\documentclass[11pt,letterpaper]{article}
\usepackage[margin=1in]{geometry}
\usepackage[T1]{fontenc}
\usepackage{lmodern}
\usepackage{amsmath,amssymb,amsthm,bm}
\usepackage{graphicx,booktabs,array,longtable}
\usepackage{csquotes}
\usepackage{microtype}
\usepackage{placeins}
\usepackage[authoryear]{natbib}
\bibpunct{(}{)}{;}{a}{,}{,}
\usepackage[hidelinks]{hyperref}
\newtheorem{theorem}{Theorem}
\newtheorem{proposition}{Proposition}
\newtheorem{corollary}{Corollary}
\newtheorem{webtheorem}{Theorem}[section]
\newtheorem{webproposition}[webtheorem]{Proposition}
\newtheorem{weblemma}[webtheorem]{Lemma}

\theoremstyle{remark}
\newtheorem{webremark}[webtheorem]{Remark}

\newcommand{\Var}{\operatorname{Var}}
\newcommand{\Cov}{\operatorname{Cov}}
\newcommand{\R}{\mathbb{R}}
\newcommand{\um}{\,\mu\mathrm{m}}
\newcommand{\Supp}[1]{Web Appendix~\ref{S-#1}}

\newcommand{\EmbeddedPDFRef}[1]{\csname EmbeddedPDFObject:#1\endcsname}
\newcommand{\ReserveEmbeddedPDF}[1]{%
  \pdfobj reserveobjnum
  \expandafter\xdef\csname EmbeddedPDFObject:#1\endcsname{\the\pdflastobj}%
}
\newcommand{\EmbeddedPDFObject}[2]{%
  \immediate\pdfobj useobjnum \EmbeddedPDFRef{#1} {#2}%
}
\newcommand{\EmbeddedPDFStream}[3]{%
  \begingroup
  \pdfcompresslevel=0
  \immediate\pdfobj useobjnum \EmbeddedPDFRef{#1} stream attr {#2} {#3}%
  \endgroup
}
\newsavebox{\EmbeddedFigureBox}
\newcommand{\EmbeddedFigure}[2]{%
  \resizebox{#1}{!}{\csname EmbeddedFigure:#2\endcsname}%
}
\newcommand{\DeclareEmbeddedFigure}[3]{%
  \ReserveEmbeddedPDF{#1}%
  \sbox{\EmbeddedFigureBox}{%
    \pdfliteral{q /EmbeddedFigure Do Q}%
    \rule{0pt}{#3bp}\kern#2bp%
  }%
  \immediate\pdfxform resources {%
    /XObject << /EmbeddedFigure \EmbeddedPDFRef{#1} 0 R >>%
  } \EmbeddedFigureBox
  \expandafter\xdef\csname EmbeddedFigure:#1\endcsname{%
    \noexpand\pdfrefxform\the\pdflastxform\relax
  }%
}

\title{Separating Cell--Cell Interactions from Tissue Composition: An Identification Floor and Honest Cross-Fitted Tests}

\author{\small Yoolkyu Park$^{1,\dagger,*}$ and Hungi Jung$^{2,\dagger}$ \\ \small
$^{1}$Department of Statistics and Operations Research, University of North Carolina at Chapel Hill, \\ \small
Chapel Hill, North Carolina 27599, U.S.A. \\ \small
$^{2}$Department of Surgery, Seoul National University Bundang Hospital, \\ \small
Seongnam, Gyeonggi-do 13620, Republic of Korea \\ \small
$^{\dagger}$These authors contributed equally to this work. \\ \small
$^{*}$Corresponding author. \href{mailto:yoolkyu@unc.edu}{\texttt{yoolkyu@unc.edu}}}

\date{}

\DeclareEmbeddedFigure{schematic}{506.177}{227.184}
\DeclareEmbeddedFigure{floor}{539.4299576001}{199.34}
\DeclareEmbeddedFigure{application}{514.9064347152}{186.564625}
\DeclareEmbeddedFigure{window}{398.438125}{161.788125}

\begin{document}

\maketitle

\begin{abstract}
In multiplexed tissue images, cell types can share a neighborhood because they favor the same compartment. For Hölder exponents up to one, under design and smoothing conditions, the minimax separation for segregation combines a floor proportional to the squared change in composition allowed over the interaction range with a term inversely proportional to the square root of cell count. Below a multiple of the floor, a random composition reproduces segregation exactly however many cells are observed, and no test adapts to an unknown smoothness bound. We fit smoothers on disjoint random halves of cells outside the interaction neighborhood. Exact null moments give a conservative finite-sample Chebyshev test and, under regularity conditions, an asymptotically valid normal test. Inverting the Chebyshev test bounds composition roughness from below under independent labeling. In METABRIC breast tumors the pooled plug-in $z$ for CD8 T cells and macrophages fell from 16.21 among all cells to 0.28 among non-epithelial cells. All four pooled epithelial--immune scores were negative across 724 METABRIC and 332 Basel images. At a common slope bound of $10^{-3}\,\mu\mathrm{m}^{-1}$, only the Basel T--epithelial test rejected independent labeling after Holm adjustment across 18 comparisons ($p=0.023$). Breakdown slopes relative to measured gradients varied by pair and gradient summary.
\end{abstract}

\begin{quote}\small
\end{quote}

\section{Introduction}\label{sec:intro}

Multiplexed imaging measures dozens of proteins in intact tissue at single-cell resolution, and a routine analysis asks whether two cell types are neighbors more or less often than expected. The answer is reported as a cell--cell interaction, related to outcome and compared across cohorts. Most software computes the expectation by random labeling, which holds the cell positions fixed and permutes the cell-type labels over all cells of the image, as in histoCAT \citep{schapiro2017}, the imcRtools workflow \citep{windhager2023} and squidpy \citep{palla2022}. Random labeling treats the cell-type composition as constant across the image. Tumor sections consist of epithelial nests, stroma, vessels and lymphoid aggregates, so two types that share a compartment are neighbors more often than random labeling predicts even if they do not influence each other. \citet{bull2024} note that correlation functions cannot distinguish \enquote{colocalization of cells due to co-intensity} from correlation, and with simulated tissue compartments a recent hierarchical model has type I error rates of 11.7--17.1\% \citep*{eliason2026}. In the METABRIC breast cancer cohort, the pooled plug-in $z$ for CD8 T cells and macrophages is 16.21 among all cells and 0.28 among non-epithelial cells (Section~\ref{sec:data}). Restricting the reference population changes both the cells that are compared and their label probabilities, and the drop is consistent with a shared stromal compartment. It leaves a harder question, because composition can keep changing within any reference population.

Local corrections estimate the composition near each cell by smoothing the labels of other cells and test whether the residuals of neighboring cells are correlated, the labeling analogue of the inhomogeneous $K$-function with a leave-one-out kernel intensity \citep*{baddeley2000,shaw2021}. Tile shuffling permutes labels within small squares \citep{crawdad2025}, and recent methods model heterogeneity across samples and conditions \citep{chang2026,emons2026} or spatially varying interaction structure across tissue regions (\citealp{aldous2026}; \citealp*{bhadury2026}). Every such correction must decide how much of the short-range structure belongs to composition. That decision cannot be made from the data alone, because a composition that varies at the interaction scale produces exactly the dependence that defines interaction. The point is old. A clustered point process can also be a Cox process \citep{bartlett1964}, and homophily and contagion are generically confounded in social networks \citep{shalizi2011}. What has been missing is a quantitative answer. How smooth must the composition be, and how many cells are needed, for an interaction of a given strength to be detectable? And which tests attain that limit?

This paper answers both questions for cell-type labels.
\begin{enumerate}
\item \emph{An identification floor} (Section~\ref{sec:floor}). For Hölder compositions with constant $M$ and exponent $\beta\le1$, the minimax separation for segregation-type interaction is of order $M^2r^{2\beta}+n^{-1/2}$ under the stated restrictions on the design and nonnegative smoothers (Theorems~\ref{thm:floor} and \ref{thm:param}, Corollary~\ref{cor:rate}). The first term does not shrink with the number of cells, and no test adapts to an unknown $M$ (Proposition~\ref{prop:adapt}), so a sensitivity analysis in $M$ is the necessary form of inference.
\item \emph{Honest cross-fitted tests} (Section~\ref{sec:method}). Two smoothers fitted on disjoint random halves of the cells, both excluding the interaction neighborhood of the target cell, give exact null moments for any number of cells and types (Theorem~\ref{thm:moments}). Chebyshev calibration controls size uniformly for every $n$, and inverting the test gives a lower confidence bound on roughness under independent labeling (Theorem~\ref{thm:honest}, Proposition~\ref{prop:breakdown}). Normal calibration is asymptotically valid under regularity conditions.
\item \emph{Benchmark and application} (Sections~\ref{sec:sim} and \ref{sec:data}). In simulation, tests without bias bounds reject a mixture of null distributions more often as $n$ grows, while the bias-aware normal test rejects none of the nonconstant bump-mixture replicates. In two breast cancer cohorts we assess which associations persist after compartment adjustment and report the roughness bounds at which pooled tests cease to reject independent labeling.
\end{enumerate}
The exact moments, the honest test and the mixture and covariance computations behind the lower bounds are formally verified in Lean~4, and \Supp{sec:lean} lists the steps that are not.

The device behind the lower bound, a random-sign perturbation of a nuisance mean that imitates a second-moment effect, is used for the effect of the mean on variance-function estimation (\citealp{wang2008}; \citealp*{cai2009}; \citealp{shen2020}). There the perturbation stays at the design spacing and the resulting term shrinks with $n$. Here the mixture belongs to the alternative class, so the perturbation lives at the interaction range and the floor does not shrink. The impossibility of testing conditional independence without restrictions \citep{shahpeters2020} is the unrestricted analogue, and a smooth nuisance field raises the same issue in Ising models \citep{ghosal2020,kandiros2021} and for autocovariances under a smooth trend \citep{hallvankeilegom2003,tecuapetla2017}. \citet{diggle2007} estimated the intensity of cases from controls to isolate their clustering, the closest classical precedent for separating composition from interaction. Cross-fitting with spatial buffers is used for nuisance functions in causal and autoregressive problems \citep{gilbert2021,emmenegger2022,lee2026}, the product structure of our bias is that of cross-fitting in semiparametric estimation \citep{neweyrobins2018}, and bias-aware critical values over smoothness classes are due to \citet{armstrong2018,armstrong2020}. Our bias is quadratic in the composition rather than linear, so their optimality theory does not apply directly.

\section{Setting}\label{sec:setting}

An image contains cells $i=1,\dots,n$ at fixed distinct locations $z_i\in\R^2$ with labels $Y_i\in\{1,\dots,K\}$, and all probability statements condition on the locations. Write $x_{ik}=\mathbf 1\{Y_i=k\}$. Cells $i\ne j$ are close when $|z_i-z_j|\le r$, the interaction range. Pair weights are $W_{ij}=w(|z_i-z_j|)$ for $i\neq j$ and $W_{ii}=0$, with $w:[0,\infty)\to[0,1]$ vanishing beyond $r$. The default is the disk indicator, for which $W$ is the adjacency matrix of close pairs. Let $|W|=\sum_{i\ne j}W_{ij}$ and let $\nu_{\max}$ be the largest number of close neighbors of a cell.

\textit{Composition classes.} For $\beta\in(0,1]$, $\Sigma_\beta(M)$ is the set of functions on $\R^2$ with $|f(x)-f(y)|\le M|x-y|^\beta$. For $\beta\in(1,2]$, it is the set of $C^1$ functions with $|\nabla f(x)-\nabla f(y)|\le M|x-y|^{\beta-1}$. In the binary case ($K=2$, $a_i=x_{i1}$) the composition class is $\mathcal P(\beta,M,\eta_0)=\{p\in\Sigma_\beta(M):\eta_0\le p\le1-\eta_0\}$, and for $K$ types every coordinate $p_k$ lies in $\Sigma_\beta(M)$ and is at least $\eta_0$. For $\beta\le1$, $Mr^\beta$ bounds the change of composition over one interaction range. For $\beta\in(1,2]$ it bounds the departure of the composition from its local affine approximation over that range, so affine trends are unrestricted.

\textit{Null hypothesis.} Under inhomogeneous random labeling, labels are independent across cells and $P(Y_i=k)=p_k(z_i)$ for some $p$ in the class. We write $P_p$ for this distribution and $H_0$ for the set of all of them. Global random labeling is the case of constant $p$.

\textit{Alternatives.} Interaction is dependence between the labels of close cells. In the binary case the class $H_1^+(\gamma)$ consists of the distributions $P$ of $(a_1,\dots,a_n)$ such that (A1) labels in sets more than $r$ apart are independent, (A2) $P(a_i=1)=p(z_i)$ with $p$ in the composition class, (A3) $\Cov_P(a_i,a_j)\ge0$ for all $i\ne j$, and (A4)
\[
\Gamma_W(P)=|W|^{-1}\sum_{i\ne j}W_{ij}\Cov_P(a_i,a_j)\ \ge\ \gamma .
\]
In the two-type case positive same-type covariance is cross-type avoidance, or segregation. The mixing class $H_1^-(\gamma)$ reverses the signs in (A3)--(A4). For $K\ge3$ and a pair $(k,\ell)$, $\Gamma_{k\ell}(P)$ replaces $\Cov(a_i,a_j)$ by $\Cov(x_{ik},x_{j\ell})$.

\textit{Minimax separation.} A test $\phi$ maps labels to $[0,1]$. For levels $\alpha,\alpha'$ with $\alpha<1-\alpha'$, the separation $\gamma^*$ is the infimum of $\gamma$ for which some $\phi$ has $\sup_{H_0}E\phi\le\alpha$ and $\inf_{H_1^+(\gamma)}E\phi\ge1-\alpha'$.

\section{An Identification Floor}\label{sec:floor}

Fix a profile $\psi:\R^2\to[0,1]$ supported in the disk of radius $1/2$, vanishing on its boundary circle together with $\nabla\psi$ when $\beta>1$, with $\psi\ge1/2$ on the disk of radius $1/4$ and with Hölder constant $L_\psi$ of order $\beta$ (for $\beta>1$, of $\nabla\psi$ with order $\beta-1$). For $\beta\le1$ the cone $\psi(u)=(1-2|u|)_+$ has $L_\psi=2$, and for $\beta\le2$ the profile $\psi(u)=\cos^2(\pi|u|)$ on $|u|\le1/2$ has $L_\psi=\pi$ at $\beta=1$ and $L_\psi=2\pi^2$ at $\beta=2$. Place centers $c\in\mathcal C$ more than $r$ apart and let $\psi_c(z)=\psi\{(z-c)/r\}$, so that the bumps have diameter $r$ and disjoint supports. Let
\[
\tau_W=|W|^{-1}\sum_{c\in\mathcal C}\sum_{i\ne j}W_{ij}\psi_c(z_i)\psi_c(z_j),
\]
the share of the pair weight carried by pairs inside a common bump, and $c_\beta=2^{1-\beta}$ for $\beta\le1$, $c_\beta=2^{2-\beta}$ for $\beta\in(1,2]$.

\begin{theorem}[Identification floor]\label{thm:floor}
Let $\xi=(\xi_c)_{c\in\mathcal C}$ be independent Rademacher signs and $p_\xi(z)=p_0+\eta\sum_c\xi_c\psi_c(z)$, with $\eta\le Mr^\beta/(c_\beta L_\psi)$ and $\eta_0+\eta\le p_0\le1-\eta_0-\eta$. Let $\bar P=E_\xi P_{p_\xi}$ be the mixture of the corresponding null distributions. Then $\bar P\in H_1^+(\tau_W\eta^2)$, and every test with $\sup_{H_0}E\phi\le\alpha$ has $E_{\bar P}\phi\le\alpha$. In particular no level-$\alpha$ test has power above $\alpha$ against all of $H_1^+(\gamma)$ when
\[
\gamma\le\tau_W\min\Big\{\frac{Mr^\beta}{c_\beta L_\psi},\ \tfrac12-\eta_0\Big\}^2 .
\]
\end{theorem}

The bound holds for every number of cells and every configuration of locations. Under the neighbor and weight conditions of Corollary~\ref{cor:rate}, the bumps can be centered at cells so that $\tau_W$ is bounded below by a positive constant depending only on $\nu_{\max}$ and $w_{\min}$. A random composition that varies by $\eta$ inside disks of diameter $r$ produces exactly the dependence that defines segregation, and the composition class cannot exclude it once $Mr^\beta\gtrsim\eta$. The witness is also a positive Markov random field with respect to the graph of close pairs, because its law factorizes over bumps whose cells form cliques, so replacing (A1) by a Markov property of range $r$ keeps the floor. Its constant is small. For a Poisson pattern with about 12 neighbors per cell and $\beta=1$, the cone witness gives $\Gamma_W=0.0050(Mr)^2$, and alternatives with dependence up to a range $2r$ or $3r$ raise it to $0.064(Mr)^2$ and $0.20(Mr)^2$.

\begin{theorem}[Parametric part]\label{thm:param}
For constant $p_0\in[\eta_0,1-\eta_0]$ and any $\eta\le\min(p_0,1-p_0)$, the mixture of Theorem~\ref{thm:floor}, with any profile $\psi$ with values in $[0,1]$ and disjoint supports of diameter $r$, satisfies
\[
\chi^2(\bar P\,\|\,P_{p_0})\le\exp\Big\{\frac{\eta^4}{2v_0^2}\sum_{c}s_c^2\Big\}-1,\qquad v_0=p_0(1-p_0),\ s_c=\sum_i\psi_c(z_i)^2 .
\]
Consequently every level-$\alpha$ test has power at most $\alpha+1/4$ at $\bar P\in H_1^+(\gamma)$ when $\gamma=\tau_W\eta^2\le\tau_Wv_0\{2\log(5/4)/\sum_cs_c^2\}^{1/2}$, and in particular when $\gamma\le\tau_Wv_0\{2\log(5/4)/(n\,s_{\max})\}^{1/2}$ with $s_{\max}=\max_cs_c$.
\end{theorem}

No smoothness is needed here, so indicator disks of diameter $r$ may be used. When cells fill space the bound is of order $(n\nu_{\max})^{-1/2}$, the order of $|W|^{-1/2}$, and in the simulation design of Section~\ref{sec:sim} it is $0.06|W|^{-1/2}$.

\begin{corollary}[Minimax rate]\label{cor:rate}
Fix $\beta\le1$ and $\alpha+\alpha'<3/4$. Suppose that $\nu_{\max}$ is bounded, that each cell has another cell within distance $r/8$, that $w>0$ on $[0,r]$ and $w\ge w_{\min}>0$ on $[0,r/4]$, that $|W|\ge c_Wn$, and that the cross-fitted statistic of Section~\ref{sec:method} uses nonnegative smoothers with window radius $R=c_Rr$, rows and columns of $|S|$ summing to at most $\Lambda$ and $\Lambda_{\rm col}$, and that the smallest ratio $\theta^*$ of $Q_{uv}$ to $W_{uv}$ over close pairs (Section~\ref{sec:upper}) is at least $\theta_0>0$. Then there are constants $0<c\le C$ depending only on these quantities, $\eta_0$, $\alpha$ and $\alpha'$ such that
\[
c\big(\min\{M^2r^{2\beta},1\}+n^{-1/2}\big)\ \le\ \gamma^*\ \le\ C\big(M^2r^{2\beta}+n^{-1/2}\big).
\]
\end{corollary}

The lower bound follows from Theorems~\ref{thm:floor} and \ref{thm:param}, and the upper bound from a power bound for the honest test of Section~\ref{sec:honest} (Theorem~\ref{S-thm:power} in \Supp{sec:statistic}). The corollary is a statement about exponents, and the constants are far apart. In the simulation design of Section~\ref{sec:sim} the lower constant of the first term is $0.005(Mr)^2$ per pair and the upper constant 12--18$(Mr)^2$. For the $n^{-1/2}$ term they are $0.06|W|^{-1/2}$ and $290|W|^{-1/2}$, so at the sample sizes of this paper the upper bound guarantees nothing and power is better described by the local analysis of Section~\ref{sec:upper}. Uniform detection of segregation weaker than a multiple of $M^2r^{2\beta}$ requires assumptions about composition at the interaction range, whatever the statistic, smoother or permutation scheme.

\begin{proposition}[No adaptation]\label{prop:adapt}
Let $\mathcal P_\beta$ be the union of the classes $\mathcal P(\beta,M,\eta_0)$ over $M>0$. Every test with $\sup_{p\in\mathcal P_\beta}E_p\phi\le\alpha$ has power at most $\alpha$ against some element of $H_1^+(\gamma)$ whenever $\gamma\le\tau_W(1/2-\eta_0)^2$.
\end{proposition}

A sensitivity analysis indexed by $M$ is therefore the necessary form of inference rather than a convenience.

\textit{Other alternatives and directions.} The floor does not depend on the mixture form or on the direction of the effect (\Supp{sec:identification}). Smooth ferromagnetic autologistic models are undetectable at couplings of order $(Mr^\beta)^2/\log N$, with $N$ the number of interaction-range disks needed to cover the tissue (Theorem~\ref{S-thm:ising}). When cells have a hard core $d_0$, mixtures whose labels are $r$-dependent only at the cells give a floor of order $M^2d_0^{2\beta}/\nu_{\max}^2$ even against interactions whose covariance stays constant up to distance $r$, about $10^{-4}(Mr)^2$ in the images of Section~\ref{sec:data} (Propositions~\ref{S-prop:homog} and \ref{S-prop:pairsign}). In the two-type case, random compositions that are $r$-dependent fields cannot produce mixing on average over Poisson cell patterns (Proposition~\ref{S-prop:mixing}), but among three or more types a third type can absorb the change, so composition produces colocalization of two types as easily as their avoidance.

\section{Honest Cross-Fitted Tests}\label{sec:method}

\subsection{Construction and Exact Moments}\label{sec:construction}

\begin{figure}
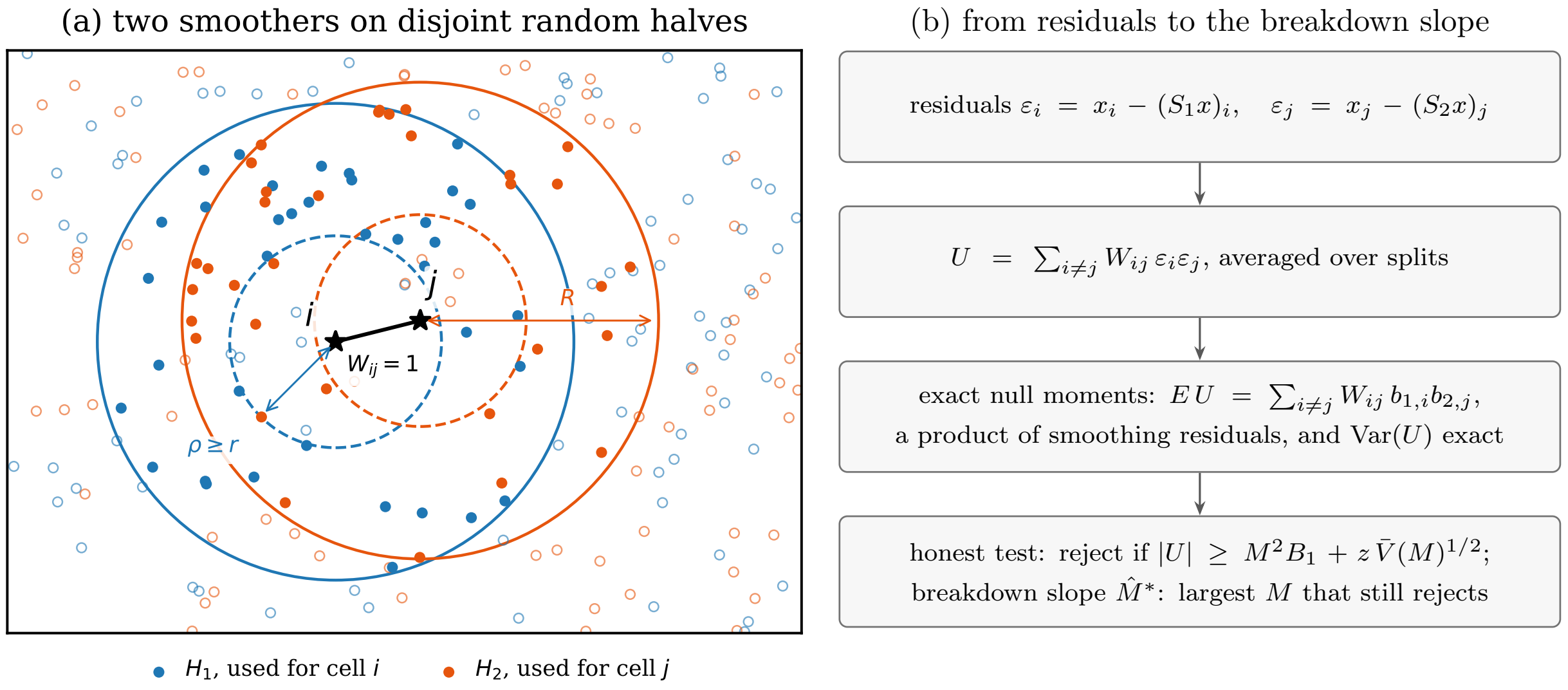

\centering
\EmbeddedFigure{\linewidth}{schematic}
\caption{The cross-fitted statistic. (a) For a close pair $(i,j)$ the smoother of cell $i$ uses only cells of the random half $H_1$ at distance in $(\rho,R]$ from $z_i$, and that of cell $j$ only cells of $H_2$ at distance in $(\rho,R]$ from $z_j$, with $\rho\ge r$. (b) Because no label enters both residuals of a pair, the null mean is a product of the smoothing residuals $b_{1,i}$ and $b_{2,j}$ of the composition, which a Hölder bound controls, and the variance is exact.}\label{fig:schematic}
\par\medskip
\begin{minipage}{\linewidth}
Alt text: Panel a marks the cells i and j of a close pair with stars. Around each, a dashed circle of radius rho lies inside a solid circle of radius R. The two rings overlap heavily, and within them the filled blue cells of half H1 are used for cell i and the filled orange cells of half H2 for cell j. Panel b is a flowchart of four boxes that leads from the residuals to the statistic U, its exact null moments, and the honest test with its breakdown slope.
\end{minipage}
\end{figure}

Choose $S$ independent random partitions of the cells into halves $H_1^{(t)},H_2^{(t)}$, $t=1,\dots,S$, independently of the labels. For each split and half, $S_s^{(t)}$ is a smoother whose row $i$ puts weight only on cells of $H_s^{(t)}$ at distance in $(\rho,R]$ from $z_i$, with rows summing to one, and for $\beta>1$ the rows also reproduce affine functions (local-linear weights). The exclusion radius satisfies $\rho\ge r$ (Figure~\ref{fig:schematic}). With $L_s^{(t)}=I-S_s^{(t)}$ define
\[
Q=\frac1S\sum_{t=1}^S\tfrac12\big(L_1^{(t)\top}WL_2^{(t)}+L_2^{(t)\top}WL_1^{(t)}\big),\qquad U_{k\ell}=x_k^\top Qx_\ell .
\]
Equivalently, $U_{kk}$ sums over close pairs the products of the residuals $x_{ik}-(S_1x_k)_i$ and $x_{jk}-(S_2x_k)_j$, symmetrized and averaged over splits.

The matrix $Q$ is symmetric and has zero diagonal and zero row sums, and with affine-reproducing smoothers it also annihilates affine functions (Lemma~\ref{S-lem:zerodiag} in \Supp{sec:proofs}). The zero diagonal needs only that the two halves are disjoint and that no smoother row of a cell puts weight on a cell it is paired with ($W_{im}\ne0\Rightarrow S^{(t)}_{s,im}=0$), which the exclusion radius $\rho\ge r$ ensures.

\begin{theorem}[Exact null moments]\label{thm:moments}
Under $P_p$, with residuals $b_{s,k}^{(t)}=L_s^{(t)}p_k$,
\[
E\,U_{k\ell}=p_k^\top Qp_\ell=\frac1S\sum_t\tfrac12\sum_{i\ne j}W_{ij}\big(b^{(t)}_{1,k,i}b^{(t)}_{2,\ell,j}+b^{(t)}_{2,k,i}b^{(t)}_{1,\ell,j}\big).
\]
With $v_{k,i}=p_k(z_i)\{1-p_k(z_i)\}$, the variance is $4\sum_iv_{k,i}(Qp_k)_i^2+2\sum_{i,j}Q_{ij}^2v_{k,i}v_{k,j}$ when $k=\ell$, and, when $k\ne\ell$, with $g_\ell=Qp_\ell$ and $g_k=Qp_k$,
\[
\Var U_{k\ell}=\sum_i\big\{g_{\ell,i}^2v_{k,i}+g_{k,i}^2v_{\ell,i}-2g_{\ell,i}g_{k,i}p_{k,i}p_{\ell,i}\big\}+\sum_{i,j}Q_{ij}^2\big\{v_{k,i}v_{\ell,j}+p_{k,i}p_{\ell,i}p_{k,j}p_{\ell,j}\big\}.
\]
\end{theorem}

The mean is second order in the smoothing error. Averaging over splits keeps the zero diagonal, so the moments are exact for any $S$, and it removes the dependence of the $p$-value on one random split. Disjoint supports of the two smoothers remove the shared-cell term, and for distinct types exact unbiasedness over a class rich enough to take arbitrary values at finitely many points requires them (Proposition~\ref{S-prop:disjoint}).

\subsection{Honest Tests and the Breakdown Slope}\label{sec:honest}

Let $m_{s,i}^{(t)}=\sum_m|S^{(t)}_{s,im}|\,|z_m-z_i|^\beta$. If $p_k\in\Sigma_\beta(M)$, every smoothing residual satisfies $|b^{(t)}_{s,k,i}|\le Mm^{(t)}_{s,i}$, for $\beta\in(1,2]$ provided the rows reproduce affine functions (Lemma~\ref{S-lem:residual}). Hence $|E\,U_{k\ell}|\le M^2B_1$ with the computable constant $B_1=S^{-1}\sum_tm_1^{(t)\top}Wm_2^{(t)}$. Because $Q$ is symmetric with zero row sums, $x^\top Qy=-\tfrac12\sum_{i,j}Q_{ij}(x_i-x_j)(y_i-y_j)$, which gives the second bound $|E\,U_{k\ell}|\le\tfrac12M^2\sum_{i,j}|Q_{ij}|\,|z_i-z_j|^{2\beta}$ for $\beta\le1$, and we use the smaller of the two. For the variance, $v\le1/4$, $p_kp_\ell\le1/4$ and $|(Qp_k)_i|\le M\omega_i$ with $\omega_i=\sum_j|Q_{ij}||z_i-z_j|^\beta$ give $\bar V(M)=M^2\sum_i\omega_i^2+\tfrac18\sum_{i,j}Q_{ij}^2$ for $\beta\le1$, and the same holds for $\beta>1$ with the affine part removed, because $Q$ annihilates affine functions. With separate constants $M_k$ and $M_\ell$ for the two types the bias bound is $M_kM_\ell B_1$ and the variance bound $\tfrac14(M_k+M_\ell)^2\sum_i\omega_i^2+\tfrac18\sum_{i,j}Q_{ij}^2$. For images independent under the null, the bias and variance bounds add, and the constants may differ between images. A design with $Q=0$ has $U=0$ and is assigned no rejection.

\begin{theorem}[Honest test]\label{thm:honest}
For $Q\ne0$, the test that rejects when $|U_{k\ell}|\ge M^2B_1+\{\bar V(M)/\alpha\}^{1/2}$ has size at most $\alpha$ under every composition in the class and for every $n$.
\end{theorem}

The normal version, which rejects when $|U_{k\ell}|\ge M^2B_1+z_{1-\alpha/2}\bar V(M)^{1/2}$, is valid asymptotically and uniformly over the class, and so is its one-sided version, because $U_{k\ell}$ is asymptotically normal uniformly over the class when the absolute row sums of $Q$ are bounded and $\Var U_{k\ell}\ge c_0n$ for every composition in the class (Theorem~\ref{S-thm:clt} in \Supp{sec:statistic}). The folded-normal critical value of \citet{armstrong2018} is slightly smaller. A plug-in version replaces $\bar V$ by the exact variance at a composition estimate, corrected to first order for its noise (\Supp{sec:statistic}). It is an approximation without a uniform guarantee.

\begin{proposition}[Breakdown slope]\label{prop:breakdown}
Let $\hat M^*=\sup\{M\ge0:{}$the honest test at $M$ rejects$\}$, taking $\hat M^*=0$ if the set is empty. For any composition with roughness at most $M_0$, $P(\hat M^*>M_0)\le\alpha$. Thus $\hat M^*$ is a lower $(1-\alpha)$ confidence bound for the roughness of any composition that could have produced the data without interaction.
\end{proposition}

For $\beta=1$, $\hat M^*$ is a slope per micrometer. Below this bound the test rejects independent labeling. Above it, the test leaves open whether any composition in the class matches the observed signal. Figure~\ref{fig:application}(c) shows how $\hat M^*$ is read off the honest critical value.

\subsection{Power}\label{sec:upper}

Power rests on an identity. Because $Q_{uu}=0$, for any label law $P$ with marginal surface $p$,
\begin{equation}\label{eq:powerid}
E_PU=p^\top Qp+\sum_{u\ne v}Q_{uv}\Cov_P(a_u,a_v).
\end{equation}
Under (A1) only close pairs enter the sum. With $\rho\ge2r$ the smoothing window of a cell contains no cell that is close to a cell paired with it, so for nonnegative smoothers $Q_{uv}\ge W_{uv}\ge0$ on every close pair, and (A3) gives $E_PU\ge\Gamma_W(P)|W|-M^2B_1$ for every $P$ satisfying (A1)--(A3). A Chebyshev bound on the variance of $U$ under the alternative turns this into a power guarantee for the honest test whose rate matches the floor, with the smallest ratio $\theta^*=\min\{Q_{uv}/W_{uv}:|z_u-z_v|\le r,\,W_{uv}>0\}$ in place of 1 and a correction for close pairs with $W_{uv}=0$ for other designs (Theorem~\ref{S-thm:power} in \Supp{sec:statistic}). This is the upper bound of Corollary~\ref{cor:rate}. Its constants are loose at realistic sizes, where power is described by a local result.

Against centered autologistic models, and centered cross-type attraction models for $k\ne\ell$, whose positive coupling shrinks at the parametric rate, the standardized statistic $(U_{k\ell}-E_pU_{k\ell})/(\Var_pU_{k\ell})^{1/2}$ is asymptotically normal with its mean shifted by $h\rho_n$, whereas the standardized score of the most powerful test that knows the composition is shifted by $h$. Here $\rho_n$ is the null correlation between $U_{k\ell}$ and that score, which has a closed form in $Q$, $W$ and the composition (Theorem~\ref{S-thm:local} in \Supp{sec:statistic}). Thus $\rho_n^2$ is the Pitman efficiency of the cross-fitted statistic relative to the test that knows the composition. It is below one whenever $W\ne0$ and $\eta_0>0$, because $Q$ has zero row sums and $W$ does not, and this loss is the price of removing the composition. In the benchmark designs of Section~\ref{sec:bench}, $\rho_n$ was 0.84--0.85 for local-linear smoothing with bandwidth $25\um$, 0.79 for Nadaraya--Watson smoothing with bandwidth $17.5\um$ and 0.63 for a doughnut variant that pairs only cells $10$--$20\um$ apart, Pitman efficiencies of 0.71--0.73, 0.62 and 0.39. In weak-interaction designs the predicted shift was within 1.8 Monte Carlo standard errors of the simulated one (\Supp{sec:simdetails}). These models are not $r$-dependent, so the result describes operating characteristics and does not narrow the gaps of Corollary~\ref{cor:rate}.

\section{Simulations}\label{sec:sim}

\subsection{The Floor in Practice}\label{sec:floorsim}

\begin{figure}
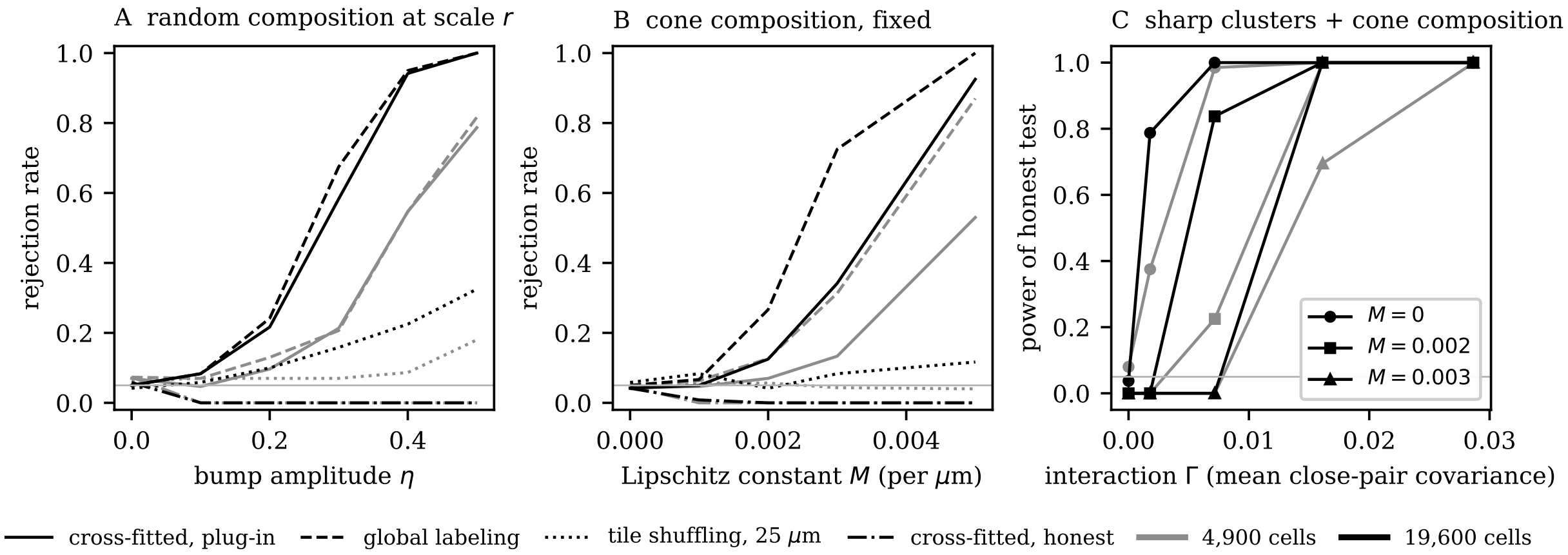

\centering
\EmbeddedFigure{\linewidth}{floor}
\caption{The identification floor in simulation. A: rejection of the bump mixture of Theorem~\ref{thm:floor}, which every test valid over the composition class must reject at most 5\% of the time. B: size under a fixed cone composition with Lipschitz constant $M$. C: power of the honest test against sharp micro-clusters under a cone composition with constant $M$. Gray, 4,900 cells; black, 19,600 cells.}\label{fig:floor}
\par\medskip
\begin{minipage}{\linewidth}
Alt text: Panels A and B plot the rejection rates of four tests against the bump amplitude and the Lipschitz constant. In both, rejection by the plug-in cross-fitted test and by global labeling increases along the horizontal axis and increases faster with more cells, while tile shuffling stays below one third and the honest test stays at or near zero. In panel C, the power curves of the honest test shift to the right as M grows, and for positive M they stay at zero for small interactions at both image sizes.
\end{minipage}
\end{figure}

Cells were placed as a Poisson process with intensity $0.01\um^{-2}$ in a $700\um$ or $1400\um$ square (about 4,900 or 19,600 cells), with $r=\rho=20\um$, Nadaraya--Watson smoothers at $h=25\um$, a single split and $\beta=1$ (\Supp{sec:simdetails}). Figure~\ref{fig:floor}A uses the mixture of Theorem~\ref{thm:floor} with $\cos^2$ bumps of diameter $r$. It lies in the convex hull of the null class with $M=\pi\eta/r$, so a test that is valid over that class must reject it at most 5\% of the time. The plug-in cross-fitted test and global labeling reject it increasingly often as $\eta$ grows, and more often with 19,600 cells than with 4,900 (0.94 against 0.55 at $\eta=0.4$ for the cross-fitted test), because a fixed per-pair bias is multiplied by $\sqrt n$ in standard units. With one-sided normal calibration, the honest test rejects in 7.3\% and 5.8\% of the replicates at $\eta=0$, where the mixture is a single null law, and never for $\eta\ge0.1$. Figure~\ref{fig:floor}B uses a fixed checkerboard of cones with Lipschitz constant $M$, the least favorable composition we found for this statistic, and again the error of the plug-in test grows with $n$ (0.53 to 0.93 at $M=0.005\um^{-1}$). Figure~\ref{fig:floor}C adds sharp micro-clusters outside the Lipschitz class. The power of the honest test rises with $n$ when $M=0$, while at $M=0.003\um^{-1}$ its detection threshold stays near $\Gamma\approx0.01$ for both sizes, as Corollary~\ref{cor:rate} predicts.

The constant attained by the cross-fitted test is set by the distance between a cell and its smoothing window. Narrowing the window from $h=25$ to $15\um$ cuts the exact null bias by a factor of two to four and lowers the local efficiency $\rho_n$ of Section~\ref{sec:upper} at constant composition from 0.87 to 0.75 (Web Appendix~\ref{S-sec:window}, Web Figure~\ref{S-fig:window}).

\subsection{Comparison with Other Methods}\label{sec:bench}

We compared the cross-fitted tests with global label permutation, tile shuffling with $25$ and $50\um$ tiles, residual products with a single radius-excluded smoother, exact permutation moments of the $K$-function \citep[KAMP;][]{wrobel2024}, an inhomogeneous cross-$K$ test calibrated by a parametric bootstrap, autologistic pseudo-likelihood with a penalized spline trend, in the spirit of \citet*{jensen2025}, and the CRAWDAD package \citep{crawdad2025}. Each design used 500 replicates, 50 cell patterns with ten labelings each (\Supp{sec:simdetails}). Power was computed under centered autologistic models, whose coupling $\gamma$ is on a different scale from the mean close-pair covariance $\Gamma_W$ on the horizontal axis of Figure~\ref{fig:floor}C. For two types at $p=1/2$ with the disk weight, $\Gamma_W\approx\gamma/16$ to first order (\Supp{sec:statistic}). Table~\ref{tab:bench}A reports the plug-in versions of the cross-fitted tests, as a practitioner would run them, and Table~\ref{tab:bench}B the honest versions. CRAWDAD was evaluated separately on a subset of designs; its decision rule rejected no balanced binary or eight-type datasets and rejected every null dataset with a rare reference type (\Supp{sec:simdetails}).

\begin{table}
\centering
\caption{Benchmark of one-sided tests in the segregation direction at level 0.05 (500 replicates; Monte Carlo standard error $\approx 0.01$ at rate 0.05).}\label{tab:bench}
\scriptsize
\setlength{\tabcolsep}{2.2pt}
\begin{tabular}{lrrrrrrrrrr}
\toprule
 & \multicolumn{7}{c}{size} & \multicolumn{3}{c}{power}\\
\cmidrule(lr){2-8}\cmidrule(lr){9-11}
 & & lin. & & inter- & & & bump & flat & islands & lin. strong\\
Method & flat & strong & islands & face & cones & Voronoi & mix. & $\gamma=0.04$ & $\gamma=0.04$ & $\gamma=0.15$\\
\midrule
\multicolumn{11}{l}{\emph{A. Plug-in and resampling tests}} \\
Cross-fitted, NW, $h=17.5$ & 0.042 & 0.052 & 0.078 & 0.056 & 0.226 & 0.626 & 0.204 & 0.426 & 0.344 & 0.998 \\
Cross-fitted, LL, $h=25$ & 0.052 & 0.054 & 0.102 & 0.086 & 0.462 & 0.812 & 0.218 & 0.506 & 0.446 & 1.000 \\
Cross-fitted doughnut, NW & 0.038 & 0.048 & 0.072 & 0.074 & 0.140 & 0.308 & 0.040 & 0.332 & 0.262 & 0.992 \\
Tile shuffling, $25\um$ & 0.048 & 0.052 & 0.056 & 0.050 & 0.068 & 0.128 & 0.074 & 0.186 & 0.142 & 0.852 \\
Tile shuffling, $50\um$ & 0.064 & 0.060 & 0.136 & 0.132 & 0.306 & 0.760 & 0.198 & 0.420 & 0.532 & 1.000 \\
Autologistic PL, bootstrap & 0.046 & 0.050 & 0.094 & 0.102 & 0.726 & 0.936 & 0.150 & 0.436 & 0.432 & 0.996 \\
Autologistic PL, Wald & 0.064 & 0.056 & 0.132 & 0.104 & 0.798 & 0.922 & 0.228 & 0.490 & 0.520 & 0.994 \\
Inhomogeneous cross-$K$, bootstrap & 0.082 & 0.090 & 0.060 & 0.088 & n.r. & n.r. & n.r. & n.r. & n.r. & n.r. \\
Global labeling & 0.056 & 1.000 & 1.000 & 1.000 & 0.872 & 1.000 & 0.238 & 0.568 & -- & -- \\
KAMP & 0.052 & 1.000 & 1.000 & 1.000 & n.r. & n.r. & n.r. & n.r. & n.r. & n.r. \\
Radius-excluded smoother & 0.640 & 0.604 & 0.756 & 0.696 & 0.972 & 0.994 & 0.898 & -- & -- & -- \\
\midrule
\multicolumn{11}{l}{\emph{B. Bias-aware cross-fitted tests}} \\
Design constant $M$ ($10^{-3}\um^{-1}$) & 0.00 & 1.14 & 4.55 & 8.75 & 5.00 & $\infty$ & 47.12 & 0.00 & 4.55 & 1.14 \\
Honest, NW, at $M$ & 0.052 & 0.000 & 0.000 & 0.000 & 0.000 & -- & 0.000 & 0.460 & 0.000 & 0.926 \\
Honest, LL, at $M$ & 0.052 & 0.000 & 0.000 & 0.000 & 0.000 & -- & 0.000 & 0.514 & 0.000 & 0.922 \\
Honest doughnut, at $M$ & 0.038 & 0.002 & 0.000 & 0.000 & 0.000 & -- & 0.000 & 0.328 & 0.000 & 0.806 \\
$\hat M^*$, LL, 90th percentile & 0.00 & 0.00 & 0.00 & 0.00 & 0.94 & 1.35 & 0.70 & 1.03 & 0.82 & 1.97 \\
\bottomrule
\end{tabular}
\par\medskip
\begin{flushleft}
\footnotesize\textit{Note:} Size under five smooth compositions, piecewise-constant Voronoi compartments, and the bump mixture of Theorem~\ref{thm:floor} ($\eta=0.3$); power under the centered autologistic model with $\gamma=0.04$ (flat, islands) and $\gamma=0.15$ (strong linear trend). Cross-fitted tests use five random splits; the doughnut variant ($h=17.5\um$) pairs cells $10$--$20\um$ apart and smooths within $10\um$. Part A: plug-in and resampling tests; dashes omit power where null size exceeds 0.5, and n.r.\ denotes not run. Part B: honest one-sided normal tests at each design's Lipschitz constant $M$ (dash for discontinuous Voronoi compartments), and the 90th percentile of $\hat M^*$ in $10^{-3}\um^{-1}$ (zero when fewer than 10\% reject at $M=0$).
\end{flushleft}
\end{table}

Every method in Table~\ref{tab:bench}A exceeded its nominal size under some composition, and every method run on the bump mixture except the doughnut variant rejected it more than 5\% of the time, which Theorem~\ref{thm:floor} rules out for any test that is valid over the composition class. Global labeling and KAMP fail as soon as the composition varies, and a single radius-excluded smoother fails even at constant composition because of the shared-cell term. Among the remaining methods there is a robustness--efficiency frontier. Tile shuffling at $25\um$ held its size best but had the least power, and local-linear cross-fitting had the most power under flat composition but was the least robust of the cross-fitted variants to cones and Voronoi compartments. Nadaraya--Watson cross-fitting and the doughnut variant lie between, with 1.8--2.4 times the power of $25\um$ tiles at $\gamma=0.04$.

The honest versions replace this frontier by an explicit assumption about $M$ (Table~\ref{tab:bench}B). At the Lipschitz constant of each design they rejected at most 5.2\% of the replicates in every null design, and under flat composition they kept the power of the plug-in tests (0.514 against 0.506 for local-linear smoothing). Under a non-flat composition their power depends on how the interaction compares with the bias bound $M^2B_1$ at the roughness of that composition. At the steepest slope of the islands composition, $4.5\times10^{-3}\um^{-1}$, the honest critical value lies 16--28 null standard deviations above the null mean, while an interaction at $\gamma=0.04$ shifts $U$ by 0.9--1.3 standard deviations, so the honest tests had no power there. For local-linear smoothing their power fell from 0.514 under flat composition to 0.000, the cost of guaranteeing size at the roughness of the islands composition. The strong linear trend has slope $1.1\times10^{-3}\um^{-1}$, the critical value lies 3.1--4.1 standard deviations above the null mean, the interaction at $\gamma=0.15$ shifts $U$ by 4.0--6.0 standard deviations, and the honest tests detected it in 81--93\% of the replicates. As in Figure~\ref{fig:floor}C, a test that is valid at roughness $M$ detects only interactions above a threshold that grows with $M$. The breakdown slope summarizes the same evidence without fixing $M$. Under Voronoi compartments, which lie outside every Hölder class, it was positive in 77\% of the replicates, and its 90th percentile, $1.35\times10^{-3}\um^{-1}$, exceeded those under interaction at $\gamma=0.04$ (1.03 and $0.82\times10^{-3}\um^{-1}$) and fell below that at $\gamma=0.15$ ($1.97\times10^{-3}\um^{-1}$). A breakdown slope therefore supports an interaction only when it exceeds the roughness that is plausible for the tissue, which is how Section~\ref{sec:data} uses it.

\section{Two Breast Cancer Cohorts}\label{sec:data}

\subsection{Data and Specification}

The METABRIC imaging mass cytometry cohort \citep{danenberg2022} contains 1,123,466 segmented cells in 794 tissue microarray images from 718 patients. The Basel cohort \citep{jackson2020} contains 844,498 cells in 376 images. For every all-cell pair, the primary analysis included all images with at least 300 segmented cells, giving 724 METABRIC and 332 Basel images. An image in which either target type was absent contributed zero to $\sum_m U_m$ and retained its geometry-based bias and variance terms. METABRIC phenotypes were grouped into epithelial cells, CD8 T cells, CD4 T cells, B cells, macrophages, fibroblasts, endothelial cells and other cells. Basel reports a single T-cell metacluster. We used $r=\rho=20\um$, local-linear smoothers with $h=25\um$ truncated at $2.5h$, and five random splits. This interaction radius is about twice the median nearest-neighbor distance in METABRIC (approximately $10\um$). If a half-window had fewer than six cells or a nearly singular local design, its target cell was omitted from the close-pair graph but remained available for smoothing.

These graphs use $\rho=r$ and keep close pairs up to the image edge. The null moments of Theorem~\ref{thm:moments} and the Chebyshev size bound of Theorem~\ref{thm:honest} hold for every such graph under independent labeling. The power guarantee behind the upper bound of Corollary~\ref{cor:rate} requires $\theta^*>0$, which failed, mostly through pairs near the image edge, in 514 of 724 METABRIC and 304 of 332 Basel all-cell graphs (\Supp{sec:statistic}). The finite-sample level of the Chebyshev tests below is therefore guaranteed, whereas the rate of Corollary~\ref{cor:rate} is not established for these graphs.

For comparisons within the non-epithelial population, we conditioned on the epithelial membership mask and required at least 150 remaining cells, leaving 698 METABRIC and 331 Basel images for every eligible pair. Under independent labeling the remaining labels are independent given this mask, with probabilities $q_k=p_k/(1-p_{\rm ep})$. On a region with $1-p_{\rm ep}\ge\epsilon>0$, the Lipschitz constant of $q_k$ is at most $2M/\epsilon$ if both original probabilities have constant at most $M$. Thus a bound on the conditional composition can be much larger near an epithelial boundary. We also repeated the analysis using images with at least 20 observed cells of each target type. This screen uses the labels entering $U$, so its results describe sensitivity to image selection (\Supp{sec:datadetails}, Web Table~\ref{S-tab:screenholm}).

We pooled signed statistics as $U_{\rm pool}=\sum_m U_m$. The reported $z=U_{\rm pool}/(\sum_m\widehat V_m)^{1/2}$ uses the plug-in null variance $\widehat V_m$ for each image. For image $m$, let $B_m$ be the smaller of the product and variogram bias coefficients in Section~\ref{sec:honest}. Finite-sample inference instead uses the summed bounds of Theorem~\ref{thm:honest}, under a joint null with independent labels across cells and images conditional on the designs. The common breakdown slope $\hat M^*$ applies to both types in every image. We use $M=10^{-3}\,\mu\mathrm{m}^{-1}$ as an interpretable sensitivity point, at which each type's probability may change by up to two percentage points over $20\um$. For the four epithelial--immune pairs, the geometric means of the two median $50\um$ RMS gradients range from $0.82$ to $1.43$ in units of $10^{-3}\,\mu\mathrm{m}^{-1}$ (Table~\ref{tab:cert}), providing a descriptive reference for $M$ on the product bias scale. We inverted the two-sided Chebyshev rule at this fixed slope and used Holm's adjustment over all 18 cohort, population and pair comparisons \citep{holm1979}. The separate four-comparison adjustment in \Supp{sec:datadetails} illustrates the effect of the family definition.

The multiplier $\lambda^*$ allows each type in each image its own bound, equal to $\lambda$ times its estimated reference gradient. Our default reference is the noise-corrected root-mean-square (RMS) gradient at $50\um$. Because the gradients are estimated from the observed labels, these multipliers are descriptive. We also report the exact standardized null mean at a composition smoothed at $25$ or $50\um$, and compare with a separately screened compartment analysis (\Supp{sec:datadetails}).

\subsection{Results}

\begin{table}
\centering
\caption{Pooled cross-fitted statistics on fixed image sets. The all-cell analysis includes 724 METABRIC and 332 Basel images. The non-epithelial analysis conditions on the mask in 698 and 331 images, respectively.}\label{tab:data}
\scriptsize
\setlength{\tabcolsep}{3pt}
\begin{tabular}{lrrrcrrrrr}
\toprule
 & & & & image $z$ (\%) & \multicolumn{2}{c}{semi-synthetic} & \multicolumn{3}{c}{$\hat M^*$}\\
Pair & images & $z$ & Stouffer & $<-1.96$ / $>1.96$ & $\ell=25$ & $\ell=50$ & normal $\bar V$ & Chebyshev & plug-in\\
\midrule
\multicolumn{10}{l}{\emph{METABRIC, all cells}} \\
CD8~T -- macrophage & 724 & 16.2 & 16.1 & 1/14 & 0.8 & 0.11 & 0.16 & -- & 0.43 \\
fibroblast -- CD8~T & 724 & 27.6 & 31.7 & 2/30 & 0.9 & 0.11 & 0.57 & 0.30 & 0.69 \\
CD4~T -- B & 724 & $-6.4$ & 3.1 & 3/7 & $-0.9$ & $-0.02$ & -- & -- & 0.25 \\
CD8~T -- B & 724 & $-0.2$ & 5.0 & 2/8 & $-0.5$ & $-0.03$ & -- & -- & -- \\
CD8~T -- epithelial & 724 & $-90.3$ & $-69.0$ & 53/0 & $-10.5$ & $-0.72$ & 1.15 & 1.03 & 1.23 \\
macrophage -- epithelial & 724 & $-116.4$ & $-92.4$ & 62/0 & $-12.6$ & $-0.82$ & 1.46 & 1.36 & 1.52 \\
\midrule
\multicolumn{10}{l}{\emph{METABRIC, non-epithelial cells}} \\
CD8~T -- macrophage & 698 & 0.3 & 0.9 & 3/3 & $-0.8$ & $-0.01$ & -- & -- & -- \\
fibroblast -- CD8~T & 698 & $-3.9$ & $-1.1$ & 5/3 & $-2.6$ & $-0.14$ & -- & -- & 0.35 \\
CD4~T -- B & 698 & $-13.0$ & $-2.5$ & 4/2 & $-1.8$ & $-0.09$ & 0.54 & -- & 0.74 \\
CD8~T -- B & 698 & $-7.8$ & $-0.4$ & 3/4 & $-1.4$ & $-0.10$ & 0.07 & -- & 0.51 \\
\midrule
\multicolumn{10}{l}{\emph{Basel, all cells}} \\
macrophage -- epithelial & 332 & $-95.3$ & $-73.6$ & 64/2 & $-10.3$ & $-0.68$ & 1.29 & 1.17 & 1.37 \\
stroma -- T & 332 & 19.8 & 31.1 & 8/43 & $-0.3$ & 0.03 & 0.63 & 0.37 & 0.74 \\
T -- macrophage & 332 & 20.3 & 21.3 & 2/30 & 1.1 & 0.15 & 0.35 & -- & 0.54 \\
T -- B & 332 & 3.4 & 8.6 & 3/12 & $-0.2$ & 0.05 & -- & -- & 0.14 \\
T -- epithelial & 332 & $-138.6$ & $-115.1$ & 79/1 & $-16.3$ & $-1.21$ & 2.01 & 1.91 & 2.07 \\
\midrule
\multicolumn{10}{l}{\emph{Basel, non-epithelial cells}} \\
stroma -- T & 331 & $-32.4$ & $-26.1$ & 34/1 & $-6.4$ & $-0.40$ & 1.67 & 1.48 & 1.76 \\
T -- macrophage & 331 & $-4.4$ & $-0.1$ & 6/8 & $-1.6$ & $-0.05$ & -- & -- & 0.37 \\
T -- B & 331 & $-8.2$ & 1.6 & 4/6 & $-1.4$ & $-0.07$ & 0.14 & -- & 0.55 \\
\bottomrule
\end{tabular}
\par\medskip
\begin{flushleft}
\footnotesize The pooled $z$ uses the plug-in variance. Stouffer's statistic combines image-specific $z$ values with equal weights. The two percentages are the fractions of images with $z<-1.96$ and $z>1.96$. Semi-synthetic entries give the standardized null mean under composition smoothed at scale $\ell$. The breakdown slopes $\hat M^*$ are in $10^{-3}\,\mu\mathrm{m}^{-1}$. A dash indicates nonrejection at $M=0$. Only the Chebyshev calibration provides a finite-sample guarantee.
\end{flushleft}
\end{table}

\begin{table}
\centering
\caption{Epithelial--immune avoidance at the sensitivity point $M=10^{-3}\,\mu\mathrm{m}^{-1}$ on the fixed image sets.}\label{tab:epi}
\small
\begin{tabular}{llrrr}
\toprule
 & & pooled & Chebyshev & Holm $p$,\\
Cohort & Pair & plug-in $z$ & $p$ & 18 comparisons\\
\midrule
METABRIC & CD8~T -- epithelial & $-90.29$ & 0.03902 & 0.54625\\
METABRIC & macrophage -- epithelial & $-116.44$ & 0.00612 & 0.10401\\
Basel & macrophage -- epithelial & $-95.27$ & 0.01690 & 0.25351\\
Basel & T -- epithelial & $-138.58$ & 0.00128 & 0.02312\\
\bottomrule
\end{tabular}
\par\medskip
\begin{flushleft}
\footnotesize The pooled $z$ divides $\sum_mU_m$ by the square root of the summed plug-in null variances. The Chebyshev $p$-value inverts the two-sided Chebyshev test at $M$ with the summed bias and variance bounds over images. Holm's adjustment covers all 18 cohort, population and pair comparisons of Table~\ref{tab:data}. Web Table~\ref{S-tab:screenholm} lists all 18 comparisons and the adjustment over the four pairs shown here.
\end{flushleft}
\end{table}

\begin{figure}
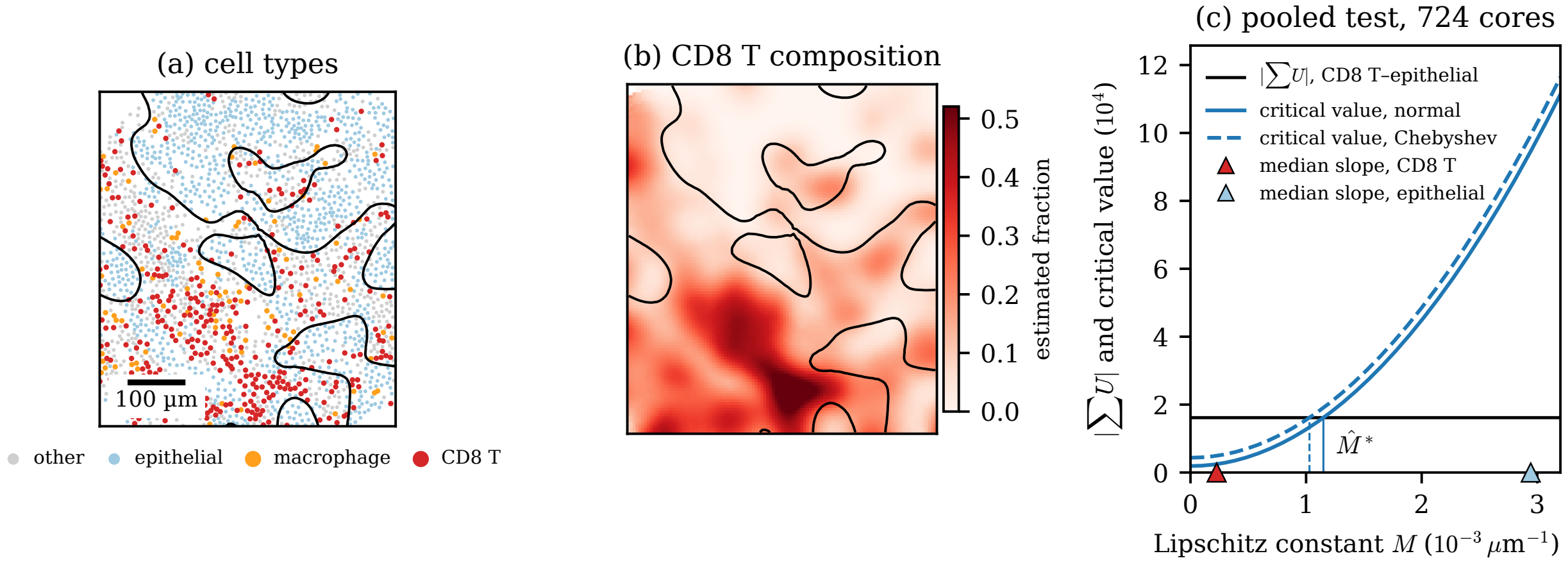

\centering
\EmbeddedFigure{\linewidth}{application}
\caption{CD8 T cells and epithelial cells in METABRIC. (a) Cells in illustrative core 361, with the boundary at an epithelial fraction of $1/2$ after $30\um$ smoothing. This core had the most negative CD8 T--epithelial $z$ ($-14.3$) among images with at least 1,500 cells, at least 80 CD8 T cells and 35--65\% of cells in the tumor compartment. (b) CD8 T-cell composition smoothed at $25\um$ in the same core. (c) All 724 eligible images enter the pooled CD8 T--epithelial statistic. The normal and Chebyshev critical values cross $|\sum_m U_m|$ at $\hat M^*=1.15$ and $1.03$ in units of $10^{-3}\,\mu\mathrm{m}^{-1}$. Triangles mark the median noise-corrected $50\um$ RMS slopes.}\label{fig:application}
\par\medskip
\begin{minipage}{\linewidth}
Alt text: In panel a, CD8 T cells crowd into a stromal band from the left edge to the bottom, and epithelial cells fill the upper and right parts. In panel b, the CD8 T fraction reaches about one half in that band and is low in the epithelial regions. In panel c, both critical value curves rise with M from well below the horizontal line of the pooled statistic. The CD8 T triangle lies left of both crossings, and the epithelial triangle far to their right.
\end{minipage}
\end{figure}

\begin{table}
\centering
\caption{Certificate multipliers $\lambda^*$ on the fixed image sets for pairs with a positive normal breakdown slope.}\label{tab:cert}
\scriptsize
\setlength{\tabcolsep}{3.2pt}
\begin{tabular}{lrrrcrrrr}
\toprule
 & \multicolumn{2}{c}{slopes} & \multicolumn{3}{c}{$\lambda^*$ at $50\um$} & \multicolumn{2}{c}{$\lambda^*$} & bias\\
Pair & first & second & normal & Chebyshev & Cheb. range & $100\um$ & q90 & share\\
\midrule
\multicolumn{9}{l}{\emph{METABRIC, all cells}} \\
CD8~T -- macrophage & 0.23 & 0.30 & 0.36 & -- & -- & 0.71 & 0.16 & 0.33 \\
fibroblast -- CD8~T & 2.02 & 0.23 & 0.72 & 0.374 & 0.3445--0.3835 & 1.45 & 0.35 & 0.89 \\
CD8~T -- epithelial & 0.23 & 2.94 & 1.12 & 0.980 & 0.9205--1.0004 & 2.17 & 0.57 & 0.43 \\
macrophage -- epithelial & 0.30 & 2.94 & 1.27 & 1.165 & 1.0977--1.1895 & 2.59 & 0.66 & 0.38 \\
\midrule
\multicolumn{9}{l}{\emph{METABRIC, non-epithelial cells}} \\
CD4~T -- B & 0.36 & 0.00 & 0.58 & -- & -- & 1.06 & 0.29 & 0.00 \\
CD8~T -- B & 0.43 & 0.00 & 0.09 & -- & -- & 0.15 & 0.04 & 0.00 \\
\midrule
\multicolumn{9}{l}{\emph{Basel, all cells}} \\
macrophage -- epithelial & 0.34 & 2.84 & 1.13 & 1.006 & 0.9498--1.0266 & 2.44 & 0.61 & 0.51 \\
stroma -- T & 2.21 & 0.72 & 0.49 & 0.288 & 0.2748--0.2929 & 1.01 & 0.27 & 2.59 \\
T -- macrophage & 0.72 & 0.34 & 0.50 & -- & -- & 1.05 & 0.26 & 0.75 \\
T -- epithelial & 0.72 & 2.84 & 1.25 & 1.180 & 1.1405--1.1945 & 2.49 & 0.70 & 0.47 \\
\midrule
\multicolumn{9}{l}{\emph{Basel, non-epithelial cells}} \\
stroma -- T & 1.81 & 1.11 & 0.95 & 0.844 & 0.7818--0.8846 & 1.85 & 0.51 & 0.61 \\
T -- B & 1.11 & 0.00 & 0.15 & -- & -- & 0.28 & 0.08 & 0.00 \\
\bottomrule
\end{tabular}
\par\medskip
\begin{flushleft}
\footnotesize The first two columns are median noise-corrected RMS slopes at $50\um$ in $10^{-3}\,\mu\mathrm{m}^{-1}$. The next columns give normal and Chebyshev multipliers at the default correction and the Chebyshev range over five noise corrections. The $100\um$ and q90 columns use normal calibration with RMS slopes at $100\um$ and uncorrected upper-decile slopes at $50\um$, respectively. A dash indicates no rejection at zero roughness. Bias share is $M_kM_\ell\sum_m B_m/|\sum_m U_m|$ at the median slopes. Further diagnostics appear in Web Tables~\ref{S-tab:ident} and \ref{S-tab:sens}.
\end{flushleft}
\end{table}

Tables~\ref{tab:data}--\ref{tab:cert} report the fixed-image analyses. In each cohort the same cores enter every all-cell comparison, including comparisons involving rare immune types.

\emph{Compartmental colocalization.} In METABRIC, CD8 T cells and macrophages occurred together more often than expected among all cells ($z=16.21$). Within the non-epithelial population the association was close to zero ($z=0.28$). The analogous Basel T--macrophage statistic changed from $20.27$ to $-4.39$. The pattern is consistent with the two types occupying the same stromal compartment.

\emph{Avoidance among non-epithelial cells.} For CD4 T--B and CD8 T--B cells in METABRIC, the non-epithelial plug-in statistics were $z=-12.95$ and $-7.81$. Basel T--B gave $z=-8.19$. None of these three Chebyshev tests rejected even at $M=0$. For Basel stromal--T avoidance, the Chebyshev $p$-value at the reference slope was 0.00914 before adjustment and 0.14624 after Holm adjustment over 18 comparisons. In particular, the positive CD4 T--B Chebyshev breakdown slope seen in the count-screened sensitivity analysis disappeared in the fixed set of 698 images (Web Table~\ref{S-tab:screenholm}).

\emph{Epithelial--immune pairs.} All four pairs had negative pooled $U$ in their respective fixed cohorts (Table~\ref{tab:epi}). At $M=10^{-3}\,\mu\mathrm{m}^{-1}$, only Basel T--epithelial remained significant at 5\% after Holm adjustment across all 18 comparisons (adjusted $p=0.02312$). The four unadjusted Chebyshev $p$-values ranged from 0.00128 to 0.03902. An adjustment confined to the four epithelial--immune comparisons gives $p=0.00514$--$0.03902$. Its familywise interpretation requires choosing these four independently of the results (Web Table~\ref{S-tab:screenholm}).

The METABRIC CD8 T--epithelial row of Table~\ref{tab:epi} pairs a pooled $z$ of $-90.29$ with an adjusted $p$-value of 0.546. The $z$ divides the observed score by its plug-in standard deviation, which treats the estimated composition as correct. The $p$-value lets each type's probability change by up to two percentage points over $20\um$ in every core, allows for the resulting bounds on bias and variance, and accounts for 18 comparisons. A very large association score can therefore support different inferences under different assumptions about composition.

Scaling the slope bound to each image's estimated $50\um$ RMS gradients gives a descriptive comparison across pairs. The RMS gradient summarizes the typical change of composition, whereas a Lipschitz bound limits its largest local change, so the two differ even when the RMS gradient is estimated without error. Across all five noise corrections, the Chebyshev multiplier exceeded one for METABRIC macrophage--epithelial and Basel T--epithelial. At the default correction, METABRIC CD8 T--epithelial fell just below the Chebyshev threshold at the estimated RMS slopes ($\lambda^*=0.980$), and Basel macrophage--epithelial just exceeded it ($\lambda^*=1.006$). Their respective ranges were 0.9205--1.0004 and 0.9498--1.0266. With the uncorrected 90th-percentile gradients all four Chebyshev multipliers were below one (0.50--0.66). At the median corrected slopes, the explicit cone and $\cos^2$ composition mixtures each reproduced less than 0.05\% of the pooled magnitude in expectation (\Supp{sec:datadetails}). In a separate three-way analysis with a target-count screen, the compartment-restricted pooled $z$ was 88--91\% of the matched control $z$ (Web Table~\ref{S-tab:interface}).

Figure~\ref{fig:application}(c) shows the fixed-set threshold. Web Table~\ref{S-tab:screenholm} compares it with the screened threshold. For METABRIC CD8 T--epithelial, the 390 images added to the 334 screened cores had a combined $U$ of $-1{,}509$. The added cores accounted for 9.3\% of the final $|\sum_m U_m|$ and 46.2\% of the final bias coefficient $\sum_m B_m$. Per 1,000 ordered close pairs, $1{,}000\sum_m U_m/\sum_m |W_m|$ changed from $-3.68$ to $-2.20$, and the common Chebyshev breakdown slope fell from $1.37\times10^{-3}$ to $1.03\times10^{-3}\,\mu\mathrm{m}^{-1}$. The common $M$ allows the same maximum slope in every core, so even a core with few CD8 T cells contributes the full bias allowance $M^2B_m$ determined by its geometry.

\section{Discussion}\label{sec:discussion}

The difficulty of separating interaction from composition is usually stated qualitatively. Theorem~\ref{thm:floor} turns it into a rate. Below a multiple of the squared change of composition over one interaction range, some segregated label distributions can be represented exactly as mixtures of independent-labeling models, however many cells are measured. Any test that controls size over the composition class consequently has power at most its size against these alternatives. A statistic whose bias is a product of two smoothing residuals attains the floor in its exponents under the design conditions of Corollary~\ref{cor:rate}, which is what cross-fitting on disjoint halves away from the interaction neighborhood achieves, although the constants are about three orders of magnitude apart. Global labeling, leave-one-out residuals and permutation within tiles come with no comparable bias bound, and in our simulations their errors grow with $n$ under compositions in the class.

The four epithelial--immune scores agreed in direction across cohorts. The evidence against independent labeling depends on the allowed composition slope and on which comparisons enter the multiplicity adjustment. Image eligibility is part of that inference. Using target counts to select cores conditions on the same labels used to compute the statistic. Eligibility based on total cell count, with the epithelial mask additionally fixed in the non-epithelial analysis, allows the finite-sample calculation on the intended image population. The CD8 T--epithelial example also shows why raw pooled totals should be accompanied by the number of retained close pairs when that population changes.

The bound $\bar V(M)$ uses $v\le1/4$ and $p_kp_\ell\le1/4$ and bounds $(Qp_k)_i$ by $M\omega_i$. These bounds can be loose for rare types. Across the four epithelial--immune comparisons, 1.1--12.0\% of individual images exceeded the normal threshold at their estimated $50\um$ RMS slopes. Omitting the quadratic variance term alone gives 3.5--19.0\%. Replacing the entire variance bound by the null variance evaluated at a composition smoothed at $25\um$ gives 25.0--46.4\%, a descriptive plug-in calculation (\Supp{sec:datadetails}). Both the linear and the quadratic terms affect single-image sensitivity. Simultaneous upper confidence bounds for local composition probabilities could tighten these terms. If a type has probability at most $u$ at a target cell and throughout its smoothing window, each probability difference in the smoothing residual is at most $u$ in absolute value, which could also tighten the product bias bound. The present results support using the method chiefly for cohort-level sensitivity analysis.

Pooling treats cores as independent under the null. The 724 METABRIC cores come from 657 patients, and the patient-level Stouffer combination in \Supp{sec:datadetails} adds the variances of a patient's cores under the same assumption, so it does not assess sensitivity to residual correlation between cores of one patient. The bounds also extend to contrasts between groups of patients under the joint null of no interaction in either group (\Supp{sec:datadetails}).

Several questions remain open. The constants of the floor are far apart, both for heterogeneous profiles and for profiles that stay constant up to the interaction range (\Supp{sec:identification}), and a uniform power bound with useful constants at realistic sizes would need control of dependence beyond pairwise covariances. Smoothing within correctly mapped compartments gives zero smoothing bias when composition is constant within each compartment. Our map used a held-out third of the cells. An interface measured independently, for instance with a stromal marker excluded from phenotyping, could preserve those cells for testing. Finally, the analysis conditions on cell positions and therefore says nothing about interactions that move cells, and segmentation errors that assign signal from a neighbor to a cell create short-range dependence that no composition model removes.

\section*{Acknowledgements}

\textit{Use of AI tools.} Large language models (OpenAI GPT-6 and Anthropic Claude Opus 5.5) were used in 2026 to assist with mathematical exploration, drafting and checking proof arguments, code development, and language editing. The authors conceived the research problem and methodology, directed all analyses, independently verified and revised every mathematical proof and numerical result, and take full responsibility for the accuracy, originality, and integrity of the manuscript. In addition, the finite-sample theoretical identities and lower-bound lemmas were formally verified in Lean~4 (\Supp{sec:lean}).

\section*{Funding}

This research received no specific grant from any funding agency in the public, commercial, or not-for-profit sectors.

\section*{Conflict of Interest}

The authors report there are no competing interests to declare.

\section*{Data Availability}

The imaging mass cytometry data of the Molecular Taxonomy of Breast Cancer International Consortium cohort that support the findings in this paper are openly available in Zenodo at \url{https://doi.org/10.5281/zenodo.5850952} \citep{danenberg2022data}, and those of the Basel cohort at \url{https://doi.org/10.5281/zenodo.3518284} \citep{jackson2019data}.

\section*{Supplementary Materials}

Web Appendices~A--G, Web Tables~1--4 and Web Figure~1 follow the main text. Web Appendices~F and G describe the Lean~4 project and the Python code for the simulations and analyses, along with their outputs.

\clearpage
\appendix
\numberwithin{equation}{section}
\setcounter{figure}{0}
\setcounter{table}{0}
\renewcommand{\theHfigure}{web.\arabic{figure}}
\renewcommand{\theHtable}{web.\arabic{table}}
\renewcommand{\figurename}{Web Figure}
\renewcommand{\tablename}{Web Table}
\makeatletter
\renewcommand{\@seccntformat}[1]{%
  \ifnum\pdfstrcmp{#1}{section}=0
    Web Appendix~\thesection.\quad
  \else
    \csname the#1\endcsname\quad
  \fi
}
\makeatother

\begin{center}
{\Large\bf Supplementary Material for ``Separating Cell--Cell Interactions from Tissue Composition: An Identification Floor and Honest Cross-Fitted Tests'' by Yoolkyu Park and Hungi Jung\par}
\end{center}
\medskip

\noindent Numbers of sections, theorems, equations, tables and figures without a letter prefix refer to the main text. Web Tables and Web Figures are those of this document.

\section{Further identification results}\label{S-sec:identification}

\subsection{Autologistic alternatives}

The witness is a mixture, not an autologistic model, and its covariance is heterogeneous across pairs. For pairwise interaction models with a smooth positive-definite profile the same conclusion holds up to a logarithmic factor.

\begin{webtheorem}[Autologistic alternatives]\label{S-thm:ising}
Let $\beta\le1$ and let $\varphi:\R^2\to\R$ be a positive-definite $C^3$ function with $\varphi(u)=0$ for $|u|>1$ and $\varphi(0)=1$ (higher-order Wendland functions qualify). Let $P_{\gamma,h}(\sigma)\propto\exp\{\sum_ih(z_i)\sigma_i+\tfrac12\sigma^\top J\sigma\}$ on $\sigma\in\{\pm1\}^n$, with labels $a_i=(1+\sigma_i)/2$, $J_{ij}=\gamma\varphi\{(z_i-z_j)/r\}$ for all $i,j$ and $\gamma\ge0$. Then $P_{\gamma,h}=\int P_{p(g)}\,\mu(dg)$, where $p(g)(z)=\{1+\tanh g(z)\}/2$ and, under $\mu$, $g=h+\gamma\sum_j\varphi\{(\cdot-z_j)/r\}\sigma'_j+G$ with $\sigma'\sim P_{\gamma,h}$ and $G$ an independent centered Gaussian field with covariance $\gamma\varphi\{(x-y)/r\}$. Let $D$ be the convex hull of the cells, $N$ the number of disks of radius $r$ needed to cover $D$, $N_r$ the largest number of cells within distance $r$ of a point of $D$, and $g_{\max}=\operatorname{artanh}(1-2\eta_0)$. Suppose that $|h|\le H_0<g_{\max}$ and $h/2\in\Sigma_\beta(M/3)$ on $D$. There are constants $c_1,c_2>0$ depending only on $\varphi$ such that, if
\[
\gamma N_r\le c_1\min\{Mr^\beta,\ g_{\max}-H_0\}\qquad\text{and}\qquad\gamma\log(eN/\epsilon)\le c_2\min\{Mr^\beta,\ g_{\max}-H_0\}^2,
\]
then every test with $\sup_{H_0}E\phi\le\alpha$ has $E_{P_{\gamma,h}}\phi\le\alpha+\epsilon$.
\end{webtheorem}

Couplings of order $(Mr^\beta)^2/\log N$ are therefore undetectable, and the floor for these models shrinks with $n$ only through the logarithm, which comes from the supremum of the Gaussian part over about $N$ correlation areas. The argument needs a smooth positive-definite profile; the disk-kernel autologistic model used in our power simulations has a matrix $J$ that is not positive semidefinite, and its representation needs a diagonal shift whose rough nugget falls outside any Hölder class.

\subsection{The mixing direction}

The convexity argument does not reach the mixing direction in the same way.

\begin{webproposition}[Mixing by composition]\label{S-prop:mixing}
Let $\pi$ be a distribution on the binary composition class under which the random surface is an $r$-dependent field, and $\bar P=\int P_p\,\pi(dp)$. Then (i) $\Var_\pi p(z_i)\le M^2r^{2\beta}$ and
\[
\sum_{i\ne j}\Cov_{\bar P}(a_i,a_j)\ \ge\ -\sum_i\Var_\pi p(z_i)\ \ge\ -nM^2r^{2\beta};
\]
(ii) for the disk weight the same bound holds for $\sum_{i\ne j}W_{ij}\Cov_{\bar P}(a_i,a_j)$, and for a positive-definite weight the bound is $-w(0)nM^2r^{2\beta}$; (iii) if the cells form a Poisson process of intensity $\lambda$ on a domain $D$, independent of $\pi$, then $E\sum_{i\ne j}\Cov_{\bar P}(a_i,a_j)=\lambda^2\Var_\pi\int_Dp\ge0$.
\end{webproposition}

Averaged over Poisson cell patterns, an $r$-dependent composition mixture therefore cannot produce mixing at all, and for a given pattern the mixing it can produce is limited by the variance of the composition at single cells rather than by covariances across a neighborhood. The Poisson assumption matters: on patterns with a $7\um$ hard core a mean-zero dipole mixture produces a small amount of mixing on average, about $-0.0002(Mr)^2$ per pair, against $0.005(Mr)^2$ for the cone segregation witness. On a fixed pattern with a hard core, the pair-sign mixture of Proposition~\ref{S-prop:pairsign}, whose labels are $r$-dependent only at the cells, has covariance $-\eta_-^2$ on every close pair, with median $1.3\times10^{-4}(Mr)^2$ in the same patterns. These bounds are not shown to be attained, and in the simulation design they do not by themselves show that mixing is less exposed than segregation: the bound in (i) is $-0.081(Mr)^2$ per pair, larger in magnitude than the segregation witnesses. The disk weight is not positive definite, and a deterministic composition with wavelength below about $1.64r$ biases the disk-weighted statistic towards mixing (Remark~\ref{S-rem:mixingtest}).

\begin{webremark}[Mixing tests, a conjecture]\label{S-rem:mixingtest}
For one split, with $\bar b=(b_1+b_2)/2$ and $\delta=(b_1-b_2)/2$, $E\,U=\bar b^\top W\bar b-\delta^\top W\delta$. With a positive-definite weight the first term is at least $-w(0)\|\bar b\|^2$, which matches the scale of Proposition~\ref{S-prop:mixing}, while $\delta^\top W\delta$ can be of order $\nu_{\max}\|\delta\|^2$; over random splits $\|\delta\|^2$ is smaller than $\|\bar b\|^2$ by roughly the number of cells per half-window (ratios 38--72 in our designs). Averaging over splits leaves the expected $\delta$-term unchanged and only reduces its variability. Compactly supported positive-definite weights vanish at distance $r$, which conflicts with weighting all close pairs equally, and the default disk weight is not positive definite. We conjecture that a cross-fitted test with a positive-definite weight attains a mixing floor smaller than the segregation floor by a factor that grows with the number of neighbors, but we have not proved it. On patterns with a hard core $d_0$, Proposition~\ref{S-prop:pairsign} bounds the mixing floor below by $\min\{Md_0^\beta,1-2\eta_0\}^2/(2\nu_{\max})^2$ for every weight, which is consistent with the conjecture.
\end{webremark}

\subsection{Homogeneous covariance profiles}

The witness of Theorem~\ref{thm:floor} has covariance that varies across pairs and falls to zero continuously at distance $r$. A mixture over random compositions that are $r$-dependent fields on the plane cannot do otherwise, and a mixture whose labels are $r$-dependent only at the cells is limited by the cells just beyond distance $r$.

\begin{webproposition}[Homogeneous profiles]\label{S-prop:homog}
Let $\beta\le1$, let $\pi$ be a distribution on the binary composition class, let $\bar P=\int P_p\,\pi(dp)$, and let $i\ne j$ be cells with $|z_i-z_j|\le r$.
\begin{enumerate}
\item[(i)] If the random surface is an $r$-dependent field under $\pi$, then $\Cov_{\bar P}(a_i,a_j)\le\min\{Mr^\beta,\tfrac12-\eta_0\}\,M(r-|z_i-z_j|)^\beta$.
\item[(ii)] If $\bar P$ satisfies (A1), then $\Cov_{\bar P}(a_i,a_j)\le\min\{M|z_i-z_{m'}|^\beta,\tfrac12-\eta_0\}\,M|z_j-z_m|^\beta$ for all cells $m$ and $m'$ farther than $r$ from $z_i$.
\end{enumerate}
Consequently, if $\Cov_{\bar P}(a_i,a_j)\ge\gamma$ for every close pair, then $\gamma\le M^2r^\beta(r-d_{\max})^\beta$ in case (i), where $d_{\max}$ is the largest distance between two close cells, and $\gamma\le M^2B_{\rm cell}$ in case (ii). Here $B_{\rm cell}$ is the minimum of $D_i^\beta E_{ij}^\beta$ over ordered close pairs $(i,j)$, where $D_i$ is the distance from $z_i$ to the nearest cell farther than $r$ from $z_i$ and $E_{ij}$ is the distance from $z_j$ to the nearest such cell.
\end{webproposition}

In case (i) the covariance falls to zero as the distance between two cells approaches $r$. In the 1,056 images of Section~\ref{sec:data} and in the benchmark patterns some close pair lies within $0.15\um$ of distance $r$ (within $0.32\um$ for the non-epithelial cells of Section~\ref{sec:data}), and in the two cohorts the bound $M^2r(r-d_{\max})$ has median 0.019--0.033$\,M^2\um^2$. Assumption (A1), like the null hypothesis, constrains the labels only at the cells, and case (ii) allows much more, because the closest cells of an image are typically several micrometers apart (median 3.1--4.7$\um$, smallest $1.0\um$). In the two cohorts $B_{\rm cell}$ has median 67--98$\um^2$, or 0.17--0.24 times $r^2$ for $r=20\um$. When cells have a hard core, mixtures whose labels are $r$-dependent only at the cells do give a floor against homogeneous profiles.

\begin{webproposition}[Pair-sign mixtures]\label{S-prop:pairsign}
Let $\beta\le1$, let $\mathcal E$ be the set of close pairs and $\nu_i$ the number of close neighbors of cell $i$, let
\[
L_+=\max\Big\{\frac{2\nu_{\max}}{r^\beta},\ \max_{\{i,m\}\in\mathcal E}\frac{\nu_i+\nu_m-2}{|z_i-z_m|^\beta}\Big\},
\]
let $L_-$ be defined in the same way with $\nu_i+\nu_m$ in place of $\nu_i+\nu_m-2$, and let $\eta_\pm=\min\{M/L_\pm,(\tfrac12-\eta_0)/\nu_{\max}\}$. Attach independent Rademacher signs $\xi_P$ to the pairs $P\in\mathcal E$, write each pair with a fixed order as $P=(i_P,m_P)$, and set
\[
p^+_i=\tfrac12+\eta_+\sum_{P\ni i}\xi_P,\qquad p^-_i=\tfrac12+\eta_-\Big(\sum_{P:\,i_P=i}\xi_P-\sum_{P:\,m_P=i}\xi_P\Big).
\]
Every realization of $p^\pm$ is the restriction to the cells of a composition in $\mathcal P(\beta,M,\eta_0)$. The mixture $\bar P_\pm$ of the corresponding null distributions satisfies (A1) and (A2), and $\Cov_{\bar P_\pm}(a_i,a_j)=\pm\eta_\pm^2$ for every close pair and $0$ for every other pair of distinct cells. Hence, for every weight $w$ vanishing beyond $r$, $\bar P_+\in H_1^+(\eta_+^2)$ and $\bar P_-\in H_1^-(\eta_-^2)$, and every test with $\sup_{H_0}E\phi\le\alpha$ has $E_{\bar P_\pm}\phi\le\alpha$. If every two cells are at least $d_0$ apart, then $\eta_\pm\ge\min\{Md_0^\beta/(2\nu_{\max}),(\tfrac12-\eta_0)/\nu_{\max}\}$.
\end{webproposition}

The covariance of the pair-sign mixture is the same on every close pair, at every distance up to $r$, so on cell patterns with a hard core and a bounded number of neighbors, alternatives with homogeneous profiles have a floor that does not shrink with $n$, for segregation and for mixing. Its constant is small. With $r=20\um$ the median of $\eta_+^2$ is 0.022--0.061$\,M^2\um^2$ in the two cohorts and $0.058\,M^2\um^2$ in the $7\um$ hard-core benchmark patterns, or $5.6\times10^{-5}$ to $1.5\times10^{-4}(Mr)^2$. This is 33--90 times below the cone witness $0.005(Mr)^2$ of a Poisson pattern and 1,600--3,000 times below $M^2B_{\rm cell}$ (medians of the ratios per pattern). It exceeds the bound of Proposition~\ref{S-prop:homog}(i) in 55--68\% of the images and in every hard-core pattern, and $\eta_-^2$ is 12--17\% smaller than $\eta_+^2$. In the Poisson benchmark patterns, where two cells can be arbitrarily close, $\eta_+^2$ is about $4\times10^{-8}(Mr)^2$ and $B_{\rm cell}$ is $0.013r^2$. For $\beta\in(1,2]$, bumps of diameter $d_0$ centered at the cells take the same values, and the argument of Lemma~\ref{S-lem:bumps} gives a floor of the same order (Web Appendix~\ref{S-sec:proofs}).

The proof of Proposition~\ref{S-prop:pairsign} uses only that each sign is attached to a set of cells of diameter at most $r$ and that every close pair lies in one of these sets. For any such family, let $k_i$ be the number of sets that contain cell $i$, $c_{ij}$ the number that contain cells $i$ and $j$, and $L$ the largest ratio $(k_i+k_m-2c_{im})/|z_i-z_m|^\beta$ over pairs of distinct cells. Then the compositions $\tfrac12+\eta\sum_{F\ni i}\xi_F$ with $\eta=\min\{M/L,(\tfrac12-\eta_0)/k_{\max}\}$ give a mixture of null distributions in $H_1^+$ with covariance $\eta^2c_{ij}\ge\eta^2$ on every close pair. The pair family of Proposition~\ref{S-prop:pairsign} is limited by the closest pair of cells. Attaching the sign of each close pair to all cells within $r/2$ of its midpoint, together with its two ends, lets two nearby cells share most of their sets. Over the all-cell images, the median of the per-image ratio of its floor to $\eta_+^2$ is 1.31 in METABRIC and 1.45 in Basel, and it gives more than $\eta_+^2$ in 83\% and 86\% of the images. It gives 224 times $\eta_+^2$ on the Poisson patterns and 0.76 times on the hard-core patterns. The better of the two families has median $2.0\times10^{-4}$ and $8.2\times10^{-5}(Mr)^2$ in the two cohorts, so the constant of Proposition~\ref{S-prop:pairsign} is not sharp among exact mixtures. These statements concern $\beta=1$ and segregation. The floors are $\eta^2=M^2/L^2$ only while the range condition $\eta k_{\max}\le\tfrac12-\eta_0$ does not bind, which for the pair family holds when $Mr^\beta\le1-2\eta_0$. For the midpoint family we bounded $L$ below by $2k_{\max}/(3r)$, its value for pairs more than $3r$ apart, so the reported floors hold when $Mr\le(1-2\eta_0)/3$.

\subsection{Several types}

For $K\ge3$ the floor applies to cross-type contrasts in both directions. With a third type $m$ absorbing the change, the mixtures $\delta_k=\eta\psi\xi$, $\delta_\ell=\pm\eta\psi\xi$, $\delta_m=-(1\pm1)\eta\psi\xi$, with $\eta\le Mr^\beta/(2c_\beta L_\psi)$ and baselines $p_{k0},p_{\ell0}\ge\eta_0+\eta$ and $p_{m0}\ge\eta_0+2\eta$, keep every coordinate in the class and give $\Gamma_{k\ell}=\pm\tau_W\eta^2$. Colocalization of two types can therefore be produced by composition as easily as their avoidance.

\section{Proofs}\label{S-sec:proofs}

This appendix proves the results of the main text and of Web Appendices~\ref{S-sec:identification} and \ref{S-sec:statistic}.

\subsection{Identification}

\begin{weblemma}\label{S-lem:bumps}
If $\eta\le Mr^\beta/(c_\beta L_\psi)$, then $\delta_\xi=\eta\sum_c\xi_c\psi_c\in\Sigma_\beta(M)$ for every $\xi\in\{-1,1\}^{\mathcal C}$.
\end{weblemma}

\begin{proof}
By scaling, $\eta\xi_c\psi_c$ has Hölder constant $\eta L_\psi r^{-\beta}$ of order $\beta$. Take $\beta\le1$ and points $x,y$. If the segment $[x,y]$ meets at most one bump, the bound follows directly or from the fact that $\delta_\xi$ vanishes outside the bumps.

Otherwise let $x'$ be the point where $[x,y]$ leaves the bump containing $x$ (or $x'=x$ if $x$ lies in no bump) and $y'$ the point where it enters the bump containing $y$; disks are convex, so the segment leaves a bump once. Then $\delta_\xi(x')=\delta_\xi(y')=0$ and
\[
\begin{aligned}
|\delta_\xi(x)-\delta_\xi(y)|&\le\eta L_\psi r^{-\beta}(|x-x'|^\beta+|y-y'|^\beta)\\
  &\le2^{1-\beta}\eta L_\psi r^{-\beta}|x-y|^\beta,
\end{aligned}
\]
because $|x-x'|+|y-y'|\le|x-y|$ and $t\mapsto t^\beta$ is concave.

For $\beta\in(1,2]$ the same argument applies to $\nabla\delta_\xi$, which is continuous because $\nabla\psi$ vanishes on the boundary circle, with exponent $\beta-1$ and constant $2^{2-\beta}$.
\end{proof}

\paragraph{Theorem~\ref{thm:floor}.} By Lemma~\ref{S-lem:bumps} and $\eta_0+\eta\le p_0\le1-\eta_0-\eta$, every $p_\xi$ lies in the composition class, so $E_{\bar P}\phi=2^{-|\mathcal C|}\sum_\xi E_{p_\xi}\phi\le\alpha$ for any test of level $\alpha$. We check (A1)--(A4) for $\bar P$.

(A2): $\bar P(a_i=1)=E_\xi p_\xi(z_i)=p_0$ because $E\xi_c=0$, a constant surface in the class.

(A1): given $\xi$ the labels are independent. If every cell of $S$ is more than $r$ from every cell of $T$, no bump, whose diameter is $r$, contains cells of both. So $(a_i)_{i\in S}$ is a function of the signs of the bumps meeting $S$ and of independent uniform variables attached to the cells of $S$, and $(a_j)_{j\in T}$ of a disjoint set of signs and of the uniforms of $T$; the two are independent.

(A3)--(A4): for $i\ne j$, conditional independence gives $E[a_ia_j]=E_\xi[p_\xi(z_i)p_\xi(z_j)]$, hence
\[
\begin{aligned}
\Cov(a_i,a_j)&=\eta^2\sum_{c,c'}E[\xi_c\xi_{c'}]\psi_c(z_i)\psi_{c'}(z_j)\\
&=\eta^2\sum_c\psi_c(z_i)\psi_c(z_j)\ge0,
\end{aligned}
\]
and averaging with weights $W_{ij}$ gives $\Gamma_W(\bar P)=\tau_W\eta^2$. \hfill$\square$

\paragraph{Theorem~\ref{S-thm:ising}.} \emph{Representation.} Since $\varphi$ is positive definite, $J$ (diagonal $\gamma$ included) is positive semidefinite, and $\exp(\tfrac12\sigma^\top J\sigma)=E\exp(G_c^\top\sigma)$ for $G_c\sim N(0,J)$ \citep{hubbard1959}. Hence
\[
\begin{aligned}
P_{\gamma,h}(\sigma)&\propto E\prod_i\exp\{(h_i+G_{c,i})\sigma_i\}\\
&=E\big[\prod_i2\cosh(g_i)\prod_i\pi(\sigma_i;g_i)\big]
\end{aligned}
\]
with $g=h+G_c$ and $\pi(s;g)=e^{gs}/(2\cosh g)$, a probability on $\{\pm1\}$ with $\pi(1;g)=\{1+\tanh g\}/2$.

Normalizing gives $P_{\gamma,h}=\int\prod_i\pi(\sigma_i;g_i)\,\mu(dg)$ with $\mu(dg)\propto\prod_i\cosh(g_i)N(dg;h,J)$. Expanding $\prod_i2\cosh g_i=\sum_{\sigma'}e^{g^\top\sigma'}$ and completing the square,
\[
N(g;h,J)e^{g^\top\sigma'}\propto e^{h^\top\sigma'+\sigma'^\top J\sigma'/2}N(g;h+J\sigma',J),
\]
so $\mu=\sum_{\sigma'}P_{\gamma,h}(\sigma')N(h+J\sigma',J)$, which is the law at the cells of the field $g$ in the statement.

\emph{Smoothness.} The null hypothesis involves the composition only at the cells, and a function on the cells that is $\beta$-Hölder with constant $M$ for the Euclidean distance extends to $\R^2$ with the same constant (McShane's extension applied to the metric $|x-y|^\beta$), clipped to $[\eta_0,1-\eta_0]$. It therefore suffices to show that, with $\mu$-probability at least $1-\epsilon$, $|g|\le g_{\max}$ and $|p(g)(x)-p(g)(y)|\le M|x-y|^\beta$ on $D$.

Since $\tanh$ is 1-Lipschitz, $|p(g)(x)-p(g)(y)|\le\frac12|g(x)-g(y)|$. The term $h$ contributes at most $\frac M3|x-y|^\beta$. The spin term is a sum of at most $N_r$ nonzero translates of $\gamma\varphi(\cdot/r)$ at any point, so it is bounded by $\gamma N_r$ and has Hölder constant at most $C_\varphi\gamma N_r r^{-\beta}$; the first condition makes both small multiples of $Mr^\beta$ and of $g_{\max}-H_0$.

For the Gaussian part,
\[
\begin{aligned}
|G(x)-G(y)|&\le\min\{|x-y|\sup|\nabla G|,2\sup|G|\}\\
&\le r^{-\beta}|x-y|^\beta(r\sup|\nabla G|+2\sup|G|)
\end{aligned}
\]
for $\beta\le1$. The fields $G$ and $r\nabla G$ are stationary Gaussian with variances of order $\gamma$ that do not depend on $r$, so by Dudley's bound and the Borell--TIS inequality \citep{adlertaylor2007} their suprema over $D$, which is covered by $N$ disks of radius $r$, exceed $C_\varphi\{\gamma\log(eN/\epsilon)\}^{1/2}$ with probability at most $\epsilon$. The second condition makes this a small multiple of $Mr^\beta$ and of $g_{\max}-H_0$. Hence the composition lies in the class with probability at least $1-\epsilon$, and $E_{P_{\gamma,h}}\phi=\int E_{p(g)}\phi\,\mu(dg)\le\alpha+\epsilon$. \hfill$\square$

\paragraph{Theorem~\ref{thm:param}.} For a cell and $a\in\{0,1\}$,
\[
p_\xi(z_i)^a\{1-p_\xi(z_i)\}^{1-a}/\{p_0^a(1-p_0)^{1-a}\}=1+\delta_\xi(z_i)(a-p_0)/v_0.
\]
Hence $d\bar P/dP_{p_0}=E_\xi\prod_i\{1+\delta_\xi(z_i)(a_i-p_0)/v_0\}$, and with an independent copy $\xi'$,
\begin{align*}
E_{p_0}\Big(\frac{d\bar P}{dP_{p_0}}\Big)^2&=E_{\xi,\xi'}\prod_iE_{p_0}\Big\{\Big(1+\frac{\delta_\xi(z_i)(a_i-p_0)}{v_0}\Big)\Big(1+\frac{\delta_{\xi'}(z_i)(a_i-p_0)}{v_0}\Big)\Big\}\\
&=E_{\xi,\xi'}\prod_i\Big(1+\frac{\delta_\xi(z_i)\delta_{\xi'}(z_i)}{v_0}\Big).
\end{align*}
Each factor is nonnegative because $|\delta|\le\min(p_0,1-p_0)$. Each cell lies in at most one bump, so $\delta_\xi(z_i)\delta_{\xi'}(z_i)=\eta^2\varepsilon_c\psi_c(z_i)^2$ with $\varepsilon_c=\xi_c\xi'_c$ independent random signs.

By $1+x\le e^x$ and $Ee^{t\varepsilon}=\cosh t\le e^{t^2/2}$, the expectation is at most
\[
\prod_c\cosh(\eta^2s_c/v_0)\le\exp\{\eta^4\sum_cs_c^2/(2v_0^2)\}.
\]

Under the stated condition $\eta^4\sum_cs_c^2/(2v_0^2)\le\log(5/4)$, so $\chi^2\le1/4$ and $|E_{\bar P}\phi-E_{p_0}\phi|\le\mathrm{TV}\le\tfrac12\{\chi^2\}^{1/2}\le\tfrac14$, while $P_{p_0}\in H_0$ because $p_0\in[\eta_0,1-\eta_0]$. The second condition implies the first because $\sum_cs_c^2\le s_{\max}n$. \hfill$\square$

\paragraph{Corollary~\ref{cor:rate}.} Choose a maximal subset $\mathcal C$ of cell locations with pairwise distances greater than $r$. By maximality, the closed disks of radius $r$ about these centers cover all $n$ cells. Each disk contains at most $\nu_{\max}+1$ cells, so $|\mathcal C|\ge n/(\nu_{\max}+1)$.

For each center $c$, choose another cell within $r/8$. At both cells $\psi_c\ge1/2$, and each ordering of the pair has weight at least $w_{\min}$. Thus each center contributes at least $w_{\min}/2$ to the numerator of $\tau_W$. Since $|W|\le n\nu_{\max}$,
\[
\tau_W\ge \frac{w_{\min}}{2\nu_{\max}(\nu_{\max}+1)}.
\]
Theorem~\ref{thm:floor} now gives the composition term in the lower bound. Each bump contains at most $\nu_{\max}+1$ cells, so $s_{\max}\le\nu_{\max}+1$. With $p_0=1/2$, Theorem~\ref{thm:param} therefore gives power at most $\alpha+1/4<1-\alpha'$ for $\gamma$ up to a constant times $n^{-1/2}$. The larger of the two lower bounds is at least half their sum.

The upper bound follows from Theorem~\ref{S-thm:power} with $\zeta=0$ (no dropped cells, and $W_{uv}>0$ on every close pair because $w>0$ on $[0,r]$), $B_1\le\Lambda^2(c_Rr)^{2\beta}|W|$, $\bar V(M)\le M^2n\max_i\omega_i^2+\tfrac18nK_Q^2$ with $\omega_i\le K_Q\{(2c_R+1)r\}^\beta$ because $Q_{ij}\ne0$ only when $|z_i-z_j|\le2R+r$, $V_{\rm link}\le n(\nu_{\max}+1)K_Q^2$, $K_Q\le(1+\Lambda_{\rm col})(1+\Lambda)\nu_{\max}$ and $|W|\ge c_Wn$. \hfill$\square$

\paragraph{Proposition~\ref{prop:adapt}.} Take $p_0=1/2$ and $\eta=1/2-\eta_0$ in Theorem~\ref{thm:floor}, with $M\ge c_\beta L_\psi\eta r^{-\beta}$. The mixture lies in $H_1^+(\tau_W\eta^2)$ and is a mixture of distributions in $\{P_p:p\in\mathcal P_\beta\}$, so any test of level $\alpha$ over $\mathcal P_\beta$ has power at most $\alpha$ against it. \hfill$\square$

\paragraph{Proposition~\ref{S-prop:mixing}.} Conditional independence gives $\Cov_{\bar P}(a_i,a_j)=\Cov_\pi(p_i,p_j)$ for $i\ne j$.

(i) With $d=p-E_\pi p$, $\sum_{i,j}E[d_id_j]=E(\sum_id_i)^2\ge0$, so $\sum_{i\ne j}\Cov(p_i,p_j)\ge-\sum_iE d_i^2$.

For the variance with $\beta\le1$, take $x$ with $|x-z_i|=r_+>r$. Then $p(x)$ and $p(z_i)$ are independent, so
\[
\Var p(z_i)\le\Var\{p(x)-p(z_i)\}\le E\{p(x)-p(z_i)\}^2\le M^2r_+^{2\beta}.
\]
For $\beta\in(1,2]$, take $x,y=z_i\pm r_+e$ for a unit vector $e$. The three values are pairwise independent, so $\Var\{p(x)+p(y)-2p(z_i)\}\ge4\Var p(z_i)$, while the gradient terms cancel in the Taylor expansions about $z_i$ and $|p(x)+p(y)-2p(z_i)|\le2Mr_+^\beta$. Let $r_+\downarrow r$.

(ii) Under $r$-dependence $\Cov_\pi(p_i,p_j)=0$ when $|z_i-z_j|>r$, so for the disk weight $\sum_{i\ne j}W_{ij}\Cov=\sum_{i\ne j}\Cov$. For a positive-definite weight, $\sum_{i,j}W^\circ_{ij}E[d_id_j]=E[d^\top W^\circ d]\ge0$ with $W^\circ_{ii}=w(0)$.

(iii) The second factorial moment measure of a Poisson process is $\lambda^2dx\,dy$, so $E\sum_{i\ne j}\Cov_\pi\{p(z_i),p(z_j)\}=\lambda^2\int_D\int_D\Cov_\pi\{p(x),p(y)\}\,dx\,dy=\lambda^2\Var_\pi\int_Dp$. \hfill$\square$

\paragraph{Proposition~\ref{S-prop:homog}.} Conditional independence gives $\Cov_{\bar P}(a_i,a_j)=\Cov_\pi(p_i,p_j)$ for $i\ne j$. Every composition in the class is $\beta$-Hölder with constant $M$, so $|p_j-p_m|\le M|z_j-z_m|^\beta$ path by path, and $\Var_\pi p_i\le(\tfrac12-\eta_0)^2$ because $p_i\in[\eta_0,1-\eta_0]$.

(i) Let $d=|z_i-z_j|$ and, for $r_+>r$, let $x$ be the point on the ray from $z_i$ through $z_j$ at distance $r_+$ from $z_i$, so that $|x-z_j|=r_+-d$. The field is $r$-dependent at every point of the plane, so $\Cov_\pi\{p_i,p(x)\}=0$ and
\[
\begin{aligned}
\Cov_\pi(p_i,p_j)&=\Cov_\pi\{p_i,p_j-p(x)\}\\
&\le[\Var_\pi p_i\,\Var_\pi\{p_j-p(x)\}]^{1/2}
\end{aligned}
\]
by the Cauchy--Schwarz inequality. As in the proof of Proposition~\ref{S-prop:mixing}, $\Var_\pi p_i\le M^2r_+^{2\beta}$, and $\Var_\pi\{p_j-p(x)\}\le E_\pi\{p_j-p(x)\}^2\le M^2(r_+-d)^{2\beta}$. Let $r_+\downarrow r$.

(ii) By (A1), $a_i$ is independent of $a_m$ and of $a_{m'}$, so $\Cov_\pi(p_i,p_m)=\Cov_\pi(p_i,p_{m'})=0$. Hence $\Var_\pi p_i=\Cov_\pi(p_i,p_i-p_{m'})\le(\Var_\pi p_i)^{1/2}M|z_i-z_{m'}|^\beta$, so that $(\Var_\pi p_i)^{1/2}\le M|z_i-z_{m'}|^\beta$, and $\Cov_\pi(p_i,p_j)=\Cov_\pi(p_i,p_j-p_m)\le(\Var_\pi p_i)^{1/2}M|z_j-z_m|^\beta$.

The consequences apply (i) to a close pair at distance $d_{\max}$, and (ii) to every ordered close pair with $m'$ the nearest cell farther than $r$ from $z_i$ and $m$ the nearest such cell to $z_j$. \hfill$\square$

\paragraph{Proposition~\ref{S-prop:pairsign}.} Write $\eta$ for $\eta_\pm$ and $p_i=\tfrac12+\eta\sum_P\sigma_{P,i}\xi_P$, where $\sigma_{P,i}=0$ if $i\notin P$, $\sigma_{P,i}=1$ for both cells of $P$ in the $+$ case, and $\sigma_{P,i_P}=1$, $\sigma_{P,m_P}=-1$ in the $-$ case.

\emph{Values.} $|p_i-\tfrac12|\le\eta\nu_i\le\tfrac12-\eta_0$. For cells $i\ne m$, $p_i-p_m=\eta\sum_P(\sigma_{P,i}-\sigma_{P,m})\xi_P$. A pair that contains exactly one of the two cells contributes at most $\eta$ in absolute value, and the pair $\{i,m\}$, if it is close, contributes $0$ in the $+$ case and at most $2\eta$ in the $-$ case. Hence, for close pairs, $|p_i-p_m|\le\eta(\nu_i+\nu_m-2)$ in the $+$ case and $|p_i-p_m|\le\eta(\nu_i+\nu_m)$ in the $-$ case, and $|p_i-p_m|\le2\eta\nu_{\max}$ for all pairs. Since $\eta\le M/L_\pm$, the first two bounds are at most $M|z_i-z_m|^\beta$, and for $|z_i-z_m|>r$ the last is at most $Mr^\beta<M|z_i-z_m|^\beta$.

Because $t\mapsto t^\beta$ is concave, $(x,y)\mapsto|x-y|^\beta$ is a metric, so McShane's extension $f(x)=\min_i\{p_i+M|x-z_i|^\beta\}$ lies in $\Sigma_\beta(M)$ and equals $p_i$ at $z_i$, and clipping $f$ to $[\eta_0,1-\eta_0]$ keeps both properties. The null distribution $P_f$ depends on $f$ only through its values at the cells, so $\bar P_\pm=E_\xi P_f$ is a mixture of null distributions and $E_{\bar P_\pm}\phi\le\alpha$ for every test of level $\alpha$.

\emph{(A1) and (A2).} $E\,p_i=\tfrac12$, a constant composition in the class. If every cell of $S$ is more than $r$ from every cell of $T$, no close pair meets both sets. The labels on $S$ are then a function of the signs of the pairs that meet $S$ and of independent uniform variables attached to the cells of $S$, the labels on $T$ are a function of a disjoint set of signs and uniforms, and the two are independent.

\emph{Covariances.} For $i\ne j$, conditional independence gives
\[
\Cov(a_i,a_j)=\Cov(p_i,p_j)=\eta^2\sum_P\sigma_{P,i}\sigma_{P,j}.
\]
Only the pair $\{i,j\}$ can contain both cells, so the sum is $\pm1$ if $i$ and $j$ are close and $0$ otherwise. Hence (A3) holds with the sign of the respective class, and $\Gamma_W=\pm\eta^2$ for every weight vanishing beyond $r$.

\emph{Hard core.} If all distances are at least $d_0$ and a close pair exists, then $d_0\le r$, so $(\nu_i+\nu_m)/|z_i-z_m|^\beta\le2\nu_{\max}/d_0^\beta$ and $2\nu_{\max}/r^\beta\le2\nu_{\max}/d_0^\beta$, which gives $L_\pm\le2\nu_{\max}/d_0^\beta$.

\emph{The case $\beta\in(1,2]$.} If all distances are at least $d_0$, let $f=\tfrac12+\sum_i(p_i-\tfrac12)\psi\{(\cdot-z_i)/d_0\}$ with a profile $\psi$ as in Section~\ref{sec:floor} and $\psi(0)=1$. The supports have disjoint interiors, so $f(z_i)=p_i$. The proof of Lemma~\ref{S-lem:bumps}, with amplitudes at most $\eta\nu_{\max}$ in absolute value in place of $\eta$ and with $d_0$ in place of $r$, shows that $f\in\Sigma_\beta(M)$ when $\eta\nu_{\max}\le Md_0^\beta/(c_\beta L_\psi)$, and $\eta_0\le f\le1-\eta_0$ when $\eta\nu_{\max}\le\tfrac12-\eta_0$. The rest of the argument is unchanged, so the floor is $\eta^2$ with $\eta=\min\{Md_0^\beta/(c_\beta L_\psi),\tfrac12-\eta_0\}/\nu_{\max}$.

\emph{Other families of sets.} If each sign is attached to a set of cells of diameter at most $r$ and every close pair lies in some set, the same argument applies with $\nu_i$ replaced by the number $k_i$ of sets that contain cell $i$. Then $|p_i-\tfrac12|\le\eta k_i$, $|p_i-p_m|\le\eta(k_i+k_m-2c_{im})$ because only the sets that contain exactly one of the two cells contribute, no set meets two sets of cells more than $r$ apart, and $\Cov(a_i,a_j)=\eta^2c_{ij}$. \hfill$\square$

\subsection{The cross-fitted statistic}

\begin{weblemma}\label{S-lem:zerodiag}
$Q$ is symmetric and has zero diagonal and zero row sums, and with affine-reproducing smoothers it also annihilates affine functions. The zero diagonal needs only that the two halves are disjoint and that no smoother row of a cell puts weight on a cell it is paired with ($W_{im}\ne0\Rightarrow S^{(t)}_{s,im}=0$).
\end{weblemma}

\begin{proof}
Fix a split and write $S_1,S_2$. Then
\[
(L_1^\top WL_2)_{ii}=\sum_{m,q}(\delta_{mi}-S_{1,mi})W_{mq}(\delta_{qi}-S_{2,qi}).
\]

Take a term with $W_{mq}\ne0$, so $m\ne q$. If $m=i$, the factor $S_{2,qi}$ is zero because $q$ is paired with $i$ (for the disk indicator, $|z_q-z_i|\le r\le\rho$), and $\delta_{qi}=0$ because $q\ne m=i$. The case $q=i$ is symmetric.

If $m\ne i\ne q$, a nonzero term needs $S_{1,mi}\ne0$ and $S_{2,qi}\ne0$, so $i\in H_1\cap H_2=\emptyset$. Every term vanishes.

Symmetry is immediate. Rows of $L_s$ sum to zero, so $L_s\mathbf 1=0$ and $Q\mathbf 1=0$; if the smoothers reproduce affine functions, $L_sz=0$ for each coordinate and hence $Qz=0$. All properties pass to averages over splits.
\end{proof}

\paragraph{Theorem~\ref{thm:moments}.} Write $x_k=p_k+\varepsilon$ and $x_\ell=p_\ell+\varepsilon'$, where the pairs $(\varepsilon_i,\varepsilon'_i)$ are independent across cells with mean zero, $E\varepsilon_i^2=v_{k,i}$, $E\varepsilon_i'^2=v_{\ell,i}$ and, for $k\ne\ell$, $E\varepsilon_i\varepsilon'_i=E[x_{ik}x_{i\ell}]-p_{k,i}p_{\ell,i}=-p_{k,i}p_{\ell,i}$ because a cell has one type.

Then
\[
U-p_k^\top Qp_\ell=\sum_i(g_{\ell,i}\varepsilon_i+g_{k,i}\varepsilon'_i)+\sum_{i\ne j}Q_{ij}\varepsilon_i\varepsilon'_j,
\]
the diagonal terms being absent because $Q_{ii}=0$. The mean is therefore $p_k^\top Qp_\ell$, and the bias-product form follows from $p_k^\top L_1^\top WL_2p_\ell=b_{1,k}^\top Wb_{2,\ell}$.

The linear part has variance
\[
\sum_i(g_{\ell,i}^2v_{k,i}+g_{k,i}^2v_{\ell,i}+2g_{\ell,i}g_{k,i}E\varepsilon_i\varepsilon'_i).
\]
Its covariance with the quadratic part vanishes because every product contains an index that appears once. In
\[
E(\sum_{i\ne j}Q_{ij}\varepsilon_i\varepsilon'_j)^2=\sum_{i\ne j}\sum_{u\ne v}Q_{ij}Q_{uv}E[\varepsilon_i\varepsilon'_j\varepsilon_u\varepsilon'_v],
\]
only $(u,v)=(i,j)$, which gives $v_{k,i}v_{\ell,j}$, and $(u,v)=(j,i)$, which gives $E[\varepsilon_i\varepsilon'_i]E[\varepsilon_j\varepsilon'_j]=p_{k,i}p_{\ell,i}p_{k,j}p_{\ell,j}$, survive.

Symmetry of $Q$ gives the stated formula. For $k=\ell$ the same computation with $\varepsilon'=\varepsilon$ gives the single-type formula. \hfill$\square$

\begin{webproposition}[Disjoint supports]\label{S-prop:disjoint}
Fix distinct cell locations and two types $k\ne\ell$, and let $W^{(1)},W^{(2)}$ be smoothing matrices. Suppose a function $T_{ij}$ of the labels satisfies
\[
E_pT_{ij}=(W^{(1)}p_k)_i(W^{(2)}p_\ell)_j
\]
for every composition in a class that can take any value in the simplex at finitely many distinct points. Then $W^{(1)}_{iu}W^{(2)}_{ju}=0$ for every cell $u$.
\end{webproposition}

\begin{proof}
Fix the composition at every cell other than $u$, and let the probability vector at $u$ move along the edge $te_k+(1-t)e_\ell$, $t\in[0,1]$, of the simplex, which the class allows because the locations are distinct.

Labels are independent, so $E_pT_{ij}$ is affine in the probability vector at $u$ and hence affine in $t$. The target equals $W^{(1)}_{iu}W^{(2)}_{ju}t(1-t)$ plus terms that are affine in $t$. Equality for all $t$ forces $W^{(1)}_{iu}W^{(2)}_{ju}=0$.
\end{proof}

The proposition is the elementary analogue, for independent draws with different distributions, of the characterization of unbiased estimability from samples \citep{halmos1946}: from one draw per cell only functions of degree at most one in each cell's probability vector are unbiasedly estimable. It concerns exact unbiasedness over a rich class; over Hölder classes it leaves room for approximately unbiased estimators with overlapping supports.

For distinct types, the leave-pair-out estimator $\sum_{u\ne v}W^{(1)}_{iu}W^{(2)}_{jv}x_{uk}x_{v\ell}$ suggested for intensities \citep*{shaw2021} equals $(W^{(1)}x_k)_i(W^{(2)}x_\ell)_j$, because $x_{uk}x_{u\ell}=0$, so it does not remove the shared-cell term.

\begin{weblemma}[Residual bound]\label{S-lem:residual}
If $p_k\in\Sigma_\beta(M)$ then $|b^{(t)}_{s,k,i}|\le Mm^{(t)}_{s,i}$, with $b$ and $m$ as in Sections~\ref{sec:construction} and \ref{sec:honest}, for $\beta\in(1,2]$ provided the rows reproduce affine functions.
\end{weblemma}

\paragraph{Lemma~\ref{S-lem:residual} and the variogram bound.} Row sums of one give $b_{s,i}=\sum_mS_{s,im}\{p(z_i)-p(z_m)\}$, and each term is at most $M|z_m-z_i|^\beta$ in absolute value when $\beta\le1$.

For $\beta\in(1,2]$, affine reproduction gives
\[
b_{s,i}=-\sum_mS_{s,im}\{p(z_m)-p(z_i)-\nabla p(z_i)\cdot(z_m-z_i)\},
\]
and each bracket is at most $M|z_m-z_i|^\beta/\beta$.

For the variogram form, expand $\sum_{i,j}Q_{ij}(x_i-x_j)(y_i-y_j)$; the terms $\sum_{ij}Q_{ij}x_iy_i$ and $\sum_{ij}Q_{ij}x_jy_j$ vanish because rows and columns of $Q$ sum to zero, and the cross terms give $-(x^\top Qy+y^\top Qx)=-2x^\top Qy$ by symmetry. Without symmetry the identity fails (a formal counterexample is in Web Appendix~\ref{S-sec:lean}). The one-sided bound $p^\top Qp\le\frac12M^2\sum_{ij}\max(-Q_{ij},0)|z_i-z_j|^{2\beta}$ holds only for $k=\ell$. \hfill$\square$

\paragraph{Theorem~\ref{thm:honest}.} If $Q=0$, then $U=0$ identically and the test has rejection probability zero by convention. For $Q\ne0$, the quadratic term $\tfrac18\sum_{i,j}Q_{ij}^2$ makes $\bar V(M)>0$. On the rejection event,
\[
|U-EU|\ge|U|-M^2B_1\ge\{\bar V(M)/\alpha\}^{1/2},
\]
and Chebyshev's inequality bounds its probability by $\alpha\Var U/\bar V(M)\le\alpha$.

The variance bound uses $v_{k,i}\le1/4$, $p_{k,i}p_{\ell,i}\le1/4$, $(|g_{\ell,i}|+|g_{k,i}|)^2/4\le M^2\omega_i^2$ for the linear part, and, for $\beta>1$, $Qz=0$ (Lemma~\ref{S-lem:zerodiag}), which lets the affine part of $p_k$ be subtracted inside $(Qp_k)_i$. \hfill$\square$

\paragraph{Proposition~\ref{prop:breakdown}.} The rejection region of the honest test shrinks as $M$ grows, because $M^2B_1+\{\bar V(M)/\alpha\}^{1/2}$ is increasing in $M$. Hence $\{\hat M^*>M_0\}$ is contained in the event that the test at $M_0$ rejects, whose probability is at most $\alpha$ under any composition of roughness at most $M_0$ by Theorem~\ref{thm:honest}. \hfill$\square$

\paragraph{Theorem~\ref{S-thm:power}.} The identity \eqref{eq:powerid} is $E[a^\top Qa]=p^\top Qp+\sum_{u,v}Q_{uv}\Cov(a_u,a_v)$ with $Q_{uu}=0$. Under (A1) only pairs with $|z_u-z_v|\le r$ contribute. On those with $W_{uv}>0$, $Q_{uv}\ge\theta^*W_{uv}$ and $\Cov_P(a_u,a_v)\ge0$ by (A3), so they contribute at least $\theta^*\Gamma_W(P)|W|$; those with $W_{uv}=0$ contribute at least $-\sum\max(-Q_{uv},0)\Cov_P(a_u,a_v)\ge-\zeta/4$. Lemma~\ref{S-lem:residual} gives $|p^\top Qp|\le M^2B_1$.

For the variance, $U=\sum_{u,v}Q_{uv}a_ua_v$, products of indicators are indicators, so each $\Cov(a_ua_v,a_{u'}a_{v'})$ is at most $1/4$ in absolute value, and by (A1) it vanishes unless an element of $\{u',v'\}$ is within $r$ of an element of $\{u,v\}$. Bounding the indicator of that event by $C_{uu'}+C_{uv'}+C_{vu'}+C_{vv'}$ and using symmetry of $Q$ gives
\[
\Var U\le\frac14\cdot4\sum_{u,u'}C_{uu'}q_uq_{u'}=q^\top Cq.
\]

Chebyshev's inequality completes the proof; the cross-type case is identical with $x_{uk}x_{v\ell}$ in place of $a_ua_v$. \hfill$\square$

\paragraph{Theorem~\ref{S-thm:clt}.} Order the cells. Group the centered statistic of the proof of Theorem~\ref{thm:moments} by the largest index: $U-EU=\sum_jD_j$ with $D_j=A_j\varepsilon_j+B_j\varepsilon'_j$, $A_j=g_{\ell,j}+\sum_{i<j}Q_{ij}\varepsilon'_i$ and $B_j=g_{k,j}+\sum_{i<j}Q_{ij}\varepsilon_i$.

Because the pairs $(\varepsilon_j,\varepsilon'_j)$ are independent across cells and centered, $(D_j)$ is a martingale difference sequence for $\mathcal F_j=\sigma\{(\varepsilon_i,\varepsilon'_i):i\le j\}$, with conditional variance
\[
V_n=\sum_j(A_j^2v_{k,j}+B_j^2v_{\ell,j}+2A_jB_jE\varepsilon_j\varepsilon'_j)
\]
and $EV_n=\sigma^2=\Var U$.

Bounded absolute row sums give $|A_j|,|B_j|\le2K_Q$, so $|D_j|\le4K_Q$, and expanding $V_n-\sigma^2$ into linear and quadratic forms in the independent vectors $(\varepsilon_i,\varepsilon'_i)$ with coefficients bounded through $|Q|$ shows that $\Var V_n\le CK_Q^4n$, as in the scalar case \citep{kelejian2001}.

Since $\sigma^2\ge c_0n$, $V_n/\sigma^2\to1$ in probability and the conditional Lindeberg condition holds, so the martingale central limit theorem \citep[Corollary~3.1]{hallheyde1980} gives the result for each composition.

The constants $C$, $K_Q$ and $c_0$ do not depend on the composition, and a martingale central limit theorem with explicit rate in terms of $E|V_n/\sigma^2-1|$ and $\sum_jE|D_j|^3/\sigma^3$ (for example the Berry--Esseen bound of \citet{haeusler1988}) therefore gives convergence uniform over the class. \hfill$\square$

\paragraph{Theorem~\ref{S-thm:local}.} \emph{Moments.} With $C_i$ and $\omega_{ij}$ as in Web Appendix~\ref{S-sec:localpower}, by the proof of Theorem~\ref{thm:moments},
\[
U-E_pU=\sum_i(g_{\ell,i}\varepsilon_{i,k}+g_{k,i}\varepsilon_{i,\ell})+\sum_{i\ne j}Q_{ij}\varepsilon_{i,k}\varepsilon_{j,\ell}.
\]
The terms of $T$ have mean zero because $W_{ii}=0$.

A linear term times a term of $T$ contains a cell index that occurs once, so it has mean zero. In $E[\varepsilon_{i,k}\varepsilon_{j,\ell}\varepsilon_{u,k}\varepsilon_{v,\ell}]$ with $i\ne j$ and $u\ne v$ only $(u,v)=(i,j)$, which gives $v_{k,i}v_{\ell,j}$, and $(u,v)=(j,i)$, which gives $C_iC_j$, survive. With the symmetry of $W$ this gives $\Cov_p(U,T)$, and the same computation with $Q$ replaced by $W$ gives $\Var_pT$. This proves \eqref{S-eq:tiltmoments}.

\emph{Joint limit under $P_p$.} Let $Z_n=(U-E_pU)/\sigma_U$ and $S_n=T/\sigma_T$. For fixed $(s,t)$, $n^{1/2}(sZ_n+tS_n)$ is a linear form plus an off-diagonal bilinear form in the independent cell vectors $(\varepsilon_{i,k},\varepsilon_{i,\ell})$. Its bilinear coefficients are $n^{1/2}\{sQ_{ij}/\sigma_U+tW_{ij}/((1+\delta_{k\ell})\sigma_T)\}$ and its linear coefficients $n^{1/2}sg_{\ell,i}/\sigma_U$ and $n^{1/2}sg_{k,i}/\sigma_U$. Their absolute row sums and sizes are bounded because $|g_{m,i}|\le K_Q$ and $\sigma_U^2,\sigma_T^2\ge c_0n$, and the variance of the form is $n(s^2+t^2+2st\rho_n)$.

Take a subsequence along which $\rho_n\to\rho$. If $s^2+t^2+2st\rho>0$, the martingale argument in the proof of Theorem~\ref{S-thm:clt} uses only these properties and gives $sZ_n+tS_n\to N(0,s^2+t^2+2st\rho)$. Otherwise $sZ_n+tS_n\to0$ in $L^2$. By the Cram\'er--Wold device $(Z_n,S_n)$ converges to the centered bivariate normal law with unit variances and correlation $\rho$.

\emph{Log-likelihood ratio.} Under $P_p$, $\Lambda_n=\log(dP_{p,\gamma_n}/dP_p)=hS_n-\psi_n$ with $\psi_n=\log E_pe^{hS_n}$.

Changing the label of one cell $m$ changes each $\varepsilon_{m,\cdot}$ by at most 1, and the other factors are at most 1 in absolute value, so $T$ changes by at most $2\nu_{\max}$.

The bounded-differences inequality in its exponential form \citep*[Theorem~6.2 and its proof]{boucheron2013} gives
\[
E_pe^{\lambda T}\le\exp(\lambda^2n\nu_{\max}^2/2)
\]
for every real $\lambda$, hence $\sup_nE_pe^{2hS_n}\le\exp(2h^2\nu_{\max}^2/c_0)$.

The variables $e^{hS_n}$ are therefore uniformly integrable, and $S_n\to N(0,1)$ gives $E_pe^{hS_n}\to e^{h^2/2}$, that is $\psi_n\to h^2/2$. Along the subsequence, $(Z_n,\Lambda_n)$ converges under $P_p$ to the bivariate normal law with means $(0,-h^2/2)$, variances $(1,h^2)$ and covariance $h\rho$.

\emph{Le Cam's lemmas.} By Le Cam's first lemma $P_{p,\gamma_n}$ is contiguous to $P_p$, and by Le Cam's third lemma \citep[Example~6.7]{vandervaart1998} $Z_n\to N(h\rho,1)$ under $P_{p,\gamma_n}$ along the subsequence.

If the uniform convergence in the theorem failed, some subsequence would keep the supremum above a fixed $\epsilon>0$. A further subsequence with $\rho_n\to\rho$ contradicts this, because along it both $P_{p,\gamma_n}(Z_n\le t)$ and $\Phi(t-h\rho_n)$ converge to $\Phi(t-h\rho)$ uniformly in $t$, by P\'olya's theorem for the first.

The power of a test that rejects when $U\ge c_n$ follows from the uniform convergence. Since $h>0$, the likelihood ratio is increasing in $T$. The most powerful level-$\alpha$ test therefore rejects for large $S_n$, with randomization at the cutoff if needed. Under $P_p$, $S_n$ converges in distribution to $N(0,1)$, so the cutoff converges to $z_{1-\alpha}$. The third lemma applied to $S_n$, whose limiting covariance with $\Lambda_n$ is $h$, gives $S_n\to N(h,1)$ under the alternative and power $1-\Phi(z_{1-\alpha}-h)+o(1)$. \hfill$\square$

\section{Further results on the statistic}\label{S-sec:statistic}

\subsection{Power}

\begin{webtheorem}[Power]\label{S-thm:power}
Let $\theta^*=\min\{Q_{uv}/W_{uv}:|z_u-z_v|\le r,\,W_{uv}>0\}$, assume $\theta^*>0$, and let $\zeta=\sum\max(-Q_{uv},0)$ over ordered close pairs with $W_{uv}=0$. For every $P\in H_1^+(\gamma)$,
\[
E_PU\ge\theta^*\gamma|W|-\zeta/4-M^2B_1,\qquad\Var_PU\le V_{\rm link}=q^\top Cq,
\]
where $q_u=\sum_v|Q_{uv}|$ and $C_{uu'}=\mathbf 1\{|z_u-z_{u'}|\le r\}$ (diagonal included). The honest test has power at least $1-\alpha'$ whenever $\theta^*\gamma|W|\ge\zeta/4+2M^2B_1+\{\bar V(M)/\alpha\}^{1/2}+(V_{\rm link}/\alpha')^{1/2}$.
\end{webtheorem}

\paragraph{Pairs near the image edge.} With $\rho\ge2r$ and nonnegative smoothers, $\theta^*\ge1$ and $\zeta=0$. For $\rho=r$, five splits and $h=25\um$, $\theta^*$ over pairs of cells farther than $R$ from the edge is 0.42--0.54 for both smoothers in the simulation design. For Nadaraya--Watson smoothing at $h=17.5\um$ the range is 0.04--0.19.

On the fixed image sets of Section~\ref{sec:data}, $\theta^*$ is defined for every graph with at least one retained pair. It was nonpositive in 514 of 724 all-cell graphs in METABRIC and 304 of 332 in Basel. Among non-epithelial cells the corresponding counts were 309 of 692 and 222 of 328 nonempty graphs. The ratios of total $\zeta$ to total $\sum_{W_{uv}>0}Q_{uv}$ across images were 0.06--0.44\% over the four cohort and population combinations. Removing cells within $R$ of the convex hull from the pair graph, while retaining them as smoothing cells, gave positive $\theta^*$ in 77--97\% of graphs that still had a pair (median $\theta^*$, 0.20--0.47). The ratio of total $\zeta$ to total retained weight then rose to 0.5--1.0\%. Among trimmed graphs with $\theta^*>0$, the median increase $\zeta/(4\theta^*|W|)$ in the required $\Gamma$ was 0.003--0.015, or 0.6--1.2 times the threshold $0.40|W|^{-1/2}$ by the median within-image ratio. The application uses the untrimmed graphs. The null moments and size bound in Theorems~\ref{thm:moments} and~\ref{thm:honest} hold for these graphs regardless of the sign of $\theta^*$.

\paragraph{Constants.} With $B_1\le\Lambda^2R^{2\beta}|W|$ and $|W|\asymp n$, the theorem gives the upper bound of Corollary~\ref{cor:rate}, and the constant in the floor is $2\Lambda^2(R/r)^{2\beta}/\theta^*$, so the smoothing window should be as close to the exclusion radius as the variance allows. The variance bound $V_{\rm link}$ exceeds the true standard deviation about 80-fold in our designs, so the guarantee becomes informative only beyond about $10^5$ cells per image (Web Appendix~\ref{S-sec:simdetails}).

\subsection{Asymptotic normality}

\begin{webtheorem}[Asymptotic normality]\label{S-thm:clt}
Consider a sequence of designs with $\max_u\sum_v|Q_{uv}|\le K_Q$ and $\Var U_{k\ell}\ge c_0n$ for every composition in the class. Then
\[
\sup_{p}\sup_{t\in\R}\Big|P_p\Big\{\frac{U_{k\ell}-E\,U_{k\ell}}{(\Var U_{k\ell})^{1/2}}\le t\Big\}-\Phi(t)\Big|\to0,
\]
where the supremum over $p$ is over the composition class.
\end{webtheorem}

For $k=\ell$ and a fixed composition this is Theorem~1 of \citet{kelejian2001} for linear-quadratic forms in independent heteroscedastic variables. For $k\ne\ell$ the innovations $(x_{ik}-p_{k,i},x_{i\ell}-p_{\ell,i})$ are independent across cells but dependent within a cell, and the martingale argument of Web Appendix~\ref{S-sec:proofs} applies with cell-level blocks. Uniformity holds because every constant in that argument depends on the composition only through $K_Q$, $c_0$ and $\eta_0$; $\Var U\ge c_0n$ holds, for example, when $\sum_{i,j}Q_{ij}^2\ge c n$, since the quadratic part of the variance is at least $2\eta_0^2(1-\eta_0)^2\sum Q_{ij}^2$ for $k=\ell$. If the composition varies at a scale $L_n$, $p_n(z)=p^\circ(z/L_n)$ with $p^\circ\in\Sigma_\beta$, and $n\asymp L_n^2$, the standardized bias of the plug-in test is of order $n^{1/2}(R/L_n)^{2\beta}$, which vanishes for fixed windows when $\beta>1/2$, and asymptotic plug-in variance calibration requires $R_n\to\infty$ with $R_n=o(L_n^{1-1/(2\beta)})$ (for $\beta=2$, $R_n=o(L_n^{3/4})$). In this regime $M_n\asymp L_n^{-\beta}$, so the floor $M_n^2r^{2\beta}$ is negligible; the minimax message concerns fixed $M$, where the honest test with $\bar V(M)$ is the appropriate procedure.

\subsection{Local power}\label{S-sec:localpower}

For the pair $(k,\ell)$ write $\varepsilon_{i,m}=x_{im}-p_m(z_i)$ and $\delta_{k\ell}=\mathbf 1\{k=\ell\}$, and consider the laws
\[
\frac{dP_{p,\gamma}}{dP_p}=\frac{e^{\gamma T}}{E_pe^{\gamma T}},\qquad T=\frac{1}{1+\delta_{k\ell}}\sum_{i\ne j}W_{ij}\varepsilon_{i,k}\varepsilon_{j,\ell}.
\]
For $k=\ell$ and two types these are the centered autologistic models with conditional logit $\operatorname{logit}p(z_i)+\gamma\sum_jW_{ij}(a_j-p_j)$ \citep{caragea2009}, and for $k\ne\ell$ centered cross-type attraction models. Let $\sigma_U^2=\Var_pU_{k\ell}$, $\sigma_T^2=\Var_pT$ and $\rho_n=\Cov_p(U_{k\ell},T)/(\sigma_U\sigma_T)$, which have the closed forms \eqref{S-eq:tiltmoments} below.

\begin{webtheorem}[Local asymptotic power]\label{S-thm:local}
Consider a sequence of designs and compositions with $\max_i\sum_j|Q_{ij}|\le K_Q$, $\max_i\sum_jW_{ij}\le\nu_{\max}$ and $\sigma_U^2,\sigma_T^2\ge c_0n$, and let $\gamma_n=h/\sigma_T$ with fixed $h>0$. Then under $P_{p,\gamma_n}$
\[
\sup_{t\in\R}\Big|P\Big\{\frac{U_{k\ell}-E_pU_{k\ell}}{\sigma_U}\le t\Big\}-\Phi(t-h\rho_n)\Big|\to0,
\]
so a test that rejects when $U_{k\ell}\ge c_n$ has power $1-\Phi\{(c_n-E_pU_{k\ell})/\sigma_U-h\rho_n\}+o(1)$, while the most powerful level-$\alpha$ test of $P_p$ against $P_{p,\gamma_n}$ has power $1-\Phi(z_{1-\alpha}-h)+o(1)$.
\end{webtheorem}

The correlation $\rho_n$ is thus the efficiency of the cross-fitted statistic at the given composition relative to the Neyman--Pearson test that knows the composition, and $\rho_n^2$ is its Pitman efficiency.

For the pair $(k,\ell)$ of Theorem~\ref{S-thm:local} let $C_i=\delta_{k\ell}p_{k,i}-p_{k,i}p_{\ell,i}$ and $\omega_{ij}=v_{k,i}v_{\ell,j}+C_iC_j$. Under $P_p$ the tilt statistic has $E_pT=0$,
\begin{equation}\label{S-eq:tiltmoments}
\sigma_T^2=\Var_pT=\frac{\sum_{i\ne j}W_{ij}^2\omega_{ij}}{(1+\delta_{k\ell})^2},\qquad\Cov_p(U_{k\ell},T)=\frac{\sum_{i\ne j}Q_{ij}W_{ij}\omega_{ij}}{1+\delta_{k\ell}},
\end{equation}
and $\sigma_U^2$ is the variance of Theorem~\ref{thm:moments}. The proof is part of that of Theorem~\ref{S-thm:local} in Web Appendix~\ref{S-sec:proofs}.

\paragraph{Local power of the honest test.} The one-sided honest normal test at $M_n$ rejects when $U_{k\ell}\ge c_n=M_n^2B_1+z_{1-\alpha}\bar V(M_n)^{1/2}$, so its power is $1-\Phi(b_n+z_{1-\alpha}\lambda_n-h\rho_n)+o(1)$ with $\lambda_n=\bar V(M_n)^{1/2}/\sigma_U\ge1$ and $0\le b_n=(M_n^2B_1-E_pU_{k\ell})/\sigma_U\le2M_n^2B_1/\sigma_U$. If $E_pU_{k\ell}\le(1-c)M_n^2B_1$ for a fixed $c>0$, as for constant compositions, then $b_n\ge cM_n^2B_1/\sigma_U$, which is of order $M_n^2n^{1/2}$ because $B_1$ grows linearly in $n$. Against alternatives at the parametric scale the honest test then has power above its level only when $M_n=O(n^{-1/4})$, which is the local form of the floor.

Theorem~\ref{S-thm:local} concerns a parametric family that is not $r$-dependent and therefore lies outside $H_1^+(\gamma)$, and its Neyman--Pearson benchmark uses the composition, so it is not a valid test over $H_0$. The theorem therefore does not narrow the gaps of Corollary~\ref{cor:rate}. It gives the operating characteristics of the honest test against a standard interaction model. For two types at $p=1/2$ and the disk weight, $\Gamma_W(P_{p,\gamma})=\gamma/16$ to first order in $\gamma$ and $\sigma_T^2=|W|/32$. With $\alpha=0.05$ the Neyman--Pearson test reaches power $\alpha+1/4$ at $\Gamma_W\approx0.40|W|^{-1/2}$, and the one-sided honest test at $M=0$, for which $\lambda_n=1$ and $b_n=0$, at $0.40|W|^{-1/2}/\rho_n$, or $0.47|W|^{-1/2}$ for $\rho_n=0.84$. The lower bound $0.06|W|^{-1/2}$ of Theorem~\ref{thm:param} in the simulation design concerns a different alternative, the disk mixture, and is 6.4 times smaller than the first of these. Web Appendix~\ref{S-sec:simdetails} compares the predictions of Theorem~\ref{S-thm:local} with simulated power.

\subsection{Plug-in variance}

Let $\hat p_k=S_{\rm full}x_k$ with a full-sample smoother whose rows sum to one. Then $E\hat p_{k,i}=(S_{\rm full}p_k)_i$ and $\Var\hat p_{k,i}=\sum_mS_{{\rm full},im}^2v_{k,m}$, so $E\{\hat p_{k,i}(1-\hat p_{k,i})\}=\tilde v_{k,i}-\Var\hat p_{k,i}$, with $\tilde v$ the variance at the smoothed composition, and $E(Q\hat p_k)_i^2=(QS_{\rm full}p_k)_i^2+\sum_m(QS_{\rm full})_{im}^2v_{k,m}$. We therefore use $\hat v_{k,i}+\sum_mS_{{\rm full},im}^2\hat v_{k,m}$ for $v_{k,i}$ and subtract $\sum_m(QS_{\rm full})_{im}^2\hat v_{k,m}$ from $(Q\hat p_k)_i^2$ (and add $\sum_m(QS_{\rm full})_{im}^2\hat p_{k,m}\hat p_{\ell,m}$ to $(Q\hat p_k)_i(Q\hat p_\ell)_i$), summing the corrections before truncating the linear part at zero. In our simulations the uncorrected plug-in standard deviation was between 0.4\% below and 8\% above the exact one for common types and 9--15\% above it for a rare type with frequency 0.1, and the corrected one exceeded it by at most 4\% in the common-type designs where we checked it.

\section{Simulation details}\label{S-sec:simdetails}

\subsection{The floor in practice}

The simulations of Section~\ref{sec:floorsim} used Nadaraya--Watson smoothers with a Gaussian kernel truncated at $2.5h$, and tests that are one-sided in the segregation direction at level 0.05. The honest test is the normal version with $\bar V(M)$, with the one-sided variogram bound in place of $M^2B_1$. Each point of Figure~\ref{fig:floor} uses 300 (4,900 cells) or 120 (19,600 cells) replicates over ten cell patterns in panels A and B, and 200 or 80 in panel C. In panel A the mean of the cross-fitted statistic matched its exact value $\eta^2G$, $G=\sum_c\psi_c^\top Q\psi_c$, at every amplitude (for example 155.6 against 150.8 at $\eta=0.4$, with a standard error of 5.4), and tile shuffling at $25\um$, although less affected than the plug-in test, also exceeded 5\% (0.33 at $\eta=0.5$ with 19,600 cells). The cones of panel B have peaks $80\um$ apart, and their exact null bias is about a tenth of the bound $M^2B_1$. Tile shuffling, whose tiles track this coarser composition, rejected in 4.0--5.7\% of replicates with 4,900 cells and 4.2--11.7\% with 19,600 cells. The micro-clusters of panel C are disks of diameter $r$ whose composition is raised or lowered by $\eta_c$, and the honest test uses the true $M$ of the cone composition.

\paragraph{Constants of Section~\ref{sec:floor} and Web Appendix~\ref{S-sec:statistic}.} The constants quoted after Theorem~\ref{thm:floor}, Corollary~\ref{cor:rate} and Theorem~\ref{S-thm:power} were computed on Poisson patterns of this design (0.01 cells per $\mu\mathrm{m}^2$, $r=20\um$, about 12 neighbors per cell). With $\beta=1$ the cone witness gives $\Gamma_W=0.0050(Mr)^2$ and the $\cos^2$ witness $0.0016(Mr)^2$. If the alternative class allows dependence up to a range $r'\ge r$, cone bumps of diameter $r'$ give $0.064(Mr)^2$ for $r'=2r$ and $0.20(Mr)^2$ for $r'=3r$. The upper constant $2B_1/(\theta^*|W|r^2)$ is 12--16 for five-split designs at $h=25\um$ and 15--18 for the single-split design of Section~\ref{sec:floorsim}, with $\theta^*$ computed on cells farther than $R$ from the edge. For the $n^{-1/2}$ term the lower constant of Theorem~\ref{thm:param} is $0.06|W|^{-1/2}$ with indicator disks, and the upper constant from Theorem~\ref{S-thm:power} is about $84n^{-1/2}$, or $290|W|^{-1/2}$. It is large because the variance bound $V_{\rm link}$ charges $1/4$ to every linked quadruple of cells, which makes it exceed the true standard deviation about 80-fold. At $M=0$ the guarantee of Theorem~\ref{S-thm:power} therefore needs $\gamma\gtrsim84n^{-1/2}$, which is below $1/4$, the largest value of $\Gamma_W$ for binary labels, only when $n\gtrsim1.1\times10^5$.

\subsection{Choice of window}\label{S-sec:window}

Web Figure~\ref{S-fig:window} shows the exact standardized null bias under three compositions and the local efficiency $\kappa=\langle Q,W\rangle/(\|Q\|_F|W|^{1/2})$ against the centered autologistic alternative, both computed exactly for one cell pattern. Narrowing the window from $h=25$ to $15\um$ cuts the bias by a factor of two to four and lowers $\kappa$ from 0.87 to 0.75. By Theorem~\ref{S-thm:local}, $\kappa$ is the efficiency relative to the Neyman--Pearson test that knows the composition, so the Pitman efficiency $\kappa^2$ falls from 0.75 to 0.56. Averaging over five splits raises $\kappa$ by 0.016--0.065 for Nadaraya--Watson and by 0.022--0.118 for local-linear smoothing across bandwidths. Local-linear smoothing removes the bias from linear trends exactly. Nadaraya--Watson smoothing leaves 0.019 standard deviations at $h=25\um$ and 0.043 at $h=35\um$, because the design imbalance of the two halves is independent and averages out in the product.

The exclusion radius is not the only way to obtain exact moments. Lemma~\ref{S-lem:zerodiag} needs only that a cell's smoother never uses a cell it is paired with, so pairing only cells at distance $(d_{\rm in},r]$ lets the smoothers use the immediate neighbors within $d_{\rm in}$. With $d_{\rm in}=10\um$ and $h=17.5\um$, the standardized bias under the islands, interface and cone compositions falls to 0.10, 0.06 and 0.65, against 0.15, 0.10 and 0.96 without the inner band, and $\kappa$ falls from 0.79 to 0.63. With a narrow outer window ($h=12\um$, truncated at $30\um$) the biases are 0.02, 0.01 and 0.19 and $\kappa=0.41$. Such designs trade efficiency for robustness while keeping exact moments. The narrow design was not included in the benchmark below.

\begin{figure}[htbp]
\centering
\EmbeddedFigure{0.8\linewidth}{window}
\caption{Window choice, exact computations. A: local efficiency $\kappa$ against the autologistic alternative, five splits. B: standardized null bias under a cone composition ($M=0.005\um^{-1}$), the islands composition and a logistic interface of width $20\um$. Solid, Nadaraya--Watson; dashed, local linear.}\label{S-fig:window}
\end{figure}

\subsection{Benchmark}

Beyond the tests of Table~\ref{tab:bench}, residual products with a single leave-one-out smoother fail in the opposite direction to the radius-excluded one because of leakage (two-sided size 1.0). The inhomogeneous cross-$K$ test uses leave-one-out intensities, and the autologistic pseudo-likelihood was tested with a Wald statistic and with a parametric bootstrap. CRAWDAD was run on subsets of the designs and replicates, as described below, and the $K$-function tests on a subset of the designs. The cross-fitted tests of Table~\ref{tab:bench}A use the uncorrected plug-in variance. Their $z$ had standard deviation 0.86--1.05 across the null designs, and power is not size-adjusted. The labelings of the last power column of Table~\ref{tab:bench}, the centered autologistic model with $\gamma=0.15$ under the strong linear trend, were drawn on the same 50 cell patterns with the Gibbs sampler of the other autologistic designs and analyzed in the same way, except that the tests implemented in R were not run. The honest tests of Table~\ref{tab:bench}B use the smaller of the two bias bounds of Section~\ref{sec:honest} and the Lipschitz constant of each composition, which is 0 for the flat composition, $0.8/700$, $0.6e^{-1/2}/80$ and $0.7/80\um^{-1}$ for the strong linear trend, the islands and the interface, $0.005\um^{-1}$ for the cones and $0.3\pi/20\um^{-1}$ for the bump mixture.

The cross-$K$ bootstrap is close to nominal two-sided but rejects 6--10\% one-sided. At $\gamma=0.04$ the Nadaraya--Watson and doughnut variants have 2.3--2.4 and 1.8--1.9 times the one-sided power of $25\um$ tiles, and at $\gamma=0.02$ 1.5--1.7 and 1.1--1.6 times. Under the islands alternative local-linear cross-fitting had less power than $50\um$ tiles and the pseudo-likelihood Wald test. The doughnut variant is not fooled by the bump mixture because its pairs exclude the distances below $10\um$, which carry 71--89\% of the within-bump covariance. For the same reason it cannot detect interaction confined to contact distances, since $10\um$ is about the median nearest-neighbor distance in METABRIC. Pseudo-likelihood was competitive in power but less robust than local-linear cross-fitting under the cone and Voronoi compositions. With hard-core cell patterns the one-sided sizes were 0.050--0.094 for the cross-fitted tests, 0.034--0.038 for $25\um$ tiles, 0.110--0.126 for $50\um$ tiles and 0.108--0.130 for the bootstrap pseudo-likelihood. With eight types the cross-fitted tests, tile shuffling and the bootstrap pseudo-likelihood held their size (0.03--0.07). CRAWDAD was run on the flat, eight-type and rare-type null designs (200 replicates each), the bump mixture (130), the mild and strong linear trends (120 and 40), cones with $M=0.005\um^{-1}$ (20), the eight-type attraction alternatives (170 and 10) and the rare-type alternative (200). Its own decision rule rejected none of the balanced binary datasets and none of the eight-type datasets, including those with attraction, and it rejected every null dataset with a rare reference type ($p=0.1$). Under the rare-type clustering alternative its $Z$ pointed in the wrong direction, with one-sided rejection 0.00--0.01. In version 1.0.1 its $Z$-score compares neighbor proportions on the union of the buffers around all reference cells between observed and within-tile shuffled labels. With common types the union covers almost every cell, so $Z\approx0$. The observed neighbor set also excludes reference cells that the shuffled set keeps, which gives $Z\approx-4$ at $25\um$ for a rare reference type under the null.

\subsection{Local power}

For the checks of Section~\ref{sec:upper} we rebuilt every cross-fitted design of the benchmark from its seeds, which reproduces the stored $z$ values exactly, and computed $\rho_n$, the exact null mean and variance at the true composition and the predicted rejection rate. The four weak-interaction designs are the binary autologistic model at $\gamma=0.02$ under three compositions and the eight-type attraction at $\gamma=0.1$, each analyzed by the three methods. The largest difference between the predicted and observed means of $(U-E_pU)/\sigma_U$ was 1.76 Monte Carlo standard errors, with standard errors of 0.04--0.05. At stronger interaction the first-order prediction was too small, by 3--10\% at $\gamma=0.04$, by 4--5\% for the eight-type attraction at $\gamma=0.2$ and by 18--22\% for the rare type at $\gamma=0.2$, because higher-order terms in $\gamma$ add to the mean. To separate these terms from sampling error, we sampled the tilted laws directly on ten patterns per design, with 16 Gibbs chains per pattern, and estimated the second-order term $(\gamma^2/2)\kappa_p(U,T,T)$, a joint cumulant, from 10,000 null labelings per pattern. For the binary designs at $\gamma=0.04$ the first-order prediction plus the second-order term is within 0.01 of the exact tilted mean in all nine analyses (standard error about 0.006) and accounts for 77--100\% of the first-order gap. The remaining differences of up to 0.09 between this prediction and the 500 benchmark replicates are consistent with sampling error of the replicates, which share their labelings across the three methods. Sampling all 50 patterns of the flat design directly gives exact means 0.05--0.08 below the replicate means, or 1.1--1.8 standard errors of the replicate means, and second-order predictions within 0.012 of the exact means. The chains mix well, with potential scale reduction factors of at most 1.004 and lag-one autocorrelations of the thinned statistic between $-0.05$ and 0.04. For the binary designs at $\gamma=0.04$, importance weighting of the null labelings, without a Gibbs sampler, gives estimates between 0.04 below and 0.001 above the exact means, and for the rare type at $\gamma=0.2$ it gives estimates 0.59--0.81 below them. For the rare type at $\gamma=0.2$ the second-order term closes 60--64\% of the gap to the exact mean. The standard deviation of $(U-E_pU)/\sigma_U$ under the alternatives was 0.97--1.09, except for the two designs at $\gamma=0.2$ (1.13--1.31).

\section{Details of the application}\label{S-sec:datadetails}

\subsection{Data and specification}

Every pair in the all-cell analysis uses the same 724 METABRIC images and 332 Basel images with at least 300 segmented cells. For the non-epithelial analysis we condition on which cells are epithelial and retain each eligible image with at least 150 non-epithelial cells, giving 698 METABRIC and 331 Basel images. Conditional on this membership, the remaining labels are independent under the null with type probabilities $p_k/(1-p_{\rm ep})$. The roughness bound in this analysis applies to these conditional probabilities. An image stays in its pool even if a target type is absent. Web Appendix~\ref{S-sec:screening} compares these fixed sets with the earlier rule requiring at least 20 cells of each target type.

We used five random splits, $r=\rho=20\um$ and local-linear smoothing with $h=25\um$ and outer radius $R=2.5h$. A cell remained available to smooth its neighbors even if it was removed as a pair endpoint because a half-window held fewer than six cells or its local design was singular or had condition number at least $10^4$. Across all cells in the fixed sets, the fractions removed as endpoints were 3.4\% in METABRIC and 3.7\% in Basel. Among non-epithelial cells they were 26.8\% and 23.3\%, respectively. We retained eligible endpoints up to the image edge (Web Appendix~\ref{S-sec:statistic}). The zero diagonal and zero row sums of $Q$ still hold, since $W$ has no edges at removed endpoints.

For each pair we sum the signed statistics $U_m$ over images $m$. Under independent labeling across cells and images, conditional on the design, the image variances add. The descriptive pooled $z$ uses the square root of the sum of the plug-in variances, evaluated at a Nadaraya--Watson composition estimate with first-order corrections for estimation noise (Web Appendix~\ref{S-sec:statistic}). The Chebyshev test uses the sum of the design-based variance bounds. An absent target type gives $U_m=0$ on any graph, while the design-based bounds still enter the sums. Six non-epithelial graphs in METABRIC and three in Basel have no retained pair. Each contributes zero to the signed statistic, the bias bound and the variance bound, and counts as a nonrejection in image-level rates. Every all-cell graph has a retained pair.

\subsection{Diagnostics}

For a specified semi-synthetic composition $p_m$, the standardized null mean in Table~\ref{tab:data} is
\[
  \frac{\sum_m p_{k,m}^{\top}Q_m p_{\ell,m}}
       {\{\sum_m\Var_{p_m}(U_m)\}^{1/2}} .
\]
We evaluate both moments at the observed type indicators smoothed at $25$ or $50\um$. This calculation asks how much residual association the fitted surface alone produces under independent labeling.

To estimate the typical slope of type $k$ in an image, let $\sum_j a_{ij}x_{jk}$ be the fitted gradient at cell $i$, with vector weights $a_{ij}$. Under independent labeling its expected squared norm exceeds $\|\sum_j a_{ij}p_{k,j}\|^2$ by $\sum_j\|a_{ij}\|^2p_{k,j}(1-p_{k,j})$. We subtract the image mean of this noise term, estimated with the $25\um$ smoothed probabilities, from the observed mean squared gradient and take the square root of its positive part. Web Table~\ref{S-tab:sens} reports how often a corrected slope is zero in the fixed sets. The multiplier $\lambda^*$ uses these slopes, estimated from the same labels, as references for each type and image. Interpreting $\lambda=1$ as a smoothness bound requires the true composition to satisfy the Lipschitz condition at the corresponding reference slope.

A separate calibration study drew 80 eligible METABRIC images at random. It used image--type combinations with at least 20 cells of the type and a positive noise-free target slope. For each retained image, type and composition we computed the expected corrected mean square and drew 20 independent binary labelings. For truths smoothed at $25\um$, the type-specific medians over images of the square root of the positive part of the expected corrected mean square divided by the target slope were 0.998--1.006. For truths smoothed at $15$ and $10\um$, the respective ranges were 0.980--0.987 and 0.946--0.974. A fourth truth used the observed type proportions on either side of the tumor--stroma boundary defined by a $30\um$ epithelial map. The corresponding ratios were 0.99--1.30. Subtracting noise estimated from a random within-image relabeling gave ratios of 0.90--0.97 for the three smooth truths. Web Table~\ref{S-tab:sens} compares the two corrections on the fixed image sets.

For a fixed placement of the lattice bumps of diameter $r$, let $G=\sum_c\psi_c^\top Q_m\psi_c$ in image $m$. The signs are independent across bumps. On constant baselines, shared signs for the two types give mean $+\eta^2 G$ at equal amplitudes $\eta$, and opposite signs give mean $-\eta^2 G$. We average $G$ over four random lattice offsets in each image to obtain $G_m$. Web Table~\ref{S-tab:ident} gives the amplitude $\eta_{\rm wit}=(|\sum_m U_m|/\sum_m G_m)^{1/2}$ that matches the observed pooled magnitude for each profile. A probability model at that amplitude must also satisfy the range constraints on the composition (Web Appendix~\ref{S-sec:identification}).

\subsection{Detailed results}

\emph{Non-epithelial fibroblast--T and T--macrophage scores.} Among non-epithelial cells the fixed-image $z$ for METABRIC fibroblast--CD8 T was $-3.85$, with semi-synthetic standardized mean $-2.59$ at $25\um$. For Basel T--macrophage the corresponding values were $-4.39$ and $-1.60$. For each pair the design-based Chebyshev test gave no positive common-slope breakdown in this population (Web Table~\ref{S-tab:screenholm}).

\emph{Stromal and T cells in Basel.} The pooled association changed sign between all cells and non-epithelial cells. In the non-epithelial analysis, 34.4\% of the 331 images had a plug-in $z<-1.96$ and 0.9\% had $z>1.96$. The Chebyshev breakdown for a common slope was $1.48\times10^{-3}\um^{-1}$. With separate bounds proportional to each image's corrected $50\um$ slopes, the Chebyshev multiplier was 0.84. These two summaries use different assumptions about composition roughness.

\emph{Pooled T--B scores.} The non-epithelial pooled statistics were negative for METABRIC CD4 T--B and CD8 T--B and for Basel T--B. Giving image $z$ values equal weight in Stouffer's combination gave $-2.47$, $-0.43$ and $1.63$, respectively. None of the three fixed pools rejected at $M=0$ under the Chebyshev bound. METABRIC non-epithelial CD4 T--B had a positive Chebyshev breakdown under the earlier count screen (Web Table~\ref{S-tab:screenholm}). In the fixed sets the median corrected B-cell slope was zero in both cohorts. This explains the zero bias shares at the median slopes for these pairs in Web Table~\ref{S-tab:ident}. The image-specific multipliers use the slopes of each image before pooling.

\emph{Epithelial--immune avoidance.} For the 657 patients contributing the 724 METABRIC images, Stouffer's combination of patient $z$ values was $-69.34$ for CD8 T--epithelial and $-91.62$ for macrophage--epithelial. Each patient statistic divides the sum of its core scores by the square root of their summed plug-in null variances. Residual correlation between cores would require covariance terms. At the image-specific corrected $50\um$ slopes, the METABRIC CD8 T--epithelial Chebyshev multiplier was 0.9805. Across the five noise estimates it ranged from 0.9205 to 1.0004. For Basel macrophage--epithelial it also crossed one, whereas for METABRIC macrophage--epithelial and Basel T--epithelial it exceeded one under all five estimates. Web Table~\ref{S-tab:sens} gives the normal-calibration variants, including a common slope within each image and alternative treatments of zero slopes. Using each image's uncorrected upper-decile or maximal slopes gave multipliers below one for all four pairs.

At the median corrected $50\um$ slopes of the two types, the $\cos^2$ and cone witnesses each produced less than 0.05\% of the magnitude of the observed pooled statistic for each of the four pairs. Here a profile of Lipschitz constant $L/r$ at unit amplitude has amplitude $s_k r/L$ for a type with median slope $s_k$, with $L=\pi$ for $\cos^2$ and $L=2$ for a cone. For an indicator disk the equal amplitudes needed to match the pooled magnitudes ranged from 0.13 to 0.21. The disk profile permits a discontinuity at its boundary.

\emph{Single-image thresholds.} With each type's corrected $50\um$ slope and $\lambda=1$, the normal test using $\bar V$ crossed its threshold in 8 of 724 METABRIC images for CD8 T--epithelial (1.1\%) and 24 of 724 for macrophage--epithelial (3.3\%). In Basel the counts were 14 of 332 for macrophage--epithelial (4.2\%) and 40 of 332 for T--epithelial (12.0\%). For the pooled tests, $1.96\sqrt{\sum_m\bar V_m}$ accounted for 10.6--19.7\% of the critical value. To assess the cost of the bound per image, we repeated the comparison with the quadratic term $\tfrac18\sum_{i,j}Q_{ij}^2$ removed (3.5--19.0\% across the four pairs), with the exact null variance at the $25\um$ smoothed composition (25.0--46.4\%), and with only the bias bound (58.3--66.9\%). These descriptive comparisons require a positive excess over the bias bound when the reference standard deviation is zero.

\begin{table}[htbp]
\centering
\caption{Identification diagnostics on the fixed image sets for pairs with a positive normal-calibration breakdown slope using $\bar V$.}\label{S-tab:ident}
\scriptsize
\setlength{\tabcolsep}{3.5pt}
\begin{tabular}{lrrrrrrrrr}
\toprule
 & \multicolumn{2}{c}{slopes} & \multicolumn{3}{c}{$\lambda^*$} & bias & \multicolumn{3}{c}{$\eta_{\rm wit}$}\\
Pair & first & second & $50\um$ & $100\um$ & q90 & share & $\cos^2$ & cone & disk\\
\midrule
\multicolumn{10}{l}{\emph{METABRIC, all cells}} \\
CD8~T -- macrophage & 0.23 & 0.30 & 0.36 & 0.71 & 0.16 & 0.33 & 0.21 & 0.17 & 0.05 \\
fibroblast -- CD8~T & 2.02 & 0.23 & 0.72 & 1.45 & 0.35 & 0.89 & 0.34 & 0.27 & 0.08 \\
CD8~T -- epithelial & 0.23 & 2.94 & 1.12 & 2.17 & 0.57 & 0.43 & 0.58 & 0.46 & 0.13 \\
macrophage -- epithelial & 0.30 & 2.94 & 1.27 & 2.59 & 0.66 & 0.38 & 0.72 & 0.57 & 0.16 \\
\midrule
\multicolumn{10}{l}{\emph{METABRIC, non-epithelial cells}} \\
CD4~T -- B & 0.36 & 0.00 & 0.58 & 1.06 & 0.29 & 0.00 & 0.37 & 0.30 & 0.09 \\
CD8~T -- B & 0.43 & 0.00 & 0.09 & 0.15 & 0.04 & 0.00 & 0.27 & 0.22 & 0.06 \\
\midrule
\multicolumn{10}{l}{\emph{Basel, all cells}} \\
macrophage -- epithelial & 0.34 & 2.84 & 1.13 & 2.44 & 0.61 & 0.51 & 0.55 & 0.45 & 0.14 \\
stroma -- T & 2.21 & 0.72 & 0.49 & 1.01 & 0.27 & 2.59 & 0.31 & 0.26 & 0.08 \\
T -- macrophage & 0.72 & 0.34 & 0.50 & 1.05 & 0.26 & 0.75 & 0.23 & 0.19 & 0.06 \\
T -- epithelial & 0.72 & 2.84 & 1.25 & 2.49 & 0.70 & 0.47 & 0.82 & 0.68 & 0.21 \\
\midrule
\multicolumn{10}{l}{\emph{Basel, non-epithelial cells}} \\
stroma -- T & 1.81 & 1.11 & 0.95 & 1.85 & 0.51 & 0.61 & 0.69 & 0.58 & 0.18 \\
T -- B & 1.11 & 0.00 & 0.15 & 0.28 & 0.08 & 0.00 & 0.24 & 0.20 & 0.06 \\
\bottomrule
\end{tabular}
\par\medskip
\begin{flushleft}
\footnotesize The first two numerical columns give the median corrected root-mean-square slopes of the two types at $50\um$, in units of $10^{-3}\um^{-1}$. The three normal-calibration multipliers $\lambda^*$ use image-specific corrected slopes at $50$ and $100\um$, or the uncorrected 90th percentile (q90) at $50\um$. The bias share is $s_k s_\ell\sum_m B_m/|\sum_m U_m|$ at the median corrected $50\um$ slopes $s_k,s_\ell$. The last three columns give the matching amplitudes for bumps of diameter $r$.
\end{flushleft}
\end{table}

\begin{table}[htbp]
\centering
\caption{Sensitivity of the fixed-image pooled multiplier $\lambda^*$ to the $50\um$ reference slopes.}\label{S-tab:sens}
\scriptsize
\setlength{\tabcolsep}{1.8pt}
\begin{tabular}{lrcrcrrrrrrc}
\toprule
 & & noise & & noise & relabel- & uncor- & one & & \multicolumn{2}{c}{zero slopes} & zero\\
Pair & normal & estimates & Chebyshev & estimates & ing & rected & constant & maximal & noise & excluded & share\\
\midrule
\multicolumn{12}{l}{\emph{METABRIC, all cells}} \\
CD8~T -- epithelial & 1.12 & 1.05--1.14 & 0.98 & 0.9205--1.0004 & 1.17 & 0.91 & 0.38 & 0.24 & 1.09 & 1.11 & 19/4 \\
macrophage -- epithelial & 1.27 & 1.20--1.30 & 1.17 & 1.0977--1.1895 & 1.33 & 1.05 & 0.49 & 0.29 & 1.25 & 1.27 & 16/4 \\
\midrule
\multicolumn{12}{l}{\emph{Basel, all cells}} \\
macrophage -- epithelial & 1.13 & 1.07--1.16 & 1.01 & 0.9498--1.0266 & 1.18 & 0.95 & 0.45 & 0.22 & 1.12 & 1.13 & 7/0 \\
T -- epithelial & 1.25 & 1.20--1.26 & 1.18 & 1.1405--1.1945 & 1.29 & 1.12 & 0.70 & 0.29 & 1.24 & 1.25 & 4/0 \\
\midrule
\multicolumn{12}{l}{\emph{Basel, non-epithelial cells}} \\
stroma -- T & 0.95 & 0.88--1.00 & 0.84 & 0.7818--0.8846 & 1.06 & 0.78 & 0.64 & 0.16 & 0.95 & 0.95 & 2/4 \\
\midrule
\multicolumn{12}{l}{\emph{METABRIC, non-epithelial cells}} \\
CD4~T -- B & 0.58 & 0.53--0.60 & 0.00 & 0.0000--0.0000 & 0.65 & 0.44 & 0.33 & 0.11 & 0.55 & 0.65 & 29/55 \\
\bottomrule
\end{tabular}
\par\medskip
\begin{flushleft}
\footnotesize The first four numerical columns use corrected root-mean-square slopes as in Web Table~\ref{S-tab:ident}. Each range uses probabilities smoothed at 10, 15, 25 or $50\um$ to estimate the local label variance, plus the $25\um$ estimate with first-order correction $\hat v_i+\sum_j S_{ij}^2\hat v_j$. The remaining multipliers use normal calibration. In order, the alternatives estimate noise by one within-image relabeling, omit the correction, use the larger corrected slope for both types, or use each type's maximal uncorrected slope. The next two columns replace a zero slope by the square root of the estimated local noise in the mean square, or omit images with a zero slope for either type. The last column gives the percentages with a zero corrected slope for the first and second types before any omission. These choices use the observed labels and are descriptive.
\end{flushleft}
\end{table}

\begin{table}[htbp]
\centering
\caption{Interface diagnostics for the count-screened all-cell analyses.}\label{S-tab:interface}
\footnotesize
\setlength{\tabcolsep}{4pt}
\begin{tabular}{llrrrrrr}
\toprule
 & & \multicolumn{3}{c}{Screened full graph} & \multicolumn{3}{c}{Three-way split}\\
Cohort & Pair & $z$ & smooth $\ell=25$ & step & control & compartments & ratio\\
\midrule
METABRIC & CD8 T -- epithelial & $-85.4$ & $-10.0$ & $-19.7$ & $-57.0$ & $-50.0$ & 0.88 \\
METABRIC & CD8 T -- macrophage & $14.8$ & $0.7$ & $3.2$ & $9.1$ & $7.4$ & 0.82 \\
METABRIC & fibroblast -- CD8 T & $23.5$ & $0.7$ & $6.7$ & $15.2$ & $11.4$ & 0.75 \\
METABRIC & macrophage -- epithelial & $-112.8$ & $-12.2$ & $-21.1$ & $-74.8$ & $-67.5$ & 0.90 \\
Basel & macrophage -- epithelial & $-94.5$ & $-10.2$ & $-17.9$ & $-62.4$ & $-56.8$ & 0.91 \\
Basel & stroma -- T & $19.1$ & $-0.4$ & $10.8$ & $11.2$ & $6.0$ & 0.54 \\
Basel & T -- epithelial & $-137.8$ & $-16.2$ & $-29.1$ & $-93.8$ & $-84.0$ & 0.90 \\
Basel & T -- macrophage & $20.1$ & $1.0$ & $5.2$ & $15.9$ & $13.8$ & 0.87 \\
\bottomrule
\end{tabular}
\par\medskip
\begin{flushleft}
\footnotesize The full graph uses the original five splits and requires at least 20 observed cells of each target type in the image. Its $z$ uses the pooled plug-in variance. The smooth and step columns give pooled null means, each standardized by its own null standard deviation. The smooth composition uses Gaussian smoothing at $25\um$. For the step we split the full image at a $30\um$ smoothed epithelial fraction of $1/2$ and assign each compartment its observed type proportions. Table~\ref{tab:data} gives the primary fixed-image results.

\smallskip In the three-way analysis each cell is held out independently with probability $1/3$. We construct the $30\um$ map from this third alone and exclude it from all test pairs and their smoothers. Both analyses require at least 20 cells of each target type \emph{among the remaining cells}. The epithelial pairs with CD8 T and macrophages in METABRIC then use 265 and 357 images, respectively. Those with T cells and macrophages in Basel use 260 and 216. The matched control uses the same held-out cells, eligible images and five splits of the remaining cells. Both fits use Nadaraya--Watson smoothing with at least three cells per half-window. Only the compartment fit restricts smoothing neighbors to the target cell's compartment. The ratio is the compartment-aware pooled $z$ divided by the control pooled $z$, each standardized by its own pooled plug-in standard deviation.
\end{flushleft}
\end{table}

\FloatBarrier
\subsection{Count screening and multiple testing}\label{S-sec:screening}

Web Table~\ref{S-tab:screenholm} compares the fixed image sets with an earlier analysis that required at least 20 observed cells of each target type. Applying the screen selects images using the labels entering $U_m$, so its $z$ and breakdown values describe sensitivity to selection. Reapplying the screen to the 9,789 fixed image--pair records recovered all 4,250 screened records. The signed statistics, plug-in standard deviations and geometry-based bias and variance coefficients agreed for every shared record. Figure~\ref{fig:application}(c) gives the fixed-set critical curve for METABRIC CD8 T--epithelial. Web Table~\ref{S-tab:screenholm} gives the breakdown slope under each selection rule.

For the fixed-image tests we specify a common Lipschitz bound $M_0=0.001\um^{-1}$, a sensitivity point allowing each type's probability to change by at most 0.02 across $20\um$. For each image let $B_m$ be the smaller of the product and variogram bias coefficients in Section~\ref{sec:honest}. Write $U_{\rm pool}=\sum_m U_m$, $b(M_0)=M_0^2\sum_m B_m$ and $\bar V_{\rm pool}(M_0)=\sum_m\bar V_m(M_0)$. With $e=|U_{\rm pool}|-b(M_0)$, inversion of the two-sided Chebyshev test in Theorem~\ref{thm:honest} gives the raw $p$-value
\[
  p_{\rm Ch}(M_0)=
  \begin{cases}
     \min\{1,\bar V_{\rm pool}(M_0)/e^2\}, & e>0,\\
     1, & e\le0 .
  \end{cases}
\]
If every $Q_m=0$, then $U_{\rm pool}=0$ and the second case gives $p_{\rm Ch}=1$. All 18 reported pools have positive variance bounds.

The primary family comprises all 18 combinations of cohort, cell population and type pair. Under the joint independent-labeling null, all-cell tests condition on the design. Non-epithelial tests also condition on the epithelial membership mask. If their conditional type probabilities satisfy the bound $M_0$ almost surely, averaging over masks gives marginally valid $p$-values. Holm's procedure controls the familywise error rate under arbitrary dependence among tests \citep{holm1979}. Strong control when some nulls are false requires marginal validity for each true-null $p$-value in that configuration. For a given pair it suffices that the collapsed labels $k$, $\ell$ and other be independent across cells and cores under that population's conditioning, with both target probabilities satisfying the bound.

At the 5\% familywise level only Basel T--epithelial rejects across the 18 outcomes. Basel non-epithelial stroma--T has the third-smallest raw $p$-value (0.00914) and enters the ordering with the epithelial--immune outcomes. All four adjusted $p$-values are below 0.05 when Holm is applied within the epithelial--immune subgroup. A familywise interpretation of this column requires choosing the four outcomes independently of the results. We have no documented prespecification of this subgroup and report the four-outcome calculation as an illustrative sensitivity analysis of the choice of family. Both adjustments apply across outcomes at $M_0$ only, with no multiplicity adjustment over $M$.

\begin{table}[htbp]
\centering
\caption{Effect of the 20-cell screen and pointwise Chebyshev $p$-values on the fixed image sets at $M_0=0.001\um^{-1}$.}\label{S-tab:screenholm}
\footnotesize
\setlength{\tabcolsep}{3.4pt}
\begin{tabular}{lrrrrrrrr}
\toprule
 & & \multicolumn{2}{c}{Pooled $z$} & \multicolumn{2}{c}{Breakdown $\hat M^*$} & \multicolumn{3}{c}{Fixed-image $p$-value}\\
\cmidrule(lr){3-4}\cmidrule(lr){5-6}\cmidrule(lr){7-9}
Pair & $n_{20}$ & Screened & Fixed & Screened & Fixed & Raw & Holm (4) & Holm (18)\\
\midrule
\multicolumn{9}{l}{\emph{METABRIC, all cells, 724 fixed images}} \\
CD8~T -- macrophage & 292 & $14.85$ & $16.21$ & -- & -- & 1 & \textemdash & 1 \\
fibroblast -- CD8~T & 347 & $23.45$ & $27.63$ & 0.43 & 0.30 & 1 & \textemdash & 1 \\
CD4~T -- B & 140 & $-7.40$ & $-6.35$ & -- & -- & 1 & \textemdash & 1 \\
CD8~T -- B & 139 & $-1.50$ & $-0.20$ & -- & -- & 1 & \textemdash & 1 \\
CD8~T -- epithelial & 334 & $-85.37$ & $-90.29$ & 1.37 & 1.03 & 0.03902 & 0.03902 & 0.54625 \\
macrophage -- epithelial & 414 & $-112.75$ & $-116.44$ & 1.60 & 1.36 & 0.00612 & 0.01835 & 0.10401 \\
\midrule
\multicolumn{9}{l}{\emph{METABRIC, non-epithelial cells, 698 fixed images}} \\
CD8~T -- macrophage & 292 & $0.07$ & $0.28$ & -- & -- & 1 & \textemdash & 1 \\
fibroblast -- CD8~T & 347 & $-4.00$ & $-3.85$ & -- & -- & 1 & \textemdash & 1 \\
CD4~T -- B & 140 & $-13.31$ & $-12.95$ & 0.31 & -- & 1 & \textemdash & 1 \\
CD8~T -- B & 139 & $-8.45$ & $-7.81$ & -- & -- & 1 & \textemdash & 1 \\
\midrule
\multicolumn{9}{l}{\emph{Basel, all cells, 332 fixed images}} \\
macrophage -- epithelial & 250 & $-94.49$ & $-95.27$ & 1.29 & 1.17 & 0.01690 & 0.03380 & 0.25351 \\
stroma -- T & 284 & $19.05$ & $19.78$ & 0.38 & 0.37 & 1 & \textemdash & 1 \\
T -- macrophage & 236 & $20.06$ & $20.27$ & -- & -- & 1 & \textemdash & 1 \\
T -- B & 46 & $0.72$ & $3.42$ & -- & -- & 1 & \textemdash & 1 \\
T -- epithelial & 284 & $-137.81$ & $-138.58$ & 1.97 & 1.91 & 0.00128 & 0.00514 & 0.02312 \\
\midrule
\multicolumn{9}{l}{\emph{Basel, non-epithelial cells, 331 fixed images}} \\
stroma -- T & 284 & $-32.23$ & $-32.39$ & 1.50 & 1.48 & 0.00914 & \textemdash & 0.14624 \\
T -- macrophage & 236 & $-4.42$ & $-4.39$ & -- & -- & 1 & \textemdash & 1 \\
T -- B & 46 & $-9.79$ & $-8.19$ & -- & -- & 1 & \textemdash & 1 \\
\bottomrule
\end{tabular}
\par\medskip
\begin{flushleft}
\footnotesize Group headings give the fixed image counts. $n_{20}$ is the number meeting the pair-specific screen. Both $z$ columns use pooled plug-in standard deviations. Breakdown slopes use the unadjusted two-sided 5\% Chebyshev cutoff and are in units of $10^{-3}\um^{-1}$. A dash in a breakdown column indicates that the test did not reject even at $M=0$. An em dash under Holm (4) indicates an outcome outside the four-outcome subgroup.
\end{flushleft}
\end{table}

\FloatBarrier

\subsection{Comparing groups of patients}

Let $G_1,G_2$ be disjoint groups defined by patient membership, with image sets fixed for the labeling analysis. Write $w_g$ for the sum of $Q_{m,uv}$ over images $m\in G_g$ and ordered pairs of distinct cells at distance at most $r$, and assume $w_g>0$. By \eqref{eq:powerid} and (A1), the mean of $\sum_{m\in G_g}U_m/w_g$ equals the sum of $Q_{m,uv}\Cov_P(x_{m,uk},x_{m,v\ell})/w_g$ over these close pairs, up to a composition bias of absolute value at most $M^2\sum_{m\in G_g}B_m/w_g$. Here $x_{m,uk}$ denotes the type-$k$ indicator for cell $u$ in image $m$. The covariance sum is a weighted average when the close-pair weights are nonnegative, as for $\rho\ge2r$ with nonnegative smoothers (Section~\ref{sec:upper}).

Under independent labeling across cells and images in both groups, the contrast $D=\sum_{m\in G_1}U_m/w_1-\sum_{m\in G_2}U_m/w_2$ satisfies
\[
 |ED|\le M^2\sum_{g=1}^2 w_g^{-1}\sum_{m\in G_g}B_m,
 \qquad
 \Var D=\sum_{g=1}^2 w_g^{-2}\sum_{m\in G_g}\Var(U_m)
       \le \sum_{g=1}^2 w_g^{-2}\sum_{m\in G_g}\bar V_m(M).
\]
The exact null variances are those of Theorem~\ref{thm:moments}. These bounds give the honest test and breakdown slope for the joint null of independent labeling in the two groups. Testing whether two groups with interaction differ in its strength requires the variance of $D$ under the alternative, for example from a bootstrap over patients. Patient-level statistics can also enter survival models as covariates, where their null variance (Theorem~\ref{thm:moments}) indicates how much of their spread across patients is sampling noise. A difference between groups in composition alone can still produce a difference in these statistics, so the floor applies to comparisons as much as to single tests.

\section{Formal verification in Lean~4}\label{S-sec:lean}

The finite-sample results were formalized in Lean~4 (version 4.19.0) with mathlib commit \texttt{c44e0c8}. The project builds with warnings treated as errors and contains no placeholder proofs. A verification script rebuilds it from a fresh copy and checks that each of the 145 declarations listed in the project README depends only on the axioms \texttt{propext}, \texttt{Classical.choice} and \texttt{Quot.sound}. The table names those that state the results of the paper.

{\small\begin{longtable}{>{\raggedright\arraybackslash}p{0.30\linewidth}>{\raggedright\arraybackslash}p{0.64\linewidth}}
\toprule
Result & Lean declarations\\
\midrule
\endhead
Lemma~\ref{S-lem:zerodiag} & \texttt{crossFit\_diag\_eq\_zero}, \texttt{crossFitQ\_diag\_eq\_zero}, \texttt{crossFitQ\_isSymm}, \texttt{crossFitQ\_row\_sum}, \texttt{crossFitQ\_col\_sum}\\
Averaging over splits & \texttt{splitAvg\_crossFitQ\_diag\_eq\_zero}, \texttt{splitAvg\_crossFitQ\_transpose}, \texttt{splitAvg\_crossFitQ\_moments}\\
Theorem~\ref{thm:moments}, one type & \texttt{quadForm\_integral}, \texttt{quadForm\_variance}\\
Theorem~\ref{thm:moments}, two types & \texttt{crossForm\_integral}, \texttt{crossForm\_variance}, \texttt{crossForm\_variance\_symm}, \texttt{integral\_centered\_mul\_of\_disjoint\_indicators}\\
Null mean as a bias product (one type) & \texttt{crossFit\_null\_mean}, \texttt{dotProduct\_crossFitQ\_mulVec}\\
Residual products for general smoothers & \texttt{integral\_residual\_mul}, \texttt{integral\_residual\_mul\_of\_disjoint}, \texttt{integral\_residual\_mul\_crossFit}, \texttt{integral\_residual\_mul\_leaveOneOut}\\
Proposition~\ref{S-prop:disjoint} (edge argument) & \texttt{coeff\_eq\_zero\_of\_unbiased}, \texttt{prod\_edgeComp\_affine}, \texttt{smoother\_weights\_mul\_eq\_zero\_of\_unbiased}\\
Lemma~\ref{S-lem:residual} and variogram bounds & \texttt{abs\_smoother\_residual\_le}, \texttt{abs\_smoother\_residual\_le\_of\_affine}, \texttt{abs\_bilinear\_le}, \texttt{bilinear\_eq\_variogram}, \texttt{quadForm\_eq\_variogram}, \texttt{abs\_crossFitQ\_bilinear\_le\_variogram}, \texttt{crossFitQ\_quadForm\_le\_variogram}; symmetry cannot be dropped: \texttt{bilinear\_eq\_variogram\_false\_without\_symm}\\
Theorem~\ref{thm:honest} & \texttt{honest\_test}, \texttt{honest\_test\_quadForm} for $\bar V>0$. The counterexample \texttt{honest\_test\_false\_without\_pos\_Vbar} concerns the unmodified threshold rule at $\bar V=0$.\\
Theorem~\ref{thm:floor}: level of a mixture of null distributions & \texttt{mixture\_rejection\_le}, \texttt{power\_le\_of\_eq\_mixture}, \texttt{mixture\_lintegral\_le}, \texttt{mixture\_measure\_le}\\
Theorem~\ref{thm:floor}: covariance of the bump mixture & \texttt{sum\_sign\_mul\_sign}, \texttt{bump\_mean}, \texttt{bump\_cov}, \texttt{mixBern\_cov}\\
Theorem~\ref{thm:floor}: $r$-dependence of the bump mixture & \texttt{mixBern\_indep\_of\_disjoint\_supports}, \texttt{mixBern\_indep\_of\_disjoint\_supports\_uniform} (the geometric fact that cells more than $r$ apart share no bump enters as a hypothesis)\\
Theorem~\ref{thm:param} & \texttt{sum\_sq\_div\_null\_eq}, \texttt{sum\_sq\_div\_null\_le}, \texttt{prod\_one\_add\_le\_exp\_sum}, \texttt{sum\_abs\_sub\_le\_sqrt\_chiSq}, \texttt{abs\_sum\_mul\_sub\_le\_half\_sum\_abs}, \texttt{sum\_abs\_mixture\_sub\_null\_le\_exp}, \texttt{sum\_exp\_sign\_eq\_prod\_cosh}, \texttt{cosh\_le\_exp\_half\_sq}, \texttt{sum\_exp\_sign\_le}\\
Proposition~\ref{S-prop:mixing} & \texttt{cov\_eq\_cov\_condExp}, \texttt{offDiag\_ge\_neg\_diag\_of\_psd}, \texttt{offDiag\_ge\_neg\_diag\_of\_psd\_weights}, \texttt{sum\_offDiag\_cov\_ge\_neg\_sum\_var}, \texttt{mixBern\_sum\_offDiag\_cov\_ge}\\
Chebyshev step of Theorem~\ref{S-thm:power} & \texttt{meas\_lt\_le\_of\_variance\_div\_le}, \texttt{one\_sub\_le\_meas\_reject}\\
Equation~\eqref{S-eq:tiltmoments}: $E_pT$, $\Var_pT$ and $\Cov_p(U,T)$ & \texttt{tilt\_integral}, \texttt{crossForm\_cov\_tilt}, \texttt{crossForm\_cov\_tilt\_symm}, \texttt{tilt\_variance}, \texttt{quadForm\_cov\_tilt}\\
Proposition~\ref{S-prop:homog}, and the single-cell variance bound of Proposition~\ref{S-prop:mixing} for $\beta\le1$ & \texttt{var\_le\_of\_indep\_of\_abs\_sub\_le}, \texttt{var\_le\_of\_cov\_eq\_zero} (part (ii)), \texttt{cov\_sq\_le\_var\_mul\_var}, \texttt{integral\_mul\_sub\_sq}, \texttt{cov\_le\_of\_cov\_eq\_zero}, \texttt{mixture\_cov\_le} (independence and zero covariance from $r$-dependence, and the Hölder bounds, enter as hypotheses)\\
Proposition~\ref{S-prop:pairsign}: covariances & \texttt{pairInc\_cov}, \texttt{pairSgn\_cov}, \texttt{pairInc\_label\_cov}, \texttt{pairSgn\_label\_cov}, \texttt{mixBern\_bump\_mean}, \texttt{mixBern\_bump\_cov}, \texttt{sum\_pairInc\_mul\_pairInc}, \texttt{sum\_pairSgn\_mul\_pairSgn}, \texttt{sum\_joins\_eq}, \texttt{endpoints\_iff\_joins}, \texttt{sum\_uniform\_weight\_eq\_one} (no pair is listed twice)\\
Proposition~\ref{S-prop:pairsign}: values at the cells & \texttt{pairSign\_abs\_sub\_le}, \texttt{sum\_abs\_pairInc\_sub}, \texttt{sum\_abs\_pairSgn\_sub}, \texttt{pairSign\_abs\_le}, \texttt{sum\_pairInc\_eq\_pairDeg}, \texttt{abs\_profile\_eq\_pairInc}\\
Proposition~\ref{S-prop:pairsign}: $r$-dependence & \texttt{pairSign\_indep} (that no close pair joins two cells more than $r$ apart enters as a hypothesis)\\
\bottomrule
\end{longtable}}

Not formalized: the smoothness of the bump field (Lemma~\ref{S-lem:bumps}); in Proposition~\ref{S-prop:homog}, the range bound $\Var_\pi p_i\le(\tfrac12-\eta_0)^2$; in Proposition~\ref{S-prop:pairsign}, the passage from the bounds on the values at the cells to the Hölder condition through $L_\pm$, McShane's extension, the bound on $L_\pm$ under a hard core, the case $\beta\in(1,2]$ and the families of sets after the proposition; Theorem~\ref{S-thm:ising}; in Theorem~\ref{thm:param}, the reduction of the bump mixture to independent signs that links \texttt{sum\_sq\_div\_null\_le} to \texttt{sum\_exp\_sign\_le}; Corollary~\ref{cor:rate} and Proposition~\ref{prop:adapt}; the single-cell variance bound of Proposition~\ref{S-prop:mixing} for $\beta\in(1,2]$ and its Poisson identity; the several-types construction; the affine annihilation $Qz=0$ of Lemma~\ref{S-lem:zerodiag}; the cross-type bias product and the cross-type form of $\bar V$; the identity \eqref{eq:powerid} and the variance count of Theorem~\ref{S-thm:power}; Proposition~\ref{prop:breakdown}; the central limit theorem; and the asymptotic part of Theorem~\ref{S-thm:local} (the joint limit of the statistic and the tilt statistic, the limit of the log-likelihood ratio and Le Cam's lemmas). \texttt{crossFitQ\_row\_sum} assumes that every smoother row sums to one. In the code the rows of cells without support are empty, and $Q\mathbf 1=0$ still holds because $W$ has no edges at those cells.

\section{Code}\label{S-sec:code}

The code archive contains the core methods, benchmark and Lean project in \texttt{colocalization}. The fixed-image application scripts and results are in \texttt{manuscript/analysis}. The original figure, table and additional benchmark scripts are in \texttt{manuscript/scripts}.
{\raggedright
\begin{itemize}
\item Statistic, designs, exact moments and bias bounds: \texttt{colocal\_core.py}.
\item Simulations of Section~\ref{sec:floorsim} and Web Appendix~\ref{S-sec:window}: \texttt{minimax\_sims.py} and \texttt{window\_tradeoff.py}, with \texttt{make\_figures\_v2.py} for Web Figure~\ref{S-fig:window} and \texttt{manuscript/scripts/fig\_floor.py} for Figure~\ref{fig:floor}.
\item Benchmark: the folder \texttt{benchmarks} for Table~\ref{tab:bench}A, \texttt{manuscript/scripts/stronger\_alternative.py} for its last column, \texttt{manuscript/scripts/honest\_benchmark.py} for Table~\ref{tab:bench}B, and \texttt{local\_power\_check.py}, \texttt{second\_order\_check.py} and \texttt{tilt\_exact\_check.py} for Section~\ref{sec:upper}.
\item In \texttt{manuscript/analysis}, \texttt{run\_unfiltered.py} uses \texttt{colocalization/honest\_analysis.py} to compute the per-image quantities for every eligible pair. Using these files and the archived inputs, \texttt{prepare\_primary.py} writes the pooled, slope, single-image and geometry summaries to \texttt{manuscript/analysis/results}. It also generates the rows of Tables~\ref{tab:data} and~\ref{tab:cert} and Web Tables~\ref{S-tab:ident} and~\ref{S-tab:sens}. The application figure is produced by \texttt{fig\_application\_fixed.py} (Figure~\ref{fig:application}).
\item In the same directory, \texttt{compare\_filters.py} compares the fixed image sets with the earlier count screen. From this comparison \texttt{fixed\_inference.py} calculates the fixed-$M$ Chebyshev and Holm $p$-values and writes the rows of Web Table~\ref{S-tab:screenholm}, four of which Table~\ref{tab:epi} repeats.
\item In \texttt{colocalization}, \texttt{step\_null.py}, \texttt{compartment\_analysis.py} and \texttt{interface\_summary.py} produce the screened full-graph and three-way diagnostics in Web Table~\ref{S-tab:interface}. The geometry diagnostics use \texttt{theta\_diagnostics.py}. The slope-noise calculations and their semi-synthetic checks use \texttt{slope\_noise.py}, \texttt{slope\_noise\_summary.py} and \texttt{slope\_noise\_check.py}. Figure~\ref{fig:schematic} uses \texttt{manuscript/scripts/fig\_schematic.py}.
\item Web Appendix~\ref{S-sec:identification}: \texttt{homog\_floor.py}.
\item Lean project: \texttt{colocalization/lean}, checked by \texttt{verify.py} (Web Appendix~\ref{S-sec:lean}).
\end{itemize}
}



\ReserveEmbeddedPDF{schematic-1}
\ReserveEmbeddedPDF{schematic-2}
\ReserveEmbeddedPDF{schematic-3}
\ReserveEmbeddedPDF{schematic-4}
\ReserveEmbeddedPDF{schematic-5}
\ReserveEmbeddedPDF{schematic-6}
\ReserveEmbeddedPDF{schematic-7}
\ReserveEmbeddedPDF{schematic-8}
\ReserveEmbeddedPDF{schematic-9}
\ReserveEmbeddedPDF{schematic-10}
\ReserveEmbeddedPDF{schematic-11}
\ReserveEmbeddedPDF{schematic-12}
\ReserveEmbeddedPDF{schematic-13}
\ReserveEmbeddedPDF{schematic-14}
\ReserveEmbeddedPDF{schematic-15}
\ReserveEmbeddedPDF{schematic-16}
\ReserveEmbeddedPDF{schematic-17}
\ReserveEmbeddedPDF{schematic-18}
\ReserveEmbeddedPDF{schematic-19}
\ReserveEmbeddedPDF{schematic-20}
\ReserveEmbeddedPDF{schematic-21}
\ReserveEmbeddedPDF{schematic-22}
\ReserveEmbeddedPDF{schematic-23}
\ReserveEmbeddedPDF{schematic-24}
\ReserveEmbeddedPDF{schematic-25}
\ReserveEmbeddedPDF{schematic-26}
\ReserveEmbeddedPDF{schematic-27}
\ReserveEmbeddedPDF{schematic-28}
\ReserveEmbeddedPDF{schematic-29}
\ReserveEmbeddedPDF{schematic-30}
\ReserveEmbeddedPDF{schematic-31}
\ReserveEmbeddedPDF{schematic-32}
\ReserveEmbeddedPDF{schematic-33}
\ReserveEmbeddedPDF{schematic-34}
\ReserveEmbeddedPDF{schematic-35}
\ReserveEmbeddedPDF{schematic-36}
\ReserveEmbeddedPDF{schematic-37}
\ReserveEmbeddedPDF{schematic-38}
\ReserveEmbeddedPDF{schematic-39}
\ReserveEmbeddedPDF{schematic-40}
\ReserveEmbeddedPDF{schematic-41}
\ReserveEmbeddedPDF{schematic-42}
\ReserveEmbeddedPDF{schematic-43}
\ReserveEmbeddedPDF{schematic-44}
\ReserveEmbeddedPDF{schematic-45}
\ReserveEmbeddedPDF{schematic-46}
\ReserveEmbeddedPDF{schematic-47}
\ReserveEmbeddedPDF{schematic-48}
\ReserveEmbeddedPDF{schematic-49}
\ReserveEmbeddedPDF{schematic-50}
\ReserveEmbeddedPDF{schematic-51}
\ReserveEmbeddedPDF{schematic-52}
\ReserveEmbeddedPDF{schematic-53}
\ReserveEmbeddedPDF{schematic-54}
\ReserveEmbeddedPDF{schematic-55}
\ReserveEmbeddedPDF{schematic-56}
\ReserveEmbeddedPDF{schematic-57}
\ReserveEmbeddedPDF{schematic-58}
\ReserveEmbeddedPDF{schematic-59}
\ReserveEmbeddedPDF{schematic-60}
\ReserveEmbeddedPDF{schematic-61}
\ReserveEmbeddedPDF{schematic-62}
\ReserveEmbeddedPDF{schematic-63}
\ReserveEmbeddedPDF{schematic-64}
\ReserveEmbeddedPDF{schematic-65}
\ReserveEmbeddedPDF{schematic-66}
\ReserveEmbeddedPDF{schematic-67}
\ReserveEmbeddedPDF{schematic-68}
\ReserveEmbeddedPDF{schematic-69}
\ReserveEmbeddedPDF{schematic-70}
\ReserveEmbeddedPDF{schematic-71}
\ReserveEmbeddedPDF{schematic-72}
\ReserveEmbeddedPDF{schematic-73}
\ReserveEmbeddedPDF{schematic-74}
\ReserveEmbeddedPDF{schematic-75}
\ReserveEmbeddedPDF{schematic-76}
\ReserveEmbeddedPDF{schematic-77}
\ReserveEmbeddedPDF{schematic-78}
\ReserveEmbeddedPDF{schematic-79}
\ReserveEmbeddedPDF{schematic-80}
\ReserveEmbeddedPDF{schematic-81}
\ReserveEmbeddedPDF{schematic-82}
\ReserveEmbeddedPDF{schematic-83}
\ReserveEmbeddedPDF{schematic-84}
\ReserveEmbeddedPDF{schematic-85}
\ReserveEmbeddedPDF{schematic-86}
\ReserveEmbeddedPDF{schematic-87}
\ReserveEmbeddedPDF{schematic-88}
\ReserveEmbeddedPDF{schematic-89}
\ReserveEmbeddedPDF{schematic-90}
\ReserveEmbeddedPDF{schematic-91}
\ReserveEmbeddedPDF{schematic-92}
\ReserveEmbeddedPDF{schematic-93}
\ReserveEmbeddedPDF{schematic-94}
\ReserveEmbeddedPDF{schematic-95}
\ReserveEmbeddedPDF{schematic-96}
\ReserveEmbeddedPDF{schematic-97}
\ReserveEmbeddedPDF{schematic-98}
\ReserveEmbeddedPDF{schematic-99}
\ReserveEmbeddedPDF{schematic-100}
\ReserveEmbeddedPDF{schematic-101}
\ReserveEmbeddedPDF{schematic-102}
\ReserveEmbeddedPDF{schematic-103}
\ReserveEmbeddedPDF{schematic-104}
\ReserveEmbeddedPDF{schematic-105}
\ReserveEmbeddedPDF{schematic-106}
\ReserveEmbeddedPDF{schematic-107}
\ReserveEmbeddedPDF{schematic-108}
\ReserveEmbeddedPDF{schematic-109}
\ReserveEmbeddedPDF{schematic-110}
\ReserveEmbeddedPDF{schematic-111}
\ReserveEmbeddedPDF{schematic-112}
\ReserveEmbeddedPDF{schematic-113}
\ReserveEmbeddedPDF{schematic-114}
\ReserveEmbeddedPDF{schematic-115}
\ReserveEmbeddedPDF{schematic-116}
\ReserveEmbeddedPDF{schematic-117}
\ReserveEmbeddedPDF{schematic-118}
\ReserveEmbeddedPDF{schematic-119}
\ReserveEmbeddedPDF{schematic-120}
\ReserveEmbeddedPDF{schematic-121}
\ReserveEmbeddedPDF{schematic-122}
\ReserveEmbeddedPDF{schematic-123}
\ReserveEmbeddedPDF{schematic-124}
\ReserveEmbeddedPDF{schematic-125}
\ReserveEmbeddedPDF{schematic-126}
\ReserveEmbeddedPDF{schematic-127}
\ReserveEmbeddedPDF{schematic-128}
\ReserveEmbeddedPDF{schematic-129}
\ReserveEmbeddedPDF{schematic-130}
\ReserveEmbeddedPDF{schematic-131}
\ReserveEmbeddedPDF{schematic-132}
\ReserveEmbeddedPDF{schematic-133}
\ReserveEmbeddedPDF{schematic-134}

\EmbeddedPDFStream{schematic}{%
/Type /XObject /Subtype /Form /FormType 1 /BBox [ 0 0 506.177 227.184 ] /Resources
\EmbeddedPDFRef{schematic-1} 0 R /Filter [ /ASCIIHexDecode /FlateDecode ]
}{%
78dacd595b6f5b370c7ef7af10f6748ccd9a44dd5bf425585760408105cdba87b403dcc4b97876dcc6ee5aecd78f928e6e89
9c346b560c481cf73b144552e427f2949173c2c88b09bbe72fc74f4638e1d438317e9eac271f084abc20416cc2a87056914f
c4c3f1272d6314c22f2ee114a1f1b7fe7eb226adba0f6535f18f273ffebce6e4a7cde470725834cfbcdad9a81bf1c39b5b83
3394394980196a803fc884bc09384bad35f88539aab8f45a0e8ed022b0c451a74193a33392a492d0d129391ede30c5dee107
9fce8410c3d9f566edbfc9e17ab1bd3cfd385f6de383dd66fc7bb188cfdf5d2fe67f9e6ea660874f57f1d976b5793f9d21b0
98be3dfa65f2fca8f2b2dd3a06a3b23f0661564521078b51a982cbd499788e4a1a14fe74df53e91cd502ac0f2ea78030100c
211843a516bc42571ed59459a36a188c44ad203dc6284755bcc238fa649401cc34bf3cc31a5dd5161370d5c2b8abc32891b2
13d79a0a63c163a34d053aa91d28302a556820635642052b06948350b2daa8c6924d277e7986b307ab1a2ece969d4a507a51
45ad17adb3c92e8cf7413ac89c0c160f4b1ac22d50c6ff6dcae3628319c3ada096958c178618b482f194f18e69928442c697
c4c624c505ae5a703cc3000cdfc5279211854f4c281e81110485fb51e02a2aba1c15d43b62fae27645e659771365ccf0b9b3
09c7af98179d4d30a845014af92017a1378cb18e31e88d74a1c27b560caf46d4d616186a83019c99a89b77d64aca7dfa16a1
cfbddd87402bb7bdb4148fa417c87a0b41b986dac7a7d31907c9ba87c335e552343a971d8b2c558f7a36cbced968aaadf856
6703fffdd92c6f51792cbcaaee1a261f6b7b5615f76332395740a56042374c5ed086c90b5c5193e254e0035563122f49e1b4
b9c1e48014a478d29a616ea946da343593e3d928cdb4a9983c432d9367b865f20c57f49c37aab06c53cbe4d98386c98bb3d5
4e3928bda8de62f26cd79d4c2e3019047c25930b45359a9a99bca103ce31f9fd4e5128e4e66fddba72cca6d2464ea82e034f
9b3ae67a5cff6bb700308e6466f1106dcbc050491d0fba53a099521a8d858b6a87b8c0b862bae05dcde33ebf77d6a2bc6ba9
b2a7cb7fb335e7742f30eaf8c359f78b34f5b8168558c3833f4cc3dd3a9f821bfe9a0a362caee7e78bd3886e2a3422dbf7ab
cbdd761ff99454e9934fc9c747251f26a8e6d85337dc93c0967a125a9512f3876d748b2151836f78eba2331836216c4b3bda
51ab2c54a4834d9d91be8f4da624a0219c04367493c04215497d4192152dd124835b9ec9ae553c9322d089df4d9649e6dcc5
310e8fc8c9afa3180714700ae9f68a022db778678e322167179fe727bb988f573e433fae56f15febcd7a71e5d375b77dd2ef
22040ccf7116c2a9a8cf520647a81e4b718eae1817cacdde4d53f65bd1940d34f5e82cf5ae631b260840a38b776d7dda632f
878521bf6487b65584fe0e3d5673610868580d8562b703e02f31c00c1f43358fb9f2feda8fce6c38fd38e5c30976b1c39852
9bb391ead62880d9b5d95d5c5e9d472c0f298936af0253aae1f5d4ba617ebdb771ec671b677a18877cd411d3ba4fada544ba
cc5aaaf0318955496a9c8ae36266b30c36c49ad1c21e0a8fd4fa5eb382800a26d58dd11c308db9861b9339208748a69bc11c
bb29507e14ce2c9f90b6994b68dbcb25b41aa9d31e15948c69f8351bdef06bf6b06c92e3d089e2ad262e197417bf4a468dfa
ca164e029eaeda3f8b1b875d6794091572b1b95a6cc76ad8e1b72753244e8ec9bf5c44de55c3e55967c88a63faf2ae02b821
ef0b6089da8dc23a00b567f253c3cbdedcc525163ce60776a0e2bec16b56491d7494692afc8bbace806d9a29d2059ec942df
f75f56483bfcdd598ef3b010beb964650a05dd51e169d0799bb3e0eb2e7f80d0fb67d5977b28c754436613034c7de13b9912
a93ec93feb5e5ed073984756af543e4dcccc28372a30b31a1fa5579748cae1d5a5e7e0f0ead2f79f61993654f857811897f1
add11fbde809e48b18bd51ea65e712c6ab7a7416f21b01d1f141e2a866a20ba3e0580eabf9f5b92f937dafabfaf1971a2f9a
8bf9585ddbdd65ea5d6275ed6bad4b0977f9bfb044c5ff5ae294ea199d61d43c3b086b03cd9bf026903309615eaf60a5a834
4c86d1fad5a806d7860b82f9b42cd453af32942341eb918a8ec9db20718a462ef16257d63908c8daaf0660be86184352c47d
b0bd33d6d9f07c959fcf8ac005392087847ca9673819838d837185324db5c29a7a905fcc5261fe0f4e69df503a2e5aa730ec
5628211ee49472542990e25bb8853f57e1f3beff27fa07c1f18638
>}

\EmbeddedPDFObject{schematic-1}{%
<< /ColorSpace \EmbeddedPDFRef{schematic-2} 0 R /Pattern \EmbeddedPDFRef{schematic-3} 0 R /ExtGState
\EmbeddedPDFRef{schematic-4} 0 R /Font << /F28 \EmbeddedPDFRef{schematic-5} 0 R /F37
\EmbeddedPDFRef{schematic-6} 0 R /F39 \EmbeddedPDFRef{schematic-7} 0 R /F40
\EmbeddedPDFRef{schematic-8} 0 R /F41 \EmbeddedPDFRef{schematic-9} 0 R /F38
\EmbeddedPDFRef{schematic-10} 0 R /F43 \EmbeddedPDFRef{schematic-11} 0 R /F42
\EmbeddedPDFRef{schematic-12} 0 R >> /XObject << /Fm1 \EmbeddedPDFRef{schematic-13} 0 R >> /ProcSet
[ /PDF /Text ] >>
}

\EmbeddedPDFObject{schematic-2}{%
<< /pgfprgb [ /Pattern /DeviceRGB ] >>
}

\EmbeddedPDFObject{schematic-3}{%
<<  >>
}

\EmbeddedPDFObject{schematic-4}{%
<<  >>
}

\EmbeddedPDFObject{schematic-5}{%
<< /Type /Font /Subtype /Type1 /BaseFont /LRRLUO+CMR10 /FontDescriptor \EmbeddedPDFRef{schematic-14}
0 R /FirstChar 40 /LastChar 119 /Widths \EmbeddedPDFRef{schematic-15} 0 R /ToUnicode
\EmbeddedPDFRef{schematic-16} 0 R >>
}

\EmbeddedPDFObject{schematic-6}{%
<< /Type /Font /Subtype /Type1 /BaseFont /ZZRNUB+CMR8 /FontDescriptor \EmbeddedPDFRef{schematic-17}
0 R /FirstChar 22 /LastChar 120 /Widths \EmbeddedPDFRef{schematic-18} 0 R /ToUnicode
\EmbeddedPDFRef{schematic-19} 0 R >>
}

\EmbeddedPDFObject{schematic-7}{%
<< /Type /Font /Subtype /Type1 /BaseFont /WGLDFS+CMMI8 /FontDescriptor \EmbeddedPDFRef{schematic-20}
0 R /FirstChar 34 /LastChar 122 /Widths \EmbeddedPDFRef{schematic-21} 0 R /ToUnicode
\EmbeddedPDFRef{schematic-22} 0 R >>
}

\EmbeddedPDFObject{schematic-8}{%
<< /Type /Font /Subtype /Type1 /BaseFont /TBPOGY+CMMI6 /FontDescriptor \EmbeddedPDFRef{schematic-23}
0 R /FirstChar 59 /LastChar 106 /Widths \EmbeddedPDFRef{schematic-24} 0 R /ToUnicode
\EmbeddedPDFRef{schematic-25} 0 R >>
}

\EmbeddedPDFObject{schematic-9}{%
<< /Type /Font /Subtype /Type1 /BaseFont /MANPGY+CMSY8 /FontDescriptor \EmbeddedPDFRef{schematic-26}
0 R /FirstChar 0 /LastChar 106 /Widths \EmbeddedPDFRef{schematic-27} 0 R /ToUnicode
\EmbeddedPDFRef{schematic-28} 0 R >>
}

\EmbeddedPDFObject{schematic-10}{%
<< /Type /Font /Subtype /Type1 /BaseFont /ZUELSB+CMR6 /FontDescriptor \EmbeddedPDFRef{schematic-29}
0 R /FirstChar 49 /LastChar 61 /Widths \EmbeddedPDFRef{schematic-30} 0 R /ToUnicode
\EmbeddedPDFRef{schematic-31} 0 R >>
}

\EmbeddedPDFObject{schematic-11}{%
<< /Type /Font /Subtype /Type1 /BaseFont /CRPSRL+CMEX8 /FontDescriptor \EmbeddedPDFRef{schematic-32}
0 R /FirstChar 80 /LastChar 80 /Widths \EmbeddedPDFRef{schematic-33} 0 R /ToUnicode
\EmbeddedPDFRef{schematic-34} 0 R >>
}

\EmbeddedPDFObject{schematic-12}{%
<< /Type /Font /Subtype /Type1 /BaseFont /TGAAIC+CMSY6 /FontDescriptor \EmbeddedPDFRef{schematic-35}
0 R /FirstChar 3 /LastChar 54 /Widths \EmbeddedPDFRef{schematic-36} 0 R /ToUnicode
\EmbeddedPDFRef{schematic-37} 0 R >>
}

\EmbeddedPDFStream{schematic-13}{%
/Type /XObject /Subtype /Form /BBox [ 0 0 267.341 223.199 ] /FormType 1 /Matrix [ 1 0 0 1 0 0 ]
/Resources \EmbeddedPDFRef{schematic-38} 0 R /Filter [ /ASCIIHexDecode /FlateDecode ]
}{%
78da2be4325430004208999ccba5ef996ba8e092cf15c805004a470586
>}

\EmbeddedPDFObject{schematic-14}{%
<< /Type /FontDescriptor /FontName /LRRLUO+CMR10 /Flags 4 /FontBBox [ -40 -250 1009 750 ] /Ascent
694 /CapHeight 683 /Descent -194 /ItalicAngle 0 /StemV 69 /XHeight 431 /CharSet
<2f612f622f642f652f662f682f692f6b2f6c2f6d2f6e2f6f2f702f706172656e6c6566742f706172656e72696768742f722f732f742f752f77>
/FontFile \EmbeddedPDFRef{schematic-39} 0 R >>
}

\EmbeddedPDFObject{schematic-15}{%
[ 388.9 388.9 500 777.8 277.8 333.3 277.8 500 500 500 500 500 500 500 500 500 500 500 277.8 277.8
277.8 777.8 472.2 472.2 777.8 750 708.3 722.2 763.9 680.6 652.8 784.7 750 361.1 513.9 777.8 625
916.7 750 777.8 680.6 777.8 736.1 555.6 722.2 750 750 1027.8 750 750 611.1 277.8 500 277.8 500 277.8
277.8 500 555.6 444.4 555.6 444.4 305.6 500 555.6 277.8 305.6 527.8 277.8 833.3 555.6 500 555.6
527.8 391.7 394.4 388.9 555.6 527.8 722.2 ]
}

\EmbeddedPDFStream{schematic-16}{%
/Filter [ /ASCIIHexDecode /FlateDecode ]
}{%
78da6d554d6fe23014bce757780f95da03c5764802558464e743e2b06d55aad55e2131dd4890a0100efdf7ebf12378d9f600
1a3f8f9f67ecc1dcfd785d4f54dd6dcd247ce4eccd9cba735f9949f673730ceeeef2ae3a1f4c3b3c1b539b7a9c3d3db1d7be
abd66660f7d92a5fb5cdf060c9abb6da9f6b33b2be2769f3d1b49e827dd8fdbbf93da90ebde093edb9d90f4d3be1e0be37c3
de72be9b66b6c66e6acc2df965fa53d3b54f4c3c72ce6da168ebac3bc0c329985e74b0e9a86cd7b4757f11c3b6901608c9ea
a61a2e23f75d1dec6160f1faf33498c3aadd75419ab2e99b9d3c0dfda753f8104c5ffadaf44dfbc1ee6f94d999f5f978dc1b
a8603c582e596d76b6a1f5febc391836fdcee095f2fe79344cbab1205555579bd37153997ed37e9820e57cc9d2b25c06a6ad
ff9b4b68c5763752134be5737c852a5a06a90c2d96310adc625bc064a8a930b785b0b438a282c5411a0b8b13e50a16076982
c92443232ed043a1875a5c77b1ba4605c97c5454fdd9f417ed3c5c6019476329790c2ca90e053c24ac8167840be08876e0c0
3161d7e7b2b35b0b9d5248d7934cc4313e18677ebcc038f7e30ce3e21ffec8296f6be009e74184380b010ff6241360495ce8
13ce03d73360e741668ee33cc81c5a055d83c2218b84ea0a784e6b1dc779e3b9e3d01d14f0229c3711c1bf205fa5e393a7cc
710ae24860f2e1f695826e1bfaa5a46bcc8113c2f01d524f054e483d4370423a230d3dd1250238938838113851417ea1212a
c9177ac69cf8d090e454073f217e8e3e09e97477a938e9c4792ae173a3a4cf8d0a7d6ed4cce746453e372af6b95189cf8d52
3e378ab2a2a0475dbc3b7eeef3a48adb3ca9f2364f9adfe6498baf79d2f26b9e74e8f3a4673e4f3af279d2b1cf934e7c9ef4
dce7492f7c9eb4f279d2dae749673e4f3af779d285cf932e7d9e32eef394099fa74cfa3c65b3eb9db95fbefba5e36dc23b7a
7df5aa73dfdb07d13db6eea1c313d7b4e6fa1e1fbb2356b98f7bc8c7bf0d8c5ecae02ff1f3a850
>}

\EmbeddedPDFObject{schematic-17}{%
<< /Type /FontDescriptor /FontName /ZZRNUB+CMR8 /Flags 4 /FontBBox [ -36 -250 1070 750 ] /Ascent 694
/CapHeight 683 /Descent -194 /ItalicAngle 0 /StemV 76 /XHeight 431 /CharSet
<2f562f612f622f632f63697263756d666c65782f636f6c6f6e2f636f6d6d612f642f652f657175616c2f662f672f682f692f6a2f6b2f6c2f6d2f6d6163726f6e2f6e2f6f2f702f706172656e6c6566742f706172656e72696768742f706c75732f722f732f73656d69636f6c6f6e2f742f752f762f772f78>
/FontFile \EmbeddedPDFRef{schematic-40} 0 R >>
}

\EmbeddedPDFObject{schematic-18}{%
[ 531.3 795.8 472.2 531.3 767.4 826.4 531.3 958.7 1076.8 826.4 295.1 295.1 531.3 885.4 531.3 885.4
826.4 295.1 413.2 413.2 531.3 826.4 295.1 354.2 295.1 531.3 531.3 531.3 531.3 531.3 531.3 531.3
531.3 531.3 531.3 531.3 295.1 295.1 295.1 826.4 501.7 501.7 826.4 795.8 752.1 767.4 811.1 722.6
693.1 833.5 795.8 382.6 545.5 825.4 663.6 972.9 795.8 826.4 722.6 826.4 781.6 590.3 767.4 795.8
795.8 1091 795.8 795.8 649.3 295.1 531.3 295.1 531.3 295.1 295.1 531.3 590.3 472.2 590.3 472.2 324.7
531.3 590.3 295.1 324.7 560.8 295.1 885.4 590.3 531.3 590.3 560.8 414.1 419.1 413.2 590.3 560.8
767.4 560.8 ]
}

\EmbeddedPDFStream{schematic-19}{%
/Filter [ /ASCIIHexDecode /FlateDecode ]
}{%
78da6d554d6fe23014bce757780f95da03c5764802558464e743e2b0dbaa54abbd4262ba912041211cfaefd7e34770d9f600
1a3f8f9f67ecc1dcfd78594f54dd6dcd247ce4ecd59cba735f9949f673730ceeeef2ae3a1f4c3bfc32a636f5387b7a622f7d
57adcdc0eeb355be6a9be1c192576db53fd766647d4fd2e6bd693d05fbb0fb37f367521dfaf9647b6ef643d34e38a86fcdb0
b7946f66992db1cf25e616fc36fda9e9da27261e39e7b650b475d61d60e0144c2f22d87494b56bdababf28615be80a846475
530d9791fbae0ef624b078fd711acc61d5eeba204dd9f4d54e9e86fec3e97b08a6cf7d6dfaa67d67f79f85d989f5f978dc1b
88603c582e596d76b69ff5fd6b73306cfa8dbb2be3ede368987463419aaaae36a7e3a632fda67d3741caf992a565b90c4c5b
ff3797d08aed6ea42696cae7f80a55b40c52195a2c6314b8c5b680c95053616e0b61697144058b8334161627ca152c0ed204
934986465ca087420fb5b8ee62758d0a92f9a8a8fabbe92fda79b8c0328ec652f21858521d0a78485803cf0817c011edc081
63c2aecf6567b7163aa590ae279988637c30cefc788171eec719c6c527fec8296f6be009e74184380b010ff6241360495ce8
13ce03d73360e741668ee33cc81c5a055d83c2218b84ea0a784e6b1dc779e3b9e3d01d14f0229c3711c1bf205fa5e393a7cc
710ae24860f2e1f695826e1bfaa5a46bcc8113c2f01d524f054e483d4370423a230d3dd1250238938838113851417ea1212a
c9177ac69cf8d090e454073f217e8e3e09e97477a938e9c4792ae173a3a4cf8d0a7d6ed4cce746453e372af6b95189cf8d52
3e378ab2a2a0475dbc3b7eeef3a48adb3ca9f2364f9adfe6498baf79d2f26b9e74e8f3a4673e4f3af279d2b1cf934e7c9ef4
dce7492f7c9eb4f279d2dae749673e4f3af779d285cf932e7d9e32eef394099fa74cfa3c65b3eb9db95fbefba5e36dc22b7a
7df4aa73dfdbf7d03db5eea1c313d7b4e6fa1a1fbb2356b98f7bc6c77f0c8c9ecbe01fc100a783
>}

\EmbeddedPDFObject{schematic-20}{%
<< /Type /FontDescriptor /FontName /WGLDFS+CMMI8 /Flags 4 /FontBBox [ -24 -250 1110 750 ] /Ascent
694 /CapHeight 683 /Descent -194 /ItalicAngle -14 /StemV 78 /XHeight 431 /CharSet
<2f422f452f4d2f532f552f562f572f622f636f6d6d612f657073696c6f6e2f782f7a> /FontFile
\EmbeddedPDFRef{schematic-41} 0 R >>
}

\EmbeddedPDFObject{schematic-21}{%
[ 490.6 632.1 882.1 544.1 388.9 692.4 1062.5 1062.5 1062.5 1062.5 295.1 295.1 531.3 531.3 531.3
531.3 531.3 531.3 531.3 531.3 531.3 531.3 531.3 531.3 295.1 295.1 826.4 531.3 826.4 531.3 559.7
795.8 801.4 757.3 871.7 778.7 672.4 827.9 872.8 460.7 580.4 896 722.6 1020.4 843.3 806.2 673.6 835.7
800.2 646.2 618.6 718.8 618.8 1002.4 873.9 615.8 720 413.2 413.2 413.2 1062.5 1062.5 434 564.4 454.5
460.2 546.7 492.9 510.4 505.6 612.3 361.7 429.7 553.2 317.1 939.8 644.7 513.5 534.8 474.4 479.5
491.3 383.7 615.2 517.4 762.5 598.1 525.2 494.2 ]
}

\EmbeddedPDFStream{schematic-22}{%
/Filter [ /ASCIIHexDecode /FlateDecode ]
}{%
78da6d544d6fe23010bde757780f95da03c51f24810a21d90991386c5b956ab557484c371249500887fefbf51b13ac563d10
4d9edfccbcf10b73f7eb753bd155b7b713f5c8d99b3d7797beb493ecf7ee14dddde55d79696c3b3c5b5bd96a3c3d3fb1d7be
2bb77660f7d926dfb4f5f0e0c89bb63c5e2a3bb27e2619fb51b781823eecfeddfe9d944d53cf27fb4b7d1cea76c2c17daf87
a3e3fc74cc1cc6be608c52fed8fe5c77ed13138f9c7307acdb2aeb1acc708ea6571d6c3a2a3bd46dd55fc5b03da44542b2aa
2e87eb1b3dcbc65d0692b79fe7c1369bf6d045cb259bbeb9c3f3d07f92c28768fad257b6afdb0f76ff45993bd95e4ea7a385
0ac6a3d58a55f6e00abad99f778d65d39f06bc51de3f4f96497a175e55d955f67cda95b6dfb51f365a72be62cba25845b6ad
be9d49ee53f687919b3a2e9fe3a174bc728041bc26c00807085413b107120048113ec5e400160034011907801a22f7800250
b8580a0fa40e9048970b424d0600e9d210403514da2a3038570066488fa90b9f41580c768c2e3249a0230123f58c040c8d2e
7a1186d398cb88309c5100e66138837463c2704e9d03f2309c418d8c87e1320940dd8673b73e5eef6c3e5e77f96fd75f9de1
6a81340e19527274e5d2e373c4bed2c2209ef9788dd85ba0a1c125514c75aef3512edda29054b320bcc0dc82f8dc50ecbdcd
114bdf9770e9fbe6b816e9fbe6a8237ddf8270dfd74dece2d4c7c4414d191bb296bc8e33f095374342bf323e86db2af3be52
9c7bbc40bcf638f1a98ea4fa33eeef0a7d63eaa524668f0b1f034f882304f0947a09050d29f5920adad2dc73705729d5e139
f0f9d54ee09a7b3db8672d825f5a06bfb40a7ee959f04bc7c12f9d04bf741afcd23af8a5f3e09791c12fb3087e65e2e6177d
59f425e19f8d2d74db19e5a5efdd3aa155456b020ba26eed6d9b9dba13b2e8476b705cba787b29a2ffd4de82f7
>}

\EmbeddedPDFObject{schematic-23}{%
<< /Type /FontDescriptor /FontName /TBPOGY+CMMI6 /Flags 4 /FontBBox [ 11 -250 1241 750 ] /Ascent 694
/CapHeight 683 /Descent -194 /ItalicAngle -14 /StemV 85 /XHeight 431 /CharSet
<2f636f6d6d612f692f6a2f736c617368> /FontFile \EmbeddedPDFRef{schematic-42} 0 R >>
}

\EmbeddedPDFObject{schematic-24}{%
[ 379.6 963 638.9 963 638.9 658.7 924.1 926.6 883.7 998.3 899.8 775 952.9 999.5 547.7 681.6 1025.7
846.3 1161.6 967.1 934.1 780 966.5 922.1 756.7 731.1 838.1 729.6 1150.9 1001.4 726.4 837.7 509.3
509.3 509.3 1222.2 1222.2 518.5 674.9 547.7 559.1 642.5 589 600.7 607.7 725.7 445.6 511.6 ]
}

\EmbeddedPDFStream{schematic-25}{%
/Filter [ /ASCIIHexDecode /FlateDecode ]
}{%
78da6d544d6fe23010bde757780f95da03c51f24810a21d90991386c5b956ab557484c371249500887fefbf51b13ac563d10
4d9edfccbcf10b73f7eb753bd155b7b713f5c8d99b3d7797beb493ecf7ee14dddde55d79696c3b3c5b5bd96a3c3d3fb1d7be
2bb77660f7d926dfb4f5f0e0c89bb63c5e2a3bb27e2619fb51b781823eecfeddfe9d944d532793fda53e0e753be1e0bed7c3
d1717e3a660e635f3046297f6c7faebbf6898947ceb903d66d95750d663847d3ab0e361d951deab6eaaf62d81ed222215955
97c3f58d9e65e32e03c9dbcff3609b4d7be8a2e5924ddfdce179e83f49e143347de92bdbd7ed07bbffa2cc9d6c2fa7d3d142
05e3d16ac52a7b7005ddeccfbbc6b2e94f03de28ef9f27cb24bd0bafaaec2a7b3eed4adbefda0f1b2d395fb16551ac22db56
dfce24f729fbc3c84d1d97cff1503a5e39c0205e1360840304aa89d8030900a4089f6272000b009a808c03400d917b400128
5c2c8507520748a4cb05a1260380746908a01a0a6d15189c2b0033a4c7d485cf202c063b46179924d09180917a4602864617
bd08c369cc654418ce2800f3309c41ba316138a7ce017918cea046c6c3709904a06ec3b95b1faf77361fafbbfcb7ebafce70
b5401a870c2939ba72e9f139625f696110cf7cbc46ec2dd0d0e09228a63ad7f928976e5148aa59105e606e417c6e28f6dee6
88a5ef4bb8f47d735c8bf47d73d491be6f41b8efeb267671ea63e2a0a68c0d594b5ec719f8ca9b21a15f191fc36d95795f29
ce3d5e205e7b9cf8544752fd19f77785be31f55212b3c7858f8127c41102784abd848286947a49056d69ee39b8ab94eaf01c
f8fc6a2770cdbd1edcb316c12f2d835f5a05bff42cf8a5e3e0974e825f3a0d7e691dfcd279f0cbc8e0975904bf3271f38bbe
2cfa92f0cfc616baed8cf2d2f76e9dd0aaa235810551b7f6b6cd4edd0959f4a335382e5dbcbd14d17fa2e482ed
>}

\EmbeddedPDFObject{schematic-26}{%
<< /Type /FontDescriptor /FontName /MANPGY+CMSY8 /Flags 4 /FontBBox [ -30 -955 1185 779 ] /Ascent
750 /CapHeight 683 /Descent -194 /ItalicAngle -14 /StemV 46 /XHeight 431 /CharSet
<2f6261722f67726561746572657175616c2f6d696e7573> /FontFile \EmbeddedPDFRef{schematic-43} 0 R >>
}

\EmbeddedPDFObject{schematic-27}{%
[ 826.4 295.1 826.4 531.3 826.4 531.3 826.4 826.4 826.4 826.4 826.4 826.4 826.4 1062.5 531.3 531.3
826.4 826.4 826.4 826.4 826.4 826.4 826.4 826.4 826.4 826.4 826.4 826.4 1062.5 1062.5 826.4 826.4
1062.5 1062.5 531.3 531.3 1062.5 1062.5 1062.5 826.4 1062.5 1062.5 649.3 649.3 1062.5 1062.5 1062.5
826.4 288.2 1062.5 708.3 708.3 944.5 944.5 0 0 590.3 590.3 708.3 531.3 767.4 767.4 826.4 826.4 649.3
849.5 694.7 562.6 821.7 560.8 758.3 631 904.2 585.5 720.1 807.4 730.7 1264.5 869.1 841.6 743.3 867.7
906.9 643.4 586.3 662.8 656.2 1054.6 756.4 705.8 763.6 708.3 708.3 708.3 708.3 708.3 649.3 649.3
472.2 472.2 472.2 472.2 531.3 531.3 413.2 413.2 295.1 ]
}

\EmbeddedPDFStream{schematic-28}{%
/Filter [ /ASCIIHexDecode /FlateDecode ]
}{%
78da6d554d6fdb3a10bceb57b08700e9c1353f24522e0c03a4640339f4034df0f0ae8ec4e409886543b60ff9f78fb36b9b6d
91438cd570b93b3b1c86779f7e3ece7cbf7f8e33f3458a5ff1b83f4f5d9c35dfb687e2eeaedd77e75d1c4fdf63ec637f5d3d
7e153fa77df7184fe2be79681fc6e1f439253f8cdddbb98fd7ac8f93427c1dc69c823ee2fe29fe3beb76c7f77af67c1ede4e
c33893c87d1a4e6f29e7a3659130f1072668cb3f713a0efbf1ab505fa49409588f7db3df61866331bff010f32bb39761eca7
0b19f10c6a85d2a21fbad3e58b7ebb5d12039b1fdf8fa7b87b185ff6c57229e6bfd2e2f134bd13c3cfc5fcc7d4c769185fc5
fd1fccd2cae3f970788b602164b15a893ebea48269f6efdb5d14f38f06bca53cbd1fa2d0f4ad9855b7efe3f1b0ede2b41d5f
63b1947225969bcdaa8863ffd79a76bce5f9e59a5ba75cd9a41fad17d5aa582a9d626508a82d80124045802d0158000e80f7
1b001e40e02d1a00eaa996b778006b001b021c004d0d5059ab8501400d2c030e00d52362aa4586410dc3353c78942ac5157a
4b99e2625921a3e20c8d1a164258c55b40cc62384bbd8d040f0b0ab662a006003dec02805b0370e0e41c0b842e0e8b0e6db5
d432011e80f759428fc93df3a059021a049b350da817eaac69408de0b3a60135c23a6b1a50af9159d3067335faa6693aeceb
a9d6ea7acadd7fdbe96208ad15ea48458205f0929ae296623e72457149f886623ef9062ca5e5bd9041b2240a4723e9a865db
2226ced51a63ca0dab449e600e2572141f89451dc52a9806f18273a0bb96ec0528ac15c7a8a335c7d8ab99434906aad92a94
bfe098f23dc794dff25e70d36b8e1788f9ac14b819ea2b0df61ae699dc9b62569a4c62582b19109389ab4016ad38a63aa495
3194cf334a70333ca3a47c36afc7ec86ef8f84810cdb5e11ce9c15f897ac89414ec5f91a56a8f8fe6acc52f1b5abc1d9f279
39f0b48e6370b3dcd7a1bebdd4414dcb7dc9c496fbb614af29bfa27cd6aac42ceee22b7070a4953730afbb6805ff38d64a83
832bf94ec04b8e7d4597dfd5ac03e507f618cec8111f6b29a7e518e7e2d61cd375dc700c3ef56f9ef192eaf00d55d9ff5e67
ff7b93fdefcbec7f5f65ff7b9bfdef5df6bfafb3ff83ccfe0f2afb3fe8ecff60b2ff4399fd1f42f67f68b2ff1b936769cadb
8c74cbe956e39f3b1ea2dbb3d19da729bd28f45ad14b81376218e3ed413bec0fd8457ff4125edf5d7cfdd814ff037389e951
>}

\EmbeddedPDFObject{schematic-29}{%
<< /Type /FontDescriptor /FontName /ZUELSB+CMR6 /Flags 4 /FontBBox [ -20 -250 1193 750 ] /Ascent 694
/CapHeight 683 /Descent -194 /ItalicAngle 0 /StemV 83 /XHeight 431 /CharSet
<2f657175616c2f6f6e652f74776f> /FontFile \EmbeddedPDFRef{schematic-44} 0 R >>
}

\EmbeddedPDFObject{schematic-30}{%
[ 611.1 611.1 611.1 611.1 611.1 611.1 611.1 611.1 611.1 351.8 351.8 351.8 935.2 ]
}

\EmbeddedPDFStream{schematic-31}{%
/Filter [ /ASCIIHexDecode /FlateDecode ]
}{%
78da6d554d6fe23014bce757780f95da03c5764802558464e743e2b0dbaa54abbd4262ba912041211cfaefd7e34770d9f600
1a3f8f9f67ecc1dcfd78594f54dd6dcd247ce4ecd59cba735f9949f673730ceeeef2ae3a1f4c3bfc32a636f5387b7a622f7d
57adcdc0eeb355be6a9be1c192576db53fd766647d4fd2e6bd693d05fbb0fb37f367521dfa78b23d37fba169271cd4b766d8
5bca37b3cc96d8e712730b7e9bfed474ed13138f9c735b28da3aeb0e30700aa617116c3acada356ddd5f94b02d740542b2ba
a986cbc87d57077b1258bcfe380de6b06a775d90a66cfa6a274f43ffe1f43d04d3e7be367dd3beb3fbcfc2ecc4fa7c3cee0d
44301e2c97ac363bdbcffafeb5391836fdc6dd95f1f671344cbab1204d55579bd37153997ed3be9b20e57cc9d2b25c06a6ad
ff9b4b68c5763752134be5737c852a5a06a90c2d96310adc625bc064a8a930b785b0b438a282c5411a0b8b13e50a16076982
c92443232ed043a1875a5c77b1ba4605c97c5454fdddf417ed3c5c6019476329790c2ca90e053c24ac8167840be08876e0c0
3161d7e7b2b35b0b9d5248d7934cc4313e18677ebcc038f7e30ce3e2137fe494b735f084f320429c8580077b9209b0242ef4
09e781eb19b0f32033c7711e640ead82ae41e190454275053ca7b58ee3bcf1dc71e80e0a7811ce9b88e05f90afd2f1c953e6
3805712430f970fb4a41b70dfd52d235e6c00961f80ea9a70227a49e2138219d91869ee812019c49449c089ca820bfd01095
e40b3d634e7c684872aa839f103f479f8474babb549c74e23c95f0b951d2e746853e376ae673a3229f1b15fbdca8c4e74629
9f1b455951d0a32ede1d3ff77952c56d9e54799b27cd6ff3a4c5d73c69f9354f3af479d2339f271df93ce9d8e749273e4f7a
eef3a4173e4f5af93c69edf3a4339f279dfb3ce9c2e749973e4f19f779ca84cf53267d9eb2d9f5cedc2fdffdd2f136e115bd
3e7ad5b9efed7be89e5af7d0e1896b5a737d8d8fdd11abdcc73de3e33f0646cf65f00f8786a779
>}

\EmbeddedPDFObject{schematic-32}{%
<< /Type /FontDescriptor /FontName /CRPSRL+CMEX8 /Flags 4 /FontBBox [ -27 -2955 1546 771 ] /Ascent
45 /CapHeight 0 /Descent -600 /ItalicAngle 0 /StemV 77 /XHeight 431 /CharSet
<2f73756d6d6174696f6e74657874> /FontFile \EmbeddedPDFRef{schematic-45} 0 R >>
}

\EmbeddedPDFObject{schematic-33}{%
[ 1121.5 ]
}

\EmbeddedPDFStream{schematic-34}{%
/Filter [ /ASCIIHexDecode /FlateDecode ]
}{%
78da6d934d6fe2301086eff915de03123da4381f405221a4101389c3b65541ab5ec11e6824e2444e222dff7e3d9e84aa2b0e
448f67de19bf930c935fef7b3f53f509fce899b30f68ebde48f0f3dfc7c69b4c442dfb0a74f70aa0408dd9f685bd9b5aeea1
63d37c2776baec9eac78a7e5b55730aa1e8b367029f5b704ef61d3037cfab282bf897feacb6b576a9fa3f6507657ab799466
36c67ec4982bf903a62d6bfdc28267ceb90d6cb5caeb0a6768bdd9e083cd4667e7522b33986127b4e6052153a5ec86937bca
cabe0c2cdedfda0eaa9d3ed7de6ac5661f36d976e6e61c3e79b337a3c094fac2a63f9cd9ccbe6f9a2ba00bc6bdf59a2938db
8676f6d763056cf668c0bbe4706b8085ee1c902b592b689ba30473d417f0569cafd9aa28d61e68f55f6ea8389d07a9585aa9
48ec8347d9dc558cb9e8ae955f47e3ad22d735d96ed69603c7c50239a4f81639a2788a1c1373e439b1ab5d90be405e523c43
4e8803e4943847ce8823e40db1bb2b270e9105b140de12c7c805dd853ee3c13f6ae2c17f823cf8c7bbe2c1ffd2728ab56118
a0b70c6b390f7116e1384ad18f089c86637f11521c7b8a8818e71531317a16737ad3aecf82d8f549895d2dce1b06e162f81a
eeede397c4adbbef88ec8db1ebe356d3ad052e44a9e1bebd4ddd6095fbb9b51fff64787a2bbc7f0dc21bba
>}

\EmbeddedPDFObject{schematic-35}{%
<< /Type /FontDescriptor /FontName /TGAAIC+CMSY6 /Flags 4 /FontBBox [ -4 -948 1329 786 ] /Ascent 750
/CapHeight 683 /Descent -194 /ItalicAngle -14 /StemV 52 /XHeight 431 /CharSet
<2f617374657269736b6d6174682f6e65676174696f6e736c617368> /FontFile \EmbeddedPDFRef{schematic-46} 0 R
>>
}

\EmbeddedPDFObject{schematic-36}{%
[ 638.9 963 638.9 963 963 963 963 963 963 963 1222.2 638.9 638.9 963 963 963 963 963 963 963 963 963
963 963 963 1222.2 1222.2 963 963 1222.2 1222.2 638.9 638.9 1222.2 1222.2 1222.2 963 1222.2 1222.2
768.5 768.5 1222.2 1222.2 1222.2 963 365.7 1222.2 833.3 833.3 1092.6 1092.6 0 ]
}

\EmbeddedPDFStream{schematic-37}{%
/Filter [ /ASCIIHexDecode /FlateDecode ]
}{%
78da6d554d6fdb3a10bceb57b08700e9c1353f24522e0c03a4640339f4034df0f0ae8ec4e409886543b60ff9f78fb36b9b6d
91438cd570b93b3b1c86779f7e3ece7cbf7f8e33f3458a5ff1b83f4f5d9c35dfb687e2eeaedd77e75d1c4fdf63ec637f5d3d
7e153fa77df7184fe2be79681fc6e1f439253f8cdddbb98fd7ac8f93427c1dc69c823ee2fe29fe3beb76c7773b7b3e0f6fa7
619c49e43e0da7b794f3d1b24898f80313b4e59f381d87fdf855a82f52ca04acc7bed9ef30c3b1985f7888f995d9cb30f6d3
858c7806b54269d10fdde9f245bfdd2e8981cd8fefc753dc3d8c2ffb62b914f35f69f1789ade89e1e762fe63eae3348cafe2
fe0f6669e5f17c38bc45b010b258ad441f5f52c134fbf7ed2e8af94703de529ede0f5168fa56ccaadbf7f178d87671da8eaf
b1584ab912cbcd6655c4b1ff6b4d3bdef2fc72cdad53ae6cd28fd68b6a552c954eb13204d4164009a022c096002c0007c0fb
0d000f20f0160d00f554cb5b3c8035800d010e80a606a8acd5c200a00696010780ea1131d522c3a086e11a1e3c4a95e20abd
a54c71b1ac90517186460d0b21ace22d2066319ca5de46828705055b315003801e7601c0ad013870728e054217874587b65a
6a99000fc0fb2ca1c7e49e79d02c010d82cd9a06d40b75d634a046f059d3801a619d350da8d7c8ac6983b91a7dd3341df6f5
546b753de5eebfed743184d60a75a422c10278494d714b311fb9a2b8247c43319f7c0396d2f25ec820591285a39174d4b26d
1113e76a8d31e58655224f308712398a8fc4a28e62154c8378c139d05d4bf60214d68a63d4d19a63ecd5cca12403d56c15ca
5f704cf99e63ca6f792fb8e935c70bc47c560adc0cf595067b0df34cee4d312b4d2631ac950c88c9c455208b561c531dd2ca
18cae71925b8199e51523e9bd76376c3f747c240866daf0867ce0afc4bd6c420a7e27c0d2b547c7f3566a9f8dad5e06cf9bc
1c785ac731b859eeeb50df5eeaa0a6e5be6462cb7d5b8ad7945f513e6b55621677f1153838d2ca1b98d75db4827f1c6ba5c1
c1957c27e025c7bea2cbef6ad681f2037b0c67e4888fb594d3728c73716b8ee93a6e38069ffa37cf784975f886aaec7fafb3
ffbdc9fef765f6bfafb2ffbdcdfef72efbdfd7d9ff4166ff0795fd1f74f67f30d9ffa1ccfe0f21fb3f34d9ff8dc9b334e56d
46bae574abf1cf1d0fd1edd9e8ced3945e147aade8a5c01b318cf1f6a01df607eca23f7a09afef2ebe7e6c8aff012d71e947
>}

\EmbeddedPDFObject{schematic-38}{%
<< /XObject << /Im1 \EmbeddedPDFRef{schematic-47} 0 R >> /ProcSet [ /PDF ] >>
}

\EmbeddedPDFStream{schematic-39}{%
/Length1 1662 /Length2 11718 /Length3 0 /Filter [ /ASCIIHexDecode /FlateDecode ]
}{%
78da8db705509c6b162d8a6b70f7c6dddddddddd699c6edcdd03c125c19de0162c58700bee0e018204b7e07039323367e6bd
aa7bababbaffb57d7f7bedaffea62657d36416b7025b0065c02077667616360180a4b2063b1b808d8d93858d8d03899a5acb
cedd11f8b718895a07e8ea66070609fcc340d21568eefe269332777fb3530683000a1e8e00764e003b8f003baf001b1b8083
8d8dff5f866057018094b9a79d15409905a0000601dd90a825c1ce3eae7636b6ee6f69fef508a0b3a407b0f3f3f332fde90e
107702bada599a8300cae6eeb640a7b78c96e68e004db0a51dd0dde7bf42d009d9babb3b0bb0b27a7979b1983bb9b1805d6d
44e899005e76eeb6000da01bd0d5136805f8a361808ab913f0afce5890a8015ab6766e7fc935c1d6ee5ee6ae40c09bc0d1ce
1208727bf3f00059015d016fc9019af24a00556720e82f63a5bf0c98007f9f0d809d85fddfe1fef6fe23901de84f67734b4b
b093b339c8c70e6403b0b673040254659458dcbddd9900e620ab3f0ccd1dddc06ffee69ee6768ee6166f067f566e0e901157
0798bf35f8777b6e96ae76ceee6e2c6e768e7fb4c8fa4798b75396065949829d9c80207737a43fea93b273055abe1dbb0feb
5f93750081bd407e7f036b3b9095f51f4d587938b36a83ec5c3c80f2527f9bbc8990fe23b301ba03b8d9d8d878f939014017
00d0dbd296f58ff05a3ecec03f95ec7f88df3a08f073063b03acdf9a0006d85903df7e90fcdccc3d810077570f6080df3f15
ff8d90d8d901567696ee000ba08d1d08e93fd1dfc440ebbff0dbf05dedbc01866c6fdc6307b0fdf1f9f793f11bbdacc02047
9fff98ff395f56250d0d256d55c6bf3afeb74e4202ec0df063e66203307370b301d8ff2019efdb43c07f875133b7fbbb8c7f
f8ca83acc100febfaa7d3ba67f55ecf93701e8fe5e0e7ac07fc75201bfb11608a0fb0fc98dd8b8d92cdfbed8ff9fa9fea7cb
ff1fc3ff88f27f23f9ff1624e3e1e8f8a79aee4ffdff476dee64e7e8f3b7c11b693ddcdf164019fcb606a0ff35d505feb5b4
ca402b3b0fa7ffd5cabb9bbf2d8238c8c6f1dfc768e72663e70db452b373b7b4fd8b2d7fc9b5ffd832473b10500dec66f7c7
b502607e1bcdffe8de56cbd2e1edea707ba3e49f2ae0dbe6fc774a699025d8ea8f15e3e0e60198bbba9afb20bd0df90d7103
fcd8df76d10ae8fd278901ac2c20b0fb9b0be0adbd008035d815e98f89f2f30258cdff10fd85f800ac16ff467fd4c36af50f
c80e6005fe03720058adff01b900acb6ff80dc0056bb7fc0b73c0eff806f891cff01f901ac4eff816ff46505fd03bee505ff
03bee575fe377ce335abf3db150672045abbff47cafeb7f42ffafdc7f7ad48d77fc0b722ddfe017900acff347eabd9e31ff0
ad48af3fe17f4dc0d2c3f52d8ffb9f3bf2369e7fe13f6f3c20d01b6889b4bc00b6140cb76f08efb8ab1327f262fe39213c4b
fd53f7133db3dfb26ba7c7032a7c1a7d6d76e8a6eb6ff1b4915ef4b51d69ba1bb115b267bfa3af8df0516d29eaed8ffe4fa6
491ad33fdb9196a67007278b8fc4bf0c902012336b89edf93fbbf8eb8438407f85fca6409defe2c187aa568875e7d52febfd
65a062752c72e1a7fa5e2d8f22f253c50c739cf607a390cf73d4051639f3f81470eecc24080c98e7de687337bf6731f3265f
c914921891028ee3384bfc0cb638e2efe77dd7abb438dcba09a8080cf049a06f30c7a669fc240ed215f016fdca4ad60697bc
bf0a9590e5a1307d5c63466739e0c8aab5d38805b5f4d57b8e2d77b1efe6a701c2698906b17653eacb5bb04d5d2954716adb
4d63b1dceb391d8024075dd612eb1dbf757aaced3f12aff0cf7c7905cca1e8db1e05b7f6f83d36baac8f8c0e32df467e6cbb
1b6919dde909d0fd2c3a204a6ce3c5ce206d6b18b3b62a9045f211b6c27d1caa5b17288f8dceb4e299c6f7046b1e21e4177a
21f140c8be5dccc666c007db0739467cca9def7f111f50dd74aed2b53587b1e8e885691cf61ac6a3f26af1ede3f97931812d
a7b45976245d419674cffb828d045a75c10660d35c5805ebf956352e8b54b90f9952f236f7d2c510550997bc6cdfcfc59a91
18a3501ffb7bfe2296539d6aedf644ee2275f2dde358e6f183a6906189b0887efedf5607dbbf2b2b254c700ec13af7e7d55e
d11223f13fb478c30bde776904e775bba966c88ac74295deeeccf7215a509e684978923bb524a0e96a093fec8c77973e7931
55c84b51d3962c2b0f281e0a45f92f1b2f4995f411d4058be1bebf6909e668e854bda38378978dbe1e4f9dbdb70a4d69da39
d2a273f1422841a4a9755748e9bb05d1b73833be43fe7d7a87cacfb1dae833577f2b4090fb4306b463120df9e24e70e6d5a6
30dd879817eb98d60d6d2ff1c615e2faacb941e0e6f1f0188c04257c27271bb6c8c4b11ca0b89c367b43a98411fb752125d7
7edb3747ec812ca155b8c6cda86103fb4628a252d8a85b1f926a534a345afb17e2d4de134f5ade8cb76c813a68a892283386
60d7ae967a2135977c711a49890302afde303713e37aa724889efc8375845e2882e501462c02c48ed03c2df6704342fa8506
961aa774017938b3403d062dc6d6cb9438aa025a2d0bec727f83a4c20484827b05044d9f1f2f5c224f3d126588ce54df54f7
d1fc82dc3c51ca552c6bf928aacca28d17850c776a82d7856634957fed6b08585d68627fe22332f2b45960b3500e293b5bb9
3f42c6ab50fba1d35963cbfd52f6c688edf9467ef866f561304253535d5d96e8088199e8156ffa112eb2e9583eac615ad715
9fb79a1017fb3cc995b0fbe49d1cc40a1a8a204fc6d7d60057450ff643cec74c6d490858bb4cdc4639f7ea86e9d980df3ef8
a40b2f8196a4d4469b2e67c1466d6d7b5caa5c9ee48659a6b7f26d8575675f6a9f7006f72c6309d936492a2f9515ee95f3f4
d9aac6123cfd7836f153178ceefa0050ae792e2399149c04c2b7d0a9395a33fa8a2c6b8af8bf46e87b98f0644715fdcca325
0d2464733b9148497c56e95063146923fb3854d966753ed6f0c8ab5151393e1e3e54b532a7e06d1976e31fcd01c74762ef86
08110814c7166938a86d522d6a7da36e15c60832983cc081ca551cfdf8be1b1ec919f951733c0c99ed0ca52803bd2895d3af
fcdf26ca3645fb8b1453712de9f1bf4fdefe303ab7c973afa6112ddf899fecbf7e04c1691d7975a6ec7550e55ba91849b666
154faca8fe5e98142bd1e5f0263be0732157b29aa9d921021f05eeb2ebb07e18961016abbf7e3821674beed12f00fbde539d
dc4fdd0e0553f7577f9b1ec1f1fab1f183c3185cd1b709a8317d1695b7fd8ee6715020e1742a038727d0c48e8ab0946e8550
3eecc3731bdabdb201209141290bc75702716b5e7ad194331fcdd6d5e6cafade95f1ffc2171ea64ddea51b7bee907fb25579
685cf0db55c504a07e628756f2b49c77c281b56b2aeda619a6503efee5d42fd6a23b1f598d5f5cedb46641cdb6fc0996e1c1
2cf308bebf20653385d5faa62fc7b1935a4d9aa3408054479c5e534289754fa802f7941bab423257bc68f336c751b0c6ef7b
37321d65db921e2c48062e99022116e21344389f9fc6123819a3a75f0b8f6ed754f71cca42fc2548d11fac8c6c8ab530fb51
1763f38538a913c56757f9e09527b832bfdea4f3d6dc3d970c414c95fda880350742e6822cb599d387943263c7124d597971
4a76c0e079dd10daa9eab5cd77f2834c101c22a663c57d2545af82480d106649472241578238605f873e11664d7378d58e80
a9e02c64e3f39906266e2b6e4967bb4cdc107445a75a0ecb41a712cc74b0f71d4c29bea9e1849e4e3406ded0d5b57a7e2d8a
96daeb8d6758fe3d12709429d9f9658530cc35c836e5a135c503f72231fc8319a3a31f21c22d50fd7621d154cc1a8f879350
6e53cddc73da97407b1cc7b42e6f1c7bd0a0929d94a16277280eb336419971c8c2393852d06fbd620c31a45a9d13f7404747
70a9a0633c7a161505b34695453577cb8ea9838faf728134ecb82f32c2264fd848784b8c4266eb4151e64eabc0e2f1535297
4e48bc239da34d4d4fce8ac54c75dc3a79eb41c57cfafc77a31fcff7e65ceaec78756381804afe6d179d92d6cf6325806437
1387934237abf82c10aa6f520f027c75d17eb1c002b152ae2f81713cc679852b724c49614598a78a5457acb22c2f3f95bc92
dd0b4d8af8920dd10f90238f0f64940262828c6ad3a59022e2cf23b5f162d5787171fc51cfc894a08d19c7429afda63bfd9f
8b8db4cb1310ba2b533fc368ba0bd94214012f043899ac768876cb331e8816b984bd92487c43fdbee27225f3799c90483e2c
9786f9de5fb7d7ddd9c53daa2b72de398b4fb2187215692e715878a5325a02dccb8b4f558084f8b6c452630debc8031419e4
a481bb95828cbcce46c55ede0498fadbb24826ebdcd235e36b07293898c479a66b5fd3659b6e7f9c1f06d3918abf736cc19a
3e65375be3f360b9f2603cdd59a678b812113c670af21d699ca64602b7d76ee67b430fb37d720bc5581826c0d46597f19e3c
9d3bf8b809b079714a8a6bf2d6f1b9695ac198cea130a6f40dc4fff8d9d9f3b2072587243229e3f7d240198bab04ff7b2a1c
50abbe64b524fb43bf7b859518e56d191fc494834ad45cac621d64feac6051f7afb914c71099073a38e4a50f2a5731a59082
5b3f4aac5eaa7a1f4f2093abd19f36193a6284064b0bd27606b0c2b12aa0ddba4bf9859a64135024a5cd22740e0b6db73a47
faa971d45326644ac06a9dfddbd31861415415899d14d39c9c1db76afc8bfcc706716efc5190b4663bd5a428c7ae2bbaf702
2aa925c7c36eb33743a92d646270d661efb24d4874d7236838b8bc30cb1d7d91f7a8e63874cfb01abcd623068f55f92d5e8b
7e52a896e6618d8eebf631c8601aafb3969af9547c50af478c1e4935524feb259261a858b3d3c2bd883b305101fcb6322d52
360b2b8ed4a4d6faa2d2a6c89b508bc0a891b1706ce85a92f534ac7f338075a2ef020ce8a85f227e57127bbba41fa5f6708d
342103a92d5d787cf5f2e17c5631ac7b1e4dee036d439308a481bad8049afcf2963b75d1ee6b7a76b761193d6f3f4dd753be
ed0428e75d3dc1a13eeddcaa796fb356df05ad9e6bd0dc66062bc8149406ada054808b8c40972e86d4b09fff7d2a13a9ba49
85c44f51a66973f5dcc421cb62a317f42b10f7c44ca9b6c81861d5bc58bb00f2eb80df5334a6e7f699d2e65adb5e2157ca35
fb27aa202c5727f1d753b734ecdec1ab4b90c126344346bb70f9eda8dbcd39380bdb42cc2fc9670119815c52922cd60fcf12
814537c8a0f9f88c748f0d0f84bc65e11c61e3a6f6c98a8a79dae877edc45c168d9290f33718cc563447ac77fe5f03b31af9
0e1e5846fad39b19ad27867c7dca3cc3c87e983648f8d68c46e5455e73887ace7520c041fdbaeab8598cead383342f22e723
982f76424a5f8dc2b7f15a26874742b8e3fd2983a82fe6a3691c7811910795517def08965838ac4a41244bcff12467bb3f68
8991490d7db7bef9705a2ebd696e181d6a7da7e6cd152c17918fe3ff0b5ef7a95170559674011b5d1f8b8253e29a7bf9ea69
efee879c11c56ca521dc1d7bfa44c68925b7ca1d7f85b30807af42bbc4f355931b2bf4694cb30ddd75b9378e3a13aabb8861
cd36df788aacdb406b8800990615bf310f364df483934f0dbd5d13418ce7c8a17def6fddb258a57e0d0913120d0973c76f47
de5f4cf317f573041ab8953fd7c26cee9dae629fc5c063c5b757a3351105953db2da17476c0d2c1790076c98a59e1e49d1e8
12070f57ec2ea8eb7da58293dbfcbc8f36e62e5f5cdd8118343ccfb2ac4f7c21b604c2a9744d4737bd3e06ad85f422a074a9
3eedb88babc0322ba180d2633f8ac29a6ee09a9846acda60273705e6e20e2608c69991d64d9d94294a446d9cf300fb9b7bd6
c86a935a9144520f1cbfa93b274512561606cdd2774736e7dc37955cb2c609c26b12d0be33f97a23b0eb0ea3506f2ff60bdb
bbf24b98341d9beca8e0441bceaaba0b547cf950aa8316edd7f35c74c15ca1441722d0d2ae4788bba392247b8a5e3f807b9c
f023924bd262bfcb4c4988e86eec942fbba426633771911a7b5f202970044a93631c99b352215d7965343ccd91ee62b6890a
81b40d38a2fe93d5c959855b911752f71918e34fdfb7bab4611e23975ebd65f075b696cb234e3be0db40ff7edf75191cb636
609afc7bba7edaf12f887bf23948798439fd9f65b379048ad4079625d2b5adc7a48f59c212cb63f2c644c19a70ecb700caab
d9360c5bd43143f54825db0a7de69fa1819aae2f3ee6dc0065357dd9f2f92941527e9838626398f2693c7d08dcd80831eee0
e4f5cd693a42a4bbec2411e49cbeb096112939fba842f41d371c579c50efb574a43d965e9195a985dada8249878f352b8efd
8234e9b59442fb2b997597c0447aebe95d19df2cb396a116d1c21d794b08314d399a74a157d88ee45318a02708820f2b2447
2694d86847a7be8273e0cabc5a9cdcc7a76347e06296113bd792bcc8fb1c363dd46252f965e1fd0f56eb00cecdf4aca3efa5
433b14b2e5e918d5fbb4d6ca6ed3f073647abadacf014fbf9b5a5af5fb9be24c9083e716a9edc787c8e5bace6255e55e1146
c6e3c94cef9b63050fa21a262e8392b6155a3d0b501b428cd72891a896e12bd05669256d4543b83384c235f19f49da4fbed7
f55c88a554e3e720669a573506336367880759c6441e3a8275211b95e92e13e05c215a3e53ae9a2d4e81f4fdec57cd38fd1d
a13c496accda68e51df39f83b6eac97f363e15dcccaaf58e1efba516c249d77e3a139e70363f30410e59b23d52758036cb69
39cdda71be7b7857cde7e1b346e57a99be7e95e681641d60e9d2c7c8dde31348febc6e3c9ff34074d77d8caacd540dd9e59e
4fbf6727207799dfa51ad1bf4a57630a540d392491d1e97a719982c315a9f238d8ea67961588fe71d1ef1b961b9125676044
9da07de2b9d40d418a32e5bb6382bff55d22ce873a58456bbbb5995318bf46ef684bb4c3095b21f90e97799633fc96c46e67
514e6c2d31c47a32ecd9a15583e528ebe6d9933678c8890593765c73a757aa7d6cd9d1e7890c9f3705b7f09dfecd82e497ea
01a7c32693e65a590f6e53d92cf8bba665f6d2473477d60f73c82f5b7163dbe7485b5cefebd9ab41d1a68f522201ce708c7b
d61536fcb654240f650390c66267caf7664b26335f65157f0b7257c2157b0d1db5c3eccabbba9c8997715a8b876eea63d879
5405f7aeeff65d764e9f1dcf60eabba21f9248ddfc1648885bb12285d728815e07a488bb63bc3b62ae4864c2a6df6e33af28
1e69b3dd9ee4379cfbecf18aa0b98fcfaf984c108ed1afa5586eae47b5eb653392d2ac1a66b80de51d4be4321456e403234f
97feda618fceadf4b942bedbb229c0df96b49b2ac9576a1b4b7bb237d1ab5d3f8eb1f3b9ca4060e26cbf7fb278209bfddb65
9363d032b09b6808d2cce34e0c8ad01fa6ea980c67612dfaa03173d0648587e89435d767fad1ccfcbc00b3741825c6b07622
06dd6ca66b4c3ab7a8d78bb007f6d89c94e85d81ef9edda59279381e6680fd68ddfe88d86351cccf8081206fb4bd33dd6658
f88c9fa028b7f84f6c2006543fd45a47518372dfabd8b451cc5891d67dbe13326766a5699ae1a247e8334533118059631975
af597ae1e9fab711b70ad136cfefed8326464fef199271169ed5477f94a958584730ae76746d84e7822910e3963f6b2376f8
4e5fd9fcb21937df8d4ff46894c99fce10eac3d27bf0596fada89a827a60e8801ab6f9fdca25aac761dc9bc8552697395e98
b5789969d233e5a2040437647fd23eb90a7c775b929c3cc52b724c3f587c9521074acc856a2bab16e7a538f4938ad41062e5
707f561e18babe77fb591afce8c02faf779d6c5d549e3d0dbdfd3af9fcc888b55caa277dfc391d7d98e5d70ff2eea93e0a06
85554d5ed4ee157dd25a04e203d833fb5514129a9965f10385778c957bd1dfbeb62360ae600adfa4ab14fca4e2f16cd024d4
d2c7784fb98a5fa3f4f1956be2f3e9be6816c424edefe3dbbefb20193c2b1d59a50526c67d5c379c2ad997e22d13d6d67297
c63e9891df90b61f06d0aaeeef0d48ab7cc3a5caddc5de71de023133b3f20e9497c444ece985746dcd76953ba2455e170b47
5e037f00b74e74a414e6fd3fb6b5e9a1054c7f96748d19c61bd41a161f3e40b6c67f86fa54949caa1620081b4331dba1f87b
967222492bbf0555e6f6c1689ff0f9d6642e059a84101a593003b1dbff970640caa83f1445ce638ee7557de079ea136d8e3a
e7b085a037656b3c19e71599e6984085bd05c6c92a75abbb42123cabc86881eaeb693689abce46482f81b771dd57b80e11fa
edb3bc9e26e360c79c7b686e7017bae17ec1948c203a930a4d192d965d798c51bd8a7a8ffca5dcdcf8f51459b7a896c089d8
92cc19debeca2b32bca65ecf3a5bbd50157dbbed96ee025c482797f55719a916db845596a52f5a50f4269ab0d3cca41120af
806726e468039556699c2949c8ebda5b647de4380d5860936fefdcef73ce8e1923a4e625f7939717fd1fdc6609783281a659
de3d7161d0dab9feb1b47ac94d4351e1b26268da0be83cf2c8466419e1fb6bbf459e0446744c1fa14644e64bf919e4d90270
679b8c28fc967010ce38b5577c072f463c7665b2ada1c7968b236ed436cb46e9a513a3f9d0429fb44bb21fbce86bce6db062
bee3b8cc096c59d953967cad5dc389b6857719422ec9ee7003a3307f154b46528496e54dd8a1a8f6579f8a82daab78f80930
b8f475f38f080d28ea579a6fd4c48a291108b6ae9877eaf8dc2bce7996b070ae89ccd12f4c90b460f8c5aa6529e54921db68
ed0b2e6935f125aee5f10a7ef0d33d9a3fdc77f72e7bf6e9c12cf17dd797189c9aeee99a17823105cf5e2645327f48c4cbc6
88c9349d8dc4c46ca44f4f59eae18d866b1b42da3d7a02c430a12baba2968cbfa9b7ea114f54dad72434b9f923efc27abe40
c655e452f42df4677fb139d03aec58df35b128d71d40805ba56b60a0027cf082abba3ba0c789f990b2136182ad42ee5adda7
ef5b2254124983e7fa5150dc4a5e92c88c353a240089d91e00c9e36bf4ddf1126dec39265b918b90613c44919960715db744
f3a75aec788d78852c6d583f6effda03de97a99931e9338ea5293935de1ebc5ad213fd5ddd11e60699114c69e5a445e3cc61
89f7db79549810c6171cd442ea5ff7b603f7a06d4d7e90ee1c8e1dd705150d51ae0cf959953a7b4d879340d83dd8fd220323
f0bba68c6bfb817cbf07f0681a5f09ab99ee9ce086bc6fd20274a86b39ab97a868187e8eb95505173febaaf254b0580dfea8
261423df5ffc78206e425883b7af885cb326d92248edbe36b610d1ce0abba7c9d8aa3b686148f845c2213a81d2497795d540
ed4bd0a62e93857031e163b05d168596a66dade3d902ae4c105ebb29ecfea7ecf64198d751fc8bf088baf5ae71cc16e2335e
5d8c8aaaaf6633ed9395ee23bfec5563656541629f4eecb8b4ef581113e45fe53c597f77a1b9b16aaf5154f51198248ab56c
741fb48abf3c4906571d3d3b978eabcfb57f849997ff9e1238bb26227285b5027ad29ead456f9685ff500a6a8dfff54118de
80f856ad6adc552e197e799864c9c233bed210ec2d24cf9bfa0d538d05e649034ec5720af6125a81d4de3d3dc0c607b1dd27
409b8fa944a88d56d7af02b0273aa148af5d3e9486ea2cb9e80167935ab8824a11f6f66fb929e6f78797ef88eca2bad2758c
8acd68e801fbfa831d19deefc5ee8ebfdbdf56cde06261faca7f562e3097d5cb6416aedec8aa2f151093fd0558df9e79fef5
2cf7233d1409cece00791274170c820699141f777a43c4c2a5eda07079519bfc26d6470a75c95c3910a3a0c77f183779b9a2
d9b16939bed9dc560599f42bd59ef0836d8b8568be618542ed318707c30ca8c0b5dc8fdee69010c272bfe06ef023ae607000
7974b1bb9ebe4fe7993c87845a248610e420ae7a19ccdce94dd8dac2f45d10bccf33fcb216b7939ed894ad42219572795950
372dba7ebb8799235b0343d780605782f038d5c7775ca939a4dc7f10162ccadaf228c0165586fceaf17d3c44c9c148c2db78
e9792bf436270c419d8534f78b7dba0466f45299cc4a9eb060dc707287df7c37ad9458a8425a3371be44a0e353073bffb4a3
6d73bea4fb27c518051cc9eba6e0777ee8024fd44fef3c43c34d4524ac5b6669d9591ebd10711f2bb713719e3543fdc361d1
aab673064a521a51dd9885571b4c2dc73f0c20a90e64b38a8297d3d9dd5f23058a6ad0963c34a9fae810fac0885cc161f871
b6a85270fe3a15425941af1887bc31085fa76a9107261e7a5b88906c2ac720772519f706c042d7b865ea2d8209e35d3ed979
727ab495e89c8c0fb0b04d67c5370359c26890aa079f238a7e831f5a9746f6b741f25da63b3727155c44ae6806fb33627a9b
724256417692809fac61cb64ae849d18f662760934e66fef32cff05a81f25a93247899ddc6266a33521db2dc2327a21e05c6
dfe169089f1891c0869b8e43fef192dd9a00230cbcceccb89443a631dc802f4eae7a26182ca00f9d1fc226997517b4e89c8c
9898e3eda9be48e04d43b615c130e40e0624ddd26756e9840acd75253ba1ee1e8cc3c3a3f5aa4f3d699540a3552a72119df4
fc5adbafdeed98522d473537862b833d63f90c4325fa25d482189e0fdddae314c153c07cc633fc526c0cf5514d520c62d2d5
821a15b9b46590d60d1bca1c55268ae9670ddae8b32f8abc393bd88af0dcaf5c16e17e32ffdaa0ea896441d274f073ef5722
1bd642e1bc28230852cf5f220f1677ed458072899376e97cdf8f8500725f6a17f53d1c6cde08fe4105b372e6dab815551f44
f436fb1964c04a057a05d4d7d2bdd363f8e794e9bc6c2430102c47a2ea27760eac2321161f5914f258c99d8e1c1c1d3561e6
de11f4515d52a25550cea6913c3ac65919d85f3e81648f1d291f2eb72659144cb45bbea136e63387a8a19fdb04e413106af1
cb250ac723ae0a959d462b6aca2e2fcb1de75d393459cfe65dc0c345f386de16c410c77e38a4a72cf0504f0f2bba956b6bab
aa5ccd6a7a49e5cdb28c0d7174c923b5c8bcb7384e7b2c2e6af8468d73903cea42a63ff6ab3f7e1feb1bbf64db8ad17b6e83
507cd1393e300e18684235d926246e8ab2ce6ac545abd74beeb16311c05d32944a534b107c724ffaa130a4d873f93c9505e9
68af13f5998cd7d6cf55a4918a4e38ee1a4aa5dae89ba849e3e5178ddfe54f411d0c4896effae5916e899a65902b9de5a1bd
e2a87faa1ccd4c13e6ca18a335c3cdbaf9cd241b6975cc957de56c319b5f9ffbb5e5c16f7ccefedee78969fbca1e20aba1d3
75db10e3f69d5a532820928a585198186de3ce8695e2f39a09722189ed62330345139a2e69106a5af5221ed15949fada3296
385bfcc294b97381448be2177200a7650355c5434d010d39af8e6279bdac1a5174f850ecf8c1d41c2c94370eb1ff3cfc374f
ef6cee8cb979b27b8ff198d5dd20fb3a60b3338fc9f209c3d083c347cd4912bae4fa0f03a4aabc8b7cf1b534bc2229d8d445
e6de5ec4d31d1ff006fc6767e8ebed969d85a8ca31e058a8eb4a290a13ac7587c05ecde3c75dae1f3b2fc4a9f52373f70649
421386052b0625f6899ce3bc5b7ed4b87949d15b20337248e764f099b7f4603c05f45ee5034d0a11ee63b915e81f72d39907
4db12df476091da58a588aa727922f54be8f2b4fd5eb1d5d1b0face7e1bfcb3c59eb3bbfc64212abc8622427d0785d170f72
1e94bff6cc01cda529a531cb03daae7ad8d8fb4c659a83dbd1eaa2b674dde0951846f5d74ba7306ef3c025a4a6dd96b18bcf
569c1a451995d842e30adc6cac55bd7d695938a9974c03fb295a706b1fb340f3e22f4d961c1fca66f1d099a37b969081f2b6
ea36c218a997c78fc7e9ba106e4f0685bcfd1f90d94b251a62e12dbb15bed57b381c9a98736a00c847c357939e853b2ae1cb
eef383248b1cabeb1da1b0e16412337f94e66c86f830d6d5d1bfccbc83e78f50d9f7e431daccb2e75892085cddb88b68ca85
0c8bd9f36222c1174d5ae25d19914104a6aadf09575852538e677fda76756e246c818d44b0c377f1fed0b8773e671bb1d6cc
80c02627f47be638256e77a42904f8cd661c871083c7287c226252c6d641c23a276af910da398db4155d6ae5863771f607c4
86dc17000c3ca3f1b3b8cdebcc0fe3a172b74659627646ef319a805f49c5bd248a0d6b258fb74dd44c50ee0348c51471f58b
a7ef3d46262fbaf4f3bc7b044344c209d98bc0d83868f3ab2913041b2d8371906022d808faeecfdadf213fe5301d6bf4820c
e32dc65a1a0d3c7fbb7154dcd8a5c60f4c4d062ed204ce555e6fead94cce0804ed735a4cdfe727ca8f97ce5a0bbc32efd853
c87f9bd99481349ffd658451de751cfe3579ce8026b86d90ebe340ca083c5d3d2d39d8dea4f09ce31d0f291c492db5a2c099
403bbec52bf58f8fba0fdc70560b50e8d11e92b753447e6a48265e72f934e90a3a30b5e1799cda6d6eb023a1d4ab1a745192
d91a4b96efa0ef99ecd7836a4bebd821e260d2cad5bafa16dba37a1de1a345680da3aea761be24442661a618963cf64e4f4e
a94b6c4401c5bec798148ee4d1371632e7664e0bd2109442875f05485d233ac5ec713cb274a0acbfd8ab2e4a5c65fae613c7
b2b9c3a110be7371d0ce4cb90c8f87d8f7de96f9d242753ea33845c8469cf37ef9cc87272140173f0881cec640838f597cc3
71f95183ca81caf372ac44ee068b8c25979f55824bd26090f83ebbe2f8ab50d02b2b2e9e0c5c25ad337da64a8d3bf6fb129e
9f99fbb7ebb2aebdfd4afb28130bc236dbf16ca6ce705e6a6182b3295ef4004202f86aea910e287a6f9c86b95b3c7e1da3b8
b9c7b320be96b0bd32aad51fd3652949f1c497169f440d65614645ade0970b481968c998645a3470e59daa8b06e6b29f1bac
e7bd615b46a45590a37f0e6cdc4c49671c2314628c1ab1de28ef0f9350a25ca4d45db8d076bc9008b02493d704f904139af5
0654991067c53be95466736b9ee9496e52e2ccb586a947cde569396cd9fd0ea4fb8ccac7be3d5253fb29589b79aa3afbe892
97a0098eae5e405ab244824a93937cbc9779920f3334a0bad0e23c6261c6d304ce8101cdd8cd45c2af2ae4cb7cbb895c9b65
4ac59408a6ef9a85f377c4320f5eed50d3988fd6db0865a29c635e099fcb9d51e418bfa44d7bace68de57c26d2b55c811706
16596ca47b735c642d9e8633941215adecec3613238b379f401f6ddbdea616051bf3dae7d5e11596b26965d79c5b6a7f52fa
6537d02c328acd89979c48caf9f0be3244b00e4d6e7f3910edbb01cffa0f71de5f089a9cb1240c405c510531542a601b27bc
7dce4b441c3b13bfcab85ca24b5dc0e135d5fcd1ab8cc9815afca332ad82d469f9a56790c7b471651f359e304e2106c0d1ef
487154a9d4e7fcddeeee6b3fad8e216ef7f7d41b3dc261baae4e07e9c6584a1b389f799ac392874e07a332a46f8f02b57168
44ebab19d5ac5fc0c105e5c19b4fd9ad49372127a36516c5a9ae757c65acfacf5390819fb9b39ca8afd463d483144b9e4f65
0e890f62632f6c6db80d3c9fda4428f65c8b3e8b615539670782af409f60228f0a82c11ae365e9f30576878fc621258c8ab2
9cb5e454373045ca0ff85050f9522fb6af9b2413c68bebf20218240c6cda5c2e0223e89544ef10f99c84f4f86f2c290dc042
806de1d5fa5ec7410bf13378d0eaaf975d57a238e2a98ccf5aef215a0f110d257c273e259ea9e71179d431d2c540c17e5dd0
0f5d23a14ec6bf55bae2a43d70418b7173d606d77d20f71d55464dbb0d7b348977b9aa307daece5c5744429fbf7b50eb7d89
9efd62dab2d177d42108c74f453a517053c61951d40929087d4ed62d8f924898d820d5247449de32c0cb3bb9dc76faf832b6
911f458cfd13a79c384286d3ef4c4766de330e016ee59cfbd3ed745db71c1eb42ea3f79cdc4f4cfebc1f48a8b4bf61bbab6e
5ffd80b3af86798135775e4e22868507287aa683633fae1da8a00fd26ae0903a274101af1089226c9f6d1c73d869f7f1fb70
d54c54a2af9db2db9c0604d69b7c5cc9d87e924bdcf18844559863b764a219563c4cfc2287f88a3815b8c2fa4dd974f4ddf0
ef84c81a1598f994c62aa75a6792833ccd5e3412ea0e2b596c520147b03f015409f67b2c26b55c73d99a38f7781969cfe81d
bb4702152b9b38edae0750a4a70b545137b60c7110278abc78237d208d1e7c9c42f8699e7847adab3809b662acdfc6ef0958
a334be730116d9e5d1552dae10f3d9b5f2f5f50b8bc44f6d66fd8e4f755eccf3b7b99f12ba664965617209b68ea23f8cd099
0425cf0e2f28225ae68f48117bb5663802e5dca448850e45f3c778f30f909a7391f49c858807ee4ecd5fe1c3fd6d3f45515e
4a24aaa0a6a7ba9e0ed362fb0525b0bf68324716d868d965dbf761e58e4441174cc5150cc1d5e70b2a61c508fac6a309e04d
740a0533a2c64684f1f87feeaac768688494961b635d7fbe4861759a78c6462033aadc4934de6be655d08f726fc653eee798
44ff3d249e53833edd384b475a8760177c12997e211f9b35c185b9365e4c8a358590912f10c3b4acd2e6f5f1e21179253919
69707f92b8393faa515a33ce30ab4357c7ccaf1543d7cbeef77a91c6b3e6b76bf4f0aa8e2c5c4ed047fcc10ddf3dafa14eba
9d5a0e695aa2c910b25aa5a31c4d114f5ab3fee152aee63346d056c26325b3b7e77b733c86f7b8b4b787694325770e95fa66
6a89a24d24a841649aace60589efd2bf94a7caec100f64b961cbf9cdfdca47fdb69729514b661b2d08d149886834426141c6
d9cd695b47b9fa79ad4862fe31ec83dce1a5dc16d1c21523de0d9effa5ad5d67fb17977256ba80765470fc5782a8ea491c4c
5d8767bcd3c0a33635e3d068637e785387fa9af024eb249a6741912cc28e3606028ed4fbf594a182aa33f79fb7bcbf88bfa7
85acd54d96fbe61c060e64d8db6747e1d540ffbaed7c67157a0fa6ce8acc1beba75f9444da56f56db846a58c362e8767a870
842cb0f42c5b633bb9b95976ee3c18dd40b6d8c19764ad68c2ef8e34d5cb35850a7c3f6427a595eb92a2023b7c8c736319ed
4f355ab4ee99a7a397d69eafa8af9dafe6b203c9af8745219dc44d15296d9fb0c5b837f4fe059e8e9d2cedf8452e8948355c
8f7b21b593739d3959b447941876af27a483087bcfa7dedb10dea3ad3a670845015a880e763457b87fbd54de8448d58a9c8d
a3e572b9812c61c981d3d68316b0a72be8d8d83339ad1e45c69e0ea29a184662defdc6ba02b1dbe248d37f24436641abb3c1
388ebdaadac3ea1bff09106c6276ee7d1270d4104ca6ba9c2473b5eff69574c3288c8fbdc07da7dce98289424a980166b354
4372a0f7bc91e6b3bc0f1d271ddd428dab8fa369b0870ccc88603c25389e26fd0522a4a2bd077be6eabc6cc99e556f530042
a6a103a5a2bab5330c190f875d4aa995290755be86acaf228454753b6040361bb50195d99db73a5658d9a9e60657dc79c3c4
adb5fa8ed1fd0b3797ead1cf1953a5414f453d82b28cd89f8f4d057cefcdb5971b2b6807626ab31dbc6f6f862c4b570cb4a7
0e1219ddd98333e49ef8be9b1052ab66a67e7244fe94ce65ca274a9a21230fbbae7c7d716ab49fba4adcc679a0a2c1754ef0
59f5cec7d8b76216b3a5388a8eb50afebc1e852caa48f587b0487b74313b7e7cef399d290df5f738e5b298828841382727c7
4aa4c4d2f71dce3f1320cdd35d5c7ab089f4b762ccf391f5d505676a55bf79620211a7e7b7c524bec59242866c9d6c47debd
de27fc121abe636282359e87f905e2bd482139c58a0d7e96c0f127c2846b888e6217dfe3754577d7ba9594bd7ad5507779d0
f9cad14b5aed351315fdc40580ca2381c45ce84ffbfe559b5cfe47c547df70142f3b532fa79b31cb907ce4e12287162441df
e16a55b0426970e34f03e304d7fdf9715aecc0a7c74adc3966833df2359214d3f81a9fe4c44f128e2dd2828f24b75c4cec6a
2db7dd4e2e567e094e3ba39153466ead25739d6c861ff401b5cf31bf6d1633dab13d4a3e1d3fa354dc95502e6bb3edde41f8
9adef0753fdf8faa373b8eac295669d3c7408c2539939d17dea5c77be3a06159a5c9450ddc5d44245817f9667f50025f0253
ec00b7fc29eb094b27b2ef449087e66d4c8876fd73ab456663372110392cc8bb442542d87619ad2e3612fc427fdd6a2c9461
bdf617aac70dfbdfd0fd000eb47c733550dfd80fc2865dd38faa92aaf04ffd104d83c430f543a47dc105855ce60e078c67bd
d546c2fb20d99dba00904fd2d8d5603bc57fe9404f4bfaccde8e7b97b45a4b1a7b34141ff36e22da3e519da5c39ba22e0d11
547da2427c2b610040c0da3acfece231f97e8be7adebd3eca48a42a29effab45136d599e575b09511c64b3115d317c6d6188
9f75565e4194cb7f1a31de0767f99264d226ebfeb163894955b38b915f8e7a916e0ba1e7d41465ef2715a799217ba85bf216
da99753bfbebcebbece803c2aa67ed5721f1272b84832af6522d94e77a11b2973d58eb9679559dc7d7c0ab864f85d1fa1d4e
7e5fe72bbc529dccac9c27003f92b9be90d6e05e32f8f031cf0925b52e47d5d991c86d09544123d9958d3212ee98612a32c3
166118d84315c443c8e82e71da5a71c0ffb87e9779482d960f9d950fc49a5f87bf143cfd748d1d0bbc46aa5057d2143110ac
adc02ba8bac80bfd41ac4c620f4b6d648bb3db495a7a5f1cd12249b6adc4c05f67f8423bc26fafd6ffda1589e0048cd22ffb
fe0d6ed65534b2b1d5f4a4ae96f4e0a3823931813cb366483da4888e99134569ffa6ff661bfe625787e82b7693c7c749c8ba
b9115dbe7e082032f7f4c630fa43cace5bcb69a9d5392df44f1e94cf58239484a9be7611aac22e039dabdf072989394803d7
45a3c82a1a35c425b6e08d9b8e2926738e2e53e41048e6f6d25d89d0c0cf4a44076c8f96f2930650faec4443798cd5134390
980ed5b014759da63d0e84503fc7f27b36ec35b7d136631b8511a8f5baf0f35f949d9da027bfe924f7ecd3ca73b769737fb9
e10839bc467fe231a29432b53328a18a8e3761ecbd4793d91b975626423208191f1d287e1aa3e7b3d0dab8efa7becff0fed9
59ea1a76e153ed511253408aab1b4a799ed8f89e7b27335e72f5acd83b8e907459b12080e743c33043987ae586e12866ea79
811886ef9ecb8b03ff727d1b38a687d901286cf328c266e25b1ddc7b4db55fa75af56e549acea9572e9fc9831b90ade594fc
91520d5d88e942cdd15fd111487074964f8f552488b0f2d48d62045dfb595f83a4d57c7f98ae3cb0d30a83f9e96ce4a1a726
2a651b7c301cc949f83ea028539f21b01d9d446460e433cf776dddce77e5b8eb42f77eb1f073780eb0277306d0ccadabef33
c25c8f0fac7de6663b30782579630f51575fc5fd18958bf2f268fb6f33ea29713ef536884a3541130fea74230b2a53c2baf5
9aa0372ea2dd2946cab27adfbe9ea7bc6f57c2bd5031463581393ca9d5607ecac6fe21466f78cde7b2f92e9ba4db85210e5b
cb896e3ba1d51f0749bd4de5a639d44c1351ec0be9833b4bc4d7e29f5c95870e4e17f97eb7c2ef66f39fa0a6096602f9f278
0bf22f95158befabfb44aace91069e879a077bed8d27b5b3acf526b4f3bceb6e11e9b2d66965fd3ffec034a0bb8c51961e7e
74ef808f96036526401d52a6fc9ab8edaa3f1642cb7640aa32f21487f6b83cac57987545ef773236b6a1cf7712ae8787cb17
21baea42a9fe78fa5bb932077e4f83b650b3c997049d2c3207536f4b93657467e0cbff01d203cb52
>}

\EmbeddedPDFStream{schematic-40}{%
/Length1 1877 /Length2 11366 /Length3 0 /Filter [ /ASCIIHexDecode /FlateDecode ]
}{%
78da8db405509c5b162d8cbb3bc11a87e0eeeeee24b836dab83b84e0eeeeee1edc82bb0609ee2e0916fc917b67e666e6ffab
deabaefafaacedebecbd0f25a9b21a8388a99d3150d2ced699818591991720a6a0ca0d6066666364666645a0a454b7740601
ff9622506a021d9d2ced6c79ffd08b39028d9cdf64e246ce6f660a76b60059171080850dc0c2c9cbc2c5cbcc0c606566e6f9
b7a19d232f40dcc8d5d214a0c00890b5b3053a21508ad9d97b385a9a5b38bf65f9f7114063420b60e1e1e1a2ffcb1d206203
74b43431b2052818395b006dde329a1881006a76269640678fff0a41c36fe1ec6ccfcbc4e4e6e6c66864e3c468e7682e484b
0f70b374b600a8029d808eae4053c06fba0045231be05fc418112801ea16964e7f8bd5eccc9cdd8c1c81803701c8d20468eb
f4e6e0626b0a7404bce506a8c9c80394ec81b67f1bcbff6d400ff8d7d500581859fe13ee5fdebf0359dafee56c6462626763
6f64eb61696b0e30b30401014a92f28cceeecef400235bd3df86462027bb377f2357234b9091f19bc15f851b012445540046
6ffcfec5cec9c4d1d2ded989d1c912f49b21d3ef306f972c616b2a66676303b4757642f85d9fb8a523d0e4edd63d98fe6aab
b5ad9d9badd7df67334b5b53b3df144c5dec99346c2d1d5c8032e2ffb2781321fc2333073a0338989999b9785800400700d0
ddc482e97770750f7be05fcabfc46ff5fb78d9dbd903ccde28007d2ccd806f7f085e4e46ae4080b3a30bd0c7eb4fc57f2304
161680a9a58933c018686e698bf04ff43731d0ec6ffcd679474b77800ef3dbe0b100987ffffe73d27b9b2d533b5b90c73fe6
7f3597495b5b55514394ee2fc2ff51898adab903bc18d838010cac1ccc0016662e6600d7dbc1e7bfa3281b59feab0ae67f7c
656ccdec003c7f17fb764bff2ed8f55fdda7f9d762d002fe3b96a2dddbc4020134ff0cb82e3307b3c9db87e5ff79ccff72f9
ff9beedf51fe2f03febff548ba80407f69697eabff3f5a231b4b90c7bff46ff3eae2fc36fb0a766f1b60fbbfa61f807fafab
02d0d4d2c5e67fb532ce466f3b20626b0efacf255a3a495aba034d952d9d4d2cfe1e95bfe51abf170c64690b54b673b2fcfd
a00018589899ff47f7b65526d66f8f86d3db3cfea502be2dcd7fa794b035b133fdbd5dac1c9c00234747230f84b716bf210e
8017cbdb1a9a02ddff9a600013a3ad9df39b0be08d9e0fc0ccce11e1773fb939014c9abf457f211e2e0093d13f881bc064fc
0fe2013099fc83d8df90a5a3898b8d19e82dc3bfc51c6f2e2676a03736ff96b0ff367c5bd17fc2fea6c964fa0764013001ff
03397f230797b731f8c78015c064f60764033099ff01df1258fc0139004c967fc0377e567fc03782d67fc0b772ff4cf446d1
e63f90f52dad8d9189e31f645858de4aff13be156bf7077cf3b0ff87f99badfddbbb6a0b029a39ff2365f997f4efb5f88ff8
8d943dc8c5e98f606fbc1cff806fbcfed172bc55ea04b4b1fcefbb66617963ebfc077c63ebf2077c63ebfa077c8be1f60f64
7d2bf7ef46fed78099b838be15ebfcd703f0367dffc67fbde540a03bd0046179c1ce84efb355fde78e5fb522046e0c7b9302
73947b1f526919bc961d3b5d1e506093686b323f6d38de8a248d7c455bdd91a0b911fe4ef2ec75d2da081bd296a0d2fee8fd
6410a73ab3d78eb0348d3338557822d23040044fc8a02ebceffdece0ad19600dd90ade2d4b99ebe0c28da29c8ff9cbad5fca
bd61a07c653c78614f65bf86530ef1a97c96214a235237a0649e32cf38eb1b1e198c330311dc7b8c4b77d4f99bdb398c9ca9
5712d9383a049fd328b6222fed4dd6e8fb6f9e6b95eaac4e3df814f8da7844903718e333545ea287c9b2b88b5ea5451b8ee3
29df703a27b66c12594087349e7b8aaabf1c8f7eea53514ef1d2e0e3314162c944446eb449c7db9490c6c3b0675daeca6eb8
cbfce46a3576a2578ab77c550271d5a318f45f99cf34a4763a71f2ed277d554ef1cf4a696ca9f271c6706fa4ef9171950d67
ebd702cb3e1dadd083f989f028b07ae65bfe85240dfb204b7adb11c4270dcf3b4ee31ea7cca43cf298f82a9f46f07368224a
ed970cb593f13a417bc1a2af5f0a018cb77237a47d128ea28912764b7595c844c9cd4d9d68ade6e04e1f19adba455cf194be
183aaa3b44321e8dea8e47127d8f132d9df44604df8e0dd8d7f7b8420782d2999d6421e3948a51ed7cdcd4c657dfe1a46030
a3b4fb927a128c874c3791f52d4f29aa5a2edbae632ddf19a1016f139d729bfbd7b702256941bc3ec4af356cbb1f11e2a091
89145c7f4196524a295e762d0ee0d2cf167732f91a4e87dd91659c42ca42a3c52ab1b2717f8fb584014432e8c9487d32d9ed
34e3da9df72f1effc9ba38f12eee7b818f3a00e5e56010b9ac4ff80a9eca1d1b8b0867a30183db02e133f6b297b9512307e1
86bcffd037376a5f2801a5d66b793c13f35fcadb75c7ae713c759263199e0fd894689fca7fa17f067a3f872aa530cd199f07
d6e087742d3d1e1f6bc25704c8216fc7ebb7b6f40773cf87b4f3d32a54f5044506eff0c341d86c404595f28d1f24c711c6a3
56f92899b2e869ab4d621d10db28079a4db0d7ceb7d112a5fa28846528178f1c9ab34ed78358cca3c6b7256f268fd2512ed6
f3ed915334c19a7d7e467a7d29e1d741c726e4b09455d22a179cf8c60189f1311c43bb8bd09fe2f4769d8523fe820b7cd3b7
150ff6503b1b0c82b0ec7bd2d459316a2183469078ef8c8d1004db64e369d9676ab8d37bfcd0ae82d6ed524c4db4293ad565
023fa967fb6c0f715c42e92957772f7b7f0c13791ca8fc3b16b78ec3af5d6bb318effab1c3e05b248f1414cd69d4c0c3b701
2142f3b450a76d427c1dc20a98e30567a28a1b95525d1f6f62c42d518fed90216d919095ab97d53abeaa7d5c1187d6b13eb9
8d8c311fd23977ceecedbbc2acd3d0c7178abb1c1d9931630d6fc91efbc175771b46eeef7ea60176d8d769d316dd79636855
643262dd48e1c2db02a35a8eb51c146e2971a0d186d6fe2d6f122d87ffb463fbcb37467d2daabad49560c258bb75021e27c1
42e4fbb1bef6aeb8b63d841a6eab2e4f7c99b99523b1144abbf6b9a78ae874255b6e829faae6e4c97ce21dc96304d9d368ac
7134b274816357decb33bd705299092587dfce324fb036ad79bed72f04625d013df2dfa13f97320df3d0d13318b3b9b5f27c
88fdc505e193b578bc0240413f3bf1fcb655d04ad852f4323157f64ec3e359ddfea1c779f1b1253aec87c153122b7be98111
e7ac53697cc8010e29ccbc5206807cdc72319262bd417753e4cade9122d68d23b1541775cc1e149a1baa65cbe252686faa73
3e7606b2ee852605a99eeebafef48f608f34700fc2e8383420a67807a3a973f1089320adfd9c7d2757f3e8a8eb7693d711ef
e742796e509f2e89ac8fd8062950b00b16778895349f351c2c166cd6994ac9b0b8d2dcdda02ad8b11dae02093189ea2c7c3d
36d4cfdf1a22e6d8f884b779ee59b0bf9b203e13944a83790bf4a939e6c5839398021392f00feaba31c7411240bfeafd7e48
647ccb4c16345deba0ae04ae270fa3ebdf858025329ef16ba61266b65f6122e146213a29c8ae8224c75dbead2b226cd6d765
84582fb0dd6066e6fb01323816cf34e348b9f05c33713a41c41093dd2ff0c16c190a029dfc552d7ae9e0708ca2ae0199ef82
ace9cb48b427b94f6d65d5b8dd458d03be9e11ee919cddc57e36094138b01f6f2535c3136fc52494dc65831f10ac246d8a3f
5bd0e27f1ce984cc23fc361d12c1fd25708828e2fb51b2d78e0b6c1f72449049c1fb3dbed231a0d077e3bd8d21309066e58b
f7dd24ba7061ff0c037a3ea1d58988d9364f5397e693f1755e033c063784fae38a69bd151e3be99166475bea2eb0fec6315c
db755e9b1ea8593c1b0c25c48ca00db176f66140aa10e6def57505d488e97b743eb81c5aabc3e1281fe2e300a77ee96a6325
b6b05761acd08fc16b9ebd75d6db17720bdfafa974826539b97f90f0bd1c95b88adc7bc6c3e68e817d82817eec4726a25990
e4adfb870794a8ce8acea6287ba7968fa4519c0e0c319f1550a9cf319ff30662be6f90b10c0dd25fb539b98f34c9a815a440
65f467fec2d07c25d33fb3e2da63cb941875d584494ba381ca57d73f97bb65c8b7168afc3172913e17cfb8c567ea45fb6dcb
1e4ed402178e8ebdf230c645031ef17d7dc521c416ef834d42111f3f11a5d3a73a520684a2ed11ee8f191374f654ecb75669
d8bbf85f6e95f43b58ea8ad4ed5da3c483f45362b1fafb614922d35790afdd29529bcee8cc40d7bbeef171f879fb5cbd3502
dbf82c6a9c8204978ed80f0a04219e00f6cd0d66bd57e49482898f694162c5eb63ab2498c7bdf0acf87ed0836d064eaf1c2e
e295fa0fd4423b1bf0cf5640079c5f87b0bd0f3725d048b51d3498b8709df77309cb2a122e6694ee5f0f696cc8e0c0c14511
698c1d1db41cf05dc0f5262ba489cc8796fbc1f00aedb97213f2ab78eb14975544b04dc13bce82bf7dd34afa1170a1624de3
8d3dfdd5e0ce4fad806978ec8edb2841f37e9892daeebb2b99066b3469e56989359105878a50cc555e39ed7e789239956c1e
c539490916bafffc18fde6d08a959a1e36d528cba9a4c3f4f8fbe1e3f72b231fd3df47c7589683427eaa99292662a6783968
13c2d750430a0f6d9a42fdb20511b5783d93c9f58b5fe783bf0ea4c32fb03bafa86bc5a0e2f2a4bf7ef461b700aa20817a0d
b88413d1b50935facbba988a444cdade4f63829589f4eab213f129ce7935c290b0bf9ee53d4171dc54eef6c8d7f42e0b906a
7230227665f93882cb6d3d12e114397cda554b837c4fadfa20908de2cfdb35e22c096ee013a0a4c980778eefbf8cf69c2e19
27f0633b9a88efbd764e7bd4e7f5dedc703429bb48e01d8244f4a238bd08c52960ea2a75fefef80cf5ad357d03f8dba85ffb
d25102e164fc79567949726b24a7c29adf31511ad2dbe0543458669fa2edab6d6e3bb180a5280882ae26d6f6630b1eb4b7bf
e5e0c09c1a38cb6655603b0ca5f5c2ca5e67e7255abb0c8b7ad493d395d796b1489fca1b9d188fca470312f7c2d2d27f04de
ec0d227e3026a9d5a35589425fb432b3858d19b3b70840db73ae4ae39bd48d5d8dbce5e63da6f70aadaf1778869445950864
d5f9502ddee6253ec0b5da4ec4d37b4b402debbf9e7d37bdf748ce723dc56295c23b634b77032d954fbf3830459f84bd87e9
07727a61eb58fb2429f3aa2ebd19234d1864180f769bee9ad9957452b6274039c768b4ef012e171c1356e5a1207a6f4bdaa9
e662ede14ac651b722ee89b050b24a40cdfe6ee46e6312ecb38a3732d54269db70cf5c739b2a9e099b35d55cc7b26c6e2e12
bbcd24f295ce528f4ec2b085f7a41e512626c7eec638028fec97e1a856e2e19b6f0b7308cc5528df7672cc7feee2f71a2010
bad96e757b1d75a8883443db429694b54a9927c8482747eac14a20dfd7a25cc5d9f90d6866bd36eff977b0f3d284c08de9ee
a8176a91c03af17737e4b66c26f729b60a6d2228e40c239d211f51d78e1a7c614857a782b211d375b5df22fa55cef1a526cd
60015f0619fd95d14544710cf76b0c7bf58cb3f0d361914b7afebea1d6c63cce70fba8c8eb96ff956d41bfc356654f4cb5d4
f84521d7147da54d3dc48e44c9710b278f60d2dc53299774f03b91946f789dc099eb2aaf9de9eadc2821cdb931c5027a7be5
5bf58475d0640ea29979369fcc4be7466f92503e59da8ae8dcc29cb8c7743d92fca9822082c41afd00bb49f142aff711c23b
9f38945014e83d8187835aac2f13eed8c672e7ac4fe7101de18ffa7be7fa1feea0344434b9aadd557c81d8e4e712b2e3664c
1ce0aaac25485619eb43288df3b6c21092a8f3e93254d8a6de6d49746f0f9ab17ef0489b62294d961ae0bb35fa4dfafe1216
ff227ef8e740e44affb67440a5c30c13345a334efc3a9bd062cc60ba6f372cf40226662181536c6d567f8c9a9f8f6626175d
fc23cac35580eb141cf76ec89af5d66ad90fd2e16ebc9cd1331f77800097f64f9f408ca1ccc6a36c3d93e4e213d2052bdff2
28211f2a2c34e38f85a92d0e8b9612116cb89098ee7136cf9331f78fda9cdf8a25aeb9f21bc4f6e878e5e2e68fc312eab6ee
d77c254da114cdcb59e0a426dbf6296f21e3c12616034a139b4e57a6d677075602593cb61b70d7f7602ab042c703dc3ac540
d067fbe59575d9e30e2c92bd2a66b92c509b03ed11493cba4917c7c25f3b320e4820beefd10c84127ed18f4c279420829a6e
78a66ca9a3413293f50c1bdfd12746cc68a44e6aa88783b86c46560534b24ec4041b7deec0190acc3fd7c2405908d66ec9bb
7ed6260c9c9b310f32ed83d7e09963daf1cebb9ed39cea5437c0c059a37a47619051891324c092f6e8893929b528b5e53e74
88c404114656ffa5b846f2eec60ebab4819ffe7d5c540deed22dea32f9a2d60fbb45de0d23aa582bb135bc80e19a81229733
bc1bb94ec3f2da3197afd1d0df6700904cb7992e0f6e4612ce51d91d007210a9bfcfa7abd80ac4ade7ec56f0c8a720aba590
e0cc509c12eaf740558b3d06d42846620bac52be280b188522bbc1a54a939a18d4a9db01f7071a924bcb1ebc7037c416ec7b
c5f11697eda425e45b62f5f31bb8588152f0364e7d80665aea4b21453d4ce4a50e61bc59eceee4d9c269f283b08ed3823598
df1dbbb56cfba53053b59a52f994ad96ac2e99cdf0bbbd734f2f96bc066882a8c4a617e8ccbdf790d0a1b24f72e7cbf1de50
345df9b65d6e62980b2d8061f6112a082b83ae4ea94af475c61aade8aa27639dcb56d97dee9224bc4baef76454a460fe9393
599518bfa2a8287e7849dd9f17c70819823ad172faca690323f9a8bc5d831550f0da607ff5f465ebde66a3a8a9c14914a0c8
db5d5a1f323e87f6cf2866134451091b82a3cdaaa547f2dc000bc66e1a937fd4060fcf4c45e0a47c4932c9ccf05bb3eccc46
1cb245ba0e07ef02efea3c27237a7ff3520ae98945a348933dd9f7580e9966789e4e830ed35e3e6b952c28169ec2f25e6c24
bd744306232b160ecbd8f9d131c40ac224e58343d79db3bf908e8c2509bf83e9d65c8196dbf92d5a7a63e32a36cb60209d70
f96ea0a14ff1ac2d9d757f2bfce99c5c06f234dd93539b131dcd23b84455d2300b5db042c2996f2de51eb55d45cc38a3d7a0
6a3cf2fb594af12a7cfac9862e8de35d8c65e5c1de1082622b984449def186b558b37750f03ad6175e86811fc69e16dbf690
09388da8a62270718ecc9c5f5cc9247194b92fbcb1e2e9203d7d1503ba56e80444fdc21470d0efdcccae61ef04391eb75e42
223989362afc44bea7cbc84650c778318880ea292cf82e6cc8ab545a8839484c4d99dfb9b0d195932c88990494df68cfa1cc
44cc7bb3748bf6ebf1eda37ed9ca45560f25ae6d9bf1e98a580be838bef27a7f1d9bf3851456497f84e30899afb887736370
be5f662c2e32cd8661d05df80a626e6534c66956a79b15ac230c3e529130466cd9383995aec072dfaf3d239f6e24f257c577
5b74179fcdac9cc8eacbc8732b50075662a27893c0880164369c4fd872ebea7b967036a16d9eb211478ed9cb9a047e358f26
5d638d111dc8534be5ae41ed1e654ecfe866b5dc59e9838043cb888dce42e66bf76aef58f5db7efc39f2c9e195c8659551c2
1a2103808b6cf18ed8cfe2164e9b4bc39f1cce33c098074cdd1802d22fddb5771d2fd71df02a81afd8f84aa0cad0d0bd7ee3
4219a5e11acce935da256ba09535b308d73bd0051a9717d3c267c469c2a40458dbcc5565fe85be5e507ed5750499dae1ad3e
6b949c5b88ebfcbd089fccfaecb2c097cf7def208052a53e8bfd2bf081cba746b3f2c9535418427181630cce480abbcf998c
3055a9f9805e909937f2d976da9628451618ad1e742606d46177954e4355e7a10087e0aeee68515df030b868fe2b61d8062b
fb08d9852479e1ba5c6695fcabc47bd1239b32c25537c4fc3513fa915a0bfa1c7e72b37677ca173ae20d790ef6dc248fc7a3
6e5530b828f2e8942b35a796e820ba69d86bab72cacec1e1ad85c7c5c33e4787bd1cd9daac80eed8fa4fb5a8bd01bb28a377
f87bb077d2a729d50186ab30e86ade4937d446d8a45c3ab70bda54fe8ca2ae723c81ec5ebe1335d41289b8a0dddbe28db57b
511e244ef3afe1b66aa8543e67e392f2191c8e2c3e1f0fb3da85f8a6a4876a36889495de4bb0531acc4d4005a87884613202
02e8583ac1032f477dd6036ef8262675d552417212c295a394b83003852a64f3b0bbeb5767949ae25765619511d16d19d8c1
b3fbad1e67a4a17ccc382556868c6524cd49c734f73cf5abdb4fa6f0a42589cc7c3131c94c3ae0fd115782e40e76dba6d17a
594a823d30029e0cb415af155f4c897e159b7f0079dbfac3c97fab87ffb1231923213d9ba1eb0ba226784599fd1c3f86d5ba
f5e362722fd6e5718753a30c583cf0e4772e552dda65db1cf3554e613c2ceb4989e831321d78fdb4069ded3f171077d2a729
6461a7cae1f159682dcd35b7bf45fb4598baf19caec7c6239563579898d73e7b167be0f208cea15dc3d063028175132eefc5
179f4a78a251d1992f3032f7e59700e4642de265ccf71bbb93de6e2dba72b9c1958c809870c9be01e720cf7ea6664f556b7f
bf7b3d4ffa6c501f0fdb57aa4644fd68a40aaca8578fc77eaa964a5d84bc904657ba02d78fc7986d272e62ac0f9032d5b928
a730be39d7efcc03d3e29d047e91d1d8efa97494dce692a9aaf41c95a839b0b8878d17d31f1cb1988d84ea487dead273a722
994842f7b3ce496de390782aa5467af28c96b4497948d3ab0abb5cb551eff39afa6491c6e2d7af1d1e974a33485c83f1ec9a
e9acd4f38e9837cf7a3a1f0f3f83ddcf0effb257f9d5773e1fa9e2b02e07c4b71d71b7d6b0f632c6326cd870ed21239c1e1e
4d6a76d92c1f7c5d22c159b088ce1f3849b7b6e396999bce64105044f751f12a587352fea8475d46486b783f13dfc59a9364
d071d3bcb4cd2831041ec7d941e01eed4041a0e645e8c1e580ddd782cbc84aeab5cc0d7c2139b870ace05b5aedb331f79237
d7519fdfe1b1016612463fb8c8cfa126562741e133c725eb682a0adbbdf9c91002dbaae56ec347ac88a90e53397145b3a05c
929a5d3f4ee6417f76f2762ace849b06769e8deaf1b9979b894d8c9d8f52d54a1aaf6a359a813c65d9d4e9549291e4971a53
5bc29a91d79dbd8e50f74045da75e792f4d18fe513e368d7d9dd605e44c65b5137fbb60d445973839b89e3cd2b24def85b62
d44ac6cffa53b25798f614e5cdcfdbbca1ac11e83e764e3805d88e47c123c067c43d9ca27eb6e380d0b62c72d5a2bbfaf226
e12cd82d9c3e717424e06a57394371cbfa696df53403393f7faceee913032cf74073d90e4a5c18ef07279f3a37f8b96274dd
d0aba345ef62d9fcb3418f7254ff0d94155e7a77b391bb1edf8f7eccd74a88c2c59960a9cf0deef22666fd79273cca56d45c
46a5a9b7e96c22f00b57153e90beca997a8565eae4da1f2e6fd27ae0ec00ead8c342ace4b8818965b8c33a530285b9bb1a11
8a6cdda3b2010a9f147747d502074fc2a60c92bc98e8e732eeaca5cece26b42976244f6aa43ecf0f8c6fe8acf7897aa97dd5
2158c7aaccd7d488b27cb4ccd6bf8b1e030942fce421547aa81263efe1585e68ddd674b8afd44a17eef3a2e0e9b1db58f114
cee62d80e3ac51aa03bec0fd30c6bafd3eb2c9c36cc938f112179f66a60d0e862f1d5374d8595c9b81839382555d9c42a133
d805626682426faab833033025bd0a0bcd8a3e7c0d1d50c90f2255681f9b891ff04708c690fa983fde6606a5b8a36c09ebd0
d68a86b3237b53629c630cd54db656f0d8e3dac7d4aa6291c5c46c19dc9c7082752d68d13e1e7f8c7d0b8dabd4e2050fffb0
ec11f3112ae6aea142f17d3030f808cc260d67f3278b9437145b55557aeeb865d12c0c52b4e4f1251cf487471f1d8ba0e388
f2e114c7c71720720ec3590ec1d482f2a559a88442188f0d5ca9c2f05e8d11fd606ba48ac9de27bc9a64a5c4bbd1612ebd04
f4b5869931b3c02006496db59be111c63d50cae0ddc4c3e2e532ee41c714c4942ef790255e2a15c65d433cfdbcadc4c34d3e
a9cae28d69bcf184762bedfd02965717614b0883e12ae54ae717b52ba53b68595a4d1e130b0994b232b8cc685f9c46f29ba5
439443f5591c64656dd6da9f19ec7a457e457b789c687757202df917c1ad4854c7a4d7d1922125c68bac0959f64b5d579da6
2d59310acf134727f925ca2ba2f84f050d09a86cd1b99d76c94781904be15cabc22e9f03a633e658fa9d343f85f0b978eb56
e584191738262890389bcb352724747ce0007c11aa91adbb76e1fa3e74b3dd2a9f791a97e61b5509264a62cf897246e85844
5146df111b794b14c1f734ed9dd19bba74b13230f7f971617486aa40342fe87b47ec96f93b10c9ade0f00c41aa4d57a48bc6
6c20071096c89096913f4a6eb10d3e0efed2c545ea86997e36d94446a6c56ce2d4be7e7234e63dc0aa9dd6fc9cf845da8580
54f8e70e7ed25a6d3b63aad63d379a3d65dd8e0c7be28f5a226d1a500e03a61af4b936a63fd4cd6356471d71b59c8e54eefe
9d56d0979a5ab32164c87d5d13d89747a944b1f45df2b335b6da182b137fb68b907608dd6c511f6cd7a4cf42cb5351d6b0dc
b4a55a2201fc37571ba6be7d183fdc117890aaa5838c49493f048f83bb4c4cdbb85bda443f205e22c6df43762596e9aaf672
278e60c78c6e9bf35a0465cec4625bbd54bad497e3b21259e7623e8851acf446533612174a424acaa1a1a94ce26ef8b4a7c8
a4e7f6151e274de2f5f3368720224f3aac56ec3dbf9e22d6822245305e04e82383230c68f2f06ccfa70bec0822c7789549e7
631b8660d0ddc0141ef831ef70bf87b053820c85d9587f8403e0c22ed18c8ef29073d8ceca4b516e638ba9ea5031d6d6ddc0
b169900703543ec092902aa327e9e4ada49529cfc4dcb713ea05cd81fc20bfa319ba67256bb64ebf204b5545bbcb0c584a0b
0971492b6cfe989c22f674efbeb687ea0a3935299d0ba6f131a56df611f3c667d730f3b3adc34c5a8874851b3a203aca632a
1572148930df3ac6147e82a3d5b7d640332b196ab3380d631bdfd29750d48835cd8d6df74b1f9049548d581aab2ffa755eb5
775dec4940a323fb95176d249b1219f9833ce0326ba854f8c35ae6f5730ef734d4b8647deadd4bb505314cce09fee294d12f
945063be1809b29492db707af697a7a10254caa6901ca7e24f321d778c71d50ca61241ac6ed2d15622c7d5d07e0f7e1097fd
98de05d1b483b8c703769d64134d7d82d7e71f3eabf4c9896622621f660118a65aeb127b40f6efa35fb911147f29cecac8ea
16dfa8ee8d0b9852ca23319df14de0e986f36062cd1d49e38e5e882b15878f19464dfae77954f508e00150133bac69552371
2f0d8480ed173fc5751111bbdb88681882650703c52a1c688e9ae84ca6505db3cb47a0f5cc0f42bc920bf76f37c882c4506d
d53aa1fbbbb705dc4682a2f121925abe8fb032d37b39ed54b73d62bf98b495d75e0593fdf2d116582d8b4e5130eb47bc15b6
e8766d89059f8f29167f7ba08a28a915f5a4f7b4ecea40d8093231bf528ab21554ae6461eb730a71e4549c1c4a0befb3e181
049db14ba529f01d4525d0175ff53e9132b1f887137b53818f7c6329518311b043a595cbe31f5db78843a861a8ac901eb5aa
4be5d4b5b96810ac917a954095cdf8058fb20ec7defa5e5983c3a1bc1b818720733f4a3d8b86a323346ba4ffd80ba2a5313a
60a36f71b400d9b715e2ab3cf80acc55a33fd1744d936d942809eace46bc1dd6edfb007498fb903419bd76dedc6313eda80f
de9f48118519704443271921948f5bace00d6166d137d7314c87100421327e3ca411ccec734e24dcea56385f72360c33a9e7
78b040c6d6538c21ad37f3f1463d3538cc81293fab19ec1596f107e8a3aefb364be22f58f36f365263dbd91d7a2c5edacf29
b95d7a1ce4e19a9b7f201e7b2809f8e8846971e554ec9ea003fc788c9ffc54bd96d7545117ea92546966045ec456604ab8cf
3a364b6323c0185dda42d5b23a2d4b626bf69c798cdc22bb8474c9996ed08061998e29c5b44234f242366e6e7039ff6e513c
b45229d9d737a98a9f2bddbe65c42cbede964234df3244976d3185d78e063edc7b49d129a04cbb064180b87b6a575da63209
d42505fefe18e1a95717b30bfd48f47d378f81f51994c4b1cfe60a3f668bc98610de57c4043903e6d760d6ef5122751a8bf3
3a34414db2ccf4d3c703b5c9a6b0ba484ef458064dd90a4c488622a8ad7b1dc1cf00fa20822c292125a36b1a7a39bd41626e
cd9c75b6eeade412e5739e4fd491d8c90722efbd4e47945885eab882d5b527576dc56027050d270fb77e760fee7f96bf662c
196edee5a87e8ae5407912fd2eae492f7394764528514f2df89085ec4d53802b99b9998862b9788424a7cbe6bdbda7f59449
00d799e45fb3bdedd3c505858f10a856dda33124e75873519492e9ab67a8de43db9f486d54c671d4dd371a3be3ffe51a8633
abba379bae18e3729226e57e092e82dfd7cc6467d98bce5780d5b86cc739d821e6d60cc62311b1f4597566c27e57c94967e5
cba3b4af873cd85afce2ecc687d472df2c72d075228b6e53820625dcbb32e89e1c7227b0bdf452db1b53ae056971828419f2
8a44f25bf7190658a66872ef663e86b936ff5a97fe39ae15b16fbf38f4cb9b0e9e6483e73e97944c34b90dd0ba3f04e0f2b3
3e460ba11050255bcd9d236939c66f76d842defb54075804fd706c93531e4bd83725369dd74e9efc7aca50b903be56f19caa
4c7eecd08cc23269064c201ce67dcd7c272739536097e1fa5957db637411aad958d9a05938c359cbc1cf49766ca432eca516
61817a86e640550afb72479364c1351e8661e17042c80d4c582b14a19e2883451807a3a7118655a6e7023ffb6cf1e813f6a5
adecf7f36a392b885db6ca6f680c305a71c1ef7ee602bee0256f99a656683325ee4fdb95f58792fdf0469be45805cb35d9bc
52b32edc29e9bf8d146417c7bbac09fa12c1c788aceacd334df419e60376c949d92b82fd4528980106647a64f9730bba097f
1864f321ddfd053ae954fec20114034e8696e7c9aadcc975f971d9a4ee47d2f9fd10144cf2ba936299e3f28027622160025c
1386cc01796d0201a90658dafad814a0c0e0d65748dbe2a0d81195b1f3631dd8de5295de4e2832119986055d88521dbf042b
1b147ba9e2619b148fa0a42d7bf18980db86e093c1787ead5cfd71f60e05234be33bdef8b8a7b3265117cba13ed071aedd99
b2fe60dd9ae28924357d56ba140d1a97d5bb34cbaca7e668626c7ecadb763c669292f86499dbcfb199fe909f8a839e3b8bf5
e71d82d4ebb959d9dfe1de93cfc1bd342397b68513bf8ed6f72dd13298c9aa4dc55310d64337c913dd6faf8919d3e2b04eee
a4bdef46d5bf41baec59a8be66914fa64cf9ac7a5652b7355688288956c7833e99c2b4a4016bc87c5f753a3f843f6385f2e4
17432b35b02013a52d2dc2fbb877f89da4e5407c535de2680c16ddc0506934b03a8a0c7471bf43ba4993c07055db57846936
385a9d9447c443ef029e53c5168960755248881a6a15ec770654f2106b6ddb73bb471017219e8ad7719d3f83c89aa0aff652
4544d34eeda4792107c9b3d3df52b3265460f3d97ed97c685cae6bc222a07f327f27cc5cd56e52f86c9faf589444d7aa671d
607af1f0f398062a8fd2a82f34dfcddb0d21bc45bf302aab0c716117474696f576d04fe9dd220a591dd29a8a1fb2347644c1
b3f049c77d4163b8fff2e92b26b1f1f841f0b65960094c95e3f129b3b321668e26113801292e65174d9d551087502f8551e2
29838f546e31c4a5e5ccadabc73736e2bbe90db86933c2314f701863a4eda3cf5054a4cb6cdfeb4ccd2012e45ba444857ca7
2e581ba54eb9b788bf6ec8c13a18ec9a45072b217033599544e9c98ab8d5f94a92054d940636157bb89324662e82a51bb69a
6ff3ca42a57caa67f35f69f4e66dcaa1863eea3aa872dab8a404a8e35b6fc7e7431368a0923aaef0b4620306557fe04f777a
86b8e182db3c78ee3b2af59a5d63ef3061f12c4f13cc24bf136cfaa060300d31bc7c3b24a34d1aa0eaf38322b2f447065600
27ab63733a4bc27801cc13e933552131bbf253b3f037cfeae2e094aae93536ede0e7682c6b8377f75853643aa77ea239b40d
d381de27614a03b23344983859eb7ca087ac5773960e2ed9b4ca8330a847fb28671e0d1703a12ff035d627d990fab86d40ae
3255e90b648101605564b43acf529d8f693786e70cbf6a6710eaacd8330e01fb42c7e6ab068275bd7e6e3c28cd99112addc7
6d5f153e7264f1e14b77e94d9dce054d2c8a91c5ebc914c0dde2568a5a872024e859646746d39111a7d295292df92bb0f4f3
cf565facb0debde7820f0bc7e11ec6c629fd7d53e897563fc326cc49573a4fa956c7dd4607209ca9c41c1b0ef15e63bdb3ac
ef0463cbe9e615380d5dd079ee54f206eaa9a8d32af4fc0fa74495bb3e00a437b4b4c69b1a4cd9c4754b116f71f8d2a4231b
52ace79a590ec0f95f807cac46c736eaa84ba3e9a9b12b1231b867f337301662a32668598b1d0b36ec821419b1fb1efdb9b9
6de101315d828cea341d61de461c0753bbee08d1b855e9a3f5956301f31cbe6c6707511fa65e3e6a1a276c4301aa497e954b
846d079f04916e925f2ba4d0e9645f40b2262524493d1566b94e7ab5b9ae5045f1f7c458e4bdceac0a26b57252813f327498
4874319006b49dbe46af5fde24fab63fda4cd8342e09c3a7b3086dd91a6811d59d1817dcb98fcdaec736ba38be0c0bfbe8c4
2ec75964d37e7d3fdd16980616ccf7386bfbf5436e4fb36678f0e48946bc8c1d6a5b463e15307875fc57878c75d7d71e994a
1804dde244a75aaf0c59f80545be0a023ef24dab8359c96db33015c68667eaeb3cf1645f99191a743a27fdf4c089f682f001
b235d4857af45d31d83c58287aab870373439d04ec5891a53b45d6c36a20d7d5d79aa574a27d137a6cf6d67e74e6e5d41506
1a71175c5060b5f1398151c706dc069b00e36d57bc60c78380b82e1ed9475dcf532fd8045fb6acd70bdffb0bc909c3439345
cffec368d3815ef0e92cbb98ea6faf0514a61721f623a56bec74a24fa682987712d3d23f7cac09783102fc1a6dac33764e4d
049df248667229bb364e661ba04679455b8aaf521a328c5e8c92a4d29e3be82120c2200998f6a1f2a50d82c51a3cc4a4b489
f52b512d5c4fa69276e64543498405132796d989b1aa137d95b1923ed562c53b5531f8d963086585ea45b88ec07df0190f1b
282cf23f53a8eb1974f258b27f8de1a036b8c254edc4b6e24291fbb0b6e5156cc774f0cc9aa89e11826d65df833b708dfa49
168aa030ae10ded0090756af6ee9e0e7eb4f3089b2c284a96cc95ebe59e7096af44173e64caaad25b6d72c055bd275e1463d
4f8ad1e1dc3b7d275e98bd998baab9a76ca3cc566d740ddad0592c817ed148f1904c68c68caac3d449dcdcf5d1455a4d5aaf
956f852131d528dead5f5083f536e1e4e357f11bcc7cd05472b48767c5dd8cfa14e084ab7eda465a04a5b0d7b39600f5a8d4
05f258af935e53f3e6e0e79699168934b409e36f93b48f8f77ba92807040341ffe9b07807b4812dabc9ed3f93c8114a6959b
5b79924b4becf0a5c16e7b976c147abc937118577cf9529a96fd861bf98d913e4bec2a51446e6ee870b6958a0e26d2e5839a
afcb13ca737c14c3356e4dc1506c12459212ad38dbcf07ab68fcaeb9cb269223969c58c651eeacfdcfb4fe619ca9702b1f50
ebbda7b4e8af5d82204b3ca653c8f744daf63a2ee9466543afa05d0ee817c16e54675c3f3ef1faf37659234599ce59ea4863
732bbe00d4d242a690b32accd9f29881825c0fd6bad422c9f70657d9306dab293594867662805fb3d518f60f47336e458915
c43cf65203d0868e1b039a11daf57235be9d659f1c42f8f0d40861f6d2767391c4359e1cb5a2060744afa64a4a974f1ba918
23baf729357f54a5e011fbf89a178944a294aa62990695d5c3dbab9468c668d0ad864e7ff47c672ade84075decb25698a747
2c3aeb0e95ad26805ef8f0cdea638ebf170eab6b2d5cfef8669d55e04fcd2a1350c64a31b673e682f0b36ab0d72276dd1cc9
c3d1b58d2463ea4059f014e339e09c01385a2f9e347a20ee9e57ca5b224d023373e72cb72e090ed1f6abe10b24fbe8ba70d1
b60c14eea1d546b4ac2762c501fdbb571af7e3fab25ea16541afc4ae77059716595a3c8cb5101ac9899ab7d6cbd22fdf7884
4884912e247c476e7cc2a120272a228b60b929f62f76a06c98ecbf1740a108957d36a68f704e7571a9495b74cfc7f892c345
c7adc295ce312c96caa22c7717d5e1883630f2f3407f95cedbf3a55277d36e90c75a6eff44ddb637b219f210f185ddc8244c
f6fe12695554e606b92c4a98891cc0dedc2159b8e7b6c93af0426befd7402f10a2d5bf6e73082baf1b4ebcf7653e2b48bad1
34aa9f23d5ebe459b642a3ec53452df493901f11ffe548f7c0a3f87d7c036f13b8c82fd8bbd4390a5d7a8039fc4ba3577a72
faf24cb686aa7713d31c42d08f8899ce11b932585ec5296bde8ce58aaaf3cc14748425beb38f21ec4146d59caf1bd1e3454c
e2c92e62c48138e71c31b0d353cddf2e191b307dc62c1c023d460aca0938ab299587bf8e5d6d227467ca76ba1f31f5a0897b
7047fcc0fc80ca96fac2af62900aa13e146668ae0ec7c8f843e2402e4cf755e332d8531735c8c389606380c655f2cbb4897b
b62e9c72b438d13027736a60149bed07467d244c79d8c8a2418cabcdcf3bac2e778dd5205f379c96af8317b52b543b925d0d
0e6b3f23e32810bd5695f76f5c27a6023104d9bd8fd06fadaf9a4e6273ee18fdcf611f3a3e86d39fdb4581c0b0b89fb73422
bfa67e5d8845b9e8dcdf63200b896b7a95cf2fd142980bbf4b9ad2de7f417f4717da4a7c6daf0b38dcf4ce7d3cafb2881824
2d8021b17928e0bf18189f1875a6ec850bdc5d1e1b9368df5b0a48b8cb9e2a3064efb5dfa25ea4ff3c85b1c12b7aee891cdf
86c3bb4225f9c940d404316d2346f36e8ce1a7cead3ed5e7c8f859f35ef2c8f3d1750cda199f2caa27258ae5689d1396eb32
d0f03b5946e4fb22ee6a3d3ce5113a7f8a6a61134a49d3fcd90d35bff66ad733d17055514809550061eb232d83c6a5f4e58e
3f7b5fcfc71ec7e57cba6f21ec9e3d9a47bf58c5720ef41ffdcde83e16cbc62e69b22e5e8ae2b734e7613be768068b3c2ced
04acad9862daae585b5eca0cf58e18874bee7e5c466a178fe2cb56adcd67fd722c72828e8a361e809687995340c1c82830ee
e6cb2354737576d154c89c28b545c5915615f1905ae1d46a1109b09adda989884e77416dbb9745f1f0b5716a667d484c63b2
085508a72187328bccd6a1af78aca7887b0fa5e58c287972f6ecd7ab00fe7f005c4797de
>}

\EmbeddedPDFStream{schematic-41}{%
/Length1 1551 /Length2 8556 /Length3 0 /Filter [ /ASCIIHexDecode /FlateDecode ]
}{%
78da8db40554d46bd7368e0812a274490ddd30437787208da4d400030c3103ccd0210d924a4b37081202d221dddd0848a708
d221f0a1e73ce7bccffbffaff57d8bb586dfb5ebde7bdfd77533d26a68734a5bc2cd210a70189213c4051401c8aaaa2a0901
80405e2e2090079b91f12514690ff9db8ccda80b714640e13091ff1120eb0c0123ef6d7260e47d9c2a1c065076b107807801
20011190a0081008e0010285ff1308771601c8815da19600552e80321c06416033cac21d3d9ca1d636c8fb63fef30960b160
058084850539fea403a41d20ce500b300ca00a46da401cee4fb400db03b4e1165008d2e3bf4ab088d920918e22dcdc6e6e6e
5c60070417dcd95a829503e00645da00b4200888b32bc412f07b60801ad801f2d7645cd88c80973650c45f766db815d20dec
0c01dc1beca1161018e23ec30566097106dc1f0ed0565201a83b42607f05abfc15c001f87b37001017e89f727f67ff2e0485
fd49065b58c01d1cc1300f28cc1a6005b58700d41554b890ee480e001866f93b106c8f80dfe7835dc1507bb0f97dc09fcec1
0005694d00f87ec0bfc7435838431d91082e04d4fef788dcbfcbdc6f591e66290b777080c09008ecdffdc9419d2116f76bf7
e0feeb66ed60703798d7dfc00a0ab3b4fa3d84a58b23b70e0ceae4025192fb3be4de84fdafcd1a8204f00385047985780110
2700c4ddc286fb77f9971e8e903f4ed06ff3fd043e5e8e704780d5fd10101fa815e4fe1fb61702ec0a01209d5d203e5effd3
f1df081b040258422d90007388351486fd6ff57b33c4ea2f7c7ff9ce5077c02be03df74000e0efbf7fbe8cefe9650987d97b
fc1bfee77eb9f51455e414b4d9ff9af81f9f8c0cdc1de0c5c9c307e0e4e10702402010102078ffe1f3df6534c0d0bfdb00fe
9bab04b38203eee3ffb47bbfa7ffb4ecfa370358fe56072be0bf8ba9c1ef690b01b0fccb7223203fd0e2fe07f4ffccf53f29
ff7f14ff5de5ffc6f2ffdd90828bbdfd1f37cb1fffffc70d7680da7bfc1d70cf5a17e4bd0254e1f73a80fdef503dc85faa55
8558425d1cfeb7570909be578234ccfa9ecd9c203e2e20df5f76284201ea0eb1d480222d6cfee2cc5f769ddf5ab387c2201a
7004f4f7e3729f0504fe2fdfbdc02cecee1f10c43d31ff728111f76a43feb9c6df1872afa7ffee431e6601b7fc2d3c1e7e01
00d8d919ec817d7ff5f7881fe005ba57a825c4fd0fb501dc5c3038f23e05703fb30fc00aee8cfdfb9a050400dc32bf4d7f21
6100b7fc3f485010c0adfa0fba5709b7f6bf881fc0adf32fbaafa2fb2fbacfd3fb07090b01b8cdff41fcf727dcbf160ee07f
2cbc7c006e88e3bde4ef57f21f1b880708e076ff1f9007c0edf907fed7f0162ececef7dbf9c3d9fbcdfc07ff798220107788
05f6dc34dc4234c8b632a8e9a2429ac28d7363580cfd20f9429f877338cf0413d9253f6eba12ab9d9e36ffa25861ae13a460
62dbaa26e37491b13473ecb55e4553edc177c649a3b06d4d63fe76faeeecc1649cd73925dd346e1d4ab65ea20cb548a17317
8a065504de172c494beb2e6d465c6fe6ca8ed9383726a2220541e5e026b5962f1545692a2494c2badf57b590f5e6bd2f37f8
4fc89833571d6202d7228dde24a9a5d0bd106bb9c48e752bc75beee9ef5dc0efa46a8bf057760dd838640841e2cb8bf652ca
f16194eb2ee43cfd9559e8ba9742e065ae48284ba2ad9a4dab4586dd2b9b135f7f4066f48b1e43df36ed8d4d1cf61e196e73
4eca56be0ddc2fe99297bf3d50282973922ed43adccb526c5141cbafa57e26b47f9cf47c0e7db97fddfef197d14696347a6f
9b8f4547af7d07a4b5236c93cb2c0c39b9f4b220d384d99d046c6e16c0c7e7f58aa02431f78caebca3e49660111d1bc0b990
b8e6b5804ad2918719b51b65d1a3185c4e0a1725bb4d7c69d7ccb6289728edc895216848b680461125a6ae4d0425f15ae849
3101f59be935721126f919e6696996c250a1f922af72b5f077be5d389318a7a5d444c23488a7e2cf332ca99cbccc57a568d6
080cfc325106e99d21cb71f2983117340c6c9eb029ce8f1d8db3ce9c7830fa0209b599e0f4ed378830e61ffe4ce8fa42c78d
5428f86427bfbea256549ef136642727d571b7b69a7d92c57fcb2b78a26b4fd7ef37bff3f6ecd21cbd788a540da24125f5a2
1143d722432b4848ce4648b878effa3427cbd1732d2b45e5205521fdb5f1d47a725393c32b0a835f67d9db9753d5c61df1e7
80e3b370d3231b5db31925d0bef2d9facdd05c71019b58a763a697abe4398e1165a3626b92f1c5ce4e9a841c008b1c39f458
a780fef64068883490a36cfe1ad29cb3adc809609d06c9def1d16f0baf4ed5c8f82615c76fcb45289963de9e3e327d2cc011
2d2eba1001e718553b55524db37d6bfacb55178eb2c1d88b6f935dd8ab097d00a84fe5d7175bea2efe681f074c51a1080cea
c41a38c36f583a88255a451703c7d65f6b88a4565f9b4b65a03ad0a3305f90ae1d2f588d544794d241d64ff969598fa06484
89cf9445d20e79dd18f096a8aef29f224213363777b79e327ae7f99d648c7d7e3f55cc78f724551d8faefcbb9bae5b2550f9
fd6dba11a6e97c857cd30a45cd0faba3a06e5c0d7cac1dfa7e4da4d3975a17090b203b43bf45a6c8fe141ffd20af16bb8def
038f07bcfc9fca6e56f9785e047da809ac3e73d8d85dd4f3e4a2c8b1dea0b58d92467b96ed3d31e21a2d9c04c147733a1b58
ecf65078e5d43e87dbdd9e846a6f1b8cbd73b651881297b64919c36a3bfebe00f150c2beb082a22891a7b2d87a6b2c19ddf1
e566ab14a62ab7d7ab46f31f2f406a7a1937cd541457263348423fb942aa3e86af47a22b9d911c1c37c1a5bb3532069afcbc
c98f7cbe680ab620b4008d03de7c54e0cefa886beaf23071a1990fa0842c8383806a00f391a8fd342f3d95e38b37cf521f8e
3db38413be8901f05079b569744b4e2fd85df6df447dc67be1d2a9c3883a4f5cf4ba915fca7444069ccc76803ee9a513db14
931a48c0d953b788aa981352026a53a178c3aff34501b8733d1e24b1abaa458af360067714c46ae994710ee6109b09a0d14a
4c186f972934783d679a5d9998555b3f790112cdd1fdd0a7477cc566acff41f2fa89e439261dca0bc3bab0c010145b294d21
e640c852237255bff1b3ff769dd802c9d3d9cb86bd04bb4d3cafb33e04a7216a66b6839353b073a71281c3e2a4130a925b18
9c19a24614f1c8dac2f080645221d4aa8c4acedbbe6857e5adc1659b8cedb975e95570456aa18f59f961c05a5a850d0369ca
26434acccbe75d9c04cc621dd6a60afa9e220eebe783e7d7adf431afd2c9456a7db1c58d8ae4cfd64f274acc7b30d07131f6
e68a57dcc817aba33d8eb36f1847dfd37a5915fbfab3b83dd3f151ed0e12f885492843f31e883a9fb3bfd7b554c194031f50
ad9001caa3b8a1907e747f1e6d4ac6c82d593bdd1161c18fd6e77ea47cf5706c4869f4e58da86fbc2813bed1d53b162cd547
9e0b76da2c610f2eebc29573c740263ab6ebe6db9040019e002af67211c9bae327e5bdbc4c77eb9c87f2cdfa47dcaf9a6e72
5ddb523d6ba55efc6c89b4a2df6dbc798d71b2c08e50782b3ae8032c7aa199596e44d6ba9063889746413b28ebbbcdc778ee
f7f5db1d1ee5b202deddb562e9189eda2f9edd673ae1feb1c2014a6f2765b6fca8f8058ee8d7ae3f2ed11b2ef133d131879c
cd97fbe007b7597f7cb5b5e5fbd47fa5aa71f5f943569f7ded89c0e9fdfd490a0fb69c6e1580589b76df682fa537ab995f97
bde28afeec65df7ae5d9517526974f1bb7d733e3bcb7d3f664653558a46385f874240c9aaef2e313bacd94cbdd06995203fb
eda85b3e94ec8cc29713373af8c023f2ee6f4b77aa16581dab9d8d41cd20939b729c19fc49cb124a03326fe6c0b8cbc012b1
5f2706f89924f278ea7962d29c0c68956f99fb1623dbf0d48d488a2e5b649464d1d6035e5778bb450aad54e9a44e7cc93ace
d392fa24c52593e95384961f299013e81fc0d39ffb3248bda8eb962280bcf6e713e3cd585db51ee97526a5c9411e2dd0f1f5
ece3a8f945a9a22f18033299e47433df65634d03d248adb3165d64037bf462f56d9e77857f54e1da5f657c69e1ec4b89e87d
259ba2a43548561b5eddcf3cad3f6b392c50c69c6872241181faeeec9dd6f21717dd1977faa334090f878886e954a6e547a7
ee0d8f320f22e2d1dfa6519f607a6485d85dd6b4dc4ddab93f1f17e4358dd89b7d33f5d904986d18f0567b84713171be443c
c9f6913bbfbe9e815c5ace36066dd8699d188bff2564daa64320ea4dee679bdaac9a19999a6b290c62f2d0ac4d1ef5dd3e2d
6216da1db217446c150be5d4d137c975f1c88e7e932d8cd9a73ff7e94469816c14fdd88f50908a55d7d82d77a4f6e542ca06
b86fbd63e769b01b4af85f084704c5501d4c112cfc3c7248c33098f8d03308230228f49e4b704ab2d6d0f5163eaa15d68def
794439c15610e566fb649270c0fa62603e813c31129ea6b2d1c668e08b6aa83a70d3061079929f51fdb66e10a7b17beb0b0a
2078e4f0bc393b922863bb895683b665c43e0a04f4526843ee32ebefe665bc6bdc7a81da29affa7046687c6503793a3d10c6
f20b25539397c6ceb0e8315672d852ea6e365f65e628d4549864849ac923370cd16f2f3dd99fc4cb16b68a2ae56ec72185ea
edfe657d813b2b686e6ca23f7186557d5a4c3ab89d1c5fff44774a37124d2438c4ded8b69dcf5ecb8ee00ad7dddfe96d469e
d1ce0d9e3dfb4ca03a45d395a442bcc11754b0ded08fe70e81e9bbd99546f619d3588212161fd8aee520f2f9d4de5856bf2a
b21e28ba3e152034420f0a1d7ddaa623b447d7dc993e4aed003a7d0dde06ded87c85c448d75ac9f64eb1a3a61de4bce55ffe
3ed0c38f9105dfe9506b497423183289b04b6b1221fca6f8e45a4c5b9bfeb95bb6ab400ff6400f4665433500148ae9ff38f2
a0294af1f14fc822fd50c01e074089a6dd74f0ddeda479b9401b3a5917f58dde0cea938c16f3e0171dc469eb762557278127
f4a1c19f09b0af1f202f57060f3697cdaa666fcea9af0811d13585565c6f4ada3160df6d2c3e0edc15e5c08ee5f9a30cf962
e819307ef9f0327f29a82ded541d6e2af00d39ca45b59e3cb3484fd6dc2eaade79c5b5f256848b8efc63506355be3b3665b8
9d61cbc9a7e55946461442530547e5420df1e287f5bdf625b8a10bfecb7bccf3dde496eff0d9bfbdaa0a2ebc290e14113d46
e91a90546ff0e4df243ac7531f65453c27ebc6d11c7524d978746b72de590b1b56a9a245c9cfb6246bf3a4148be7a5698f9e
b5100c3c6cd9627c8a2dabfdeacdcfb91218b9dc1ed651c4358fe02c13cf473acf97c2deb69f54066ab61c498cafb4e63f8d
90652baa2c7de28a101a75308c3b34bd0995aef5e1f45c387a12336f79202aea7e2ff9f78e493d47a200dfe964b1f191ef8c
adab21ceee5cb32dc205ec8b2ffd4bee8abe17cc7b4977a3f0919f7b46b1665a2480f35f87a8006c19451a1e72845a4a16a1
c0396792d195607d341e4274064eb069ba3bcc6fa31717b51472a3d309178daa9a53b64f9b42efbac96cb9a213afd2cea4d2
12def9681edef6b03b642ad0084fc94d650f7c382c5bc77a12658f9ed34a42fb617ec340bbea2604a6f6307f3d2072733591
1ee4c6c02550ade49e8aff70d09cc66f506440ae6f762ce88378d32ef39d26c9fef7100bcc69b697a676af79be21772be621
158cdeaf6d9f07e5d179bc83cc0139def18fe9153e39e71a38bed4194cbf8a490b158b327a5af0826fbbab4fe988e881ae51
ac4e2ebaaf6f897e38e9dd4f21dab633dea8cebc2fe7a5b911c2b22e8cedbe6a7e1dbd352c9adb749eb457562b67abf9f5f8
88d7725f945ac8c7d301f4cdf9093fb9f18be7fa270c731053b2d896c4a7f4f597ee1399cf965447d75ae2b802456be3d87d
b1ec70f03cca3d37d43b877187bfba7e02ae81ac2e0da73815464d35965fe4bfe9537901900a60d0697fa458af4cbb98a1fa
f0cb79d4264eded342d2e5c25317e6866187e2eb071bd6a806d5945c3b2e7db9aacf5ca90c230bbee75b0d32a95d2ff40729
565d78a0a04aed2c48ca5dffd4f5cd4d276784973be5fd48e42345f42fbd9fd04cbcda1895726de9896070a49d2c25d50e10
825f8d273056f34882bfd94e704eafa0fa8ad8c4bf4fb9040d11d6f2bd18c86592db72f1fb3a49b58a91804aae1d6eb08cb7
8c6be33593113cc2d87ae4fece38d0f51d9b583c4447d649aa2b9269674d96eb9a19a69717f8ec4857c288f41a08a647a1c8
7812d61a4efec3c7d8fd5780ab96552d2b32172f17a132bc9d77bd288b488f32d5937a4ddfb9fd395b10aa9ed8d2daaf5f38
b71c8f64f2c2f033878ee734361ac074db9cd0ea0206b39669f503c9635ec1f787d072dd7e99fe00d095482daf8f111a8fe1
6629f36585ee8e3b6bcb7f1ac2614729e3f84c004c1c12fe589028a777ab504917122a7b3724055dd87c35998425f570df57
c83ffed9993bd2f56d46315738a1954e15c393c726ee9d5e0e3b28556621d1e15734f4dbfe865f3738867ede685edf6a5566
958eccd7a0753f79e317865b064867f4a70ed2eeb5d807ae342ae6e46c72cd4f6c8a7e153fff11bf5b217486f0d86bad613e
3634c69376afea77339cdeb5a550fc9ca1aed41d22185ec031c8bf01df3cfbf0adf553bbd8b89d54012342c9c8cc4c5ae935
e62705ce84e77ba86b8494e6661f57f6607e635742408296fc9f8c5f35eba34c1c5347ba3b5d0210c6f155da7ca9dc5514a4
73312eadb2183b55c77cec118f2289b40dfd9c0e1dcdc6cac8730ffdc3ede5ca96e3c7df03f938b5dc16188009223704b3bb
edc4e3b7f4239e876de11c648a2a432a98986e18bfc6b3535e984cab8d46df159bff4c605d6e8dd9d64c8627f4f3579eace9
bc4e170cd85cbbc31b08631b91562ce26b0739bf7588deb97bb23f94266e36fa355049ec188f4db7549c654028e7829efd82
7dc3437047122639ca2827ed87c8572c5d4a16e5cd205eb9e57f68c266db9bf67cef52271c848102a3eee24ec6ab7043ad32
d3ae498db52e96a2007f64468c533e776d72c792dfec689a37614d68700c401b8f99384e3b2c6265b84aaaeaa222dd3174cd
5690695079aa9824e4ca53d24a0fe6d7c7de1ae9e19b0fca3811beb054eb93e24d3512d64366c4a4944fbf5439c77f7dc828
b2bf4e329e4a369b3b3aa9ff50f8149fd338c68bc9f47213f4cc5cd0c79329a4849ecfd8ea1951df337e960a7bb10f388f86
b0bfc831711a13563cc89a10fdf44ebd2b594be0a8256ee4246957061e229a1b9b0c084c8c6624e0023fe0e36e6ce37ce579
e4686a9954b5aaddb3ceb3f1cd0edcac110d67f504fb8448bc9edf0a4ab6533c98f5954be66ffdc93f099740953296566a71
301daf89fb8875d0f983b23caf5839af2200fe3ddc04ad2fd6a17624841a1ed11221e67880e1afac27555f3688c2d895b4b8
9861870227727919b2be3011e547817e1a045dc02991e7eead0eafdb382de3b3856d8469b6223b9b67d5157e49d00be06be8
483fa0d39fdb4dad1bbc0e5be09631b96c637918139920f514c6401f3e8e4b7af0bc6d69713b2f1e88b956dd68d536edcc31
bd6e8942bd127534b840f72968efc0a9f9823d19f14bf2312caf88a0d937b2f7522e163f5925b2fc08cb3356760e509a1671
bd1a1b1de951242b3f70fb53f95bceeca87bc978798a32e23c95b6f068fb17cb2855c264b2c4eabc35c2f87c5888ca4c3495
e0d16429892e8a78dd628c8d5d65707a5ccc4b0ce58b337d5502b604a1419339cf3d0bc8fb2882cab16079d662ad049c56da
2401b42eb79a9fd9c1ab3f692852277913f19b999905e8107d2d0e3b81544a66e2400b46eb51f7ab40ab43d2773c2cd6a07d
bc5bf44d4d2652627afcdb14c730f5c89ae387ae065259cfc9ef26cddfbdbe9e0a537c757ec0309662a4c2df9f9348347c35
7f3d8512866d958fcadd8e2b80af7e76e611ff6eec4de0cbb541e164efa86b63f569f41eb378dbd29db92a101e7b58688a1e
7120b365815acada5c074b64a1fdb72e5071994cdfbc1c9989f14a6e9327e6117b9b0a5f052829248740f3bd0e8113762475
f12991624a9c0c1975d3dd9385385d69e6ec1fdaa785bf9c26f013706c5d372eda0259a349aa08f87e54fc7c3af17434a289
fc31c47b8a12dad54a616fe6821fc81db1198f0e6f7491309dad7a89ca6aa28d3ec6491d0c73f3b9e1781c66a856274f3c2a
fbe0a4fcfcb1c1e3282d74c867cff6c9b343dea60b440d95a3dfe8551703ce8f1b9ee021ade792a45c806cc7e0824cfa3676
35bc76d2a4ab458884c651ae33d9d763a55e460fb2bd4b2934940f30ce928afcb2ce556dbe4f73ed35935eb5c3d351f90687
afc8483bd47fd804e5cb7ff8068dccdc31c227672390717f832a3b3dfa9efec3208bef38bd7877dbb391774f4ee33abaf58d
0a7ba73d9635876bf877be179bde383269f1b7bfd1569e3601a40caf4fd63a9f24e8b76ebacc9087e7442f3d498b5d4bcf61
cd20e88e077de4902914fc81ab18f4497d4aae07571eb0316cd7323c605435a81edc5b7864401e297561034b4e1061fb59b8
591d2ea529999210f859439e4cce4ffbd8a83ce2eac32e6071366b3d33bf22a85cb68aecea38f858d54653332e73dc5d7524
e9bd9cebacf14e6463d6ccd955d97841c98a72315bfdcbd51cfa05bae3719baf14c1333fa17aa6fc84fdef139e91815f0cbd
2c2a7e76cc30431f8a125acf85befb03e0df9d74f8d8e02118dd61cbaf431cf5ed88d591de2a7d270961cc0eb2db96c9d33e
5869595f8a1c99eadff5ad6eb926fef533ff0ab6ab831fcda8cc1266ceeddcfad45429789f7559f61f22e0c50f831583dab5
f251827f7ec19c823057b7af5e65e57164243b094443441e93f0db2f5fdf8e69fc686e1f7cca9420f8e6929bb82c8dbc32e5
ec3bb67c9c3b9b412a01c6f94111c3ec467bd762c19795a8899c82acfd8ce31faf8b430e40f1fc77bd64694612c405136736
15f49ba03185c128a30c33e9523b8c1e6d49dbbae7ab89b322ecda3872b522c6eccff35f76f19312b5475811bf33d28d1131
eab72e3d05f61e947d5805c42b8d57227db03baae1721d76b847883b09f6c93e12ea4eb9d91ba1852bfe6f3e09f62b91fb01
da1c520739465c6d67fa594b8bbdbb223b23a204612935438f977ce8b13c93442d25c5b7779cf565705373ba355c322d6ccd
bf8977afcffb334425e3c8735884a7f0bcf5ea2ed9ebe1dcac0591738c0f22d374d06d3ff2f221f6cb831e8396fd405f6d1c
344be72f7ef2b57615856b77baaba47f19c55789500ad54cec55bd49c1b409ea2e25fc3855f42e499ffc655a082f0b6f19bc
6426241cd8971b978ef525efa881309fe4cd3cbec12766f1a49c77d6b569ca125adb74d706c189dd91cff7beba78a78b337c
4817660628a3352a444073d09d713f183b773790d709684901e10b4629eccc53a3c46f048f6b340a7f31a8b4ba060048beaa
5457e55f867e8c19796e10b4feb2ea8dce3ad933d523df66dc066bfc278d4edf255395dcace32f0495b70772718dc8a9cd37
9f225064d03a20d16973075c76ddee589bb125f49e9cc26f41883174abba85bc2938717a97da70579babf3abb1fdeab81a5e
02ffef4458d5c28b214f2bbb3f5a12b77d1c4ff7c30c95f0b226cd91bd9966b358b72a3b9630fb90e5e21e2db9876ddc4de3
ef375028af2b57d6ef19de849d9bd6f22b3053d72f2de9110168eb46db8f8f083124b6514c9d4c14c3a4f2edc69f026da4d1
2f5b409ff1898176fbab629e81ec3a2766f385e1c375d5aa58f829f323accfa43ce4fad771768c17188d937214a18462c494
31e7657527d0fc0f69717db3d1bf50bf61828f29758fd16b059c3b3ebcd002140a7ce78441914b9ab3fb7a198f3d1cfb43f4
58a64f0a923f33b5d6abe4ac2fe93c3b69656bf2fec9566a9802dae3c48577e211b0f078f2c95fb27292701b331abb3cfda4
9c32ddbab3a885d9da111a90fdc35f317d7e9253cbd7de9510f32bf92714074f4fa7148a8ceb758914dbeffc51967ba024f4
ecb747c1cbae7d9bdf79a993d1e69d862b91e0a4daf209dcdcf12a70e7fb731a7199c3c77e2d577cb5cefa4992b38a62cab5
2605f57276f354d226bf7ae43c83be3e1a6cd2fe06eb08702e461523bd15f472192a4adee1bba00ece9b2c80ca33bd71adda
48e3aebe3ca0263bc1f0fae626735468232e71b51b24c3f9cc8d0c2f743d5cb9dad3704a144dbfbfc9dd3b3fa3ffe88e4640
4aa61d4cd6165c11648cb370ca027bdd5c93f759baab90c673213546bda7cc87084b990b5e69a169a2d972f473e0c9c8dae6
2ba2f8609061f86c0d16c94d2efde07eb9b987e0f94e23b5ae7797899d2e09df9edbb9167a37c205bdd6affb058a149fdfde
8abc572d7d41147b983465ceaddbcce641729ea3ffd96850c9c1af3bedba9a62f4413cee29274d0cbe228df31e1ebacf6b69
a5f50c1f7a1efb259b1bc80c8dee20f008ce0d0cc11e98efbab87b5c39a21db5dfa7bdb014d5e5e3f31e3717c775e5852baf
89496e00ee22b02b8d13c4adcfb3c52474bc5d66260ec1f8bcb9c3252dc547225ca5b6d71cc227f40905582b530addb8fe95
538ee2a831a7e4090580ec97467b1b8c290ead247d20062fbe56048d4cd8b704ef40d6a875368d7de973e20c49fda7206af7
cf646dfe272a04822c1246c1ac6d142c47428b18b3a77c5592cad38237ce78b819de0d1654b33fad8475fbcccd7061338a56
5bc38f6e38272383a4ccfa211e73f95fd4d553d8541c237ebedfce0aad8b2214df1ffec186d9764b17575189b6775151cf48
5d3ee6dcdecbae83889af7a9ee14c97e1510dd6990c696f8c10d65c7e1f89a0176d8d682f3504ba46c8cf6c13c1596886248
564f4c03495f934d2d8acda94329723be98caa43ff0d696c605386ead7cf9f0830ed755b274a87a7db0bd59bbffb11636f5a
86bf9f798503d32d766e60a676fe5ccbdc9c8473f475a889fcc558cb17bb1fde200d0bb3d6b9764525a222fe698afd09ffa3
374f7d178a0ab69ffda86c655bc0c6cccc0e36da7ee05c5f934520343e25967a271f2b4596d5e20dfd6e9a183ed51dabb21b
faa554e6439becc5a9e6d20d2531ba6e2cad5eb8f3551ddfc937afb4986f7627f2548db6b8fb23b1fcde619d6675f469b698
49f8715246d0c256c31d1cea8ef0ce2a150981d475a86bd293fa742b22680e9f73c4f104701752d437e09a3c5c86c6001cc0
58c87ed72a1463050a13d15356fdc979fd54ece5c0709430b359cc0937e2db73ffa66575ba912dd1022d067250737aee5099
3ee5a6554cfd47e2fa391b9bb0e5ae8efaf69a9517270a3c8859fe35cb69b296583669cc4747e35b0495982c3f9ad00ba783
80bc6e6d5f04832520fad7dc4a985cb5c3fd6b920bdb6a7d36d524bc2eefd417c96c6a7ceae98ccdb2a50fe9571863f545dd
5cd4cb31613d431c830e9174873e1a13e0324142136f14bfcdb490077d75d2e1a89f27fc4947c10dc110fd629c6af72edb77
b6e9e3dd9bd19e2f0edff6ad0a8961ed6e197ff5f768e41eae557ebb7ecc9436fcabdf50c4ba58c2f3cd29c984e7d635202b
3693599388bf9347c4e4b2c2149d778cb5fc813a275024ee8042ed2a4efa43951af3c1c9997288efed7580d60a055f43f165
cb94ce970b03d81387a88eacdb42194f1ba69838c8af574b923b5bf40e30a609df059ef6fe3b3671f6bee43ca3a4cd2e9a05
fe40c7dbc1936bd5f3c16dcc95754758d75d71d70e6a20a5dd1171ab2d55be84459eac3c192005af57b1f1aec8208c8e23ef
5958f005f57e1adb7bde0aea67a758b61879b4381e746875b60645a55c34e1a847d7e59133a13e82323c7e4add25c6ce09dc
5a6b6c49b4165607f0f5d4bb318d1151075e12336e8b08278385ccb2d3351ee323bf3ae3a8f3701836bbe60545ccc3ce6700
c04385b736f9cddf4937799c9dd587caaa26c55708d71d83a823ade5df6fa3687f1bbeab88f86ec7aac77cb0e3029d7e4ffd
50d4f922372ea023b8734f9db4a1f0d29c454de4323e8318f775c3ed5c341d659bd80226d135e9180151a2b44d74e36755c7
9196ef0d0cdd24b951c81d5f9a5a79f342799d2aeaef139a66bc1ceef4c13dd05f748bef438405b7e5cf0a7006ece532a1ba
b6d006db5031b310f2ed690a980ea4b66ad0b0d185fb53e900476338aedb036b1753228c586c1a2cfc8fadb10c48f62ca329
077b4841f32e5a273978eaf935aa42e567720601064ef0f06b147fab9837d1a086aff69e521b9fc3b491d41105785496e713
fd77bda88fb501cb68ac729f4eb0c7852dac918fb75a192a3f331e30bc26c0aabc3399bc4d6bb01621ad26a455948ff58cf3
6eb82ea6c990bf26c87a184de4b19d6ef0905ba1e428bf13976a1f886af4fa93f796660411d6ae09e68efdaca893de065b5c
9ea695f575e27a96b600421aa9cee54db32b63ac9f7de8ba8a6689fb74d7484c9c716c614e58f311cacbbaa787a76d7974ea
5b79a710076816fff227b1c7cc0af502a28d8a264273ea5cf178f6687641e7913355e936feecae9f1d1a31c8a9fb483e1a5a
ba9a9354befad0bef43d494fceee4282787338424ce54d4fdd6416f51e3b6773d147672ad6e7a54b76628b7b25212ba889c3
4c792bd31a7513d62b9876ced3c34d26d1d9aa284ca759f64b7b970a52b87ce4cdb4ce371e262b7b9842a4ec11fb440e7c17
1b83dc5f0baadb51e3b93bcafdf79ba5edafe84ca7427b7f84cba4650acea12e7bc2ae7bb2a86698cba0aedf62a9ef887091
62ba7449db492bf51b6c8dcd883e07b321aea1b188d86195477af9c4ad678486d267d2b7cab158af6ed7ddc53348d0abe7a2
948ca4f7e7da7610f13f15cf647e61f2332e1d5a3c493c1e6a3032289c7137d02d958b6e34d479fd8a98964035aec3a3a5f1
a3c0567bf943a05445c61e6dde27bae2668d403ac2674f9339a81fe104c3654606b91d7ad25dbef398f3468e0556ca6adcaa
1ce0cce1c8de4af432b5ee546ba93a5ea825d5d33970ba25ae78cdcd79bb790111906dd77056e024bedb0446564131c29622
ca7bf5118755bd649c2f69beddad5fa0827b325a2dedf67167d3ba62bf577abb07d1d6f385924f16f6719f30d347c1b92ee0
276d1edd192b0bb90a967539a6a1dcef942333551ae3cddaa60ebbd11e3e23d1ae7ffbbe2a7d692ea77fb059d5dae2dd5abd
1f1738720bc30a3a15d2768c4613f8a195ef7dd898769a68747b6f39aa99ec9ac469ac877fb64b87ed107b1e3c9c00493ff5
7a0dadc7de23eaa461a579f3919899d1ca6ac4b1c5d9b9ad539c8f55ab8317c479ca61144d22f3dc13c7585438e58cfc5b5a
727dce12d4c7fc6b9552b055a3cd4b5fe63c0dde577d05e4ed4cba3ea43dad4a96ab6d5181ab5d433e741cc52ed4addb6cff
076eacaa68
>}

\EmbeddedPDFStream{schematic-42}{%
/Length1 1436 /Length2 6505 /Length3 0 /Filter [ /ASCIIHexDecode /FlateDecode ]
}{%
78da8d740734dcddd6be96e891e8dd6841b419354a441d21884e108c99c128338cd143d444efbd97207aef2dba8812842044
49b41042f428df2479ef7def7bffffb5be6fcd5abf39fbd9e5ec7dcef31c6e766d3d410518ca0a0e462131822021a0344049
53534d0200048a0a01812224dcdcfa088c03fc2f9884db108e7641a090d2ff11a084864330584c1982c1c669a29000755707
00481400929006494a038100112050ea5f8128b4344019e2868001348500ea2824dc85845b09e5e48946d8d862b0dbfc6b09
e085f2014052529202bfd3010a8e7034020a41023421185bb823764728c401a0878222e018cf7f94e095b5c5609ca48585dd
dddd85208e2e4228b48d1c9f00c01d81b105e8c25de06837380cf06b608016c411fe673221126e80be2dc2e50fae87b2c6b8
43d0700016704040e148176c862b12064703b09b03f4d434008f9ce0c83fc11a7f0204007f9d0d002404fa77b9bfb27f1542
207f2743a05094a31304e98940da00ac110e70c023b08610c6032300802061bf02210e2e286c3ec40d8270805861037e770e
018015740010ec807f8de70245239c302e422e08875f230aff2a833d6515244c09e5e80847625c487ef5a78c40c3a1d863f7
14fe73b3f648943bd2fb2fc31a818459ff1a02e6ea246c804438bbc2d594ff0ac142247f6336700c401c785752f42e080077
06c03da0b6c2bfcaeb7b3ac17f3b7fc3d8097cbc9d504e006bec10701f84351cfb47e2ed02718303306857b88ff77f3afe69
9180400018028a0158c16d104892bfab6361b8f51f1b7bf9688407c01488e51e0800fcf5fbf7ea09965e3014d2c1f3eff0df
f72bacafa8fd48f531ff9f89ffed5354447900bcb17b0a8a880301201131104012bbf0f967156d08e2af2e807fa7aa21ad51
0010f04fb7d863fa57c76e7f1180f72f71f001fe594c0b85652d1cc0fb37c9cd80e24028f603fa3f53fd77caff8fe1bfaafc
6f24ffef86c0ae0e0ebfddbcbffdff8f1be28870f0fc2b004b5a570c56009a28ac0c90ff1d6a04ff235a4d380ce1eaf8df5e
350c042b0405a40d96cc82203121a0d81f1ce1024678c061da080cd4f60f65fee006bfa4e68040c2b5512e885f6f0b360b08
fc2f1f565f507becfbe182e5e51f17c4052b36ccef6bfc65c3b172fa671f2a48280af64b7722e21200081a0df124c15e3dd6
12c732012b5018dce337b301c2424814069b02c0ceec03b046a1497e5db3b81440182b5e47c82ff83702028a038411ff614a
0084edfe6d4a8000c22e0e1017dbdfc83f5a81baa2d1d85e7f3308dbe7bfecdfef011cee018792cc4ea3a032417635416d27
550a4cee82aba3b2d776534e8c4504475f9a1361fa54262c96e3f43233e61e9680677b416073bb4e2d45e793ac4f1f7e787f
a965abf3143b1264036fd8b059c54c5f1de1be8ff73e66e698a668c2c9354a5264952e42f7e168b384df7c4d7c1f66d3a7c7
4df194a7a66726defd36f52bb0a47a709b56c7ebaa57191ab4cc5286db2bba9866ab41fd55f1037a9eec15c7e8c0cf1166a1
c95a691c0f653b4e49e2dc2b6f2e0d0c0dcedfea65e90af757770b58fdcef51c734b45669059598cb0d2703eefc6797691db
561aa5b7952a9512ad9e662ebb2e3dc9a0525e42f32ebdd93927a1b15d46a86d3cc9163d457b5eda7a812dca2ff95454bc3b
f06e72f67b8e173661de30d905b0ae5f47f38717431374c7b3d79686be3890be7ed7ca9bc1f9d4b6ecd5fe33dfb70a7ae176
291550134121a31cf834556e2fe51d772890f4b85915942ceb91d5f7723fa52358dac096c7ec2abb84ad995a17c1c23287b2
228b9932e5dd06ec7b9cc612b6e3fba74e8926630cdbde943ceb982682145c30dc0872dd6f62b567c597a793292fdcb7f821
9a4476fd8cc6792c9bd4f4dd069ebb2965f7b345f456bb325f6071d4d01dbc4d9dfdb1d68f44f35c2a7bf43b050b05af22fd
1a4781c79de2e5f22425e91f5390ea6450ab4fe9dde6a51344f1289df534c294ba409e215390bae2933a8ed918d69b810a88
50cd129c3b54bc927ef66479cfdd36a73ee8b829a8b0e9358306186694bc6f272d7e2f9b7b3b186436177e5b9844e1c19b72
456ba2278f4025c5f286f17943b32b2810467a84b1a2dd52b9e6f44be42e3745f02a4e2074f0c713b37cb6235f1bdbad4a1a
1809dfe9a38f02c3a279cc3a89c59b3fc164a1d116b2e5052d0bf964e5a4801495a42b45886cd2227dc3096f73e87084cb93
1e6f668c06df02e0e5deea878bb8894f9a3c30a6f9ed7bf9d29142eea45cc1e67caf792992bd42f868f85b12f0a389d446eb
c838d3ae19e9c88bbc496452c595679cc39d13ddb772d877ef569fdc1a68bcc8cc14be0d69aff0c25fbcd51619d9699793d4
f2335830ce8d2f3201803dd5f4ad4ead1250fac704ca7eef81e3065106e46af5b41081c28b81a88ddebe166467f4fd1d70fb
d98560289bf406cccca7386274b052ac7a472af981322733db54f8fda81842e2c89af8c81753c38c15eacc5d9962477e879a
a30c3b13bb2ecaf7e692e2caf5e0473c16f792b08d364599c804d3eebee1e312a412f1f19491a2ebdd76eabd762d325dd23d
3cb78bcf8bf99de90d7e10653757c9db9bb8f7945a7d033aad981fca68c393f37d28ccd1fe0f3e01e9e24e09b24784ea054b
d5082c47976b06678f5d24b412ead7a41e6002a07345b8d43429e752915f453c659ddfd6ad06dfab3b429a1f69f5a35c347b
17c3b9e6dfeb24c4ea3471dc952e8cb03871eba44e93f11f66cfbd859f85d919173158a9a2691456d824b1bd0af1d076d7c8
f5da8878de75d9cb9532fae8ebf9329ebc7980e00c55e53681b38861cdc901cd3ba89d7bd0c84e5e00aeb689f17df1dd0802
8eafa1176d7afdc2179d1e71ebee689df42382a220c412d3c015ee66161ffcebd1295a873fc9e12e75da3ace931a7e091b81
31e1076f73d2d86f5ab70356d4eacf734663e2cc4e417bd4dc0fe248f5715c51d9a13fe757cdc35cd8182bdcd0d502119b1b
544fde336046201d451956419459b21aa689ce0a541dfd4c2ad29b15376b4d3e92c1237d9864f1de4436cef68ffb4682337f
14d40f951b0e51c73b05820598351f728f11e20d2bb9bf4ab5b52270a53f6eee14d8a45b467d1e346fdf2f0cbeb164665827
b2cbcbeda9c1f44a4b4df2e517b906a3698d1eba35cc10eea1dd5acaf32bd57132bef79fb3181dc4c97ad368c4b63625724b
6cacb5afe8da9a0e7feaa42bc7694877edaa939cf0d5ea0bbc2b7236dc3e2fb0ee7bb0013758fd16355fa23a9454c219177a
b73e7a3efc98e9ea9b907e6ce1aa80b60b81c02623fb8d7ee258e6f6a0daf2a268cc7e8e66ca244bc5beb3cf431689f6397b
8c7be5e65748a69469fe3bdf7039adb2b38c3887ed9143564f87ebd726eccce886cb1e1c2f6b541ec5ec43d3b6fa5bf177a7
62f03ee300efbf43459ca5047167d28687c895145d9b604bd5f0bf60fcd6cce4faac0e2ab2e64a396370b9c5906bff707f0a
b8505be41192344ccedaba984aed355a42960784b83f8f98125747e13624528fa573599c1e7c632c984e35aaded734f1b7fd
c95da32a9f97a0e12bac2af226e319acc5463c8bf150cb15e998051ca0626720b73c4fded202dd6bb6d8cbbc699ef122bbdf
be45ea7bcdcfd82b3079339788fca0bbfd154e30012c9841798f325b135ee109fd19307fed3d60c76f61bb7022e4362c6732
65f7f6199571d66389e4af6b24e75df58b512cb7fb46e78ef306c72f8635c4b999ce6c5582ece8c4fa97385b190ceee45ff0
c0deb41e0cd930a51dd50667d5e4edb0d9721a2eb598a58b3a2e9a3a6b2ddc909500b25f65be152995f3b566330dc0db1334
9d49efd4d98e70b75f724a76916acc9d52d200eb1d06f7b6ac1e3b0feaf514527471c654a00b6fd810dcf2e6cc60b5ceb16a
5afbc8f0bec8d73842592d91bb4cda92c571a91bad293ca3b664373a807f03f53aaabf2d3c61a8d8f34c63df715ced5d3d41
8cb0f1e683e17591f6c9b6a5db17ec95dfeb19f87a2f4ce14582be8613b52d9b455cf73f2c73a1ba38690cbf29d99ce93144
3f4a513db0d579cac0f8339640b73a57d2957a67dd3582ac224eb9cf795bd0234296496fff4cdff1fbaaf6bec489da4e683e
ffc14cef74346d516e28e58325bdee77b7e97f8c2b5809f94fc7bde02c2702432e8e241e868ccba90695e88dd1d79792a6af
74211a2a973edde9adc64fb19d6bf71c168869d3fd41f5a5fecb720173910d6c328e3a8a51ffa3cc54ab42b7f79312a154b5
83c3d12fc546b51c9f2ba7acfaa4ade94a8d6f6976f5d31ca8c89dc7cf6a59f84b092787f6b3d3e05452273f5a2ae7e348d5
dc21f643d6ab27278dff9061cabb2b2560b29c451e5636d7464143ae50181da01cf3324219c3a396d6d12307553b9f12283b
9e181c0a5cda61ec4c58a57999b6f22ac83b80fa510bb36f227e9b11db89d2457c9702a762c9046589017fd39098a2d9402e
0b019e4173428fe094198ef9eaf49eeeaadcf00099e9749ad1dbd8f5ebfbd9eae7a0f3915816273a43ff35364f68f5c24016
91d26198d7e97a0ddd1c23e3e4fb6f058d6c7ec2f7b3608782e7620d115eae5935835ceb1ac3aabb02d7646376f9430d5ebd
9a249ae14e11d773f4db1bf291ed7e41f38814ed77fe22e1942d8baf979ca7bdc3a8ba6ee91bbca6d099f66226054679ead1
d3ad755829f1f03d8335c3ae2a31a7e636dc37b1b8f57119e9a6e244178dfdbbc74365c409f6d9f7edddbf26a2a7c37b7b09
a7a87f0e56d8157248d02d3da40023a8b9ef57cfdf2b399e12b07c40b21eebb106ff16b655c097ad3fdaa451f1201d2e93cd
38de5ea7d01759b4b64d2a66f7228495d080c89ec2bb20abde69f64805c2d32cd3052ca66b203f8c0f92740efb16061fbfd5
3067afae6973964f9374203cd7233931ab04a7871296d4427d3ab85e057f2e077b2485c4c22cba6361fc3c637e90b5d7de4a
9aa7147bab1ecbdf5ef37d18f3e67c7932a993de13e313eb3c795ad6d9b1e596e2d5c43e1f9ac6199b353ad241c7c5717a99
d3a3b0314e55986e34cc7f1d03b691b277c99e15d62a4c7254b42c8be2d3fffa14fe5c2a6e168a171772dbaa33aa80d1282f
cd3bb6c080893befd8c0e9d95ab1523a99d585aa5780ec9a30111f3d1866cc3f994fbc4613476709df8f16d818297d11e0d1
a0846660a386ad136b7f9577130f1d53b122ed3bb524ac9b73cd872d210a58b7396f9a5c9bcc391ccf0ae34955c7333e0de7
03be332cb6d4f974bd04a3edd41b763e9f20ef603be43ad303c8c705d23c53bff61487f56371566490408345a6863e54e42c
7df709ae16b04d2586b877a353e6f04d2bf6f568de342891b68c5d6b951e6c319e557a5ae2e40050f413da666a5ef2be0a55
3bd360499b7a25b19cac06282109230fa17dccaab3327f7c8a1b515bc7c6621af3681b1c95a4a0c0c22514a1faa929d6b39b
1332275c8a5717aa0d26d28eb9458930e9818d35eb757f88aef2af02db17d402c7a9fca9752dc76e0905aa4dabec0fe9949d
d80811227b5a95e46dd88b9ffd18721bb95257a176e98eaf09675d06ab66618c0346c7f9ef4ee665533c7fbe5592f8a646ec
0d1d6145590055fc2b4669d47c560e688f2aa873ffd6a8d7b9085eb4e92043eabdafed25f9d4db1e7ac51b2bcc4ad2869829
d3cfac5b749f1dd9f43a103e3fbf0d86e10ba9c04a4aad8e194b5b0cfbc20857bbd8c58a67841c2d3e7a7ef4efa9a539975e
e869d0af8a41e488d20452c5258a4a1b84eaa5702d920b3dff49a3392b55bf6ba1c1c94218f249fc6aa2d6b2c06eb1b1aedd
4bd1c7c6dc48f419ddd6a8aa67c1b1854cc9d2203fe295f0ca7d22791af019a8ff8e267aec4e009f786069e0fb0755d4baa4
08a4bbbf7b5d60b5a63c71ab7d2f974403c7e29a8cb774a6da9d7a2724516afbe5507088d03cef6ea5040f615dc89863dbd5
4bc78283f8661d4613f37c6045b07dbd0dc377df22d1880ad853819582c7c781648c3fa8d6ca13d3f62cdf9d8974a95826a5
7d5ec4e1c23d69336a5ed7e4768c5e2752bbc78c7c39fae1c748a81e1d6ed6a3d74564558215b4c44efae033e7d2f7519f97
dbbada24fc1afd38e2dea337fc1e928c6ea75f5024591ba40be4cb09d39a45d0fb5e1f24191472ebc96a26903f1d3e4ad462
32b887bfeee1c795e97184b8c4a179663bb5a5421f56d703609e2f49aa239624ef657bd796d291e589174d839d7ca7c076d7
78aca7206bdda943e4b8c494a979b2dddef745e27af5e33b01a3d7cb49dbbecf4e684215c7d0865787ac635983d7f338e33c
eeef577d370988942b172884792a509cbcef7aba49c28631220d6f86292d22669acfb3551562f1f471fb8276cf63d77dde4e
5b2835a31c42aa2311331f3eedd1388a5a0405e94de46d214ba0944f0ef41abad8a535c097bcdea3b4c73bb336da5f9d7246
b360dfe5f8facf836327890c325c226bcf71b278bf8dc7198110610d42e31e25a795b85f6f4606d6e503a41650b46595cef6
0e6fe2421f868b5d4c918e5406a46429247ecf208d835f4a04ee0ef6d3c0efb906947ba9f386135875bcf0206840c91fcc6b
b05101ba2feb1aaff74ebd8bdd80d37992e429e7cf2e8bc004cb75be2c7c25d8c6d9e591634979562dc52ab4e1c64e3b322a
b95c1d327de798d2f3b6f7c68746a35b52693f183db7124f723453877164ef1e5d71430fb83e92ff0c2ff84c6ff2ae23591a
efe0b4c1eb1dab7d26d249a2db90512ffee66a9fcc1814e7a95b81dcb3681ad9cc441dcebd7da16174cbf30207074017cb59
d3a516c970c675c1e7e7092ccd8aea0dc754a5de1de2e4ebe152c8d3dac53a19ba365836cedced9f2688833c1d9006a5a334
4dcecb73912df92ffca49ebe0ff2d27ea6d63c74bdd1106de446ab05d5e60aa1b3c295bc1d3a5c4cf1038fa646f47389e3c8
a40ea4b3bcc0d87fa17b633c3087a3219d31ddc303323840f8ec2db9972c9e12d30ee7c6373aeece50323baddce031db2674
54ac84c462916ae9308b052d213dab44e48700824197854b4e76e7bbb5b755eb4b71532e7e4e5c6beb1be8d2682d58970e58
49b50f07b4df31b465feac97f17067a62f5e343be0f414d908ee0147596eaf88396dc9b5b66fd13ed9e763368f6ad3595008
cd7ff5b438676f2efdfd59d3802d9824405560c7cca2b5ab78f023035dad5d918440ef61c96ab1e9461fc6b2c82ace5caf54
2a87e869b1e247e2e7b75d1d697bea78f805274efac5afc5dedab7e361d6b7a1cf2b5cd5b46d1a088691857c2629d599d2cd
1ce7d4cd3b57cb1438b9041b0ff064e7836548dfec93744b7cdd8bb78eaa59b171f0946988ee598a50bb3f184aab19ea1289
9913821e50f24b8a1689e324aa7939e5cff0cd273204cdbe95bce66f65db316c4af6818519dccfdf7c44c9735cd9f6f8497c
6c3ba5cea141d98044688ccd7c43d741d7096e7a52ef352b5dfcc65cd54b9fe46138a9599a020339e8dd94152807f1969c75
6538c1cdffa1c2654e519577918d35ab2bc551536fe469a16931d2ce07da31be5075f14e42f7fe51db388b89bd0151eb12fe
7268eac8b4c54542b6b7e9fecc50d9c961e42c8e592ee3b0b49ca60989db42774af4cbc4bad9113c05b0ea97b6b76406b8fc
3552a60828f763cc0aa62f666eb328592527f1bea97ab8bb9667be53ce6218ea591b079e6eb1b8e53a83dde5bd914fb7df19
fab71d34959c58bb1ed500a38bdbece1f8d0c4c759b4c527cf3bf9990a7334e076871d7345ace8d1008d0dbc39fa06b08aa5
613d3bff8b4777f4eb0bd8f0ac26f214830d55d3fb066d7ecc60861f070df1b858460a92866c2a9868285feb7c42b52d662a
5159f0cd8b883aad4f77ff99f6d3d6bbb67cae96e93f0bc9081012abfa918a5ecb6b993a866475aebedf0f2c5e3f384f7213
eeae733aad592ee726a01c246c99eb9c7c1445a19f640a362376a38f8a5313e776eb495fbec9c0bf7b1408cd8c39cf4f11bf
8ae5d32b423d14bd1be26184ea6ebd552af93663b3723a23b2ba537ae6f184dcc088e4b1ea93f695935221c78df3301105eb
691ff8f29a6a37c5cd65ebc02f4fd92a2f17d7656455d5d9fc3667eaf480dcef9d4b20373fb16f596fc9430f47bc5e79b076
36337ddca8d2d8188d5fd85a0c1842343713d109936e333e8111521e7b10f8dee57d2a185bc88dff2a797ec42d9b14584154
999c575c7223d2dd32caaed86802185ad570232f26f85c303370d39c4dedc467395a5ae4735d4144b6fe0fa14862cc8c0cd7
c72fe356556f1aed6f54bb06d44c336816848407aca2d65fa55ecb8596d16636b90c8297ebd4966f52677af973436573f49b
985c5a5aa7150f8a2d0a32a653016bd5eabdf461f477742b4dadd1ef2f94539ff298cf06a5b51f4eaa7d11fa4e4e4463d20d
c2ef3405b677f4a2366ace9972d94694719db8ede141cfe5ad2ff81a8735899519b8f5de5f8ecfb7474118b3c45c49d40dbd
ca4611097a96d0847719078fa590d3f7524f9ead25c52a0ab7e0e2ebf3157164178cbc28d461352492b92b685a33acd1b73f
2d363ad8c87d733623d1391aa776e37c27b536f05085d31f80a349b5ad73f78503a84a45f998fbee7079938298d218453ef5
97d98564cebca6cceb3f8e82c7ea5101b9bc126229cfcc711c98cdc25937c6ba27f179335873a8dba5246f741259caf9beea
e69d317ef5c0a27f5730a895bd90b2fe812063a0e4ea731196e1794729c59268ce235034c908b87bc22c5965e9947a4fc2ee
44f3a1c916873fe7c39ce837f75414287e283ef22f11453feadfc6cbeb18369103bbe588cd2f8fbeb576ef95350e6e374678
a2855ea009bea115873bd3c2b70a9e2edd676a743430aed48767dff6b1e08cfcac1cba3e25a76833f7323f6b534129474b4b
accd07d4224364ab2a4330d440305aa7e04cf912bfc0ab75cac637428818baf292951f2f602e7622e93256d672965d2dbf27
fdedf0f64de410ddcfe23da1c2b71f8a95f559c1150f2275e929bb28054c3e27fc9011eb41ce09c8ef86dd1a5010e18de979
ef4f1f3b48f49d90bd44e33d593e4df59cd9231b7fc7d79195039002dc6472aa271ec6e2df9c3571e815169a93f5a60164a2
c25b3815a53390153323f998099297052402e9f83e8462fa3bcf4fe49a72e23f9d32b9d029ef7712289b36b97bd96544e6e0
05c4edff6887df50137a441c8627204ad02c4e7044ed7ae33ceb0bc2382fa8b5747ef95efaddb1d6276f16af2b9b00a30927
8e256abd1f9fbca6d9175c7bc375afcc876ab3a2c3bd463f995d64f5e972ef515760f8c5654be6c7f3593c59f16d8f369afb
64cbf5a6c2a6f8e61f88e836dea4569cf0c5e7f554a13e55de715d19eb3b9a1cb76b5eae4fd5ff5ef603a0730fa4bfe47e37
ef73a6dd1bc56c46f1f8ec69a6d5be173330d787ee9acd23dd1f3f5532dda7b6b5fcfe14a956fae5b19664fbebd76ea1628c
cf3bdab9efb7db0f7029234437683379be158c9e3fcc62e21f4ba27b1d287883b6ee2c7c35eb1bb0b68a5e3fb9366b9222ff
a524872a314310372b3b912fd8e186f531f944954797ad9fb8dcb856426de51a313af7a33a23d7a27c276d0aa7aaeac11058
b89140477738f6c2f6b1d333724f1a863d2e9cc4e80c133b4d516d02d1095eda9f0a62c50d83963a973b5b24ad45c7de16e7
4e4d49463c650b2f5b3e792bc7737ce2e13f73f2bbdf0c486d07e5e6af6e35c0410c9f1ad8b799c2c66c8bbe144fec754bb4
bb17ad2a0510b191b7b17592555dbce01d53cd0d9c96f69040adf6adedf40655904c79191e5f4fc94af4ef36fd245a7d7e36
3010d3ffa4d2d5b9ed6673b9eb93c2f207fa388272d5a0e3991f2cf5238bb64b6eb8d4758eabc772d50fbeecb6263c3a42ce
c6f5f36cfea4db2ed695fa52bd1312eb706dc96e64d87f9d54250d5ac7447cd9f8562ae3437238f92e01987eada98a4d248e
791a5cb5d3eda7bdc94c685d98678b60f0a82b73e67fbddad75ae599b4bfada56c3f7e4f34dbd43788a5cdc4955c668f8138
2f8863fced1ed260637ffa70a7313b828ef2a74e6f79f95aa0be1aa4a8ecea9444ab6182a9a5cd60a2ee43e224458ccc98a5
e0ddedd2eec0d4b840f20f422b85ccabee3bc924fce6996cb682107ee7168f7c91462ae8e8121fb026e9945b3174e59ec8c6
8b36458e8884c1d375d7c7cd940a10b1176e5cbcf57c1b329eded9afcf0698759565332c2da4985b892f173cf398c597792e
9c0ebed3af3c4e9e5f4468d5de546a13bc5d07ea4c327361d0c5d55e4d30237e5cc1592e7105950cdc5323ee76fcd04113c6
e17c9426c75c617521d99faaef5e9c733d7369c55ba7aac998e7aaf44369aa48d5a050074b0a14c728206a047e5010603072
c07141d19157ec4b64dab7aef81df39815e7bcdcf79be5b4af4135e5ddfe91cb5a9c5c8a323c26d6aad1fc70103989fa6c40
32f770ce1abd6cb2e41791ad5661d0e7c6eee5cf066957eb27e247e677bcfe07acb79444
>}

\EmbeddedPDFStream{schematic-43}{%
/Length1 1413 /Length2 6326 /Length3 0 /Filter [ /ASCIIHexDecode /FlateDecode ]
}{%
78da8d750754535bd72dd2c42822205d25f41e7aef4dbaf41a40431220940492d07b074140a44811a508484740447a07e945
3a088214a94aaffe51effdee7fbff7c6786f9c3172ce9eabecb5f69a73878d49cf904f1186b281df4721b17c82200129a0b2
8ea1b904504040182420200460633342609de07fc1003613381a834021a5fe9783321a0ec1e230150816e7a783420235dd9c
8082c2404131294171290101a0908080e4df8e28b4145005e28e80017540404d14128e01b029a35cbcd0083b7b2c6e9bbf3f
819c502ea0a0a4a438efef70a0a2331c8d804290401d08d61eee8cdb110a71021aa2a00838d6eb5f293865ecb15817297e7e
0f0f0f10c4190342a1ede4b878811e08ac3dd0008e81a3dde130e0af86810f20cef03f9d81006c40237b04e60f6e88b2c57a
40d070200e704240e1480c2ec20d0983a381b8cd81861ada405d1738f28fb3f61f075ee05f6703140409fe27dd5fd1bf1221
90bf83215028cad90582f44220ed80b608273850f7be3608eb89e5054290b05f8e10270c0a170f7187209c20363887df9543
80f715f581105c837fb58781a2112e580c088370fad522ffaf34b8535645c29451cece70241603f8559f0a020d87e28edd8b
ffcf641d91280fa4cf5f0b5b041266fbab09989b0bbf3112e1ea06d750f9cb050701fec1ece058a0a880a49898882410ee0a
847b42edf97fa537f27281ff360afe82711df8f9b8a05c80b6b826e07e085b38ee05f0c140dce1402cda0deee7f3bf0dff5e
010405813004140bb481db2190807fb2e360b8ed9f356ef8688427102c80e39e2050e0d7f39f2f2b1cbd6028a493d73feebf
e7cbafa3f8404fcd9ce74fc7ffb12929a13c813e7cc202403e495151a0a0a08428505c5c12e8f7ef347a10c45f6508fc13ab
81b4450125ff548b3ba6bf2b76ff8b009c7f89830bf8ef5c0f5038d6c2819cff90dc524054008afb11fcffa6faef90ff1bc3
7f65f97f91fcbf0bbaefe6e4f4dbccf9dbfe7f9821ce0827afbf1c70a475c3e204a083c2c900f9dfaea6f03fa2d581c3106e
ceff6dd5c04270425044dae1c8cc2728021210f9832330f7119e70981e020bb5ff43993fb8f12fa9392190703d1406f1eb6e
c1450908fc970da72fa823eefec0e078f9db04c7c9e7dffbaa22a128d82f9d09898a01216834c40b2080a393106ede3e8238
41c2e09ebf990ce4072151585c0810d7a31fd0168506fc1aaba0801890df0682fe85fe06840481fc76bf6e31381aeeea861b
cadf060120bf3302e986f90dfcab06a81b1a8dd3e26faae00afc7bfd5bf870b8271c0a981a4741a5431dde86d61f972b3278
f07ded27febcd818156fde11298a659f78e563af7d354d6dc455e911ac84b637456ff24df8f0b82f03f7f6f180a7654b46d0
7cb2fa4b2c9e6ac792d22e5f56cfc95e8b1dfc1e0c456f37289e12b544bef354ff2ed923302198e399e9f8799878abf06ae9
4aee452366d5427203e80ccc3bbcc594cc807da298d9b8811f6dcec4ba88fd686e792b96020d7eee66a4613c78811f8d883d
52aa8fa13bf15a02c7a67e9c9bebcee8ec4811fe91add5562656124e4977db5b9329b337a18b8eea6986660d068f5d49f450
f94d523c33dd6be770f679024847db1458b2b6791d0ba1cbb8fa64e7e9fefd9afb4c840a0ce0c6c9114d3e989fcd6ec7db39
dd6b8e411482f55b9159d084d3b91dcd75ce6ae2af0a8046b158df1e119ddcd8fabd05115b933b4fe5ba94c94848b3e8a5a7
fb6a3deb5a1c0c1a14ed60910a7c2c018552453d25b3b1e95177bc15db8edc3d2b0e88c9ca29c3a3e50dcce046e165d9e4ec
0c759f201329c5534d0e157d74c437bda40f2c400ff63dab8bc8934d22b4655c5c9def1c403d62d6dae0fa845ecc5fa2da4e
7b84376618f156620975fd44912c868196ad82d4b1bb6f9c541c7ae905430ef38b5b123ad71ad85f7dedefccb7948da9759f
73487f1d5bf31de638d1bbdd7b3d59b6937f7194d9a28a2b3b49f43285fade743d815f1b8d75ff42125f80fc37efa1b6c9eb
a1cf7ea0bdcb36765caf3ce4e787cec890d64e1cf3ccbf8e21eb7fe80eb954308f5e9816e25d377e1f48d456e11560d7aa48
91df18c1a9449cf100ebfd04ffedd3fcc3dc4e8161a0f797d06cf556fb7e365146f6edef9d41c5018f8982ede3853ae46424
b6f9f14b840e463e7e507fb5a7dc5bf1b9e72163e5d13a4365f8d7fc6dd047e31e1ec20164fe68184b3a963c76ea64892a0e
5f7583e20dd86d17f93a0bf9c364c85e135f522483d4ef221c3f95c025813d418d73d3f9f18a78c6141138d899c95a363146
6e035af518c01c2ccb7c7dfa934b6760e9e4aafabdf30b79b43cd0da3f3e3ae921b88caea1893c3544c53bbdc51d90de60b8
65cc6855158c2437e1e5b86d02e9c91208f96e9f90db216c255d675fe9717019c7085b6b7872e29c261ae59d1fe9a837708b
3e57446a319186fc507c305cec004412bca324d640a53f164937703e45182473e6ba78a7e5d6638ed823ca45e780b7814ee2
5e0276ac95df82a4d58dbb41274e3de4dc6cc90222233ce982600e250f737c093be11b56d68fc66e0ca490077ea427715197
67f50aa93cd0d70b6c1766b43b87a1334337e3f5ee4989a5e9fc4c6bba2a19f649b9dd843140f9b34540ac00c81e4deba021
4a8972520a7c99dc4d45e403243b0c7ed6ee3c96b6ec9b4d531c29bedeed0af1f182c14787aced8f74448f906a1868158b51
fe3bd033ba3cdf3d305db2dedc5ec86a9d89da52e7158ae06ccd69f20a198fd5a0ef48c90c6e079d400c2158e6262d6dbb95
15761d934390a5ef76e2f342d3aff019897f1645f1d40ba6825597a2d5509d426251e37e375e3cb92844db15b05d90b1827d
9f281c6459bb0078aa7582a79f6f2086be6e31fb98e5bcab77c3bbf3d4a86d9f492682db8bda4a6932eb9de759af4081d9bd
7a0257568bc6fe63c64ce83ec94c7e41556edce21bf76bea816982dc71f82df1e8af1711f3a11cc18b04d60d62e383a6bd03
4d49145ba496e6cedfcd0de45f7bd5a531a90145616c5909e49b5a330736e33ff30faadf8209d4f1a506d4e8abfc8156e7ed
f8814d017825da7469d7f70ba1990f46ab6356492d8272bd876e08828fa44596835cacd7aa1fecb89c15ba2aa90029c83826
a38e67a8389e44ccf7a707909b259cb43f71acf2cd5d51c323d44fead4ab0a5c9fbe286f576510896cb28e236b58ab1d7e1e
bd20fbd3976dce6294ebd16339685b54aeef73cdb506ae874e1e11ac60c6af3b59264f8dc8cb8ec5ee46da5202de77c8cebe
7c41784eb0b4c41df7e5eb0d844b89b62a1ad06e201fbcec9ae88e6d18da7d92781d6d78f328cd688f25a46e82a2f8a44160
39a98ffec22eb74ac7ea2b9bcf5d3bcac682e7e62fe93e85174284bf10a4539a66bcefedfafc43e5d59b095ef6b7f64c9d53
216fa2622687e8b75c7f5a032b88000cd55a67a48a4baf89fcc54d7ac7c63a9295923c8f4882477a08755484c486ef9c2de7
d27b95a352df4b65860d3045c49033ab2c01a5299d597eeca5b572e0d5a59a2a03682cbd30ec77871cfbfc2c0a0314d5f696
a2696a1d664f8296379717c2530c0e8318cb4a77df3befb2c7bd7ddb69b5119187340d09a6ed82e62cbf1665afbf29c67cf1
73e2e9faa1ca44e9e19940ded1e47085f9f79714ac9d933d33a5993f54a8e53ca60cce83ba022d22201f326f7f95685f6ad1
c314cc48711dc730bf5215eb176379efd15a6057ef7f9cc9ea372376ae8161282075ee90306ad60a6d6725e64a5818690de1
bcbc6ebe4be4a1f0a5ad51d81edc6e9cd859db708d25bc693ff4f046bc0d34b506c0d5c8045ab85b47bff669d834ace6d1e2
6332e800714f8ae27c6b302b8147be3ced767914b5e21d400eebe99be258f212dd56e26a8555a8dc48bcaad0cd99718d74ab
7706cb5d4d3cddf16ddacf6da5ae650b62ae31e8368a224c81da733d0417b76d652cc8124fd3ee2712c233d4c3e9380ef56f
d13a485ab67ab84ebc7c3947fe72e870f9d015ee911076670592ba13a3049aa2b8b7c47c83cba08d55b03973496431ea6569
d8ebd214a95633b78b91e5a2f295e6f7ded5f799efccd43153bfe6183611e6fbe682303c6ae2e34d1dbae4fd5e69151acc90
1cd9cfb9442827e8ae778f5dc829f7cee31f2ed4421ad477b52eb29438663cd4423ae9231ed5a013359256f66fc70d81cfb6
0bf7b41d3e71ae6d7be113db50c14b93f55a879d3b3977ded6a548c3eb5f5861253256988c5923d6235b2028656adfb31034
51b8bc836883bdb3d0a084aedb5035acc8d9bde261caf7029a79c1ce779510d5d4b7b76d26ae576078df1e41c513c3bd26f5
e829278fe25321660b8dfac9ba2f39ef45f75e9265791e463bb6b8c218f4c54fa0eedf01054f69f649acea7dde2794c8ac0c
92fadeb55ba3dbd0321bba872a7009db9a789c531e3006f54e9ff2a1683325be01967a1d6c90cf94a499bd2fd089ef0215cc
86be506a37d12230614a27312a695d710a690f337f97774b039ffc2c3ad2e9dd372fee28a16ba3ef86358366e1e1b0fa0d31
e5a179ce84528ba3d3456dc9d1870ae474e6ae72af02e83cdb09eff26726ad65a7e8ce4a649415de2366964bfb72ba1f585c
6868cbf75637908392965d87c63cdde0fc23ccf4bcb4f2808921737348e1907f9b822912dfbfd3d59fa5c0ac83afd14ded45
ed9311ece7973b88ef84bc2d08763c945995e4771e881cc652f0503f1c1893153ffa606eeb9b487bd8003c4940bdfacb3136
bd7b61333aa1e9d1ed55750061dca174d9e0a39f1925c5e1aad038a6b3945bc59f9f8ae9916f4c1df653e9b0946649d0a7fa
0f9ab6a0ab9a9a298feddca8dd9455616d83a5db469c4372aacf4f4162cb1bf65763a9e785e78ab5512c608edadb4b2d364f
e264d7ae18ca15ba4c06ab2ee53ebe472e92303d56ff90851b2cfde2cbeaae7abfd8dcb95dc9799ed169ee3dbfe5e1ad08af
be04a7853ba8e8d709f1325fa26b1af0341672666453f5510303116f7c6105aaa7ad9b1c314bdc09c556768f7afd145ee909
10c505161a855cbd9154522061177f812f053c881d5aae1e7fcba732279e7d32d28f5f31a5c57193e9dbf8a31f814d273a29
1196ab353912a8f5d027f4fc2d97f5b66a07f6134b8d91f11599fa8159b1a41471845497b3deb972e7bdef0642d59b12d272
0fc1c860e2677504a6955d84332539fe1865d048aa42b9f1697d858e2365df279e57767b9f230e1d2a57ecf4f194f1aa5d90
163311969f0bbccbad51f88335a2c332c68d5f68864b8665ee911a58671be3fef02e7a117b25a45a6a27d5891f7d522299d7
c143aa07e989d6f6f8bc5be71af09c18760a26240d6bd0c59e899c95c76ee3556be69225f6b66c15efecc7934e85e317ac61
30286b3b07bcf3734060503144cdcb5c51d635f28d1949a6b89651a4fc2297f55491afee11e7b28810e304d2cc51d2e5bbea
5b595dc4027fb597288fc46a869896a975a4987be1ae22e6cb0afdf850327f515b6d77fef32a0603cf9e303d47856d1529ff
4f5a44771e02c532b543e15351637164faa4594ade3dacda916f5cfdd48c1fa7d82d21bbcf018b35676b079f558db664767a
ddcf64955e2227f7d1f97ecc00fcf0153d19920ff3a98b47196ef2a1733d7b5c7a548756bd6b7936e9ef4b1b908b915f69ee
be8e1d39b8c2f079a2a3a3709fda48c9657413201484c27819888bd2a9900f1f68e94a9c96b26de958da74ceafd54f7734f2
a6d51cc3d7ae33b9b91761927d0a2931478cc9d1a553e6669d2d6f78d45602f172ecc8ed555206d8aa2b15efe3253c1362fe
c42370b7fa76c6f5c69350b9d4dc89631283d56519328f91fa97beccee5d780617465b8380cb9f2994ce97755769ee5f4d79
59db5f5ba6ce703e7e945ef984424565b6b243fefdf3fcaa57eeac8c754b04ed3407e0fcc9e5cfea23e643f496de7a3fc341
889a9004ce9aa4b3e01763f16390fa34f1db15661d9b5f3e0bff3c13aec87aa24861340eb95966ed9522658bf232e6f1f828
3d4c16b792773dbc051d6e140c3ead4b15a13aa2bcf3908605717ae16eae74d74c7a7cf38143a67b87e30da930a3a599de5c
fc9b3fa91f4be95b76fbce902e454e1b870897f565eb23aef892862f6846662fa96a745543f0e7ed6a09c286215c4de4146c
1bfc7b2b5f556ebe96f71bb420ba567528d791ec96f530b85eecbd1defb5a5da5a1d63978e5e7abe0ad0747e98738e8473ce
db4b2149a933c8b6cf684571c0bec7e08eb97a56f942585542f07c7256bc308b689d7628eab6a3318b69ec20943291aa2158
41f7292b3efa25f16cc58bb2a61e3e16979dcebecdd214cdc2db408a55552692ca339ec99f22e6cfa70739f8442cade5afcc
d926c7ac931e3990b03c4dac73f2f0497aa33748698aaaf6cec80a217801673f50e8e4cd21656e3dbc88ba718b6210bd05a0
877d406ba0b5eabed225767e17e63f1b785a00665f6a1096003d276057ed4c2eaff7b7b80c340c0405d9046dd6566c290094
c4c63c258264ab99cfb312122603fb7c2525444c3eb4fdd02e7a12951bef4ac618e676725673f050c8e0f2bcf711a26e461a
b87026a0d06a77cc29c0dd3e14499fb7de94fdfaf471e8c5a125c654c43ac4f8a3d4c4cd38bedba98a01d0c7e16de5c91b8b
998af4fe645a84090f880ce103baebd0db2f82c1dc0acfd7438ab916fcaadcad7dd2680901391dad4d992f6a1e41863d57c9
e4e070f3b08d8c70529f16365d13d5a3a5f0f83597035e1435522cb257aef1e6ee2d3f22c2222aeb3dd7467217937717c677
6e0328baaec576751baa0ef55e4137936c7f33449512dc08174dd3ecd8e7ba924cac752c1e1d27a9c1d759cded74dece9a6d
f426e612523948837564f9be7f8b6d4fd4594907767b602ae9344eb2580cca9162b22ecdf7b364ac2be9840f412d63ddba6a
734b57bdab3c78893e3ee73db86bd070a5c0e24be406bd02d2b11096f3fafe0ff9f6730611bcc6a4894c29bbef24c7f17251
e4dbb2e597699bce0bd98c397dbd09db25613d017c6245e274b0f98a33f61a2d48463bb843974d742ddc7e20bc69bccf608b
47530a443a63751417cd0d532d5c880ceeae4a1026e7cf50b63bf4a18f9377c41e173f9d4b88de21fe8a6dd97f563fcd2af6
7efac862275aa941d21c3136b127785d7be026867d7cb4b9e6a271593ac26bec638cd51dba6673c6708e6be207dd7c14cdee
2eba57f58f98019610afa2e981ad2eabb5674a257d29d2f7a32ee894960ecb1b3f3c34fdf4385bc49a75639ef60d73c5567a
8952bd9201d3289494e6391d3478754acdb537ebc83d2784e6aad5987f65fb3375ee3e494072935517b17254c778365fcaec
bdb13706030ee04fb48e418efe54c54c89fccc44eb6309c75704235d186ec4c45cec47bdd8ed3161bbf283bc087032d1f850
579878f96042409d5be529c997fcfd8f2031326f16628c03f7d00c1367f7885f494c603d8b7afd997b6cf09eb75f6c1911c5
684eb9fa8b2493fefdcbc055bc13226cc7dd8fc3bc20261ea7c90c83e1ee592008c56f241198b21ea1fff30c2ee70d66b27c
f6e8f99284a391725e42e0a6b5ff6b507b7d65d6e1e6513e1a540e8d605d4b8a5ad2c862eb3e7f315ee070a1d8169366bc9b
54c79ac3a4c0eb67c01c4456ead33447e5c3ebd819a6dd265457b5a37968a2632a44447a45735931ced6f7ae549e463b8512
551f37ed24915a1b86288b4e65648e9780d1d21ff423487242e55519b89d922e19fa8eb5f90e110ff7c91efb38d0dfa38f9b
e42148dbcda2b9ecca88c6286fd6884a17752155c6667d64a8cc3c206422bdba9528b30f3162969735b41251d6ca6f75b2d0
eb3ba18fd9aae10a4025faa52e67e9f2cc7a58ddaaa1a19cbaee5656fcaaa08a3f1b20a310846f51dc3db7a5a834b873bfbf
aa0f8365cc5d551cf023a1eefed69828992b753c4ffa13365f58768a134deb85c98f891d050f21681b68e49d8996218327b9
d9807145f5b5218076dd4f8ff629aacf7e1f20b7b7d23ea23eed683dff68df4c73cc7b5a3ba66db47143ecaa75bff2879e7d
ce55785857aa0d0b3689e0e3eefa1b790ee7ec200e38d6894855d8a3a349fb79b4b870084cbb49cb91c32d7e6f1cff0b95fb
f75a319282fd536dab2b5504898e92fef2cc7ef56c0ce5e2802b6587928a10113cbae163829cc9603f2fd93353d9e20537ba
9f5f4f89ef654f909ff4b1d8ad66ccd0b037b9bdc014986b6022e35a039e88a2390db759f8a146f78895974a66de310ef6c8
1e48b45c29d77d189962ee1932ea4e239ac89a7f63ff39c04d3192ff7c5daeafa0a31482465043acb5e72db21fec6bdcaa96
cbb7ee2c0e5c7fce1379eb9224bce2c6012f916f6e911c17c94acb808d49eb87aa6f1adaed6e6a9b52ad095ea06c11391ec0
f82565c727bff4443cefeaed7aed5dd94b92c7db1e893702f5e676d6afda28d6ef55e713e3a55e5708316b6046f95454bd6d
3eeb1436d4335d983c903471689dbb055e50ddbff3bc92ac340022d16c7edaf6ecc9f0f53c5a89db8435019a32df7d5a4f69
3b85b3ee2d185d308b6e86426f4dee2a7ea3b4bd695add7e15a3e7959d90df6233473e9433aa64c7457627416b00142ca5b1
3a6ca51f532acd11f21533c29cc3525eeab8eb5924f3ee55f5cc84eec3e94fdc73eeef60555b2b1fb7fbbb72d783be746d47
d171b30a0a88133cf0537d74a284a7c02a366e7a38fa3ef4dd9795efe8fbd2165fa50c54c94beeb9c3048bb2fd34deeb8593
78fa7667db9d609ddfb6aed776741918cf1a375737bb83b449f2ca8ea8c6b2442d949ed1dae21ffc206de88998447dab2665
65d19a835d53c2778989cdabac7e1a84a56c9cbb198ab9bc99e97a3c012545905ea59b7f9eade42c4d2d3a8114afb965765c
eb4b1f958c6a9f5b4a753cdd87e6b9a9dca2d277ba72f7a081d324e44760f56000c7acdeb31b4127a1f787d5b5a9fdcefa2c
15e9c9dc8a76e34964a42dc376f9d2cfbdb9602591de23e524256a145c9e9ed3994f638cc16dbc545ed14143a1a50c1ce086
b365e78ededa9972820adfcf667aaa113e36f9287d4fe190e684ae39f343e5bbc7d9ab3f6efc90153372fedae762339aa2a3
1e9ce51956eefb43b3c4a1f6c9e8da8f2b4d99a5ab44a46be0b293e21b6e7b4577fcad47729956c91eb72d8d76ad2c91001f
06cee3ab342aad15ade49d0d0b2e22d212802cc47b8e9d4ebab95b8bec4190da57b2b3ba799a0b3c67c6442a8ea3694cd51f
12089a2ac7afd4a729298a1c8f37b6bd93b14ea5bb1dfacdcc4a4a5f2fda0aa4f582c284e42ea9af6ff3acb0343b09419028
0f650e7bc0c798b41ee380144fdd463cbd1f547a649c8c33219b41b493787b6ff3abbf112c3a02356ceaed4bf995fdf54ba5
1f357c279091cae2f4f882689f027fe1f8d0be436ef18e1c613a9d32776a3398d6dccc2d25ff8484bba3a559c97be4016faa
bbc6f5fbefba7b2c694ba421595650df854988ffcc333ac4f0226927c0e2cdc432fbac74dbb3b95ca650f7354d21cf9317d5
bb1b44e368caa72b24bea66b6b602fea9cc8e6113a9bc44b1ab9d75ec5b39f06aa360189ed516bde9ceef174a67a7e5aa18f
518004c01c5148742aa55d507602c5cff3a57e4f73d6d7d8d99e1eeba6e9a78c16e6e474a74babd7c0e97de914a405120ff0
dd066f91e1677c7ab107a31ea3f17108c237c2733a0e9500bed506969e33cc9d36220891e5e5eab68bc20eee1a647d37cb0f
9c5e2b58dd6756bdca314eb555be16e2ec0410edd93549c93b96275bb16fe4eae8de2571b961f3be08feaae5da8a5d3fcb46
a63f5beb64ad7d6cab3cf57d91f259bc6e319316a5fe7cecc6702ab446810dbb05dc9e919e55d437f9201ab640647217a4bf
a9cab12f766528bb86954536e1dd805502f9c665197332d51cfb6d8de1ef7aa6f09d064074f223e6fe47aac65c3afa2735f6
5c228174b9f73753aed1ced85bca4e33ea7c3562074b35a501c8e4e2cd1ee609b8f5ecc2916a9f8919af7b268c7e10629a66
01ee9234e891d1594f154a4c4af89f17da810248e16b2c19e62cef6819fcd44708dd13ebf346aedea1363bee52b8568d081e
7b28d654f4de8c54259d813c9d8af8538986a1f57379320470ee0d5797ec377c5a1d4650604416ba8cd695703ae04ca18d19
30b0a171fac15c39d03c3a4b90ae2df055cf5d37b5eb863ea300bd554d894d153ceec60a94df072b93a582d5ff01487c19ae
>}

\EmbeddedPDFStream{schematic-44}{%
/Length1 1391 /Length2 6192 /Length3 0 /Filter [ /ASCIIHexDecode /FlateDecode ]
}{%
78da8d770554136cdf378d8094342a8c6e18dd5dd20d4a08630c98c00663c046778374091212d2215222a14888b494a441b7
8084c43bf57eeefbbd9fef3be7fbceced9ae7fc7f5fb5de78c83c5d04440d9016e0fd180c39002c282401980aa9eb1040008
14150402458839384ca14857c81f2d31873904e10985c364fe975d15010121313a351012e3a6078701b4bd5c01c2a2006109
1961491920102002044affc7118e9001a881bca10e003d4180361c06f124e65085bba31150276724a6ca7f8e006e300f4058
5a5a92ff773840d90d8280824130801e08e90c71c35404835c01267030148244ff2b05b79c3312e92e2324e4e3e3230872f3
1484239c1478f8013e50a433c018e2094178431c00bfc605e883dc20bf071324e600983a433dffa84de08e481f100202c028
5ca16008cc1313e0057380200098da00132d5d80813b04f6c759f78f033fe0afd500840585ff4ef757f4af4450d8ef601018
0c777307c1d0509813c011ea0a011868e80a2251487e0008e6f0cb11e4ea09c7c483bc415057903dc6e177e3208086b21100
8499efafe93cc108a83bd253d013eafa6b42a15f69304b568739a8c2dddc2030a427f1affed4a0080818b375b4d0ef6b7581
c17d607e7fce8e509883e3af111cbcdc85cc60500f2f8896da5f1e1815f13f3a270812200e040225a5a401100f0004057616
fa95dc14ed0ef96d14fea5c6f41fe0e70e77073862468004401d21981f623f4f9037048044784102fcfeb7e1df12b1b030c0
010a4602ec214e5018f13fd9316a88e31f1973f308280a6005c4004f1800fcf5f9fb6483c196031ce68afec7fdf7e50a599a
a9eb9aa8f0fd1ef86f938a0a1c05f0131001020444c48100616169518024e610f0ef2c8620e85f5d00ff89d58239c201d27f
9ac56ce93f0d7bff75fbdc7f118307f0ef5cfa700c622100ee7f006e0d140782315fc2ffdf30ff1df27f43f7af2cff0f80ff
773f1a5eaeaebfaddcbfccff8715e4067545ff65c7e0d50b89c1be1e1cc300d87fbb5a40fed0550fe200f572fb6fab161284
e18032ccc9f5ef25423d35a028888321140976fe03953f7ab35f047385c22086704fe8af070520200c04fe970dc32ab00be6
d1f0c4e0f1b7098221cdbf4baac3c070875fec12119700801008109a1873c518491ce0278ca1a10304f51bc1002141181c89
090160c60b0038c211c4bfee5342182004f1f0c26c1da3fead11930608611eaabf654c6521a40ffcb7fcafd2602f040243bc
dfd0c0f4f51ff937cb211014044c3c3b0507cb863f6a086f3fad53beed23b0322c3fc1b16291c523e0378b78ed754e4698ce
539b1bba84f8a19c3ef08662feab3af7b1d227e64bbfadd646c2a8b654a3573ffd2f6c938dc7565e11cf8cd2f68e146f29bf
787797e88e80a9d2aaffa587bf79880b6e2b76a7364781879714996111d5a94f8f26eac5bb8ab90f91532b46abb5123a2417
15e3020966f1d621651f390aedf326e959099002776ff0deda47917f3cfe31712b7fe49a593b998f38603b41b4c4cf725924
f16cd277a1ca54c4b38b819dc192fe2eeef1ad0f639c7e2aeb19da74d37ecf4b96101f3227695f0f7d764b13765de7f65dd1
373e456c7c7fc8c93122c3cd402f844bad1517bfd4762fc5ad8c2585402c6f7f5e7b09a5f55db2d5de93df20057a6de02ad9
80573551cab26a4f7d6797f6e1d6d2ea58e3b0c0a6ba02996819525693305e6e45d6f540e62dd1698981493a45d167135e03
c5b7d2ed05234bb39f97338fea6ee59fc6338fdd11b88ac2ff594e70ea03403bd14b61e97a3e404eb5a8a79178dcab730716
2b0c4691f5ad10aff37b2c6fcec5b6a6917b24d545521355b9515aaad948be6f6b357011d0b45db0fe6232485587d6cc3dd6
7d7a7499961399a4cf2875eef4645bc488204066e36d31c5aaffb9aad07541417ee307c69f1f07925d6f8ea5f624a8ee2f88
d8750527ea7a890e2d902fbfa2551bc6bfc14b3ea9da572ecef47a2c7bc5f3e3986e9819830bbddbbdca9b1a1e73a68eef2d
d9d78ff6abc7fdb3e7575b443ea05c3fd51844794da33a529c34d785fcb10964b639f56141fbecda03ea46ab7be88bb5c530
b1ebc38f0ae5ee68ae6f4ca25961d78fa00b827317138c7da5f7f106a50363bf0e0fd90349bc56f3ec65352aa4b80d351ee0
f84f53ab9970a4a0e5af229d602c4621961951a3594feb2a0ca0cfa3237fac563f4fbac71fb90692b43bce4858b5b5943ba0
72ca3bb0ecbcdc1a1abad55891052158cacf0f0a9acc0ff5488f160a2478c465a52ebc40747bc0c789bf5fa73a38ed880cee
36a5a027d5285b872807c4aadea7c6bddbc49414654ee94243ed342320145e75b469ac34ce3cebef7a608fa51a22d43ee21b
5aa753c8648b57518303e5e79a121e313ac8967c982adaa6de7a3a51b1b71d1477579a1db7cbf23bd5a403dac615c99488a0
28a094cfa7cfc979af104f45ad13ccf1611c4421fcda7a971b49fe9556509dc9f7b10577345ba89efe87e426894f81bcd3ea
b2539dd7a3aabd0c45b5a087c22768f1b2854e2ee756b42371bd2725c52741a990913af19a2e309452935d7cd009b8a763ea
7a85eb7e31ad34ab60c370edf86516c7627c2d7c4ed2e8b1a6f956da7215ac982270305fde067fc332c92e9eb26837359277
15275092649df52979c18dc79fe2c1ab33273e77812d853865b197667e0ff26082ef6d515a9d8605d5650f4a8d6c18ed5e26
97e3f5eae9b38c12a4d21e777fbeefb321b3bffac2583612599d33439f9bca67e3945cb74965f9917a20e28af412eee95391
ba6e6416c293db2c1557d30f8afda2fc10d4e7ad80ea4de819bc17da6e0b7b3f1b8815921d17c0b0c77c3d1f146782cc79af
99172b5975735ca79d5e325aa4c331af5af5e90c14a849e326eca5181a46372916c39961b3b5f629abd661e85a2378dedd6d
938ab0c5ec562db9734bc144437e3703bb4f8975adb603a1145f974981ba7c59d7fd986b58f3a07932c7c49ce278ed54bee6
7b2bb1b0f8b76a233c733be38f09920336c3e4125ef26c73b85f27a97178205cc6e0dfc3ef7373122477f43581f1626992e4
1e7e6c330893d5d0c22d2e6c3656cd5ed33917c0713c7ed47ebc9b39ef3b4cf8aab9706c94691e90bbb57de1fc35de936f99
3795c14c20b7967989dcbae792fb4eb7bc9a95038fd2c522071e32f8388387edf96c4a8cde6ca6b2fc99b83efb737a61a34c
976a72e7cad59e5bfe776a835ef358f683f7e669a6dbe3e40392d24724ba1ee27cf029e682787d74c4bf574beace82123cea
2264f564399b3279ee3ed3e0ad8f75a33f132d951ef88323e76954c4282db5ed8edb4735a9be9668c446ddc00be7381f0b2f
dbeeaf93126fac689b52e49b8813cdfb6cee3ca86c1151e46827ae6e7bb7c9132ba4639cd61ae40f91df9d1bc2afc02d8cb1
c7b60dbf6e029aac87bfcf106f8b8d10159ee9741b40ad2a64dc9ccc6a34c83ba7a4a9eb62bacf62f5357476a6002fd1ee6c
c01bcff1bb7fc067e3f3242231d6385a82401e4a022f9a1def9b4beaec6695d91dd6712f2d64b7e29f328eeb00ab6f4abc62
9f2a295b8d4baa3674517c37847e85be8e168b68a392c3ffb1b0b83f98e221afaac90ed8a7cb68517298edaec76f0e3c1c50
1636920a5516f05f8ab117b6f3adae7f8405684cca982bbdb0b4a62812ec43ec95df2c08fcd6e7d999cdfad27e93cec6189d
9d9dcd89e857035238ebe6984cf26c56f94ae8adca7fc9a4ba62e56a353058cfb8eac5ec9fc84f842f9d9d62e228a04e6e7a
f955caf7f716763f1ace3ebe251a106e90e7da4fcae59f5a702e7ffb3dcca5f7f1ec96aaa37a6e88f8b9864dec681c3a03e6
ec7fe783c30d3a42825121eece40ceabb52ec3ba1215e997aad7a8e07869e2b3c54da2c71cf49f4ff8a5f309b129470c6ee8
6c4fece8d267cde379dfa1bec71f4fad309e1871147f875fa2310d87954fadb7e510e1ac2b6743a8cc27b1aafd46fba5f4da
ca5c67d22e4d90c8c9233916215791ca67996fbe6d768a33f6a6d777395d9b63a7f411d1adf012512fe64a693dd519015b1f
e56e1b305fe495454e24050498666c9ed02cfca492cebba336cebd3108c69990de2bccf2844c3344e4f9ec184749492eb034
0cd7c1933d535ae9ed91693703558af48b0d994fbbf7efdd77ea1f22645379c1314cc3942c7516599c0efd929fd69ab67d74
e1543247aa9996962c91267945219c2967cc3a95c0e508ab1610c93585f034f7ad3e53b6e8f50f11953b79c0ff34bde94ccd
fed2a726266824582402786dbe3b3fcd516f599343227f97b1ea4aab3fa2afd0e55cf562674faa2e5449c12880c93e42a3d0
0a6f93eab0dcf790cbbd80d3e43516af59233bd38b990ca6675889a1203be3312df9c9a70e3bde69b6cc33d64c731679e527
2acb6dc7f5b654afc59bcadec1acb20f9638bf9b6d5e36dd8756ef7f49d67f550bf61aee44ec7bdfd2cab431d77abe541369
6db2bbcbaa955843b0ea4d1694ef494490e2f1c03d9760c4d0e65adcff1ad71f61f029063b181b0cde5001d21c26e898fb4f
176c458a9d2c0255418fdcbf946e790e4829f7aebfcacb5c23fb3ab4f9f2baf0018e98c0f6c4966ee3cadacd1743b53dfb38
5092bd855d0df37a3f8e5e14ee6d757cdb9bd5b9bc76a3ba066fcdb60f89fbd2743aab35df0fdcf5f40faa70d37ef686439c
61a1366564afd441d99b4a8ee1d1724d476803be8005ded31d9925727e791a95079b55576e8515cb59db1056843892ac9abc
a6287860e7fc5d5903de9d6d05fe659f5994962de018f2c233a699f2f3875954863112aba8b1c40812d38eef2c7b0d10b062
ccd574bc9b361e8ecd6d57c87b7620f4b55dabc1cf675f0c9a650062fd79dd93b0674d95bfdccae12770ca57786e5729b5f0
5949f6ad836561d001633f13ceedc5dba8fc430b2efae57d2765ece465765cef2e8f510e1caa54795aa54a1280219adad6a7
6c74a3ac23d82fb6aad5bd5b2277696cfb8a4c17a7482fe3d22d498fed92d8983862e3c193a0c6f6dcca31f6d182edeae0bd
c823bcdbbdfea512add9fecc2bd949fd8f351f536b6536dbf538ba78dc060528215bf02d39f3b892dfc87fbb215a12fed11e
4fabcce98ed953f11b6190e0eafedecb72e5d7b5d8df0a026ef15f947de759a3958132dce805ef3b5bcab8766ed48e87de94
d3c891fc8827c4277558db695db8d56d2ef822f731dcdae57e312104bbbde06d0d1edfad67bd438c123e252a4d50d56e7783
2724f5b7c835de7767eaa3d46c86f2e598e8def0113db004cea086b3eff66fe2455959064e85a6957d8be8c876573c3db46f
96904e0a5139d178ce4faf992809daa5eb36277830359ff2730b7987f40d6ef40fbd206fb08208ad05db73d12ae0da16428f
bd1a44744daa45ebe6fd64f4fbb9c16607eb737e62aa3019ee020ed5adb6c5bd723d68b3d2de90f5719e553dc111f1fc7142
0c327d7c99467e3dc14cef0066a5624bd32f92618b9b40a604f3ca52d41de7a751b330a3fe7015ff11d582c520ef1590cee6
d6942e717702a54c7a17f510dfc8f2271f0735d1e3ad185e3fba91d8c96604c0f81e9115ef0374661b7005da570d04d9e35f
de36942c1f404f75c4b87b33b994997efa306d6accc7f14e126bf7e20673c33c4b4cfd277ac7d3460d728093e6e232569e61
579c88a20907957f933d47ddd7abcbcca26d7d4e64af8c02bdbac1fb07dada7bdaf8d7c5299689a9151bd72af70458b002d1
ede30cee931f62c7abb0bf1684472f4fd53aec9052fa0a6d169f2f89162f3fa2492ee6ede2beab8d6df8e41d3b576d71b5f2
37d3d3a6e6db831528d6a67747f65962816236f7c926db56533a4eee6649d7489bcabdfe98fffe89ce26eeb388fe7c695a2b
2dbf2cc3738530c6307132632f3027f76e05ed8f09e07c85771e65c34141beecbedf86e8250a2fe0952d8760079e7d3ec4a2
6097b2a248046282da56c30e95627a56791bbe9dbdfed96fee55d746f37a6ec00d824e1f18bf85e17e01758def139c8cf85c
d9293ee3686484f9d0fd83aff24324d6e35ea56d2ae984ef760ecdb855bd7b717d14ef4d133d425e0762a9e517d3a3903be7
13616de7b1ea862956fd2872acca09a18486f04291e6509e9a31bee5c6e66cf27cb2b1f7f503a8880f6249cd27e56449125a
f79ceae0ea535c514533b006fbd5db3dd6aed51f74539cc1385b17f64264153a012e9be639578b6a445e0564126c7c731bf4
d27042fcaadd3743d017b4b7b89dd0e2ae267107512d3ffd3afa08492c5f0f331bc590c63703f35e3a1cbbc7593809e826d9
97e4bb9ffc64405bcff912703b199749dc5886ecf50d7755e924bcbcd2a42b9597361359285bfd1c46fba255b5eeb294e930
69f21ec9976eb87b2282ebec139e86c9d6001fa9d9eb81c179f6d3a6bee1362a06925667d959a2512decc240c290e18ce250
1926b8da69bfd9eb6a7af5ca338d16345c16f8e92af8eebee2e2dcf89c9be1a80223eb205b5f7aef791d2975a2a9b8360b2f
e369d00add99099d5299deaa249b14e7d4b64069d959013a0476de23fd64f1e82c755cbcb94e33bcde7471f2e340b87910e1
d12de293cacc1d3e0bac7a132b46834d37519a64433787980d7eca4abe3493e051faeecbb203d3edc2489753f43185b11c3b
09cdd0b78560a2cf22d7a01b9ce33f5a68f34366e44b543f16fcd8abfbac3f4c2a4d2bd9c456d1f025f07daf5edcf171eebc
697d9762a38a2fcdf769e3b1a1d6822223ea06dfcaf2699b5b549e7115a65bfb4713864a86e1d96444119a027d8b006369bd
cee015dfea95c3e0151e51f2a49c5dbdf61f8425f7aa9f12a26ebfb936e7224c2911e49ef0ac3b975d1ef45fb2253030cd9f
7c4a787f6b5cda8e49133ffce605bba1c56c31e4c15ed8c1dc98bca33ac7c762034686447707b59205b723df7efcf2a91e9b
cfad34910d2fa74af07dbba274d2f8f0505a34e62ec43e84f53247477a6d84056e63d3628317150fcc3bd80acfbeb4b5e4b4
ea5d14eddb949ef2f6154d3f93fe4a198d873ca3176743b87dfac2a53e0d2183b30fe71495a52f8cf3f94ac184e415618fc6
5ed03f1de3ae269cee19e4e6a7cd29350811f121d9909328d480810b845956945b7d6f9b8f1adf3ea7e9d1ca7eb539d3c067
ead205a793aec99a93215f1ab0b03e8bf64e12a4c041072b8bd3a4adf6be71fa8ee568a3f84882e7ba17f5a4023dc7a2ff58
bc28ce897f4246242f2e277a5723a19d9263cd6b52dd6eb766076b42651d15e47372837ba176203287b391ec61cee150685e
28759c7ee9b1f9c87afddcd3e211e57e1552173e4e9c374a035c236359919e6fc222db226d65db08b3ab9aa9633d8eb96dcb
2f5bf26476da0567657876436999acc6881815f1ecaf90ef04f5f74b09fcbb3e1acebaf506bce9128af6aa5d6693546635ab
7ccbd3146e6a3d6d89daba478a8007c045e6eee78ff50c8284cebc0f5e87a889ba98a51f59758d4ff331f26f55364cbd65a2
a7e71fc4bd6140b223c6b114c58e4f758febe25dbd2a01a5d0adc6aba6b864bc9bf11745dd93fefc45def2ad7b011b91b9cc
aad115b9b9683b89076486bc2dab478f7153118b5d52841cd47e7031709af06a9b81c9814a138b8332e10d89ad6a1b0b0ea7
2ef08a69c1f7dcc5fca67207bc6b05bd6891eed687fa8f8276c9baadf42795d20b782d5ffeb8bb2acbe6ec3caf23b9c62622
5babbad05414fab35e84d447f5d6e3dac464d609e65eab46b45aa5763f3fdb907dd09b18cbf068735e49008a72294be79bec
d3b8c4d134f4a07770bab484168f162b113b4b72cfa4d0f3b397521dc7f0146ec322b4f1667723659899c9333b327239e633
6a163f86d83ccdd2d16e483b5f68d82ab76ae80b2fe5ba371e780ca3aeeba6602fab0c92d3f44753f28fd573080e26f31944
4e55c538060f1b83aaea6844f81c1046bb6de3b445d442b2ddc905499f92cf02547fff09550af2bfe60487bfebdc4b10fea9
d58f73c720a00d5bfcf32e2769288b1c42cf36b5eec055756cf6c91a7930e3003ecdb43e395b695f52e20b31b234056f7e3e
bdc503e1cdb52e77f718dbafb589dbd5c07da861f0d5507afa5e7c687bd54154528b534987a570f46399e088c05528778a6b
7cec5779178f1a506dd5be609730f9b7dc99452e19599ea9451b31e67dfe2d093a2f14784d7756b91e1436fb90a5fc48970c
ab212685b5c08bf629f3932fc403b981f22523aaa3a2e1561d004161a94becf9475a67e7a6063346a4c44661d734cc4fd3d0
f3dc25ea34bc5d3f153aabb896a296c2c4a349999c4983c2cd833317ecf8d78fbd11a2a677134d27fc96ae5f8bd21874cc5d
eee5d9362524ca6d0d874fa86df311955d632f3ad32c32c634883251bd0fa592fbbc0215dcdcf7b9c728533731ccde7e4344
8d3ba4f9b254beb27ab7728249075e95e7c29ed77fabac935736467f742815c066a141c24940cd587ec236fef3c53a253dae
58025940912fc25271806e80ed900d7c705c70eb60e409733f60051fb4b29a12ef7e1218e79673e49bfc725b87bb815e1d72
c411e9386bb2e2e370e6ed9fa9dd911772b31c12954a756100577d13bc77c10f34422641b1d8e303a64259fd599d7a0c9e68
d5100bd547c96537a81b56f0324f2f71aefa3aaf3e236fce516bd78e5ea0114cfaf13e0cccc0a8d66779f08538bcf5da124a
31bc597fb294face39e76781951c93692fa4c11f10fe29c194679835045f79f3f5cc67610b5ecd657f2489f98e35934dfeb3
f43cec8e12c557797a779641d114c7044ca75e2b1f0cbc226aa463183e157cdbd8b8a8fc7eb52e35b6a91bc895f068eec75c
d1ee552fde1385627793cb200a73b6d46e2f1985d78e762364acb62a7970581a8b8f7e4726c9085a4855fcd908c25ef113d6
63c99fc34acf14f922ccc5e3269946b66ecfd4541712abc44748b4bdffb1037ff0f6c023a8e761b5789aa53ba019c13faf92
42f7f08b0caefe3b6d3ceaaec6ac926be0a1a4f248ec52192279b22c405dac60c057b97a7d91cce231e2c3370ebba97bb3ee
966d27eef2037ef9f0a0b7825d7e6b6c5c15c72fbcf428d7134deb97bfbf4d7f876715ea74a8e5f7d64157593fa5378828c0
1ef70e643b5248439154b9c5c6de2b6fca7ba03ba4b2d8632135c94a546e444c4fc5b498c61dd4bc8cd07f6d550d8c58ba30
08a612b39e2bc659af8d5481f7b67cf89aa35145f2f8390e56dee5dcbe42ca20d73b42871a18aa5d3f33890d4a51e062abbc
f8a469f245585d91279b6cfb279ffe1f43c9d8d822255314b15d74cf43b1521aca6c7238e1ece0b5e7584534760e51e8b447
dac1e40b52f7273d77b395564be626b74c58dedcfc6292627ed5787ad5719453f5744bd34604a70214514a1949e88f7ba80a
e4ffe16a3cd25e43dfbc230b0e2ced02c7d38f463d6fbfbaf5f5a0c5166095544520b19d6d70ce7930936701ce0e2a507b38
74eed94988237912957788c8b832ac3d77a337bac6cf9f09916e19ddcbbabbfeacda7b39d13e3741f025711a8eb4c805cf0c
bf1943ac41f6654769ffd5aeee01731a85cb6246443d6d664f595e87bd109b1fb6f2c8b5549f89667a73ac6ecaa2edb7eab4
03190358bc85df5407ab71296bfc2173fe3756eea9a19e2d7d97c21c2e30cd63cfb9045ccfcbd3e811c7ece64307f01b55bd
b267e1a931ab460e4749df821cb1ed13e8ac216bdb87a177b328bbc0af2e37c4f5eb75da75f6fac7262121767462b941f545
23b08e94da21301f51c41312320de36845146220b6b0421bfb9d6cd628ab03842d7e793569c26e4e59ab5ca957a1953435f3
75beec35b38276d3819ad4689854f22b8f87b3b594cd568d6fe30484fabf9e70ea2576eebd25e58a5e10d8f9811eff3c3449
b4884f87a36f4dce9557c6527e9f2c543c4d67c5c363ad77e5d8a342b670cde5fdc724e45111c3092ed3ed21e374ff9052be
1108d7fa0e3e3f0e6369b26f4a46a6951ec578e1a7d579be96a860129f16b2c0d51e7d9562ecff014a23cae6
>}

\EmbeddedPDFStream{schematic-45}{%
/Length1 1376 /Length2 5906 /Length3 0 /Filter [ /ASCIIHexDecode /FlateDecode ]
}{%
78da8d74055454dffab6748774281c4a40892186ee469a21a4196606181c666060484941949212a5410169a51b4140425a14
0911440449e9963fc6bdf7ffbbdfb7d6f7adb3d63967bfeff3d6decfb3f9b88c21c2aa708c13420b83f616161301c901ea06
9ab7640010484204041227e7e333437aa3107fcde47c1608ac17128396fb5f00752c02ea7d61d3807a5fe00c3068e0260e05
88490062527262d2722010200e02c9fe0b88c1ca011a501f241c3010016e62d0082f723e758c873f16e9e2ea7d51e65fbf80
004c101093959516fa1d0ea8ba23b04818140d1840bd5d11ee171561501400c1c090086fff7fa4105070f5f6f6901315f5f5
f51581ba7b8960b02e4a8242802fd2db1530457821b03e0838f06b60c010ea8ef8339908391f60e68af4fa6387609cbd7da1
5804706140216108b4d745040e0d4760818be20044571f30f240a0ff80f5ff008480bf7b03888988fd3bdddfe85f8990e8df
c150180ce3ee0145fb23d12e80331285008cb4f445bcfdbc8500281afe0b084579612ee2a13e50240aea7401f8dd3914d052
3501a01703fe1dcf0b86457a787b89782151bf4614fd95e6629735d170758cbb3b02eded45feab3f0d241601bbd8767fd13f
277b1b8df14507fe5d3823d170e75f43c0711ea2e668a4270ea1abf117726122ff8fcd05e10d8041b2e2d2529200c21340f8
c15c457fa537f3f740fc768afd325f4c1014e881f1009c2f864004219d11171ff2402fa80f02f0c6e2104181ffdbf1cf15b9
98180047c2bc0127840b124dfe9fec176684f39ff5c5e163917e800de8827b6200e8d7f3ef3fbb0b7ac1316894ff7fe0bfcf
5754ddd41862aa7fe3cfc4fff6a9a961fc804061716940585c160c06c4c0925280b4b41810f4cf3cc650e4df3e40ff09d645
3b6300d93fed5eecd3bf5af6f9cb0081bfea1004fe99cb1073415b0420f01f96db82c020d8c54becff9bebbf43fe6f14ff95
e5ffc5f2ff6e480b8742fd760bfcf6ff1f6ea83b12e5ff1770c15a9cf785020c30173a40ff37d412f147b506083812e7fedf
5e5d6fe8851254d12ea87f6f23d24b0be987801b23bd61ae7fe8f2c76efe4b6628241a618cf142feba5700613110e8bf7c17
da82ddbeb83bbc2e38f9db85b890ce3f4b6aa26118f82f8d8983a50028160bf527075d5049fce2a803c52ec40847f8fd6631
202a82c6785f840017e30501ce182cf9af13950101a25e3877f7df579b37c2cffb979bfc1f2560382cf64266bf497051ff5f
ebdf9a4620fc1030f2c90f18987c845b5544cbe14b55765fe1c5217149aa37dbd1bba4965f35597d964cc2d5158e76249ec5
7c546d1fff16733447b72a5e745e7ffb1613e8b6fde73b6931724c93352d44ce4e0b427746e8148f57bdf471e2dff45ebdb2
769440c35d98361b4c7b7d7755de09320b796e2dbecffcaead70ada1e2a0dc759384bf6e4d84127f282e868445464682123f
a25535b3b542ad52f5dd789b2eafd7b6cad07af255adef4769dd6fc5c15fde4c9508e121e7d3d85756a83cf4763357949e2c
bf9662d3731c01def17d7dad49463e57aaeb88d19955a0a9b9c9ae911b7592b9d075b6c7187ba7259d76618547e2696e8029
fb483bd29f5c9f5b52dc45f3c5a647686766e8e77d9b2caab3ebd0c75db84b01657e92372f532a3c8cf97a4869d3c129e587
1656deb5bc9283ed323611cb48ecfe59a49b834ab010e92954f2ed16ee8eb58e41ef7646cfbdd2074cfc68f186aba673b797
22dd863afd3924a7b2827deed164e37a62ac5f56e50c8e4e0b1120686547aada7850b1ceda9ff6888c485f37732d342c35cf
9b2bc71ed9eb729da97ed82c8a53a6678f160bbc66b51f229ce92ae08e9c77174169e16d1f2aee101c0682a73ebfd00844af
ca7054d673a29e06bc9b086decae0ecd39bbe7daeb083bb9726369d9ec5379d90083e2b9823c6fc94aaaa69084eacb71ee9e
9cc915037a96b37d7b5977d60de333aaf7cc97390b0bd3a7ce76ce265e25e9c893d306ea9f8413ca8272625e4c448f2b25db
d58de9526c4b056c9ef51d9753e2e61874029b2458254df4c00c77523137226bd929722c8b8405840433da76b6acd138db5d
bcb49e84e47da97c43ce04946745b08078eddec663775f6be78a611e4a216736c5e5477ef1c17b2f44750a3242de04063479
71279245d2339b0cd5bcc400d59e451b4362b32c4197251c36c36f3434be9a74859035d3b11a17d9424b1dd65559a16cadd7
8763b44b0aeaefa85d2e8d77f35d215b681a3ecea5f7f80c69375e39e5c1fb39744df196affc5b06c0d5afe9bca1efbbe09b
b123043766b7930a705b9f0f3dbb53e93cf04c1b3e4e0de1f002604e94bdfddfeff17045acc80f84e07b8ebc21b24848681a
3250e1e8e20974870b4d4d6654811db21dbe17be612547f9e07d1dda164bb4da5f59f2139e114e3f52bd27c3ffb65ef6f6d9
ad15aac7d1fa1272f50aaae324e3b5611b94017832f5a33c46e0943b990d8ef4958b522991e6f7c3490fee84663397bd0e98
b8479834f456cc185d0f6f2d7a3a04ca70e1b440b206279c66aa50414c1f783db5fe2938babefa0076774f47a85986416411
3186ea1a8dca4215fcb88beca19aada67a67f704aca4654cfb95ea5e9bdebe09a34db78e41f1e7a3b2b48446dd651ad97747
e3b48bf69b3cbc4e48a3c45d6ddf67f74deae93e6773f8dd25cddc31379666a7f30c1d6889e153e359d369bcff8925d95f30
a433bb6dc4df303b13a7bc60dd79dff7baa8d775fcd7da389441dfa96c219816f6ad2fd886bad0c9f01aa1f7a0d4eddd4b9d
4e493af4d3f4b84c41b8f03d5b1d2fc7980e8e1b11a45cdcade8e9f589514266c2b89d8e7a0d42c70cecb5a3afdeb9e4c72f
249c2daa068a17ad18f7377e423e362ba412296236059ca79c703f091fd2efe0b108aa1dc90d7f91c2b81db16f7a4290c515
b5f45b729277f9bc6d219bc3e0bd50135196b0f8ecca19ecd476f6fa2c9559c487c9d7d1e44852359ad32d0199d2de1e48d2
8f03bf88c6d0fed2d4b6041672a76db327451654069d45f5cb6b889eabd3bb490462c5a7dc2d44a5b61dc6a4726f981f5990
1875104781022c3e6be7777d8d3c5620275b52b6ebb22a3f13fe482c44705bab729ab77c2caef31e3bf57321a726f35271f6
fbeb5134e63392aa7d638e5555fb75cd929e9dee9ca84be5f2b9a538d95a596a480867864333f159201bf88e317dfc449c48
dc3578ff90db4b3a6be4a6bc219feae9f1187e62b9fe3ef2e8fa0362e7b45d596272dd3e17eb1068dff0cce72b8a3a75ecf7
9f3bbf3ddb7dc4fa5485b581daa2307033222cec98f02375ff9b67911620150b57d8bc9a1d504cd49f403a89f782b28a0a5f
8fd3a4cc3825fa70df446a2c79e15aab0da37698c6e95d11bcbbfddf322846ef10becb970b44370e7bf0065af0e860fb9f1a
ecab9229432dc62f3dbbc356126d4500a38e275f781a5d9932005a379f28ef7ff2d29e6ee1b9ef7bbe9b8b122bb9251c7146
b4ac899b231261c8f87ab9deb4e0c47e6a5d5b26b2b9bc78da5bd6e72b1fde315eb76aca51da734e30fab8715dba9d737a8b
0ba6b55497693dc5a9a8f7cd308b8bf63c818959903e7889a25b42b472757ca3b9e17563b80991ed816e4e81e251b0554ec1
31e718673e07f581fd17218b11bcde3863f13c03127e750642e3025b2312b16c692b15d76f50daf21e6da10fdc2d5b1c6391
bddba6c79b5e4aa77efbbe4c226849ab864236af1e3ae4e6741701d76ab62b34aa5fa07f52cb2ddc69347033e97eda002584
31b29caf99ce6aaefc26c1de762d96dfc034bc1cd71a1b2b633f083b7c10c43cff606976ff768b05ffd5e31358052823a71e
b45057d99cbaeea251acd8cee666a0b9e6446a648fcf1223732584806a3976326d48ff56f0355e60d0e3d300537333db6577
1bc6ee446ba762470ad3570186b6360a379d0f93cc93245aeaf886183952dee3818ff397c7e79cf3af017b77d550dcfe18bb
14e6d0749c83aa4a92497f9928adfc94c34e4833b3ccb1814bef73cf829f824acfcbdfa9b0f9134ab2865980bc1ba3252221
4775d84ff6da8f4e377c94cbe734a4bf9230baa7f7501f5630e542c891b401310f7c4d7f10933f485c9377b3c9f4a0949a8e
596d201f6b6959eae9b5e369ede235a3d0cb12653c084d4dc6492dc829428a73124e9404c5a3b5e53749d07cc44437c2a12c
9ced7dd39b4758ee893d4a4ca76316e446b6d15569bb4cb2b0d400459a6a92f776f0ea7dc82084e0f61543818efb31bb4968
87c7d77c69963c6669151cef6f77d7ddd1b96f92cca5be103c9978f483ae91bfb8e0e57d06ebf92567aef8f8a5483fb93c3c
57870eed91447c2e9f30daafcf77d3aa881bacf88ab8d53055d9860fa23832eb4c0288b2a1cb9b0a65c115cf88afb0de06ba
0ef60fd6357466d8b17d6e4a81d305e21ef58b1ae9dc39b38c69545d44a4dee7db1a265f6a7542c414764acc57685d965b93
48c2745ed52c58c7d8f233aac0ed7466eaa1c988f764fe8c7bbdb2d854ca55653424965d3c619707449aadddd221cd7d5b7e
dffe2d05164fab54267129445e5cede4874ddc72312745db66a53fed7149b90deffcd4cf73be0178b33903e4dec9036ae3a5
38ae166752d3b1f1570f93826e1fa6a659975c2d3c6c1dd6ab62ea923819b2ce7d3f68d4e6971472d0157d7683eb41583afb
9c2e4eb1cae47ba8505bba1983e59d57c421b452c5776b8a1f746f757e9fa91f7b36a2b1dcd8f4d214e493624bec8466b3d6
7854bc1fc7782f97a1f82682f38ab58d34490c4caac33dd136229c7222a31ce01e1bd4a855b7b3f50083fcc0e5ab5d988e10
81125031f3f6f548ec7be412450eeb54b1b53ddf22e6fe62f25a6ced3573e5fe229f4b6b7987990f407c16398f7be21783f3
cb1f06f0470726492d64a4e47970cbafb37ea489db608bf7c19f7fcffe89f04991bb627bc0ce7211a617c674ab5def399227
f7acaf9f58e774b043e40daf03dc92ebbcc852116f81e299a2515e8dc581e7e24f9ad4a03107ff528a9538f5c026e9edebca
5112e6cac149713d9aa4fef75b5dcc74cd87270a03f0e75c4eecdff2e10156cd03eba31bf86d65a384d946b3a885b8551a48
e4b9ad914fce9e35ac7d31506dd84d3b822dbe4e26a7cf0614bd9f1d99617066a6edf03e32c2d6c836957b7cd2f10e298346
dc493e530f2fbfd4bdf9abd7bad36898c0571fddebec0a96aee39d04e52bf1ef6c4419e976d1ae7b268866bd63a7b2b4c903
8df6d7f25126f6815f3d25c5b8f93c7702cfbaefec2368cb129caab287bfd106967e9cf9e88e2917e5e93c352cd27d29a530
5f4b5d28f202715243a3ae88f6eb0fbde2d4646aedde8e8b42e5058929eddf5a48aa4015681e2463a3d322d87bb84d2ca45e
7e6986c00c4df38178324e0e6ee56bcf8263861fb7ec96eb537eeb11cd22cae0b1bc4c9e78253d7d738e8e2724c90e3e73f0
ade20b640c3facdaea9de377ab5e9dca6372ffe8f8d60d0222053ff37847ac453b43daf15b0495a0297fb8cd2b6256c5f225
92f888f47932c54a7d7a0fd129e159c1dc3209c62107fa3a99f9a0985217fb6b14cb6cb24c0d02025aeb2ea6cbf7f022421a
236613df42db04ba257f6c09cb27e696d010352f9baa00038121d470ef9a6dbacef65c97870c55469f788dbe67d77fdbea93
fb316243c92e9273454fcf7136550b64cbe1c1630d23a5ced1de8901ef28f96a14e8bf6ef2121928bdf4e5bc2eb42ab114f5
f60ac707979b390daeb86b205351bd349da4d57d6daff9ce6f37af7f6de4a222ee0e5a4b7e8bdb348ccdea6253b94625251b
a14414d83102deb3682633de5760cb65811280487fb0d2a780e3bd70648f23ae37bd1e78b2187ae954409c2b99afe4274c5d
a79a9d8dfe1634952930af20c8e7edf736d9c22d08eb9c79c15a4b94fc65f2d099efb6962b59ee79aef2429fa8ee66cf8a0f
0ba8645b7f6af797d8d2d2d0586b3438fcae12c94f75f02a6025efb30e7d1da2cb054492809de9b84b7d0332cc4f8f38d373
d7375f4cfa2927a930fbfeadfff5ef1ee70725fbd37d36f08d2e5f02d1dc907be7acad533c1f6bac2cb5aedf061f346d9d11
81e153e9f2e5592fda51beda89b04765d0db8e3eeb3cf04bab036ac66179c52e8ccd67721c98b66612ea8a1977df17f15e75
0f2414cb85c3ed4e9554c1376c5c549db516783f97849debd4b0cecd83faf8c13cfc445b92cbcfb6024c4ae6372eadae62bb
4e8d639269dca72e316e98e07f06a802dffb110f925e4e97f5854b03aef8bc67c374534114b34d447849d351adf57083ced1
cffac701a181f413f95f848864afbcfc14d03b882d828e3fb9be1675b37a57a6d8acf5d3e2bc89f71491a93072f49d3c0c35
030f4aa40adb36aa107ed207889b029d042d9d4df1214f429df504d5c81f461ec5128072bee4169ab6a7d755b0b2f950ea8d
3eccf0800b8e3b120dfd18ca37a466a64f78399985df597053eabbe20865e4b547cb950fa9d4e34b624c3ea1f7ed941916a8
c61912df65de6deb554ad4f7bb9cfa7535db2e3a75bbc0c4408bd7f116b2ccc6ec2485a2fddcebb104b7dc54b5963abb7df8
cd76bdb21c21a7bc055ddae97ce79456e76ab8c83716c13293e95712ca431f1e39934fa056f29702b807fd766a8839d57115
976e5a10d13da1371ff9da507f77b34c141c4ec5e3a932289d8d3bc1ac72bd883b5d37fc1c23143be2e178801f54e8f7cdb8
7c6b754dbb6598ff957cf597e170652a7603b0790a2b6459d03f794e766ee56e037de9fae74ff9bc0f350fcd509e3219d28a
00689fba31c262b557d032de9c6a52484ac833b53cff537d21761947aa224e593e9666ed5a3764c626c7acaf33c2b1b19513
fe042d919156ef40a6067150928766a494c32a9ea15770d43a234784ef2297495f0aca65d9e7b4806bf8a9c22a4f8bdacc16
7dba134759dd67df49a87db6b235d6930fac65874b7d2f501ced49dd823f2fc42f4e1f00f968923a5c2ddada90dc3a03f53f
9d10ea623f65fdc8dff9fc38e37c3468aca28b2d9775a232b2dd8d35eaaa8bee1356a1c7cab7c87a36f5b91d54430f3a8a4a
bfe63fc9223e715b03d1b43e4eda7be7b0dda906545d96ff36321496270594102a247178081805a2c1d139185d9da75d0a62
7bb1196d62ee37840f67f2794a791a549c0336a7a3d8a75953925dcfd3323f04155d451828b1f557c7053a511d76090ac0e8
5d773d795ac482ecdfdb11b0f0e104f460e4d34a9e227a5707fca395ef350cae9cdedfaf5091ad88828c5c5e4c7732e3af7b
a0db9be54396e8b5f269bbfa9c36cce46badb09dcbeb9b42df85fbf85405f0a76fa0e2db78bf1dafffa038da4d0ccf13d821
b3bc6e3ed0353925cf1cf074ab50b0d18bef90a5aba9a5e875f1120dbd4ae2a70e3764092eda1bd4aa3388aa4996dd7e28a0
c42147d4373d7735e2083e00baf5817f102f716a41a23a7a56aed6cdebf446b0694d832460d4a25aeaf7222a9928f7699588
aa1a8d8ee3dc1519f5351fa1ea31187bda577cd53b47cc1e0b854e41d9ae4fd467d774c6ee3fa2293a537e7aa271e4c8e050
825b5ac6b3f4059fd09884abb7c299ec8e219a68775693ad008952d1847be1b401c87cc7270dc678b2da6692e1e9d0939982
64841bb4a722769f9d5f92adef608f2a40b7a0e6d24743c1f6118d7b3f92ec7dc3f8ba8e9b26add389a3f46d0e367947d36d
f55d9d66823689e2fa04dc34714792335b7c376af7a2e0462efe23e1cc71fd3f642454c12db42900de3a5a97a4012712c3eb
eff0d09e51a9b7559d8aace0be75bf14ece3b771d457bf03b5396430aae0299525398f20272629c187f392f575c5abea085f
c2b2bb96c5b94f4ab101282512355f71b18735b8e1eaabe7e3d55556a23773684a65d78d02356b1b37a4b85f61a237aab70d
e79e9f756725357dff90c5ffe84505c93daea5660313ab9fd798eeba4fa84c74ab65950a2578d64ea317bd9e8e0db468716b
15499fda4bfbf04ca6a87976e448c4f7b670630675d2ad5e27714dd289c053677a72a2ac198387abdfb0d7ddd0ef60817688
91ee1f81e9d4728b351774d4ea33ce6852e5f3276e9d3785642afe7096c4e7f2f2322516085297d2f474324ca920f04e5d2e
33ac6a66b6638a1e028b980cdeb8536a3f8eb384285d790298512cac05b8c4f963c5831a3d0e3e2e34b47ec828faa40472e8
5d447adfb4ecb2a0b6c47b304a194ccb236ccb57b7fa81247ee94adce6c372fb92a753e78a710405d5b1ad2a9de4592fb3f1
3932e201aad63b159bfa862e990fc1d18461da0b31b16c1485181330dc9268ab886f4d4b735a7dfa19ea75a34041fb11c58d
83ed7aedb20e0893f0648ca7fd8ec9a57553f12f14316123069b1da8e41faf6d534b2796de4a3fc6f99f3f03f9ce3746bc58
6c6d8dcb7f161df780e25ce5654288fc4c3928fba37ee4feee5db69e662c366be660def312c75cff5afb0223cfd54fdeb1b9
c25cd5df7124465549c36d1a04217d03ece2d8738b7d88249347e4e261469e2af322be67d7910995222b6141de8a2a7e49df
a10164dc895fe81b448983c2af307ae8360dc80072c8da59a7dc0bca538a9fbe14db96cc63bc7d96e0e58ab9a4f4c291d4a2
bf48e88c4461997373e4a529ba7da346af4d37a45285e2944e6cb7de28939a85804edf9901331fbe057d01dbd3de3fceaa15
f33eac81b9709644d21e5fb9fffab565efc35a72956b1debe781b93d3c75abfb30f39faeac4f8eccaaa7abbf54e1d4de3dec
9c3c09be1de631ca6ac00e5f912970594c16ae67bba3bdd1c1e29cfa6359d9c3a5a43bdd2d80bdeb69e0ec20e7c2fd056dae
472c2eb9276c918d4fc2f5f64d38d92ae63eb63ebb85ca276690dece171faf001db3b4d3e449145862debd9d0cab15f0c5e8
f996f012dcf8688675e18af3998438bdb5a0bd65498ad564de6236a2a1aff76d626a4ceaefce8cdf5086a32b80c749ddc36f
0ac82c26e6b8b9846f9e297bcc699d595666f32e57c817295f97122ae0f0be8352b16d1df564257ddf785d5e98612565ca37
2ae3c822771963af50e3a240c3b2870bad13b96d477cbaa731146b4b7a7b6d71b600de5cfbd5e290a5159b116ca7f6f33527
3bd7542501f606d5dee163db130766270d7db395cd2ca5a5c865d862c6714a1e19be76d7adfab9f3416f5fbd2542c187f08f
beaf3f282c3c5663ea323413e93294b78db53bea582ca29d004bb69d97dff9d2edb64902f17c7dcccdb0cf799a576611b2df
7654b6c96c5ff9dc31cb8bd731954a5089e2c3ea2483196aad466f4749ff337d77a2943e384524bca88f27b43a9abd6a3275
f0fe61682b2847b9ebfb33859e6359716a12f7d0a5d81d2b0ff1ba422e4d371b3a9613b9a18f1c9c2fbed93e8bd5ba6a7426
12b6e09a5f59b7942dbee2f7d6ccee95e1e367761dcd1ccf92a014f00d05ce062c634e68c669381928b9292f053ff6e748bc
e9d0c2e21b3a25a862ac9530c2f1cbdb21b2068405cb1599a987859553a6d858b16a3917e2e9e50c5567ed12443ddfb95081
aa823cefcb6370b64faf174be075e4dbe51d50df23aec9b396747ef761a90a35422312a65619c536d9fc21c9a4ff011ce223
5b
>}

\EmbeddedPDFStream{schematic-46}{%
/Length1 1403 /Length2 6180 /Length3 0 /Filter [ /ASCIIHexDecode /FlateDecode ]
}{%
78da8d76075493ddd22e02d2ab341190d03b24f4aef4deab80422001222401124a288222bdf70ed2952e457a55a4231d4169
2a1d14e922e58fe59cf37fe7deb5ee5def5a79dffd3c33b367f63cb357b8d80c4d849420487ba83a12811602090365012a7a
26f7240140a0983010284ac2c5650a43bb42ffc2245ce6500f140c8990fd5f062a1e50301a8ba982d1583b3d2402a0ede90a
0089014092b22029592010200a04cafccb10e9210b50057bc120003d6180361201459170a920dd301e30276734769b7f7d02
781df80020191929c1dfee002538d403e6004600f4c06867281cbba303d8156082748041d1987f84e0957746a3dd644544bc
bdbd85c1709430d2c349914f10e00d433b038ca128a887171402f85530401f0c87fea94c98840b60ea0c43fdc14d908e686f
b0071480055c610e50040aebe18980403d00d8cd01265aba00033728e28fb1ee1f0341c0dfb301808441ff0ef7d7fb572018
e2b733d8c1010977032330308413c011e60a0518a8eb0aa37dd082003002f2cb10ec8a4262fdc15e60982bd81e6bf03b7330
405dc90800c616f8b73c948307cc0d8d1246c15c7f9528f22b0cf694d5101015241c0e45a05124bff2538579401db0c78e11
f9d3591704d21be1f777e10843401c7f1501f174133143c0dc3da15aaa7f4db010c97f3027281a200194919414970240dd01
501f0767915fe14d316ed0df24e8178cad20c0cf0de90670c416010d803942b12f123f14d80b0a407b784203fcfe37f1cf15
09080480c01cd0007ba8130c41f29fe85818eaf8678d6dbe07cc07600dc46a0f0400fe7afefd751f2b2f0812e18af98ff9ef
fe8a986a282969a908fca9f8df9cb232d207e027240e10921197c6ca55540620252d0908f867144330ec6f16c0ffb86a211c
9100993fc9624fe95f097bfded3fefdfd9e003fc33963e122b5a2880f73f1ab7014a001db03fa0ff6fa5ff76f9bf09fc5794
ff97c6ff3b21754f57d7df34ef6ffeffa0c170982be6af0156b39e68acfef590d82940fcb7a905f4cfccea4121304ff87fb3
5a6830760e94104e582d0b81c48581e27f70184a1de6038518c2d00ece7f14f30737fb3569ae3004d4108982fdba5ab05e40
e07f71d8f17270c15e1f28ac2c7f5350ecf4fc735f35840312f26bcc44252401600f0f308604885593a88404c00f849d4708
d4e7b7900122c208241aeb02c0d6180070447a90fc6aab1840048cc2d60c43b960bbe0fc8bfc8d4b8803441050a7df971eca
158cfac3fc636b074f0f0fec04fe560836af7fad7f8f3b14ea037520999b413ac83d7d58fbb4edb44689c95b6875846069a5
2322fe5e6fb8049a7bf6999fb32e61a6c684bbb21da4f2e6509ae1fb17a1e333fe4cfc5f4f477d6cbab31f2fa66ae6a171d4
7a3f2bef09e50ffc38e87682de86206f39bd934a8bf84cfd2dc18885d2ce1adf9a27c962e63c44aa476cbd6aade8a203b56e
25b30d80034a8ea9d85299d0314ab91ddbb851f7d83857d083f76ca8626f7858a77b9a6a99bdbbc08d82c59e28b74533fec0
7cb68ecd185c58e8cf7edb9b26b65fa0f3ba5ab232948691d6579b2d7728b18f912e215bbb1185c3ad2c71acf222259e9db1
181ecabd8807af9d02313373c6be6104c201a5cc0f879460e1d7123d349f29dcde7958a5d0269f300c9f797ee89bb2b93996
e3ba441bf17a3e23c4cae1a2d8dbef3e8d81f521ee2a805ea5eb8e3551fa38c33d18bd7a35c7f753dfca40b7a3549cacf15e
aaab6daa667019773c7aee229e1a81127e22ff68a461c880c0d7b3433facb69b8058277bddd97665fcd8fcea90a765de9dc0
71953bcbfb39bf0fa26e0b6df5a66ae64656d3deab225c498e4734a96dc6e5cfd6e0a71a2bc95dccc2f9603df294a8c616ff
3e9b453c5ee24d73d0c2b4b3f0a3eba507aad1a39ac77a9721d35284d977ac24145ea4d8e8e8a5eb5b57d496a0249f2f6d58
ca52868fb53f97087936a2b522affcaad53e8d875c8d3ecca0b325bda38d9fb54d5465478e68e30e97bf476730e676e96e86
163fcb277cd0b8496c4c658a4b2f4516ba366e3bcac493ece88500ae4cfe18e9728aa416fc893545be900b81f0130bc228aa
a47dd9d14f43d52f186b29be66b3391ee2dc48d991d1eb67104485410e6f769683cd98391735b97577566cbb66e46f860f80
457f06f5d2c831ecf21d490b4e71c94abcfc5aacff856c1b616d3e43dda81c11fb05affd4eb0389d186bbd8baa600553e71a
4b8febd6be077cb8676b647f7fad4de6673a57e54582f1b3bc212bbc02f9657b03df40f1aa1a73fe88b71da3b72b34587c3b
35d8edb2578047596dec939e3bc96e8fa8f81ef3dc2021515a82327fefdbd353bb96112646838be7103f1fafa3b6a444e5a1
4edb31554b4051f94aed3c7d19b47dd0bac44b265d7946d54f7b29b33c34de56533ac30322612efe781b2482515efd18e467
eb0c0939bfffb2d2d4ed5a2426bd3fd13ee5dc78859f4358234993ea0b75385029731b079798c6dad69986f71ac3e06be3f3
08373f1ab63d32336d7ad58d7b44ab5763bc9a1e7d4c4c6d933a670176b4129695b2196f051e86b7b38a4956beb84ed3fc5d
a1cf65fc69b705bd67ddaeb825aef3b775f77157b4a607d14a73dd068b8efb434d72ff9ff6f0921f79a1e770b86240fe5e5f
0ea95c527a684fd3bc3ccae853ebab5a886a4ede53b6e6cb710e32e29aac48a5f60aee7c99e063fd9ab9ad503689ecb4441b
654d7ecdd075e105b133eac81e2e329cabb228a75725063bc1d339f23af6467371e659addf59b78c8e7cf4804f32bf347c7b
463597ab4b59edda4f757834b0331ab4c046da2272af63bd467ed3832f56331fa254524f2a273653b85d1e694e7ef93a558b
051e5a14abcc16f0ed04efa942e2d95a779497ad78662d1cb5b226555eb1b2797d0c977152efa9bd982da0aa18453fcef021
828cea763dcc0674650861f230eb6b7de0d9039af7d8fbd62cb8191e29bff5fef5932c5218d56d8ce54f0bf483d44f866af1
e027ec8528962c4dbc6807f27587972e6de8a19faceb86323bdf8752df4284632a1ae7b4d09c7df749754e15375722d74fa5
ed5d3f5c67de3cacfa1829a957bdc3c8ad936f86128e21e58b7d612839a71647309de65e1037208184e2336ab697861ef7a2
664143be4cb312674f1f8375808d73e873e9373b2cb755eb1df09e8cabf26613a5bb05093edd899b09627833daf5b19e212c
5f4cf2c4e695fca325bfe88b011241891a0982bd42c540ffa160e6de81640dda8fba57892b5592aa5b3ee72e8f0213aaa763
28bf84f22572ccf2f79a24e38254bf3d7cfde867ebd43cb97596cf5d4aa69a6e12d1c2949b02c379bcf616d531ec51731c99
c45b05b49f5d15f3c841e27803b3457e7015ee89574a6b12b1300f62d984083ebfaf67df360a7f3a35c36ce905f4428f8c88
ee2b0d68b08d2a311f92b4ddea19ce99f8a65a3b94f7515a3ce026bbe25e99f263b8fc06e474ef3240f85ee86ea96ba13028
6a4a47051472aab4b5b87f2db889397b55bc981faa6e1994ac38e3181b24cee7224d9a32799e72433f96d34eeae7abb4f407
94c9c534ba8e57f9b367b2c951697665aea53748df3c318228dc64eb6bcbb57573ff981ee973897bdcdc984d3c67c314b541
51cf2c43edf79cdd6042c24fa66f49bae29446b0c07b448d9bd1f45a0972ad574beb4112c10a2f3dd3c9a6161dd9e7ed78e2
7d7d673bbef7cb772fedfd03c983ae6924da7a12b5ec1ff9e72919b17f7aec9d4e806cd22504097536deccd6e68ef9d4c939
fa8d19466b7ced7560fefabaa6ee94d754d371d8715fafb922934dc7d1661808df06885f492e223f81d1dcf1975c2abdeb97
9173f1265aae9961f7da273f75726e3e6a3a207050485582f3416f6d57c634ff24bff63e6793ad74ce7e31c7c96d3116e266
586908556f333dcb4aa2698a9ac33469a56063a8fdb59a76734c6a819bcb187f960058aac2b0bd76e79a420b87f918ce4da2
a330253a2eba91459bbc6e1ae67851a6c5a19f73cc07f2c0e2fbbe4881b6ed93f399c6f626e0f1c486e7c095acf56565c175
e6a68604ad65bd2f67faca7855d52f93685ebf4a5dd030ea19dcfc9affb4b635fcde0cd8a221dcdd5f67e695d4f1a63a5105
db05f0c165c81716b9ae116e8e808aabe15122ad52ab8122da7daaee18863cf2a1ce9ba1893d0e3aef8dd4dacb0f7a89982b
5b83f065960a07c31a97682839b8634ebcc96c0e867c03903206fbd23f07ad5850ea8ab9eaaf63e4bffa7fcced8ab866326f
f9517cfb722aeddc628059d1ff74e8bde758fa67f3fde52d42690b6471c7b4b445aa4bdd5034a42d5a4e82bbd4ce6736f110
c908ce97b2825354e7661f0a186691246f26112a50cc39135090283cd70107875b5bb5b80f24bb13aa8739dc56e2d27cff0c
ef15c4a146b2f5b1e0d6e02dbcbe9a900f0c21917b3e278b6f1bb70a634c1f72b8163c1bab37b50b2b8d4928626dd82fe7a5
1db1b8bf8f821cd292daeae8ff645af01a5193bad1406c5fc3cd5618149ef4054cb4ae00f2d4bde6bb469dc8de2970945e28
8ac1eb2eb6cb55e012f6896b1e3f753dd3737a1bdcc0a8311f3039277a9733f7f0ced327d28e9bfbbb59cb44b475071e414e
adb4d3314b3da9c111f7ead922edca024f08e4e1292d9e383861f4f93f8a1f64e2cb2cd6e203393a0b1f92be9937e0ec1a42
e4bae56b34caec973590b1073eaff04f16fb591e40c05bac89679d13fa106fa1cabe40d863eec75ac32d1cd6798bb77856da
1a122c0e297b8e9e728d6e1c2279cd3db433abcf9de4286fd987863af36d67001c77c99aed68538c327c0b4aa37f0e233491
6fe45f6ce023bf542d58165b849ccce17dbf32eaaa935f7062a5b35a9ed5ef3808d36bc46f62639d6653957c52bec175f116
736be161c1a6fbfdde6874b54e1e0df83beebc210d7b2b35fdc6bde015957053de174f8fe4a8e20a223b02bc337bce0653bc
67472cc3876d0492a13504659c0d75aa2a161581d5d7d9556a4c0f3ff7efa1e5c58f783a670993075d21387786be93bc2f8d
64d85015656b1240ae5392c92bc6b44c7dcf3f8fdf9d7a7ac3971afd4855376dde78929c9a522a94d49d1daf577c1d47d42d
d6fca1199014ec4f76b2c259b0c08123b85c0d2e7df6d6539d5aff7ca7126f53c2b7b7a085fa837171907a9e280b87e8f09c
2a49c7f5870f7c545527ceb54edfc1f5fdce36e509222b39faecad47098197a4435c91c79b5ad331b55706cc4bc97d720571
37fbada659e7d3be9c08c2171b5d0927f785a7d9e8ca23b8c32970c813561966ac0e1332ea854c442bd4ab7c2a875fb639d6
34a24213af1210cd00c0735cfcb6fcbeb342939afaa4dcca6591c3f24b619c40c9f976833c619cc96766b8d16314ee810a45
febd90cd050d0e97f80ae18b4e3b1b9bbaf1d0a0ac611cafada5e6be86a97beee7270140f76da02de6d6cb92d613f809faf5
954984d0b786d825a3072fba90df8b6bf28b0772884ca07d64e67e847400ca712f015f98b79470bd55956ca24034ad4b6b4e
76bf446b60e2d27d8c4c620ec3a3d53b1620398d39fface4139168bc33174b0b1b9a09b039a2a14e5c566a79d4887417a941
e337268068ad7c960949814b2d98cc5d6ab25e0793ffe41c8979f24a8fbfd8220bb5d69d5bf10ebf485719e981137bac4088
d3e278a89b9cb91c69632bd4f5b07275b9b36f94f79975ef7d399e9e60d21ba9c29bcffb7f1cd46fbc5e8eb9d211e2757806
0f41179768cc74d92d297cf55270d0fdd9a2fb2e1cf52e2363e223fc667c1bea0785b071879cdf8dc899543abae55646c6d7
f266d32432a94117175ff9a2b4d59b02688cd3535ca6b223da859de75d350b1fb78a8f1ce8bbb32069689b7db53f0c426ded
65bff7f8a9cf21729e9357df5305f906d55a632ae446ebf1afd9278f244ee6555e24b7451a3311ccb04992864544e20b3594
9f371cae56043c0c3db7968860fb9444d65654f8f599d8ebad1c9fe80deaa87b1715a7f9ce5a6e12d17d48748300cb428550
7f5734f07062d09b34d73f6586dd902cc73828b3d2c234a02eb84c3b6dd8db9d9b31f3ad609e9b661a840bc4ad57bd352d3d
a19bb3b7d0b5e117583af849a887b1837322f5d2ff9c377dc1304c5a3b035a9e6a58d245f9c15e20ec27647130dcf20af551
e5a60e43f88b4fea186bc24a2d491efad3ade4759f763a82b18077dffa86bd1e885db88510751a595ede33d440e42d6ce317
e8760e0d7f69780627394d294b3125a1ded79a1a72e6a328e0aa7cf0c975d896e4c6526a4180cea6312db45ff25b4f4a79d9
45d6c4e517dffd947062c91373b7976e5b0d3c4b91071c6f56c724c61743eb6c78f81766cfa444db1e68138ae771ee5f17b6
5c8c5978ed3a473ca6752dff112be5157cc6a53443817011cd75e726aec4f319aa07f3e42a8db3ae22281b01ce59137a9fdc
85ae19427efa242ae1b5958f8eb1a929289c3eab135357a1b9a367fdcfd658827ff00321665fb73161655994b55bd551dffc
a33fdff2e586c7f776e6d9a59e537406e9b4ad920ee4c78e8dd16a557d4a699f50f1e21df54be55bbdff44709737b1276f23
1e768ec8911de4322163c9af7c3413650f54adcaf6318a2b86f1f81653beced960a3ba3c2a066ea7eb5915acd6df73e8a9f9
aa2f317b50412ac7145ac13cd4bd7e6eba328f61f089526217a2f0c157152cf5abf29a7ca752d66d699a4a27e03bb0c6f563
fefb4ca1fe0d49bb64c20caec41f146d7ac22e53aed2ec0af7ba273b348f2b227bef34907311861047fade25b63d36e9fbe2
a80ffc7e6c57f8914866a16ae68b64b3934c80895ab6cccd66ab407881d7174c39e5c9c8d1f71cab59896af063bafc4084d9
c20fe3db1115c572bc46ab45fb9aa85009b3187ae9be2172fead6ebb7cd9a707dc8592c92ebd564156951f465aaa0b7a5276
824bca1e928308a1f0fede45a2c1170f30a5f591d20d9a2e718807299a9f195aa62d0d0ab5af3f7e29f4fe7adaa2f2ad3d1f
e63375ae3a3d7b4065844d1a5e178e59f2c12da6b612fb3bbc030e30dd4e7270c577f1a3fdbb00352e37620f67836ee09a29
e5bbdcf87042a42f46fcb1c999717fe65bfdd30be125d30b178b1e34a3b143f224d702bf387551037727d34c9478f49930f3
d7b60cf56a53f57c286ffb371d4fb3998866fabbd23812473fa464823765b7a2d6e52bf5cd52ab31814ff5b392d767cc30ca
248374e7a71c5bc432690f3eb8b2de2670af6fa50f639b2b9419f2f01b916aefdb88b5d5ca7c7209601449f8012997227a6f
3236d4130832af011973c1a7709384c0adeeed44eebc4126e95509b82c9ec29fc371a9e34c6f643cae752b29dcb982d35b1c
e325becd17e8d69333423e6949304f0abebfe9b9c06dd129ff7ef98d90d565a2ed74c8d4f03ad40f3626beac3db9c47f7bf2
4db6e49736211b10ebf1e7a78f9d368a4c697165675b5489d2e40afa6da8f2ceba68ded0533abd8b4036e517be9f9f045b85
3ec1db19534305df95443bee53e4f8d9c97ec2414fb489967773081d9440489902628bc2c3baa2fbf435cbe946014d0f8fdf
84f46c36d5b413e24a3b6578a5232f77f958c99a92d4dcf9bb1fbf7cab3a31892746b848c395d3cc41b6ab28ecec3bb4f6fa
f6b022be6b563605aa3b156475c57e00e4fe60ac1748763e3d27ac43937490b5b8172ec573efbd7e7a7e02c570208606c8f0
313917979fba0a901424b02592381f564261b9007ba3af39f3bc099cff835834017e94ab0e68ab29346da4cd791347f6d991
228a5ab0a959ebb98dd588ea435d220a59d84cbcbf6282eff7b0e30d92b51d9620f55a747cf27a15abd35922ddaef180fdea
434819867271ec181e63fb338ead758327223b88768a1f607a76833118f008c31770dcdf065a57614d9b6de548dd94dda3e5
11f56a881462c87224f84650fba2cccde54ef29bf851ba99f7d4a3bee73769dfad85ec2cbfad2fe0cfc2a8806dd7a8c3153f
6c25f7ec98b31f8d191c0b104c192069bd6e0132bdc47cbce3de78930e46415d55b553759c252e353a2c0298291a62550403
6c4e8bb656b90b0cdfc7bea3a4543f74f235d40eab398eabf81eb344fde152cad37cb02c97aa2428765e97c16f8c4349220b
3d9ad5f2e89e2657ec4b65dd09bd97375044f814d5fa6cc3a8419d8a5ba8223cc067dd241300318995646e882dcf7b9df6bb
84071589b24bdb4a199da41f5bf59a6e4a38a6273cc519a5ca155c650faaad691393c5fe252191d1ce97404a8d7004330799
5a4ccf068d662bad28b5e5d596401502bed23aab9b6bd3d7912796e6f40285da06893608a20a7eb4981ef14801bd31ed6a7d
d21d9603c31132fe85b4d429b10a5e378864ede156ceb711142f2d8dd245ac85ee88023e25b1ef957e3578e972bcda01b5ab
236588c187346f2dbf2c2f7dcfbfe121ac788b3108d180d60eab7bac98177dae53c894ae1277b240f0be3842a13f6f7406cd
e6fd622424578c47ffe77eadd3461654905735cb2d7fffcce06b21fabdf5b821df10c550cfc0c9999caf05f58bfa5dbad3fa
e3bb90e652535fe5dd83ed01538f45ed5d6373a4931c75654dd914dfdab137e2cae532713349974b413c07dec7583a3d8913
9e49612d3adcc4b63d6d19c4e89b531cff797cd063db6459f799203a2690fec3e77dcf0af28b5d4278d8e5b10f29a9ebf554
07915b38351354365b90f5c20f5df29cc622dd3b66c40d63ed9fd6b33648ed7c773edf71f7ec7afeae8438a08e7168a1cead
82b9537d70b03693f5ce7913d98a3b63169f4224fd82ec47db15136fa0016f9d484887bb614fb6f1f4e5db8bec8e93e89c06
f1dea775e8c29880e32893223793aeed7875c6bdb2fa83fe8e55bab6f185d23db3dadd2ba98a360d3b8636489e7119b3a8da
f25a9eb1e92bebeefbfe3829fce334cb0f921e8df0d1d3ac4c35de179d6be5fa2a3b5a1f667d1131f0232ec5effa92aea45b
da4e74abf5a365fb025115cfdcf16c8e3c07e7ea2f17c9638fc327d84af6b482241cea7b7c938cd4d8f9421de33734624920
043dae4e88190b468777160bab4f97d1e51876aab7781b7a7ac13a6ba9df708af75e5871fe601a402d51c832b18adddfc464
fb8788f0b32f1ddd8e204bc084fafac51bdf67460fe78c776a969f3404bde5fefa9df7e0d89db5be0895e6247b08db0b6b0e
6fe9bf0802e96a24193ae3dcb5b0c88c547421b1e836dbaa69fc6ec7c4b7e3e6f3c9a4dcbe559a31ac69eabd3846a2221e3f
ac80473ae50e28738341b06b77d28f5f5049e81cd751661467376f4abf16e3f2007fd6a2227ff1516143929cd32343039481
33629f27668d3e47496c14141832fd560c9f973d5180a2677a744a5373bd55267058d789af1fe7673269d7264a74f5135045
4f2985ab6a9e059f35598f917cebba5b1a4a010d89699b1b69f12021eaa06d44fb468b74e8f39dbd8399d87c8d80afdee408
e5645053d81cbbb0a01e6d24af146c3aeccbba3e85d6d9a70f6296337a61c1443ceb2c3a6f2c8920fe80a12210c7e37b5a7b
a9b2dcdca61847789de5a45ad43693de3e7ad953ed1c3a2545e36077879ba791d914ee49b082a6ebf49bef0a2327a1a3eb55
8aa8ca30e238e5549d93f58c89e2ebcc54f6d2ac9c0753927ec55ccd4d88080520384b3ccc8e06593212afa951d6050c0da8
b4e180efec1f441625c96b134f7da06e3fc6dc9997c599c689a1d615475eb7947fa79299e15f3e7f927ff0451e455a62f548
56d0c26dc7813ec4df929e30ce3b7861f0104f5d398ae6c09c4ffbaa4e3561253bd7e289885cec10c19580bc8b586ecd98af
387f93c6b9f8593b6778c0f2a3e8d3d929fc6b577dabbc2f43d9b8a34e4bbb42f502c64f046e1c962935347c547c11f64c9d
40064816a6320f0d528ea33a0e2aaab91dada99f261f98c4dc836f78f395a44dbbedf2c0c964e72395e92ad72ca762aec486
b3fd0ede228fc4ceb6c8832404ade05121a5a8cb4e766cd0c6ab1f9e1f886f86773a3d5e1edc0fbbcbfc3feaddc81f
>}

\EmbeddedPDFStream{schematic-47}{%
/Type /XObject /Subtype /Form /FormType 1 /PTEX.FileName
<2f55736572732f796f6f6c6b79757061726b2f446f776e6c6f6164732f6d616e757363726970742f666967757265732f6669675f736368656d617469635f6f7269672e706466>
/PTEX.PageNumber 1 /PTEX.InfoDict \EmbeddedPDFRef{schematic-48} 0 R /BBox [ 0 0 518.4 223.2 ]
/Resources << /Font << /F2 \EmbeddedPDFRef{schematic-49} 0 R /F3 \EmbeddedPDFRef{schematic-50} 0 R
/F4 \EmbeddedPDFRef{schematic-51} 0 R /F1 \EmbeddedPDFRef{schematic-52} 0 R >> /XObject << /M0
\EmbeddedPDFRef{schematic-53} 0 R /M1 \EmbeddedPDFRef{schematic-54} 0 R /M2
\EmbeddedPDFRef{schematic-55} 0 R /F2-DejaVuSans-Oblique-epsilon \EmbeddedPDFRef{schematic-56} 0 R
/F2-DejaVuSans-Oblique-rho \EmbeddedPDFRef{schematic-57} 0 R /F3-DejaVuSans-greaterequal
\EmbeddedPDFRef{schematic-58} 0 R /F3-DejaVuSans-minus \EmbeddedPDFRef{schematic-59} 0 R
/F3-DejaVuSans-notequal \EmbeddedPDFRef{schematic-60} 0 R /F3-DejaVuSans-uni0302
\EmbeddedPDFRef{schematic-61} 0 R /F3-DejaVuSans-uni0304 \EmbeddedPDFRef{schematic-62} 0 R
/F4-DejaVuSansDisplay-summation \EmbeddedPDFRef{schematic-63} 0 R /P0 \EmbeddedPDFRef{schematic-64}
0 R /P1 \EmbeddedPDFRef{schematic-65} 0 R >> /ExtGState << /A1 << /Type /ExtGState /CA 0 /ca 1 >>
/A2 << /Type /ExtGState /CA 1 /ca 1 >> /A3 << /Type /ExtGState /CA 0.6 /ca 1 >> /A4 << /Type
/ExtGState /CA 0.85 /ca 0.85 >> >> /Pattern <<  >> /Shading <<  >> /ProcSet [ /PDF /Text /ImageB
/ImageC /ImageI ] >> /Filter [ /ASCIIHexDecode /FlateDecode ]
}{%
789ced9d5d931dc7919eefcfafe84bd2f61cd4f7871dbe90cc151d1ba1f052e2ae2e567b010d2112d80146044071bde11fef
e7ad3edd5d7da61a9821214b7658bb520c72baeb2333ebcd37b3aa7a9e7df1e2cf2f6f5ffce6cb5f4effedb7a767dbbf6edf
9decf48aff7e3b99e915fffd71b2d397fcf7db93e15faf4fd19673e0a7bbcb4fcef9b3e35f66fde9bbd3e98fa767bfe09577
3cf7e5e964dd39e458ca64f339e744232eb973b139d55576d7cb9cb1672fd9fae62a698d7f3f5d7e916b4a352d6db868ce3e
d832ffdf648b3fc712a6b72fa6df4d6fa667bf706d44e73afd783267eb6c4c25b91c118574f94fe61fd9d0a3f3d1d5e9375f
4e7bdd9cac8de71a4d60943ef05e0c313221ebe3b9a45c5deac4368673cdcea53a0577f6b964ef279bead9c4646c9e62e249
178a9d6e4f1aacf535a73865738ec8e9de56cbb3de043f957a4ed1c4ec7aa135f5cc24f47a2774f51c4a0c6ed6c0a5511b0a
caa839db7e04169da7eca3ad6a631b2e6638673a431fddcc6c093c6d98dbd4a9a193de9e6a3a1bef4deea5b4555dc92ef7cd
cad821d45cf76360943959e7fbf106c3a3b946d74fad13ae4ad8649bb6d60637ad6e5df706d846b9d96a9d4d67d391fd6fc7
6ed1bcf5b7a7afa69fefafff3cf9b39ce71cc2f42facaa6fa64fe1c1ccd9966aaa9507bb702ec6a5d28bad6749571f5141ae
349bac61101e85d376b453a1a91463c85241c051d0748d537567e7425503a19ebdab411e6118988de54ab87a702714b0b0e2
f18dad4deb3090a591da0f4003b4d984e8d5c4365a3d125cc581a66e6216d3d690aadae81c7893d286c967538b373b792de7
647228b56f9859fa508c56433f8a82d394c2f2ea875cdc396af2b59b5c275bb5d0cb16756d0d6e7addfaee4db00db3b3d636
a1ceae231fb81dbbc627f462fe978933c4024ef1d6fa1ffe6122a334be3a37f2594c91a24f2e4ec19e6dc0251446041da80b
e0ebc48c023c899e1106015d8836009a11e4a996e5938489c525d9caa1f8029c54f021a35bec6227a789c564811e00c0f364
0cbdd0da74760188706aa1937b7406d6d0c2d6ac8df1ecb00f43ecc62008f6a17a3343ff3a60292f80386186c5cbdc04ee3e
242f04dd14d14969c3b973f6c9cadfbba78d3f276f41ccbe65c031827926ed87912320127c2cfd985366a5592d8675729ba8
d7c32add54b635b8e976ebbb374337cece64db943ae38e1ce176ec1f7f61f8fdf9aecc622f291069e4cad19e8958c5f6621b
ebb932661f04029e5116ec9e507b54cf4281142ca8291564fc05cbd12930508dc96a202b82b94803c2015f69d1eda4bd0d7b
b98ba046ace21b5bc3c2b9283260fb5108120b71d4b740b80dd9065c06a2a147b6d909e7934886dd39f3269d71bc188bebec
9e66481ecfb169d7b295c358ab81eec681c91967160bea46cdc2743964b5d1cdb093eef4d1c93bed6d2d778ade86b133ca36
e6ce80dbf43a538fdce276ec2d17c7fe1ed7fec9ceede5dc560cff1c7f061fb693b202100f2a557810af272679060d7e1945
20d4f07a7af60f66fae27e7dfac629ba64a65383af0a61867ffa1c4d1a3d9ecf0a7026109ed274c3745d228a466ce7074f63
c16c9c8bcc0410406b3526a655600b8391247952867790bd90669c99a5abc6c59c474fcb2360280461235fbe71728a0a2d4d
a0dc6028be404832960e4206866e4428ab29f8631a76606dd38ca5176c099ff13918a0268e1463bd85a1084e230c612a5a2a
10e920963b9aeb39553c1ed68047108c203768310e757e63131047b4526833523bd981b10e9653458dc793cda04432b1c265
2a0bb7a61c820b69f4f88dcd227c0e2049deb0ac02ac9ba89560706ee404b0388f1a4ccb076f0847050748a9409246cd8365
40732da58482eab51c683c1553eb70f458d6174822f661b1a572c6e159d9acff91a18874675c2a6325c09ec910ab4b2daeca
45470e6f0d112e451632417dba21e433f39cf1ff381abed57801d7ecb066d282f25e990d2b6664ade895b218d4c74a5590ad
c136d38dbdcc11c84c220962813b74c9db069f87d49aa1db3bc56717e55722973749d19065c51b69347a0680ab247e0bbb50
0a16430831435c479605795076f6263acf8abdc18fc85941734b12387c9e259b985a7417220fe54f01d436c325eeb055456f
604c32f3744d45b515c6317405109c450139a17fc00ce0263fad70e8307664d62cbe08e18b560b054405142a0bcb8ca65b89
2421c3bac835831e57e60e483b08cf68fc3e337a16358e069d8ff02c5e1515abc3d9f21b31a84cbe1078e286a0083c55865f
ddd0d308ef2802e73122b1ca64f09b00f5f743df51488c055b99860a841b668d5f17543cd43ebcd3283a4197183ef0c67a2f
56d0321a0d2880e52b96d532f18a4b004480400c750fb304f05c91b3c51982188d7319d41dad131838d06d8024df1c011fc6
c8986ee4f722a505e28ba76bc902af907703adf3a3b66f6c11baa376d64968791e29185341b5234de2e5600bea57ba852241
5b523aabd98fc642d64814c8e03c23173627e27a242e0f5d38680d251cc6a29a49509832c9b4f8c950e9b8238bdc88ce852c
529388b16ee8bf2cbfc84a0ed9178107ee4e768d3da5f8b15a4840594b30c8e2f3247f4c91d551fc70b90a0c682f2970c89b
31420db93a006d040637a1013b8187d8e4d47a62e2210a40865325a4f2a063a6ada014ce8ed542d01a0e86e6d04430625261
222b0294e086646ba3a1f87026dee052b8988c84490b73e6fffd68e8000b1816c56dbc141915be996b3a843dbc8990449092
c3645612f3c0c6c3d605d2202ef12c918d4c9589023b0440776026fa37190c52adafce3892f19e9c01ef9147b64c2d0a79c5
b88d323c108dc98f41928517446c18c12579205c051b8656955d08979e7057429eb9020e9908f866dcbe46032b8ccafc5a06
94c10c8590329eadea82104440518b553107d692781a050c492528899f447c8d1c10dee55b4821a71da370264402328cd7cf
84381a407fcef287eac150c43fc812bacfe55c08df15dfab070b9678833724711b6bf03b163731dfe521a812424aa84abf5c
6e1902eb059012971b721b2d11b507e1823965e9520091c358970c57011904ae20a3d245727db88819526e1e06b9aa5708e6
65205e000c86c721c7c5c5f1630f361ad554a33413e57307e10f47e759b893dca5d13a021b998d807918fe48c66008b013a3
ecc22bb14f419e50c69642d9f0b09a05670d7ef021629612cc212800dc9ee483b411f5dda84e8f9fcaf587d19e7c0f564516
92557e01fc88b41a7d1da91e048198048771559dc02b6563dc08181a8da504300368acaad068c9434ac95449bdc63c4ee115
5533126f155e992c5a01de0e12352f8444ed41983a43089645b379e8f5a23e6d4983c6acea1c156f09cec4b4de545fb54cf7
53e5b93fad6cb34527c50f4c4fe42dadc008a2452dbe8b3eecce9882f488dd094e45da29ac7130c42e2988dd6bbb9e957000
1b0d9308557855b0e0ce62fdfdf3011a6f5b7269626e3516189f092c94e006cf47d4a7d045806455939912b151a8c9a3a10b
54bc9c3017dfc28d11c7657db71ac86030d01ead7bc60fe1d164a22a3e51f98e2ba3175007cba0007151fb2a41343ee30e70
b13c1a5052c2cacc4899b2e29349304f32ae8cd1c61d1064206fb0091c967828f729e43901d234329657a5000ae495c355a3
d23deb3ad80537ec150493ce398fcb94063b782304cd04339eade820f4cecb77f095aad84f4400531718d83f5f0593c09e70
5e4b950c2c1af15c1fd3d819702e489623aae29749f9209893e51983d103aa560124c4a61b427f898c0d923872869bc69358
7ad8d2e0f637da1961e505af2adcd07b50b6658d40db94d2b14e58e12003a6184d176e6d556b50bc64f4a195905b841ab5de
28a547f78c16eea3da1cb0447784d0d16c6171c097535243ec6626059f8749d5a16ab44c33e986c3f345cb52d0a68a11152a
a3179c60cbc2c382009ea7f58387e196b1576ae7265407ad643020b287c079e56ae36595b45d0413af465ce08665a2a2037a
cd696caaa43d228841514d4afb4ab80db4238cf4ce18c451ad688cca1830d84ca0527c3023afbf296d8fce8bf1094220d151
e57a3f7449916d9bda562a0faafe14b3085dcb8f468dc3e1847e047bed4d03b5f025d55548784796223ab13a816d3b675000
1c29acb2dde1f3c0a52959191c169a12d06d0a0b8434666428abb482a6add61b3e81bbd4c63a861e8959c43948b38d9b979f
53983858dcf034ac48d40e2a7fb5d4922480a437b9a1daa548a3fcaaaa64d7b83f26559aa44dfc81c3e303156615c5f25a98
8daabe8412c6cb09ce548da8beb30d5aab32c022704e43af31a5652e05b61f05f5c444285c28ca07461db0ea591545f52896
462b44da6894e88fb1550d3ad2695cd82993a967edc22778c17005923a492141e974d5610b954d31352a8ae3d8a67a10fc8b
c8961a16b643265679ec50fb458418b60d3e8656f44d99f4924038721c97cea8021651a0082ca9ac623f581284b4c355a21d
449ff174fc4b7e0ce024789385aa8c0295159c11a532d97ceb8e69784cebcd7095607d1609d80e04cc9b0fc4d882fbd5347c
1e0e17602d209499e33e5c89401bc7481fe40adef27c54e2a2ad22d63781278e290e9c463eef202da1d5318867bc5fcbc850
11286ec5f0a84a5d14f6c0a4ab192e6f080ae140bc5f2dc3e3557f24c4da3202569504337ec97ad62612d080fbcb81871ec9
8a1350439fd08a9debb7dada54a9dd8f7c26e8cc0116ac20bb174d24fe93dc03cb4394cf6a4e7529965c94ffc2af61f57e2d
9d5ffbfb59999f57f4ce7248b043913e0f57130c39a5c06a554967a639704525bbc3c65501248d060cc81440f90aaa02924c
7ecc0d8a4e66b1805a26323395d26a8fb5f821b066c613558383858bb103064cbd947587e66ab23418b4dc0a695fe365f074
4ffa35e4ad64d264b51057bc97d1d7aa4d3ea5bd6b21e09a97790511ab6d9fdc76beb54d10b37143d5473228f82c1152d115
43f03638596d19222b9cd324fc17c5eb6496f2018000fe598640a61302a4ce10371767d22d6693012717f2d8138062a19136
ebdd7c72260325caf487ba177d508623fb2bfb8eaa3490c0386d900e3d13e70a01ca21a42798c38a554f6a09db509df206c5
f86ada0c540d1747267085313c456de5e20c4e78a47c1d60966f9a309a81aad5385acad213aeaf2a18105295910f07d44209
19230ecd8499804a4255c9f8909e11ba415395b26ad0e621400503b4ebd682bd64a43feb48c1b6d77a3ade6b7dfbed256f7d
e4ceec63f770df7e4b3edd2702f816f4b5ed03b6b378aa4007bb70a85f3fa82aaa5ec06aaccb5e0a600e474b173aba7f3e6b
75e33a29ab24a1ad5bfcc20aa4dda875b0094336dae25ab9c2295a04689a1f3d9e95001b31406dee2b01ad724fd674193c4e
73067e96a07ca12519406cabf799d1c8db0ea381336975d8a945592845ab6c0c9e06e221d38409abecba34582081bc24eefb
879d0a70049f8212f99700195e2abd8c94129500b422ac6da90ccb1b4a5d2e61e76a182c3722bcc1188d2842ecf0d0882f86
4b05e67a92a03b0c5dcb99499274b20c6d85579a91c2a3536a2422c45a2620163109e2f15023c9aaa44a0aabf3954c5965ed
2afb1c0d3c4460096236ef8a41feb3f24d97cbd0556049f8b487f510e7b552033e00842c25d0fde33ab481a5f15ba35db61b
1d246565e068cac246f614d2d5aacda7aa144f09a05176327ad8c92a585e564a8d4169ab1967a9233fc4a0caf9a0b65e2187
704bf8c66acbd6eef54449e16c9c95d3a69d13545d0162a897aa9de8a2fcb2ce9b8b41b9313c68687fe7e0923517d2ee7610
4689004cc5a0faf160c49b34926aeabcf9939a56b377a3a902b7f2255708820ca78a696354ed990f152966eeb5914b8c8f53
d591354572d43e6abd2d33a8b5718d4caa42e573db8318cdd5db33849c584cecf6ad3e05a7000c5c8c238f89ed68052483c4
a23445653116525937f647c68e1364e575a59d0257e1af54954e07cfeb5c93f6381b278b3a7b90759ac08cbd9730d2ceb479
dad701972267211b1a0f451b4b707011ca181b0772da410525c72ec3c0551e6171ce87c78c0155735c0ad45763d1166ad4e6
606d08435f001766c843ada7a433e5b86c52c9b81544f1f4a4bdf751e33a84ad34111098dd212b2f5651ab8f465ffdfc437d
6b15f8745c055e23f0236bc68fad2eef237092a738153615964ad27e02c9bb350bd65c51ca0021839aeb3442c22f31974f38
1f11d80d9e2faa5ce8881ae864dba162b127164cbd543dae9ac79ec133899c0ce99f8c4b7e66756e68f4344a0f3a8033d3c7
1b9d1be59fa059b8a451fbe79da69a75241a4050e08b5e009c2ec9dfd5c8058c46a705a24bedc8988e4c78b1873c6a5b1be8
248ab6edfc8673abac5a90c60eb5e81b2e821c425d780466138e243f1a37a12c45301b9db9d8a6990d7183a839d4b98ca2fd
a4566f6c87779c7676619d71ac45a32323b0f6d8604f857f0d1dbbd1e5e005b17c904b1c5eb5725ddb001e420ec3c16b3791
dcc9e830d49cb3a8a09888b3cb699ffdf30080534eee8dce460af81a3ff4290d0773d38e4cb2ec7801d86e911346a9edfa50
46e3c93a13e78c8e33116374ce3cc2570059371cbdd3816116b96de54fdc9318cf7004f5c3c9e67680be5a1193299ab63a48
1cc74e2365c078486da175baf6728efc54830975e4363a5ac8ea0458b4a63db1907129cec7d1d31e45142689e324df8a4eca
cb88b565e8346a3c62230870ab09826725e9805e8943205028d6590014415eac4c332a12a3f751f3ed640c200d34927ab210
2114d0dbb2d082bdbf9f8197a4c89d54ac238c24cd380fbd57ab0193a8e0e666e264540c840c87d140aa6213880d220515af
701d95cf731ad9479b3c461520df4ed9b68df12c5646dbc3c164d3d213c65efc7c6ad7a970687cf4a3896a735bfbd4a43356
e70c183c8c40e7e49704fdca77814fdbea6155aad1515f520a188431e3e188036b7b10823417fbab7669dc1edb7f6e3873d3
dfb74065cfeef4231d3fd8bceceeba5c6e9abcee4f1677879eef9613f35dd3f0aa69d8ec836b7d0fafebd1dac3bb7eaf8777
fd78f4711705fb072f6f1fb66834fe698ee3b6cd6289bd372d68a8b21c480309779117f3b9b9c9e9975f9f9efdaaf9eaf4f5
1fdbc5c7afbf39fdf3f4d9ef3f7bfefbcfa7f73fde4fef5edfdfbfffeec5db77d3fd9be99b97ef5eddbf7cf37e7afbfccd37
f7afa7ef9edffdf9c5bbcfa77f99befefbd3df7dfdf3f9ca368393ecb0cd620e0717f3ea8c73d82e13c9bfdc55659adc57e7
a474cd2de9d2496eddbaeee1af7af3db8b6741f3a5bc4797221e5c390b3af5d02e66192db1aa3b51afdb65ab00e689d4ea18
646611ce379f4a4881055943cba3e37c7c896c394f45a5fe2468698ea00c44a73363c30a9d7bf4ade5618f77ed5a80eebb34
70d04e60d04daee62a95c758f85e9b91e0625633836eef4e3542f4c969ca54ad8e31457f59378f278aedb29cb6faf7d76874
a369a60e7527d7959ee2758ab2973a05af96abedaf2d482d3a73bb341f54aabe343fecf66effbc0e0b6d60e074ba75e965fd
150b6ed8fbddfef9b5eb6d313eb64835992b8ffb40a9ec375f3ebeddd329a9e4a5e34a936fa94c73c564da8539507a1526dd
245241615259aad8590419d20519ed87ebb4b7ae3ed973d6496764316b0f4d8fb118db8941b269cf539d84144ee7414f9d08
62ab00b236a4dd7bd559bafedaa586b68374dac695b56b81c2a76df40a7830e026bacc7215dd9e483473c92a79ac42e59eba
07b03586c49072855da751a75f45bfb7c1218aed18e836835eb4ccb3975db4b135b6aa6cebb4d3ec3ab6d500db0436333d34
e7ed7c4d7c86cd39745d013fcb2ce9a519f8bd765fec82fc4670473a2ecc7ff62b3765c580cffefbe7d3d7af4eba18e7613b
faf58deca0f570795259476dcfdaf6ac4ed5b7231c3ad377f5acbdb4fa9fda93ae1168ddd3269f68b1e6b3e9e8173fb45fa0
3753da795dbbfce2dda5d3a8adc768b7375e5cde1067d2f3f122ffe6220ff0c0d802c3c7fafee3e51751bbc57a6169eafed2
94d37124e60b36cebf78bb68425bb3a4731f9fdfede50d5204d5763f3e8dbba5258f316bdca671f88ba3be3763bf6c4fb4c8
fde89cfe0258074f9faecf973dbe56d0ee06a6078845d6ad63973bc8922c5e61966d55db6bd0d2ce95dda3969d6f91f4b0d5
8b96f5dccb2ecb7e6b6c0587aed70e45baf1ad70d3cd6305a56ec61d7a496aafe04bbb9b798f5f1285070026a9b94230dd2e
ac5710b693ad73ee85ab6ed6063b15ae5def74bd8cb133ca3a95ce780fcdfc7124d33d2df3978332f7ffa1ecff6e287bb541
d9b35f8499d46fdf7631976fbb9cc49ed73bb0464b97a4a451f43a90df35e26d1ec8f56358a456274c1f485da3f9b67df2a1
933b5d10f0aefa4bdbab5c80d1e47d8f4e5f6698a5fd8dde557abb9bcf26a76d13fbb6d75662df761c48d7f1dd9e7af9369f
bbbd7c9b7dd763a7a9a1be6f978fe75ca77dc395bf36a113848d6d83538cbd65921b00c8555704b0e63ac07dd02bc81774e9
a87d56a6c5a5347b452803f99c47d407f2f6698f4d1a06d2964825bb7e08e42257e9746bbb93fbadedad47efb7b6b7f16dd2
dbdd7c36b9da0e7ddb6b2bbe6fdb0ca4ae1ff726dfe673b79777b30fbd4ed6b647fa7e9a57ac4d787de1a11ddc552b8f708a
c74285ad734938eb12b12e10cf1fe308eea1789ebef1576209bd5d84c53f10266d49a5cbb71d56b196d2457e772daf5bcb97
eedafa589a5ec7d6496ffba97472215099bfdcb26f456795fc831ef7d26ddc9b7c9bcedd4ebc4d7debaf53d248cf4ff385a5
055dc9f0ba923d0fc327df3384527b6798c3c6ef9a2f683bae5d6db1ed30cf3ef0bb9524cc68a2aa9bf5737cba84a75797f0
546dd2212caf4f9f5c338db9bfffda9ed47742aa763db6e86a1fe794499b9cd99bb69903ff2a561f352af1a1f44ed2e2aea4
a57d07e022f3eea14cf7c81b0beda46591def5d2aa1353edfdad1fd256b7c89611adb2db7ef4abf4ee947483f44ada8e835f
f5d3c9ca36ce4e6ab771eea47e6d339b2b7d8cf479e57b8faee4ac1efa17a8e6f439fd65c845a7a427ddeb6a93b6a17d4f40
ce6e7ddb0ef26ee65c72ee07e4f8ede26fdf2b1163eda976bafeb4bca27ae9afdccd172f5e3dffa71f7efbfccdbb9bfff187
bb97dffff0e2e6ed77f7d317f7d35783f77107ddb02cb377ef9af27d53dfbe7df1fcfd8bb72fbeffe1f9dda524fb6140c6f9
7418585f7bb2babd9b5badafda0752e047df90ca7b69bbf01e2f32dd3b79288366b42fa274520726e9e675bcb4baca1324a8
c9fbbe744c395ea4dbb836e9ed6e169bfcaec96d7cd08a16577ed0632f5dc7a7b637f9329bbb6be93cefaeb7553f03fd5eaf
85c71609b6b5f0a823084f2a14f4c8bf0e796601867c599ff6f908eaff66035bed02e0aa3aae0d4fd1fdea75cb828818b485
af5b9e075b16a776c4fe326ef9ec39b74977a36f33eb9e399d9cc29cf589c67430d5d9790f28e8a351be96da8bef4ea15dd1
d2addd4e1c70ae64828eafb5b3adae7dd2a2972a5d6e9b81a75eaa4a7e5249ff6e2f863610fdb28e706fbd557d40a2b48f0b
6e235b85b7bb79ace23bc41074e2a34a14abd8e99013ac3dc7beb79d74199a1adec4cb3cee76d26dce7d6fab7a861a6eaefc
cbbfb6c1b5b7e074a3626ff055bc37f82aeeada52feb05f2b7bdc1e140496738af2c1e9574c147f6068f4e76519ab7732fc7
62322eec0cbe0af7065fc57b83afe29db5d6de76d2cbc8f6f65ea7b137f836e7beb7553d430dff4d185c97b782b932f722dc
197b11767682e7185de8db595adf08c829852b43ab98507469756f6912f48cf2edded2a4f3a84340d2fbe022dc5b7a15ef2d
bd8a77665a7bdb4997a1ed4cbdce6367e96dca5d678b6a067afd9b30b26ea4c31df7465e843b232fc2ce4424d75637ae7646
d6b78a8ab936b2ced041fbae6c9cf4f1507f65e2c4ac5ddc5b7895ed0cbc4a77f65da59d6db68e7ae132a69d71d7f1ef8cbb
4db5eb6951c9409f9d711f97097a7d9fcead55625d67d2f5dcd0f10157b74ab1beb9163b3a2cd7d0d1771569af6a98c197d0
0a8f4b117329de161ddbea93c1979764109da762538098ecabb7c5e89e9a92bfab0271f146f72975ae77fec5f3a5f37ca9d2
d6ab42e9833e0e47355d125ddd2ad78d3dab63cd3ac0dd72840b3b22698b5da6eb494fa2d3d92c2c76f5b05f55b7a4b439c4
4bd2d1d5662fcffcdb432d7eb4f7562ed4a741da05c3c3ee7fffd9927cbbb4536bdfff6f2ffd677d37d62d7b01572daefd2f
fb9a31e8db8bb96d645c0ff60333330f86f8f932447d93a20659e5e3d3d76d3660abda608f673fef46e8538dda69d67dcaf4
a1a65ffd55edfaeaaf6c57f7d7b5ebabab943b6e3973d4294a637403c2a97ab4c214b8759080bff8d3bb9777f76f7649f8ae
45dd68bc8caa5501d616aff3f0d72fdffcf0eeb89d9a74edd9e80b81b17c9291b5dc34270275d477311f37b4af7638efb041
87f3bab8ae4fac7438dfd5ef167bfde352ec8b26151ff5518507361d78fd4dd6a7b0fbfdc5aba5aa5c59777b9daea1f715c1
cbad1421b442cd552b4b97bfbb341248c8b3aa86e131e8d0b5dc77a9e3e39194385b17760129ea335726d876fb613fc863a0
b1574043c429419f8e34578165f08bc3e0f5e765ebd10303a6fbc561ac3d8cce877d7c7b146b0ffb388cce87135cb650a38d
701d086ef8f4133cecfc30d0ffe9681e879ce190b0bcbff461f40583f679a95de70748a65bdbba91aedb5ef5b210dbca0edd
cafee2e5bb3fdd3dff9f37ef7e78fdfaf9fb97f76f4e1d5a44b735a89ff50db8accf9a346c64394aa763c07873ffbed5134f
87e0a3ef9ee8c285d775ec7e781fc2b2e3e602545547ab6332fe69ad7df501e2aaf3ffc5ba32c2b36d4d1ebad312bc9cabd9
a72d741faf97dbe50da30b146173a7431f38f4cc37476cf690e61e7ae6e12f0e3b7f7dc15232d1f905f7b1c57af8c6a1760f
27f891f53258acff79399e31b3b1ed14c6d104fbb8f177175caf3185766f7219e0e3e39c0daa8f5a3fef7c3de8e193462677
2e3ab9ae2f1ef963def5879fc4d47d30ced4aaa34473974bb84ada92337bf2f804a2fd53c7e57ed6b89e1e978fa0d8f0433b
c333f3aca740f157fb0b043bbea5aafb153c8dd3ea43bc395cc087e1eb30441eaeeca7e7db9f10050f47b59cc8d2e72e72a9
0fcf573d3ed43f1dec0e7f7138c1ef8e7475c81a0ee1f190931dcefcfff9aacc2764d5876a7f32b1edc3c43f5d628cf566cd
4d77e3d00791f4bdc5b85fa7fa8c90b544151b1e86957fbc341a4ad547cde2d6f1861c870af83f477adafee08ea3b99ea3e9
e46f893b1034610482874be870351edaf2e97eff74e43a7ce31376fe64def313a060611d5e1fe1eaad7bf8c62784ff43b879
32fcf7abf17f1da8ed718bebba1d6dd678cf08db31a5070dfd7a619144767d552b4c3aea3eebeb88ef7495b59baba7b7867f
b9600aa4b51dcdae1fa7781142a38f2b78fded81c39ae17fbc04c40f952cff7de95d7fac415f5af317031c41dd938b928be2
42363b68fc58e9f058bd8734f7d98529ea6f25cc65fdcb2f1e61900da0fecb8738a4bee4ecd37ade6dc1b80f1fe739ce97f5
916edd5cb7497f1ac65e0e683e6ced87372f8d376154fbd3771e3718d63717af60380fb9e8c2e01fcf2c0f81e230a2fceb82
e821a49cf3162e0f83f0610cf871f1227deda76fea30383c9d401e1296c3511dd2f3435d3d2693fdf565dd465e2c7a3f7d60
31fc87653691d51b0bf0f501d77e7aa039d4c9a1d10ffde7e945c84f17c61f07488fe05b4fcf0c0e15f5f430fa747d1c06de
4f5864fadbe4221f2cca667dd749f553857a7f88b76ec15bbb7d10617f0ae2eacabbaf6e0663ddebc2975052d01f0adec993
6e00e976f94eaa7bd546df9ede49f58771da5ffe6817bcbd8e0974cdcfbfba6a7e7de50eb9ce9b5e3fdf6e65efde5145d986
94ae87aaaf3e45fd89a5dda074bfbbea28ee5eda6eb9cd6750ae86aae6e75f5d35bfbeb21feaf6fc83a1ea4f093b7d337cdf
54d647858cbeb6d10bf5c7b7f4777c7a61d2979342ae0f87996dfb5b29f56a94db0bbb416e4f5f7d71627810c3557d428405
62dd633f39f187df7f3efdf1edfdebe9ed8b772fbf8144bc9bdedf4fefbf7b31fd0166f1afdfdcfff8667a7777ffa717bb0f
4e9cfe3725d44f62
>}

\EmbeddedPDFObject{schematic-48}{%
<< /Creator <4d6174706c6f746c69622076332e31302e382c2068747470733a2f2f6d6174706c6f746c69622e6f7267>
/Producer <4d6174706c6f746c696220706466206261636b656e642076332e31302e38> /CreationDate
<443a32303236313132353030303334372d303427303027> >>
}

\EmbeddedPDFObject{schematic-49}{%
<< /Type /Font /BaseFont /GCWXDV+DejaVuSans-Oblique /FirstChar 0 /LastChar 255 /FontDescriptor
\EmbeddedPDFRef{schematic-66} 0 R /Subtype /Type3 /Name /GCWXDV+DejaVuSans-Oblique /FontBBox [ -1016
-351 1660 1068 ] /FontMatrix [ 0.001 0 0 0.001 0 0 ] /CharProcs \EmbeddedPDFRef{schematic-67} 0 R
/Encoding << /Type /Encoding /Differences [ 66 /B 69 /E 72 /H 77 /M 82 /R /S 85 /U /V /W 98 /b 105
/i /j 114 /r 120 /x 122 /z ] >> /Widths \EmbeddedPDFRef{schematic-68} 0 R >>
}

\EmbeddedPDFObject{schematic-50}{%
<< /Type /Font /BaseFont /BMQQDV+DejaVuSans /FirstChar 0 /LastChar 255 /FontDescriptor
\EmbeddedPDFRef{schematic-69} 0 R /Subtype /Type3 /Name /BMQQDV+DejaVuSans /FontBBox [ -1021 -463
1794 1233 ] /FontMatrix [ 0.001 0 0 0.001 0 0 ] /CharProcs \EmbeddedPDFRef{schematic-70} 0 R
/Encoding << /Type /Encoding /Differences [ 40 /parenleft /parenright /asterisk /plus /comma 47
/slash 49 /one /two 61 /equal 86 /V 97 /a 114 /r 124 /bar ] >> /Widths \EmbeddedPDFRef{schematic-71}
0 R >>
}

\EmbeddedPDFObject{schematic-51}{%
<< /Type /Font /BaseFont /EGWXSV+DejaVuSansDisplay /FirstChar 0 /LastChar 255 /FontDescriptor
\EmbeddedPDFRef{schematic-72} 0 R /Subtype /Type3 /Name /EGWXSV+DejaVuSansDisplay /FontBBox [ -46
-463 1794 1233 ] /FontMatrix [ 0.001 0 0 0.001 0 0 ] /CharProcs \EmbeddedPDFRef{schematic-73} 0 R
/Encoding << /Type /Encoding /Differences [  ] >> /Widths \EmbeddedPDFRef{schematic-74} 0 R >>
}

\EmbeddedPDFObject{schematic-52}{%
<< /Type /Font /BaseFont /BXGEVL+DejaVuSerif /FirstChar 0 /LastChar 255 /FontDescriptor
\EmbeddedPDFRef{schematic-75} 0 R /Subtype /Type3 /Name /BXGEVL+DejaVuSerif /FontBBox [ -770 -347
2106 1110 ] /FontMatrix [ 0.001 0 0 0.001 0 0 ] /CharProcs \EmbeddedPDFRef{schematic-76} 0 R
/Encoding << /Type /Encoding /Differences [ 32 /space 40 /parenleft /parenright 44 /comma 58 /colon
/semicolon 97 /a /b /c /d /e /f /g /h /i /j /k /l /m /n /o /p 114 /r /s /t /u /v /w /x ] >> /Widths
\EmbeddedPDFRef{schematic-77} 0 R >>
}

\EmbeddedPDFStream{schematic-53}{%
/Type /XObject /Subtype /Form /BBox [ -1.732051 -1.732051 1.732051 1.732051 ] /Filter [
/ASCIIHexDecode /FlateDecode ]
}{%
789c6d904d0a03310885f79e221778217f927a855e23300c85de7f3b9936584bdc88f8f47b6a0e2f4ae1493320c75e4be21c
de946263a98d4d31c5878854b94bdc846729c7525aff76ad6c90aad0096540b93fec67443d545e10c3580ec6ca88dabef335
1de4f86387c0f3c2be12f6bdf177179cc3e1fc07ee23e17c1cae13cc8927d1411778df59b7
>}

\EmbeddedPDFStream{schematic-54}{%
/Type /XObject /Subtype /Form /BBox [ -1.732051 -1.732051 1.732051 1.732051 ] /Filter [
/ASCIIHexDecode /FlateDecode ]
}{%
789c6d904d0a03310885f79e221778217f927a855e23300c85de7f3b9936584bdc88f8f47b6a0e2f4ae1493320c75e4be21c
de946263a98d4d31c5878854b94bdc846729c7525aff76ad6c90aad0096540b93fec67443d545e10c3580ec6ca88dabef335
1de4f86387c0f3c2be12f6bdf177179cc3e1fc07ee23e17c1cae13cc8927d1411778df59b7
>}

\EmbeddedPDFStream{schematic-55}{%
/Type /XObject /Subtype /Form /BBox [ -8.007505 -7.558336 8.007505 8.162278 ] /Filter [
/ASCIIHexDecode /FlateDecode ]
}{%
789c6d8e310e42310c43f79c221768e4243f75ba32721284e0fe2b154c485d2cf9c94eecfa14e85db6a4f98c60eb5b068c58
8b53618bf4d5fa929106b0507fd0cdafe62cdd9d6446cd1fed6a5e3ac2aa3af30ba19b07d89ddb9e22c763a7b7a77d0f919b
7c005a09283f
>}

\EmbeddedPDFStream{schematic-56}{%
/Type /XObject /Subtype /Form /BBox [ -1016 -351 1660 1068 ] /Filter [ /ASCIIHexDecode /FlateDecode
]
}{%
789c2d923b6e04310c43fb39052fb080f5f1ef3c0b042936f76ff3a4a430c4b12c0e45e9b1bbe577eae7b1e58a11b2d80adb
32e73b8ede8fd9504c322315dbfee2b5ca80321735437987eca42675efc77d6ae691efa3b94c91def1fdc45a7f37979b7995
7689934c31cd69ca5db59cc1410d19b8732e7d9e5ca03595e1bce38c4d5cc57ba83ca5949b53dab3235ae8af90afe06dcae9
aa78c85435ca0c2de9a363dcacce0ac1eb03469c715b30770dbc11abd9c22f9e594714f0bed0e7e98a992047f93fa2a3423e
ae0adb0979d7593283518e5f197e75cca84c21c77114db40e73c3a3d12dad930db1156faea509d3608df5aa8c4b13dda9a29
b4a5b94e597774db5526655172cb5eda34bdcab7d0ab0dc5fe17a3f7bd8988653415f9fb8e46566b227c8ffa0b1b32cb3ed1
58dcde1f48aa259662b78358ebd9fabd78698cc1fb81e57f0ddfcff7f3f50b95ff7aa2
>}

\EmbeddedPDFStream{schematic-57}{%
/Type /XObject /Subtype /Form /BBox [ -1016 -351 1660 1068 ] /Filter [ /ASCIIHexDecode /FlateDecode
]
}{%
789c4550b98d03310ccc55c5346080bfa47a16385c70ee3ff550bbc64524f8cc37742d446cbc87692096c13290ea70156438
aee1ae484e7db29620fcaed7883d4f97914877e49cc4506e4a886b0ba509cf8ddc1b361b2dc9a18b7fb1a0cafb5da86cb008
34ef4abc0eff3ef51a36efcea2f04a9809b4d97507d5426712045a0e42732c3c3399f81bfadf09f96b7567c4a62f4d3a11d6
27816bfc8ee46bab7c8fd40577ea6efd5b494012ebab68ba2afe70ba7a1ba73227ea3b13b92f8c79dd3fe675504ceda09ac8
934627d10a75d354d21c9d752afd23982dda1922d3d02e2461986c42ce363c9ef3a0b906886d07f06ba56dfd7c004eed5acb
>}

\EmbeddedPDFStream{schematic-58}{%
/Type /XObject /Subtype /Form /BBox [ -1021 -463 1794 1233 ] /Filter [ /ASCIIHexDecode /FlateDecode
]
}{%
789c35ccbb0dc0300804d09e2918819f812c14b970f66f632c53f1c4e90e981ced71fca034527041a8a0915ea9d056a56c2d
91d81a31509defaf5616ccd349dd83852ef4cd5e3d98f0fe37dc18df
>}

\EmbeddedPDFStream{schematic-59}{%
/Type /XObject /Subtype /Form /BBox [ -1021 -463 1794 1233 ] /Filter [ /ASCIIHexDecode /FlateDecode
]
}{%
789ce33234305330363555c8e532373602b372c02c237323200b248b60416433b8d20015f30a7c
>}

\EmbeddedPDFStream{schematic-60}{%
/Type /XObject /Subtype /Form /BBox [ -1021 -463 1794 1233 ] /Filter [ /ASCIIHexDecode /FlateDecode
]
}{%
789c4dcebb0dc0200c45d19e293c027efe40168a5224fbb7c1e088144847ba065cb83aa9293d45bb4cddc56ba738437e9059
c8ba666d829fa4212ab7940204b3ac5b3cab8c7b4b40273e06d81b7900f64dc54e5beb0d414d295bfe15756bed7495f305f2
a928fd
>}

\EmbeddedPDFStream{schematic-61}{%
/Type /XObject /Subtype /Form /BBox [ -1021 -463 1794 1233 ] /Filter [ /ASCIIHexDecode /FlateDecode
]
}{%
789ce3d235b23055b0303050c8e5d2353234063373b8742d8d15cc0ccd402c43334318d3c8c452c1dc18cc34363687899a18
98c21540cdcae1ca80a8363583180c6582c4d300ebf51632
>}

\EmbeddedPDFStream{schematic-62}{%
/Type /XObject /Subtype /Form /BBox [ -1021 -463 1794 1233 ] /Filter [ /ASCIIHexDecode /FlateDecode
]
}{%
789ce3d235b6345630373155c8e5d2353430043373204c3373631013a40089095190c19506004b060b74
>}

\EmbeddedPDFStream{schematic-63}{%
/Type /XObject /Subtype /Form /BBox [ -46 -463 1794 1233 ] /Filter [ /ASCIIHexDecode /FlateDecode ]
}{%
789c55cd410e80300804c07b5fc1074c8016810f190ff5ff5717ed416f13b2bbb41e242c4e574bf697f361c48074d8927350
8fac9b266d8222826a3faa33888d8f64af4993aa1b8474e4c2fa77b6e306138a1a9e
>}

\EmbeddedPDFStream{schematic-64}{%
/Type /XObject /Subtype /Form /BBox [ -2 -2 2 2 ] /Filter [ /ASCIIHexDecode /FlateDecode ]
}{%
789c6d90390e03210c457b9f820b7c648b00c315a6cd1168a24873ff76620820e23496b7ffbc887b13bb933e06e2a3bb887d
28f9e0d863f639979035080f29519c784e9c9266ba5369d630da558a496aa0d6a7c099fc8a96a6f316775566eb46545b699f
03238205c30cc7b61bc6c6d80f81bd147ffe01fb35fcb2d1977f113de906a3ed490b
>}

\EmbeddedPDFStream{schematic-65}{%
/Type /XObject /Subtype /Form /BBox [ -2 -2 2 2 ] /Filter [ /ASCIIHexDecode /FlateDecode ]
}{%
789c6d90390e03210c457b9f820b7c648b00c315a6cd1168a24873ff76620820e23496b7ffbc887b13bb933e06e2a3bb887d
28f9e0d863f639979035080f29519c784e9c9266ba5369d630da558a496aa0d6a7c099fc8a96a6f316775566eb46545b699f
03238205c30cc7b61bc6c6d80f81bd147ffe01fb35fcb2d1977f113de906a3ed490b
>}

\EmbeddedPDFObject{schematic-66}{%
<< /Type /FontDescriptor /FontName /GCWXDV+DejaVuSans-Oblique /Flags 96 /FontBBox [ -1016 -351 1660
1068 ] /Ascent 929 /Descent -236 /CapHeight 0 /XHeight 0 /ItalicAngle 0 /StemV 0 /MaxWidth 1350 >>
}

\EmbeddedPDFObject{schematic-67}{%
<< /B \EmbeddedPDFRef{schematic-78} 0 R /E \EmbeddedPDFRef{schematic-79} 0 R /H
\EmbeddedPDFRef{schematic-80} 0 R /M \EmbeddedPDFRef{schematic-81} 0 R /R
\EmbeddedPDFRef{schematic-82} 0 R /S \EmbeddedPDFRef{schematic-83} 0 R /U
\EmbeddedPDFRef{schematic-84} 0 R /V \EmbeddedPDFRef{schematic-85} 0 R /W
\EmbeddedPDFRef{schematic-86} 0 R /b \EmbeddedPDFRef{schematic-87} 0 R /i
\EmbeddedPDFRef{schematic-88} 0 R /j \EmbeddedPDFRef{schematic-89} 0 R /r
\EmbeddedPDFRef{schematic-90} 0 R /x \EmbeddedPDFRef{schematic-91} 0 R /z
\EmbeddedPDFRef{schematic-92} 0 R >>
}

\EmbeddedPDFObject{schematic-68}{%
[ 600 600 600 600 600 600 600 600 600 600 600 600 600 600 600 600 600 600 600 600 600 600 600 600
600 600 600 600 600 600 600 600 318 401 460 838 636 950 780 275 390 390 500 838 318 361 318 337 636
636 636 636 636 636 636 636 636 636 337 337 838 838 838 531 1000 684 686 698 770 632 575 775 752 295
295 656 557 863 748 787 603 787 695 635 611 732 684 989 685 611 685 390 337 390 838 500 500 613 635
550 635 615 352 635 634 278 278 579 278 974 634 612 635 635 411 521 392 634 592 818 592 592 525 636
337 636 838 600 636 600 318 352 518 1000 500 500 500 1350 635 400 1070 600 685 600 600 318 318 518
518 590 500 1000 500 1000 521 400 1028 600 525 611 318 401 636 636 636 636 337 500 500 1000 471 617
838 361 1000 500 500 838 401 401 500 636 636 318 500 401 471 617 969 969 969 531 684 684 684 684 684
684 974 698 632 632 632 632 295 295 295 295 775 748 787 787 787 787 787 838 787 732 732 732 732 611
608 630 613 613 613 613 613 613 995 550 615 615 615 615 278 278 278 278 612 634 612 612 612 612 612
838 612 634 634 634 634 592 635 592 ]
}

\EmbeddedPDFObject{schematic-69}{%
<< /Type /FontDescriptor /FontName /BMQQDV+DejaVuSans /Flags 32 /FontBBox [ -1021 -463 1794 1233 ]
/Ascent 929 /Descent -236 /CapHeight 0 /XHeight 0 /ItalicAngle 0 /StemV 0 /MaxWidth 1342 >>
}

\EmbeddedPDFObject{schematic-70}{%
<< /V \EmbeddedPDFRef{schematic-93} 0 R /a \EmbeddedPDFRef{schematic-94} 0 R /asterisk
\EmbeddedPDFRef{schematic-95} 0 R /bar \EmbeddedPDFRef{schematic-96} 0 R /comma
\EmbeddedPDFRef{schematic-97} 0 R /equal \EmbeddedPDFRef{schematic-98} 0 R /one
\EmbeddedPDFRef{schematic-99} 0 R /parenleft \EmbeddedPDFRef{schematic-100} 0 R /parenright
\EmbeddedPDFRef{schematic-101} 0 R /plus \EmbeddedPDFRef{schematic-102} 0 R /r
\EmbeddedPDFRef{schematic-103} 0 R /slash \EmbeddedPDFRef{schematic-104} 0 R /two
\EmbeddedPDFRef{schematic-105} 0 R >>
}

\EmbeddedPDFObject{schematic-71}{%
[ 600 600 600 600 600 600 600 600 600 600 600 600 600 600 600 600 600 600 600 600 600 600 600 600
600 600 600 600 600 600 600 600 318 401 460 838 636 950 780 275 390 390 500 838 318 361 318 337 636
636 636 636 636 636 636 636 636 636 337 337 838 838 838 531 1000 684 686 698 770 632 575 775 752 295
295 656 557 863 748 787 603 787 695 635 611 732 684 989 685 611 685 390 337 390 838 500 500 613 635
550 635 615 352 635 634 278 278 579 278 974 634 612 635 635 411 521 392 634 592 818 592 592 525 636
337 636 838 600 636 600 318 352 518 1000 500 500 500 1342 635 400 1070 600 685 600 600 318 318 518
518 590 500 1000 500 1000 521 400 1023 600 525 611 318 401 636 636 636 636 337 500 500 1000 471 612
838 361 1000 500 500 838 401 401 500 636 636 318 500 401 471 612 969 969 969 531 684 684 684 684 684
684 974 698 632 632 632 632 295 295 295 295 775 748 787 787 787 787 787 838 787 732 732 732 732 611
605 630 613 613 613 613 613 613 982 550 615 615 615 615 278 278 278 278 612 634 612 612 612 612 612
838 612 634 634 634 634 592 635 592 ]
}

\EmbeddedPDFObject{schematic-72}{%
<< /Type /FontDescriptor /FontName /EGWXSV+DejaVuSansDisplay /Flags 32 /FontBBox [ -46 -463 1794
1233 ] /Ascent 929 /Descent -236 /CapHeight 0 /XHeight 0 /ItalicAngle 0 /StemV 0 /MaxWidth 1000 >>
}

\EmbeddedPDFObject{schematic-73}{%
<<  >>
}

\EmbeddedPDFObject{schematic-74}{%
[ 1000 1000 1000 1000 1000 1000 1000 1000 1000 1000 1000 1000 1000 1000 1000 1000 1000 1000 1000
1000 1000 1000 1000 1000 1000 1000 1000 1000 1000 1000 1000 1000 1000 1000 1000 1000 1000 1000 1000
1000 1000 1000 1000 1000 1000 1000 1000 1000 1000 1000 1000 1000 1000 1000 1000 1000 1000 1000 1000
1000 1000 1000 1000 1000 1000 1000 1000 1000 1000 1000 1000 1000 1000 1000 1000 1000 1000 1000 1000
1000 1000 1000 1000 1000 1000 1000 1000 1000 1000 1000 1000 1000 1000 1000 1000 1000 1000 1000 1000
1000 1000 1000 1000 1000 1000 1000 1000 1000 1000 1000 1000 1000 1000 1000 1000 1000 1000 1000 1000
1000 1000 1000 1000 1000 1000 1000 1000 1000 1000 1000 318 1000 518 1000 1000 1000 1000 1000 1000
400 1000 1000 1000 1000 1000 318 318 518 518 590 500 1000 1000 1000 1000 400 1000 1000 1000 1000
1000 1000 1000 1000 1000 1000 1000 1000 1000 1000 1000 1000 1000 1000 1000 1000 1000 1000 1000 1000
1000 1000 1000 1000 1000 1000 1000 1000 1000 1000 1000 1000 1000 1000 1000 1000 1000 1000 1000 1000
1000 1000 1000 1000 1000 1000 1000 1000 1000 1000 1000 1000 1000 1000 1000 1000 1000 1000 1000 1000
1000 1000 1000 1000 1000 1000 1000 1000 1000 1000 1000 1000 1000 1000 1000 1000 1000 1000 1000 1000
1000 1000 1000 1000 1000 1000 1000 1000 1000 1000 1000 1000 1000 1000 1000 1000 ]
}

\EmbeddedPDFObject{schematic-75}{%
<< /Type /FontDescriptor /FontName /BXGEVL+DejaVuSerif /Flags 32 /FontBBox [ -770 -347 2106 1110 ]
/Ascent 929 /Descent -236 /CapHeight 0 /XHeight 0 /ItalicAngle 0 /StemV 0 /MaxWidth 1342 >>
}

\EmbeddedPDFObject{schematic-76}{%
<< /a \EmbeddedPDFRef{schematic-106} 0 R /b \EmbeddedPDFRef{schematic-107} 0 R /c
\EmbeddedPDFRef{schematic-108} 0 R /colon \EmbeddedPDFRef{schematic-109} 0 R /comma
\EmbeddedPDFRef{schematic-110} 0 R /d \EmbeddedPDFRef{schematic-111} 0 R /e
\EmbeddedPDFRef{schematic-112} 0 R /f \EmbeddedPDFRef{schematic-113} 0 R /g
\EmbeddedPDFRef{schematic-114} 0 R /h \EmbeddedPDFRef{schematic-115} 0 R /i
\EmbeddedPDFRef{schematic-116} 0 R /j \EmbeddedPDFRef{schematic-117} 0 R /k
\EmbeddedPDFRef{schematic-118} 0 R /l \EmbeddedPDFRef{schematic-119} 0 R /m
\EmbeddedPDFRef{schematic-120} 0 R /n \EmbeddedPDFRef{schematic-121} 0 R /o
\EmbeddedPDFRef{schematic-122} 0 R /p \EmbeddedPDFRef{schematic-123} 0 R /parenleft
\EmbeddedPDFRef{schematic-124} 0 R /parenright \EmbeddedPDFRef{schematic-125} 0 R /r
\EmbeddedPDFRef{schematic-126} 0 R /s \EmbeddedPDFRef{schematic-127} 0 R /semicolon
\EmbeddedPDFRef{schematic-128} 0 R /space \EmbeddedPDFRef{schematic-129} 0 R /t
\EmbeddedPDFRef{schematic-130} 0 R /u \EmbeddedPDFRef{schematic-131} 0 R /v
\EmbeddedPDFRef{schematic-132} 0 R /w \EmbeddedPDFRef{schematic-133} 0 R /x
\EmbeddedPDFRef{schematic-134} 0 R >>
}

\EmbeddedPDFObject{schematic-77}{%
[ 600 600 600 600 600 600 600 600 600 600 600 600 600 600 600 600 600 600 600 600 600 600 600 600
600 600 600 600 600 600 600 600 318 402 460 838 636 950 890 275 390 390 500 838 318 338 318 337 636
636 636 636 636 636 636 636 636 636 337 337 838 838 838 536 1000 722 735 765 802 730 694 799 872 395
401 747 664 1024 875 820 673 820 753 685 667 843 722 1028 712 660 695 390 337 390 838 500 500 596
640 560 640 592 370 640 644 320 310 606 320 948 644 602 640 640 478 513 402 644 565 856 564 565 527
636 337 636 838 600 636 600 318 370 518 1000 500 500 500 1342 685 400 1137 600 695 600 600 318 318
511 511 590 500 1000 500 1000 513 400 989 600 527 660 318 402 636 636 636 636 337 500 500 1000 475
612 838 338 1000 500 500 838 401 401 500 650 636 318 500 401 470 612 969 969 969 536 722 722 722 722
722 722 1001 765 730 730 730 730 395 395 395 395 807 875 820 820 820 820 820 838 820 843 843 843 843
660 676 668 596 596 596 596 596 596 940 560 592 592 592 592 320 320 320 320 602 644 602 602 602 602
602 838 602 644 644 644 644 565 640 565 ]
}

\EmbeddedPDFStream{schematic-78}{%
/Filter [ /ASCIIHexDecode /FlateDecode ]
}{%
789c3d514196c52008db7b0a8e201010cfd379f3fea273ffed04fb5f1736a92824312b658a2d7ed242966df9d1a1b90ffd1b
303dec1ea8672f3065694a2c486e956be42cc98cd320d9aa315676a519efa44ee9fb512a886025c2a47b87423059cd25bebd
2b93ac5cc289c93941e44c561619cf479558cd07bdba42a6c94a4ed1c93b5cfcbd063886537c55bb2c48f7396eefd7e43d3e
4339da5134ac50a1ca7b38553d247613cc904db194a566dd9bf7159bc65294021a1f39873100ac109f4ecb10b78e0a6ee2be
69b5ce3c0f3d78bd0a5a8d71378f1a63b4b03a2ae2cbc0809a8127c05ea039a47da34375e4c660d7838c8f9566d4186a7c0c
9ad944eb27c22263bce0ea99bed7c1eb55d18a7eff013faf684b
>}

\EmbeddedPDFStream{schematic-79}{%
/Filter [ /ASCIIHexDecode /FlateDecode ]
}{%
789c35cdbb0dc0300804d09e296e04f331847da22885b37f1b9c88069e0071ae8201892aae03218993893d3f3ed4c345ce06
372fc9e4d648585d2c9a335ac6508b12a7b64c70e85eb2feb0dc91f5e16f1db8e8a6eb0533901c98
>}

\EmbeddedPDFStream{schematic-80}{%
/Filter [ /ASCIIHexDecode /FlateDecode ]
}{%
789c3d8db10dc03008047ba660048301dbfb44510a67ff366f12a7004e9c5edf5cb9b036aca68e197c08498cc49b347ad224
95c2560bc8f17b29343ebbc3b0bdf272d62dafa3a15a03c9b04dc82d97bdf3af9b74d1f90001361c40
>}

\EmbeddedPDFStream{schematic-81}{%
/Filter [ /ASCIIHexDecode /FlateDecode ]
}{%
789c3d8e3912800008037b5ec113e490e33f8e6381ff6f453c9ab093498030c105d95b420c9d133702b21c3c41c8870a7461
6489260b7dbdafd25e6aaf285833663a339a4caf33a4b727ad0fb13a9a5213b14c7e3ea8ff70c101fb052a8a1e30
>}

\EmbeddedPDFStream{schematic-82}{%
/Filter [ /ASCIIHexDecode /FlateDecode ]
}{%
789c3d503b7205210ceb39858e80bf98f390c9a4d877ff36c29b49e19558db0829776042173f3927966e7cc9d050a4173e43
a7c26ae319c6d61f5b2f7361b1ef31e13671861759147c27cb1bc3ea762e5b06af68214f22f7d9714772c6c55bd356369eff
573ce367b8198c3a9fe141e6015f02b3f746e3eed5d850aa860434aeb241f6d58876f970425e3487d4ba8c6f508bf6a1b9e8
cba0ac332cc9b86d9cb559e4d1788654367b8668f67d9d1f8fb93b414644a197dd382e0b61ba6208a7bd1688a23d9ab971e4
dc8d41cb67349b822846188b7f897a030ea579ee79dd3818ff5f2c87117dff024bd05702
>}

\EmbeddedPDFStream{schematic-83}{%
/Filter [ /ASCIIHexDecode /FlateDecode ]
}{%
789c355239ae5c310cebdf29788101acdd3ecf041fbf98dcbf0de9410a438265898bdc515868bc2cd02b30e9f863cf4d57e1
ef533b5938f83c5586f646d946c741ee46e7c6fbc9e29b1aa43bba1d71bef1fd44c4cdfcbe3df03b235871e3a433b073501d
dfe8c68ab264c5899fc31ee2658f7ac2202cef44a62154715562886e4e9699417cd6cf428c936d23b2f9a6282d7ca1aac827
be91f25961666dc46fd872a26d7292305bbc40445e833cd78d64b2ec66466daf03a3a497f4d22c92d6347a1abad148b67f1e
be34339c0db2d3d5d6209b85316a1dfa4400d9253d2b9544710d877a8c34a4337ddfde4c3ac75645b7544519db939bf12609
a74f7ba9c2edf929c8a3601eae73b98db40dc93955079d35ca8e113f6d5188c24dba76a4cb6f38c263526428e5fa14469a3d
77282d1d2e26d8a8dfa4318a24c2f5294bf60d69d6a2f6900b4549c345561ff60efe7fbff7f3fbfcfc03c88a7db2
>}

\EmbeddedPDFStream{schematic-84}{%
/Filter [ /ASCIIHexDecode /FlateDecode ]
}{%
789c3d50416e04210cbbf30a7fa052620881f74c55f530fdffb50eb3bb07644348623b3b6188852fefc83adcf8f6e61187fe
35c638ec6e3e17b8e661039c133e098e2d34b0272ef5e90f070e6adc41ebafca3969f0fab1036eae0a69c80423a199dd66c1
d5bad6890c4fac8e11847bad182b35c811aee5b6107c645d6dfa5beadbc9dd2603cc51ccb430033125d8a8fee23570ba1ed0
d7930225acb084f1b0b2ab5edd46e9cd5d72cb0a1fd8a6d763d4717c0f5420d2254d95105181a954094accf5c9f76ebfede7
1f4e0e48ae
>}

\EmbeddedPDFStream{schematic-85}{%
/Filter [ /ASCIIHexDecode /FlateDecode ]
}{%
789c358cb109c0300c04fb9f4223486f4756f6092185bc7f1b19ece6390eee3dbaa88c58d32983b73c06aa979828bf4cc2c6
b5896162ca226fc79d32d16815e63e487c787f989812ec
>}

\EmbeddedPDFStream{schematic-86}{%
/Filter [ /ASCIIHexDecode /FlateDecode ]
}{%
789c3d8db111c03008037ba660042036867d72b91464ff3698386ee04f02c9cd91d0350793100e713c195298f400bb140588
8c3c99d4e9d79437355baeaf94801d987635e4eb906f1f864a3663b597c28d6bafe6801bae1747661e66
>}

\EmbeddedPDFStream{schematic-87}{%
/Filter [ /ASCIIHexDecode /FlateDecode ]
}{%
789c3d923b6e03310c44fb3d052f6040fc4a3acf06410afbfe6d1e6527c52e0972341a0e559e32c44b1eea926bcaac215f7a
c54e710f79bdb3a5122b247449944944ca7d450c894947bb0b6afa89f7e5f44f45378825064754d0b1c6e710e30b5731876d
547710e060752ff198a2887198ef4b6789cd14e5ac71e789234f474511ace8534baa2e7b379985a0c6d2a5006fed802ef490
b4bc82bdb6ac53de1cebe9b80315e12eea4d13d9fcadbdd055ef59504c676db1c28d766798fcf9755f3f4c5267dad765de88
0099923a11b2250fca6149d81d1dc90431e24498f977965d09050f82b9ee2b27b866c299d038d1b3cf74661bdc2446c1dc3e
ceeea0407133715261097cacfc8cc68c8151c6034064c3bd5c1ec5247a1e84b18b8e18cad64f05dd8f361a43ce6e16bb6459
5cd85e85c96ab85af1ac9ef09da0bc9d7e584ff6ff9f7d5c7ae2d8f72f657578a1
>}

\EmbeddedPDFStream{schematic-88}{%
/Filter [ /ASCIIHexDecode /FlateDecode ]
}{%
789c3d8cb90dc040080473aaa084e3a720cb7270ee3ff521841376b40f1c890bc5cee1100c5f78115036be30e6063642573f
44fe534efa0029a369d4469a4ecef57883b44ca5eaf707907e178a
>}

\EmbeddedPDFStream{schematic-89}{%
/Filter [ /ASCIIHexDecode /FlateDecode ]
}{%
789c3d8f4b1203210844f79ca22f4095208a9e6752a95998fb6f034ee28aa7687fd4070a5844c15a1cea0eef052f21b18666
8e0f697d6891a883e7869ebf92c4c05e2165eb348c16b34e5ce479d305366386b04acc911bee6917c778b77d39726c88dd09
b388a7a758fab0dbc13a1ee4f416adbb4249e1bc98868c19c675a7bd4e954537c9781a46ad5fd745da1abaf56cd30f8dfff6
a6f7179b4e374b
>}

\EmbeddedPDFStream{schematic-90}{%
/Filter [ /ASCIIHexDecode /FlateDecode ]
}{%
789c3d904b0e03310843f739858f1030e4739ea9aa2ea6f7dfd664a46e6207e4e44198a183a9230691a3e3652d62e81af8b6
60b98570e97484853471b5e8bacd096e45e706e54baf46dfc7f952450f7a9af2a18e5b826bc1b683dcb0a5eec99817c2ddf8
88852363ca39ff4ee1187e6a02da848f813421f48e647d406a067d4721d72cd1c751e12a7d2a02c914204b5775a29c50b3b2
f9ec41fefaafe16e9ff6fe0193053826
>}

\EmbeddedPDFStream{schematic-91}{%
/Filter [ /ASCIIHexDecode /FlateDecode ]
}{%
789c3d8c410ec0200804efbc623fd0041115ffd3343dd8ff5fbb46db0b4c76614a37280e2b9c5515c51bce241f3e92bdc35a
c790a29947439cdddc160e4b8d14ab5816169e6111a4b435ccf8bc286b604a29ea6967f5bfbbe57a01ff9f1c2c
>}

\EmbeddedPDFStream{schematic-92}{%
/Filter [ /ASCIIHexDecode /FlateDecode ]
}{%
789c358c410ac0400803efbec20f1456d7ac7d50293dd8ff5f1ba47b300c210e1c3af4700662f2522f13b3d5f8ca2e4be0a9
b1403283a613021bc22828690f0753cfae298fec97e19b7e75c923f707e2671826
>}

\EmbeddedPDFStream{schematic-93}{%
/Filter [ /ASCIIHexDecode /FlateDecode ]
}{%
789c33b330513050b000623373330573234b8514432e230b33a0402e9705582087cbd0d010ca3236315230343405b24ccd8d
a162308d40594b90413950fd395c195c6900743212a1
>}

\EmbeddedPDFStream{schematic-94}{%
/Filter [ /ASCIIHexDecode /FlateDecode ]
}{%
789c3d924b6e03310c43f73e852e10c0fad99ef3a428ba98de7fdb2725e98a1cd916456a96ba4c59531e903493847fe9f070
b19df23b6ccf6666e05aa29789e594e7d01562eea27cd9cc465d5927c5ea84663ab7e851391707364daa9fab2ca5dd2e780e
d7269e5bf6113fbcd0a01ed4352e09377457a365293463aafb7fd27bfc8c32e06a4c5d6c52ab7b1f3c5e4455e8e807478922
639717e7e34109e94ac2685a889577450341eef2b4ca7b09eedbde6ac8b207a998ce3b2966b0eb74522e6e1787211ec8f2a8
f0e5a3d86b42cff8b02b25f448d886337227ee68c6e6d4c8e290fede8d044b8fae6821f3e0af6ed62e60bc533a4562644e72
ac698b554ef75014d3b9c3bd0c10f56c4543297391d5ea5fc3ce0b6b966c16784afa051ea3dd264b0e56d9dba84eefad3cd9
d0f71f5ad7741f
>}

\EmbeddedPDFStream{schematic-95}{%
/Filter [ /ASCIIHexDecode /FlateDecode ]
}{%
789c4d8e4112c0200803efbe224f0014d1ff743a3dd8ff5f2b52a7bd2d84842811089920ada218c18ae0e0e458a9e34ed215
ca0563ed0a77a779948d2789b5650bf29049223f0a55db6bc83b4368e7e678b58eaadaebd3390579a5c8ff2854ef118edd77
a42b9d0fff63292e
>}

\EmbeddedPDFStream{schematic-96}{%
/Filter [ /ASCIIHexDecode /FlateDecode ]
}{%
789c33363657305030343257d03532365530323450303733514831e4823173c12cb06c0e175c21840992cf81abcce1cae04a
03006b3a0f87
>}

\EmbeddedPDFStream{schematic-97}{%
/Filter [ /ASCIIHexDecode /FlateDecode ]
}{%
789c3336b4503050303757d03534345530323250303432514831e4323434073373b96082396096890190610824c11a72b860
5a73c03a20b250ad395c195c690071a21267
>}

\EmbeddedPDFStream{schematic-98}{%
/Filter [ /ASCIIHexDecode /FlateDecode ]
}{%
789cb330b6503050303430533034375230373652303135514831e402098198b95c30c11c30cb18a82c072c8b60416433c06c
235353a81e100ba2c710ae12c182c86670a5010051be1916
>}

\EmbeddedPDFStream{schematic-99}{%
/Filter [ /ASCIIHexDecode /FlateDecode ]
}{%
789c458cbb0dc03008447ba660047e26f63e5194c2debf0d10256eb827dd3d5c1d0999296e33c343069e0c2c865d7181f496
300b5c47508e9deca361d5beea048df5a652f3ef0f949eecace286eb0102421a66
>}

\EmbeddedPDFStream{schematic-100}{%
/Filter [ /ASCIIHexDecode /FlateDecode ]
}{%
789c3d4f390ec3300cdbfd0a7e20807558b6de9322e890fe7fade8a01d0411a078c8b2a363050e318149c71c8997b41ffc34
8d402c839a237a42c5302c7036498187415649573e5b9c4c21b5055e48f8d6143c1b4d64829ec7881db893c9e803ef525bb1
75e801ed1055245db30c63b1adba703d610435c57a15e6f530638992479f554beb010513f8d2f97fee6eef767d0160c5305d
>}

\EmbeddedPDFStream{schematic-101}{%
/Filter [ /ASCIIHexDecode /FlateDecode ]
}{%
789c358fcb0dc3300c43ef9a820b04d0cfb23c4f8aa08774ff6b25a7050c9830c927d91683918c434c60ec9863e125544fad
3e2423b7ba493910a95037044f684c0c339ca44be095e9be653cb77839add4059d90c8dd596d344419cd3c86a3a7ec154efa
2f7393a89519650ec8ac53dd4ec8ca82159817d472dfcfb8ad72a1131eb62bc3a23bc588729b1969f8fdefa4375d5fdc7830
d5
>}

\EmbeddedPDFStream{schematic-102}{%
/Filter [ /ASCIIHexDecode /FlateDecode ]
}{%
789c458db10dc03008047ba660040c31305014a520fbb7015b32cdff4927fdbb38120ed24c134665c37bc0a5b4f05b247362
40e926364e2adb44d9627e7a9b1a6fda0b659bea29ce67c00bcf0fa13f1d59
>}

\EmbeddedPDFStream{schematic-103}{%
/Filter [ /ASCIIHexDecode /FlateDecode ]
}{%
789c459039120331080473bd8227485c82f7accbb5c1faffa907d647a2e902341cba164dca7a14683ee9b146a1bad06be82c
4a9244c116921d50a763883b28484c48639248b61e83239a58b35d9899542ab352e1b469c5c2bfd95ad5c8448d708dfc89e9
0654fc4fea5c94f04d83af9231fa1854bc3b838c49f85e45dc5a31edfe4430b119f6096f4526058489b1a9d9becf6086ccf7
0ad738c7f30d4c8237eb
>}

\EmbeddedPDFStream{schematic-104}{%
/Filter [ /ASCIIHexDecode /FlateDecode ]
}{%
789c333636573000425d4b23056320dbdcc85221c590cbc8d404cccce58209e67059188355e57019406998a21cae0cae3400
fb840e1f
>}

\EmbeddedPDFStream{schematic-105}{%
/Filter [ /ASCIIHexDecode /FlateDecode ]
}{%
789c2d514972034108bbcf2bf48466a7df63972b87e4ffd708ca07060d8b40745ae2a08c9f202c57bce591ab68c3df333182
df05877e2bd76d58e440542049635035a176f17af41ed8719809cc93d6b00c66ac0ffc8c29dce8c5e1311967b7d765d410ec
5daf3d9941c9ac5cc42513174cd56123670e3f99322eb4733d3720e746589761906e561a33128e3485482185aa9ba4ce0441
128cae468d367214fba52f8f7486558da831ec732d6d5fcf4d386bd0ec58acf2e4941e5d712e27172fac1cf4f53603175169
acf24048c0fd6e0ff573b293df784be7ce16d3e37bcdc9f2c25460c59bd76e4036d5e2dfbc88e3fb8cafe7e7f9fc035f605b
88
>}

\EmbeddedPDFStream{schematic-106}{%
/Filter [ /ASCIIHexDecode /FlateDecode ]
}{%
789c3d52496ec4300cbbfb15fcc0005abdbc2745d1c3ccffafa594a6878084259b14953c138214bcd4917323ddf1a5c3cf86
4ec7a7992dc77bd8b687f96a66123076695522700d9d02b380f2493d7963ceaa14337e537194571c4b786e92084a78c26f25
c2355ca58887200e9c22cbea784ff0d9b6677fc8da357e466cb653fad32c8d3e7ba08708b1fa1fa42b925597e049ab4146a9
5211c7ab46b4cec568a49043ec3b29750e0e15a9cbd79813d32a456520ff1393981ccc03eb12a7db35b1c6448d67a2bc9e2d
5378b5ad62b7416752378b12e2e4e1442a457616b4158bae291a9befb0bf90995a36d3b3d831a1fc62b676b2979bd2606f6d
c3e96557e0479abc8bc4a92dab2987e1ed242a7113ad6230ba4f66657efa87b1b91acbd37de27c21952e9c3aebb012cc21c2
d16b3a86675db5baef5f801678c9
>}

\EmbeddedPDFStream{schematic-107}{%
/Filter [ /ASCIIHexDecode /FlateDecode ]
}{%
789c4d51496ec3300cbcfb15fc40007113a5f7b8087a70ff7fcd0c9db4056c702c52b3d033860cb12d0f75c93da4e6902f3d
5453d2e4a7418d25d781a13f80298091ff507837cd240a9411123bc4ca2475c8896b1b9429ee50c2eb75d7f308af46b13031
80a11059e8e404e2099c7968579b6423d25ac20955eb3bd34806127c939399a8c27a1ed425a28f47b6b14d5b306a78e87b09
832ce531c0b8c3be0bd6717df6721ddf3d611ed81191968a69411a0a70ba9b64627ab5621802240b9ca009105806b2446cd9
0c1bb0aabd06d0add5f50e4be4cec522d09a7d25acd3822320d19455b746d57bdd44f41073b4a9b0eec0a6f3d7c0b6477535
289cbf9118eff90230aa64da
>}

\EmbeddedPDFStream{schematic-108}{%
/Filter [ /ASCIIHexDecode /FlateDecode ]
}{%
789c3d51396e04310cebfd0a7e20800ecac77b2658a4d8fcbf0de5206946b4c4c3f2d43418caf0e18972a232f1e9a3a1d7c4
f72873b819b8166a8374a4e119b482a05a2d4db1ba3e23a22ef2a9b9a47e305dfda596ce8af27632c46c1b81644053dabe6c
caf2192d2f75daaeefd4015d9f91fadece9ca838c873543b82298ea239b587057816b8574f84b288f720eb0fe5be5a861429
762f59374117e4a22e46e9c56f2f552da779a3f0a994093f9ac4ef7d9570b45c899fbc35eaf444c8b73a62b8b8ad39bd7a9b
54a03d25e81015a5d3ee79b91e5b9ec4ec367d412ecc7ec7d228ee0f7afe7fd57b7c8dd70f9fd9560c
>}

\EmbeddedPDFStream{schematic-109}{%
/Filter [ /ASCIIHexDecode /FlateDecode ]
}{%
789c3590416ec4300c03ef79053fb08024da71f29e2d167be9ffaf25e5f410489049691872219031f04aa23830f4fde4e1d1
4cfc7673dec80c5c0359c4bdf03e925336cd25ca3c91ad39fd72adee2aaa1595735baa9657f8c8797599e1b11a16fc9aabd5
d3d3205e9a6899c9bcde55eb65ea896c1654b5de7cb29b968587fe7d7c9b9f72ef246cb8c0309cb28c9c4f98e1fbf238bfaf
b9ee30ee8c6385f1b6c7bc23f63ff3d6aef23d81446905ed959ec2dca16850e919fb8eab53e59e88a4156263f109c6fed9de
9af8cfe36c9f3f97f45222
>}

\EmbeddedPDFStream{schematic-110}{%
/Filter [ /ASCIIHexDecode /FlateDecode ]
}{%
789c358e4b0ec3400843f79cc21718690c613ee749157591de7f5bd32a1bb08c79105ce88881c648b81324f1a295b5073e36
2466606db463823d149d388daa0cd013c7feb51d654b14e2b60726a523297a4f6867112dbca2398556c85dcc1ae6ff8df3b9
7edbdbae2f93ab1e3b
>}

\EmbeddedPDFStream{schematic-111}{%
/Filter [ /ASCIIHexDecode /FlateDecode ]
}{%
789c4d51396ec4300cecfd0a7e60019e3adee3204891fcbfdd196a37496170445173d02355544ae56121c34ce650f9b0abbc
a45c7e2ef600be1b286a46fdd66504362597c42ef194184bb6dc5780f851123a9bdac7a9f7e51a8d0ca2e662a6321cfd8923
ceeccec5e270725f0091b0a592a0e274d6449fcf0b1dd2551c015628eb4174c0187454462e5acc9dd28ec98808190b374453
17d20478ffa1c1a864f943bd8daf7ee2915851a355af9d4482dc210fa1d81b987663426a689bca09dbb43279e3b94fc760ac
96d8c68473233686c41ad81458c3bb9ea510d9c22c260c51f96637194830d19c7e44fcf5330068a256bbdac6d4b069f868db
a6c93bd28d789f4f05376445
>}

\EmbeddedPDFStream{schematic-112}{%
/Filter [ /ASCIIHexDecode /FlateDecode ]
}{%
789c3d514b6e053108dbe7145ca052cc2fc979a6aaba9877ffed33cc6b573816d886c451991253be6012ae1266f28d515049
bf06221add0ff2fc20ac14e416a8098ecb5e720d558a81fd47ccc5802ad730ced77babb88bcf2d19a49d5e3bc43901a478b2
46c9548842f7081cc19ce2cc49c2c3454f8d8209c4d23bb84ded5a01b2116880923d42d56b2c527c3307d6aaa239cba872a9
2ced50c0a6c3268fe435c8a83e17d1935db9caf2466ed11d4e397fb4e8e560bf3fa27f27bcc62f37cd0ef922e21d927e8af6
b4c3793bad4cd5acc3715d9ea53cab72296edd0cd354070e7dad73ae3ae7e2bab355eb63cae7fa77bce9fef30673f95a51
>}

\EmbeddedPDFStream{schematic-113}{%
/Filter [ /ASCIIHexDecode /FlateDecode ]
}{%
789c4d904b7203310844f73a0547e00f731ea7525e38f7dfa691ca1e6fa6df805a82b66262b2c4c78da992e947d6605ad1df
b2b64d2f9052a6905550f678f07f193d9679537190e945254ac6b6f5b1344f453d71026af0b44f471ab72529ba69b6353a77
0724d7bc18f5459e33c5746f0a9da6d60d3c9a6fd96551bde178d1fac079611fba09331e4a5712ec9f1d24b0940466940be4
81bd6447667ca24316f0ec8a238368a414d0d9d81a9e40cc8cb4dc4fdc361bbfd37eade7fafd0739ac4bcf
>}

\EmbeddedPDFStream{schematic-114}{%
/Filter [ /ASCIIHexDecode /FlateDecode ]
}{%
789c3d524b6e05210cdbcf29728191c88fc0795e5575f17aff6ded30af8b110682edc43363c8901c729ba94c554977f9d22b
2d2566c96f239cbf1bdcd3511d72ab0ddc3bc01cf2bac20c146388cf385cb6d601afcbf27386ef365da29b8a6ab8d464dd98
a206a01b0034b7aee0256125c435ea8392aa638bce0900bb4a298d4d29e5db996201292d870f3be075b9c7815ea480b06fbe
84f0d301b6e1e832b557a57b82c5fe434b826fc88e469750d131c13bc547d18cd83c2bac0c6fa4982e3a55389b6cb8b0c59e
a7b5b8584f10c0e119b7012a564756cf87f994908ed950802b94c74174c06ce828955cb4183ba41d93112d84afa799d48d6e
3aeb7fc4a8dfffa1bfaf9faeb495f801881ce30f8316587d6f607af362fd680751f048dde28d218d3e51b8c81338278c7e90
9a2fa49c6075ebf54c808851b2a263c59bdd6420414573da11b167f20034d1516ed99d176c2abe0eb04e90e6f134cf96d8de
f71f71218d5c
>}

\EmbeddedPDFStream{schematic-115}{%
/Filter [ /ASCIIHexDecode /FlateDecode ]
}{%
789c4d90418ec4300804ef7e054fc0d060fb3d59adf690fdff751a349acc21e94a10a64c02a2e2c957ce94952a3f7360f2fb
bf224cee31cd1e58ba49ec78804df7b0695f045b45b60539c542057b8bad94d0946bb8d691107797e0e3793aaf01cb26e496
a0124eb017ac04a723e8c49950ed74d7aa905ab0eef0016dbdfdc92e00f94059bc2939c569cb4ba1e6b8950b5d1121be0e7d
9638b4f31a76769365b693718fb0591576fbe11f6ec1b13a6d9faaccf71aede80365d7fbbec7dff87d01cc004b50
>}

\EmbeddedPDFStream{schematic-116}{%
/Filter [ /ASCIIHexDecode /FlateDecode ]
}{%
789c45504112c42008bbfb0a9e405051dfd39d9d1edaff5f3750677bd064204462351595eabc6c0d19641f14329f2a7792d5
056a32740a506560c951608d6215d46708bd26b2e3bb4283542cec19db2ec613ae89544587cc9b84c23b38436c08b709f170
712aace73b81ecf0e5ac70935470b73db35d62799aee3047398bc1a41b734556922b8912f30712b20cb3973426cad69f74e6
b91eb79785fc2cdf1fb5233fa0
>}

\EmbeddedPDFStream{schematic-117}{%
/Filter [ /ASCIIHexDecode /FlateDecode ]
}{%
789c45514b6e443108dbe7145c00099390cf79a6aa6631bdffb626c96b575831d886549898e80a517788c365d42e5f283093
3e4d7e0e62072c64d814a0cbc09257810fb69ba0d91e43f45dc98cfb32ebe958edcef8557160abee4a1f32897a93ece801ce
b036a4da6cd25365b0c363fb642543e7fdc224bb83d9eecca392e9a9faecf32a6fe676697d70b73a36f82408e6fb943cc13f
d2b007c12a237582463f2a2ac68e504966e6b9ce116b3d35331ca4dc4c1d537473a8e4345a8e0d51aa28d63abf809947da70
3aadb5ad3f94de9cd54a21f0b8ea1922d249d30a3d6465ac2a354e7d957ed11cd9c0152cf3b4a480e49c6fbc872e3f356cff
acdfc3bccbf72f0f666585
>}

\EmbeddedPDFStream{schematic-118}{%
/Filter [ /ASCIIHexDecode /FlateDecode ]
}{%
789c4d8f4b0e03210c43f79c2247207f729f51d5457bffed243062d8e0271c0763dda003451e860c6e1d2e6c34eaf6df5852
7e254aa988fa82f79194c917ac86a9eb41649a242420e673551ca418493ae4a0e54ae89e43e0cca7cb633d5f4d37cc82ea5b
a7218c0bc8ede9506d30e2a165e6373754fedb3e37de8b348e
>}

\EmbeddedPDFStream{schematic-119}{%
/Filter [ /ASCIIHexDecode /FlateDecode ]
}{%
789c33363250305030b20413060ae666060a29865c4606a60aa6460ab95c2031202307cc3000d3300a2c6c68688a60981b58
40a4100c33b062a0690816587906571a007ce51753
>}

\EmbeddedPDFStream{schematic-120}{%
/Filter [ /ASCIIHexDecode /FlateDecode ]
}{%
789c4592cb75c5200c44f754a11284be508f7372b248fadf6624fcec059e01197425d8b68849039f2d935c95bee6f0b9c830
fe86ab9379907b922d23afc146d78889df45294c7a5be46abd46ea59c954f219b498b15710591264c8b80c39f7adca158173
a1dfd1141fc3d04879b403e9fc1a15fe38c4132c3699b2f21c16cccc9d62a18e48d0ee56f0f3710e6ef3792a44ce0b3507e8
0a036ad92a6b5704ae53075af09822505f8f76c0345ea3e8d3ed121c52dd9533b47aa93bbacb9ab39954fda6947d9c8435a5
98de9422d29482538ab2f45096ebd4b2f9354567f323bd3c455e63b18a3f5ee373d7de3aed712659ae2a88a2410fd7224155
ce5195709d6d7dabf50634f47e15c6c79dc6402bc21dd9f53e6a563781c6deafef1a3fe3fb1f968c7eb8
>}

\EmbeddedPDFStream{schematic-121}{%
/Filter [ /ASCIIHexDecode /FlateDecode ]
}{%
789c4d904112c3300803ef7e054fc00213fb3de9747a48ff7fada069d2438c32026b4db88b8a058fe821c34c1ebd79e7ff3b
cb801cad03b7f0d8a8387189d117053afe94234d60b2ad0b868acf29d8d8ad217b33cd2b5d8c81196ab1aaeecdf1c5f098bc
8f316b70d6e98c4c1f64ca70d5aa669a4e522460bde127b4f0e655cb708f5b24c5a98229465ae59739866421ab0fbadb2a1e
73adba37ac590a911d10708f394787a72dbe995b30dfaa62ae74fab9462cbd45d2d5be8ff66acf0fd5654b5c
>}

\EmbeddedPDFStream{schematic-122}{%
/Filter [ /ASCIIHexDecode /FlateDecode ]
}{%
789c3550cb6dc5300cbb7b0a2df000eb6b7b9e3c14bd64ff6b492505028891458a544d932939e5a32e9926e92e5f1d3e553c
e41e1e87c077e24942a71c956b047a6a47229628de582d0f5f801c229cf0339b12b6f1428da8124a063077b05ec3280d645a
3da1c75e8ed66a15cd6c55d6670f113773824ec8696b14811e3569e109728ddfcec498f730cb465ae02bbe2345ea420bff38
87aec56235d106f0305930353726b6443ee602435b28c7bb71012ba292de47583d11db5f4ef268e8f0d614657db700f55a86
516d4e3db75e74494d9afe8fc1483f7fb155530b
>}

\EmbeddedPDFStream{schematic-123}{%
/Filter [ /ASCIIHexDecode /FlateDecode ]
}{%
789c55924b6ec3300c44f73e052f6040fc99d279520459a4f7df768672d11408c0678b1a0ec7b962c8105b72da28c93524dd
e54b0f1b293653be3779c8bb494bc5b444cdc54265a93c0ebb5272426649987824cbe3701c02c252ae21114b56e175948b62
4a405ee7ec6a68c009c81d37d0e1f3ea2b6194a24660444b56ed1945b51e0aa287800a4d85f5096cfa0aa16d8f92df951ec7
eb50657771bdd5f026a4ae7bcd3f0a9f24a32baa870435cbd033eef969f4ed9d9dd7aef0ecd514131d4c97a392cef202dd79
3bfcb2ee04485a53d8a16a7de7ea002082676a9ebaa7b02264df441f67b6b1455b306af8d1f7ecd5677f29c0a999bdeff887
fc03740a1fb48f19d607eeb85ec7f30708556b75
>}

\EmbeddedPDFStream{schematic-124}{%
/Filter [ /ASCIIHexDecode /FlateDecode ]
}{%
789c3d4fb90d03310ceb3d05173840cf49b2e74910a4b8ecdf86f205290cd1341fcb9740500b8746c075a152f0d4d170739f
61de90776d5d3a94843a1e632dccd56e9dd9c3c5c8129c3ac1c7d0738b6329f9b6674e745ed9f96f7becb252c3352ca9a9a2
c69146ad2a22b3dd6c0a61794dc6c79e775f239d7df81191edf1c9870e3958d4a14719ee9da47e8d7bbd6bbcc7eb0b2d6c31
38
>}

\EmbeddedPDFStream{schematic-125}{%
/Filter [ /ASCIIHexDecode /FlateDecode ]
}{%
789c3d4f3b0e433108db730a5f2012260192f3bcaaead0de7f2de4a99d7030fe646c8120884e330c12e18207db6ff5b99104
de8d9a3014b440d709c6c658b89a0a411128935cebcc215a4ca249435d988c2331f764cac475a14c3da23a044b1277878cb3
6c96314a817bbae6db36cb355b1be7e93bb9cebcf30a7139ea62ed23e1383d04dd333f179ddcff2f5fedd59e5fc6543073
>}

\EmbeddedPDFStream{schematic-126}{%
/Filter [ /ASCIIHexDecode /FlateDecode ]
}{%
789c4590411242210c43f79ca247a069a0701e1dc785de7f6ba8e377c1cf1b5ad2fc3297758ba90f8523c2eede0ad1ed5d14
bbdbab11f3221a7deaa43186b1eb70d9adc58671ca2f439a16dca5aa9c1e11d28dc30dd45bb82a40942f1cea9fa558fb5444
031a181a764125f09fd4b5037fe05ca773fe61f81694db45441ec2c9170625a26622b78dfecd9bb2a412792d25d657cf3f8e
22fab9f1721a3879c910656de0bcfd6df1d69eedf101f3c04403
>}

\EmbeddedPDFStream{schematic-127}{%
/Filter [ /ASCIIHexDecode /FlateDecode ]
}{%
789c3d924b6ec4300c43f739852e3080f58bedf34c5174d1de7fdb47a7e822a020c51249a93d6d58dff622a83bac33edc32f
32b1ed47e83decfbf2b1fea36dee6e1e6e2bccabec4e7b5f7e4fabb4186e5916750bde57ac13a453a592cb7a92cede36c372
0edb75c0435d4e546d0917bfa9d4349f4b9564e60e786d864c2680e78d2fcd6b714b727476d5e7b435cfefef0b6e1943a3d5
50b09c3441c5b259686f93b262021d11d5becf3c39125d0725c74f944e372cca12c61ff3f6b61abc1920aa6b6b8aa244c3f7
952bfea3811f187233331f5a85dbf409dece8638b8409104990d2345be794b179f7e78c3b789f8bc826e7f384e8528579281
ab34cad33e1ae990f28bcd66b06fb2e94709aec6d6ae931585dc919f0d1f1447ea1f4d0bf9a53d49cbfd9c021522565adc02
7ac422cefa58c160c0736ac1290925eac968892f1c87c5eb2c9575a52d54963dd7f8bebeaecf5f630f7fec
>}

\EmbeddedPDFStream{schematic-128}{%
/Filter [ /ASCIIHexDecode /FlateDecode ]
}{%
789c3550410ec3200cbbf30a7fa0124ea094f7749a76d8fe7f9d0dea855889e3d8640e54e48983d911022d1b5e2c6ecd13bf
720a8cc43571b401d61475e02ed4cb04a3a3cd5566ba2d4012df12c107f142977aedd0ce451c19a6f621699122a4e961df36
eee7fab77c0aeb297f534e16baa4c040abb14fd3742a450b4d7a2eff94aaab26f26d14b52d46c8f5de89984bc599adbaaaee
6862d40933d2bfa29d0c87b34ab2c2aa59f71d5767c9dd9193c590b7bd63b75659eea5fae4b995edfd070c8946f2
>}

\EmbeddedPDFStream{schematic-129}{%
/Filter [ /ASCIIHexDecode /FlateDecode ]
}{%
789c3336b4503080c31443ae34001de60352
>}

\EmbeddedPDFStream{schematic-130}{%
/Filter [ /ASCIIHexDecode /FlateDecode ]
}{%
789c4d90bb0d03310c437b4fc1050eb0247fe7b9204871d9bf0de943e2547aa62c8252c98e0c9f382c10b3a08d8c8725cb03
a575bc137b824b506d12d4dba401d2fca7bb1b1cdb749ba8bbc9e24b33e08cd20adc0cd57126e76861b6a808ca75a8501eb6
de322265a6aa9483bfdb4478c5a4ec0de672d14e2286a0de0ba20fd906a738c20fe1389c7e7dddc06b5ef55cb144d68c15a6
2c26dd684099bb0f5f456b9cbf935de9959e1fd17040ac
>}

\EmbeddedPDFStream{schematic-131}{%
/Filter [ /ASCIIHexDecode /FlateDecode ]
}{%
789c4d904b0e03210c43f79cc217a8447e04ce3355d5457bff6d4da6bfcdf8910c89cd704787262e62183d11b2709566e185
cf16aa458f939450ff7da053ddf4ab6b375c0291b06530818d04c71fcd5c7071967aedd31ca54753994592017e7dc2826551
c1bedb2784164ae7da0d92b3f368b4fe85d3a6acf9472af926096eecc1490cccc133f7624b0cd687ef44d64bb6d151e7ccbd
dfd602071ccd794d94ab9953d24a95018f4a7efad82ff7a3d3c8bddd5eac6143d2
>}

\EmbeddedPDFStream{schematic-132}{%
/Filter [ /ASCIIHexDecode /FlateDecode ]
}{%
789c4d8e3b0e80300c43f79c2217406ad238a8f74188a1dc7f25555a60b19e6ce7030717de3404ae0c697c08a9ed61dc0467
f39d3b45e185d1e9a4d57f94a1a04ed2d658a40499617a35f4a39c5d37933245b1d593ca6347bed3e9a2f301930121f7
>}

\EmbeddedPDFStream{schematic-133}{%
/Filter [ /ASCIIHexDecode /FlateDecode ]
}{%
789c4d8f390ec34008457b4ec111608001dfc78a5290fbb76136cb0d3c7d3e5b5847421e2154d0f8c29b418326fea0b320b3
62820ba1762fea662f1abe84d3bc68553d1e9f702d48b0b92d415ba0441409d1545a2923bbee963ae98135b819bfe8147d53
8b73a8b26fdf7923e10b9f3f99f32d4a
>}

\EmbeddedPDFStream{schematic-134}{%
/Filter [ /ASCIIHexDecode /FlateDecode ]
}{%
789c55904112c3300803efbc424f3032d8c97f3a9d1e9cff5f6be290a617b3830468eccd50a09c8f3be1bae3a5c25d51b5e3
102b05d63a86d4faa4300ef13f5aaa354d1f09768fde6698dba72d6e2494f0e8af9e42eddb0236074becd11ce23c75434c91
59cef64cbde46aa045acdeae2c73fd0d2b31cd1e74895bdaf207867ce4fd0501183836
>}


\ReserveEmbeddedPDF{floor-1}
\ReserveEmbeddedPDF{floor-2}
\ReserveEmbeddedPDF{floor-3}
\ReserveEmbeddedPDF{floor-4}
\ReserveEmbeddedPDF{floor-5}
\ReserveEmbeddedPDF{floor-6}
\ReserveEmbeddedPDF{floor-7}
\ReserveEmbeddedPDF{floor-8}
\ReserveEmbeddedPDF{floor-9}
\ReserveEmbeddedPDF{floor-10}
\ReserveEmbeddedPDF{floor-11}
\ReserveEmbeddedPDF{floor-12}
\ReserveEmbeddedPDF{floor-13}
\ReserveEmbeddedPDF{floor-14}
\ReserveEmbeddedPDF{floor-15}
\ReserveEmbeddedPDF{floor-16}
\ReserveEmbeddedPDF{floor-17}
\ReserveEmbeddedPDF{floor-18}
\ReserveEmbeddedPDF{floor-19}
\ReserveEmbeddedPDF{floor-20}
\ReserveEmbeddedPDF{floor-21}
\ReserveEmbeddedPDF{floor-22}
\ReserveEmbeddedPDF{floor-23}
\ReserveEmbeddedPDF{floor-24}
\ReserveEmbeddedPDF{floor-25}
\ReserveEmbeddedPDF{floor-26}
\ReserveEmbeddedPDF{floor-27}
\ReserveEmbeddedPDF{floor-28}
\ReserveEmbeddedPDF{floor-29}
\ReserveEmbeddedPDF{floor-30}
\ReserveEmbeddedPDF{floor-31}
\ReserveEmbeddedPDF{floor-32}
\ReserveEmbeddedPDF{floor-33}
\ReserveEmbeddedPDF{floor-34}
\ReserveEmbeddedPDF{floor-35}
\ReserveEmbeddedPDF{floor-36}
\ReserveEmbeddedPDF{floor-37}
\ReserveEmbeddedPDF{floor-38}
\ReserveEmbeddedPDF{floor-39}
\ReserveEmbeddedPDF{floor-40}
\ReserveEmbeddedPDF{floor-41}
\ReserveEmbeddedPDF{floor-42}
\ReserveEmbeddedPDF{floor-43}
\ReserveEmbeddedPDF{floor-44}
\ReserveEmbeddedPDF{floor-45}
\ReserveEmbeddedPDF{floor-46}
\ReserveEmbeddedPDF{floor-47}
\ReserveEmbeddedPDF{floor-48}
\ReserveEmbeddedPDF{floor-49}
\ReserveEmbeddedPDF{floor-50}
\ReserveEmbeddedPDF{floor-51}
\ReserveEmbeddedPDF{floor-52}
\ReserveEmbeddedPDF{floor-53}
\ReserveEmbeddedPDF{floor-54}
\ReserveEmbeddedPDF{floor-55}
\ReserveEmbeddedPDF{floor-56}
\ReserveEmbeddedPDF{floor-57}
\ReserveEmbeddedPDF{floor-58}
\ReserveEmbeddedPDF{floor-59}
\ReserveEmbeddedPDF{floor-60}
\ReserveEmbeddedPDF{floor-61}
\ReserveEmbeddedPDF{floor-62}
\ReserveEmbeddedPDF{floor-63}
\ReserveEmbeddedPDF{floor-64}
\ReserveEmbeddedPDF{floor-65}
\ReserveEmbeddedPDF{floor-66}
\ReserveEmbeddedPDF{floor-67}
\ReserveEmbeddedPDF{floor-68}
\ReserveEmbeddedPDF{floor-69}
\ReserveEmbeddedPDF{floor-70}
\ReserveEmbeddedPDF{floor-71}
\ReserveEmbeddedPDF{floor-72}
\ReserveEmbeddedPDF{floor-73}
\ReserveEmbeddedPDF{floor-74}
\ReserveEmbeddedPDF{floor-75}
\ReserveEmbeddedPDF{floor-76}
\ReserveEmbeddedPDF{floor-77}

\EmbeddedPDFStream{floor}{%
/Type /XObject /Subtype /Form /FormType 1 /BBox [ 0 0 539.429958 199.34 ] /Resources
\EmbeddedPDFRef{floor-1} 0 R /Filter [ /ASCIIHexDecode /FlateDecode ]
}{%
789cd55b4b73dcc611bee357e02855c4d1bc1f871c2c3b56952bae94225572b07da0562b8aaa252991949dcaafcfd70016d3
83c58058aee572ac122df60e7afa31fdfa30fbfcbbedaf979bed3f5fbe68bf7ddd3ccfbf6dee1ad57ec4df8b56b61ff1f7b7
56b52ff1f7a291f8edaa712609ab537211bfeef8af2a25612c6832fff343d3bc6f9e7f83c7efb0fc65d3582f8cb126b8d605
a12218aa60845736853490769ca442128a38e50747d2c05cf7cc2f202404161122632ba2344e0ba5527461dc8d51b0076dd6
bc80b6bf359ff153b66790bdb54e04a7bd4f36c9d45a2b9454111b6fae9a176f9ae7dfab36b66fde77f678f3aef9a97d2285
7cdafed2bef9a1f9db9be655d3c9d0a428ac344199716f46a9ee9db490c63bedac8a7adddefa706f85073d1ef3299b9991aa
bb2b1385d1d24465ac73ebb6b77c7bc62b689182f5011b1a25ac1978e0519174c7141c468e4fde3e6ddf7c6c9c505ac9446e
ee7678f2654f76baf77e4fbeeac841b8e8d8e24ff33cda8eac85c34125130ce4f38e6c05247446e947b1deed5997ab2fe7c9
f71dd908ad6cb40feaf86e7ecbed5e6cbeb657112ef80cf34a49b129877ff92862d4b153706ff9cd55fbfc7b7df6ddf6e3f9
bfbebc3ebfbe3bfbc7dbdde5e72fdbb3edfd79fbdd4dfb6a3c4a2c6093c0b934d1e134592de2943a77a0b417be755ed8e88d
d64a397b6414e5ed239da7a48d2ab6cfd4eaf64817c94ae5a3f6491f19482ceb4868618c0ca1d89f91ab022869842111ac25
41d7c6d281043ac193d0c1941264725d02fc8c36442f1dc5e382047e5102640e6325b12824c8e4ba044e0a2355903e181cca
0509e2a204c18b90ac57e52164e4ba04282ac1e13f1da3f5750954790c3f53396bcf889786345df824d849f639a3cae676fb
71bbb9bfbcb96e6fcfefb705c7765a04950e4247e99178a54416d64a24dfde6edb7fb7d7ad6e7f40e0221bbf6ccb3acd6a99
0f940efad8084e68af0202d25b91101c96dc926b4f801628ae1430b491b3cee39768293d46dd9173a9501ab5c2682c213ab6
891afc02dca9440801a5980cfeba79d51ea9d54f2dbcd12af8e51798ecdda06153d730220f3aa563a9e1a877a1219263907d
4ae22a6247afa4f66b55c43609b9cd3f5a45450abab51a7a81ddacf12b34e454a66141ce1afa847fc5d0b75659c188536c0c
c5c263f543eaeb5ca8b2239b078feaac237915616d12a332350b326b6b3839ab5914a847a849f157d7c60913bb7c52382de1
e4843e98b23608b164cce086ac0d6d9e4ce7f8f25c820b1a93a426c7b24875a747dec2813c4637c8e79ded520dd70dfd24ea
ae521357a10e40e7ae1df8bd952b622eebd64c74b3c22a051ea56ed8ca7b39c9974a68346672e2b61005ba32a94ca9597422
a15df3ba542c05c45a888f3e84f3b1b6944b7a45fe8f620cd98826379ccd6992ccacf891644362268fdb771cc779b0c66f3f
1ccd0c97e07330995ecd4ea658b96aaa2dd60d4f5739ca4efe7ea655dd447b518ca899114e9cb20f0c57df0ca3044af6e1d4
70301855c8b7c3f412b5edb7581ca3ae8f1a6a6e062651792bfd03b35845becdc0c3466919f928d69531afc2e36e202be9f8
8ec74d7f95d5952d2b663d6ec4ad48526152d1b262eeca9695417961989dca81817538cdb7fb61f7d51e5ae270951ce0aa46
2b83245f842fa61401b95c34635432d2187fecc923c026ad501a258ca1c6fd38a90aba68496d1b7a3fa351ecd7c14d726654
d64883e44f95d12e4eaaefefac901a9d09650eb35680996119051d6e8fc868d9e08c5415c0489a0d83364ea1f8ac15a0863b
69e585a4294f062a346bb0a7bf0f41c5ce6b252a2bf9e1b810f9301fc3c7a58dff0ebcd7c5f071b9b192642a5a56043cae2e
1c9592722af8b1fb1c9393e4dcb2672be6f8f9c9b0ddc48f15f7ce66a94a396c87ba8291323e042afefcb40edbc1c8621562
f7e5fa529a17df96a81dcf7d0cb6d31233c5943c0b04278ad6b15b3b05c163a270088f89b28ce14d443905cde3799dc3794c
9607f0bc893027217b5c1a0eed716996b1bda934a7a07c5c1a0ef371699671bea934a7207e450566901f976619f39b4a7332
faa768c64dc9b89ee169186037a2ec758c7d8c1d03054e473fd66860b4f5287dae0b32ab050142311560206b0b386404d308
afbdb6aec5d4a8bcd291c6580c1a34d3aa411c1fe2101daf1fa9c61cf6d71c40462b541a11b342a560057a191b43a952f218
8453b076a292a799371590c363543ac0fa1ed2087d0a25d3a94614301d38cb9dc4c959a36c9642a3ccfb24850e0087e65d05
872ef51a4e59a1172f36bc27e5e4ac172733bd664086c785d0a2f8232cc4ddc2f085f2a0a968b1cde4a0414e64ec308d1d84
5a4ac9f6b9fc778a9da53cc0b012ae4bc0bf0cfe2b75413a0b69c0f3992e04fd38eb5c17f05c9919a0eed4a859d225e3595c
150645f268199d58460b5bcd34415dd02851e94445e6a2e5f87386f8929036fe29c2a40ec8b1eacc4f191ba58f44e46646f5
b9191c9c0e26f8abd9091e2bd78dffe5c21195abf07c00959bd8b980e666da83176dbbb9b9dee2c7d5a79bbb4b6a149eb5ef
2fffb37dc75b85258803ae13b6b4862334217815fda823238dcab0278f80384c44118de89133d0c949f5093f20eb78cce3e8
df56431c33ef81e945aed65eea8cdc3252fd428f8201647404b2a8b500839a7b0f8d8e0365d824861c67d2c285a228a493d4
39438095fbcfcc2fe448239d0fa170f79e54df3f458c29e809acc1d8b4727f53c1570c9451f8112d34d2abf0950ab2711c14
70cc105ec11e2a28c8d70766c3323a538323e6518359431c87b61c871255b0dbe340f159a9cf06ce4132992b204c45c38a97
16d19913e1b15f07b253852415012b9254e43ece8f15b9676dbd0436d9807cb7dfb0809c0c879c5e9e5f5d4dae87b13ae2e1
476d7c7781c1a03f4c53f26c6540cf83d2a4911b65dfbc78249680aa6216a6f459a0898912e1b314e90665214a26af10252a
01cf28950c3d701cd0c4ab2b12b40c1a7f4a59187d85304a5a6464f4c036d013c7014d5c1a8d1ed48285994893e96ba441d1
d512adb4d4f4c471401397c6a178e0b0f58d049326d3d7486351dda20ece797ae238a0a9e8830c06379c3b3b9126d3d74813
68480e462a454f3c026832ce09199331ba630841e866d622d0f4e9e6b7ed6d7bf3befd807ef2eebebdc78f29dc346adab33d
096e624d5f50304e775b18464bb0148e1266163ab046cad45da6b0cad179b1dd7b6f45b760fa8366e167b485c184d2c84e62
519241da89f1870e9c64ba18fe37b8a2f3c45e2a8bb81d05a33f7d32fb5122778daba13822db691bd1171baa41a1be98702f
d8ce05eb5d37437bd3e5b4dec5d3d5c62351279ad2acd134107a4396a0def770b193d4c0261fac9694abf38aafeab7225b67
bff16ccddc16a17138dd63dd40b1e8b5717fd31700d84255bd266716949ed2188bad2f3ca5ea9ea2cb3b7a61f5acabd49fcc
550599b9cb2411a2ec8bdfd774977ec85dbaeeaeb915a58b304862da717da4ceac2e5d640cf56e5448fa1abe5f7d8abb16fc
e4458c6aa81ff0130101da752f42a4713915a260617d6979f2144c458b0e022bd8a4298fcc79aa73d3827f0691d8893613ff
288fc265bdec2ec8aac1b27bdb9aa9af9c047398c6b61ad2e0419e32cdc46f3262a54619187d6f26be72500abae9ae1bcd2b
be927f267134f8671a477bf7a00cbb90c81427bae7e3b28b0e43c8565d246716946e491a6dab4b7dbd99595dba454541ef52
fb0899593deb22fb6771d134d5fd711e720f79c8d53d34b762e21585e6b49b0d74dfed4f97cfbac59d5e886a802f9ff2a2d0
7e182018b098c9ab00df19e0720e9104a7033cf36a16cf24687815185a2e1c01df0acf0700df899d1f027cbf6ddbbb0fe7b7
9fdacdeecbddfdf6f6aefdcb01025c42bfa6807e61c801f8b5d6093b585e12ec471a20f23989ce3eea9f4c4e0f1487a65c0f
4b1424f51e4d1d276a416fea1b46a274d2b3ca242aaadd1ba4cc3fe153e23f8ad013365cd09e8448b5fd774d9c1948449053
de9cd8c9b06918a997745790468dd806bdde07d6da10d6fd628f75d341edb0ee83579bd6fabd429186cc0e20b942ae318764
fa8aaa9c234f4a75f13495e7912959d824249c4c26a2a76b0b9e5a7069950ab62326e3d1ed11113900bd139987beb0823c6e
3a32c23cd057384134c3664124d84417b451cefef94c46af984ccb79222ad0842557ec9f04da32ba5dbe69b8ace8353542ca
b75cab94b5ca1618899bce3036200deac95a99e80a61c1154702055afb89045a387a97e60b69f58c5a7ade027ac6589927b3
6adebff0409695792b6be5e72ce0b305ba53595cf3f6c3c1b584b5d37c62d8951382ef428f630eb7e202bb1517e88a3f7d9b
7784119f7f6f86157f1d56a05596d224e9f58012ca7cb7f6b052b370401f1e67c221938b70e0e499c2cb0d1f3c8f337748de
15e4c4c97a8efca1c76c3e2ed83538baf3f507d8d50ad8caabee6ae0f0811870608d939d82770f3f51fd40aff41d5d0198f1
5d2617bee3e419df7106919f89d119deb133915dc7c92b7ce4ed1f74f6bfae8fcc928f14f4a667baaf85d0358fee975da31d
dd2e187e999dfcfa6e849a3ad67d84b2fbd8dcdedcdd9dbdbfbcbfdfbe7bd67eda7db938bb2cda8de9bdaca970c89e84e0ed
2554e41f37caa5b4158afddec979d1feb4bf1255ca4a8073c76d41e08bddcddbf35dbb3b7fbbdd5d5e5f94a256bee7d5494a
9759fa7718a3b0f485274e8251a5a7574c71b5c89aeeb20f0f68760a3d7bdfd99fb0fbe1bcc0eda9ffee7d5bbef4d1260e2f
78daf215dbc107fb5739ca6a17d885e2b67624ef86d74d2e6974402a7ff061ffd60a838694210bf5a5f6c1fbfd2baa24230f
a5fa075535aa8a5fd736bfc8afb692ebde0f0e1f3cab695e3589ae05a3ab7dd00ed7ca7d8861e68decf80e2db4fb7768813a
4b9bc6f770fda958736b9b46b666f6fb7dd34b32e8835579ac75429f549e618341c61d73ac8d3223dbd5d9a37fcfc02392b2
99eec6c7992b8286fa39f24816dd24c8c949280e92beab58908a112cdf4329ee76c891f582f8f619927dbbd9ee767773524f
6a2586c5e18def5e5c4b379a3809e206747b25e940dc52d41047b60ba2aaf4cccfc9faaaf91f9729c82d
>}

\EmbeddedPDFObject{floor-1}{%
<< /Font \EmbeddedPDFRef{floor-2} 0 R /XObject \EmbeddedPDFRef{floor-3} 0 R /ExtGState
\EmbeddedPDFRef{floor-4} 0 R /Pattern \EmbeddedPDFRef{floor-5} 0 R /Shading \EmbeddedPDFRef{floor-6}
0 R /ProcSet [ /PDF /Text /ImageB /ImageC /ImageI ] >>
}

\EmbeddedPDFObject{floor-2}{%
<< /F2 \EmbeddedPDFRef{floor-7} 0 R /F3 \EmbeddedPDFRef{floor-8} 0 R /F1 \EmbeddedPDFRef{floor-9} 0
R >>
}

\EmbeddedPDFObject{floor-3}{%
<< /M0 \EmbeddedPDFRef{floor-10} 0 R /M1 \EmbeddedPDFRef{floor-11} 0 R /M2 \EmbeddedPDFRef{floor-12}
0 R /M3 \EmbeddedPDFRef{floor-13} 0 R /M4 \EmbeddedPDFRef{floor-14} 0 R /M5
\EmbeddedPDFRef{floor-15} 0 R /F2-DejaVuSans-Oblique-eta \EmbeddedPDFRef{floor-16} 0 R
/F2-DejaVuSans-Oblique-uni03BC \EmbeddedPDFRef{floor-17} 0 R /F3-DejaVuSans-Gamma
\EmbeddedPDFRef{floor-18} 0 R >>
}

\EmbeddedPDFObject{floor-4}{%
<< /A1 << /Type /ExtGState /CA 0 /ca 1 >> /A2 << /Type /ExtGState /CA 1 /ca 1 >> /A3 << /Type
/ExtGState /CA 0.95 /ca 0.95 >> >>
}

\EmbeddedPDFObject{floor-5}{%
<<  >>
}

\EmbeddedPDFObject{floor-6}{%
<<  >>
}

\EmbeddedPDFObject{floor-7}{%
<< /Type /Font /BaseFont /GCWXDV+DejaVuSans-Oblique /FirstChar 0 /LastChar 255 /FontDescriptor
\EmbeddedPDFRef{floor-19} 0 R /Subtype /Type3 /Name /GCWXDV+DejaVuSans-Oblique /FontBBox [ -1016
-351 1660 1068 ] /FontMatrix [ 0.001 0 0 0.001 0 0 ] /CharProcs \EmbeddedPDFRef{floor-20} 0 R
/Encoding << /Type /Encoding /Differences [ 77 /M 114 /r ] >> /Widths \EmbeddedPDFRef{floor-21} 0 R
>>
}

\EmbeddedPDFObject{floor-8}{%
<< /Type /Font /BaseFont /BMQQDV+DejaVuSans /FirstChar 0 /LastChar 255 /FontDescriptor
\EmbeddedPDFRef{floor-22} 0 R /Subtype /Type3 /Name /BMQQDV+DejaVuSans /FontBBox [ -1021 -463 1794
1233 ] /FontMatrix [ 0.001 0 0 0.001 0 0 ] /CharProcs \EmbeddedPDFRef{floor-23} 0 R /Encoding <<
/Type /Encoding /Differences [ 46 /period 48 /zero 50 /two /three 61 /equal ] >> /Widths
\EmbeddedPDFRef{floor-24} 0 R >>
}

\EmbeddedPDFObject{floor-9}{%
<< /Type /Font /BaseFont /BXGEVL+DejaVuSerif /FirstChar 0 /LastChar 255 /FontDescriptor
\EmbeddedPDFRef{floor-25} 0 R /Subtype /Type3 /Name /BXGEVL+DejaVuSerif /FontBBox [ -770 -347 2106
1110 ] /FontMatrix [ 0.001 0 0 0.001 0 0 ] /CharProcs \EmbeddedPDFRef{floor-26} 0 R /Encoding <<
/Type /Encoding /Differences [ 32 /space 40 /parenleft /parenright 43 /plus /comma /hyphen /period
48 /zero /one /two /three /four /five /six 56 /eight /nine 65 /A /B /C 76 /L 97 /a /b /c /d /e /f /g
/h /i /j 108 /l /m /n /o /p 114 /r /s /t /u /v /w /x 122 /z ] >> /Widths \EmbeddedPDFRef{floor-27} 0
R >>
}

\EmbeddedPDFStream{floor-10}{%
/Type /XObject /Subtype /Form /BBox [ -6.5 -6.5 6.5 6.5 ] /Filter [ /ASCIIHexDecode /FlateDecode ]
}{%
789c6d90390e03210c457b9f820b7c648b00433b65ae4113459afbb713430011a7b1bcfde745dc9bd83de963203eba8bd887
920f8e3d669f73095983f09012c589e7c42969a63b95660da35da598a4066a7d0a9cc9af68693a6f715765b66e44b595f639
30225830cc706cbb616c8cfd10d84bf1e71fb05fc32f1b7df917d14937a3cb48fa
>}

\EmbeddedPDFStream{floor-11}{%
/Type /XObject /Subtype /Form /BBox [ -6.5 -6.5 6.5 6.5 ] /Filter [ /ASCIIHexDecode /FlateDecode ]
}{%
789c3350c8e23250f0e2d235d433550013b95c70660e980961e92298195c5c4e5c002efd0a5e
>}

\EmbeddedPDFStream{floor-12}{%
/Type /XObject /Subtype /Form /BBox [ -6.5 -6.5 6.5 6.5 ] /Filter [ /ASCIIHexDecode /FlateDecode ]
}{%
789c3350c8e23250f00262433d53855c2e5d10052672b89098195c5c4e5c00b48107ef
>}

\EmbeddedPDFStream{floor-13}{%
/Type /XObject /Subtype /Form /BBox [ -6.5 -6.5 6.5 6.5 ] /Filter [ /ASCIIHexDecode /FlateDecode ]
}{%
789c6d90390e03210c457b9f820b7c648b00433b65ae4113459afbb713430011a7b1bcfde745dc9bd83de963203eba8bd887
920f8e3d669f73095983f09012c589e7c42969a63b95660da35da598a4066a7d0a9cc9af68693a6f715765b66e44b595f639
30225830cc706cbb616c8cfd10d84bf1e71fb05fc32f1b7df917d14937a3cb48fa
>}

\EmbeddedPDFStream{floor-14}{%
/Type /XObject /Subtype /Form /BBox [ -6.5 -6.5 6.5 6.5 ] /Filter [ /ASCIIHexDecode /FlateDecode ]
}{%
789c3350c8e23250f0e2d235d433550013b95c70660e980961e92298195c5c4e5c002efd0a5e
>}

\EmbeddedPDFStream{floor-15}{%
/Type /XObject /Subtype /Form /BBox [ -6.5 -6.5 6.5 6.5 ] /Filter [ /ASCIIHexDecode /FlateDecode ]
}{%
789c3350c8e23250f00262433d53855c2e5d10052672b89098195c5c4e5c00b48107ef
>}

\EmbeddedPDFStream{floor-16}{%
/Type /XObject /Subtype /Form /BBox [ -1016 -351 1660 1068 ] /Filter [ /ASCIIHexDecode /FlateDecode
]
}{%
789c3d50c10d033108fb670a16a844309030cf49551fedfedf9a5cd54784851ddbc9885502a87c862f9787e992f7c0fe43df
5b60071511d55e213e97f85671875cc3939b2483de9aa0ee790d78dd1bdd54a418331c41c69cb7d5c5e88434e934689199be
449917f7980909ef02167fd441bc4494342cdaf0c4ac5335901d1d25115d864c320c389375f98046012a5823e811da7563ef
e31b0596a3eaf739d7788de7170fc43896
>}

\EmbeddedPDFStream{floor-17}{%
/Type /XObject /Subtype /Form /BBox [ -1016 -351 1660 1068 ] /Filter [ /ASCIIHexDecode /FlateDecode
]
}{%
789c3d52498ec44008bbe715fe402476aaded3d2a80f3dffbf8e49327d323185c190e354c369d2f83dd403198dcf61f61f69
2e4c76a282ee0dcd242a31a0cbf162c6a14d76b0ea46d63d195b640d2af35eb1266122688359a302563ef03a5c75028f8d6e
78b39b24f910466e0823b62342583b4229f64c9afb1ba9b06b4d24ec6d82a403d57850ae3a465237331ac4bd9e04474c765c
2c5447d7d0da68451a27dd48ce4220ed797fef0b22e711e95470cccc3dd5d9325a4317d68c155c07b5164e3ed2c4d98f9593
0b88b58934c9150cd23e577e31ce8ae2160a53107444734193ec3f50f3dab77075f0a27742b0c5b5da9e36466f2365bcde2d
6eb43491ee1a71edb8c435791e5e799466f902def075cc4cf7fff0fd713ec7fbf8f903c1076c35
>}

\EmbeddedPDFStream{floor-18}{%
/Type /XObject /Subtype /Form /BBox [ -1021 -463 1794 1233 ] /Filter [ /ASCIIHexDecode /FlateDecode
]
}{%
789ce3b2b4503050c8e50252e646960a395ca6a646482c33133320cbd0d21c896500a4c1ba72b832b8d20099ae0cbf
>}

\EmbeddedPDFObject{floor-19}{%
<< /Type /FontDescriptor /FontName /GCWXDV+DejaVuSans-Oblique /Flags 96 /FontBBox [ -1016 -351 1660
1068 ] /Ascent 929 /Descent -236 /CapHeight 0 /XHeight 0 /ItalicAngle 0 /StemV 0 /MaxWidth 1350 >>
}

\EmbeddedPDFObject{floor-20}{%
<< /M \EmbeddedPDFRef{floor-28} 0 R /r \EmbeddedPDFRef{floor-29} 0 R >>
}

\EmbeddedPDFObject{floor-21}{%
[ 600 600 600 600 600 600 600 600 600 600 600 600 600 600 600 600 600 600 600 600 600 600 600 600
600 600 600 600 600 600 600 600 318 401 460 838 636 950 780 275 390 390 500 838 318 361 318 337 636
636 636 636 636 636 636 636 636 636 337 337 838 838 838 531 1000 684 686 698 770 632 575 775 752 295
295 656 557 863 748 787 603 787 695 635 611 732 684 989 685 611 685 390 337 390 838 500 500 613 635
550 635 615 352 635 634 278 278 579 278 974 634 612 635 635 411 521 392 634 592 818 592 592 525 636
337 636 838 600 636 600 318 352 518 1000 500 500 500 1350 635 400 1070 600 685 600 600 318 318 518
518 590 500 1000 500 1000 521 400 1028 600 525 611 318 401 636 636 636 636 337 500 500 1000 471 617
838 361 1000 500 500 838 401 401 500 636 636 318 500 401 471 617 969 969 969 531 684 684 684 684 684
684 974 698 632 632 632 632 295 295 295 295 775 748 787 787 787 787 787 838 787 732 732 732 732 611
608 630 613 613 613 613 613 613 995 550 615 615 615 615 278 278 278 278 612 634 612 612 612 612 612
838 612 634 634 634 634 592 635 592 ]
}

\EmbeddedPDFObject{floor-22}{%
<< /Type /FontDescriptor /FontName /BMQQDV+DejaVuSans /Flags 32 /FontBBox [ -1021 -463 1794 1233 ]
/Ascent 929 /Descent -236 /CapHeight 0 /XHeight 0 /ItalicAngle 0 /StemV 0 /MaxWidth 1342 >>
}

\EmbeddedPDFObject{floor-23}{%
<< /equal \EmbeddedPDFRef{floor-30} 0 R /period \EmbeddedPDFRef{floor-31} 0 R /three
\EmbeddedPDFRef{floor-32} 0 R /two \EmbeddedPDFRef{floor-33} 0 R /zero \EmbeddedPDFRef{floor-34} 0 R
>>
}

\EmbeddedPDFObject{floor-24}{%
[ 600 600 600 600 600 600 600 600 600 600 600 600 600 600 600 600 600 600 600 600 600 600 600 600
600 600 600 600 600 600 600 600 318 401 460 838 636 950 780 275 390 390 500 838 318 361 318 337 636
636 636 636 636 636 636 636 636 636 337 337 838 838 838 531 1000 684 686 698 770 632 575 775 752 295
295 656 557 863 748 787 603 787 695 635 611 732 684 989 685 611 685 390 337 390 838 500 500 613 635
550 635 615 352 635 634 278 278 579 278 974 634 612 635 635 411 521 392 634 592 818 592 592 525 636
337 636 838 600 636 600 318 352 518 1000 500 500 500 1342 635 400 1070 600 685 600 600 318 318 518
518 590 500 1000 500 1000 521 400 1023 600 525 611 318 401 636 636 636 636 337 500 500 1000 471 612
838 361 1000 500 500 838 401 401 500 636 636 318 500 401 471 612 969 969 969 531 684 684 684 684 684
684 974 698 632 632 632 632 295 295 295 295 775 748 787 787 787 787 787 838 787 732 732 732 732 611
605 630 613 613 613 613 613 613 982 550 615 615 615 615 278 278 278 278 612 634 612 612 612 612 612
838 612 634 634 634 634 592 635 592 ]
}

\EmbeddedPDFObject{floor-25}{%
<< /Type /FontDescriptor /FontName /BXGEVL+DejaVuSerif /Flags 32 /FontBBox [ -770 -347 2106 1110 ]
/Ascent 929 /Descent -236 /CapHeight 0 /XHeight 0 /ItalicAngle 0 /StemV 0 /MaxWidth 1342 >>
}

\EmbeddedPDFObject{floor-26}{%
<< /A \EmbeddedPDFRef{floor-35} 0 R /B \EmbeddedPDFRef{floor-36} 0 R /C \EmbeddedPDFRef{floor-37} 0
R /L \EmbeddedPDFRef{floor-38} 0 R /a \EmbeddedPDFRef{floor-39} 0 R /b \EmbeddedPDFRef{floor-40} 0 R
/c \EmbeddedPDFRef{floor-41} 0 R /comma \EmbeddedPDFRef{floor-42} 0 R /d \EmbeddedPDFRef{floor-43} 0
R /e \EmbeddedPDFRef{floor-44} 0 R /eight \EmbeddedPDFRef{floor-45} 0 R /f \EmbeddedPDFRef{floor-46}
0 R /five \EmbeddedPDFRef{floor-47} 0 R /four \EmbeddedPDFRef{floor-48} 0 R /g
\EmbeddedPDFRef{floor-49} 0 R /h \EmbeddedPDFRef{floor-50} 0 R /hyphen \EmbeddedPDFRef{floor-51} 0 R
/i \EmbeddedPDFRef{floor-52} 0 R /j \EmbeddedPDFRef{floor-53} 0 R /l \EmbeddedPDFRef{floor-54} 0 R
/m \EmbeddedPDFRef{floor-55} 0 R /n \EmbeddedPDFRef{floor-56} 0 R /nine \EmbeddedPDFRef{floor-57} 0
R /o \EmbeddedPDFRef{floor-58} 0 R /one \EmbeddedPDFRef{floor-59} 0 R /p \EmbeddedPDFRef{floor-60} 0
R /parenleft \EmbeddedPDFRef{floor-61} 0 R /parenright \EmbeddedPDFRef{floor-62} 0 R /period
\EmbeddedPDFRef{floor-63} 0 R /plus \EmbeddedPDFRef{floor-64} 0 R /r \EmbeddedPDFRef{floor-65} 0 R
/s \EmbeddedPDFRef{floor-66} 0 R /six \EmbeddedPDFRef{floor-67} 0 R /space \EmbeddedPDFRef{floor-68}
0 R /t \EmbeddedPDFRef{floor-69} 0 R /three \EmbeddedPDFRef{floor-70} 0 R /two
\EmbeddedPDFRef{floor-71} 0 R /u \EmbeddedPDFRef{floor-72} 0 R /v \EmbeddedPDFRef{floor-73} 0 R /w
\EmbeddedPDFRef{floor-74} 0 R /x \EmbeddedPDFRef{floor-75} 0 R /z \EmbeddedPDFRef{floor-76} 0 R
/zero \EmbeddedPDFRef{floor-77} 0 R >>
}

\EmbeddedPDFObject{floor-27}{%
[ 600 600 600 600 600 600 600 600 600 600 600 600 600 600 600 600 600 600 600 600 600 600 600 600
600 600 600 600 600 600 600 600 318 402 460 838 636 950 890 275 390 390 500 838 318 338 318 337 636
636 636 636 636 636 636 636 636 636 337 337 838 838 838 536 1000 722 735 765 802 730 694 799 872 395
401 747 664 1024 875 820 673 820 753 685 667 843 722 1028 712 660 695 390 337 390 838 500 500 596
640 560 640 592 370 640 644 320 310 606 320 948 644 602 640 640 478 513 402 644 565 856 564 565 527
636 337 636 838 600 636 600 318 370 518 1000 500 500 500 1342 685 400 1137 600 695 600 600 318 318
511 511 590 500 1000 500 1000 513 400 989 600 527 660 318 402 636 636 636 636 337 500 500 1000 475
612 838 338 1000 500 500 838 401 401 500 650 636 318 500 401 470 612 969 969 969 536 722 722 722 722
722 722 1001 765 730 730 730 730 395 395 395 395 807 875 820 820 820 820 820 838 820 843 843 843 843
660 676 668 596 596 596 596 596 596 940 560 592 592 592 592 320 320 320 320 602 644 602 602 602 602
602 838 602 644 644 644 644 565 640 565 ]
}

\EmbeddedPDFStream{floor-28}{%
/Filter [ /ASCIIHexDecode /FlateDecode ]
}{%
789c3d8e3912800008037b5ec113e490e33f8e6381ff6f453c9ab093498030c105d95b420c9d133702b21c3c41c8870a7461
6489260b7dbdafd25e6aaf285833663a339a4caf33a4b727ad0fb13a9a5213b14c7e3ea8ff70c101fb052a8a1e30
>}

\EmbeddedPDFStream{floor-29}{%
/Filter [ /ASCIIHexDecode /FlateDecode ]
}{%
789c3d904b0e03310843f739858f1030e4739ea9aa2ea6f7dfd664a46e6207e4e44198a183a9230691a3e3652d62e81af8b6
60b98570e97484853471b5e8bacd096e45e706e54baf46dfc7f952450f7a9af2a18e5b826bc1b683dcb0a5eec99817c2ddf8
88852363ca39ff4ee1187e6a02da848f813421f48e647d406a067d4721d72cd1c751e12a7d2a02c914204b5775a29c50b3b2
f9ec41fefaafe16e9ff6fe0193053826
>}

\EmbeddedPDFStream{floor-30}{%
/Filter [ /ASCIIHexDecode /FlateDecode ]
}{%
789cb330b6503050303430533034375230373652303135514831e402098198b95c30c11c30cb18a82c072c8b60416433c06c
235353a81e100ba2c710ae12c182c86670a5010051be1916
>}

\EmbeddedPDFStream{floor-31}{%
/Filter [ /ASCIIHexDecode /FlateDecode ]
}{%
789c3336b45030503034300792468640969189428a21174800c4cce58209e6805906401aa23807ae26872b832b0d00e1b40d
98
>}

\EmbeddedPDFStream{floor-32}{%
/Filter [ /ASCIIHexDecode /FlateDecode ]
}{%
789c3552396e04310cebfd0a7d20806edbefd9204891fcbf0da9d91403717452943b5a5476cb878554b5ec74f9b495da1237
e4776585c409c9ab12ed528648a8bc5621d5ef9932afb7356704c852a5c2e4a4e471c9823fc3a5a70bc7f9b6b1afe5c81d8f
1e581523a144c032c5c4e81a96ceeeb0665b7e9669c08f78849c2bd62dc75805521bdd38ed8af7a1792db20688bd99968a0a
1d5689167e41f1a0478d7df618545820d1ed1c49708c895091807681da480c397b2c76511ff433283113e8da1b45dc418125
929839c5ba74f4dff8d0b1a0fed88c37bf3a2ae95b5a5963d23e1bed94c62c32e97a58f474f3d6c7830b34efa2fce7d50c12
34658d92c607adda3986a299515678b743930dcd31c721f30e9ec319094afa3c14bf8f25131d94c8df0e0da17f5f1ea38c9c
36de02e6a23f6d15070ee2b3c35eb981154a176b92f701a94ce881e9ffaff1b5bed7d71f1ded80c6
>}

\EmbeddedPDFStream{floor-33}{%
/Filter [ /ASCIIHexDecode /FlateDecode ]
}{%
789c2d514972034108bbcf2bf48466a7df63972b87e4ffd708ca07060d8b40745ae2a08c9f202c57bce591ab68c3df333182
df05877e2bd76d58e440542049635035a176f17af41ed8719809cc93d6b00c66ac0ffc8c29dce8c5e1311967b7d765d410ec
5daf3d9941c9ac5cc42513174cd56123670e3f99322eb4733d3720e746589761906e561a33128e3485482185aa9ba4ce0441
128cae468d367214fba52f8f7486558da831ec732d6d5fcf4d386bd0ec58acf2e4941e5d712e27172fac1cf4f53603175169
acf24048c0fd6e0ff573b293df784be7ce16d3e37bcdc9f2c25460c59bd76e4036d5e2dfbc88e3fb8cafe7e7f9fc035f605b
88
>}

\EmbeddedPDFStream{floor-34}{%
/Filter [ /ASCIIHexDecode /FlateDecode ]
}{%
789c3551390e03210cecf715fe40248c2f784fa228cdfebfcd8cd1561ec15c86b4942199f25293a821e5533e7a992e9cbadc
d7cc6a34e796b42d7380b8b6bc2fad92c089668883c16998b8019a3c0143d7d1a85143171ed0d4a55328b0dc043e8ab76e0f
dd63b58197b621e789206228192c41cda94513166dcf27a435bf5e8b1bde97edddc853a5c6961881060156381cd7e8d7f095
3d4f26d1741532d4ab35cb18091345223cf98ea6d513eb5a34d23212d4c611ac6a03bc3bfd304e00001371cb02649f4a94b3
e4c46fb1f4b30657fafe019f475496
>}

\EmbeddedPDFStream{floor-35}{%
/Filter [ /ASCIIHexDecode /FlateDecode ]
}{%
789c4d8e310ec43008047b5ec1072201c680ff738a5238ff6fcf984b740d3b6261c14590f0e8ab78137419f86110221453bc
412d364d684dd198173dee840bf6e69dd265357a94368e9d3541d77091199599875ea09c317fb562f497a3e1289cc4410ff1
2893c71fe4fe7e26bf3abf3175292e
>}

\EmbeddedPDFStream{floor-36}{%
/Filter [ /ASCIIHexDecode /FlateDecode ]
}{%
789c3d514bb2c33008dbe7141cc1989f7d9ebe79d3457aff6d253ccd2668821192280b1912fc64b9d4dcf2a7d7f49298f2b9
6c1bc17d7928816f977420953de475852f510d899ca271ea54658728f1072f6c448f98b1430e8b14525a6ee10ed657af25ba
7f02eeebddd0310c31a58da0667a234f1307a9afd57458ab1059d48a95c34f759237a21dbcc8993d939e1404964cfc016b16
24609ef50822ba1f1554d4697d585aa2c2e103ce6bb41ec048a1581f84bd44e126a550b8b06d39b6a5623ea0c1b075ecae47
7b238498f44d771b7e3de8aae057c1863b3afcfa0a04bafa3470822305e3aed19ecdd8c9e049b20f3ed34feda311a96133b2
0451f26e9d6a6cd1eac80648a7f4ed1903f3f8ff027f296dcc
>}

\EmbeddedPDFStream{floor-37}{%
/Filter [ /ASCIIHexDecode /FlateDecode ]
}{%
789c3d514b6e443108dbe714be40a540f824e77955d5c5f4fedbda19b52b1c02c686aec444163e6ca167a2c3f16943d0cec2
cfa8bd609ea868f4429ec63a7846462110db6f6bb05ef1192bfba265fc49783bcc98f7657c903491135601b63ec394b30d9b
4a4e257d163f58e0e992b6c8adc00cc7726a37f6668a5d19a8b5456f87fafb16cb43c84f6842e41b0527f72a644cb46b44ee
405b9141de276ad366cb9c5052f66bd49a7f882a936e926b296b248d565c1e37f6258591e3ccbb0c452e83ee849c5aab68cd
19a992bee5208d9ef7ed537c3b147272abc2ba6e8ff95da1581222a5674db9d6238f404e5ec790bc4e6b427670abbad88105
15dafba4cfff715fe37b7cfd02ed3d5ff8
>}

\EmbeddedPDFStream{floor-38}{%
/Filter [ /ASCIIHexDecode /FlateDecode ]
}{%
789c4d8d4b1280300843f79c8223d0caa7dec7715cb4f7dfda8863dd90c78404776561b3395c85a3ee7c147af601b1ca9d8a
b6051e31695a1f20d4697be34969568d3f09ee9b2c280dad78bc48b21472d1790397771edf
>}

\EmbeddedPDFStream{floor-39}{%
/Filter [ /ASCIIHexDecode /FlateDecode ]
}{%
789c3d52496ec4300cbbfb15fcc0005abdbc2745d1c3ccffafa594a6878084259b14953c138214bcd4917323ddf1a5c3cf86
4ec7a7992dc77bd8b687f96a66123076695522700d9d02b380f2493d7963ceaa14337e537194571c4b786e92084a78c26f25
c2355ca58887200e9c22cbea784ff0d9b6677fc8da357e466cb653fad32c8d3e7ba08708b1fa1fa42b925597e049ab4146a9
5211c7ab46b4cec568a49043ec3b29750e0e15a9cbd79813d32a456520ff1393981ccc03eb12a7db35b1c6448d67a2bc9e2d
5378b5ad62b7416752378b12e2e4e1442a457616b4158bae291a9befb0bf90995a36d3b3d831a1fc62b676b2979bd2606f6d
c3e96557e0479abc8bc4a92dab2987e1ed242a7113ad6230ba4f66657efa87b1b91acbd37de27c21952e9c3aebb012cc21c2
d16b3a86675db5baef5f801678c9
>}

\EmbeddedPDFStream{floor-40}{%
/Filter [ /ASCIIHexDecode /FlateDecode ]
}{%
789c4d51496ec3300cbcfb15fc40007113a5f7b8087a70ff7fcd0c9db4056c702c52b3d033860cb12d0f75c93da4e6902f3d
5453d2e4a7418d25d781a13f80298091ff507837cd240a9411123bc4ca2475c8896b1b9429ee50c2eb75d7f308af46b13031
80a11059e8e404e2099c7968579b6423d25ac20955eb3bd34806127c939399a8c27a1ed425a28f47b6b14d5b306a78e87b09
832ce531c0b8c3be0bd6717df6721ddf3d611ed81191968a69411a0a70ba9b64627ab5621802240b9ca009105806b2446cd9
0c1bb0aabd06d0add5f50e4be4cec522d09a7d25acd3822320d19455b746d57bdd44f41073b4a9b0eec0a6f3d7c0b6477535
289cbf9118eff90230aa64da
>}

\EmbeddedPDFStream{floor-41}{%
/Filter [ /ASCIIHexDecode /FlateDecode ]
}{%
789c3d51396e04310cebfd0a7e20800ecac77b2658a4d8fcbf0de5206946b4c4c3f2d43418caf0e18972a232f1e9a3a1d7c4
f72873b819b8166a8374a4e119b482a05a2d4db1ba3e23a22ef2a9b9a47e305dfda596ce8af27632c46c1b81644053dabe6c
caf2192d2f75daaeefd4015d9f91fadece9ca838c873543b82298ea239b587057816b8574f84b288f720eb0fe5be5a861429
762f59374117e4a22e46e9c56f2f552da779a3f0a994093f9ac4ef7d9570b45c899fbc35eaf444c8b73a62b8b8ad39bd7a9b
54a03d25e81015a5d3ee79b91e5b9ec4ec367d412ecc7ec7d228ee0f7afe7fd57b7c8dd70f9fd9560c
>}

\EmbeddedPDFStream{floor-42}{%
/Filter [ /ASCIIHexDecode /FlateDecode ]
}{%
789c358e4b0ec3400843f79cc21718690c613ee749157591de7f5bd32a1bb08c79105ce88881c648b81324f1a295b5073e36
2466606db463823d149d388daa0cd013c7feb51d654b14e2b60726a523297a4f6867112dbca2398556c85dcc1ae6ff8df3b9
7edbdbae2f93ab1e3b
>}

\EmbeddedPDFStream{floor-43}{%
/Filter [ /ASCIIHexDecode /FlateDecode ]
}{%
789c4d51396ec4300cecfd0a7e60019e3adee3204891fcbfdd196a37496170445173d02355544ae56121c34ce650f9b0abbc
a45c7e2ef600be1b286a46fdd66504362597c42ef194184bb6dc5780f851123a9bdac7a9f7e51a8d0ca2e662a6321cfd8923
ceeccec5e270725f0091b0a592a0e274d6449fcf0b1dd2551c015628eb4174c0187454462e5acc9dd28ec98808190b374453
17d20478ffa1c1a864f943bd8daf7ee2915851a355af9d4482dc210fa1d81b987663426a689bca09dbb43279e3b94fc760ac
96d8c68473233686c41ad81458c3bb9ea510d9c22c260c51f96637194830d19c7e44fcf5330068a256bbdac6d4b069f868db
a6c93bd28d789f4f05376445
>}

\EmbeddedPDFStream{floor-44}{%
/Filter [ /ASCIIHexDecode /FlateDecode ]
}{%
789c3d514b6e053108dbe7145ca052cc2fc979a6aaba9877ffed33cc6b573816d886c451991253be6012ae1266f28d515049
bf06221add0ff2fc20ac14e416a8098ecb5e720d558a81fd47ccc5802ad730ced77babb88bcf2d19a49d5e3bc43901a478b2
46c9548842f7081cc19ce2cc49c2c3454f8d8209c4d23bb84ded5a01b2116880923d42d56b2c527c3307d6aaa239cba872a9
2ced50c0a6c3268fe435c8a83e17d1935db9caf2466ed11d4e397fb4e8e560bf3fa27f27bcc62f37cd0ef922e21d927e8af6
b4c3793bad4cd5acc3715d9ea53cab72296edd0cd354070e7dad73ae3ae7e2bab355eb63cae7fa77bce9fef30673f95a51
>}

\EmbeddedPDFStream{floor-45}{%
/Filter [ /ASCIIHexDecode /FlateDecode ]
}{%
789c3552418e243108bbd72bf8c04a814048ded3abd15ee6ffd7b59dea43095702b681acb96cd86afbe3d36a1deb0cfbeb4f
ae657e8efd0a45b9654d8b83186d33867d9e1c6e338fcd95f7f3adf879a25d28662923c6796b7c5f16ef215646eae08628db
98e163a9a69b6420c96de27c45a821c5349ac02d5d299b3659ae0640f76de4f3fc7b329715ee7f2f3ab805c39a61e961ab58
3ff7b1b52155dbd6b97a8c6ceaa2402d3322f2ad0978208b9f215645e8a029a201f374d9533599744396c41f59a94e1d4638
a8f7044e98417c6be8962c720fd66f3fec6d9e898aadde30a13d2c0f9700fd0057d067698577cf1c3ee39d8dd0c409260643
3520442fc97a9b67eb7dd0232377725f8c77e1de3941a6439f436a7131881c806a3b24ee24921dcf9441df2dc391b7058e0d
e8607185213905d04e72094e671d86de0a0f84410300588ee7848f3b71e82d9c91caada340de7adbeca165965244246646ee
7c6beadd6c658995f1ca0841b7a6cb468d7a8d651f59cd7459ffee83bbf9f90ff09ba266
>}

\EmbeddedPDFStream{floor-46}{%
/Filter [ /ASCIIHexDecode /FlateDecode ]
}{%
789c4d904b7203310844f73a0547e00f731ea7525e38f7dfa691ca1e6fa6df805a82b66262b2c4c78da992e947d6605ad1df
b2b64d2f9052a6905550f678f07f193d9679537190e945254ac6b6f5b1344f453d71026af0b44f471ab72529ba69b6353a77
0724d7bc18f5459e33c5746f0a9da6d60d3c9a6fd96551bde178d1fac079611fba09331e4a5712ec9f1d24b0940466940be4
81bd6447667ca24316f0ec8a238368a414d0d9d81a9e40cc8cb4dc4fdc361bbfd37eade7fafd0739ac4bcf
>}

\EmbeddedPDFStream{floor-47}{%
/Filter [ /ASCIIHexDecode /FlateDecode ]
}{%
789c4d51496ec4300cbbfb15fa4000edb2df3345d143faff6b290533e8212063531229a72531eda04b8c220e951efa92156c
437f87a51fba97e4f9c7dcb9194a3d949481c9a4ba8149afa5c1e4859b4a6090b10fbe966d1de68e9330f203251fdc04dc58
c61831ce4135e91b3081b615e729893eef1e9bacce04d09383989efa9cc0cf8552d45cad97284a121592ddb1ad0d01a5aad3
20d39b1989e12f361d87d8a9b8fb0abc26a963a8cf3c00227935b19d6dcbc5bb77bb5312c1ae0233b3069f3cc3d0b81586cf
055e2a673bd0618a61846bcd8cc6cef430854b470245049749d51a462674b04ed3fbcbde83480dbb87f58bde9fb7bdd7cffa
fe0331f55ffc
>}

\EmbeddedPDFStream{floor-48}{%
/Filter [ /ASCIIHexDecode /FlateDecode ]
}{%
789c458ecb0d80400844ef5441092c3fb51f633c60ff5701cdee055ef8cc8c8b23a18c2cb63b6eca780e103d9075c3a7c9c5
306010f52ce636e00673cdd70758387b7437fe8f268ca314d26502eb5e60d496019a7a8b3ef50ab4e87badeda236a80813a8
535d2f57f72994
>}

\EmbeddedPDFStream{floor-49}{%
/Filter [ /ASCIIHexDecode /FlateDecode ]
}{%
789c3d524b6e05210cdbcf29728191c88fc0795e5575f17aff6ded30af8b110682edc43363c8901c729ba94c554977f9d22b
2d2566c96f239cbf1bdcd3511d72ab0ddc3bc01cf2bac20c146388cf385cb6d601afcbf27386ef365da29b8a6ab8d464dd98
a206a01b0034b7aee0256125c435ea8392aa638bce0900bb4a298d4d29e5db996201292d870f3be075b9c7815ea480b06fbe
84f0d301b6e1e832b557a57b82c5fe434b826fc88e469750d131c13bc547d18cd83c2bac0c6fa4982e3a55389b6cb8b0c59e
a7b5b8584f10c0e119b7012a564756cf87f994908ed950802b94c74174c06ce828955cb4183ba41d93112d84afa799d48d6e
3aeb7fc4a8dfffa1bfaf9faeb495f801881ce30f8316587d6f607af362fd680751f048dde28d218d3e51b8c81338278c7e90
9a2fa49c6075ebf54c808851b2a263c59bdd6420414573da11b167f20034d1516ed99d176c2abe0eb04e90e6f134cf96d8de
f71f71218d5c
>}

\EmbeddedPDFStream{floor-50}{%
/Filter [ /ASCIIHexDecode /FlateDecode ]
}{%
789c4d90418ec4300804ef7e054fc0d060fb3d59adf690fdff751a349acc21e94a10a64c02a2e2c957ce94952a3f7360f2fb
bf224cee31cd1e58ba49ec78804df7b0695f045b45b60539c542057b8bad94d0946bb8d691107797e0e3793aaf01cb26e496
a0124eb017ac04a723e8c49950ed74d7aa905ab0eef0016dbdfdc92e00f94059bc2939c569cb4ba1e6b8950b5d1121be0e7d
9638b4f31a76769365b693718fb0591576fbe11f6ec1b13a6d9faaccf71aede80365d7fbbec7dff87d01cc004b50
>}

\EmbeddedPDFStream{floor-51}{%
/Filter [ /ASCIIHexDecode /FlateDecode ]
}{%
789c3336b65030503031513032365030b23451303630534831e43281b072b9606239601648550e970912032297c195060006
c90e35
>}

\EmbeddedPDFStream{floor-52}{%
/Filter [ /ASCIIHexDecode /FlateDecode ]
}{%
789c45504112c42008bbfb0a9e405051dfd39d9d1edaff5f3750677bd064204462351595eabc6c0d19641f14329f2a7792d5
056a32740a506560c951608d6215d46708bd26b2e3bb4283542cec19db2ec613ae89544587cc9b84c23b38436c08b709f170
712aace73b81ecf0e5ac70935470b73db35d62799aee3047398bc1a41b734556922b8912f30712b20cb3973426cad69f74e6
b91eb79785fc2cdf1fb5233fa0
>}

\EmbeddedPDFStream{floor-53}{%
/Filter [ /ASCIIHexDecode /FlateDecode ]
}{%
789c45514b6e443108dbe7145c00099390cf79a6aa6631bdffb626c96b575831d886549898e80a517788c365d42e5f283093
3e4d7e0e62072c64d814a0cbc09257810fb69ba0d91e43f45dc98cfb32ebe958edcef8557160abee4a1f32897a93ece801ce
b036a4da6cd25365b0c363fb642543e7fdc224bb83d9eecca392e9a9faecf32a6fe676697d70b73a36f82408e6fb943cc13f
d2b007c12a237582463f2a2ac68e504966e6b9ce116b3d35331ca4dc4c1d537473a8e4345a8e0d51aa28d63abf809947da70
3aadb5ad3f94de9cd54a21f0b8ea1922d249d30a3d6465ac2a354e7d957ed11cd9c0152cf3b4a480e49c6fbc872e3f356cff
acdfc3bccbf72f0f666585
>}

\EmbeddedPDFStream{floor-54}{%
/Filter [ /ASCIIHexDecode /FlateDecode ]
}{%
789c33363250305030b20413060ae666060a29865c4606a60aa6460ab95c2031202307cc3000d3300a2c6c68688a60981b58
40a4100c33b062a0690816587906571a007ce51753
>}

\EmbeddedPDFStream{floor-55}{%
/Filter [ /ASCIIHexDecode /FlateDecode ]
}{%
789c4592cb75c5200c44f754a11284be508f7372b248fadf6624fcec059e01197425d8b68849039f2d935c95bee6f0b9c830
fe86ab9379907b922d23afc146d78889df45294c7a5be46abd46ea59c954f219b498b15710591264c8b80c39f7adca158173
a1dfd1141fc3d04879b403e9fc1a15fe38c4132c3699b2f21c16cccc9d62a18e48d0ee56f0f3710e6ef3792a44ce0b3507e8
0a036ad92a6b5704ae53075af09822505f8f76c0345ea3e8d3ed121c52dd9533b47aa93bbacb9ab39954fda6947d9c8435a5
98de9422d29482538ab2f45096ebd4b2f9354567f323bd3c455e63b18a3f5ee373d7de3aed712659ae2a88a2410fd7224155
ce5195709d6d7dabf50634f47e15c6c79dc6402bc21dd9f53e6a563781c6deafef1a3fe3fb1f968c7eb8
>}

\EmbeddedPDFStream{floor-56}{%
/Filter [ /ASCIIHexDecode /FlateDecode ]
}{%
789c4d904112c3300803ef7e054fc00213fb3de9747a48ff7fada069d2438c32026b4db88b8a058fe821c34c1ebd79e7ff3b
cb801cad03b7f0d8a8387189d117053afe94234d60b2ad0b868acf29d8d8ad217b33cd2b5d8c81196ab1aaeecdf1c5f098bc
8f316b70d6e98c4c1f64ca70d5aa669a4e522460bde127b4f0e655cb708f5b24c5a98229465ae59739866421ab0fbadb2a1e
73adba37ac590a911d10708f394787a72dbe995b30dfaa62ae74fab9462cbd45d2d5be8ff66acf0fd5654b5c
>}

\EmbeddedPDFStream{floor-57}{%
/Filter [ /ASCIIHexDecode /FlateDecode ]
}{%
789c3552496e03310cbbfb15fc40006bb7df93a2e8a1fdffb5940601125090b590d4a42536d2f01243e445b9e24b96e78169
e16fb9276c2b5c37f438ec2ab414ef65e9d0dc30632618df1a7c2f559f4892990a88064c2e5f8ec1b88d3f571db8876906c1
9927981208ffc9b6f7924a141bd56a9819593672f98d893cc85904b13935365fc20b71cea8892d8376aa5f1829eb837d92c2
dd1b57f8e076a08216d43e28eb1a29c41f675492b8219cfbca2646dd2f1da694c72d32743ba0a3bf4b223f112b85e484faea
40ae239ba6b2833c7a83d1adba0dd4d52ae9f221cd1ad365f768a73609fa9509e5613e077aaf9f65fb2239f76f697bcba82b
b2687ebbe2364e1eeebdd49d7444061fef27e2caaeb093d363a173c6a2e3b48653fb13e83d8d6419f164e8605738c5583c56
b6dbe4c9096efbc1aba34011d1bbc876e7f4a4cbdc929f202fdd539bfd474f6bfbfe07e9627d73
>}

\EmbeddedPDFStream{floor-58}{%
/Filter [ /ASCIIHexDecode /FlateDecode ]
}{%
789c3550cb6dc5300cbb7b0a2df000eb6b7b9e3c14bd64ff6b492505028891458a544d932939e5a32e9926e92e5f1d3e553c
e41e1e87c077e24942a71c956b047a6a47229628de582d0f5f801c229cf0339b12b6f1428da8124a063077b05ec3280d645a
3da1c75e8ed66a15cd6c55d6670f113773824ec8696b14811e3569e109728ddfcec498f730cb465ae02bbe2345ea420bff38
87aec56235d106f0305930353726b6443ee602435b28c7bb71012ba292de47583d11db5f4ef268e8f0d614657db700f55a86
516d4e3db75e74494d9afe8fc1483f7fb155530b
>}

\EmbeddedPDFStream{floor-59}{%
/Filter [ /ASCIIHexDecode /FlateDecode ]
}{%
789c4dcd410e80300804c03bafe00994c2d6fec7180ff5ff578118f50213925dd0c1c24d35a64de361ca7ba36679b86abbf2
22c5fc00df421972f447d011ea225511c2f8a9a2d9ff4232575f169d74dce05d1a1f
>}

\EmbeddedPDFStream{floor-60}{%
/Filter [ /ASCIIHexDecode /FlateDecode ]
}{%
789c55924b6ec3300c44f73e052f6040fc99d279520459a4f7df768672d11408c0678b1a0ec7b962c8105b72da28c93524dd
e54b0f1b293653be3779c8bb494bc5b444cdc54265a93c0ebb5272426649987824cbe3701c02c252ae21114b56e175948b62
4a405ee7ec6a68c009c81d37d0e1f3ea2b6194a24660444b56ed1945b51e0aa287800a4d85f5096cfa0aa16d8f92df951ec7
eb50657771bdd5f026a4ae7bcd3f0a9f24a32baa870435cbd033eef969f4ed9d9dd7aef0ecd514131d4c97a392cef202dd79
3bfcb2ee04485a53d8a16a7de7ea002082676a9ebaa7b02264df441f67b6b1455b306af8d1f7ecd5677f29c0a999bdeff887
fc03740a1fb48f19d607eeb85ec7f30708556b75
>}

\EmbeddedPDFStream{floor-61}{%
/Filter [ /ASCIIHexDecode /FlateDecode ]
}{%
789c3d4fb90d03310ceb3d05173840cf49b2e74910a4b8ecdf86f205290cd1341fcb9740500b8746c075a152f0d4d170739f
61de90776d5d3a94843a1e632dccd56e9dd9c3c5c8129c3ac1c7d0738b6329f9b6674e745ed9f96f7becb252c3352ca9a9a2
c69146ad2a22b3dd6c0a61794dc6c79e775f239d7df81191edf1c9870e3958d4a14719ee9da47e8d7bbd6bbcc7eb0b2d6c31
38
>}

\EmbeddedPDFStream{floor-62}{%
/Filter [ /ASCIIHexDecode /FlateDecode ]
}{%
789c3d4f3b0e433108db730a5f2012260192f3bcaaead0de7f2de4a99d7030fe646c8120884e330c12e18207db6ff5b99104
de8d9a3014b440d709c6c658b89a0a411128935cebcc215a4ca249435d988c2331f764cac475a14c3da23a044b1277878cb3
6c96314a817bbae6db36cb355b1be7e93bb9cebcf30a7139ea62ed23e1383d04dd333f179ddcff2f5fedd59e5fc6543073
>}

\EmbeddedPDFStream{floor-63}{%
/Filter [ /ASCIIHexDecode /FlateDecode ]
}{%
789c2d4e5b0ac3300cfbcf29748182ed24757d9e8db29fddff77923b0828e8654dbf60a885c3272216dc4fbc7d90d98eaff0
2cb8192e69814abc86cb68e49775c0773552c97c988a7684ed27129eaad009160ad84f9a9fc9cb54f9e4de6a61fa1093bd4b
f5420a3cd80c63327050fbb98f718e9d81fa577fc6fd03b96828c9
>}

\EmbeddedPDFStream{floor-64}{%
/Filter [ /ASCIIHexDecode /FlateDecode ]
}{%
789c458db10dc03008047ba660040cb661a0284a81f76f03b6649aff934efa3731246c34235518272b3e0dfab08d6b930c41
87d445ac3d286d11458bd1ed6372bce82ca42dca27bf9f0e1fbc3fa7b91d60
>}

\EmbeddedPDFStream{floor-65}{%
/Filter [ /ASCIIHexDecode /FlateDecode ]
}{%
789c4590411242210c43f79ca247a069a0701e1dc785de7f6ba8e377c1cf1b5ad2fc3297758ba90f8523c2eede0ad1ed5d14
bbdbab11f3221a7deaa43186b1eb70d9adc58671ca2f439a16dca5aa9c1e11d28dc30dd45bb82a40942f1cea9fa558fb5444
031a181a764125f09fd4b5037fe05ca773fe61f81694db45441ec2c9170625a26622b78dfecd9bb2a412792d25d657cf3f8e
22fab9f1721a3879c910656de0bcfd6df1d69eedf101f3c04403
>}

\EmbeddedPDFStream{floor-66}{%
/Filter [ /ASCIIHexDecode /FlateDecode ]
}{%
789c3d924b6ec4300c43f739852e3080f58bedf34c5174d1de7fdb47a7e822a020c51249a93d6d58dff622a83bac33edc32f
32b1ed47e83decfbf2b1fea36dee6e1e6e2bccabec4e7b5f7e4fabb4186e5916750bde57ac13a453a592cb7a92cede36c372
0edb75c0435d4e546d0917bfa9d4349f4b9564e60e786d864c2680e78d2fcd6b714b727476d5e7b435cfefef0b6e1943a3d5
50b09c3441c5b259686f93b262021d11d5becf3c39125d0725c74f944e372cca12c61ff3f6b61abc1920aa6b6b8aa244c3f7
952bfea3811f187233331f5a85dbf409dece8638b8409104990d2345be794b179f7e78c3b789f8bc826e7f384e8528579281
ab34cad33e1ae990f28bcd66b06fb2e94709aec6d6ae931585dc919f0d1f1447ea1f4d0bf9a53d49cbfd9c021522565adc02
7ac422cefa58c160c0736ac1290925eac968892f1c87c5eb2c9575a52d54963dd7f8bebeaecf5f630f7fec
>}

\EmbeddedPDFStream{floor-67}{%
/Filter [ /ASCIIHexDecode /FlateDecode ]
}{%
789c3552cb8d1d310cbbbb0a35b080f5b7eb99c5620f49ffd7909a17600071f417e9f2922dd5f2a52ed92e1d26dfbadc5a3c
e4ef72f800621fc92de12e27e459912daa47a2b768d558f3cd08d1756186874e897732821eb1afb065580967d03ecbce1e64
119361b691d588e8451ebc8aa976fbb5ce1a226d1566a8296a54ce348b94c4ff512efe39e459bf4b8bf31547e9e5bc14d394
40770bee3bc5e818c8f36db809e57ec6627bdd83188d348cd8c3cab3127f5e31f4f9ceb12f17448abe09eab07c32e8c3de15
f070efd0eee663b9ba0e52dcaef8405d73c2b9c0973219cea997d86701701b909d808abbcb5fd25a7a2319c7504f4794f659
7e5f850304373a461cd811a70b68cf8e8d8512de9ea58812757f5640a60f427681c02897e2bb40cf2aae1a90a14e0a27d505
817806b438eee3b1402d1f09c82fb38f94096a2950ee2dff85a2683fff001e7e7dcf
>}

\EmbeddedPDFStream{floor-68}{%
/Filter [ /ASCIIHexDecode /FlateDecode ]
}{%
789c3336b4503080c31443ae34001de60352
>}

\EmbeddedPDFStream{floor-69}{%
/Filter [ /ASCIIHexDecode /FlateDecode ]
}{%
789c4d90bb0d03310c437b4fc1050eb0247fe7b9204871d9bf0de943e2547aa62c8252c98e0c9f382c10b3a08d8c8725cb03
a575bc137b824b506d12d4dba401d2fca7bb1b1cdb749ba8bbc9e24b33e08cd20adc0cd57126e76861b6a808ca75a8501eb6
de322265a6aa9483bfdb4478c5a4ec0de672d14e2286a0de0ba20fd906a738c20fe1389c7e7dddc06b5ef55cb144d68c15a6
2c26dd684099bb0f5f456b9cbf935de9959e1fd17040ac
>}

\EmbeddedPDFStream{floor-70}{%
/Filter [ /ASCIIHexDecode /FlateDecode ]
}{%
789c3d523b6e43310cdb7d0a5e2080f5b7cf93a2e890de7f2da5041d1e444b3625522f2db1518987182215e58a2f59b790f7
e07789054a1452825283ee4499e0b9d499b1032d9f57b663e27319f91ab9914f135e4ab66425c894190856527562e4e94a23
75841cf829f8ddf0e88a732a5766d9ddf78195c1ee548ce85cf25f58169addac27080f2835b4220d9d283cb3d2482f3b5f54
b72169ab71e76db2e938a1f7ed08557286c9e8c6e3428ef3c8bcf84142da99d3fed1a6e7621492bee89afc239f7e121b97b0
787b372f7df484f2e30da51e065ae703ac1da064c91eb2a7e3d37e1164f24f1c471b695084fb4cedc2d7da15e3a816d4447d
c69edae7776f3b835e835c078dcb8d4cee20eb0c75d86ca2e579ef865eb7efc13e13533f9bc8cd8c0ad26af693b9e75f7024
5ded29f2fa6ca5633b6bef0c59f318f5f0e6b009674bde9748a46c466e7026ef95cee09ffff3b57ed6f71faee18468
>}

\EmbeddedPDFStream{floor-71}{%
/Filter [ /ASCIIHexDecode /FlateDecode ]
}{%
789c4551498ec3300cbbfb157c82b5dbef4951f490f9ff752867d03924242c9922e5b4c4442efec216ca152f19a20b11819f
5176c8dd24571399859a06616b49418ad7b4700d15435941fd11d2a507af61f19cb874afc153a9d677622632389c13733e18
762a649e01a7bead0923ea9256ab09a51f635569dd5288ce8a5481adf770367e8954b38ef7cf1af30b61445d9bb24e663460
aa1cc92fc9372dcfc5012e0ef70d571adb7af0b17bd84a466434755aee60277cf68219801db9fd8468e4c2cc0f93cdf05c96
f4926377102f249f468c6a9311fe1ee41a9ff1fe0511b9524d
>}

\EmbeddedPDFStream{floor-72}{%
/Filter [ /ASCIIHexDecode /FlateDecode ]
}{%
789c4d904b0e03210c43f79cc217a8447e04ce3355d5457bff6d4da6bfcdf8910c89cd704787262e62183d11b2709566e185
cf16aa458f939450ff7da053ddf4ab6b375c0291b06530818d04c71fcd5c7071967aedd31ca54753994592017e7dc2826551
c1bedb2784164ae7da0d92b3f368b4fe85d3a6acf9472af926096eecc1490cccc133f7624b0cd687ef44d64bb6d151e7ccbd
dfd602071ccd794d94ab9953d24a95018f4a7efad82ff7a3d3c8bddd5eac6143d2
>}

\EmbeddedPDFStream{floor-73}{%
/Filter [ /ASCIIHexDecode /FlateDecode ]
}{%
789c4d8e3b0e80300c43f79c2217406ad238a8f74188a1dc7f25555a60b19e6ce7030717de3404ae0c697c08a9ed61dc0467
f39d3b45e185d1e9a4d57f94a1a04ed2d658a40499617a35f4a39c5d37933245b1d593ca6347bed3e9a2f301930121f7
>}

\EmbeddedPDFStream{floor-74}{%
/Filter [ /ASCIIHexDecode /FlateDecode ]
}{%
789c4d8f390ec34008457b4ec111608001dfc78a5290fbb76136cb0d3c7d3e5b5847421e2154d0f8c29b418326fea0b320b3
62820ba1762fea662f1abe84d3bc68553d1e9f702d48b0b92d415ba0441409d1545a2923bbee963ae98135b819bfe8147d53
8b73a8b26fdf7923e10b9f3f99f32d4a
>}

\EmbeddedPDFStream{floor-75}{%
/Filter [ /ASCIIHexDecode /FlateDecode ]
}{%
789c55904112c3300803efbc424f3032d8c97f3a9d1e9cff5f6be290a617b3830468eccd50a09c8f3be1bae3a5c25d51b5e3
102b05d63a86d4faa4300ef13f5aaa354d1f09768fde6698dba72d6e2494f0e8af9e42eddb0236074becd11ce23c75434c91
59cef64cbde46aa045acdeae2c73fd0d2b31cd1e74895bdaf207867ce4fd0501183836
>}

\EmbeddedPDFStream{floor-76}{%
/Filter [ /ASCIIHexDecode /FlateDecode ]
}{%
789c4dcdb111c0200805d09e2918011410f7c9e55298fddbf84da10dff711ce0a5b1b0094a1776ed7c29adfe4558e141359c
2dda944a1eaa0df2d8c0fe204b3d646ba898e298d5dc500b687ede12e41f0fdd1fab7d1eff
>}

\EmbeddedPDFStream{floor-77}{%
/Filter [ /ASCIIHexDecode /FlateDecode ]
}{%
789c3550cb6dc5300cbb7b0a2e50c0fada9e2745d1cbdbff5a514e4f6622fea4b4c44426bec4106b62b9e25b86c986393ec3
b281cf400adc264412cff07048f13d1734ac5f4be7a490af92142362b5264539a14965d1338f83217c9fa1ef1f25b7183af3
d5c89676913cedcaf7e61031990c36a1e6766b1781deeaef2acff8edadb8e8a72817c96218c4149bd2bd21a13c88baa3bb52
5ac0b7a1a65102b2d3bb5d7dadb9413b5e8e01ab277656238fd38c98f66ac257bbf0da34e57b5388184b065b50b3efb10f5b
d293a5ffd7e04a3f7fc84753fe
>}


\ReserveEmbeddedPDF{application-1}
\ReserveEmbeddedPDF{application-2}
\ReserveEmbeddedPDF{application-3}
\ReserveEmbeddedPDF{application-4}
\ReserveEmbeddedPDF{application-5}
\ReserveEmbeddedPDF{application-6}
\ReserveEmbeddedPDF{application-7}
\ReserveEmbeddedPDF{application-8}
\ReserveEmbeddedPDF{application-9}
\ReserveEmbeddedPDF{application-10}
\ReserveEmbeddedPDF{application-11}
\ReserveEmbeddedPDF{application-12}
\ReserveEmbeddedPDF{application-13}
\ReserveEmbeddedPDF{application-14}
\ReserveEmbeddedPDF{application-15}
\ReserveEmbeddedPDF{application-16}
\ReserveEmbeddedPDF{application-17}
\ReserveEmbeddedPDF{application-18}
\ReserveEmbeddedPDF{application-19}
\ReserveEmbeddedPDF{application-20}
\ReserveEmbeddedPDF{application-21}
\ReserveEmbeddedPDF{application-22}
\ReserveEmbeddedPDF{application-23}
\ReserveEmbeddedPDF{application-24}
\ReserveEmbeddedPDF{application-25}
\ReserveEmbeddedPDF{application-26}
\ReserveEmbeddedPDF{application-27}
\ReserveEmbeddedPDF{application-28}
\ReserveEmbeddedPDF{application-29}
\ReserveEmbeddedPDF{application-30}
\ReserveEmbeddedPDF{application-31}
\ReserveEmbeddedPDF{application-32}
\ReserveEmbeddedPDF{application-33}
\ReserveEmbeddedPDF{application-34}
\ReserveEmbeddedPDF{application-35}
\ReserveEmbeddedPDF{application-36}
\ReserveEmbeddedPDF{application-37}
\ReserveEmbeddedPDF{application-38}
\ReserveEmbeddedPDF{application-39}
\ReserveEmbeddedPDF{application-40}
\ReserveEmbeddedPDF{application-41}
\ReserveEmbeddedPDF{application-42}
\ReserveEmbeddedPDF{application-43}
\ReserveEmbeddedPDF{application-44}
\ReserveEmbeddedPDF{application-45}

\EmbeddedPDFStream{application}{%
/Type /XObject /Subtype /Form /FormType 1 /BBox [ 0 0 514.906435 186.564625 ] /Resources
\EmbeddedPDFRef{application-1} 0 R /Filter [ /ASCIIHexDecode /FlateDecode ]
}{%
789c84bd4bb2264976a437cf55dc0dc45f6e6fb321c816c1880380032ca0052c12d20d0a001172fbd4ef1cbb11950175b004
d55d99f9a787bb3dce538fea5ffedb3fff3fffd77fffe77ffcfbffe5eb7ffddffff8cbafbffaeffff147f9fa17fdf7af5fcf
d7bfe8bfffef57f9fa7bfdf7af7f3cfaabfff9c728fd739ed9dbd05ffe8fbffdcbb2e767cc3eebd0df7ffefc97ffe71f7ffc
1f7ffce5eff498ffd0bff6f77ffcd19fcfe8cf9efbabcfcfe9fbe9430f2f6d7cd62acfb3fee66fff8fbffddb65f4cfa87537
fded5fcff8dbbf1b7fd4bf7dfdfc87f5ec5f4feba3b4fe75c6a7b6a7b4dd1fbd5d79d667acbef638abd5af7fffe7af7ffafa
d7afbffc5dd5cbfef17c76296bf6f58ca33529acc9dffeadbffff35ffef58f7fd3dab152e56be9cfdc734ebd456f7ac5a66f
e8a59767d6affffe3fbffef2bf3d5fffedfffef9ebfea94f7d4e6bab2d3db39d3df437e6d9e6b74dafdff4e3ae3ff1c7f3a9
a7cfd9169fff9f7fabfdd9656981663ff1ebe71965e9fff457ee2de63aadf67d468d47b747cfed67b9d7289f473fde7beabb
5886fdac36c77cfc6ffb2e7d2d7d911e5c3eda86ca93d7303fd6c11973e825fbd029fb743db7ecb5ed4bfcd02bf7b6fa3346
29f1ca659faeff96ee7e5df767b7b1cf73e6d8fcbcf553b4edc3be752d9f3d46d7b1a96bc5afcfde6d96598b7b78399f3ac6
39bb9c5cebb2b4903a64dd7d65991fbd45d396d739e2e13a8da54cadbffb75d502efe79993673f9f55cf384f3fdd6dfa8f3a
3e65e8843fdf2f5e8a16e4d95a16fbf3f23975e8594f392d16f1d166b134eee76d7ffad1811fb5b4f8f58e3fabe906d855d9
9f39eaa965ecc5d19e5afca263bedd8aeba2163d6a569dbcdc9f3afaaecfb2bfe63bdbd47e3cfa491cd8a6ed69e7a9f6de8c
cf900d686bccdccbb5ebd2c6dbf38a59ebcff39c31e3d7e5e10f5add5eb21fdacbd9abf652473c7efec8a2e8d470daed5d2f
b5ad39ce28dcf5a13ba95379ecf2d5f9593a265a8f2716bbe90cb6d3fb72f7f7870ca5de78e9b6b7a947eb44b5d1fa78d999
f5a9b28f6bb71d5747d69153b0ba7be91f32677a63adf48a952e32117d9e35dd7b0c0ea81ea645e97113fad3f6d1f11df6ad
f7e714190ebd79498ba695d3023df5656b74db7bd7e13fb99195833bebe3b786efd247d5f274565b86e5b4a2df2f7fcbeaa3
dfeed947acf694c5962df2fbc8b367df530f968dd0b38bce9e5ea6d835299ff5e8fb86fedf9e27aae0b39ee25ee447d52169
556fb966bd37587fccaabdd837a92b1ce038d8077e3e6aeb720ab3ba57919daaa78ddef4ae795a771fbbec6a2d8fd65b866c
ea754fecbcdca89c69ddf64ddad17e6cbd7a5d3dac9a4cc329aca7fdccb131825397b6e5c3077fd43a6b58eba0bdd45aafe7
e99c6e5d8afde8b6fbc3fd51ec21d7a8ad899ba05ba6d56bd6726b3d64b5b5792b9dc2d48dd46b786339864eab96b9ca18c7
ce70ca6b91139c6ef5f49adac3e73c2deefaa97b357955f7db1f4b518a764516aac662e80f9115eed870e725658de6def293
6972745d96e28de11edd964c63796a91556b9cd4ad6047b7a8d9fbd8cfe7e102d6a53ba9c3c5699a5ca2ed5e4436b8ce3d15
e1f45c3d7999ddcab2df58f5e8ba8f7ebd66ba1a2d9ddedaff5a7746ffe829f24744020a07e536088ddc5bb7c9e59585920f
c8d5939faeba312f3186a2045d12d992dc72eda93cf0768fae33eef9e19f6f6c489b4bf778d9482eaeee1eba50ba258783a8
2fd45f4c453bee13cb27fc7a91f1ed79a865dae4025fa2b9b364fcab3e72e782141d8f69038cde7405f48347df19d77c1619
4145abf69affe81d2b8c639199d6b36baf445dc3ae08be43fe7fe81c696fe2e138f8fa6285e5d67596b696b8e4c93e72800a
44b7b7945b96746b6b64c276bc89c245d9ed662ffa541cda09178a0cde170667748569dd0780250223ad8b7e158e6911b18d
63adb0f666ca57e8c4e979614414aa8d59a77d742309d1a62bc868797de5d4ba62efed7e5eb8633be2ac4a58d415f7edf4db
e600b2938fbeb1a7515d0aa7b64ee4f4e981de50a64c7127eba190b42a455b36c2fd51b862b28cb3904df06b9936c54ff6d7
32ee53b15627928bb594a17a145ebc18774545321ab250588f2f827ec5a38a29fdb3a7f64291a462e099078ae44d46ebb867
57f92cd9b2cabbc682e8244d1d6d1b9ee938698bf573193f1215f905591f1d306bdef58d3a3ff2163a196157f55885c6721e
6fa1333992562d7dc13365a2ba0f2a1514e9542bf3c8ec6a2b9a1fe44f36ca965d584531d38e20518f54bca5af780b449400
751d38d98378e91327518bf4b28dba1edab8936f2d3fa344ae76ef49a70cc39659522cb947583f45883a26d677e836ca1d9e
a92fdd19b7c81ccb8dc965bfd8ca2e0f268fa7dfe57d5c7a31258b6e2317be45b74617325d9e4e936edbb097575f59b07825
02da888565ab6cd8f7435e894d961569699f6424955614eb4b89c8e5089446d6dc47d25afc8775d39c10621abd4886e43b7c
c34bfca9d45e29e36cfa3f765dd640f7a5bf38bcfe59b2bf0ff58599e7a928af5768669dfa643d94281f0537b97aba608b94
c5ed4c5f7a6b85e10ae077c45a4591a7ce63752bd2eb077bae1ff45cbecd4b37c5046e176bc4b2e403957dd90400c31bc91f
436b2d13d322fa8df351a7520ec59eee1bb7123c12bab3c36f60e6e582ab0fcc14f4151cb95c5cdc815570bcc53b02854f5a
b7d9673a24252ad82a7b7175a2b5aeb24c143738ff4a1d158958cb5465b81446e8d8b59ef5161d12a5dddd876532098ae174
f59e8cb41e1cea52286723713d6eca3a4faa50bcb4be903fca1f3ca258059e93f2134f5688aac5983e39214bd78555b4505a
962e74b094472e5bbad0e990d3547af1d49bd3cbae2b9d5dc36df990b1965fd40588421849a4b669fb3ac7a6ae448ad69fb4
643a818a7674cf7dc4ac7fac0fc42146f0a4bd572ad3dd2ecac1c886693d9e7c67ca4b8ae175157dba2b0bfa0c056d73e73e
caa4eafe78a3a7f846879ef2831c11774b8e7d34ad8fdf19528da8ccd4f8ad62b887eb655f5af6904ca7c5cd925752cca44b
603d6821081bc46f37632c3a2d727cc7069e5a6819de434924936ec563b2a6db47592d4c93d20b1203deba53279ce571e75a
3e407f6ae51c9f34634b31a86ecc4b48b6e45de47d4638a3ae8f6c4d8ecca70443c1a31eb5646b46ba17b9489d439fe92ab3
d3020cd2ecf84805944a05dbb266ecc758a4763a9e85e8304a22da512d6a71f6666d4cccc69eb670a2da5499271f4aae859f
9069e6d4dd5c464650a7cf57163ac5bd4a1ac08dd1f9587121cc49d5cd5d24efca2b23a3924b5772de6dc947d6499fd48869
6e8152f7b89db71244899dabdaf2b4a8cabc09cb9b77023c4e57511b9d119cf685f8b6d88717bd8a7c804259999d588fa9d3
a535f4e5b8bab14f0a249a5e9862695f3ae93284eef2ca9dcb4b2856dadfd527995f7df7b4d79167cbc5365dc71ef91a5644
37e1f8484beffdc8af2837ce25ec8d3a4bf35e46f9ffd4719599d6dde26c2b697c9467dafb4bcc429c35b01e515397c596c7
7b7c4ef5a347395a89c3d87171148f5011b1de60e8a8ca401114661d5be65897e0b14e97ac5b56f7c86b6539a453d49abe52
2acfa1eb44f5a16459ab135936854fee1e281a2a872aa96c435a914a24e5eb93ca369afe19d1691a61624e1247bb7a6d7d8e
3645f65d9e2b4a7787368a56dcad47fb10f99c8539cd72e66edc4e5b21a2542fd7a923a1ece7d616f4a7c846d9f05a0907c5
625dabf1c49bc84495a64fb62b321402283896f52d2b4ba53a898a447d76471d96ade88ab0d3f52a9da547e3bb3f250a0a2d
1c5f440194659498beb831bd0b3e4cb158f601782b799ce6f7122bafa34fb41fe1387d2659409f85f5cd97d12d93198c67eb
c94776edb1868ac2d9a3c44a0b1e77527f8eee822eb4af9b3d5ba9a052a4f1dd109b9177da7372e83fe968285faaf7c0ea7f
6a2fedf11e3afe32a945913686875de24d7c62a5cdc4571f85d925d74f664507c7eee550b2ad541a5b9c8130a53c592cdbe9
a843e90e269e6a449aee292f55ec4b2b1018a4ce07d7cfa6ebc9da57bb8b3296f459170e648789d252e806f93ae26a9fb8ec
ca0539214531a81c9e122d1bdf56ca05cffcd928d231d5f66bb3ec59550c752acda1db715122846b9fd57701f46773262a8e
94cea5c2afaa84ef258dd8da5e39c5d3b262a4d07f602aaddd19913bea7ce0047036f2db7a8beafb5bb463755db1ac59fea4
ebac4db41f294bf2c8892accee19614cec2196c517d3153be36f65aa5a8f1863e033ab6da3e8ce14629d411017e57299e123
eff1d89c5709829ca9a201a2545e448e6653f8b6a9a622fea283410ad83232927790b1f58d50dd757ca4163c82451920c510
c3e6128dbad083b994358e42a2622ff9876677bd53b7d5f236851698d673964cdadcdeb42a34d269a2539a693d3d7279b2c7
bec88f4ab04d13652d42128e14f1ccf007b07e26e9b12c4dcf0459d10bd5191b074c224fe5484521338755ee57af421e675f
a4d1af906fa2779cb58bfdf4a7fa94a9cbde6981f7e0e8c7472aff29addaa89f10b75088a5791fb50b7aabdbd728b57e3868
0a5ce9f6a24057c35ed942ec8aee6ba93d000f4f0423c3d72e1a79b4028ba6cc37adb0565b1ec486c3839a425caa27033aa5
61f4169bad1f6bf964847563f5e30cff9440e9b5b16f2ee3e44d00924c8c36165e0648abb37d12a950b1d3b9cd6ba02d9491
d5bfe81e3d09b75b99a35e9b2d0bdec1d6f85e51a3a2aee4238c7ac4cf4da15d5bbe5c4a2b40d14da37a16b1d120dbaacb97
acf6fc1cf264599e4cc2f546d119b0278aa89604b51192e07ceb50a0346dbdb4297a96af69daf6f01f4aa8fb0a3cc68bffa0
60ac1f9d6c7ecb640397f04eaf9f8fd2c0aedb7dc1008f227922121b33fc90b7a15b7bbe61347a11258a7bf968bb57c01f0a
04d6ed8e548577fbf874251ac33afe7adb1ab601a0111008db4bd98706acc26b5d87e8cd117010d9fbc055b1a802f14af96c
6541602ba27f7bed4d834d01a5fe73235745decae36cd1aff40f7dbb02f626a345997905d01ec550285745ef6486b7912994
0d5fb627f16374999a4791ceb82e0143c1355a3e24d15ee80e3e3a7661771e8ed8b1c641368ae2eba310377343858a0fe822
6fd24afb5032d555b8201d3d57ff2249fc4be0a00cb513a9663d7b1722cb336c69580b486f41273a1aa771739a5ce6b0a103
3e9542732470b48be8e98ce17daa3e93eebe42c49a1dcb4ed190eaa6fb4c5c08ee54cb98c978d5813d9449fcee5027a95cf2
fbde4a7014214dff709af6b28f3ae02b7aeb4af87466c85c5f767e710fe51bd3af2aa6d32d268c7e594139328067d460b4fa
0a8fb7fe15dfeaaaf82432e41a0d627df3a34359ecd56985085b478e1d8c7ba6e70235b22610901da59172ab0253c9225d5a
5b68da3c3a606a7d95b4c69d0e9c96e8e5e18013960cd98a366ea1b57eec9ddc944b75faa93d677b1804cbb659ad6e7b0358
77caed409258cb496c8f096834ee412eb48c059e3822c7bb901fa7e884ea2ccbeea4319653c0d40e6b1de86468d375d1c281
e8b9b223db6e62a1bb4971a62b34890458496721fc7107bb7d88f3153de91759b73f4051bab77fda18c2d045e56a454974d1
c9f2057605884b8185f2a404161de541a494d63951885cb497949406200ff092fe0deff6b85ffb00f74bf0cfd4b5571c6aa3
861fca22e79e9cfb125d7e2c84eea6efa6ac25f7ab184dde495720602fbbcb4abcf46477a5b6a368e791ade1c6288e7e487e
7d1de8d0f93ba7a563922b56d6eca1677279879b3565ff228dac3ab64b46d3a691942ce7a21691955685ce93fe9b37c20a2d
27f53c25f83ffb0d1118d9bed8d4c9ec5abd9a7e496111c5530f578ab2f678cacec60e992fe7dfc6f0853d947da98436d1a0
968b01dbe31d8772a075f0fbc408513850bca1f7b2fbc2826c2d1df5bc844dd11852eced011284da55494ddc731aebf2bd0f
dd2ef7e3c57b6f4a50e706730a2bf11c2f65c24a5f5dc1c5ba682f65fbd835f7ec439d55970f9c55dc18bdd3245f75e75a6e
4311bf9cad8e461447aa92b75a9bc500ffd87204ba21c0f112afc94dd34ef9ae519c3eb994396f979543ad2cd17e24a53f7d
9552ce9dd03080622fc904f6038812b8ac089cb5c8f2dafebe50035f7a673d8ade66940ee4274fadd6dec8d729f000e115d7
eb8045a25fe7d6630d52356a9027c3d51ea522aea48d9ca9a5d03da017c95a9397527bf32db7a6005261242dfd88153a48f5
c796cf88880edd7a70b4195a1007cb937ae077b4611f59f6712b90b2210f3510eb92641803a6d7b2224662aac5e9b6db055c
8ecb35bff1ab94c30192dad35781e3cb61757c5286667c44dfbe703f9443ca9a47cf3b4c7080fc5f5c016506020faa4669dd
a3f7a57fddbe76542b2b70ecb4945410c0afdbd63af8f6b9e92646de418278a2a9676fef8ad5a6c3ff645203f45229bec5ce
cf49cc497f228c9f8c6b8f4ebc3f510a7114d5ecb1d76db5b64174d96cb4b03f73761d3a3997db5cacd46d96fd4665f73a4d
ec62f9d9b77c228a76fb783ed4cef495649011ee83dcd09bd87dd4855d5142cb721ba0f58561783122f37028da8a7291f2fa
3a0002d9bbae4864d03fd8b21f598a974bd3611d1e24b97100bb019acb2058be0ee0837fb6564c1609ccea751e07d83cc811
f7911d9bad7f56685fe0d3099adf3a1fa31124ca199ce72274e5d16455fc2cc620a090853fdaeeec550fc03fabbe44d8a0e9
40b4a59b5eba8fda2d1f9be91e503da3c75a5a56e2099a69d8facda1bbd81edd9bef4b83d1f2812da9fa5884884fc0ecc6db
c041959b964527b4c89cf7e14f20cef0f8fa4d78aaa31758a8e876e99037dd0c6bfc067815051fb59d2896f7f1685bbdc9ae
115fcc7a88183254a562f0581746ecae3f77af7d7b2fd47248343dc6b4515f53b0429e1645abf1908179cc4d573c4e86a190
6cdea18a4206576cd5857c5a6e595b572eba83412d86933c26f0d09da1a7d7328b286436816875166a7d369d339a13d9beef
caaf5f3a243fa68c3628030a502cf7d2178feeeba075c76aeb948c4c95642895c2d173b7fbfea1f9071c2fb30e8539da4805
11b601084e8df8a9e75840a5bda213650d6b8b0348aba6f6eb97b6a2d0eee355babeba230a12ca735da4ae220350cd83d529
3c5772ef27cf898c147fbded652713eb60ade850478f58096104836e05e5c61a35db5d339198b2667bd974baf11a4006e7ca
062af007f241db1205764ba003b66d11458d260f4803d65e1c0e895c69a7c239c1cb54fd2f2579b61ca13721992a335adf3c
5bf762533ff7a9d8d2cf7390a28669c5cef6d27c67314e2bc0db935500055d0a32c64b20c5d0e2693a460cbb10ff55fa75cd
960ce8292b35797413a276ae8f2ba05ebc25a1bb4a8fa430074554ae7bb3cecb78071ef2108adfcaba8e92ccc4b2910e08e7
03acbdf5efa9284610bcb1aced43557300540ee812a370f21f1e24f689048541b275c344e5ca14336c42b8e42ce89d3e5994
1b3a4d135ff9b276851647a1870ce090b20c45066b17aa7c293d6d1deb283d9ea2fd5648e09e2c2b4cc39cde5f6ef822c2c4
b0be78309af5f509f4d9a2f84c05caa761b2d860d514054501195c1b834bbecfc5e4c3a2581f1d3662874dbdd7b6ede5c02a
2051face1988e873ebb635f289ff0a04ed9dc6eb94de2b3db297a8658129638bb356758033566b9a4e00951412f50ce3180e
90ebf7730f64d314df1af331719680254538e55e7bf355dabff15cb01083b480f16d18a2a5deeb7bd48f505477fca153ed3d
ba1c0647553b923baec8ac0324b1d1a78e8e6eedee27a1a03298a7bf00744a8c219709662e333cc518b23ac507008af92865
31b3376f358e89cd6d6b00a09686028aada42760fe7d10671cdfb68a6e79c079b5dc1914d1c6673b5f86e6c0958260c99281
fcb5ecdeb2d98fde9be1293dab7e57c841163dcdcf332afd690c7e9ca31bcc122aee037cf35648942dda0c66cd6c2e4d3cfc
b41093ca8f63f63181a6b4352bf0a297f129190d103de1916452b5f031d3e373a5cefc145ba34c362c5f650cc7c75bf4915b
a523731b791bf775863d7f80b8e5e80883723d14f863247c0da576ed4c8c4cd5ecfe3074d54f542c6d1ad6086709c8134778
e89dd9060d43013137f200d4cc74b0325f566d6635005829221ca45f715a3b2f7eb6350d5d6e540e4e2f3d1329c4c91dd6f4
112a2c72509cdd4d39a8925818d78f45d9a403d3ec991928d49fe0736c60a6cca7d2027d4e5c1999941333a7f6c703975bb5
d24c183e81615616d67deb31124d996bc669334d6f31e8717cf95366b810f716466fe2b5cf00b9dfbccd59a4a05b163d50df
747995aed802d4d41da0d6fe009e894d5c4cf482acb475553ca76e6f8fb3c4b838c5711faf7c4070009ecdd1042668368334
be15bb3f20fb94dfb6c06b0277796278ca7a3b6df2009d7d47f718b693dbb58d19bd730508c5f84fa0fc74bf9962b069b4f6
30b8008ecc68d6c1b42c404d3c908216c41a4c8962213195633f36e824a12ac0675a84cab2adb311c9bd440914ee300891e7
ee4a86d25e40288799b4422b308ba432930b008b77d06057e5461fe0f297df61e864548b8b9d80fc95a881729c519ea4dc38
3c3e8847eb3e87631cd9b79b74b18bef0ef50f51812ecc881bce94e1d12e5980863246200271d6b203ab801c349b472512b3
d0fdd09f9e1bd341fe4e0f5a1e1c69d0bf65655edc3a6989bdb50c3d00c26ef8e50c2b0086e1826d78a323178dba1c558a00
955186b7945b5908c5ed5a33bc39d15db39fd898c79a0f8899a082d02954a67e7c2e4563a1c12b72e7e406d36f4adb7c0037
a39724eb52c38a115b40e16287a609c9c8b202ec91a6b78285f1b87dca76dadf015426ac29f54f2802dc17ce0f956b52987c
ebb1a04bd82fd3284abb0059eed9bf93732a89f27bafd10a8994ae60ba5b229755fd8a8c98a8a05fad743b6b32050c8f4f5b
0775dd47b9f59321198667caf7db5dacd11f5da02da2f3048892a4df3b8ca3c81a82953df3ec018d5d01827fc95a89afa017
49529446898122957d6fb9d0d9b1fb617c7b836b4761b6670ea01d339fef81a51ddb026acebe76237f6a34d2236aa78f0918
f40d743e19c7a9e77b1ea591d6efb77e199380fac46cfc171aeeadf9a1dfaabc9cbaee796e8daf4578bafd64f88f51080c48
aefbc9d1b7c8687ccedfe9a7ea74e8966411734779e185e5a76025e58080d7df6aa32ef921b2f4ce0b2004bc30519f4f8046
7bc5b510dd6f9ae8b71cbd18e9f223233fc299d325b803d92d08749ac718528f791aa37727839bb39e607c79a92f2b9266a4
64677003cc7b2ba8b09121b82aacf909f689b88fa583e2f23054dd74857b0a8512b1a0ac8151641fc029c227ab7b36b30c11
0ae116caf332e55a699d1c66d46e618d7c5061881f5f6954762bdc0cb709c66994cdf18167d5722b428102697f47cb4af5e4
785f786a80799f3662f4976608204c9bf3c84211a6c359d25bc250397cfd250218814967ea273393e0a5a06761bd635304a0
207c64fafa2c19e0d37c5d320a65ca40c79391241d5060d03e0066002eb8b06a70b2d40dec5121bfed40c8bccb650118bc23
237a0f1d5d1f4a32e054e50b9f08f2a28f2ea3422cfee2d287ce35b3d02d91a27a87f2f8d495f94b109cf2cc2dc1c74f2407
72afd64d9f4fe0ab63d23a0e94d25ec0ef2f10d7f69951ead65db923dc74aeb6af162839d11d64567e6798d302b6acf36bcf
9faca5cc3f744425505215fa9617be2172ffb0ab17b700b87d52d4f405b0a3f8fa61d6f14ed102c12f2f7359f4432638bde0
44d05be9a5b68cdff29331b2960a3ff5f4fa0de80ca0c1db751fd4d4a8e2af401417ca8d32f9fed9a017379584d3230465d0
0ae0981d06654920b461d62bdda90ccff1e3bfd1c539ca409fef81f60928d757a361676326b5e0fcbffe7f7a995891b9a1e9
09d05a347de894bfe0d9a9cf032fc05ac71101a0460bd7e6d15ba1a16e0ca6378e08e563851b363fd736ceca94d50a560c02
e2c17cbb87161e60632d4a6cb7a78faf56cee2578f914ae50209fc9457a23f68f3c0262332745319628d5d91c59c9711c29f
eb4307b607091a5d4106c5e0ba7ba980158678a8c7e564b69204586b2cbc2ba6952ab53d1826023e5998283abe732c1b853d
63b425cb324d6685d0dfc78991d071b4c360cbbf43d16517bbcb86c8512c5803a2914eff9d11f8b7aa319eb18d68674407ec
50517f09841faabaedd4a48c62be1de4c5ab9b819ba91002a4a3069a51a31c662d6587255176324d8eb2c1b3d9793fd5c154
0ccfdb33ba54f4d417bec9f6f40b9412d040ad9dfd0d6db992fcee49483ab7a483f54dd4e733a9319324da7a92de84216f6d
7b3a046a275aa4e25331f23600e4e51aa84efe08c0d42ff8c00ecbc8e49a288b6850b3f8d32df77b600a2194cc788eb9abf3
7850c994915272c2b06e5c328e36e1f34b2d33286d1466b50bdf7d0238e367b9798d0631d5ddf785e504a6e47342f8802616
32ca3e05f28bc793a72ceac65c83d1d3e890abbfb0b2407645d16b8c7e47fcf4b1d4c13ce6a83190001bdaba0583aae0c517
82cba45a3be1f0ca9e96ee0fcc9a3ea684092a0870464b541fd9ef8a916afbf341c440156c669d68756aaa1eb8b01934634a
e81618f4b5305f4c3b1c0be340ef40ab76a027e4a407c1a7cf63412a6c9d537880e2545380a56ae693bc1565e3d1410c646c
8b93297efa923067039e487fa74000de191f8ac8393e4c62d779a3fdfac0caf346dcb33f1368a3fce3c8f7980046609972a6
52379de3a4f82f4ce5434aab94cf7a5eda9e31c739325a7dc0826d3b43cb5c18443de0e76e4cd99206d49689a05a1b60faa0
234a100c38bcb78998d52005a2f6d972c64afff340dd6961fa84b38c67b44c08211aa1f2f752cd0479d869c78569820b0562
c3b7634d6f43673331e9d196ee2f758e16b307d14848eeaa605fdc2f93089d46f09181e444e59be8e8f5ed7da3d2aa4350b3
6b8e7e50166638c6236c26e61a9802159a3465bcd9f4f49860fa13e93423242a4b4b07eae36d8a5bce148a864bcd420d9e76
a6357ce02c8ace2625efac3453cdde3eba907ba47f44189747a4c44cb6b54fa0d2ba0c352db3c4d7c84181c83af6d194950a
537b234912f1c167f9dcbb05264df1e94c3b292345af6dd9fac2520aaba5623438c249cee8088a5e178874d234104137ce97
53a060b05e181816fd9256a19dcdfe50f0b8bc11ec14edf6808b227987756d57f5141d3aaacc91c8daf480f43325ae30ee8d
5b041e52589dc60dfa181b38bb780b3ce1387988ce62484ec78e05f29020e05a9d39fabaa2c95703007a862f6e41a838747c
00c745617505958087421cf04205a2b044e3327d077cd70fd47d063f659037667e07070b9e2c7bf094eeea3d9f596632e632
10416a63979acb7598c89e519a1970429e48829c1deb01c118518560065421c2cb7866906e53273db78a53176f056382af05
b7a3d4158683f3ddf1013a64ab2132ec8db8e01949a758a3495596e7a52eba2d036ee2c4643e505ec81e5b3c0bb82b46d802
8e1c913845d6681db895a611c1f8665977786203ca8481d467543021442099fdf6010ae5b10dc14e9b872931dc4a5e729814
9f622d24696083c6fb4450a1588813624fb432d7278cd6b32e3dadf2faf9c2b0da82f2b882460e46c75e89cdea7c69a62a12
ea3530f6b91ced611ae625fed08fe1b7843636627058510168d871aaf301d448e8336e05077faaa4d4ee8bce34f7108cca37
6dca430fc7334c539351d8544155a415d32e32b16aeb2cb25c6074a038cc720f6dcc694b4350ea29b0812129c2155883c012
d91841c6e309a4e7532f674a7982c0d24f9332ca076dcffa0e561802c037d9e59b208018bd519e1bd57fc2f517a427799a5c
71a3447a4b71e411abfa284b193743992bc67502a928e7b9b6671fc5ae43287a982a8c810f1d7fc06636d3655bd628f7300d
46116002f02877dabf7490c2a883598358c293610cf61006deb6e2923348d4a968796673d0e5fa3e3816fa4d2d697c7a2ca1
cc1eb3c932d6c97c471c57b520e3055eb9c98780af646ad9e9d3be4ce731f1bee03d85cd2959f52a7be4e7203f14af78953b
a30aab6d8c36bb279fe08da3049df43e87592c2addbee644e7a906997e141ab581db0326401340a1337678445008531ec13e
3788524e49731130aa0ee8cac39106ec0974e699afcba48b96557d89ed61e1d74260cb6fe0d63a3c0d3e5b0c6633c5533c3c
fbcb72bc0f21d1cbe415a741f146ac1d83c40ab05ea89194c3d728e7413e1eb745df375fa6922760f1083efa25c8a1f353e7
f1e431f51390a220bfff4a980f64b2be9850fa2738f414658dc0359680389ee1533aa67088dfe40aa2d1cd70596512db2796
8c8a81fe5ab7853e68e3d90a66c5a9c82ed53aee9c1657e06d5c77c6d883d2b9933310c46ed1d27a498741648f5abe09361e
8c84077090bfc0fedcfbad8dd2f404556559a86af42507c8e6f4b7fa5c88333de2afe93d65ba603a4eca9105392833a576a9
a3600ef16836866ad4da1f3f933402fba5533262ae419bc904fdf120343aa40a0c2031e9df93d450634ddb2a0bbd921394b0
09331a67777a2876d33b4cf65854a67088d889f61eb00036125af4634e60bd09240f208ae21bb524f172cfe7c6414f0c9969
493c0281604f36a6457336ed0254b2cdd652b90594b83b89567af3ad7d5450ebb312b012e0429ecbf34ee9ad2f5f7f9b90b1
af24bdc91f53bc8458dabaa4410706af7be12114c8271ede0621b8c6b8054f1045d500f772177ce2456f3614313239d20d60
34ed256965cc9e8c389a76fc4fdcaaa7d28921232409eefc61006708432d970e5ce295f6fd9dc1c9e671f1782a6665437ce2
b90cc340e3947c7bf662ed234c7d938e574f3026e1dfe32b048bc975e5a24199118bcd2c70f1c41d50c8cb08af92d4c54aa5
41f5d93b304b8c02d118bf870f586a9b1e0f0464ac932c1798bba278a3f85457dd37ae15c3f5c94030472e0c65659a04ee27
e77899c8ac4cca7c23e868c755c8b05e6a26345f99922e1159e0759a0f9fe0b682527af60b550c5afb136418eefa06291919
5456c361916504c2e32be93a35d026b37c13d432b2ea0d1acc064abc0a68d7c8e48987caeb09e9f2b55146bad518253520c6
7ddf3fd8af752ab2d11d33b34f7b0981e97fc1dc294f7407af98e28569d5b3f5add076a2c499cdc3c13100fffc12bbeba002
d17a722658afa49f0e3f9cfc03ea2ce45d568f5c86b3d795a0f83c09e0183cc69344fd56dc65b78b9f22a1510b3709f4b079
bccbeeb45ffde6ac4f8c67233dd3ef1ac2d774fc09dc1d12a032d77381ee83ae88fed22e39e81dc67598ef087e4eda9e8c4a
d9c008be9607c6dcbcc31151c142600b0001776298b4e647c2ae0242c98f70828ad75297d3f6d50e088aec6299a286eccec3
892b979c98205adefbf8a0bcb1d701e5289912361add0a62dfc65a37dccec18470735e46de5f3a1c9bc8818e386176460e14
e8ab2de6e880e369084abeb98ba8a7320ae30351fa6561332323644a030b618d20a46af46252cf8338305a18be733ce0be0b
144c26df0c3b809ef243fd3aaeb44acb5a17e125b30378d5c2d8060355494a9c0dc4a6d463f869a31f51ad7d20e388440c7d
13edd4f1d37903e636a6216ea7603361a868f3a579c2901d7bf89cef5e70a34a6a7d480bc18d32c10667498922ca0b87488b
895a386d5bf21e6fba543631adfcb1cccd5ff91b90bb343b6ce487901bc58259c645442e26b57d8999226947510a74613826
fafad92cf505281a69e36483f4a195033bb6b59495b613988f5aaff417b20e2f246f41adc6513bfbe28201dcc32c66bb32fd
130c33434624297e49acfbf0f357b4d1d1e492b90bcf44b3afcfe1811654503a880998d8b29c3982da62f837a1c917e7f8f6
2072f88e11211bebac18c9533894633b1be61b20fa2f53d2b2e923e8514642d81a07bbbe28b8c8668788d6a901da6eb83574
266da77b2a4493dd2bebce6bcbb16337bdd85a195995c1840448793029f2da06fb40ec0c54e0f9265ba289e821440d7421c4
ec57338a41d90996f825ea622c0e46ecb88d3534b2a256ee16a47d289a9c193da1c022221316fed2bdf6fec88400cca817ac
021e105d167bd76b985db41c2e16bb04e2d2d3665480b634db4734f409fc022cf74ac17ce8b1cc5c6deecd4a750a1b0630f0
42813509e899717c1002b358fdf699e0e766a88345883175bcf7b6831f851a9092af9a7892153300fbc51b349052908573eb
0384259b4971c9cb36d258d589dbf55b7e7364a5d2f7500844e9798f1d0811388a1b2d345f015204c5e0c2b9b31c907740da
f602d92a2bd4e3d6939c2340559b1f47000a5be094a2aa1ae958e3f53d489432031877dacc6959c186acc7568b882e5ac0fe
e7f7cfe10c6cafa8525a9ffc0bbdddd98f7a30b4cb9a62e59ce05e6565fa85122b7e1a2f074a2e81c1b1bdc225d0fa0166ad
48de638ed899aeefba4c6950c32914b4610bd0cc803bedcbd4ac85a3046aeb34f8b1418b5b66fdf6182a7283dd7b260550b0
12572a07d967660ce4615cd095854105c14f7f452c661463eb9bd5d162c744dff33d554d92bddee4ae2a260df5d49ee25824
391d9097971218d12602577a4386190a3b5ea7037bbec23b473b87d064db8b5e614aab0c16e6a02541336a6af62e2269c35d
85982c2155cc7e8e667d81ac13eb7ad005beb3522ba420df8423fbc2f631e210edbe19181aef39202160d81ff07f068974a8
ab2edc8b7f1c934181711507a866c0c16dab232bf013e8296670f6fce4117971d44aa798af99b7620f233ba0d917c2aab383
ef646553f31005bc198649ef1d758293f341a167c9f0b44f67193c214228b7154b390c224f7f562100003f89985be6067b21
d5eb61b648ed124c9624ed0d69c3ad50e78559bc9171420b7af1e6346e5bf3a23ce1209114867e229b4b255cf51baf7d54f8
279211f83c6630405779b60574b3e8e55f89229d7299ef620f55474ba83011d4529b0d00124cb1fe451405408146a9f5bbf1
113ad6e755ab389487912f8976837c6b799151596028b1d9376d9b94b9e4e3ad4768c8e01d92d7f6abd0505163f045b4130c
6c0d958b9a2fc28ca94f4efb87d2fb032c38b97a0eb2ae2fbc668d99ed5d27733c49fcd0c08f571b31448640a9be94efa9ed
9001599e174987bb44f20f35e36d515391f4e5d08046d4fd934d0c321858f3fc183b0a0f8c37305b97a8f71594e4d6921ca0
7a32928af62efca36d1a69beb5c3a677aa77e999ea08e5b7173eead0371fe1bb92907f05966679a6e1b12121e2ae27f6b847
74b6df62e7064fca03a553465c2700abbee0fb0995abe0d10e89f210c17a514046c2a7b5896acab93a820bc909af932d6309
440b5cc7376b65289916af39b026109111496a9ca807eb47e7c35f5fdcc142e226575b1b8ed0b8b50c347714a3c2037467ac
14758366f4108946eb8399c7dd6e510248cf093ca35b950a072f43a5f369694ab44a05363cff2ecabcf9b65a52f3025178e8
aedd6752f23d38affaad5fb2d7f27d6ac6f5c15ad0f2cae80fce5c740d7d9d8b0a062333893600ec0a7fd09b00f2021cadb4
20c7c2c10d360fd66316800e1efc0c19bcc00f0aecc72ec741e01c8d9d1e96926c9c412fdb1398a114437c1f002b82722608
7c93890217ed03aa01292f4a3cb27d23a323a386aa2cf2c1b11a85391ecf1f432940065ad9c3b8e474817af4984bb0fd0f30
0608b6771a4a9993e0f4349bb2986184cbeed4cb200395f1737cdda0300eb31b65be2c836af1c66bfb6aa50add33d273c05a
ce7d79615ba27b4a27a3d4c880263ae9e745c785492584314ea885c79b905053a6f09c3a70434fe01731c4c3bff978020fc0
1705767794c2724e7ec04be4a7d9b58dc1eb8bf466a2dea97dbfcc453069c31906c070d71ad2015fa69c00f575e02a04bdc9
f7bd9331d7d95fcc7300d432789f545068b9bd94239014618b2fe908447dfc51ce78d00b50541192651909333b83d48daf42
ed117d89be92da175d37fdc7af87b6a5caa30fa6f433a4ac0d63ed8dfb8769ed75498088802b45114b84013e8d3f7aad6f9e
08a0b91e3b0ea32ed3f9a8f5ec3b8d3081c77b925c64454074f7722189709bc385e9991f63885ceb9b44290bea0ae5bfbe0f
0fb9bf025b1a6179f2069ab18886f9b22dd263df0c36102241b8e523fdce98b75e7bb45bd42c8401c5e7c73f429815f0ca38
e71b45b3e0abb465223a90700c20009df001e8bcc9ad7d4f91c17048f401c344dba5c6dbbcd04af50fdc4c88a1641807c6ea
19b403dcab44ef84f394836ffc7f909fbd8c90d15c0f94f7ba12928de195735eaa8940730b1dcedb0a007d40bfd3de043c3a
453bede0ceeed213e82fef0d5a431b96f6c549030f2804d9a997de7381d26ca7ae1cefcd69b4318eec1322e09dd0f69a916c
04bce51c305031e41842cc40ca00e7be093c058686f1a10c88905c53046087788899653360bebd1813a0d963fab59ea11bc4
f05b24eb6384337fe178036d4e72d22e8792fe049076be2d51a18519b06d8f3b2a0a82b1d09a7d33514cda9e6fb861b0b6ee
c743f13a63726806d43b93f86447d10bfd9c606c89da77fd6e44d5c00cbd5c4a5d02287bb3860cfc12049027f3a01a56e992
5e35506a3b08a9d8c0ac6ca66b21d5a9df1064aa96d52ee0d0ce3c30736714871701b7f6924d7f227dae7d640f88e09cf9c7
97eed220ea44f0e8c4ecf1d85407fb8bf368ccc022eb5c2241a15ad5e67f31c8866c630d81ce1fa95daef3e2077301e1d3ed
6e25e94183bb857aab1f165d7062073d4ebd423f70244e0f42667a9b288e72fd4560d0ea6ab615c027d2a74420350a6d65bd
aa69431841d806462e964f1e3b8a463e4f871516360904dc2f960145f06efbea212e06ca90d65fe068d0607dbc24b4ccea82
3a8c79a584e87066012af8521b320e83c9f872856be14d0f895cf726d0623023c45c5a5645c0884171631f3ea23a19e092ec
0c3f4c9d0380f5f547a6939922ba2a92b81a002d2fad61a8d51ed83392811c429c97b1cb1f88921e06d846e4604fa0f74288
cbd96168bf22276efd46ceb2124a4cedf5c56a03179037cf3ade1378bbeda7011bdaa1036aac0b3b24d51c2120fa373fff87
3ffee1ebdff0a7ac163a0a323372f4d052761428fbd7194141488554e18c9239c8113104684cfffb3f7ffdd3d7bf7efde5ef
ead75fffe30feada8792530fcd74aa9d83aa70707d42c65f43b2f6dfffaaf7fb97afd75ffff1a75fffe3dfbffff2f7e7fef1
6fbfcc57d01b33340138ad50144142a3fc9ac12ebfb92daadc258b810d4c011425e6971d3e295476485882f30cb98c9f9c20
bf3d15a6f086414e5b5e208a4017c4fcb861ea6b341532216702973eba7bdfca903d98bf125121110a8c0bc3fc74842419b3
54272d51a1b345096cdab7e04e4383189101d94285a8cc3df907790e8688f31ef7a8f19fb9bef3f13fff5a81c401a52f83b1
ef84a5ae9192c0edd6597783ca0775ed0410d130c6257f5fe9df5e9b8e13ac50c1e94a641d14d2f6b5a1c01a24ce7a9b8406
8f20eefad9defbd3cf275af6003266ba6f288b11429f6ef92a8020858adcd30b7b7a4294aeb94757a89610922f29514ff4a6
cce0b8b304d1a38e4641b02007d191b4db7f23d5f2fb82a0eea9c02ab530217decbfc2d23f1fd30aa915188b99168e31631a
b5f6a69c90c2dc0064669a5a0461fbcfa1a7f29b1d07a3276b9ba90ef3848c567fbbb63f7f220829e5729da1a31a30e16823
95e57e3da9bb401cf124f1ca0a619fb3ed2622a8d603ced013211af2b7ed2764e7b7cb85f07988b5a5a7ea0fa5d7615fbad2
00811f99fc2fd107cca2f90da7b5a7d55014f0f4b4f8a8cf8ffa0b5bf3db8b40c6588212e51e269928b88a9d610247d24225
3d067661f6d03affa409fc6da957e8a30e72e7e4add73b74a688dc65d43985346785785d813aeb99981bf78d277c75678c2a
75a1a012a64fe0dea384e637c235dfbadcb281b2c38f7d7410af1464e01870cbbe7cd2a2bb7b0e4a70ac18704baadd120de0
eeec075048f4b402b3734b36b5c6a4a87d0fc8730bd5d095347080c16a3038bb87b77838c0e71cdc01a18468877b768f223b
08d92bab5565db314efec68453d9889c65ef77831efa25c3fae75dd42500c2debe859080dad65f0ce3ffe9841460aa28875e
8400036cf6a5f72798e57a4f0193122765fd9468fbed2c153efda9172453148c6a1f8b3b77e4ee1040235674c9982001646c
c65edc2790b80b8da5a45de985fcd07b0138387536cfaab7f85111202ac3fa97424502f19719636e2992ba7fce1bfdf9c728
2641f7124162ec218316e367c6f7db42933da1281dc536066264b9b1f46ec70f5a001d21dd73b5493b82aad5463503fa4686
f2cf9d12c4503788b6dc7b10a1e38ee119fb160b5e10e2b805098e07c83702f4022f3a5a05bd58075339c5a4068454012386
1672da43aa5bdb624405179a574b919f3244eff72b3a90bab5f5ea2552b9a110f978571e3e107e9d287b101ae309acad9995
7a583ff55cda01e43282efcc7d21b3afbad7eb2af21dda8abf44ec7f37a60bfb85cb8d57661e87abf872f0e46941816a35e0
428a6ee8cf14f53f5983ce7521bfcf049faaebf30bf7f5e7950ed648b4bdc7958f5ab456b68d27e4e62afc631b150b0ee91c
35000b760f297311c93666fd62580162c19f9880df167a13c56f3c4af05d132f2f860adc7be0b6980938693cc89807349fce
7a04b98e5cb71cca77eaabace5f1717db973d3947b2342607c7ead659fdc12c7562e502df06fbff4a97f8f63774aebb67c0b
1a63419c6e5dc5444a0b21d9962056c0625a8a67d82b0b8115f35f952a4e20e68807d971f7eb15c71f2dc54b4da4d58b3779
f1cc48bcc9ce8df63d074cb9c307eb2d083de4107728be0c90af7b751f412e266398556b59eb8db96faac3eed79825f04fb5
df5156b28d33a8b33aef59a1cca19fb412df89b618acaa2f170622b70e4be189ee5cd5d60ce22b1b437e68a3173ae9e5b6fd
a039ab36412b94203ae3dc21a5c93437d50c1b2383be841d3ee6a2606a1874abdd86f7c8cc8128957b3ae01f3ec56e20a8ce
88bba8e787b6698ca9be4456d4abe41be86b4646f9005aaed607f5f989790b0af6494ea0cb1bf24a6f8f7e009a9ff36da3a1
a35ee057ad836bd46e5af099509a01c9c638c16b3ca873d3ca95426d9c6fcafb3e4a193515d7a36985e407666ad8b01ec56b
ed5d2dd7d1fef4b336980839aaa747a416d13454bb4835d89d09d574145e6b4ec24deac18f8d80603c82992b6430d8444456
c7782962303e0d256c44616173cf1ef63047e5189279fd1b7740f1d078f6110d896a87f6fea2f621a2a4e2ed3de780001258
43d0353c49ed81deb55d0c64e418f6394f3090c8d953a2fcc9dcf67b983e12af576e839c9083cabdbddea15f843ed359fd0a
6835e40cec37f6f086906705f312335c2710c0f6c99b31de1a4950b6172a2ac665d94209fd02982a01d60526714792f392b3
3c70128f76b96c641c1114b551c70aa4142961ce2153a408b5691f75200bbb93c2f80995f3d0e7322f3182efb8f620fd7862
964da19bf5296d867426814606c6652073f0536ae6f72082561a4c5851ef78a2a4a363ea8b674fe7d4a0ed1b117723ba9c3e
8de4c754f0fa1d421928136869aa37fb343756d0d245b36c4410614f051b024bcb1c375d41ea1b4953fbcaa81c3cdce4cb4b
4b0a1e2075f768663469ab303cf61538fa19b809bb25b0a41ec021e7d2449718d4ab76af17dc8030058e93e78d3a1bb4b32f
6539b0cba43e51fbdc0c573fdba641548a464bdd983be25aa100fad903feed7040a7333770f7f5ad3eb963d6cba7147a6968
b07b928370a368b75bf3ac0b17d442354ac2140629edfa0a0d4a4110edf5a45567c41bbcaa7be519aeba3052b3bfbb148784
f9b166bfed18220a56b98b1f26d856c4618f3fa7944d44542e9b8cca6cc95cde02309adc2b9482c20391ad52a6b451741089
07e743fe1c164a44b3b6ad1f6b51264a454f461d8ce257943eecb126f92db2b2487b7d2592b9d084b39b0ea503307a39966f
a1d9cae0f68b1f422a1089a2a04f0815c2278880bc49006607c2322b1270bcb09a3edc5570f0a0a15628844644cfac35838b
3e81534c33f9b21ccec48eca4c2f9b8530e0c06094fc7680ba620c6575bf206d7cd2ee2a944fa7051975e8cedb6f6cc44abb
257873cc20e25f6f4784d2480d45f708da20657c7e8abbfcbee5556b7cdabcf868a438f62f16fbdfb7fc54f2526071d9c7a7
1b3c6df18ce8078556f0e7117928e20c6e27bfd21fb9a9356b501728b165cadf2743f0d2400933d0048d3472b6f91ad1704a
b5b0409132de650c5121987f34bc78cce2227f183d258a35b3bf24ec20a8a091c8162a3df61537dd9b48984f90babd700902
14bee1cdaeb790f4ca7616758c18e3719b12e2ebed616a3d20eb0486bd37eb5c4efc6362ae84a04df4309810f675a2d67180
a842c600c9a31bf6d2d30a05b387faa4ee6df64337b401c5564a1962dec12159bfc5d717ad137bab609e42ae8be480d30c98
0b4a57ebb666d0a3836e8d823ea011b98061ebe88380f52184b9e45a5a9ac128dddb6220389d935f43563176c71aaf1daaa9
74bcc665eb94936d3e0003fc0ba70c2db5d6ae324f09b2737f5b81edcb598eec51a1d841e3c0ba4e74e881e740801528b101
6cb8fc14a7fe3db5c17a43f27b394619d03b3fe7137e5b6780fe7cd577bd561b3969370ef7dacc7201e2e93b7605f5186a51
360d6a60bed83f253759fbe70b47cc6bdb747683923bfdd605270365cd6f780cf03d21d49914cf9dde964cb47f9116659680
e9066b428d8d6cf6ae74d061cc189c8c80b039f36fe6587fb3753d52ce2c023075c00c84ad99816e9aa1ce7e092a2a94ea5a
1fdb66aca13ea708a8d7db745aa1026703bccef11f94676e205143baae58f35c3e584dad44d9898d58eca8e27f9f80b74f40
73f6b9ad0d0841d72f3cfbefa69104a9477a13b0b31677c1a6d4a1dd1b889cabb413f45e10a3bed45098d58321258b6c9562
ba2df4cd18fa806a22046ba8262ac93ac72758c8726cd4d872ec3b480821a5b10bcdac854efe667e16a702db2bd3c0d60dc2
69581851fa9e8c658d89c37ced292acae7f2c3565a2d518eb31fa8440f7ef279bbf2d19ff73db21f94e6774bf45fb6c2295c
50baf6deaad306ac57df090802d82ddf7ecb056087039238c9644ff3fdc8860e0900e46f092bfe1c05dd3e7f5b709d81c3aa
89767ce0447b8abf2bc1e9b9a2f0b52e1118ac17cfcfe9addf434c846d76498e3d322eb4d1de1a8c1d4b400fed4e133d518e
f3bd3a19e918d00d02b7540b63cf6d519c82cb0a35ea2789879ade5e36c99781b1e8542e28fc5fa2f08804bd6b01cf7442ab
31a3f9d09603b26973329d7550be35a5171b458f35ec42ebb2c077a4382395a0e8908148b54694263f8c3c095f8948bb4f5b
20828b0316b276abd6fb00839509f64dac41063b1866cf819c076008d4caf6403322271f0220312795016ea1dce41e4ea79f
24932c2ce3af89d6af0da0957c0018a804ba09e741ea1b4a416b4797a219825ba806529260438f556cd811384ec49c43f634
5a974ceb8f620d4d34ab49a946eb394e0c8b265a61f6027414d619c16e2bca12640ccf0ab9705f0fa0265a4858ee0d285081
7a937ea03880b831273f462409f5a70ad9ef276f21cc10722690f852acec2168ec321004f39e199d3fee2dac0e057abe974c
41b72570cab9eb35e9ed7f2a0dfffe8d3339c4ebb861ca64d6e1b1af1dd3b3e8c2eb345390dd418d3a7c16093eec1086d52c
4f221407cdb5fd4492bda62daca9c8822a219a9fd605a0375b20621b575c1af00bb4f3fed26c2242768269beece6cae269e7
7d31b8320ea430974bc85a0fa085e7b1dfd860fd9c0d52be2c793c28fc8d698f2a3a284ca19513f2aa50925566b9dffa6ab0
1cff247a67e68ec57e39202d60ffbaed17fcd0424372d8a0172a998acd43612a0ad88369c1658b7eccab5049efedc224ce7e
200db7aeab4767800efc4878e844b7b2f8bad58c2a03a4b7a93843865a4321cb051f0cc72d64225b704554e2c3c57893bf01
9be6d4faa61f47b50c4a025b7c8461bd14123392a148f950f17b0b6ca2e203a500e2cf412d0a52fa19be058757ec74a17792
c492c221db67fd4645dc8cbc6d8c8483847a2949a0cd27e126260279221941ee826972bb7ea8d34ceae723dea2865aba2275
9b9b817963a2b4dc2162d41cca4bb71340bf5e1630734dd30eb3143c3cf6e2ea35e48d65fbb3fc489399b0dd163106cd4b2c
59dddff075224edb39893973e6f4911bfbeefeee323df02c527c703701fdd161814a466f6d3118902f06d096de6f41881d8c
d1f286e97c304ada886c2a2bce8bf4cfda52c03f189ab92ff5728587b7afea8bb17331fdb7c1978e2c4c504d789a6d52f16c
b061a40149b3da426de0a5998ba004b471399b4c7598691bb71a1089713c9e1b133e092169bec70113063022063ca265873a
e62faeed3faf337d37daa8907266c6806f29bea0895147a788da6af2def5508fdb367f2a91e443d47cc1e2340f8f6f764255
4e0516eacfcc3de11f60cccc9b69a80cd056ab573a2e78a08eaf24b64a5201b9d61d9661886e52e3b55839a57d3ad48a8452
a399f18a7ea8e3be5831744ca92d6792d820c96f8ffdc80ecaf58cb87bf1e4c1f1f6c5c40647231316a1d71da92a7abdc3f6
b799010314c0f47ad0df52590cfa2a0bdec67f2a546b892238b03af797561ca2c198d9fd3ca15ac480253868bb2d300ed415
d392f9c64c25816e726e798674285091087cf5eafcdf7ec50d6ea69bb0cce95b064473d040d9c3d47a806e76a4ccd059847e
b7efba06652a02baf5f69310819c213deb336cf813d1688841e32758148a8f819443052924042357fbac9252558bea99cc30
41b078d5ee834f1636a4974217836593f5bbe2cf0526200b2d92d78f06c003f2208a08c1afb26c451891ed8716b7dcdc4d8e
a09da3ef63d78fa44521c10e8232bc27be71f9363ef126f8c9e803644784415cc52bd6fa9266a34a0a8a31e1649d0457e18a
8740e0c41954eba9b34127bd557ffc3e079a13605109e9a1ecfbcc634b144c92d22e6960fa930088a1ae976bb09068e83590
efd17da200912867efc9b515407977a6d94800c6fcb507098f4f4cfd7fb333c273ad1466db9ceec70c31cb1d42f05f31514d
49cd37fbd81a39897946968ef43f74548b5de91f2db82c0b48a4cc2ca9583c7ed681d8ed806a3d2bf93875679815b3805198
431a734ed972a75f0071972fd940045768e4b472f9d7e4310060da2861d090a471776a021b6a489f7920aff67047d9f8b97c
27b4399003f4cd6099801a629cb75e89a91ab6ae024706109a68bcde77a6a7e43d06f5fd0d6e97c674d65961af5bb674846c
31114569d1bf96958162d083eb989225b02340cf6cf54f9384bf27abf81ef8cf3346182b79fe6dfa3efa275a713388ba583a
9813f778e9ecaef681c5e8201317497634b956b56fd2e88c9fa8ffa5857cc051c869bc0084a150d1b5bd7a60721d7c862daa
a356cbd87f707186ed8093d1e3261ac07484a4cabc5d9419336036649a21001a338d694d2155639ac2632f47b09cad90f748
80247c956bbe21271427858c4a490d71e638fab6090608c9d0be5c373d8377996a99f7e5f0bee860c6e447e0c94e0d9d576b
d5234f7882642fce3464fa5c33dbecc02bc202fbac96e47c1d459ad37d39345a7d4074daa5c19e3d9a2e3e4c883f9a98b366
31b95395e7a4be604607c92c64b7c1635a2a94957e740c084290843c57f96fc1658bf0a85decf1d1e21135e568d0e0c9305b
bc1848384fe9fb97f42e703d165fb269b042308b1d7123a5c8d0a3e8de4e43404869f5d46f483db2e49177d86f8cc189090c
2875ccd0f2933df5811351f80ed5c92b2aa133f234bb335a6a98735195cbf47607731d7ec1af5e90402158db2e4c2bf30c5f
76880e6940bd13e91d6c2d1e46d85630b942a49113800d8a7554475e20c8d42d81776729b2467f73767f0d6873350a5c3d2b
3c8dcbf8345f25261252288414463e598b0dd5bf0fc940e96fe6e35676d0e0a58a1e93370b80bd47903c2536445f014baebd
e9e30301fb81f82aa36519066db9b53815b8375842d47fe384b0e7653fde5acb21c202f97c0f5d4cb2348fe8a56fcb24660c
0365a4ac88b6179bb586a6c481d2aed5e412aa40c5db7c7c06087c6882712edfa4d6488429e3b409f4081c07e46f7772371e
0db997bd6224fd85eec793201fc520f4996c24d979b45e7b7f0380a1816bcb0f0f28e844dd83582981d91539f05fc219bf07
b438a11205fbe8ed5014f405931fb20ab50129bbcd3c0807ceea76b42a846b11cc5d7b3d39a6a4e49b3e8c2f09d5d0d9ebd1
588c1729876a818d39e9223f48988c4b5d3682e60479389bd60d4e8812f718c422bb25c9f3958a11452312c4365204886a72
abc7d679e8acc0b8d221a80ae80e442dd30f3f23de35d1c1ad31ce4ca50900bf3f1e0d96ba91cc98612515dd045ad5c786cc
83210ff9cde44246c830e2dba96672ada06b98de0e520aebbd3a7a963ac3cab96ba8316b6148a36cc1703c21775a43642b11
99349d9b2ff35005847421346f2e16ed3c41dced6f0082486d07f9228fa695f14bd9f8b7d51bdc3e4ed33cf7e2c29d44eaf1
b2d80463cf4dea22de799e6a8bb3d07acc1355a0e87e3ff83268355fdc39250165257913016832e8edeb4db2a884a558ec5b
ae03aeb11e8bd6d2a60757dd894311f6a387c29487cee8362ac80264d9fba54591f5d12db3013353e98c8828ef0bdc75327c
0c0f25a51402713eef9a591da0c2ed2b4335f8fc27a3e3644825f8d79812f246642398c99c36b79113523123d59714fa937c
ccf8ac547ba8fce57c83ef63c7c08065535441ea43a66e13bbb63f39515a3376d70e5dd2a2b7ead713cbd00296b60eedc062
2b4333ea9b10533dd7dbc54052f1af41dfadc38c956a0f4ceac5017c29e82ae959f097b59cd718a1d76eebdba1f5d5621a98
8d8b8d393107e5eb09d4e1f9c357bb5d9886ef9fb6640d813ed4360cc705f448678b32ad9d6663d2be927fed5b67ef8b9eca
4b17778144ee9104e63ddf085ceec7f7a5160c994a5e83a83e72e231ca2fdac13fdf2f844c951277857039dd3a9319cfde2f
9a9cc421a5f5a428a2585b57f1e83b2405069cabe869c765440983d2b43dd703a53e726782fbcc211e2053c7772a10b62c4a
316a76b68904e0d2b3213363298b835da3528d33003ccc0ebc4c8f433fa66d5e37b07d3067db0fa276ca1f0fc227515843d2
11a93a6f5681af50a0c805994014f8175f7c81020918157b7a52f4666164792958303b36fb65061a30d0024974cbb1a9272f
9aabb7bc411b935cdab6bc90dc3dd0d63fe5b613c8d69e97a657876b4a77b0fdace04c68090873edc6ec4fd8487de7254585
2a4bcb6e37bda31e3790a5b8e25621d40efbdbcb115158438f75640160211938972d79720ff6081c5d161ae97ac2fee17d41
0f928d11ae366f0dd3f5dbaec981e921dc72c62dd08745c0ecb374461682e17745c39a91b21073f4ebf704aa5a2f3dae021a
67042c81050141bff530e797750b66c6c67e5e0a8813554e263bf7beaa9ce89f1dd20a67e1273020def57e26f33e218aedab
f9b4ff1038d46b9790d139efc871bab2da9c0a436ad48c179978f1f35c4c0431ce0f6d4b86b634461e3f735ba16c0288f124
829d2086ecdb5a1165c705ed337418129a35424bd736ca366da401b2ba5d4dce8afe9d3521a89e64130e5afe08401f868abd
790a2ac106cfe315e366f40a9e345fb98329affd24240794075fa2af3ded4f40b6e119bfd844ead2be0e212fadd0a2610c66
8e31416a0812df7b0e25894462c91301b8013d068bef08220c4afe6be4ac40b9723a2f7d0d7888b0cef702e8dd29645afbcb
d884fe5981bc32d6e381ca0f224cb787cc03156a754f4cca34b2f527c675dd7551f6cf18177175745810ca2bf41adda3db07
818e106f487b53b82b9e678671aad1501f4c38c346bbae7b2c3d038a0a24106fb8a8361a663e21a032cfb8f2fa66b608c6c9
f30201853992b9c71605432e56092a009f7f51c66cc4093882cc03a103f70414519259306104ff3c2e3d2856ec7e83c160d0
7780620e02b70ec2bbbae32fef4243f600100e9b04ffce4b2844c324c498f60d957188e517c3ee6f8697aa252354252d290c
2be5f1c03a10373c09f1b7c06fe54cde3c1e1840055d71cc088a11fcc51300efc7469da81502bb387d8ff55deba15ae51d17
f4052d92be956c59f0806e266ddc8b43abbca1301e23f55c0a7e4b56cfc330e4f7178df5cbe6c05cf8e32bc03f02d0c600c9
93b4194f9ce917a844e0e9513d3870f66799a04386653b1b03f18d0795f85b1507a0059edc9fea1aac0245517ebb0cfb0c05
f8b13c45648cbdc3df10435741f4713c53440b5100581366f4bd1e457a0420f62a56f8f483ca312b1b0fc33eeda519c3500b
4a80f414126f0bc41f7d1bfb8103c52de080bb5fcb3b8299cca725749053319e538dbb98f06f7aab44b719ce850bed382ba8
76dd3b9f985b656e29985da268387ab550b918aba459d2bfcb5eda4fc4df6cfa37686af0f31eba5b547d39ce2f8b01f72923
e9fd628564cbb497c78fd6848469a12db612be12d03745d5de9cea231555a3c173a241806628ea861e875419cda5b8b9c655
fc430c0ab089fb75d0ec2c8492932ebf519e2d7eb12bfd789ace901c7c85869f5e62f8b9fa4e861b50fddb80a3fd4a07ce5f
4466da4b07d99c6e2bdaa2109abc110b85d81b6aa05fc9c40ff3a60dc06b50e6c983ae04e93c9508bcfbe23943f0203aa07b
481a0a5d2d5018d5d646f70701e1074cf615d6c68c3dd39f555867415b222118658d507ce976f950750f3c0341441cecd404
f0a5735d3e2eee68096b96dde1520c7f56a383d4b0ebcf9db009469b97711cea9d488a3033b632845c882c37ef9da1357a90
c10b1006754f001f7ee4b545fc05a7c3734b048cf5768f003d606948309ed44f83a109025affce1f6e628726333d2e84c6ba
992f555aa66c94fc7d2bb8c820c843fe528efaed3d804b079e388baeb074788eaa9acaad60b7d7cd58d74c9c807b8b900a2b
28db65710a28ade7e59339250fd1e25e66b916fdc05fdcc8bf57019fe811f71a254398f51f925d776bb1639314f1ac34d4c0
d29b4f6d7f44cb46e1942e6a6639e8b28c563ce017fe48723ce6bcd3223480fadbe3468034c170ddcead7a315df878cad71f
51fe784811b25e0d69d8f378a83e95049de1d89a6fa53052b9ee3bd072b707a8cf8ab314fd7bc88e7db9a4205f3949654782
5d22807a6c3b2814126996a4ce00a31667bdf0ee14b0b605468c9638ec4ea3f1f15d07eeb702e20e4e28913fe068517fb767
149dc5c51b87ffdc38c7f982ef8e42a56ef43709dd84731c928437a304fa3979480bff038a53ffc250ab32b87c117832fbe5
9740fc6f163a8858192ce899c231e2a8b8e3ed60ccb882577d0f31de7a3c1318625510de5277fc8a717d604aeb8588621317
4f286b6e756997fd8b8bfb3f957b01e701f1be9d0ca0b389aa770723aa6d8cccdd1f23d23e7c63878eeca48a3cb3c23fc375
d92b456df821bad38295440102e71fd3772b1b05800248e33254317ab720d27c59e930f42bc5ccfe2b383312629dd2d28dfd
67708af697ae30fc36f08c6460cca82e2a5fb69c4369087477096d1286b2abcce4335f72a7c36408da403b07382a2597e5f1
aa72284d1b48e8df03c929df2d1ff0029e61c09b7837852d113b643ada1fa4d02738688844442533c63cd25bb1194ac4d895
0cd65628b8db9630884562ca2bc25e1f9d36104b2f5174a57c434c1039f5a15b4a93c1c6ba3bae331cdc49c601c517d1f10b
ae5b567e4711f8f24d1d900cd59a98c247cd18428af5e8c82d3c6f9419c4e44b06eb6ae228b94d02d29749a727a87fcac97a
30f5a7ced37d65707e42e324e678bf82318f29b46581cf2d401605b1b00bb9800703a8d50bd307149170b624d14c488ec824
f966580915fb0aeae272bf333bdf3db5102a304180148415c8d1b4a87a7bf83a95554615b5ebbab4a1ded0103fb731c78c71
9e3ea3b51175ec81dff4260c32370a9e27a8a4c23ef2c9ba362f30a5a19d7ba8bc245b29ea354cb8bc8cece0d8e03c09e422
c83166227d9512741791c3ce61b81eb85c4fcb0da70f63516d26f3424458a16661ed9d92b28d8005539099b1c004bc5fa23b
6adefc62f1d4687c3f319b6aab4b30de83d2daf5521101bddc2fb89902fe3488312f092f839940125f20c4540314fdac3b6a
274bb6a87a795e6ee8cc89d25a92bfc14a1131968b8b2b744f0813449615cdf267783cd8a2cd41a1793fd9921b8dee3b9cd8
d694d511234b572b0b3e5986a4ed8f3b6ebc609b561e3d187659781b7a30e3893006745dd1b7009e547c993e5977007ae629
55943c15f9fa9c7dd3f7417cb0ac08a49948c568da32fd93fa4ac4dcd9a0045cabf4cf46dd4099513621fe89e5506a8640f4
e339f9006b1373c7d04ee8370740ff25a8419c013daf80f4d103adcd83660a52be4cd343d470e1e58978b4e5adc26c8ff6ef
72ea06ab3a42e6f68822f717c33c275dc6d41ec1bafc52d40c06f61ae58e27085f8274d523f31b093b085b30c6d856106380
6c2db6e044bfbbc11e3e727e03f8d14b16a94421681575f8c289326c0dfb986732089a41b4c677bfe5d281ca982f9712206f
68b991560e274ab926025fbbdae070c75372926a0c78ccc64bfd076c6a493c61d02f94186ef1730d9496c27dc3851a076432
38fc36fe1850bfef00190e7626e32cbb1c950086c729ec47e9bdf4d0b1f2d82eaa3fe48e2c42421c305fdd4299e41143b767
228017790884a204c9d62c1d183cbbb6e4da52ad0d7cc6db7776770299f18401208eb986653d334c4b0b76b37a8526912555
a4e341e320aa0b4cd13bb418a265107c81de35ff189f5543b37c2508a6d4d4527f991783831506f82b4506cf0543c3eed1d0
0f2aa8e1f4254a5589e4d92fc87fbc09430ac021b169703e5136f2ef0c833d4877f989acc66e98c26128706b5d828f85b982
8ca983eaa2785280162a7241b311753652ebe2b9bea3f08e7043cf62a9223c92b4edc7427a0ba40e988534bda81d1cc0f13e
530d7a0e66a75283e9d10d67dad8f6aa68a11fc2e239be8320dd0800146f832491b46beb925c74b660d0733f96b755b2f900
680830244c4b54ce6c1ec7743dcdc9675ed1075acbd0c7f9a6b87691ccb77307bf82f218c98f668bab34a0e00b4507e6726a
1eda51f6c78a2039f0b8cf8c37896bd7f1063276069182d03be1136b28e8f93cff8486357213f7e41539afe7455e03d12d98
5b424328461ba8705aa40230f74589afae2b81f13ca948ede99988c5f0ddaddf5fd3f6070af892c054e52021059d582d5e5a
8ff7a490f3134261edaae8c08155317efe2eee0fc8e28a1046ce7996c380a807f9a29e867c24022571acf5e0e7799193f8b1
379d21667b4776e47ad0902343e8c38af0458c9fa57e8d62a23df61be5cb838e311627cb62489397f6321027ef055f15f227
d73c0502c5426b64af01b4a2121ee25f91bc3cf5e51b17629de80db692110b839e44432f8f26d0d595daf7162c88dc15a5fa
fa07a60b85f1e8fd33f511c5fa17af01ed1a03544f648e4f435f8ca1045fd28025145df499104418faaad7fe399fb0343364
32e29dd12f23beb5059e19c3424ff694e8822a21b548d91f91828e6cd1c71d280db071b331c5820c3882ff7b943a9d72921d
6fdb812bc10194fa6d15d2f2595f7bf40a64019966ad1e7e4ad418adad060dbc51ea5bfd7ca3e206026d3e612d4c51036deb
0151694a0295fc6c7b3ae884ed18faea495bd582cc1122097fcda10460e62fe52bd11ddc703abe72a1c892ed1e73a9489bcb
587b93b0c9f707befe8ec4d1158fa0c59a3264813bbbb87722e2c0426c3f3211c78359bedb6e694cb18eee99203649396689
2245fc3ad5673d41deae983d02b8537e4e0e6df40b7c0124943199977fa2a0465d1f89060ffcef8c776c9cec5599c18966fa
efeeed4457b487fc3725ed07d5d2c7a60e045990721c54ca0358a6f8fdbcc8c5c17884d8e7a1109450591d94e9298c60351e
7b535b3a69ab096fc6cb00bf72d6c2282fe0c76cbdeb8f508c7aecaf7f6c3acd100c0239f9caf16c24fa6c8511ec099c9e77
7e6e725b940eda26142338b246f49de66d0c6ad361c3b18d662a8c15485c7427d9415949ebcb4fc8f60108c91d8480817116
9fee6c00b83859588062a5999f5f7ec48844aac3bda9e54b43b66993742fefb22a9d9698106859b8c6df2206617771833c19
fd201d1d9bc8100973a1b6e4b501e6c31a1ee33a0c05efe24b47159a565c3d05f6c4a1155cd8db782083cb287adde1ec0906
f67898e20f065ff553d4f54ed25247b8ef6f4ba3737ee0504d4ce30494797c4534fcf881306d3ed77ee0f36083b6753a58df
d12c683b265f0a3341d07cfa6a2bd411a0cfc3a9a4413d4c6c59cb0efc8ab495b8a384779e41c9f006462bf0385658176fdf
a38cc0a2fbd586ce067da05fea3144922fe49d33244223fd8c421668c4b6bd1c4c2044ea092ab444d8c054c6ec8cbd33905c
516d4b2e0d12203a4e3e04a950ee863af353d2a28648fbb2471520ffa0dacf7c6f46e2c10845a7c25df4cdf4619200d66f68
63c3c5dbb81dcd14e03bed84e84090a4d49762ee064604217b82374934799317690ec8c54b0456b97a8bbaccb1a1504b8e59
a54727275f60358ad905bf7c875002daa016379d916319acf136cd023c1b335292eb194aa5e28748e4181b4c24f37a8d492d
0b71c6975d44df17d1d8c4bea24551c072789a591d1058802e612bc3488a1b8ebde9a1b14e07f22703426c8b1f7646275316
6105f351404f5062a99eba1cdd22c81fe5b2ae9e01ea8c2f643c382f19043a09971c573e51510b746feed9a03e06da4febb9
8343812af7e44ef4e70b5dbd8473cb913185e3c5b87e6ce68099041a972e55362184463d63f144db61eb7a5d2e4044cf9ef3
822859b0cb02772a39b4ce3c12341fbeecda3e0124e746de9224c849fd3bd6832dc88569889d8ccb74062b0a205e186cd388
57765121f38d7a05b39bb5f90901588a1024020b94424a3194eec1e534c8502a599741e541012ac4e0ecd1ae501b54ea10df
c42f8ada65747c8e84f6aa0255640782b61ecc08f8a67744a4be68d065857d96125bc8fa9a9d7c0211196cb6f73e96b818b6
14b2921276ee8b247d0a04483e3761c81fbab724bf83d8111cc3eb48486104f92443fa03d43858faedd95b647cb03c5feedd
41f375f9ba2bb0fc129eb15c0eecaa580b0a8e974a2a4b81ed4a6747a316e2bc976ec2443dfc9c7e07b44b18776b80210304
f05d739d951ac0a651fd1475672c8f7a4c79faf7fd62d2f7f16dcb4f1ad1756f80d2064a431ee31215aabd389cf3b2715772
9ee13b800a10cfc0de7ecf5dad30af3ee41b402f40b6a640f0497e081fe333c407df2962a029d94601f381c1e9050f060818
98414af7902e95e3277598eda529dce735eed012322860abd64805b26f4f1258f13f3a3d5d2fca056526d3983f153e0f4e7a
7bec35d3df0daae3ebcfe9ae0f24655ecbb48aae99487d12abc41722eae133ee462854c78dcb160d74da2d6f7549266a1470
0663d384ac087cfecb282b55da901cce2d273fda9e7b0e121a84dc01a504082344b0b7e7359d07c6348a87276b870dd54ae0
cfde25cda0a192cf8f965adda1e9f726751e334040ca83b89209aaf552ec81f233b8aeea37af07a4f1737b3ef3d1a840c073
3b33470a517bb92fcf7ebe9122d48d62e0ea9bd96eadee3b3d114a7279bf47ed1f54009ec753c777aa523da0cf3766d90c9d
34eb055a8dd10768fe2e8360257d7851a800f5c9980cb5c224e4ab3127e2d9e7207466d07566b11d0e7dc6b33c85660c68a3
6d3fe6e534222f3c9efd19f155e476117ab9f00e1caadcc7cbb45389e6f74e1534f2ed0e66cf1fa7c2f80bbdc05e330c3990
dffbc06ca17c0de5454b24128904f8525b49226261740a6eb66c65f590097e9119a94169d3026c13215fa5fbcbcc9b3dd8ba
2c0bbdad68ee0563a88c937f34dceacace48aabe15f81e343c3c4a6761130a83bd79520172bef5f650654014b1c0899ddf08
3cac798df6860098dc6685d5258fea423fa0fb91d08d8ac384fba325935b61f88bda8bbb8f03d6d940f05e36864649f8f1d8
22da480a15ceadcd47256ec4c8a73dd86887330ff26dfb1e80a3ed85710de49e1cd17c7acaa347aa561f2f421c833894bcda
774a1a7297e0e37d460f9700e2aa97566680112b2fe0bd1832c6519f202a809f05e548db92517cddc1a1420894f811ca9270
64dab76e499d421a918c9768d456dff01c0795eaa84ee66452aaf8f9055914df62ea7adf897598208f9774fcb1e8b140a918
2f1aa04a853c32e0963381006a45211c4e05c265581074e75e4821c84967c077d3420d16d3870b61e0d13638cc2c27824fce
80ebfe32be0661fae835218ddc027cbc77d32300599455effc2d86e14df31c52b40e916bba52c0bae5854a137ebb0395260c
4951f261a6f085067dc1dcdb2867ac043fa23004279887501106833ed2edcd71647a0a607fdd559f8193a336799324328427
0651ac9b9185247581dee4478ac04122edf54c987edc5c838420f3db1a421a2f5966309beb002586fc168dbbd76151b83582
65fdc934a27007a024f389014aa687bedabc2d63b99deddb43e0c498baa2bb90650b1644498a47a880a96f3b128f3b9f0431
c87c013482fe09e048bb3ca78740b7bdb48da30c4b54089029618a48764dcf5510c3d4329564b111541e327cc4c35ec46119
1c93aba92599e75b634ed9b7cc180166eaba27a1a0fcba0c05c9b18ff6b147102be729d9f4b297a7a7449c069c4479dabd36
417ea600dd0371404e8462d6130146536436104cb33732043783172ab0e7ca39d1acf2dc7231fd011b14858e6c223211dcaa
3fdc8dd1bb9400a81767b182e7dcd35f05733f222bb49da26540b5f765008429f00e48b65f8770501043cfe32510005fdcf7
9d7bd02578026df672a2d03da3f05f57cefc14b8f95e78ac4764404c8d5ffc95c22f14296c4b6e7568aa10d862f29599d9a0
97d92ffb3ea27f3783af26aa7294759667ede064438ff5cdd386f1a44fe223dcca3cdd5e40bb660ec564c3cd96946accc593
134625112e7274413da36bc8aac83a5ea6910ece826ee81b95eaca225b4956dee8d856cbf514d13ec8de75291619e35f6f3c
37c8b92ac267e422fbdd2be0a6cb777763186b55a86e621496b23a2523cf7227873028297d0bc84fa868946efaee021246cc
3f7deb6284741b6cee2fce74931a203f75b594d094395e9f40579d89c1894c59cf09248a1ec3c305e42395c7cb3c7d072f24
eef059dafc60c4fa11e15eb816094e7da375fdd0a06756e35aa8109da93e76a1307d02ff9ed86f5d1c59420f3c01345621b7
29410646e11331a8173d6384a80039f4abb288ccb5bc9eaf7d92c79615ea90f3d6dee18ff6410e1db9c19b9e54cf8553336a
a12f9318fa9307e3a9578d3dc8fbdf406348473f945bae3a700de4e4f1325e3f2206803efb09160e39ddc5b5b5c10893d51b
9ee6f38d1427945bbeaa5ac83a20b480d82da23e9d548456fccda5560af1dab93ae1d07a5389b54536da9f03c6aecb283c90
a5dad5e7113f42e1017532a00b59b1022bf842463723d8873f35fb608d8944642e5f670976101b5d711d88112952db4b20b3
c09947f8fb72dc20b3e0df1ac25306792e7a82127cc6676e412603f44871d654a9811b0739281fb49c0f53ebd450eed02595
9f518eaf96c2a0a70bc644f33794a9a4d299f5a41de38eea41cd29465a2f80e4fda0824c02d8db722bd30c7e86baf0cb7dc4
4dd025b8a5e98654ebcb6c8a7c4181018c69ca68b37514078b873960cc649c98b7adc932de496ad196f72811f46338fdb77e
3c83eb63da8fd48982ad6933131f8dc4839cc47c7c041f8a652d14a62f440f410b5891ec7b903cca57f4a07da060755e9a7d
8c8a85f023b4320911c11c830af7370cc96bd82c5646cdca7d2092f4508e0df3c91305bcec3e3107e0036cea1c6098732c3d
e6a268e384b13285e9450b93b6ea933d830277f17a7c0d0546e127b89d9fe4f920a40b72130fb59c51c9661cf0eaca17c8a1
1f4f73db10ad8518b0274970072fea9b007205a869209d988769e83e36e01cbe4b44afaf855cc9578a6f6ce4295ea89e1783
4eab5ed63a05301ca6f582e8db81155cb8d9a85c1c24d75f70770cc3c1ef418413a59c1ee4f18fbd0140c61596ea4e95cba5
ceb8ca7a3c27079f3821ee9b095c008d08d3818fdc29b3f5504d4f087d907730d3e2cf7509a433e8db980ca94f24285ec16d
d4a0c250aa74055a76df34c7ec8ecf4f50b484121a4c041ce9bd2de8631387296340d6353e90ce0bef6ccfe9e24490d2b7d6
eff98066a6fab1f0416d06bdb2b9725e01f78139b6676fd39e62367dc75cc8fc3929682f0c015f8d1e65be0893a68a43bb75
5fe743fb698686446e39bbf878ee6608091e995ed81b336c417570ca34d8832db70179ead24dcc717d18337aa850dbaa6d74
038127efb47b089b56cf5c4b918869975aaf5ec84698e27802d809f17aa3a07499a111356d2fc27a038a8d09c1d9ed2f407e
aa30a67adc5d70ed849c4192a36f86f9f64bdf8c381f49160ab1576a1c1c2a89a1dd49f0fc3434a3b01fae146bffaad205dd
f5618af61bd42763bf3ca41a869c15c321d74bebdfc441bdc09fa67c12a09c72624ca5772a4472d22fbcae157bf620101ad3
108403dd5778d1ad87100123157108ef1c32b93e7a2fb8a5196c74295231a8e20c4fe5cf88146404adc4a0165ab5b26c2f6b
1d23542b828375e349c571c887f8ebdb514bd3d9c80405dc7b85b8e185ea0800f18cb9a0af509d08b18f971871a1e6780047
5f5a13a5c6ca3c1f0f0400180ccbe9a8771018c00fec44be225ce114021d7fd97c51076c8f278d1d0887330cba6356004e0f
2a391ea38ff3d0d9ab6456336bf507451ff7e0092bfe41f3e2ca26c01708abfe0b2017ec3e0a64517c427aadf51794f9a0e9
2d973e76626c60304298c91600da51dc4699e25c507a8c4be1d45e4a941d8c209623f06e74d8aae7e3c20857968d0240b9cd
888ed09907237e20d8ab3dcd3bc69792ed1b2f7487996fadbbe15099055cd51f6bccd3d53588a304ebeff416985ee9a659b0
cb658c250e5106e6e1382455b0fa00bcbe44620be5613f3449811263b44342eb479a0508d95ec8c6cf07fe83d0ee4e66ed41
81a4d9f8b355669918219a773e0f0ac750aaf5710bc26f60392f372441aed7ef63e07f013eccbb18c0b34dedca935b0d663e
68e544521a0a629982f981d6f08e705a0708abc3f280d6bcbdb99017413004c35cd8c9ad0c9ccab4b57c9c6ca6f6f15aa9cd
ddb90836dd6df0323e507c5d81bb07896468862c0a70c21b44d3ecbb70419ebe971f528500b084e5dba98707d3a29cefb140
47bdf4d39f9c868f4eec03d9e87a615e851e0d95a079a5f39e0a79e0f0557a1cafa2c901e7d1578ef6e1e5bdc2b47208aabb
f4696f90a30c5307cc92dc56ce3d51cea5b9a7ba54bbd792a57d2c1f8a81ba31336281182cfbe490d9dbb4b44a4eaaf4e053
7941f09c0fe81e5dbfcb8748205758158b58a11a2e171771ce572433c880bd68b84615a2d1778c7da9816e1d3e4404178cce
13c34ed72ce8daeee2fbb0606cd70c36fe6f2811a52adfa29471479485a1f89dd07bc6dcdfb4233753eed881abfc4dcc1ab7
e0edeacec03c8d4be5f88404119cccfe3c0d6ac53d06cae3acca4604d78dfb48bcc584cd663c299d7790e1ac7e3803da951d
44d999b1a1a53d42bbc097c11ee4b995fc24435e003f80d4bdc8240f50473a51a1a5064f7a4c6cbf881a36e0930cef86e35d
4f54d0ec6fcf0756e8e74e1c315ed3b416fe7ac1ae7682d9335963e9c2c64bd858814a70f051f73b711482e108d7d86d241f
6e544fc655315bfb21a9f083740c865466ee2f2a077b7d1000f7fc55bc66630c3db38805e59be7a445887187e8407c2231ed
ea8fbf5f2173ca88ef068b956119458e6da35ac49b14692e82e0ac53815083abc06b3d00ee7d76a8af64044039606daf9455
8226729454374f09024a4b7e04856184109a3feb7b9ce9040dd20b536a0f807c0da023ef4b7a6c03444e088227ba03418958
016f377b3efa07a02ae34bf5a2b0d0843adb7a515069237046515aab0842a3b1602b273de88c152fd5923de3bac1d5d9b750
88ca44dc0cddcd02952f57ddd7d07f4cc859360c2323d12407690f1d0f9b7bc5207a8da1f9081380ec514ef5651692079005
27e6fca39554881b3caf1f3399753219b76fa560833af27a6e72309bfa17dd5a3645069bc988171d7470e00761f85aae70c8
a4d28630aefdf924be013351136a86865839dd4bba34ad20f35c2df8de82d60b36028f0c8667f630ca3293bcf30915b1fa42
bb02600224c6bad3f9686bb1dededa70016a10205dd51ae0cfa0b03cd703086db62f93ffc2ac0d5cd42ffdd442a376dc912d
45e43092db9ca0e193998a42873c4a273a5cb5f8518b509c00290d5631285a2089c65ad9eb95c221f0cea470bae291a74f8f
580882e31375c99808ae8c92541ae3d6d63014dfc22e6672524e00067d6c5d21b8c5a55cd6e20772d755bd0a1d2e46ab05e1
c9b3bfe11072aa2183694e08b70a3c978eff6503d95050bcc8e66e66e80988be3596684eca36d867339801cbd59e259368c8
2174007cdd1361c3124a1af5b2a855720da657dce64087798007e478012dd2353da52fa3dd33c45192b98c98ab42fbe73919
06d01b658af15cd86da0a5f2015fe195676081d38706a003f88235d713a6dbde6fce18175ef7ebc59b13d34eea98b78b4e3b
5a41bb875581716c5180ae173946ff6ebcdc5dcad790fea6f4b1724708a23ca1c500098608d2b75ad74668e145c6a29e6048
2fc8a05fd9307df27aa3ae6c903dac10d7c9b6135077c68eece51d61901a34b05730111e4f884cde1ece3c506118279c2eac
650a066c06b13fa16acefc46da32f8e54ff14e0cfe0b086061fb3c09528104c8eba84071cc56285208fd1caa7c88213fd5de
de96fa55105caecb79b26b9e7177fce84d62fe4104d17a5a20d848845e56a423244385ed04ec3326b6bd0ccd894202f63fb9
bbc1d58f173561e45f111845aba8dd2c892ce81599444fb2a4008712e88ec8d87919ddad7145fa6542a59eb399547d495d4b
290ca93e19e113648190f46d850d1cf201a3979d35929f363df52edc4d78ba32ef78b25c3f3ad52fb48a8a96077eb97cd7ed
50172a1e8310fadfb0b850090e3a15f81ddb0b67122b8d7e7c79ee703727f56104f485e8bb42ca4abfe472d452b8ae9e5b34
5dfa4275be458ca8488a89ecc777d7a038d51fbbbeb9f00a6d4ffa012feab9489c285c5d019406350868d44efac2aef100b3
401524eb2c1403c8f4dca3a15349918f9290775865a185f1ae7406a7c0c9a4074662b2bc579a5a68128f5630195240188eed
db0a680452f685682a6df0133094e241d8c1104bff64de79269d6b081ffdd9d6472a8e6332f4a64910a0314be6eb321561f8
3e51dccd8deca43ecdeb270064e2b8ddd171c8ad65a0b675bb852619fdd49d3aaae8aa2083e8bb719b316944a1ae9a045538
70e7be6ea7a879205e55474df209da65cff685bbc8a3f9448e2a56f5c8055392f5d55d722ae87e5a8cd822130f67fb8bca93
82258cf4ce3d9f210beea74436a546b85a6ebd221a71cf8b1623b4e38070940926ed09ab51034be993ffa04e608f33a08404
510ecd777651c7a8a92e92fde80d9ff0dbe86eb0078638494271a03c1d5e63bc7e605b61882f50852477f00f5bf806f7e541
55262065818580f3637bc90cae1743465a912718981fd49696af3a919e538ec7cf5e267d8aa9117e7aebb461b402769a28ba
1a4ee625e108393ce67aafb00bdc590cb6be6143778176e8bba52a631c8c89befcdaa88b1f6a71f5b2ff309becf5704bd013
1ee453c3fb9728a00f2f54c1ac189e23f7f07abb1755436a320702b8da2fa9ff1934697c0aa1d813380b172cab1b8c9342f9
e035742896a10f7f56445a909b53f87c61b10765811cc7bea5eb312935f8cae481c1a933799233fa347bbcf908574769410b
3d6e6d1e97be3dcf23e59b19d265fb32ec562a892fac64a8a8d3e66e57ada820be02439e2f1d2a6300527051c9153c61f1b8
5a682f433d925241f9ae612226f2d64c052b3c9918dcb72ec9b4b2a786aaa02c43c6f0ea0eca54b546466b53e30d373e0c36
e312f4300255eacb180773e30a51f95792bf06931303f8f63242244e72b4477692d06e24d7736fb25a80144e87153bbef201
6cec3b38ba8cf2289b160527a4cfcd74a9efc8c042cd6037f0ff002635a4b3bc304e90dc2936056e9f45196e256c692fe10d
e47d13f5921cd662707cbc88ae056111765fc7a4dd9a31b0a4e7850d7aa0c558427722a1eb3d5aeaddbef860e8aa80074e25
c6453d647928a4ceaa7c7309a99324b99b08dd2bf0f261fe937377cc7813c405a7726f7ec4563eb702c63f97411d6203c618
3d93308ce8a1c47595407b7082140b8beb51bc7c104cbc82f57a2cf0191bdcc87f0e68a8b3244e63e389c113b7e734860e14
be675c26bad073b2dba29b8ba8a28c68b9d23f540a6418fc967708e56860ddb1a746d727c52edc4a2b3da1760db94036fe63
a6dae7315a0a19ba96f6a350ee24a3b28009d87d99bcaee55b16956e04c2ddbebc0c003704b3d271411d57fe7401fee18f7f
f8fa377a09839293b644bb20dba09d2164969dfa3aa001e09fd0551a7a834512226f1b73fefffecf5ffff4f5af5f7ff9bbfa
f5d7ffe00f666e9655c921641c2b65689d997fffabdee85fbefeab9ffce3df7f953ffeab27fcf16fbf6e49a041600f90f595
f360f24e076e42fb9a9f57ff762952fce29b3101e20e32296a8ee6c7baaf5dc7203110412f20ab34daaf3b587f3334c09a72
4d32865c1b79c7657e9c4803447c76cea540748d74ef7722faa75f137f3780ae210650c2014487debd470eb9372498eff421
5563c603eccf53cc0746d675395a025af9135efca71fc31acd5ce7848ee107f3430d56ad9fd4b67f5e9050fb806df3a2db19
0fa46538dc3746d57fcafb80ec480197ced8e2fa0e84fefcd684108bd16166e72a113631f64ffed9df1ecd44d666b08979dd
009fd4e04ab62f52e918c1b21d1909c6a313bccded9e3d752b1638e0103124be402ea5e2acddb3698c2a2947b2ab942c5cd3
673f7e0531d4cc97409fc77786222833e43ff547febc28600f5ab4428215321a76fb17cfe39f9fdd36d33d8885b5f23d64f9
704abe275a7ffb795c7fb834bfe7fe9abc1d1388eee71d1d5cfd72207ff0953ce775ffaaa2fff9d9548c179c672d8a2168db
3cbcf8586eef416fcc1144a3e1bd50e30ca5047b1ba82a8ce0490c3da24acec8f875f3af42134cdb316138e7af201a18bdfd
84a8fcf95e5e7a47845bb2cb8d6af3cfaac29f7e3b4a14ff91904af0cb8a82fae9eec1415f8e8e26da3479242b287e3c9ebb
69b075d35b5c7104e18847e87bd847333f57420e801027628512d4a4d51d130ae784aa4ca025128b96edfca971f2db6b1f6a
030dd7d46f2d491e98f8c9de4bc54f4fcff10d4aaa809a28f302ea70fb4eef09b5af905ea0388981fb8568f94f57e704edc3
0976f4403d110e2ceb17062ab744bdd015454acf200e31a87df1605543d2ecccbc0c935a7ae3e9ee3234b6beb66f6a9060bc
47e6c01ed80ac7b7f2e8ba52a068057f286d42b7260d991d12d1ebdb695ed3067aeca20c1a3985767f4be6e4a1b00e7e33f7
da0c2d1cf46a02675982ac72f550a5378f3e507b9620e72fb705b098d2afcec6521239c857ad9e643afa37d16d7bbb92f009
30fcd76ed2414a53ec7d0702038bcdffc7dcbb2c5daf25c761678ca7f886d2a07fe17e99d2b23b4233d33db33460b4294a8e
6e5ba61c56f8c1fc7e5e590bd8a82c6471c361f2006404e3fcc5f5a156e606d6b52a6bb31ced3fd42a5a7d790526f9c2228a
b37087f274b6dc47b535e40dead1d842f4b02a419a07c67a8cddb8f15564a3e6040ad2e1e2e4a796e4692da1483e1a6aa63d
6a7d8df518acbc8788b5fc4861c55fddced211c85a679132560c105b51ad11708c0bb6b21e9eebf13b6e4a71ef213f05cba6
1fb1ededa6aa4a0bd56e84ee8ac6abcd7e188b77b50a44042f5332e3587925e44beea11623166ab8c5553f24469edee45287
3d2868b36c1eb94242dee2865a061d24536ce01e2d4478d43325d41bb17eae9b4d147c9e11b09d7dec58c0b7b5b0604d5bc4
d9985cd841c679446d997eaf618ec44124f2caaf17cbfc1e379e58f7d96d1ec464874e7e349ded64cb4c36ee471d2da4f4fa
f153402dfc325df9ba3b2844d70566d92c60721ae4a460216cd0c05eb77ac6b05816c3f8c9aa8e8b9e11b5cd70f06afb475c
6ce1ae598e0c56520eab51e894da91378a7b99e04c32874cb532e1582fa19092d5a1fe957c5b717a866b6b2c02ea85698bd8
9841ae5fb17d43d431f200f7e860ec39cb74b9a9e6233245a049bfd69481059a3addbcea111063715946979fda56c788c543
b9d0f9a3141d163d133ef96535d519bb81b6f665e0978be90553e5084d7d4b48807241f9d07ab94282c4001282474bd8ab94
a34afd324956c6b1ec3d4d2bbaaba9a256540af9c77abf30db4b0d910da044261e724746f95dae5069c449e8b45f68af7692
dbe94d406731e6288131546d3d5cad8c88bcd41f7d599e59287b3de8c2aaba5dcf6c87300a5a61f332870cf59a6bc1e90856
3eead10b340690418cd4373b3442f067bfe9b57157ab15b5888bb5db14dc78d4caa1b2d7057c8bca34db50cb0bf550ec1fe5
6bd2d9ddcc609b86a9c63c8ca8abb069fea0c38b37aaad673558fcb73862518c20a313012885f17acdda6117784ae55c7a3d
57a5e623b314d7a6d9e662c6821ea5ecc7655f46a32218ea57cb2f07ab05882a5601bc11b14ef3a0477a88ccb765ecc0a762
4f5e2063387c4e3e78a0da3048a282f61eab8cf02ae40e243f3aeaf19515e3b855b53cc84a0cb3dec92176ac4c0a73bb071a
40fed3b672aaf182425f505b1cfa9d8f0d22d0482b4f6660e45fd99eddd63f0b0a8f4c9b5e3b94bd02ce02a0f3b3edd1a528
d360e29ef2852a2f10922276bd08cc55085896b370192f7147b8d4d3de1e01d350a0974b98c59256a1125dd3c210098a746e
bd382a2b6064ba0fd3ae79842d733fca5f711a7e59160eceb82c526eb684d75eae8d104d8caa83268362673c038403b6b6a3
83847fd6f3ae7d8e32159cfd946f445a6495129f11018a43effde82b6ddd50ebbffd63de323ed71d882d10fa1c507b682dfb
cc0e126bd0096e4fa1ea21ac57006ade13d7b0531b2c08eada14d78d6539553e6228d0176237dba22f9fb82e6afd87cdca9e
cf283ade55699f71342160f56c842c20c5ceeaf3d8194de906ca827f3216a9f9b420f604d9c66b8dc54165f3b5ff94d9e367
2fb83259a00fd5ef9a4f080fed874f681c3332217419921736d241027a3411f6413e1cf7550b0e6a0f6d66c88fa136dfa0e8
5e0753ccc06d6995dcc1880b3958f9ec010207cb824cd57a0a80e97c453087e4107266b8c641048a9dff0cb61cffd4c4e39e
cc98b927940d99fb3abc6ff87ac74191f287cda4d4905930d588cb193204081f94bf8f4d2d35e07d0fa6472df1f553608b5a
23c20b71919de5f021ce71c31a6b6d1583269a019d05cbc3b67178306d93edc811091c225b0585bcdbbd1ccc8aebf77ed6af
0a527de65ab5b7de338c76b38fbd8282f90b1252636b7a99fb2df900bd5eddf1b217b554bade66390beb44f5f94e34ee71d2
b25868694d89c0a136ee7f66f9ad41d2a4ec0520335cc7f4a920dcbaf6388e885f2696bc784febe0bbe0b19d5d4a5d5bcfc3
2ffb6dca2fd219485b634dad647b3021652b16ba1c9b0ec45df4b21f33f2d851e4d618abb5241177d4aaf71552a51b3414ad
9e1096cba8df525e5ffd2dacb8539d501badab153f10a3d47fb2112eef48d9866122acabbd11226c735927aa6e0fa60783e2
43b55c2ce279ad28a7fad1112337d7ead816793c58b2d7aac7c0de540391c2d4ee5a6910293b0b3d8711b3ec7821bf8f139f
9ad6d7f716e4dca9e6a6b9688529ed3c160198403b4a8418767a88cdd51b3154e883aa5dafdebdbe37011d08334d36142307
0ff13ebd7aaf91613bd5c20f7ba828f4ec901aa37bf2cb4ae38dd3528f2a7b484b0f8524f5ecb2a1c20a1671397b146ad958
2dc87a97fc0dab952746ddc6a382a8ad99f42736afbf109058a6dfad9e4215c065ae6c7b3df959fd87653dd43e909bd84152
4f4e4f1ba4212c29670f8881b22dea524806714a30a100f15a8f1ecbe887d96d68d5c3cb7c83149471ddab71e270b60cae9f
0dc16502b164996e3daac6a2e5820b33f94a21f404220a7b9c01ca7d977142be260384c770cc87a281962f8550904ffe421c
2951857b827cf01e66808484bedbe40fb940eab0459eed6c69b6d81ee05e6a54dffa028575c828584eaeed4c505ef8233c12
46287b47cacf888c4f6bbd22867cece5e887520a560bb94cc175c76469fbfdaadf11a42594b160860c4a8d74417d0cac8155
73949141ce1d1c8c5516a02c5db16797431a4e51b000060d35621497fc085253f31ecaf0f4a35531e8ec58784231a556f30d
8dc10e9ff0566faa517ed14e6fd56f897339486d9ade4dbdf65c717430cac91d2a28659b3476164c691b69242f7e4a93c78f
0c87bc4ba1aca6b940880c89547200b4387c44a32243a3665f203cd7c25764735c0874b807aa218f38d981a6b29cf6d65fa8
1e8df0f81ae20f25b26e983f251db9758f83a4d184a06bc4578f6dd047f92eb05d852bcbb265bfa643b81a9498f4a8564b1d
0f88f7da2f7f9121dc7ece8978ec9e7f2112bcdd105e5f1984021f92e7e50b8b9acba5d70398a9e31a0e684d9b5f8c98788f
20ad67578416e65f2043b044757c808cc8629bbafae3ac38fd9d5bb926c1b17d59386df8e4eb545966b3d6cac2ea89d5f21c
3bcb69de97a3a87c34b672cc1c468c4ea68555ef544a27068417c95dd182ab8c693531687b25dbd1eead4739c24ec8c56f51
bcd0aa785bb82bce743f8ad73ca0a08ec08245f9580b964ea64d3076728030e560a8efd4ea3d83a5247fd2f6c398593e6264
9bd99195fd3816e1837c5d3de714467ad41caa51870b761548bc56fc4deb2fc8b2e2e4a4fe3858d574a69babc7b5193f332e
ad6b79115cdf74b31ca72ca0714596ee712b806ba47e5817b920808c16bed9b1661d76abc9974e7a1a1ecb9763b2e9c8faae
4bffd1eed2e5b3a14e88484cc4f8d5b3785b976c7acfd29b1c1a8a07547d76ac465b287aca21130bae32bb4f7b40232a6f60
3c5484209ea0b0857890694f7f40d809f4b593f709151bca7a6dd965feb6b2821ef4a602caa2d09842acc27c1c822248b195
13fc6cc1344885c318d89b40f6662193c9dabf4353e4075a57b02b3a4b4dc50113f40e16580346569c9ee9fd38d4bf20dfb2
ee3b21cc7f086895c31f4a198ed0e4982c43b7a0c5fd7c2ff77a185991a10351895ddaa10c3ee56b5fe48fb85951aa15f1d9
f5d5c3f60252e4aa1f88db283dc1ce71bf695d6a70857c9966a8d4ae56bc68cf86860c84c9d0c87e58546559e5546576ec7c
111d2a4f782031854dd654b374a18a86d367f90be2c988905fe723937c458cf89cccbe2827d29635c82e2583407be836670b
c0d1b4860f5d4c082b60ff26095946a84a43e5adab87bc28240b2d3579e0f58765c60a70b3a8cefd75821cd0a992cc1f18a2
817aab05d01e723210105fe4d67a4469b6f2cbe07eb526bc6cc8b3c34b957d61f358e6255cd5d83dda8a53d179958b8cb2f1
84901be46f8e61151bba5ec3b40212383140652a6b6e91bbed2727efb21beaa1b452f634b51e6ff9e9a1a5a917f3bf70518d
2af5f54b80bc122a9fc965031630907e4229463bdd297ba8cee5618491b2330143c8fdec72bcc8685ca64d2ed037ac99370b
b2ab292f056b070cf2ad42399cb1c57a7f5deb79ea880ce34dee6e70845566c7f23b2f358bb50c0eeb645b33c95f8f9a7588
1f69fb3d800482bcdb275026bee0536737a8fb6f89a26b28632267bd1972c02b4e192d781e91ade5651de5278cb30c7b2b20
0f684fb6e3da711e646b7bbd51dbd62e6f2d8c1f6a4628bdad3e06db048f5045ab9a171da2085af9ba22a4c1feff907adb93
1abb32336d1ff181b87a4130540bb95f7bff5a08f8965586eef6848e400a6d5a6c23809fbcdbba49ae8cfade76718859b650
ea09e3101278f5aed6246871825bbfe0f24196b6edacb759d82620f4cf0e426ddd3acf132e5be5e4d4632d5fbeca61dcbb3d
d47269831a4c3a8c07b6aa28ef08ea9677080fde1635be8ec875ead6a3662df22b70859a6ce0ca220ac111b650acaf08a276
66d9babc2128058db6474403f2d1ca78a9d84370d980925a48bbde6f9790a9ba7572cd854583edbfc7aeeab35810f3fcb9dc
8c677aad09c54247ad26642cebd6affa8a008118a6b1b6cc7559b9e1b8bb4f8e9926c836a086ce7edb8b1a8ed39c0c23b8f9
40851dc4c5d5553992d2864d4ea9dbaf052f5f79debce7fb5972dbd2cbf7695eb0888742d6ae2a57d600658e1de5d73be0f4
0c055f2c30c576411067df74bf511612612e1065b58f00d50c7a7d2685048105023ee33e470ee5975950a645bf50654a45d5
1f08704d35f0b4c3c1a5fcd171685ede1194dca82b92f2234dae0e0d9f5ffdb2840a6c966b95789c047593dcb7d9c7d821da
e6a8cc8798abb2aa4a8ec13b3b2a825c7937eef50f1074b76c72ed6c95bd31980e7bae0744074d6759f51b417cbd15649f2c
f01f85c9513846bed85859e21e61da8b26a2221ec27c925362c8db9896d29edc800ad90b043be5066145b15054489bf6e0c3
b91bb05f58e54a14caa898e36ad27a799b30e7e9b70fc3080a4eb7d3aeba8a3c693b3d504fdeb0f444f80062da2cf6dfe2db
3abd5a9c37948b29af673f54a507a416429d669367b3a83c80106cc43bd50d3bd233d7511ec923bd1cf99788deace76e50dd
43d88e9caa9160667a177865ab3e050a48c84d21de40e8230d658f5fa391b034c109be1e8871d388429acbb06775a384deb8
6cb235746ecadacc0a52d8a331e70c5db259597e41ab1995686ab4cb8ad35454fa503f254aa90eb8521faa4e3612457a9388
d487c4134aba228fa92e8cac1c51d7cbdfdd72a420390dd5c4b56eab7be8d66c7217091d8e0dd21ae3baafb65142cf6ad8a8
a53c4e0bca37d62d559604d9aacb3ae9b9a6b3f91402a75dbde5995040741e36b9a0c3a132aa524337caeea6b04c84cca1e4
045546202f638521ea2d234e7dcb182ed723c82c44741e8ab360542bcbe832072607912392e831854e9f6248eb645a1e12e5
68ea7c28c0699c20d1010786728581c50b764d6b9dab4d7c675e3abdc0c5e1374e5ae7650f22c56886aa41eac956a61213d7
ba8bfeae90ad43bd1dd9784544cf82c46e8baf446c8a95d9165f815d5e9a64e751ddb874629b467d803fe1be1df50f51ccac
ae757052bde929010505502d663bd651163130e11444bdabd09cc481a5555fc3ca08a11910694ff6cb65aa28efded8ed527e
f85190e5253fe07ec0f51fea696c556809bb780458ab9ea032d90af1b07ed7ed409599093aafc94dc2325888f87e03029508
94e749967f287d5cf6b0d35c63b39039d90f7a85365b2c37a4498779d792420cd0b28dd9ce705cad046bade3585ebe0135fb
f4bc67398e13e236867dcf8420fda5d73bacee175255acba72dd8fad16afd5c9874f883945d00c1433f7632314d6d5976a7f
b00ba711d960163e3ca064eab8b6fa72b947c03ae6b1ae1e9a4f16f03b0cf2671faa9455790ba73a48959d9895b8d6e72408
e6c292ce44e84cd762861cdc36ea3d023641c878aec181a88a83124d724381b844bcdb766afb875ab510a9aa9bbcf85aed22
7f81c69eddc0414066d6b7867f58b19ac452a85fea75a75d979549555ead9471b82c85569c71d9d554a11d795e83de85a328
2d6a5f95f16f570a44a1ed1e79aeea735f70e651765f489cfda9d54610842f17f233b45a9185b1b47678d0232ecc062a39ee
f426208a88b6ba6b42a957849ee8896cc0610dca6fcc638d52e90cf430c845f16267d33db287f6db0f08dae1fa5a8f81a8af
0175a6bd4a741985b0bb4ef7142609861fbb265721d7bcd5112d48db5a26949e1a8eb39a42773be8d3782b29d943026697bb
47bd40bdc6c01a1707406558aaea826504c640dbcae10ff329d26acbb8b017b52a10a065aeb8c6ad154a2b1cd2bf284e8f2c
75b931b4fc5dcc635023dc6fdf16c472cc72f6458e578f02db8863a9d3de66117c7aaffc0b01d9eb0a31223bfc1b11203a6f
7a1580fc6068884167b49efee1660de136facd9e7092825899bdf6140a458f7aff816099a5433ad89e1fd2a27ac280db7d35
23a0a41fc6cab1af7113a82536a19ea2be45e82cb7af7c646d2d31dca12618666f39a7627f5c7a5d3ee27aa0379b42632f51
96f55f797f565c874e55e873461e96fc22476cc83044638f626b45e4836e3a92a41062a732d3b497d1c1c46182b73a4c0ff5
49ca761957187569840a723dca22e9735c54ecc6adf62e945aa634c8c8cbfb5f1489401d5a8b36b242a2134ee8e51277b455
ea065dad3a2c74185a7b7d853a2030b487ba65bb6b17ac90df2de3a6ecf46c7ab848cad833b6a082b524b146705df61a8864
9af6330f2c14938b8415178008d1abaa23c859c54a54ae2bb11b43542346913d766d5c4d86541f82ff32c562cc08fbc6d0ca
237472aa41b5540861961f7aab9a8865aec119aedcd84cb8981b51e4bbee823abb81eee57866194438a9dea6650fc680e4cb
a64fc5caae10b257a31d90e0dcbe7c2ef3204309902c384ea6cc8310ba5a02abc5b9830e56ed111a8824ca5d626b405a7319
19647cc584111be501214b57776ec850c789a51ccd105a8439cf028dec1e03f7fffa861329cd76815f6bc797110465c793c3
115c902056a86ea951fd79859a83fc0290650c41c8c1ded2c20492cf75f821169438d19aad088dadcc46440c2ef2101e5a70
d0b2409df1fad996af6bb68a85d9067c306d88b5ce60900cde505d401e405a425a6f2955f6995b7117287f65bf226aeef526
c38244052cd097551eac632c8358453fefdad32bd68808d592d33f6aba238261996a69bc0deb80551f004de50bc158d4eed3
c680e2eded32c98d1bc414562b943dd618a3ceca492f7afd84b0a1d1b6bc739d49cbe4bf94d77096cb668c0a0b24dd709357
b55230808c3a3414754fcb4c80fa6cf5fa0a05ed91a6a5371218facae78773c2a9ae40cb1bb8411c34f921cb467eb41b7b5b
522eb63893eb2784a2a194c9540580cb1839e1b84d6f1f4b37204e5976e75d1d54171382ef5a1d63849b31a8776dbb1c5187
2c991e170af294756a91d3bdeedbc7de4af6e9e05044708e48294078edae4e8ef1182b063ddbe1ce7cab7ac2b622c274ad93
112cc2a8bc7eed74049aa0147579153779340255695ca52c47f81252bbbab94b36a7286782826e88e7eff78b5f68be28ba67
6ca556a8fad792993d2a6c21f621497580fc2f5496977ad90f61cc71d2d707f82131a92032a2c6cfa152156e0be5ab3d629e
418ec032ec1827142cefe591446f2b3d8b8ea99ae3a882d1ae8bfc6970f707cd19c404d63d0a8a51e1dc5edfb995c1155285
653c3d8258508172d892cf66c4a14f671974c31ef3826bf4458e52bd053377a8c3513389cb648090d8e41ef297851294afdd
664800c062441e80a32432ca6facfbceaacc67d8dbf772e18722dba61edb1d31ad33cea5ce8ccfcb8201a2ed6b7ba806a382
1dc61dbd1286a4c3800d751d2ead044437ea675b6a304aafd80121b4d59049a3dfeb1ea12850349f97bd1b65d8b4658fe2ae
ac115000ae381e8f1a1138db1ce4621fd2fb389e9c8f356567258c119aa14fc44c5b6bae85b0b028c29579d7ebf8cd1e21ad
d091b4682f0b51c06ebcd7712f580820ea1ed5c7edddc65204112272b93f61382d631f024f8d13c47094b95b6f2450e77d46
763e8ee2ebbe03b1f783dec96276b27abb169965810448f35ff4e20f0264cb8cb2e37b31472ce82da5539ff297f77a41acbd
e9d317bc659e429e83fcc6b033813ec7608522ac2e2bb2f5e5d45e9695d86d2298dba631ecad9742b6bc04b76a8f568fb86a
672f16072e9723263b0a5d883de7b48cd8e398ee9091976fa926dd5eb7a8c7be3db91d2e8036e4e38dbbe857e93ef294f5e1
6d1957115eb7a172423ddf3285897595cb4a54a54352fad69bb26f798f30b3cef2981c2b068ba2c738b30fd9a80cb30d3204
6380f273f949902968b301061b14b2d13b14dcacae2848bc57ef2aefe88acb29f5b3a092b345cdef97c923c422cb972e637a
30a6a2a224b219ab6aeb8adbce562f13c7deb2102cfeb7d257562d28c527436aa1265a4676e4e3d4983cc45f4cf399f21c7e
4954684448efb09f5e20911dfa55722b0105430873a39aca5eb7d5b208e56b6d5944389c1e8fdd2696ab50269093de0c853a
2be8d8d6f32d541d1b5a399be2180ae5b85035b64640a2e8060e23b29bae0d09ff7bb5cfb24b456cba0c8ec1d501922171b0
5b87ecd6ceb7f4970be5f10e01a30815abb7620baa844dfa9542c881c97b982c917d0888c34c0215a1258250c3bedb578a23
f2c5b1234f2648dc3e99d8ebf191f54be99bbe0ec005276a884087bf9e31745895ca196f44aef6d8efc1b71382626c45af0f
7eaceeee865369cbee42c4eb36d256ec9f59d1cef2767ffe5bf1ffc79f7ff36ffffefffacf7ffefbbffde3dffcfcf9bf36cd
86db039cd777c86e82ccc6cf5f61c4e0d357eb804c8ce9e72f4d21ee97dd922f969b3b5b91a36a9e70263699d9f44260b5ac
2e94752ba376f9849023b4db111554c6e3ce948b4cce077684738d285e077bd91a40dfb1da4d9db52ff6ce0e6dc7da1e053a
ca0cba987dc6fd7c6d8f8df4820354d82105b23f7ffc855dfbba9a1d57d1e368f6c94ab5597b139b2883d86e5f4c64bada71
2d504998cb3abf2f8f32fb8ae0b7c1ece52780eae76e379d64b3230f0792eeb5b99d07c2bc219da4abadb7ce6a6257336a49
f573e84d679f02249dd07bc4e1afe6d5aa55998834ec888c6ae7dd6e03f716584331c7b254596a7b54422df300ec162a553e
a6faab608d55ed03ae16da2efe8a9d95b8c7fd337e73e875aebb7945c0c1c66f08a44ba0be37b8d7ac980784c24152baacec
70e680b24be5f5ebac221ee254ab1d938a3d7d80d8c786138fce36d9a830b8dbb1125c615f6d07d859e7b181335504b30f8b
69ae563baeb0f17e4380c0747960eeb15742d154336f462beca897d0a3102eec138452edcd192c2317657dadfd804964b7a3
8204de2872db6f08a1437910b3b7ed663f55bffe82ec7ed9f01b2a8812da6320f18cd0b66a2f4304a260aa7d46347e1fd8c1
626db238cbb2e147bc6d995976fb5ca618438bc33c13bea9f616a16e683f953f2de3667dce8430421c6ec0bef4167455edc8
cb5fcd8c75d3bc5b5b8806811c1c334141c6eca39dddf4666eade2d96e4666d284a70c503943123cec036e03b1fb85bd5b10
c2b1dbb1d7b49f7640b8505b3f1f0834e2d208ed7b8ba4ad5f610f91bcb2e798cc0e81c0fad5621934237beec73e0da461f4
bb1da195c362f63255969e9a1d737107810cd8cbdcb5d5af199ade1bcec27ef049a2c042df1ff6b2826b27b3e3eeb61d767b
dbe102fa079f360a580cb5bda9b6637b699f3cd27586dd0e85e7adb6ef0f58569c1941423f184f66ab5dbfdb4d167baaf65a
f20df6b9bc775024b0a901c3dbfab18fab0d17c50e71f1bdfb50456a71b90a3b7446fb6db70f559008e31e420fe66effa051
c2601acd6e0778d3f1a1236b70323b62b0da2e0c17655cb5bbca71b743376fb7236c6c0ca3518b97aaec958e51ed18bd5a53
e0453ac3616ffb69d8edfd3e1a9d83638b0c9971da1f6f5954b8ef851d69cb750070832cec189a8e41791fab8b193994c3c0
237b8b046ffb439e0860ef4d9988278eb6afa9e871a241fbc9caa8f384d5e2be0d137f98e03007ce76f2c513a2f5bedd5f71
d83b93af61b46ec26d11158f30f73041173b0e3dea27dab6a6e737af4cb29bfd8b19fbf8196edd1a023fa11d0e9bb5458da9
9e7eef6d85a22472c5612d43d8684c6e0b6a7f97e1d0ccb8a3b2817143c51e4b48a6776cc35e7598ece7802218ea4f553374
92da89dfe00df9fbddb2d60fa10ce818e2caa26a447458f83ab0d6ea90d9651f194affda10b7a1e0b2a55ed8b73721a3a19a
cb9a1bf204f4a96e030a2f202f1aaf53d94e8fdd6eee16bb4aa67160c31c8c64591b35a05a6d8f8674054a5cdb9b8a9766ae
d60965727a1e7a8a7928ab01fcbc7e64dbb03681146b1d08717e3757330e58c79ec7cd75b3b5c95487d905f9356335a3facc
b8f1a8bca20aa5d5d9b00f1e9263b0aebfa0281787fc15e1d6c89be4896385fa011685668656f56c6654521a3b5b12e0c41c
3b6098a18108b1219eab8a79d820936b535e8b48465811638bd3ed6ab5006c98478be8dca6dd6c0919c53ca0f05187a1d5ca
3e94df7437636739aec1630f0d342c0cac7f6562b0d5e8da5999be7d76dc3ae47054f30a39fe2160b7648075aa665c24db12
78b5b8b2766f8d0da5cd45cb86410f159789d762c67125d6cb169c0b295b9851edad85420afd640bd63f1036b61f786b2dc8
a8986d3fb11aaf50a52ccb3e6b6d3ad16d9d5ca7ceb42f8a75b21966dbad33a685621da1dc38ede6b200b151678192483fee
afc8d90d44a8a0444b9dfe5b28e4c08cf1cd126961c684b0187214a3eb075bb978e438141eeb4f833cd8ae4e860b125b7bfc
1e366de1d46daee6ce44acf93758ac72d4d2d7170d77e1f6b32f16a219dfb3b90c7208aadddfa819796dd58c62c861c10519
86099a9b757dd623a4b09a5bcbd2e5e5dcbc40f502da2d30cfb59e6a31a33432ee35eb1a126ff961468ce5c64bced9aa5195
363f5888e02ed2689db1dc43069b997bdcf96cd58c3de53c987945aebb757044eb11e30bcee8a16eb85533f213313af756f4
15ebef621e70656ea3736fb1cfb6eeb3bc3f845d987580c4be518274722c716045e5c2ad1a918f85a1b9472ae636dab03f43
8a11a5797eeae26fdb6c793b633734e39a14e60565ce0c2224485010d4cce53b41d98cbf34c8c11db0b8c27ab366439a1129
117813fac1320d6c682e66c4878f662dcbf3d6d898107082ab3c2c6521116b8345b5a2485535e3b214e6050a823822c43a19
358aec658204d484b457332382db66fc69b6cf1503653fdbc9d156ade5bd472e34acc8093646213a35b4389bc08600315436
af4d58c8a11e60356302b65ea3b8cc88b3366c3790a7dcede61952828b99cbc60a5921662eaba66dabe60de1d766c61a0a82
3b3fb69541a649b5965da48d653d2a940ed3b09b5185074bb8010b0768abc1dc231608cb396cabf07a6c5b3523a3bb336b79
5c3bed8d5b08499737a1eccd7002de2dbb1913c76466a497d96275c20b525eff421f0e86b05a5eabb9bcb236f10e569f6518
7673e92fcaa760bbd9e113d80e7381be9ba775b5d30c3b01e86d256967b59092aae615a5a46633a30a7b5fadc808c7bb372c
78c1ab43ec28107a54ac582e209a767ff286c0bc6286ee05d21c433f0aabb3d565ab6020ac5bcc63cdf2ae1e31cdd84e6e44
2ac85c37258e2728e5a3844ff809460478b6637d737a08f5631034b3492bf28f3e6235dbaf36e0e015c1755de9f68823826d
b3b9debd50508eeed63af9a0a81b821506339b9eeac8ef30c4aba039b39b71e53598d556a2137f1d63f93ba8d22cbb79415c
aa9991bbd0efade709bbc2623669b1cd462798472491143334fb11f759cdb806b18eac56b5a75d82cb1583c8feb2ce76b781
996ddc4c4b7ede5b2315cd8e624c91b2b5e5c564f97e4b385e72d0eddc66828074256a467d043eb629b4020b066c7f0ae3d8
2e3f13aae4d8cc865bee02ad1be810661aadac03de4a7706f33963b4256f792e3415fe0a33eac0e31df12be1d182040d798b
42c77d9d78cb8b8175ebbaed6b6c1cb31ee605350469e93da250e3dcd505390e29ea3350d0021a93bcaab7db2de4c3db1660
295ff252ad282bb754337246ed352b2f00aaaaf41ded2e8a1595e2877edf895811f86246bced613d76332829d4a1ca62b5a2
049bb164e72f9bed2f91cb627a0f6646d0f5baf0ce6a34514be422d8c6cd72e18bd5a2ffed4528d632486f36848cc8b65feb
521fd718ddb637de96d54ec65ae835592d261b2cf6cc65dba50d38eade87966db675373669d8047d86b83a86b416d83cda91
8dbd6f3d527dcb6e0c576e0b0faac56a6548876ac57d68f753b6681bc298f701045be9d6acc8d6987858d9c0d6eca68101e5
2f37442ca3e676b5226f1de7d873fdaf6326e911298ad36ddc9cd99671b28c6debee089197ae3bacb8ce587eca0e0aeffab4
0f569395ecf9d9b05bd9eae203de3abb59db3a8b761cf6d1719b91d665f95d563f8f51203b68eaeb84686b60dc646e35f5ba
db474c48df96e7968d4fbf7c9aee3cc288404afe29cafe06a27bfdfe4bb45630a018919bbfd14fbce2e0069532ebfb309918
2eac53e1bce777a758cbafbb1909e7eb876a2a7d5d8d9f6f307652b5feb17fdd61c599eb401f0c1e803bf1fd3382bc71f9e6
a80b9f2f1156eccb17ffd5ae183c90e6e9bf7b6cf110f530d631a2908da9cb71708e27600bf2dc2b0d3eeb6a95d26dbd3e42
fc0c8a58fe473847b5d5aab7f5f564e63c6fde3adb656e76dafc395586d5028d7ee85019efc78a8a8a3f74760cf30c4da61f
3a3bde2c5bfc78c8e7e818cf1edab59e5e9e27c730e3ea311c1017738f33847aa6f93920c69e7e41ede11f3a1fc6b940874b
dbdd5c8665dc4efab7dd9d0e6f26505a577a305b7cf34a5f8c3b1bde064b54ab47583d2aa8e27c86be3a984d6468a54fd496
bca381749fb335862c7947dfbe2d84ad40380d149d658cdaa5961f6b7cff566c1cea5ac096cdf8f8061aaf3a935c1a91a8e8
47bc4a896d8f69d8f407e9665f10695d991d2cdb830765d897f21e0c238fecfe00bfd8675369dc7f4fdc2160574607723dc4
8706dbadf90336773f800336ec7086fd6de94dda88cf4edd2d034e16474485ee76ab4c1826587b173b2c52c311afbbddc091
30ea1174fb2b5ddef96e5bc311b2bb256901d08e20aabdac01d6e388fa7374ed6e5bda0d7708ed7c7c4928823cc693717769
d3221fed7883a12d32b6f5c0c89fbcfb4b21d4994622fb6e879803b64274e2ef2f9720723d4de33e6694cd210ebae93ec15f
5d95d1aa6cb96c8859edbbc505ce715d8b0badf31468c4dd4d5787e4f3c868b464807a69751e3b61818a92ddf6dcd1d2accb
477d3ed74e42175c3f0ee40dd6051a7778028e0bdbbded6a5af6c87842c6842d15cf23562469a0cee07ef0bae108c9346110
c8664d57ab6cbd5acade682111d58a056b996c262bc6b58ebbd546c619aa789022af47b19f635ed47241d8a01de7a2faca82
c2e5d36032dd36d89fc7c750d0c629d154ad4876ee070478b7e86e3dc94650460b61432c9421ad7198f793e90967516b3dda
72e7db26dc3cd403509837d3a78319c193f5d0fe73488e5d49f968f64be9cf59fb84739f761feecfa379b44661f3c30c79b3
f29bd2a33fe7fb13ee3b1166fa43d70193d5e71af66760b0b4c2d68cf173a96021ef43bf5533de9d1ed2614cd4e76a02c9a5
10b0aa371628bb89f848fa09cefb0d2c8af09071b7e2d72abfa8ff6deba136c24de84538af4eb0d699a7bd775badd8b1d1ab
6447dd38cda5d7ce5dca60058dea4cc36ec66134be34f73edba78efb037af7612d332f542efce7d35989326be3bf1f33e3b4
973fa0ce94f8c69efd59394a3b45a44ef716105836678c10b78ad06b9a98a4f3b2111bf90d91afbb19a70ba553c434cc9027
0bbf566f6ad2b89fa59fdc6a49adc8e9a2f7c6ee3dfb32a2fb37b2b7f84aacb66c9fbcd401bfc741698f90a40935f44c5877
6f0c71c705a279d0da18ce0bd5016f18ae2f51e3fcecdc507efed98ad21d53d80738e2a1976dbf58077b48f847c413c247bb
a33164272d0e6a83c071bdcbfb5cc8ce2691d4572838d744bd84d9c4974d13e787ae636764395adc2dddbae2a0b245f0fe0f
ddaeceb6b1d8431f50400b1a14c588c0c1fa639d37ae30979fa7af179f9f8b558851f7ddbcf7fa737f8af35248b8d697fa73
4d8a944d489ad6ee7dae4371e88a68b223f663bf0d9d47cb0cdc832d3ac422604d51cc93291cfcd065e85cd319f7d69fcf6b
b69bd67d49f7f96e1107d54d7cdf8983624437f63ca6cc38bc18c14c352330acfc1aa8b8838baf99c72b285fe08c310c7a33
fe6b583eb13cfb1059cceb68f2da34ccda493642dc7850c6b9770b6d4a1ec2116ed72f6112c0233614f0e329c342f3f60be3
7332c2834764a4542bc460c7115624d72e334d7233de132b42eee7d979c1ad478d79b29b4c685dc00a856f3b5528d67683e2
9e05132219c9ac9feb4d08fb6235b1df6e228a7d986185cedbb472db0d994fe3b4cff5a6fc6dd675b1c234669d20fdb02035
b77c19d3b2517f714dd2ee87cb36ad63a90b2bc2f4f9f216c6b2261af9fad76e659031b0f30b5dffc2c9828ddfd631e94b6f
a7767bdbcf4f04e163d30bdfcde8581993177cab288bc3bf72316f7829762b0287cb7a7c19abce407853162b5e361f116565
00edcb50b24c26435517aae7db56cca805b91d2f61f94acaf0b7cc28aebac74be18469c1ed2ecc1b8afa84bb7c98cb7a6137
7f5efb6581c0fab42dfb4762eac813cc90989a07fe76a0f051e691314ceac58c13cd8fb5ecc4f0bbae5873eebb83cf37bcd8
9afb888d3bbe76e87943d63e844314f368d918bb197534c134e67194afe6f1a5985161613b6230f6416ac1ed4019f5361ed2
562b8ab27547fcd83e00daa5e732d6108673b85cb1c42fefdb11563663c560adad5af7c463ee6ac56dbaba733f47e8158178
fdfeab9f83fcdad5a55cc75302d6f4961abe9b51b7b110b4da01dcbe783ce7951515ed904dc893d35a45ecfb105cb4f626ff
567fdfde4a20e144bf982d5c77decd7be812cce5d5aa03f7190155cce3be98767154c56ab564fadd8c72b3dd1a1e7d0469ad
b869e98f99fd08f40244487a1f3bf16eb32d0821b70d7a8f4d31fd06e7061d6f43f94ec64fccd96af529e8ddc1c5d662f10e
f4b2dabe1d392ffc1d9806cd682fb1ffc62c957e458552fa22adf58625157dd578f6884a5d3430a0239d6de56970c1b9c28c
92b47ed8ea6d2ee0c1d0e3c695110ec40e33ee6a7a1a65cd8cba793d0dc9ee21b860de4c4860ef9c693bd360ef8e2b6608ff
acfb1605d77f483a1c691a71871265cec10fd0862044d432b6ec27336fb84a8076a99fa1dc1943317758ebd843ce2303dc47
5b1946337f4e068a1929aef5d73d0f0066841574f3fe90cf3e1f13e88c8337337fb6f3c56c05fbbadd8c4489b6a349d86de6
61c6455f357f76ed908d5a972358f68c4ec3351e94913b6c04602f7f3b0f74bde7a3d3507b7038ce005d741a2e098fbbd71a
8b6b17647479e8c3d370ca00e5dc63802da32bb2bce86ad287a701f6820222fbe88d905b5c5af81b4e17b606f286797fcc19
b666098e7b38326e852dcecfdf9dfa58b6015719167d445311ec9389e4ef76bbe7dff8c6d6c7b8e16747d9c399e745d8a12f
d48539176fcf868f64e289dbce9546d36dadf3fc5aa32ffdfdb10f71c34b88f3ed919716b0238ea53d02dc8f58365c5aa3f4
c911e18ee37563d85d66bb1837bcfae356afc35c381b52e190e5b2d09a0adf4fd962d497d2c5adc18cf292c30f05a855f366
cb0e178b06b99376dc8f4b6c8244d62677f00c511b10650e99c81f0b3a83dab5fd4a8e2e17a206723b9308b25834e817efd7
ed9fdff40c51837eac259afc58d419128c7a7e8fcec0b50162f9a842f16331676590b0a015ffdeba08b5621e50e9d082a3ce
50b4c1a4eaf6502a84a2ad63b7f1c7e242d1ecdbeaa67aba764697157399682af27529ff3d8fedcc5fa80b3a432675ad474a
d165a8e73d99aa1d8597c18cca92663ec3cbacf532db06c9859721abb6b3faef7b7819641d434fcef0b2a1a6ae1bdbab45bc
e0a291b0a3d8d61e7fe078c55aa33c0e012dfeb7c1d40f75c2857fdeda1a09b2f48ed8b32d3d935eb4d52ec8505d8f82405c
b7edb4baad5be815576fb5833ebe64455239040d3948054c7508d0a48816c79fcd7473bdb02c6c43197e0dd132ee47b08b06
0c97bb7986820307e2945fb2bc0b35e2c905f9b81fb83795efd14ed0572c0e50288ea287dc4bd25bc5c0610d718cbd05bad5
ab1fdcf5b5888be32826f7baf6167037db0acdbddc3d82dfbb7a7ce0be1084c1cffb767983d096c523fa282bf795152b8279
ba3dc4b4f8b3486a17be757ebf7690d3cdf53416a501a0bbcd8161e7c0d021251e597cfb28d299082bc59c61ccc11157cf81
6b6e28321550abb8646648a3836b1f2be78639141a1e515cbd9acb0bb259289f8bc37343280ad36cddfeec73c0b558be698c
61c5164abacc036f906146458e3644327758bc4e35c6cfc29eedd8807b72ce1f10b65bf7b81d3f0d9914fad82fc711755fd6
7143ccfc70d3193245e675decf60cee9af5bad8e607b9c88a062f42541c5cdae90cb2e7ba2ba804118f880f437feedfc2cdd
61adb10ccb71de7dcceabdedeee6e99382b54204885f18bf3ac0e1e6668a31d58e6265888d722fa32d3edac96248dd0bedd7
24881b2c1d188f787ce8e3d967ebbe2d3b23df4c6f9d3e5cbfe681922156cec78ebaec8afa79e351c16ebbda3d6ccf0d226e
9d854b56d3d1d8ad93d53ea7016a0f219fb0d6f4c39c9dab970d49c763a25bdbf5761c749c776cf614ccd57e64f54b472ce9
add279588262dcb6c2f4d58ccd4f1b62fafc0ab7b70d4b3b8603ca23f9690edbf96a2fb35238fe841d311f61e36eeb298803
d7bdf899cd32a0382bee8b767bd968af3675b9c9c867c50c365e4d6bd8d3db9cf6b990edac14bc45d5d1f2d125ef60758a78
ad635bdfe32d0d2ba29a04646b56bf8676294358fd948dd3729c6723a624aeda3b5338b700645efdbbc74cb687dfad28aabe
2f23dcdec2751d2c6de37e2e8ef30b5442095b1ac74ccdf9afdb2bcf24767afbf2dcff204b810a6dc6f80362a243c99af87b
e3e58644507c3f3e1bafbf34ffc9059ae3eb1a2c2ce1af3e2abd98ad90b405789f31ec304f9082fda140787c749b4d10144d
0f336a06d8e2e38cbd87cbc25d5d619d01fcd822a17a6f588fe1127cabe718b02e93cdb4fe6adc2d555a64ff42f695d72418
72cab7643b74b7ca8019a73661d10933aa57afbc9ec0b3cbfeb8f6ef5c38a0274bbfeecb897eb2b987fb772e2780117ab73d
2f2760ded67a74e7160e8818849cce484b84fa1bd4a5f2669b3d8b3ba68c9a33bdc5deaa65d9170ee57193251ef9541d9739
8360cf711b77f3349be025a701a1c6256ae07286d188d5a705eabbd00a9fc1633be4faba6233b5d90a9be23336cc5ed83671
9887cf0e2abb26548b1d76f368e2d9142cb221fbad462bf898930dd71ba8674b112a5892608aea299ce55ca820f6050510f6
b60324f239280fd94f580f2f1455b3e1ebb2632c1f8283bb847ee9470ebfc3fa07778c1c7eb7d92936a2507d20905917dc0b
f9602258716a3e5140d286f2090844a3f03bf40cb1561305da9d8857abdc331cfd2def210ec15ce4d7f96be0ad5e96997f67
b49c2159b45bcb7bdf77f4f79f976dc5a9cad2adb4d7dd6c76abc90bc5382c1651417d3dbe830d7a75a876598dbd15e365b0
9f2f09715dfdbcd424af2a54d373fcd6f9316eb65f5b37fe723714a96eb7bab55cabfe32ff0ce790b021a7761b6dbd8f00bd
de3632fe873c47950ddfc5b0a777e124157a1ff42a9c0313be1b28b46d348a6d903d5cec54784566d486fdaa7fc3ce71b058
b7799d8e38986db1201cff8e9e43e96671df75a1788ec6b65a87aa75b562595428f1effe39a0172b2a83d48de067f4c7c27e
41b67f9d2a10a750faeebfa97302d96a2097917ece361ba626ecf3696ac2170c3182ad5a51d7ad8cc3febb3e67b70d6f396a
befcb8942b6c1650aea3daca5a67e070b6330babc54b0af9a21fcac26aabb4bbbd38c55c7649bde513bb61c8a561b55870a3
aae68fa5610d5887773c96b9ecacd612dad66d4fc3c295bd85d5b97172311d873a0bba5175b1eaea43087ac3a347e82ad228
ee1d4e90fa9f6dc65c0653e18c33810733ff329da13d03adcc81966fe567194f942d7a6af4836315a91cfdb41e196f65e659
c244aa562d58cd34ff93e979543d8e9ffee7dfa57a1c67e0aa8f487141fa2e74bf3cd51d6853048b8b4ee33fd0cf978f09ed
bd3e48fefc16d87efee1079a239d21fc87e6ff30ed933f9884124626dc0b148ae6b6fe613b5669dae66ffed4fc9bffa1bcfc
3f7ffa8f0db452fef4bf34fff3cfbffaeddfffabdffeeeb77fffaf7ffbf9edcfbffdfd6f7f29fffbf3dbfff9dbfffddb7f29
fffaaffffae73ffcfce9df35fffd9f9affb1d9e9ed8afb169227c57df9ef0671c7c38e19056d76cccb786cbe4eeb5f9c1511
1f5b9dcacf073863f955ffe3e1b222ae6811c00d2180c5a257466bfd5764f3cd4897984fe35fbc11c5866ae8b4fbf3d3589c
fd0d745c3c972386d08f3a8c09af9475f966ea30279b4b60b3fbad2dfffbf3dbfff3db5f3d77b810451add82bce57f30eec8
f447fe67818974e01e9a9113c251ecba0429280bd6c7a7b5d8665ce29a6d99ad2ea49579d8daf201627fd35971fb3f37500e
9851d613d6b1a61b96c74c16da8bb35adbf138d38c387a13846ac80a9146cbd8f93c6fb1ff5a56ef198529b08fc45f9f7d5c
217632d92ee883c5f2e29084bc23fefcfbcfe0a01f5062d2b7ea8013fb18f73444aea35233fb45a03834917d0f11a8d64324
ce21f1b613f369fd90733ef043e1e9d9737df6f1fc554e2ce7af177ee33fd7b7fe7cddff183f701c03d95fac88125f5192d1
bd8c737819fff7f221ffa7f219ff23bf8a285c8cbca1015a6f0b4abd43e3008a570531aabb2375fe1f8fb73469dd50ebbffd
63de323eb7b13c22880b4ff47e5b7ad90855edd33aa2142852e81c93c83783ced9409c172b6a52f5fdf9e38c0b64ed70abf4
f9099dc9fdd664dddf0af7bccfdbe33cbbf7ccf5f1f3463a2c9f37d7a176efb855fade9058e65a8eb5b0e4ec9f38d4c5eac2
be71d3541645abef25c429cab279f568bcedc47d5a3f049d0ffcd0787af67c9f7d3c7f9913cbf90b8adffafb7b8ee4bccf5f
dd7ad7ffbe4c58ff797fdfff52feebef7efb8b7feb4d3d7c7f093b2b2a8b0d37f4f7e7e355ffa79a94f7bb6bfea92734d82b
acd09b9fac1a67675526fe6aba0f8bc5529c564b3ec1318a9593860cbfa598e0c067e4e100dbfb16c156a0beb5f2ced8738c
105cec605ba0cb4736f7f33aeb62e70b65ae734f4470fb8a484ce7dbbd98ae9738748768f28f43b3e1ac1e0b0d87fb63fb33
d828a341190e7ccb1585a55132f37c62b1a1cc4659727adf96ebd4619777f6127ba6d5f2354f34cee6703beb8721f7c40f93
ceb7e3dcf5f2f3eb38349f5f51fddedf5f6c6cabf6bfbaf962ffb5bcca7ffeed1fcb60fe5fcacbfd77bffdc36f7f1f5616c3
3ed4d6baa5f67e2ebbf6a80dc0388fff0ce749eb865adb707ef3b965a952468875ae5fab7bf5915d8c02ad9b7ff7bb3a1e2c
ab7ff96b66711d80dd2f81cce2b2778340c4f9a3c188c0efc9bfffde48eb96d3fc7967fc533f2f97ef815f80b8de7e5e598f
ebf36e7b0edc6700338aab4d0bb5b5a43ca8c3fba7f6d05db5c1987ad0e12a04f5357d6f3b042261b3eb719d46e2e0343bbe
cea73a66cf1ed0af70f6d6fd62272ef7db8af7e0c69a0621fcf3ffa7c1febffbeddffeb69655f69f787fd2c5fd09bcd91ea5
efaca081edc0b0a15a81f2afcd809488111a8aa7f52fce5a5e74bbeeb41340f70cb21b3eec3f3f0d909b337c1e8972c9c34f
bd8b2c3b710ba6c28dce58b6fcd0932cebc95350b23104ffab9292d4feff5aec1df20011a41cfa4b765c90da7566784e66ff
3ca7f9df9ac4077c5b18037698c137d99d8fac4f1a43f5ad7cc03752277bbb5461df64773eb23e690cd5b7f2517d6f2bb4b8
aebe9dddf9c8faa43114df588e1cebe9018be83afe8e880bc568feab8c2475d192b5da172df9439a0407f0e18ab9451c42c0
4776c291e1563c95ed76e203be71263456bd53f24d76e723eb93c6507d2b1ff03dff82bc5e77f14d76e723eb93c6507d2b1f
f08df38ade2489d837d99d8fac4f1a43f5ad7cc0372e9566bb7063df64773eb23e690cd5b7f201dfb8316f11e0117c93ddf9
c8faa43154dfca47f18da4b37534c507f2cd76e723eb93c660bea50ff8b6b2c228861d7c93fdf491f529c1507d2b1ff08d30
f5b213b8f826bbf391f54963a8be950ff84604f2dc9a5a2df926bbf391f54963a8be950ff8c67dcb664a23ec9becce47d627
8da1fa563eaaef0dd932c2b7b33b1f599f3486c3f7d5077ce3447da99a35e49bece423eb93c2507d2b1ff08dd8ab0efbcbe0
9becce47d6278da1fa563ee01b09e1d3bc5c7c93ddf9c8faa43154dfca077c433116baa9d137d99d8fac4f1a43f5ad7cc0f7
86f89b69b8f826bbf391f54963a8be958fe2db6acf2fa3292179df6c773eb23e690ce65bfa806fe4f8588991e09beca78fac
4f0986ea5bf9806f0b0ab5d033f64d76e723eb93c6507d2b1ff08dfdd866b1a2ec9becce47d6278da1fa563ee01bb9b4a6f9
c5aebdd979c87aa41154cfc2031c4fb6a5be60f666e720e98fec7d752b9e5fdd6e65b314a74e32bbe727bd917d2f6e216286
38980d61d3282782629db315bb40fef63aa1a05fdd01e8860d35b4a39c7b4f6c343420461aecb82d11b13713b4840841db81
7899866db492e71b2ac62fedde59d4ed8206d4f081ac5a36d4f2c07ce7998d4607d00bee2d11c81e5093dde14bd890dc1db0
51cf74b25f625cca0eb46faddaf556c67c53043830ffd3cd9a03f0d7a735092cc05d2d0db2bfc025bbc395d1a0693b007783
553eb13ef6f3b622131bff40500914c6970f66d5b2a19607ec3bcf6c128440be59bace74414e768730634433b823771f20ca
4d21ddd23ad9962ee35c6f3c80ab860d35dc71df796293c02bb0217089f26e1136db1dbc8c0e4ddf011b51aab5ec0df20521
19db76762588ea2f2d42f20edcaa65432d0fe0779ed92408811ce1d756852a2027fb89306324617047be40841f47bb781311
3930b59d5d8622877f4341e503b96cd950cb1df9ad6736094220ef11d63e8d17e46477083346348307f276442176f47245f0
29d45a3110adb8ca9c5009fc835cb56ca8e581fcce339b0421900fb87f87b45b404e768730634433b8239fc7b27442e49f09
e2ac107eb09ac86506e8503178bf89c95a36d472477eeb994d8210c8c75fa806de5e9093dd21cc18d10ceec8275497d9ecae
08793608b119517f6945f95ee8317c7e73d9b2a1963bf25bcf6c1284403e59e9be21ae5bd8ee10668c680677e4236a9d4cf5
cd845e00b4de3007a3d0d652af0276e4b265432d77e4b79ed924082bf26dc1347045eeec0e61c6886670473e9437b0ed6b84
076a946c43bd195f26d30e85f8c58e5cb66ca8e58efcd6339b04219023ebc2ca2906e464278419238ac11d798f32f410ce43
d8d66a31b556a07c194754a8dd96cfdb2e5b36d472477eeb994d8210c897aa3077414e76873063443378209f5bab5f6bbdc4
15568bf812c4b8ae28f68ba0a703b96ad950cb03f99d6736094220377105241505e464770833463483b617943ee07bfb35f7
1bd4af836fb23b1f599f3486377c690a61413e217675406c352367bb439831a2197cc1e82a1102394a012feb78414ef61361
c648c2e031a3d6ffb1090fb7e3ab69d4e31341e98d0d4944c78caa5a36d4f29851ef3cb34910023972322ce9312027bb4398
31a2193c56512874be747549bf8e63bfa1a629563ccb3a948f64f820972d1b6a79aca2ee3cb34910023992e7a7a5bd2027bb
439831a2193c56ce1d6a706c76abdbd67d8dbda553f9206748367f16cea261430d8f75f38d2736093cc046be7f4dc564d864
77f0323a347d07ec0d1574ade8ead67d628bb0ada917d7e781976cd950cb03f89d673609422047b023f29a2372b23b841923
9ac1637b0c5ddece3ab641787041f52edbe3941f6ddc4ee0b261e31b1e9be31b4f6c1278806d1a7ee3e56290ed0e5e4687a6
ef80bd96be0dc36c671748335a5082b0351d1628ae2f276ed5b2a19607f03bcf6c128415f9369a34d605b9b33b8419239ac1
e300ac9f51f5d94e6d4c42d6aa86e01d85104b07a195e3004cb56ca8e5710076e7994d8210c8abd8c274414e76429831a218
3c904f2d32b9ad6328ec88631c7b4991a4b141b0fc002e1a36d4f0c07de3894d020fb057ab8eb45e6093ddc1cbe8d0f4bde0
645b22047228080efde53698ed0e61c68866f0e95b0c09afc0464a549910c7089bed0e5e4687a6cf7288da7f2a68ebff6fb4
9604006028515896701760643f01648013826c03247dc037d411a7b6bdf826bbf391f54963a8be950ff846a6edba5d6eb9d9
ee7c647dd218aa6fe503be4728b5409a2af826bbf391f54963a8be950ff886d2e20cb58ce09becce47d6278da1fa563ee07b
86c0ae0501b36fb23b1f599f3486ea5bf9a8be4d6923def8b2ddf9c8faa4311cbeaf3ee01bc58856289d04df64271f599f14
86ea5bf9806fe475f6d3e5a69bedce47d6278da1fa563ee07bfbb50cf3385e7c93ddf9c8faa43154dfca47f1bd588effe5fc
96ccce43d6238dc03c2b0f70dc59b19de8d759cfc727bdd17daf4eaf4f874f93bab944c591d93d5ef745f5bbfa144f87d3e1
574d610a4ebdd93d3ee98bec79752b9e0fb79607dc5eee1ed9ee3c24fd91bdaf8e9507789e4c5da0bb7826bb7391f54823a8
be950ff84605bbc154fcd837d99d8fac4f1a43f5ad7cc0f7826cfd75b9f826bbf391f54963a8be958fea1b45642f37506c77
3eb23e690c87efab0ff8462dab09caafc137d9c947d62785a1fa563ee01b2581b679bef826bbf391f54963a8be958fe27b6d
4d606e8bbed9ee7c647dd218ccb7f401df2662819a2dc137d94f1f599f120c2fd8584984400e3dfc6e9ce28cc57687306344
335859573ee01bd23ad3b05e7c93ddf9c8faa4311cac4306074538c1d0d00e280d675b95a18506673b7fce6665cb865a1eac
df7966932004f211ba3726ddc6c8c9ee10668c68069f3ebf90f0001bf58b70231f6193ddc1cbe8d0f43d17a226810170997b
d6a55be26a85ed0e58468426ae7e5dca077c2f5649a0bdf826bbf391f54963a8be950ff85e21ced95e12fcd8ee7c647dd218
5e7032291156e410a0bca417b2dd21cc18d10cbee0345a220472d4e0b380a7809cec843063443158df37e5a3f8deec12d874
73c937db9d8fac4f1ac3d30191121e60a3be4ebb5c721ed97ec2cbe848e83b60431715f55431eaa108f10af15793e5457817
a4680fdcaa65432d0fe0779ed9240881bc4cf928a677414e768730634433f8f45d978407d803e4d34c7c946193ddc1cbe8d0
f41db0bb699c503cd6a639a81da34a6d0de62bbfccb87e2e7565cb865a1ec0ef3cb3491002f96895d7c70b72b23b8419239a
c117c4fe4a84403e41ed70bc24bab2dd21cc18d10c1ec85b144db7c845c84a8ef3801d940921757d0f89cc0f72d5b2a19607
f23bcf6c1284400e95ec76bce4adb0dd21cc18d10c1e57f90b146617fb24bb118257484cc5b5fb08a94754ab3faef255cb86
5a1e57f9779ed92408817c31fdd0211eadb1dd21cc18d10cbe20de5b220472682aaf78690272b23b8419239a415b45481fd5
77f95a5167efe2dbd99d8fac4f1ac3e1fbea03be21e63d7797c354b6938fac4f0ac31bde7585108a2616ccd54ee16a23d81d
c28c11cde0f3e39b4608e490b11ab6f5829cec1f8419231983cfcf691a219043e3156f4a444e768730634433f8fcc2552304
f201e581d798961dec0e61c68866f0e19d8a8607d8a8470bf5ec089bec0e5e4687a6ef3810285b668817d9e6bded2c9b6ed8
672248f54cf3e74040b56ca8e5712070e7994d8210c827932c6c2fc8c9ee10668c68069fcf52d408811cb5748679b82027bb
439831a2197cecb44f030360a88a978df40530d91db08c084ddcbf78b48e060060ebafaeefe20d1a9b5df733b89a1e2c8ba4
87aa76d60ed375b0f466e720e98fec7d752b9eef45d62e6e4fb37b7ed21bd977bcb53fffed8b52dafe908b52dab9d8f176c4
ad8efd54edee3999dd3d4728a539df37d4c7c847d6278d4128a539df37d4c7c847d6278de19f45adec9023bb76f422b74600
3e76ea68064c135125a2948f8b149af3edecc147d627854148a139df37e4c5c847d6278d4148a139df37e4c5c847d6278d41
48a139df37e4c5c847d6278d4148a139df37e4c5c847d6278d4148a139df37e4c5c847d6278d4148a19dbeefc88b918fac4f
1a83904273be6fc88b791f599f120c420acdf9be212f463eb23e690c420acdf9be212f463eb23e690c420acdf9be212f463e
b23e690c420a8d7c7f9517231f599f34062185e67cdf90170b3eb23e290c2f889390082f426c0eb9b313c28c11cde00be272
24c28b0c9c43eeec84306344332864e09cef1bd26ae423eb93c62064e09cef1bd26ae423eb93c62064e04edf77a4d5c847d6
278d41c8c039df37a4d5bc8fac4f09062103e77cdf9056231f599f3486a7936b24bc8b029d83edec042fa343d32714e89cef
afd26ee421eb91467055a0738ebf8abb9183a43fb2f7fff2d944aaf351de8e301d66ea7c025512f3f8bb2ba045793b87f834
0768091182b617cc890add45decea17676c297b021b97b3a645282bb88dc39d0ce4ee8323234792f88a493082f22770eb9b3
13c28c11cde00b22e924c28bcedd89dcdb0961c68866f0e9e02609ef2272e7603bbb8797d191d0f702e50689f02272e7903b
3b21cc18d10c1ec8c76e867688bd8973fd22eda21ef7d64b99833e839b6cd950cb03f99d675e44ee1cf21bd26d8430634433
78c8b3a0aaee60d776e55dc45df5584564cab8d461f2f9fce6b265432d0f79963bcfbc88dc39e437a4db0861c68866f0405e
469e16896eb89bc71aa3dd75f9eaaded70ca30c9960db53c90df79e645e4ce21bf21dd4608334634836f90f753082f227784
fc63278419239ac1170850498417913b87dcd903c28c11c5e0213ab60e435958d8d03b94096098da1e5d5ea04db7aded39c2
c9960db53c44c7ee3cf32272e790df906e238419239ac103f9b42eedda5595cdbeea6cdaef33f7e3d695e5ee6784932d1b6a
7920bff3cc8bc89d437e43388e10668c6806dff09b2b8417893d87dcd90961c68866f00ddfb9427891d83b917b3b21cc18d1
0cbe606c97082f127b0eb9b37b8419230983c77c6ee7e9f59b2c1fe532cd35340955b3ebd6f298ce55c3c6373c66f31b4fbc
e8eb39d83754e3085e4687a6ef80bd9515de54d75788421a96c576d45580b44c079fc02ed9b2a19607f03bcfbce8eb39e437
84e30861c68866f058b4db91c95263edf043adb6a39a973a3a7fee6444bb86da1d0bf6efcfbb88eb39cc3724e3085bc685e6
eef9585d8df022aee7903b3b21cc18d10c3e1fb1aa115ef4f51c7267278419239ac1e7433735c28bbe1e21ffd80961c68866
f005676f12e1455fcf2177f68030634431f8f461ab8477d1d773b09d9de0657468fa9ebe4c90f02ee27a0eb6b313bc8c0e4d
9f5d47491f1785bbd3b7b7938fac4f1ac3efad70e780dd508df30032c0094142e1cef9bea11a473eb23e690c42e1cef9bea1
1a473eb23e690c42e1cef9bea11a473eb23e690c42e1cef9bea11a473eb23e690c42e1cef99ebfabc6918fac4f1a8350b823
df5f55e3c847d6278d4128dc39df3754e3828fac4f0a8350b873be6fa8c6918fac4f1a8350b873be6fe8c6918fac4f1a8350
b83b7d7f178f230f598f3482abc29d73fc453fce3f3ee98deefb45e1cef9fc2a1f478fd77d51fdbe2adc39a75fa5e3e8f149
5f64cfaf0a77ceed0ddd38f290f447f65e28dc39cf3754e3c845d6238d4028dc39df3754e3c847d6278d4128dc39df3754e3
c847d6278d4128dc91efafaa71e423eb93c62014ee9cef1baa71c147d627854128dc39df3754e3c847d6278d4128dc9dbeef
a8c6918fac4f1ac3d3db0b09ef22aee7603bbb8797d191d0f782802589f022aee7903b3b21cc18d10cbe20705b22bc48fb39
e4ce4e08334634834f1f204878175d3f07dbd9095e4687a6ef41813b05eca2e8e7003b3b01cb88d0c4bde0884c22bc48fb39
e4ce4e0833463483cfe7756b8417614187dcd90961c6886650080b3adf37e4f2c847d6278de105418112e1455890907fec84
30634433f8b4c49d8477511574b09d3dc0cbe850f41db0e7719e97b1f6d2ae4b963a22b6f12a5db66ca8e501fcce332f9a86
27f23b627d8430634433f8821b2789f0226be8903bbb4798319230f874fcab8477d13474b09d9de0657468fade20eea7105e
640d1d7267278419239ac11744fe4a8417594387dcd90961c68866f00db25f0ae145d6d0217776429831a2197c347c4262bb
081a3accce4ed8322e34776f90f553082f82860eb9b313c28c11cde00b22dc25c28ba0a143eeec843063443328040dc9f757
9140f291f54963108286cef70d99bee023eb93c2f0ecf7adb05da40c3f7d253b61cbb8d0dc3d3f9a6b8417294387dcd91dc2
8c918cc1e703c334c28b94a143eeec8430634433f8f06255c3bbe8183ad8ce4ef0323a347dcfefcb34c28b94a143eeec8430
634433f8fc11844678913274c89d9d10668c68069f3ff2d2082f52860eb9b313c28c11cde0f3c79c1ae145d3d02177764298
31a2197cf8205fc38b8a870ef569267019199abcdf41c851743fca293a54a7997a9f8095d45ce514c96da29b7898e9f9496f
64df6fc8297e1e12e414fd82cbdbc7f29bb5435fedee3999dd3de722a748be6f4814928fac4f1ac3454e917cdf5017241f59
9f34867f1939c500e0ab4421753403a689b8c82992ef1b1285c147d62785e10599ba1261107324e4ce4e08334634832f48ee
9108839424217776429831a2197c74f72ab105094bc2ecec842de34273f782f36789300868127267278419239ac1e737331a
6190ef24e4ce4e0833463483fff2e1f71240d006f5c0bc9d0064803541176d50f27d436fd3fbc8fa9460b8688392ef1b7a9b
e423eb93c670d10625df37f436c947d6278de1a20d4abe6fe86d928fac4f1ac30b42ad24c2a04c1a907fec84306344337851
2625df37f436838fac4f0ac3d3615e125e902525d8ce4ef0323a347d9572e5230883926f67271f599f348637bce80a619025
25e4ce4e0833463483175952f27d43ea937c647dd2182eb2a4def71da94ff291f54963b8c89292ef1b529fde47d6a704c30b
627725c2208a4ac89d9d10668c68062bebca47502625dfce4e3eb23e690c176552f2fd55f2933c643dd2089e0e5157e05814
95309f66c296502189ab5c8be7b36e69707b98e9f9496f64dfdff06109742c5d4aa04f734097702198ab5c8be707f150f2eb
ece421e98fecfdd3eb14092e8887126867277419199abce7af2d34c2201e4ac89d9d10668c68065f205a211106f1508fdcdb
0961c68866f00567191261d00f25e4ceee11668c240cbee0fc4a220cfaa184dcd90961c68866f00571651261d00f25e4ce4e
083346348307f232c90ddd6667556d2d76693fcf548686b91d87cf4fae1a36d4f0c07de3894d022f8887126c672778191d9a
bee360dee2db5afb3db6aeed11fd6677096337f5a601f83998572d1b6a791cccdf796693200ce2a184dcd90961c68866f005
97311261100f0dc83f76429831a2193ca4f5eaffd85dc9bc968e958fd26e51bab2c85c375cc01ed27aaa65432d0f69bd3bcf
6c1284413c94903b7b409831a2183c90b7f35086228bde2e034f59e42eb68758b63d29eeb37c932d1b6a7920bff3cc264118
c44309b9b313c28c11cde02124b976cbb0d8b551d92c95a1685d87cd7ad9cfc330b7e77e4cb66ca8e5212479e7994d823088
87127267278419239ac137fce60a61100f25e4ce4e08334634836ff8ce15c2201eea917b3b21cc18d10cbe606c9708837828
2177768f30632461f005f3b94418f44309b9b313c28c11cde0b180ebb761d84f9bdba146f92ef5033589db738b2a5b36d4f2
58c2dd796693200cfaa184dcd90961c68866f0f960718d30a888127267278419239ac163a3f65cdd7b8d30a8881272672784
19239ac1a7937825bc20214ab09d9de0657468fa9e0f1cd60883846840feb113c28c11cde07382101258500e25c0ce1e8065
4428e25e703d291106f15042eeec8430634433f8f4d5898417c44309b6b313bc8c0e4ddf453cd4fbbea399493eb23e690cbf
af782801bb21c8e901648013822ee2a1e4fb862027f9c8faa4315cc443c9f70d414ef291f54963b8888792ef1b829ce423eb
93c670110f25dfd377414ef291f54963b8888792eff9bb2027f9c8faa4315cc44383efaf829ce423eb93c670110f25df3704
39838fac4f0ac3453c947cdf10e4241f599f34868b7828f9be21c9493eb23e690c17f150effbbb2e2779c87aa41144f15072
fc459ad33f3ee98dee7b100f259f5f9539e9f1ba2faadf513c949c7e55e5a4c7277d913d8fe2a1e4f68624277948fa237b7f
110f25cf370439c945d6238de0221e4abe6f0872928fac4f1ac3453c947cdf10e4241f599f34868b7868f0fd5590937c647d
d2182ee2a1e4fb862067f091f54961b8888792ef1b829ce423eb93c6f0f4125fc20bbaa51eb6b713bc8c0e4ddf0b629524c2
205d4ac89ddd23cc1849187cc16e56220cd2a584dcd90961c68866f0e98829092fe896126c672778191d9abe07053c15b0a0
584a809d9d80654468e25e1022261106e95242eeec8430634433f882434989304897127267278419239ac1a74fa125bca05b
4ab09d9de0657468fa2ebaa5e4fb861a27f9c8faa4313c2d6928e105d1d200fb632778191d9abe17dcf4488441b794903b7b
409831a2187c4130a24418744b3d726f278419239ac1e7a5a034c2a05b4ac89ddd23cc1849187c838ca54218a44b09b9b313
c28c11cde00beeb225c2205d4ac89d9d10668c6806abc49ef211c443c9b7b3938fac4f1ac3b3127b0a5b900d25ccce4ed832
2e34772f88179108837828217776429831a2197c3ad459c20bcaa104dbd9095e4687a6ef0d3fb84218944309b9b313c28c11
cde0e33fb88217444b03ec8f9de0657468fadef0832b84413295903b7b409831a2187c8348ae421884535d8fc94e08334634
83cf27b06884413895903bbb4398319231f8fc9a51230cc2a984dcd90961c68866f0f9bda14618b45309b9b313c28c11cde0
f3951d34c2a09d4ac89d9d10668c68061f3e79d2f082702ac176768297d1a1e97b3e1b55230cc2a984dcd90961c68866f0f9
e3658d3008a7127267278419239ac13709a712ecafa2a1042e234393f7bb0aa712aaaf9aa4d4fb04aca4260aa706b78942ea
61a6e727bd917dbf219cfa7948104ef50b2e6f4734fe5c2662b3bbe76476f79c7f460151d5d1a0be4a009c9d3a9a01d344d4
cf715dfb61ea4b977e70df3baf53f9bd3bfc636ecb0a61dcb64f56956ed950cbfa41de7b6693200cdaaf84dcd90961c68866
701f88ba3d050abd2ccbdeb22c1a2d937f292f1c7401c7cf79ac6cd950cb7d28baf5cc264118446303f28f9d10668c680677
e4e5031b2d1503e95e435f16bedd8c7f2c65f41c4a87d7cfae42b66ca8e58efcd6339b046190ac25e4ce1e10668c28065f90
3d291106c95a42eeec84306344337864d28d4369d11dd72213e2d62dd37118bbb15d70617664d2a9960db53c32e9ee3cb349
1006c95a42eeec843063443378ec24db71b14b192cfad76559866db07cbfa99fb6e264fa7ce7b265432d8f9de49d673609c2
205c4bc89d9d10668c68069f3f3dd00883702d217776429831a2197c7e0fad1106e15a42eeec8430634433f8740c83841764
6d3d6c6f2778191d9abe1704ad4884415497903bbb4798319230f85c64960416947c09b0b313b08c084ddcbf7c7293041064
820998b313800cb026e8e9d042092f2814136c672778191d9abe43a7f7ea23680407df1f3bf9c8faa4313cf8f12860419898
003b7b009611a188ab642b1f411d987c3b3bf9c8faa4315cd481c9f70dcd5bf291f54963a8be958fa0cf4bbe9d9d7c647dd2
182efabce4fb86e62df9c8faa4313c3ea62878411ad8c3f6768297d1a1e9bb480393ef1b82b7de47d6a704c30b62e325c220
0d4cc89d9d10668c68062fd2c0e4fb86e02ef9c8faa431bc20174322646162027e9a095fc687e6af722e3cb03a30393ecde4
20e98fecfdd3193f0a1a0b1307c48799a0254448dadef08209742c4c4ca04f7340977021988bc2c4e4f686e62e7948fa237b
ff864154e10bdac484dbd90960c687e6efb925a90416248909b0b313b08c084ddc0bb6ee12619024f6c8bd9d10668c68069f
3eab91f0821e31c176760f2fa323a1efe99c0f092f8811136c672778191d9abe17c4ff4b84418c98903b3b21cc18d10c3e1a
8f2db1052562c2ecec842de34273f7825b0789302811137267278419239ac11768364a8441893820ffd80961c68866f0406e
eb0b9b68d675e89769ae7148eb70ec930fe0aa61e31b1eb86f3cb149e005196282edec015e4687a2ef805d5ecb79a9292865
8d511676ad09aaf6fb85c9671a530d1b6a78c0bef1c4268117348809b6b313bc8c0e4d9fad8da58fa0024cbe9d9d7c647dd2
18aa6fe523e8f0926f67271f599f3486c7df72052f88f07ad8de4ef0323a347d2f18d624c220c24bc89ddd23cc1849187cc1
54261106115e42eeec8430634433f8e8c245620bf2bb84d9d9095bc685e6ee05cb548930c8ef127267278419239ac1e7c3cf
35c220bf4bc89d9d10668c68060fe46bd9360d355878282f69b774abfda3bcabd33ae2553990ab960db53c90df796693200c
0abc84dcd90961c68866f005870f126150e00dc83f76429831a2197cfa6648c20b3abc04dbd903bc8c0e45df0b0e1625c2a0
c34bc89d9d10668c68069f3eb297f0820e2fc176768297d1a1e9bbe8f07adf77e467c947d6278de1f7d5e1256037b46d3d80
0c7042d04587977cdfd0b6251f599f34868b0e2ff9bea16d4b3eb23e690c171d5ef23d7ed7b6251f599f34868b0e2ff99ebe
6bdb928fac4f1ac34587977cdfd0b6251f599f34868b0e6ff0fd55db967c647dd2182e3abce4fb86b66df091f54961b8e8f0
92ef1bdab6e423eb93c670d1e125df37d46dc947d6278de1a2c3eb7d7f97b8250f598f3482a8c34b8ebfa8dcfac727bdd17d
0f3abce4f3abc82d3d5ef745f53beaf092d3af02b7f4f8a42fb2e7518797dcde50b7250f497f64ef2f3abce4f986b62db9c8
7aa4115c7478c9f70d6d5bf291f54963b8e8f092ef1bdab6e423eb93c670d1e10dbebf6adb928fac4f1ac34587977cdfd0b6
0d3eb23e290c4f2fb325bc20014cb09d9de0657468fa5e101e23110615608fdcdb0961c68866b0ca272a1f4187977c3bbbf7
91f529c1f0f4265ec20b22bc04dbd9095e4687a6ef41355a052cc8ef126067276019119ab8e7e50234c2a0c34bc89d9d1066
8c68065f70342911061d5e42eeec843063443378207fb060b94418747809b9b313c28c11cde00b0ee225c220c54bc89d9d10
668c6806dfa08eaa1006216042eeec8430634433f8826b278930680107e41f3b21cc18d10c3eaf74a511062d6042eeec0161
c68862f005eaa81261d002f6c8bd9d10668c68065f90992d11062d6042eeec1e61c648c2e0b3daac0a5b500126ccce4ed832
2e34776fd02f5408830a30217776429831a2197c8366a54218348809b9b313c28c11cde0e322a50a5e902126d8ce4ef0323a
347d55f659f90842c0e4dbd9c947d6278de105714a1261d00226e4ce4e08334634839575e523a8f1926f67271f599f3486ea
5bf90892b8c1f7c74e3eb23e690c8f7f640a5ed0c325d8ce1ee0657428fade30aa2a84410fd7f598ec8430634433f8e4ea41
630b4ab884d9d91db68c8b8cbbe757c91a6150c225e4ce4e0833463483cfef8635c2a0844bc89d9d10668c68069f3f01d108
83122e217776429831a2197c93122ec1bea1ef4af0323a347d2f53c225e437f45d0961c68866f0f9a36d8d3028e112726727
8419239ac1c76e313430d6c025bca79960653468da9e8fbc94f8582497609f668297b021b9fb1d847f45e7598137603accd4
f904aa24e62a195bf6406b594ab4760e312e85f17e583f31d3ba65432daf92b1f93383daefa7c341edd72f27bd1d690ffd58
06ffbff07332bb7bce15f95c869871c0392bba5c4f5b07a5957b366c7cc32beef4894d022f6804136c672778191d9a3ea111
dc2f6d5fbeabb9f678ecc6b29ada5681fb6cd9504ba1119c3eb34910068d6042eeec8430634433b88f592d84b4badee45c97
a14c9bddbc602299d7d1bef9ee73a4205b36d4721fb36e3db34910068de080fc63278419239ac1431d79bf2ac402b1b574cc
615becc72a6be6b2339d3ef3936cd950cb431df9ce339b0461d00826e4ce1e10668c280677e4d80f62c8fbb13c3c7b1797d5
7eac711acb52e2bc89922d1b6ab923bff5cc264118348209b9b313c28c11cde0a18ebc8b38db1678ecdb32fb8c9677db963f
6f2d82f15047562d1b6a79a823df796693200c1ac184dcd90961c68866f048ee2c5bc26ee9aa9c6bd9fdf65b198ef00f1cb1
7773fb19d975cb865a1ec99d779ed92408834630217776429831a2197cc159924418348209b9b313c28c11cde00bee5d25c2
a0114cc89d9d10668c68069f166091f08246b087eded042fa343d3f782502289306804137267f708334612065f90e5281106
b16042eeec8430634433f886df5c210c6ac284dcd90961c68866f0f913258d30080a137267278419239ac147457dafc0828a
7100fcb113b08c084ddcf347681a61903326e4ce1e10668c28062f72c6e4fb864430f9c8faa4315ce48cc9f70d8960f291f5
496378c380ae1006316542eeec8430634433f886775d210c52ce84dcd90961c68866b0be6fca47d053f6bebd9d7c647dd218
1e14a554c08288330176760f2c232221ae0a9f2a1f4146997c3b3bf9c8faa4313c9dd421e105056782edec042fa343d3f7fc
fd8046c80ace04fc3413be8c0fcd5f547026c75f158ac941d21fd9fb17246c29742ce21c401f664297702199ab5c8be7b38c
32b93dcde1f9496f44dfa38c32b9bd210f4c1e92fec8de3f3e942870414399403b3ba1cbc8d0e455be958f20674cbe9d9d7c
647dd2185eb0409108839cb147eeed8430634433f8821da74418148d09b9b37b84192309832f48d5920883a8312177764298
31a2193c903f58684d220ca2c684dcd90961c68866f005c7c71261903626e4ce4e0833463483cfca392b6c41d498303b3b61
cbb8d0dc3d1d692ce10545e300fb632778191d9abee35eac7c80fdb499322b8e038665b1a9bfecdf6afcdc6748972d1b6a79
dc8bdd796693200ca2c684dcd903c28c11c5e00be45e25c2a06b4cc89d9d10668c68065f70172a1106556542eeec84306344
33f886df5c210c9ace84dcd90961c68866f040beb65d3fd4f884cdc6a2b17ea065b5d961f3f699cc64cb865a1ec8ef3cb349
100659678fdcdb0961c68866f0e9815dc20b9ace04dbd93dbc8c8e84be67958d15b6a0e64c989d9db0655c68ee5eb0629308
83a633217776429831a2197cc12a5d220c9ace84dcd90961c68866f0053b338930683a137267278419239ac117ecc625c2a0
e94cc89d9d10668c68065f70022311064de780fc63278419239ac1a78ffb24bca0e94cb09d3dc0cbe850f4bde0285b220c9a
ce84dcd90961c68866f069b139092f683a136c672778191d9abe7f795d65092008467b60de4e0032c09aa0aa7daa7c044d67
f2edecde47d6a704c345d3997cdfd049261f599f34868ba633f9bea1934c3eb23e690c174d67f23d7ed749261f599f34868b
a633f9bea1934c3eb23e690c174d67f27d4327997c647dd2182e9acec1f7579d64f291f54963b8683a93ef1b3ac9c147d627
85e1a2e94cbe6fe824938fac4f1ac345d3997cdf504a261f599f34868ba6b3f7fd5d2e993c643dd208a2a63339fea298ec1f
9ff446f73d683a93cfaf82c9f478dd17d5efa8e94c4ebf8a25d3e393bec89e474d67727b4329993c24fd91bdbf683a93e71b
3ac9e422eb914670d17426df377492c947d6278de1a2e94cbe6fe824938fac4f1ac345d339f8feaa934c3eb23e690c4f2f75
25bc20274db09d3dc0cbe850f455ca958fa0e94cbe9d9d7c647dd2185eb0af920883a6b347eeed8430634433f88258348930
284a137267f708334612069f3e4290f082a234c176768297d1a1e97b4140b54418a4a509b9b313c28c11cde00b326424c220
2d4dc89d9d10668c68065f70362c1106696942eeec8430634433f87402a0841774a509b6b313bc8c0e4ddff37a521a61d095
26e4ce4e08334634832fb8fa920883ae34217776429831a2197c81d2ae441874a503f28f9d10668c68061fbde295d882a234
6176f6802de34271f702a55d8930284a7be4de4e08334634832fc8e3970883a2342177768f30632461b00a902a1f41d9997c
3b3bf9c8faa4313c1d3d22e105596782edec042fa343d3f706a55d8530c83a137267278419239ac11704444a8441d999903b
3b21cc18d10cbe20384e220cbad284dcd90961c68866b00e30ca47507626dfce4e3eb23e690c2f08cc930883ae3421777642
9831a2197cc397a6100655eb80fc63278419239ac1378cae0a6110b626e4ce1e10668c28061f9f4e15bca06aedba4b768297
d1a1e97b7ed5aa11066d6b42eeec0e61c648c6e0cbb4ad09f90dc566429831a2197cfe2c42230cdad684dcd90961c68866f0
e1c3270d2f085b136c672778191d9abe37095b13ec1b72cd042fa343d3f77c18aa461884ad09b9b313c28c11cde0f361a81a
6110b626e4ce4e0833463483ef10b626c05f359b09564683a6edf9cb42898f85ad09f6692678091b92bbdf55d83a604a14ac
0f33753e812a8911f2ce73194996c9321aa6695ca6699d3f451764c3861a0a79e7ec8941d5fad3dba06aed9793de8e949775
1aaadd3d27b3bbe708d86b593596fd608f2f6c5aca1f96776f1914ee4fcb865a0ae0e9339b046110b626e4ce4e0833463483
07f23260ae83ad919661ded787e8f250a6cf615d4f2173d5b0f10d0fdc379ed824f082aa35c176768297d1a1e9db616f386f
5826cc22e5ff6e65f3871851bca26d7947e6732da25b36d472077eeb994d8230a85a07e41f3b21cc18d10c1e7adef3300e4b
67bd5cb0f9afd541e7cdce40f08a1e72dea261430d0f35ef1b4f6c127841d29a603b7b8097d1a1e8db61ef1accb617e887be
fcbfbbd95ed1a96c81c732b57c70cb960db5dc81df7a6693200c92d684dcd90961c68866f010f31e7a3be1b0439efddac294
d7d778c72f5b36d4f210f3bef3cc26411824ad09b9b313c28c11cde021e6bdb6dddcd9e0b32e18fd21bd857f749f61e910f3
562d1b6a798879df796693200c92d684dcd90961c68866f00527d5126190b426e4ce4e0833463483cf1f24698441d29a903b
3b21cc18d10c1ebbeb07cb474b8441d5da23f776429831a2197cbe5498461854ad09b9b37b8419230983551053f908bad2e4
dbd9c947d6278de105b1631261d09526e4ce4e08334634830f9f63697841549a603b3bc1cbe8d0f4bd204c52220cead201f9
c74e0833463483cf9fe06984415d9a903b7b409831a218bca84b93ef1b8acde423eb93c670519726df373493c947d6278de1
0defba4218d4a509b9b313c28c11cde01bde758530a84b137267278419239ac137fce60a61d0b6f6c8bd9d10668c68065ff0
9b4b8441e49a903bbb4798319230f882a4078930486c137267278419239ac1e7ef6634c2a0b24dc89d9d10668c68061f5472
57c0585c9bf09e668295d1a0697b3a9d4981635d6fc27c9a095b428524eef9db28898e75bd03e8c34ce8122e247351d79bdc
7e15cc0ecf4f7a23fafe820449852ea88a136a67277c091b92bb37bc620a5f101627dcce4e00333e347f8f0f270a5ed03427
d8ce4ef0323a347d0f4e1a0a589032f780bd9d80654468e25eb03292088394392177768f30632461f005fb00893048991372
67278419239ac1171ca44a8441ca9c903b3b21cc18d10cbe401e52220c52e684dcd90961c68866f00579911261103427e4ce
4e08334634832fc88b940883a67940feb113c28c11cde00b2ec924c2a0694ec89d3d20cc18510c3e1d5e2fe105417382edec
042fa343d3679b03e923488a936f67271f599f3486ea5bf908a2dee4dbd9c947d6278de10529a1126110f5f6c8bd9d10668c
68069f55b756d882a2376176768f2de322e1ee055399441874bd09b9b313c28c11cde00b962f1261d0f526e4ce4e08334634
832f58b24a8441d79b903b3b21cc18d10cbe609b2211065d6f42eeec8430634433f882ada9441874bd09b9b313c28c11cde0
73a710125890f30e803f76029611a1897bc1399b441814bd09b9b3078419238ac1a7c5fe24bc20e74db09d9de0657468fa6c
b5287d044d6df2edece423eb93c6f0fb6a6a7b607774aa094006581374d1d426df3774aabd8fac4f09868ba636f9bea1534d
3eb23e690c174d6df27d43a79a7c647dd2182e9adae4fb864e35f9c8faa4315c34b5c9f70d9d6af291f54963b8686a93ef1b
3ad5e423eb93c670d1d40ebebfea54938fac4f1ac345539b7cdfd0a90e3eb23e290c174d6df27d43a79a7c647dd2182e9ada
e4fb865235f9c8faa4315c34b5bdefef72d5e421eb91461035b5c9f117c56afff8a437baef41539b7c7e15aca6c7ebbea87e
474d6d72fa55ac9a1e9ff445f63c6a6a93db1b4ad5e421e98fecfd45539b3cdfd0a92617598f34828ba636f9bea1534d3eb2
3e690c174d6df27d43a79a7c647dd2182e9adac1f757b168f291f54963787a992de1054d6d82edec015e4687a2efa2a94dbe
6f2845938fac4f1ac30b424624c2a0a9ed917b3b21cc18d10cbe60372b11064d6d42eeec1e61c648c2e00ba20d25c220ab4d
c89d9d10668c68065f105b2b1106596d42eeec8430634433f88204218930c86a137267278419239ac1c7c5a515bca0a94db0
9d9de0657468fa9e1774d20883ac36217776429831a2197c5ec44b230cb2da84dcd90961c68866f005a104126190d526e4ce
4e08334634832fb85b960883ac7640feb113c28c11cde00bd41925c220ae4dc89d3d20cc18510cdab259fa08f2d6deb7b793
8fac4f1ac31bd44015c2206f4dc89ddd23cc184918bcc85b93ef1bbacde423eb93c6f006f5598530285c137267278419239a
c1cabaf21134a6c9b7b3938fac4f1ac30bf42b24c2a0314dc89d9d10668c68060fe465b1d321c7034b9c32c7974f63b62f12
136237b7c3678b285b36d4f2407ee7994d8230684c137267278419239ac18bc634f9bea19c4c3eb23e690cd5b7f211549ec9
b7b3938fac4f1ac31bde758530a83c07e41f3b21cc18d10cbe6146530883ca332177f68030634431f87418aa8417549e5d77
c94ef0323a347dcf6f1634c2a0f24cc89ddd21cc18c9187c7e83a81106956742eeec8430634433f8bc2c914618549e09b9b3
13c28c11cde09b549e09f60df9628297d1a1e97b93ca33c1bea15d4cf0323a347dcf9ff36a8441e599903b3b21cc18d10c3e
7fb6af1106956742eeec8430634433f8702eac86c75acf84fa3413b88c0c4ddef39758121f6b3d13ecd34cf012362477bfab
d673c094883a1f66ea7c02551223448fcbdb55fa04c58cf6d78415d4b09d3fa46ed9504b217a9c3e33c83d7f3a1ce49efdba
d2db910332ce6db5bbe76476f71c817c2a8be5719d36fc34d350164e2bf28205f04fc3c63714b8b3273609bca0f54cb09d9d
e0657468faaeb0cb585ebeaab2863445ea6dea07e8bead02f7d9b2a19657e0f9339b0461907b26e4ce4e083346348302795f
e5a96d6e2993cc30f7ddbc8e0af9a765432d05f2f4994d8230c83d07e41f3b21cc18d10c1e42d775178c5562f918db7a8f80
5e2e7d3b8fd3d61eab50ddb2a19687d0f59d673609c2a0f84cc89d3d20cc18510c1ec8bbaad5fb53059991b0d45a2f4bab65
eb87cf6da46ed950cb03f99d673609c2a0f84cc89d9d10668c6806f719dad2d246f46bf1ffdd7eded87d7e16ed1a6ab7cfce
379ed724d882d633617676c29671a1b9db31f7f580a74a319701772a73b9fd4ce358dfd0cf3653b66ca8e58efbd6339b0461
d07a26e4ce4e083346348387bef7b474651b58af09dbf2db2cf5861c193d6df9a373f92d5b36d4f2d0f7bef3cc264118b49e
09b9b313c28c11cde0717e587e14bc84b6f347985d3bd643afba2d1c3ef3b86ed950cbe3fcf0ce339b0461d07a26e4ce4e08
334634832fb881950883d6b347eeed8430634433f8825b778930683d137267f708334612065f105d231106a56942eeec8430
634433f8743895841764a609b6b313bc8c0e4ddf0b529b25c2a0344dc89d9d10668c68065f1039281106a5e980fc63278419
239ac1aa79ac7c04ad67f2edecc147d62785e1715973052fc84c136c672778191d9abe37bc6c0a6110b926e4ce4e08334634
836f186014c220724dc89d9d10668c6806eb67a67c049969f2edece423eb93c6507d2b1f41e8d9fbf676f291f5496378c1bb
2e1106a16742eeec1e61c648c2e0d3839b8417549e09b6b313bc8c0e4ddf0b444224c2a0f24cc89d9d10668c68065f201222
11b2dc33013fcd842fe343f3f7fc6d9cc4c78acf04fb3413bc840dc9dd83eaac02140b3d07ac879940251448c21e17e015d0
58639a109fe6002d2142d056a5adc5f383ca33f97576f290f447f6be3a561e82ce327976767291f54823a8be958f20764cbe
9d9d7c647dd218de2176ec01df91f0256019119ab8aa51a87c04b961f2edecde47d6a704c30b262a8930c80d137267278419
239ac1172c4e24c220374cc89d9d10668c68065fb0f3920883dc30217776429831a2197c8152a14418e48609b9b313c28c11
cde0f381901a61901b0ec83f76429831a2197c812ea74418e48609b9b3078419238ac1e7c37e35c2a0384cc89d9d10668c68
06df20aead1006bd6342eeec843063443378d13b26df37547cc947d6278de10def9b4218f48e3d726f278419239ac1178cae
1261503d26e4ceee11668c240cbe6046950883ea31217776429831a2197cc12a4a220caac784dcd90961c68866f0e9d34c09
2f481e136c672778191d9abe176c9524c220794cc89d9d10668c68061f14fe55c082d231017676029611a1897bfa944bc20b
7ac701f6c74ef0323a347dcf07d76b8441ef98903b7b409831a2187c5a884dc20b7ac704dbd9095e4687a6efa2774cbe6fc8
fc928fac4f1ac3efab77ec81ddd110260019604dd045ef987cdfd010f63eb23e25182e7ac7e4fb868630f9c8faa4315cf48e
c9f70d0d61f291f54963b8e81d93ef1b1ac2e423eb93c670d13b26df373484c947d6278de1a2774cbe6f6808938fac4f1ac3
45ef38f8feaa214c3eb23e690c17bd63f27d434338f8c8faa4305cf48ec9f70d0d61f291f54963b8e81d93ef1b2ac2e423eb
93c670d13bf6bebf4b099387ac471a41d43b26c75fd484fde393dee8be07bd63f2f9554c981eaffba2fa1df58ec9e9572161
7a7cd217d9f3a8774c6e6fa8089387a43fb2f717bd63f27c4343985c643dd2082e7ac7e4fb868630f9c8faa4315cf48ec9f7
0d0d61f291f54963b8e81d07df5f857cc947d6278de1e965b68417f48e09b6b30778191d8abe37680e2b84416d99903b3b21
cc18d10cbe6047291106b5658fdcdb0961c68866f0b93313092c882c136067f7c0322212e25e703c261106916542eeec8430
634433f88248028930882c137267278419239ac1179c814b84416499903b3b21cc18d10cbe201f4a220c3acb84dcd90961c6
8866f005b77c1261d05926e4ce4e083346348307f2e989aaa6125b505826ccce4ed8322e34772fc8f5940883c23221777642
9831a2197c81f6aa4418149603f28f9d10668c68065fa009291106856542eeec0161c68862f0a2b0ec7ddfd10d261f599f34
863768af2a84416199903bbb479831923078515826df377483c947d6278da1fa563e82c631f97676f291f54963b8681c93ef
1bcabde423eb93c670bc6d9046dac505caf262c2358d8d44c3d89539a9fcc9e76d532d1b6a79bc6d779ed9240883c6312177
76429831a219acac2b1f4165987c3b3bf9c8faa4315c5486c9f70de55ef291f54963b8a80c93ef1bdab9e423eb93c6f086b1
4d210c2ac301f9c74e08334634832f531926e437e47303c28c11c5e09b54861dec5bdab9042fa343d3f7329561427e433bd7
21cc18c9187c7e63a81106956142eeec8430634433f8fc61804618548609b9b313c28c11cde09b548609f60df95c8297d1a1
e97b93ca30c1bea19d4bf0323a347d2f531926e437b4730961c68866f0f9bc668d30a80c137267278419239ac137a90c13ec
af02ba042e234393f7fcb595c4c72ac304fb3413bc840dc9ddefaa321c302572c287993a9f4095c43caeb5aba48183c4b05f
543a3b524dca47b954fbf99cccee9f2360977df758f649a375726ea7a10c34e3a4707f5a36d452004f9fd9240883ca30213f
ed8c30634433f80659698530a80c13f2d3ce08334634836ff8cd15c2a0321c901f76469831a21914c8375c944ef3862523e6
53da72e9960db514c8d3673609c2a0324cc84f7b449831a2187c83b2b44218548609f969678419239ac137284b2b84416b98
909f76469831a2193cf495779158f4729b9020379afc55f9347b48b68e9f1b39d9b2a19687c6f29d673609c2a0354cc84f3b
23cc18d10ceec8cb92a96b7b5b2c2245103b42db1a9665dc38f7f8f90ee4b265432d77e4b79ed9240883d630213fed8c3063
4433b8239fda7918073b0629cb8d5defdb7eac3d16e8b30a972d1b6ab923bff5cc264118b48609f969678419239ac117dc53
4884416bd8237776469831a2197cc1fdb34418b48609f9692784192309836f5059560883d630213fed8c30634433f802dd02
8930c80d13f2d3ce08334634832f50e9900883dc30213fed8c30634433f8820c7a8930c80d07e4879d11668c68062f72c3e4
fbbb8e6ef491f54961788162834418148709f969678419239ac1e72f2b34c2a0384cc84f3b23cc18d10cd6f74df9089abfe4
fbb4b38fac4f1ac345f3977c7fd7d1651f599f34868be6aff77d43c9967d647dd2185e30ba498441f397909f764298319230
f874dd00092f68fe12ecd3cef0323a347d6f1079560883e62f213fed8c30634433f88234008990357f09f8c7ccf8323e347f
cfdf0a4a7cacf94bb03f668697b021b97bfe4250a263f1df007a3733ba840bc9dca342c717502cfb4b583fe6082aa14010f6
b4048c8216148709f2696770091592b8e7af3d35bea0774cb84f3b03ccf8d0fc3dfe6b2b78416a99609f768697d1a1e9ab32
cfca47503df6be9d9d7d647dd2182eaac7e4bbfbaae54b3eb23e25185e307a4b8441f598909f76469831a2197c4196a64418
548f09f969678419239ac1172ccf24c2a07a4cc84f3b23cc18d10cbe60492e1106d563427eda1961c68866f005f9a9126150
3d0ec80f3b23cc18d10cbe60df2d1106d563427eda23c28c11c5e00bceb724c2a07a4cc84f3b23cc18d10cd61955f908bac3
e4fbb4b38fac4f1ac34577987c7f57d3651f599f348637bceb0a61d01df6c89d9d11668c68069f3e6392f082e830c13eed04
2fa323a1ef05139a4418448709f969678419239ac1172c6224c2203a4cc84f3b23cc18d10cbee09449220cbac384fcb433c2
8c11cde00b362b1261d01d26e4a79d11668c68069fde954b78417d98609f768697d1a1e97bc1318c4418048803f2c3ce0833
4634834feb834978417d98609ff6082fa343d1576578958f20014cbe4f3bfbc8faa431fc0e32bc0a40d0172660a79d016480
35419554e52348007bdfcece3eb23e690c170960f2fd5d56977c647d4a305c2480c9f777595df691f54963b8480093efefb2
baec23eb93c670910026dfdf6575d947d6278de122014cbebfcbeab28fac4f1ac34502987c7f97d5651f599f34868b0470f0
fd4d56977d647dd2182e12c0e4fbbbac6ef491f54961b8480093efefb2baec23eb93c670910026dfdf8575d947d6278de122
01ec7d7f55d7650f598f348228014c8eff69815d7a7cd21bddf720014c3ebfe9ebf2e3755f54bfa3043039fda6adcb8f4ffa
227b1e2580c9ed77615df690f447f6fe22014c9ebfcbeab28bac471ac14502987c7f97d5651f599f34868b0430f9fe2eabcb
3eb23e690c1709e0e0fb9bb62dfbc8faa4313cbdc497f0820430c13eed115e4687a2ef6512c084fcbbb02d23cc18d10cbe60
372b110609608fdcd91961c68866f01d12c004f8bbb02d01cb8848887bc1199d4418248009f969678419239ac1179c484b84
410298909f76469831a2197cc1e58b4418248009f969678419239ac117a4ac4884410298909f76469831a2197c5ee34a230c
12c084fcb433c28c11cde01be47015c220044cc84f3b23cc18d10cbe411456210c42c084fcb433c28c11cde00b525025c220
041c901f76469831a2193c90db2acb62b9d675281b9ab94ec4eb705c9d1cc055c3c6373c70df786293c00b2ac004fbb44778
191d8abe03764d86b73771c56ea2dd7fa92ac5359c79e6b265432d0fe0779ed924088306b147eeec8c30634433583542958f
a0024cbe4f3bf9c8fa946078fa6593f0820031c13eed0c2fa343d377112026dfdf457dd947d6278de122404cbebfcbeab28f
ac4f1ac30b46558930081013f2d3ce0833463483170162f2fd5dd4977d647dd2182e02c4e4fbbba82ffbc8faa4315c0488c9
f777595df691f549637899007140fe4d569711668c68065fb0629508830031213fed1161c68862f00532121261d020763df6
76469831a2197cb25089c616d48709f369f7d8322e32ee5ea63e4cc8bf6bea32c28c11cde0f301cf1a61501f26e4a79d1166
8c68069f3f7bd208830031213fed8c30634433f8fc79a34618348809f969678419239ac1e7236035c2a0414cc84f3b23cc18
d10c3e9fbaa511060d62427eda1961c68866f04d1ac404fb9bbc2e83cbc8d0e43d7f6f26f1b10631c1fe98195ec286e4ee77
d5200e9812b1e1ddcc9d4fa04a62843ce93c8f65356513ca3495f1755ae741aa931e0d1b6a28c449b327b206f1d95bd620a6
45a5b75bc24d19f7fec28f49ccee2957d065365dca1cb2da39ef38f7fdb0cc6dab9497cf960db5bcc2ce9fd924f858819871
3bbb0798d0a1d9133f768743e5d208a3ccb8e1bfcace482a4e7f5a36d452fcdce9339b041feb0f336e67f700133a347b4a68
7b1edb3275da52b96cc9bbad2b3f9452a23d5b36d452496d67cf6c127cac3e1c717fec1e60428766ef1daacb577cac3dccb8
9d9d01267428f684f270dbe1447b5b700450f6fffd00b1bb55e03e5b36d452280fa7cf6c127cac3cccb89ddd034ce8d0ec1d
b8c72a898c3e96ad7fdd07e2871aec147f5d3eb055c3c6373c50df7862938063d16106edec1e5dc285a6ee0d62cb0a1f4b0e
336e67f700133a347b87d4f2be25c2f6af6dc7b2091a4ceb7fc18eb8efa06b78482dab960db53ca496ef3cb349f0b1e030e3
76760f30a143b377082daf65661febc9e63097b7702a5b01fba10a55ff2f73679224bbb2e4d8b9ad225750e27db3a7dcffb8
d804494079f09c93bcb41429a97f21f64cf5c019ec09fb0e6daf27dc38b2d9c82568f9c89c2df079dcb0738bae80c10e76af
879865e2f3b061e3565d01831decded98fde10ce73861d5a74a10b5e04eb7ac857263e4f19766ed11530d8c1ee75705b1cf9
3c63d8b94557c06007bb77e64d7126f36c6127165dd18211ecdbf9b7c499cf93852bf7aa2b60b083ddabb9c25ef9775a6e29
111aa2fe7bc8ef263e4f15766ed11530d8c1eef5b0a5119f670a3bb7e80a18ec60f77af8bd89cff38c9d5b74050c76b07b35
cdd82bff4e08b612a121eebf872d8df83c49d9b85557c06007bbd7c1cbc6c8e739cace2dba00063b827b1de4b4239f07293b
b7e80a18ec60f7ce7fb4c97c1ea3ecdca22b60b083ddeb20b10df92c44d9b13759f18219ecdd427dd27ab28866f9c94ebcc9
8ac63ea0691d7c28446c969e5c911759e1d80874ad83141f62b3106547de64876323c0b513e3a201c9c3931d5574a56207d0
aeb3a38a10cd73931d5974850b4eb071737c3055f0e862af2cba96080d71ff35b8d82a1f0803b612a121eebfc6167be5df61
bc5a223414fa9f2b53058f0df6caa26b89d010f7dfc1be0bf93c34d8b94557c06007bb77fe1b37cce791c1ce2dba02063bd8
bd0e0ecfc8e781c1ce2dba02063bd8bdf3df2d633e8f0baedcabae80c10e76effcb70999cfc3829d5b74070c76907bf31e95
2a7858af57165d4b8486b8ff0e2e7390cf83829d5b74050c76b07b3526d82b7f7f86df5a89d010f7dfc3364e7c1e126cdcaa
2b60b083dd3bfbd212e13c22d8a14517bae045b0ae831d39f27940b0738bae80c10e76af831316e4f37860e7165d01831dec
5e0727a8c8e7e1c0ce2dba02063bd8bdf3aeb011cb13811d5774e50a2eb069e7bf12cd7c1e09ecdca22b60b083ddeb203c0a
f93c10b872afba02063bd8bdb3b3c210cee3801d5a74a70b5e9075350cd82bffcec0b512a121eeff9f46013bd6f767bcaef5
1f68d99c1a046c950f84eb5a89d010f75f6380bdf2ef685d2d111a0afdd71060affc3b58d74a8486b8ff1a01ec957fc7ea5a
89d010f75f0380bdf2ef505d2b111ae2fe6bfcaf57fe1da96b254243dc7f0dfff5cabf0375ad446888fbafd1bfb5f2af385d
2b111ae2fe6bf0af57fe1da65b4a8486a8ff1afbeb957f47e95a89d010f75f437fbdf2ef285d2b111ae2fe6be4af55fe99a6
6b05423bdc7d09fcf5b2ff1da7abb3732bdcb7c7fd7ac55f59ba363b36423d97b05f2ff92b45d766e746b0eb12f5eb457f07
e85a016e063baf41bf5ef77778ae5508ed70f735e6d72bff8eceb512a121eebf86fc7ae5dfc1b9562234c4fdd788df5af957
76ad95080d71ff679fb4239c07fc3ab4e84e17bc20eb3ab842433e8ff7756ed11530d8c1ee9dfdd81be13cd9d7a05557bae0
055bd7c1dd26e4f3805fe7165d00831dc1bd0eeeaa229fc7fb3ab7e80a18ec60f73a7858837c1eeeebdca22b60b083dd3b3f
5886f93cdad7b94557c06007bbd7c15bddc8e7c1bece2dba02063bd8bd857b38c48c5f564f7f7ef3dba1d34ff37c0dd79f97
c77dfdb96960b3810bf581195b80f34c5f87165de982176c5d079f81219f27fa3ab7e80a18ec60f7cefee60fe13cccd7a145
57bae0055bb7404f1fdcceeff6bf3faf61177c9912fb6e7f6f46af0f436960b3810bf481195b80f31cdf0abdea4a17bc60eb
96af5a87adf1759d0e359ff718d931be7e3dfee3ba7e6dbe7cd54a239b8d5cbe6a3d32670b7c1ee4ebdca23b60b083dc9b2e
81b08207e95a65d5b5446888fbaf31ba5ef9773eac96080d85fecfdec011ce43741d5a74a50b5eb0751de435239f07f83ab7
e80a18ec60f7cede85239c27073bb4e84a17bc60ebe6bf2daae0c9bd5e59742d111ae2fe7b482c263e4f0d766ed11530d8c1
eed5cc60affc3b87d74a8486b8ff1e527b89cff38a9d5b74050c76b07b1dacb5807c9e565cb9575d01831dec5e5f59c5cefd
3b81b700063bc8bd0e221990cf938ab563d31530d8c1eef59253ecc4bfd377052d18917c3bfff365e6f39462e7165d01831d
ecdef99ff1329f67143bb7e80a18ec60f716eec7e5f2ba4e5f557f6fefebf057397e1239dd16185afe6e0f437060d3810bf5
81195b80f3786287165de982176cddf9af5a339f87133bb7e80a18ec60f7cebf3fce7c1e4decdca22b60b083dd3bff7321e6
f36062e7165d01831decdef9cf8198cf92891d7b93152f98c1de9df6ce3142591cb1b36eb242b10368d7f90f7591cdd28a2b
f2222b1c1b81ae419ce9e532ff0ac3ff7e2c3b1bcc2a5e06361b0861a669c69255bc345ba28ae5b452e5e97ba3e1b4e37f7d
92206f934c57b7347f490ad6b222ebfca11bec1d8c1e1afc7ea7a3cae5ff3d2fb7f1a5cec7e5454eaf239b8d04abe39c8de9
4a4cb0428bac74c10b740ea01fd7db7bbafb313638aef578ff6eefeff3c86623013aced998ae64043bf42a2b5df0029deb23
197947570282155a64a70b5e807300fd194fdceecff128366c909fe162e52631b930b0d940404e3336462bd1c04a2cb2a205
23d03608441e2e3cbf434f738cefb8837ddf3f4ffa83de46361b0981c871cec674251a58a14556bae0053ab740bf5ecfcf70
013a3638be9ef37d5e2eb7f92f71d8d30f87b40d9a46361bb9401f99b3315dc905566891952e7881cefd9d16fc05ba8e7f6f
efe1ffaeaff91af4f59d1ef33e6eeb59010d6c36f0efa4e0c88c8dd14a22b0128bac68c108b4ed8f78cadefecb061b4e41ef
c345ef730a6ebe7feec321fcf15eef7ce3c86623ff980fcdd998aec4012bb4c84a17bc40e77a084106ba92052cd02a2b5df0
029d5b9e150fa7878fefb4f50d57dcb7cbebf27e4c77c72fc3b9e9657a737a79564c239b8d5c9e151f99b3315dc902566891
852e78c1ce9dfd1c8dd04a0ab0128bac68c108b4ad83975e88ae44002bb4c84a17bc40e73ab8cb4e74250558a14556bae005
3a77f63b5d8456f27f9d7895152d1881b6f5f033035d891e5668919d2e7801ce2dd08feb6b7c803935f89aef1b4dcf00c69b
e2e383bff56085239b8d5ca08fccd998aee40e2bb4c84a17bc40e77af8a581ae840e2bb4c84a17bc40e7e6e85d98bf64fe6a
599175fed00df6be4bfcd5b2bf936c75fed00df63e9785f94bdeae945559e70fdd60ef1dfc39115d09db556891852e78c1ce
75f0da33d195a45d851659e98217e85c0fb9ca40576276155a64a50b5ea0731dbcd84f749eb1abcc9baa6cc109f4ad836f38
80cd3376957853158d7d20cfce8e7d022c4fd775d845552cb680ec3aff01249179b4aef26eaa93b10b7bc7ce7ff048642560
5781455636f6813c3bfb3b43022be9baca2bb292051bd0b45db6ae96fd9d58abf3876eb0f75db0ae943d1057abf3876eb0f7
5daaae96fd9d552bf3876eb8f70e02a588aee4f92ab4c84a17bc40e7cefe5b22b492e4abc4222b5a30026d3b2f268da04a7a
afb28aac50c10234ac836303d195e85e851659e98217e85c07b9874457727b1d7a95952e7881ce7570d6437425b457a14576
bae00538b78becd5b2bfa36875fed00df67efeab7d4857e27a155a64a50b5ea073bbb05e2dfbfd1942abf3876eb0f71eb66b
a02b41bd02adb2d2052fd0b90ef6604457827a155a64a10b5eb07373321fcc5f7272b5acc83a7fe8067b3f313515a04a36ae
b28aac50c10234ecec933d422ba9b84a2cb2a20523d0b60ecee989ae64e32ab4c84a17bc40e7ce7f5515e94a30ae428bac74
c10b74eeeca429422b91b84ebcca8a168c40dbe61d26cc5f4269b5acc83e7fe8067aff07c1b0d07c89bb552691b5f9808ac6
cc56c2fc258e56cb8aacf3876eb0f75d18ad943d10f1aaf3876eb0f75d12ad96fd9def2af3876eb8f75d0cad96fd1deeaaf3
876eb0f75d06ad96fd9decaaf3876eb0f75d00ad96fd1debaaf3876eb0f75dfaac96fd9de9aaf3876eb0f75df4ac96fd1de8
aaf3876eb0f75deeac97fd95e6aaf3876eb0f75de8ac96fd1de5eaf3876ea0f75de2ac96fd9de3aaf3876eb0f75ddcac96fd
1de3aaf3876eb0f75dd6ac94fd19e4aab3875eb0f31a34ab35ff3bc855a6e63eb0e79232abe57e85b8ead4d805f45b2366b5
deaff4569d9abba08e6bbeac56fc9ddbaab37327d4f52e5c568bfe8e6cd5e9432fd8f92e5956cbfece6bd5f94337d8fb2e56
56cbfe0e6bd5f94337d8fbd9e7db8456d26c9d7895152d1881b62d29babbf94b9eac9615d9e70fdd40ef1d5cc2125d099355
6891952e7881ce9d7da782d04a92ac10abac68c108b4ad83473544576264155a64a10b5eb0731ddc4927ba9221abd0222b5d
f0029d3b390602d14a7aac128bac68c108b4edfcc00fa42bd1b10a2db2d2052fd0b90e5e1b25ba921babd0222b5df0029d5b
be25f9de864be2f9538ff101c37dfc806dfcc7e333bf1bb63eaec191cd462edf921c99b3315dc98d556891952e7881ce7590
054774253456a14556bae0053ad7439028d095d058851659e98217e85c075fc4115d098d75e85556bae0053ab740ff7d91fa
3fd3870fc32ff27d4f27e6efefdf03ae75f3c691cd462ed047e66c4c571263155a64a70b5e8073bbbc58297b208555e70fdd
60ef1d6c604457926a155a64a10b5eb0730bf4fc7f534fafe19a60f8739ff7f6d76147377de8be42d3c8662317e8237336a6
2b49b50a2db2d2052fd0b9d3c37901ad64d42ab1c88a168c40db7a48e605ba9251abd0222b5df0029d3bfb7362422bd9b84a
2cb2a20523d0b60ece8688ae04e32ab4c84a17bc40e77a08a705ba92c9abd0222b5df0029debe0c371a22b81bc0a2db2d205
2fd0b9390019e62f79b85e769575fed00df6de591aae42ff4e7975bae00538d7c137bd44579270b77e4d56bae0053a77fe97
974857c270155ae48d2e78119ceb2c0a57a17f47bc2a5df0029d3bffcb4ba42b31b80a2db2d2052fd0b99e427095f877b8ab
a20523d0b6ce127015fa77b2abd2052fd0b9933f3945b4127dabc4222b5a30026d3bff4d75a42bb9b70a2db2d2052fd0b9f3
df54473a0fbd55e64d55b6e004fa76fe833a62f3e85b25de5445631fc8b3f39f3f1399e7de3aefa22a19bb408e5118eb7055
7f1f131cfe670c4e9d721cee184bba0e6c3a90a258c38c25f576e9b5a4deca39a4cad3c74a9f69250a9d24c8db24403c7e88
f31cfedac7bdccd0dc7c08bd13f23ab2d948608e7336a62b99bb0a2db2d2052fd0b93df473f839eec3e5f5633c517a8c57a0
e35f24416f239b8ddc43e7391bd395cc5d851659e98217e81c253a0fa7fdc36f31e52c3e3ec3ef33fc1dbc9ef44baf239b8d
a44ce7346763ba92b9ebd0abac74c10b740ea0efe3322097cb74fffbb95d1400f43ab2d948808e7336a62b99bb0a2db2d305
2fc0394a571ece202eafc77451f0bc8f81a2b7ef1b23a5d791cd4652ba729ab3315d89dd556891952e7881ce01f47bb810ba
be87416383c386791f0eaa0fdcbcd791cd4602749cb3315d89dd556891952e7881ce41d6f0658c0e1df67bf31fe3f52f3116
a0b791cd4642d6709cb3315d89dd556891952e7881ce2dd0c3d9c3f7395d00bc4763aecfeff3396d97eff945af0d9a46361b
b9401f99b3315d49de556891952e7881ce2d59c3c36ee6719f1eeabc87b3c5d767d8d1ced1c0afebfb73ffded6334f1cd96c
e492367c64cec674257957a14556bae0053ab7c40dcfaf79ccaf017cfeb6c5e9277a3c2ee33dbbf50a0347361bb9c40d1f99
b3315d49de156895952e7881cefd413f9eefebfdfa99ef4b5e9e8fef7bbe8b337e527a1936cced1a0347361bf9077d68cec6
74257957a14516bae0053bd7c11b254457c277155a64a50b5ea0731d3c3822ba12beabd0222b5df0029d5ba04f5cd290e84a
f8ae428bac74c10b74ae83a7654457f2771d7a95952e7881ce75f03e20d195fc5d8516d9e98217e05c0fbf34d095fc5d8516
59e98217e85c0fbf34d095fc5d851659e98217e8dc2e7f57cbfe8eb6d5f94337d87b0f1b18d095f45f851659e98217e8dc2e
fd57ca1e88b6d5f94337d87b070fde89aea4ff2ab4c84217bc60e7ce0e8a27b412fdabc4222b5a30026d5b88cf587d16b94a
e8afe28aac5cc105f4ac87a463a0f3d05f65de54650b4ea06fe73f6d27360ffd55e24d5534f6813ceb21e4784fe6b9bfcebb
a84ac62e90631d7c4b06649efbabbc9bea64ecc2deb1f31fbd1259c9fd556091958d7d20cfce8bea22a492f8aba4222b5330
00ed3afb0358422b61c34a2cb2a20523d0b65dd8b0943d90e3abf3876eb0f7f31f6f235d893a566891852e78c1cecd618e30
7f091bd6b222ebfca11bec7d2e0bf397c45f2d2bb2ce1fbac1decffe5b22b492fbabc4222b5a30026d9b8d86f94bf2ae9615
59e70fdd60ef27462b035409dc75d65556a860011a76fe6b4048570277155a64a70b5e8073bbc05d2d3bfdefff4c93d5f943
37d8fb5c16e62f91b75a56649d3f7483bdef226fb5ecef3c579d3f7483bdcf6561fe123a2b6555d6f94337d8fb8941a80055
b266955564810a16b061bbac592dfb3b5155e70fdd60ef671f8808ad24ce2ab1c88a168c40db3a38bb22ba123aabd0222b5d
f0029deb217f15e84ae8ac428bac74c10b74eeec3c28422b89b34a2cb2a20523d0b6257f75377f897df5b2abacf3876eb0f7
7f90bf0acd974c596512d99b0fa860cc6c25cc5f625fb5acc83a7fe8067bdfc5be6ad9df89aa3a7fe8067bdfc5be4ad90389
aa3a7fe8067bdfc5be6ad9df89aa327fe8867bdfc5be6ad9df89aa3a7fe8067bdfc5be6ad9df89aa3a7fe8067bdfc5be6ad9
df89aa3a7fe8067bdfc5be6ad9df89aa3a7fe8067bdfc5be6ad9df89aa3a7fe8067bdfc5be7ad95f89aa3a7fe8067bdfc5be
6ad9df89aa3e7fe8067adfc5be6ad9df89aa3a7fe8067bdfc5be6ad9df99aa3a7fe8067bdfc5be4ad99fc1aa3a7be8053baf
b1af5af3bfb355656aee037b2eb1af5aee57b4aa4e8d5d40bf35f655ebfd8a55d5a9b90beab8c6be6ac5df99aa3a3b77425d
ef625fb5e8ef44559d3ef4829def625fb5ecef44559d3f7483bdef625fb5ecef58539d3f7483bd9f7d9a4f6825f6d5895759
d1821168db2ef655cbfece34f5f94337d07b0757ce4457625f155a64a50b5ea07367df2021b412fb2ac42a2b5a30026debe0
292ed195d857851659e88217ec5c078fea89aec4be2ab4c84a17bc40e77a8a7d55e2df99a68a168c40db3a78c988e84aecab
428bac74c10b74ae83cc1aa22bb1af0a2db2d2052fd0b90e5e4825ba12fbaad0222b5df0029d3b3bda8dd04ae6ab128bac68
c108b4ad83744aa22b99af0a2db2d2052fd0b9e55baccff57d7f4f6f720ee7a3c3b8e1a799be111cce2186e3cd65db8be1c8
6623976fb18eccd998ae64be3af42a2b5df0029debe00334a22b99af0a2db2d3052fc0b95de6ab943d1068aaf3876eb0f70e
3630a22b99af0a2db2d0052fd8b90e76254457325f155a64a50b5ea0731d7ce1487425f655a14556bae0053ad759ecab42ff
8e3555bae0053ad743082ad095e457851659e98217e85c079f75125d497e556891952e7881ce7570ea4b7425f955a14556ba
e0053ad7c1c77e4457925f155a64a50b5ea0731d7c0946742577d6a15759e98217e8dcd95f82115a099d5562911d2d1801b6
75b06e0bd195d0d9ad5f93952e7881cef590bf0a74257456a145dee88217c1b9f3ef4f215d099d556891952e7881ce9dff19
1cd295d059851659e98217e85c4fa1b34afc3b5155d1821168dbf95fff215d099d556891952e7881cef5143aabc4bf135515
2d1881b69dff2405e94ae8ac428bac74c10b74ae87a45da0f3d05965de54650b4ea06fe73f2624360f9d55e24d5534f6813c
3bed630402f2ac59c75c5405627832aa871052488a2d71b372f6a8f2f485d0f735c9324990b74900faf31c4e98c637eea606
6fb7fbfb75b97c097a1dd96c2440c7391bd395b859851659e98217e85c0fbf34d095b859851659e98217e81cc5cd5ee63fb7
b1bfe5387205e67560b38114361b666c8c56b2669d7895152d1881b65194f2f84ce93afe57e3a7b8e3ff1a2ee769dbde4636
1b4951ca69cec674256b56a14576bae005384701bbafdbfd76bb4e57b7cfdb7c8ebcded6e691cd4652c06e9ab3315dc99a55
6891952e7881ce01f4f3f5b80c6747d3d5edf331dedc18763f982abc8e6c3612a0e39c8de94ad6ac428bac74c10b740ea03f
e30214f76985bde17c7cb80078dc3e6f625e06361b08c869c6c668256856894556b46004da06e9bab7f770f4bebc5e7390fb
e33a14185f708274dd7564b39190ae1be76c4c578266155a64a50b5ea0737fd0f3a9e07d3c337cbfe644d4e9b267d8014c99
cf9735471947361bf9077d68cec674256856a14556bae0053af777ca39dd9c79cc0f49e57f5fd65febef9c13c6351bf777d2
7960bec65c256256705556aee0027af6877bbf3fae97dbf432e4e736fcb57dc718f3ff991ed90e274de3cfb500e3c86623ff
900fcdd998ae44cc2ab4c84217bc60e79607b0272e654b74256256a14556bae0053ad7c10358a22b11b30a2db2d2052fd0b9
0e1ec0125d8998556891952e7881cef510260c742562d6a15759e98217e85c074f9d89ae44cc2ab4c84e17bc00e77af8a581
ae44cc2ab4c84a17bc40e77af89b06ba1231abd0222b5df0029ddb45cc6ad9dff9a93a7fe8067b9fcbc2fc25e455cb8aacf3
876eb0f75dc8ab943d9060aaf3876eb0f70efe9c88ae84bc2ab4c84217bc60e73a787385e84aceab428bac74c10b74eeecf7
1908ad44bd2ab1c88a168c40dbce4db6052e0f7955da4d55aae0013a76fe237d62f3905725de5445631fc8b3e5f9df709935
1cb5a6857d6fc3dff9b4f94d7b9cf16bb7cbfdb59edbe3c8662397e77f47e66c48e621afcebba84ac62e90630bef796ba912
9987bc2aefa63a19bbb077ecfc479d4456425e15586465631fc8b30e72f588ada4bd2ab2c80a179c40dfcefeee95d04adaab
128bac68c108b4ad83cf9b89ae64cd0ab4ca4a17bc40e7e6643f98bfa4bd6a599165fed00df7be4b7bd5b2bf835475fed00d
f6de410c1ad195ac59851659e98217e8dcec35cc5f025fb5acc83a7fe8067befe0cf89e84adcac428bac74c10b746ef61ae6
2fc9af5e769575fed00df67ef6ce9ad04aecab128bec68c108b06d17fbaa65a7b759fe335155e70fdd60efbbd8572dfb3b51
55e70fdd60efbbd8572dfb3b5155e70fdd60efbbd857297b20dc54e70fdd60ef73181dcc5f1258b5acc8327fe8867bef6087
457425ff55a14556bae0053a377b0df39710562d2bb2ce1fbac1de3b3810135d0961556891952e7881ce9d1d544468258155
894556b46004da366f5d307f8941d5b222ebfca11becfd9fe491ee9a2f19abceb4cada7c4045639668d7ddfc250655cb8aec
f3876ea0f75d0caa96fd9d30aaf3876eb0f75d0caa96fd9d30aaf3876eb0f75d0caa943d9030aaf3876eb0f7b3770c845612
58955864410b46b06dbb04562dfb3bdc54e70fdd60efbb04562dfb3bdc54e70fdd60efbb04562dfb3bdc54e70fdd60efbb04
562dfb3bdc54e70fdd60efbb04562dfb3bdc54e70fdd60efbb04562ffb2bdc54e70fdd60efbb04562dfb3bdcd4e70fdd40ef
fff787386abec4bb2a93c8da7c40456376f1ae5af67776aace1fbac1de77f1ae52f66780aace1e7ac1ce6bbcabd6fcef0c55
999afbc09e4bbcab96fb15a1aa536317d06f8d77d57abfe253756aee823aaef1ae5af17776aacece9d50d7bb78572dfa3b39
55a70fbd60e7bb78572dfb3b3955e70fdd60efbb78572dfb3bbe54e70fdd60ef4bd8e96efe92b1ea655759e70fdd60ef679f
93115a89775562911d2d1801b67570ff83e84abcab428bac74c10b74eeec3ba78456e25d855865450b46a06de77fb7897425
de55a14516bae0053b77f267d88856b25d955864450b46a06d3d65bb2af1efe052450b46a06de70786205dc976556891952e
7881ce7596edaad0bf834b952e7881ce75f07e35d1956c57851659e98217e85c07df4c105d8977556891952e7881ce75f075
0cd1957857851659e98217e85c1f49a73bba12efead0abac74c10b746e81be0cbfc2e5f2f727377d73f89e3ed47a3f9e8fe1
e4f8b29e96e0c8662317e8237336a62bf1ae0a2db2d3052fc0b905fa7e9b52d5a606fff6aed3a6f8a9ef0ee2c8662317e823
7336a62be1b202adb2d2052fd0b90e366fa22be1b20a2db2d0052fd8b9cec26515fa7772aad2052fd0b9cec26515fa7772aa
d2052fd0b9b333d809ad24cb2ab1c88a168c40db3af85485e84ab2ac428bac74c10b74ae87bc51a02bc9b20a2db2d2052fd0
b9e95e1ccd5fb25db5acc83a7fe8067b3ffb0b19422bc1ae4a2cb2a20523d0b60e2eeb88ae04bb3af42a2b5df0029deb21e6
14e84ab6ab428bec74c10b706e811eae3986ab91e9d396ebf5f57d3e2fd7e957795f2f8fef733cbc2fd034b2d9c805fac89c
8de94ab6ebd6afc94a17bc40e7ceff260ae94ab6ab428bbcd1052f8273cb1db9618ff3bcbde6ef5ae63fbfd7741a336c1bb7
dbedb55de1e0c86623977b7247e66c4c57b25d155a64a50b5ea073e31103e72fe9aa5a56649d3f7483bdf710290b74256055
a14556bae0053a77fe977748570256155a64a50b5ea073a7e7e8025a0958556291152d1881b6f590a30b74256055a14556ba
e0053ad743aa2cd079c0aa326faab20527d0b779d7b99fdd234eb5e6a6eae4dc09757dfe878e44e659a7cebba84ac62e9063
148df8ba3c9fc3d6371e4a87ff74fe5ba704cc6d64b391148d98e62c59a74bbb25eb54cea65c1e764f8377936eb3247d99e6
f40c4c802b51a7ca6cb2b0252fd0bae9ca952a94b8512d6cb214480d61ff94b2fa1c4e469f97e96ae4f1be7f2eb7fbe74e01
85dbc86623296535cdd998af648e3ab6c88297ec40f77af8d302be923aaad8261b5eb203dca3dcd13187e272996ec93eb787
c7943bba8c6c36927247d39c8df94aeea8629b2c78c90e748f92478773c4c7e7f91d2fc49ef7f7e3f2797d3062761dd87420
e58e86191bc395d851653659d89217681d30bf5f734ed378ca34dd9d9df223097a1dd96c2450c7391bf395ec51c53659f092
1de81ea48f0e1be37738b44ea78dafcb6b3863bd7fd6f8071ed96c24a48fc6391bf395f451c53659f0921de8de923f7a9d1f
3c8c2dfe5df85ca6acd0f9f1f17dbda0e491cd462ef9a347e66ccc57f24715db64c14b76a07b7f67847f8fcac7b3b7f77bbc
1b3ddf6b181a1d53941e6b44380e6c36f0ef7cf0c88c8de14a08a930bb2c6cc90bb4ee8f79cc32fcfcdd3b1ace4feff7cb70
ce3ab63b1e58dfcfc77bbdbec291cd46fe511f9ab3315f8921556c9337bc6407bbd7c1b368e22b41a48a6db2e0253bd0bd0e
1ecc125f8922556c93052fd981ee75f03611f1953052c53659f0921de85e0f798dc057e2481d5b64c14b76a07b7386215428
91a05ad8642b901a82fee7c250a1c4726a6193a5406a08fb3ffb293cc195544e653659d8921768dd2e97530bff8ebcb402a9
21ecbf879d09f0955c50c53659f0921de8de2e19540a1f88bdb402a921ecffec8d9be04a30a8329bbcb1252fd8ba0ef6dcc4
57a24115db64c14b76a07b3d24a2025fc907556c93052fd981ee9dfdc607c17948a822ab2a64c90934ae83f70080ce634295
5955610b56906fe73f0f27360f0a75e24d15b46004b976f6373900e639a18aabaa710517f6969dfdd11580959850e53559c8
820f645a079fd6115d090a556893852e9981de2dd8ff7c6544c42a21a14a6bb2502517d0b40e3e9325be12132ad82e0b5eb2
03ddeb20b18bf84a4ca9629bbce1253bd8bd5d50a916fe9dc269055243d8ff5c182a94b0502d6cb214480d61ff677ff84e70
25a954994d16b6e4055ad7c3c60d7c25ab54b14d16bc6407bab74b2bf5c2df1f699c56203584fd77b00f25be1259aad8261b
5eb203dcdb85966ae1df79a056203584fdef624bb5f0ef44502b901ac2fe77c1a55af87726a815480d61ffbbe852297c2015
d40aa486b0ff0e7625c457a25315dbe40d2fd9c1eecdc97950a1c4976a6193a5406a08fb3ffb704570253b55994d16b6e405
5a379b0d154a7ea91636590aa486b0ffb930542831a25ad86429901ac2feff41a227b45f524a95ca64e93ed1a2394ba6e7ae
428912f5c2224b81d410f6bf0b13d5c2bf733a4b81d410f4bf8b13d5c2bf933aad406a08fb3f7d37017025cb54994d16b6e4
055ab74b3395c207d23aad406a08fb9f13f5a0420915d5c2266f055243dcff2e56540bff4eecb402a921ecbf83a33ef19558
53c53659f0921de8de2ed8540bffce0cb502a921ec7f176daa857fa7865a81d410f6bf0b37d5c2bf7343ad406a08fb3f7b07
46702559d5994516b6e4055ab7cb56d5c2bf93454b81d410f43f17860a2500550b9b2c055243d8ff2e02550bffce17b502a9
21ecff1fe4c942fb9eb02a50a64aef8915ada919ab5af5bf834c75eed00af65d5256b5e0af24539b9b1b819e6bceaa56fc95
636a738746a8eb9ab4aa357fc798daf4a119ea7c97b5aa657fc798dafca91dec7e97b6aa857f07995a81d410f6bfcb5bd5c2
bfa34cad406a08fbdf25ae7ae15f89a256203584fd9f7d8c23b892b9aacc261b5bf202acebe05628f195d455c53659f0921d
e85e4fb9abc27c2054d4d89217685d078f8789af24af2ab6c91b5eb283dd3bfb3500822bd9abca6cb2b0252fd0bab3dff420
b892beaacc260b5bf202adeb2c7f55b17f878b1a5eb203ddeb20aa87f84a02ab629b2c78c90e74afb30c56c5fe1d306a78c9
0e746fc11ece5e5eef79137c7f5eb7fbeb327dd971fbcbc958b7711ad86ce0027d60c6c670258255994d16b6e4055ad75908
ab62ff4e1835bc6407bad7c10751c45762581d5b64c14b76a07b9d05b12af6ef94d18297ec00f77af8b581af44b10ab6cb82
97ec40f73a0b6355ecdf49a38a97ec60f73af8ea8ff84a1cab629b2c78c90e74efeca317c1953456653659d89217685d071f
8e105f8964556c93052fd981ee75f0e108f1955056c53659f0921de8de72363e6c8fd3be6e6c71d824dff7ef9411f29976c4
d7e773fdc3c691cd462e67e347e66ccc57625915db64c14b76a07b1d5c84105f8985556c93052fd981eef590180a7c251b56
b14d16bc6407bab7604f0f86dff33d8ff1bbbdcf7427e0f5be3f3fefcb67bd6308e39a8d5b907fcfd798ace4c23ab0c80296
8c40df3ab8b5407c251956b14d36bc6407b8d7c13761c457b261b77e8b2c78c90e74afb37458c5fe1d7d2a78c98ee0de9cf7
07154a42ab1636590aa486b0ff5d46ab16fe1d406a055243d8fff95f00225f4969556c93052fd981eecd7e43859294aa854d
9602a921ecbf87545ce02b71a98a6db2e0253bd0bd1e727181af04a62ab6c98297ec40f7e6cd0c2a7868a9d65555a64fed60
f78bdbafc7709ef4989fce4ca9b2eff96bbe4bbd9182239b8d5cdc3e3267433a0f4d556655852d5841becd46ef67f7e052af
b9a932796885fa3e3dc393124777c1a5db915ef5f15ba4f7f877f2bf3e4fd2659e1ed2348970175f2ae4a21b6172841d8400
53a97d209bd36aa49e9861ae4d357631a2567bd5ad46ea8919ba0b1215f2031999853039420e02f9ed751d4edabfd3cbc0cf
f94c765d558707361d08dc69c65d90a8601f88c934bc6407db07d88f35f6724cfe1cb302870b630a98dc46361b09e071ce5d
9ca8901f48ca34c2e4083b4889a2c369d5f53dfc976397c3267affdc86f32c225f47361b4989a269ce5da2a8901f08cb34c2
e4083b08a1a297d7306cd819ce07a2eb5f2c24906f239b8d8450d138e72e5454c80fe4651a6172841d5cc85fafe7e7fb1e7f
92f7f89ed270f678b94d9be9753804dfefb2b5d3c8662317f22373ee724585fc406aa6112647d8c1255a74cc8cbb4ee718ef
f7f077f8b88eeb5b0d5d7e1ed399d675bdf8c391cd462ee1a247e6dca58b6ee44782338d3039c20e2e01a3e36724b77939b3
cbf3721b7a9eee450c9a2707f3c866239780d12373ee024685fc4076a612264782837fe4c3c9d4704938afea767f7d8633e1
cb6d7a3238fe7d0e97eedbfd5a1cd96ce41ff9a1397719a3427e203ed3089323ece0f29cf5737ddfdfd32dd6e1aa71f84f86
1f68fcc7fb3bfc40f7d765bba785239b8d5c9eb31e997317332ae40712348d3039c20e2e8f5a4f5c0412097751a3422eba11
2647d8c10e9e3c22e12e6dd4c857dd089323ec20e48d4aed03419ea546ea8918ce0e4744bc5ddca9608b6e78c90eb6af870c
4e22dc859e0ab9e846981c61077bd8c110e12e7555c84537c2e4083b78f6eb1c88b74b5d156cd10d2fd9c1f6f5f08313e12e
f6752357dd089323ece0d93f38e2ed925f055b74c54b7604fb3ad8b721e12efc55c84537c2e4083bb8909fb8b43212eef25f
855c74234c8eb0830bf989eb2b23614d8115f04d36bee407fbd7c12b2dc4578360057b930d2fb881de75f07e07d1d52c5883
5e64a30b5ea073e73f9644ba9a082bd09b5ce88217e0dcf94f07916e970b2bd4a21b5f7003bd5b3efdbb5c8756ee73aae9e5
3a7d9c37bfa2707d8d8f7cd655937864b391cba77f47e6dc25c30af781f453034c7eb07f0bf9796b4a32e12e2656c84537c2
e4083b78f627ae88b78b89ddb05537bc6407dbd741140e12ee9262855c74254c8e0407cffe7c1df17639b5822dbae1253bd8
3e48aa95da077258ad46ea8919e6da546317192bb545b71aa92766986b538d5d6eabd416dd6aa49e9801925badf6f75728a9
d5483d31430f7fe044b84b6f1572d10b6172841c84fc56a97d2018d56aa49e9801225ca5f6816c54ab917a6206487195da07
e251ad46ea891920c875ab7d24a3d46aa49e98e1ec1d3ae2ed925c055b74c54b7604fb20cb556a1f482ab51aa92766986b53
8d5da8aad416dd6aa49e98e11f848c12c02eb455c044378004cc06416cabd43e10596a35524fcc30d7a61abb7455a92dbad5
483d3103e4ab5aed9fc1a55623f5c40ca7ef3b086f17ef2ad8a217bc6407d90701af52fb407ca9d5483d31c35c9b6aec9256
a5b6e85623f5c40c1d9c1421e12eea752357dd089323ec2084bd4aed0329aa5a23f5141820ef556a1f8832b51aa92766e820
b80b097799af422eba112647d841487d95da07e254ad46ea891920f8556a1f4854b51aa9276680ec57a97d20d9d46aa49e98
a183fd0b12ee22608d7cd58d3039c20e4208acd43e90ae5a6aa49e88017260a5f6818055ab917a6286b34f20106f17432bd8
a21b5eb283ed9b2da71a352c762b2db255481d31c1fffde93fb55fc268056a55b5f780cac6ece268a5e6cfc0579b9e7ba1be
f789b452f467deab4d1f7ac1cef7a1b452f640deab5508fd60f7904b2b950f04be5a89d411134034add43e90f96a35524fcc
00e9b452fb40ecabd5483d310304d45aed9ff1ab5623f5c40c67ef96116f17522bd8a217bc6407d9d7c1f90712ee726a855c
74234c8eb083673fc643bc5d54ed86adbae1253bd8be0e9edb22e12ead56c84557c2e44870707945e1f31d76f97363f76113
bdbeaf9fe91f97e553d8f515051ad96ce4f28ac291397791b5427e2090d5089323ec605fa9b5827d2090d5f0921d6cdff979
234cb84bae1572d18d3039c20e76101b8584bbf05a2117dd089323ec6077f1b5427e209cd5089323ec60072fd322e12ec556
c84537c2e4083bd841e82312ee826c855c74234c8eb0831da47c22e12ecbd6c857dd089323ec600fb1ae44b88bb31572d10b
6172841cece13727c25da2ed46aeba112647d8c10ef2aa9170176a2be4a22b61722438d8c1973148b8cbb51572d18d3039c2
0e9efdd51fe2eda26d055b74c34b76b07d1d9cc620e12edd56c84537c2e4083bd843d02b11ee026e855c74234c8eb0831d9c
b423e12ee356c84537c2e4083b786eee29b1ed026e855974634b5eb0773d648012e12ee356c84537c2e4083bd8c1474148b8
0bbb35f25537c2e4083b38dddfc61abbc459a92d7aa9917a22861eb63722dc85ceae1d9b6e84c91176f0fced8d0977b9b342
2eba1026479283e7ffe64cb80bbe1572d18d3039c20e9e794461b65de6ae308b6e6cc90bf66e8eaba41abbe05ba92dbad548
3d310344df4aed03a9ae5623f5c40c3d6ce544b88bbf1572d18d3039c20ef6b06723c25d02ae908b6e84c91176f0fc5bfb4c
582378057c938d2ff9c1fef590f40c7c358557b037d9f0821be8ddf91f54235d8d0136e84536bae0053ad7433e2a65f8962c
603dff527dfcb46e7c0c3ae9324fd2651e221f6f7e5fc784e4b1cbf17f0d17eb5f245f46361b49e469ce16084b16b0918b6e
84c91176f0f4f067c22b31c4862dbae1253bd83ec21e8eadc3c6f81c4faf1e9f61037d5f5f2fcc475d47361b49e069ce1608
4b0872215f75234c8eb0834b00f3be468921b6daa2971aa92762a018e2ebe5fb9d4eb0c6d0e0dbf85efee38279bcebc86623
298838cdd9026149223672d18d3039c20e5212f1f5f69eb2f7c62ec7f894fb777bc39f47361b4949c469ce16084b12b1918b
6e84c91176b0b32462233f90af6b84c91176909288afe3b3ecef7b3cf43dbfcfdb7d8c7afc00f936b2d9484a224e73b64058
92888d5c74234c8eb0830bf9e3fafddce7c779f7f5d1cef063dda727f89f2d599f06361db8701f98b105bc12436cd8a21b5e
b283edfbc39e574cbf4fab91be1e97f97ef7b48dceb1b9977553c791cd46fe811f9ab305c21243ace4aa1b6172841dfc3b5d
befee5e68c67b3972906eefe9d32a3c773fbe1bf7daed7c338b2d9c8bfd3e54373b6405862888d5c74254c8e04079718e2fb
e37ab94d2f5d8ee7f7e3cb68d35b69c3e5d4e3751b7fbe851c47361bb9c4101f99b305c212436ce4a21b6172841dec2c86d8
c80f84eb1a6172841dece0d138129618622317dd089323ece0d98fc611af641017ec5537bc6407dbd747faf29eb024201bb9
e8853039420ecee9cb54a3c4105b6dd1ad46ea89193a78bb10094b0cb1918b6e84c91176f0ec6c56c42b19c4862dbae1253b
d8be7963a31a2508d86a8b6e35524fccd0c33e9d084b10b092ab6e84c9117670721d6b94345eab2dbad6483d0586b33774c4
2b51bc862dbae1253bd8be0e8e274858a2788d5c74234c8eb0830bf9894b8923a147f11af8261b5ff283fd5b1e1e5d878ba7
eb777ac47599effc4fbfd473389f1f36d9fbba77a181cd062e8f8e0eccd818ce73788d79938d2d5881c69dff8814e93c87b7
402fb2d1052fd0b9f39f95219de7f01af42617bae00538b7403fa7a0dce9dbcbdbf7fdb97c9ef7f9bdc7d75f8f2b348d6c36
72813e326763ba92c36bd4a21b5f7003bdebe0db3ee42b39bcc62dba01263fd8bf0ebee744c292c36be4a21b6172841d3cfb
035ec42b39bc8aadbae1253bd8be0ef28990b0e4f01ab9e84a981c090e76f0ad3a1296285e2317dd089323ece0d9e1048857
52800d5b74c34b76b07dfff7c92d085022860d4c740348c06cd02e62d86a8f6f81fc88edb51aa92766d8450c97da3f8373ad
46ea891996dafb1a25e4d76a8b5e6aa49e886117f26bb50f04e75a8dd41333ec427eadf681e05cab917a6286d3f71d8457f2
850d5b74c34b76b07dbb7c61ad7d2438d76aa49e98e11f04ce124049103630d115200107837609c256fb402aafd5483d31c3
2e41d86a1f08ceb51aa92766986b538d12f26bb545b71aa92766d885fc5aed03c1b95623f5c40cbb905fab7d2038d76aa49e
98e1ecfd16e2957ce182bdea8697ec60fb966ce37d8d12f26bb5452f35524fc4d0c10936129688612317dd089323ec600717
554858028e8d5c74234c8eb083675f45235e4937566cd50d2fd9c1f62dd8f7cff0473fafb1f4fefbd47fbea9f7bedd2edb76
fedfc3da82fc73b616c04aa4b2018bae60c988605c07f78990b0043a1bb9e846981c6107e7e451aa512295adb6e85623f5c4
0c276e660456729c0d5874034b46b071bb1c67ab7d201bd96aa49e986197e36cb50fa4135b8dd413332c69c6fb1a2549b9d4
5e75ab917a6286b377e18857429c0d5bf48297ec20fb3a3866236189903672d18d3039c20e76709e86842545dac84537c2e4
083b78f64501e27980b5528b6c70c90c366f8ec5850a96316d755755a70fdd70efff20357bdfba07581bd1265bef0c4aa6d4
006b2bfa3319daa60fbd60e735c0daca1e8885b60aa11fec7e17606d950f84425b89d41113ec02acadf6815068ab917a6286
5d80b5d53e100a6d35524fccb00bb02eb57f26335b8dd413339cbd6b44bc12606dd8a217bc6407d9d7c1b110094b80b5918b
6e84c91176f0ecd33ec42b01d68aadbae1253bd8be0e2e6491b004581bb9e84a981c090e76f0920b1296006b2317dd089323
ec60076f352161c9b03672d18d3039c20e2ee4c3f6f7bcbde6f5bfe7fcb6d7f40ef3e776bddd6ec369d64a4e239b8d5cc88f
ccd90261c9b03672d18d3039c20e2eef2c3eaeaffb7b4a28fc5e5ef31fe5f42af1d8ee9864b8fee638b2d9c8e59dc52373b6
405832ac8d5c74234c8eb0831dbc978b8425c3dac84537c2e4083b78f6274d885702ac0d5b74c34b76b07dcbf70ef3ff4def
c6bf8623fe70089e7f9ceb65d81d7fc78890e57b071ad96ce4f2bdc391395b202c01d6462eba112647d8c10e227d91b00458
17f25537c2e4083bd8c1179b485802ac8d5cf442981c21077bf8cd89b004582bb9ea46981c61073bf83b47c212606de4a22b
61722438d8c1e783485802ac8d5c74234c8eb0830bf9f0e35cc74f14c62e3f8fc760d46bfab1c6f390ebeb725f7f731cd96c
e4427e64ce16084b86b5918b6e84c91176b0872467222c19d6462eba112647d8c11e929c89b064581bb9e846981c6107cf4d
7226b6925e6dcca21b5bf282bd3b3f8591094b86b5918b6e84c91176f0fc1446262c19d6462eba112647d8c11e7e73222c19
d6857cd58d3039c20eee32acadf68164e65223f5440c1d7cc6888425c35a3a36dd089323ece09caf4b354a8ab4d5165d6aa4
9e12c3f9376298b0a4481bb9e846981c6107cf3f8e3361c9923672d18d3039c20e9efc8934e395186bc316ddf0921d6cdff9
a7ac4c5842b48d5c74234c8eb083f30e866a94186bab2dbad5483d31c3e91b1be1950c6bc316ddf0921d6c5f0ffb5522f40c
6b03df64e34b7eb07fe71fc591cf33ac0d7b930d2fb881decd7f5f30bfa74897b28b6cf3876eb0f71ea27529fdb9a448eb19
90eae367a1dfdb63d6659ea4cb3ca76729135e8990366cd10d2fd9c1f6edb187a3cb70d1f4bc4c7b83c7fbfeb9dcee9ffb1b
b8b791cd46eec1f39c2d109614692317dd089323ece0e93f38e19508e982bdea8697ec60fb7a484b27c212606de4a217c2e4
08394801d6af61c478eb60dc46e71b08774c715e07361d48f1d561c616f04a7ab5618b6e78c90eb6afb3f46a233f90c96c84
c9117610c85fafc7f73d1ce4c7269f8ff7f3f979ddf12f7c19d86c2070a7195bc02bd1d5862dbae1253bd83e8aae7e0dc386
43c0bc4f9a9fe23fe807df46361b49d1d569ce16084b74b5918b6e84c91176b0a7e86ac33e10c86c78c90eb6efef0479f805
6eb7e9cb9ef7ebf3ba0ea7d2d343aee1fcfdf21d6a5cd7eb021ad86ce0dfe9f191195bc02bb9d58aadbae1253bd8bece72ab
8dfc401ab312264782839de5561bf9813466234c8eb0831dbc0b808425b7dac84537c2e4083b7876da26e295d06ac316ddf0
921d6c5f07699b485872ab0bf9aa1b6172841dece04500242cb9d5462e7a214c8e9083bbdc6aab7d208dd96aa49e98e1ec37
0a11af84561bb6e88697ec60fb3a78df06094b6eb5918b6e84c91176f0f4fd2ae195c86cc316ddf0921d6c5f6791d94a7e24
86da089323ece02e32db6a1fc882d61aa9a7c070f6c686782532dbb04537bc6407db77f64e15f14a5eb6618b6e78c90eb66f
c19e9ea05ca653eaeff4d7709bee830e1795cfe110f8bdae172838b2d9c805fcc89c2d107a5eb6816fb2f1253fd8bf8ef2b2
8df96718b4b1052bd0b8535f63232e4fca2eb88b6c5cc105f4ac838f8c88ce93b20d7a930b5df0029cebe09b32a22b49d946
2dbaf10537d0bb0e3e2244be92946ddca21b60f283fd5b3e1cbdde3fc34e6efa711eb7d7bc98dbb4793eef8fcbedb93e1ae3
91cd462e1f8e1e99b305c292946de4a21b6172841d3c2f8508c14a46b602ab6e60c90836ae836818242c19d9462eba122647
828367a700205e09c8366cd10d2fd9c1f6cd69af54a3a4545b6dd1ad46ea8919fe6d4ab5818d6f7afc487e368004cc06ed52
aaadf681e467ab917a62865d4a75a9fd33f9d96aa49e986197526db50f243f971aa92762d8a5545bed03c9cf5623f5c40cbb
946aab7d207ed96aa49e9861ae4d354a54b4d516dd6aa49e98611715adb58f24245b8dd41333cc71af54a3a4395b6dd1b546
ea2930ecd29cadf6818464ab917a62865d9ab3d5beff4e48b61aa92766d8a5395bedc7ef8464ab917a62865d9ab3d53e9090
6c35524fccd041540e12962c692317dd089323ec6007e7834858e2a40bf9aa1b6172841d3c3b2408f14a92b5618b5ef0921d
64df79573c085602ac0d5874034b46b0711da422216109b03672d18d3039c20e7670598f8425c35ac95537c2e4083bb8905f
aec335f07dde2e2fd7a9b7f97edbf535be20fc7cade434b2d9c885fcc89c2d1096306b23175d099323c1c1396d976a943869
ab2dbad5483d31430737d090b084591bb9e846981c610767d7a946c995b6daa25b8dd41333ec72a5adf681ac66ab917a6286
1efece88b0a45a1bb9e846981c61073bd8b72261c9d42ee4ab6e84c91176b083e3291296586d2317bd102647c8c11323eb09
aca4691bb0e806968c60e3ce3e4546bc12a56dd8a21b5eb283edebe09a08093d4d5bc14536bee407fb77f61d7282b3206f43
5e55250b46b06d7326f17e76cfdab6a29b6cd3732fd4f73f080f87d63dc8db8836d97a0fa0684b0df2b6b207e2b1ad42e807
bbdf05795be503e1d8562275c404bb206fab7d201cdb6aa49e986117e46db50f84635b8dd41333ec82bc4bed9f09d55623f5
c40c4bed7d8d92a66db5452f35524fc4d0c12d42242c69da462eba112647d8c10e0e87485802b5955c75234c8eb083e79df0
2158c9d13660d1152c19118cebe07a0e094b8eb6918b6e84c9117670b977317f0d35fff54daf2abde79fea7219dfa6b86c6f
30e2c86623977b1747e66c81b0e4681bb9e846981c610717f2ebf3f17c3ebff3a678191fc87d960f46aec37ff6595f62c491
cd462ee447e66c81b0e4681bb9e846981c6107cf8fdb60c292a36de4a21b6172841d3c3fcd87094b8eb6918b6e84c91176b0
836f8d90b044691bb9e846981c61073bf8f005094b94b6918b6e84c91176b0832f4090b0446917f25537c2e4083bd841ac32
1296286d2317bd102647c8c11e7e73222c51da4aaeba112647d8c10efece91b044691bb9e84a981c090e76b06f47c212a56d
e4a21b6172841d5cc88761c35176fa331c1f1adcdfefe91eeb70c1f1bd8c6fd4ade7ed38b2d9c885fcc89c2d1096286d2317
dd089323ece0f951714c58a2b48d5c74234c8eb0831d9cbd226189d23672d18d3039c20ef610ae4c842550dbc84537c2e408
3b38470c538d12696db545b71aa92766986b538d122a6db545b71aa9276658a295f7354aac73a9bdea5623f5c40c7d045aef
094ba8b4918b5e089323e4e0b991f5c456e2a4a557d38d2d79c1de9d7f3c61c212666de4a20b6172243978fe3d01262c61d6
462eba112647d8c1f3cf9e98b084591bb9e846981c6107cfff6e9b094b9eb5918b6e84c91176f0fc6b05262c79d6462eba11
2647d8c1f3738698b0a4691bb9e846981c6107e7a461aa5122adadb6e85623f5c40c3dfca511a1475a1bf8261b5ff283fd9b
3d870a1e2a6d8537d90a847eb0fbc5eef3168a443a8fb42ed08b6c74c10b748e128e2fc3dfdef08ff14ff231763ebe64b76c
633cb2d9484a384e739648ebb5e11269ad6761aa8fdfaf3eefc321ea7f7d9ea4cb3c40fe790ed772e3fb8b5397b7dbfd3d46
1413f93ab2d948208f73b6405852ad8d5c74234c8eb083d39511d628b9d2565b74ab917a62861eb637222cd1d2857cd58d30
39c20e7690a8bdc72bb9d2862d7ac14b76907d94a8fdba3c9fc3e16e8a3bff4ecbb30c1b2971af239b8da444ed34670b8425
5adac84537c2e4083b787aa236e1955c69c316ddf0921d6c5f4fb9d2867d202dd9f0921d6c1f600fbfc5eb369cbf3fa7e3c0
b033ba0c7b645c1b621dd96c2480c7395b202cb9d2462eba112647d8c19e72a50dfb4060b2e1253bd8be25577a4ce8bc4e7b
dcf77bb8547b5cc70bf4a1c9a1fbf11cf3badefbc091cd462ec9d247e66c81b0444b2bb9ea46981c6107ffc86fe3e75cb7f9
32f5326c8743cfd37def41bbdf2eafc7737d5a88239b8dfc233f34670b84255adac84557c2e448707089969e6e3acec17af7
db7ca13afd58afe9d2e9b15d22e0c866239768e92373b64058a2a58d5c74234c8eb083cbb3f1e132e2f19d02c83ec3bee0f2
babc1ff38ff5787d2fd3ebf2cbb3711ad96ce4f26cfcc89c2d109668692317dd089323ec6007cfc691b0a44b1bb9e846981c
6107cfbfdfc584255dba90afba112647d8c1f3efbc306149973672d10b6172841c3cff260813966c6b2317dd089323ec600f
5b3b1196786b2317dd089323ece0f9777799b0c45b1bb9e846981c6107cf8ede45bc126f6dd8a21b5eb283edebe190468425
de5ac95537c2e4083b787ac434e195646dc3165df1921dc1be0e7e70242ce1da462eba112647d8c10ef66d4858f2b58d5c74
234c8eb0831d9cc320a1e76b1bf8261b5ff283fdeb285fdb987f46481b5bb0028deb255fbbe0fe088f36aee0027ad657beb6
41ff8c8e2e74c10b706e817e5c2eafeb744be47b7b5fbfefdb7d5a80ea33dff0de5e5fc4814d072ec807666c8c56c2b50d59
74830b56a0710bf31604308c797c86139a799b7ccc8798f5604d039b0d5c980fccd8025c49d63668d18d2e99c1e675101383
842559dbc84537c2e4083bd847b2b6021fc98b36b064041bd7c167d2485892b58d5c74254c8e0407cfce0d41bc92ac6dd8a2
1b5eb283eddb256b5bedf1bd8c1f81d25623f5c40cff3659dbc00ea4551b4002668376c9da56fb405ab5d5483d31c32e59bb
d4fe99566d35524fccb04bd6b6da07d2aa4b8dd41331ec92b5adf681b46aab917a62865db2b6d53e90566d35524fccb04bd6
b6da07d2aaad46ea891976c9da5afb485ab5d5483d31c32e59db6a1f48abd61aa9a7c0b04bd6b6dab7df69d55623f5c40cbb
646dab7dff9d566d35524fcc70f63112f14aa8b7618b6e78c90eb6af839322242c91e2462eba112647d8c17963a31a25d4db
6a8b6e35524fcc70766e1fe29544ef82bdea8697ec60fb3a48a644c212ea6de4a217c2e40839d8416e0f1296746f2317dd08
9323ec600799c34858d2bd8d5c74234c8eb08367df5141bc12edadd8aa1b5eb283ed5bb04f5c9c0e094bb4b7918bae84c991
e0e0d9b70d11afa48a1bb6e88697ec60fbe67c6baa5172bdadb6e85623f5c40c3d6c6c445852c58d5c74234c8eb083bb5471
ab7d202bdb6aa49e9861ae4d354aaeb7d516dd6aa49e98a1831c38242cb9de857cd58d3039c20e76700e818425d7dbc8452f
84c91172b0839bf4485802be8d5c74234c8eb0831d9c312361c9f83672d18d3039c20e9e7d8984781ef0add4221b5c3283cd
ebe06a9cf82ce3dba85755e18217ecdcd9f75d00cce3c58d77938d8c6d20cbe6846498dd13c0ade826dbf4a117ecfc1fa49a
43f3255edca044b7f6032c5ab38b17b7ca0722bbad44ea880976f1e256fb4064b7d5483d31c32e5edc6a1f88ecb61aa92766
d8c58b97da3f23bbad46ea891976f1e256fb406876a9917a2286cee2c58dfc4068b6112647d8c11e0e4a4458e2c5955c7523
4c8eb0837dc48b1bf081d06c054b4604e33a38d146c2122f6ee4a21b6172841d3cfb661de2956c71c316ddf0921d6cdfd9b7
ac10af048b1bb6e88697ec60fb3a082844c2122c6ee4a21b6172841d3c3f3e8d094bb0b8918b6e84c91176f0fc1031262cc1
e2462eba112647d8c10e3eb642c2122c6ee4a21b6172841d3cfb9323c42ba9e2057bd50d2fd9c1f64de7c958a3e47a5b6dd1
4b8dd4133174f00d0812965c6f2557dd089323ec6007df802061c9f53672d1953039121c3c3f9992094baeb7918b6e84c911
76b083c32912965c6f2317dd089323ec600fe9d6445872bd8d5c74234c8eb0831d7c12838425d7dbc84537c2e4083bd8c14a
44485872bd8d5c74234c8eb0833dfce6445852c58d5c74234c8eb083bb5471ab7d20a9db6aa49e9861972a5e6affcccab61a
a92766e861ff46842555dcc8452f84c91172b087631a11966c71e9d874234c8eb083277fcdca782558dcb04517bc6447b2ef
fc1357262cc1e2462eba112647d8c1f32f5698b0048b1bb9e846981c61074fce6660bc922a6ed8a21b5eb283ed3b7fd52926
2ca9e2462eba112647d8c1f333c598b0a48a1bb9e846981c6107cf5f638d094ba6b9918b6e84c9117670cef6a61a9e2a6ea5
37d92aa48e98a087bf33e0f34c73c3de64c30b6ea077a7ef5301cd03cd0bf1221b5a30026d8380e9f773f8031c1352867f3c
9e97d7fbf6d9ded2e291cd4642c0749cb3049aaf0d9740733df5537dfc5afa7affccbacc9374998702cd87dddef3333df979
3c5fc3d5ecedf2a500da7560b38114671e666c01afa4991bb6e88697ec60fb7669e656fb4046b7d5483d31c35c9b6a943cf1
527bd5ad46ea89192849fdf9797f9e97e974f6f1be7f2eb7fbe7fea6df7b1dd96c2425a9a7395b202c91e2462e7a214c8e90
8394a43e5ca2bec60cd5b1cbcff3fe7a5f5f2f8c5a5e47361b4959ea69ce16084ba4b8918b6e84c91176b0a74871c33e9095
6d78c90eb60fb0877dcfe57b794f4b9f3d1fe385ec6538b723ee7564b391001ee76c81b0a48a1bb9e846981c610729557cfc
a6ecf9fa4e3fcfa7dcf8e091cd4652aa789ab305c2922a6ee4a21b6172841dec2955dcb00f64651b5eb283edeb2c555cc98f
64651b6172841d5c52c5efcf61eb9beeba7f2ee399f5f73ebdaaf41e43b1bf97efba57e791cd462ea9e247e66c81b0a48a1b
b9e84a981c090e2ea9e29fe1acee31bfb8707f0ddbe2f3729bb6d2d7b05ff87e2edbdd4d1cd96ce4922a7e64ce16084baab8
918b6e84c91176f0fceb60262ca9e2462eba112647d8c1f36ff3316149153772d18d3039c20ef690304d842555bc90afba11
2647d8c10ede4940c2922a6ee4a217c2e40839385d27618d92eb6db545b71aa92766e8e0253b242cb9de462eba112647d8c1
f31fa23061c9f53672d18d3039c20e76f07a211296686f2317dd089323ec600fc715222cd1de4aaeba112647d8c1b35fac44
bc12ed6dd8a22b5eb223d8d7c10f8e8425dadbc84537c2e4083b381d54b04609d7b6daa25b8dd4133374701283841eae6de0
9b6c7cc90ff6af835318e2f37c6dc3de64c30b6ea077bde46b17dc1fe1d1c6155c40cf3af81285e83c5fdba037b9d0052fc0
b9b3bf3a22b492af6dc8a21b5cb0028d3bfb0333842bf9da062dbad12533d8bc0e424b90b0e46b1bb9e846981c6107fbc8d7
56e023a9d106968c60e33af84e18094bbeb6918bae84c991e06007df86236189d83672d18d3039c20e9e9dde817825dddbb0
4537bc6407dbf76fd3bd0dec4062b6012460366897ee6db50f24665b8dd41333ecd2bd4bed9f89d95623f5c40cbb746fab7d
2031bbd4483d11c32edddb6a1f48ccb61aa92766d8a57b5bed0389d95623f5c40cbb746fab7d2031db6aa49e986197eeadb5
8f24665b8dd41333ecd2bdadf6f57762b6d6483d0586b3f7d7885782c50d5b74c34b76b07d1d1ca091b0c49a1bb9e846981c
61073b382943c2926c6ee4a21b6172841d3c3b400ef14aacb9618b6e78c90eb6efbccb0e042b59ea062cba812523d8b80e32
6b90b0c4a917f25537c2e4083bd843b436119638752317bd102647c8c10e426091b0c4a91bb9e846981c6107cfbe75867825
4bddb04537bc6407dbb760ffbdf03e35395c8d4e77ecc7263fe39ef8727fad39a838b2d9c805fcc89c2d109638752557dd08
9323ec60075f1e236189533772d1953039121cece0d638129644752317dd089323ece042fefe3c1faff7f42739ec03c78ff8
eef34720f37b948ff566318e6c3672213f32670b8425cfddc84537c2e4083b38278b538d92a86eb545b71aa92766e8617b23
c292e76ee4a21b6172841dec611f4384254ddec84537c2e4083bd8c17105094b9a7c215f75234c8eb083679f48205e899237
6cd10b5eb283ec3bfb311ce2951c79c316ddf0921d6c5f078fe190b0e4c81bb9e846981c61077b48d027428f925770918d2f
f9c1fe9d7dfb83e02c47de905755c982116c5b0737ba80cda3e40d79930d8e9d20d7cebe974b609e626fbc9b6c64c106346d
ceff86f94bd0bcd515dd2a847eb0fb7f109f4fed97147bc312ddfa4fb86ccf2ec5de6a1f4886b71aa92766d8a5d85bed03c9
f05623f5c40cbb14fb52fb6732bcd5483d31c32ec5de6a1fc8662f35524fc470fa3e83f04a84bd618b6e78c90eb6af83e73f
485822ec955c75234c8eb083679f06205e09b2376cd1152fd911ecebe0ac0f094b90bd918b6e84c91176b083337d242c59f6
462eba112647d8c10e1e0a206189b33772d18d3039c20e7690cb898425cedec84537c2e4083bd8c19d52242c71f6462eba11
2647d8c185fc361c68fe2e472ef7f906dbf4a1d873cc991c6fb9ace434b2d9c885fcc89c2d1096387b2317dd089323ec6007
9f45206189b33772d18d3039c20e76f0210c129644fb42beea46981c61077789f656fb404e7ba9917a22861eb637222c89f6
4aaeba112647d8c193f35019afc4d91bb6e88a97ec08f69dfa0912b295207b6316ddd89217ec5d070752242c41f6462eba11
2647d8c10e1e7e216109b23772d18d3039c20e7670da888425c8dec84537c2e4083b38078b538d12256fb545b71aa9276698
6b538d12e66eb545b71aa92766e8615b27c212256fe4a21b6172841dece0e13e129620fb42beea46981c6107cf3d9a115b89
b03766d10b5bf282bc3bfdc485f04a7ebdb46bbae1253bd8bef3af0f98b044d81bb9e842981c490e9e9c47c17825bfdeb045
37bc6407db777ea4171396fc7a2317dd089323ece01ff9e3737ddfdfd35ee8334c7b19f642f7f11fefefb01fbcbf2edbfe1c
47361bf9477e68ce16084b84bd918b6e84c911767021bfdfa68bc4f96f70be3133a50d7eea6d4e1cd96ce4427e64ce16084b
84bd918b6e84c9117670211f8e3ab7cbe52f01e9319c677edfe3b1e73d1c771fc3c5dc76f682239b8d5cc88fccd9026189b0
3772d18d3039c20ece41ee54a384c85b6dd1ad46ea89197621f256fb673eba55481d31410f5b3af07988bc616fb2e10537d0
bb05faf91eae933ef36dd7cbf3316c8e738aeae7f5ba3cafcfedbe2e8e6c3672813e3267633acf912fd08b6c74c10b74ee2f
7f767a847cff8ec79c65c7fb19f7bc8fdb737c6bf5715b4ed87864b3917f09b487e62c39f26bc325475e4fc1449ff210861d
c5ac6ff3245de759c89fefc7f476e8f3f1ba0c5be3f33185becf6b693cb7e05d18d76cdc42fd7bbe16d84a88bc316fbab325
2fd8bb8579f82bbc7ca7cb86e7ebf2194e672fef29e1fffebcbfc7dbdf6bce328e6c3672e13e32670b8425c2dec837dd0993
23ece092b37c790d9d4d2b5f3d87e3dc6738b3bb8eff783c2eb7ef7035f35cf6683cb2d9c82569f9c89c2d109600fd42bee8
4e981c61074f5f2f81f04a7abe616f7ac54b76907d9da5e71bf9ef5878274c8eb0833da5e71bf6ef4c78c74b76b07d803df4
f8bcdf3ed37e683899fcbeef8ff795168ad846361b09e071ce16084b7abe916fba132647d84120ff8c2f4ddca76bc8e13f1f
4e3686a32d822f039b0d04ee34630b78253adfb037ddf1921d6cdf1e7bd8fb5d6ed3ee78da44ff2e273ec0bd8d6c36720f9e
e76c81b0a4e71bf9a63b6172841d5cd2f3af8fdb6dfa1cfcfdfa3be08eff78bde7e4f7f5288e039b0d5cb2f30fccd8025e89
ce576cd11d2fd9c1f675169d6fe4bf03e18d3039121cec2c3adfc87f07c23b6172841d3c7ff13c262cd1f946bee94e981c61
073b48bc45c2129d6fe49bee84c91176b0872873222cd1f9857cd19d3039c20e9efdf404f14a6ebe616f7ac54b76907d1dbc
51878425b5dfc837dd099323ece0bc6200d528b9f9567bd3bd46ea89194edfd808af84e61bf6a63b5eb283edeb61ef428425
34dfc837dd099323ece0f90f2999b084e62bb9e84e981c61073b582801094b6ebe916fba1126478283737a3cd528c9f5567b
d3bd46ea89193a589803094b6ebe916fba132647d8c10ef63148e8b9f906becace97fc60ffce3ea4109c87e61bf32a3b5bb0
028deb2534bfe0fe7722bc730517d0b30e5e63233a0fcd37e855ae74c10b70eeecc00e422ba1f986bce90e17ac40e3ce4eeb
40b8129a6fd09bee74c90c36af836ff890b084e61bf9a63b6172841dece0bb4d242ce9f94a2eba132647d8c10ef22b90b0a4
e71bf9a61b61722438d8c1c7d94858d2f38d7cd39d3039c20e9efd353ee295f47cc3de74c74b76b07dff363ddfc07e27d23b
4002668376e9f956fb7722bdd7483d31c32e3dbfd4fe9548ef35524fccb04bcfb7dabf13e96b8dd41331ecd2f3adf6ef447a
af917a62865d7abed5fe9d48ef35524fccb04bcfb7dabf13e9bd46ea891976e9f95afb4022bdd7483d31c3d9fb4cc42bc1fd
86bde98697ec08f675709044c292dd6fe49bee84c91176b0831323242cd9fd46bee94e981c6107cfce6e41bc12dc6fd89bee
78c90eb6af836b00242cd9fd46bee94e981c61073bb8ee43c212e26fe49bee84c91176f0ec0b7dc42b09fe057bd11d2fd9c1
f62dd8cf295366da03dfbeefcfe5f3bccf9900afbf0f66576e1ad96ce4027e64ce16084b82bf916f7a254c8e90831d04ef22
6149f037f24d77c2e4083bd8c197a8485842fc8d7cd39d3039c20e2ee4afd77001f57efeed7d86738ecb757a84f2be5e1edf
e7f82ecc424e239b8d5cc88fccd9026109f15772d19d3039c20e2ee48feb6b7c2037ed805ef3f1674a4218ef328f8f92d6bf
731cd96ce4427e64ce16084b88bf916fba1126478283673f8b41bc92e06fd89bee78c90eb66fceb1a71a2543df6a6fbad748
3d3143077935485812fc8d7cd39d3039c20e9ebeb1115e89ef37ec4d77bc6407dbd7c3de85084b7cbf916fba132647d8c10e
8e284858e2fb0bf9a23b6172841dece011281296047f23dff44a981c21073b387344c212e26fe49bee84c91176f0ec8b24c4
2b09fe86bde98e97ec60fb3ab82a46424ff057f04d76bee407fbd7c17d10e2b3107fa35e54830b5eb07367dff102304ff037
de557632b6812cebe0a62eb17988bf21afb2c30527d0b7b36fe0135a593fc09037dde1821568dcbf5d3fc0b07e67f27bff09
97edd9ad1f60b57f67f27b8dd41333ecd60fb0dabf33f9bd46ea891976eb0794dabf32f9bd46ea891976eb0758eddfc1f8b5
46ea89184effe325bcb27e80616fbae3253bd8be1e76d44458d60f5072d19d3039c20e76f0f40d09cb120246bee946981c09
0e2ee4ff7e2165042b2b0718f0a63b5832828d5b80afe72da48c8465e50023df74274c8eb083cb85d5890b2923615939c0c8
37dd099323ec60070fa090b0ac1c60e49bee84c91176b083f05b242c2b0718f9a63b6172841dec20ea1909cbca0146bee94e
981c61074ffd4c02d9ca9a01c6bce9ce96bc60ef3ab8398c8465cd8042bee84e981c6107776b0658eddf49f8b546ea8918ce
ddd288adac16a0cca23b5bf282bdeb60bf828465c10023df74234c8e04073bb8198f8465d90023df74274c8eb0831ddc8c47
c2b26c80916fba132647d8c139c69d6a94e07eabbde95e23f5c40c677f0a867825b5dfb037ddf1921d6cdf2eb5df6aff4ec2
f71aa92766e8614327c2b26680916fba132647d8c10e4e9491b0ac1960e49bee84c91176b08337b590b0ac1950c817dd0993
23ec6007278d4858560e30f24daf84c91172b0836fe691b02c1e201dabee84c91176f0fca427262c8b0718f9a62b61722439
d8438a3e1196f5038c7cd39d3039c20ef610364d8465fd0023df74274c8eb083e767d93161593fc0c837dd099323ece0423e
c7ce4d7f93b7fb6d38e45c5f5397c339c7fdf1787cd65bdb38b2d9c885fcc89c2d1096f5038c7cd39d3039c20e2ec98dafcb
b0ff990fb7c3ef741f36912938f73dfc70c399d8e3bdfee638b2d9c825b9f1c89c2d1096f5038c7cd39d3039c20e9e9f59c9
8465f50223df74274c8eb083bbd50bacf6af607eaf903a62821eb636e0f3d50b0c7b951d2fb881de2dd0f339cdf46af0f533
eef76ed3fef03dec042e63e6dd7a270447361bb9401f99b3319daf5e50a0ff64a70b5ea0737f11c0aff7f5f6b94c0fb99edb
fb636394ea70493d9ce57f9773461ed96ce45f04f0a1397df582ad615fbdc0ce01551f9338eecfdbaccb3c4997799624ffcb
f08b7ca793d9e7f646d5e0d0e5323fd95cc07160d3814b8eff81195bc0f3050c1c5b74c34b76b07d0bf6e379198ee753e4fe
f63aced0e4f5f97749b772d3c8662317f02373b640e80b1838b9e846981c61077b58a883087d01834abeea46981c6107979c
ef71b39c76b6cff14d8dfbf33dddb8793c4a6a210e6c367049f93e30630b78be8081638b5ef0921d641facdbf0bd0cd76dc3
3fa62ec7c8c5f17dd26dc50a1ad96c24acdc10e76c81d017307072d18d3039c20e429efd755a41e4316da3c3de7ff85fcfe7
fb0be4dbc8662321d13eced902a1af61e0e4a21b6172841d04f2c7eb7ebbbc1ed369def37e7ddd5eb7ef7a0f8447361b09e4
71ce16087d0d032717dd089323ec60476b1838f6ef647ec74b76b07d7dad61e0e4bfc3f99d3039c20e76b48681611f48e677
bc6407dbb79ca68f5f2edee6a0d8cb73f8713ef7e7fc4bad7f9acb693a8d6c3672394d3f32670b84be8681938bae84c991e0
e0f96bca31a1af61e0e4a21b6172841d5cee720e07d7d7757a2debf31ecfe9c7e4d8e99aeaf5f7307dbdcb49239b8d5cee72
1e99b305425fc3c0c94537c2e4083bd841ee3212fa1a064e2eba112647d8c10e3e0546425fc3a092afba112647d8c10e9e60
21a12f63e0e4a217c2e4083978eaab47c8e60b1838b3e8c696bc60efea02065efb7732bfd7483d314307df2023a1af61e0e4
a21b6172841dec61c10c22f4350c9c5c74234c8eb08367070d209e2f6060d8aa1b5eb283edebe0e13812fa02064e2eba1226
478283750103affd3b96df6ba49e9861ae4d357c0901af2dbad5483d314307271048684b0838f8261b5ff283fdeb60d7467c
b68a80636fb2e10537d0bb0e8e6444676b0954e84536bae0053ad7c1bbbb44676b0938f42617bae0053877f60ba484e66b09
38b2e80617ac40e316e6d763bcff367f62397da1f47e4edfe35dea953f8e6c3672a13e32670b7cbe9c80738b6e80c90ff6af
83000d24f4e5049c5c74234c8eb0831d446820a12f2760e4aa1b6172841d3c3b4903f17c2d01c7165df1921dc1be0e3e5746
425f4bc0c94537c2e4083bd8c127ea48e8cb0938b9e846981c6107ffe972020ef63ba2df0112301b549713f0dabf23fabd46
ea8919ea7202b5f6af887eaf917a6286ba9c80d7fe1dd15f6ba49e88a12e27e0b57f47f47b8dd41333d4e504bcf6ef887eaf
917a6286ba9c80d7fe1dd1ef35524fcc7076a808e2f94a0686adbae1253bd8bec972ace1cb09786dd1b546ea293074709c44
425f4ec0c94537c2e4083b78f68911e2f95a028e2dbae1253bd8be0ece8491d097137072d18d3039c20e7670f58384be9c80
938b6e84c91176b0832b5e24f4e5049c5c74234c8eb08367dfdf403c5f4ea062afbae1253bd8be0eee6521a12f27e0e4a217
c2e40839d8c1ad4b24f4e5049c5c74234c8eb0831ddca946425f4ec0c94537c2e4083bd8c1b30924f4e5048c5c75234c8eb0
83cb73b81397b446425f4ec0c94557c2e44870b083577990d057147072d18d3039c20e2ee4d3c9f5149ef619b68cf7f3359f
7f7ceecb29e7024e039b0e5cb80fccd8029e2f66e0d8a21b5eb283ed9b1752a01abe9c80d716dd6aa49e98e1fc2fa299d057
147072d18d3039c20ef6b08321425f51c0c94537c2e4083bd8c14105097d45814abeea46981c61073b78630e097d45012717
bd102647c8c10e723e90d057147072d18d3039c20e76906a8384bea880938b6e84c91176f0f4b51408cf5614306a910d2e99
c1e67590f04a7cbaa28053afaac2052fd8b9f3527c01c9d61270d24d36263680cceae08e2eb1d95a028ebcc906179c40df3a
78e84974be9c80538b6e7cc10df4ee9f2e27e058bf23fabdff84cbf6d4e504bcf6ef887eaf917a6286ba9c80d7fe1dd1ef35
524fcc509713a8b57f45f47b8dd41333d4e504bcf6ef9cfc5a23f5440c736daae199fe5e5b74ab917a6286b39f3a229e07fa
1bb6ea8697ec60fb3ad86522a107fa3bb9e84a981c090e9efde40ff13cd6dfb14537bc6407db77e23a0604e669fe0e2cba81
2523d8b80ece7291d0d3fc9d5c74234c8eb083675fd4209e47f93bb6e88697ec60fb7a58c480083dcadfc94537c2e4083bd8
c1f53b127a94bf938b6e84c91176b083479d48e881fe4e2eba112647d8c10eee5621a107fa57f25537c2e4083b5803fdbdf6
efb0fa5a23f5440c3d6c6f44e8b1fe46aeba112647d8c10ef63148e8b1fe4e2eba12264782831d1c5790d063fd9d5c74234c
8eb0831dbc3984841eebefe4a21b6172841dacb1fe5efb775ebdd7483d31430fdb1b117ab2bf938b6e84c91176b026fb7bed
df79f55e23f5c40c67bf2b86781eebefd8a21b5eb283edeb2bd6dfc97f87d53b6172841dec619d1a22f458ff4abeea46981c
61073b78570c093dd6dfc9452f84c911727079a1e17319a69aded11dfe0ac7b38dc71c39f028f18d3cb2d9c8e585862373b6
40e8b1fedab1e946981c610717f2cb6338bacedbe27b38b3bfbfa658d1cfedef146cddd86960b3810bf781195bc0f34c7fc7
165df0921dc9be25e7eb737ddfdf536acd6749151dfff1fe0e7f1c43c7db6e1d47361bb9e47c1d99b30542cff47772d18d30
39c20e2ef9ee532ccffc5ad718c53dbe5634a591bda6976c1edba68e239b8d5cf2dd8fccd902a167fa3bb9e846981c61077b
48dc2642cff47772d18d3039c20efe915fff4eaec72e2fd39fe7fd3bee84dfe34178d8293ed7f3751cd96ce41ff9a1395b20
f44c7f2717dd089323ece0427e1dceec3edfe927f90c47becbe3393dd77c5f861390d7e7715fb7761cd96ce4427e64ce1608
3dd3dfc94537c2e4083bb844577e3e97c77ddaf5be877defeb33fc914e4993c3b1e03dec986eebc11c47361bb984571e99b3
0542cff47772d18d3039c20ed64c7faffd2baede2ba48e9860ae0c152c55df0b6fb21508fd60f7cb463ede817bcc871af9df
97356375d9c4f7e39a8d5b36f0dff335e6b23cfd8abbc8c6155c40cffee2689f633eef770ec37e0fbb9bebe53dad26721f5f
d39e36d13f601ed96ce45f1ceda1394b9efeda70c9d3d7932fd5c7108eef733803fd5f9f27e932cfb292c0fd717f5fc75dee
63bb3ebf8cdff5ddbfe373d5e597c681cd062eeb081c98b105bc92a76fd8a21b5eb283edeb610105222c79fa462eba112647
d8c1254f7f38b80c07d3f15cea791fce2e1ed7cb1417fcb83c87e3cc777b6b8b47361bb9e4e91f99b305c292a75fc857dd08
9323ece042bebd2031de32785c8626a7851ecaf70c3cb0e9c085fbc08c2de0953c7dc316bde0253bc83ec8d37f3f8763fb98
bd3636f9bc0cffcd677b0b8147361b0979fa71ce16084b9ebe918b6e84c9117610b2d52fcfefeb355e558c5d7e9ef7e18ff5
f57a02f936b2d94848578f73b64058f2f48d5c74234c8eb083407e1fae976fb7ebf416fcf336df277c50bcfa36b2d948208f
73b64058f2f48d5c74234c8eb08394a77f7dddc770f4fbb4990e07c2e19ae64dab476c239b8da444fd34670b842552dfc845
37c2e4083b4891fad7efe7fe794e1fcabfae9fe1e7ba0e571140be8d6c369222f5d39c2d1096487d2317dd089323ece01ff9
fc6ecc7dba7a7acd0b344def8c0cbba22905feb2fe9de3c86623ffc80fcdd9026149d55772d58d3039c20e7696aa6fe407b2
e2953039121c5ceeb65ddedfdbe5f2f739d7b4bcc37bfab18613fbc7e7f6dc9e8ee3c8662397bb6d47e66c81b0a4ea1bb9e8
46981c61074fbebdca782552dfb04537bc6407dbd7c10782485822f58d5c74234c8eb083a7473f135ec9d32fd8ab6e78c90e
b6efdc5479622b49fac62c7a614b5e90771d3c2844c292a76fe4a21b6172841dece01d33242c69fe462eba112647d8c11e7e
73222c69fe462eba112647d8c173ffc289ade4f81bb3e8c696bc60efe67523a84609d3d7daaa5b8dd4133374f0301e094b98
be918bae84c991e0e0d98750c42b39fe862dbae1253bd8be5d8ebfd53e904f6f35524fccd0c3c646849ee36fe09b6c7cc90f
f6af835003e2f31c7fc3de64c30b6ea0771d1cc388ce73fc0bf4221b5df0029debe0056da2f31c7f83dee44217bc00e7ce7e
7791d04a8ebf218b6e70c10a34ae83779391afe4f81bb7e80698fc60ff16f2cff093dce788cdfb707a757d5f3fd33f86b3ac
611734467f2ee434b2d9c885fcc89c2d10961c7f2317dd089323ec6007e9a5485872fc955c75234c8eb083677fb48a7825c7
dfb04557bc6447b0af83340b242c39fe462eba112647d8c10ebecf46c292e36fe4a21b6172841d3cfb837cc42b29ff862dba
e1253bd8be7fb07c01019425040c4c740348c06cd06e098152fb672cbfd5483d31c36e0901ab7d2096bfd4483d11c36e0901
ab7d2096df6aa49e9861b78480d53e10cb6f35524fcc70fadf2fe195d50b0c5b74c34b76b07db3e554a32c21a0b555b71aa9
2766e8e030898465010323175d099323c1c1b3cf8b10afac5e60d8a21b5eb283ed3b2fcc03c1caba05062cba812523d8b81e
d2fb89b0ac5b60e4a21b6172841dece03a1709cbba05462eba112647d8c10ebe3847c2b26e81918b6e84c91176b083e76e48
58962e28e4ab6e84c91176f0d4a74fc856162d3066d10b5bf282bceb206b1809cba205462eba112647d8c10e5e1f41c2b268
81918b6e84c91176f0ec1877c42b2b1628b6ea8697ec60fbfeb09f9fcbf5759d62083eeff10388f159d5f88febebef01c2c2
8d239b8dfc033f34670b8465c50223175d099323c1c1857cf8791edfe919c270f4bd0dd7acefc7fc36dfe3f5bd4cb9b50b39
8d6c3672213f32670b8465c5022317dd089323ece0427e993f3e99f63edfe1b7f9bea7cbb7f7f7ef54733d47c791cd462ee4
47e66c81b02c5a60e4a21b6172841ddc2d5a60b50f44f15b8dd41333f4b0bd116159b4c0c84537c2e4083bd8c33e8608cba2
05462eba112647d8c1b30f2a8857562c28d8ab6e78c90eb66fc11ecee487e3ced4e4f47ef8fb3d5dbe0f17b57340e17a6d88
239b8d5cc08fccd9026159b1c0c8452f84c91172b083d71a90b0ac5860e4a21b6172841dece00a0909cb8a05462eba112647
d8c1b39ff5239eaf58a0d4221b5c3283cdebe09137f1d98a0546bdaa0a17bc60e73a78d60d6cbe7481216fb2c1b113e4dad9
f77209ccd72d30de4d36b260039ad6c1435ea22beb1618b5e8c617dc40efce7e4284706555038316dde892196cde3f58ae81
00ca92090626ba0124603668b76482d53eb00c81d5483d31c36ec98452fbe732045623f5c40cbb2513acf68165084a8dd413
31ec964cb0da07d602b01aa9276638fdef97f0ca92098aadbae1253bd8be0ef6d64858964c3072d1953039121cece0f93212
9655138c5c74234c8eb083679f93205e593bc1b04537bc6407db77e262110456964c3060d10d2c19c1c6f5b0580411965513
8c5c74234c8eb0831d5c5a21615935c1c84537c2e4083bd8c19375242cab2618b9e846981c61073b489846c2b26a82918b6e
84c91176f0ecbb278857964c28d8ab6e78c90eb66fb76482d53eb01040a9917a22861e3636222c4b2628b9ea46981c61073b
d8c120615932c1c84557c2e44870b083830a129625138c5c74234c8eb08373843dd5288b16586dd1ad46ea8919768b1658ed
0351fc5623f5c40c67af39857865c502c316ddf0921d6cdf6ec502ab7d208adf6aa49e98a1831d2b1296450b8c5c74234c8e
b0839d2d5a60e407a2f88d3039c20e2ecfbd86aba7c7b0f79b2e23aeafeff379b94eaf3abdaf97718ad77a48c191cd462ecf
bd8eccd9026159b4a090afba112647d8c1b3d30210afac5860d8a217bc6407d9b73cdc1eb6c9ebf8b1f3b8257e1e8fe1c03b
a7808dcf68afafcb7dbdf582239b8d5c1e6e1f99b305c2b26281746cba112647d8c1e53d8ef9ffa6c65e9f618f3f6c9ad3f3
f8eb6538f27fc7d4e0e53d0e1ad96ce4f21ec791395b202c8b1618b9e842981c490e9e1fa3c68465d1022317dd089323ece0
129d37bf5f31bf5a33fc42cfeb6dda4c8713de39ee6f25c791cd462ed17947e66c81b02c5a60e4a21b6172841d5cb2bd5f43
6fe30bc9e34b6ccff1ddf4c77492ffbe0ea73bf7a1d7951c47361bb9e47b1f99b305c2b26881918b6e84c911767009b0dfb2
a9dfeff1359b3969e1f59d165219433d97fc7a18d86ce0125f7f60c616f0ca8a05862dbae1253bd8bef3f33199b0ac5860e4
a21b6172841d5cc85fc309c67d7a73f0fd94dbe1afd770e8795defdfef4a4e239b8d5cc88fccd9026159b1c0c84537c2e408
3bb85bb1c06aff5c0ac02aa48e98a0ae5860857f86f25b81d00f76bffc6d0f7b9adb74b3fffd1a2e1d8706a78cded77061f1
1dfe3eaeeb150a0d6c3670f9db3e306363345fb4a0102fb2a10523d0b6bf8d7bb8587bdc6fdf39707cf8ff8796a67f0cbfcf
63b8aa58f7663cb0e9c0bf4dfbc88c65c582b5dbb262819e7ca93ee6ac8caf764ebacc9374996759abe1fefc3eded30bd98f
e7fb797dddefd389d5fd3eb4f8ba6ea7e73cb2d9c865ad862373b64058162d3072d18d3039c20e2ee4dff7e3759beeff3eb6
3ca0714589d71ccdbb2e5a80239b8d5cc88fccd9026159b4c0c84537c2e4083bb82cd7f0fc0c274fd3fbc06bc2fcb496c667
38a57f0f146bb6358e6c367259aee1c89c2d1096450b0af9aa1b6172841d5c162db80c7d7da703cc737b6e3b2ead70991f6a
2ee038b0e9c065d1820333b68057162d306cd10b5eb283ec3b7dad06c22b2b1618b6e88697ec60fb60ad86ef73e8f979995e
8f7ebcef9fa1dfcffa6c9847361b096b35c4395b202c2b1618b9e846981c610769c582f1dbc9cb653ac57a6ed16fb462c132
b2d9485ab120cdd9026159b1c0c84537c2e4083b08e4efebf3711d2e27c63df0f37579de87cbe507ae52b18e6c3612c8e39c
2d1096150b8c5c74234c8eb083b062c1f5f5bd7f5ff7e931cb70aef51ccee9dfb4b16f039b0e84f50ad28c2de095e50a0c5b
74c34b76b07de75f94306159ae40c95537c2e4083bd8d97205467e20845f099323c1c1e53edbfd367d0e3ddd13fb4b2b9d6e
967ceaf3591cd96ce4729fedc89c2d1096e50a8c5c74234c8eb0831d7c23878465c5022317dd089323ec60079f9c236159b1
c0c84537c2e4083bd841f62f1296450b0af9aa1b6172841dece0a1191296a50b8c5cf442981c21073bf84c0e09cbd205462e
ba112647d8c10e728091b02c5d60e4a21b6172841d9cc3eca946593cc06a8b6e35524fccd0c3f646846509012317dd089323
ec600761db485816305072d58d3039c20e76b07745c2b28081918bae84c991e0e0a96145c856562f3066d18d2d79c1de75f0
c9391296b5138c5c74234c8eb083e7fedac4e6ab2618f2261b5972829d9b57aa800abe6e8115de642b10fac1ee3bd895129d
af5b50a017d9e88217e85c07e72b44e7eb1618f42617bae00538d741c21fd195a50b8c5a74e30b6ea0771dbc178e7c65e902
e316dd00931fec5f07319e4858962e3072d18d3039c20e7610e7808465e9022557dd089323ece079df12225859b4c0804557
b0644430ae83af6591b02c5a60e4a21b6172841dece00b69242c8b1618b9e846981c6107cffe241ef1caa205862dbae1253b
d8be7fbb6881811d5808c00012301bb45bb4a0d4feb91080d5483d31c36ed102ab7d6021805223f5440cbb450bacf6818500
ac46ea89194eff1b22bcb25e82618b6e78c90eb66fb69c6a94450bacb6e85623f5c40c3decb089b02c99a0e4aa1b6172841d
3c3bd801f1ca7a09862dbae2253b827d272e1c406065a5040316ddc092116c5c07571c4858964c3072d18d3039c20e767095
898465c9042317dd089323ec6007b71590b02c9960e4a21b6172841dece0f90b129625138c5c74234c8eb0831ddc9746c2b2
6442215f75234c8eb0831dc4df22615938c1c8452f84c9117270217fdc6fd7c775f94af7397eea35b57c1ffe9bcb7b0ce558
c86964b3910bf991395b202c0b2718b9e846981c610717f2cb63d8ddce37b08703efb027be4c3fcfed6f2fb5de21a781cd06
2edc07666c01afac9a60d8a21b5eb283edeb204a1e09cbc2094aaeba112647d8c10e5e5543c2b27082918bae84c991e0e01f
f9fd73bf5f1ff3c7caf7d7e77b7f5e6e63cbeff14bceefe7b2edd87164b3917fe487e66c81b02c9c60e4a21b6172841d5cc8
875dcee5365d448c671f63a0d0942cf47e0e6761b7f163bf959c46361bb9901f99b305c2b27082918b6e84c9117670b77082
d53eb01c80d5483d31430fdb1b119685138c5c74234c8eb0833dec6388b02c9c60e4a21b6172841dec61b10c222c6b2714f2
5537c2e4083bd8c192117bbcb27082618b5ef0921d645f07e7cc4858164e3072d18d3039c20e9efa160bb29525138c597463
4b5eb0771dc40521a1af9aa0e0221b5ff283fd3b3b938be06cc904435e55250b46b06d1d84b8029b2f9960c89b6c70ec04b9
76deed5c42f2c5128c74938d2918807675f0589de8ca6209462dbaf10537d0bb39e89d2a94150bacb2e8562275c404673f91
43bcb29e81618b6e78c90eb66fb79e81d53eb04680d5483d31c36e3d8352fbe71a015623f5c40cbbf50cacf68135024a8dd4
1331ecd633b0da0782faad46ea8919e6da54a32c2aa0b555b71aa92766e8e0092c129645058c5c74254c8e0407e74c62aa51
62fdadb6e85623f5c40c673ffd45bc12eb6fd8a21b5eb283ed9b2da71a2561df6a8b6e35524fcc70e29202045682f50d5874
034b46b0711d9c8d226109d63772d18d3039c20e76f0cc1d094bb0be918b6e84c91176b083dc6b242cc1fa462eba112647d8
c125617e5fa3a4db97daab6e35524fccb04bb7b7da0732db4b8dd41331ccb5a946c997d7daaa5b8dd4133374f0778684255f
dec84557c2e44870b0837d2b12967c792317dd089323ece02e5fde6a1fc86cb71aa92766d8e5cb5bed03c1e95623f5c40c3d
6ceb445822e68d5c74234c8eb08367dfc543bc926e6fd8a21b5eb283edeb2cdddec80f64b61b6172841dece05340242ce9f6
462eba112647d8c1531f4e205bc9b52fccab6e6cc90bf66e61be3cded3117edc0c3feff7fb3e5ca68fff989e9b5e9fcff5d7
c691cd462edc47e66c81b044db1bb9e8853039420e9efdec11f14aaebdb46bbae1253bd8bef35f6262c2926b6fe4a20b6172
2439b8bc5a31fdd5cd3fcffd36bf503775f99a429286d1ebab1534b2d9c8e5d58a2373b6405872ed8d5c74234c8eb083e767
ac3161c9b53772d18d3039c20e2e09d09fcfe5719fde257c7f8603cdf0df7de630f6d7f5fdb90f87e1851c47361bb964401f
99b305c2926b6fe4a21b6172841dec21e09d084bb4bd918b6e84c911767021bfce597063977f2fc75fa62ee7fdd27d7ddece
239b8d5cc88fccd9026189b63772d18d3039c20e2ee199af39c078ec72bcef3bec982fe3dee8f5b87e6fb7fb7d7daf844736
1bb9c4671e99b305c2126d6fe4a21b6172841d5cc8876e3ef7e91588f77d0dfc1b335e87e6875f678b0da5814d072edc0766
6c01cf53f58d7a930d2e99c1e6f5f073039f67fa1bf6261b5e7003bdebe1af1be83cd6bf402fb2d1052fd0b9197abcd77dbd
bca63738efd30a9ff7fb14c27f1b63da1f8feb7a639747361b39431f9bb324fbaf0d97647f3dfd547d4a5f79df675de649ba
ccf3f773df2edfe18af1361e6c8753abcb7538dc4e9f09dc3ec31160cca95f5734c091cd46fefddc87e66c81b024fb1bb9e8
46981c610797a51c86338d295971e8f2315c4cdcde7397f7eb7bfc89b6a8501ed96ce4b298c391395b202cc9fe462eba1126
47d8c1654d836945a8e993afc71cc436858cdc873fcbc7b0d96e6b39c0b8a6e396f50c7ecfd7025bc9f42fccab6e6cc90bf6
6e59cd605d33e8b1dd0d1aff32aff7eff80c7db92ac381cd062e6b191c98b105bc92e96fd8a217bc6407d9b764fadf2e8f61
ec94c07fbfce9f074c5be7e5f979dcbfdb5b553cb2d9c825d5ffc89c2d1096587f2317dd089323ece072f0be4e879db1c9f1
0589fbf33d7dddf9783c07cb8673bc2de11d06361bb81cba0fccd8025ec9f4376cd10d2fd9c1f641b2fd757ca0371ce7a71d
f177fc5fcfe77b0d3ac791cd4642b27d9cb305c292e96fe4a21b6172841d04f2e770f5f4bdbca70cb8e7635c05f2325c5312
f93ab2d948208f73b6405832fd8d5c74234c8eb08390e97ff95b016f6af972fd5b120dc8b791cd4642aa7f9cb305c212eb6f
e4a21b6172841d5c4e56c7dbecefe7b414dc73da14c7af5fc608fecb70c2f3da9e9ef0c866239793d52373b6405862fd955c
75234c8eb083e72f28c88425d6dfc84557c2e44870b0832fb890b0c4fa1bb9e846981c6107cffe2c18f14aa6bf618b6e78c9
0eb66fc11eae1d6ff7796dd3ef14ccf0983fc3a987721ed96ce4027e64ce16084ba6bf918b6e84c91176b0b34cff42fe33a9
de089323ec6007a9d3485832fd8d5cf442981c21073bf89a09094ba6bf918b6e84c91176704eb6a71a2555df6a8b6e35524f
ccd041a60c12964c7f2317dd089323ec600fdb1b11964c7f2317dd089323ec60072f622061c9f45772d58d3039c20ef610c1
4d8425d3dfc84557c2e4487070ce3ca71a255bdf6a8b6e35524fccd0c1310d094bb6be918b6e84c9117670769d6a78c6bd95
de64ab903a62829a716f857f06b85b81d00f76dfc31f18d079c67d815e64a30b5ea0731dec4f89ce33ee0d7a930b5df0029c
ebe0c481e84ac6bd518b6e7cc10df4ae83f441e42b19f7c62dba01263fd8bf0ebe0141c292716fe4a21b6172841dece0ad7c
242c19f74aaeba112647d8c10ed2fd91b084dd1bb9e84a981c090e9efd2525e295a47bc316ddf0921d6c5f076104485892ee
8d5c74234c8eb083676702205e49ba376cd10d2fd9c1f6cd29dd54a384d15b6dd1ad46ea8919fe41ca3e0194a4fb02b6ea06
9080d9a05dd2bdd57efd4e8f2f35524fc4b04bbab7da0722dcad46ea89194eff1b22bc92746fd8a21b5eb283edeb20eb0009
4bcebe918b6e84c91176b087c30511969c7d2557dd089323ece08981f3045612f60d5874054b4604e33a380946c212b56fe4
a21b6172841d3cfb4345c42b39fb862dbae1253bd8be05fb31ec6dafd3a717c345dab0cb1f5fe5fc9fe9b1e1f5fbf86e3b73
1cd874e0827d60c616f04ac8be618b6e78c90eb6af87a879222c21fb462eba112647d8c11ea2e689b084ec1bb9e846981c61
073b88cd44c212b25fc857dd089323ec600f51f3445842f68d5cf442981c210797cf15e7ff9b7e92d767d8eb5e5ff38f751d
5afc7c9febbb633cb2d9c8e573c52373b6405842f68d5c74234c8eb0834b1ef2f37d1d76bcf3bb0f97e7e3fb9e1f528e271e
97e7f0f3ac37757064b3914b1ef291395b202c39fb462eba112647d8c1f33fd464c292b3afe4aa1b6172841d5c3ed49c773b
d36b3fd7cfdf6b8d63978fc7fcb9dd7af686239b8d5c3ed43c32670b842567dfc84557c2e4487070211ffefdfe4c8fe487f3
ace7f0ffeed399d77bfcb1bec399d7fa9be3c8662317f22373b6405872f68d5c74234c8eb083e77f9ccb842567dfc84537c2
e4083bb8cbd9b7da07d2e3ad46ea89197ad8de88b0e4ec1bb9e846981c610797bdeb1c0c3f1df76f4367efdbf5351d039ec3
4e62d8277cd6ab621cd96ce4b2773d32670b842567dfc84537c2e4083bd8c30a03445872f60bf9aa1b6172841dece02c0a09
4bd4be918b5e089323e4e072e63cb4787dcf6f4e7f3e8fc76db8be995a1e2f77aeafcb7d25c791cd462e67ce47e66c81b044
ed1bb9e846981c61073b783917094be0be918b6e84c91176b0837b0348e881fb0a2eb2f1253fd8bf0e32ab88cf32f78d7a55
152e78c1ce7590cb066c9eb96fc89b6c70ec04b9d6c1b37e62f3f07d43de64830b4ea06f673fe627b492bc6fc8a21b5cb002
8debe0a10df295dc7fe316dd00931fecdfd9cf2711afe4fe1bb6e88697ec60fb76b9ff56fb4096bed5483d31c32ef7bfd4fe
99a56f35524fccd0c17359242cab0e18b9e8853039420eee561db0da0792fcad46ea891976ab0e68ed2359fa5623f5c40c73
6a31d528b9ff565b74ad917a0a0c1dec5691b0ac3a60e4a21b6172841d3cfb208a7865d501c316ddf0921d6cdf6ed501ab7d
2054df6aa49e98a1839335242ccb0f18b9e846981c61074f5c6f81c0caaa03062cba812523d8b87933a31a25f7df6a8b6e35
524fccd0c366468425f7dfc84537c2e4083bb8cbfd2fb57f66e95b8dd4133374f0aa0d129655078c5cf442981c21077bd8de
88b0ac79a0e4aa1b6172841d9c33e8a9465975c06a8bae35524f81e1bc5d2a8295c5060c5874034b46b071bbc506acf68100
7fab917a62860efeb891b02c7560e4a21b6172841dece0961a1296a50e8c5c74234c8eb08367bf41887865a903c316ddf092
1d6c5f07b78d91b02c7560e4a21b6172841dece08b50242c4b1d18b9e846981c6107bb59eaa030ff0cf037b6e4057bd7c1eb
834858963a3072d10b6172841c5cc8e7dcace915a0cf78eff3f2f7dedfbc17ba7fd6f3351cd96ce4427e64ce1608cb6a07d2
b1e946981c61079747fe9f61dc7bda2e3fcfe77021f1f9dcc77f0c87e1d7fdfeba6cf7927164b391cb23ff2373b64058563b
3072d1853039921c3cff350f262cab1d18b9e846981c6107cf8f9e63c2b2da81918b6e84c911767059ede0ff33776e49afdb
d80e7ee72832822edd25cfa9e7ff7c48ca948065a07fbd9c88a94a5536f60ac90fb6752589b26b533ecb9481ed73beb41acb
0168f8cf762c7543edf1229795892a5bdac1933693210c6907440e3a113a47b483ef6fb4a80943da0191834e84ce11ed600f
9bc02bc2907640e4a013a173443b28b6159da732f769af27e2b21c2cdf5dada3dc56f4aa4c5429b615b56d264318d20e881c
7422748e6807cb8db1ee832307a8eb5ba61edc8834c1d9b3e88137fda78e6f993a30e391a3efe14b2ee878d3ff00dd64a233
5e48e74ee8751fce3be53ca6ecce3e4ec35ca357a67c9d57966b5f57e9ba3251e509fdaccdb0e9ff35e0b0e93f5e85b19eff
bf7c1b54756ac7e9573bdfb88373c272ddea335f4fe57b8bb2c56e19653e03941bd96ba2a6ae4c54f98d3b78d466328461d3
7f22271d089d23dac12f799d323ad587047379fc5a065747b95f0b18bee4b23251e597fc519bc910864dff899c7420748e68
075bdcc136d51f611e58d93ceaf89439d3f56b3a7d530bda4f5c5626aa6c71074fda4c86306cfd1fc8410742e78876b0c51d
6cc3bacef5cbb8dc37cc79944b5baed1c86565a2ca1679f0a4cd6408c3eeff444e3a113a4794832df4a01e80eb63c192cf30
1de351ef21e723dfcfed25b8a191cbca44952df6e0499bc91086ddff899c7420748e68075beec1ba2f75cfac75d986cfbead
4b4da798867c68ca9d5c7145a22e515dcb3cf8bbbd64d8c2d6ffc44c3ab0392fb477bfd76beb6738e625ffa1fe34cb66c7e5
05d02caed7eeca4495bfd76bbecd6408c3d6ff444e3a103a47b48362ebff7ca9b1d7a59765c8534cb99095892ac5d6ffb6cd
6408c3d6ff444e3a103a47b48382fc53b60d5bb74f4da838c262655d99a85290db3693210c5bff1339e940e81cd10ebe7fc9
aa09c3d6ff48ce3a103a47b483edf94bbed8588efa91e43be6fc792c6bddfc275f82e5e351fe725ecfdc6465a2caf6fce549
9bc91086adff899cf49bd039621c6ccfdc6ad64c9d0355f6482c1b26d49d13f2ed54be1429f189d733375599a8b23d737bd2
66328461eb7f22271d089d23dac1b6c0a8e6a7d5f73dc75eaeeccb6eabe50fe3159dd71618a9ca44956d81d1933693210cbb
ff1339e940e81cd10eb667ebf59033d4cfe753c738d57d148e655ca7bab6f87ab6ae2a1355b667eb4fda4c8630ecfe4fe4a4
03a173443bd8c10223491876ff0fe4a003a173443bd8c1f6b99230ecfe4fe4a413a1734439d8c11b53491876ff2772d281d0
39a21dec60235d4918b207889c7420748e6807dfde6148e285e001c2261df09c1ddabe33ee41f511b6fea7be49873edc9834
430fa1078a306cfd8fe4ac03a173443bd8c101461286adff899cf49bd039621cec60d5aa240cc103444e3a103a47b4831d9c
4e2561081e2072d281d039a21dfc091ea0beffdcd19f7a7023d2043d843d083e8e3d206c9401cfb821bdebe18b26e838f620
40df32d0192fa4733d1c57041dc71e1034ca4467bc10ce7570b5a6e842ec0151930e7cc60de9dddb976a122e641e1034e940
e7ccd0e6f590f6a00843e60191930e84ce11ed600f690f8a30641e2039eb40e81cd10e7630415f1286cc032227fd26748e18
07df5b0722c142da0101930e60ce086d5c074b762561483b2072d281d039a21dec6069bc240c8107444e3a103a47b4836fef
0121f142d60261930e78ce0e6d5f8b24f8ed23c42184be41873edc9834c3bf92f3f00b10b216088c740270c0caa0d7bf4b0a
2fc43c1036e980e7ecd0f69ddf25d547c85aa0be49873edc9834430f072e4518b216889c7420748e6807dfde6541e285a005
c4661df09c1ddabe0e96044bc290b840e4a4df84ce11e36007eb252561485c2072d281d039a21dec60e1a0240ca10b444e3a
103a47b4831d3ca59684217781c8490742e78876f0d58583922d242e1033e9c0e6bcd0de35e661d9eb8e00e56b78ecfb3ee7
534ef943ddf9755cd7ebd39695892a1bf7933693210c890b444e3a103a47b4837de40efc1286c485400e3a103a47b4831dec
982b0943e20291934e84ce11e5600fb9038a30242e1039e940e81cd10e76307d4b1286c40522271d089d23dac10e76439784
217101c9590742e78876f0fdc5a29a30242e1039e937a173c438d8c1344d49181217889c7420748e6807db32d96d5ef2c55d
fd48f6b260ffbceadaf255f7928f55d73c745998a8b02d927dd0623278216e81b049073c6787b6af611fc7b0cc7505d49e6f
abf28f623dce51e67be963ce97db17b7aa4c54d9c09fb4990c61087b2072d281d039a21d6c5ff5f2467239d7e8c37f0fd74c
f2f645ffad4b54d7bee67fb7970c5b88792066d281cd79a1bdebe0902e0943cc0391930e84ce11ed6007a7714918621e0239
e840e81cd10ebebfdb85260c310f444e3a113a479483ed46a5becd39e39df2c97e2f2b69a77a53d11e98b5fb145598b0b0dd
a63c683119bc90f140d8a4039eb343dbd7c19da9240c190f444e3a103a47b4831dcc5f92849cf180e024039ff343fbd7c1c4
16c547190f440dea0d67bcd0cebdbd2d9b00e38007e24519c8b40dcab20e263628360e7820649401ce38217d7b6f4e83820a
d10e044b3a601913a4656fbf23927021d781a049073a678636af8397829230443b1039e940e81cd10efe443b50df0fe212a8
0f3726cdd0c1ec19491882250239e840e81cd10ebefe335378215582b049273c6787b2ef275582fa7e90d4407db83169869f
5409ecfb495203f5e1c6a4197e5225a8ef075909d8871b936138fb567d845c07ea9b74e8c38d4933bcb8d5be020b890e044c
3a803923b4713f890ed4f7839c02eac38d49339c7dab3e42a602f54d3af4e1c6a419cebe551f21de80fa261dfa7063d20c6f
1f4b255e4856206cd201cfd9a1ed3b2d577d846c03ea9b74e8c38d49337470e2968421592190830e84ce11ede0eb5f368517
6215089b74c2737628fb7af8c0156188554072d681d039a21dfc8955a0be1fa406601f6e4c86e1ed2f9bc40be10a844d3ae0
393bb47d3fe10ad4f783c800eac38d4933bc9864a1c042a60201930e60ce086d5c07cf742461c8542072d281d039a21dec60
8192240cb10a444e3a103a47b4831dcc8e9484215681c8490742e78876b083dd0b246188552072d281d039a21dec26562130
ff1916406cce0bed5d7bf5760ce334d7e9c9c7a7be3a58ce596dcb3a8de529dcf50b9795892adbcbb7276d2643186215889c
7422748e28073b8b5500f247610144e81cd10ebe3f4750138658052227fd22748e38073b8b5520f207610144e81cd10eb649
34df15d16594c3b0ecf920f4a973bbcad138ff46d68b5c5626aa6c13699eb4990c6188552072d281d039a21d7c79be98c60b
990a844d3ae0393bb47d9d652a10f983a40022748e6807dfdfbb5313864c0522271d089d23dac1ce321588fc415201113a47
b483827c2abbc50fdb56b7989d9671c9979e9f43915f95892a05b96d3319424e7420709481cff9a1fd8b890ed4f19f7105d4
81198f1c7db33b7fed8ff93cc1cfd77e9ed99ab95ee5dc3b60cbc28485cdec072d268dc6710e81f89601cd18216d3b89f379
aefe530e00d306df8cb1ad9c6e87535d99a8f2647ed6e64f9cc377c03f710ef7d51fea7543a5239fa3ffcbed381ddae921c8
4211fec4390039e844e81cd10e9ee41b2c6fca27fc259ff68f9aa494ff97e1dcd1f44bae2b13559ee4cfdafc897300f20721
0544e81cd10ef61064a1087fe21c88fcd289d039a21d6c4116dfdbf37fea6e7fc73c6e439d383a978dc1ca0ab7eb142a2b13
55b6208b276dfec43900f983908240e81c510eb6208bb201fea7ce235cf67c093fe62bba1abdb08efb58570a34725999a8b2
05593c69f327ce01c81f441510a173443bd8e21cf297eff8d427feebfd943f0f7918cec7d00d5c16262c6c810e0f5afc4974
00ec07390584e7ecd0f6b5937819e25af7de2ff3aee6753f3fa8f8e44716262a6c27f1072dfec43900f6839002c2737668fb
44a8413ef5e54bf9bdeeffb70e53596db50c9bb854bc2b13558a5003dbe64f9c03903f08292042e788765090efe3bae4df60
0d1a297133f934b42eab22bf2a13550a72dbe64f9c03903f08292042e78876f0fddb034df813e77093a34e84ce11ede0fba1
929af027ce01c8414742e78871b02da13a537ccfcc816f6c52fdb096efa5d6452e2b1355b625544fdafc897300f207210544
e81cd10eb6d5efdf47926594db3745a82e5b1f96ed33d499fa6df5bbaa4c54d956bf3f69f327ce01c81f841410a173443bd8
c18e0792f027ce01c8412742e78876b0bb380722ff33a480089d23dac10eb60596843f710e400e7a20748e28073bd8c74612
fec4390039e844e81cd10e76b08e4a12fec4390039e844e81cd10ebebd47b0c4fb8973006cd009cfd9a1ed13710ed0f78390
02eac38d4933f4f06553843f710e3739ea44e81cd10e7630154412fec4390039e848e81c310ef6b0edba22fc89730072d089
d039a21d3c430d541f3f810ad037e8d4871b9366e8e0542e09639c0380df32f1393fb47f1d2c87567c31ce01b06f99f08c1b
d2bb1e3e6c4117e31c08bac94467bc90cef5705c117431ce01a06f39d0192f84731d5ca72aba9f3807a0069df88c1bd2bb0e
ae1b24df4fa20370834e80ce0fed5f07936725e14fa20390834e84ce11ede0dbbb1d48bc9f38871b1b75c2737668fb3ad845
5712fec4390039e848e81c310e76b073b224fcc9750072d089d039a21d7c7be58dc4fb0975006cd009cfd9a1edeb601b0449
f813ea00e4a013a173443b78ee48affaf8495680be41a73edc9834c3eb21000aef27d481b02f9df09c1ddabe1628f1dbc74f
ee02f40d7ae8c38d49319c7dab3e7ec20fa06fd0a90f3726cdd0c38f4c11fe442f0039e844e81cd10e76b08a5412fe442f00
39e844e81cd10ebe188ca4c07e42176e60d409cc19a18deb6059a124fc095d0072d091d039621cec6059a124fc095d0072d0
89d039a21dece0ce4812fe842e0039e844e81cd10e76f0b84712fe842e0039e844e81cd10e76f06a5612fe442f0039e844e8
1cd10eb6d7f1db5eaeaceac03ee3306df966a6ce2058c635dfdf7cee64475999a8b2bd8e7fd2e64ff402903f08142042e788
76b091d7e550e777b12c9299b7bacb72beab3b9fdf5c5f765598a8b0713f68f1277781b0ff4c13203c6787b6af832d6c25e1
4fee0290831e089d23cac1b6b2b0cd612e1343ae69bd79945bfd4e2ef73a5a5999a8b2ad2c7cd2e64fee02903f48132042e7
8876b083adaa25e14fee0290834e84ce11ede0ab1b934bb69fc4859b197562735e68ef5edf7d5fe1fdc42d0036e888e7ec30
f6b5d594f542baae51d9b76ba54699f33a0ddbb27e866beaacac4c54d956533e69f3276e01c81fe40810a173443b581f03c8
3e7e320fa06fd0a90f3726cdd0be6cf9703bd5fbb67d3bb6b1ac3eac5f8dfd3c208cd72f4c15262a6c5fb6072dfe041e00f6
83cdfc09cfd9a1edeb60ceae24fc893d0072d089d039a21d7c37ea41b1fd041e0033e8c4e6bcd0deb5f367be8439be6f6cc7
659ce77c45538f09f3912f7bd665bfa86565a2ca76fe7cd2e64fe00191ffb98d3f113a47b483ed6a31df9a4fc3f0bd7aaf73
e6f73aca7d5997231f16aeefb9ac4c54d9ae169fb4f9137800e40f76f20f84ce11e56007f96492f027f300c8412742e78876
b0835b4249f8937900e4a013a173443bd81e03e4436e59fd576fd9b7f3b55ebd9d2b53170ac0f53b9795892adb6380276dfe
641edce47f6fe84f7cce0fed5f07b33f145fc83c00ea4b4538e38576ae83a77c822dc61e00f22d139c7642b9d6c1135dc516
630f00f99609ce38217debe001bea2fbc93f006ad089cfb821bdeb60ca83e4fb8940006ed009d0f9a1fd3bf747577dfc8410
40dfa0531f6e4c9a41841040df0fb6d6a73edc9834430f5f7445f8134240e4974e84ce11ed6007139b24e14f0e0190831e08
9d23cac11e3e7345f8938200e4a013a173443b285210eebe9f240b501f6e4c9a41a42040df0f9205b00f3726c3205210a0ef
079bfc531f6e4c9aa183738a24fc89430072d089d039a21d147108d0f7838801eac38d4933883804e8fb41c400f5e1c6a419
5eccbd50603f190c000c3a813923b471a7d9aa8f9f2004e81b74eac38d4933bc3d5753e2fd64300036e884e7ecd0f69d96ab
3e7e5210a8ef4ba73edc9834430713b924e14f100290831e089d23cac10e26ef49c29f20849b1c7522748e6807451002f4fd
60877fecc38dc93074f07d93843f5908400e3a113a47b483220b01fa7eb0d53ff5e1c6a4197ac94200e0073bfc139833421b
d7c16d8124fcc9420072d089d039a21dece0119724fcc9420072d089d039a21d6ce4e736cfe7f1b63e85dccf2fe930ac7557
bfeb7b2e2b135536f2276dfe642100f9831dfe89d039a21dec6001bc24fcc9420072d089d039a21dec280b8198ffdce19fd8
9c17dabbf66eea332dd37abe34ab5333ca6c94f287bae9d3b45d1b47ebca4495edddd493367fb21080fcc10eff81d039a21c
ec2e0be1227fb4c33f113a47b483efcf58d4843f5908400e3a103a479c836dc6e2bc8cc354efdacaaae2b2dd4bddf7255f63
2edb5466275d33165565a2ca3663f1499b3f590840fe60877f22748e6807db8c8b32db69daceedcbd621dfceccebb98ddf36
9f53afae1917aa3251659b71f1a4cd9f2c04207fb0c93f113a47b4836d6651d97275ae1387f7639ab63cc8e39c6c976fa9f2
9774bac86565a2ca36b3e8499b3f710840fe60937f22748e6807df9fc0a7097fe210801c7422748e68071bf9b81ec358b7a6
dcbf7b040c7594f98c9c8f48f3750da32b135536f2276dfec42100f9834dfe89d039a21dec21084211fec4210039e844e81c
d10e9eb100aa8f1848005ddf32f5e046a4095e4f061070310b01986f99d88c15d2b81ebe65822ec62110749389ce78219d3b
a1cb83a072e95e7ef3d3746cc738e52360fe43be819d3fe5f95c7ba2ac2b13559ed0cfda0c7108d780431c025e00a25eb761
fa2ca70eed381ddaf90641acdfb7476594dbb4acd367a8d73b6339ecede586ea4bae2b13557e83201eb5990c6188432072d0
89d039a21dfc0641acebb4ece59aaa84364cf9bc5eb69a2e432e6fd9f257f29ac0a62b13557e83201eb5990c6188432072d0
89d039a21dfc0641ecf39c2f60eb463bf3f2c997f3fb5ee75e4de509cd706cd7b6b2ba3251e53708e2519bc910863884407e
e944e81cd10e7e0f6ec3967f8a6557edb2777d09f5cb9f501964d986a63eb068c7365598a8f07b687bd2623278210b81b041
0f78ce0e655fcb42f83e912a835cf3edfb3aae6bf941ce53db3dad71cbca44952d0be1499bc910862c0422079d089d23dac1
2ff979dea99353d621fb738c47bd6b9ff3e1385f6a1dd7ca455d99a8f24bfea8cd6408431c0291834e84ce11ed604b8128db
81ccf5cbd86eda8f722e5aa6b52c5858a6eb3a595626aa6c39104fda4c863024221039e844e81cd10e8a5c807c17b36d65e5
6119e5b1e60bae71db642ec05599a852e402d83693210c8908440e3a113a47b48382bc1c8d3fc35e1790e753eef81907c83c
9195892a05b96d3319c2908840e4a013a173443b28121186ef46f6e787357ed7dd09f2bb3251a54844b06d2643181211901c
7522748e6807dbfdf0b60e59afcf2b5698c6b06d5bbe8fcd979a9feb7e585526aa6cf7c34fda4c863024221039e848e81c31
0ebeffdc4b1386440422079d089d23dac1f694f7bdc45b4d181211881c7422748e68073b4b4420f2e5ef7dfe89d039a21d6c
6fb08665af3335ca9b977c75bdcf9f9a3776dc175fed1d96aa4c54d9de623d693319c2908810c82f9d089d23dac157dfdd49
b6908540cca00736e785f2ae838d752561c8422072d089d039a21dec609b594918b210881c7422748e68073b58662509431c
0291834e84ce11ede04f1c02f5fd60937feac38d4933f4f07d5384210e01c9512742e78876b083d9209230c4211039e848e8
1c310e76b086531286380422079d089d23dac19f3804eafbc136ffd4871b9366e860cb3449c87108047ecbc4e7fcd0fef515
8740d8f515c8ffdaeb9ff08c1bd2bbf38b26dae74082d06d93a97d331a39f6d6ed4ffb1c0940ddde7268df8c468cbd83cb35
45172201881a74e2336e48ef7a388e2abe100940dca013a0f343fbd7c15583240c9100440e3a113a47b4831dcc9e95842115
00c9512742e78876b08319d39230a4021039e848e81c310e76b09c511286540022079d089d23dac1b7579a49bc900a40d8a0
139eb343dbd7c1ea1b49185201881c7422748e68073b58ed250943260191834e84ce11ed604f990401fbcf7dfe09cfd9a1ed
ebe1035784211181c8410f84ce11e5e0eb1fb8c20b610c840d3ae1393bb47d3d7ce08a3084311039e844e81cd10ef6705457
84218c81c8412742e78876b08f3006047e12314060ce086d5c071bbd48c210c640e4a023a173c438f8766499c40b490c840d
3ae1393bb47d1ddc904ac290c440e4a013a173443bd8c1e33549189218881c7422748e68073b78252c0943120391834e84ce
11ed609b06504f3ae756f2c73cedeb761e978eb9dd51b55900aa3061619b04f0a0c564f0420c0361834e78ce0e6d5fdb6735
0f72f9d4676365b65dbeaadd9773aa4abec618eafe8d6d9f555599a8b2edb3faa4cd640843124320bf7422748e68073bd85b
571286240622073d103a4794831deca72c0943120391834e84ce11ed60db3dfbbbe4b60c2c1f73f31968a98f08f7ec559908
b85c9fb9ac4c54d976d07ed2663284218981c8412742e78876f0ed54028917c218101b75c2737668fbde5fd4a809431e0391
838e84ce11e360232f7b36e41f5e19e59a2f30d612895346b90dc3b06ff755abae4c54d9c89fb4990c61c8632072d089d039
a21d7c7f39a7260c6910440e3a113a47b4833d7ce68a3004421039e844e81cd10ef6f03b5784211082c8412742e78876f0fd
05eb9a30c4421039e844e81cd10e7630595b12865888407ee944e81cd10e7610a52509432c0491831e089d23cac10e6e5724
6188852072d089d039a21dec609aba240cb110440e3a113a47b4831d6c3a2409391602c141263ee787f6efd549ea8a8c0221
88f75211cbb8a03deb60e6a860e3400842be6582d34e28d73a989badd8381082906f99e08c13d2b706bd0cc336d653ea27bb
f3d9a7b96e7f54f641ca07e1fb05a12c4c58d8901fb498345a48832064d009ce58218debe183567c210d82b8412740e787f6
af83895d923064511039e844e81cd10efe645150df0f1216a80f3726cdf0f6eb2a8917822802f6a5139eb343dbd7c3974d11
86200a22073d103a479483e7974df511a220a86fd0a90f3726cdf01305817d3f8957a03edc9834c34f1404f5fd205e01fb70
63320c3f5110d4f7838003eac38d493374306b541286280822079d089d23dac10ee62048c2104441e4a013a173443bf81344
417d3fc859a03edc983443077375256148a42072d089d039a21d7c3182438185180c02069dc09c11dab8f36ba6fa085914d4
37e8d4871b9366e8602e99240c491881fcd289d039a21d6c291cbf7d842c0aea1bf4d0871b9362e860eea2240c5914488e3a
113a47b4836f4f569578210683b041473c6787b1af830f5c1286180c22079d089d23dac10e0e30923084701039e844e81cd1
0ebebde644e285280ec2069df09c1ddabe0e2e5c246188e22072d089d039a21dece042591286280e22079d089d23dac10e1e
0348c210c541e4a013a173443bd8c123744918a238881c7422748e68073b98af2b0943204720bf7422748e68073b98af2b09
43200791831e089d23cac1b7e7eb4abc90c601c3259df09c1ddabef75f806bc290c641e4a003a173c439d8263d1c733ee0d4
ed12f319f7f8cc6b89092913148e75fc1cc3b5b9b2ae4c54d9263d3c693319c290c641e4a013a173443bf8fe7c5d4d18d238
881c7422748e6807df9fafab09431a0791834e84ce11ede0fbd3ba346148e32072d089d039a21d6ce4f97398ea83b1fc2bcc
c7c161af0949db7ee4caa1fc581bb8284c54d8b81fb4980c5e88e2206cd009cfd9a1ed7b7fd3554d18a238881c7422748e68
07df9fb5aa09390884c06f99f89c1fdabf338244f4c0691cd4f12d5307663c72f43d7cd1041da77104e826139df1423a7742
0ff352afe8cec888639d3fc354e74b8ce7bbade14a5dd195892a4fe8676d86348e6bc0218d032f00512fbb9f6df9f8557568
c7e9d0ce3787a4dda2fe533243ca0dcd74deb5b75b9a766ba40b13167e53489eb4980c5e88e2206cd009cfd9a1edfb86906c
e73fe5c3c927ba7bdbedf193af6ecb6e57eda0a22b13557e43481eb5990c6188e22072d089d039a21dfc869094ebf963ab71
7a73be63cf07bbb1ce679e5a78487baead2b13557e43481eb5990c6188e208e4974e84ce11ede0495e5ee38dc356afe6e73a
0d6c2e7b16d6c498755c96f17a2fae2b13559ee4cfda4c8630a4711039e881d039a21cfc1ed3976daa1948ff94e0843d1f8f
a7b94ee4cd97d74bbefe81252fb23251e5f798fea8cd6408431a0791834e84ce11ed60cb21f9e47bd5a9dec72cf7766e7994
f5ee7dd8963b87445526aa6c39244fda4c8630a4711039e844e81cd10eb6348e7cf2c9a7da72d3b6cedb372ea30c79cc1738
ebf0b95e4febca44952d8de3499bc91086340e22079d089d23dac1df64866cce9ebf1d47fd7cf28756139186eb0a465626aa
fc4d66f06d264318d238881c7422748e6807451ac752d3bfcacea879c8f9e67dc917209bcaa4b82b13558a340edb66328421
8d83c8412742e7887650907fb67d9bf20deb5a3faccfb20ef9624026b05c95892a05b96d3319c290c681e4a813a173443bd8
c83fe3f859eb84cbbd1c0cc67c81bdd60f6b3fafb4efcf5c5526aa6ce44fda4c8630a4711039e848e81c310eb6a73f6547d2
e5cc0586ff1eaecfad3dfbf9ad4b54d79efcfcdd5e326c21878398412736e785f6eefda79c9a30e4701039e844e81cd10eb6
07fa2f266c4bc290c341e4a013a173443bd8c102274918723802f9a513a173443bd85e5f2d633ee2d78790f94473be4daf7b
6d972d7af763babfedb23251657b7df5a4cd6408431a0791831e089d23cac10e762e9784218d83c8412742e78876b0836da5
246148e32072d089d039a21d7c7bf193c40b511c840d3ae1393bb47d1dac7f9284210884c8412742e78876f0f50f5ce18514
10c4469df09c1ddabe0e7ee19230a4801039e848e81c310ef6b07bbd220c2920440e3a113a47b4831d9ccf2461c8202172d0
89d039a21d3cf321541f9c02425ddf32f5e046a4095e5d20aec838ff83806f99c08c0fd2b598ff11bafd235a83da37a39163
8ff91fd4ed9fd116a17d331a31f60ee6b429ba90ff41d4a0139f71437ad743ec89e20bf91fc40d3a013a3fb47f1d4cdf9484
21ff83c8412742e78876f0edcb248917c23f101b75c2737668fb3a58c6290943f80791838e84ce11e3609b9d3ece473eadd5
abf765dace8db6eaafb3cc299fd67b12a3ac4c54d966a73f693319c210fe41e4a013a173443bf8e2aee10a2cc47e1030e804
e68cd0c6bdbdf244e285cc0fc2069df09c1ddabe0e165949c290f941e4a013a173443bf8f6aa3a8917323f02f6a5139eb343
dbd7c307ae0843e60791831e089d23cac1d73f708517323f081b74c2737668fb7af8c01561c8fc2072d089d039a21d7cfd90
aef042e00761834e78ce0e6d5f078b09256148fe4072d489d039a21dec6031a1240cc91f440e3a123a478c831d2c26948421
fc83c8412742e78876b08367b89230847f1039e844e81cd10ebefa544db285d80f62069dd89c17dabb0ede404bc210fb41e4
a013a173443bd8c1ac034918923f881c7422748e68075bfec531eef933a99fcfbae65fe271cce7dc98699be76db82fdc6465
a2ca967ff1a4cd640843f24720bf7422748e6807db1c9b791987a95e629443527928549f0eed6b3e344f65dfef6b8e8daa4c
54d9e6d83c693319c290fc41e4a00742e78872b0832db3256148fe2072d089d039a21d7c7f3d9d260cc91f440e3a113a47b4
832fafa7d37821f903b151273c6787b6effd55659a30247f1039e848e81c310ef6907fa10843f20791834e84ce11ed600753
65256148fe2072d089d039a21dece133578421f983c8412742e78876f0fde58c9a30247f1039e844e81cd10ebe1de924f142
ec0761834e78ce0e6d5f9b15febd81aed71bc3926f99f3d7b15e6f0c53be1829534ddbbc705599a8b2cd0c7fd266328421f6
23905f3a113a47b4831d5cb74ac210fb41e4a00742e78872b0837b154918623f881c7422748e6807dbfd69fd369eb39ecad5
461e58bdd29ebecf52ae6955aa305161bb3b7dd062327821f383b041273c6787b6af61d799da7505563e0f0d5399c7fd4d55
5ca71a487371abca44950dfc499bc91072e60782834c7cce0fed5f7bf834eec33cd615a6f9a05b1f90d527076bfe2de60f6e
bee680a8c24485edf1d383169386a3d80f42be5424334668db3a580020d838f683906f99e0b413cab50ee67529368efd20e4
5b2638e384f4eddc255eb41fb237a85fd0a907331e39fa0ea69349be90bd41dca013a0f343fb775aaefa08e917d437e8d487
1b9366f849bfa0be1fc43a501f6e4c9ae1ec5bf511222842df974e7db8316986d6f76f1f210482fa063df4e1c6a4187e4220
a8ef07c10ad4871b9366e860dabb240c1114488e3a113a47b4833f1114d4f7835807ecc38dc930743061541286000c22079d
089d23dac10e5e4e4ac2108041e4a013a173443bd8c15a2a49180230881c7422748e6807df4e359278217b83b041273c6787
b6af83b9179230646f1039e844e81cd10e7630df461286100e22079d089d23dac1175347145848fe2060d009cc19a18d7b7b
4699c40bb11f01fbd209cfd9a1edeb208f401286d01122073d103a4794831d4c9e9484217404c9512742e78876b05e30ca3e
42f207f50d3af6e1c66418cebe551f217b83fa069dfa7063d20c6fcf129678217883b041273c6787b6af839fb8240cd91b44
0e3a113a47b483ef9dc4245888dc2060d009cc19a18debe04a4d1286c80d22079d089d23dac10e66c94ac210b941e4a013a1
73443bd8c18da8240c911b440e3a113a47b4831dacf79784217223905f3a113a47b4836fbfa29178216f83b0410f78ce0e65
5f7b23b7aeebb89f9b881df9a3c8d71ae79cbff2aa65dc86f93a93c9ca4495ed8ddc933693210c911b3062d289d039a21d6c
911b75164c7d3e72ec65a7e5b21944f9c3b87dbf988d5c5626aa6c911b4fda4c8630446e1039e840e81c710eb6d7eeeb3ece
63bdb838d6615d3efbf934b09c8187755cef5b705999a8b2bd767fd2663284217283c8412742e78876b04db5a873f3cf2c98
7cfa59c665d86b62c4565f142f9feb7a555626aa6c532d9eb4990c6188dc2072d089d039a21dec216c441186c80d22079d08
9d23dac1f7271669c210b941e4a013a173443bf8e6469b9a2de46d1033e8c4e6bcd0debd3f215c1386bc0d22079d089d23da
c197e70d6a3c0edb20ea5b2638678636af878f5bf071d40761df32e11937a47767c288689fc33642b74da6f6cd68e4d8abd7
e5b1cc50164e9773dbb84f7b3ef69f977563bb8cff7ec34c65a2caeaf5c33643d8c635e010b681977ea8970dcea6e1387568
c7e9d0ce3766e4bc65a93f84a99ef286f5534759ae65f35df4954aa72b13557e63461eb5990c61c8db2072d089d039a21dfc
c68c7c5f1b96cf679af33763c8177a47fdb04a7a60bebe6de4ba3251e53768e4519bc91086bc0d22079d089d23dac16fdec6
387cd6f216ab8cb23e785cb67a14c807a3f1fccb2fb9ae4c54f9cddb78d4663284216f23905f3a113a47b48327f9b61e6b59
755443219679c967817c142ca35c96fc1dfdecd7d30f5d99a8f2247fd6663284216f83c8410f84ce11e5e077d1c1b04dd331
d7cbba65c83fcf7c3157e350ca3be4facebcad395085890abf4b0e9eb4980c5e08db206cd009cfd9a1ed6b611bf3fa59f6fa
8026dfc6ade336cf757def5c4261f2edecf5ec435726aa6c611b4fda4c8630846d1039e844e81cd10e7ec9b77dcc1f45dd5d
6abd9fbd96ddebb771fd8cf98cdfc86565a2ca2ff9a33693210c611b440e3a113a47b4836d4dd5581702954196376af3bad7
4793b9f9fc1bc9d7d777ee82284c54d856543d683119bc90b441d8a0139eb343db27f226ead7b0ccfa2c433ee77ece3268e3
2a4c5828d2265c8bc9e085980dc2069df09c1dda3e817d941bd7b97e25f39d633e082dd31593260b13150a6cd76232782163
03b151273c6787b6af612fe3e798cfe750f3f5042e8f78aeebf5015b15262c6cd80f5a4c062f046c1036e888e7ec30f6bd7f
53a40943cc0691834e84ce11ed60074ba9246188d92072d089d039a21d6c4fb3df0b90d684216683c8412742e78876b08335
459230c46c04f24b2742e78876b0bdb29aca97f39c2490cbea64eaf3017cbda29eefe7f8b23251657b69f5a4cd640843cc06
91831e089d23cac10e5e534ac210b341e4a013a173443bd8c1ec7049186236881c7422748e68073b78292f0943d20691834e
84ce11ed609de025fb085917d437e8d4871b9366787b13678917b22e101b75c2737668fb3af89949c2907541e4a023a173c4
38d8c1a21f4918b22e881c7422748e68073b38a948c2907541e4a013a173443bd8c1e25549c8491b047ecbc4e7fcd0fef592
b441c07f86581098f141bad6c1aa6c45c7391f01bac94467bc90cec59c0feaf6cf088bd0be198d18fbbb5f2dc115123e8817
7422333e48d77a388628be90f041dca013a0f343fbd7c319531186840f22079d089d23dac1b72f0e255e48f8406cd409cfd9
a1edeb6062b4240c091f440e3a123a478c831d4c86978421e183c8412742e78876b083658b9230447d1039e844e81cd10ebe
bd8a4fe285a80fc2069df09c1ddabe0e96f848c210f541e4a013a173443bd8c12a3e4918d23e02f9a513a173443b58af1565
1f216f83fa063df4e1c6a418de5e4927f142de0661834e78ce0e6d5f0f5f364518f236881c7422748e68077b38c028c210b9
41e4a013a173443bd8c3b9541186c80d24479d089d23dac10eae9f246188dc2072d091d039621cece09a591286c80d22079d
089d23dac10e1e254bc210b941e4a013a173443bd8c1fb4949188237881c7422748e68073b78272d0943f00691834e84ce11
ede0fbabea346108de2072d089d039a21d7c7f559d260cc11b81fcd289d039a21d6cb36ece692267164a3e10efd3b8d551e6
9bbb7c05b21cd7b75d5626aa6cb36e9eb4990c6108de2072d003a1734439f8feda324d188237881c7422748e6807dfdea55c
e285d40dc2069df09c1ddabeb62dfd997954e77f6dd7d4f512927166a20cd7bc595999a8b26d4bffa4cd640843f00692a34e
84ce11ed600f51048a30046f1039e848e81c310ef6103fa10843f00691834e84ce11ed607d1e22fb08d117d437e8d4871b93
66e8e1fba60843f40591834e84ce11ed6007c7184918a22f881c7422748e6807df3ea948bc107d41d8a0139eb343dbd7c155
84240cd11781fcd289d039a21dece0ca511286e80b22073d103a479483ed6e61cebf837cd0af3770df7bd7ba6bc4119f87c8
ca4495ed6ee1499bc91086e80b22079d089d23dac10e66aa4bc2907e41e4a013a173443bd89e0ad487b3e74e21f9426b5fb7
f3b1dd31b79760eda1802a4c58d89e093c6831193c8ebe406a9009ce99a1cd6bd0f95b397e83320f58067a9c3ba59477fd8d
5a5526aa6cd84fda4c9a8fd22f88fa5211ce78a19d6bc8475dea59bf8d9f3aa57839f352e3e23a5d99a8b2213f693349364e
bf20e45b2638ed8472ad11bf1834abd838fd82906f99e08c13d2b7f678339f56eb86a0658047beba28cbd7eb939a72101ed7
6b85b8ae4c54d91e6f3e693369ba90bd41d4a0139f71437af7f64669122e046f1034e844e7ccd0e675f0305b1286d80f2207
9d089d23dac1d73f7085171247081b74c2737668fb7af8792bc2907712c82f9d089d23dac19fbc13eafb418648e8c38d4931
bcfe655378216a85b041273c6787b6efb45cf511c24eb06fd4a90f3726cdd0c1914d1286a81522071d099d23c6c15727694b
b610b242cca0139bf3427bd7c12a1749184256881c7422748e68073b98a72d0943c80a91834e84ce11ed60074b272561c859
2172d089d039a21dec60da89240c392b440e3a113a47b4838dfcf8e4bf3de740cdf9c034eee351ff908f4febb194391b8d5c
5526aa6ce44fda4c8630e4ac1039e844e81cd10e7630bd4c1286c01522079d089d23dac1b7a76a4bbc10b812b02f9df09c1d
dabe0ee6904ac210b842e4a00742e78872f0ed49c3122fa4ad2036ea84e7ecd0f6fda4ad50df0f124cb00f3726c3f093b642
7d3f8811a13edc9834c3d9b7ea23449e50dfa0531f6e4c9aa1831f99240c9127440e3a113a47b4836f1f55255e083e216cd0
09cfd9a1edebe0342a0943f00991834e84ce11ede0db9984122fa49e1036e884e7ecd0f6bdbd9a53e285c813c2069df09c1d
dabe7643fc6296b4240c912781fcd289d039a21dec603f004918524f881cf440e81c510e76f0be4a1286d4131831e944e81c
d10ebeff3a561386d41322071d089d23cec136efe0fca77e24db918fbae3767e58631e62bed25caec77db23251659b77f0a4
cd640843ea0991834e84ce11ede0fb33d33561483d2172d089d039a21d6cf38b8e79cedfcc7ae89db77cda5987a98e325fe5
8e9f63b88f70b23251659b5ff4a4cd640843ea0991834e84ce11ede0fb3b616ac2907a42e4a013a173443bd8926ece03eeb9
40e4281fca543fac7df9ee357e3de69595892a5bd2cd933693210cd927440e3a113a47b4833d64fc28c2907d42e4a013a173
443bd8c8e7355f5cd4f570f9da7acdffcef56a7b2f814c9f7cb57d5dc3c9ca44958dfc499bc91072fc0981df32f1393fb47f
671288e8810348a8e35ba60ecc78e4e87bf8a2093a8e3f09d04d263ae38574ee8c3fd98e21dfb52f7512edb82dc367d8e77a
40285b7e8ff3676debb54d65a2ca33fee4599b21fee41a70883fc16b40d4cb6e7fc7b8ae5587769c0eed9ce49f61fc6cc35e
2fe3c7f2da7cf8d46bbdb16e003eeeedc7a50b13159edc8f5a4c062f649f1036e884e7ecd0f69ddff2e93b41a0a6759c0f42
f3657cfda4f63a8fa39d417561c2c2f33bfea8c564f042f00961834e78ce0e6ddf89bde45f5ffeedd539a5f9b67518cb754d
fd70da9bfb762cd595892a4ff0676d264318824f02f9a513a173443bf80d3e293995ebf9622fff1acabada7c9d5b863c5f6b
6fbfc127b23251e537f8e4519bc91086e01322073d103a4794832779be7e5eb625df2e9751e6bb9665cf173b7bfdb0b25bf9
82e3cbadeb12d49dd44fda4b862da49e1033e8c4e6bcd0deb5d49371ccd7b375b2c6f25d1f510f42f3348df594732db39195
892a5beac9933693210ca927440e3a113a47b4832df5643df2156b9d28be0ef93b912f69cf511ee3942f2f8fb53de0d49589
2a5beac9933693210ca927440e3a113a47b4835ff2fd938f439f3a2b67dd86631ba733ab7399d77cc21dc6f14af89195892a
bfe48fda4c8630049f1039e844e81cd10e8a0490b233cb9aafa5cb95c6f2d9ce87b1edf58dae4c542932406c9bc91086ec13
22079d089d23da41419e7f8df9e739d66351feb0d6f998d66555e45765a24a416edb4c8630c49f2039ea44e81cd10efe92ef
d3995b525fbd6cf958b4e55fcab108f2bb3251e52fb96f3319c2908042e4a023a173c438d8c1124a49181250881c7422748e
6807dfbf13d684210185c8412742e78876b03de53cc63d9f62eb28d7351f788f633eb36aa77c58da86fb3d8eac4c54d99e72
3e693319c2908042e4a013a173443bf8fe337d4d18125002f9a513a173443bd8c11c7d49181250881cf440e81c510e76b0dd
b2240c0928440e3a113a47b4831d4ce39584210185c8412742e78876b0839dc4246148402172d089d039a21dec6002b3240c
f92b440e3a113a47b48367f68bea2384a060dfa8531f6e4c9aa183ef9b240c2128440e3a123a478c831dcc099184210485c8
412742e78876b083e521923084a01039e844e81cd10e76704695841c8242e0b74c7cce0fed5f07b380141f47a110f62d139e
71437ad7c19a3745c7512801bac94467bc90cec52814eaf6cfac8fd0be198d187b0f5f3041170251881a74e2336e48ef7a38
9e28be108842dca013a0f343fbd7c115ba240c8128440e3a113a47b4831d5ca74ac2908982e4a813a173443bd8c1423b4918
3251881c7424748e18073b5868270943260a91834e84ce11ed60072b042461c8442172d089d039a21d7c7b4988c40b992884
0d3ae1393bb47d3d441628c2908942e4a013a173443bd8c1ea274918325102f9a513a173443bf89389427d3f08fb087db831
2986b36fd5470826a1be41a73edc9834c3d9b7ea23448350dfa0531f6e4c9aa187df99220cd120440e3a113a47b4833d9c52
146188064172d489d039a21dec2c1a84c81f045e20a173c438d82ea0d67a6d578f83d3673f86633d77effc6cdfa782d70594
aa4c54d92ea09eb4990c6188062172d089d039a21dece046491286681022079d089d23dac10e9e0948c2100d42e4a013a173
443bd8593408913f08bc2042e78876b0b36810227f10784184ce11ede0fb0bb03461880609e4974e84ce11ede0fbcb903461
88062172d003a1734439d8c1a6de923044831039e844e81cd10eb65ddc8f6358e661ab5fc67cc639e6f538375dcff738c7fc
b926ceebca44956d1ff7276d264318d241881c7422748e68073b98762609433a0892a34e84ce11ed60232fdb0aec6b9d1596
8f456bbee6d9ca65673e16e5f3d276ef99a02b135536f2276d264318d241881c7424748e18073b4b0721f207891b44e81cd1
0ef690d0a10843360991834e84ce11ede0998ba2fa08e920d437e8d4871b93666847d77cc699ea354efe099c93cdeb81703f
a7655e33d86561a2c2766c7dd0623278211a84b041273c6787b6af83938a240ce920440e3a113a47b4831d5c4848c2900e12
c82f9d089d23dac1f757726bc2900e42e4a00742e78872b05d36d7fd62cf5c8732c9ba4cb9ad491e5bbd9b5bee04015999a8
b25d363f693319c2900e42e4a013a173443bd8c12c6e4918d241881c7422748e6807dbedf137aba88ef2f33de1d551e63bba
6958efc760b2325165bb3d7ed26632841c1082e020139ff343fbf7febe2c928f024288fa5211ce78a19d6bcf80f2916738ea
59369f82ea32e073a6fdb91bd87cdf15cbca4495ed19d093369364e3801042be6582d34e28d71a713dc99c93feca657cfed5
d50f67fa4e6eb86615aac244858df7418b4983713a08f1de3291191ba4693d7cc6822ea4831035e8c467dc90de9d1bd9ab1e
424407f50c3a75e146a409cebe551f212483fa069dfa7063d20caf7fc1155e08c9206cd009cfd9a1edebe099ae240c211981
fcd289d039a21dfc09c9a0be1fa43f843edc9814c3eb5f3685174232081b74c2737668fb7a38a42ac210d181e4a813a17344
3bd8c10a3f4918223a881c7424748e18073b785d260943500791834e84ce11ed60076b1b256108ea2072d089d039a21dec60
72be240c411d440e3a113a47b4831dcc19978421a883c8412742e78876b083b566923004751039e844e81cd10e7630735a12
86a00e22079d089d23dac10e664e4bc210d441e4a013a173443bd8c134374918b23a02f9a513a173443bf8f6cc698917823a
081bf480e7ec50f675307f581286ac0e24479d089d23dac1b7234a245e8809216cd011cfd961ecebe003978421a484c84127
42e78876f0f50f5ce1857c14c2069df09c1ddabe1e3e704518f251881c7422748e6807df3ea44bbc908f42d8a0139eb343db
d7c1395c12867c1422079d089d23dac10ea6aa4bc2109142e4a013a173443bd8c1547549185252881c7422748e68073bb83f
9384212525905f3a113a47b4831dec882109434a0a91831e089d23cac10e9ec348c290920223269d089d23dac1f6d4f118f2
5fd503d0f1a90bae97f3fd713e288fe5e47b5dc4c8ca4495eda9e3933693210c2929440e3a103a479c83ef3f69d684212585
c8412742e78876f0e5972a1a2f44a41036e884e7ecd0f6b53778f3f0593e75bdd0b12dd3b00dfb524739e40bcca14e61686f
f05465a2caf606ef499bc91086881422079d089d23dac1f7a71a69c2109142e4a013a173443bf8fe7a144d182252881c7422
748e68071bf93cd5571de771f73ce5d40951477cd82a2b135536f2276d2643182252881c7422748e6807cfa810d507879450
d7b74c3db81169821852421dff99c0411d98f1c8d1f7f02517741c5212a09b4c74c60be9dc19d5b1e61bc57a0bf94f09d358
9732aeba0b7f3ecf0ef9e2636b37e3a63251e519d6f1accd1052720d388494e02518e8751fc2320bbdea773b4ec7767a0a29
21ecbfd33718cfd9a1ed3bbfe56515ffb6cff51e6a9a8efc594cfb56477ce6c78c57148fae4c54797ecb9fb5990c61c82921
f25b6742e78876b0b39c9240fe57fa06133a47b483df9c92fce7e9337dca1555be6fc8d79565294319f2b95ee6b8eec57565
a2ca6f4ec9a33693210c3925447eeb91d039a21cfce694d4d3e9541746cde55d4bbec4aaf317a6fd9addfe4d2a9195892abf
59258fda4c8630a49510f9ad33a173443bd8595a0991ff9dc1c184ce11ed6067692544fe770607133a47b4839da59510f9df
191c4ce81cd10eaab4923aaaa51e9a964ff9af75dddbef5c5726aa546925aecd6408435a0991df3a133a47b483827ccb3f8a
793aeae7b32ef91b3b2ffbb82bf2ab3251a520b76d264318d24a901c7426748e6807455ac9782ed2a80fc5b7fc03cdc33cae
b70aba3251a5482bb16d264318d24a88fcd689d039621c6c0b69b735df24d62724fb0aefbeb67c53316ff99aebfab6cbca44
956d21ed933693210c6925447eeb4ce81cd10eb6157e6575f1b49dcf86d6613a8e793da365b6f95cde7eadf05395892adb0a
bf276d264318d24a88fcd699d039a21d7cff4184260c6925447eeb4ce81cd10eb6e7ac751fb2731de2314ffbba9daffd8eb9
bde96dcf595561c2c2f69cf5418bc9e085a89280dd74c6737668fb3a98ce2b0943540991df7a24748e28071b799dc4502f33
3ff583aa571ff516b35c5f5fcbfb445da2ba46fd777bc9b085901262be7566735e68ef3a7869280943480991df3a133a47b4
831dec69260943480991df3a133a47b4831dbc22978421a484c86f9d099d23dac10eb6bd9684212205c9416742e78876b083
cf5c1286881422bf7522748e18073b8b4821f2bf833f98d039a21dece0d82e0943440a91df3a133a47b483af9ec9251b87a3
10f225339973423bd7c13e958a8fc35108fb9219cfb821bdeb60de8fa2e3709400fd9599ce78219debe19316741ccd42d097
1ce98c17c2b9331146b41fc251a8df5be71ecc78e4e83b58e826f942380a71df3a033a3fb47f3d9c371561084721f25b6742
e78876b083ab444918c251901c7426748e68073b98362c0943380a91df3a113a478c831d2cf19384211c85c86f9d099d23da
c136497e9c8f7c5d513f9f65dace6d3ceb0fb44c6d9fd67b5e9dac4c54d926c93f693319c2108e42e4b7ce84ce11ed60239f
8fe538979b9e934586ba7d47be97d9f3edccbd0ce67f97a5c6fc676bc98085581402be7506734668e37a886b5084211685c8
6f9d099d23dac10e163d49c2108b12c89bce84ce11ede04f2c0af5fd77d448ecc38d4931fcc4a250df7f478d701f6e4c9ae1
271685fafe3bec83fb7063d20c3dfcce1461884521f25b6742e78876f0c543aa020b6928080c3a833923b4711d5c3748c290
8642e4b74e84ce11e36007d78a9230a4a110f9ad33a173443bd8c19b034918d25088fcd699d039a21dece0d983240c692844
7eeb4ce81cd10eb697c1f9ae6d5acf7550759ffb7dafd314ce5d7af309f05a4a2a2b1355b6d7c14fda4c8630a4a110f9ad33
a173443bf8f61e6e122f44a110f6ad339eb343dbf7fea21b4d18a2500279d399d039a21dec604f674918a25088fcd623a173
4439d866f9e46fc37ed4175ac750b6772cdbca975196217f86cf7691cbca44956d96cf933693210c5128447eeb4ce81cd10e
b6bddbc769cb479d1ade716cf9a0bb947bf4b2e15dbed0da8e65bec86565a2cab677fb933693210c5128447eeb4ce81cd10e
b6fdfacb74907c8a2d03dbe771cb67a3bdce3f3b96ba64e79ac7a82b1355b6fdfa9fb4990c6188424172d099d039a21dec20
04461286281422bf7522748e18073b98c7280943140a91df3a133a47b483672088ea23849150dfb7ce7db83169869f3012ea
fbef940deec38d4933f4144642d87f476c309eb343dbd7c1615d1286301222bf7526748e68073b0b2309e47f456c30a17344
3bd8c1458c240c6124447eeb91d039a21cec2186451186301222bf7526748e68073bb8649784218c84c86f9d099d23dac146
7eccf3b8d48fe498b77cfc5fcb5d543d28afe327ff4caf23bbac4c54d9c89fb4990c21879120f82d339ff343fbd76e4f877c
e41b86ef6ed0cbb2e6cb9d7a48de977539f249f8faaecbca4495edf6f4499b49f3511809513795e08c17dab90e562208360e
2321e44b6638ed8472eddcba5fb4ce8920d4e92573f3662c72e467b7a2fd90c941fdde3af760c62347dfc18f59f2854410e2
be7506747e68ff7e1241a8efbfa32eb80f3726cdd0c1c9431286501022bf7526748e68071bf98ba1bf9230848204f2a633a1
73443bf8130a427dff9d7611fb7063520c3d7cdf1461c80521f25b6742e78876b08793b6220cb920480e3a133a47b4831d3c
4d9784211784c86f9d089d23c6c10e52e32461c80521f25b6742e78876b083141c4918724188fcd699d039a21dec210b4611
865c1022bf7526748e6807db0be231ff14c64f5d1e329c935cebc7b3e653df966fa3af5546aa3051617b3dfca0c564f04228
0861df3ae3393bb47d1dcc129784211484c86f9d099d23dac10ee64248c2100a42e4b7ce84ce11ed600f71288a30848210f9
ad33a173443bd8c1b6da9230848204f2a633a173443bf8def4360916e24008f8d62398334219d7c14c464918e240901c7426
748e68073b98312c0943220891df3a113a478c83afe76228bc100742d8b7ce78ce0e6d5fbd25967d844c0eeafbd6b90f3726
cdd0c3974d11864c0e22bf7526748e68073b38c048c210cb41e4b7ce84ce11ede07be7120916d23808f8d619cc19a18deb60
a191240c691c447eeb4ce81cd10e7630495c1286340e22bf7526748e68071bf9920ff6635de399873b7ef669ae4f43cb63d1
cff2b9cf25b2306161e37ed0623278218a2360379df19c1ddabe0e76db9084218a83c86f3d123a4794831ddc094bc210c501
23469d099d23dac1f73723d084218a83c86f1d099d23cec1f7e7c66bc210c541e4b7ce84ce11ede0fb0ffa346148e320f25b
6742e78876b087f8154518d23888fcd699d039a21d7cffb1b6260c691c447eeb4ce81cd10ebe1e3ca3f042140761df3ae339
3bb47d672c84ea23846150dfb7ce7db83169869f300ceafbaf9409eec18d4813c4300ceaf8afa407eec08c478ebe875f97a0
e3308c00fd9599ce78219d3b2321f669d93f5519fe336ec7fcd9f6b20cad0457940bbe925e71429bca44956728c4b336390c
e31e308761d04510eac77fb67c82fa9c3ab4e37468e7fcb88773ffb0b11c01f225edb1acc35af7ba1eb7619e720bd726abba
3251e5f9713f6b331942cec36072d089d039a21dfca6424cd3bc6fc350131cd6125b509f578cc77a3eeaff62ff55f6bdfd7f
d05a32601c83c1c0a0139833421b7702afe7ee70531d62599ab1d7399df90fd746df27b22e4c5878423f6a31193ccec088d8
974e78ce0e6ddf3703237f14e3278f358f6b2e393cf57d46fd98c66519f7b25fe33703435626aafc66603c6a331942cec060
72d003a1734439f85d65330d9f9ac8930796afde4a74c5514ff5f927b9ad9f7db96e057565a2caef2a9b476d2643c819184c
0e3a113a47b4832d0363ce47d8bd6e92b8acfb3a6ef35cef64e6391f90b6f17e0dae2b1355b60c8c276d2643c819184c0e3a
113a47b4832d0363cfd5437d1eb7de6feecaa6f2dbb87ef2bdfbb537beac4c54d932309eb4990c2167603039e844e81cd10e
f69581c1e47f273b30a173443bd857060693ff9decc084ce11ed605f191844fe20d981099d23dac1be323098fcef64072274
8e18073b58352b09390383c9412742e78876b083257592903330981c7422748e68073b58972009390383c9412742e78876b0
cd5e3d86311f85ea0b96bddcc7954d97cb1fc6edbbe1f3357b555526aa6cb3579fb4990c21c76044f24b2742e78876b03d88
782f595b13720c0693831e089d23cac1b7e7704a3c4ec2606cd009cfd9a1ed7b37fd43b171060633834e6cce0bed5d073bc4
4b42cec06072d089d039a21dec60deaa24e40c0c26079d089d23dac13acf49f6c12914d437ead4871b936638b7f1567d700e
04f70d3af6e1c664187ad8cf5a11720e0493834e84ce11ed6007bf7249c839104c0e3a113a47b4831dcc4690849406c1e0b7
4c7cce0fed5f07fbf4293e4a8360ec5b263ce386f4eefd0918928ed22022749389ce78219d3b8fa8a27dca63e06e6f39b46f
4623c6fe6ec08ae0e21c08e6059dc88c0fd2b51e8e248a8f7320981b7402747e68ff7a386f2a42ce816072d089d039a21dec
2b0782c81fa41b30a173443bd8c1da2649c839104c0e3a123a478c831d4cd595849c03c1e4a013a173443bd8a6674ff96fa6
a97e3ecb797a3f873ce633df382ef7e9535626aa6cd3b39fb4990c21e7403039e844e81cd10e769103c1c07fa71b30983342
1bd7c1a20b49c839104c0e3a113a47b4831d2cb491849c0311c92f9d089d23dac1b75775493c8ea0606cd0039eb343d9d7c3
07ae08390083c9412742e7887630066070df7fc73a701f6e4c9aa187038c22e4000c26079d089d23dac11ecea28a90933088
1c7522748e68073b58152e09390983c9414742e78871b0af240c26ff3bdf81099d23dac10ed63949424ec26072d089d039a2
1d7cf5018864e30c0c66069dd89c17dabbf7d7f96842cec06072d089d039a21d6ce4f5d4736e2f76ccd3be6ee741e9988765
29f7f417b82a4c58d8b81fb4980c1e67603036e884e7ecd0f6b5b7ffc3362f737dc87fac9f2d7f45f77a89b77fbef735d781
4d5626aa6c6fff9fb4990c21676044f24b2742e78876b0839dda242167603039e881d039a21cec60136549c819184c0e3a11
3a47b4831dcc6f92849c81c1e4a013a173443bd8c136f192903330881c7522748e68075b2ec07930aa7b99efe5d473bebedf
3e756ace325d1fb92a4c54d872011eb4980c1e07603036e888e7ec30f675107a2209390083c9412742e78876f0f5200885c7
d91b8c0d3ae1393bb47d317b83fbfe3b5482fb7063d20cafffc6141e676f3036e884e7ecd0f635ece3c8a7bcba9e75cf77ef
f920b41ee728b7713fe67c577771abca44950dfc499bc91072f60693834e84ce11ed603b9d941778cbb9152bfcf7704df06e
2793dfba4475ed54f2777bc9b071ea4664be7462735e68ef3ab87890849cbac1e4a00742e78872f03cb4a83e38f782fb069d
fa7063d20ced62b51ce8f3f8aa43eb30e533507d8f9cb5f99c3a7f5dacaaca4495ed62f5499bc91072ee0593834e84ce11ed
e0e9baea839227a86b90a90737224d706e1d2e7ac0ec07eef752b179331a3df60e7e5e828db21f18f996094e3ba15c0bd90f
dce95fa10adcbc198b1cf9abc76ec5c59913cc0b3a91191fa46b1ddcfb483ece9c606ed009d0f9a1fdebe002451272e20593
834e84ce11ede0db57a4128fe32e181b74c2737668fb3ab8eb93841c7711c92f9d089d23dac1d73f7085c7491b8c0d7ac073
7628fbde3d982b36ced86066d089cd79a1bdebe14a411172c60691a34e84ce11ed60074f6e2521676c3039e848e81c310e76
90282309396383c9412742e78876b0bda758f7711e8f734afa50f602395f96968967c33aaef774035999a8b2bda778d26632
849cb1c1e4a013a173443bd8c1ca5449c8191b4c0e3a113a47b483ed65e4baaee37eae1c3eea0638c379b75a5667d44d8eae
b791aa3251657b1df9a4cd6408396683c9412742e78876f0ed95a9128f3336181b74c2737668fb3a587b2009396383c94127
42e78876b08319f892903336981c7422748e68073b98872e09396323925f3a113a47b4831dcca093841cb6c1e4a00742e788
72b08359939290c336881c7522748e68073b98a52b09396c83c9414742e78871f0ed69d9128fc336181b74c2737668fb7af8
c01521477d3039e844e81cd10ebefe812b3ccef9606cd009cfd9a1edebe10357849cf3c1e4a013a173443bd8c8d721df40d5
498de502e39cdc57ae3db6fc3f7dc6e37a0aa30a131536ee072d2683c7691f8c0d3ae1393bb47d1d9cc62521a77d3039e844
e81cd10e7630115e1272da0793834e84ce11ed60bb68bd0f44f3f2598ee1b39d139a9773a7e56b9f095598a8b05db23e6831
193c4efb88d8974e78ce0e6d5f072127bf781cf5c1d8a0073c6787b2af83c9ff9290a33e70c4a413a173443bf8fe862a9a90
a33e981c7420748e3807df5cf6a0d938e48399412736e785f6eefd7d453421877c3039e844e81cd10ebefcbc4de371c20763
834e78ce0e6d5fc31e96bddeaa9603eeb1effb9c3fac3aca69fdace3ba5e2f526465a2ca06fea4cd640839e183c9412742e7
8876f0fd153e9a90433e981c7422748e6807dfdfee511372c40893834e84ce11ed608c18e1beffcaeee01edc8834410fdf36
c14701278c7dcb8467dc90def570701174147012a19b4c74c60be9dc1970926f5ff2df0cf55319cbe4887c65b997018e75d9
dbb02ddf4fda5426aa3c034e9eb519024eae01878013bcf443bd6c2199cfcca70eed381dda393feef9bb72b7dcb84d73f667
c837aee51267ac9f512e6e1fb7ae4c54797edccfda4c8630049c1039e844e81cd10e7e034e8e71c99717f505fa94af60c7b2
b1710da0695738ed88aa2b13557e634e1eb5990c61483a2172d089d039a21dfc467e7c07592e72e6691bd6cf549f517dbfa2
e3d40e6bb23051e137f0e3498bc9e085a493807de984e7ecd0f69dd87bd99b7e299b87e741eee51e629f6be2d074aeeb1cae
239bae4c5479823f6b3319c2907442e4a00742e78872f0bb27fcb2e56fc4a71e7fcb9ddbf1c9b76e359b639896738adc975c
5726aafcee09ffa8cd640843d20991834e84ce11ed604b3a29992b537d50b1dcbbdc96581a3eb09bca44952de9e4499bc910
86a41322079d089d23dac196f7311ce3f1a9d716ebfde0b5c4d20ce7564f0d5c16262c6c691f0f5a4c062fc49c1036e884e7
ecd0f6fd063facc7b0677bea7debb26ecbbae6d3f1957ea00a1315fec63ed81693c10b1927840d3ae1393bb47d22e9a3a6ed
941d0ccb88cf7d0caf2937ba3061a1c8f9702d268317024e081b74c2737668fb04f6be9d919ce588d47e89ab8c76b92a1355
0a70db66328421e004c9512742e7887650049ce4d263ced7d2e52cb48dc79c8fccdbb108f2bb3251a50838b16d264318024e
881c7424748e18075bc0c9b60ee3bcd429fc2bbc6adab66dc837b12575aa059ca8ca44952de0e4499bc91086801322079d08
9d23dac10ea60e4bc2107042e4a013a173443bd826d0ce4b56ea5bc4323daabc66a8ef1bf2ad7b3e15977c9a6b02adaa4c54
d926d03e693319c2107042e4a013a173443bd8a6919effd44704dbb1aff9cb713e3c1887fdcc20baa691aaca44956d1ae993
3693210c012781fcd289d039a21dec601aa9240c0127440e7a20748e2807df7f04a30943c60991834e84ce11ed6007fbc24b
c2907442e4a013a173443bd8c11c5a4918924e881c7422748e68073bd8bf5b1286a41322079d089d23dac19fa413ecfb497a
08f5e1c6a419de7f45ac0943ce0a91838e84ce11e36007bf3449187256881c7422748e6807cf2c06d547483aa1be41a73edc
9834c3bb810c8a8d334e08f99689cc39a19d3bdd163d70ca08757ccbd481198f1c7d07a75045c7292301bac94467bc90cebd
3f0542d271c60941df72a0335e08e77af8a4055d483a216ad089cfb821bdebe1f8adf842d20971834e80ce0fed5f07574a92
30249d1039e844e81cd10ebe3d8351e2859813c4469df09c1ddabece624e88fc417807123a478c831dcc51968421e684c841
2742e78876b08379e99230c49c1039e844e81cd10ebebd1041e285b013c2069df09c1ddabe0e56d549c2107642e4a013a173
443bd8c19a1b4918c24e02f9a513a173443bf8f6222b8917c24e081bf480e7ec50f69d911faa8f1037427d834e7db8316986
9fb811eafb418806f5e1c6a4197af8892bc2103742e4a013a173443bd8c3d94c1186b81124479d089d23dac10eae60246188
1b2172d091d039621cece0aa551286b81122079d089d23dac10e561c49c2103742e4a013a173443bd8c1e36449184247881c
7422748e68073b7859270943e80891834e84ce11ed60072f682561c81d2172d089d039a21dec606f274918724702f9a513a1
73443bd8f6f39aa7fa1dfca7ae12391f84d65dc78e786c9795892adb7e5e4fda4c8630e48e1039e881d039a21cec600f3749
187247881c7422748e6807dbbe7ddb909b3a9f0ae6abab791ea63a9d7f9ff3af725f97fdba319795892adbbe7d4fda4c8630
e48e1039e844e81cd10e7630e14a1286dc1124479d089d23dac177b7c8576c21748498414736e785f1ae83fd862561081d21
72d089d039a21d3c376b577d84e40fea1b74eac38d4933bcbdf5adc40ba123840d3ae1393bb47d3d7cd91461c81d2172d089
d039a21d7cfd035778217484b041273c6787b6af870f5c1186e891407ee944e81cd10e76b083bd240cd123440e7a20748e28
077bf8cc1561083e2172d089d039a21dec602b774918824f881c7422748e68075f0ff052789cb982d420139c33439b173357
a8e33f3245b079331a3df60e16820836ce5c21e45b2638ed8472ad118fe753823aba732fd8a15c477f534a611da7ac4c54d9
889fb499341b27be10f22d139c7142fad6563a6ddb7a948ddccbaeee75c9e930946340fefd7fa6699eaf5d187465a2cab6d2
e9499b49d385dc17a2069df88c1bd2bb86bd9c2bb0ea4df1f508280f311fef3ff953d92f6a5598b0b0413f683119b810fa42
d0a0139d33439ba7d6b5e5b3f9316cdbb906f11b497b7dc56565a24ab5aecdb5990c61087d2172d089d039a21dec60459f24
0cb92f440e3a113a47b483e7d943f511925742df974e7db8316906e1fa5c4eadd35ed74295377bf9a4b78e6ae5ec5d99a852
b86edb4c863084af1039e881d039a21c7cfd00a3f042fe0a61834e78ce0e6d5f3b87968d31f7b5de2caecb675df772a35ace
f0c330e40bac6b13485d99a8b29d439fb4990c61c85f4172d489d039a21dece09a5c1286fc1522071d099d23c6c10e9ee34a
c290bf42e4a013a173443bd8de5a0cf9a05b5ee09545aef334e7b3c0b67e2ff5e665598e8b5c5626aa6c6f2d9eb4990c61c8
5f2172d089d039a21dece04d95240cf92b440e3a113a47b4831dbc9d9484217f85c8412742e78876b0bd91ae4b14ea662cf9
bea68c71aacf46f2c1689dea2bb5eb8db4aa4c54d9de483f693319c210c142e4a013a173443bd8c1021849182258881c7422
748e68073b5814210943040b91834e84ce11ed609b6db4d6895075aaebf4f96eba56bfa6db7772d035db485526aa6cb38d9e
b4990c61886009e4974e84ce11ed600773eb246188602172d003a1734439f8f61c798917f257101b75c2737668fb3a98292e
0943fe0a91838e84ce11e3e0db33c5255ec85f216cd009cfd9a1edebe1035784217f85c8412742e78876f0f50f5ce185fc15
c2069df09c1ddabe1e3e704518f257881c7422748e6807ebd32fd9470841a1be41a73edc983443737d3ef277e03ce795f8cc
b267653df897edfc86dbf2ff5d969adf7fb6960c58c83e2160d009cc19a18deb603d82240cd927440e3a113a47b4831d5c33
49c2107f12c82f9d089d23dac10ea20a256148402172d003a1734439d8c14a0c4918125060c4a413a173443bf87e30a7260c
0928440e3a103a479c83efaf41d184210785c8412742e78876f0fd27019a30e4a01039e844e81cd10ebebffa461386281422
079d089d23dac1f73703d184210a85c8412742e78876f08cc5507d843012ea1b74eac38d4933bc9bb2a4d8420c0931834e6c
ce0bed5d0fdf3445c82128047ecbc4e7fcd0fe9ddf33d103c79050c7b74c1d98f1c8d1f7703017741c4312a09b4c74c60be9
5c0f611c2a3e24c490e0f517ea6567cba9ccb9fd2fb7e37468a787300e45186248881c7422748e68074ff2b5ec7f7b6c738d
4ec897567990635df1384d9f7c2fbb4c4b23d795892a4ff2676d2643186248881c7422748e6807bf791cf90359f6a15e55ce
db96bfa39fa5cec8cbf7cfdf19345f725d99a8f29bc7f1a8cd640843124920bf7422748e6807bf7350e6755fd7a59eef96fc
f7e723f832ca23ff5559cd70cd419195892abf73501eb5990c6148222172d003a1734439d8f238ce4fa4c646e4018fdb3cd7
a95173f99e9694816b7a9fac4c54d9f2389eb4990c6148222172d089d039a21dfc929f134eebb392fc71ecd3311e355a703e
f285f59e29aecf5c5626aafc923f6a3319c2904442e4a013a173443bd89248d672f42d635cb6211f7af3b7b47c54d3b0d534
8d6b9e99a84b54d77248fe6e2f19b6104342cca0139bf3427b2762483e794cc73ad45903cb3e1fc35416b98a69857765a24a
114462db4c863024911039e844e81cd10e8a488ee54acc18ca549d6ddaa60f4cd75695892a4524876d3319c2104642e4a013
a173443b28c83fe581e0bad5054ae57130edc7a92b13550a72db663284218c04c9512742e78876b08349b4923084911039e8
48e81c310eb6d9a46788635d87b62dc33933a68c723fbfa4c39537242b1355b6d9a44fda4c863084911039e844e81cd10eb6
5d01ca72e2e9dc9c6358877a9971ce330fd13bba325165db15e0499bc91086301222079d089d23dac10e76829084218c84c8
412742e78876b0cd2cfc4efdac0f67be27dcf3c35ab6fcb51caeb39aae4c54d966163e693319c2104612c82f9d089d23dac1
36b37099f3f5ddd81e7a974ca8e5fcb0ea475536926d330b5565a2ca36b3f0499bc91086301222073d103a4794838dfcbde0
614d18c248881c7422748e68073b9853290943180991834e84ce11ede0ab9bf54bb6104342cca0139bf3427bf7fef35d4d18
6248881c7422748e68077f6248b0ef27e11ad4871b9366384312541f210884fa061dfb7063320cef465228b6100142cca013
9bf3427bd7436680220c0124440e3a113a47b483dd049010f39fc91e44e69cd0cef5f0590b3e8e3f21ec5b263ce386f4ae83
93b6a2e3f89300dd64a2335e48e77af8a4051dc79f10f42d073ae38570eedd1fb4e00ac127c40b3a91191fa46befcff2d07c
21f884b8412740e787f6afcd691ae6713d3775fa8ce7ef6fab1fce1dbfdde634a9ca44956d4ed3933693210cc127440e3a11
3a47b4833d059f20f693380fc2737668fb3a98b8280943f00991838e84ce11e3609ba63bce473ea5d739ebcbb49dcf02eb77
b44cae9dd6fb118bac4c54d9a6e93e693319c2107c42e4a013a173443bf8de5c740916f24e08187402734668e3de5ecb25f1
42de0961834e78ce0e6d5f072b3d2461c83b2172d089d039a21d7c7b698fc40b612701fbd209cfd9a1ed3b9337541f217184
fa063df4e1c6a41872dff51497cfe3e7e4ae7a2418cff921f515e4fffaeb6c6f7943e1ffef6400429c0981814e000e581bd4
c3af481186301522079d089d23dac1cec25488fc414408113a47b4832f9e211558c8504160d409cc19a18deb60cd92240c19
2a440e3a123a478c831d5c004bc290a142e4a013a173443bf8feca1d4d183254881c7422748e68073bb8bf9784214385c841
2742e78876b091e7534dbefbaca31cce4fa7be155df38df136dc93026561a2c2c6fda0c564f042800a61834e78ce0e6d5f07
dbd548c210a042e4a013a173443bd8c8c7fc291c751bdbe32817adc3f73d78be8eab7bc55d3f725999a8b2913f693319c210
a012c82f9d089d23dac13685e2bb5979fd327ebe9bacfe536211bea7e0ebae4f5626aa6c53289eb4990c6108502172d003a1
734439d836e41af2a886e1fb65acfb6eee758ed3bee43be47c3ebc9e6cc8ca44956d43ae276d2643180254881c7422748e68
073b9830240943800a91834e84ce11ed600fd1318a3004a82039ea44e81cd10e36f27ceccbe7defaf90c65fbb4b255771d65
feb03e43fe955ee4aa325165237fd2663284214685c8414742e78871b06db6f83d10d703d05aaeb7967acd99ff87a9cc9d5d
aedfb9ac4c54d9365b7cd2663284214685c8412742e78876f0ddd01cc516e25b88197462735e68ef7a0877508421c185c841
2742e78876b08770074518125c881c7422748e68073bd848561286101722079d089d23dac1332849f511625442df974e7db8
3169860ee20e246188512172d003a1734439d841ec81240c312a440e3a113a47b4831d641f48c210a342e4a013a173443bf8
f6eee4128f6354901a6482736668f33a5847a3f828c485a82f15e18c17da39819ccfed7938634d7fdd86f19bdea590afca44
9502d9b699241b87b810f22d139c7642b9a69689e593dc944fece54b99bf85cb3ae4a1ae82f8ae4c54a99689b9369366e310
1742be6582334e48df04f4519e42e75b9932be3d9f71d6653ad472c8ab3051a140762d268d16125c08197482335648e304f3
be9d0b3cea87b2851552ba3251a5a0b66d26c317425c881b7402747e68ff14f9b82e63be4fdbeb2887753ef27d8cfc8a5f95
892a15b96b3319c210e242e4a013a173443b28c8b76df9e46b8bba5e775df6753db6592e016d85890a05b76b3119bc90e042
d8a0139eb343dbf793e012fafe339784fa7063d20c3dfccc146148702172d003a1734439f8fa4155e1850417c2069df09c1d
dabe1e2e19146148704172d489d039a21dece0fa501286041722071d099d23c6c10ed6594bc290e042e4a013a173443bd8c1
e3544918125c881c7422748e68077bc8ae518421c185c8412742e78876b0bd2a3bc67ddeeb5284635df311e838e6f33170fe
78e66db8670acaca4495ed55d9933693210c092e440e3a113a47b483edc5709d8775bed63ae67cb1b59d53758eb94d676def
855561c2c2f65af8418bc9e085f816c2069df09c1ddabe3601225fdfd4f083f2d6fed8f7ba375bfda44adcd0b8ae17b7ac4c
54d9a6403c693319c210df42e4a013a173443bd8c1b25b4918e25b881c7422748e68073b98e82409437c4b20bf7422748e68
073bd8905b1286f81622073d103a4794836f4ff9977821be05b151273c6787b6af8329bb9230c4b71039e848e81c310e7630
4d5b1286041722079d089d23dac1b7173b48bc10df42d8a0139eb343db77a679a83e42860af50d3af5e1c6a419cebe551f21
c584fa069dfa7063d20c3d7cd11561c8502172d089d039a21dece1e0a60843980a91834e84ce11ede08be9310a2c64a81030
e804e68cd0c6f5901ea30843864a20bf7422748e68073b58962a0943860a91831e089d23cac10e2e562561c8508111934e84
ce11ed6023bf4fb7f3f2598ee1b39d9b232ce78cea6b39822a4c54d8b81fb4980c5e0850216cd001cfd9e1ec7b7fe98d260c
012a440e3a113a47b483ef078a6ac210a042e4a013a173443bf8febdb8260c012a440e3a113a47b48367bc83ea23449850df
a0531f6e4c9aa187a82245180254881c7422748e68074fd7551f21c884fa069dfa7063d20c67dfaa0f8e12a1ae6f997a7023
d204314a843afe3327833a30e391a3efe1b026e8384a24403799e88c17d2b96fac44bd8a18ea4d43fed6e793ed329d192053
9930b1ced773555d99a8f21b2bf1a8cd1025720d384489e06508ea6547ce631db6aa433b4e87764ef26d998f1a73514639ef
5b3edd7dea2de4744ef23caea38aae4c5479923f6b3319c2102542e4a013a173443bf80dd458f6b1ecda55ae32e6bd5ce1ec
f930544679de4c0dd75b045d99a8f21ba8f1a8cd640843940891834e84ce11ede0f7f57079da3f9debb897f118973279a106
9ee4bbab355f79ddaf876565a2caefebe1476d264318a24402f9a513a173443bd843a086220c5122440e7a20748e28075ba0
c6b918a47c33977cb39cafb0a76feccb92afb4e76bf9852c4c54d8e2341eb4980c5ec811216cd009cfd9a1ed6b691ad3b07c
a6fa6828df38cef3320e75cdeb32e4e3f3fcb91f37e9ca44952d51e3499bc910861c1122079d089d23da4191a9b1aff9d2aa
6cf659bfa32dfb44cc7eb92b13558a4c0ddb663284214d84c8412742e78876504cf4ca67bfcf673d7f9deb30956578cb2033
35aeca449562aa976d3319c2902642e4a013a173443bf8f67c4a8917a244081b74c2737668fbd414b7b1ac66ffece507b97e
f275675d002ba7b8b5ca44956a8a9b6b3319c2102582e4a813a173443bd8c18a174918a244881c7424748e18073b58552709
43940891834e84ce11ed609be2f67d3c56ee9f8661d9b77dfed40fabdcbfe7efcb7a3d719395892adb14b7276d264318a244
881c7422748e6807db14b77aaaad6feccbe3a2b20767dd8c33df42e78b8f3243f19ae2a62a1355b6296e4fda4c8630448910
39e844e81cd10e76162512c8ff0cc82042e78876f0ed296e122fe4881036e801cfd9a1eceb60b31b49187244881c7422748e
68073b98dc2709438e0891834e84ce11ede0dbfb3a49bc102542d8a0139eb343dbd7c176ec923044891039e844e81cd10efe
448960df4fe239a80f3726cdf0ee7ef08a2d44981033e8c8e6bc30de75f04d938421c884c8412742e78876f00c8e517d8428
11ea1b74eac38d49339c7dab3e38d283babe65eac18d48139c3d8b1e3854833abe65eac08c478efeec56b4cfb116a1db2653
fb663472ecaddb9ff6395882babde5d0be198d187b0753d2155d8897206ad089cfb821bd6bd89fe1c857cf759ad9b81ddbb4
7f37d9fa6e817ccfd89495892a1bf6933693e10bf112c40d3a013a3fb47fefbf87d684215e82c8412742e78876f0fd57a39a
30244c2039ea44e81cd10ebe3cb949e3857809c2061df19c1dc6be0ee6f149c2102f41e4a013a173443bd847bc04013fc84d
2030678436ae8379c99230244c1039e844e81cd10e7630175d1286840922079d089d23dac1b7175d48bc903011b02f9df09c
1ddabe9f8409eafb41b042e8c38d4931fcbb091304f6203781001cb036a8875f91220c0913440e3a113a47b4833d1c391561
48982072d089d039a21dec236102819fe426109833421bd7c1f590240c0913440e3a123a478c836f5f004bbc102f41d8a013
9eb343dbf7fe4c6f4d18e225881c7422748e68073bb8bf9784215e82c8412742e78876b083a7a49230244c1039e844e81cd1
0eb6976ec7304ef3b93ffea77e2d973a93ef58d6692ce7dceb339795892adb4bb7276d2643181226881c7422748e6807df7e
cb2af142bc44c0be74c2737668fbda6bf5f39ffad67b3bf25169dcce0f671cf6311f8eaebdb17465a2caf65afd499bc91086
780922073d103a4794831dec99230943bc0491834e84ce11ed600fc11a8a30c44b1039e844e81cd10e76307d461286780924
479d089d23dac10e8235246188972072d091d039621cec2158431186780922079d089d23dac10ea6c949c2103241e4a013a1
73443bd8c11e789230844c1039e844e81cd10e76163241e40fa21388d039a21dec603aac240c2113440e3a113a47b4831dec
75290943c44520bf7422748e6807dbe4ef7c81fd59ebc0f6b20c645c3ff5ae7a5bf6739fab7b7f535599a8b24dfe7ed26632
8421e282c8410f84ce11e5a098f03fedf93c3c6cdbb934e37bf52527fc5f95892ac5847fdb66328421e282c8412742e78876
50908fe7edc419cb30e41f48bef458d5da96bb3251a520b76d264318222e881c7422748e68073b58e4210939e502c141263e
e787f6af870dbb051fa55c10f5a5229cf1423ba796318dcb3a4f474debcbf752f9f3296b6ce53aa65699a8522d64726d26c9
c62917847ccb04a79d50ae09e27c72cd279d329b2e8f6e8acb146565a24a416cdb4c9a8d532e08f99609ce38217d13d0f396
7f70d3581f0aaee5bde9b88e8bfa45df95892a05b46d3369ba107441d4a0139f71437ad7c1f244c917822e881b7402747e68
ff04f9f8dd03a12e9cfd6c9f792c7b3928f2ab3251a520b76d264318822e881c7422748e68071579d9e5365f3e6f7594e5bf
d6759707f1ab3251a522776d264318b22e881c7422748e68077bf8cc156148da08e4974e84ce11eda0fa9d977d00f673178c
f57c0934cb43fa5598b050fdca4d8bc9e085980dc2063de0393b947dea54765d4ed4a5f279d0d3679787b7ab3251a53a95b9
3693210c491b440e3a113a47b4836fafbe9678216603b151273c6787b6af83842e49186236881c7424748e1807df4edd9378
216383b041273c6787b6af3d6b1a977c9b36d78723c736961494fa406c3f9f048fd7c3745598a8b03d697ad0623278216083
b041273c6787b6af3d4a2fcf9aa7ed5c229bbf8447fe4a9e4f0bc28d98ae4c54d91ea53f693319c210b041e4a013a173443b
d8c12b3349180236881c7422748e6807db0be29a7a5a7f82c75e76d12aeb9cca1fc6ebc7d95e10abca4495ed05f193369321
0c191b440e3a113a47b4831dcc859084216383c8412742e78876f0d59591922da46b1033e8c4e6bcd0de75b0a6471286748d
407ee944e81cd10ebe3db14fe285680dc2063de0393b947def4d5d956021540381512730678436ae8359ca9230846a1039e8
48e81c310e7630335d1286500d22079d089d23dac1b7d77748bc10aa41d8a0139eb343dbf7ffbfc6420284c40e02039d001c
b036e8f5cf53e185b010c2069df09c1ddabe33a844f511e23aa86fd0a90f3726cdd0c3c1431186b80e22079d089d23dac1d3
75d54748cea0be41a73edc9834c3dbc157122fc46604ec4b273c6787b6efc584140516d2320818f400e68c50c6f59090a208
435a068c987422748e6807dfcf09d184213083c8410742e78873f0f59c108517d232081b74c2737668fb7ac8095184212d83
c8412742e78876f0fd5d23346148cb2072d089d039a21d6ce4cbf9d2bd8e721f3ff9eb5127871fe75bbafb7a4d16262c6cdc
0f5a4c062f04751036e884e7ecd0f6bdbf684e1386a00e22079d089d23dac19fa00eeafb41fc05f5e1c6a4197e823aa8ef3f
1330a80737224d10833aa8e33f5328a803331e39fa1e8e2b828e833a02749389ce78219dfb86369469b9f97757a3258e737e
725d63d6160e5f9b24eaca4495dfd086476d86a08e6bc021a803af80502f9b7d2ef9a4547568c7e9d0cef765e3347cf225dc
546778e41ba66cce5117224c473e147df6e5da674b5726aafcbe6e7cd466328421a883c8412742e78876b0c5557c9f95ff53
37f7cb1fcf36d4a584f9739a97310ffc9a49212b1355b6b88a276d264318823a881c7422748e68075b5c457d1356170c2fe7
8e68f5ba3e5fe3e65b9af99ef1aaea12d6b5a88abfdb4b862d447404e64b2736e785f6ae05550cf917b92e758c47beac9d97
f37894cf78eb3a4ff7ab085d99a8b245553c693319c210d141e4a00742e78872b065550cc7787cea7168bdefcbcb4f73386f
5a1bb82c4c58d8922a1eb4980c5e88e8206cd009cfd9a1ed6bd89f251ff3eb4bb1751bbeafffebef729df73aafede2569589
2a1bf8933693210c111d440e3a113a47b48362e6ccf589e441b6db553961a815262a14f3665c8bc9e0857c0ec2069df09c1d
dabe0ee6fd4ac290cf41e4a013a173443bd8c15c7e4918223a881c7422748e6807d5ba95ed337fb6b93e9bc9bfcc35df4fec
6a46ff5d98b050ad5a312d268317f239101b75c2737668fb3a58a22509433e0791838e84ce11e360070b122561c8e72072d0
89d039a21dec60e1b1240cf91c440e3a113a47b4831d6c302009433e0791834e84ce11ed609b21b7eef9deb8de4ee4bf5897
7ceb7c4ee2cc57f643fe85de0ff26565a2ca3643ee499bc910867c8e407ee944e81cd10e7630434e1286880e22073d103a47
94836fef9923f1423e0761834e78ce0e6d5fc35ed7752c7b9097411ecb327d8673feea7d1bddb85565a2ca06fea4cd640843
3e0791834e84ce11ed600fc9248a3044741039e844e81cd10ebe9e4ca2f0423e0761834e78ce0e6ddf4f3e07f6fd247b82fa
7063d20ccdf2a99cfacee964f9f0bb2cf9b6e63c15d4cd08e6fb8c222b135536d39fb4990c6148e92072d091d039621c3c73
0c541f212783fa069dfa7063d20c3f3919d4f783f409eac38d4933fce46450df7f0650500f6e449aa0831dff141fa77410f6
2d139e71437a17533a42b77f046050fb663472ecefe6ee082e4e0721dc5b0e5cc605e159c3dd8f75d9ea5ae4cfb84ce525f9
5cbf06df0bb9e59aee202b135536e4276d264d17d241881a74e2336e48efdabbb97c725ba79aadf719cf03ed563f9723df3e
e5cb8c7bc58cac4c54d9decd3d693319be900e42dca013a0f343fbf7fecb6f4d18d241881c7422748e6807df9fe1a209433a
0892a34e84ce11ede0fb93b9346108082172d091d039621cec60436c49180242881c7422748e6807fb080821e007b11704e6
8cd0c675b080441286801022079d089d23dac10ee6804bc2101042e4a013a173443bf8f682038917024202f6a5139eb343db
f72f84742880903e4260a0070007ac0c7afdf35478219b84b041273c6787b6af871fb0220cd924440e3a113a47b4833d1cb4
1561c8262172d089d039a21dec239b04819f246e109833421bd7c1a598240cd924440e3a123a478c836f2f619678219b84b0
41273c6787b6af833b4d4918b249881c7422748e68073bcb2621f207891b44e81cd10e76f0a45212866c1222079d089d23da
c10e5e3e49c2904d42e4a013a173443bd85e389e3b8fd781952d5aa76139df159dc7a2f99e55202b1355b6178e4fda4c8630
c49304f24b2742e78876b0b37812227f10ba11089d23cac1cee24988fc41e806113a47b4831dcc249184219e84c8412742e7
8876b0cd1e3ae6795cce4900f3967f846bf948ea96607984c7701fdb6565a2ca367be8499bc91086781224479d089d23dac1
b6b358beb8ced756e71497f297e35487bc2fcbb9f9fc756c9795892adbce624fda4c8630c4931039e848e81c310ed69770b2
8f1010427d834e7db83169861ea251146108082172d089d039a21dec6086a2240c0121440e3a113a47b4838d7ccc0de66b9d
f3e097afed96b53ea6da877c1bbb1d65a959235795892a1bf9933693210c0121440e3a113a47b4836f6fda28f1423a086183
4e78ce0e6d5f4b0719cfc9846594df3bb8a12679acf9eb997faef76a225999a8b2a5833c693319c2900e12c82f9d089d23da
c1b737279578211a84b0410f78ce0e659f8a89f89ef6cee530e776838bda76f9ae4c54a962225c9bc91086681022079d089d
23da41b1b8e42887e2b95ee5af7bb9da5ba6438626b4c244856269896b3119bc900b42d8a0139eb343dba7d6d48ceb32e68b
8d72f85df3a5ed9cefe517b9fdf05599a852ada9716d2643c8b920080e32f1393fb47f3d4446083eca0521ea4b4538e38576
ae87e4845f36ce0521e45b2638ed84724dad105c3fdb96ef60ca217739d67c04cee73df9e5be2a1355aa3582aecda4d93817
84906f99e08c13d2b75fe8351fefe625ffa19c7496b27f6cd9e0484506dc95892a7fa17d9b49d3855c10a2069df88c1bd23b
819d4febcb7a9c0bb5f397635da7e1a33eeaab3051a180762d2603174241081a74a2736668f31af6582fa8ea7965289badef
f5f1fb12f7199685890a1bf6831693c10b8920840d3ae1393bb47df5ce57f6113239a86fd0a90f3726cd20be697bf6682c93
92eb1763d8f6e9b8f797d395892ac577cdb6990c61c8e408e4974e84ce11ede0ebbf3185173239081bf480e7ec50f6f5103f
a30843260791834e84ce11ed60076bcd256188e54072d489d039a21d7c3b8d44e2854c0ec2061df19c1dc6be0e0212256188
e52072d089d039a21d6c8f3aca5ea6fb5a1f37aef517589e0595cf6a18867dbbe701e8ca4495ed51c7933693210cc91c440e
3a113a47b483eda95e793dbd9cef72e0bf87eb736bcff47eeb12d5b5277a7fb7970c5bc8e42066d089cd79a1bd6b6f4beaf2
8e73ade87ced59549ec54df9083c2f773e83ac4c54d9de963c693319c290c941e4a013a173443bd8de8ace43beb0a9eb208e
6d99f295e5be9c012acbf619eacd637b2baa2a1355b6b7a24fda4c863064721039e844e81cd10eb637e1753ac250dfe37dea
aae8e9f806a8ac53fd88ae37e1aa3251657b13fea4cd640843260791834e84ce11ed6037991cc4fc206982d89c17dabbce32
3902f99f611344e81cd10ebe3da14de2854c0ec2063de0393b947d1dec062d0943380792a34e84ce11ed6007d374256108e7
2072d091d039621cec606ab6240ce11c440e3a113a47b4836fcfc59778219c83b041273c6787b6efdf0de720b007a91304e0
80b541675e82ea23246450dfa0531f6e4c9ae1f5ef92c20be11c840d3ae1393bb47da7e5aa8f108f417d834e7db83169861e
0e5c8a3084731039e844e81cd10ef670b25284219f23905f3a113a47b4831d5ca048c210d441e4a00742e78872b091e7abd5
fddc8c297f41c7bafce35c883f6e65dfe2eba992ae4c54d9c89fb4990c6108ea8011934e84ce11ede0fb6bbb356108ea2072
d081d039e21c3cb775577d84b40cea1b74eac38d4933bcbf9780260c6919440e3a113a47b483af674628bc109541d8a0139e
b343dbf7fe62264d18d232881c7422748e6807cf9f99ea23e45550dfa0531f6e4c9ae1fd472c9a30a4651039e844e81cd10e
fea46550df7fc650500f6e449ae0fdad71241f677510f62d139e71437a77da2ddae7b48cd06d93a97d331a39f6969cf0dd33
f79f32fb61ca87fbfdcc77c867dc699bd6fb81b1ae4c54d992139eb419d232ae0187b40cbc0c613d1f9ff2e9aaead48ed3af
765a72c236aceb5cbf11cbfd5437ff2f4b5b0cd1c86565a2ca969ef0a4cd6408435a0691930e84ce11ed6023bf7784ce5752
53be7fd8b7f287b9ecb0391cc770bd069495892a1bf9933693210c6919444e3a103a47b4832d39623d3efb5e2fa8af492775
9447beb8dd4bfa45239795892a5b72c4933693210c9919811c7420748e68071bf9a7b8520779ee833dd7956df3671bf24dc6
7c4fef5085890a1bf7831693c10b8119844d3ae1393b947d2d39a23ca299ebac93b6a0ef28a35ca6b52c8458ee80145999a8
b225473c693319c2909941e4a403a173443ba8e631ad7bae3c866faa4bfe5a96e8073991a95526aa5433995c9bc91086cc0c
22271d089d23dac10ee6004bc2109b41e4a403a173443b28c8d76d19f2dd72ddf6705dca46ba031cd86565a24a416edb4c86
30c4661039e940e81cd10e0af24f7958b06e75b955795444abfe7565a24a416edb4c8630c4661039e940e81cd10e8ad94cd3
b9faa63e2adbc63cccfc7d39d4677e5726aa14b3996c9bc91086e40c24671d089d23dac10e663349c2909c41e4a4df84ce11
e36067c91944fe200f82089d23dac1368febbb2d4319e5672dcf2c96fab82a1f84a7f213592e725999a8b2cde57ad2663284
213983c8490742e78876b0cde63aa71f9d5b50ccf9749b7f87759dddfa0d2bbb9ef7c9ca44956d36d7933693210cc919444e
3a103a47b4836dbf8361ff4cc3f09d7294cf00eb67af4b69f7e54cabbb3e735999a8b2ed77f0a4cd640843724620071d089d
23dac1368f6dd8e665ae171ac7faf92e1ead5fd3ef065bd7bc455999a8b2cd637bd2663284213983c8492742e78872b087cc
104518c233889c7420748e68073bd8cb461286f00c22271d089d23dac11e6243146108cf2072d281d039a21d6ce4f99b39cd
e7d1e8539fc12fe78e3b718593aefc3fe6de25dd555dd9bacdab14bb04f7f313db75faeb9fbe0286207ad0fa99a41671527b
c6d192686dd89887a4dea4e5203fd3673384293f43c8a51e089d113638c8fbcfede3bd1ed82fa4ecae5b933ce64b90414e2d
9bb41ce467fa6c8630a5774472ad074267840daee1023446cacf90b1a5be8fe18ec9301cf23364ec13991432863b266628b0
4b1912a6f40e21977a207446d8e021bd43c6fe6730858ce08e8809ae0ee721388dee10e6580e6c46058aab908d03741a1c92
a0f772a0332ed05c95e010c1fd672646e23216c059ade010813e1188217cc606babbfe8d34f3a5e010e1967a00743ed8dfc5
133f182fa58608b6d4039ed3c1faae9fe9c384293524926b3d103a236cf0fa99654c985243845cea3ba133620c1698478984
293544c8a51e089d113658233544804f6461089813c1e20a4c1646c2941a22e4520f84ce081bbc7a4e3ee2a5c810c1967ac0
733a58df721f8063a4dc8e3476a88731dc3131c37f9bdb21602702291280032641977f96082fe57608b6d4039ed3c1fad6cf
128d919233646ca98731dc31314385131711a6e40c21977a207446d8608de48c087c260f42c09c0816572c3943c84f444244
4267c418bc7aa131e2a5e40cc1967ac0733a58dff593cd9930256708b9d403a133c2060b3c5540c2949c21e4520f84ce081b
2cf0cc0c09537286904b3d103a236cb0c0fb17244cc919422ef540e88cb0c1026f1b9130256724f2500f84ce081b2cf06e1d
09537286904b5d089d113258213f8208537286904b3d103a236cb0c04c12244cc919422ef540e88cb0c1027b4121614ace88
e45a0f84ce081b2c306f0a09537286904b7d2774468cc10a99214498723b845cea81d0196183c5723b84fc441a85103a236c
b0586e87909f48a3104267840d16cbed10f213811442e88cb0c14ab91d827d228d42f09c0ed6572cb72391ff339042089d11
364853fefb67b15f644febca94bf0b10dab1766fd9a4254df9777d364398a23b845cea42e88c90c102cb3c9030457708b9d4
03a133c2062b645810614aef1072a9074267840d528ac5f4ec27ded7f22bdcefa6a6fe6bf8fbe072aead6593969462e1fa6c
8650d33b22b894039ff3c1fe6847eebf75c5cbc2e2dfb4ce4ea08c83bd659396b423b7ebb3319fa4770875a8ee70c6059ba3
58877e77d16f25d755e49fe7f7d6cfbe4f0aa8d95b366949b10eaecf866c9ade21c8b11ce0d80459abb0cf3eb0697a8720c7
72803326d0db5894fb7bf53ba86532593fd5fc5d60cc07f8ec9757cb3ef1dba25c6ad9a4e558947ba6cfc67429bd43a8a51e
f88c0d7457602d32f2a5000fe1967a00743ed8df20ef1fbddff7b62e1bddde1af7a3bcbfff72a437726ad9a4e5203fd36733
8429c343c8a51e089d113678c8f090b14f2453c818ee989861587f7f5eebf6fbafe9d67fd2decb8ffbdfd6d5ef3dec0bda35
69378cffbbbf66d8527a47620ef5c0e65cb0bbc1bcbfa89fe733be6efd147f5f4e03b7d76b7e03b54153c3161b0eea133d36
8397d23b045bea82e77490be028bfd9130a57708b9d403a133c2062b5c2511614aef88e45a0f84ce081ba4c5fef77e1df97d
2f8f65faefddabff007e31b0656bd862435aea6f7a6c062fa57708b6d4773ca7c3e8a35bc0e9d3cff4dfe541ecfce6fc7d7b
dcef7823b4b56cd2926e015d9fcd10a6f40e21977a207446d8e0209fd6ad66e65bf4d7fd31bf6eba2d6bd35ff75fff4a3ec3
8d10b56cd272909fe9b319c294de21e4520f84ce081b1c4fb8e6fd839ecbcf7b3f1df7fbc5e77bb986ef7fac7bff53fd1edb
1b2b6cd9a4e578c675a6cf6608538687904b3d103a236c70bcbd78beeeb7c76379afd67f03e679e9cb04f54fbfa3981edfdf
fe7e165b366939de5e9ce9b319c294e121e4520f84ce081b2c96e121e427922984d0196183c5323c84fc443e85103a236cb0
4c8687309f48a61036e782dd1558ce838429c32391877a207446d8e0d553eb102f657808b6d405cfe9207d0596762061caf0
88e45a0f84ce081b2c30611809538687904b7d2774468cc10a5be11361caf01072a9074267840d5e3d2f1ff1528687604b3d
e0391dacefbfcdf010b013b91802e08059d021c343c63e114e2163b8636286cb3f4b8497323c045bea01cfe9607d154e1e44
981244845cea81d0196183157e308830258808b9d403a133c206af5b61826029382401877a007322585c81ab22244cc12142
2e75217446c86081b53548988243c211a77a207446d8e0f50bdb9930058708b9d4374267c4191ce4efe5d096bf4f3f137c6f
dff73a9bf437fddda36de4d4b249cb417ea6cf6608536c89904b3d103a236cf0fa85454c98624b845cea81d0196183d7dfe8
33614a2e1172a9074267840daed1063446ca0e91b1a51ec670c7c40c15b2248830259708b9d403a133c206c7c3b4dfedfbfa
2dabad7ef7bf57e2cbb7e2efd662bf5cc6964d5a8e076a67fa6c8630259708b9d403a133c206af5fc2c7849a9b22e0b11cf8
9c0ff6b77ecf60044d2e918163390c608e078fbec2070de834372541efe540675ca0b991a7b0ce009f3f02af7e1298df9c2c
e907d37c1bf77a3ef63c0568d8a4e1c85338d1e32134e5ef680fa129fb2558ac2fdbecce9b21fe3fedc7d5433f03fb737f7c
6fcb85e57b0fa19b37ab9eeeeffe97fa6def20b165939603fc4c9f87d094407e220a44089d1136386696dcbef7ef6fd948b2
9ffbe777a5f344bafeb7baddd6fba86d6609356cb1e1985972a2c743624ac03e910322784e07eb1bd8cfded57b7950f07e4e
7fb3cae6835c025e6ebfed5e945b366939c0cff479484c11f27f468108a133c206af9e4a847887c494801dea09cfe9207d05
66a422e12131259087ba103a236c106696dcdebf699a5738ce47d96fa4e61e269c59b2b56cd2126696d83e0f892981fc440e
88103a236c90a6d8df1f9fe935ef613a9f96defde7f6b73f67e3964d5ad2147bd7e7213125909fc801114267840d0279ffa1
fd7ddeeb0cc7f7ebf37e7fa727ae27190d9b34046ed7e3212e25609f0801113ca783f5159844858487b894401eea42e88cb0
41583bf59ce788f7f3ffb2c2eb313f2efcbdeff451df5b3669096ba76c9f87b8949dfc4c0888103a236cb05c5c4a203f1102
12099d1163b05c5c4a203f11022284ce081b2c179712c84f848008a133c206c712f0ded3f7ef9ef1feba3f9fb7c7f2c3fb99
57fcf4cfea677bc28a2d9bb41c4bc0cff479884b09e4af7f878008a133c2060b4c1944c2435c8a906f75217446d86081ed3d
90f0109712c8433d113a2364b0c066364878884b09e4a12e84ce081b1ce4ebff2db338a7efe7dd4f85eb364bfd5eb2df5dcf
8f6f0639b56cd272909fe9f3109712c84f848008a133c206c734d1652b86f519e4fcb3fb9c96c9ccdfc7dfdb90ed2127356c
d2704c123dd1e3212b25609f4800113ca783f50dece5f5fdbaebccf7f9e867a069b997fca6db726ed862c3817da2c743504a
c03e11ff21784e07eb1bd8f3edd3df3e5a73d8eebc13cbbaf3ceb2efcebc87f9e0a6964d5a0ef0337d1e825276f233e12342
e88cb0c102b1404878886909e4a11e099d1163b0c0b6744878088909e4a12e84ce081b5ca352688c434c4b183bd4650c774c
cc00312d61ecd7bf4248640477444c50200c89f872524bc0decb82676ca0bbcbb369002dc7b408f1280b9a1181da0671bf57
9ccf33cbf14debbca4e547774a9b9870cb262d07f3993e73584b80fe670e49a2332ec0dc80eed78cfd54b7ccd1b8f7eba9f7
fb765f6e273efdbfe8b791d336d7075b366939a0cff499c35a02f4891012e13336d0ddf5d30f98ef10d612b8435d009d0ff6
572bac25609f8820113ca783f55d3fa78d090f612d3b79ac0ba133c206c73cc67e9afdac57cefdaef9becc347cfe7d35ffde
e98e798cd4b249cb318ff14c9f87b096407e228224123a23c66081b9ab4878086b09e4a12e84ce081b1ce4efdbe7fb5d2e28
e6a976ebddd2fcf5dc5e670f7068d8a4e1e03ed1e321b225609f0822113ca783f51598978e8487c896401eea42e88cb0c102
6b1190f090da12c8435d089d1136f89f24a71c010e913002b6d505c001b3a01147731ce390da12c60ef534863b2662b87a45
0fe21d525b0276a80b9ed3c1fa2a7c8b88f0901913c8435d089d113658e1cc498487cc98401eea42e88cb0c12a99313bf099
2414017322585c81eb41243c64c604f2508f84ce883178f5c276c43b64c604ec50173ca783f55dbfc081090f9931813cd485
d0196183051e6e20e12133269087ba103a236c70905f98e68d8487cc98401eea42e88cb0c1022f6490f0901913c8435d089d
1136385eb92e5b642f0bcdbe9f79aef8fc307dfec77dfa5b0cb1bd72a5964d5a8e57ae67fa3c64c608f93f935084d0196183
055eb323e121332690877a227446c8e0985af17c2cdfbeff2d6f8ad6eca6e528bff9b116b66cd2724cad38d3e7213326909f
4842114267840d169854828487cc98401eea42e88cb0c102c9294878c88cd9c9635d089d1136b8dc12e31887d4963076a8c7
31dc31198672a92d81fc441689103a236cb05c6a4b203f91452284ce081b2c304d13090fa92d813cd485d019618305767644
c2436a4b200f75217446d860add496807d228b44f09c0ed637e6dff79b99df7bb97b9d378dee97f3bf65e1fef4faac8ffef7
f9f7d4b249cb31fffe4c9f87d41621ff671689103a236c90525beebfefb3dfcfcc5fc8fe3d7d4efd67f0bbcdbfc7964d5a52
6a8bebf390da12c84f6491244267840c964b6d09e427b24884d0196183572f2f42bc43644bc00e75c1733a581f60cf33c5e7
d8b06599ebbe721eb8b7964d5a02b8edf310d9b293ff3b8f44f89c0ff657600521f1a5c89640bd55239c71c1e60a2c1705b6
1cd91290f7b2c0b109b25621bd04d872644b40decb02674ca0b76bb32480eb10d61278435dc88c07b436801fb757ff7f2e6b
b59ff775fdcb920170cb4b62b165939603fa4c9f87b096c07d228244009d0ff677f546078877486a09d8a12e784e07eb83a4
9630f689141219c31d133354f8b011e121af45c8b7ba103a236cf0ea8c9a23db21ac2530877a62732ec85d859f4c223c84b5
04f250174267840d5258cbf2b3f65aa2ab5ebff97fbddf1fcc2cd95a366949612daecf4358cb4e7e268544089d11365820c4
10090f792d813cd423a133620c16d86a00090f792d813cd485d0196183e5f25a02f9891412217446d8e0789e79bf2e8b1909
0f792d813cd485d019618305165f23e121af259087ba103a236cb0c0425c243ce4b504f250174267840d1658a6878487bc96
401eea42e88cb0c142792d81f9440a89b03917eceefa0d6c99f090d722e45b5d089d113638c8bfbffe2d5857213cfb9f6a7e
eabafca3ffc5fa3dc53c396a9053cb262d07f9993e0f912d81fc441049227446c86081051e4878886cd9c9635d089d113658
60be2e121e225b0279a8474267c4182c30471b090f912d813cd485d01961836bc4078d71c84d096387ba8ce18e8919d6b169
8c43b449183bd4650c774ccc00d12661ec13991d32863b266658c7a6310ef92261ec509731dc3131c3d58bb810ef102e12b0
435df09c0ed67761c406811d32450270a80b9813c1e20a4cc847c243b888906f75217446d860812b17243c848b04f2504f84
ce08192cb01c19090fe122db114b5d089d113658216283080fe122813cd403a133e20c5e7f87c284877091401eea42e88cb0
c1eb775b60c243b848200f75217446d86085c803223c848b04f250174267840d5ebff686090fd126813cd485d0196183573e
7d61b643a849600e7561732ed8dd1a7141631c6245c2d8a12e63b8636286ebf7ed61c21c2b12c0f7b2f0391fecefdacf1990
e5389300bc9705cc78406b15fecc4097e34c047a9485ceb84073e33dfcbcbef4b9bc38ede7da5bff933cbff31fe8f597cef0
186bbab8659396e35dfc993e53a2c976c029d1245ef7c5fab207f0e7fb59eaa11f570ffd1498ce858429d144c8435d089d11
3638de4fdea67e64cbe7f0dd7f65befdbaeabec470bc6e8f5fbfa3786fd11ed8b249cbf17ef24c9fcd10a65013210f752174
46d8e0f2b405c748b12269ecad2e63b863620698fdf15d26a52c2947aff7f4ea5f8bdb8f5e876f0d9b3484b91faec766f052
ac8860877ac2733a485f81a9b148986245843cd485d019618305264323618a1511f250174267840d0279ff30de7eb7cf7203
f17ecddb33debe3724df5a366909e4b6cf660853ac889087ba103a236c10c83fd3ba927b99a0326df374807c6bd9a42590db
3e9b214cc922421eea42e88cb04158e572fb5b69b87e41d777d82fccd7d85a366909ab5c6c9fcd10a66411210f75217446d8
60b16491487e262f43089d1136582c5944c84fe465444267c4181c8bf8d6b5fbcb517ee69d75d70745d3efef73b93d7ba086
4d1a8e457c277a6c062fc58a0876a80b9ed3c1fac6fcae398071d999e213fff76dfbb88ed95dc7764dda8db95dffeeaf19b6
142822cca12e6cce05bb1bab93e785cb8f7597f0dbfbf6e87733ef75dd5d3aa573cb262dc7eae4337d3643980245843cd485
d01961830556a423610a1449e45b5d089d11365860361f12a64011210ff544e88c90c102bb5e20610a1411f250174267840d
0ef2ded5f3b33c9dfa8ec716eb99a97f299ffd4bb2bd2bc4964d5a0ef2337d3643980245843cd485d0196183cb5d318e9152
3d64ec509731dc31314381d9a34898823d843cd485d019618305765342c214ec11c9635d089d11367875800de2a5540fc10e
f588e774187d05360e43c294ea21e4a12e84ce081b3ca47ac8d8cf7f6765c818ee9898a1428a0c116aa68880ef65e1733ed8
dfe05e5e2fdc96cb8ddf7222787cff7e81dffd1ae0b74f00c1964d5a0eee337d36e6d34c11c1decb82676ca0bb02bb12129d
c68a24e851163ae302cd0de8c77c6db9cea4edd7b4cb4c85e54ff4b7b5da3e49195b366939a0cff4d9984e6345047a2f273a
e302cc5dba0281b852a088f086ba90190f686d00ff6efdd4befc117ef7a9ffc67dfe0229ffb617df2f4fb165939603fa4c9f
cdf0a54011e10e7501743ed85fa54011c13e119321784e07eb2b30830f0953a048248f75217446d86081599b48980245843c
d423a133620c16586f8284295044c8435d089d113678e12ee90496a2440438d405cc8960710516d820618a1211f250174267
840d1658608384294a44c8435d089d1136f8df468924b07fc673088003664187281119fb4446461ac31d1331548a1211ec13
011982e774b0be0adf22224c5122421eea42e88cb0c10a674e224c5122421eea42e88cb0c11a512211f84c40868039112c6e
00df9fdfcfefb17c20fb09e0b724992ed72ef37aa3c7bc11c460a6964d5a0eec337d364398a244843cd423a133620c5ebf8a
8509539a889087ba103a236cf0fab51c4c98d244843cd485d0196183d74f4066c2942622e4a12e84ce081b2cb0930612a634
11210f75217446d8e0205f37eb5e8ef23b5f68dcfe9ee63f967c9fe7fea6195b366939c8cff4d90c614a1311f25017426784
0dd6c8d43812a6349144bed585d01961831532358830a5890879a8274267840c8e7915dfe7f3fe5a5f9c3ca7fe517ccfafc0
977dcbdff77e96decf70d8b249cb31afe24c9fcd10a63411210f75217446d8e072878463a43c0f193bd4650c774ccc50601e
0f12a63c8f481eeb42e88cb0c1439e878c7d22a5228ee18ec9300cebd3ad4b5a9fa7f7cfe7f3d93f96cbac97fe81ed5700af
cff659c7964d5a0eeb67fa6c8630e5790879a80ba133c20657eb34464ad490b1435dc670c7c40c6386e2df23eed9d06d9915
d07f0366438fb4b6855b36693966299ee9b319c294a821e4a12e84ce081b2cb0f71e12a6440d210f75217446d8e0988c3bcf
66ebffc17c609fe77dea97bc9ff909773fc065b5df7d3bbf61cb262dc774dc337d3643984235843cd485d01961836302f67d
9df1f3bf3548695e8e7f5ba689af573ecf2d79915b36693926609fe9b319c214aa91c8b7ba103a236cb0c0a47b244ca11a42
1eea89d019218305f65545c214aa21e4a12e84ce081b2cb0b8060953ae869087ba103a236c907611eeffcd32ed6b3eca47de
271c5b366949bb08bb3e9b21d45c8d081ecac2e77cb03fda377abaf5cbeb69997bf3fa4debabfa27716f2d9bb4a47da35d9f
8df9245743a8b76a84332ed81c2d137d7f7eb7f924381fdef4783c3fd3ed461ff2bd659396b450d4f5d9904d733504792f0b
1c9b206b051662139be66a08f25e16386302bd55c9d510dc13791142663ca0b50a5107c4977235843bd405d0f9607f957235
04fb445685e0391dacafc21f9c0853aa879087ba103a236c70903f1f73fedc72f2794e7f9b5ecc47d9ef98fb89ffb7bda9e3
964d5a0ef2337d364398523d12f956174267840d16d84d040953b68790877a227446c8e0d5bb3b205e0af610ec50173ca783
f5150bf688e467e22a84d019618305f63840c214ec21e4a11e099d11639056fadfe7174cbfcf7c127effde8fe7f2841257fa
8f964d5ad24a7fd767338429d843c8435d089d113658e0710b12a6600f210f75217446d8e078b87861503112a6600f210f75
217446d8608157564898823d843cd485d0196183572fd343bc94ea21d8a12e784e07eb2bb086080953aa879087ba103a236c
f0fa1d359930657b0879a80ba133c206afdf39970953b64722dfea42e88cb0c141be4f757dbe7eafefed37ad6ba05eeb6491
ed4b4e0d9b341cdc277a6c062f057b0876a8273ca783f4159858888429d82392c7ba103a236cf0c2ddef092ce5790870a847
3027c288bb3ce580f05298876087bae0391dacef10e621639f08c89031dc3131438119da4898a244843cd485d01961838728
1119fb443c878ce18e89192efc661358ca2f11e05017302782c515582d878429c244c8435d089d11365821d1830853968990
87ba103a236cf0eaeb15c44b4126097bab0b9ed3c1fa0a2c0740c2146422e4a19e089d113238c8fb87e1fd5ef26dfbb9e736
af15f98eed4efa21beb6672edcb249cb417ea6cf66085390493862a90ba133c206af5ff1cf8429c844c8433d103a23cee0f5
37a24c98824c843cd485d0196183d72f0161c2146422e4a12e84ce081b1ce4f7cf6d7ecabd1ce57ab3bc7c25dfd3e33bdd5e
cfed6b4e0d9b341cdc277a6c062fa5980876a80b9ed3c1fac643b6dfe3f578af0b53e6fbc5e7bcf5e7fc8f657af563dab2a2
b9659396e331db993e9b214c2926421eea42e88cb0c1eb372862c2946522e4a12e84ce081b3c6499c8d827323a640c774ccc
b08e4d63689a880cbd97650477444c70fd465cc8a7992282bd9705cfd84077d7eff986749a2992a04759e88c0b345761cb7f
ca02499922f1122cd697ad8fbfefb51efa71f5d0cf9a2f4163a4540f193bd4650c774cccb08e4d63a45c0d193bd4650c774c
cc70c8d54863ff333042c670c7c40c05a62020618ad610f2504f84ce081904f2c774ef57b7bf7582d07a8dffc449c15bc316
1b02b7ebb119bc94ab21d8a12e784e07eba3c9d0db2ced6546583fe87e273b11f7d6b2494b9a0cedfa6c8630e56a0879a80b
a133c206817c9a5ebfcf7b9df2f97e7ddeefefb43d04c0864d1a02b7ebb119bc14aa21d8a12e784e07eba3150ff3c65def69
5992343f9a939b226ed9a425ad78707d3643984235843cd485d019618330d5e6be6e5dbc3caf996e53bfbc7c7edff451df5b
366909536d6c9fcd10a6500d210f75217446d86081f53d489842352279ac0ba133c2068b856a08f989b48848e88c188305d6
f12161cad510f250174267840d1658bb8984295d43c8435d089d11365860621912a6740d210f75217446d860855c11224ce9
1a897cab0ba133c2060bec478084295d43c8433d113a2364b042ae0811a6740d210f75217446d8608169944898d235843cd4
85d019618305767b41c294ed21e4a12e84ce081b1ce4efcffd79ffaefb86dffa75c7677dcd364f27b8f58bfdfdfd3cb66cd2
72909fe9b319c294ed21e4a12e84ce081b2cb0b71112a66c8f481eeb42e88cb0c102792e4898e23d843cd423a133620c5e3d
5d1af152b6876087bae0391dacafc0066648989245843cd485d01961836baa098da1d91e32f45e9611dc1131c1f56f15904f
b33d047b2f0b9eb181eeae7f5f88749aed91a04759e88c0b3457603102d169b68740efe544675c80b9eb2741205d4af810ea
50173e6303dd5d3feb85f952c2877087ba003a1fecafc0f20b244c211f421eea42e88cb0c102f3fa9030857c44f258174267
840d160bf910f213d11591d01931060bac3d41c214f221e4a12e84ce081bac11f221c027a22b04cc89607105164220610af9
10f250174267840d5e9eca407829e143b0435df09c0ed63782308e63a4108e34f6569731dc3131c3218443c63e912e91c670
c7440cebd834464ac290b1435dc670c7c40cc5923084fc44be83103a236cf0eaa56d88976230043bd405cfe9607d35623022
f0997007017322585c816b25244c3118421eea91d01931060b5c1f23618ac110f250174267840d8ebbc14fbfaa9896c7bdbf
fbeb31cf9b7fae8f02d720e7d7f63c115b366939ee06cff4d90c618ac110f250174267840d5ebc1280f15206866087bae039
1daceffaf9d14c983230843cd485d0196183573f3947bc148021d8a12e784e07eb1bafc7bef7cff3b344ce7ec7ce66f33ffa
7ddcd44fccb7fd1a195b3669395e8f9de9b319c2148091c8b7ba103a236c70bc0c5e3275efebe3e5c7fa80797995352dcf5c
5f7bdc0bb66cd272bc0c3ed3673384290043c8433d113a2364b0c00400244c0118421eea42e88cb0c1417efbbd9633c07c94
fd0ff479dca7e5c56dbfccecb7b0afef76dd862d9bb41ce467fa6c8630c56f0879a80ba133c2060b044120618adf88e4b12e
84ce081b3cc46fc8d8274225e218ee980c438149464898e237843cd485d019618387f80d19fb44a8848ce18e89190a7cd691
30c56f0879a80ba133c2062b048f10618adf10f250174267840d1698bc8a84297e43c8435d089d113638262cf7dff9c7f2b4
e033fded313eff63eaf794fdb7e9b56d5c8e0d9b341cd3954ff4d80c5ecade48d85b5df09c0ed637e6e64fd3fb3b9fece633
dffd313f34ba2db3c95ff7dfe3f17c860538d4b249cb3137ff4c9fcd10a6ec0d210ff544e88c90415a8f31fd9efde2637968
d2ff52ef4f3ffa6ddb766cd862435a8d617a6c062f056f0876a80b9ed3c1fa6001ce773e213d97c958efcf3cf9eef5f812f6
d6b04943587ee37a6c062fa56e0876a80b9ed3c1fa68b9d5bd9f81fa712d5b8ff733d3e7d97f77907b6bd9a4252db8727d36
43a8a91b113c9485cff9607fc03ddf4bccbf3dcb5f677f6604dc5bcb262d81dbf6d9984f5237847aab4638e382cdd192e5b1
89feedff7b8d67a1b8c26c346cd290162c9b1e1b8269e486f0ee6521630da4ecf2ddca014cf33684772f0b99d180d206f1ad
7fdd1ecb949ff7f49e9bfdeecbfef9af7479c62d9bb41ccc67fa6c4c975237843ad485cfd840771536e427be94ba21dca12e
80ce07fbbb366085d852e4863087bab03917ecaec25f9b0853e4869087ba103a236c7090ef1302e68dbd5efde2f1b924a23c
c74bcc014e0d5b6c38b84ff4d80c5ecadb48d85b5df09c0ed607e7f1fe01bddde7d9cacb69f7367d1edffd0d30b76cd212ce
e4b6cf660853de8690877a227446c8e0e53fd88497f236043bd405cfe9607d7469363d9e8fc77dfdfd993facf77e3145d90b
7bcb262de9d2ccf5d90c61cadb88e4b12e84ce081ba4f0bffbfb75ffdc9733f07bbabd9fdfc7fb859bca6c2d9bb4a4f03fd7
673384296f43c8433d123a23c620dc74f6ab8a7efe7d2ffb93f62f47ff73dda72f5d8eef2d9bb484db4edb673384296f43c8
435d089d113678f5f315c44b611b821dea82e774b0bef110b5df0c7fbecbfe23fdb0deeba1fd6fd9b4e0f9fadd7edbfe49dc
b249cbf110f54c9fcd10a6b00d210f75217446d86081b5924898c236843cd485d01961830596512161cadb10f25017426784
0d5e3ddd03f152d8866087bae0391dacafc0fc1e244c611b421eea42e88cb0c102ab3d9030856d24f2ad2e84ce081bbc7af3
6ac44b611b821dea09cfe9207d05762b47c214b611c9635d089d11365861877a224ca91b421eea91d0193106d720041a2345
5fc8d8a12e63b863628643f4858c7d22d041c670c7c40cebd834460a9f90b1435dc670c7c40c87f00919fb44b6828ce18e89
190a2c6a42c2944221e4a12e84ce081bbcfaa48e78298242b0435df09c0ed65760663612a6080a210f75217446d860854402
224c2914897cab0ba133c206af4f2460c2944221e4a19e089d113238c897e5409ff56bd8af31ee4bb4fc77faacd34eb7a9d9
d0ae49bb41fdeffe9a614bf913e158a52e6cce05bbbb7c477ec24be113821dea01cfe970fac69df8f7d68d2c1793fd1e793e
c6d7badb4cbfd8bacfcbbfb60f39b66cd272dc899fe9b319c2143e21e4a12e84ce081bbc7e050213a6f009210f75217446d8
e0205f96d4ae93c7bfcf47bfae5d6f2dbee9f511376cb1e1e03ed1633378297c42b0435df09c0ed637366b5a928797dfdbef
67ded67cdeed66fec77dfa7b6230b8b165939663b3a6337d364398c227843cd485d019618383fc6feaf872d2fd7b89bbee9a
f89a7eb7659ac820a7964d5a0ef2337d364398c227843cd485d0196183d76f4dc68429fa42c8435d089d113678fd76744ca8
c11b02be9785cff9607fd7af33423e0dde10ecbd2c78c606babbfe4d02d269f046821e65a1332ed05c859df1293023056fc4
6bcf585f7685fe3dd67ae8c7d5433f87e00d19fb4498858ce18e8919c87aff8b2c3ba9cd86e62da69ebf7d930f6ed9a42559
777d364398623f843cd485d019618305662320610a1d49e45b5d089d113658e16f4e84297444c8433d113a2364b0c2398608
53ee889087ba103a236cb0c07470244cb923421eea42e88cb0c102736f9030458f0879a80ba133c2068b458f08f989400d21
7446d860b1e811213f11a82184ce081b2c163d12c9cf046a08a133c20607f9ef7effbd97970df39ce6fea5f82dafd4a77912
e4bc31d04e4e2d9bb41ce467fa6c8630458f0879a8474267c4181ca12b6b50e6b2de705ad7f22e87dc7f06971fa3db7686c3
964d5a8ed095337d364398a247843cd485d01961831542578830458f0879a80ba133c206c7bad5f9e5c26bbd6b0dfffbb67d
57c7aad563bb26edc69ad57ff7d70c5b0a1d11e6501736e782dd1558a58c84297424916f75217446d8608149954898424784
3cd413a13342060bacc947c2143a22e4a12e84ce081b2cb0e70812a6d011210f75217446d860811d769030858e0879a80ba1
33c206c783bfdbe7f7b8ddfe7671ef57b7efdf67f964f6ab8f57bfafb86dbf67d8b249cbf1e0ef4c9fcd10a6d011210f7521
7446d8e0f5cf799930858e44f258174267840dae490c34468afd90b1433d8ee18ec9305cff3e850953f2879087ba103a236c
b042d40b11a6e40f210f75217446d8e0f5ef4d99507347047c2f0b9ff3c1fe0a2c52203ecd1d11ecbd2c78c606baab10b602
749a3b92a04759e88c0b3477fd2418a4d3dc1181decb89ceb8007357af4421b4143a22c8a12e7046058abb74621b92a5b811
210e75417326d85c81698c4898e246843cd485d019618363eaea6b7d89361fe5e373ff7d1ecf65abc879cfc87e5bb38363c3
161b8e89ab277a6c062f658d44ec58173ca783f55d3d451bf152d08860877ac4733a8cbe0273f29130058d0879a80ba133c2
062fdf189ff052dc886087bae0391dacaf58dc88909fc8d1104267840d56c88020c2943822e4a12e84ce081bbc3c6885f052
d849c2deea82e774b0be43d8898c7d2240248de18e89180e612732f689080f19c31d133354f8a013610a3b11f25017426784
0d5e7e4e27bc147622d8a12e784e07ebabf0234e8429f52492c7ba103a236cf0eaab36c44b9127821dea11cfe930fa0a2c2f
43c2147922e4a12e84ce081b2c1080898429f244c8435d089d1136381eb7fcfa95dd7b7d1ab24c1af87c96d3d2badff863da
26d571cb262dc7e396337d364398524f843cd485d019618317afbc61bc147922d8a12e784e07ebbb7efd0913a6d413210f75
217446d8e07833f97df613ceba20ecd9ef5c9feff925d2b21fd3fbfefbdef6705f6cd9a4e5783379a6cf660853ea4922dfea
42e88cb0c1026fa39130a59e0879a8274267840c169881808429f544c8435d089d1136b8dc2ae018297744c60e7519c31d13
331448c040c2943b12c9635d089d113658e0f3868429f544c8433d123a23c66081bf3912a6d413210f75217446d86081b32b
12a6cc15210f75217446d860854f3b11a6cc15210f75217446d86081397d48983257843cd485d0196183631ee7df6bf3f9c0
fa5dcce335f58fe3fc8ffe1b3c4f6c7f6ddf736cd9a4e598c779a6cf660853e68a9087ba103a236c70cc57fef6ebd9e712e7
f7f93e1efd57e0fd5db6f0ecf7389feff3f7d8e62060cb262dc77ce5337d364398625712f956174267840d1698a38e842976
45c8433d113a2364b042e00c11a6e415210f75217446d8e0e5813384979257043bd405cfe9607d05965d21a126af44f05016
3ee783fd155862487c92bc22d45b35c219176caec01a6260d3f01541decb02c726c81a10cfcffddfcfe939ffdaf44bcd75d2
c49388b7964d5a02b1edb3319be6af08f25e16386302bd415241bfaaea5758b7e527f6d5af286f8ff9aa12a0f7964d5a4252
81edb3315dca5f11ea50173e6303dd51d04e3fbfdee6dff7f910a7fe1bf3997306087b6bd9a42545edb83e9be14bf92bc21d
ea02e87cb0bfcb2386082f45b00876a80b9ed3c1fa96073e38460a4191b1435dc670c7c40cf41deb5f8557ffc77bf9607cfb
e1ceb736f81d1b2d9bb4a4ef98ebb319c2948392c8b7ba103a236c90d240debf699a17abcc47f9ed27c27e9933e102f5ad65
93969407e2fa6c8630e5a00879a8274267840c56c86a23c2148522e4a12e84ce081ba46ba5e9f5fbbcd778a6f7ebf3eea786
275e228e864d1ad29592e9b119bc948312b1635df09c0ed6077704b7fb3c63be5fc5adbf80fd42677e6c43b7045bcb262de1
9ec0f6d90c61ca4111f2508f84ce88315860753a12a61c14210f75217446d8e0a52b95912d85a00873a80b9b73c1ee0a3ccf
45c2148222e4a12e84ce081b2cb0a20c0953088a9087ba103a236cb0c0521b244c2128421eea42e88cb0c131f5e5759fe61d
51e743be4debaefecb82fa293df0e0964d5a8ea92f67fa6c8630e5a00879a80ba133c2060b2cc640c2948322e4a12e84ce08
1b2c968392c8ff19f02184ce081bbc7a661fe2a51c14c10ef584e77490be62392891fc4cba87103a236c704da6a031521289
8c1dea710c774c86e1904422639f48f79031dc31314381c9bb48987250843cd485d0196183053eeb48985258843cd485d019
6183579fdc102f05c00876a80b9ed3c1fa2ae48110610a8011f250174267840d16f81d47c2940123e4a12e84ce081b2c70ed
8684290346c8435d089d1136786d1e08b1a5f497c4bcd585cdb96077572f8f47bc14fd22d8a19ef09c0ed2376e4a977ba5e5
0dcbf7b76ccbf1f8feed04f47e2c3b1f6d37a5d4b249cb71537aa6cf660853004c3862a90ba133c206afdfee830953068c90
877a2074469cc1ebb7b661c2940123e4a12e84ce081bbc7e1329264c1930421eea42e88cb0c1eb370e63c2140323e4a12e84
ce081b1c8f5a9faf7e0e5ad63eced71df3d60dcb1e0efda6ead5cfc4bffd7b8e2d9bb41c8f5acff4d90c618a8111f2501742
67840d5e3f5d9a09530c8c9087ba103a236cf0faa9b34c986260843cd485d0196183d76f81ca841a0323e07b59f89c0ff677
fd9461e4d31818c1decb82676ca0bb0a7f6ca0d31898043dca42675ca0397833fe9d8ff4b95c75f4b3606ff77a7c7112e168
d8a421bc19773da60c98ed6853064cbcfc0cf57563f27e765cea7b3fae1efb21ecfbd42f745ecb75e6fbd3ffc3fe13f8c179
755bcb262d09dcf5d90c614aa011f2bdae84ce081bac90cf4084298146c8f7ba123a236cb0c08459244c0934897cd495d019
61836be2118d91326064ecbd9ec770c7440c574fbd41bc140023d87b5df19c0ed657e16b4684290046c8f7ba123a236cb0c2
a9950853008c90ef75257446d820ad44983ed3e3f35d1e41cf8fdddfb77e5ec6bff9d6b2494b5a89e0fa6c863005c008f95e
574267840d160b8011f27fc79a28a133c2068b05c044f2f51fff67ac89123a236cb0c0422b244c013042bed785d0193106c7
e2baf92dc7e7bddcbcbc5fbff7fb332fbd9c7f8d6efd7a73da5fdb71cb262dc7e2ba337d3643980260847caf2ba133c206c7
82ca7ede7d2c5b4df7439aeef3677359fbf859efdfefdb7b3b6ad8a4e1584e79a2c766f052fa8b60ef75c5733a585f8175a4
48983260847caf2ba133c206afcdbd21b694fe9298475dd99c0b7677fd7300264ce92f42bed733a1334206af7fd0c78429fd
45c8f7ba123a236cf0fa87bb4c98d25f847caf2ba133c2060bcc9d46c294fe22e47b5d099d113658607729244ce92f42bed7
95d0196183d7bfbe62c294fe12c9435d099d113678fd2b4b264cd93342bed785d0193106d7dc1b1a23a5bfc8d87b5dc770c7
c40c05562920614a7f11f2bdae84ce081bbc7e7200136afa8b806f65e5733ed85f85d41be0d3f417c1deca8a676ca0bbebe7
c0209da6bf24e8bfb2d21917686e40f78fe2fdb3fee27dbfafd7e3779b968fe6bc9ce43edd9edb271c5b366939a0cff4d998
4ed35f047a2b673ae302cc15d88395e852008c50ef75e53336d0dd98d5f798e76aac5bfaf71ba8d76b7a2e932fbfeb5e40cf
fbf601c7964d5a8e797d67fa6c862fc5c008f75e5740e783fd0df27e0bf39a965f9bdfbeb2793e11ad679ed736531b5b3669
39c8cff4d90c618a8111f2bdae84ce081b2c3063190953124c240f75257446d8e0205f6fd9d7b9f4cbe4f2cf7b594f70bbcd
ffd56d9fcc892d9bb41ce467fa6c863085c108f95e174267c4182cb014050953188c90ef75257446d8e020ef85effa17797d
fe2e24974fe9e3f378dcf6bc84ffbb591bccffecad19b0140323c07b5dc19c0816572c0646c8ff1d6ea284ce081bac908e41
84290646c8f7ba123a236cb0520c4cc2fe57b48ae2391dacaf632f3fb2d3ed6f5dc8f285bcdf3f53ff4f972bb5ffebffdd11
e75f31ff5f3703903266046caf6700074c822eff7b125e8ab711ecbdae784e07ebabf00526c2146f23e47b5d099d113678e1
af1481a5601b01deeb0ae644b0b8022b44913005db44f250574267840d16b80845c2946d23e47b5d089d11637090df9ef7f7
635aff24eb7de1b43cfbf9f6efc5e331ed0fa9b165939683fc4c9fcd10a66c1b21dfeb4ae88cb0c1021b7c2061cab611f2bd
ae84ce081b2cb07e0e0953b68d90ef75257446d8608107a64898e26d847caf2ba133c206c72b99f7e7de6f50d6673eb7f7eb
37cf13ffdf3c2b62eabf07f7f7fe50095b3669395ec99ce9b319c2146f23e47b5d099d1136385e402eeb75d7f4a1f96df8bc
c7d0f20a635a3e96af3d381b5b3669395e409ee9b319c2146f93c8475d099d11365860a20112a6781b21dfeb99d019218383
7c4ea578ac4fb36fefdbe3fb7d2e4bd67bedb96ea3be9153cb262d07f9993e9b214cf13642bed795d019618305a6d520610a
d711f2bdae84ce081b5c837d688c146f13c70e751dc31d133314f8bc21618ab711f2bd2e84ce88315821ea840853bc8d90ef
75257446d86081c94c4898e26d847caf2ba133c206976f1a8e91026664ecbdae63b86362860a9f37224c013342bed795d019
618315ce3144980266847caf2ba133c206af5f27cb84296026918fba123a236cb0c09468244c013342bed733a13342062b44
eb10610a9811f2bdae84ce081b1ce4f7f58ef27feb4ccef93ded6d39ca75f2d5734b1be6964d5a0ef2337d3643983266847c
af2ba133c20661954fbfacfc3ebfefc7fc9dec5fd5e7f4b84f5fda397d6fd9a425acf2b17d3643a81933117c2f2b9ff3c1fe
6803f1bf1ffd7593fbfbedfeea5f125cddb4b56cd2923610777d36e6938c19a11e5581332ed85c81e58bc0a6193382bc9515
8e4d90b502eba2894d336604792b2b9c3181de00fab55d52dce65dc6a77e91f9fbd08ac5bd659396006dfb6c4c973266847a
af2b9fb181ee2a2428115fca9811eebdae80ce07fb03f27efff47ace9b07ce47b96e21f8a433d9deb0c586c0ed7a6c062f65
cc08f65e573ca783f51d326664ec7f27a7e818ee9898a1c2878d0853c64c221f75257446d82090f7f3eeedd76fd597a37ccd
5b18f6cb6a4c5ad95a366909e4b6cf660853c68c90eff54ce88c9041daf7605af7d05fd6ab4fe95e905b366949fb1eb83e9b
214c193342bed795d019618315ae958830c5cc44f250574267840d1658098e84296646c8f7ba103a23c6e0b80b5ed7a02cb7
e79f79bef6fa6e79fafdfd75b647dad4b049c3710f7ca2c766f052c68c60ef75c5733a585f81d73748989266847caf2ba133
c206c7cbcae7e3fd78ff961518afbf19eacb939a6f9e8e812d9bb41c2f2bcff4d90c614a9a11f2bdae84ce081b1cafe697d7
2bebfcf4796f82e7b43ca8e937cdeb5cee6d023c356cd270bc983fd1633378296646b0f7bae2391dac6f607f6ff7c7735dd6
f85be6a9bfd6251aaff7e33e4f0adb3eead8b249cb017ea6cf660853cc8c90ef75257446d8e0a5db56235b0a9811e6bdae6c
ce05bb2b30e7080953c08c90ef75257446d8e0207fadd7cff3513e3ef7dfe7f15cf660fdae91b0fb1c526cd862c3c17da2c7
66f052ba4cc21e75c5733a585f81e97548980266847caf674267840c5e1d3d8178295d266287bae2391dacef902e2363ff3b
b145c670c764180ee93232f6bf3353740c774cccb08e4d63a47c17197bafeb18ee9898a1c0970c0953be8b90ef75257446d8
6081c5734898225e847caf2ba133c2060bfc9222618a7811f2bdae84ce081b1cd74dd3f4e8b770cbdfe77e9f7eeff7edbe4c
44fcdc6fafdffb376d9f766cd9a4e5b8763ad367338429e245c8f7ba123a236cf0f2d00fc24bf92e82bdd715cfe9607d056e
119030a5bc24f251574267840d0ef2659ddf3ad9f4fb7c7c960772f33ffaadfb6b3e516de0d4b0c58683fb448fcde0a59417
c1deeb19cfe9207d0536f540c294f2128e38d695d0196183d7cfd566c294f222e47b3d123a23ce6085dc0b224c292f42bed7
95d0196183d73fec63c294f222e47b5d099d113678e5e687cc96f25d8479af2b9b73c1eeae9fd7c58429df45c8f7ba123a23
6cf0e2bd4d192f85bb08f65e573ca783f58d495d53bf967cbe963fce7bcdb15fb69d99fad9b75f783c7fdb0b3b6cd9a4e598
d475a6cf660853b88b90ef75257446d8e0f5d3d99850c35d047c2b2b9ff3c1feae7f5d877c1aee22d85b59f18c0d74373fff
c0fe355e250dfb57d6fecdd1e0b1d3c6e8d3efd92f70bfcb37a1df39f7efe487a656ed0d5b6c48dba29b1e355e653f5a8d57
91ebbe589f373c9f5f372ff5d08fab877e2aec074f841aafa2e4a12e84ce081ba40c807bbf747f4fbf6552d037addde5964d
5a520680ebb319428d5751f250174267840d16983889841aaf92c9b7ba103a236c30c7abe8d8ffce0dc963b8632286756c1a
43334e74ec509731dc31314385cf3a116ac6899287ba103a236cb042d207116ac6899287ba103a236c70fdbcd1189a32a263
87ba8ce18e89192eff31253c8d1851ec50173ca783f5d1e283cfedf1bd4dd33ae9f46f9f659a46b9b76cd292161fb83e9b21
d48811218f75217446d8e0207fadeb22962703db9e05f3cccf65ebd3104e470d5b6c38b84ff4d80c9ee68b2876a8473ca7c3
e8bbfec68409355f44c9435d089d113678fdcd28136ac4889287ba103a236c703c7d984319ee8f65ded7e7d9fddce70d5fe7
cb9ed77253b32dc2e0964d5a8ee70f67fa6c86502346943cd485d0196183d73f7162420d1ac9e45b5d089d11367875b8ca91
4d23469439d4139b7341ee0aac1546428d1851f250174267840d1658958f841a31a2e4a12e84ce081bbcfedd09136ac48892
87ba103a236cb056c48892ff3b3843099d1136582b6244c84f046728a133c206af7f47ca841a31a2e4a11e099d1163b042b8
0a116ac0899287ba103a236cb056c08992ff3bb843099d113658602e04124ac08982ef65e1733ed8df1aa7032348c4880ebc
976500733c78f457afc22034c917c9c4a32c6846046a1bc4affed1fbcb3cf9ae6bff5eebee76cbde76f304f3814c2d9bb41c
cc67fa6c4c27f9220abd97139d7101e60a4c29233acd1751ea50173e6303dd5d3d8110e1345c44a1435de89c0c96572b5c44
c9ff1d99a184ce081b2c303f1a09355c44c8635d089d113658604e3c126ab88892877a2474468cc10a7b7913a1868b2879a8
0ba133c20607f9a3ff7f1ecb263ffd8a7ed982f9392d47f9eb27a9fe83b47ddab165939683fc4c9fcd106aca889287ba103a
236c7090ef8f5de72fe13a9377fe7e4ef36491fb77fb0da7864d1a0eee133d3683a711238a1dea82e774b0be0a090544a811
234a1eea42e88cb0c1cb2329084f234632f656173ca783f5fda711230af6efec8c0ce08049d0e57f4fc2d38811c50e75c173
3a585f852f30116ac4889287ba103a236cb044c48802ff3b3843c19c081657e1b28408356244c8635d089d113678f5d26fc4
d37c11c50ef588e774187de3aeebb7eed3b5fc3da6eff4f8fc6db4fb770edeb9b1659396e3aeeb4c9fcd106abe889287ba10
3a236cf0eadb6cc4d37011c50e75c1733a585f81274948a8e1224a1eea42e88cb0c1f190f679eb5fbb65e16cbf477cf40b98
cf6b7d5bda7f6f6ecb5698e3212db56cd2723ca43dd36733841a2ea2e4a12e84ce081b2cf0020e09355c44c9435d089d1136
58603b7224d470914cbed585d01961830536e546420d1751f2504f84ce0819bc744209b269ac883287bab03917ec6e30dffb
2dff779923dcaf2cfb8dc4ebbddc3f7d6efdca72ea3fbbdbcf38b66cd272709fe9b319428d1551f250174267840d56f89c13
a1869a0879ac0ba133c206d798071a43634574ec508f63b863320c05263021a1c68a2879a80ba133c2060b4c6042428d1551
f250174267840de658111dfbdf61193a863b266658c7a63134d843c70e7519c31d1333ac63d3181aada16387ba8ce18e8919
0a5cbf20a1466b64f2ad2e84ce081bac708621428dd650f2504f84ce0819bcf4fa05d934544399435dd89c0b7677f516b288
a7891a8a1dea82e774b0be0233fd91501235043c9485cff9607fd72fbe46be98a8a1d45b35c219176cae40780ab049a28622
ef6581631364adc256e0c026891a8abc9705ce98406f055683129d266a2875a80b9fb181ee68d9f33cfde2b94c5c7c7fe629
baafc717c35346c3260d69d1b3e9b119388dd350e850173a2783e51558ed8d849aa8a1e4a12e84ce081bcc891a3af6bf7322
740c774ccc50e13b46849aa891c9b7ba103a236cb0c2299508355143c9433d113a2364b0c2cf27116aa2869287ba103a236c
b0c03a5824d4440d218f75217446d86081359148a8891a4a1eea91d01931060bdcfc22a1866a2879a80ba133c2060bbcaa42
420dd550f250174267840d16584382841aaaa1e4a12e84ce081b1cafa4fb21de3febdfe7fb7dbd1efde67d39e467ff46defb
5deb468e2d9bb41cafa4cff4d90ca1e66a2879a80ba133c206af9e8380781aaaa1d8a12e784e07eb2b30d51f09355a43c943
5d089d1136386619dddfaffec95c1201fad5f53c1de93ba64ddcfbb961fb31e7964d5a8e794667fa6c8650a335943cd485d0
19618357cfaf423c8dd6c8d85b5df09c0ed657608503126ab48692877a227446c8e07287886368be858c1deb32863b266618
d6df8b966576e9e3f7e93715ef6517a97eb07fdf8acd3ab56cd272583fd36733849aaea1e4a11e099d116330a76be8d8ff4e
acd031dc3131c3d55f71c4d3600fc50e75c1733a585f85900722d4600f250f75217446d86081df7124d4600f250f75217446
d8e0207fdda7f94a7239eb4eeb8aa3e528a7f4088a5b366939c8cff4d90ca1067b2879a80ba133c206c7c5faeff17abcd7bb
887ecffc7bceefd6e67f2c53181ed3f6ea865b3669392ed6cff4d90ca1067b2879a80ba133c2060bcc9c4542cdf650f25017
4267840d8e5bd3e561d1f2d3f7fdcc1bdece5b212cb791dbe671e3d6945a3669396e4dcff4d90ca1667b64f2ad2e84ce081b
2cb09907126abc8792877a227446c8e07810b3e411ae1924731ec1fc715c26c44d8fd7effe7ced8befb1659396e341cc993e
9b21d4788f78c452174267840d5e3fe7860935de43c9433d103a23cee0f5b3489950e33d943cd485d019618397c71e109e66
7b2876a80b9ed3c1faae7fb7c0849af0a1e4a12e84ce081b84f749cff98fd3af3497a37ccc5fd6dffb8e6fd2b6964d5ac2fb
24db6733849af0a1e4a12e84ce081bbcfe4d1a136ac8879287ba103a236cf0fab7a74ca8211f4a1eea42e88cb0c1eb77ea66
4209f950f0bd2c7cce07fbbb785a0ac249c28732ef6561332a505c4af8c8c39a288f5196fecdd1e0b15f7f3ae5708e14f211
affe627dde69fd797faef5d08fab877e8a857c08f989e80a217446d8209d4eefdfdbd43f11cb44a5dffbf15c2e78f1743a5a
366949a753d767338429e443c8435d089d113678fde9940953c84722dfea42e88cb0c10253af9030458c0879a8274267840c
1e224664ec13c11932863b266658c7a63152c8878c1dea32863b2666a8f05927c214f221e4a12e84ce081bac707e23c21431
22e4a12e84ce081bacf09b4684296544c8435d089d113678fdcd1113a69491481eeb42e88cb0c14a2923827d223b23e2391d
46dfc0fef5cbd9fe83331fe47c0ebcf76fe27cc4d3ebb3ae9dd92f5ca965939603fc4c9fcd10a69411210f75217446d86081
b54748985246843cd485d0196183173fec63bc143122d8a12e784e07eb2b309d160953c44822dfea42e88cb0c1321123c27c
223823b13917e4aec01462244c1123421eea42e88cb0c1022bf49130458c0879a80ba133c2060bbcb343c2143122e4a12e84
ce081bac10b44184296244c8435d089d113638c89f8ff7a3ffd62e47397f12a7dfb2d55effe949130cb165939683fc4c9fcd
10a68891481eeb42e88cb0c102efe69130458c0879a8474267c418ac10b44184296244c8435d089d113658600f37244c1123
421eea42e88cb0c102737090502346047c2f0b9ff3c1fe72c4880cfccf080d19c01c0f1efd98ecb5ee00b5d8981fd33c6e7f
b3b41ed367b97ddd2e9fb065939663b2d7993e1bd369ca48821e65a1332ed05c81853844a7292302bd97139d7101e66aa58c
08f489000de13336d0ddd54b8f102ea58c0874a80b9d93c1f20676efabffcc2db918f77bbfa27cdfee4ba0d8a77f3e7eeff9
67607053cb262d07f8993e9b214c2923421eea42e88cb0c1024b8f9030a58c44f258174267840d16580d8284296544c8433d
123a23c6e020fffe1efd877dc987e81fd0e7fc766df947bfb5797f5ff31ec1839c5a366939c8cff4d90c614a1911f2501742
67840d164b1911f213f11942e88cb0c14a2923827d223b43f09c0ed637b0a77e197fbf2febe0fa15fdf3d9bf9ccb513e6ff7
dbb35f6e6dd72ed8b249cb017ea6cf660853ca889087ba103a236cb0423c0311a6a09144bed585d0196183ff6dd088809d48
d048000e9804550a1a11ec13f11982e774b0becb4f5d84975246043bd405cfe9607d157eab8830c58d0879a80ba133c2068b
c58d44f233391a42e88cb0c102997f48981247843cd423a133620c16b80343c2943822e4a12e84ce081b1ce4b7d76759b23e
df267c3f9fcfb3ff9c2e77c8f3c3f67b6fbe9153cb262d07f9993e9b214ca123421eea42e88cb0c1024fd39030858e0879a8
0ba133c2060bbca440c2143a22e4a12e84ce081b1c2f239faffbedb1fcdcce27a679c9feb276fff39e5e53ffc8ee7f736cd9
a4e5781979a6cf660853e8889087ba103a236cb0c0ab77244ca123897cab0ba133c206af8dde20b6143722cca19ed89c0b72
37a6d5cc5330ee8fe55dd9e7d9af31eef37af57979c86b59a671dfceead8b249cb31ade64c9fcd10a6d011210f75217446d8
e0d59b76235e4a1c11ec50173ca783f5159847858429712492c7ba103a236cf0faf5c14c98f24e843cd423a133620c0ef2bf
3d9796d36fbfe49afa75d67294f343efc77d9e7836c8a965939683fc4c9fcd10a6bc13210f75217446d8e0f5abe19930e59d
0879a80ba133c2060bcc2343c294b622e4a12e84ce081b2c70e9868429eb45c8435d089d11365860061d12a6a419210f7521
7446d8e09a724363a4ac9734f6569731dc31314385cf1b11a6ac17210ff544e88c90c10ae718224c892f421eea42e88cb0c1
0abf2b4498425f843cd485d01961830512eb9050435f2278280b9ff3c1fe0a5c37139f84be08f5568d70c6059bbb7aa50980
69e28bf0ee6521630da4acc0922262d3c41741decb02674ca0b7022bc8882e25be0875a80b9fb181eeae5e2f887029f145a0
435de89c0c96b7e69ed018297345c60e7519c31d13335448cd22c294f822e4a12e84ce081bacf01d23c294f892c8b7ba103a
236cb0c229950853e28b90877a227446c86081e7aa4898125f843cd485d0196183973e4547b694f51299635dd89c0b7657e0
76000953d68b90877a2474468cc1024bd7903065bd0879a80ba133c2062b249e1061ca7a11f250174267840d16d84a180953
d68b9087ba103a236cb0c01c08244c592f421eea42e88cb0c1025b8623618a7b11f250174267840d16d8201f0953dc8b9087
ba103a236cb0c09a0b244c712f421eea42e88cb0c1abb320102fc5bd24ecad2e784e07eb5b434f688c14b82263877a1ac31d
1331149852888429ee2592c7ba103a236c700d3da13152e08a8c1dea710c774c86a1c0571c0953dc8b9087ba103a236c70fc
a0fd6efd0a635ae653dfa7eff4f8fc5d6bffcdaadfbfe3d8b249cbf18376a6cf660853e28b9087ba103a236c70902f8b2396
0978bf7981c8fdbb9c88a6cffaaa6fbb6c85764dda0dea7ff7d70c5bca7a11e6501736e782dd0de6c7fc4a6d5dfe727baebf
b8cbcbb5bffd21f649ead8b249cbc17da6cf660853d68b9087ba103a236cb0c00261244c592f421eea42e88cb0c141be2c7f
5adf5c7f9f8f7e165ce7d57f9f6329c500a7862d361cdc277a6c062f05bd0876a80b9ed3c1fa0a6c32818429e845c8435d08
9d113658603319244c412f897cab0ba133c206c71484f5cfb34e207f3efa2fcd7d5ade44f4cb9ce7ebf5fa6ee4d8b249cb31
05e14c9fcd10a6a017210ff544e88c90c102b395913005bd842396ba103a236cf0e2f9ab8c97525e043bd4039ed3e1f48d57
0af399eff35e0ef2fdfaf5fff73c1b6a7efc7fbbdd3e53ffff6faf14a8659396e395c2993e9b214c292f421eea42e88cb0c1
eb77a967c214f422e4a12e84ce081bac107d418429e845c8435d089d113678fd8ee54c98825e843cd485d019618340fe99d6
27c2cb2edb533ab173cb262d81dcf6d90c610a7a11f250174267840d0279bf9f793f1fdf65c6e5fbf5fd7d9eafcffd43e45b
cb262d81dcf6d90c610a7a11f250174267840d02f9fbfe9c5efd2f34df51bd9f9fd7ed3b7df14fbe356cb12170bb1e9bc1d3
9417a1decb02e764b03c80eef7719fe54dc87c88f3d3f0e76fdffa815b366909d8b6cfc67c1af422d87b59f08c0d7407d0cf
a95f5e3eeecba3fff7637d4cfaa2dde9f7964d5a02b4edb3319dc6cc24e851163ae302cd5d9f57c6f1302966265e7bc6fa9c
33f07bdcd67ae8c7d5433fd75fba30618a9911f250174267840d56d8929f0853cc8c9087ba103a236cb0420c0311a6989944
bed585d019618315a2378830c5cc0879a8274267840c56885b21c2147223e4a12e84ce081b3c84dcc8d827a25b640c774ccc
b08e4d63a49819193bd4650c774ccc50e1b34e84296646c8435d089d113658e1fc4684296646c8435d089d113658e1378d08
53cc4c248f75217446d86081eb18244c4933421eea91d01931060bcc6146c2943423e4a12e84ce081bbcfe812313a6a41921
0f75217446d8608139cc4898c266843cd485d019618305167b21610a9b49e45b5d089d11365860791f12a6c819210ff544e8
8c90c10273b8913045ce0879a80ba133c2060bbc3d44c2143923e4a12e84ce081b1ce4dfe7f3fe5ab7777ff6d3eef33d6f3f
b5ac4e7cf78b90db3eab155b366939c8cff4d90c618a9c11f250174267840d8e77e5efcffd79ffaef3556efd37e8b34ed2ea
3f44d3adff12ef1362b0659396e35df9993e9b214c9133421eea42e88cb0c102f123489822672279ac0ba133c2060bcc0c41
c2143923e4a11e099d1163b0c05a0d244c9133421eea42e88cb0c10ac12b44982267843cd485d01961836bf40a8da1a12f32
f45e9611dc11314181fd22894f2367047b2f0b9eb181ee0a2c0a223a8d9c49d0a32c74c6059aab153923d0ff4c534974c605
982b30a394e852e48c5087baf0191be8ae52e48c409f0852113a2783e55d3a5b1cd952d88c3087bab03917ecaec0aa08244c
6133913cd685d0196183c5c26684fc44844a2474468cc1626133427e224245089d113638b6b1efffea1733cbdfe7f598e61d
169629eebf65f3f9c77bbf2dc0964d5a8e6deccff4d90c610a9b11f250174267840d0ef2e7f7d53f9dcb417efeeed4d6efe7
a77f45f7e091ffbb591bccffecad19b0143323c0a12e604e048b2b163323e427c25384d0196183c5626612f93fd35584d019
6183ff6dcc8c809dc84f49000e9804558a9911ec13f92982e774b0be4a3133827d223c45f09c0ed6572c6646c84f84a708a1
33c206c795c9fea1ec3727afefed37ad57cfaff5deffb55d9840c3260dc775c9891e9bc14b1933113bd605cfe9607d051684
2361ca9811f2508f84ce8831582c6346c84f24a708a133c206c7a385fe5779bcd73bfff98af139af425a1e037cd70fe796a8
842d9bb41c8f16cef4d90c61ca9811f250174267840d0ef2650bf565f793cf3c19e0392d6f8afbd969fdc46e5f726ad8a4e1
e03ed1633378296046b0435df09c0ed657e0351412a68019210f75217446d8608197ae48980266843cd485d019618305d208
903005cc24f2ad2e84ce081b2c30ad040953cc8c90877a227446c8e0d55b09235eca9811ec50173ca783f52defa0708c14f4
226387ba8ce18e89192ecfb721bc94f212b1635df09c0ed6572ce545c84f6497444267c4182c30750b0953ca8b9087ba103a
236c704cddeabf359feff26cbd7ffbdfeb19e07fcb7cda7e7fd3ef6eb6bf39b66cd2724cdd3ad367338429e545c8435d089d
113658e02a020953ca8b9087ba103a236cb0c0d42d244c292f421eea42e88cb0c1025bae22614a7911f250174267840d16d8
e0050953c64c22dfea42e88cb0c1916f731c23a5bcc8d8a19ec670c7440c153e6f4498525e843cd485d0196183ab751a23e5
acc8d8a12e63b86362860aa94244a8392b113c9485cff9607f053285884f7256847aab4638e382cd1598790e6c1ab522c87b
59e0d804592bf00880d8346a4590f7b2c01913e8adc0f539d1a5a815a10e75e13336d05d811b32e44b692bc21dea02e87cb0
bf43da8a8c7d224344c670c7c40cd7a630105bca5911e6501736e782dd5538ab1061ca5949e45b5d089d11365860173c244c
392b421eea89d0192183152e178830e5ac0879a80ba133c2060bdc0023614a5b89e4b12e84ce081b1c3724dffba75fc42c8f
66deeffbedde3f9aeb99a97f349fd36d9ff9812d9bb41c372467fa6c8630a5ad0879a8474267c4182cb09a030953da8a9087
ba103a236cb0c0f202244c692b421eea42e88cb0c102efc49130a5ad0879a80ba133c2062b648e10614a5b11f25017426784
0d0ef2cff7fd9a96f584bf7e253f6f8cf95cdfe4afdfc8d7361d005b366939c8cff4d90c614a5b11f250174267840d164b5b
11f213192242e88cb0c1028b0e9030a5ad0879a80ba133c2062ba5ad24ec7f2698089ed3c1fa2afcc1893065bd0879a82742
67840cae393334464a5b8963c7ba8ce18e89192ae47e1061ca7a11f2508f84ce883158e047050953d68b9087ba103a236cf0
dac413624b292fc21ceac2e65cb0bb0251754898b25e843cd485d0196183e38279b98ebb2d6f397ecbedcce3fbf7a2e9fd78
bc7efb5b426cd9a4e5b8603ed367338429eb45c8435d089d113678f52452c44b412f821dea82e774b0be023b7a2061ca7a11
f250174267840d1678358b8429eb45c8435d089d11365860e20b12a6ac9744bed585d0196183055e232161ca7a11f2504f84
ce0819bc7a4623e2a5a09770b852173ca783f55dbf6d3813a6ac17210ff540e88c3883b4d1e8e7f6f8dea6e590a7c7df8909
3797dd5a3669491b8dba3e9b214c592f421eea42e88cb0c1ebb7d565c294f522e4a12e84ce081bac90fb418429eb45c8435d
089d11367879fa05e1a5a017c10e75c1733a585f852808224c412f421eea42e88cb041207ff4bb99e7fc1c613ecaf569c213
833fb6862d36046ed763337829e545b0435df09c0ed607d8f3aad7776f399f87fa11ae8fe79ec4bdb56cd212c06d9fcd106a
d08b80ef65e1733ed81f70dffa3d453fb2654a56ff563ea77eed31e1e6f85bcb262d81dbf6d9984f835e047b2f0b9eb181ee
08fab6ee85311fdfd80005f3ab46c3260d09d9f4d8184d535e12f1280b9a1181da0aec178de92c29e5255e78c6fa1c31f07e
7cd67ae8c7d5433fe36ffd5a376e9f8ff2b9bd3c9e4fc1cb0e75df3db78a1ab6d870fcad4ff4d80c5e8a7811ec50173ca783
f515d8141e0953c48b9087ba103a236c902210faa9a7df3c2ffb1bf55b9977ffd9fb5050d9deb0c5861480607a6c062fe5bb
24ecad2e784e07ebbb3e8e900953be8b90877a227446c8e0f5b7254c98f25d843cd485d0196183877c17197bfec73f325364
0c774ccc502155870853ba8c9087ba103a236cb0c2e78d0853ba8c9087ba103a236cb0c239860853ba8c9087ba103a236cf0
f21f15c24bd132113bd605cfe9607d05ae22903045cb0879a8474267c4182c162d23e427025384d0196183e356617adfeecf
65bff4cf3becc235f5ebfae7747ffeb6473ed8b249cb71ab70a6cf660853b48c9087ba103a236cf0ea47f988977265043bd4
05cfe9607d05d6c22061ca9549e45b5d089d11365860850212a65c19210ff544e88c90c1022b149030e5ca0879a80ba133c2
06c73bda6536cc7d9d52fd5827552faf18a765c6c46bdf281c5b366939ded19ee9b319c2942b23e4a12e84ce081b2cb0501d
0953ae8c9087ba103a236cb040c60812a65c19210f75217446d860818c11244cb932913cd685d0196183574f3b42bc142a23
d8a11ef19c0ea36f602ffb1eafa7a27e2efa2c6f1397f3d2d843756053c3161b0eec133d3683971265043bd405cfe9607d6b
ae0a8d91325d64ec509731dc3131c321d345c6fe675c898ce08e88092a247d009f66ba08f65e163c6303ddadbaa17f4d5549
c38eb2f46f8e068fbdc0044ea2d3541581decb89ceb80073b5525504fa446088f0191be8aec03465e44bc12ac21dea02e87c
b0bf02abdb9030c5ab0879a80ba133c2060b2cc140c214af12c9635d089d113658604f6b244cf12a421eea91d019310607f9
ba8dc2bae7f89207f359a3326ef98e085b366939c8cff4d90c618a5711f250174267840d168b5711f213e92142e88cb0c11a
f12a027c223444c09c0816572c5e45c84f848608a133c2068bc5ab24f27fa68a08a133c206ffdb7815013b911b92001c3009
5a2efe718c9471226387ba8ce18e89192a659c08f689e40ec1733a585fb18c13213f11de2184ce081bac947112b1cf247708
9ed3c1fa0a5c0f2261ca3811f2508f84ce8831582ce344c84f247708a133c2060bece7828429e344c8435d089d113678f573
62c44b1927821dea82e774b0be621927427e22b943089d11365860912612a68c13210f75217446d86081ad089130659c24f2
ad2e84ce081b2c967122e427c23b12a133420607f93c05e0be2c46f87c9ef77e61399f77fb517e5fcbfa88fb76a38f2d9bb4
1ce467fa6c8630c59c0879a80ba133c2062f4f7721bc94b022d8a12e784e07ebabf00727c214b212c9635d089d113658e04b
8e84296445c8433d123a23c66081d5f748984256843cd485d0196183c5425684fc447488103a236cb0c256fc44984256843c
d485d019618305f61741c214b222e4a12e84ce081b2c70d18e84296445c8435d089d11365860371d244c212b897cab0ba133
c2060bcc5d43c214f122e4a19e089d11325860ee1a12a68817210f75217446d86085bf3911a68019210f75217446d86085ef
39116ac04c040f65e1733ed85f81ddc3894f0266847aab4638e382cd15988e0c6c1a3023c87b59e0d804595ba337a0778d78
9141f7b2746f8e058fbcc0f512d1a58817a10e75e13336d05d852802e24b112fc21dea02e87cb0bf43c48b8c7d22be44c670
c7c40cebd834468a5a91b1435dc670c7c40c15ce284498a25612f956174267840d16b83140c214b522e4a19e089d113258e0
52090953d48a9087ba103a236cb040ec0612a6a895481eeb42e88cb0c10ab11b4498a256843cd423a133620c16988f8d8429
6a45c8435d089d1136385ecddfbb96fb6f99bd745ba70f2c7fabf7f4f84eb7d773fbb053c3260dc78bf9133d368397725604
3bd405cfe9607d976e128e6c29614598435dd89c0b7657604b78244c092b421eea42e88cb0c10a392344981256843cd485d0
196183156227883025ac0879a80ba133c206d7d8091a23659cc8d8a12e63b8636286756c1a23058da4b1b7ba8ce18e89192a
7cd68930058d0879a8274267840c5638bf11618a3989e4b12e84ce081bbc36f282d852c08930877a64732e8cbbc1fc985fab
adb3e36fcff543b9de53ad3793fbad19b66cd272709fe9b319c2147022e4a12e84ce081b2cb08a0e0953cc899087ba103a23
6cb0c0ba5c244c3127421eea42e88cb0c1ab6753225eca3811ec50173ca783f515786d818429e644c8435d089d113658e0e1
3212a69813210f75217446d8608179a44898624e843cd485d019618305e69122618a3949e45b5d089d113678f50e61889732
4e043bd4139ed341fa0aec1e8d8429e6241cb1d485d0196183d76f80c88429e644c8433d103a23cee0f57b7d32618a3911f2
50174267840d5608fb20c2147322e4a12e84ce081ba4b08fe975fbdd3ecb5d64ff29b8ffeeb7efed45e45bcb262d29eec3f5
d90c618a3911f250174267840d52dec77c1fd97f83e6cbccf77e334d791fa365939694f7e1fa6c8630259d0879a80ba133c2
062b045f10614a3a11f250174267840d5688be20c2147622e4a12e84ce081b3c92f7bbe6efe7fbbe2db3da5f9fe7f7f6787e
b797a6dcb249cb23b9efb319420d3b11f0bd2c7cce07fb23ee7e78fd26fa359f855ff33bb07955327dd6f7964d5a12b7ebb3
319f869d08f65e163c6303dd01f4f7fdf9dde6c9a6cb013efa0dc69c5642d05bcb262d01daf6d9984ef34e12f4280b9d7181
e6c615ebbae274b96b9cd6dd479795d6d367fdb5bd6ddf6d6cd9a4e5b8623dd367ca3bd90e38e59dc42bd0587f2def22fb5f
f2ff693fae1efa29b0713512a6c813210f75217446d8e0e5492f8497f24e043bd405cfe9607d9447b81eefb2b1c274ff3ea7
c77dfad2a5cbdeb2494bca23747d364398224f12f956174267840d5e7f8bc28429f244c8433d113a2364b042f80611a6c813
210f75217446d8e021f244c69e7f01fe11e42163b8636206b27effdcbffd94b804d97dd3e678dcb2494bb2eefa6c8630459e
0879a80ba133c2062b7cde8830459e0879a80ba133c2062b9c638830459e0879a80ba133c2062b459e44ec33411e82e774b0
3efa39fd6fe37d9930459e0879a8474267c418bcfafa09f152de896087bae0391dacafc0c36d244c7927421eea42e88cb0c1
025b0520618a3c11f250174267840d8e3758f3deabaf753154f8dfb7ed8b3ade5f1ddb356937de5efdbbbf66d852d84962de
eac2e65cb0bbb1547e4e11fbdb7ff6feba3f9fb7c7f268ea333f9be99fd2cf468d2d9bb41c4be5cff4d90c610a3b11f2504f
84ce0819ac10f9418429ec44c8435d089d11365860450c12a6b013210f75217446d8608119194898c24e843cd485d0196183
05d68520610a3b11f250174267840d8ec947eb8e25cb81cd118a8fdbdf828e7573c5e7be2506b66cd2724c3f3ad367338429
ec2492c7ba103a236cb0c08a18244c7927421eea91d01931060bac8841c2147922e4a12e84ce081b3c449ec8d8278244640c
774ccc50210c8308357045c0f7b2f0391fecafc2a70df8347045b0f7b2e0191be82e07aea461ff912622fd9ba3c1632fb025
29d169e08a40efe544675c80b90273a6892e05ae0875a80b9fb181eeae5edd8770296d45a0435de89c0c965760bf61244c69
2b421eea42e88cb0c162692b91fc4c8688103a236cb0c0eeda4898d256843cd423a133620c0ef2f71287b244273c7efd76f1
fb5e673af79fc07549de464e2d9bb41ce467fa6c8630a5ad0879a80ba133c20607f9f7f7e8b74dcb0ef7fdabf99cdff32cff
b8f55ff7ef6bde9a7a9053cb262d07f9993e9b214c692b421eea42e88cb0c10a393344986257843cd485d0196183c5625784
fc449888103a236cb058ec4a22ff67da88103a236cf0bf8d5d11b013512609c00193a043ec8a8c7d224f44c670c7c40c15be
4844989257843cd485d019618317465511580a5c11e05017302782c5550a5c89d867624404cfe9607d052e09913005ae0879
a8474267c4182c70038484297045c8435d089d113658e0e90612a6c015210f75217446d86081a7f448983257843cd485d019
61830536f944c294b922e4a12e84ce081b2cb0593912a6cc15210f75217446d8e078eb3eef5dff5857eadfdeb747fff3bcd7
7940693119b76cd272bc753fd3673384297325916f75217446d8e0a5732c902da5ad0873a82736e782dc5dbd3e18f152d48a
6087bae0391dacafc04422244c692b421eea42e88cb0c1e5c608c748792771ec589731dc3131c3f882dd1f53ff655fce79df
e9db7ff3decbade3a79f156f5dd373fb39c1964d5a8e2fd9993e9b214c7927421eea91d01931060b6c40808429ef44c8435d
089d1136387e4ed66bb975eec5f76feef07c94afd7ba2269fba661cb262dc7cfc9993e9b214c7927421eea42e88cb0c10293
b89030e59d0879a80ba133c2060b4ce242c2947722e4a12e84ce081b2c30890b0953de899087ba103a236cb042f60511a6bc
9344bed585d019618357ef20857829ec44b0433de1391da46f602f0f24d753513f177d968d2196f3d2c8851ed8d4b0c58603
fb448fcde0a5a413c10e75c1733a58df72fd8463a4ac11193bd4650c774cccb08e4d6368da471c3a94650477444c50e01108
f149da87506fd508675cb0b9abbfd600a6511fc2bb97858c3590b21cf52183fe33c542ba37c782475e60ee2fd1a5a80fa10e
75e13336d0dd6a9b4648611b3272a8cb10ee8898e0f20f36e1a59c0fc10e75c1733a585f85531811a69411210f75217446d8
e0d5bfd48897224612f656173ca783f515d8d3140953c48890877a227446c8608149bf48982246843cd485d0196183055e52
21618a1889e4b12e84ce081bbc7a422ae2a57c11c10ef588e774187dd7c64d105b4a1611e6501736e782dd1578078d84295c
44c8435d089d11365860122e12a68811210f75217446d86085f001224c1123421eea42e88cb0c10a411b44982246843cd485
d01961836bd8058d91423e64ec509731dc3131c321e443c63e119c2163b86362860a9f75224c1123897cab0ba133c2062b04
6d10618a1811f2504f84ce0819bcf6779cd852b848648e7561732ed85d85a00d224c1123421eea91d01931060b5ca623618a
1811f250174267840d1658878a84296244c8435d089d113678f56327c44bf922821dea82e774b0be02719f48982246843cd4
85d01961831582368830458c0879a80ba133c2062b046d10618a1811f250174267840d16983c8884296244c8435d089d1136
78e9e441644be1228979ab0b9b73c1eec62cba7e3df9585ead7cfa8df37d4e8658a6fa7dbeb7fe799d53c7c7243a68d8a4e1
984377a2c766f052b88860877ac2733a485f810d8b9130858b842396ba103a236cf0faad069930858b0879a8074267c419a4
5d35eff3bb9edf67fe64f63f533f1bcf9b8a11f9d6b2494bda55d3f5d90c610a1711f250174267840d160b1711f213911942
e88cb0410a17b93fa7573fcef922fbfdec37cedfe98be05bc3161b52b488e9b119bc942c22d8a12e784e07eb2b962c22e427
f23284d0196183c5924584fc445e86103a236c9092456eeb0acbf920c7ca7cdc277a346cd29072454c8fcde0a55811c10e75
c1733a585f85780d22d4581101dfcbc2e77cb0bf0a091bc0a7b12282bd9705cfd84077047d9bafb2962763af77ffbd793f6e
3fda117c6bd8a421219b1e1ba369a648221e6541332250dbd58b7b300824058ac4abce589f63043eaffb5a0ffdb87ae8a742
940a11a64011210f75217446d8e0f8884ffd626a5efe30df45dc1ff3dcf1dbb28b73ffaffbb7f0b93d57e5964d5a8e0ff999
3e9b214c9922421eea42e88cb0c162992289fc9f491942e88cb0c10afbfd1361ca1411f2504f84ce08195c9335688c94ea21
6387ba8ce18e8919aebf1f64c2942922e4a12e84ce081b5cadd31829d543c60e7519c31d1333144bf510f213591542e88cb0
c10a9f37224ca91e421eea42e88cb0c10ae737224cc11e913cd685d0196183057ed39030057b0879a8474267c418ac14ec21
d827e22a04cfe9607dc5823d84fc445c85103a236cb0c07a7c244cc11e421eea42e88cb0c132c11e89f99f7115c2e65cb0bb
62c11e42fefe775c45227446c8e07833fcedf7adaf75fdc0b37ff99eef79bf9be56dd7bb9fa06e5b201bb76cd272bc193ed3
67338429d843c8435d089d113658602b21244cc11e421eea42e88cb0c1024bdc9030057b0879a80ba133c2060b2cb942c214
ec21e4a12e84ce081b2c106f818429d82392c7ba103a236cb0c0da23244cc11e421eea91d01931060b4c6a44c214ec21e4a1
2e84ce081bac90014084295644c8435d089d1136b8264dd0181aec2143ef6519c11d11135cbdde8be034d54398f7b2b01915
28aec2a70ce8345324418fb2d0191768eef23f33a069a08810efe584664480b60a7f66a04b8122421deac2676ca0bbcbffd0
04970245043ad485cec9607903bbdf95cf6ba797839c27aef7bbb66546f79412aab965939603fc4c9fcd10a64011210f7521
7446d86081559c489802452279ac0ba133c2060bace244c2142822e4a11e099d1163b0c05a46244c8122421eea42e88cb0c1
0a512a44980245843cd485d019618315a2548830058a0879a80ba133c2062b44a910610a1411f250174267840d56488020c2
142892c8b7ba103a236cf0bf0d1411b013211d09c00193a043a0888c7d222643c670c7c40c15f2698830058a0879a80ba133
c2066b048a08f089a40c017322585c855f0b224c9922913cd685d019618383fc75bb4df7e5a2ad1f6eff2a3c9ecbeef8df75
9af47e51880d5b6c38b84ff4d80c5e0a1411ec508f784e87d15760c93f12a64011210f75217446d8e0d577bc8897d244043b
d405cfe9607d055e0d20614a1311f250174267840d16580a8d84294d44c8435d089d113658e0b52f12a63411210f75217446
d8608157fd4898d24412f956174267840d16d8f21f0953a68890877a227446c8e0b5392ac49602458439d485cdb960776302
d3b79f719ecbfe139fefe3317d9fefefba32a25f5a7f9fbfc7f6aa1b5b36693926309de9b319c2142822e4a12e84ce081b3c
048ac4b1cfc464c818ee9898e1d24f1ab2a52811610ef5c8e65c187705b65840c2142522e4a12e84ce081b2cb0ad0612a628
11210f75217446d86081eb07244c5122421eea42e88cb0c1625122427e222043089d1136582c4a44c84f046408a133c2060b
4c9843c2142592c8b7ba103a236cf0ea9dba112f45890876a8273ca783f415b82546c2942622e4a12e84ce081b3ca489c8d8
27323a640c774ccc50602b3a24d42c93081ecac2e77cb0bf023314894fb24c847aab4638e382cdad7113c7de35504406ddcb
d23d1f0b1d770e149141ff999521dd9b63c1235f8785fe53a4878c1bea3282391e3cfa02d33f912f058a0877a80ba0f3c1fe
56e534464af590b1435dc670c7c40c05668b21614af510f250174267840d16d83d160953b04722dfea42e88cb0c14bf70a46
b614e921cca19ed89c0b72b77ec7688c14aa216387ba8ce18e89190acc5443c214aa11c9635d089d11365860a61a12a65c0d
210ff548e88c188315f66027c294ae21e4a12e84ce081b2c30470f0953ba869087ba103a236c7090ffed9bbc1ce56b5a5fa0
2e47f97bdf5eb7e7b4fda260cb262d07f9993e9b214ce91a421eea42e88cb0c135ed80c648f9163276a8cb18ee9898a1c0cc
48244cf916421eea42e88cb0c10a9f37224ce91a421eea42e88cb0c10ae718224cd91e421eea42e88cb0c10abfa84498b23d
12f956174267840d5e9b70416c29d54398433db13917e4eeea293e8897823d2276ac0b9ed3c1fa0a3c6142c214ec21e4a11e
099d1163b0c0935c244cc11e421eea42e88cb0c1cbe32d082fa57a0876a80b9ed3c1fa0abca542c214ec21e4a12e84ce081b
2cf066120953b0879087ba103a236cb0c07b78244cc11e421eea42e88cb0c131f7e2f99e3edfe51aeb7b9bdf29cea104f351
f6237ef55fe1693bb163cb262dc7dc8b337d364398823d843cd485d0196183d7c65b105b8af410e6501736e782dd1598d385
8429d823916f75217446d8e0d8886c7acf95e528df6bfec4f2de619aa67e797b7ffeb6fda1b16593966323b2337d364398b2
3d843cd413a13342060beca08b8429db231cb1d485d0196183d7ef36c88429db43c8433d103a23cee0f53b4c3261caf610f2
50174267840dc27eaadf796271ff2deafffbddaf37bfef7e7ff301f0ad619386b09baaebb119bc14ec21d8a12e784e07eb03
ec7e79f77ef6e39abf90ef57ff9a3e5f9f3b726f2d9bb40470db67338429db43c8435d089d113608e4fd54f459eeade6a39c
b74878fef685a9dcb2494b20b77d364398b23d843cd485d019618340fee87fa0e7fcd4683ecaf5d9d113c1b7862d36046ed7
63337829d843b0435df09c0ed6479126f373b97ef299bf90afdffcbfdeef0f86d86c2d9bb4a44813d767338429db43c8435d
089d1136489126fd36b2df4a2dab445edff773fadca709f78ade5a366949a126aecf660835db43c0f7b2f0391fec8f324dde
df7e6b715bae305f9fe7f7f69897c800f7deb2494bca34717d36e6d36c0fc1decb82676ca0bb0a412e40a7f11e097a9485ce
b8407397de93612647caf688579d5abff5dfc2ef5a977e5c7deba758b687909f48ac104267840d16481d44c294ed21e4520f
84ce081bbc3e75900953b647220ff540e88cb0c1ebefc99830657b08b9d485d0192183d7df9331614a161172a9074267840d
56c8b820c2942c22e4520f84ce081b3c248bc8d827f232640c774ccc40d6e70d32fa5971b998ffa67de8b865939664ddf5d9
0c614a161172a9074267840d56f8bc11614a161172a9074267840d5638c710614a1689e45a0f84ce081b2cf0bb8284295944
c8a5be133a23c660a56411c13e919721784e07eb2b962c22e427f23284d0196183576720225e8a15116ca9073ca783f58d7b
c3bf7df2979bb9f73c15eef55a8ef8fe7acc27c4d7367f175b366939ee0fcff4d90c610a1749e4a11e089d113658e0ed3412
a670112197ba103a2364b0c08c04244ce122422ef540e88cb0c102b3509030858b08b9d403a133c206af9e768478295944b0
a51ef09c0ed657609e1912a6641121977a207446d86081b9854898924522b9d603a133c2060b2cb044c2942c22e452df099d
1163f0ea69b488976245045bea01cfe9607d05563822618a151172a9074267840d1e624564ec7f8666c808ee8898a0c27efc
c0a7c922821dcb01cfd84077153e6840a7c922097a2f073ae302cd5d7e5201344d1611e25816342302b415582b4e74295944
a8a51ef88c0d7457e1d34d7c295c44b8a51e009d0ff657e1644684295c44c8a51e089d113678e96223644bb1229159eb81cd
b960770596a923618a151172a9ef84ce88315860511d12a6581121977a207446d8608125a448986245845cea81d019618383
7cdd0976ddb27f59edfb5977b6bfe5873dd8b249cb417ea6cf660853ac88904b3d103a236cb0c216f944986245845cea81d0
196183230de1d1bf078fe5adcaefb5fef4ac877cef3fc2f7fb6bff25c7964d5a8e4c84337d36439862451279a8074267840d
d608543912a6dc112197ba103a2364f03f08542180146a2260520f000e9805150b3511f213991d42e88cb0c10a1f66224ce9
26422ef540e88cb0c10aa76e224ce926915ceb81d01961830576f940c2147022e452df099d1163b0c0c52912a6801321977a
207446d8e0d54f58102f059c08b6d4039ed3c1fac673faf7fb7dffac1b4b7fbfafd7a3dfb82ccfd4e7dbe67e6ed8965a70cb
262dc773fa337d364398024e845cea81d0196183e335e4fa7fcb814ddf7e13d1ef1897570bf75b3f29fddedb84226ed9a4e5
780d79a6cf660853c089904b3d103a236cb0c0ab67244c0127422ef540e88cb0c131dde0f9badf1e8fe5a140bf719cb3a596
90a9cfbbdf4c3de6b9e5db74036ad9a4e5986e70a6cf660853c049220ff540e88cb0c1b1187e7d49b6eea2fdfd9beb361fe5
ebb54ea2df1ea762cb262dc762f8337d364398024e845cea42e88c90c10ad12e4498624e845cea81d019618305a61521618a
391172a9074267840d1e624ee2d867223c640c774ccc50e0f3868429ec44c8a5be133a23c660810d379030859d08b9d403a1
33c2060b4c6343c2147622e4520f84ce081b1cd712fdc6ed71bbfd6d1bbf2cc6f92cbf7efd22fff5edffdd768ec1964d5a8e
6b89337d364398c24e845cea81d019618305aea29030859d08b9d403a133c2060b4ce043c2147622e4520f84ce081b2c1676
92c8ff99e22184ce081bac147622d827223c129ed341fa0a4c5d44c2147622e4520f84ce081bac10384284296a45c8a51e08
9d11365860d226126ad44a049772e0733ed85f811937c427512b421daa3b9c71c1e652d48a8cf9cf1c13e99e8f858ebbc03c
1762d3a017418ee500674ca0b76b5320802b05cc08afd40399f180d6d64f168d90225e6464a98721dc11314181b935489802
66845cea81d019618383fcc27c76244c0133422ef540e88cb0c1022f6e903005cc24f2500f84ce081b2c107e8184296646c8
a52e84ce08192c309f0a0953c88d904b3d103a236c7090bf6eb7e9be7c187ffdc7f6f7793c9767eef3c3f7fe0dd93fecd8b0
c58683fb448fcde0a5849b88adf580e774b0be35f982c648193332b6d4f731dc31198602efc4913065cc08b9d403a133c206
0b4cde43c2943123e4520f84ce081b1ce49fe937e6e4bc7eafefed37ad575cf304a8f7be0e1d1b366938b84ff4d80c5e0a98
116ca9073ca783f555f88313618ab71172a9074267840d56f89213618ab71172a9074267840d16f82d45c2146f23e4520f84
ce081b2c70fd848429de46c8a51e089d113658e06e010953bc4d220ff540e88cb0c1ab273b215e4ab8116ca90b9ed341fa0a
3cd943c2147213c9b51e089d113658e0393612a6901b2197fa4ee88c188357efb9807829e146b0a51ef09c0ed637de54fdbd
3e5e9e8abd1eb77eca5d76f3ffde5ed3efb63c101d6faaa8659396e34dd5993e9b214c2137422ef540e88cb0c102ef659130
85dc08b9d403a133c206c72c84e5d9e7fa466d9e1e317f0b97a39c96772baffdf4862d9bb41cb310cef4d90c610ab91172a9
074267840d8e9927d3ad9f7cd68753fd94f47cde1ecb56cf9f79b7e3cffbf5d94e6fd8b249cb31f3e44c9fcd10a6901b2197
7a207446d86081d9464898426e845cea81d019618383fcdead7c973d1afb59f8fbb8f51bbae5286f8f7e6afabe9edba71d5b
366939c8cff4d90c618aba1172a9074267840d8e5dd2e64da5fa75f47c609fe77d7adde7ad4fe60c87d7b2e9f69688c02d9b
b41cfba49de9b319c2147593c8433d103a236cb058d48d909f08704984ce08192cb0a13412a6a89b70c4a91e089d113608db
5f3e1ff30495cf72813dafdf794cbff71d37fedc5a366909db5fda3e9b214c5137422ef58dd019710681fcdeaf38faa5d7b2
58aaab7a7ffa9d266503ec0d5b6c08dcaec766f052ce8d604b3de0391dac8f767abd7f6f53bfde9aaf39fba7f3f15c7e9589
7b6bd9a425edf4eafa6c863045dd08b9d403a133c20629e1e73e754ff306957320cfe37eeb637d2818616fd9a42565fcb83e
9b214c5137422ef540e88cb0c10a213f4498a26e845cea81d019618340febe3ffbe5c77bb9d8ea171eafdb77fa62e4cbd6b0
c586c0ed7a6c062f45dd08b6d4039ed3c1fa00fbd9efa11f8ffb3263e5fd58dfc3bde84bbeb76cd212c06d9fcd10a6a81b21
977a207446d82091cf1357fa25f692bab53f2e24f2d1b2494b22777d3643a85137021ecb81cff9607f97673a019ce6dc0873
2c0736a302c511f1fdf6fb2d2781f9f81ef31a8ad76d42e4d1b2494b62767d36a6d39c9b04bd97039d7181e60ae5dcfc1dea
21e766bfe48cf539cea45ff8aef5d08fab877ecae5dc04f213e92d42e88cb0c102b7254878c8b909e4a12e84ce081bbc7a6f
7ac43b84dc08f656173ca783f515082340c243c84d200ff544e88c9041ba1b5b1773af7f9edbd42f73fa31d3d97c6fd9a425
dd8fb93e0f213781fc44748b103a236cb042e808111e426e0279a80ba133c20621e4268c7d22ba45c670c7c40ce5426e02f9
89e816217446d86085cf1b111e426e0279a80ba133c206af7ff4c18487909b9d3cd685d0196183173feb62bc43c24dc00ef5
88e774187dd73fdc64c243c84d200f75217446d860b9909b407e22bd45089d1136582be426609f886e113ca783f58d77b4f3
f2f9c7ba11d0ad9f72bffd04bc5ef24dcff596667b474b2d9bb41cef68cff47908b911f27f46b708a133c2060bec0e808487
909b401eea89d019218305e66220e121e4269087ba103a236cb0c0fc1b243c84dc04f250174267840d8e0957eb0631cb51ce
571b8fdbdf1cb17586f7737ba2cd2d9bb41c53aecef479c8b909e427d25b84d01961830526d921e121e7269087ba103a236c
b0c0c44a243ce4dcece4b12e84ce081b2cb0681a090f3937813cd423a133620c56482a20c243d44d200f75217446d8e0b5fb
f513db21e4263087bab03917ec0e426ec2d8ffcc6f9111dc11314181f5f2c497436e02f65e163c6303ddadbaa1ff1c3323c3
8eb2f46f8e068ffdea75018496636602f15e4e68460468ab70fe06ba43cc4ca00e75e13336d0dde57f68823b64cc04e85017
3a2783e5ad5f2b1ae390f212c60e7519c31d133354f8bd24c243d6cb4e1eeb42e88cb0c1026770243c64bd04f2508f84ce88
315860270e243c64bd04f250174267840d16d88f02090f592f813cd485d019618305569222e121eb259087ba103a236cb0c0
ba69243c64bd04f250174267840d8e9088fbf3dbaf609735ceafc7b4cedd583ea6efe7ebf678ef4f7bb06593962324e24c9f
87ac1721ff678289103a236c70903fbfafeffa17e91f8b75abc6e553faf83c1eb73d15e4ff6ed606f33f7b3b44bc04e013c9
2609cc89207103785f21307ff7d69d0ae7439c5ef7c7effedd7ecba861938603fa448f870098807d22d644f09c0ed6572106
85080f0130813cd485d0196183cb35238e718860096387ba8ce18e891986f57e11fd59b7c5fcbd6ef7e59bfffcfbe19fe764
6d734eb8659396c3fa993e0f112c3bf9996011217446d8e0d53bfc20de217f2560877ac4733a8cbe02974d4878c85f09e4a1
2e84ce081b2c70938084870896401eea42e88cb0c1020fcf91f010c112c8435d089d113658e05511121e22580279a80ba133
c2060bbc1e44c243044b200f75217446d8608157c24878886011f2ad2e84ce081b2cb03c1f090f112c813cd413a13342062b
84cf10e12182259087ba103a236cb0c0c6044878886009e4a12e84ce081b2c100682848700989d3cd685d019618305fee648
78088009e4a11e099d1163b0c0f71c090f0130813cd485d01961830582c590f0100013c8435d089d1136582e0026909f8835
114267840d0ef2f7e7de6f21d7f72db7f7ab1fe3ba78e73bf51bfa7ea7b9fdcdb165939683fc4c9f870098407e22d644089d
113658e0ea15090f0130813cd485d0196183e5026084fc9fc92642e88cb0c102f76a48f8ff337766c9b22a59b2fc67143982
2bd13773aaf97f3f023660b6507b873f56899448a595278eea8e1341ebb6eb801172c90b613242061b3cd68784bb0e182197
dc08931136d8ae0346c80f749b186132c20617f2ef6dfccd9b2f994cefe1fe9eaa9ff6f233af55bc5e78c591838d5cc88f6c
73d701b391ffbbe0c4f8920ff6d7e02a1cf1950e18a15e53850b2ed8dcb9d5247baada3d23b05b6c58ec807c3558f196d86a
f78c206fb1c10513e8adc13575a2db35d008b5e4c6176ca0bb06eb3a23dfaeff46b82537c0e483fd357802030977fd37422e
b91126236cf0ecbb6688b72bbf116cc90d2fe9607d0d96ad47c25df98d91afb91126236c70ae99a23976f53332b7e4658eb4
4fc4d0e0711f24dcd5cf08b9e446988cb0c1d93acdb1eb80d9e6d6dce648fbc40c8bf5cff8eff07e9f9e16b98fc71abfb7f8
a7ff301e728cc755bf9b8c8b751a39d8c8c5fa916dee1a6884fc40af8a122623c1e0dc7e4373ec3a60646ec96d8eb44fcc30
cf4d73ec5a58646ec96d8eb44fccd0ab8545b00f748b185ed2c1fa1a7cbd20e1ae8545c82537c264840d36f83943c25d0b8b
904b6e84c9081b6c70e48684bb16162197dc08931136d8e0cc0c09772d2c422eb91126236cb0c1250724dcb5b018f99a1b61
32c2061bbc7b8484bb22162197bc1026236470b9bcf67c3eafefebfccf70fc1cdec6bfd27429ecf747bbbe2eeb62903c72b0
91cbe5b523dbdc15b16ce447ea458c301961830d2ea323e1ae8845c82557c26424186c70d30809775d2c422eb91126236cb0
c18d4224dc75b108b9e446988cb0c1e5b6f8e73e1ed7ceff26efaff1a3f8bcdca6bdfcddc7ff7e2edb773b8e1c6ce4725bfc
c836775d2c427ea061c408931136d860cd1724dc75b108b9e446988cb0c106ebfc20e1ae8b45c82537c264840d2e4bdffe7d
11ff76ecfbfc1d6c3ea6538df778f0f55bf2edb17ec3e1c8c1462ecbdf1ed9e6ae8b45c80f348c186132c206db75b108f981
8611234c46d8e0d98b7921deae88c5b0d7dcf0920ed6d760c56324dc15b108b9e485301921830dd6ad43c25d11cbbac7961b
6132c20697b50a5faff168f23dfd1bfcbd99369e3e5fa6bd7c5cbfb7dbfdfede16c0a591838d5cd62a3cb2cd5d118b901f68
1811c26424193c79ad67c6db15b108b6e4869774b0bef3d77a66c25d118b904b6e84c9081b3c7fad6726dc15b108b9e44698
8cb0c1f3d7de65c25d118b904b6e84c9081bec50be4384bb2e162197dc0893113608e4eff158e73a1e7d4c8531e399e47d3c
9d7fe0df7c1d39d848208fdbdc75b108f9818e11234c46d8e0efb60ecf51db5064ea2db619d21e3101151ebdc6d3f7e77451
faf97cbc9fcfcfeb8e8b7b2f03071b487547618bb50d4598ff59f5616c41058aeb50f10474b50dc5a097d8e8820b34d7a09e
13bb4c4a278a1efd69fefb3f9fe3b1fffff966422c5b695051897ca511c5b82557c0a083ed35383b40bed28762dc922b60d0
c1f63af4c0105f294429dc6bae804107dba343e4f7e5f6b98cdf43bf7dbcfdbdb7820bf6af23071b4987c8699b43e02b7528
c62db903061d64efece5fa11ae34a118b4e44a175cb03a6a67f87b136dfac35cae7fd766887a1d39d8486a6748db1c025fe9
41316ec91530e8607be79f13305fe960316ec91530e8607bcd1a588cfb40af8802061d6c6f6ebda1194a038acd2cb94e1176
88f7bf59ff89721f29f650c0a083ed9d7fbd81f94a018a714b2e804147b07776891ac295ee1383965ce9820b56d7e01811f9
4af389714bae804107db3bfb5609c295de1383965ce9820b56d7e00410f94aeb49e15e73050c3ad85e83a50090af749e18b7
e40e187490bd06afc3235f693c316ec91530e8607bcbd33d7fff167ffbf8fcbe9ee327f23bdfb4fe5bc6717dd203470e3672
79bae7c83687c057fa4e8c5b72050c3ad85e83e7b990afb49d18b7e40a1874b0bd064ff0215fe93a316ec91530e8607b0d16
6b43bed274a2dc9a2b60d0c1f61a3cab8a7ca5e7c4b82517c0a023d8ebb06a3bf1959613e3965c01830eb6d7e0a96ce42b4d
27c62db902061d6c6fd7736233ffb3c2432708bbc37bdfe1730674de7262d05bac74ac02bd75f888019b37ac14e425563816
81d64eafdd0030ef5731de2d7632b600cae67f4db0f5d26f62b34aae13f0cee09e3758bd03e94ac189514bae7c4106bbdbd5
9bd8cc074a3b748ab043bcff0d0e8b90af949b28b7e60a1874b0bdb3ff41235ce9353168c9852eb808ea1afc56215f693531
6ec91530e8607ba7ae148364a5cdc4882557b42082bd35583a05f94a9789714bae804107db6bf01e2af2952613e3965c0183
0eb677f6abe608574a4c0af49a2b5d70c1ea7a94b7ecf94a9789714bee804107d99b8b2d6886522762334bae53841de2fd6f
b08605f2953211e3965c01830eb6b77cd26ee34fcb6d7adef9fb984f4ee61dbe8ebf79d7eb63fbe1c491838d5c3e6947b639
04be526462dc922b60d0c1f69ad59828f7917e0e050c3ad85e83252c90aff49818b7e4021874047b0bf7f5f978fe92df4fdd
efa6d4777e39f56fa589f10466e5a691838d5cb88f6c73087ca5c5c4b82557c0a083ed9d7a808864a5bdc4882557b42082bd
35b82c8c7ca5bbc4b82557c0a083ed35b8fd817ca5b9c4b82557c0a083ed2ddcd3ede6791191df2306f7d765dac5dbdfc1ed
7a804803071bb8501fd8e210e04a6989414bae74c105ab6bb07201f295ca92c2bde60a1874b0bd06eb16205f292c316ec91d
30e8207b0d562d40be525762dc922b60d0c1f61a9418205f292b316ec91530e8607b0dd66a40be5255a2dc9a2b60d0c1f6a6
13409ca15485d8cc92cb146187c2fe37f8a4215f290a316ec91530e8607b0d2aa090afd48418b7e40a1874b0bd06a55fc857
4a428c5b72050c3ad85e835f50e42b1521c62db902061d6cafc11113f2958210e3965c01830eb6b770cfff33edd6eb331ebf
5e5ff32dabf11b617e047be5a691838d5cb88f6c73087ca51ea470afb902061d6caf472dca9eaf948318b7e40e187490bd06
4fff215fa906316ec91530e8607b0d4e7f91af148318b7e40a1874b0bd060f16209fd78228b6c48a1764b0bbb36f72139a35
8218f09a0a185b60654dfa408cf49f75174a8502c85583bbbb44e66d2006bcc5cac61ad059832bd3c456ba408c5972c56315
e8edf45608422b3520862cb9c205132caec1bd17e42b2520c62db902061d6cefec5bf808572a400c5a72a50b2e585d831bca
c8570a400af79a2b60d0c1f61adc5845be523e62dc923b60d041f6e6020e9aa1947fd8cc92eb14618778ff1b3c2c827ca578
44b93557c0a083edcd0518344329feb099259729c20e85fd6ff01947be523a62dc922b60d0c1f61a7ca7215f293c316ec915
30e8607b67ff80215ca93b3168c9952eb860750d8e5690af949d18b7e40a1874b0bdb30f4d11aef49c18b4e44a175cb0ba06
cd7cc8575a4e8c5b72050c3ad8de72bef97add1ee3d7cedf17edf8ffbc5ca7cbd8efebe5f17dfeaeed2ee79b3472b091cbf9
e6916d0e81af749c18b7e40a1874b0bd06cfcb235f693829dc6bae804107db5bae969d581f8c7ca5dfc4b82577c0a083ec35
b82a8c7ca5dd44b93557c0a083ed35b8fb817ca5dbc4b82517c0a023d86b70970ff94ab389714bae804107db6bf09c0cf295
5e13e3965c01830eb6d7606d13e42bad26c62db902061d6c6fe1be3fc76fdfe942e7780cfd1cfff73e1d55bf7f8f1d7cc7a3
eaf5680d470e3672e13eb2cd21f0954e13e3965c01830eb6d7a1d783f84aa389714bae804107dbebd0ea417ca5cfc4b82557
c0a083ed2d0b558dbfb08ffbf418ea78b0717b7deecfcfbcaed478bafeb97f6febf7398e1c6ce4b254d5916d0e81afb49918
b7e40a1874b0bdb317e842b8d26552a0d75ce9820b56d760a562e42b4d26c62db903061d646fe11ebf686ed30315e34febb8
cdcb7b3ab81cbf8ca7afe0ebfadb4d03071bb8501fd8e210e04a8989ecaee54a175cb0baf36b6b98af549818b7e41b60d091
ec2ddcaff1b4f83e3d2ef57ece8b274fcf18bcc683c9fbeb7aff7e576e1a39d8c885fbc83687c0573a4c8c5b72050c3ad8de
c27d9d9f22fadffce0edef25dbcbf4b799d712bbafb77b79e4602317ee23db1c025f6930316ec91530e8607b1daa7a88aff4
9718b7e40a1874b0bdd38b7a08ae549718b4e44a175cb0baf3d72067be525c62dc922b60d0c1f64e5e839ce14a6789414bae
74c105abdb3596d8ccffece3d009c2eef0decff3c2f6bd34c4a6dd62dd3eef0ceef9f9cb22239b578614e42556381681d696
73ecdf82d47f8f4c5d1fd7fbf845ff9c0e2dee9ffbfbfd7cbcd7e3321c39d8c8e51cfbc8364b61c8b2bfa52f448eb4349efe
8bef29956d70ba6da2c1e935b195aa10459658e058045a6b70d2416ca52544912516381681d6ce3f1443b65210e2c86b2c70
2c02ad9dbdba3981956210e595d8c8d802283bbf2f10d94a2d88224b2c702c02ad35ab0451e47f575d081c8b406b1d3a1a80
ad948128b2c402c722d05ab3221045fe77bd86c0b108b4d6a1f305d84a0189224b2c702c02ad75f8b70c6ca57b4490351638
1681d6ceae7122b0d239a2bc126f646c81959dfd8b4c60a56e4479251632b680ca1a1c74115b691a516489058e45a0b50687
d6c4567a46145962816311686d3981fabd56f3987af2f4ffbeacff1697d3a7fdb8c1c62d274fffdedec054a55cc461d758b0
5801fa6a70824c6ca55744912536381601d61a3c67416ca55244912516381681d696476ac6ad8ceaa7c7029ecff1b7725433
7f20c7d3f7fbebb23d188a23071bb93c5273649b03b3953611459658e058045a6bf00e35b1952211459658e058045a5b1e94
bb4f2f444e7f91cffdf67ebee667f23fe3e1cae3b760eafa9c1c0d1c74e0f298dc812d0e0c560a4494576221630ba8acc103
91c456ba4304596381631168adc11bd3c4566a431459e20d8e45b0b5068ffb125b690c516489058e45a0b579e179d87aa9eb
d0492596cdf3aee07eefaa3a74d27ff550c8c6794770afcf7e131eb0bca14361b754b0d000c96ab0b22370793987d32ea980
a102b2357f96f65bf6720c9d6f4b6dd3b813fbfdedf0b5b8e72aad1c8a2bb1a0a104f2d5e0978fc84a2187024b2c6cac019d
35582385d84a1388224b2c702c02ad2dc88ff1abfaefbd84cf73fc6d1acfaae675c9a655c97e6f652dc83472b0910bf2916d
0ecc564a4004596381631168adc15f99d84a0588224bbcc1b108b6d6a0eb85d84afb87224b2c702c02ad35f83124b652ffa1
c8120b1c8b406b0d8e6289ad347f28b2c402c722d0da82fc1e65bca6bed1eff571fb3d5d779f17269ecf951febabc73872b0
910bf2916d0ecc564a3f14596281631168adc17a38c4562a3f1c798d058e45a0b505f939bd0b3dadb670fbbe3f97cf735ea6
75fc129af77045a691838d5c908f6c7360b6d2f6a1c8121b1c8b006b0dde2e27b65233a2c8120b1c8b406b73b9096cbd747c
e8a412cbe6795770bfcf5eba80c04ab787f24a2c646c0195cd9261eba5584326d55836cfbb82fb7df6920104560a359457e2
8d8c2db0b2065fd1c456ba3414596281631168adc1c274c456ca3414596281631168adc1e116b1951e0d459658e058045a6b
700d84d84a8586224b2c702c02ad35385b24b652a0a1c8120b1c8b406b0d164320b6d29de1c86b2c702c02ad35580781d84a
6d86224b6c702c02ac35b85f4b6ca53143912516381681d61aac694f6ca52c43912516381681d616e4cbe8fe72f9bbc83abd
07f49e9e93187f4b1f9ff13066bd0c8223071bb9201fd9e6c06ca52743903516381681d61a3c7b416ca5a0439125dee05804
5b6bf05726b6d2cda1c8120b1c8b406b73130a6cbd1463e8a412cbe6795770bfe74961eba5954227955836cfbb82fb3d4f0a
5b2f95103aa9c4b279de15dcef065f95c456da2014596281631168adc19a40c4568a201c798d058e45a0b5e5b0e76fb587f9
0ce3ef71d0e9f77adcbfdbe5b95d58c391838d5c0e7b8e6c7360b6d201a1c8121b1c8b006b0d8ef488add43f28b2c402c722
d05a83e379622bcd0f8a2cb1c0b108b4d6e0e635b179e983104b2a68ac019d35b8dd056456f9a0b86bb871a102b4d5a4ee41
21ffd5672044840e8e1a5c50032a2f7a50d62d152e1440ae962ba67fd5a1bf3dbb3d5ef35355d377cbf7ef8585f58a298d1c
6ce472c5f4c83607e42a1d0f8a2bb1a0a104f2d5e0f60e91958607059658d858033a3bfb4e0b81956607e59558c8d8022a6b
b0083ab1955e07459658e058045aebb0d23eb0954a07475e63816311686d411ecf52c7c390e943f8180f086ebf9741a7ddfb
5eeed7eb633b69c091838d5c908f6c7360b6d2e6a0c8121b1c8b006b1dfecac0566a24145962816311686d2eaf80ad970e07
995463d93cef0aee77832f10622be5118a2cf106c722d8da5c59015b2fcd0d3aa9c4b279de15dcef66ad0d8afcef42028163
1168edecdf62022b7d0dca2bb190b10554d6e00122622b550d8a2cb1c0b108b4d6e01904622b450d8a2cb1c0b108b4d6e061
40622b1d0d8a2cb1c0b108b4b6203faeafdf4de3e99fda6b3e569876ef5596b9e091838d5c908f6c7360b652cfa0c8120b1c
8b406b0dae74105b696670e43516381681d61a3c83406ca59441912536381601d61abc5a4b6ca58f41903516381681d61adc
9027b652c5a0c8126f702c82ad35b8554b6ca58541912516381681d63a3411005b2960506489058e45a0b50e2504c056ba17
14596281631168adc10289c4566a1714596281631168edd4254d88aa742d28acc482c50ad05787be01602b350b8a2cb1c0b1
08b4369d9bd3d64bc7814e2ab16c9e7705f7bb53bf81f3fe6bdd7e21630ba8acc3e70ad84ab381224b6c702c02accd9f2bd8
7a6916d8fe1b16cbe6795770bfe7b59161eb65597f9d54e275f3bc2b61bfe74961eb654d7d9d5462d93cef0aeef7c9ff8210
acaca5afbc120b195b4065b364d87a59c85e27955836cfbb82fb7d723307829535ec95576221630ba86c960c5b2f0bc8eba4
12cbe6795770bf3b3481005b593e5e912516381681d6764bc7eba4ff5a955d36ce3b827bdda177644fe64bd62bef960a194a
205f1d3e537b2e5fadde699754c05001d95a4eb7a7e759e6a5b67ee749bfd518a6d389d7f4f8d2635b0600470e367239dd3e
b2cdb252fddfde9685eab79f6e49479cc7f5b706e9ffd92642bc6da3c1d936b09575ea85784b0d2d88206ba79e86015559a0
5e58b7d4a08202f275febf5e622bebd31bf1921a5a1041d61aac0e0b6c65857a21ded2821644ecad9dbdfe2f8095d5e90577
4b8d2b5820651d16e4dfb395c5e985784b0d2d88206b1dd6f0deb395b5e985784b0d2d88206bd3192f6cbd2c0e2f736ea96d
3cec0aed77870a803d5b591b5e88b7d4d08208b2d6e173b5672b4bc36fc4921a5a1041d63a2d0d2fb8d77f2c7eae5cc1022a
3bbf448dd8cadaf042bca586164490b585f87bbd7e9fd3e1c8fbf76d3a3551fdf6eef19ecfaab68f348d1c6ce4427c649b03
b295a5e185784b0d2d88206be79f8d125b59195e88b7d4d08208b2d6e00a30b095e5e18d78490d2d88206bcb39da188dc77e
d3c30797e7f86f6edcbdf91fe0eb3e1f73afe7683472b091cb39da916d0ec856568717e22d2d6841c4deda721e3e9e265f6e
b7e94da2dbfbfd5bb4665abde6fd7c3d5eb7df29c87a1e4e23071bb99c871fd9e6806c65717821de52430b22c85a83d74981
adac0d2fc45b6a684104595b1e60b98fdf68dfe9caf8673c2abbbc2eefc7fc717c8ce757d3ab4dcb032c3472b091cb032c47
b639205b591a5e88b7d4d08208b2b63ca574798cdae787e3de9fd7edfe9abe71c62f9cf93d9cf5e93b1a38d8c0e519a5035b
1c10acac0c2fb85b6a5cc102296bb0722ab09585e13762490d2d88206b0d561105b6b22ebc106fa9a2051168adc17a47c056
968517e22d35b42082ac3578ba12d8ca9af442bca586164490b5dd9af432e7ff7fe975db74d811daeb0e9faa1d992f4a2fb8
6b685c2c017c75f8d2d871f99af406fb171a162b005b0d5e3bdf73f982f802bb86058b15ec6c35782a79cf55d6c317da2d35
309600be1afcde0359590e5f78b7d4c8820672d660dd08602babe10bf1961a5a1041d6ce3e8405b0b212fe862ba971050ba4
6c5e0c76bff5b214bdccb9a5baf1b02bb8dff39cfbad97b5e065ce2db58d875da1fd6ef04f17d8ca52f042bca586164490b5
068718c056568217e22d35b42082ac35588d15d8ca42f042bca586164490b58578fc1a1bcf5ca6f52daed7f1b4fc79b94e7b
f71e7fafbfcfdf75828598460e3672213eb2cd01d9ca3af046bca486164490b5066ba2005b59065e88b7b4a005117b6b0dde
e204b6b20abc106fa9a1051164edecb5c201acac3f2fb85b6a5cc102299b7f0ef75b2f6bc0cb9c5b6a1b0fbb42fbdde06d68
602b0bd06fc4921a5a1041d61a2c2f046c65097a21de52450b22d0da427cb98fa787af796d87f997eb357de83fe307f1767b
6d2b82e2c8c1462ec447b639205b59815e88b7d4d08208b2d6e0ed6f602b0bd00bf1961a5a1041d61abcfc0d6c65fd7921de
52430b22c85a83833c602bcbcf0bf1961a5a1041d61a5c8f05b6b2fabc106fa9a1051164adc10543602b8bcf1bf1921a5a10
41d616e2b33a9501ac2c3c2fb85b5ab88285bdb2b32fb2005859745e70b7d4b8820552d6602503602b6bce0bf1961a5a1041
d61aac2a0c6c65c9f98d5852430b22c85a83a51b80adac382fc45baa6841045aebb058f69ead2c382fc45b6a68410459ebb0
c6fe9eadac762fc45b6a68410459dbad762f73fe6b2979db78d815daef06af28015b59695f88b7d4d08208b2365bde6fbd2c
752f736ea96d3cec0aed7783755780adac746fc44b6a68410459ebf0af77cf5616ba17e22d2d6841c4de5a876fe83d5b59e7
5e88b7d4d08208b2d6e0913b602bcbdc0bf1961a5a1041d61adca105365fe57e03de42030b1ac85983e72bf664b6c8bdd02e
9952b102b2b59cf85fdf97fbf53b5d7abacc1717a7bfc2f335fe8e5c1ef7f5ac81060e367039ed3fb0c501907c997be15c43
4342fabda60637b1f654becabda0aea151b10070d5e06ece9eab2c722fb45b6a602c017c9d7d2b07b0ca02f702bba586151c
90b0b317d505b0b2bebde06ea971050ba4acc1f2d4c05696b717e22d35b42082accdab80efb75ed697b73997d4361e7685f6
bbc362eb7bb6b2bcbc106f69410b22f6d666cbfbad97f5dd65ce2db58d875da1fddeadefbecdf9cfe5cb6de3615768bf1bfc
fb01b6b2babb106fa9a2051168adc342eb7bb6b2b4bc106fa9a10511646d5ecf7ebff5b2b6bbccb9a5b6f1b02bb4df0daa65
80adacee2ec45b6a684104596b70771fd8cae2ee42bca586164490b50677f781adaced2ec45b6a684104593bb5f30ba8caa2
eec2baa506151490afb34ffd00ac2ce82eb85b6a5cc102296bf01e01b095f5dc8d78490d2d88206b0d6e09025b59ce5d88b7
b4a005117b6b0d6e09025b59cd7d2396d4d08208b2d6e08233b095c5dc85784b152d88406b0dd65a03b6b296bb106fa9a105
1164adc1ca05c056967217e22d35b42082ac9dbbc8f79eaaace12eac5b6a504101f95a58afb7d7f8e99bfe818d1fbddbe5f1
3b33feeddd6dfc437c1ef7f57405470e3672e13db2cd01d9ca12ee42bca586164490b5066b8f005b59c75d88b7d4d08208b2
76fafae27bb0b286bbe06ea971050ba46cb786bbccf9af55ca6de3615768bf3b7ca8f66c651177235e52430b22c85a872f8e
3d5b59c35d88b7b4a005117b6be7fe2ceca9cad2f1eb7e6a6a504101f93abfda83d8cabaf542bca5821644b0b585f8eff2cf
7448323d6a391e6a4f872497f160e1fa5bd56921a691838d5c888f6c7340b6b268be106fa9a10511646d5e777bbff5b270bd
ccb9a5b6f1b02bb4dfe7b7e4105b59355f88b7d4d08208b276fee133b19575f385784b0d2d88206b1dfec67bb6b268bf106f
a9a10511646dfe97b4df7a59355fe6dc52db78d815daef0e9fab3d9b2fd92fc06b68604103399b1defb6ed6be6cb846b685b
e6dd803d9ea7db6dd917adb7e9fe42db30ef04ec6f836715e79df115e77dd57a3d6ad0f8f3dff372ff3ca6583612e26d230d
aea5109d2f5c6fd0122b5d7081e6ce3fce403a5fc1dea02556bae002cd9d7d064a68be827d215e63450b22505b83355d89ce
17b13768899d2eb80073cbd2bdafd7f3f3fb27f6fbf776bdfdae5e5f7e87fbafc7f57bbbddefeb821c3c72b091cbd2bd47b6
39309d2f656fd0122b5d7081e66085e6fbafc268fc484ccd02e3bede5edff10b6885a691838d84159ae33607a6f3d5ec0d5a
62a50b2ed0dcd9ab70139aaf666fc4122b5a1081da3aacb40e74be9cbe414bac74c1059a9b17f187edfb9af636adc4bafdb0
37b8eff3b4b07d5f585ea7d558b71ff606f7fdec0f35a1f9daf2462cb1a00511acadd7eaf206fdcff5d38d2eb840736797ad
109a2f2f6fc4122b5a1081da968391dfa5f2f7733a587a3ebecfe7effde1e9c0e1321e9abfb6a77679e4602397839123db1c
98ce5798376889952eb84073cb31e76f5df2eb6dbadcf8be8f9f85eb6ff1ae7107c7bfd6efa4efba9e5de0c8c1462e479d47
b639309d2f325fa0d758e9820b3477fe755ba4f375e60d5a62a70b2ec0dc721ef96b61f95bd06a3ce1bddfc7f3dce91ec278
06fc1e3f9ceff5e419470e3672398f3cb2cd81e97ca979839658e9820b34d7e01d52a2f3d5e60d5a62a50b2ed05c83d7fd88
ce179c376889952eb840730ddef8233a5f76dea02556bae002cd9dbd7e06a1f9b2f34aacb1a20511a8adc1fa9d44e72bcf1b
b4c442175cb0b9068f45129d2f3e6fd0122b5d7081e61abce04974befebc414bac74c1059a6bb0e211d1d902f8c6bca5ca16
4ca0b70e7fe73d9bad816fc45baa68ec819c75f806db93d932f8857749958c2d90b1b397b4022c5b06df60b7d4b158c15e57
83b7f081cc57c2376089958d3d90b3060fb0139b2f866fc8122b5c3081de16e8f97fa6dfcbd767fc2b5c5ff357ea75fc57f0
f93ed7cb9c3c72b0910bf4916d0e4ce7ebe11bb4c44a175ca0b90647d844e7abe22bb4c64a175ca0b906cb79119d2fcb6fd0
120b5d70c1e61abca24174de0b60d0122b5d7081e6ce3e972234ef05306289152d88406d67ff32139af70218b1c48a1644a0
b60687d744e7d500062db1d2051768aec1bab144e7ed00057a8d952eb840730d1644263a2f083068899d2eb800730d56d021
3aef08306889952eb840730d5e80263a6f0a306889952eb84073b52bc0a6fde76af8b6fdb037b8ef0d562d223aef0b50688d
952eb840730d3e6044e79501062db1d005176cae41bf0ad1796b80414bac74c1059a6bf0fd49745e1c60d0122b5d7081e64e
5d4e81b8bc35c0702556ae60019df5ea0d30e87fae8c6f74c1059a3b7b190942f3e200239658d18208d4d6e0a097e8bc3ba0
40afb1d2051768aec1a574a2f306018396d8e9820b30d7e07a3ad1798f80414bac74c1059a6b709d99e8bc4ac0a02556bae0
02cd35b8d44c74de26a0d01a2b5d7081e61a5c75253a2f14306889852eb860730daeba129d770a18b4c44a175ca0b906cbc4
109dd70a18b4c44a175ca0b9060f68119df71a18b4c44a175ca0b9057adca7cbed36fda6dedeefdf92b5d3dab5efe7eb311e
3f7db76f6f1c39d8c805fac83607a6f36a03839658e9820b34b740cfdf39f32fcaf86779dfaeafe909b2e778b8f8783c3eeb
c1098e1c6ce4027d649b03d379b782414bac74c1059a6bf0f21ed179bd42815e63a50b2ed05c87bf34d079c382414bec74c1
0598ebf04506745eb260d0122b5d7081e6969facfb6d3af7997e5efe4e76a71dfcd4054671e46023979fac23db1c98ce7b16
0c5a62a50b2ed05c83bba34467550bca2ca9b20513e8adc1e906b069db8201afa180b10534d6e02ee19ecb3a178c764b150c
1d80ad0637ca80cb6a178c764b158c1d90af85f7c46e5820f3e605039658d9d803396bb02632b179ff82214bac70c1047a6b
b0f435d1790b83414bac74c1059a9bd7ae87ed7b15824d2bb16e3fec0dee7b83e5eb89cebb180af41a2b5d7081e6960688dd
f6bd10c1a695d8b71ff606f6bd5622d8b4ffac1cb0ed87bdc17dafad083aedbfd7fdb7ed87bdc17def50130074de8c60d012
0b5d70c1e6e6a600d8bed713d8b412ebf6c3dee0be9fdd5943685e8e60c4122b5a1081da16e2e7f4253e959fdcbee3f9c5e7
399f608e476bf3c3072b328d1c6ce4c27c649b03d3793f82414bac74c1059a6bf00807d1794582414bac74c1059a6b70639f
e8bc25c1a02556bae002cd35b8054a745e9760d0122b5d7081e61adc02253a2f4d306889952eb84073673fba4c685e9a5088
d758d18208d4d6e01a0cd1796f82414bec74c105986b708991e8bc3a41a13556bae002cd35b8824e74de9e60d0120b5d70c1
e61a2c7f47745ea060d0122b5d7081e63aac950d74dea160d0122b5d7081e61aacbf42745ea660d0122b5d7081e616e8ebe3
769b1eb378bfc631bfa5dba66509dff38ab5d7f5cb9b060e3670413eb0c581d1bc4cc1882556b42002b52d6b0b4dc58af7e9
73f89abf63a673a1d77b5e31f6b27cb679e4602397b5858e6c73603aef53306889952eb84073a7ff9901cd2b158c5862450b
22505b877fcd40e79d0e062db1d2051768ae57ab4381fe476f81d1051768aec38f15d079b383414bec74c105986bb06e16d1
79c583ecafc54a175ca0b90e4b88039db73c18b4c41b5d7011cc9d7fd48d745ef460d0122b5d7081e6ce7f6e05e9bc69c2a0
2556bae002cd9dff2c1ad279d984414bac74c1059a3bff4c1ae9bc6fc2a02556bae002cd9dff7c29d279e584414bac74c105
9a3b7f2d3ca4f3ce0b839658e9820b3437d732c0f6ad78c266dd52dd7ad817dcf3523d6173fea3e1c136ce7b427b5dda27ca
8cdc34b1a4ba6dde0fdae7f3ff0961774429a090431a8ddfff3d3fbfa3cb2997ada47cdb4c83a54389af545028b6c4869774
a0bd0657e188af945028b6c4869774a0bd4e3514cefcaf8e05634b2e505d83eb33c4578a28145be282977480bd05fb3abfff
f47740fa7b6eeb32ede27cf3e3bede0ce691838d5cb08f6c7360be5245a1d8121b5ed281f69a955128f6bfcb160c2fe9407b
b48efdbc16fbedf76f723c03bfbfc6a93e0fc25e470e3692d6b14fdb1c98af345228b6c4869774a03d2a2c9857529bd79dbf
bcc6c3c671875f84bd8e1c6c245516a46d0ecc573a29145b62c34b3ad05e872a0ee02b9d188a2db1e1251d68afc35f1bf84a
2787606b6c784907daebd4caa1ccffae9c50b6e482d535f83e23bed2cba1d8121b5ed281f61afc7a115f29e7506c890d2fe9
407b0d8e5588afd47328b6c4869774a0bd0647a6c4570a3a1c7b8d0d2fe9407b0d6e2a115fa9e8506c890b5ed201f61a9c72
125f29e9506c890d2fe9407be7df6e40be52d3a1d8121b5ed281f6cebf0c8d7ca5a843b12536bca403ed35786b94f84a5587
624b6c784907da6bb09207f195be0ec1d6d8f0920eb4d760d516e22b8d1d8a2db1e2251d6cefec677209ae147628b3c4c696
5ca0ba06af09135fa9ec506c890d2fe9407b67afa34c70ded8a1c85b6a64c9048a9b5e5182ed7b6b86cebaa5b6f5b033b4e7
a7ff83da83796986e32ea971050ba4acc3d7e69ecd9b3394784b0b5a10b1b7d6a15461cf56ba33145962830b2ac85b83f718
88aeb46728b4c4469764a0bbf3efca235fe9cf506c890d2fe9407b0d6e1f135f69d0106c8d0d2fe9407b0d8ef789af746828
b6c48a9774b0bd798d6e98a1f458e8c412db04698770ff1bac9e477ca5cc42b12536bca403ed35f8fd20bed268a1d8121b5e
d281f61a1c74135fe9b4506c890d2fe9407b0d3a1189afb45a38f61a1b5ed281f616eceff88fe0397f06a78bb8bfc729a62f
c1cffc5efdbad6068e1c6ce4827d649b03f3955e0bc596b8e0251d60efec05ac09aed45a28b3c4c6965ca0ba53176627b252
69a1c0121b581281de1abca54f7ca55243b12536bca403ed3558918af84aa986606b6c784907da9b0e8f6886526ca1134bac
13a41de2fd6ff03123be526da1d8121b5ed281f6ceed7900b2526ba1c0121b581281de1aac6c477ca5de42b12536bca403ed
352bb850ec7f1738185ed281f6ce3e4220b85271a1cc121b5b7281ea1aacf0427ca5e4c2b1d7d8f0920eb4d7e01898f84acd
85624b5cf0920eb0d7605952e22b45178a2db1e1251d68afc1891ef195aa0bc596d8f0920eb477f64d25822b4d17c2acb1b1
2517a8aec13d0fe22b5d178a2db1e2251d6cafc14d00e22b6d178a2db1e1251d68afc16561e22b7d178a2db1e1251d68effc
176e91af345e28b6c4869774a0bd068f78115fe9bc506c890d2fe9407b0bf678ec369e8b4f5fba97df97d16f9d92ff4d4fd9
de1fe379d36bfdd1c691838d5cb08f6c7360bed27aa1d8121b5ed281f61a3cc7487ca5f7c2b1d7d8f0920eb4375dd3a0194a
f7844e2c719920ed10ec7f878f19f095f609c596d8f0920eb477fe8a0dc857fa27145b62c34b3ad05e83fbc8c4e70d14422d
a9c12519e8aec1e39340671d148abc864a1644a0b5b34f00f6585e41a1b05b6a58ec00743538a307322fa150de2d35b2a081
9c35b838096ca58642912536b8a082bc2dcc27764a135d29a25068898d2ec940770d56dc26be5245a1d8121b5ed281f616ec
c7e5f2ba4e7f8ceff86ffffbbedda77392dfc9c938c5f6c7c681830e5ca00f6c7160b8d283a1cc121b5b7281ea3ab48e005f
69c270ec3536bca403ed9d5d1d4070a5864399252e6cc905a8db1571e8c4ffeeb8b009d20ee1feefaa3864e2033d133641da
21dcff061f6ee22b651c8a2db1e2251d6caf434b04f0953210c596d8f0920eb4d7a0358af84a2388624b6c784907da5bb02f
f7eb735e9ee67b9d6f74bea6df98cfed7abbdd5edba9158e1c6ce4827d649b03f3954e10c596d8f0920eb4d7e09908e22bad
208a2db1e1251d68efec9bc604574a41945962634b2e505d87860ce02bb5208a2db1e1251d68afc163b2c4578a41145b62c3
4b3ad05e8367c189afb48338f61a1b5ed281f6960b839ffbf8153fdfe0bdbfc63fc5f3b757bfb7a03fcfebf885bc9d5ee3c8
c1462e17068f6c7360bed20fa2d81217bca403ec3558a88ff84a4388606b6c784907da5bee36fccef31ff34750feefcbbaaa
c972af613f6eb071cb9d867f6f6f60b2d20ea2c0122b5812c1dece5ea28fe04a3988324b6c6cc905aa6bb0100af1957a10c5
96d8f0920eb4d760d91be22b05218a2db1e1251d686f59e4681cf47d4e7f8cf76f3dbbebf33b9d37bc1eeff15ce1fdf96c8b
1cd1c8c1462e8b1c1dd9e6c07ca52544b12536bca403ed75f86b035fe909516c890d2fe9407b0bf66f4be36fc9b48be3b9c1
f3f73af66f175f97f1a8f2b55d41e291838d5cb08f6c7360be5216a2d8121b5ed281f63a7ca5015fa90b516c890d2fe9407b
1d5a5280af148638f61a1b5ed281f69a558628f6bffb300a5ed201f61a3c0a427ca53464db5f8b0d2fe9407be71f8a235fa9
0d516c89052fe908f6ce7fba0df94a7188624b6c784907da3bff7c13f94a7588624b6c784907da3bff114ee42be5218a2db1
e1251d68effc077691afd48728b6c4869774a0bd9357e041b8d21ea2cc121b5b7281eacebf6c887ca53f44b12536bca403ed
9dfcf011c2797989226fa991251328aed697e8acff2a08b1ad879da13d5f2c8fdf75d7bfbffd67fcbcff1e689e176e9c966d
fc3deab068a691838d5c3c1fd9e6806c5e9fe2c44b6a684104596bf0e422569f9406153dbcd3fcf5df6f719bcf9ccb76522e
db69f0fc3b129612152397dc08931136787e811e13961e152397dc0893113678ea7569642b3d2a8579cd8d2db960770d4eee
91b054a918b9e485301921830d2e62216169533172c98d30196183cbc5dad76b8cded319f8f817f91dcc5ea6bd7c5cbfb7db
fdfede3a1768e46023978bb547b63904c252a862e4921b6132c2061bacc58f84a553c5c82537c264840d9e5d388178a553c5
b02537bca483f575a81721c252ab62e4921b6132c2064fff83135ea955516ccd0d2fe9607dd3e3963847a937b1b925d739d2
3e0586796e9aa3748cd8dc92db1c699f9881ca55dee311d6e535dd427addfe8eb13ff4f75e470e3692ca55d2368740586a46
8c5c72234c46d860835f33242c4d23462eb91126236cb0c1110c1296b29142bee646988cb0c1e5a8f5fab8dda67501c65d1a
bf7fc65fdde9f187f7fcd6e3753d54a781830d5c8e590f6c710878a56cc4b0252f784907e9ebd0b24284a56fc4c82537c264
840d36b82b8584a572c4c82537c264840d9e7f638a094beb88914b6e84c9081bec50b74284a578c4c82537c264840d9e7f9f
86094bf788926b6e84c9081b3cff560d1396fa1123975c0993916070ee68a0394a0d88cd2db9cd91f689191abcab8d84a509
c4c82537c264840d3658c20809bd0fc4c0b7d8f8920ff6d760f16de2f34212c3de62c30b36d05d878f39d0792d49815e62a3
0b2ed05c876f54a0f3661283dee242175c80b90e85154057ca498c5a72e30b36d0ddfc1b42339486109b59729b22ed111334
b8558884a524c4c82537c264840d36384e45c2d213a2e49a1b6132c206e7ca0c9aa39475d8dc92eb1c699f02c3aeaec3e63e
d0446173a47d6286069f75242c951d462eb91126236cb0c1fb6948585a3b8c5c72234c46d860838307242cc51d462eb91126
236cf0ec2637c42bc51d057bcd0d2fe9607de73fedc384a5bbc3c8252f84c908196cb072311296060f2397dc08931136d860
856e242c551e462eb91126236cf0ec57eb11af547918b6e4869774b0be734b5b88adf47828b3e6c6965cb0bb066b99216129
123172c995301909063b742d1061e9123172c98d30196183737b0bcd516a3d6c6ec96d8eb44fcc30cf4d73946e0d9b5b729b
23ed1333cc73d31ca5e0c2e696dce648fbc40cf3dc3447299ab0b925b739d23e31c3d95fe788578a260af69a1b5ed2c1fa66
e53447697bb0b9252f73a47d2286068b622261297c3072c98d301961830d8e1791b0743e18b9e446988cb0c106e70848589a
1f945c73234c46d8e0d9278588579a1f0c5b72c54b3a82be06570190b0943f18b9e446988cb0c106f7a491b0f43f18b9e446
988cb0c1f3df9165c2520161e4921b6132c2063b2c534f84a505c2c82537c264840d2ecf1bfdbdf4375d867cfe563f7c4cab
5e8e5f0cb7df23b78ff5df398e1c6ce4f2bcd1916d0e81b0144118b9e446988cb0c106cf982161e98228e46b6e84c9081b3c
7b9526c42b4d14862d79c14b3a485f873f381196320a2397dc08931136d8e11f3911963e0a2397dc08931136d8e0ed4624f4
4a0a0597d8f8920ff6d7e0c61df1592b8551afa9c205176caec1112bb0793785216fb1c1b109b276fe1a00c8e6f51486bcc5
06174ca0b706e7e044571a2a8c5a72e30b36d05d832bbac8574a2a8c5b72034c3ed85f8325a091b0f45418b9e446988cb0c1
0e051d4458da2a8c5c72234c46d8608345de91b0145614f23537c264840d9ede4e4278a5b3c2b0252f784907e9dbb556d8dc
07ea206c8eb44fccb02baed0b98f7432d81c699f98a1c13f71242ce515462eb9122623c160874203222cfd15462eb9112623
6c70f9417b7f9e8fd7f450f1773c59faad61769faf06cf0fcd3fd64bca3872b091cb0fda916d0e81b0545818b9e446988cb0
c106073148585a2c8c5c72234c46d860837b6648588a2c8c5c72234c46d8e0427e624d3a12963a0b2397dc0893113678f69d
23c42b7516862db9e1251dacafc193b348581a2d8c5c72234c46d86083e7e391b0945a14f23537c264840d36b87f8284a5d7
c2c8252f84c908196c705119094bb585926b6e84c9081b5c1665788d27cdf7e972d8f8df98d6a99f7e7e5eaff10b783cbdfa
2ee43c72b091cba20c47b63904c2d27161e4922b6132120c76e80020c2d27461e4921b6132c2061b2c2b8584a5ecc2c82537
c264840dd2b233d7efe7fe79de7edf46e329edfd75bbbe3e0f225f470e3692969d49db1c0261e9bb3072c98d30196183d3e9
31ce514a276c6ec96d8eb44fccd060b11f242cbd13462eb91126236c70217fcc1f88e95fc5fa3af5b897f7e92afc675bd58a
060e3a70e13eb0c521e095de09c396dcf0920ed6d7e1ab95084bf584914b6e84c9081becf0734a84a57da290afb91126236c
f0ecc71110afb44f18b6e4052fe9207d0d6eca236129a0903db6dc0893113678fe4a3f4c583a288c5c72214c4692c1f3cf10
99b0d45018b9e446988cb0c1f3172a60c2d24461e4921b6132c206cf7f2e81094b1985914b6e84c9081b3c7fcd19262c7d14
462eb91126236cf0fcc7e599b0b45218b9e446988cb0c1f31fcf60c2524c61e4921b6132c206cf7fa59609bd9ec2c0b7d8f8
920ff677f21b5808e7e518c6bcc5c61654a0b8f3df2b453aefc728d04b6c74c1059a3bbf498a9b2d4a43861e7e6afefc1d08
8d4703532edb49b96c67219f7f6be7a7e0c623acf7edfa9af6f2397e16c7e3ebcf7a828223071bb9901fd9e610084b438691
4b6e84c9081b6c70591f094b4386914b6e84c9081b6cb00c2b12969e8c42bee646988cb0c1066d9048587a328c5cf242988c
90c106ada748587a328c5c72234c46d8e0427e62bb2f12969e0c2397dc0893113678f66556c42b2519862db9e1251dacafc1
7d1c242c3d19462eb91126236cb0c11d05242c3d19462eb91126236cb0c11d2c242c55194aaeb91126236cf0ec8610c42b2d
1d862db9e2251d41dfaea5c3e63ed03d6173a47d6286061f36242c2d1d462eb91126236cb0c1170c1296960e2397dc089311
36d8a19f84084b4b47215f73234c46d86083e327242c451d462e79214c46c860836366242c451d462eb91126236cf0ecbb96
88575a3a0c5b72c34b3a58dfa9d591c856fa398c5972634b2ed85d83a250242cfd1c462eb91126236cf0fc3bb54c58fa3994
5c73234c46d860838b7e4858fa398c5c72254c4682c1f3ef51336169073172c98d301961830d5e9a47c2d20e62e4921b6132
c2061b3cb38f84de0e62e05b6c7cc907fb9b7b046006efe7b089b7d82608fb837b7ffec31048e7fd1c057a898d2eb840730d
d6d5223aefe730e82d2e74c105989b3f60b0fdd29061f34a6e3384fdc1bddf3564d8cc077a1f6c8ab4474cd0e11784084b43
86914b6e84c9081bec70d44084a52143c93537c264840d36583b0e094b3f87914bae84c94830b8ebe7b0b90fb44ed81c699f
98619e9be6280d1936b7e43647da276668f07d8e84a521c3c82537c264840d36f8f946c2d29061e4921b6132c206cf7f9691
094b4946215f73234c46d8e0f9cf32326129c93072c90b613242069727faeed3fafed32feff8d33b7ef9bea67f939ffbe5f1
f8adaab23ed04703071db83ccf77608b43c02b0d19862db9e1251dacafc17beb48581a328c5c72234c46d8e0f98f70326129
c93072c98d301961830d566940c25295a1e49a1b6132c206cf7f7895094b5586914bae84c9483038afa44f7394b20a9b5b72
9b23ed1333ecca2a6cee0305103647da2766d89555d8dc070a206c8eb44fccb02babb0b90f1440d81c699f9861575661731f
6861b039d23e3143a7b28a82fdcf0208c34b3a585f872f17222c5519462e79214c46c8e0fc61a3394a5985cd2db9cd91f689
191afc98216129ab3072c98d301961830d0edd90b0945528b9e646988cb0c1062bc22261e9ab3072c995301909061b9c9922
61e9ab3072c98d301961830daec32061e9ab3072c98d301961830daeb62261e9ab3072c98d301961830d5e4f40c2d25761e4
921b6132c2061bacb78484a5afc2c82537c264840d3678d40a094b5f45215f73234c46d86083a53190b0545618b9e4853019
2183d3f123ce514a236c6ec96d8eb44fccb0fc4bfb3df477bd4d5ffbeff110e771fd1dde8d863e8fe9d5c8ebfa8b8a23071b
b9fc4b3bb2cd211096d2082397dc0893113678ea336ec8e675118a2cb19125136caec18b7ec467751146bda60a175cb0b906
37a981cdeb220c798b0d8e4d90b5e510f9ef01d3e92cf9fbf71cf1f4a1fcbc7e1bb8af1f6c1c39d8c8e510f9c8360766f3ba
0843de62830b26d05b833322a22b7511462db9f1051be8aec1357ce42b7511c62db901261fecefec6b6c8857ba220c5b72c3
4b3a58df827d4abf37b29596086396dcd8920b76d7a11984084b4b44215f73234c46d8608325e491b0144518b9e485301921
830bf978c43a9ea04fcbbd5f5fe359e2fbef51f2f7e7719f0eae56721a39d8c885fcc8368740586a2a8c5c72234c46d8e0ae
a642e73e52be6073a47d628653bf5d90ad145418b3e4ca965c0477e7aff2c284a5a0c2c82537c264840d36b8738084a5a0c2
c82537c264840d36b8738084a5a0c2c82537c264840d36384e46c2525061e4921b6132c20697d3a2f97fa63fc96bfcfefd8c
bffad3c7f47a795f3fdfdf7a48cb69118d1c6ce4725a74649b43202c0515462eb91126236cb0c19d03242c1d15462eb91126
236cb0c19d03242c1d15462eb91126236cb0c19d03242c1d15857ccd8d3019618367bfb08b78a5a0c2b0252f784907e96bf0
6e3a1296820a25d7dc08931136d8602506242c0515462eb9122623c120905ffe0a8aa7bdbc5cff3e9a44be8e1c6c2490c76d
0e81b0145418b9e446988cb0c13df96b3ca31b4fe53ed337f278b8f1785ec65d7e02f93672b0917bf2bccd211096820a2397
dc0893113608e49fdfb7f17d3afc78be7f2f128cff4868919b75e06003813b6d710878a59dc2b02537bca483f5edda296cee
039d0b3647da2766e8f06123c2d24e61e4921b6132c2063b7cc1106129a83072c98d301961830d564e43c2525061e4921b61
32c2063b5473106129a828e46b6e84c9081b6cf0e40512968e0a2397bc10262364b0594785901f6a5e30c264840d9ebff209
13968e0a23975c08939164f0fc1be34c583a2a8c5c72234c46d8e0f9ef753261e9a83072c98d30196183e7bfddc884a5a3c2
c82537c264840d76686a20c2d25161e4921b6132c206cfbfcacf84a5a3c2c82537c264840d9ede584078a5a0c2b02537bca4
83f59d7fbb9c09bda0c2c0b7d8f8920ff677fe2d63e4f38e0ac3de62c30b36d0dd1ff467fc3c3ee7eb91dfebdc8afe9abe8a
c783bddb6d3c12598fdc70e46023ffa00f6d73603aefa828d04b6c74c1059a6bd651b1ee70e9a8d0c34fcd1fbf2fc8f16f39
e5b29d94cb7696c3d5f91b76fe7dfdfc9d3d4f67558ff9e462fd73e3c8c1462e87ab47b63904c2d25161e4921b6132c206cf
3f5067c2d25161e4921b6132c2061b2c4b8984a5a3a290afb91126236cb0c1692912968e0a2397bc10262364b0c1eb2f4858
3a2a8c5c72234c46d8e0423efe1d6ed309c4f8cd3b9eb68f1f8f6927dff383e5d7f59f390d1c6ce0c27d608b43c02b051586
2db9e1251dacafc1ed3b242c1d15462eb91126236cb0c1f546242c1d15462eb91126236cf0f45212c22b0515862db9e1251d
acafc1a575242c05154aaeb91126236cb0435b0211968e0a23975c09939160f0f46a0ec22bf518862db9e1251dacafc31f9c
084b3d86914b6e84c9081b9c4b49688e525061734b6e73a47d6286060f84206129a828e46b6e84c9081b3cfb1705f14a3b85
614b5ef0920ed2d7e01002094b3b85914b6e84c9081b5c0e1b7f77b0c733d5e910ef31eee1fbf742fdeff5e6cbe5321eee7e
d72f181c39d8c8e5b0f1c8368740580a2a8c5c72234c46d860839bb548586a2a8c5c72234c46d8e0d90f38225ee9a8306cc9
0d2fe9607de73692105b69a75066cd8d2db960770d9ee245c2d24e61e4922b6132120c3678531d094b3b85914b6e84c9081b
6c708917094b3b85914b6e84c9081b6cb0b638127a3b85816fb1f1251feceffc273190cfbb310c7b8b0d2fd840770deee110
9d776314e82536bae002cdcd6d0db07d6fa7b069b7b86c3fec0dec7b6da7b0690f743ed80c617f70ef3b7cb289af746318b7
e406987cb0bf5939cd51da296c6ec96d8eb44fccd0e11784084b3b85926b6e84c9081b6c709310094b3b85914bae84c94830
b86ba7b0b90f343ed81c699f9861d74e61731fe85cb039d23e31438337fc90b0b45318b9e446988cb0c1063fa04858da298c
5c72234c46d86083f75991b0b45314f23537c264840d36e86f43c2d24e61e49217c264840c9effe42a1396820a2397dc0893
1136d8e08d7d242c0515462eb91126236cf0f44612c22bed14862db9e1251daceffc479599b0b45328b9e646988cb0c1857c
5a1be5329dac7fa77f8eb7cf5fa3e6f3361dddafe43472b0910bf9916d0e81b0b45318b9e44a988c04830dd67143c2d28d61
e4921b6132c20677dd1836f781be099b23ed1333ecba316cee037d133647da276638ffb17c262ccd1c462eb91126236c70d7
cc61731fa89cb039d23e31438f8a8a3d6129e728e46b6e84c9081b3c7bbd44c42bcd1c862d79c14b3a485f873f3811965e10
2397dc08931136d8e10b86084b2f88914b6e84c9081bec70184184a51744c93537c264840d9e7dc48c78a514c4b02557bca4
23e86b560a62e407aa2e8c301961830d2e85206129053172c98d301961830d2e72236129053172c98d30196183e72e5b4f6c
a50ec4982537b6e482dd9dfd5600e2952e10c396dcf0920ed6d7e0f12e242c5d20857ccd8d301961830d1ee943c2d20562e4
9217c264840c9efd0c27e2951a12c396dcf0920ed6d7e00518242c3524462eb91126236cb043ed0d117a1989824b6c7cc907
fb6bf08e1bf159198951afa9c205176caec1ab9cc0e6652486bcc506c726c85a83bb94c4e6652486bcc506174ca0b70e0d2c
4057ca488c5a72e30b36d0dd72f6393d4d3a7709fc0e2aeeafe91fe2f8eff0fd1e7f6e9f6b59010d1c6ce072f679608b4380
2b4d24062db9d125192cafc162e24858ca488c5c72234c46d86083bb1648582a498c5c72234c46d86083cb8a48582a490af9
9a1b6132c206cfbe8e8c78a58fc4b0252f784907e95bb06fbf4789a7f5bdbe97fbbc28cabcd8d77c81645b330c470e367201
3fb2cd2110963e122397dc08931136b8eb23d1b98f746dd81c699f98a1c1356c242cad24462eb9122623c160831f15242cad
24462eb91126236cb0c1f32e48585a498c5c72234c46d86087860a222cad24462eb91126236cf0fc151999b0b49218b9e446
988cb0c106e74748585a498c5c72234c46d8608367b591b0b49218b9e446988cb0c1e502fef5f61a7f6c3ef387f1337e193d
efd31b7897f158fbf579dcd7c3661c39d8c8e522fe916d0e81b0b49218b9e446988cb0c1062fed2261692529e46b6e84c908
1b3cfbaa2ee2955612c396bce0251da4afc102f248585a49945c73234c46d820f5345c5ff7dffda6dff1c7f37dbb5ec6f3db
f71dc8b791838da4a686b4cd211096561223975c0993916010c85fe3c1ce7dfc52f85da97a8edf4feffbf81d8c1515ebc8c1
460279dce610084b2b89914b6e84c9081b04f2e77826fb18cfe47e5fc2e357f1e3f279ad47ae3c70d081c09db63804bc5249
62d8921b5ed2c1fae66e0e9aa3f482d8dc92db1c699f9801943fd60bde3f41bf550c6edff50c91470e3612a4c76d0e81b0b4
9218b9e446988cb0c10effcc88b0b49218b9e446988cb0c10e5fad44585a498c5c72234c46d820fd9c5e7f67b2e39fe5b797
dfe7ed3e1dfce2cfe93272b091f4739ab63904c2d24a62e4921b6132c206cf3e7e42bc524952b0d7dcf0920ed6d7e0de3012
964a122397bc10262364b0c1491212964a12d963cb8d30196183e72f7dc284a592c4c82517c26424193cff620813964a1223
97dc0893113678fe8b9d4c582a498c5c72234c46d8e0f9977b99b0549218b9e446988cb0c1f31f5367c2524962e4921b6132
c2064f2fe620bcd24762d8921b5ed2c1fa16ece7fdfbb84cff20bfd353f59fe9d5e3d77b5ed264bdba0fe3061bb740ff7b7b
43602b6524c62cb9b12517ec6ee9a7f87b52e3b78fb7c76bfed34cff2ebf7f4fe42dd43872b0914b3fc5916d0e81d0cb480c
7c8b8d2ff9607f0bf7e372795da72bd1dfdbfbfa1dbffba775a77e0b507d1fdfedfb1c070e3a70a13eb0c581e1bc89c498b7
d8d8820a14b710cfcbcafd8e2abee351c5ebf1783f6fd347b3be8a8023071bb9301fd9e6c074de4452a097d8e8820b34d760
f9116c10294d247ad4a9f97df4399effdca65cb69372d9cef947a94c589a488c5c72234c46d8e074b90be7285d2036b7e436
47da276698e7a6394a1b47997bcd6d8eb44fccd0e0e155242c6d1c462e79214c46c860837351242c6d1c462eb91126236cf0
ecb56611afb47118b6e4869774b0be05fbb7d4f4f5363d5ffbbe5f5f8febefb1a0dfa5c1c7f403b85e5be591838d5cc08f6c
730884a58dc3c82537c264840d36b8438b84a58dc3c82537c264840d3678d70a094b2187914b6e84c9081b6cf0ae15129642
0e25d7dc0893113678f65565c42b6d1c862db9e2251d415f830616242c851c462eb91126236cb04335041196420e2397dc08
931136d8a18a84084b1d88914b6e84c9081b9cab48688e52c851e65e739b23ed133374f8bc1161e9e43072c90b6132420667
eb344769c5b0b925b739d23e31c3e95feb84572a310c5b72c34b3a58df82fd7a8d7bf39ece22c663cadf3b6397e968e3f75f
bfddeff22c048d1c6ce4027e649b43202c9518462eb91126236cb0c1b11b1296560c2397dc0893113678f6c20888578a3114
5b73c34b3a58dfd9e7a588575a310c5b72c54b3a82be06172290b0b46218b9e446988cb0c10e4d2844585a318c5c72234c46
d860830bad48e8ad1806bec5c6977cb0bf0697d689cf5b310c7b8b0d2fd840771d2a5080ce5b310af4121b5d7081e63ab444
009d777218f41617bae002ccd54e0e9bf640d784cd10f607f7bec3bf2be22b9d1cc62db901261fec6fd7c961731f689ab039
d23e314383b79d90b0747228b9e646988cb0c1b91d82e628ad1836b7e43a47daa7c0d0e0b713094b2787914b6e84c9081b6c
f07943c2d20862e4921b6132c206e7cf1bcd513a396c6ec96d8eb44fccd0e0bb15094b2787914b6e84c9081b6cf01b8e84a5
93a390afb91126236cb0c1129c48583a398c5cf242988c90c1066f6d2361e9e43072c98d301961830d167842c2d2c961e492
1b6132c2063bf4b01061a9e53072c98d3019618367af6e8578a59343b13537bca483f59dbda434e295420ec3965cf1928ea0
af41090b1296420e2397dc0893113678fe63e94c58ea408c5c72234c46d8e0ae0ec4e63e50726173a47d6286064ba92161a9
033172c98d30196183bb3a109bfb40c985cd91f689191aac2f8584a50ea490afb91126236cf0ec65dc10afd48118b6e4052f
e9207d0dd6ed43c2520762e4921b6132c2063b7cd489b0d48118b9e446988cb0c10e3f684458ea40945c73234c46d8e0d9c7
6e8857ea400c5b72c54b3a82be06eba82161a9033172c98d301961830dae452061a9033172c98d301961830dae772261a903
3172c98d301961836dea408cf940cf85b12517ecaec153ea48581a418c5c72234c46d8608367b591b0348214f23537c26484
0d36789412094b2388914b5e08931132d8e0d15d242ca520462eb91126236c703a37c4394a2d87cd2db9cd91f68919ce7e88
14f1bc9343a92536b82483e535789092f8ac93c3a8d754e1820b36d7a18664cfe69d1c86bcc506c726c85a83bbe2c4e69d1c
86bcc506174ca0b706372889ae747218b5e4c6176ca0bbf3179661be52cb61dc921b60f2c1fe1a9c052261a9e53072c98d30
196183e72fa9c384a596c3c82537c264840d9e7dc103f14a2747c15e73c34b3a585f839b5548586a398c5cf242988c90c1b9
9c82e628c51836b7e43647da2766d81563e8dc47ea1e6c8eb44fccd0e09f381296620c23975c09939160b0c1a31748588a31
8c5c72234c46d86083c76d90b0146318b9e446988cb0c106c76e48588a318c5c72234c46d8e072a8fe39afe518094b318691
4b6e84c9081b5c2e6bfe7d1d4dff0c9fbf75be1ed33a50efebe3f65bc9e0b192e3c8c1462e97368f6c730884a518c3c82537
c264840d36587c04094b3186914b6e84c9081b5cae3e9cd8018b84a518c3c82537c264840d36b8c48784a518a390afb91126
236c1016f31e3f86d7cff335bd8bf81c7772bac97a05f26de460236131efb8cd2110966e0c2397bc10262364b0c102ee4858
ba31945c73234c46d820908f1fc5fbed76fdfc3e99cfdbbcc4e1832ee56f23071b09e4719b43202cdd18462eb9122623c120
90dfae97f1d0f33d1d763e2fb7dfedecc705eb0ad691838d04f2b8cd2110966e0c2397dc0893113608e4d7df37e1f8abf3fb
051a7f7baed345612cc758470e3612c8e336874058ea318c5c72234c46d8e0ae1ec3e63e50fa6073a47d6206b23e1ed13fef
afe9d9a4d1d3bc6c285625ac23071b49d6d336874058ea318c5c72234c46d8207dc7fc96c6bd5ca625efc76f82db6f31ea3b
fda26e23071b49df31699b43202cf518462eb91126236cf0f4161ac22bdd18862db9e1251dacaf432b0811966e0c2397dc08
93113608078fd7f932f47c5ff1f27a8fbffae3d10f1d3cae23071b09078f719b43202cf518857ccd8d301961830d4e1890b0
d46318b9e48530192183672f808278a51b4376d772c34b3a58dff9b72e99b0746318b9e442988c2483cdba318cfc40e38311
26236c70b9e2fbf706ec6f2f5f8fdbe575793fe605831fe38773baafbf5cf1a591838d5caef81ed9e610084b3786914b6e84
c9081b3cffbe0e13966e0c2397dc0893113678fed3ea4c58ba318c5c72234c46d8e0b92511c4568a318c5972634b2ed8dd52
1d301edc3c9fd337f0f7f65b7ce93bdf85f87d40afe337d4fa33ce23071bb954071cd9e610084b3d86914b6e84c9081b5cc8
dfafdffb23bf9d1c7f6b1ee36fcf6bfe4b3de69b6eeb4ba53470b0810bf7812d0e01cfbb318c7a8b0d2ec960790bf4e77b7b
dfefaf691fc7cfe7ef0addf41fc68fe9781efd7b3d6fa1a691838d5cb08f6c73603eafc730ec2d36bc6003ddfd418f8751cf
f7fcecc0f8eff17a7dde1fd39b79e33fc9d7af6a603d48e791838dfc833eb4cd81e9bc1ea3402fb1d1051768aec1aa6d586b
51ea31f4c0d3f3cbb8b5d79cdb7652be6ea7c1ed2b242cf518466eb91026236cb0c3df9c084b3987915b2e84c9081bec5190
b1272cd520855c72214c46d8e0f9a7654c58aa418cdc72234c46c860874f3b11966a1023b75c08931136d8e11b8e084b3b88
915b2e84c9081b3cf58d2b642bbd20c66cb9b02517ecaec1cab348587a418cdc72214c46d86083c73190b0f48218b9e54298
8cb0c1066fd92161e9055172cf85301961830d5e8341c2520d62e4966f84c94830d8610d79222cd520466eb91026236cf0ec
f7bd10aff48218b6e5829774b0be0e5d3044587a418cdc72214c46d8608367ae90b0b4921472c9853019618374c3f0f5bd7f
5ff7e962c9eb7a7bbec77fa8f4e0d13670d08174bb306c710878a592c4b02d37bca483f4ed2a496cee035d1b3647da2766e8
f06123c2d24a62e4960b6132c20667eb3447e905b1b92d9739d23e31c3e9dfe984574a410cdb72c14b3a585f83450290b0f4
8228b9e742988cb0c106076e4858aa418cdcf28d30190906cf2e8241bcd20b62d8960b5ed2c1fa1abcbd8d84a517c4c82d17
c264840d36b8fe8284de0b62e01a0b5ff2c1fe1a5c6b243eef05316c8d052fd840770dde03223aef0529d05b2c74c1059a9b
fb1a60fbdecc61d36a6cdb0f7b03fb5e9b396cda038d133643d81fdcfb06abc7235f69e6306ecb0530f9607f1d3ede44587a
418cdc72214c46d8e0fc61a3394a3387ceedb9cc91f689191a7c932261e9053172cb37c26424186cb05a1c12965e1023b75c
08931136d8e16f4e84a517c4c82d17c264840dee7a416cee036d173647da2766e8f07923c2d20b62e4960b6132c2061b7cb3
2361e90529e4920b6132c2061b54b92161e9053172cb8d30192183e73faacc84a517c4c82d17c264840d3658a201094b2f88
915b2e84c9081b6cb0ce1412965e1023b75c08931136d8609d29242cd5204aeeb91026236cb041170c1296761023b77c234c
4682c1b3bb6010af548318b6e5829774b0beb397d742bcd20b62d8960b5ed2c1face7f0d83094b2b89915b2e84c9081becb0
743e1196561223b75c08931136d8a18f85084b278a915b2e84c9081bec50914184a513a5904b2e84c9081becd48962d8079a
3e0a5ed241face7fd78a094b278a915b2e84c9081becf05127c2d28962e4960b6132c2063b7cbd1161e9445172cf85301961
830d7ecc91b0d4a218b9e51b6132120c36385c47c2528b62e4960b6132c2069bd5a218f981b20f234c46d86083cb6f48586a
518cdc72214c46d8608307d691b094a318b9e542988cb0c1068f6d236129473172cb8530196183673fee8578a519a5604b2e
784907eb3bfb714ec42bb528866db9e1251da40fb02fd7df8ac5dff7b438cff779bb4f7747887b1d39d848008fdb1c0261a9
453172cb853019618344fef7053cad5174b9fe2dd682e4cbc8c1461279dae610084b298b915b2e84c9081b3cfda34e785eca
a2d4160b5c92c1f21a3ca24f7c56ca62d4926e70c1059b5b7ec3c6ffd7f53e2dc9f27ececf974ee78fafd7eb329e5adcbfeb
22663872b091cb6fd8916d0ec8e6a52c86acb1c0b109b2d6e0414662f3521643d658e08209f4d6e08d69a22ba52c466db9f0
051be8aec1ad61e42ba52cc66db900261fecafc199181296521623b75c08931136d8e006291296521623b75c08931136d8a0
b101094b2f4b21975c08931136d8e0d63012965e1623b7dc08931132d8e16f4e84a515c6c82d17c264840d36780404094b27
8d927b2e84c9081b6cf0dd8e84a593c6c82ddf08939160b0433f0911964e1a23b75c08931136d8e0c146242c9d34466eb910
26236cb0c1c22f48583a698cdc72214c46d8e0d94bcf225e29a4316ccb052fe9607d0d5e2d44c2524863e4960b6132c2061b
5c7f40c2524863e4960b6132c2061b5c6843c2524863e4960b6132c2061baca68e84a590a6904b2e84c9081b04f2c7753cb7
f9dddcfdede5f82bf4b88fdf55f437df460e3612c8e3368740580a698cdc72234c46c86083b208242c85344aeeb91026236c
704ffefc5e3ef7c7f81f7ec79c8fdf330ddfdf1b2e40be8d1c6ce49e3c6f730884a590c6c82ddf0893916010c83f97df9bfd
afdf11c7e3f97a3c9fb7cb7ae48a03071b08dc698b43c02b6d34866db9e0251dac0fb0c773b9cbf5f79ef5b49797d77b3ce6
fd50ebd43672b091001eb73904c2d24663e4960b6132c2064fff83135e29c2316ccb052fe9607d1dfe85136129c23172cb85
30196183f0ad7e798edffdbf53cbdf5e7ec6efe2f158fb457ff16de46023e15b3d6e730884a508c7c82d17c264840d76a800
22c2d28563e4960b6132c2061b54cb2161e9c23172cb85301961830d1e0a40c2d28553c82517c264840d3658b90f094b178e
915b6e84c908196cb0422712963a1cd9e3920b6132c206cfbf7fca84a50ec7c82d5f0993916470b9d87c797f6f97cbdf8b23
8ff148e3fb9e2e128ec71f8ff160f7b2fecd71e46023978bcd47b63904c2528763e4960b6132c206cfbfa9c484a50ec7c82d
17c264840d2e8f6e8fbb787dcfef007f3e8ff170eb327f4cefeff1dfeaeb725fc971e460239747b78f6c730884a50ec7c82d
17c264840d2e2f2adc7ef7f7a71b40e337d3f7f178dd5ff3dda0f996c8765309470e36727951e1c836874058ea708cdc7221
4c46d8e042fefe3c1fafe9a6df773bc4fa3d6d3e3f8bf9585fb5c491838d5cc88f6c730884a514c7c82d17c264840d362bc5
31f2037d2f46988cb0c1857cbe0b3277d8dc7ecfc5bd9f53c7c9a5de48c491838d5cc88f6c730884de8b63e01a0b5ff2c1fe
3a54c4009ff7e218b6c682176ca0bb05fafe19cf1ae7fd7bffddc89f3f9eeff113bafea5ff316c5870ffb9b581a1bc0ea7b0
6eb140050528acc193f95863b3abc3d90e3c35bffd77bdbceef73997eda45cb673f63d43c4db75e108b6e4869774b0be062f
a120e1ae0b47c82537c264840d36589d1209775d3846bee646988cb0c1f3cfc99870d78523e49217c264840c36585b0c0977
5d38422eb91126236c705a5b0ce7d8b5d1c8dc92db1c699f98619e9be6d8b5c2c8dc92db1c699f98a1c1433048b86b851172
c98d301961831dbedf8870d70a23e4921b6132c2061bbca88084bb56988d5c73234c46d8e0d94b6c23deae1246b02557bca4
23e86bf02e0e12ee2a61845c72234c46d8608775f48970d70a23e4921b6132c2061b3cf28584bb56182197dc0893113678f6
1b8688b7ab8431ec3537bca483f511f67cc9777ed1f9f21af7f1fe79be907b1939d848024fdbdcb5c208f981a69542988c90
c106cff821e1ae9346c82537c264840dcecd2c34c7ae1546e696dce648fbc40cd00a23731fa83bb139d23e31c3e9ffc4096f
d70a23d8921b5ed2c1faa882e83d9e445e5eafe94379fbbbd985b7c6d791838da40aa2b4cd5d2bcc467ea4ebc40893113678
f6ba1088b7ab84116cc9152fe908fa1a3cb18e84bb56182197dc0893113678f6913ae2ed2a61045b72c34b3a585f83cb8d48
582b61047c8b8d2ff9607f0d2e43105fad8411ec2d36bc6003dd35b8da4674b512c6a097d8e8820b34d7603d2ba2ab853402
bdc5852eb80073fb421a99f640d18acd10f607f7be5d218d701fa85931c0e483fdcdca698e5d258ccc2db9cd91f689193a7c
c689705748b3916b6e84c9081b6cf0362912ee0a69845c72254c4682c1065fe648b82ba41172c98d301961830d6e9721e1ae
9046c82537c264840d36386a41c25d1d8e904b6e84c9081becf03727c25d1d8e904b6e84c9081b9c4b88688e5d218dcdbde6
3647da2766e8f01d4384bb421a2197bc10262364b0c12f2a12ee0a69845c72234c46d8608395a69070574823e4921b6132c2
063b941011e1ae9046c82537c264840d36588106097785341bb9e646988cb0c1066b6c21e1ae9046c82557c26424183cffa5
0026dc75d208b9e446988cb0c1d3ab78086fd74923d8921b5ed2c1fa4eafe221bc5d218d604b6e784907eb3bff151026dc15
d208b9e446988cb0c10e553c44b82ba41172c98d301961831d5a3a8870574863e46b6e84c9081b3cffb51f26dc75d208b9e4
8530192183d37912ceb16b8591b925b739d23e31c3e9354084b7ab84116cc90d2fe9607d1dfe9911e1ae126623d7dc089311
36d8a00c0709779530422eb9122623c1e0423e3f703aede5efbdf6dbe5afc5663c939e6e64af5f303872b0910bf9916dee2a
6184fc40d1891126236cb0c1a90212ee2a61845c72234c46d860834b0248b8ab841172c98d30196183cb05d7fb7b3cc89aff
22f27f5fd6875896cbadfb71838d5b2eb6fe7b7bbb3218613e5071626cc905bb6bf0ca1912eeca60845c72234c46d8e0d90f
fd20deae0cc6b0d7dcf0920ed6d7e0414e24dcf5c108b9e48530192183b0e8cdf7f57eddc6d3b7dfdf6afce33c9e97dbf58a
0b1dad23071b098bdec46deefa6084fc40cb891126236c703a5ac639768d2c32b7e43647da276698e7a6396a2dca36b5c436
43da232668f04927be528b22d46baa70c1059b3bfb3b0dc06a278af06eb191b10652d6e0014662ab9d2882bcc506174ca0b7
b35f9526b45d218a204b6e7041058a6bf0361df2ed0a51845b72034c3ed85f839bb248b82b441172c98d301961830d6eca22
e1ae1045c82537c264840d3678ac0e0977852846bee646988cb0c10637659170578822e49217c264840c76f89b13e1ae1045
c82537c264840d76f8774e84bb42948d5c73234c46d86083ef7624dc15a208b9e44a988c04830d1e6d43c25d218a904b6e84
c9081b6cf0763c12ee0a51845c72234c46d8e0d9efdc20deae0d45b02537bca483f53578471c09778528422eb91126236cf0
ecd728116fd78622d8921b5ed2c1fae832d3f57dfd8c7fa8cf7449e83edf4fa7a5e3b791838da4cb4c699bbb3614213fd0f1
6184c9081b6cb0aa3412eeda50845c72234c46d820ad24fe1a8fb56ed7cfef87e8399edebcaecfeb835e9ddd460e36925612
4fdbdcb5a118f93f3b3e8c301961830d568f47c25d1b8a904b5e0893113278765102e2edaa50366ccd0d2fe9607d7fd8d313
41afa958f1b752e8e372b9dd7f7bfc280f96f1c04107fe611fd9e2ae0745b00f147c285ed211f42dd8cff7e332fd651eafcb
f7fd7a3ea63fcdedf29afe501bf57edc60e316e87f6f6f578222cc07aa3d8c2db96077d3a57b9c63574322734b6e73a47d62
86d33f6684b7ab21116cc90d2fe9607dcb97ca753cb69bbeec9ee3df653c817a4f07768ffaa4090e1c6ce0f2a572608bbb0e
12c13ed0ac61784907ebebd0be4284bb0e122197dc0893113608bf9fb7f15331fe452ed31febf2797cc753582c135b070e3a
107e3dd316770524827da056c3f0920ed6d7e0501109770524422eb91126236cb0c15d6824dc159018f99a1b6132c206cfbe
398978bbf611c196bce0251da4efecfb7588b7ab1e5977d772c34b3a58dff917379970573d22e4920b6132920cb6ab1e11f2
03851a46988cb0c1e52da3f197f5759dfe097edebfd5be7f6bd1fcfec375fdc7b9bc654423071bb9bc6574649bbbea11213f
50a86184c9081b3cffed0b26dc558f08b9e446988cb0c1e5b1f0c7f5f57b11e6b7cb97df2efe2a4b7f7bf92a97bb78e46023
97c7c28f6c73573d22e4070a358c301961834b0dc5e53efe215e735bc2fca2c46bfa988ebf05b7dbedb5fd3bc791838d5c6a
288e6c73573d22e4070a358c301961830bf963be04f5dbcbdbfbfa1dbf12a6c7793f7365ebf66c340e1c74e0c27d608bbbde
11c13ed0a761784907eb5bb0dfafdf1b0bbf7d1c7f6e1ee3cfcf7409fe7b79cc8faeafaf12d2c0c1062ed807b6b82b1d11ec
7ff669185c92c1f296fe8debfd33fef6fcfe18e3a1c56b7eb677fa78feaa426ebfc5be960a0e1a39d8c8a585e3c8366be988
60ffb35ac3f0820d74d7a474c45843bbc8121b545080c29687e8aef34fecffe607032ef7f138727a5afb79ff8e7fa4fbda23
c423071bb93c4477649ba57464dde1523aa2879c9a5fffbb7ec733cc3997eda45cb6d3e0d977242cbd23462eb91126236cb0
c15d68242cbd23462eb91126236cb0c1938448587a470af99a1b6132c206cf3f2163c2d23b62e49217c264840c3678c20809
4bef88914b6e84c9081bdcf58ed8dc07ba3c6c8eb44fccd0e08936242cad27462eb91126236c70b64e7394de119b5b729b23
ed133374f8ac1361e91d3172c98d301961830d9e2943c2d23ba2e49a1b6132c2061b2ceb8b84a57ac4c82557c26424185c8e
634e6c7646c2523d62e4921b6132c2061bbcfa8284a57ac4c82537c264840d9e7d1b01f14aef88614b6e784907eba3d5f9e7
fdbdfdfe52e3b1e7fd35fe207dd63f388e1c6c24adce9fb63904c2523d52c8d7dc08931136d8a18083084bf588914b5e0893
1132d8e05e211296ea112397dc08931136d8e0214a242cc527462eb91126236c70577c62731f68f4b039d23e314387cf1b11
96ee132397dc08931136d8e13b86084bf789926b6e84c9081b6cf0bb8284a5fec4c82557c26424185cc85faff143f89e8e6a
c743bbf1f7e772990e777ea5ccb7fbfdbdbd1f4e23071bb9901fd9e610084bfd89914b6e84c9081b6c70f08884a501c5c825
37c264840d36386140426f4031f02d36bee483fd9dbab40d9179f789016fb181050f68adc102e244e7dd27057a898d2eb840
730dda1188cebb4f0c7a8b0b5d7001e61a5c6524bad2bc62d4921b5fb081eee6fe119aa1749fd8cc92db14698f98a0c12546
242ccd2b462eb91126236c70b64e7394ee139d5b739b23ed133334f8d78d84a5fbc4c82557c26424186c70e30a094bf78991
4b6e84c9081b6cf0338684a5fbc4c82537c264840d36b84d8b84a5fbc4c82537c264840d9e7aa4866ca5f5c4982537b6e482
dded5a4fcadcffecf2b039d23e3143874f1a1196d6132397bc10262364b0c10d33242cad27462eb91126236cb0c32f0a1196
d6132397dc08931136d8e0d80d094beb89914b6e84c9081b3cffa97c262cad274aaeb91126236c7021bfdfa6e5b4a7bdfc5b
c57dfa6a1a7fffc77fa097edd9741c39d8c885fcc8368740585a4f8c5c72254c4682c1068b382361693d3172c98d30196183
e7bf89c184a5f8c4c82537c264840d3658b21c094bf789914b6e84c9081b6cd67d62e4071a3d8c30196183e7bf7dc384a5fb
c4c82537c264840d76286520c2d27d52c8d7dc08931136787a0b07e195e213c396bce0251da46f577c62731f68f4b039d23e
31c33c37cd51da476c6ec96d8eb44fccd0e1834e84a57d44c93537c264840d36f87243c2d23e62e4922b6132120c9e5ddd86
78a57ac4b02537bca483f535386a45c2523d62e4921b6132c2069b558f18f9815a0d234c46d8609bea11633e50a8616cc905
bb3b7b4148c42bbd23862db9e1251dacefece75711aff48e14ec3537bca483f535686340c2d23b62e49217c264840c367896
11094bef88914b6e84c9081b04f2cfefb2f07d6a8b7dbe7fedc08f9b7cd461e06003813b6d710878a5f2c4b02537bca483f5
11f6f575ff7d2ddda7bdbc5d2ff7cbfd4dab596d23071b49e0699b4320f4c2150597d8f8920ff6d7e0834e7c56b862d46baa
70c1059b6bf0dc28b079e78a216fb1c1b109b246cf8bbe2fb7cfe5f59ad71afbbb024e4b9d6e23071b49cf8ba66d0ecce69d
2b86bcc506174ca0b7068b38135da95d316ac98d2fd84077cb73a2e307ff7a7f4caf743fe79dba4e4f36be5e97fbeb7aff2e
c7e33c72b091cb73a247b63904be52bb62dc921b60f2c1fece5eb40bf14ae78a614b6e784907eb6bf026211296ce152397dc
08931136d8e03100242c9d2b857ccd8d301961830d1e0340c2d2b962e49217c264840c9e7ad28d6ca56dc5982537b6e482dd
4dd770718ed277a2736b6e73a47d628606ab8e2061e93b3172c995301909061bbc768084a5efc4c82537c264840d3678d504
094bdf89914b6e84c9081b3cfb3a0fe295be13c396dcf0920ed6d7e03a0f1296be132397dc0893113678f6d50ec42b7d2786
2db9e1251dac0fb0c7b3c6cbf7f2be4f3d158fdfcdd4cbf8f54cdcebc8c1460278dce610084bdf89914b6e84c9081b3c7bed
6ec42b6527862db9e1251dac8f166b7f7ede9f71ef7e3fb98ff15fe2e5f67b7500b8b791838da4c5dad336874058ca4e0af9
9a1b6132c20617f2df7589dbdcc2f27afeae167ca713bac7a31c2af3c8c1462ee447b63904c2527662e49217c264840c2e75
0c8fe7e5fbb9ccff06afe33f88c77b3ae67a5c9fe337d37832bdbe198e23071bb914321cd9e610084bdf89926b6e84c9081b
5cc82f5383c2f4bd3b9ec85e6fe3c9cd541b71b9cc0b38aee03470d0810bf7812d0e01aff49d18b6e48a9774047d7fd8afef
6f0dbd69f9fcf960f23edd2fbe7fc763aceb7802bf5ef9a181830dfcc33eb2c521e095ca13c396dcf0920ed677fa5f9bf04a
db8a614b6e784907eb5bb0efb7df45c7e9a7e6feba3d7f376da72fa1eb783c3dfe2b5d973ce591838d5cc08f6c730884a570
c5c82537c264840d2ee4dff1ec71fc169ebe7e2f7f378d7f7b791f3f92bf1fc1f59e058e1c6ce4427e649b43202c9d2b462e
b91126236c90faab9ee381c6efe6ed6f2f5fb7dbfd3dfea3fcd2cff83a72b091d46095b63904c2d2b962e4921b6132c206e9
b8ed7af97ea725be7f4759b7dfb3488f0b9da26c23071b49476e699b43202cb52b462eb91126236cb059ed8a911f281331c2
64840dd28dbaebef21ee71e7e6a3cddb7d7a0f166fd42d23071b4937ead2368740586a570af99a1b6132c2061b2c738284a5
79c5c8252f84c908196c70bd11094bf98aecb1e546988cb0c1f35ff362c252be62e4920b6132920c9effb2131396f2152397
dc0893113678f2d3b28c579a570c5b72c34b3a58df82fdb98cff0ae677f0bed343fb8fef77fe4b79651e8f1c6ce4027e649b
43202ccd2b462eb91126236c707924fe799fde40fcedf2e53b1e554e479c9fd77bfc4afadd745a1f88df8f1b6cdcf238fcbf
b73704b6d2b962cc921b5b72c1ee9a75ae18f9813211234c46d860a7ce15c33e502762784907eb5bb03fe3beddefbfef9def
7dfc84fe1e019bfec3f8411d4f687f6f012edc3472b0910bf8916d0e81d06b570c7c8b8d2ff9607f4b03c9ed751dbf68a76a
98c7dc8034eff0753c71be5ec7e3ac851b470e3672e92139b2cd81f9bc76c5b0b7d8f0820d74d7a276a5b0867e952536a8a0
0085d17db2d065bf3bcb58070e3a90ee92852d96ce95756f4be78a1e696a7ef9eff61eff81dca75cb69372d94e8327e990b0
74ae18b9e446988cb0c106cf002061e95c3172c98d301961830d9e2a43c2d2b952c8d7dc08931136d860a111242c9d2b462e
79214c46c8608357b990b074ae18b9e446988cb0c1857c3a5a9efacc3ebf3fc9f5315ffd7dbfa6172b1fdb09098e1c6ce442
7e649b43202c8d2f462eb91126236c70ee3da1394ae78acd2db9cd91f6891916eb9ff147e7312f54707f8d1f8de7ef9dc6e9
45b0e7753cc4df6a2871e4602317eb47b63904c2d2f862e4921b6132c20667eb3447e95cb1b925b739d23e314383d53791b0
74ae28b9e646988cb0c106dfec48583a578c5c72254c4682c1539f9a45b6d2b662cc921b5b72c1eece7e5515f14ad58a614b
6e784907eb6b70b08e84a56dc5c82537c264840d3678cb09094bdb4a215f73234c46d8608337da90b0b4ad18b9e485301921
830d5e5f44c2d2b662e4921b6132c2061bbcb08a84a56dc5c82537c264840d02f9fb357e1d7d6fd3f3d5cf57596197470e36
12c8e336874058ba5e8c5c72234c46d86087bf391196ae172397dc08931136d8e1df391196ae1725d7dc08931136d8e0bb1d
094bd78b914bae84c94830d8a1e586084bd78b914b6e84c9081b6cf0f8071296ae172397dc08931136d8e0ed6524f4ae1703
df62e34b3ed8dfd9ef2e139cd7bd18f3161b5b5081e21abcbc4b745ef752a097d8e8820b34d7a01c80e8bceec5a0b7b8d005
1760ae410b06d195ba17a396dcf8820d7477fef35ccc57ca668c5b72034c3ed8df5cb9427394ba179b5b729b23ed133374f8
b01161299b5172cd8d301961830dae6e2261299b3172c99530190906e72a0a9aa3d4bdd8dc92db1c699f98a1c1cf181296ba
172397dc08931136786ee909b195a2176396dcd8920b76d7e0e964242c752f462eb91126236c7057f752e6fe678989cd91f6
891996b9f77394c2159b5bf23247da276268f01c041296c2152397dc0893113678eebf6f622b552bc62cb9b12517ecaec36f
281196aa152397dc08931136d8e06815094bd58a926b6e84c9081b6cf0ec071296aa1523975c099391607079efe2f9be8e3f
f0d3b339cfcbf8eff1b744c5fca5347e4eaecfed79551c39d8c8e5bd8b23db1c0261a95a3172c98d30196183cdaa568cfc40
81881126236cb059d58a911f281031c264840d2ee48ffbedfab7639ff90ecf63fe073afdf3fc3d54bf90d3c8c1462ee447b6
3904c252b562e4921b6132c206cf7fdd86094bd58a914b6e84c9081b5cc8c7ff7c7bce3b367e4cbff7df85d0692f7f3f4cb7
d77a219d470e3672213fb2cd211096aa9542bee646988cb0c14e552b867da0bea4e0251da46f57b562731fa82fb139d23e31
c3ae6ac5e63e50206273a47d6286796e9aa3949de8dc9adb1c699f98a159d989911f68f150c26424185cc8efd33b77d357c1
e8e73d3d2d307dfd5f1e8fefe7b61e33e3c041072edc07b63804bc527662d8921b5ed2c1fa96a3a7bfc73d7ffbf57a8ca7ac
97f763befdf0787d2fd3badbcbd1138d1c6ce472f474649b43202c6527462eb91126236cb0c113cb4858ca4e8c5c72234c46
d86083156791b0549e18b9e446988cb0c1b39f2345bc527962d8921b5ed2c1fa1a3c478a84a5f5a490afb91126236cf0ecb7
fc10af549e18b6e4052fe9207d0d1eb142c2527962e4921b6132c2061b3c508884a5f5c4c82537c264840d4e878d3887f78e
e8d412db0c698f98a0c1a38cc467bd2346bda60a175cb0b9b3d79a05302f1d31de2d3632d640ca1aacde446c5e3a62c85b6c
70c1047a3bfbb783d04ae388214b6e7041058a03e6fbef60e6f69eae8efcaeb98fdf02cf2b3ea1ba8e1c6c2450c76d0e81af
348e18b7e406987cb0bf06c74848584a478c5c72234c46d86083e73491b0948e18b9e446988cb0c1b34f0710af348e14ec35
37bca483f5cd1d143447e9fcb0b9252f73a47d2286e5c3365f679dee97bfe63b87d3bf8fd77b3e92baac0b19e3c8c1462e1f
b623db1c026169fe3072c98d30196183bbe60f9dfb489f85cd91f689193ab440106169fe3072c995301909061b3c7e8f84a5
f9c3c82537c264840d36f82947c2d2fc61e4921b6132c2061bbc668384a5fcc3c82537c264840d3638ef47c252fe61e4921b
6132c2063bb4601061e9ff3072c98d30196183a7b760105e29ff306cc90d2fe9607d1d5a3088b0f47f18b9e446988cb0c11e
2d187bc2d2ff51c8d7dc08931136b82c16ff7c3f2ed35fe7f1ba8c0779cfc7b4b8f97c87e9b95d748171838d5b168afff7f6
86c0569a3f8c59f2c2965c90bbb38b2010afd47e28b6e6869774b0be05fb3d7e1c7fd7802ebfbb97afdfaa6ed3f5b0fb7856
f5fc5e5fdff5b80d470e3672013fb2cd211096e60f23975c0993916070219f8e2da73bc5cfcb780af3b97ea68733ee9feb6d
fce37cd6858479e4602317f223db1c026129ff3072c98d301961830bf9f4b8fcf40eeff3fafa3dfcf5988e34eed38fedafbc
6125a791838d5cc88f6c730884a5ffc3c82537c264840d36ebff30f203ad1646988cb0c1063f664858fa3f8c5c72234c46d8
60830318242cfd1f462eb91126236c108e587f4b08fe1e1898ee755c1fb7df3ae3773c565f470e36128e59e336874058fa3f
8c5c72234c46d820f57f8c3f40efe77cf8311e783c9f9fdf1acc54fff13770b081d4fe11b63804bc52fe61d8921b5ed2c1fa
1a9c90236129ff28e46b6e84c9081b6c705701094bf987914b5e08931132d8e0652b242ce51fb2c7961b6132c206cf5f9491
094bf987914b2e84c94832b8bc84f2194f26de5383c1e7f91c477c3ef7f925b1f1dfc7fd75d916c9c791838d5c5e4239b2cd
211096f20f2397dc0893113678f2d3a38c57ca3f0c5b72c34b3a58dff9cfe4336129ff3072c98d30196183cb33f9afd7ed71
7ffffe0d8e7fa6f1947afcdd9d9e707e5f2f8feff3f7f4ebf2503e8d1c6ce4f254fe916d0e81b0548018b9e446988cb0c1a5
11e2af7de6b797b7c76b7e7b60fa948ebf3e8fcbfdb5be758423071bb934421cd9e610084b0588914b6e84c9081b5cc8dfaf
f1ff37d537dc1fe317d3e5fb9acb5a1ef3cb516bc30f0d1c6ce0c27d608b43c02b1520862db9e1251dac6f6985b85c9feff9
cbe7fbb85cafcffbe372fffb8cfe5d43598a2168e46023976e8823db1c02a1578018f8161b5ff2c1fe7a558018f63f8b2e0c
2fd840770bf4f6acc6ef9fdefc4bf3fb57b956a72ecc3070b0810bf2812d0e8ce645208578890d2d88406d673fca87051ea5
08448f3ab77c3cfebcbf1ee3f9d614af9b09b16da5c1fabac8576a408c7bcd1d30e8607b0d9e7542be520262dc6bee804107
db6b7036867ca502a470ffe50e1874b0bd068b40205f290031ee35af804107d96b56ff61dcff2cb570c0a083ed3578a90bf9
4af98771afb903061d6ceffcda4de62bc523c6bde60e1874b0bdf3af38305fa91d31ee3577c0a083ed3558ee03f94ae58971
afb903061d6c6fe1be8c7f85cbe5eff5e1e9a8e23d7d07bdc7fffe7834bdf54fe2c8c1462edc47b63904be5278a2dc5bee80
4107db6bf07d8e7ca5eec4b8d7dc00838e60afc122abc8572a4f8c7bcd1d30e8607b0d16a743bed27962dc6bee804107db6b
f0a032f295c613e35e73070c3ad8de723ef69817309f0e23d735867ef732a74577e42494060e3a70391b3bb0c521c095b293
02fd973b5d70c1ea1adc1644be527562dc6b5e01830eb277f61507842b2d2706bde64e175cb03a7a31f7fa7c5cdfd7e969bb
e7786c711f7f651f4fa25e470e36925ecc4ddb1c025fe93831ee3577c0a083edd12dfff100fe3efe35a6e7b0c653e7f7fdf1
bee25f7b1d39d848bae99fb63904bed27062dc6bee804107db3bfb5107842be52606bde64e175cb0ba0e1f72e22bd526cabd
e50e1874b0bd066f80235f293631ee3537c0a023d86bf05e34f2955a13e35e73070c3ad85e873a17e22ba526c6bde60e1874
b0bd068ff3209f579a18f6123b5e90c1eece7e4d96d0bcd0c48897d8d1d8034a6b70b19cd8bccea420cfb1c3b108b4d66049
2862f33213435ee20ac722c05a834b0bc456aa4c8c79cd1d8f55a0b70e951a44578a4c8c7acd9d2fc860770dee03215f2951
31ee3577c0a083ed75f890135fa95051ee2d77c0a083ed35f83e43be52a062dc6b6e804147b077ee42ff44566a5b8c78cd1d
2d88606f0dae94225f296d31ee3577c0a083ed2ddcbfcb4bd7db7443ee7dbfbe1ed7df9391bf73a5c7757ead6fe5a691838d
5cb88f6c73087ca5b8c5b8d7dc01830eb677f6c128c295ce16835e73a70b2e585d83c75690aff4c514eebfdc01830eb6b76b
8bb199ffd98152a7083b44fbdf609d1ae42b5d31c6bde60e1874b0bdb3cb2b11ae94c518f49a3b5d70c1ea3a147f115fa98a
31ee3577c0a083ed35382a46be5214a3dc5bee804107db3bff0d1de62b3531c6bde6061874047b0d9e1242be521263dc6bee
804107db6bb0c839f2958a18e35e73070c3ad8def27ace7496329db0bc7f37bdefafe998763ca49d5fb7585f58a081830d5c
5ece39b0c521c0957618835e73a70b2e58ddd90bf9235c298631e83577bae082d535ab8531ee7f969d3860d0c1f696d7b0c6
cfe3b486c86f1fc78fe4784a7e9f7678fa4eba3e9feb5f1b470e3672790debc83687c0574a610af75fee804107db5bb86fbf
c7b7a667f5c6b3b6efe3f1babfe607f7e66fe0edf93f1c39d8c885fbc83687c0575a618c7bcd2b60d041f6769d3036f33f7b
567c8ab043bcffbb46189bf99f2d2b3e45d821deff5d1f8ccefcef96139f22ec10effff9af77325f698331ee3537c0a023d8
6b502f867ca50ec6b8d7dc01830eb6d7a04a0ff94a198c71afb903061d6cafc153c5c857aa608c7bcd1d30e8607b0dee0621
5f298231ee3577c0a083ed35b8668c7ca509c6b8d7dc01830eb6d7e0e95ae42b3d3085fb2f77c0a083ed35785c05f94a118c
71af79050c3ac81e2d073bfeb28effffe902c5efcdf2e7e576bde2e368ebc8c146d272b0699b43e02b3530c6bde60e1874b0
bd068fa3215f298131ee3577c0a083ed3578fc10f9bc8046b1d7d8f1820c76372fac0edbb702189bf52fb5adf3aef07e9ffd
702b6079fd8bd12eb183a103d2d5a05289c8bcfcc58097d8d958033a3bfb097d022bd52f06bce6cec61e505a83275991ae14
bf18f59a3b5f90c1ee4e2ff921b8d2f962d06bee74c105ab6b500d807ca5f1c5b8d7dc01830eb6d7a00c02f94ae54be1fecb
1d30e8607bbbc2179bf99f3526758ab043b4fff3cc3443295cb199d7dca7083bc4fbbfab5bd199ff5d22e253841de2fd3ffb
5d3e842b4d2b06bde646175c04750d4eb190aff4ac18f79a3b60d0c1f6cefefd40b852b162d06bee74c105abeb5033427ca5
60c5b8d7dc01830eb6d7e0fc0af94abd8a71afb903061d6c0fb8c7dfb8f7742dfbb78fb7e7e771ff5e5ed831b28e1c6c2470
c76d0e81af94ab18f79a3b60d0c1f680fbf21cbff0bff7e95ae6e3f31cbf86afaf17febdd791838d04eeb8cd21f0957615e3
5e73070c3ad81e2c4dfe99aa30a61b818ff114f1f9bc5dbe84bd0e1c6c202c4c9eb63804b852ac62d06bee74c105ab5bd6e0
ff2dc5387edb4e67ddf3aa2f9fdf01ece36fe9fced0c08470e36725983ffc83687c0576a550af75fee804107db6bd03d807c
a55ac5b8d7bc02061d64af41c308f2956e15e5de72070c3ad8dec23dbdc430ad3af9d4b3f2f14ce57e9d9ef958b969e46023
17ee23db1c025f695631ee3537c0a023d86bd6ab62dcff6c0b71c0a083ed356b5531ee7f76853860d0c1f696efb5c7f845f3
b9cc856dd7f1a7eef19eeece3daee367e4757b6f076b3872b091cbf7da916d0e81af74aa18f79a3b60d0c1f6961fefebe77b
9d0e29c6bfcbef8ced3dbd4efca84f82e1c0c1062e3fde07b63804b852a762d06bee74c105ab6b70a4867ca54cc5b8d7dc01
830eb6d7e0c81cf94a958a71afb903061d6cefec0bf408578a540c7acd9d2eb860750d16ed45be52a352b8ff72070c3ad85e
83777190af94a818f79a57c0a083ec35788915f94a858aecb1e40e1874b0bdf3177165be52a062dc6bae804147b2b73c7e37
ffcff434e4ebf37e7ec61386e911c1f1ac69fc097eaeeba1f1c8c1462e8fdf1dd9e610f84a7d8a71afb903061d6c6f79dc72
dcc1eb7bfedb7c3e8fc778203fbf5d7e7f7fc6d387cb5a61cb23071bb93c6e79649b43e02bfd29c6bde60e1874b0bde521ee
eb783075fdfef6f17b993b2fa6c7439faff174f9f2b8af4f5bd2c0c1062e8f701fd8e210e04a758a41afb9d30517ac6e817e
8f3fa7e3e1d36f17b76eb6dfc3a1f36d8dc7fa520a8e1c6ce4827d649b43e02bc529c6bde60e1874b0bd66b529c6fdcf3e10
070c3ad85ea7d21483fe671988d30517ac6e69d1b8de3fefef741de83beed57cb0317d3a7f4527b7ed9c9b470e3672e9d138
b2cd21f079658a612ff1ff63ee4c9265c7916c39c72a620521ec9b3de5fec70580040d8a7bac9ca3204abefcca676509a7ea
75670be210cf91a1dd15eaf988d74657a64b5a347a1af24cef739af6697a6e23fe6a0b85f7e7684143312705aca54c286d40
eaea2825a5e175e250ae32c9b405a9ecebf96e2adc04112938adaccac7bfcb3ac43d622edb204eb91aa483b569151df25108
6d65d0392ea4b90e96b75374084721b49541e7b890e63a781940d12119a5852e65d0392ea4b90edefc5774884521b4951b3a
c78530d7c16b3e8a0e992884b632e81c17d25c0f0919820e812884b632e81c17d25c072ffc2b3aa4a110dacaa0735c4873df
df3e9074884221b49541e7b890e6f2343f353e9248f8b156c6f8ced6c86defe15725e8104302e8aa0c3ac78534d7c1f2218a
0e192484b6724de7b8d0e67a485e1174082021b49541e7b890e63ab8f3ace8903e42682b83ce7121cd7dbdb29c4243ee0889
ad0c344784d4d6c17c784587dc9116ba9441e7b890e63ab8b25374081d21b4951b3ac78530d7c13b5c8a0ea12384b632e81c
17d29c808e7bd6e11cf6f99aba91ee910f691e87807e3a033a05b43b66d074481c21b49541e7b890e63a9811a0e810374268
2b83ce7121cde5b340353e023ff8b156c6f8ced6c86d57ae9f2f7d9af39856399cce6affa93a033a956b6fcca0e990f601e8
aa0c3ac78534d7c1af4ad121ea83d056aee91c17da5c07fb4f45879c0f425b19748e0b69ae837742151d423e086d65d0392e
a4b90e26d928ba3ae183cc4f156c8e09e9adccaf19af457afeb956ab4ccf0d873c17668d67e8e714cf299ef935aa33a0b3cc
af793366906c75c207899f2ad0b407e5eceb137d8155677bb4b07715585a81d2d54798494b56077b90f7a93664dac25f631d
ace52bc890ea41602b834d7b50ce7ac878106c88f420b29501e79890def2199f1a1f891afc582b637c676be4b67f9aafa0b8
10a401dcaa0c2ec78274f6f96ae0020d111a24b6728de688d0dabe3e362834446890d8ca407344486d1dccae5574c8cf20b4
9541e7b890e63a38e5517408cf20b49541e7b890e6be5e9840a1213983c456069a23426aeb60a1134587d88c16ba9441e7b8
90e63ab8ababe890d841682b37748e0b61ae83a92b8a0e711d84b632e81c17d25c0fb130820e711d84b632e81c17d2dce7e7
24020d411d24b632d01c11525b0797348a0e291d80aecaa0735c4873e5aae65a10fa5a2af8b89fa9e45fe2729d503cd0b233
a0b35cd5bc1933683a447410daca359de3429bebe009bba2433e07a1ad0c3ac78534d7c15c0a4587700e425b19748e0b69ae
83a5b6151dc239086d65d0392ea4b90ed655577408e720b49541e7b890e60a745ebd68c8933dcefc37998e7bbae03ac57da0
fda665674067817e3366d07448e620b49541e7b890e6bacae568a1ffffbc09d2392ea4b902bd8c5b7abd2c6ded90fe24e9f1
68dac0ad59164b77067416e83763064d874c0e425bb9a1735c08734d22073ff657ce05c777b6466e7b13c7c18ffd1572c1f1
9dad91dbde6471e0637f664c707c676be4b677f0735274c8e120b4956b3ac78536d741a691a2430807a1ad0c3ac78534f775
64974243fc0689ad0c344784d4564effd67d9cc7e34a1a4a0bc9a40517f2f969bce28cdf438b32929d019de5f4efcd9841d3
217b83d056069de3429aeb208150d1217883d056069de3429aeb6056a9a243ea06a1ad0c3ac78534f7f503058586bc8d96b8
9481e68890da3ab8e9aae810b641682b37748e0b61ae8359a58a0e491b84b632e81c17d2dcd7abf02834646c90d8ca407344
486d1dac00abe8ea800d305b156c8e09e94dbdfb3b2eeb1cff0a79c1bbe538f779d947f9577e3a033ad5dbbfde9841b255d9
1e042ec51a4c5b90c6ae65c3ff8c5ca76bf0f39e2a86965b21b6f7fab8bf23d7e116fcbca78aa1f556a82dfefa556e818564
0bd25a19605a8212d64114906243ac0591ad0c38c784f4d6c1fadc8a0eb11684b632e81c17d25c07476045874c0b425b1974
8e0b69eeebe5e7151ad22c5ae252069a23426a6bb22cf8b1bf1222daf19dad11dbdec1b4604587140d425b19748e0b69aec9
d0c0c7fe8c87e0f8ced6c86def214b42d0214283d056aee91c17da5c07bb2f4587fc0c425b19748e0b69ae83370c141df233
086d65d0392ea4b9afaf9a141a9233486c65a03922a4b60e2e21141d6233086d65d0392ea439019ddeef4eb7a7f25fc556a8
13d04f6740a78076c70c9a0e991984b632e81c17d29c5a8779b8d6ea8aff394d983a967d5fe58967690c6854ab303b23068d
86b40c125b19688e08a9edebe5b6151aa232486c65a03922a4b6b2b0faba2fd715deb20de7beadf96ec5bd22ca6abb2fd117
d0571655ff3d5ed05c48c868714b195c8e05e9ace00ef90f927f66b61a5bdcc0e7db58785563a81b0bf08b11834643360689
addca0392284b6ae923100fd33f181748e0b69aeab580c42ff8a7b009de3429bebe02fade8108941682b83ce7121cd7dfd83
566808c320b19581e68890da0a719a2cb49df9f4285e56c6cd9ae6bc7d735969b210abc6503716e21723068d86180c125b19
688e08a94d2558adfb39a4c75569fbb6698abfbef877112722d619d0a932acbc3183a6430c06a1ad0c3ac78534274ebea671
38cfbc1871daf74c698ec532c8e5089ece804e71fae58e19341d3230086d65d0392ea4b9afef672b34a45f90d8ca40734448
6d1ddc29507448bf20b49541e7b890e63a78bf43d121faa2852e65d0392ea4b90e96265574c8bd20b4951b3ac78530d7c1c4
7f4587d08b7a7beb32e81c17d2dcf773e0251d122f086de58ace71e1982b73ddae5750f25f259d1c4fc33d8ffb5ac77db6a5
f56467406799edf666cca0e9107741682b83ce7121cd15e8735aa6f59a74985f924d2f1ca57fe46f68dccaed81569d019d05
facd9841d321eb82d056069de3429a2bf357a7b4e6619e8f770ef3b92cdbbc5d93f3aef7516c8e9fec0ce82cf357df8c1934
1db22e086d65d0392ea4b9021df7b5cb99b7295e1b5cf17069a7134f9e9739bfecff40abce80ce02fd66cca0e9107441682b
83ce7121cd75157341e85f090ea4735c487305fabad0bf2229a6349366bf820b86f6b52dd919d059a0df8c19341d522e086d
65d0392ea4b9027d9cd33ee79b1b7187b3ce694e40fec7506ecd3ed0aa33a0b340bf193368ba3ae282cc4f156c8e09e9adc4
3e4cf1f032e5e786e772bd8b716deb18f740e3b8d86e4c76067496e087376306c956075d90f8a9024d7b50ce3e0cf4f80b54
e75bb4987715401a5e89ea60193f954dc1808bfa1cb22aef2959399ebde7b20de294ab41d41ca1788630c593ffbcfadc11bf
8643fc243993f1e90ce8547384bc3183a663c005a0ad0c3ac78534d7c1d2088a8e011780b632e81c17d2dcd72fcc2b34a65b
34c4a50c344784d4d657ba05a07f463734748e0b61aeaf740b40ff8c6e009de3429aeb21e841d031dd02d056069de3429aeb
e0dd7145c7740b405b19748e0b69eefb9b05928ee91680b632e81c17d25c9b6e818ffd19dd80f19dad91dbdec3174cd031dd
a286aecaa0735c487305fa3eaa5db710ee45bbd2beeebccf559f37b564674067817e3366d0744cb700b4956b3ac78536d7c1
4143d131dd02d056069de3429aeb60c14c45c7740b405b19748e0b69ae9c05a6688031cf9ddaf77811b28ce9b6609a30b8e4
ebbf670e9eee0ce82ce7816fc60c9a8e011780b632e81c17d2dcd72bde2934a65b34c4a50c344784d4d6c18c5a45c7740b40
5bb9a1735c08731dbc95a7e8986e01682b83ce7121cdf510f420e8986e01682b83ce7121cd7530a956d131dd02d056069de3
429aeb605e87a263b606a0ad0c3ac7853477457a88f1996e517f6c55c6f8ced6c86d575fb02deedee299724ea248f39dc675
5cd4fed33a033ad517cc1b33683aa65b00daca359de3429b53911ee3bc2df16435efebe27f6338b643865b3c8da16e54811e
ce8841a331da02c456069a23426aeb21cf43d031da02d056069de3429aebe0dc40d121da02cc4f156c8e09e9ad83d7f6041b
a22d40fc5481a63d28676545d8346b6c5ff38dc97539d735adf893b66d1be2f8f177f71c1f656740675911f6cd98419221dd
a2e1bdab20d31694b10eae6a0419d22dc0fb541b326de1afb10e66840932a65b00d8ca60d31e94b32be8418cce7c097ca895
31bcb32d72cbbf7e68a2d0186d01622b03cd1121b5759003a0e8187051435765d0392ea4b91e56d01674ccb800b4956b3ac7
8536d743f881a063cc05a0ad0c3ac78534d7c19980a263cc05a0ad0c3ac78534f7f5f2100a8d191720b632d01c11525b07e7
b88a8e311780b632e81c17d25c077339141d632e1ae852069de3429aebe12f2de8187301682b37748e0b61eef31fb44063c6
0588ad0c344784d4d6c1941d45c78c0b405b19748e0b69ae83733145c7980b405b19748e0b69ae836b1b45c7988b1aba2a83
ce7121cd75305549d131e602d056aee91c17da5c59fd76cdafcc5ecbf44fd742fdf9727bcb6bd72fb6b2b3ec0ce82cabdfbe
1933683ac65c00dacaa0735c48731dac73ace8187301682b83ce7121cd954915d7ffe4c7e2dbb1aff14ae87a4d651cf6f138
d7e75d45dd19d0592655bc1933683ac65c00dacaa0735c487365aad23c9ce3787d15e37771cf3780f3f792af5bebc6503796
894a2f460c1a8d191720b632d01c11525b5f191780fe19e0003ac78534f7fd9b5c928e19170d742983ce7121cd95979ac67d
4899ec696b87eb4d9c3c636edda6638b87da6766986a0c682caf34bd18316834065c80d8ca0d9a234268eb21d543d0315e03
d056069de3429a6be335f0b13f932b30beb33572db8beb75ced3c7b297b4c73bf27cc16dbf8eeecfc142f405f415cfbfc70b
9a8bb11e356e55069763413afbfac7a4d098e901622bd7688e08adadaf4c0f40ffccac009de3429aeb20f64ad131d603d056
069de3429aeb6066bea263ac07a0ad0c3ac78534572e6c8e791e97bcb4c0316fb16b4d6fd9a78b90239e8b1f831d2b646740
67b9b0793366d0748cf500b49541e7b890e63ab88455748cf500b49541e7b890e63ab859a1e898ecd1409732e81c17d25c07
8bb0283a267b00daca0d9de34298fbfa96ab4263ac0788ad0c344784d4d6c1e30445c7640f405b19748e0b69ae8339c48a0e
c91e35b355c1e69890de3a589657b0d5c91e002ec51a4c5b90c6d424c76d19ce78b69667612ee98439eef4e4d4e1a733a053
4d73f4c60c8a0bb922a07daa00930e84adaf67740a28449a00f5a9824a0b50b2ae65eaff8ecd5c117ca49531bade12b5d53d
7c9b041b7345806c65c03926a4b7afd7da53680c1501b19581e68890da3a9827ace8182a02682b83ce7121cdf5101f23e898
2bd2409732e81c17d25c07273d8a8ea92680b67243e7b810e63a584a52d131d504d056069de3429a5327f4d703a3eb15be21
5eecc5ffceaa5ea7b1ce804e754aef8d19341d33556ae8aa0c3ac78534978f916a7ca69ae063ad5c8fef6c8ddef60ebe608a
8ea92680b632e81c17d25c07bb1245c75413405b19748e0b69aea3541310ff8cec009a23426afbfa6c48a131d204c456069a
23426aebe0955345c74813405b19748e0b694e459aac71ff9aa69ac47fc40ba0395ed56f9bfc353f9d019d2ad4c41b33683a
a69a00dacaa0735c48736afdf41cc391d79c5ce265c9ba4ec3a9989fc68046b57aba3362d0684c3501b19581e68890dabe5e
235fa131d8a4212e65a03922a4b6423c0d4bbcbccddb37c7ab81651cf25238cbd0fe9a65674067617e3366d074cc3601b495
1b3ac7853057122fceb44e5adebeeb5dae39af893c9f5bbc045ce7fdb99e538d018d25efe2c58841a331d8a426aeca407344
486d85383f46c80b4caee396e62c2df9ca6bceb714e6f57910a73b033a0bf39b3183a663b009a0ad5cd3392eb4b9cf333e04
1a534d406c65a03922a4b64f53991417234d806b65703916a43371808a9b3a8ce9ade77c3c89575cf1f4fc50675fd619d029
0e51ee9841d331d504d056069de3429a13a722633a94c5fd4dfa42c6bdce985fcc9577a39ece804e712ae28e19341d534d00
6d65d0392ea4b9af6f652b34469a80d8ca407344486d1d5c3f2a3aa69a00dacaa0735c48735f2fd5a4d0186902622b03cd11
21b575f0eebaa263a449035dcaa0735c48731dacbda6e8186902682b37748e0b61ae8355f6141d234daaedadcba0735c4873
df4ffd97748c3401b4952b3ac78563eefba97e928e912680b632e81c17d25c99c63a2c7bbefa4813308f3dee78ce397f2fd7
693dd7319e433cf3585567406799c8fa66cca0e9186902682b83ce7121cd7d3955597231cc04b856069763413a2b690fc33c
aed78a15e77885ad6c7903e34e6f9aa6cd7661b233a0b3a43dbc1933683a869900dacaa0735c4873057ab916e84a1b38ede3
197f0263dec0942eb39c16e0221b43dd58905f8c18341a934c406c65a03922a4b6be924c00fd33a803748e0b69ae40ef5bda
dfe4ed5bcee518ceedfa192ed73ad0cbc32c1a031a0bf28b118346438c09809f2ac01c0d525a89f518c6287ebe9238e215df
3a2fc3f5237ce25a4bb287ea0ce82ce11e6fc60c920d3126207eaa40d31e94b31e625bfe9221cfa4e1bdab20d31694b10ed6
30545124cc33a94f20abf2f6ef364d6995b1ff6110a75c0dd2c1a41545c73c13405b19748e0b69eeeb8b6685c63013105b19
688e08a9edeb75b9141ac34c1ae252069a23426a2b57cc69d9913be52dee57e6785a9a9f15eef331ef7180fdb9a4909d019d
e58af9cd9841d331cc04d0566ee81c17c25c0757cc8a8e612680b632e81c17d25c07af4c293a869900dacaa0735c48735fbf
2cafd0986402622b03cd1121b57d7f8f40d231c904d056069de3429a2bd06bbcb6dcafb7d8e2416699e27e377f0fd36bd0e3
36cccf05b3ec0ce82cd06fc60c9a8e392a80b632e81c17d25c0f9935828e392a357455069de3429aeb602fa6e898a302682b
d7748e0b6dae83e395a2638e0aa0ad0c3ac78534d7c19989a2638e0aa0ad0c3ac78534d7c1531b45c71c15405b19748e0b69
eeebeb0c85c61c9586b89481e68890da3a7806abe898a302682b37748e0b61ae83979a141d7354006d65d0392ea43995aeb1
8df15af01cf245fe95b43dcb39ad4f63a81b55b6863362d0680c5101b19581e68890da3a98d1aae818a202682b83ce7121cd
890943f1cf31c7b3c525fdf896b47c48ba9f370b68eb0ce8141386dc3183a663880aa0ad0c3ac78534a7a0d7633fe257304f
c78c87d3b8ab3f6cee9fec0ce854d0de9841d331c2a586aecaa0735c48731dcc9252748c7001b4956b3ac78536f7f55e4ca1
31c205c456069a23426afb7a5e9842637e0b88ad0c344784d4d6c1cba88a0ef92d607eaa60734c486f1dbc3f26d890df02e2
a70a34ed413913bcd1ff1eff4b79824b5a40768a07b751f23e9d019d82d71d334832e4b734bc771564da8232d64114812043
7e0b789f6a43a62dfc35d641de8420637e0b80ad0c36ed4139eb60296bc5c6f418205b19708e09e9ad87182641c70019405b
19748e0b69ae8395da151d03646ae8aa0c3ac78534d743ac88a063800ca0ad5cd3392eb4b90e8eca8a8e013280b632e81c17
d2dcd72f382b34a6c780d8ca407344486d1d2c89a5e8182003682b83ce7121cd75b02896a263800ca0ad0c3ac78534d7c18d
3f45c7009906ba9441e7b890e67af84b0b3a06c800daca0d9de34298ebe01a52d1314306d056069de3429aeb60c51745c70c
19405b19748e0b69ae83d33145c70c19405b19748e0b69ae83a7738a8e1932357455069de3429abba1c77b4e40dac033fef8
e26e3ecfb68e87d6294d4d581e68d919d07943bf1a33683a66c800daca359de3429bebe0e1b3a263860ca0ad0c3ac7853457
a619cc537effe59fbcba79beeccd539bf6a39d9e2f3b033acb34833763064dc70c19405b19748e0b69ae8309258a8e193280
b632e81c17d2dcd7d3e2141a3364406c65a03922a4b60e6688293a66c800dacaa0735c48733d04e7083a66c834d0a50c3ac7
8534d74770ce1f3ac6c800daca0d9de34298fb3aeb43a1314306c456069a23426a2bc4d339cf4b0e278be750d7fb7d5752d9
7590b1d76e64674067617e3366d0744cb001b49541e7b890e6aee01c313e9364ea8fadca18dfd91ab9ed1dfca4141dc36400
6de59ace71a1cd75b0f354740c9301b49541e7b890e63a983daee8182603682b83ce7121cd7dbf9680a463980ca0ad0c3ac7
8534d7c1c9bea263980ca0ad0c3ac78534571274aef3f26b46fb3ccdfb346e7992e81a8f68f112f478a06567406749d07933
66d0740c9301b49541e7b890e63ab88057740c9369a04b19748e0b69ae8307c38a8e613280b67243e7b810e6be9e48add018
2603622b03cd1121b57570ef55d1314c06d056069de3429a53cf0ac778e53d2ef98d9d35ee7a8679987735d9d43a033ad5d3
426fcca0e9102653335b156c8e09e9edebd59005589d2403da52aca9b402a9ebeb99877fa1102303d4a70a2a2940a8ea6025
01c1852419d03e55806907ca97e24d4bf00c43be2d18b76a4a0b01cdf239e8d319d0a978bd318324638e0d80ad0c36ed4139
ebe12f2cd8986303642b03ce3121bda975cc9f679479c9e22d5e889cbb3a26596740a75ac7dc1b33683a46d900dacaa0735c
48731d3cee56748cb201b49541e7b890e63a98c2a2e81865d3409732e81c17d25c07671f8a8e513680b67243e7b810e63a78
614fd131ca06d056069de3429aeb60229ea263944d0d5d9541e7b890e6da281b7ceccf9c967a7c676bf4b65f1f2bc667980c
3ed6ca18dfd91ab9ed1dac6ba4e8182603682b83ce7121cd5daec5f84c74c1c75a19e33b5b23b7bd83c9518a8ea12e80b632
e81c17d2dcd757330a8d892e20b632d01c1152dbd7efca2934c6b980d8ca407344486d1dbc07aae898e802682b83ce7121cd
15e8346d7ebaae3cb6355df29d7995af6519a67839945efc2fd0aa33a0b340bf1933683a86ba34d0a50c3ac78534576220d2
c293737e37755da39be3482bc2a4e08635dd735fa6e77a4e7606749628883763064dc75017405bb9a1735c08739f665f282e
26bad4b855195c8e05e9ace0c693cff318aec9d463fcc92f7b0e8d5fc6f801f124d52e6a64674067417e3366d0744c7401b4
956b3ac78536d7c1175bd131d405d056069de3429aeb21e544d031da05d056069de3429aeb60790a45c76817405b19748e0b
69ae835b528a8ed12e80b632e81c17d25c07372a141dd35d006d65d0392ea439f5c6e0f5ec2e2f23bc8df1ea6b1ab743dd63
b6ce804ef5c6a03766d0744c7701b49541e7b890e63a78ef59d131e005d056069de3429aebe0b1bea263c04b035dcaa0735c
4873650247cefc9bf204aa69dfd3c2f67985fb78645fb6293da17f2670a8ce80ce3281e3cd9841d331e005d0566ee81c17c2
5c999f742f22fc4f9e89782f3490ff44c7b64dc3fa3cfdd49d019d657ed29b3183a663c04bb5bd7519748e0b69eefbd52d25
1d035e006de58ace71e198fb7ecea1a463c00ba0ad0c3ac78534f7fd4c5e49c78017405b19748e0b69ae40ef47bcfccbbfbf
d31ed5c60dbce7d82c4f2c84ec0ce82cd06fc60c9a8e312f80b632e81c17d25c89c28847b314569b36709a86f4cad951a685
8ef3b83cebb1e8ce80ce1286f166cca0e918f302682b83ce7121cd15e8784c598f75cd1bb86cd754fabc81e7bd7ecc03ad3a
033a0bf49b3183a663d20ba0ad0c3ac7853457a0d71cc572e40d3cf76388574257dad276ff611e68d519d059a0df8c19341d
935e006d65d0392ea4b98e925e40fc23cb04608e0629adf01ee7b4cf5730c9bcc753863d1e5ff3e6c593c763492fdc1560d5
19d05988df8c19241b925e40fc5481a63d2867e9b9911a1b592bcd27de558cadb7436db3ba6ef7e2ebff5e8d97ce804e75dd
ee8dc9ac95677399b5529fcb55e5f5df7d889f72956d10a75c0df2f552280a8d412b20b632d01c11525b072fd32b3a66ad00
dacaa0735c48731d65ad34c43f824480e68890da3a78a95cd1316b05d0566ee81c17c25c076bd72b3a66ad00dacaa0735c48
731dbc51a4e898b502682b83ce7121cd75f0ee98a263dc0aa0ad0c3ac78534d7c17be58a8e712b80b632e81c17d2dcf7b729
241de356006d65d0392ea4b91ebede828e712b357455069de3429a2b3bb2bcd669bee63af674ca96d60fce3b9de765a0b223
539d019d6547f666cca0e918b702682bd7748e0b6dae835723151de356006d65d0392ea4b9727272bd787dfd218e7b66663e
795cae57dd9ee3b4ec0ce82c27276fc60c9a8e712b80b632e81c17d25c39073d8eb879792a42dcfb4cf1da6b3dae53c66d8c
fbda737aeecdc8ce80ce7216fa66cca0e918b702682b83ce7121cd75b00ab4a263e24a035dcaa0735c48735f5f542a34c6ad
80d8ca0d9a234268ebe0de81a263dc0aa0ad0c3ac78534d7c1ec0e45c7c415405b19748e0b694ecce339f21cb2fcac6359b7
786514777d8af9690c6814b378bc11834663dc0a88ad0c344784d45688c7bc0e7bfeb90d6952fd9eef85c7d3d578313aef15
b1680c682cc42f460c1a8d592b20b632d01c11525b0773b5141db3566ae8aa0c3ac78534a7bed8eb1eafb78f7c81b0c4c679
df8641cd1eb7ce804ef5d5f6c60c9a8e592b80b6724de7b8d0e63ad885293ac6ad00dacaa0735c48735f4f9a57688c5b01b1
9581e68890dabe7e3142a1216b05c04f15608e0629ad83b75f041bb25640fc5481a63d2867eae5b931cd3639f73c53f08cfb
bb7cd756be3c573a033ad5cb73de984192216ba5e1bdab20d3169431b51247dcbf9cdb9c1fdc6de314f7b2fbae4eb0ad31d4
8d6a1d0e67c420b110b402d8a7da6069057f7575700925c818b402602b834d7b50cef23b736a74469de043ad8ce19d6d915b
dec1745645c7a813405b19748e0b69ae83f76e151da34e6ae8aa0c3ac78534d7c19ba88a8e512780b6724de7b8d0e6be5ed6
46a131e704c456069a23426aebe0ae84a263d409a0ad0c3ac78534a7ce6ec7e13cf35cb1742e3aa5d5279741be7ef1740674
aaf35b6fcca0e9187502682b83ce7121cd751094a9e8187502682b83ce7121cd7570f1a6e81875d2409732e81c17d25c3e37
50e3336c041f6be5667c676bc4b6f7f0ab12740c1b01b49541e7b890e6bebe6c54684c1a01b19581e68890da7a383110744c
1a01b49541e7b890e63ab8c050744c1aa9a1ab32e81c17d25c076b762a3a268d00daca359de3429b2baf718d53fcc9e56fdf
1e7f6dd3b0c41f5e3e4b9fe216c6f3976712adec0ce82caf71bd1933683a268d00dacaa0735c48731dacc3abe8983402682b
83ce7121cd956905eb3ec673c4eb3594618de7e6d79b0af14f147703e36aefb9c8ce80ce32ade0cd9841d3316904d056069d
e3429aeb60269ca263d808a0ad0c3ac78534f7fdbb6b928e612380b632e81c17d25c0733e1141dc3461ae852069de3429a2b
d0c7304ef33513f5cc6fe12cd757b17d12a93b033a0bf49b3183a663d808a0addcd0392e84b90ee220141df346006d65d039
2ea4b9cf1356041ac346406c65a03922a4b6423c2c7b5e2426fd498e7ddfe773ce5fca3c916d8c57db0fb2ea0ce82ccc6fc6
0c9a8e5127357455069de3429aeb606aaba263d409a0ad5cd3392eb4b90e76dd8a8e512780b632e81c17d2dcd719680a8d39
2720b632d01c11525b079379151d734e006d65d0392ea4b90eceba151d734e006d65d0392ea4b90e96c95074cc3901b49541
e7b890e6ca0ce6f4539fae9dccb00e53fc0daed743be668681ee0ce82c3398df8c19341d734e1ae852069de3429afbfa2d3a
85c69013105bb9417344086de52ed178ed73ee4bfbf4b2fb90efe85c6b98cccf2202ba33a0b3dc257a3366d074cc3901b495
41e7b890e63a780eaee8987302682b83ce7121cd7570e757d121e7a466b62ad81c13d25b0f2bcaff65aba34e005c8a3598b6
208d7dfd30e32f14a24e80fa5441250508553d447ffce542d409689f2ac0b403e5ebebf99c028b3927a0b532c0b404252c3f
0155a33369041f6a650cef6c8bdcf21ee255041d9346006d65d0392ea4b90ed6b254744c1a01b49541e7b890e6d49e725b86
335ed45eabe9a63b48435a5a57ed2a4b6740a7da577a63064dc7a49106ba9441e7b890e67a38180a3a268d00daca0d9de342
98eb60d55245c7a411405b19748e0b69ae87781541c7a4911aba2a83ce7121cd7530cd53d131e704d056aee91c17da5c9b73
828ffd19e281f19dad91dbdec16c7c45c79c13405b19748e0b69ae835f95a263ca0aa0ad0c3ac78534f7f58bbe0a8d112b20
b632d01c1152dbd79388151a2356406c65a03922a4b60e26112b3aa6ac00dacaa0735c48736a3e6d3c5b59e72d3f3a8b573f
693dd3e59427064f6740a79a4feb8d19341d5356006d65d0392ea4b91ea265041d53561ae852069de3429afbfa7d6e85c688
15105bb9417344086dd75981189f5127f5c756658cef6c8ddcf6fcb16a7c868de063ad5c8fef6c8dde76f5a58edfbd25fe23
bf6c9e667e9ce739a95fb2750674aa2fb53766d0740c1b01b49541e7b890e63a9819afe8183602682b83ce7121cd75708346
d1316c04d056069de3429aebe04d5f45c7b011405b19748e0b69eeebd75f151a9346406c65a03922a4b6f21c2e8d932a6957
975fbe490fa3d319e3300cfb66cb1debce80cef21ceecd9841d3316904d056069de3429aeb60b6b6a263d208a0ad0c3ac785
34d7c1d2608a8e49230d742983ce7121cd75b06ea9a263d208a0addcd0392e84b90e6608293a268d54db5b9741e7b890e6be
5fd851d2316904d056aee81c178eb9efe7f04a3a268d00dacaa0735c48736546eb3ae7df60dada3c55efc8dfca6d8fa78d7b
8a3d2af359fff605f495d9acbfc70b9a8b1923c0b532b81c0bd259c13daf67a369f3c63b6731cfc7dce341659cf7e93958c9
ce80ce82fc66cca0e9983102682b83ce7121cd955082611ed76bcad2395ec12f5bfeabc49dde344d9bedb765674067092578
3366d074cc1801b49541e7b890e67a08561174cc1801b49541e7b890e67a08561174cc1801b49541e7b890e67a0856117488
1901f353059b63427abb6237fe8e8ea00f7ce653c5e07a4bd45617c9db92eef35fb92bd9cdbe4e7927df2e002c3b033a8be4
3763064986989186f7ae824c5b50c63a9852a112421833529fc855e5e5dffd5cd215e1ff308853ae06e9e091a0a263d208a0
ad0c3ac78534d757d208a07f266980ce7121cd75b0cab3a263d848035dcaa0735c48731d5cb42b3a868d00daca0d9de34298
eb60e57645c7b011405b19748e0b69ae838b7645c7b011405b19748e0b69eefb8b7649c7b011405b19748e0b69ae83b78d15
1dc346006d65d0392ea4b9fc904e8dcfb80f7cac9531beb33572db3b782752d131eea386aecaa0735c4873057a58e22eedca
0c4db7dee72d1fe1e201eeba2c7a120c55634063417e3162d068ccfa00b1956b344784d6d6c11143d131eb03d056069de342
9aebe0dc40d131eb03d056069de3429afbfa4d3985c6a00f105b19688e08a9ad8345a4141d833e1ae852069de3429aebe0e9
b3a263d607a0addcd0392e84b9af5fb751680cfa00b19581e68890da3a880650740cfa00b49541e7b890e66ee8d8360f677e
c8b06ec3b18de9079836705ee37e282d8759a0656740e70dfd6acca0e998f501682b83ce7121cd15e814813ce7653ee3d9d3
30c69fc091f638cbb4a6f7df17bb4f233b033a0bf49b3183a663dc07a0ad0c3ac78534974f7ad5f80cdca83fb62a637c676b
e4b617d7699d92ed9cb297f55c86786419f397816b62e8c6503716d32f460c1a8d691b20b6728de688d0dabe4e9251688cda
00b19581e68890da3a9852abe898b601682b83ce7121cd75b056aca243e006989f2ad81c13d25b0737f5051b023740fc5481
a63d28675fbf0621b090b6d1c0de556069054a57076f090a32046e80f7a93664dac25f631d2c0521c818b801602b834d7b50
cebe7e8f498131eb03bc560699a3414aebe13b2de898f501682b83ce7121cd75f062b7a263d6470d5d9541e7b890e63a787f
4bd131eb03d056aee91c17da5c07ef6f293ac67d00dacaa0735c48731d5ca62b3ac67d00dacaa0735c4873057ad8e265f4f5
52c6b6a62552ce7c93705986291ee1d28dab02ad3a033a0bf49b3183a663dc07a0ad0c3ac78534f7f5a5a34263d60788ad0c
344784d4d6c18d0945c7ac8f06ba9441e7b890e6daac0f7ceccf208b667c676bc4b677107faae898f501682b83ce7121cda9
57e686739ad21a80e9a23a5e622ffbbeca73ded218d0a85e9873460c1a8d591f20b632d01c11525b07d7ac8a8e591f80b632
e81c17d25c07d7368a8e591f357455069de3429a530b1aeec3948f6ff94ecafd628c9c7df7740674aa050dbd3183a663d607
a0ad5cd3392eb4b91e024e041db33e006d65d0392ea4b9f2ca5c7a2920cfa3d9ebff3c3c7fadf2c2dcdfbe80bef2badceff1
82e662ca0770ad0c2ec78274569eadc7abeafdc87b986348f3a8ce34d7223f2d8d97fac3f9bce7ac3b033acbb3f53763064d
c7940f405b19748e0b69aeac377cccf3b85c135be678fd33afc39417444ef300ce63787661ba33a0b3ac37fc66cca0e998f2
01682b83ce7121cd957977f7f3ffb481eb792f159bbf9647bcde1f569b8a2f3b033acbbcbb3763064dc7940f405b19748e0b
69ae873c1741c7948f06ba9441e7b890e6be9e16a5d018f101622b37688e08a1ad83e9958a8e111f80b632e81c17d25cbeb6
51e33367031f6b658cef6c8ddcf60e2227141d73366ae8aa0c3ac78534d7c1174cd1316703d056aee91c17dadce79113028d
211b20b632d01c1152dbd77b4e85c6900d105b19688e08a9ad83d9f88a8e211b80b632e81c17d25c99457b8cfbbce717a18f
72dfea3a438fffbdf817b2378d656740679945fb66cca0e918b201682b83ce7121cd9593defc92f7f5db4b27e2691f93cfd5
b6bc835d6c36beec0ce82c27bd6fc60c9a8e211b80b632e81c17d25cb9bc492fb1ddaf7cc77ded3cc7bf43bef64c933cf675
d99ffdb6ec0ce82c97376fc60c9a8e211b0d742983ce7121cd75b0ea8ba263ce06a0addcd0392e84b972b3625b87f88dcb6f
b8adf1c2643f8ebc9ac3b66dc3bc8df3f93cba919d019de566c59b3183a663ce06a0ad0c3ac78534d7c1946945c79c0d405b
19748e0b69ae837562151d72366a66ab82cd3121bdf5b0b4f85fb63a6703c0a55883690bd25807930cfe72216a03b44f1560
d281b0d5c19a7b820b511ba07daa00d30e94af0ea6360a32a66d00d8ca60d31e94b3afdf7650600cfa00af9541e66890d2ae
7c11313ea336f0b156c6f8ced6c86defe1a724e818b501682b83ce7121cdf5b0b314748cda68a04b19748e0b69eeebc9b30a
8d391b20b67283e68810da3a986da8e898b301682b83ce7121cd7570a6a7e898b351435765d0392ea439b534ebf53c21df4a
d98678911b45adea00699d019d6a71566fcca0e998b301682bd7748e0b6deeebf568151a233e406c65a03922a4b60e221215
1d233e006d65d0392ea4b9cfffcc028df91e20b632d01c11525b07cb36293a467c00dacaa0735c48731d4c1152744cf900b4
9541e7b890e63a78935bd131e503d056069de3429aeb60069ca263ca07a0ad0c3ac78534d7413aa6a263ca47035dcaa0735c
48731dbc0fa0e818f401682b37748e0b61ae0dfaa83ff6778606c677b6466e7b1bf4818ffd9962518fef6c8ddef60ebed78a
8e411f80b632e81c17d29cba7efe2f437d151a533e406c65a03922a4b60e2e25151d533e006d65d0392ea4b9afcf3c151a23
3e406c65a03922a4b6f240ecc3e07945c7940f405b19748e0b69ae83b533151d533e006d65d0392ea4b90ee6f02a3aa67c00
dacaa0735c48731dcce155744cf968a04b19748e0b69ae8339bc8a8e291f80b67243e7b810e6caa4acf1d8d274f9b481e962
771aee4973d3b6aff18ff41ca2756740679996f566cca0e998f2516d6f5d069de3429afb7e96a5a463ca07a0ad5cd1392e1c
7325016258f67c4e9c36f08887d0f9ccd703475e9f705cd767c6a1ec0ce82c09106fc60c9a8e291f80b632e81c17d25c818e
bfb2f8fbcb6bf88ff144625d8731ff89f67158cef88b7cfed2b233a0b340bf1933683a667d00dacaa0735c4873053a1e5b96
2def744ebb12481b784d105c9e9996b233a0b340bf1933683a667d00dacaa0735c4873573280189f691bf8582b637c676be4
b6f790ab22e898b601682b83ce7121cdf5103122e898b601682b83ce7121cd5d5f30313ef22ef0a94f15a33bdb22b7bcc9bb
c067fec874c0e07a4bd45697bfed723d7bcf7f877d3ce3be75cc938daf9b46f67d968da16e2c7fd91723068985b08b06f6ae
024b2b50ba3a78beaf722a1876519fcf54e5399d559ddb55b6419c723548074f4d141dc32e006d65d0392ea4b90e2ed8151d
c32e006d65d0392ea4b94f5fb7555c8cb968704b195c8e05e9ac5cb5ce4b3cf19ef2b9d8144f58ce744f34ff71b6659bd2db
d1cf55abea0ce82c57ad6fc60c9a8e311780b67243e7b810e63ab86a55748cb900b49541e7b890e6be7e9548a131e302c456
069a23426aeb218240d031e302d056069de3429aeb2be302d03fe32340e7b890e6be4fe394744cd800b49541e7b890e67af8
7a0b3a266cd4d05519748e0b69eeebbd984263c20688ad5ca33922b4b6be123600fd333e02748e0b69ae8327088a8e091b80
b632e81c17d25c7956342ed3945f4cddb77b21f9fc5867bf36f5599d5e3606349627452f460c1a8d091b20b632d01c11525b
07ab18293a266c34d0a50c3ac78534f7f55c7185c6780d105bb9417344086d6a21b67872bca5f774e33f96639db77854dde4
b4e9a733a0532dc5e68d19341d1336006d65d0392ea4b9af57c657688cd700b19581e68890daca828ecb3a9cc770ddb819c7
7d8b87f174285dc6f5be235b906567406759d0f1cd9841d3315e03d056069de3429a2bd043feb3e42944535a39f74c0b1fc6
0d7cd6072cccaa31d48d05f9c58841a3315b03c456069a23426abb89a38b23bdd29f362f5e088cfb9cd7159863ff30aef17b
5888556340e34dfc66c4a0d118eb51135765a03922a4b6afffc60a8db11e20b6728de688d0da3af8292b3a267b00dacaa073
5c48731dc403293a267b00dacaa0735c48732ace6438e678c9b7e4a34abae83fd359b338365b6740a78a33f1c60c9a0ec91e
607eaa60734c486f1dcc1c166c48f600f153059af6a09c7dfdbeafc042b247037b5781a515285d1dbc732bc890ec01dea7da
90690b7f8d7d3d5d586031d603b456069896a084e599e86a74666be043ad8ce19d6d915bdec14c7445c76c0d405b19748e0b
69ae87fda4a063b6460d5d9541e7b890e63a381e2a3a666b00daca359de3429bebe0cc47d1315b03d056069de3429a2bd073
fc99add7712beed5af80807c3e3aec5b3c551d9ebfb4ec0ce82cd06fc60c9a8ed91a80b632e81c17d25cb9785b8f73dff3cd
de6799fef4559c8fb8c3d9cfe3b981ae3b033acbe5db9b3183a663b606a0ad0c3ac7853457a0af25c1f21fe21ccff4143a5f
5f6e69d6fe324f76c52a1a031a0bf28b11834663b60688ad0c344784d4568887337edff2f2494bbcd85ce665c9efd7cff174
255e09d86c33dd19d05998df8c19341db3351ae852069de3429a2bd04f9c743c9adc73e0d2f6edf1826fddc667693dd918d0
58905f8c18341a633d406ce506cd1121b415e27d9c8e3451289d358c5b9ac89397819ee3ee763dc7ed7cde76959d019d85f9
cd9841d331d603d056069de3429a2bfbed359e87e663e9527f25e30f72d897d582dd445f405fd967ff1e2f682e667a00d7ca
e0722c48672a022a6f519ee5baacdb120f2fc3a9ee9d3f8d018d2a00ca19316834067a80d8ca407344486d1dacc8a5e818e8
51435765d0392ea4b90e56d453740cf400b4956b3ac78536d7c1d2328a8e811e80b632e81c17d25c995699de9bdbd7fce879
5dce75ddaf7bdfdb360cf12c31f5966995aa33a0b34cab7c3366d0748cf500b49541e7b890e63a780f52d131d603d056069d
e3429a2bb318d2baded3358166887b9a236ee1f55cba59814377067496590c6fc60c9a8eb11e80b632e81c17d25c99513ac4
abf86dbe5e5e9be305c0346ef99dc5f53e657ce6abc8ce80ce32a3f4cd9841d331d603d056069de3429aebe0e54f45c7588f
06ba9441e7b890e6cacca4359e078f479ee7bb0e71b7b35faf9fc47d4f3cc08c71239f9949aa33a0b3cc4c7a3366d0744cf6
00b4951b3ac7853057e60edfdfc5b481db7d9278fd1897ed1cf20ab265eeb0ea0ce82c7387df8c19341d933d006d65d0392e
a4b90e16e55774cc1501b49541e7b890e6be9e6aa8d0182a52135765a03922a4b60efecc8a8ea12280b6724de7b8d0e6f213
05353e933df0b156c6f8ced6c86dbf3e568ccf780d7cac9531beb33572db3b78f541d1315e03d056069de3429aebe040a5e8
18af01682b83ce7121cd75704aa2e818af01682b83ce7121cd9533ee6b76feb58339d2564d796bf7789298b31d9fa3b3ec0c
e82c67dc6fc60c9a8ef11a0d742983ce7121cd9557d6eec523f24f6e3da7655bf29de9b8c79bd2cb03cb032d3b033acb6b6b
6fc60c9a8ef11a80b67243e7b810e6ca05659a6c3ae6f8dd7d9fc76d19d31b1969b5a765bc26913d1794aa33a0b35c50be19
33683ac66b00dacaa0735c4873e57ec9763d6bc9dfbe714abbde215fe62fe3394df35cad6ea83a033acbfd923763064dc778
0d405b19748e0b69ae839b448a0ef11a35b355c1e69890de3a58114db0d5f11a002ec51a4c5b90c6d484f87159e7e9c84f8a
d7d8b6cfcb3eaab5d0ac33a0534d89f7c60c8a0bf11aa07daa00930e84ad0e1628155c88d700ed53059876a07c7530734690
315e03c056069bf6a09c7d3ddd4e81315e03bc560699a3414a6be335f0b13fb32330beb33572db7bf86a093ac66b00dacaa0
735c48733dec3f041de3351ae852069de3429aebe10821e898b001682b37748e0b61ae87930041c7840d405b19748e0b69ae
af848d1afa777c04e81c17d29c5a5de6bec0bc5e6cbb5eb55f646aded319d0a95697f1c60c9a8e091b80b6724de7b8d0e67a
c89b10740cd900b49541e7b890e63ac84754740cd900b49541e7b890e63a78cb57d1316703d056069de3429aebe1eb2de898
b301682b83ce7121cd7570a9aee898b301682b83ce7121cdf5b0f71674ccd900b49541e7b890e6d4c9c9b6edcb9c2709ae7b
ba0dbe4c873c212b8d018dead4c419316834866c80d8ca407344486d5fbfb6afd098b0d1109732d01c11525b0767dd8a8e09
1b80b67243e7b810e6be7e9f4aa131dca326aeca407344486d1dfc99151da345006de59ace71a1cd75108ea4e8182d02682b
83ce7121cd757071a5e8982e02682b83ce7121cd757022a6e8982e02682b83ce7121cd75b074a7a263c008a0ad0c3ac78534
571e09daeb1afb9e566b9bb66b3ae87dcaf44c3c548d018de581e08b11834663ba0888ad0c344784d4d6c1045345c7741140
5b19748e0b69ae4c6c58f3f219791dafb474d7b85c2f16c67d4f8a0c586c8543d919d0592636bc1933683aa68b00dacaa073
5c48731d4c6151744c1769a04b19748e0b69ae4c56bafe277f15b7635f8f71bbd6638cdfd0eb4af099aca43a033acb64a537
63064dc77411405bb9a1735c087365dae1324fe31dfe715c2711cbf5bdccdfcab428789977a83a033acbc4c33763064dc774
916a7beb32e81c17d2dcf7ab964a3aa68b00daca159de3c231577210a6b45472fefd9dc37c2ecb366fd78ff15a91d17ed3b2
33a0b3e420bc1933683aa68b00dacaa0735c4873057a19b7b49868dec014f832ad792d8d78bdcb2856dd19d059a0df8c1934
1dd345006d65d0392ea4b9021d4f91e2ef2c07350ce7989e2be6cddbafe3ceb31b137d017d05f8f7784173315704b8560697
63413a6b7345f0b13f433330beb33572dbaf8f15e333d9031f6b658cef6c8ddcf61ea263041d933d006d65d0392ea4b936d9
031ffb233903a33bdb22b7bc87e098bf6cc81501f153059af6a09cf5f0cdfa4b86689186f7ae824c5b50c63a780756a58230
5aa43e99a9cad3bff1db139172d90671cad5205f07812a34e68a80d8ca407344486de5e6c4b60ef19a2abfb7b99eeb102f3b
f24a885bbcec88e769f3f9dc7b939d019de5e6c49b3183a663ae08a0ad0c3ac785345726691fdbb2dd2fff0cf93da039df05
de536e53bc16589fbb13b233a0b34cd27e3366d0744c1769a04b19748e0b69ae837710141dd345006de586ce7121cc75f086
a0a263ba08a0ad0c3ac78534d74126a6a263c008a0ad0c3ac78534f77d26a6a463c008a0ad0c3ac78534f7fda5baa463c008
a0ad0c3ac78534d757c008a07fa66780ce7121cd7570274ad13160a486aecaa0735c48731dbc03abe8983102682bd7748e0b
6daedc5d1ef6b8771deefd4c9e3fb5e7c4ae7831b4c493e4e13964c9ce80ce7277f9cd9841d3316304d056069de3429aebe0
e189a263c608a0ad0c3ac7853457cebdaf5d4dfe436cd7eb8b5b7e88b75f9b373c9758b233a0b39c7bbf1933683ac68c00da
caa0735c48731dccd052748c1969a04b19748e0b69eeeb892c0a8d312320b67283e68810da54cc4859477f88fbfab8035ef6
7d9533b44a6340a30a1971460c1a8d192320b632d01c11525b590a2f9d2a6ef965fb9403bc0cf1079036768907b7e53c265b
0b4f3586bab12c86f762c4a0d1983102622b03cd1121b57d1e3e21d0183002622b03cd1121b57dbdaaa54263c00888ad0c34
4784d4d6c1e2a58a8e1923357455069de3429afbfacfacd0983102622bd7688e08adad83a588151d3346006d65d0392ea4b9
b2c79e86259e25e60d9cc72b2937af7f3ec423ca7c0e9b65c9a8ce80ceb2cf7e3366d074cc1801b49541e7b890e60af476c6
be334fe75ee3e9513c7e1f691fbbdc0b645b7ab6ec0ce82cd06fc60c9a0e1923607eaa60734c486fe55c24bd8f325d2911db
9a5e8e3eaf9fe2d2dceed59d019de56ce4cd9841b2216304c44f1568da8372a656a95df77348cb95a46ddbe2a9f99e4e9ec4
c997750674aa756abd31832443cc48c37b5741a62d28635f9f6b0a2c648c00f6a936585ac15f5d1de4f70932c68c00d8ca60
d31e94b3cfffbc028c0927e0b532c81c0d525a0f3f6041c78413405b19748e0b69ae83dc0b45c784931aba2a83ce7121cdf5
907b21e8987002682bd7748e0b6deef3080881c67813105b19688e08a9ed26be6eb26ef7ef2d9e865faf28ccf3762ef1cff5
9c72a9be50f7ddbc2fc60b9a8bc126c0b532b81c0bd2d98dbbdccb77e5f3a1296edc7ee407e3f3b84ff10b69cfe2746740e7
8dfc6acca0e9186c02682b83ce7121cd15e8744be69abfbcc48bad2ddd734e7fa0e9dcd73cd1e7b901243b033a0bf49b3183
a663b609a0ad0c3ac7853457a0f3572f3f5858e2e5dd92de16ddf206c62b906ddbed16b6ec0ce82cd06fc60c9a8ed9260d74
2983ce7121cd15e87b72673e90ecc73cc6ce297f159fb7150ab4ea0ce82cd06fc60c9a8ef12680b67243e7b810e6ca2e6c5e
e3be25cf555bd6788ab4cd73fe5ecef334e475906c27a63a033acb6eeccd9841d331de04d056069de3429aebe0ee97a263c8
09a0ad0c3ac7853457ce49be4ad855680c3901b19581e68890dacaf9f657f1e00a8d0927357155069a23426aeb20d645d131
e104d056aee91c17da9c7a0a178f68e7b0cf5702f23c9ee390e290d563b8d219d0a99ec37963064dc78413405b19748e0b69
4e3d6f1dd3d4ad73cfab4c9d6b3c97d986492ef8f2740674aae7adde9841d331e104d056069de3429aeb60691f45c7841340
5b19748e0b69ae83a9bc8a8e092780b632e81c17d25c0f5936828e092780b632e81c17d25c078b4c2b3a269c34d0a50c3ac7
853457a608451bf192274f501bd234ae73cef9f57b7a47f88cbdcf5c47d919d059a608bd1933683a269c00daca0d9de34298
eb21c047d031e104d056069de3429aebe0a57245c78413405b19748e0b69ae8397ca151d434e6ae8aa0c3ac7853457a08f71
9ff7fceee4519e1c9e790f14f739f336d84b64b233a0b340bf1933683a869c00daca359de3429beb21f643d0316205d05606
9de3429aebe0450c45c78017405b19748e0b69ae83dfb4a263c00ba0ad0c3ac7853457a0e738523c57fc27bf3270bde99ebf
8ac7909e830cf62286ec0ce82cd06fc60c9a8e012f80b632e81c17d25c072f57293a06bc00dacaa0735c48731d9c9c283a06
bc34d0a50c3ac78534d7c14c7545c78017405bb9a1735c08735fafe6a4d098ee02622b03cd1121b575b06897a263ba0ba0ad
0c3ac78534d7c1f26c8a0ee92e35b355c1e69890de3a58864fb0d5e92e002ec51a4c5b90c67a58c6fe0f17d25d40fb540126
1d085b1daca529b890ee02daa70a30ed40f9ea218de22f19d35d006c65b0690fca590fa123828d012f40b632e01c13d2dbe7
9136028dd93220b632d01c1152db156923c667ba0b3ed6ca18dfd91ab9edeaf734c6def4ba6cd232b5d3c3656740a7fa3d79
63064dc7749706ba9441e7b890e67ad8690a3aa6bb00daca0d9de34298eb60ad1b45c77417405b19748e0b694e9df9c4ad3b
d6edccbffaa359fc4d770674aa331f6fcca0e998ee52435765d0392ea4b91e4211041dd35d006de59ace71a1cdf590f921e8
98ee02682b83ce7121cd15e8e5bad8481b383faff6a7837a5eaea40a45508da16e2cc82f460c1a8dd12e20b632d01c11525b
3e4caaf119ae828fb532c677b6466efbf5b1627cc69be063ad8cf19dad91dbdec1237e45c77813405b19748e0b69ae875fb2
a063bc09a0ad0c3ac78534d7c15d0945c78413405b19748e0b69ae836512141d434e1ae852069de3429aebe0668ca263c809
a0addcd0392e84b96bff29c667d248fdb15519e33b5b23b7bd8379618a8e492380b6724de7b8d0e63a38fd53744c1a01b495
41e7b890e6be3e1352688c1901b19581e68890da3a9809a7e8183302682b83ce7121cd7dfd8c46a1316304c456069a23426a
ebe0519ca263cc08a0ad0c3ac78534571ebae62505a6fcfc7fdae3956eba4999bf94e9adb7b4fed3f3d0557506749687ae6f
c60c9a8e312380b632e81c17d25c07730a141d6346006d65d0392ea4b90ea6cc283ac68c34d0a50c3ac7853457d639cc4be7
5e8b60efc736cd5b9e877bc46fe7be1eebb30e896c0c682cab1cbe18316834668c80d8ca0d9a234268eb60e15245c78c916a
7beb32e81c17d2dcf70b974a3a668c00daca159de3c2315796901f63db78a6c3493c84e62dcc7b9f35ee63e3b7f479834e36
06349605e45f8c18341a0346406c65a03922a4b62b99408ccf880f7cac9531beb33572dbaf8f15e3336a031f6b658cef6c8d
dcf6366a031ffb33c502e33b5b23b7bd87e81a41c7a00f405b19748e0b69ae87e81a41c7a00f405b19748e0b69ae87bfb4a0
43cc08989f2ad81c13d2dbf59bfa3b3a823ef0994f1583eb2d515bfdf95efa2f16523e1ad8bb0a2cad40e9ea6042900ae860
ca477d3a81729ad03e5ce57a10a7fc0cd2c1a30345c7a00f40d7e58ace7121cd95db0369f18c31af73b4eff3b8c57ddd9eef
5fc473c26badcde7fe80ea0ce82c3708de8c19341d833e005d972b3ac78534d7c13b638a8e411f0db4952b3ac78534d7c102
e28a8e411f80aecba0735c08731dc4532a3a067d00ba2e57748e0b69aeafa00f40ff4cb1009de3429aebe07a59d131e803d0
75b9a2735c4873053a7e0fc7fd7a17e43896658a27a8f97b696bc51468d519d059a0df8c19341d833e005d972b3ac78534d7
c35f5ad031e803d075b9a2735c48733dfca6051d833e6a68942b3ac78534d7c1ad4e45c7a00f40d765a3735c68731dbc33a6
e818f401e8ba5cd1392ea4b90ea2d7141d833e005d972b3ac78534579e47a6071bfb9acf12d7259e57ecd722f7db36c45dd1
967217cbf348d519d0599e47be1933683a067d00ba2e57748e0b69ae83f9a58a8e411f0db4952b3ac78534a7de48d8a639fe
18ae35a25392c3b88e8b9a55619d019dea8d046fcca0e998f501e8ba0c3ac78530d7c172b58a8e711f80aecb159de3429aeb
60f17845c7c40f40d7e58ace7121cd7dbd72ab4263e20788eb7285e68890dacab287e99c7ccb57fb4bbea9739cf94f322f5b
9a659ece2fcab287aa33a0b32c7bf866cca0e918fa01e8ba5cd1392ea4b90ed67a54740cfda8a151aee81c17d25c9e87a7c6
67f2063eb62edbf8ced6e86defe00ba6e898bc01e8ba5cd1392ea4b90ed6d554744cde00745daee81c17d25c81ce27a9737e
99b09e1d190f67f1723b3dbc7ea055674067817e3366d07448de007355add81c13d2dbd74b890a30c46e00b7aa565c5a8212
56609775388fe15afd711cf77804cb334de2b5f69d65ffd0aace80ce82fb66cc20c910bbd1f03ed58a4c5b50c67a4892f94b
86e40df0565590690b7f8d5d0788bf6333fb021f5997abd1f596a8ad2e8ed77d19f249f052efdfaebb19ab4ddd157d017dc5
efeff182a662f00560eb7285e53890c63a38a956740cbe00745daee81c17d2dce76963028da9173531ca159a23426aebe09c
47d131f502d075d9e81c17da5c0fabc60b3a065f00ba2e57748e0b69ee82def344da2927d0cee90598b8539fd35f65da9f05
192f68dd19d07941bf1b33683ac65f00ba2e57748e0b69ee86be668de7e3f41cf7abf1ca67cdb7a1a7b4a5e9a4a57cbd7567
40e70dfd6acca0e9187f01e8ba5cd1392ea4b90b3aee66c6f13cf3fdc839ad4a378e5b0e138f6786cb32eeebf3844a770674
5ed0efc60c9a8ef11780aecb159de3429abba1f7794e8b7ee79deb12cf448f3d7e07d306a614a9e1d89e6c2add19d07943bf
1a33683ac65f34d056aee81c17d2dcfdf5bece42d7fb36cd7d5d9fff2ae5ae5df97acbce80cefbebfd6acca0e9187f01e8ba
0c3ac78530d7c38e4cd031fe02d075b9a2735c4873e590b5c571cefc8ac4721f51f3ef6f1ea63b8ee73964a9ce80ce72c87a
3366d0748cbf00745daee81c17d25c0f2727828e091880aecb159de3429a2be7a0798df3fc0af43a6e6985f9253f58997386
414a4b7b4e4255674067390b7d3366d0740cc1a8a151aee81c17d25c0717b08a8e211880aecb46e7b8d0e6c4c39a33ee668e
784cc99743fb1c7f13f331ab057eac33a0533cac71c70c9a8e211880aecb159de3429aeb60292745c7100c40d7e58ace7121
cda9558db6f558cf29bfb2bb6ecdeb91ba33a053ad6ae48d19341d4330005d972b3ac78534d7c1ba098a8e211880aecb159d
e3429aeb6042b3a2630806a0eb7245e7b890e63a58a155d13104a381b67245e7b890e60a747aa5ee9cf3d4db789170bda391
7f8cd7afcff2f664674067817e3366d0740cc100745d069de34298fbfacd7685c6040c10d7e50acd1121b5955952f31e7ff1
576a41f59f87e74759e648fded0be82b33a47e8f173417b32f805b972b2ec78274d6c1347d45c7ec8b1a1ae58ace7121cd75
9072a2e8987d01e8ba6c748e0b6dae83398f8a8ed91780aecb159de3429aeb60f50245c7ec0b40d7e58ace7121cdf5957d01
e89fc10ea0735c48731dbc7aa3e8987d01e8ba5cd1392ea4b91e7ed3828ed91780aecb159de3429aeb60191645c7ec8b06da
ca159de3429a2bc7e9b484d792d7c339e25f679ee31f256f60bac4dfd7e57954a93b033acb71facd9841d331fb02d0751974
8e0b61ae83844145c7f80b40d7e58ace7121cd75b08296a263fc05a0eb7245e7b890e6ca65f4b6adc7999f3bec4bbe493d0c
794afd329ed334cfd502faaa33a0b35c46bf1933683ac45fd4cc75b562734c486f6a41bc780a716ef391ff2663bc40899bad
6e065a63a81bd57278ce884182d5d917a0b5a2516905525707ef23fce542f60568ab6a05261d085b6a19f72d9e28add7ac9f
b83f5dd7639b256b690c68548bb83b23060985e00ba056d58a4a0b50b204ec3ac6ef7abceccb4badcffb321cdb21ef733e8d
a16e54a9cece88416231f502b475b902d31294b00ede31516c4cbd00725daee01c13d25b0f511f828ec11780aecb159de342
9a6b832ff0b13f531d30beb33572db7b089211740cbe68a0ad5cd1392ea4b9cf7722028da91720aecb407344086d7da55e00
fa67a403e81c17d25c0f6701828ea9173534ca159de3429aebe0a198a263ea05a0ebb2d1392eb4b91ed6ca17744cbd00745d
aee81c17d25c0797308a8ec11780aecb159de3429a2bd0e7389e6b7e669526b1af63bcba5cf306eed765a67dbd5567406781
7e3366d0748cdd00745daee81c17d25c078f3f151d433f005d972b3ac78534d7c35f5ad031f403d075b9a2735c4873574080
189fb11bf8d8ba5c8def6c8ddcf61e7625828eb11b80aecb159de3429afb7aad7c85c6cc8d86d8ca159a23426abbbe5d627c
a65ee063eb32c677b6466c7b9b7a517fecef48078cef6c8ddcf60e76208a8ea91780aecb46e7b8d0e63a587e44d131f502d0
75b9a2735c4873e5b67dbc889fd674533d6ecd36a6194cf90efb7ecd04189f4715aa31a0b1dcb47f3162d0684cbd00715dae
d01c1152dba793461417f32e805b972b2ec78274d6c1a41145c7c80b40d7e58ace7121cd75306944d131f202d075b9a2735c
48731dcc9f50748cbc00745daee81c17d25c074ba2293a465e00ba2e57748e0b69ae83f4873f68ccbb6888ad5ca13922a4b6
429c668ddfabf11dd77ceae5fa25e6dfe1f8bc41a63b033a0bf39b3183a663e405a0eb32e81c17c25c07ab592a3a465e54db
cb7245e7b890e6fa8abc00f4cf48878ace71e198bb966f17e33377021f5b97abf19dad91db5e5c9fd312f7ee59459edc9ce6
f8662fc739a457fcb6c7b5ea0ce82caedf8c19341d532f005d972b3ac78534d7a65ee0637f064a607c676be4b6b7a917f8d8
9f910e18dfd91ab9edd7c78af1993b818fadcbd5f8ced6c86defe1572ce8983b01e8ba5cd1392ea4b9cbb5181fc90ff8d4aa
5a8dee6c8bdcf2ef57ff556cc89d007155add0b407e5ecfb83a22243f444c3fb542b326d4119534fe2fee3b057951ad1464f
d861bd2a8fffa625cdf6ab6c8338e56a900eee3f29ba367aa282b632e81c17d25c073328155d1b3d51415b19748e0b69aedc
9eb8ce0bf2de653cee07a4f94fb45cef583dd76fb233a0b3dc9e7833661b3d01e81fb90aa0735c48731d4cfa57746df44405
6de586ce7121ccf5163d5141ffcc55009de3429aeb61997a41d7464f54d056069de3429aebe0625dd1b5d11315b49541e7b8
90e6f2931a357e1bfe507dac9531beb33572db8beb78f2308ed75dc9639ef6fc8679f632c403cd313dbb4fd918eac662fac5
886df24345fc33d600688e08a9adec47f2b3f7fc70f0d8d329445a4a34ffe69fc786653fa23a033acb7ee4cd986df28341ff
8e35009de3429aebe088a1e8dae4870adaca359de3429bebe0dc40d1b5c90f15b49541e7b890e63a380b54746df243056d65
d0392ea4b9afe71328b436f6a122b632d01c11525b07ab8e28ba36f601d0a50c3ac7853427a0d31a4bebbce595bc96332768
c60b2305fd7406740a6877cc36f6a182fe9969d0d0392e84b9b27cd2197f7a713bf31f62b89f88a70d4c4b3ba52936cf545a
d919d059964f7a33661bfb5041ffcc34009de3429a2b0b65e5f94bf9bc611de21541dcd7e673b7f918a7793f8fe71195ee0c
e82c0b65bd19b38d7da8a07fc61a80ce7121cd75b0249aa26b931f2a682b83ce7121cd95c5efee7ccbb4814b3c5b9cf6233f
479cc7781d38ad76ff5177067496c5efde8cd9263f54d03f630d40e7b890e63a58f14fd1b5c90f065d9541e7b890e60af41c
ff124bbeffbec49fdeb2c5f3c4744a319df16c79dbd24a53055a75067416e83763b6b91315f4cf50859ace71a1cd15e82b5e
25af9d9ffe77bcd6ceff9887b81318a6c9beddaa31d48d05f9c5886de84445fc335101688e08a9ad831fb4a26b43272a682b
83ce7121cd7510e5a2e89ad0898af9a982cd3121bd7570b4126c4dee4445fc5481a63d28673d84a9fc256b7227c07b5741a6
2d28639faf97ff17ab099da8609f6a83a515fcd5f52774a2fac49fc90a185d6f89daeaaf53be14589b3b51f15a19648e0629
adec32eeebabb47ddbbdfa725e477f29b7529e5d86ea0ce82cbb8c3763b6b91315f4cf5c05d0392ea4b9023d8e715bf6bc4f
bb4333afcb8ef4f2775a7be539ed919d019d05facd986df48441ffce55009de3429abba1a7e18c5fc02967e80c79f9edeb80
3dc5d3d4f58c274bcfc141760674ded0afc66ca3272ae89fb90a359de3429beb219040d0b5d11315b49541e7b890e6eeb5f9
e379c830dd67a053fc2d9ed3908e60f174e93cf639fe916e66d918d078afccff66c43677a222fe19aa00344784d47611afe9
4e609ae3912f33d28bd779a19bf88f7189c7ece30918d18da16ebc885f8dd8864e54c43f131580e68890da2ee2e59e879776
abe99c2cbf7a923776882766c3343cf3fc746740e7c5fc6ecc3674a282fe99a8003ac78534974f09d4f86dec033eb69431be
b33572dbef6fd7389ceb7e2de71ecf85d7638ad7b4f9079f1f52a6ff63f97ac9ce80cefbfbf56acc36f6a182fe9969d0d039
2e84b91b3addf08da787f964255edcecdb315e97f1f13afe4c0f2fcb3152770674ded0afc66c631f2ae89f9906a0735c4873
3d44b908ba36f6a182b632e81c17d2dc7d6230c473d463cec7ed254fc79cf222fed331dc9704e5bc40350634dea7056f466c
331f2ae29f8106407344486d1d9cff29ba36f3c1a0ab32e81c17d2dcd7178f0aad0d7ca888ad5ca33922b4b60e6229155d1b
f850415b19748e0b694ea45cec6b3c274f37c8d2063e0f58c40339eb0ce8142917ee986de04305fd33cd00748e0b694e3d85
cc0fcb96bcc78997dc637e45573e7a7d3a033ad553486fcc36f0a182fe9966003ac78534d7c162558aae0d7ca8a0ad0c3ac7
8534d7c1d24d8aae0d7ca8a0ad0c3ac78534d7c1d24d8aae0d7c00742983ce7121cd75b07493a26b031f2a682b37748e0b61
ae873c1341d7663e54d056069de3429aebe0d50045d7263f54d056069de3429a2b93c2d2c3b371ca6f5eef71efba8ce93dba
b881c7925f94199f9970b233a0b34c0a7b33669bfc60d0bf630d40e7b890e6ca420df744e7b481f172305e8c2df9d6781c7d
4a934397075a76067496c51ade8cd9263f54d03f630d6a3ac78536d7c1720d8aae4d7ea8a0ad0c3ac7853457966bb8deb7bf
26b1c78ddaa778d2f8cfb5ba567ea4f6bcc3263b033acb720d6fc66c931f2ae89fb106a0735c48733de40108ba36f9a182b6
32e81c17d25c6fc90f15f4cf5803d0392ea4b97c27528ddf662f541f6b658cef6c8ddcf6eb63c5f86dfa013eb69431beb335
72db7bf8310bba36fda082b67243e7b810e6ca6e3beddba6eba5d161bd1efa5ccb5f36d37975674067d96dbf19b34d3fa8a0
7f2eed0f3ac78534f775e098426ba30f2a622b03cd1121b575b04498a26ba20f8cd9aa60734c486f5f4fcd17608c3ea8684b
b1a6d20aa42e751139a657ebce3d7dfde2d5c034e700167911593a033ad545a43766505c4df44145fb5401261d085be236c1
118f21cb9c5fa35cf7f416d0321d2ad2e2690c68143709bc1183846aa20f2ad4a70a2a2d40c952390fcf43d97457f6887f8e
651f25edd319d0a9921ebc3183246bd30f2a602b834d7b50ce3e5fa75d80b5d10715af9541e66890d23ab8c3a9e8dae8830a
dacaa0735c48737fa30faa8ffdb9ae3fc677b6466ebb72fd1f07452bba36fa00d0a50c3ac78534f7f94f4aa0b5d10715b195
1b344784d0d6c36e53d0b5d10715b49541e7b890e63a78334fd1b5d107065d9541e7b890e63a586546d1b5d10715b4956b3a
c78536a7a20fae73d23cc726fe1ee62d1e570f1596639d019d2afac01bb38d3ea8a07faeeb0f3ac78534d7c1430345d7461f
54d056069de3429aeb61917641d7461f54d056069de3429aeb2dfaa082feb9ae3fe81c17d25c81fe30f95dd1b5d10715b495
41e7b890e6fe461f541ffb735d7f8cef6c8ddcf6eb63c5f86df840f5b156c6f8ced6c86dff7ac97285d6860f80b89481e688
90da7af8520bba36faa082b67243e7b810e60af487d9aa8aae0d5e30e8aa0c3ac78534f7f5775ba1b5a90b15b1956b344784
d6569ef28e53bceec9b349f678c9330df178968fa9c3f532d4fc3cf0949d019de529ef9b31dbd4850afa67aa00e81c17d25c
078fb6155d1bbc50415b19748e0b69ae3c23f930165bd1b5f10b15b49541e7b890e67a482210746dfc42056d65d0392ea4b9
0e56ab52746dfc42056d65d0392ea4b91e9208045d1bbf50415b19748e0b69eef3240281d6662f54c456069a23426afb7ac1
3d85d6662f80b89481e68890dabe5f615ad2b5d90b15b4951b3ac78530f7fd02e692aecd5e78b6b72e83ce7121cd5d0b988b
f1dbf483ea63ad5c8def6c8db3ed7fd30faa8ffdb9b43fc677b6466e7b6fe90715f4cf6001d0392ea4b9efbfd792aecd5ea8
a0ad0c3ac78534f7fd1e4cd2b5c90f15b49541e7b890e67a58285fd0b5b91315b49541e7b890e6be5f7a58d2b5b91315b495
41e7b890e63e3e2591684de44505fc5401e66890d23e3ee954604dde4585fb54c1a5252861df2f77acc89abc0bf0de559069
0bca98783e356ee71cbf8b47be4517af36e3cf7e578f9cad31d48de2e9943722c32e9e6d65d8457d1e5595d32be96b7a1dfd
7f18c42957837470a353d131ec02d056069de3429aebe00e98a263d805a0ad0c3ac78534576e06c523c610f72979d7127734
e73c8ef96ed59a967a4ab7e49f9b41aa33a0b3dc0c7a3366d0740cbb68a04b19748e0b69ae839b418a8e611780b67243e7b8
10e6fa0abb00f4cf2407d0392ea4b9bec22e00fd33c901748e0b69eeebfb410a8d491720b632d01c11525b9b74818ffd99e4
80f19dad91dbfefd8990a463d805a0ad0c3ac7853457a0ef3722f28eeebc6759e4e35bdcd34de9dce981569d019d05facd98
41d331eca286aecaa0735c48731d3c3351740cbb00b4956b3ac78536574e888e78bcbeaf4186e15e14266de0344c67f4b53e
f7286467406739217a3366d0740cbb00b49541e7b890e6be7edcadd0987401622b03cd1121b5a9cb9beba87dbdfc326cf15a
643e56354bdd3a033ad5058e3766d0740cbb00b49541e7b890e6d4cce96d19ce619faf4571e6f11c87b4428e9a3a5d3a033a
d5dc696fcca0e91876d1409732e81c17d29c585b271e68f678a27a2d801dff22d1d331ab2b59eb0ce8146bebb863064dc7b0
0b405bb9a1735c08736515a578be120f30796d90799bd6f44e6c5ef3299ea3c64bd27378d60d959d019d6515a53763064dc7
b00b405b19748e0b69aeac10662b462c7bdcb7e6edfa27ade036ee63cede29d0b233a0b3ac10f666cca0e9187601682b83ce
7121cd9565e28f6bb5de7ff2eaf9e5ee6f5ad2fd39ba9665e2556740675926fecd9841d331ec02d056069de3429aeb60595c
45c7b00b405b19748e0b69ee5e16b77a8c15f73079d1907c1e356dcf1cc27b595cd919d0792f8bfb6acca0e9187651435765
d0392ea4b91b3abd88b28c793f13bf89e9d6dd3cac7903f329f2f02ce3a93b033a6fe85763064dc7b00b405bb9a6735c6873
37f43cc57dea365eebf2c5818f79cecbf8a68c866bfdc9022d3b033a6fe85763064dc7bc0b405b19748e0b69ae83afb7a263
de05a0ad0c3ac7853477439f69dddc23bf5f37c7f3c5653f87bc8ad9148f357b3c692cc8b22f547d37f08bf182e642d20568
9f2aa81c07d2580fc1357fd9907401e2a70a34ed4139fb3cb3e62f16622e1ad8bb0a2cad40e9ea21aee62f19922ec0fb541b
326de1afb10e96a31564ccd900b095c1a63d28679f7f9d05187336c06b6590391aa4b40e4e33151d7336006d65d0392ea4b9
1e4e3e041d73366ae8aa0c3ac7853477af219e26d0ac7b7e4a39c7cd5be235df99ff2af333bfe85e435c760674de6b88bf1a
33683ae66c00daca359de3429bbb578b9fd678417be43727e38fcf5e7b8fbbbc6b6d98f29c48770674deabc5bf1a33683ae6
6c00dacaa0735c4873770643dcabecdb305c8bb90f6955f974651b071f86f4f5b8897fb5853b7de1f7684143316103ac5606
94a3400abb58a735fec8b7217f03d3d215f9a95ddebe6ddfe379e1938ca41b43dd78f1be1a316834266c80d8ca407344486d
17b16d46fac68dd7e3e9336fdf328cf3913ee342d69d019d17f3bb3183a663c206a0ad0c3ac785347743578f51aa248ab481
e945f7f9dc1e68d919d07943bf1a33683ae67b34d0a50c3ac78534f7f9775ba031dc03c4566ed01c1142dbbde7ba97e8cf7f
92ebd2269e2ee6bdcdb19ddbb63e732a756740e7bdff7a3566d0740cf700b49541e7b890e67a887111740cf700b49541e7b8
90e6eeb391653e72b84a3e73d8b778ea725ed14dd78bfec7b36280ee0ce8bccf465e8d19341df33d006d65d0392ea4b90eee
f4293ae67bd4d05519748e0b69eeeb6b2a85c67c0f105bb9467344686d1d3ca45174ccf700b49541e7b890e64a8ccb99e622
e6e595ae25bde72b4fe98ce78ee33a5bd2856a0c682c312e2f460c1a8de11e20b632d01c11525b7900b99dfb30e7dd4d3c88
0cf17892a688a548edebb9a1ade1283b033acb03c83763064dc7700f405b19748e0b694e3d6a1e8e7989ffc85fcaf466ce99
1600978f9a4b6740a77ad4ec8d19341dc33d006d65d0392ea43915e3728788e63b76e776c613e5e554d0d619d0a9625cbc31
83a663b807a0ad0c3ac7853427a0d344df34ad277f2fc794199fce9e15f4d319d029a0dd3183a663b847035dcaa0735c4873
1d4c1f51740cf700b4951b3ac78530d743768da063b807a0ad0c3ac78534d7c132748a8ee11e80b632e81c17d25c072bb229
3a867bd4d05519748e0b69ae83c5c9141dc33d006de59ace71a1cd95e98ee99470ceaf96edc734453beb71ad6fbe8df15b79
3e370f7467406799f0f866cca0e918ee01682b83ce7121cdf5b034bea063b807a0ad0c3ac78534d7c15a378a8ee11e80b632
e81c17d25c07af37293a867b00dacaa0735c48731d4c5657748c1601b49541e7b890e6da6891e6637fe466607c676be4b6f7
f00513748c1601b4951b3ac78530d7432294a063b408a0ad0c3ac78534d7c1aa708a8ee92280b632e81c17d25c07cb3c2a3a
a48bd4cc56059b63427aeb60615ec156078c00b8146b306d411afbfaddf6bf50481701ea53059514205475702127b8103002
daa70a30ed40f9ea61d1f8bf640c1801b095c1a63d28673d64030836668c00d9ca80734c486f1ddc895274cc1801b49541e7
b890e67a085611744c3801b49541e7b890e6fa4a3869a07f247880ce7121cd75947002e29ff11d0d9a234268eb612f26e898
7002682b83ce7121cd75706b59d131e1a486aecaa0735c48735fe79d2934c69b80d8ca359a23426b537176ff7126baa263bc
09a0ad0c3ac78534d743a68ba063bc09a0ad0c3ac78534f77546a34263b60988ad0c344784d4d6c1e30345c76c13405b1974
8e0b69ae83fb038a8ed92680b632e81c17d2dcd74b442834c6aa80d8ca407344486d85f8ab585d85c64417105b19688e08a9
ad83e7628a8ea12e0d742983ce7121cd5de939627cc6aae063addc8cef6c8dd8f60eeeac2a3ac6aad4d05519748e0b69ae83
d506151d9355006de59ace71a1cd75b0a290a263b20aa0ad0c3ac78534579e965c2b6d5f8b781df7196adac0f8c7c947b7e7
2f2d3b033acbd3923763064dc76415405b19748e0b69ae8367908a8ec92a80b632e81c17d25c07eb4a2a3a26ab00dacaa073
5c48733d848c083a26ab00dacaa0735c48731dacfca7e898ac02682b83ce7121cd7dbdaea44263b20a88ad0c344784d4f6fd
82f5928ee12a0d742983ce7121cd7d9fc720e918ae02682b37748e0b61eefb3c0649c770956a7beb32e81c17d25c1bae828f
fd995b528def6c8db3eddf7fc1241da35d006d65d0392ea4b9efbf60928ed12e80b632e81c17d2dcf7fb4f49c76817405b19
748e0b69ae87c00f41c76817405b19748e0b69eefb85b6251da35d006d65d0392ea4b9cf533f041a735d406c65a03922a4b6
ef17919774887601f353059b63427afbfefa46b121dd05c44f1568da8372562e628f719ff7bcf2d251de2d4cffd8cf78be18
bf8fd37374969d019de522f6cd98419221dda5e1bdab20d31694b1af9f54a96016a6bbd42772563ece7fe7718c87ba5c7e06
71caf5205fdfcd57688c7601f153269a23426aeb60c2b2a263b40ba09f32e91c17d25c07b7e0141da35d1ae8bb4c3ac78534
f7fdde4bd231da05d04fb9a5735c0873df1f95251da35d00fd9449e7b890e63e3ef99268cc7501f153269a23426aebe0a6a3
a263b40ba09f32e91c17d25c1bed828ffd955bc2f19dad91dbde43ca89a063b40ba09f32e91c17d25cd9790e718386e1be85
10cdace79e9fa0eecbba1cd3fa2cfda73b033acbcef3cd9841d331daa586b632e91c17d25c07ef452a3a46bb00fa2983ce71
a1cd75f0c85dd131da05d04f99748e0b69ae83195a8a8ee92e807ecaa4735c48731dbcf6a2e898ee02e8a74c3ac78534d7c1
9a248a8ee92e0df45d269de3429a2bd063fe36a64dda87340777bf96816aefaacbc680c682fc62c4a0d118ed02e2a7dca239
2284b6b296d43e4e47dcd3e6bf42fcd1c7abbff55ae07e1bd7339eb03eaf81c8ce80ceb29ad49b3183a663b40ba09f32e91c
17d25c59356c5ecf65cf67114bfc838cf112249f98cff3144fdec6e50907d09d019d65d5b03763064dc76817403f65d2392e
a4b90ed6cb57748c7601f453269de3429aeb60f96945c76817403f65d2392ea4b97bc1c77811925e39cb1bb89c7137bbeff9
d832c5efe73a1cdbf3f444770674de0b3ebe1a33683a46bbd4d056269de3429abba1e32f2c1e36afc509e3416d8d67cc7941
ca794a5953f1385798556340e38dfc66c4a0d198eb02e2a70c344784d6962febd4f84c56c1c73e658eef6c8ddcf60e163757
744c5601f453269de3429aeb600d5545877c1530972ad91c13d2db8d1c4f5dc6789e94ff2669ce5c3e92a7cdbb7772cf8367
dd19d07923bf1a334836e4ab80b85489a63d28671d1c2a041922561adeab4a326d4119eb60855c41868815f0966a4ba62dfc
35d64322d45f3246ac00f829934d7b50ce6ee26b05ad7c0376de86f8eb4ab7adf2af2deedca76d79e6cbe8ce80ce9bf8d598
41b3316505c84f99708e09e9ad873db5a063ca0aa09f32e91c17d2dcbdaef93e0ce3b8e60506e6e188975fc39c5ffe9ad2f9
4bcef0baa1756740e7bdaef9ab3183a663ca4a0d6d65d2392ea4b97b05fb753c87d8ba5cfbd378417d0e79c1a731dd33dad3
84d41b5a770674de2bd8bf1a33683aa6ac00fa2983ce71a1cd5dd0f3950492cfc3e385bd5d7ec5abea751dc6bd30ebc65037
5ec8af460c1a8d112b207eca447344486d17f1302f795af3f5578897d9f12f92af3cc6f5b8ae3dca714a7706745eccefc60c
9a8e592b807ecaa4735c4873193a9e26a57b434b7e6773dce32ff1486f41a57f2cf1742ded83efbb6f4e674067867e3966d0
748c5b01f453269de3429abba0b7b4e8e3be9c5730c8bd644afe265ab0c405ad3b033a2fe87763064dc7b815403f65d2392e
a4b91bbabc6b9ebe7c675af1fdcc2f9f8d4b7927a930abc680c61bf9cd8841a3316ba521becb447344486d3d7cb7051de356
00fd945b3ac785307741c713867549616d798fb3dcf710f29fe85ef7e9be3fe17406745ed0efc60c9a8e712b807ecaa4735c
48739f1fac041ab35640fc9489e68890daee7392edfa9f9c51b6558f3bc6339e2c1fcb3c96a726ba33a0f33e27793566d074
cc5a01f453269de3429abbcf3e9f6366cad25c9673596243fe0badf76afbe5ec53760674de679fafc60c9a8e592b35b49549
e7b890e63ab8b852748c5b01f453069de3429bbb9f21e4c757cb15291a7f01673cb61ef99ae0daee79791e0acace80cefb19
c2ab3183a663dc0aa09f32e91c17d25c795a54be70292663bfa6efe5fcfae610ed74067496a7456fc60c9a8e892b807ecaa4
735c4873e5b9e0331b7549b7edd67884bdee62e593a6f199d1211b031acb53c11723068dc6b815103f65a23922a4b6923133
e467b5f97974faad4de778e4a779c3704e535abbb844cca8c650379684991723068dc6ac15103f65a23922a4b61e5275041d
b35600fd9449e7b890e644aa4ebc021cc674cf3f6d60dcd7ecd391fe422255e7e90ce814a93aee9841d3316ba581becba473
5c48733d4409093a66ad00fa29b7748e0b614e4509a5656ae2c5400e4a3ed37f5ad75d2e7cf8740674aa28216fcca0e998b5
02e8a74c3ac78534d7c1b29e8a8e592b807ecaa4735c48731dacd3abe898b552435b99748e0b694e2d3b7d4d27bf5e9189fb
9c739fe3298680b6ce804eb5f0b43766d074cc5a01f453069de3429beb60fea1a263d60aa09f32e91c17d25c070b69293a66
ad00fa2993ce7121cd75b0ce91a263d60aa09f32e91c17d25c07eb1c293a66ad00fa2993ce7121cd75f09295a263d60aa09f
32e91c17d25c813ee6795cae57f0e72d7e1fd6f40e545ebc701dcff8ed7ca6cccbce80ce02fd66cca0e998f4d240df65d239
2ea4b91efed2828e492f807eca2d9de34298ebe1372de898f402e8a74c3ac78534d7c3de5bd031e905d04f99748e0b69eeeb
9782151a625e6ae0a74a30478394564ec6d232577b9ecd15ffff74f57365086fdb300c71a7733e77036567406739197b3366
906c75cc0b80ef22c0b40569ac8340c7bf5c487a016da9124c3a10b63ab8b8105c487a016da9124c3b50bed42b3ee33ec68d
c96f61c593e26b06aaba2b649d019dea151f6fcc20c998f402e0a74c36ed4139eb212e40b031e905c84f99708e09e9adafa4
1740ff8a31219de3429aebe09690a263d20ba09f32e91c17d25c5f492f0df4ff9f64423ac7853497df0850e3336e051ffb94
dbf19dad11dbdec3174cd0316e05d04f99748e0b69ae878c1941c7b8951adacaa4735c48731ddc6855744c5c01f453069de3
429bebe0dc40d1317105d04f99748e0b69ae83134045c7c415403f65d2392ea43915339366630c5bbe2dbc4df7b20b87827e
3a033a55cc8c3766d0740c5d01f453269de3429aeb20ac53d1317405d04f99748e0b69ae834b5745c7d015403f65d2392ea4
b9727fe27a4325df1ddbe37f6119e30e376de0b1e4a5ec9e4396ee0ce82c7728de8c19341d735700fd9449e7b890e6bebe29
a3d098bb02e2a74c344784d4d6c1c2248a8eb92b0df45d269de3429a2b7759d3d3b3e95ac57658af8974d73eb73933d19d01
9de52eeb9b3183a663ea0ba09f724be7b810e6aeb019313e7357ea8fb532c777b6466e7b07cbfd283ae6ae00fa2983ce71a1
cd9567171fa6d42b3ae6ae00fa2993ce7121cd75f0944ad1317705d04f99748e0b69ae83f50f151d735700fd9449e7b890e6
3a58184fd1317705d04f99748e0b69eefb15b7251d735700fd9449e7b890e63e8f2011680c5d01f153269a23426afb7e097d
49c7dc15403f65d2392ea4b9027d4ecbb45eeb8d9ed57bead793e3697b2eeb74674067817e3366d074cc5d69a0ef32e91c17
d2dcf7b118928eb92b807eca2d9de34298cb67816a7c269f54ff8daaccf19dad91dbde431a87a063ee0aa09f724de7b870cc
5d712f627c269fe0639f32c777b6466efbf77b3049c7e413403f65d2392ea4b91e4240041d934f00fd9449e7b890e6be3f25
91744c3e01f453269de3429afb3c0444a031f604c44f99688e08a9ad87440c41c7e413403f65d2392ea4b9ef2fac241d924f
c05caa64734c486fd7f1e2efe8c81ec167962a07d75ba2b6ba489e63613dd7eb0c25077fe793f1fdb85f9d7e0ecbb233a0b3
487e33669064c81e6978af2ac9b40565ac833702547208e247703655958f7fe7f4c0eb2adb204eb91aa4833702141d124808
6d65d0392ea4b9ae124808fd2b5e83748e0b69ae83db8e8a0e09242d742983ce7121cd7dbf0f9374482021b4951b3ac78530
57a03f4c3e5674482021b49541e7b890e6be3f0f9274082121b49541e7b890e67a08c61074082121b49541e7b890e6be3ff9
9374884021b49541e7b890e67ad891093a44a010dacaa0735c48733d1cb2041d2250005d9541e7b890e63a78e145d1210285
d056aee91c17da5c390d1d97783a9e76b4fb766c637a81219f31eed779d4f8ecc75463406339097d3162d068c83f21b19581
e68890da3a9821a6e8907f42682b83ce7121cd75f0aa80a243fe09a1ad0c3ac78534d7c1ea118a0ef9272d742983ce7121cd
75b0228ca243040aa1addcd0392e84b9b2d4d13585e95a99683cd3012507776ce9096a3cda3e7f68d518d058963a7a3162d0
68c83f21b19581e68890da4a14c83d6f2f6ddf32edebb41ff928338f7b5a99dc1edde8ce80ce1205f266cca0e9907f42682b
83ce7121cddd0bb6dd8b16e735a7ce7559f673c87b9c783a11ffdff9ac762efb42d5772fd6f662bca0b9907c425c2b83cbb1
209dddeb60570f37e285c1122f0f8e3c4b73badf10decad75a770674deeb60bf1a33683a249f10dacaa0735c48731dac83ad
e8907c02e8aa0c3ac7853477438f43dcc879ca6b56c7cba0635ab6fc339cf2f3acf47f2cd0b233a0f3867e3566d074083f21
b4956b3ac78536772fb479dce92cf907b70c639a3d965770be2e88c6e749bfee0ce8bc17da7c3566d074885e21b49541e7b8
90e6aec417313ec24ff8b156c6f8ced6c86dbf3e568c5fc78ff0539f2a4677b6456e790fdfeabf6c75fc08899f2ad0b407e5
ece68d2743f110722d0239e4b481e5dad031ee63e77d7ba65aeace80ce9bf7d5984192d5f1232def5d0599b6a08c75b068ac
20abe347c8fb541b326de1afb19b379d721f5bbe3d9512e1e2be65bca2d2cadfa5ecb0746740e7cdfb6acc20c9103f42602b
834d7b50ce7af8460b36c48f10d9ca80734c486f3d1c97041de247086d65d0392ea4b97b9defb21e4ddac0f99ce38e7dceef
ad8efb1396722ff42d3b033aef95be5f8d19341de247005d9541e7b890e62ee8b15a6ba83a7ea4a5c8d3f2fbf182b740ebce
80ce0bfadd9841d3217e84d056aee91c17da5c0f6bf70b3a249010dacaa0735c487317f4baecdb1c773c5740c8785fcfa67f
cc47fe361666dd18eac60bf9d58841a3217e84c456069a23426abb88e73d5ecae6b77f866b898523ed83f23f863d1f66efef
b6d319d07931bf1b33683ac48f10dacaa0735c487317743e7c0c67bedf3cc6a3f83cc41d4efea35ccf5bc6f2cab3d319d079
41bf1b33683ac48f10dacaa0735c4873f9dc5e8d8f0c90f6634b19e33b5b23b7fd721df772d72fe0facea7dbbd290b2ffda3
2c947dbbd69d019d97eb7763064d870c10425bb9a1735c0873f77ee41ccf29ee62f206a625efd393d91ce4b00c43bcdaddca
d2d34e6740e7bd27793566d074c80021b49541e7b890e6ee23c6f54cf20ad14a4bbe2fc79e6fff8e7635761f31646740e77d
c4783566d074880121b49541e7b890e6ee73836dcf7bbb7cc6368ed72c8cbc812942773e96e78448770674dee706afc60c9a
0e312084b632e81c17d2dc7dea3becdb382df97c25feaf7158e3ee7ee39fe83ef5959d019df7a9efab3183a6430c08a0ab32
e81c17d2dc7d9193de624b73e7d306a6558bf7bc3860fa1395e55bee6b1cd918eac6fb12e7cd8841a3210384c456aed11c11
5a5bf83e5d5ad1210384d056069de3429abb1f1e5c3b9c7c8e3aa7b7b153204bbe73b43feb8fde8f0f646740e7fd00e1d598
41d3210384d056069de3429a2b8f89ae875539b222fdef61b9d3b08778881b260b0d908da16e2c0f895e8c18341a3240486c
65a03922a4b612f532a5c52ef359c3729fa0e57dcf3ced715bcff57cd642919d019d25eae5cd9841d3210684d056069de342
9a2bd069cad699ef28c49fdb348c71a79bfe31a7d783f2ace0075a75067416e83763064d871810425b19748e0b69ae3cf4cd
ebd3e489534f4847dec0235e84ef11e081969d019de5b1ef9b3183a6430c480b5dcaa0735c48739f47dc0834648090d8ca0d
9a234268eb604283a2430608a1ad0c3ac7853427d25e8eb8675d8f1c1eb7c4436cdcc2e154d3759ec6804691f5e28d18341a
0240486c65a03922a4363159275e0e2df1d796039ae3b9c392ce2ee4628f4f63a81bc5541d6fc4a0d190fe01e2aa0c344784
d4a6961fdc96339e26e5bf4ebc045ce35f67966bd295c68046b5f8a03362d06888fe20b1956b344784d6a656a4dbcef8039c
f3a540bc4058f7b8df55eb2d5a63a81bd57a74ce8841a321f783c456069a23426aeb6091324587dc0f425b19748e0b69eeeb
f5ba141a423f486c65a03922a4b60ede6b527408fd20b49541e7b890e6ca24f1eb1d8deb35e4b851fb34e613c4f4beddbc2c
cbf1fc9d656740679924fe66cca0e910fa41682b83ce7121cd95d70186f87f18867b45833cb777cfb3d8e3117539e24fe279
db4576067496d701de8c19341d423f5ae852069de3429acb7798d5f888dde0c75ab919dfd91ab1ed1d2c5da5e810bb41682b
83ce7121cd75b0329ba243ec06a1ad0c3ac78534575ebdb8d73748db142fa7a7250e9bf775f1a220bda3b23cbf2ad919d059
5ebd783366d07475f20698ad0a36c784f4d6c14a9e82ad4ade207029d660da8234d6c179d05fae3a7983b44f1560d281b0d5
c11b2682ab4ede20ed53059876a07c09de782d3d6ce79943ecd7738dc7f0b4ef51bc4f6740a7e075c70c920cc91b04b632d8
b407e5ac8345d4151b9237886c65c03926a4b7ae923708fd2b5682748e0b69ae87100a4187e40d425b19748e0b694e40cf5b
fcbd4de391ff2a51d336aee3a27ed0d619d029a0dd3183a643f2460b5dcaa0735c48730a7a1c8ff4be66be4118cf17f63caf
4b4297ce804e05ed8d19341d723f086de586ce7121ccf5f0971674c8fd20b49541e7b890e6d48e6c8c6768715f9b5f5c8c67
85c3b11d72e7fd3486ba51edc69c11834643e80788ab32d01c11525b0749278a0ea11f84b6724de7b8d0e63a38482b3a847e
10dacaa0735c48733d249d083a847e10dacaa0735c48733d249d083a847e10dacaa0735c48731d84192a3a847e10dacaa073
5c48731d5c472a3a847e10dacaa0735c487305fa5aa937dfb1d99ed734d31ee8dae50ccfd340d919d059a0df8c19341d423f
086d65d0392ea4b9af1f222834847e90d8ca407344486de526d8bc4feb722d7e53fde7e1f951965b607ffb02faca0db0dfe3
05cd85b88f16b794c1e55890ce3a7866a2e810f741682b37748e0b61ae407f9824ade8103602e8aa0c3ac78534d7c16a428a
0e612384b6724de7b8d0e63a7878a1e8103642682b83ce7121cd75b0009ea243d808a1ad0c3ac78534d7c102788a0e612384
b632e81c17d2dce71114020d492324b632d01c1152dbf70bb84b3a248d10dacaa0735c4873df2704483a848d10dacaa0735c
48733de46e083a848d10dacaa0735c4873570485181f711fedc79632c677b6e6ff987b936cd9799cc9b68f517c2388a5ba98
53cebf9d2c456e5d431eb5d2f9aff756c6452070b8cddd2552226172ecf5cffe531f861bfcb32dfcaaef8c468cfd65b8d1ff
d9bfdd2458df198d1c7bee252eeac3f2827fb685bbface689cb1fffe572ce960b841e816069da385546e04ef094107c30d42
b730e81c2da472f90b26eac3f2827fb68551df198d1cfb50961784fecbd281748e1652b93aefcbff97aee2c715c6341ff9fa
3e876f427a60f8ccfb54a621b3cefbbed4344d07d70b42b730e81c2da472bf9fec4a3ab85e10ba8541e7682195fbfdb246d2
f5ae17647ea260739490bafd7e9fa362eb3d3748fc4481a675509ad5e57a3daf172fe65d0f8fd8c921dccbb6663b26330d99
75b9fea5a649b2de73e3cd5ba220d32a28c5eadb831f5a862bbb0c7a6ef413aa2e7cfe6f0b77da3d875b1127dc15f9f53365
8546c30d10b730d01c21a46c033c64557434dc00740b83ced1422a37c043564547c38d17740d83ced1422af7fb0b98a4a3e1
06a05bf845e7682194fbfd5d4ad2d17003d02d0c3a470ba9dc088e04828e861b806e61d0395a48e5d2825dd5a7e505fe6c0b
a3be331a39f6fc67457d9a4ee0cfb630ea3ba391631fe1c72ce8683a01e816069da385546e847b95a0a3e9440fdd8541e768
2195ab2f7de7e50877962bcffcafb0c8ddd734819a7203f2f581969986ccfae2f74b4dd374349d00740bf7748e165ab90176
71283afa4e00ba8541e7682195fbf5e1578546d30910b730d01c21a46c036c185774349d00740b83ced1422a274eee870b4e
b8f84cb9a37e480bff936b559f73cb34648ab3fb6e4dd374349d7841d730e81c2da472b547c31aa6a57b3e5dbf1ea589421c
606a1a123ea4a7cf8ccc3464d61e0d5f6a9aa6a3e904a05bf845e7682194ab4d57ee733b9674aa2e9a0884dfc2911e87af75
3368eb34a3320d99b5e9ca979aa6e9e83b01e816069da38554ae36148aaf62f333c92d35a79bf634c0e53ef7f38c6d312bb4
cc3464d696425f6a9aa6a3ef04a05b18748e1652b9d23a6a5dc297ee48ebbdf55cd6f95ad7d4f328f6fc0c2b84d8a3b0b48e
929986ccd23aea534dd374749f00740b83ced1422a3740b76b4547f70940b730e81c2da47203f47e5674749fe8a1bb30e81c
2da4720374fd5374749f00740bf7748e165ab9d2f07a2a0bdbf4ed0b69fbb2a5357f9848ed7bf890ea1d4b275a9f58da5d7f
a9681a8dd613206e61a0394248d9ded613f8b37ffa3aa0be331a39f6b7f504feec1fe60aa8ee8c458ebc7cbae7355dfb9a9d
03d6750ff3c42d9d780f9fc21d9f0f3ffb9374a621b37cbe9f6a9a6483f504889f28d0b40e4ab3117ec2ff92c17ae2c55ba2
20d32a28c50aef7a866ff99e16784b981a9dd33ecfe9b7765d47eec3587965a621b3f07eaa69920cd613e07da22f32adc2bf
8ae51fd1bfb569fe803fd9c2a8ae47a2463d82258060a3f903905b18708e1252b70c1d3ef8f0259f52a3887828386d3b4803
7c7af766669d687d6246fe54d1341a9d1f40dcc240738490b2e5eece41f37d0b2bf8d48738beb15ae2d22bfee3b58c71320d
99b9bbf3b79aa6e9e8fcd0437761d0395a48e532745e626c773666284f11d200db552043eb4c436686fe56d3341d9d1f00dd
c23d9da38556ae746cdfd3c7925b8b573b9df88f75dbc2826fade72175a221b1f46bff52d1341a6d1f40dcc240738490b265
e239683187a97eba4b877b595864e51d6bf312be89d7b1d4859b9369c8ccccdf6a9aa6a3f303a05b18748e1652b9041da744
e7325de9357c982f2cd739dff91a3ba79fe1bc941fb49369c84cd01f6b9aa6a3f303a05b18748e1652b90c7da553e66b7a51
37ddf3742ec7912eb9f3fcb4f7ccd03ad39099a1bfd5344d47e70740b730e81c2da4726fe787d79ffdc3d600f59dd1c8b197
2f583e015c74d9c3057ede532f9a79decaabf1fa05939986ccf205fb54d3341d9d1f00ddc22f3a470ba15cb994e4297e6a46
37875b77f836ae47ba062c75ef67bd94c84c4366b9947caa699a8ece0f806e61d0395a48e546f0f81074747e00740b83ced1
422a97a1cf644f36e74734c9b1e98cadc6e32d6edff73b9a7214689d69c8ccd0df6a9aa6a3f303a05b18748e1652b93cef9d
d62d9db94937b23b2c73ef6959b3b7d1e3c89727be3ad3909967bedf6a9aa6a3f3430fdd8541e76821952b2b9c679612a7e6
e12a937af1a4d5489aafcd4faf3d9d69c82c2b9c4f354dd3d1fc01d02ddcd3395a68e50670e55474347f00740b83ced1422a
571c2fca23d8f4b47d09bf81b0084b06c02ba7443ad19058fc2ebe54348d46e70710b730d01c21a46c85385cd3e7fb4eefc3
d7b8bb38dc60d25185fa76e139aea6330d9985f9534dd374347f00740b83ced1422a3782c787a0a3f903a05b18748e1652b9
fa4aae3d6a0e73a5305bbc8efc52e50e538bf41ceb7925a7320d99f595dc979aa6e968fe00e816069da38554ae427ff1aaff
cfcd346456e82f354dd3d1fce1055dc3a073b490cad537ced502370cf0285fbfe4a35ae7c84f731f9969c8ac6f9cbfd4344d
47ff0740b7f08bced14228576d3ef2a6a5f4eaff0e2befed48c64aeb71142b9acaac120d89d5e4e34345d368347f00710b03
cd1142ca5677532cd37687d9e97f71e7c39c3b1da4cf677af71c949986ccba9be24b4dd374f47f00740b83ced1422a27f6cd
8439e334c7174a71808f0f8bda37f3641a32c5be19b7a6693a5a40f4d05d18748e165239657a11e649773a6b162d2a96f807
b64935ae6a99864c657be1d5344d47170840b7704fe768a1951ba02f9ba2a31104a05b18748e1652b9017a022b3a1a4100ba
8541e7682195abe76dc2353d4cd8fe4b1d4a8e3936144ba763cebc0f777e4ee0aa4443623d6ff3a1a269341a4180b88581e6
0821651ba191b9a0a31104a05b18748e1652b901cedb283a1a4100ba8541e76821951be0e889a2a311c40bba8641e7682195
1be0e889a2a30d05a05bf845e76821941ba09393a2a30d05a05b18748e1652b9117ed3828e3614806e61d0395a48e5ead5fb
0a6b8123adf962a7e6f0ed5bd3ebdc30790e0bdf396e7eae576f9569c8ac57ef2f354dd3c186a2676e51b0394a48ddea2d3a
1a39aed3917e704b9027ac00f309d6633eaff57e96573ad390596fd25f6a9a64eb6d28005c833d9856412a3682e9c63f5cb0
a100ed130598d440a835c0116fc1051b0ad03e5180690d945e0374011664b4a100700b834deba034132ba92b5e64d63d7597
3f632f896db9d4519327d19028d6515e45d360f4a0006f0b83cc91418a3640237345470f0a40b730e81c2da47263795000fa
4f8305d0395a48e57eed07aad06840f122ae61a0394248d9067818a4e8684001e8167ed1395a08e5d21616559f1610f8b32d
8cface68e4d87ffeed1268b480e889bb30d01c21a46c03387d283a5a4000ba857b3a470bad9cba723e4f02d323f0f0212df7
297f524fa621535d39bd9aa6e9680101e816069da385544e3d5f9df76d0ebff46446724c7bde70239fafd64c43a67abeead5
344d470b0840b730e81c2da47223f85e083a5a4000ba8541e76821951b60c6abe8680101e816069da38554ae4287a5c6b5e6
5e696b98a19edb9407b8a6ae80dda457255a9f58913f54348d46ff0710b730d01c21a46c03b44b5074f47f00740b83ced142
2a37c0525dd1d10202d02d0c3a470ba9dc00dd40141d8d205ed0350c3a470ba9dc007d5f141d8d2000ddc22f3a470ba15c7d
a8fc43f36c454723881eba0b83ced1422a37c03b134547230840b7704fe768a195fb7df3504947230840b730e81c2da472b5
4d6c698c9806187e6de1bfbff30bcc30c225cc199f772632d39059dbc47ea9699a8e4610806e61d0395a48e52af435cdc79c
a60ed759b6f6e7971dcfa4a242ab4c436685fe52d3341d8d2000ddc2a073b490ca8d608b20e8e80501e816069da385546e04
870041472f0840b730e81c2da472053a9d56bdf28bdafb3acf74ec3dfd18e3ebdb797f368eeb4c436681fe54d3341dbd2000
ddc2a073b490ca65a300519f6e0cf8b32d8cface68e4d8abd6f339ad7977f93dcdeb11a68c49a37298787dae242ad1905895
fe50d1341a8d205ec4350c34470829db08df2e41471b0a40b7f08bced1422837c27544d0d104a31b6f1f069da38554ae4287
5bdab2e619db9dbe885b1eedb62ff3ba9eadb9a5cc346456e82f354dd3d18203d02ddcd1395a38ca8d65c101e83ffd2540e7
6821951bcb8203d07fba5b80ced1422af763672c8946f70f10b730d01c21a46c95386d6b493b5ccef818693dd2de8f6b59ce
337c42fbf6108b44436225fe50d1341aad3f40dcc240738490b28d65fd01e83f7d2d40e7682195fbfd9a4ed2d1fa03d02d0c
3a470ba95c5dbd5ff319eea0e98308bf8074ee3f5baa86794cf85a2ecfcd4a661a32ebeafd4b4dd374b0fe00f313059ba384
d4edf74f69141bac3f40fc4481a675509afdbed9b62283f5c78bb74441a655508a0db02f4cb976d0faa39f4676e1e37ffb71
ef4b0eb7224eb82b32c0eb044547f70f40b730e81c2da472f575c24f1ca515177d3f80dbc2e07254909a0df02241d1d1f7e3
055dc3a073b490ca0db0515bd1d1f703d02dfca273b410cad54bf60f1da5151d7d3f00ddc2a073b490cafdfebe2ce9e8fb01
e816069da385546e80c3178a8eae23806e61d0395a48e546f87a0b3a7a9e00ba8541e7682195ab17b230f1bdb6d4fa3c7c1d
e635acb2537bd533f6b73ff7ed69c6a2330d99f542f6a5a6693a7a9e00ba8541e7682195fbed1d5a70d1eda4c7edc2e07254
909afdfa40a842a3d509885bb8477384d0b20d30d95674b43a01740b83ced1422a27f6a585d5ed74842943fc54f6f0bf5ed3
6f52ed4b7b320d99625f9a5bd3341ddd4e00ddc2a073b490ca0db0d754d1d1ed04d02d0c3a470ba9dc586e272fe83fac3c40
e76821951ba03f87a2a3db09a05bf845e7682194abbd67e274e548dfbeed0cf388306148035cb7236e3a8ff6bdb5f78cca34
64d6de335f6a9aa6a3db09a05b18748e1652b9015a2b293aba9d00ba8541e76821952b4db4e6f30877f0bdb4c4d9d7fdded2
bd264c9dca4fb340eb4c436669a2f5a9a6693aba9d00ba8541e76821952b5df1e24e9ab8d1247d1071737b3a231cfef11c08
2b4df164a2f589a527de978aa6d1687502e216069a238494adb4d00f17f6f398a6dcbf6d8a3d40534bce70b54de7ee0bef5f
69569ae7ff5dcd34141d4e7ad62e0c2847022958e9ea9edad6c587a051fe309e70fb4c7d568fe70562e9eaae120d89a5abfb
978aa6d1686f02e216eed11c21b46c9978ee9e39774d18e30097dc80b44ec174a62133337fab699a8e0e27806e61d0395a48
e5066862aae8e8af02e816069da38554eeedaf823ffb878308aa3b639123cf7ff4dfea7038c1df7ca228ae47a2463dc217fa
5f32389cbc784b14645a05a5d8cf2f59ff62c1de04b04ff485a525f857aef2e13e5618716873dedd9ccc238e6d9ac3e2fab1
1ad19986ccf2e17eaa69928ce62a006e61b0691d9466235caa041bcd5580dcc280739490bae5ceda57da89b5a556aa73f874
c234615bd257302c7df6b082bfcaccd2c93464e6cedadf6a9aa6a3bf0aa05b18748e1652b90cbd9776e7f983984b2bd1ff52
f3f3e47a539975a2f58919f95345d3683457e989bb30d01c21a46c9978edb65da6f3fd57b4c04bff98ce3b7e176bd7789d69
c8ccccdf6a9aa6a3b90aa05bb8a773b4d0ca65e829eda15dee64f832ade1dfcb7265b388b3ec3c2dd03ad39099a1bfd5344d
477f1540b730e81c2da472d909227a731ee79c7e6fd37597fd3b698053b81a85854779a0eb641a32b313c4b79aa6e9e8af02
e816069da385542e43ef61097f2cc9196cba52ebdc356d0198c3c7759fc7e380aa130d8919f95345d368345701710b03cd11
42ca9689b7fd5aa62b5c5cd20097bc80bdf39772deb67bdbca77dbc9346466e66f354dd3d15c05d02d0c3a470ba9dcdb5ce5
f567ff700e417d673472ece5db150fbe849b5892259f7cc937f4e446b7851979fd76a9444362f9767da9681a8dce2a206ee1
179a2384906d848b88a0a3b30aa05b18748e1652b972bb8887bac26d3c4ec6a7f40227beba49038cc79b96a7d59c9369c82c
b78b4f354dd3d15905d02d0c3a470ba95c868e1d97ef397b662efbba4693b4e4f1b3842fe876ef77759a938986c48cfca9a2
6934daaa80b88581e6082165fbf5e44fa1d153a527eec240738490b295397eeed999f6197446a169b0d524a4ccf165a621b3
ccf13fd5344d474f1540b7704fe768a195fbb56ba442a3a10a885b18688e1052b6010c58151d3d5500ddc2a073b490ca9597
049d0bde72cfe16a73eda997e712eeafa98748fd9c75a621b3bc26f854d3341d3d5500ddc2a073b490cad9ef7de0151d3d55
00ddc2a073b490ca15f79c730d75b6e4e5bd6e77f80286d556bce42ef941caf17cbd75a621b3b8e77caa699a8e9e2a806e61
d0395a48e5ca8bbfd8ea709b935bd57ac6a36ce79a4cca967c14637a8e60ea4c436679f1f7a9a6693a7aaabca06b18748e16
52b9f28a372daad774d6684b6708c317308eef9aa2dfd1de6c2754a221b1bce0fd52d1341a0d5540dcc22f344708215b7da9
9d4d4ee2f730fdbfd3566caca630459f96a539e7a844eb13eb2bed0f154da3d15005c42d0c344708295bddbb10375adec97d
2ffcd696699eced418652d4ba3e969a024330d9975efc2979aa6e968a802e816069da38554aeee5299d2e438b56fadce8ae9
139aa6b02c880ddeeb261595687d62dda3f2a1a26934baa9f4c45d18688e105236b119e99ae2238623de4943ce1666ccd3ad
fade3d898644b115c9ab681a8d562a206ee11ecd1142cb3640675845472b1540b730e81c2da472aad5df11963ce795eea3fb
756f7b587bceb2bfe19369c854adfebc9aa6e968a502e816069da38554aef6809b732798ff723ba7693deeb4d9a21c995caf
674ba5cc3464d61e705f6a9aa6a39b0aa05b18748e1652b901daa1293abaa900ba8541e76821951be01493a2a39b0aa05b18
748e1652b9df9fc1947474537941d730e81c2da47263b9a900fa4fab90179da385506e80534c8a8e6e2a806e61d0395a48e5
06e8f6a7e8e8a602e816069da385546e840b99a0839b4acfdca260739490ba0d60ff25d87a371500d7600fa655908a8de01d
f30f17dc5440fb44012635106a0d3001135c705301ed130598d640e9a57a4b879b48b8d4a4ad63c71cd6daf779aa831d2dd1
fa44d559daa968128b562aa06d61806911946003f40f576c745301720b03ce5142ea36969b0aa0ffb40a019da385546e0023
044547371540b730e81c2da472027a2e8ecfe904ca1d7e88736c2ba2a09f4c43a680766b9aa6a3a1ca0bba8641e768219553
d0b1e9f6bca527b161f63fa796e7d268f6c934642a68afa6693a1aaa00ba855f748e164239013dedf771c475501ce0153e9f
70dd3de4d39127d39029a0dd9aa6e968e702e816069da385542ebbc888faf454e9ff6c17467d673472ec03fcaa141d3d5500
ddc23d9da385564e5d3fe7f98aae86e99170988f9e694fbdbc7ed64c43a6ba7e7a354dd3d15305d02d0c3a470ba9dc000f5a
151d3d5500ddc2a073b490ca0d303d5074f45401740b83ced1422af76b4f3d8546431510b730d01c21a46cbf9ee92b341aaa
80b88581e60821651ba02983a2a3a10aa05b18748e1652b9ba80cddd80d373b123b7714aa7268e33b71568679965a621b32e
60bfd4344d47431540b730e81c2da47203bc41507434547941d730e81c2da472033c785374345401740bbfe81c2d847203bc
4150743454e9a1bb30e81c2da47203bc415074345401740bf7748e165ab9da87b63cf98e033cb6f0939bce3459bcc29c315c
71a6a7e788ce3464d63eb45f6a9aa6a3a10aa05b18748e1652b91ff7589668745301710b03cd1142caf6fbe6e1928e6e2a80
6e61d0395a48e57edf265ed2d14d05d02d0c3a470ba9dccfcd2e041aad5440dcc240738490b255e225f6e34bfe4df7b4dedb
76ac473673cad7dbe60925330d9995f94b4dd374b45201740b83ced1422af7b652c19ffdd32a04f59dd1c8b1e73f2bead3cf
e4f5676b18f59dd1c8b1fffc4b2dd0686602e2167ea1394208d9b2d0a23eed44baff451f467d673472ecbff7509174b41301
740b77748e168e722378a8083ada8900ba8541e7682195fbbdd598a4a39d08a05b18748e1652b9df4f83241d1d4500ddc2a0
73b490ca55e8bce5217d15af7cf8298f36fcf49227ef737f949986cc0afda5a6693a9a8a00ba8541e7682195fbbd7da0a4a3
a908a05b18748e1652b9df2fe6241d4d4500ddc2a073b490cafdde1254d2c15404cc4f146c8e1252b7df6ffb536c301501f1
13059ad64169f6fbbee58a0ca6222fde120599564129265e222ce7b4842b6cdad7742ce593912d7f9f4c43a6788de0d6a4a9
c8335c9a8af493c92ebcff2fdefeef33855b1127dc1519604f98a2a3a908a05b18748e1652b9baeb2f3b36a587bf679c172f
47dea097e68bdb737c4c261a12eb9ebf0f154da3d15704c42d0c34470829db00bb3a151d7d455ed0350c3a470ba95c7d8910
cfdf2c792e3ca5465aeb9e7f894f1feafa1241651a32eb4b842f354dd3d15704d02dfca273b410ca8de04120e8e82b02e816
069da385546e04071941475f1140b730e81c2da47203ecc85774f41501740b83ced1422a37c2d75bd0d15704d02d0c3a470b
a9dc08576f41475f1140b730e81c2da4723fbf490b345a8bf4c45d18688e1052b63a173bf6695ed33aefdcef30f7bfaef4c8
ee08cb9ff598d7fbd9ec29330d99752ef6a5a6693aba8b00ba857b3a470bad5c9d75a70e21799db73e2bfab8892e3dd1e836
2ba944eb13eb9cfb4345d368b41601710b03cd1142caf6ebed590a8dbe22206e61a0394248d9c6f21501f49fa619a073b490
ca8de52bf282fec33403748e1652b9b17c4500fda769c68bced1422837406f0e45475f1140b730e81c2da47263f98a00fa4f
d30cd0395a48e54a7ba1bcde4b9b4cd6633af7354e5fe200f77d8eafca9fbd0d3ad39059da0b7daa699a8ebe22806e61d039
5a48e506f21501f19fee1940738490b265e2b5aefde290d63b2c83d6356d9a4bdd09d778c2b320eb4c436666fe56d3341d0d
467ae82e0c3a470ba9dc08ee04828e1e23806ee19eced1422b975b3ddedb11be7a5b6e4b78c595e0bd67d797b0eacbf6ed19
dac93464e6568fdf6a9aa6a3c708a05b18748e1652b901fa5b2a3a7a8c00ba8541e76821957b7b8ce0cffee1e181eace58e4
c8b3d2e771371fa1e7f97ffc475e943c5da79d4c436656fa5b4d936c703801f113059ad6416936c237eb5f32389cbc784b14
645a05a5d808978f7fc9607202de27fa22d32afcabd8089fefbf64343901700b834deba0342bbfe0653bc3423ecdc3f31deb
3c53f3c7305309fff32d1e9029bf609969c82cbfe04f354db3d1e404c82d0c384709a95b864ea75eb674906fde963d9fe08c
ff58b72d2c08d6fade44271a1233f2a78aa6d1e87002e216069a238494edd76dbd151a1d4e7ae22e0c34470829db08dddb05
1d1d4e00ddc23d9da385566e843efd828e0e27806e61d0395a48e57e6ec720d0686f02e216069a2384942d13af6104e14349
67cee303ef7b39d77cb9c9dfc8b5ee5c76320d9999f95b4dd374743801740b83ced1422a97a1e7eed5dd74a6adf347da4a14
56b9cb1e5bf5957bb39369c8ccd0df6a9aa6a3c309a05b18748e1652b9b7c3c9ebcffee1e081face68e4d88bd667b1e94aba
ecfb1926c4e945f174c7adbf618d5ddd6474a621b368fda9a6693a9a9c00ba855f748e1642b962a1531ce896f4b32fdb2092
9553b8c41dd319be20055a671a328b85cea79aa6e9687202e816069da38554eee74e49028d0e27206e61a0394248d932f195
6dabd37c7c0a0b81336e0a48d398396e549cceda42ddc9346466e66f354dd3d1e404d02d0c3a470ba95c99ff65e7dff45a61
0e8b9f755ab3ede29c0f92cefb3301949986cc3203fc54d3341d7d4e7ae82e0c3a470ba9dcafa7f90a8d2627206ee11ecd11
42cb36807ba3a2a3cf09a05b18748e1652b9e2d37996e7b0ffc587b071a34bec7f94d69e7952be3c3e9d32d390597c3a3fd5
344d479f1340b730e81c2da472e511fe544e7ac7012edd7ba539e4edd37cd617353ad1fac4f200ff4b45d368343901710b03
cd1142ca36829d8da0a3c909a05b18748e1652b9f26eea794414cd29f6fd0a37f3f442bc7bdb505e4ec94c4366793bf5a9a6
693a9a9c00ba8541e76821952bd0e12233cffb9a0c49a62bac3da735f52f5996f98e53e5eb792327330d9905fa534dd37434
397941d730e81c2da472c5ce26af32d316b535acfcb679bfeffc32ed391c5eec6c64a621b3d8d97caa699a8e3e27806ee117
9da38550aebc6f0edfc2302f4cef46c3e4210c715dd353a4253f239ecffa925d671a32cbfbe64f354dd3d1ea04d02d0c3a47
0ba9dccfcd5d041a7d4e40dcc240738490b2d50d2461d115d58fe3bbc2bc7c598ed4eb205c00f2cea6677f94cc3464d60d24
5f6a9aa6a3d5490fdd8541e7682195ab5b85c22f2fdd4df7ed59f984b1e6b36e7bdb0527f20c79759bd0dff54c73d1e704b8
2ddc73392a68cd06e8e4a9e8e87302e816069da385544e40874f241dd34d9e24c7eb0486ce34640a68b7a6693afa9c00ba85
41e768219513e7c8c288ce6539d321f4b8102b6ff3c439b227d39029ce91b9354dd3d1e704d02d0c3a470ba95cddad1d37c7
cde949c679aef3b1cdb1c340dc8fbaa503864fa3459d69c8acfbb5bfd4344d479f1340b730e81c2da472031c305274f43901
740b83ced1422a578f9da40d98e959ceb59c67f842cc733a2db1877bca12fbe93dc74e54a621b31e3bf952d3341d7d4e5ed0
350c3a470ba9dcefcffe4a3afa9c00ba855f748e1642b9f4705dd5a7d308fe6c0ba3be331a39f601dae0293a3a8d00ba8541
e76821951be05c93a283d348cfdca260739490ba0d70f5146cbdd308806bb007d32a48c546721a01ec1f7e1a00931a08b506
38c924b8e03402da270a30ad81d26b80d601828c6623006e61b0691d94660338182a369a8d00b98501e72821751ba09bb8a2
a3d908a05b18748e1652b901fac62b3a9a8d00ba8541e768219513d0cb1c3733c58e476180d31217f8db244fac3d99864c01
edd6344d47b39117740d83ced1422aa77c37aa1f6e3c7f74874f254c18a47f504d34242ad70da7a269343a8d80b8855f688e
10423675466fbad62dfc231db98a8de4eed8325a9ed1ab99864c7546cfab699a8e4e23806e61d0395a48e5de4e23fd9ffddb
4603f59dd1c8b1fffadba5d0683302e216eed11c21b46c4968559f461ff8b32d8cface68e4d807b876293a1a7d00ba8541e7
6821951be02ea5e868f401e816069da385544ecd478eed3ef7fc003c2cedf6fd8a9ddbd474a4241a12d56cc4a9681a8d461f
206e61a0394248d90698762a3a7a7d00ba8541e76821951b6075a1e8e8f501e816069da38554ae2e216313e6734fcd3af6f4
32e34afb10c26a6f9ace23fcf7cf1252651a32eb12f24b4dd374f4fa00740b83ced1422a3796d7c70bfa0f230bd0395a48e5
0678f4a5e8e8f501e8167ed1395a08e57edf5c51d2d1eba387eec2a073b490cad5a7e8d77c06f9d353d87d0f3f83eb5ae33f
cef079adeb11475b9fa2ab4c43667d8afea5a6693a7a7d00ba857b3a470baddcef7be24a3a7a7d00ba8541e7682195abdd8f
d7295c78f2f3f66b5dcef472367d2fa76dbbaf67cba14eb43eb1f63efe50d1341abd3e40dcc240738490b2fdbe9db9a4a3d7
07a05b18748e1652b9ea10306d67fab1c7015ee779ae773a2173951dcecff63b9d69c8ac1e015f6a9aa6a3d707a05b18748e
1652b9119c2f041ded3e00ddc2a073b490ca55e86d3e623ff934c068bf12d7407180c7cb5c5a671a322bf4979aa6e968f701
e816069da38554ae42ef6b7a4b9b0618af3ee9cc5bdac91e5f273ebdfa459e21af02ff5dcf34176d4680dbc2e07254909abd
6d465e7ff60f1b0dd4774623c73ec2574bd0d16904d02dfca273b410cafddc5b45a0d1e4a41b6c1f069a238494edf7f70a49
47931340b77047e768e1283782b38ba0a3c909a05b18748e1652b9df3b80493a9a9c00ba8541e76821951bcbe404d07f3a78
80ced1422a3796c909a0ff74f0009da38554eef70b3a4947931340b730e81c2da47215ba34cfcd73d2d2c93f2db3734f9ae7
59b3ce346456e82f354dd3d1e404d02d0c3a470ba9dcef9f57483a989c80f98982cd5142eaf67b3b56c5069313103f51a069
1d9466bfdf6ca8c86072f2e22d519069159462bf6e3eaccc49e870d2cf24bbf016db0484ba29dc8a38e1ae887a7b923b43a7
b37de13eb3a636849b406e99864cf5f6c4ab699a8e0e27806e61d0395a48e546b0751174343901740b83ced1422a57a17f68
0fafe86872f282ae61d0395a48e57ede265fa0d1e104c42dfc42738410b2d59764ebb9ec5bbe6776ff797a7e94f515d9bf79
86bcfa82ecef7aa6b9e86d02dc160697a382d46c04f30741476f1340b730e81c2da47215fa3ab6a3587b4fc9e53b5c63d2d5
765aee70c5dd9fafb5cc346456e82f354dd3d1db04d02d0c3a470ba9dc089fb4a0a3b709a05b18748e1652b901f6c52b3a7a
9b00ba8541e7682195abd041fd252df0ce30679ae381be34be3023bec31f989f6b984a342456e40f154da3d1dba427eec240
738490b20d30215174f43601740bf7748e165ab90136ee283ada9b00ba8541e7682195535bb4e673bec212244e9fc24f908f
f375a621536dd1f26a9aa6a3c309a05b18748e16523901bdcfebb1c5f3f77180e102305dc725cf393f89d6270a64afa26934
da9b80b88581e6082165fbf5765e85466f9317710d03cd1142ca561b166cfb745fa983ff1e2ea8613970a65bcc36efe192bb
84ffba22cb4c43666d5af0a5a6693a7a9b00ba855f748e1642b902ddb752b8ce6202f75fb4aa98f77d5db676dd969986cc02
fda9a6693a7a9b00ba8541e7682195ab1d48c21a7eb9d303ab2dacbbae702f496b82755ab67c2cbd42cb4c4366ed41f2a5a6
693a7a9b00ba8541e76821952b0d76ae6b9ec3f852339ce8bd3c87eb7b72e8285d611e6b7a9d69c82c0d763ed5344d476f13
40b730e81c2da472a595d2bea766dc4bfe20c2c4304e54e300e7306f8c0dfcea654c671a324b2ba54f354dd3d1de04d02d0c
3a470ba9dc58f6263df4dfde1da073b490ca0dd0054fd1d1de04d02ddcd3395a68e58a3fc13487efc099763ec4395398bda4
576973bacfced151aad813a84443627127f852d1341abd4d40dcc240738490b28de0b821e8e86d02e816069da385542e431f
f1a8fbb9ddd959679bc21733b70e9e5b03bf0cad330d9919fa5b4dd374705601f313059ba384d4ad20d7c7ecf1ab77c7ce97
f9823b6fd5aab712ab44436201fe52d124186c5580fb44c1a545508265ef9a7f6bc3d8e4f5174b14b5f538d49847b0bdf897
0cc626e07da22f32adc2bf8a65defd9eefe598f307be17a395dcc6360c6b9d8eab7e9f74a62133f37eab69928cc626006e61
b0691d946685381a3584994abe78872b776cff977ad8ae57ba67d74eb83ad1fac4c2fba5a26930ba9a80b78541e6c82045fb
b9c78740a3ab09885b18688e1052b6dccafbce0f4a4ac7fe7dbee3aa320df0b9a8656427d390995b797fab699a8ec6263d74
17069da385546e048f0f4147631340b7704fe768a1952b6deaf3d76d9dd300d735b7588f039cb6b0ea9caffa44d9c9346496
46f59f6a9aa6a3b109a05b18748e1652b90cbd942eb5d992a37e48e97b792f616e7e1fd5d944671a3233f4b79aa6e9e86d02
e816069da385542e434fe99dcd9c3d76cea56c284903dc96f0fb8b5fd20cad330d9919fa5b4dd374f43601740b83ced1422a
97a0c377fe9cef754dc723e22f207aafecd965e8f1034ed04ea62133417fac699a8ede26806e61d0395a48e5dede26af3ffb
877107ea3ba39163af7ff69ffa7417c19f6de1577d673462ec237caf051ddd4500ddc2a073b490ca8d70051374341801740b
83ced1422af7733f2a8146771110b730d01c21a46c2358aa083aba8bf4d05d18748e1652b93cdd9edbd18c395afb2ccb91da
e0ccf39dbb8ed705864c3424e6e9f6a78aa6d1e82e02e216eed11c21b46cc54f652e2ba0bce659f66d8fcdd4e33fa28d779c
3215649d69c82c8e2a9f6a9aa6a3bb08a05b18748e1652b9019e14283aba8b00ba8541e7682195fbf573558546771110b730
d01c21a46c033c3e5774741701740b83ced1422a575e943cae45d107234d18ceec4e969f80ce4f670a9d69c82c2f4a3ed534
4d47771140b730e81c2da472c547663a8f79d9d20d34fc3fe11e1a2ebae95369e6bcc54746661a328b8fcca79aa6e9e82ef2
82ae61d0395a48e50af4356f6b98b0a44f25da08c49619e9f7578f5fd777fb3ad39059a03fd5344d47771140b7f08bced142
2837828f8ca0a3bb08a05b18748e1652b962a992f758a4dfdf7a1c21ebdecef4bddceb069402ad330d99c552e5534dd37434
1801740b83ced1422a5737349433e6ffa59785d79aaeb071b4f3733ca56e68509986ccbaa1e14b4dd3743418e9a1bb30e81c
2da47275eb4a7c6237a587767bda78b01e6981bf5ec7bc8729dcfdf4c1939986ccba75e54b4dd374b41901740bf7748e165a
39d1d132ac01a639be348c1fd13e1d677c60adf665b54c43a6e868e9d6344d479b1140b730e81c2da4720334ff5374b41901
740b83ced1422a273697862f5e9825afe99e1aeeb4fb798795bfda5bfa245a9f28b6967a154da3d16304c42d0c34470829db
00e77b141d3d4600ddc2a073b490ca0d702840d1d16304d02d0c3a470ba95c3d95d97ca7c3977259cf255c73fecb2f25d26e
c1e7f887cc3464d653995f6a9aa6a3c7c80bba8641e76821951bcb6304d07fda77bce81c2d847203f4c15374743801740b83
ced1422a978d55447d7a8ce0cfb630ea3ba391631fc1b947d0c163a4676e51b0394a48dd86f21801f1ffd34aa307d32a48c5
06681ffa2f173c4640fb44012635106a0d301b105cf01801ed130598d640e955e77b5b3e681ec7b63e5d41c2d8d6d415a53b
b9af12ad4facf3bd0f154d62d16004b42d0c302d82124c4c6fa7393682b9cf781bdbef3ddcbd637b5735bf7d320d996282eb
d634cd46831120b730e01c25a46ebf6ee8add0e82e02e216069a2384944d10af47f81496393d04dfc3ff222ca9e74d7dcc2d
d3902998dd9aa6e9e82e02e816069da385546e002f504547779117740d83ced1422aa7bc36f6ebbcf629bda4decef59a96d8
97593d9978320d99ca6bc3ab699a8e0623806ee1179da385504e405fc987f6480fccc24d6d0f5fc75bb5ab7f120d8902d9ab
681a8dee22206e61a0394248d9deee22fd9ffddb3d03f59dd1c8b10ff0dc4bd1d16004d02ddcd3395a68e57efded5268f436
01710b03cd1142ca36c04544d1d1db04d02d0c3a470ba9dc08862e828ede26806e61d0395a48e57e7da25aa1d1db04c42d0c
344708299b6acd351f6b6ce99066a767f880d6b0e45366542dd390a99a7379354dd3d1db04d02d0c3a470ba9dcaf9f622b34
1a9b80b88581e60821651ba01386a2a3b109a05b18748e1652b95f77a9526874357911d730d01c21a46cf5c9533e0f9b9b78
961364e9f3d9b6fc60e1799a2b330d99f5c9d3979aa6e9e86a02e8167ed1395a08e5067884ade8e86ad2437761d0395a48e5
6a6bd4f266210ef0d896b0b83fb7fc70700bdfc6f4acaeb646559986ccda1af54b4dd374743501740bf7748e165ab9df77fe
9574743501740b83ced1422af7fbc6d6928ec626806e61d0395a48e57edfb75dd2d1d804d02d0c3a470ba9dc081e1f828ec6
26806e61d0395a48e52af4712ce19e1a3f957b0e53a77d9fe634c07a9cf1f97acb4c436685fe52d3341d8d4d00ddc2a073b4
90ca55e87642e29e636796b87d290db09eee7fa055a621b3427fa9699a8ec626806e61d0395a48e52af43d5de1f692be8a73
6934945e979ed7b6a61e980fb4ca346456e82f354dd3d1de04d02d0c3a470ba9dc085f6f4147739517740d83ced1422af75b
0b1bc1455b15e0b6f08bcb5141683682978ca0a3b34a37de3e0c3a470ba9dc8fbd64241a6d5540dcc21d9a238423db58b62a
80fed33304748e1652b9dfcf3e251d6d5500ddc2a073b490ca55e83d9dee4b3792ab1cee4b9f4adb725ca155a621b3427fa9
699a8eb62a806e61d0395a48e57eec9e28d1e8a902e216069a2384946d2c4f1540ff6918023a470ba9dcef1f1c483a7aaa00
ba8541e76821951bcb5305d07fb88680cd5142eaf6fba7628a0d9e2a207ea240d33a28cde22b60551bae26afbf58a2a8adc7
a1c63cc02e12e5494263937e22d785d7b46372cde156c409774504f419963de196967aaaeec7b4e7a3c20afac934640a68b7
a6693a1a9b00ba8541e768219553dbc2ca13f0dc31782e0ffee5b6b09a69c854dbc2bc9aa6e9686c02e816069da385546e80
3ee28a8ec6262fe81a069da38554ae428709cbbda7ed97711fc23eef773a3e79c486d0d3795ded93569986cc0afda5a6693a
7a9b00ba855f748e1642b9fa72f0d8c37a2bddb1cf3d0f292d428e234c4b8f79bd9fd7de32d390595f0e7ea9699a8e0e2780
6e61d0395a48e5eacbc1d87c7a4ecdbe43fe7c84fffa4c030c77b6788f999ffba3cc3464d6d7835f6a9aa6a3c309a05b1874
8e1652b9117c2f041d1d4e00ddc2a073b490caa5cd61aa3e3d46f0675b18f59dd1c8b18ff0051374f41801740b83ced1422a
37c2a544d0d166a487eec2a073b490ca0d70d35074b41901740bf7748e165ab9011cee141d6d4600ddc2a073b490ca299b91
f0d35bc2052e7e2afb756ffb14667072f6fb641a3295cd8857d3341d6d4600ddc2a073b490caa94d71b52bf37fa927d17ccf
53349250bbe26aa621536d8bf36a9aa6a3d308a05b18748e165239e534b2dfc7110d36ff8bde11fb1a1629c7213fe927d390
a9bc46bc9aa6e96836f282ae61d0395a48e5aad9c81d6ee4777ab912d65ec5d1290e70ddd7336defafd032d39059cd46bed4
344d47b31140b7f08bced14228573b36841f5a488c637ab658c7dfdf7acdcb7aded7fe7cbd65a621b3766cf852d3341dcd46
00ddc2a073b490ca15e8a7475df8208a295d1e607ce0b09e712b4a81969986cc02fda9a6693a9a8d00ba8541e76821952b5d
58b6738ea1d426e68cef79ce3575f85ab2f1f3f46c2ad19986ccd285e5534dd374341b01740b83ced1422a57faed2417a329
6d0059c31afbdeb62535760bdf8f70110a179efa0247671a324bbf9d4f354dd3d16c04d02d0c3a470ba95ce9ac942d4fa6ec
fad27c14ae7d4ae6cf85f8af342b3d95feae661a8a1e233d6b1706942381142cb3ceb14fdd762e697473dccd3fafe9954b34
5098c3fced3156d19986cc4cfcada6693a7a8c00ba857b3a470bad5ce985b71dc9cd277f0ef7bd84ef7ffe504acbfbe96986
27330d99a51bdea79aa6e9683302e816069da385546e04c70d41479b1140b730e81c2da4723fb70c1068f01801f013059823
83142ddb6efc5b1d461ff89b4f14c5f548d4a847b0a1f8970c36232fde120599564129f6f36fd4bf58f01801ec137d616909
fe95abb4878ded03b6227dbd3ba7913d2d214b77589568482ccd61bf543489458311d0b630c0b4084ab09fdb6d08301a8c80
b78541e6c82045cbc453da19b5148f8435fc7b597243d7f92c7b9c0bb2ce346466e66f354dd3d16304d02d0c3a470ba95cb1
db58afb0b4b8d25223cec7f6d8d133b7684edb0d8ebd7a8ce84c4366b1dbf854d3341d3d467ae82e0c3a470ba9dccf5b980b
341a8c80b8857b3447082ddb089dea051d0d4600ddc2a073b490ca65db8978c8ec8e2f65d300abc14d1ae0ba4ecb7ed64bb6
9369c8ccb613df6a9aa6a3c108a05b18748e1652b9e2b5910e0f5ef9ab7874fd2ea72b35bb8c8d7b8bd786cc346416af8d4f
354dd3d16004d02d0c3a470ba95c863ec2bfc265265d70e215763d538fa32968947a34577b119567c8cbc05fea99e6a2b508
705b185c8e0a52b3b7b5c8ebcffee1da81face68e4d8dfd622f8b37ffa66bcea3ba311631fe11b2de8682d02e816069da385
546e043f1541476b1140b730e81c2da47223dca5041ddd4500ddc2a073b490cafd7c3222d0682dd2137761a0394248d98a9f
4a798c9d3de87233be2d4f8af3decf7af8d4c93464163f954f354dd3d15d04d02ddcd3395a68e506585d283aba8b00ba8541
e7682195cbd06b775034b5d3ba6257fdf48ff0d38b0fbfaba58ace346466e86f354dd3d15d04d02d0c3a470ba9dcaf9f8a28
34ba8b80b88581e60821651bc14446d0d15d04d02d0c3a470ba9dc000fce151ddd4500ddc2a073b490ca8de02323e8e82ef2
82ae61d0395a48e546f0911174741701740bbfe81c2d8472e5c5df3edfd33ae74f255c61f6308d486d82e7d88fe68ca774eb
bb3f9969c82cafff3ed5344d47771140b730e81c2da472e5156fdcce711de9f0c91abe8b676cba74a48fe85eeeb841b1becc
d79986ccf28af7534dd374741701740b83ced1422a575ee617639738290e5fbdb044b8a7746820fcafc3ff773faff2659e75
79e545fe877aa6b9e82bd2e3766170392a48cdea868db605700b239ac2643135925fcbb4b96dc2929986ccba61e34b4dd374
f41501740bf7748e165ab9ba1fe9b8c3fa2f5d56eb09abf478630b97fb259e347b8e20c84c4366dd8ff4a5a6693afa8a00ba
8541e76821951ba0f19ea2a3af08a05b18748e1652b9011a2f2b3a5a8b00ba8541e76821951ba0299ba2a3b508a05b18748e
1652b901dac52b3a5a8b00ba8541e76821951bc01140d1d15ae4055dc3a073b490ca8de1a7f20f1dad4500ddc22f3a470ba1
dc087617828ed622806e61d0395a48e5ded622f8b37f3a67a0be331a39f6fc67457d987bf47fb54551dd198b1cf9003f65c1
d69b7b00b8067b30ad8254ac5ead630ba52d1f6beffef3f4dc5bebb5fadf3c435ebd52ff5dcf14116c3dc0f9448124e9854e
f5ac56584f2ea94f4358e8e42deae908db794de192b33dfbde65a221b19ed4fa50d124143c3d80fa4441a5055062d5b9d691
b7a7c7b14537e77b8f8fa592e3dcbd2ceb7ab609a6ca3464d6b9d6979a26c968eb01e016069bd6416936c0f96dc5465b0f20
b730e01c25a46e03d97a80f84fcf0aa0394248d90668c8a0e868eb01e816069da385546e80834a8a8eb61e2fe81a069da385
544e7910ece73dc559681ce0f1ecd557260435d390a95c08bc9aa6e968eb01e8167ed1395a08e506709b507474f600740b83
ced1422a57a1638788fc7e703ff6784bbbd393d86d9b9670b98f07d92ab4ca346456e82f354dd3d157a487eec2a073b490ca
a589adaa4f670ffcd916eeeb3ba3d163cf7f56d4a7bd06fe6c0ba3be331a39f69f1b8a08347a6b80b88581e60821651be09a
ade8e8ad01e816069da385546e80bbb3a2a3bd06a05b18748e1652b901da25293ada6b00ba8541e76821951b60baade8e8b0
01e816069da385546e8026108a8e0e1b806e61d0395a48e506e827a3e868b2f182ae61d0395a48e5c632d900f49f0e122f3a
470ba15c6d27b82ea94dfe7ff90850ea1d9fdad05ce90cecd4fac0ca4c43666d27f8a5a6693a9a6cf4d05d18748e1652b9da
2b33ff5f7a2e778439ea351fb955ef3c9d73f8df3c17329d69c8acbd32bfd4344d47930d40b7704fe768a195ab7d60c3c732
170f8cf88a387e37d300e36e9730c6fd791826330d99b513ec979aa6e968b201e816069da38554eef7bd9d251d4d3600ddc2
a073b490ca8de03721e868b201e816069da38554eeb7bde90517ed3580dbc2e07254909a8d60ba20e868af01e816069da385
54ae405fd33aefd99aeb9ef377f0489fcab5cc4b9852b4deb732d39059a03fd5344d477b0d40b730e81c2da472237cd2828e
f61a806e61d0395a48e54630521174b4d77841d730e81c2da47263788afc4347930d40b7f08bced142283796c94607fdc144
02748e1652b9814c3640fca7834487e608e1c8562762d714b4c82dc6ef34bc2dcfa0b67d99d7f57c9e06ea4c43669d887da9
699a8e261b806e61d0395a48e546f09b107434d900740b83ced1422af773bf098146870d10b730d01c21a46c23f84d083a9a
6c00ba8541e76821951bc16f42d0d16403d02d0c3a470ba95c7d5a923a665d79e133eddb7d5ee9bc7d3c0e33ed61b9fbacac
64a621b33e2df952d3341d4d3600ddc2a073b490ca8de03821e860b201e6270a364709a95b369df8b73a6c2ef0379f288aeb
91a851d7fd8557f8916ff9b2be1ee176b64fe984dd199f1986454a9b06c94c4366dd5ff8a5a64932986cbc784b14645a05a5
d800ef7f953f064d36fa895c175ec21c7a8dbb07fe0f8a38e1aec800ef07151d4d3600ddc2a073b490caa936c361661a0696
dad2ed6bb88d5dc7a5b674b444eb13559361a7a269343a6c80b88581e60821651be0e5a0a2a3c3c60bba8641e76821951be0
a487a2a3c306a05bf845e76821941bc14b45d0d16103d02d0c3a470ba9dc00c75b141d1d3600ddc2a073b490ca55e8fca83f
cd518efcea2ebdbf0ddf8cb4917b7a7ed332d39059a1bfd4344d47870d40b730e81c2da4726f870dfcd93fed2350df198d1c
7bfeb3a23e3d2ef0675b18f59dd1c8b18ff0bd1674f4b8e8a1bb30e81c2da472f50ab6653f8638c0f5595ac66d377b98305e
d7f95cc054a2f589f5faf5a1a269341a5c80b8857b3447082ddb58061780fed3bd01748e1652b901eece8a8e0617806e61d0
395a48e5d44ef970655f972b59148415661869ecf02eb7cad74c43a6da2befd5344d47830b40b730e81c2da472027a2e5ddb
e3a7b2dde92ded76af0afac934640a68b7a6693a1a5cbca06b18748e1652b90afd3473ddcf296e843df3e9f7f7c365996848
acc81f2a9a46a3bb05885bf885e6082164abee1669fb54bac8ecf3115be16ce99cec9a5c56a2074745969986ccea6ef1a5a6
693aba5b00ba8541e7682195abcd12d6fddef209f72d7c2073f816a607fe6b6c65117e8dcf7b139d69c8accd12bed4344d47
770b40b730e81c2da4723f6d88a1b8e86b01dc160697a382d42ce31ea987d892fa38c4e365cb1a770ec6e16d4f97c40cac33
0d9919f95b4dd374f4b500740b83ced1422a57dadb946d44c96be38c4fe6a6bc1258a6750bff93e9d9a6a4330d99a5bdcda7
9aa6e96870d1437761d0395a48e5066864a4e8687001e816eee91c2db472a565d5baa53758b9cf50b8f8dce1369e3ba93de6
2ba56595cc34649696559f6a9aa6a3c105a05b18748e1652b9d2912dffc672efb0e79946fc47dd7851bb0dea4c4366e9c8f6
a9a6693a1a5c00ba8541e76821951bc17e41d0c1e302cc4f146c8e1252b75ff71a1460b0d700ee1305971641095660a7355f
6d92fc473e289dec04e2e4780922550b069d69c82cb89f6a9a2483bdc68bb74441a655508afddc71e25f2cd86b00f689beb0
b404ffcaf573bf897fb168af01da1606981641099671e7ae89e974cd7398906ee91ff37d862b7bf89f155c996848ccb89f2a
9a06a3bd06785b18648e0c52b4d2f5371ed2bc8fd47b65bad32f6b4d53c230d4695aa7b37ac6399986ccd2f5f7534dd374b4
d700740b83ced1422a97a18fba788d030c17972d1e184dbfb8b0c24ddfcff2b576320d9919fa5b4dd374b4d7e8a1bb30e81c
2da472197a0d9fc3bee5e715f168d5bd9c6bbeccc47603b18745ed5dae330d9919fa5b4dd37474d800740bf7748e165ab911
1ab60b3a3a6c00ba8541e76821951ba135bfa0a3c306a05b18748e1652b9df9a4d082e7a6b00b785c1e5a82035cbb861c11e
a6f647dacf3cede712d636c79eaf3dc9267d5dab8b8ace346466e46f354dd3d16103d02d0c3a470ba95c86ae5b1e9351407e
f8baccf94b585cfeca52c2c9346466e86f354dd3d1dfe3055dc3a073b490ca8df0490b3aba8b00ba855f748e1642b90cbda7
b9d19a66864199e9b89729f5df9e2e3e0872320d9919fa5b4dd374741701740b83ced1422a97a1c317f15eefdc93783a1ea7
d964d715967961e4e566e5641a3233f4b79aa6e9e82e02e816069da38554aed8416df9105dbab89e537e2637e79963fdbc8a
1d94cc3464163ba84f354dd3d15d04d02d0c3a470ba95cb15459f2fb932d0d301fc598f274f9cacd780ab49369c82c962a9f
6a9aa6a3c1480fdd8541e76821952b962ae59ccd9c7f7279199f4d326247b5f9aa5de89c4c43663155f954d3341d0d4600dd
c23d9da38556ae38f9e5a35675aa74e7fd0e692df47c2f8b939fcc34641627bf4f354dd3d16004d02d0c3a470ba95c715589
3bf6cf23cf9053abbcb81848032ce6018fab8acc34641657954f354dd3d16004d02d0c3a470ba95c7954927b68a52e88f332
8775ee9a0da6e7dc0460de9f672532d390591e967caa699a8e1e23806e61d0395a48e5cac3b0b93c98fc2f3d8a8cdb2fe733
7bc02c47fe05d6a76132d390591e877daa699a8e1e23806e61d0395a48e50678a2ade8e83102e816069da38554ce7e6fcead
e8e831f282ae61d0395a48e58a9b4c7a303ba5f7e07358ddafcb7da5ad3d735a334cc7567fd33ad39059dc643ed5344d478f
1140b7f08bced1422857de474ea53bc97ff1dd613c331b5f0fa78f289f9aad1b6774a2f589e56de4978aa6d1683002e21606
9a238494ad188cccd3bd9f79c347b8d4ed5798a6a743b34b6add11ffcb82ac330d99c560e4534dd374341801740b83ced142
2a57f653cce7b185ab7adae3711c65fb471ce05ef739d51d1532d39059f6547caa699a8e36233d7417069da38554aeec9cd9
4ae38affd2bbe97a50348c369b1d4de7d3c54a661a32cbce994f354dd3d16604d02ddcd3395a68e5aacd4858e8de4b7a45bc
af73eeb69eb6ad4dfb1526aed3f140cb4c4366b519f952d3341d6d4600ddc2a073b490caa9fd7fb1c3d7bca58d2fdb1dffd3
be9ff2bccd9369c854fbffbc9aa6e9683302e816069da385544e9db8397223a23875aab7965d423f99864c75e2c6ab699a8e
3623806e61d0395a48e57ebd755ba1d16304c42d0c344708295b6d4b77ac7186b4a5d560ecba94fbb61c776915f8f4c75089
86c4da94ee4345d368341801710b03cd1142caf6d33efd8a8bd6222fdc1a0697a382d46c00130645476b1140b7f08bced142
2857a18fe433988ee9cedb9c9cdbd399a0f55ac39a7b7bf6c1e94c436685fe52d3341dad4500ddc2a073b490cabdad45f067
fff4ed407d673472ec6f6b91fecffe659f81eace58e4c807f8560bb6de5a04c035d8836915a462031841fdcb058311d03e51
80490d845af5cc5890615e53938a73cfbd5b532fc0e30893e1635eef67e221330d99f5ccd8979a26b9e03102da270a30ad81
d24b9c235aa3f9c972a6157d7cfa5dfc61c539a227d39029ce11b9354d92d16304c02d0c36ad83d26c8093538a8d1e23406e
61c0394a48dd06f21801f19f061a40738490b20dd0c55cd1d16304d02d0c3a470ba9dc006b4545478f9117740d83ced1422a
3796c708a0ff34d078d1395a08e50668b7a2e8e83102e816069da385542e4df6547dba7cf47fb60ba3be331a39f6fad8e9de
d6e94e5bc6f67081cd2661519778822efd229ec74e2ad390591f3b7da9699a8e1e23806ee19eced1422b37c0174cd1d1e104
d02d0c3a470ba9dc400e2720fed3be03688e105236d5c168bac3542dbd8c8ebe05e9ac919c18d44443a2ea5fe454348d467b
1310b730d01c21a46c6a6210bb404f539a9aefad85ac9a18d44c43a69a1878354dd3d1de04d02d0c3a470ba9dcafe77f0a8d
de26206e61a0394248d9c6f23601f49fc61da073b490ca0dd0f144d1d1db04d02d0c3a470ba9dc00cf27141dbd4d5ed0350c
3a470ba95c6d2ff843fb6245476f1340b7f08bced14228571b575e73f877da2276ed7bf8465cd79a1f7d2fc7ba1e53eb9c2e
330d99b571e5979aa6e9e86dd2437761d0395a48e56afbddf93ae2e3fc38c02b6ff6cd9da097e34c6f719e4f5a661a326b03
de2f354dd3d1db04d02ddcd3395a68e52a74dbf6192e40b157ee7295dea27b3ea9f240ab4c436685fe52d3341dbd4d00ddc2
a073b490ca8d60f321e8e86d02e816069da385546e841ef9828ede26806e61d0395a48e546b0801074743801740b83ced142
2a57cd3ee63dac79d29e9ab02a98e2f6eaab1a0dcd6181f24c4e74a621b39a7d7ca9699a8e0e27806e61d0395a48e5d2230b
559f1e23f8b32d8cface68e4d84730561174f41801740b83ced1422a3796c7c80bfa0f030dd0395a48e57e6b9324b8e82e02
dc167e71392a08cd46b8470a3aba8b74e3edc3a073b490cafddef44cd2d16004d02ddcd1395a38cafd7e0a28e9683002e816
069da38554eee75e1b028dee22206e61a0394248d92af1b49dc7992fa961a9bd842557fa1a867568588aecfb73cd568986c4
4afca1a26934ba8b80b88581e60821651bc16843d0d15d04d02d0c3a470ba9dc08461b828eee22806e61d0395a48e54630da
1074741701740b83ced1422a579f4ca57942bee0accfc6f1d859a4749878d63632d390599f4c7da9699a0eee22607ea26073
9490ba55e4fce3cb4f50c2f8ce653ed217719fb675dbb6ebb955c94c436645fe52d3241bbc4d40fc4481a675509ad5fd6fbf
33565664f03679f19628c8b40a4ab1da9b353ff44daf2ac3d2273e0d2d2743c222605b1ff36c996848ac9d593f54a4b1c933
561a9bf473c82e3c0745efb0f24fe156c4097745ea2bef6d9fc2822fbda9dde67c1a266dfa98f7f268e179e5ad320d99f595
f7979aa6e9686c02e816069da385546e808d248a8ede26806e61d0395a48e5c49bc1d41ef7ccad54f7fc346955afc95aa2f5
89e2cda057d1341a8d4d5ec4350c34470829dbcffd6b041a5d4d40dcc22f344708219b200e5f83735b93ad7998ab5da90d90
6c9b5e130d8982d8ab681a8d9626206e61a0394248d9c4dbdf39bbb5e7832fd311666beb55cfb03b99864cf1f6d7ad699a8e
9626806e61d0395a48e506d8cbabe8686902e816069da385546e8057de8a8e862a806e61d0395a48e5de862af8b37fba85a0
be331a39f6fc67457d5a9af47fb60ba3be331a39f65f9f8b5368b43401710bf7688e105ab6b12c4d00fda75f07e81c2da472
e22e75cfe71c2ea86903c67ebd8ca375a62153dca7dc9aa6e9686902e816069da38554eed71bb3141afd4c40dcc240738490
b209e22db573cf4debf6e57d885b661a3205b35bd3341dfd4c5ed0350c3a470ba9dc004b0b45474b1340b7f08bced1422857
17916b982cec79ee1f2e317b3c069c56bcf3741e612d383d5b8865a621b32e22bfd4344d474b1340b730e81c2da472d5d224
4e54eeec2d739ee10213c71597f6d1e47bbae2fab75a9aa84c4366b534f952d3341d2d4d00ddc2a073b490cad56e14ad15ca
365fb9d3795c0f2c618eb8870b6d33d3939986ccda8de24b4dd374343601740b83ced1422a578c4dce352c7ab674715db77b
0eb386333dae5bb247c3f1ec9dd69986cc626cf2a9a6693a1a9b00ba8541e76821952b1d6696dc6533f542b9bba5ee12be87
6952552f643ad390593acc7caa699a8ec6263d7417069da38554ae3412aa0f27e398d67b5dd6753db34fc5e33d533a09c94c
43666925f4a9a6693a1a9b00ba857b3a470bad5c869ebbc7ec5dab9fd8eb284e16d7fba8d03ad39099a1bfd5344d47631340
b730e81c2da472b94fd695db3ba52db073de4074a6d6c9736b8394fb64e94c4366ee93f5ada6693a1a9b00ba8541e7682195
2b1dd1eaf43f0ee98ead10efdcbe6dabcfe10bb34c3424967e685f2a9a4683ab09809f28c01c19a46823187dfccb06631310
3f51a0691d946603f4361464303679f19628c8b40a4ab1ccbba6fe83cb9aad39eefc14250d346eb18e8f73ca030327d39099
79bfd53449066f13f03ed1179956e15fc58ad9c71ddfe32c47b6e30893956d5d8f3cd07cb99dcbeccbc9346416bb8f4f354d
92d1de04c02d0c36ad83d26c80beac8a8d0e27406e61c0394a48dd06e8c0abe8e87002e816069da385546e805ecb8a8e0e27
3d7417069da385546e84aeda828e0e27806ee19eced1422b3742ff744147871340b730e81c2da472c5eca3752b9ba24f69fc
d5252f83abdc442bb34c342416bb8f2f154da3d1de04c42d0c344708295b7181c8cfe6b66c5351779ea60186fa61b258d752
4ea621b3b8407caa699a8e2627806e61d0395a48e546f0fb1074343901740b83ced1422a97a1c37c709ae296e634c03287c8
bfc4c7d23243eb4c436686fe56d3341d4d4e5ed0350c3a470ba95c869ee213ca79498bda296eaf3d62fba734c0b374c32ed0
3ad39099a1bfd5344d47931340b7f08bced1422897a1c3326bb9ce3b3d9d9bf6f850f23af3efef2cfd8faa6991ce346466e8
6f354dd3d1e404d02d0c3a470ba9dc08ce2e828e2627806e61d0395a48e58ab34b3e1f5bdca4b6fc65cc7e3b679839ceb1fb
647176919986cce2ecf2a9a6693a9a9c00ba8541e76821952b33933065b9e3fea9f4ed9bcb438c348b486d18cfdaeecfc934
649699c9a79aa6e96872d2437761d0395a48e5ca1c34ac0ac235eeca8e78dd29a8b0360adfc3787ea4ce4165a621b3cc413f
d5344d47931340b7704fe768a195cbd0db1e66c5d14e28cf1dc294625fb34dc5b4cee19bf9f81a3b9986cc0cfdada6693a9a
9c00ba8541e76821952bebcaf50a3fb2ab183e968393d9e7339d1c0857a5baae949986ccb2aefc54d3341d4d4e00ddc2a073
b490ca65e8bb7be7107e72e156137d4cd3c277c9b7dbea9ca9330d9919fa5b4dd374343901740b83ced1422af76b57588546
871310b730d01c21a46cf663db5f85467b1310b730d01c21a46cf663576785466f9317710d03cd1142ca3686a1cb3f74f436
01740bbfe81c2d84722318ba083ada9b00ba8541e7682195cbaf249bf6d188236d0108bf8338c030910a4b85fd79e3ae330d
99f995e4b79aa6e9686f02e816069da38554ae6c3358c254713fd35191355c52b779bfeffcd6f8798a57b619c84c4366d966
f0a9a6693ada9bf4d05d18748e1652b9baa1a42d03b6e41a157f8f7180f7b9a789e3b3e753661a32eb86922f354dd3d1de04
d02ddcd3395a68e5ea31a3fdbacf339dcedda7e38cb79a34690a93c7653defeb390bab330d99f5a0d1979aa6e9686f02e816
069da385544eec8cbbe33a37fc236d740a9f4fb8afdf4beb33a9320d9962679c5bd3341ded4d00ddc2a073b490caa92328f5
fef25f5a99cc7798494db207e19369c8548750bc9aa6e9686f02e816069da385544e9dc9c86eae69fd7bc4675bf779aa5328
2dd1fa447522c3a9681a8df626206e61a0394248d9ead984d89fe5dc9319c99e7a255fe946731c53b8c41ed7e34ca5330d99
f56cc2979aa6e9e87002e816069da385546e80767c8a8e3e272fe81a069da38554eed7b62eff70d1e104b82dfce27254109a
55dcd2402e8d6fbf97edd8d26422dcd9967878e4b93deb4c436645fe52d3341d1d4e00ddc2a073b490ca55e82bcc804b6389
29f59858d302e15c5e1d9175a621b3427fa9699a8efe2a806e61d0395a48e5defe2afd9ffdcb4304d59db1c8910f7009116c
bdbf0a806bb007d32a48c57e6d82f52f14cc5580fa4441250510520d65ae02da3fec4300a635507afdfae09ac0a2b30a685b
18605a0425d858ce2a40fed33904708e125237e55438efdb1c7e77e9ecf731ed7963a49aca3f99864ce554e8d5344d477315
40b730e81c2da472039ce15274345701740b83ced1422a3796b9ca0bfa0fe710d0395a48e506589e2b3a9aab00ba855f748e
164239011d4a4e73dc3a1f07f83c4251d04fa6215340bb354dd3d15c05d02d0c3a470ba9dc00861f8a8ed62e3d7417069da3
85542e4d6f557d9aabe0cfb6705fdf198d1e7bd5fa798914be0931edcc87fbd216cdf5ec7c4644a221b12afda1a269343aab
80b88581e60821651be03aa2e868ae02e816069da385544e99abece13b181b99c5015e7b98bccdc721e7064fa62153d9ab78
354dd3d15f05d02d0c3a470ba9dc00c66b8a8efe2a806e61d0395a48e50698052a3a5aac00ba8541e7682195fbf5035d8546
7f1510b730d01c21a46c033cd05574f45701740b83ced1422a37c0c33e45477f9517740d83ced1422a575b3a5e6bb8aca70e
84d71a3e84758f4e20b1e3e4b58725f8f43423d19986ccdad2f14b4dd374f45701740bbfe81c2d8472bfef4c2be9e8afd243
7761d0395a48e56a03e27d8f6d14d2a7725ddb16df8da701aee7b5a6b35f155a661a326b0be22f354dd3d15f05d02ddcd339
5a68e56a1bf539cc11e7b473e99ee6f508bffed469743fc20570dad6c78a41251a126b13f50f154da3d15c05c42d0c344708
29db6ffbe40b2edaaa00b785c1e5a820351bc16144d0d15605d02d0c3a470ba9dc080e23828eb62a806e61d0395a48e52a74
696b9e06b81df9a797067897de750fb4ca346456e82f354dd3d15605d02d0c3a470ba9dc089fb4a0a3a90ba05b18748e1652
b9b14c5d00fda76309e81c2da47223f843093a9ababca06b18748e1652b90abdcd479c2aa49b4b34a988cfc0ffcb9d35d00a
4f671a322bf4979aa6e968ed02e8167ed1395a08e52a74db4c7eddf1b4d77aaf69dab82ffbbdcf6161f440ab4c436685fe52
d3341dad5dbaf1f661d0395a48e54670391174b47601740b77748e168e723f77391168f47501710b03cd1142caf673971381
465f1710b730d01c21a46cbf5f4b4a3a5abb00ba8541e76821951bc1e544d0d1da05d02d0c3a470ba9dcef1f95483a5abb00
ba8541e7682195fbbdffaea4a3b50ba05b18748e1652b9117c3f041dac5dc0fc44c1e6282175abc8eb1eeea66951704df71e
feff35f5ec3ba30d4d58f83faf7074a621b3227fa969920dd62e207ea240d33a28cde20b51551be62aafbf58a2a8adc7a1c6
5c8f7a1c61a571a7295af457bcee2842f847b8bd6cf73d750ddc65a621b31ef5f85293fe2acf70e9afd24fe4109ec207b8e5
705fc4093f4506e812abe8e8af02e83edcd1395a48e56a3fe029bda54e4611cbd38e268c769ac2823beeacaeed8055a2f589
b51bf0878aa6d168ae02e23edca1394248d906d853a1e8e8aff2826ee18eced1422a37c05b6f45478b1540f761d0395a08e5
d48eb87c5a325de0f735eebe5dee53198eb44c43a6da11e7d5344d47971540f7e18eced1422aa7de7a1fa164b89befe97bf9
8c56bdf5ae99864cf5d6dbab699a8e2e2b80eec31d9da385544eed722df395fcbd9ccbf66ab9cbb5661a32d52e57afa6693a
baac00ba0f77748e1652b911ac65041d5d5600dd873b3a470ba9dcdb65057ff64f0313d4774623c7feeb5de30a8d062f3d31
c21d9a2384944dd99d9cd3926e71e99b501c2ee5c6f127d390a9ec4ebc9aa6e9e8f102e83edce81c2db4726379bc00fa4f03
13d0395a48e5063822a0e8e8f102e83edcd1395a48e594fdda7cace192bea40186cf28fc24d653ed7d6c99864c65c0e6d534
4d479b1740f7e18eced1422a37c08448d1d1e6e505ddc21d9da385544e4d7deb12242ec7c22a3f3a3848379f9a684854135f
a7a269347abc80b80f03cd1142c856d774f14dc371a7fe02e1a21a165e4b6aacbbbd1ee5eb44eb13eb9aee4345d368347801
711feed01c21a46cb54b433ab0995ae0ecfde5663d8ff035bca63615929986ccdaa5e14b4dd374347801741feee81c2da472
f5214d7e6a92cc58b6e5dc97f34aef1ed639ac8796bdbda6d29986ccfa90e64b4dd374347801741feee81c2da472193acc93
e6f8d2348de98c6fb3ce754a662cc7b14c7bf8cff5cbad330d9919fa5b4dd374347801741feee81c2da472c5e0e5ca4e3367
fa20527fba2535ba5f6aeba7fa49eb4c43663178f954d3341d0d5e7a68843b3a470ba95c692cb4cfe1ab38a7c728e1abb785
2bee94fd3aa2d9e319f72614689d69c82c8d853ed5344d47831740f7e146e768a195cbd06189196f63697f7558f1e62b4d1c
5f5974d7569e3ad19098913f55348d46771710f7e10ecd1142ca569a666db1adec963aaaced7b3c88c0dae5217dda5da5439
9986ccd234eb534dd374747701741feee81c2da47219fa5cc25ce19ad363e1ecb1739ee98a336ff5f65aa075a62133437fab
699a0e062f60eea21d9ba384d4ad18bc146f9234a27d2eb793ff925349b257aa7d0e75a2f589c5dee54b4593607077016e17
edb8b4084ab05f37fc1358b07679c13ed10e4b4ba0e41acada05bc7f9897bcc8b40aff2a3694b50b78ff742e019bd6416936
96b50b90fff42d019ca384d46d8016bc8a8ed62e80eec31d9da385546e8066cb8a8ed62e3d34c21d9da385542e43879bc539
858f23dbb0ecfb192eeb69efc3141be6c60756155a671a3233f4b79aa6e968ed02e83edce81c2db4726359bb00fa4feb12d0
395a48e58ab5cb7c1ce14797bbda8725d77ae6b186e2cb755c75c12cf30c79c5d8e5433dd35cf475016e1feeb81c15a46623
7840083afaba00ba0f77748e1652b9e2f6b1e4755dfab1e529f1b6e796fe47ecc1bdd517174ea621b3b87d7caa699a8ebe2e
80eec31d9da385542e41ef57b21f9953bfd8291f6ddbcf7ced291e15e5d7ec641a3213f4c79aa6e9e8ebf2826ee18eced142
2a97de05aafa7456c19fedc3a8ef8c468c7d842f98a0a3b30aa0fb7047e76821951be15222e8e8ac02e83edcd1395a48e546
b0931174745601741feee81c2da472c54e66cf67fcf34ded79169826ac6112b73ecea44ea621b3d8c97caa699a8ece2a3d34
c21d9da385546e003b194547671540f7e146e768a1951bc14e46d0d15905d07db8a373b490ca0d605ea9e8e8ac02e83edcd1
395a48e532f4513b0ea5011ec7167b20a5c747d3b35b3343eb4c436686fe56d3341d9d5500dd873b3a470ba9dc087632828e
e62a80eec31d9da38554ae3caa49c7fa97f2ed5bafb42c499edff9eebadecfa31a9969c82c8f6a3ed5344d477f1540f7e18e
ced1422af773471981467f9517710b77688e1052b65f3f4d5768345701711f069a2384902d135fe95ab31d6948e5d94ab248
9eb7f0a39c96ab6ecd72320d9999f95b4dd374345701741feee81c2da472f975e05a4fe2fc176d40c2f5660d035ef2688f10
b8ab31ab9369c8cc2f04bfd5344d47731540f7e18eced1422a37c0db6e454773951e1ae18eced1422a57f635dc716d7fa5a7
65ebbd6f610d38a5c662cb19e654e7fdec6a9079d6e5953d0d1fea99e6a2ad0a70fb70e37254d09a0d70d646d1d15605d07d
b8a373b490ca0d70f244d1d15605d07db8a373b490ca0dd09658d1d15605d07db8a373b490ca09e83b7cfbaeb0004e7d40af
57af089d69c814d06e4dd374745601741feee81c2da472bfde23afd068ab02e23edca1394248d96aebc1ec6b7ee5f5503a19
9b2613c799b7c14ecfa92a9969c8acad07bfd4344d475b9517740b77748e1652b96a8d10afa46b6a6c10be754bb8a5ed5776
3238c27a204c2f9e53d632d39059cd11bed4344d47731540f761d0395a08e5b2e586a84f7b13fcd93edcd5774623c7fe6b1b
0a85466f1310f7e10ecd1142ca5689e3eed339ed04082bcb395cde8f33751b0dd782783e7a7e8eb3cb4c436665fe52d3341d
9c557ae63edab1394a48dd2a7258f2844566fac15fe1661e7e78e92339f321f4f969d5a0120d8915f843459360bdad0a685b
b0516909a45c23988dfcc3056715d076d10e4c6a20d4aab0f39eb6b9fc971f11c60e6d53ba92ef6b989f2fdd217a9969c8ac
b05f6a9ae482b30a68bb6807a635507a0d701e5190d15c05c07db863d33a28cd063839a5d868ae02e43edcc1394a48dd0668
abade868ae02e83edcd1395a48e5063839a5e868ae02e83edcd1395a48e504f432c75d7a677a97b34f4b5cd76f93847e320d
9902daad699a8ee62a2fe816eee81c2da472bf3e2ea6d0e8ac02e23e0c344708219bb081b8f6f39ee2dc3f0ef05896f58c43
16c82dd390296c20dc9aa6e9e8ac02e83edcd1395a48e514f4143729a61eba5bb899effb32ddea22f6241a1215b253d1341a
6d557a62843b34470829db001fb3a2a3a90ba0fb70a373b4d0ca652f19519fce2af8b37db8abef8c468e5d39abece1bfdaa7
7ca62e7c1ba7304b5b950d43cb34642a6715afa6693a3aab00ba0f77748e1652b9016e178a8ece2a80eec31d9da385544e40
ffff768d5674745601741feee81c2da472033cde5474745601741feee81c2da472032c6f141dcd5500dd873b3a470ba9dcaf
9f4d28343aab80b80f77688e1052b6df3717947474567941b77047e7682195ab3d52d725355ffe2f1f358b8f9c53abdef0a1
c4631353eb782c330d99b547ea979aa6e9e8ac02e83e0c3a470ba1dc8ffbff4a34daaaf4c40877688e1052b64a1cbb16ac69
97f17527ff802d5d6eaeb7c79dce346456e62f354dd3d15605d07db8d1395a68e546e8db2ee8e8ac02e83edcd1395a48e52a
74db7979cfb1dd663c4898be97f925daf67cb965a621b3427fa9699a8efe2a80eec31d9da385546e04030a41477f1540f7e1
8eced1422a3782d588a0a3bf0aa0fb7047e7682195abd05bf6184c034c7b30d7d49539b667bee3e6ad8759255a9f58913f54
348d46731510f7e10ecd1142ca36c2775bd0d15c05d07db8a373b490ca8d609824e868ae02e83edcd1395a48e57e6b8a25b8
68abf2c26de18ecb51416af673d73381464f1510f761a0394208d9469885093a7aaa74e365b8a373b490cafddc5e44a0d150
05c47df84173847064ab661bb9c741fab19da5bd43eea8ffec64aa661b2ad39059cd36bed4344d474f1540f7e18eced1422a
57a1cbfe92f47bbbcb5be9b8e8bb73df9df5996ccb4c436685fe52d3341d6d5500dd873b3a470ba95c5d3d87abeb324dc5ce
2735323dd3fead706bd9c2d2a07980c94c43665d3d7fa9699a8eb62a80eec31d9da385546e04b30d41475b1540f7e18eced1
422a3782d986a0a3ad0aa0fb7047e7682195abd07143f392bd52a7742c62dd7337db5747719d69c8acd05f6a9aa6a3ad0aa0
fb7047e76821952bd073e9ff9806981eeb84cb4e1ae06bafb6ce34645683e52f354dd3c15605cc5db463739490ba55e4df79
4a2b36d8aa80b88b76685a07a559e58d93fd644172f6ff797a1eca57da7ff30c7995f5ef7a269960e7f2227da21d93e6575a
95c325eb12aeab47baaaac6777007dc97df2e6b3ce3875a621b31c30f954f36de75286fbb6736973c82e1c4f242dd39cc3ad
8813ee8ad4f6a8711278a459f016a6c647bccea413306145941e653def6065a621b3b647fd52f36de7d241ff6957023a470b
a9dc00e76a14dddbd1a5836e61d0395a48e56af7df74cc3e3527cedb79d7d46960bd8f292cfde211e5dafc57241a126befdf
0f15df762e20fec3ab04688e1052b6fa2ef2399eba9f53dcfd70e68fe7bd82948986c4fa26f243c5b7974b47fca751c90bcd
1142c8265e39cff1ed5df858e2ec217c38e13fedfb296d4d9e4c43a678e5ecd67c7bb974d07f1a9580ced1422a37c0ae4345
f7f672e9a05b18748e1652b901f6972abab7974b07ddc2a073b490ca0db09358d1bdbd5c3ae816069da385544e40cfb9434c
ea2071cc4bf8ed9da7da24d512ad4f14c85ec5b78d4c47fca74d0ad01c21a46c03f88b28bab7934c83eec2a073b490ca8de6
24d341ff6993d2d3395a68e57efddd56686f1b998eb88581e60821651bc0834bd1bd6d643ae816069da385546e809d708aee
6d23d341b730e81c2da47202fa48ed5cafd44162dfae7087896e0a0afac934640a68b7e6db4606d07fd8a480ced1422a37c0
ee5645f77692e9a05bf845e768219413fb98c30a779a6347fb3843dea73051bfa2b1a7d8c7fc641a32c53e66b7e6db4ca683
fed32905748e1652b95f5ba22ab4b7934c47dcc240738490b2d5a706ad23dfd6de0b87dbdb561f483f4f0d54a621b33e35f8
52f3ed24d341ff6993023a470ba95c810e97f47bbe97b4e00b2bde70b9cb5637cbb51dfb7d6e4f9b199d69c82cd09f6abe9d
643ae83f6d5240e7682195cbd0e18358e38ef2f4d42e2c73afeb0c1f4d1c60ee8f733cc79b74a62133437fabf9769269d07f
dba480ced1422a571a0ae5cd69a9e7f8123bd29ce9f079f443a9f799d24f48265a9f58da097da9f8b691e988ffb449e9d11c
21b46cc53b2736e75de2318df829ecf3b4870b6beacd97f7f59ccfcb399d69c82cde399f6abe9d643ae83f6d5240e7682195
cbd053de9098cd7dc28a20dc4ff6f438a71a0a3c7b0c74a62133437fabf97692e9a0ff344a019da385546e80b6688aeee524
d3313f51b0394a48dd32f25157b5b151dd1ddb7cdee9c632bf6e563ad19098813f553409f67292e9709f28b8b4084ab0d2e8
6f5ab3c34b1ace711dc77ea5c5c01c6eadb9d168a1d59986ccd2eaef534d93642f3319f09628c8b40a4ab16226734ee1627a
a55303e960f2b55de9d737af533e2f5b1e16389986cc6226f3a9a649b297994cc7fb445f645a857f152bad2bf329c9740466
2e2ea1d91c231f9e98f7a777a5cc346496e6959f6a9a247b9bc974c02d0c36ad83d26c3433990ef94fa714c0394a48dd4633
93e9a0ff744a019da385546e34339906fdb7530ae81c2da472a5a9f6925f9965a384fcbc66caed82af7c26baf697d69986cc
d254fb53cdb7994c07fda7534a4fe768a195cbedd3c32d25cc86e7d4b46c3a3ab3d8f04d5ce3d1857a96cac93464e6f6e9df
6abecd643ae83ffd5240e76821951bc11d40d0bd2d653ae816069da385546e041f0841f7b694e9a05b18748e1652b911dc55
04dddb52a6836e61d0395a48e532f499363d6cd9b1a0f621cc6dfd97304d0c632cef619d4c436686fe56f36d2903e83fdc5a
40e76821952bd071ef7c6c38980688cb6ef92fd60a2d330d9905fa53cdb7a14d07fda75bcb8bced1422837c2d75bd0bd0d6d
3ae816069da385542e5fc8e633cc12e3698574d129968bf9a273edeb3d85497bb990e94c4366be907dabf936b4e9a0ff746b
019da385546e845b96a07b1bda74d02d0c3a470ba95c31b4994352983266f3996e07df74c5fdb7f35e4d959d4c436631b4f9
54f36d68d3a0ff766b019da385546e007b3f45f736b4e9a05bb8a773b4d0ca1523c775bab7648c91ae335b3e8c9d06182619
b16f7199863a9986cc62e4f8a9e6dbd0a683fed3ad05748e1652b9b2ca9a633f9e79cd3fb9b2ad3e3ddd081fce113ea7b342
eb4c436659657daaf936b4e9a0ff746b019da385546e04171f41f736b4e9a05b18748e1652b90c7dad61bd7bade5db573e94
ecc3f25c8e32b4ce346466e86f35df86361df49f6e2da073b490ca8de0e223e8de86361d740b83ced1422a3794a10d88ff30
6c019a2384946d28439b8ef84fb796179a238490adbcb9c94d70d261b7ee155a1aec74eff17c457d7323330d99e5cdcda79a
6f439b0efa4fb716d0395a48e58aa1cd7985c89aad57d690136eaef94b195da4e279d6fa8e4e671a328ba1cda79a6f439b0e
fa4fb716d0395a48e5cafbe7f02d8c9e9e69f7433ce995fcece368673e1274320d99e50df4a79a6f439b06fdb7670be81c2d
a472e5cc51343188fde5e300afdc803d9dd55e8ec7e1a09c39929986cc72e6e853cdb7ad4d07fda7674b4fe768a195ab1b69
e20eef235d66b6d4ccedcaaf61c2543936e6bcdb9e129969c8ac1b69bed47cdbda74d07f7ab680ced1422a57f74b6dfb745f
53ee50182eb4c776a6b657dbbc97fe1acf86299569c8ac3ba6bed47cdbda74d07f7ab680ced1422a37c0764045f7b6b5e9a0
5b18748e1652397534e3d8c37fbba46779a9f5427f1258671a32d5d10cafe6dbd6a683fed3b605748e1652b90a9dde8fa7c9
e2191702f1f562bcec1e5b7cc9b8aedd211c9569c8acd05f6abe9d6d3ae83f6d5b40e76821951bc0bb40d1bd9d6d005dc3a0
73b490ca0de04ea1e8dece361d740bbfe81c2d847223781809bab7af4e07ddc2a073b490ca8df0490bbab7b54e07ddc2a073
b490ca8df0490bba97b54e636e51b0394a48dd06f89c051bdd753ae01aecc1b40a52b17aa7bae7f9de93fe71a3c4736afb88
5603719ced048eca3464d63bd5979aa6b85eee3a1ded130598d440a8f56bb73901f5b2d6e9509f28a8b4004aac5f9f261358
6f5f9d8eb68501a64550820d701a56b1bd7d753ae416069ca384d44d4caeafd82666ddd3819133fcc8f66db9e4a1aa9a6848
14536bafe2db54a723fed33106688e1052b601ce9129bab7a94e07ddc2a073b490ca0de087aae8dea63a80ae61d0395a48e5
d46a397c2c6becc29956b6a917e72a999f44eb13d55ad9a9f836d5e988ff748c79a1394208d906e857a1e8dea63a1d740b83
ced1422af773ef2481f636d569c45d18688e10523645bc87ac7b4dcf9eb76b0f5388b0be96cd2a9e4c43a662f66abe4d753a
e83ffd6a7a3a470baddc00cfbe14dddbd2a7836e61d0395a48e504746c4b183bb8a5bb696b852ea09f4c43a680766bbe2d7d
3ae83ffd6a40e768219553ee3673586685ff3ef54d59c3e4f83a2e79117b12ad4f54de364ec5b79f4f47fca7590dd01c21a4
6c6a16361f6bb42b49d3c4b39ca458e534ac661a32d53cccabf9f6f3e9a0ff34ab019da38554eed7cb0b85f636f3e9885b18
688e1052b6019e6f2abab7994f07ddc2a073b490cafdb2e1a1e47a3bf974b82d0c2e4705a959ed50db766e85112d618073ea
5b173f9c75dbb6eb69e129330d99b543ed979a6f271f40ff6153033a470ba95c6dc05c461807786c4b586d9e5beeb2ba1de9
58d0d3c453661a326b03e62f35df4e3e1df49f4e352f3a470ba15c6d2e1ed73bdb5cbd0be22ba4f455bce2d1dde99c9f89a7
ce3464d6f6e25f6abecd7c1af4df4e35a073b490cad5c6f9ed54e375c7131371b743fc4739e4b73f9fb4cc3464d6d6f95f6a
becd7c3ae83f9d6a7a3a470baddc6fcd1104d7dbc6a7c36d6170392a48cdaaebc7b4cefb927e79f77ce7dda76980d7322fcb
72b4cee232d390595d3fbed47cdbf874d07f7ad480ced1422a3782a38da07bdbf874d02d0c3a470ba9dccf1d6d04dadbc3a7
236e61a0394248d99235abaaff36d2e9fe6c0ba3be331a39f6117e5282ee6da4d341b730e81c2da47223d88109bab7914e07
ddc2a073b490ca8d60fc26e8de763a80ae61d0395a48e5ea54285cdec2bd3acd54ee6ebf5c3619588e67b78ace3464d6a9d0
979a6f479d0efa4fbb98179da385506e80f99fa27b3bea3ce3edc3a073b490cad5997efebff4411c57b8bbcdf90852a87fce
e1b6f6f45bd49986cc3ad3ff52f36daad341ffe918d3d1395a38ca55ab95782c25ac70d39c740a0bce335b969dd7714c6182
3a3f73039969c8ac562b5f6abe4d753ae83f1d6340e7682195ab0bd93417cd4e47eb92bf7fc951e158b6b0f0de9a6194cc34
64d685ec979a6f539d0efa4fc718d0395a48e5aaebc891fa2da5495af81dae213b3d053fa3d3f9b96fe7f39b969986ccea3a
f2a5e6db54a783fed33106748e1652b9115c4704dddb54a7836e61d0395a48e52af41c266bb90351f8c9857be8b6a73386e7
b42ce146b7adcfd75b661a322bf4979a6f539d0efa4fc718d0395a48e50a74f1b44a5e2867bc9fe6e9e37197beb6cf07ad12
0d8905f94bc5b7a34e47fca7610cd01c21a46c95386fd84f1799739d8f30933ad323e16b9bb3c9c283ac320d9995f94bcdb7
a34e07fd87670cd81c25a46e15f947eee90aec65a7d3e13e517069119460713da76abf6c6df0174b14b5f538d4984b33cb79
0b17f06cf41877f084295ada02b0ac3c19a8130d89a595e5978af4b479c64a4f9b7e16d7c2e7fdbf70c399ae1c7e8a38e1be
483954938f77a5dd786b7c181aee24e9f9dc723ef7f372a846661a32cba19a4f354dd3d1d306d04f98748e1652b96ae4531e
c7fd973a5e86df7bece915473b3ff3966ae4a3320d99d5c8e74b4dd374f4b401f413269da38554ae7adae4fb54bc9a87cbd8
1dfdefd22774c4e71a5bec79553d6d44a221b17ada7ca8681a8d9e362fe212269a238494ad1e235a97b8a5346de2599fb3f5
f1c84ff41b8a5d9f9e63442ad390598f117da9699a8eb636807ec26f3a470ba1dc000db6151d6d6d00fd8449e76821951b60
6f94a2a3ad0da09f30e91c2da472bfde4aa1d0e86903e2274c34470829db00a7c5141d3d6d00fd8449e76821955386106188
4bb8b9a4015e77f8fd851b9ddc10f6641a3295218457d3341d6d6d00fd8449e7682195fbf5fe1185464f9b9eb88589e60821
651bc0c847d1d1d306d04f18748e165ab9f4c646d5a7b10cfeec13667d673472ecbffe762934baca80f80913cd1142ca36c0
4544d1d15506d04f98748e1652b9016e178a8eae322fe812269da385546e80d3008a8eae32807ec26f3a470ba19c98f4de61
117685197af20939d72b7c39ae55fda25ba621534c7add9aa6e9e82a03e8274c3a470ba95c5ddeec67f6c8d8b763bac39731
191096dd547b3bdd23f20c797569f3773dd35cb49401ee132697a382d4ecd7eb5785463f19103f61a2394248d906784ca1e8
e82703e8274c3a470ba9dc005d5e141dfd647ae816269da38554eed7cf1d151afd6440fc8481e608a1652b7e325719601c53
bcec9c47de4b35e70ebef3f30652671a328b9fcca79aa6e9e82703e8274c3a470ba95ce955b5e55f5f1ae012d6d3e9561a07
7886ffc57ecccf1b489d69c82cbdaa3ed5344d473f19403f61d2395a48e54630d11174f09301738d92cd5142ea368285cebf
6cb09401718d124deba0342badf69243d99c1d6ef6306db88e23cd0fc3fd644aa6f6d54047671a324bb3bd4f354d92c152e6
c59ba324d32a28c5326fea42b195d12cc5962efea3fd1833ae4c342466da4f154d62c14f06b035fac6d212fc2bd7507e32e0
fdcb2e856c5a07a559f19339a66dbd96d2a9354cf6c33c315b39651f81a765bb9369c82c7e329f6a9a66a39f0c909f30e11c
25a46e63f9c900fa2fb314d2395a48e50668faabe8e827d343b730e91c2da47203f4b45674f49301f413069da38556aef8c9
a4b373d795bd5f8e735eae29dd3ea7eb0edfc3231a0f163f199969c82c7e329f6a9aa6a39f0ca09f30e91c2da472197a3bcb
96d338a6bd6eb44d035cb6f0e538a7b29672320d9919fa5b4dd374f49301f413269da385546e041b0641473f19403f61d239
5a48e54630dc1074f49301f413269da385546e2c3f9917f4ffdb2c85748e1652b9f4ca44d5a7a30bfeec137ed7774623c63e
c2174cd0d1d105d04f98748e1652b97c299996a0c912b774c4019ef735ef5bdac91eb77a9c715f77bd7eea4c4366be947cab
699a8e8e2e807ec2a473b490ca65e87539d3bb8c7c81dfb38167beab8539db19a6ccd5e743671a3233f4b79aa6e9e8e802e8
274c3a470ba9dc588e2e3df49f7625a473b490ca0de0b1a7e8e8e802e8270c3a470bad5c9efd4e5d4fbea9fa8ca78b4e7c6a
149d0baafb83ce3464e6d9efb79aa6e9e8e802e8274c3a470ba95cb1900c73d3703fb9f28dac3b6637dd77b8cec6f358055a
671a328b85e4a79aa6e9e8e802e8274c3a470ba9dc0066a18a8e8e2e807ec2a473b490ca65e82379c1ad65c97d84b9fa7e67
4ff5296ebcbde28f2243eb4c436686fe56d3341d1d5d00fd8449e76821951bcbd105d07f399690ced1422a37c07339454753
9717740993ced1422a37c0936645475f17403fe1379da385502e43df5b3434dd92055ef8b585c1864f228f360fb6b6597632
0d9919fa5b4dd374f47501f413269da385546e8077658a8ebe2e807ec2a473b490ca155f97739ae6794f87ded770f1c977d8
34daf98eebe0a71d89ce3464165f974f354dd3d1d7a5876e61d2395a48e5ca1bff328af4aafadeb7b0289ad2ce93e50c938c
f3aeaf07759e7579e56dff877aa6b9e8e802dc270c2e4705ad597574c96f77d256a2d61526baafcce12b197e85cf067d9969
c8ac8e2e5f6a9aa6a3a30ba09f30e91c2da4723fddaba4b8e8e502dc274c2e4705a9d900e711141dbd5c00fd8449e7682195
53db2dc37d245c4ed3a712bf9ceb156670728fe99369c854db2dbd9aa6e9e8e502e8274c3a470ba9dc00dbb6151dbd5c00fd
8449e76821951bc01640d1d1cbe5055dc2a473b490ca0dd0e051d1d1cb05d04ff84de76821944b4f9b557dbaa9e0cf3e61d6
774623c7fe6b2b0685462b15103f61a2394248d946f0841274b052e9999f28d91c25a46ee9bb25aaf76626f8932588d27a1c
72ccf9effd53197622f87b35cad2721462bcbffe0e0b28d88900b54649a5055062a9dbe0392dd7741cb9f37658d56efb7e4b
7f8d27d390a96e835e4d9364741401f013269bd641693696a30890ff72cc209ca384d44d1d230a73b1b0f0481d51f6ebd500
4d671a32d53122afa6693a9a8a00fa0993ced1422a37c0312245475311403f61d2395a48e5d4e1e2e8ed1c1676d93b739def
798a469aea7471cd3464aae3c55e4dd374341579419730e91c2da472eaecd4f341c4f3cfb1f7cd729fd286e0c93464aab353
5e4dd374f41501f4137ed3395a08e5940dc1b124fbf6f4a92cb9dbe7a62e642dd390a96c08bc9aa6e9e82b02e8274c3a470b
a95c9e0b88fa34f7e8ff6c0bb3be331a39f6010ee7293a9a7b00fa0983ced1422bf7eb3e050a8dce1e207ec244738490b229
1baa63bbc3da2e7d22fb76eefb15f7862a17aa9268485426544e45d368b4f500f113269a238494edd756630a8db61e207ec2
44738490b2a933e4b993659e8c4fc719666dd7aeee152dd390a94e917b354dd3d1d603d04f98748e1652b9011efc283a3a7b
00fa0993ced1422a577bae5d6166b226b3f0301d0dcbf8b0dcca7df0c215e00a0b84a743a8cc3464d6ae6b5f6a9aa6a3b307
a09f30e91c2da472bf6f9629e9e8ef01e8274c3a470ba95ced053b9df732a5bd79575963a65dec67f8626ed7b2b74ede32d3
90597bc17ea9699a8efe1e2fe812269da38554aeb63a4e1d3c739fedf30a93b5237d2dc3b73237b87f1a79ab4443626d74fc
a1a269349a7b80f809bfd11c21846c95f89ae665cd1fc99dfaf66fb92771dde8f2b407959986cccafca5a6693a9a7bf4d02d
4c3a470ba9dc08bdcb051dcd3d00fd8441e768a1951ba14bbda0a3c507a09f30e91c2da472d58f206e29ded322f35e624bde
3b779f2e07b0b6e74585ce3464563f822f354dd3d1e203d0ff97b977499b1cc7816ce75c45aca03fbd1f2be8f95d47ef7f7c
f91460d261fe1a952b07551908249cc75c2e511209bbc24ad7d10295fbb9db05a0a9bf87105f6145eb0881b27dc1d404e8d4
e243a0afb0d275b440e5ee161ff2b17ff95768fdce6870eced3cb26dd332ef599771dce20475187377f7bcba376df7bbce23
941924b39d47ded40c4ca7161f027d8595aea3052af75b6724e052730fc1bdc2cad5510135fb82fb15d0a9b9c70dba8695ae
a3052af7059f0ba053730f81bec277ba8e16a0dcef7d2e904ecd3ddc785d58e93a5aa072edde662e6b78cb9cb4ec6bcb5dce
8f61cd0d9baf53186606c96cf7366f6a06a653730f81bec29eaea34547b9e673110fb921de76e4bb8f38713b9327eebf3c63
5cb6f8c3bc9e3d726690cce673f1a666603a35f710e82bac741d2d50b976eb9e56e14d5bf153ccad41e6b5bc44bdbdafe2cc
2099edd6fd4dcdc0746aee21d05758e93a5aa072bf7f4883746aee21d05758e93a5aa0726d4952eb5e957769d4ddaa69807b
f94e866bc51d6606c96c4b92ded40c4ca7e61e027d8595aea3052ad7a0d384655fb32bc59a173ea6a579e9b9f8300cfb16ff
fe82a6cc20990dfa4dcdc0746aee21d05758e93a5aa072bf7ff08a74eaef21d05758e93a5aa0723f5e41836862ee21c02daa
601d1950b4e277f1ac2e0e1bf2992daac5792434ea9bc3c6ed13d94ba344b5368f83c6fc81be6fe48f21261b329b72e1e3ff
cc4b9cd09e396c453a6157a4b6f7cbad7fa6bc5a3badba8b57eeb374e35bae1622b5bf1f6606c9ac0dfe5ed50c4c27261b0a
6d61a1eb6881cad55fef349c7967551c533cb7264390233f138e7703db7aeecb75cfcc994132ebeff755cdc07462b2a1d016
16ba8e16a85cdbdc31afe7b2e78783cbbaafe336cff936608eb3a4293565bc1653606690ccb6b9e34dcdc074e2b371876e61
a1eb6881cab5ceac974dd4726ef589601adf7da7162606496c9d595f540c8c26261b4a6ce11b5a470890ad6d6759d6e14ceb
95d2e2a5319e70963d4f13e37fbcc599c36e5b1d30334866dbd2f2a666603a31d950680b0b5d470b540e9a2c1feb1e138f21
efbad9a6ba7114de785b66904c68b2dcad19984e4c3614dac242d7d1029583850df9c7b697d3ec3a1cf1d0dc675cb0732506
9f080b1b7a1503a389c986125b58d03a42a06cb478256f703ff2c6e77539ce7d5ef611d7725c99413269f94aaf66603a31d9
50680b0b5d470b54eed7eb57084d1c3694d8c282d6110265fbc0fa65a213930d817661a1eb6881ca7d60d10ed189c986425b
d8d375b460e57eee760168e2efa1c41616b48e10285bb11581fa62b2a11f6b61a9df190d8efd033f29a213930d85b6b0d075
b440e5e8cc19e7a6c94f320f70af0d24663c75b6cc209974eeecd50c4c27261b77e81616ba8e16a8dc0756c7139d986c28b4
856f741d2d40b90f6c6a263a31d950680b0b5d470b54ae418f7993663ef886b4c679cf8f1696fb52214c0c92d8905f540c8c
263e1b4a6c6141eb0881b2b5bbb933dddde6d95931bc8db79be9deeb8cb3f1719ded5ba6c42089ed6eee45c5c068e2b3a1c4
1616b48e10285bbb694f93e12d2f6e59f6987a1ea505cebc6c6929bfbddde7cc2099eda6fd4dcdc074e2b3a1d01616ba8e16
a85c3317b186744bee72933af1a54729e7bee6254ed7b400338364367391373503d389cf8640bbb0d075b440e56aab91f2e2
2adf56cfdbb0af739a1fa701268bda695bae97a09c1924b3b61b795533309d586d28b4853d5d470b56ae369539967d8f3797
e98ba84dcaa635fd18a7e96a1b529bca606690ccda54e655cdc07462b5a1d01616ba8e16a85c7dcebc957ff2d1b7b9abcc18
7f0ffb112fa9ed0acd994132eb73e6573503d389d586425b58e83a5aa072b567d2505fbbe72f22ad2d4a0f83d300ebeaa2f6
457362f089b563d29b8a81d1bccf86025f5101ebc880a2d5c658d93461c8e79b31f9d1e7ce6ce90f6d1b656b81c69941326b
63ac573503b2799f0d25bea282c63a906685772fcff54b9fae6b4b74fa435b79da8c0938334866e17d57332099f7d9b8f3d6
a890b10aa4d807babd0199b7da50de2b7a2363159e8a15de390f6ccaefdec7e92c4fabf240a7bafeb25e933b9941320befbb
9a01c9c46a43812d2c6cac03695688475b1f17ef70c631ce4c97fc87badc766e4639981824b1f0beaa18184c7c3694d7c242
d69101452bbd2a8fbc1776aee6086551cd528c40ae7ec1a557256706c92cbd2adfd50c4c273e1b0a6d61a1eb6881ca7da015
2bd189cf8640bbb0d075b440e53ed06998e8c46743a12dece93a5ab0729ff2d950e8bf4c2494aea3052a577d3686395e27b7
bc947b286d63cb9d5f3c01b5dbddeab381994132abcfc6ab9a81e9c46743a12d2c741d2d50b92f34c7073af1d950680b0b5d
470b54ee0b360840273e1b0a6d61a1eb6881ca559f8db49f2dad77cc039cca83aa32c0fa17f5ecddc90c92597d365ed50c4c
273e1b77e81616ba8e16a8dc37cc451e74e2f2a1d016bed175b400e5be7078039db87c28b48585aea3052a57a0e37dfd98f7
e2e601d63d4fe5025b5a2bb4a39b13834f2cc8af2a0646138b0f25b6b0a0758440d9caa93bde4d4fc77ee6dbbc619dd76241
97c757df2a55e44e6690cc72ea7e5733309d587c28b48585aea3052a57a0b771db8eb5ba714c539c2e1707a7fc327e3bdab7
8c7941f20af09b7a81b9c4dc43705d58b83a2aa0661f9888119d987b28b4853d5d470b56ae409ff1e492df35e4e36e1c4bf7
e33c3dcecd06f6b6f5a1931924b340bfab19984ecc3d14dac242d7d10295fb809f1fd189b987425b58e83a5aa072057a4eab
e3d6233fc6c9aba7d2bbc3ba64aec0fe57c6fff77fffae1198424c3c14cec242d1614685aa2d653c7b8e43f9991ded9299c7
b78e719634b657709c1824b19a52bea918184d1c3c94d8c282d6110265ab4f45ca3bb1a9b86d14cbccf2958ca97746bc6eee
eda9086606c9ac4f455ed50c4c270e1e0a6d61a1eb6881ca7dcac1e30efddff6144ad7d10295fbc0335da213070f85b6f08d
aea30528177eef814e74e2e0a1d01616ba8e16a85c7d1b95aa1c675ed03cc5fbb9698e775ae535c3b55da1be8ec2cc2099f5
7dd4ab9a81e9c4c143a12d2c741d2d50b90fbc6c253a71f010681716ba8e16a8dc07b67a109df87828b4853d5d470b56aead
1ab11d73f1fb9986313da44a5f51bdd6da1a30cc0c92d9568dbca919984e7c3c14dac242d7d102956b1b1fe6296d0eccab97
e66b2a983629a4912693bf6be3036506c96c1b1fded40c4c276e1e0a6d61a1eb6881cac1c687f86bdc8f38d86cb5b2cff106
28352fa38d0f5766904cd8f8d0ad19984edc3c14dac242d7d10295fbc01247a213370f85b6b0d075b440e5a8bff5162f29fb
91ef81e27d7ebcbe4ee38816265766904cea6fddab19984edc3c14dac242d7d102958365cbf334e4d56a79c7f774f982c1b2
e52b3348262c5beed60c4c276e1e77e81616ba8e16a85c31b680fae2a7a11f6be15bfdce6860ec1fe8934f74e2e6a1d01616
ba8e16a85cd11aea8ba5867eac85a57e673438f6f2b150df9b5ac8a75a54aa77c68223ff801f10b0394b0d056e410fc62aa0
6262a9a11ff7dfde115a1a4701e32d1ff7acec7d2df4f3aea894e651d0887fbd7706b0c4d442692d2c602c0209463b588a71
dc984f25436968b9e0a5e0ca0c92493b587a3503b389a985225b58e03a4aa06e9f32b550e8bf1c1b94aea3052af7eb0d8f84
268e164a6c6141eb0881b27dc1c603e8c4d1e20eddc242d7d10295fb75277a42133b0b25b6f00dad2304c8f681edcb442776
160a6d61a1eb6881cafdfc6b063471d210621716b48e1028db077ecd44274e1a0a6d614fd7d18295fbc0958ae8c44c43a12d
2c741d2d50b90fcc49884efc3414dac242d7d10295fbc03d39d189a586425b58e83a5aa0721f70d8243ab1d450680b0b5d47
0b54ae352eb45e3771f2bf2753a8d263b01e88575b4e4a0c92d8da16bea818184dfc3494d8c282d61102656bfd29e3c4703f
f262d66338e35f9e737ea113ff837989dfce76f5eac6cc2099ad3fe59b9a81e9c44f43a12d2c741d2d50b9d689f498e38f3d
9f64e289f538e735adf64b3d16e3d9e73c066bbf8f9941325b27d2373503d3899f86425b58e83a5aa072add16e5d469206b8
5573eed242748947e37aed67e96406c96c8d76dfd40c4c277e1a77e81616ba8e16a8dc075a2a139d586a28b4856f741d2d40
b9d63c7b8cb396f1cccdf2075bf078e4456183b53dc3c42089ad75f68b8a81d1c44f43885d58d03a42a06c5f68090f74e2a7
a1d016f6741d2d58b94ff96928f45f66114ad7d10295fb949f8642ffe517a1741d2d50b962b400f5c5d4423fd6c252bf331a
1c7bf958a82fb612fab11696fa9dd1e0d8db717d96e66bf918acddf6f265356feb9e77b30fc1cc2099edb87e5333309d985a
28b48585aea3052af75b7f07e0123b0bc5b5b070755440cd7e7d89223471b450620b0b5a470894adcd44d6bcee331f74475d
f699a74ac9b3292f13bb6622941924b3cd44ded40c4c278e1677e81616ba8e16a85c9b73e6dbbdbcf7e2d8534bded42e21fd
61bcfa3db739276506c96c73ce373503d389a385425bf846d7d102946be60eeb1eaf1a7985d8b1a67d727bb99ca49590c31a
a78757f77bcc0c92d9cc1dded40c4c278e167ebc3e2c741d2d50b9761f599602962fe2a84f71d20097a56caebaee2e303348
66bb8f7c5333309d385a28b4851d5d478b8e72157a1ce3f9fcc84fe8e217714cc3b2e67d27fb10a7115bbcb25c572acc0c92
59a15fd50c4c278e160a6d61a1eb6881ca7dc1e700e8c4d142a12d2c741d2d50b92fb4fc073a71b450680b0b5d470b54eef7
cb27904e1c2d14dac242d7d10295fb7d5752a413470b85b6b0d075b440e57edf2c13e9c4d142a12d2c741d2d50b9dfbfc041
3a6f6aa1cc5754d83a4aa06e6a6aa19ff9dfb6115a9c4742a356538bfb27b27d458d4a6d1e078db9361bbb7a95a7ae6fc3d5
4e294e9b72fbacfa9dfe95166a9bb1bfaba997c5354af5b2f0f32817deff4f32cb984ad88a74c2ae48dde0b0c693e77ee6bd
36f13f384b4ba134d4dc66280ebc1dc29c1924b36e70785533309d7a5908b48585aea3052a57fb06ced3320e5b9e09cea58d
d09c6ffba6b4ef39e6ee6d02c29941326bdfc0573503d3a99785405b58e83a5aa072ad4364bd75ff979b0bb6bbbbb407652a
8f9daf09086606c96c1d22dfd40c4ca75e1637e81616ba8e16a85cdbd5d176d4240f82ebc945eadbd96e07ae5d1d941924b3
edea785333309dda5908b4856f741d2d40b9d6f2353b9be59fdc3a5e9d565287d6b4287f5e6d83036606c96c4d5fdfd40c4c
a77616026d61a1eb6881cab5ad2c674c3cf333aaf8051df12e73c88b7a9634b6dc8af8daca42994132db5696373503d3a99d
85405b58e83a5aa0721fd8ca4274ea6821d01616ba8e16a81cccbae28c3f774bcbc7e5541c9e179a6a5a66904c9875756b06
a653470b81b6b0d075b440e56881d4b8c6ebc9b8ece5b85ce7635a17dcca726506c9a40552bd9a81e9d4d442a02d2c741d2d
50b95faf6a253475b4f0c42e2c681d2150b60fec58223a75b410680b7bba8e16acdc079684119d9a5a08b48585aea3052af7
fba70448a7961a026d61a1eb6881cafdfe2901d2a9a586405b58e83a5aa072bf3e8b119afa69dc885b58d03a42a06cbf5ebf
4c686aa621c416bea1758400d9684e925e950e435ec6b4da12089a93b4cc20993427e9d50c4ca7661a026d61a1eb6881ca7d
60f64974eaa721d01616ba8e16a85cbbcfd8ce78f5cead39d6f5eac79b1c12ca4e773b85616690cc769ff1a666603ab5d410
680b0b5d470b54aedd512ef92c9307e8cfb1f31e8fcced186cfd326606c96c77946f6a06a6534b0d81b6b0d075b440e5dab3
83296d0bcdcbb796da6d2777599aa7b67eeb7a7640994132dbb383373503d3a9a586877661a1eb6881ca55e8dafc2e376c88
5f417a7e97bea1e918aa736763a6c4208915f94dc5c068eaa721c416f6681d2158b6421cc7308ef14296c637a7d7e2639cb0
9587b5e5b1ddb5ac823383645637df573503d3a99f86405b58e83a5aa072f5b16f5989975b494da925ca9e775da747b46d3b
547dea8b89c127d687be6f2a064653330d21b6b0a0758440d9ea43fd6a369b1bedacf1cc935ba5fd4b0fe28f7857b4add73a
5ece0c92591fedbfaa19984efc3484f98a0a5b4709d4ede7fe214f3031d310dc2b2a5c2c0209567b72cdcb114fa2a50d5e32
593ee36c213b0558bfa5da930b338364d69e5caf6a062413338d1b6f8d0a19ab408a55f390a934d4cec3d98e3879d853f7c1
d4242d2d444d9dd2fe55f310cc0c9259cd435ed50c4826661ac27b456f64acc253b1c29b97702cb50d5c6b9495fe10f3f24b
88ba248a13832416da57150362a99386d05a58c0580412ac3a692ce7907c886a83c3653ce6217b6f8d53f58fae0f823a9941
32ab97c6ab9a81d9d44c43902d2c701d2550b76f996908f49f4e1142d7d10295fb96998687fedb2942e83a5aa0721f68724b
746aa621d016f6741d2d58b9d2ce78777776c3e6ccf5e275643ed303f8d6ce983383649676c6ef6a06a653330d81b6b0d075
b440e5be65a621d07f3a45085d470b54ae9a69a42e78a9014c1a53f94e8eb3749cde6b43d30acd994132ab99c6ab9a81e9d4
4c43a02d2c741d2d50b92ff84a009d9a6908b48585aea3052a57a0b7b2e823bb3eccc7be2ef3b4160b8874c797baa857664c
0c9258905f550c8ca64e1a37e21616b48e1028dbb79c3404fa4f9b881b5d470b50ee5b4e1a02fda75384d075b440e5be6009
04746aa621d01616ba8e16a85c814ef77ba93f489e3ae5653ef1df4ba3f5e47e9b9ee55468ce0c9259a0dfd50c4ca7661a02
6d61a1eb6881ca15e8d5ad578b3ffc613ba778b6cd039cebb158a139334866817e5733309d5a6a78681716ba8e16a8dc07a6
6344a7961a026d614fd7d18295ab3e227110e73ce7157b699a3a65d7b9726f34cff1d2d31a5474328364561f91573503d3a9
a586405b58e83a5aa072d5ba6f8eb703f1fea09c67e219288e2b3f8e8d370cd9dca5bdb3e96406c9acd67daf6a06a6534b0d
81b6b0d075b440e50af4b48f53bc841fe532ea76650f67fc25eee7b9b59b69ce0c9259a0dfd50c4ca7561b026d61a1eb6881
ca55ab8dbaafbedcf8a6d36cb289c837becb389de3d156017632836456b38d573503d3a9db86405b58e83a5aa0721f786c42
74eab621d01616ba8e16a8dc071e90119dba6ddca05b58e83a5aa072d56d63d9e321b04ee519f438ee6971797eae331fd932
a1f9eb7262f089d56be34dc5c0686ab521c416bea1758400d9be65b521d07ffa48085d470b54ee771b50084a1d3684d5c202
d5910005abaf5f97f9c85b44fea55d2a97ebc79076fe377bb2fafa15338364d6d7afaf6a06a653870d0fedc242d7d10295ab
6fd9e7353fb3c96b3cc6a96ed64c038c67d63535fcba960b616690ccfa9efd55cdc074eab021d016f6741d2d58b96f396c08
f49ff61142d7d10295fbc07221a253870d81b6b0d075b440e560355cb2af9a52d7fef40d55f31b5cffd7128324c25ab85ec5
c0686aaf21c41616b48e1028dbb7ec3504fa4fef08a1eb6881cafd7a612ba1a9b786105b58d03a42a06c1f58bf4c74eaad71
836e61a1eb6881ca7d60793ed1a9b387405bf846d7d10294fbc0f27ca253670f81b6b0d075b440e57e6d8f40686a2a22c416
16b48e1028db17be66a0134b13cf6c5161eb2881ba15d78d67756f2a221fd982be348f03c7aca622f2717f187748691c058c
f7662a229ff787738694e651d088cb273e6babb3877ca485a53a8f8446fd817332b1a9b387205b58e03a4aa06e1fd81c4774
eaec21d01616ba8e16a8dcef7b8c209d9a7b08b48585aea3052a577e52505fed356e1fdbc252bf331a1c7bfbd8477df5b890
8fb5f0ad7e673430f6f2b1505f5d26e4632d2cf53ba3c1b17fe1e704746a34e1a15d58e83a5aa0721fb8e0139d1a4d08b485
3d5d470b56ee03367544a74613026d61a1eb6881cafdbee11ad2a9d184405b58e83a5aa072bf6fad87746a3421d01616ba8e
16a85c6ba298baefe697afbbfff7e13a445b0bc5675e90bcd640f1ef7a81b9d46242702d2c5c1d1550b3d62873cb2f2dd28f
ed189778111da6357f396977e4be2efb058c9941325ba3cc373503d3a9cb84405b58e83a5aa072cd7021b7d9194b97fcabb9
503a22ebabf7ab51266606c96c860b6f6a06a653970981b6b0d075b440e55a1fd863dca3fc7980ed496f392c75a95d273348
66eb03fba666603a759910680b0b5d470b54aeb5399e73b7e93cc0a32d9dca5fd19017d45c6d603131f8c4d6e4f845c5c068
6a3171236e6141eb0881b235e2735aa635af132aa7dae4f292fe909bd74e5b5b7dd5c90c92d998dfd40c4ca71613026de11b
5d470b50aef5ef5ec62db50c4ba31db6780d8f279cec13b0ddee6b38334866ebe0fda666603a7599f0d02e2c741d2d50b96f
b94c08f49f160a9eaea3052bf7856efc40a72e13026d61a1eb6881ca359789611ed7e2d4748e67d98d921b8e1fd3384df172
7b9dc6303348667399785333309dba4c08b48585aea3052a777799908ffdd3c041ea77468363ffc2010674ea7121d01616ba
8e16a8dc174e2540a71e17026d61a1eb6881cafddcf401d0d4e642882d2c681d2150b62f983e009d3a5d08b48585aea3052a
f70dd387079d3a5ddca05b58e83a5aa0726db23f4fb923c8bfb237396f81c9772347dd747ffda0313348669becbfa919984e
9d2e04dac237ba8e16a0dc07ee65894e9d2edc787d58e83a5aa0725f307d003a75ba10680b3bba8e161de5da93a9748adb73
4b94f8bff1d2b21fb971c6b60dc3b06ff1efaf2753941924b33d997a5333309d3a5d08b48585aea3052af7fb07af48a74e17
026d61a1eb6881cafdfebd02d2a9d385405b58e83a5aa072b4162ade850e6372664d4b97a67188ffe9d56c85338364d26aa8
5ecdc074ea7421d01616ba8e16a8dc8ffbfc219ada5c08b18505ad2304ca46ebfce2747139d633b7cc9de314e2d80eb429bf
12834fa4557e9d8a81d1d4e342882d2c681d215036205e2ee3f434bee4441ce771f4c2cc3283640273b766603af1b810e62b
2a6c1d2550b79bc7857ce61f2e12529c4742a3fef961f5c4127b8d1b6c8d0a164b4072d5b654cd952b8e265e38e729de04e4
c7ee79bf4959aa5ffb52616690ccda98ea554df5d9b886ab3e1b7e16e7c25b6a7b1fa775396c453a6157e4d70dd7084d4d36
84d8c282d6110265ab261b43dd781087346ff1b736a73edf697cf1229aeef0affb19ce0c92594d365ed50c4ca7261b026d61
a1eb6881cad51ff234c4a978bc03f9979a1dc66bc8b41d794f7f3c3cb7f8035caea7329c1924b3fe945fd50c4ca7261b37e8
1616ba8e16a85cdbd831afe7b2e78da0cbbac7ff689ef38b87399ed9a778cf7d3d96e1cc20996d63c79b9a81e9d46443a02d
7ca3eb6801cab596a8736aff97bb5ad63e967987686d09385e37ae981824b135447d5131309a3a6c08b18505ad2304cad6b6
b2cc7957515e28365f2d95d26ea3b4e966484d1dda5616ca0c92d9b6b2bca919984e1d3604dac242d7d10295831ec7fb1a7f
7da9c1643e288778641ce341735ccb0c92093d8ebb3503d3a9c386405b58e83a5aa072301d19abdbcd5c8ecb731ed32480e6
235766904c9890746b06a653870d81b6b0d075b440e57e3fd1453a75d810680b0b5d470b54ee0bb62240a7261b1eda8585ae
a3052af7eb7d4b84a60e1b426c618fd6118265fbfd2319a453870d81b6b0d075b440e53eb0598be8d46143a02d2c741d2d50
b92f7cd340a70e1b026d61a1eb6881cafdfc070d68eab071236e6141eb0881b27dc8614388ffb48fb8a1758400d9e8f23c4e
fbb694865a6b5ae3339fc3463b5e2c3348265d9e7b3503d3a9c386405b58e83a5aa0723411cb86734bdef6bd9ce9dfd675c7
07435766904c9a88f56a06a653870d81b6b0d075b440e53e30e5263a75d810680b0b5d470b54aedd5c0d7110675ea2b7a607
19b9b150bee36fbbdddbbd1525069fd86ead5e540c8ca6f61a426c6141eb0881b2b51be8214e7fd73c335ce260977959f2db
f9791be3cdc1b45c8d3f38334866bb857e5333309dda6b78681716ba8e16a85c7b54328e739c10e76e0dcbbe95a6706980a5
dd525a4cd61e95506690ccf6a8e44dcdc074eab021d016f6741d2d58b90f3c14233a75d810680b0b5d470b54aef6b5598f35
3d94cecface271386cf1d6271b452cd7168adad706338364d6be36af6a06a653930d81b6b0d075b440e57efd949bd0c46143
80afa880756440d17ed7a80990c45b4340afa810313e495530c7e4031195cf031bd3d6a371ceabe9e38dd730c6bbceeb8d0d
6706c92cb0ef6a0624136f8d1b6f8d0a19ab408a95366be7523cbb8bfa6941d1b996afa2f4d99eda2eaa4e6690ccd268ed5d
cd8064e2ad21bc57f446c62a3c152bbcc9f176dcf24af9319e3be319b498b78f37936c4e0c9258685f550c88a5de1a426b61
01631148b0da3af0accbffd280e271382df39cd75137d3f2f16a1d889941326bebc0573503b3a9b786205b58e03a4aa06eb5
4964da127bc65bf77fa9a163e961d91a1d0ec33c24979bda2412338364d62691af6a06a6536f0d81b6b0d075b440e56a3bd0
da052c8def682db6d3f8866bce54bb81526290c4da0cf44dc5c0686aace1895d58d03a42a06cdf32d610e83f5d233c5d470b
56ae76725e4b6799d275b93af9e41eb5a9335a9aa0b546e59c1924b376727e5533309d1a6b08b48585aea3052a578d35cad3
8a7c4b3bac6d8f571ee0b4c483636f17e54e6690cc6aacf1aa66603a35d610680b0b5d470b54ee5bc61a02fda76b84d075b4
40e5be65ac21d07f1a47085d470b54ee43c61a37e23f5c2304ad2304caf6058f09a053630d81b6f08daea3052857a0e3b478
dc97353fa44967ba6b03ce509f3ab7439b13834f2cc8af2a064653570d21b6b0a0758440d9bee5aa21d07f5a46085d470b54
ae40c75b9f78af7ee65d46c39a9ab31e7bb9b0ecf5c559b31de0cc209905fa5dcdc074eaaa21d01616ba8e16a8dcb75c353c
f4df961142d7d102952bd09bad621ad29aad31de08e45fdf713d822ecc981824b120bfaa18184d2d3584d8c21ead2304cb56
2d35d674661d737f9421ef324fdf4a9e1e0f73dab4bad6297727334866b5d4785533309d5a6a08b48585aea3052a572d358e
653ce344b89c5893dfc73294af28b54f4fbb45ebb1ddc90c92592d355ed50c4ca7961a026d61a1eb6881ca15e839dedead4b
59b6949afd64a793dc383e356f9ba6b92d74ec6406c92cd0ef6a06a6534b0d81b6b0d075b440e57e7d0f4d68eaa721c41616
b48e1028db071e95109dfa6908b48585aea3052af72d3f8d1bf41f7e1142d7d10295ab7e1af915d2581e4dd67bbd6a083c0c
d33c6ced6ad5c90c92591d355ed50c4ca7961a026de11b5d470b50ae3ecd779d8ea6f460728ce7d7fcade467b6f3b95d4ff3
313348667d9affaa66603ab5d410680b0b5d470b54aebea98907638ce4370cf16818c63d5e63f3c3dab2297abcfa47716690
ccfabee655cdc074eaad21d01616ba8e16a85c7d07d92ea2ffd2fbc238538a13e43c759a8a7bf97635e2e7cc2099f51de4ab
9a81e9d45bc343bbb0d075b440e5bee5ad21d07f1a4778ba8e16acdcaf7723109a1a6b08b18505ad2304caf62d630d81fed3
3842e83a5aa0721f580f4674eaad21d01616ba8e16a8dc8f3747229a1a6b08b18505ad2304ca46abf2ebaa8fbce076bbf610
d0aafc9619249356e5f76a06a6536f0d81b6b0d075b440e57ebd8897d0d458e346dcc282d6110265fbc05a6da253630d81b6
f08daea30528977b84517db5b6908fb5b0d4ef8c06c7fe81cd0044a7ee16026d61a1eb6881ca15ada1bef84bf84fb5a854ef
8c0547fe056b82279b77b710e016f460ac022aa6ee16f2717f384848691c058cf70387317089b786d05e5101630d48af9bb7
867ce29fd611529d4742a3fec2010c6ceaad21c81616b88e12a8db170e2ca0536f0d81b6b0d075b440e5cad105f5d5dd423e
d6c252bf331a1cfb170e30a0536f8d1b740b0b5d470b54ee03c60b44a7ce1e026de11b5d470b508efa7f95971bd3926fb48e
b64292fa7fb5cc2099d4ffab5733309dfa8a08b48585aea3052a07d0f1d6224e7ef733fb054cd7b25780be3283640274b766
603af515f1d02e2c741d2d50b92fb84d009dfa8a08b4853d5d470b56ee0b6e1340a7be22026d61a1eb6881cad5ee8d833d19
dbf7d455bd7464dece7a9ebdda74526290c4dabbf14dc5c0686a2a22c41616b48e10285beb57796c4bba7d8e633a86a16e31
caa79f618a93a8b445b5f5aba4cc2099ad5fe59b9a81e9d45444a02d2c741d2d50b90f7426253ab51611680b0b5d470b54ae
b96c1c733c0a4a67e0793bce791db2357c7ab3112f358335e5c7cc2099cd65e34dcdc0746a2d22d01616ba8e16a85ceb365c
7f7ffff2dbbb6d8d07625e6bb19f655fc6b5999a338364b66ec36f6a06a6536b1181b6b0d075b440e55a5fe9fc9bcb3fbf3d
754d9db7dc55eb8817d63d5e6ada9a154e0c92d8ba4abfa818184d7d4584d8c282d6110265fb40276da2536b911b740b0b5d
470b54ae354c4f7bac8ed200fa4cebe3e6332f3539725be8715d2f1b02cc0c92d95aa6bfa919984ead4504dac237ba8e16a0
5c839ee2809623f7ee1fe66294b0975f627d317b41536690cc06fda666603ab516f1d02e2c741d2d50b92ff801009d5a8b08
b4853d5d470b56ae41efc7ba6cf9347b8ecbb4e785087980a5a5fbf5ae9733836436e8373503d3a9b588405b58e83a5aa072
5fb0bb003ab51611680b0b5d470b54ee0bdf34d0a9b189405b58e83a5aa0720d3ace1b96212f3b38b385555e7470c4b97276
2ebb2e58901724af01ff5d2f30975a9a08ae8585aba3026af681f336d1a9a589405b58e83a5aa0721f9896109dba9a08b485
85aea3052af705830fa053571381b6b0d075b440e5be61f0f1a05357931b740b0b5d470b54ee03b7d144a7ae26026de11b5d
470b50ae3d223a8e21c6b6fc454cd376c4f36b79a253f6d64dd7890c338364b687446f6a06a6535713375e1f16ba8e16a85c
7b1838965f601ae01a7ff8f376e65616f5e677b67e3e981924b33d0c7c5333309dba9a08b4851d5d478b8e72f456a3ec183e
d257b4a50da567da8f442f355a62f089f44ea35331309a5a9a08b18505ad2304caf6e3653a88a67e26426c6141eb0881b251
afbdabed4cda11749cfb1ccf36887c6506c9a46e7bbd9a81e9d4cf44a02d2c741d2d5039809eb729f7924a73e3356d451ad7
71a1f7c196192413a0bb3503d3a99f89405b58e83a5aa072003d8d439c53940ee4eb30a585b7cb802d80afcc209900ddad19
984e2d4d04dac242d7d10295fbfd9a52a453571381b6b0d075b440e5007a58cf6d4b66d06980c71aef0fc66dc3a577576690
4c80eed60c4c27ae26c27c4585ada304ea46c8ad05601cdd1e67ce69ed3eb60c6d89411209b85331209818aa08ee15152e16
8104fbc2f7fb24134f951b6f8d0a19ab408ad5ed2d6927ffb4e42d37f1ffc638f99ff2ac614ccbf697b4fbac6d6fc1cc2099
757bcbab9aeaa9720d573d55fc1cd285d7b4733d4ee972d88a74c2ae48ed36b7ae53daec9e8eb939bd353c5373f734daf128
add4da21cd9941326bbfb9573503d3a9ad8a405b58e83a5aa072dfb25511e83f3d4384aea3052af7810e9244a7b62a37e816
16ba8e16a8dc077a85129ddaaa08b4856f741d2d40b9b691a9d855967d4763fcf5c5bfce87657a7eb6ccd79d332606496c1b
995e540c8ca6b62a426c6141eb0881b2fdbaf32fa1a9a78a105b58d03a42a06cede29c1c2ba732f58f73f1785d3dcb46b3e5
b69e84338364b68bf39b9a81e9d45345a02d2c741d2d50b91f4fbf104d0d5584d8c282d6110265a35bc8f4ae6118967c4cda
bb16ba856c994132e916b25733309d1aaa08b48585aea3052a479bd5b66538873dbf535a97e4323e1c03425f99413269bb5a
af66603a3554f1d02e2c741d2d50b90fb8c8109d7aaa08b4853d5d470b56ee039bf4884e3d5504dac242d7d10295fbf5263d
4253431521b6b0a0758440d9f21e01aaaf9626f2b11696fa9dd1e0d8cbc7427df515b97d6c0b4bfdce6870ec1ff21511e23f
4d336e681d2140b60fd87e119dfa8a08b48585aea3052af7fb079d48a7be22026d61a1eb6881ca81af489c22edc73ae43768
cb3e1fb142bcfd0168cb0c9209be22dd9a81e9d45744a02d2c741d2d50b97677712ef370e6751ef1ca79c43972e9d2124fb3
f39e9fcc5eb717941924b3dd5fbca919984ead4504dac242d7d10295fbf52d15a1a9af8827766141eb0881b27dc04c85e8d4
5744a02dece93a5ab072157aa92b97f2a177ad6c8903b47644151a33836456e8573503d3a9af88405b58e83a5aa072f5f95f
5ed531e52d31735a009b1ed1e567757511d7b5488a338364d6e77faf6a06a6535f1181b6b0d075b440e56a23a325cef8d3b3
d834c0f972f44ecfa5f39280e33abc393348666d64f4aa66603ab11611e62b2a6c1d2550b7fa707b5af769387243d378b2b1
1e25f1d25afa24b6ef99338364d687dbaf6a0664138f1121bea282c63a90661f78830164e23172e3ad5121631548b1c23bec
b5695a1a5b5ece945e29a5b1a513ce943a5e565ece0c925978dfd50c48261e23c27b456f64acc253b1d26a2f2a3ec779433e
c78cfb3ec4df576aa794dae2c55f407a01511f8c7432836496567bef6a0624539b1101b6b0b0b10ea459218edfc3becdeb9a
1b01962716635e4a5f5ea94c972b4127334866217e5733309bda8c08b28505aea304ea567a679efb78acdb59ed21d6314e8d
d7bc7c6bbcac5f0a7427334866e99df9ae66603ab51911680b0b5d470b54aeb648ad9b21d298e2f47f2eed3dd300873a53ac
93ec4e6690ccda24f555cdc074ea34e2a15d58e83a5aa072d56964afbfb87fa5b5fa1ecfec4bee627ba6c56371aa5a9fe277
32836456a791573503d3a9d388405bd8d375b460e56aafe76b6a90fa326ffb381dc3541ad96607817479a9bd9e31334866ed
f5fcaa66603a751a11680b0b5d470b54ee0b7dcc814e9d4604dac242d7d10295fb96d38840ff69a321741d2d50b9ea3432ee
a96949e9c0ddf6de97ced4d336cfdbd0b6167532836456a791573503d3a9d388405b58e83a5aa0721f721ab911ff61a32168
1d2150b66f398d08f49f4e1a37ba8e16a0dc879c4684f84f1b0d41eb0881b27dcb6944a0ffb4d110ba8e16a85c819ea73dbf
2ccb1796ecd915ffbd5c4d87351e8b67db70d2c90c9259a0dfd50c4ca74e23026d61a1eb6881cad52bf416e786f1fc56aea6
e3b86fcb9e5773a5fefae59f7685c6cc2099f50afdaa66603a751af1d02e2c741d2d50b90f398d08f19f361a1ead2304cbf6
2da71181fed34643e83a5aa072df721a11e83f6d3484aea3052a57a0277164dbcaf752bc2374696b27334866817e5733309d
3a8d08b48585aea3052af781db68a253b31181b6b0d075b440e5be653622d07f3a69085d470b54ae3c1aab3ed753fe22a631
0e692e4ed0e3547b88d4df74273348667934f6ae66603a351bb941b7b0d075b440e50af45eeef28be9f33ea545d76557f6d8
ba0dd4196827334866817e5733309d9a8d08b4856f741d2d40b9f2987bcacf2f86bcfa769adc7dbef97f97a7dc9c187c6279
c8fdaa626034751a11620b0b5a470894adbeb8c9d62773d93410b38e3847cf4b029c2f4a7d7183994132eb8b9b573503d3a9
d388405b58e83a5aa0721fd8954074ea34e2a15d58e83a5aa072edfd73593f9ed701a4ff8f9794fc8778865da678a7742d63
c6c4e013dbdbe7171503a3a9cd88105bd8a3758460d93e643322c47f7a68085a470894ed03eb67884e6d4604dac242d7d102
95fbc0f230a2539b1181b6b0d075b440e57ebfcd1be9d46944a02d2c741d2d50b95fafe72534b51911620b0b5a4708948d4c
37c6f26a31ff128fb9dc13e112d72b334826996ef46a06a653a7911b740b0b5d470b54ee5b4e2302fda789c78daea3052877
771a918ffdd34643ea774683632f1f0bf5d5eb433ed6c252bf331a1cfbddebc37fec5f7e1652bd33161cf9175c039e6cdeeb
43805bd083b10aa8d8077e454f2e711a11da2b2a60a801a855bc119e95c5eb433eef8a4a691e058df80b07d3934c9d4604d8
c2c2c63a90664563a8ae5e1ff2a11696f29db1e0c8bfe03f0174eaf521d01616ba8e16a8dc17fc27804e9d4604dac242d7d1
0295fb82ff04d0a9d3c80dba8585aea3052af705ff09a053a71181b6f08daea30528f701ff09a253a71181b6b0d075b440e5
5acbb978a91ee7bc01785fcb90f2168d6ddb86791be7f3da82899941325bcbb9373503d3a9d388877661a1eb6881cafddc7f
02d0d46644882decd13a42b06ccd7f221ef5db7094fef9eb392ddb921fffc6214f691ff6723531c7cc2099cd7fe24dcdc074
6a3322d01616ba8e16a8dc071a67129d3a8d08b48585aea3052ad7fc27e6651ca6fce2ea88e7bbe34c7bfcf32f31b5ed48de
300d1a33836436ff89373503d3a9d388405b58e83a5aa072ad2fecb09fe9bddcbfb2c07b59d7b2d679cf8b8ba6780a6ad098
1924b3f5857d5333309d3a8d08b48585aea3052ad73a00cf439cb3e4f34c32568e73d53d9f738f6189e79d7c07d63a005366
90ccd601f84dcdc074ea3422d01616ba8e16a8dccf4d37004d6d4684d8c282d61102656bc4cb3cc5334dee4f7c947e064b9a
4f1c6533dd78b53de1cc20998df94dcdc074ea3422d01616ba8e16a8dc374c371e74ea3472836e61a1eb6881ca35e8315ec0
cbdbaa386fb8d6421ef575ed7c75f4a6c420890df945c5c0686a3322c416bea1758400d93ed0ae9ee8d466c443bbb0d075b4
40e51af4326e695f6c1ee016e72e7106914f3fdbedf1396706c96cd06f6a06a6539b1181b6b0a7eb68c1cafdd68701b8d460
44702d2c5c1d1550b30f1cd844a70623026d61a1eb6881ca7de1c0063a351811680b0b5d470b54ee03572aa2539b1181b6b0
d075b440e5da9cc4963e1e67b689988e3a515ea76939cd8701338364b639c99b9a81e9d46644a02d2c741d2d50b92f386e00
9dda8c08b48585aea3052ad7eeacd67d9cc7a39c67d2e684f40e3add051ddb36c443d14e64981924b3dd59bda919984e6d46
04dac242d7d102956b0f0e8ee53887a57c11477d149f06b82c65fbd5f59bc6cc2099edc1c19b9a81e9d466e406ddc242d7d1
02956b8f88c678e41ff9a57bfc22e24f31de6d673bd1a174ff9c2f4f2ccc0c92d91e11bda919984e6d4604dac237ba8e16a0
dcef1d37904e6d46dc787d58e83a5aa0723f77dc0034f51811620b3bb48e101dd9688d4e9c500ce3925739af7b5de63c03b2
6506c9a4553abd9a81e9d46644a02d2c741d2d50b9dfb765433a751a11680b0b5d470b54eef7ebee904e9d4604dac242d7d1
0295fb7d5b36a453a71181b6b0d075b440e500fa18d2dbb57c37b4acf1905ca7e1a495245762904440ee550c8ca63623426c
6141eb0881b201f1bec6a4d4ec290f70d8e2fdc2b55daf9319241398bb3503d3a9cd88405b58e83a5aa072bfef748d746233
22cc5754d83a4aa06e694d0b5517af0ff9cc2b2ac5792434ea26f298672df92018d25bc43dbfa52ebbdde20169bf25480c92
d8247e51312096187ddc606b54b0580292abb6c9daca3fb9a5d5e68efff1dc86ecd5dade21706690ccda26eb554d35fab886
ab461f7e16e7c24b9c64a40efcff4f6a70d495a8cddfc6647652f7414cc979652a7ddae6a9f4d36817454c0c92585bbfbda9
18184c3d3e84d7c29e8c5540c9ea5e9ab2d57aad2605ebbc9e4b69d9b6b6135edb4b83994132eb5e9a573503b3a9c387205b
d8c3b110a8dab7fc3d6ec87f985778381602556b9b87ce73c83d89d37965dad7692fc39bc7587c5aed29146706c96cdb87de
d40ccca6ee1e826c6185632140b50f34292536b5f710640b7b381602556bc8673ae6f3968f74f9cabd70d2e8e2997518d7d9
bcd528314862037e513130989a7b08af853d19ab8092b5ed52d640605de34966887f5f76b30d4bea7874cde53131f8c4b659
ea45c5c060eaeb21bc16f664ac024a46f3f8758f07c2911f022fdb543724e344be6506c9a4997caf66603635f610640b7b38
16025523bfc0adf643f997f7efa5d738f1aa4637a95766904cf20becd50ccca6b61e826c610fc742a06abf37c3443635f5f0
c82eece15808548d7c4c5cabd878865986633bd0dc6275dd672d915c4c3a150383a99f87f05ad891b10a2cd9cf8d0f004cad
3c84d7c29e8c5540c9be60d9026ceae521c816f6702c04aaf6f3af18c0d44444782decc9580594ec6e2172fbd03f2c327c79
1e0a8efb03fe47c4a61e22826c6185632140357a543d4e7b5e9f96cfe46bbc793e878df61b5866904c7a54ddab19984d1d44
04d9c21e8e8540d5e84175ba9dde4beff47588f720db3e23f195187c223da6ee540c0ca6e621c26b614fc62aa06434e7f9df
7a24239b5a8708b2853d1c0b81aa7d60664b6c6a1c22c816f6702c04aad66e5ed63813c9835bfc6d655948bcda5e78c80b92
d76e5dfeae17984a3d433cac0b7b2c9600f5faf59d2981a95d88f05ad891b10a2c597353b5fd4bcbbe4fc3983a56a6e1ad71
9a321ceec2849941329b9bea9b9a81d9d42c44902dece1580854ed038f98884dad4204d9c21e8e8540d5eab3d333b5f93df2
a5723ed76589e798fca39be2d7b3ef67fb09735e7079f5b9e98b7a81a9c4224458afa887620150ad2f78a23cc9c4204478af
a827431148affade23f5169ef7dc3c65dacf638825b774769906ed41dec90c9259df7bbcaa19904bec416eb435eac1500252
abd0cec3b0664f9e3498299ed3d6f405fc4b3dcee2d96618af151f9c187c62617d55312094388308ea15552ae47f4a55bae2
9d437dcb94c69296f60f675ee53fe689c298cea7a5291e2606492c3df15e550c08a5a620c26a61cf850a905805761de67959
ca8bf9786239b66d2debcbc6d253786f1b563b9941320beebb9a81c9d41244802decd95806d4ac200f79a7c6547b6ecef1cf
5379771dbf89ba38b822736690cc82fcae666036350411640b7b381602552b2d3db7f69af85f6e1d1c4faaeb596c8786cb46
aeb4f4e4cc2099a5a5e7bb9a81d9d40ec423bbb087632150b50ff4ab2536350311640b3b38168255fb40336662532b1041b6
b087632150b56f198108f29f2e171e8e8540d57eee1a0060ea0122bc16f664ac024a561d40ce389efdc8376ec322bfc0fa17
0d98338364560790573503b3a90388205bd8c3b110a8da87fc3f6ebc7f985b7832560125fbc2570c6ceafe21c81656381602
54fbf9af18c0d4fb43782decc9580594ac9ca8c77d1d86b4823e0d681db73dfee6f3d685382d5ee73326d759752733486639
51bfab19984d9d3f04d9c21e8e8540d50a72bc499e8efdcccdaf87353d804933c13cbcdaeaab7ec79dcc209905f95dcdc06c
eafb21c816f6702c04aaf685cb31b0a9eb874776610fc742a06ad5f363dce224e9c8a79934e99ff77208e6cea2dbd17ec798
1724af3a7ebca817984a0d3f04d6c20e8b2560bd3e6036476c6af721c816f6702c04aa5690532fa2fc14b95c2dabbf639e00
e76635e9eea320736690cc82fcae66603635fb10640b7b38160255fbc0ad13b1a9d587205bd8c3b110a8dab78c3e04f94f17
0b0fc742a06ad517d5592fc7e32e5e56e2953bbb168cba8ab9931924b3faa2beaa19984d6d3e04d9c21e8e8540d5ca939f78
7d1cd3d1969f454d6759509687977687a7ad85f5697c2733486679f2f3ae66603635f9b821b7b087632150b5f07b93636253
8b0f41b6b0c2b110a0daaf1f5513981a7c08af853d19ab80927dcbde4390fff4aef0702c04aaf6810d09c4a6e61e1ed9853d
1c0b81aa7de0d529b1a9bb87205bd8c1b110ac5a5b1010e7c1fb9ef7cf5ddbe7f2cbec638c5386f3b8ba93716690ccb624e0
4dcdc06c6aef21c816f6702c04aaf6fbed79c8a6e61e826c610fc742a06abfdf628c6c6aed21c816f6702c04aaf6eba5cc04
a6ae1ec26b614fc62aa064b4b4772b0df2724bf9eddaf34e4b7b5b66904c5adadbab19984d6d3d04d9c21e8e8540d53ed099
9fd8d4d4e386dcc21e8e8540d53ed02d9fd8d4d243902dac702c04a8f6856f19d8d44e44902dece1580854ed6e26221ffaa7
53872fcf43c171dfad44fc87fe6597e18bf34070d41f389a81cc1b89086e0b3a2e9400d5fac061fca4121b1161bda29e8b04
00a56e2622f2697fd864f8ca38041a6df9bc676535f1900fb4b02f8ec3a0117fe1c00532351011600b7b36960135fb829306
b0a97d88205bd8c3b110a85a39aea0bada77c8875ad897e7a1e0b8bf65de7143fec399c2c3b110a85a43deca1d631a5e7a6c
7bc6db9874f3b82de3394d6949ea854c9941321bf29b9a81d9d4ba43902dac702c04a8f6050f0b6053e30e41b6b087632150
b5e660919e828c53f65ed8e7317e11f18a9586772c63d939f0af5958506690cce661f1a6666036b5edf0c82eece1580854ed
0b1616c0a6be1d826c6107c742b06a1fe843496ceada21c816f6702c04aaf605fb0a6053cf0e41b6b087632150b5d662752e
f6ea69784bede59c8fc8a33ec5bf9ae9626690ccd662f54dcdc06cead821c816f6702c04aaf605eb0a6053bf0e41b6b08763
2150b59f1b5700985a7508af853d19ab80927dc1b602d8d4ab43902dece1580854ad211fc338cdc531e6cccbd996ec7076dc
dba8716690cc86fca666603675ea10640b7b38160255fb40f77362539f8e1b720b7b38160255cb3732545dad32e4432dace5
792830eef2a1505dad2afc87bab02fcf43c171ffda8f84c0d4a542782decc85805962c8b4cd5d52b423ed4c2be3c0f05c7dd
4e1a71d6198febfc9baebbe9f28d4699a64ddb75bfc6994132db49e34dcdc06cea1421c816f6702c04aaf681f324b1a94f84
205bd8c3b110a8da07ae86c4a62e11826c610fc742a06a5fb04b0036f58810640b7b38160255fb825902b0a94384205bd8c3
b110a8da076e51894dfd2104d9c21e8e8540d5da8388798fc77cb674f5ff3e5c0f06db6388675e90bcf610e2ef7a81a9d417
e206dbc21e8b2540bd3ef06c8dd8d41542902dac702c04a8f68187c6c4a69e106ebc3eece1580854ed0bfe08c0a6a610826c
618363213aaa7dc11d01d8d41242902dece1580854edf72d98904d0d2104d9c21e8e8540d57e6e120060ea0521bc16f664ac
024ad6fa2f9dcb3c9c796fcaba0d71369a8cd6d3f0e675def3d7d38031334866ebc1f4a6666036758210640b7b381602556b
c8cb1a4737946e766359149b8fc031dee86dd36e6d1831334866437e5333309b5a4108b2853d1c0b81aa35e46958ce29ef23
8ca796f2783baf891deeedf230334866437e5333309b1a4108b2853d1c0b81aa154f04a82e460cf29957d417e781e0a86f36
0cf2897fb80df8da380c1a71f9bc67657142b87d5e8dfad238081a6f5d86bf0ff1b7bbe61d4e715c7bfc59cf7957c894fafa
e79e4d6d193e6606c9accbf05fd5541f84365ab541703317179db3fbf698a35681a356a1767b8a378ca9f5656e44152f4fc7
b1ef7981cfb4ecfb1acf6bd73d386706c9acdd9e5ed50c48a636089ed7a29e8c5520c5ea9e83651f937b66fe2ef63411dae3
59258d6ddba6618dff7eb5f2c2cc209975cfc1ab9a01c9d403c1f35ad493b10aa4583d2b668f9125f77749fd3ecb8d451adb
11ff6ad8e6e59ab7636690cc7a567c553320991a20286f8b7a325681146bedf8c678133596b1a53741f3b9979d01a90ff6bc
a7e78dad1d1f6506c96cedf8ded40c48a6ee079ed7a24ac62a3c156bbcc9712275724b3e05f14898a62defc89cdb0aea6bc6
8e9941321bef9b9a01c9d4fac0f35ad493b10aa458db4d120fb1a36c695ac72d6d045d7257aa3937ec4d93af6b37096506c9
6cbb49ded40c48a6d6079ed7a29e8c5520c57eda3b1498d4f4c0935ad433313f69f563bb25c252bb030f6b518fc512905cbf
37a72332f53af0bc16f564ac0229f605738727991a1d385e17f564ac0229f6f346d64f2c7539f0b01675582c01caf5011b0b
20539703cf6b514fc62a9062bf7f4a46646a71e0792deac9580552ac34fc7fd6569301ff8916f5b5791c34e6bbc5807ee27f
77d0f7b5791c34e62f34db7f92a9bd80e7b5a892b10a4fc5e8373bc67bb461c8ef1ee22f6b4acf0d67f453b8328364d26fb6
573320997a0b785e8b7a32568114fbc0fe5420536f01cf6b514fc62a9062c03bace7162f897926b41cf132b98fdb869b81ae
cc2099c0dbad19904c8d053caf453d19ab408ad1842a35958a7fc893bdb406e23ccf892618961924932654bd9a01c9d455c0
f35ad493b10aa418f0fe8f6d6f894c8d051caf8b7a32568114fbb5011c60a9ab8087b5a8c3620950ae9f7b283cb1d44fc0c3
5ad463b10424d7170c149e646a26e0792deac9580552acb50089d78ab45b28df76efe7114f23796dc33c4cb59347e3c5cc20
99ad05c89b9a01c9c44fc0e35e41cfc51a905ef579ab5b6817ef40f35e9a31f7c2dfceb20aafd1726690ccfabcf555cd405c
6226e061afa0c7420540abf0633fe12792d80828670d7a24840799aa65c23ac6236acccd47e29dc9122f1943de4632c6a32e
77a56d975ace0c92592d135ed50c442536021ef50a2a15f23f942aa84399b596610dd3112f846b7e913f6e43bc233dd7ebb2
c39941320beabb9a81a8d446c0b35ad473a102a0553511c8aec663def735a6e7fbc7b6e5db95315f1a87ad4d293a994132ab
89c0ab9a01b9d443c0d35ad483b106a457e18d97c1ece8907b9e8dd3321e73696618cf26e58576b388e0cc209985f75dcd80
646a20e0792deac9580552ec5bf6018ef7cfeef89e8c5520c56a03ccfaa2b574c8def7635db67c7b166fd4a6f82b3cb73a8d
ea6406c9ac0d305fd50c48a6d6019ed7a28e8c5540c53ed0d615c8d437c0f35ad493b10aa458e15df6bad63b0d676d4bfaf3
d8a6251e11f100acbc9c1924b3f0beab19904c4d033caf453d19ab408afdbcd9f8134b1d033cac453d164b40727da199fc93
4ced023caf453d19ab408a55de7c47b694fec1653df77696b3ccb4cdf3364c6be3c5cc209995f755cd80646a17a0bc2deac9
580552acf01e75aa9ee55f9638dddbd6bd7c17b51d789d6c7432836416de77350392a95780e7b5a892b10a4fc57efee37d62
a9518087b5a8c7620948ae6fd90478debfbae07b32568114ab57a27883bdc4894056bff44eded63225ba9eb5d42b116606c9
ac57a257350392a94580e7b5a827631548b16f190438de3fbbe07b32568114fb8a3b8027fdabf9bd63627ed4ea0373482053
6700cf6b514fc62a9062dff205f0bc7fb5bdf764ac0229f6817b2220534f00cf6b514fc62a9062d511a03eefcd475b9b12e4
dbd36b4d5f3504a0c42089d50fe04dc58058ea06e0612deab1580292abc01ea94dd0b94da56f7f312a684f2286618ef382fa
b4aa931924b3e0beab19904cad003caf453d19ab408a7dcb084079ffbbcfbd27631548b1c2bb97d3689e028df16717cfa243
f6101acb2edf656996979c1924b3f0beab19904c4d003caf45958c55782a569ebe4eb501427e523ab929419cfde57618ed35
1827069f589ebdbeaa18104b1d003cac453d164b40727dabffbfe7fdabbdbd27631548b1fa6268a8abc5ffa546fdf13c32a7
b583696cf10c9b36935f0d9e383348667d31f4aa664032edfdef785dd493b10aa4d8b73aff7bdebf1adb3b32560115fbc042
6d20d3aeff9ed7a29e8c5520c57ebd7c19b0b4dfbf87b5a8c7620948ae0faca602326df6ef792deac95805520c78b7715993
29525e4cb11ce73ea7fd3fc47b6506c904de6ecd8064daecdff35ad493b10aa4d8b75afd7bdebf3ad97b32568114fb40b765
20d336ffcadba29e8c5520c588b7f61229cbfa4a8fa08556bb5a66904ce2edd50c48a63dfe3daf45958c55782af685eff749
a60dfe3daf453d19ab408adddbfbfb4ffcab7bbeafcde3a0317fa0f93c9089b180c3b5a0e7620d482fa21df7f158b73c1189
b393e28a48ab2f2d334826d1f66a06e2f2ae029eb5c51c14f293521ff85a1f4c6229e041afa08722faa74a5ff8421f4ce266
e041afa087427ad089dadac4812cf39a17e9efa987d6321d389569894112a9a94da76220247551f0a016f550880f427de14b
7d72a98582a7b5a807630d48af0f5ce5804cfd133caf453d19ab408a7dc130e249a6d60d9ed7a29e8c5520c53ed0830bc8d4
b741795bd493b10aa458e33dc7f18cbfaf34b6c52de9de96da73c066a9941924b3f1bea919904c4d1b3caf45958c55782ad6
dacab5d5cb79655d5db0916f97cafdd1705d6c313348666b2bf7a6664032756cf0bc16f564ac0229d6fc1ad242fc393fae8b
67cea93cdd29b382bc1ce77ac0cb9941329b5fc39b9a01c9d4aec1f1baa827631548b1d616323d3b9ab6d27f7d2d0f2bcb0f
ef7657cd9941325b5bc837350392a95783e7b5a823631550b1665bb01e69156869bf3a9506acb9dffe362de7382fd6d21633
836436db8237350392a95183e7b5a827631548b12f78163cc9d4a5c1f35ad493b10aa4d8170c0b9e646ad1e0792deac95805
52ece76e054f2c3567f0b016f5582c01c9d560e317318ee587167f697b7e4c977f75bae18e13834f6cb02f2a06c45267060f
6b518fc512905c0d364fddf34afd3841480d94a77cd53896719df289e5a2a5cc20990df74dcd80646acbe0792deac9580552
ec03cdc6814c3d193caf453d19ab408a357b827b6db544d04f6c515f9bc74163fec211f524533706cf6b512563159e8addbd
18dc27fee936e06bf33868ccc598e0595bcd10fc275ad4d5e671e098db77bae6d53f633e81d5c53ff9743dc79965debe757d
a7941924b37da76f6a0624531f06cf6b514fc62a90628db7f880e543ec288b254a63fd698b57a7785371cd5731334866e37d
533320999a30785e8b7a32568114fb4273fe27993a30785e8b7a325681146bbca957df9c5fb21e6b3c12e7733f8bebdb513a
175e5ddb31334866e37d53332099da2f785e8b7a325681146bf3f3218e6518ea29334ef3d673cf37c6791dfd146fffaff939
6506c96cf3f337350392a9f782e7b5a827631548b12fd8103cc9d478c1f35ad493b10aa4d807cc10814c5d173caf453d19ab
408a7de0791190a9f182f2b6a827631548317afe59972a96e522f59787cfb7afcc2099f4fcb357332099ba2e785e8b2a19ab
f054ec03ef2f804c2d176cb43eeac95805520c78d771de96d489348d2dfece86633b70d1c995187c22d0f62a06c452b3050f
6b51c3620958aedf77442432755af0bc16f564ac0229f673d38127967a2c78588b7a2c9680e46add8ab6731fe6dcd621d61b
c6e398f3769f655ad392abc5dec4616690ccd6afe84dcd8064eab1e0792deac9580552acf1b6cdfea9bf50ea7577a66651a9
4ffe704e535a83d4702931f8c446fba262402c7557f0b016f5582c01c9d556b6da16a6751cb725ef62fa977ad16ce31a877c
5eaf1d313348665bd9faa6664032b556f0bc16f564ac0229d678f3847e29e79164aa36e63e5931369feb16af21172e240649
6cb42f2a06c45253050f6b518fc512905c0d768af39c6329631bcfe4859c1b296d698ebdcc93b5de82c420890df645c58058
62e6e059afa087620148ac9f7faf0f283191f0a457d033213e0875b390d00f43af881af475710430d62ff8293cdd1fd43fc2
cdc35c74fa3ff10e673d72d42a70d42a7cc14fe149a6fe119ed7a29e8c5520c5ea8f26ce2b724fa37c298ae7cb692b5b2ee2
7df196fbe65fd336cc0c92597f36af6a062453ff08cf6b514fc62a90621fe8560764ea1fa1bc2deac9580552ac75239cd773
291b0f96755fc73881ce6f31e6791a72dbed8b17338364b66e846f6a062453ff08cf6b512563159e8ab58b8033018893ae65
5e96fcb0728ef38e759e16fb7e31334866bb0cbca919904cfd233caf453d19ab408a7d605b1290a97f84e7b5a827631548b1
df1b811199ba48785e8b7a32568114fbbdd71b91a99184e7b5a827631548b10ff40106323592f0bc16f564ac0229468f11d6
3dfece8ffc42345e53a679dfe2dd213e4768994132e94142af664032359270bc2eeac95805528cfa78b75bdfd4d5f65c8f34
fdc347802d31482275f1ee540c88a546121ed6a20e8b2540b9bed0b4fc49a646129ed7a29e8c5520c58acdc1b3b65a39f84f
b4a8afcde3a031fffea92391a99184e7b5a827631548b1bb91847ee27f9b25f8da3c0e1a3329fcbff59325323592f0bc1655
3256e1a9d8178ea827991a49785e8b7a32568114fbc2f7fb24532309cf6b514fc62a90625f38273fc9d448c2f35ad493b10a
a4d8079cb9804c8d243caf453d19ab408a7dc138e349a646128ed7453d19ab408afdba4f0360a9918487b5a8c3620950ae9f
ba070293ba4878528b7a26e627ad3e705f0f64ea22e1792deac9580552ec0bae194f327191f0b857d073b106a4d7079e423e
b9c445c2c35e418f850a8056f5817a3cdec6611bcb7b13d7143335f41997656c5de93b994132eb03f557350351899184a2d6
a0a7427e50aab6bc3acafb8b6c2d1b87712ecb947badc5f9579df45554ce0c92595b5ebdaa19884a8c243cea15542ae47f28
5550e7b66e2e8d643ee778b0cf7be9b9b66f3170b65ee59dcc209905f55dcd40546a24e1592deab95001d0aaf4a92b16087b
b66b4ceb8be2ad685ee23be639c1b8b719232706492c5dea5e550c08a52e121ed5a29e8a0520b16a13c2dc37eec8a791bc71
377e27f90de8380fe5894afd663b9941326b13c257350392a98b84e7b5a827631548b1da64b29a5ee7069867d97b98df6d8f
e3b526a53699c4cc2099b5c9e4ab9a01c9d445c2f1baa827631548b1c23ba787dbeb719456c4694cc3f5c4bb70fe57c6fff7
7fffae119040dd223c97451d01d3a232b5f3ed128fab65ab2db4872137591ef317705d206ae75bcc0c92593bdfbeaa19904c
dd223caf453d19ab408ad59ed5db79c6f1146787cdbdb4485d5ccb3f9597338364d69ed5af6a062453b708cf6b514fc62a90
625fe8c1fe2453c308cf6b514fc62a90623f6fb0ffc452b7080f6b518fc512905c5f70c77892a95b84f2b6a827631548b19f
7fb94f2cb58af0b016552c96e02957f9e50e53f13fcbfdb497786332aecb5a9a895ff7a0e597cb994132cb2ff75dcd8064ea
16e1792deac9580552acf0c65b90e9d8cfbceb6d58d3bde6b117cf9dbdf69869fdf539334866e17d57332099ba45785e8b7a
32568114fb965b84e7fdcb11c193b10aa4d8177c9a9e64ea16e1785dd493b10aa4d807661a40a69e119ed7a28e8c5540c5be
e519e179fff245f064ac022956ee0886bceb2b5e31f2e964aa6d47f22cb7d877b4c5fd9dcc2099e50ee85dcd8064ea19e179
2deac9580552acf0c6689ccb1fc752af8efb100fc57c2139d3668f783aade7e74e6690ccc2fbae664032f58cf0bc16f564ac
022956789731b53b19e772b756dbebe4bbd17842dfe27dcd5e7fbf9dcc209985f75dcd80646a1be1792deac9580552ec5bb6
119ef72f6b044fc62a9062d5f7f3acdbb0f370ca4dcc56fc1e4a3ba1d6d3b2931924b3fa7ebeaa19904c6d2394b7453d19ab
408a55db88a94cf8f2d8b6388f4f5e07f90f4bf33dafbc9c1924b3da46bcaa19904c6d233caf45958c55782a566d23c6312d
cdc8fe3bd3b8ee71325fbe8ada737ab96c2328314862b58d7853312096da4678588b7a2c9680e4fa966d84e7fdcb1ac193b1
0aa4d807f6570099da46385e17f564ac0229d6de7ca5ce48db5936fa25c3e461cd6b13a7735fb3d7d2f55e13338364b6375f
6f6a062453db08cf6b5147c62aa0623fde7f4558ea19e1612deab1580292eb038b77814c6d233caf453d19ab408a7d60ad18
90a96d84e7b5a827631548b16fd94678debfac113c19ab408afdbab53460a9678487b5a8c7620948ae0f745b0632f58c50de
16f564ac0229f6816ecb40a69e119ed7a24ac62a3c15fb8247c6934c3d233caf453d19ab408a7de1787e92a96385e7b5a827
631548b1e291f1ac2d9e11ee032de82bf32868c41fe8e6f2e4f29e119eb5c51c14f29352bfbe0a3c80c430c2535e414f44e8
4f89f2d1f3ac2a960dfea3aea02f8b9f0f232545c76d8ee78adc416fdda77198e3b780f3a62b33482669daab19884a5d1b3c
ab453d172a005a7dc07b0ab8d4b5c1d35ad483b106a4d7cf7f2e4f2cb56cf0b016f5582c01c9f5818b0c90a96583e7b5a827
631548b16f593628ef7fdb127832568114fb966583e7fdcb9640c95885a7621f68c107646ad9e0792deac9580552acb5581c
e30de091e718fbb1c509c6b2e61540fb509eb2cc57e368cc0c92d95a2cbea919904c2d1b1caf8b7a325681146b2d34d3e923
45d243fe65cc06ee79a069d3cebe2e57931cce0c92d95a68bea919904c2d1b3caf451d19ab808a7dcbb2c1f3fe654be0c958
0552acb5844d6fe5c6bc5cfc58d33a9874892c0762fccfc775bc5a1c636690ccd612f64dcd80646ad9e0792deac9580552ec
5b960d9ef72f67024fc62a90621f68610d64eadae0792deac9580552acf19665317938c7f5762afdf0ea6bc8eb7c85994132
1bef9b9a01c9d4b8c1f35ad493b10aa4d8075ab203991a37785e8b7a32568114fbc2f7fb2453e306cf6b514fc62a90623f77
217962a96784c2b6a8c7620948ae2f9cac9e64ea19e1792daa64acc253b1bb6784fbc43f5d197c6d1e078df9d70e3e80a576
151ed6a20e8b2540b9dab5befc9387b3c519fd319677bec7187fd4f959fb75ada7cc2099ed5affa6664032b5abf0bc16f564
ac0229f6053b832799da55785e8b7a32568114fb829dc1934ced2a3caf453d19ab408ab57b9332cd2c67af7823b14f635e70
b5afb5c5dc65a983994132dbbdc99b9a01c9d4aec2f35ad493b10aa458bbf79cd76d3ff232ee63484622e79c377beee926ea
1cceabf3066706c96cf79e6f6a062453bb0acf6b514fc62a9062edd942547fcd338fddfffb703def6a4f169e7941f2da7385
bfeb056452a30a4f6a51cfc4fca45533be1c9769ca3630fb56dbc9658fcabd7c1f578f3a4c0c92d86c2f5f540c88a52e151e
d6a21e8b2520b9da23c1a53ca3cc47d9657a13c736e7f9877bbe4d89c127b607822f2a06c4528b0a856d518fc512905c1f78
be0d646a51e1792daa64acc253b10fac700132b5a8b0d1faa82763154831e0cdad5bf733af6b5e8778cfb4ed33ad09b0c4e0
1381b6573120965a5478588b1a164bc072fdbed72291a94585e7b5a827631548b12fb8363cc9d4a5c2f35ad493b10aa4585b
6b79a6aecff9a7355ddbc4527b9bf8358ceb6c9dcc283148625b6bf9a262402cb5a8f0b016f5582c01c9d56097fc26347f11
fe2c3a97e5dcc378fd7431334866c37d53332099ba54785e8b7a325681146b0d8252e3f029bf658813f7719ae29522774569
efbcafe55a981924b335087a53332099ba54785e8b7a325681146bbc75ed673e97ece59d433a10e7a5d9945dbc941924b3f1
bea919904c8d2a3caf453d19ab408a35de7c3bb6d57367bc6c9497c0f3bc9d4bbc6a589b72c80b3eafb1fe5d2f2093b85478
d02be889989e94fac2f7fae012a30a0f7b053d162a005a35d4d4d5baecdb5df6a9f6a4cac3daca7eaa6b990f6606c96ca86f
6a06a2129b0c45ad414f85fca054ed725536ec8ec5e22675168d9aa4314edbf5f6bb76b9c2cc2099b5cbd5ab9a6a93d106ab
36196e2ee8a263c44e9d29ff9f54e0a855f8826dc4934c6d323caf453d19ab408a7da0611b90a94d86e7b5a827631548b1f6
d31dc778f2d8f3cf2c3d109ecf3d3f66af9d8dd203cbf6d3a5cc2099eda7fba6664032b5c950de16f564ac0229f6fb531591
a94d86e7b5a892b10a4fc57ebd2d09b0d423c3c35ad463b10424d7077a6902997a64785e8b7a325681146bf77df3945692e5
bbee79ab3766e97bc90d308773b80e66cc0c92d9eefbded40c48a61e199ed7a29e8c5520c57eda151698d41dc3935ad43331
3f69d548e34532bdf4cc77df655f7276245b6eaf8f3931f8c4c6faa262402cb5c6f0b016f5582c01c9f5ebcec680a5be180e
d6453d164b40727d60eb2490a93586e7b5a823631550b16264f0acade614fe132dea6bf33868ccf4f4ef7f68d84a58ea8be1
612deab1580292abc8fbacadce14fe132dea6bf33868cc77670afdc4ff765ff0b5791c34e6f689f7daea0de13fd1a25a9bc7
f11cf3177acb3fc9d41bc2f35ad493b10aa4d817bc039e64ea0de1792deac9580552ece7e6444f2c3586f0b016f5582c01c9
557e3ecfda6acde03fd1a2be368f83c6fc05eb8d27995a33385e17f564ac0229f681d30590a93b83e7b5a823631550b10fcc
61804c3d1a3caf453d19ab408afdfefd1a91a94783e7b5a827631548b10fdc470399783478dc2be8b95803d2eb038f849e5c
e2d1e061afa0c7420540ab0f3cdd7c52894783a2d6a0a7427e50aa3ea81fa6f88b5af39ad639deceaf73bafd8c7f883faf31
ade46aad773b994132eb83fa5735035189478347bd824a85fc0fa56a2badb48a28adf6ff973b218df1805bf32680e95a8352
3b696162f089b591d69b8a8190d4a0c1835ad443213e0855dd288621b5bbcf8f6ca6faaa2f2ff91ce3e4671dc66b3b392706
9f58bd28de540c08a5060d1ed5a29e8a0520b14afbbbd2dd2ae99046b30ce7b097569369a94df6a56fedef38334866697ff7
ae6640323568f0bc16f564ac02295678876d58e6f8377938f1b4316cd3548c334a13e4ab377827334866e17d573320991a34
385e17f564ac022956da57aec372c6796c7e0817e7777369369907bac4136a9cf0d4595427334866695ff9ae664032356ef0
bc167564ac022a56dbcfbad58f43eb5c97fba89eebb4c7019ecd9882338364d6f6b3af6a062453e306cf6b514fc62a9062a5
bdf05616e0e7d124e3f47118f272ed214eeeca77517039314862692efcaa62402c756df0b016f5582c01c95560db7ea4d2d7
3ab7329bc6d2f7b89e711a2d6706c92cb8ef6a062453d706cf6b514fc62a90623f6fecffc452d7060f6b518fc512905c05f6
a8eb10f28157b615ae7be9809c8c52867dab5f6e27334866c17d57332099ba36286f8b7a32568114fbf997fbc452d7060f6b
51c562099e727dcbb5c1f3fee54ce0c9580552ec0b67aa2799ba36785e8b7a32568114fb828bc1934c5d1b3caf453d19ab40
8ae5c79d505b7d13dc27baa8afcde3a03157df04674c184f2965af6a69983ed77359559833836456df8457350392a96f82e7
b5a823631550b1c21b8ff5738e93ccdce67f734b1287f40c2b3d426b47146706c92cbcef6a062453df04cf6b514fc62a9062
d52722fe7ccf79ce7d31d2d476ca13cd3cc91ce7791987f678b7931924b3fa44bcaa19904c7d133caf453d19ab408a55df04
b7b7309ebc962db7e4cb63bbd68354df04cc0c92597d135ed50c48a6be099ed7a29e8c5520c5fe874e7940a0fe089ecba29e
80694999c2951a626707923ca6b36e02ce37bf97c55de1e3cc2099e57b7c57332099fa23785e8b7a32568114fbc003042053
7f04e56d514fc62a906285778db3d765294b11e2d4fd88ff45694635960948bcf5afbc9c1924b3f0beab19904cfd113caf45
958c55782a569efe0d6937506aedf12f75f44f9da8d27bc034b6d4da644a77fcf5f11f6706c92ccfffded50c48a616099ed7
a29e8c5520c50aaf3d924b4f98e3f972480b6bd3d88e78f714fffbeb35356706c92cbcef6a0624538b04cf6b514fc62a9062
d522a1de0496dd1be3b18ceb791693e0d213f6ea0bc0994132ab45c2ab9a01c9d422c1f1baa827631548b10facad0732b548
f0bc167564ac022af6fb9d5d44a62e099ed7a29e8c5520c57eba5c1598d41fc1935ad433313f69f581051f40a6fe089ed7a2
9e8c5520c58077dea6bcb121bfd94e6b6cc7755c68b7bc6506c904de6ecd8064ea8fe0792deac9580552ec03bbe5814c2d12
3caf453d19ab408a7da06f3590a94582f2b6a827631548b10f747f0032b548f0bc16553256e1a918b5b4dff66dda8f7c6b1a
27bb7106388d239eafaecc2099d4d2be573320995a24785e8b7a325681142b9601cfda6a52e03fd1a2be368f83c67c372970
9ff8473f7e5f99474123fe7933f207947728f0a02de688109e64fac079e9c12426051ef40a7a28a27faa743329f01ff5dfdd
f87d59fc7c1869f9b0475db509f09f66515f19c700a3259b80715dc6b4be26e91dd3e6635a173cf75c994132c926a0573320
97da04785a8b7a30d680f4fac231fb2453a700cf6b514fc62a90625fb04b7992a95380e7b5a827631548b10fd82b01993a05
286f8b7a32568114fb755b30c0529b000f6b51c562099e72fdbae11b60a9478087b5a8c7620948ae6f790438de3ffbe07b32
568114fb409f4620538f00cf6b5147c62aa062ad0f677eb196bd9d8f78d4c7dfd698dfd2ef6bdac49eec2b1a2f6606c96c7d
38dfd40c48a61e019ed7a29e8c5520c5bee511e079ffea83efc9580552ec037d83814c3d023caf453d19ab408a35de6318b7
315f2a8ebdfa78e706b8e3751169bc941924b3f1bea919904c3d023caf453d19ab408a7da1cdf8934c3d023caf453d19ab40
8afdbc87fc134b0d023cac453d164b40727de1607e92a94180e7b5a827631548b1d243fe595bbbf4eb27b6a8afcde3a0317f
a0cd389069977ecf6b512563159e8ab5cbdf31eef39e57271c6d8b5d69aa3c6df3bc0dd335bdc1cc2099edf2f7a6664032f5
0870bc2eeac9580552acf1ced33aad7923d6b114f3adfce38e33ebb2c7e56a338e9941321bef9b9a01c9d426c0f35ad491b1
0aa8d817dac83fc9d426c0f35ad493b10aa4589bae1ff1466258ca09acb6fbcb77c8cb52ccd62e5ecc0c92d9a6eb6f6a0624
539b00cf6b514fc62a90628d37bd579d8a83cdb096bd77c53bf0e6a7c99941321bef9b9a01c9d426c0f35ad493b10aa4d86f
dbc83f99d420c0935ad433313f69d59e2ad846967d4f9d0fa7bcd6733bab97f475a2a2c42089eda9c28b8a01b1d41dc0c35a
d463b1042457735adcd6789accc662fb5a7c1f73abc72dde30cddb389f570b36cc0c92d99c16dfd40c48a61e019ed7a29e8c
5520c57efd7c0cb0d423c0c35ad463b1042417bd38fd1f7bad0399da04286f8b7a32568114a39717ff638f6320539b00cf6b
512563159e8a016f3c9d957d0f696c53bc6ccdf1b4837df3afcc2099c0dbad19904c6d026cb43eeac9580552ecc78d71084b
6d023cac450d8b2560b91aecb04519ca229b6d4d6f07cf7c0589539f3831195333aa464b9941321bee9b9a01c9d426c0f35a
d493b10aa4585b8f360d4b1c435e64335fdb0cd242c1fbc18c994132db9ab437350392a94d80e7b5a827631548b12f74ce7f
92a95380e7b5a827631548b1b6bed22ca6e32f6b1a72c7ce34b6e40a39c443f3f2ecc6cc20996d7de59b9a01c9d429c0f35a
d493b10aa458e39dd773d9f3c387658d6792b4273a771e89978dec4862bde429334866e37d533320993a05785e8b7a325681
14fb7daf662253a700cf6b514fc62a9062ad5b4d691f934f27cbb4afd35ed632cf63bc5d9bd6f37aa0c19941325bb79a3735
0392895f80c7bd829e8b3520bdd20341a82c1dfbfdc75d415f18c700a32d1ff6a82b3df3f5c36ad0d7c511c058ebf798dbaf
2ee5181b7377e178bff22fad49cfdd74e6e53a2f616690ccfa3dbeaaa93df3db60b567be9b18497498e34c25477d058eb60a
edb84d8fd2cb6ba278e73994a62bff722b9d35efa0ba6e81303348663b6edfd40c48a63df33daf8f1a19ab408afdfebc4464
da33dff3faa891b10aa4d82f1d4c8849bbe52ba9458d89f949ab0fecd80032ed96ef797dd493b10a4fc57e3f6324326d98ef
797dd4c8580552ec038dfe804c1be67b5e1f3532568114cb2fdda0b6b6acf79fe8a3569bc74163fefd3d179169e37acfeba3
46c62a90628d775987f3c84b7ed6e5da029d2ed2eb766cd36eab5b31334866e37d53332099f6aef7bc3e6a64ac0229568ea8
676d6d20ef3e51a2569bc741632eed7c9fb5b585bbff441fbd6af33870cc3fddc7064cda3ade93faa831313f6955b47dd6d6
feedfe137dd46af33868cc1f682d0b64da3ddef3faa891b10aa4d8af8d2d004b1bd72bac450d8b2520b91aec190771e6fb92
751bea6ad73cb478154c2b43aec7a9981924b3e1bea919904cdbe67b5e1ff564acc253b1f2f379d6d6c6f5fe137dd46af338
68ccbfb60e012ced5aef617dd4b0580292eb036dc6814c1bd77b5e1f3532568114fbc0ae6220d3b6f99ed7478d8c5520c580
77cc3fe5259fc49733fddbbaeef872edca0c9209bcdd9a01c9b46dbee395a891b10aa4d807be5f20d3b6f99ed7472f325601
15fb80cb069069db7ccfeba346c62a9062bf36e2012ced99ef617dd4b0580292eba7936460926ef91ed4058d88e949a9f6c8
e047dedb0025adf23da90b1a13e28350edd1d7f633fb5ea09256f98a7a058d0af941a90f3caf7e5249ab7c8fea829e0af91f
4ad51643657549f12758e665d8e6232f4798ea72c7ad1dbe9c1924b3b6187a5533109576cbf7ac3e6a5ca8006855fb471d63
bc55a856c2e964b2c7632ccdedc6b28e77bc4ec09c1924b3f68f7a5533209736ccf7b43e6a60ac01e955faa19df9b1f292db
64c649fc190fbf353f241be3d15696a317de4e6690ccd20fed5dcd8064da30dff3faa891b10aa458e19df761d9f35af0a1ec
ce3dd2b793ff3094d535f5d94127334866e17d5733209936cc77bc123532568114fb403f4320d386f99ed7472f32560115ab
7d29e335388e6ece6d50f7a33409cdbde9d2cef3295ea5ebf7dbc90c925978dfd50c48a60df33daf8f1a19ab408ad53eab6b
6a043ecef9bbd8dc639d643a9b564daff53d7127334866edb3faaa664032ed99ef797dd4c8580552acf60d8e97c77888952e
ca9b7bae3dc4ab67f9a7f2726690ccda37f855cd8064da33dff3faa891b10aa458ed443dcc7110f1b79d86b3eed370ecc521
6938ae695fed448d9941326b27ea5735039269db7ccfeba346c62a90625fe834fe24d3b6f9ca6b5123631548b1c23beeeb30
a485a5796ce3b6c78321ef758f279a785a8d77a3f5fadbc90c925978dfd50c48a69df33daf8f7a3256e1a958e18d1383e9d8
cfbc996458d382b8d411368fadb6b86a7da139334866e17d5733209976cef7bc3e6a64ac0229563aab3f6b6bef7aff893e6a
b5791c34e66ff5aef7bc7f7587f764ac022976ef5def3ef1cffeecbe368f83c6fc816b109069ef7acfeba31719ab808afdda
9407b0b471bd87f551c3620948ae2f34ea7f9269e37acfeba346c62a906285f78c5f7a6a2d5fce97d7fc3d4df62e3b85c2cb
9941320befbb9a01c9b471bde7f55123631548b1daa83f5e1a86a87bb1cd5963d259dcdc8633ed0188d78b7a01ea6406c9ac
8dfa5fd50c48a60ded3daf8f1a19ab408a7dc0310dc8b4a1bde7f55123631548b1c27bd4c58ec527a16c5c5d4af3f9ebad7e
e1e5cc209985f75dcd8064dad05e792d6a64ac0229561e6824478cdca3210fa7ccf5b6dc7c3eb7681ff6d696b3931924b33c
d0785733209936b4f7bc3eeac95885a76285370e267b23e5b16d473cadeee9779e3aefb7950c959733836416de7735039269
437bcfeba346c62a9062e581643cea86b4c1223f1c9d63ca702ef9e5dc987c14d2d7d41e34736690cc6ae0f9aa6640326d68
ef797dd4c8580552acf0960d6e436e96348f4b3ce697a90c74ba2e9cd59915338364566bd65735039269437bc72b51236315
48b1c2bb97b974de893c1f698dc8589f1457b3b76b430e6706c9ac9ebbaf6a0624d386f69ed7472f32560115fbfd062422d3
86f69ed7478d8c5520c57efdee0fb0b4a7bd87f551c36209482e787bfdbf36b90732ed69ef797dd4c85805528c569fd45349
7e5379e6f73acb897d5aafcc2099b4faa457332099f6b4f7bc3e6a64ac0229f681addd40a63ded3daf8f1a19ab408a7da0ef
3090694f7be5b5a891b10aa4d807fa0e0399f6b4f7bc3eeac95885a7621fe83b0c64dad3def3faa891b10aa4d8177af83fc9
b4a3bee7f55123631548b15f77b8072c69e6ef587dd0a0580012eb033fdd2797efe7ef592d7641213f29f581762a0f26e9e7
ef415dd0a088fea9d207ae3d4f267113f0a02e6850480f3a9155ceb69cf1662dcf00d6a55a3d13694b0c924846399d8a8190
d4c8c083faa841213e08f5852ff5c9a546069ed6470d8c3520bdbef0637d92a99181e7f55123631548b15f5f70004b5d0c3c
ac8f1a164b40727dc00109c8d4c540792d6a64ac022906bcf10a35e5be45e97b996adb4be4bd328364026fb76640323532f0
bc3eeac95885a762ad0b5f6a57b7afb941e09a373e94c70edb360cc3bec5bf6fbc981924b375e17b533320997a19785e1f35
325681146b2d16536b9e39bf768d17aaa96cca2ed3da314e6acfe96a9e899941325b93c537350392a99781e395a891b10aa4
d8b7bc0c3cef5ffdfa1d19ab808a7dcbcbc0f3fed5afdf93b10aa4d8079a1b03997a19785e1f3532568114fb42b3ee27997a
19785e1f35325681146bbcd5203e8d6dabbb5c4a97de653b877cfbd47829334866e37d533320997a19785e1f3532568114fb
8257c5934cbd0c3caf8f1a19ab408a7de1fb7d92a99d81e7f55123631548b12ffc7e9f646a67e0797dd4c8580552ec0bcdf6
9f646aa6a0bc1635325681146bd7df638ea7eedc97f798e3546f5ed370b2a9d53a9ec7703d6de5cc2099edfafba666403235
53f0bc3eeac95885a7628db7fcd48a354a9cffecd398d754a40534f3b22cc775fdc5cc20998df74dcd80646aa6e078256a64
ac0229f68566fb4f323553f0bc3e7a91b10aa8d807e6cf40a6660a9ed7478d8c5520c53e707f04646aa6e0797dd4c8580552
acddffb6c512e97472ed4b4c4f9dca63a6e17a5887994132dbfdef9b9a01c9d44cc1f3faa891b10aa458e31d8bf94f3d9dc4
93783ce0d2d8ca55733eaee71b981924b3f1bea919904c2d153caf8f1a19ab408afdbcebfe134b2d153cac8f1a164b4072d1
c3ab32cadc6f6e1be375335e448eeb610e6606c9a48757bd9a01c9d452c1f3faa891b10aa4d8071e4e0299ba2a785e1f3532
568114fbc0937620535705e5b5a891b10aa4182d45b98693bbfb6dd316a5220b5dcb0c92494b517a350392a9ab82e7f5514f
c62a3c15fbc05223205357051bad468d8c5520c57edc0885b0d455c1c3fa68c3620958ae5f36422126f553f0a43e6a4ccc4f
5ab5e59069e3f890f78eafe3b8a5eec86b59a7ba8deb39c669dfb51e92328364b605916f6a0624533f05cfeba346c62a9062
8dd7c97fecf51e2d8d2d0e6d9da7c5e654981924b3f1bea919904cfd143caf8f1a19ab408afdb2cf3331a9938227f5516362
7ed2ea0b9df69f64eaa4e0797dd4c8580552acf1c6195d3c63e74677d31e67f6e986348d6d98eab02f5eca0c92d978dfd40c
48a64e0a9ed7478d8c5520c57edfe187c8c449c1e3baa071b106a4d7cd49c17fdc7f5b16f8c23806186d93765f96793af37a
fdf4ff43bc6d1af3d7bec76bd4643d193131f8c426ec8b8a8190c4c44139afa021213cc8d4388fd2b12a8f6a3fe22c72c82f
cbe6b691ea5a2e8d9941321be99b9a77138732d8bb89c3350f73d1b4e3eb88d7b21cb6129db0d5f882adc193ed6ee360c416
15b48e10a4da077a9e01dbddc8c1882d2a681d2148b5df5b2611dbddd0c113b7a8a0758420d5da246e8a33e7236ff95ccef1
4c6ff6f3556adbea79e79ac3416290c436857b51f1eee760b87f7916dcb83a2a3c256bb8eb71ee7b3eab5c7b6cf29771c473
cc9ee664172f6506c96cc06f6ade1d1d8cf82fd70241eb0841aafdfeae84d8ee9e0e466c5141eb0841aa7dc1c5e2c976f794
30628b0a5a470852addd630fb94d755e3c395d7d2692034473586c37d994187c62bbcb7e51f16e2961b87fd9260857470592
ecd7fb0e01ecee2861b81615ae8e0a24d9178ee727dbddd1e222765141eb0841aa3d3d2dec33ff726ff0c53b43c1717fe03c
096c776f0b23b6a8a0758420d59eee16f6997f393848f1ce5068dce5339fd5ef0e13f6991695e29da1d0b83ff0fb01b6bbcd
84276e5141eb0841aa7dc06508d8ee4613466cd11b5a4788a76a4fab09fbccbf1c15a478672834eedffae03ca9ee4e13c66a
5181ea48407ab55716bff34626b6bbd784115b54d03a42906af49266dde3fcfa18eaa3da38ab4fded5f896a66506c9a4d734
bd9a77b70923fecb5141d03a42906a1ff05f00b6bbdfc445eca282d6118254cb670da87ef77cb0cfb4a82fde190a8ebb7ce6
b3fadd77c13ed3a252bc33141a377db3ff431b7702bb3b2f18ae4585aba30249f673a38927d8cd7ec168afa050753420c17e
eda0f5c4ba193018eb1514285600c4fac03ceec975b360f0b03528582c01a8f5cb978d4074b35f30cc2b780362f8874e1533
4e1ba763ce2f02e3dc279e38a76c4714ef9aceedc89d1e2a27250649aca06f2a0682babb2f18aa45858a1500b1aad9c4321f
715e934733c7ab7d9cee9c43f6a82e4b078feb2d236706c9ac6613af6adefd178cf72f8f0121ebc8409a55c789618f13bcb4
a1290d681d5347ee2377978f13ea7189279a3687e2cc2099d571e255cdbb038311ffe53220681d2148b5d2c2efc89e704b5e
f63ad6ce99f9f67f8cf74eeb301d6d29622733486669e1f7aee6dd83e122fed36740d03a42906a85b8ce5bf3ade0585b67e4
e7fd63d94c37aef5b94927334866217e57f3eec260c47f390d78b48e10a8da177c279e6c771f0623b6a8a0758420d5bee03c
f164bb3b3118b14505ad2304a9f681d6c1c076f76230628b0a5a470852ed03cda181edeec660c41615b48e10a4da077a9d03
dbdd8fc113b7a8a0758420d5bed03fffc976776430628bded03a423c55fb82c7c893edeec960c41615b48e10a4dad395c13e
f32fe70129de190a8dfb0bc7d593edeecb60c41615b48e10a4dad399e1facc3fdd07a478672834ee0f1ccbc076f76630628b
7ab48e10a8da078e2b60bb1b3418b14505ad2304a9f6816b12b0dd2d1a8cd8a282d6118254abf30e67483f6c6e11693c14f3
6ab8a51dd59c1924b3ce3b5ed5bc9b3418f15f460482d6118254fbc0dc12d8ee360d466c5141eb0841aa7dc198e2c976376a
30628b0a5a470852ad102feb11cfa5735e32134f2cf17cb3cee50e76a85dda1b316706c92cc4ef6adead1a3cf17fdb11085a
470852ed1be61477b6bb5983115bf486d611e2a95af8bddf26b0dded1a8cd8a282d61182542bc4e739accb34e453e9782ef1
56f23cf226d1b1ed9a6ccf7a3833486675547d55f36ed860c47f9912085a470852ad3ec1dcca3ff969e3e6de028de736c44b
78fce1b52798981924b33ec17c55f36ed97011ff694b20681d2148b5fa943acef9d6adac4398e314f038f63db7e4988ad1c4
76bd2ce4cc2099f529f5ab9a77d30623fecb98c0a3758440d57ebffb87d8eeb60d466c5141eb0841aab537699747d9525f8a
e4177c3116af9cdbb85f6f5d283148627b8ff6a2e2ddb8c170ff322710ae8e0a24597b45baacc379e45e23ebe226c04b3cc7
1c5bbca05d2f84313348667b49faa6e6ddbac188ffb22710b48e10a41abc018f978dfd88bfb2bc586b9fe34dfc7cccd4acd5
328364c23bf06ecdbb798311ff655020681d2148b50fec2107b6bb7d83115b54d03a42906a1fe812006c7703074fdca282d6
11825403e2743d39e3d9adfc00e3b791ded7539f0fcb0c9209c4dd9a770b0723fecba6e086d611e2a9daafbb8703d8ddc1c1
702d2a5c1d1548b20f34fa00b6bb8783115b54d03a42906a1f689a0e6c3723870bd88202d6918134fbf501fdc4521f07436d
318fc4fc241571e63bf563cf0baa96e38c27987dc46ee95766904c22edd50c4075737230d42b285428c053a90f9c949f5437
2f0743bd8242c5028056f9b1ffb3f2dd52c13ecfa2529a8701232e1ff8ac7db735b04fb4a8d4ee0c8446fdf3dfe813ecee6a
60b81615ae8e0a24d9072e3ac076b73630628b0a5a470852ed035649c0763737f0c42d2a681d2148b50f748803b6bbbd8111
5bf486d611e2a95a23decab6e434ba253f8a4856af695a5f56d7b8bb03ca0c92d988dfd4bc1b1c18f15f4dfc05ad2304a9d6
5a781637d0dc01798ff72bcb180fc334ba6319cba6ce7fad85276506c96c2d3cdfd4bc5b1c5cc47fb6f117b48e10a45a251e
e397bf95f609fb194f31cbb6e4c7e371be37a5bda4571b17ce0c9259895fd5bc9b1c18f15f8dfc3d5a470854adb5a54d8d08
9735b7441e9778821ba6bc9d649f73b78ce57a72c59941325b5bda3735ef360746fc572b7f41eb0841aa7dc1d8e1c976373a
30628b0a5a470852ed0badb49f6c77ab0323b6a8a0758420d55ab3f4613fa7f4e832194f146b9cfcce6bdf97b26cfffa1d63
6690ccd62cfd4dcdbbd98111ffd5d05fd03a42906a8d78ddc779cc9d7e8e35edb9488e4f6974f11233ac63fc4e2e62ca0c92
d988dfd4bcdb1d18f15f2dfd05ad2304a9f685a3fac976373c30628b0a5a470852ad11e7de3ed99cfb98af064e697453bc70
cecbb50d8b33836436e23735ef960746fc575b7f41eb0841aa7de15cfd64bb9b1e78e21615b48e10a45abb1e8fe5ad7c1add
90f722cc6b9910dfee9d38334866bb1ebfa979b73d30e2bf5afbdfd03a423c556b73ae71da9623ef118b2796631a9635afba
d987d2f668be8e6acc0c92d9e65c6f6ade8d0f2ee23f9bfb0b5a470852adcdabedddde1eff492f024a17fb214e80d31bfb6b
5a0d894112dbacfa45c5bbef81e1fed5dbdf73755440c95aa7f86d1de2992477ed5fe31d4ebc8ae436385bbc8accdb389fd7
e300cc0c92d93ac5bfa979773e30e2bfbafb0b5a470852ed035e78c076f73e30628b0a5a4708520d1e07c473ca1ecfaf7901
4a5a7a3b6de73a92d7bb6506c984c701dd9a77f70323feabc3bfa0758420d580782cbb32f2aeb92d96dfe32d3d3d8eb6c4e0
1381b757f16e7e60b87ff5f817ae8e0a24d9173ae43fd9ee0e08466c5141eb0841aac133bdffa93b2b80dd0d100cd7a2c2d5
518124fbc07b3360bb5b2018b14505ad2304a946ef59c6394e18e21d5d1addbc2fc3b11df894fa4a0c3e91deb2742ade1d10
3cee7f77f917ae8e0a2419e04ef19c36ef6563f33a1c4b5a504667684b0c3e11707b15ef060886fb5793ff1b574785a7641f
58b0036c770b84365a1f15b48e10a4daef3b0a11dbdd07c1882dead03a42b06a6d195afc639cfce5636edea6352d65caa3cb
1d61ff7fe6de255d521ed9b6ad5b2bfe16dc8f374e0b4efdb663f7bf7cf4441ac4b4b3286d9485cc0c0b0b738de93808906c
4ed77413cb4c43665d86f6a6e6d30fa111ffd5f31f688e104ab5efbbad2ab6a72342236e51a0394228d5be6f20acd89e9e08
8db84581e608a154abc4eb7e6d6772a1dcf6739f8f754df3de354c82d369b6b52a519986cc4afca6e6d31ba111ffd5ff1f68
8e104ab5ef5b9f2bb6a73b42236e51a0394228d546f00bf897ede98fd0885b14688e104ab54a9c1ae4a4c7e75b38c16c7106
9cdac8847bb93d4c869b3f9ecc346456e237359f0e098df8ff6d4400304706a559e15dc3cf2c489ebe8d395d52c22c3f8e2d
3e7d8bcf6beef396cc346416de57354d913d1c1a1aee1d04971641e835c297fb0fd7c3a6a1872d41606909845a2334f1172e
0b746ae8678b2d7c5ce1480da7c41cbe8b38e1bec8f7d644928e660d80bec3a473b490caa5b562aa3e1d13f0b17798f59dd1
c8b17fbd9944a1d131e1415cc244738490b27d3fad9574f44d00f41d7ed2395a08e506e8bea7e8689d00e83b4c3a470ba9dc
e79de6051aad13407c8789e6082165cb2711519ffe05f8d83bccface68e4d84738ba041d5d0c007d8749e76821951be13c22
e8e86500e83b4c3a470ba9dce7970b81462f839eb88589e60821651b6032a2e868a600e83b0c3a470bad5c9d765ee7762ce9
c5fdf60b13886539d2edddfad82aee641a32ebb4f34d4dd374f45300f41d269da38554eefba73f928e860e80bec3a473b490
ca3d2d1df0b17f9916b0be331a39f6117e55828eb60e0fe812269da385542e6b2dead35b011f7b879ff59dd188b17f7dce56
68747600f11d269a2384946d80f988a2a3c503a0ef30e91c2da47203ec935774747900f41d269da38554ae425fdb1a4698de
a01c53f8174b5e52b8c52f242ef5bb0f6e9969c8acd06f6a9aa6a3d103a0ef30e91c2da47223985b083a7a3df4d02d4c3a47
0ba9dc00fe168a8e661380bec3a073b4d0ca3ded26f0b17f792ab0be331a39f601da31283aba4e00fa0e93ced1422af7b9cd
864083f104806b94608e0c52b401ce22820df61320ae51a2691d9466f552715ce1262f35dbdff7aec7e056debbb7556532d3
90592f156f6a9a248303c583374749a655508a0d608927c8604501de1a7d926915fe55ecfb57d98a8c861400bec364d33a28
cd32f1791ee7146eb1d280e282f77008a653ea72de3b2d32b1ce346466e277354db3d19302c87798708e1252b70cbd871f56
3881e617565372cfd8d2a47c997ff91c53cfd33ad39099a1dfd5344d475b0a40df61d2395a48e532f43a4d7bfa62e29896ee
6565f8c5edfb349ff57aac13ad4fccc8af2a9a46a32d454fdcc244738490b2e5668567b63f498685f3b99ce19c33a537b2f3
766ea1c2dd82d3c93464e66685ef6a9aa6a33305a0ef30e81c2db472197abee23bd1253d200a67d7337c2feb91bc42ca3292
bab5dfc9346466e877354dd3d19c02d07798748e1652b9110c39041dfd29007d8749e76821952b1d66e7396d785cd200cb96
a9d429340ced98cee92ccf1a9d4c4366e930fbaaa6693a5a5400fa0e93ced1422a374023614547970a40df61d2395a48e506
e817ade86854f1802e61d2395a48e532f4efb6af8a2dbcbb4963f831c61963f47ccad03ad39099a1dfd5344d47af0a40dfe1
279da385502e431feda9e814e789f154933af3ff6e07f6cc2c130d8919f95545d368f4aa00f11d269a238494ede956818ffd
cb8f81f59dd1c8b19736fbf7411e75496687d396cf7ae99730d5ee284ea621b3b4d97f55d3341d1d2b007d8749e76821951b
c03f42d1d132a3876e61d2395a48e532f4ba9ce9e560fa56f213a365cebff9b8de34dc16d5df94ce346466e877354dd3d135
03d07718748e165ab90c1d4a2d219066a6d31e5f14feceec717196bd6ae532e9641a3233f4bb9aa6e9689c01e83b4c3a470b
a9dc08dfb4a0a37706a0ef30e91c2da4722338a4083ada6700fa0e93ced1422a572e93f371fcf6e2ee91cc3ef33c26ed4a3c
e20bbd7295147986bc72917c51cf3417bd33807b87c9e5a820351bc0994dd1d13e03d07798748e1652b93cd39f7e5b6ea996
ce30a57d7c9ac684b3cd394f5b7df6e8641a32f34cff5d4dd37474d078409730e91c2da47263b886fc4347130d40dfe1279d
a385502e435f4b9e37a49fdc157b9a86a9627ac032272b8879a9dfb4ce346466e877354dd3d14703d07798748e1652b9fc9c
660f27d975df53b79e799fe770a3bda479d31cbea93376b3298f6974a2f589f929cdab8aa6d1e8a301e23b4c344708295b7e
00391f67f23689435aca7386d4c7678ebfc2308bac572a27d390991f41beab699a8e561a3d740b93ced1422a571e35e78782
7980b1e499da40c7d1d6f6bde549b34cb43eb13c687e53d1341aad34407c8781e608a1652b6f14d6659ba3cbd87f71314f77
4b140fc770be99cffa925d671a32cb1b8557354dd3d14d03d07798748e1652b9baa1a8748afc2f6e350e73c6254c8e93bb7a
f919cef7db6e9969c8ac1b8aded4344d474f0d40df61d2395a48e5eadbc12d3585486f2efb7e186b3811cdc76f6a4d5464a6
21b3be1d7c53d3341d6d35007d8749e76821951b608994a2a3b306a0ef30e91c2da47203bce7577434d700f41d269da38554
4ead60496bb6b6f4967ebbe2ffdbf75336dcb8330d996a058b57d3341dfd351ed0254c3a470ba99c808e8de67ed394ba5685
d3ec1212e35a6f017d671a3205b45bd3341d2d36007d879f748e16423965a4322ff9f6370e304c68b6f50af7840aface3464
2a2315afa6693a1a6d00fa0e93ced1422af7753b198546a30d10df61a2394248d9d2a366551f6e17fda7de515677c622473e
80498f60eb3d2f005c8200d32a48c5d4e9e358927d63baac2d5398bf851fbc6abbd5320d99eaf4e1d534c505db0bd0d628c1
a40642ad6ccdf06f65784fe0f36a94a5f528d488bfee8924b0e87c01da3b4c302d82126c84a349b0d17a03c87798708e1252
b7012e468a8e061c80bec3a473b490ca8de02123e8e8c101e83b4c3a470ba9dc002dec141d6d381ed0254c3a470ba9dc008d
19151d9d38007d879f748e164239e53e52168de5536ee9cc2ea1ef4c43a6721ff16a9aa6a31907a0ef30e91c2da472153adc
d35c7b5a061a1f42df9b188fad74826927329569c8acd06f6a9aa6a31f470fddc2a473b490cad556baf97559eae41cce397b
3812d3688f33cfd25a67199969c8acad74dfd4344d474b0e40df61d0395a68e54668892de8e8ca01e83b4c3a470ba9dc089d
cf051d8d39007d8749e76821951bc17e45d0d19b03d07798748e1652b934cf57f56990818fbdc3acef8c468e3d7faca84f97
0a7cec1d667d673472ecf963457d5a45e063ef30eb3ba39163afe790f2e8371dfa61fa10ee3dd2c383708e6337509d69c8ac
e79037354dd3d12d02d07798748e1652b90abd9e614a96ae66fdff9fee6b7b45fe37cf905781ffae679a8b56110fdc122697
a382d4ac1a09ccdbb2ec6bba8efdf29ed8d4f6fffc4dd73a6df7465b996848ac46022f2a9a46a3550488eff013cd1142c856
a703f3fe9be6342f3bf7694beb4fd3a53bad265f3a67799969c8acd38137354dd3d12da2876e61d2395a48e5eac4afadb238
d7f0f34b3b23d2fdf6fe0bd398ae13b94ab43eb14efb5e54348d46c30810df61a0394268d90670175474f48c00f41d269da3
85544e390acce158d896fca54cc719eebf7ebb7ab2d9320d99ca53c0ab699a8eb61180bec3a473b490ca29e7cc70731d4694
e6a87bf852c2596799e416df3bd390a99c33bd9aa6e9e81c01e83b4c3a470ba99c7a50f1bf6c90aae8e81f01e83b4c3a470b
a99c80fe5f35c15568f48f00f11d269a2384946d8087708a8e2e1280bec3a473b490ca0df07e46d1d14802d07798748e1652
b9011eac2b3ada493ca04b98748e1652b9013a52283a9a4a00fa0e3fe91c2d847203acdc5074f495e8c6db8549e7682195ab
6b747e4bda7e934eaefbb54dd3929a4b6de19620361dbf27a032d1fac4ba42e74545d368f49500f11deed11c211cd92a71c8
0b77b44b3ec964cbc0741c4ecf974632d3905999dfd4344d476b0940df61d2395a48e5be5f7f26e9e82e01e83b4c3a470ba9
dcf76d15251d0d26007d8749e7682195fbbe2f87a4a3c704a0ef30e91c2da47263b94c00fa2f1f05d2395a48e52a74b8b58d
bef27180cb79fdae30aef4ad4c4b590f7b43ab4c4366857e53d3341d9d26007d8749e76821951bc18e40d0c16c02cc354a36
4709a95b7ccaacaac3f1019f59a32cae47a2463d8285c9bf64307d78f0e628c9b40a4ab1efaf8ed2b101b60f98cb75e15f90
2e76df4ae156c40977454668a82de860fb40e816069da385546e802e5d8a0ea613846e61d0395a48e5ea8c770adfc3954c74
f7e5ee651067a7b54b5e9df0aa44eb13eb7cf74545d368309d7812d730d01c21a46cb931bfa80fdb077e6c0b3fea3ba31163
cf1f2beac378811fdbc2a8ef8c468e7d1ce30512ffe5694034470829db001d87151d6c1f08ddc2a073b490ca8d70c21674b0
7d20740b83ced1422af775376d8506cf0712b730d01c21a46c03f4c55774b07d007417069da385546e00e32d4507db0742b7
704fe768a195abd0f31cc6923ddfca8a8bfce38bcf83e37a9efb5e46661a322bf49b9aa6e960fb40e816069da385546e80ed
9f8a0eb60f846e61d0395a48e5be7f3e21e9603a41e816069da385546e846f5ad0c174e2095dc3a073b490cad5dff412b737
a51e4fdb1e0614cebb7b7a3e989b66eced0643661a32eb6ffa4d4dd374b0bc20740b3fe81c2d8472157afb2d71fa90ceb4a5
bd691cdf7a5c5b985cb4278e22cffabc0afc773dd35cf0bb206e0b83cb51416a36c0155ad1c1ef82d02d0c3a470ba9dc00cf
48141dfc2e08ddc2a073b490cad53be7ef5cc8251dfc2e08ddc2a073b490ca8d60f221e8e07701e82e0c3a470ba9dc002f7f
151dfc2e08ddc23d9da385566e04930f4107b70d42b730e81c2da4726a63e83c850b797eb3b54f4becd3b24d7215cb9d69c8
545b43bd9aa6e9e0b641e816069da385542e3b9b88fabde5053ff58ea2ba331639f201569108b6def282c47714685a07a599
3a874c71d3559ec2ecc716cef2d3a57e4d77a221519d419c8a26b17abf8b276c89024b4ba0e4aa5789fddcf29abcad77825b
a6238db3ad1bf937cf9057af107fd733c9d43b5d90f48e3e9834ffbf5ad5994fde45959f18cd5798b4a48e52eb1117076eb1
415c9df8884443629df7bca868120b3617a46d61806911946003dcab2a36d85c10b98501e72821752b9db8b633cc7be7340f
0be38bafccd6294ebf97f001d31efe7f3d9e75a621b374e27a55d3341d6c2e08ddc2a073b490ca8de0ed21e8e07401e82e0c
3a470ba9dcd7de1e0a0d3617246ee11ecd1142cb56bc3dd21732a75bcb7053b55eb11160fac356d7e7fd57bc3d64a621b378
7bbcaa699a0e3617846e61d0395a48e5327438a9e4168f714c7933ff9c1ea0cc6b7d5f59a075a6213343bfab699a0e361784
6e61d0395a48e53274dce9148798da4e873968ec20b1e44e9ef17cbbdc1b039d4c4366867e57d3341d6c2e08ddc2a073b490
cae596a84777018d5d4cc39df59196e7c5b77647b8d1ac1b8a9c4c43666e89faaea6693ad85c10ba8541e76821951bc1d044
d0c1e6e2095dc3a073b490ca8d606822e8607341e8167ed0395a08e54a6fe770f31e2612f9f77786ebcb35e71579d3951b6e
d7a7ab4ea621b3f4767e55d3341d9c2e08ddc2a073b490ca1543937ddab6655eb3f948b7a03c3e1d8b7bafeaab6e27d39059
0c4d5ed5344d079f0d42b730e81c2da472a575796768381ddd72b970064a6ff1b66a39a1330d99a575f9ab9aa6e9e0b341e8
16069da385546e04731141079f0d407761d0395a48e5327498354c53dc9111c7b41797b06c0ff1dbd76b5a6a131027d39099
a1dfd5344d079f0d42b7704fe768a195cbd05338cfeccb9566c6539e3d6e7b3e03dd8b6d32b4ce346466e877354dd3c16783
d02d0c3a470ba95c820e33e1f0c56c7bda971ba7aaeb117e907bb9d6c46d8ce58ae5245a9f9890df55348d06930d12b730d0
1c21a46ce949a3aa0f9b0b7e6c0ba3be331a39f67274e54ef5e9b5e8944f73bf2b1f0967dc341a574696a34b661a32cbd1f5
aaa6693a985d10ba8541e76821951bc1ba46d0c1ec82d02d0c3a470ba9dcb78626820b36174fdc1a0697a382d46c00433b45
079b0b42b7f083ced142283782b787a083cd05a15b18748e1652b962471add197e69ebcd1cefbf96e5d8b28de895fbf8579b
0b9968482c66a46f2a9a4683cd05895b18688e1052b64c7c5463a8f414a96c784f4fceb6b21ab33e99d29986ccccfcaea669
3ad85c00ba0b83ced1422a971f402ec93b714a0fff97a59b9d36838efc0052275a9f981f40beaa681a0d3617246ee11ecd11
42cb36c0736645079b0b42b730e81c2da47223787b083ad85c10ba8541e7682195abef8ef2ebabbcdd3bfcefb4cde90f6bf8
492e61b6de56fca944eb13eb9ba317154da3c1e382c42d0c34470829dbf77b8c251d3c2e08ddc2a073b490ca0db02c4ad1c1
e382d02d0c3a470ba95c7dbb5f9ffcc6254c7b6c3b7ee55d04dba3eb9ece3464d6f7fb6f6a9aa683c7c513ba8641e7682195
fb7a49834283c105895bf881e608216413c4d7fe3b7fe1984c0b4ccef5372db1cfa642be330d9982d9ad699a0e0617846e61
d0395a48e546f06f1174f0b820740b83ced1422af7f0b8c0c7de8e0df8d4164575672c72e4d908e1dfea9dcb043fb206fbd2
7a1c72cc6a895fdd9c198fbc701c6ee7b9cbd65c35d190a816f839154d41f5161344bda3a0920208a906f8e10aaedee082b4
7714605a03a597e28d8fb6c31fd2f2f1b870fabaae45adae6b99864cc5ebd53449068f0b02b730d8b40e4ab30196ab2a3678
5c10b98501e72821751b6011a5a283c705a15b18748e1652b9cfcd5b041a0c2e48dcc240738490b20dd05b4fd1c1e0e2095d
c3a073b490ca8de1eaf10f1d0c2e08ddc20f3a470ba1dc08ae1e820e0617846e61d0395a48e54670f5107430b8007417069d
a385546e00570f4507830b42b7704fe768a195ab6dbcc354745e9369e1b9e721a53e21c7714ceb31c7c661b58db7ca3464d6
36de6f6a9aa683c105a15b18748e1652b9daad3dee8c9cd33cf83cc3d9759b8f330df0b7cdb981c67fb55dbbca3464d67eed
6f6a9aa683c105a15b18748e1652b911ac4c041d0c2e08ddc2a073b490ca3d0c2ef8b17f7947b0be331a39f687c1053ff62f
ef08d6774623c75e8febd62ce30cff89d386e41910cd04ae6bbb6f3164a221b11ed52f2a9a4683b706895b18688e1052b6cf
7d17041a8c3548dcc240738490b28de0bb20e860aff184ae61d0395a48e5c47c648d5f49b892a7012ed164e7da677587d132
0d99623ee2d6344d07870d42b7f083ced14228a7669eb903ca9c6f7ce7f2eb9333cf9a69c854334fafa6693a386c00ba0b83
ced1422a37820b81a083c906a15bb8a773b4d0ca09e830555ae33928dd0f9ccb3cadd37aca870677a6215340bb354dd3c164
83d02d0c3a470ba9dc0077938a0e261b846e61d0395a48e5463022107430d920740b83ced1422a37c0c3124507930d42b730
e81c2da472231811083af86c10ba8541e768219553d0f3fc8bde64a95dc81cdb13c7dda312ba661a3215b457d3341d7c3608
ddc2a073b490ca0dd0a140d1c16783d02d0c3a470ba9dc00af20151d7c369ed0350c3a470ba9dcd76fe7141a4c3648dcc20f
34470821dbd70b0a141a1c36fac1f661a0394248d9c671d820f15ff6113d9a238423dbf7cb83241d1c3608ddc2a073b490ca
7ddf595ad2c16183d02d0c3a470ba9dcc7ad31241aec3548dcc240738490b2d5f57eb13fef9166865bbac38f16097180db11
5f045cc77d6ccb4c43665deff7a6a6693ad86b10ba8541e7682195fbbea9a1a483bd06a15b18748e1652b9effba0483ad86b
10ba8541e7682195cb5e13a27e6f70c14fbda3a8ee8c458efce355c30aacf7d620ee1d059716410936c261f52f59efadf1e4
2d5190691594629f9b0f084f0c1a6bf453b82e7c8693c2394d39dc8a38e1aec8e7edf8051a5d3540dcc240738490b27d6fa9
26e9e8aa01e816069da38554ae426f61da199f85c50763f37c8609ca946e2ee7fd085fd6d91e37ca4c4366857e53d3341d8d
351ed0350c3a470ba9dca75ddf14170d3d80dbc20f2e4705a1d9d3d0031ffba76105ea3ba391631fe1f724e8e8e901e81606
9da385546e2c4f0f40ff6958013a470ba9dc589e1e80fed3b302748e1652b9af1b2a2a347a7a80b88581e6082165fbfeae51
d2d1d3a387eec2a073b490ca8d606f21e8e8e901e816eee91c2db472df9b324a3a7a7a00ba8541e76821952bd0eb9e1a62a4
6f655eca42c4f087e517fe2aaef8b9cf6232d39059a05fd5344d474f0f40b730e81c2da47205ba748b48e79c299c59a725bd
5b587e53b115aacc2ad1905890df54348d46430f10b730d01c21a46c6936a4ead352e3f1b1358cface68e4d88bd0e10b08e7
f425cd83a733087364cf8f25fc30d256d9fbe429330d9945ea57354dd3d15203d02dfca073b410cad5f3c8773ea7928ec61a
806e61d0395a48e5be7f3622e968ac01e816069da38554ee5bf314c1454b0de0b630b81c15a4665f1bba2934fa6980b88581
e60821651be06e4ed1d14fa387eec2a073b490ca0d6022a2e8e8a701e816eee91c2db472036cc95474f4d300740b83ced142
2a37c0ae4c45473f0d40b730e81c2da47203ecca547470f300f31d059ba384d4edeb3d99020c561ec0bda3e0d22228c10658
8d24c8e0e6f1e02d519069159462e2cc75ee61421c8ffc38b6fb09923a73dd99864c71e6726b9a2483a70778efe8834cabf0
af62035c9305196d3d00dcc260d33a28cdeabcab7bbcfe3ba72ddc40a726a2eb318733cdb2b5071332d39059675e6f6a9a66
a3ad07905b18708e1252b7019ec6283ada7a00ba8541e76821951ba0f398a2a3ad470fdd8541e76821952b3de662d2bca765
eaebf43bc3d4704dadbe9765be621bfd7bafadce3464961e73af6a9aa6a3b307a05bb8a773b4d0ca65e870f31ae6c05b5a36
13fe679ee223e1784599e32d50f845feea554a671a3233f4bb9aa6e9e8ec01e816069da385542e434febf65b97330d70beee
ded6537c53bb25ab86fa9e48671a3233f4bb9aa6e9e8ec01e816069da38554ae74ca5cc349e7484bf3e7fd8acda673ebc3f9
618fad130d89a54fe69b8aa6d168eb01e216069a2384942d13af6142183b39c6dfdb9cbb3b6ed979256ef48a5b0acbd7ec64
1a3233f3bb9aa6e968eb01e816069da38554ae18b81ce11a1aceeaa95f69b892a407c2e92bca6d98d77a5be1641a328b81cb
ab9aa6e968ebf180ae61d0395a48e572b3df331cf369a1eb7fb131ef75e53da26980776be2dcec57671a3273b3df77354dd3
d1d603d02dfca073b410ca65e83d2f714c0faac24d7db8cb3caff21595054d155a671a3233f4bb9aa6e968eb01e816069da3
85542e432fe5129adaf59ff54bfa2f190b2d5318ec51ae574ea6213343bfab699a8eb61e806e61d0395a48e50668e0ade868
eb01e816069da38554eed326ed8a8b861e3d6e170697a382d46c8446fc828e861e806ee19eced1422bf7b9b78540a39b0788
5b18688e1052b64c7ca68597db2f0d29dcd55fe1eef7ca279ee558d7f01b2c77564ea6213333bfab699a8e861e806e61d039
5a48e532f4916ffcd290d6df190e87dccc623aef0d299959261a1233f2ab8aa6d1e82502e216069a2384942d1387019ebfe5
4ccd7652e78a78415bf2a9f638e2cfb23c3570320d9999f95d4dd374f41201740b83ced1422a37c2b12de8e82502e816069d
a385546e044f224147479107740d83ced1422a572cb78e302b0c772069be70748b8ca7df56fe53a075a621b3586ebdaa699a
8e8e22806ee1079da38550aeccc2ae292fa2cfe79972979b664ceb1a8ec3f854a9ccc264a621b3ccc25ed5344d47471140b7
30e81c2da472e526234c1ae6e9f72bdf43f8db233fe298ee6574e51e43251a12cb2dc69b8aa6d1e82802e216069a238494ad
78a86cd7345f477a8b34cf4bf846d629b5ce9f97b282be9cc59c4c436671517955d3341d1d457ae82e0c3a470ba99c7d6ff0
abe8682a02e816eee91c2db472f931e07afec2ec7f4d8f6f96351cffd3b5a557c4737ca2b1ced3bdaf5a671a328b93f3ab9a
a6e9682a02e816069da38554ae3ce50e3ff7251ad2c6015e617ebc45af88f887f0a3dce62b7c4ff529b7cc346496a7dcaf6a
9aa6a3a908a05b18748e1652b90c7d84995298b6ece9317cb87c9e6927741c6079f370bf79d69986cc0cfdaea6693afa8a00
ba8541e76821951b6049ada2a3af08a05b18748e1652b94f97c2292e3a8a00b785c1e5a82035abaf64b7f0575bb2e1ddfbae
bdeb19ee75e3f29efb25b4cc3464d657b26f6a9aa6a3a3c803ba8641e7682195fb7a77a442a3a308885bf881e6082164fbbe
1d86a4a3a308a05b18748e1652b901d6ff293a3a8a00ba8541e76821957b3a8af41ffb976707aa3b639123ffb8bf8c02ebcd
4c405b833d959640ca95cd53fea90c47117cde1d4569390a31deaaec719dd39ad668851b9fe6a4bd95df5bbb20c94c4366d5
f64d4d935c701401ed1d0598d640e995e5fdb7363d3df0912d8cea7a246ad49f1fbd028c861ee06d6190393248d106586aa7
e868e801e816069da38554eeebb65f0a8d861e206e61a0394248d9065816ade868e8f180ae61d0395a48e50668d7a8e868e8
01e8167ed0395a08e50668c1aae868e801e816069da385546e800ec38a8e861e3d7417069da385546e8406da828e861e806e
e19eced1422b3780758ba2a3a107a05b18748e1652b911ac5b041d0d3d00ddc2a073b490ca8de00420e868e801e816069da3
8554ae42c736bc677a1218fe3b6ec3c8cf128e639ac2bd74f8fb1b5a651a322bf49b9aa6e9682702e816069da385546e0493
1e4147331340b730e81c2da472f5377de4ce28e9f79796f0c7b718f1f7375fcbb2ae5dd7709569c8acbfe937354dd3d1cf04
d02d0c3a470ba9dc084617828e9626806e61d0395a48e5047498315efbb6a4454547b81bbac28466570db45ba6215340bb35
4dd3d1d2e4015dc3a073b490ca29a38be33cc2fc2dad610fc767b80d0cb30a79c371671a3295d18557d3341d2d4d00ddc20f
3a470ba19cf27c083fc53da4a6011e770b7fe5f950330d99caf3c1ab699a8e96263d7417069da385546e801b0e45474b1340
b7704fe768a195539e0f73b8f7ded6b4e4755f9e3bf664a62153793e78354dd3d1d204d02d0c3a470ba9dc089e0f828e9626
806e61d0395a48e5bede68acd0e86702e216069a2384942d3d7855f5e928828f6d61d4774623c73ec026764547471140b730
e81c2da47203ecec5674741401740b83ced1422a970f30519f9e1ef8d816467d673472ec033c035574f4f47840d730e81c2d
a47203797a80f84fcf8a079a238490edeb87fb0a8d9e1edd60fb30d01c21a46c95383eafcadbe8f6638fefd3aedc80689bc2
b56d8e8dcc2ab2ca346456e637354dd3d1d603d02ddcd1395a38ca7dff3a54d2d1d603d02d0c3a470ba9dcc7edcb251a3d3d
40dcc240738490b28d606422e868eb01e816069da38554ee73231381464f0f10b730d01c21a46c75ad5dec72185bfdc44be7
322fcb72a455cd6bbdcd6d9e1e2ad39059d7dbbda9699a8e9e1e806e61d0395a48e52a749991c601e6b68053daf5be3ef6d1
3b9986cc0afda6a6693a7a7a00ba8541e7682195fb7255a5e482970868ef28a81c0da4621576ddc328d20ee06d0f53a370f7
919e4daee1f84bb77e0d57651a322bf09b9a26d9602702e23b0a34ad83d26c8423fa5f32d8893c784b14645a05a5d808ae0b
c20c848e22fd04b20b1f41c735ae55fc1f1471c25d916c4320ead3d6031fdbc2a8ef8c468efd5bb307c145430fe0b630b81c
15a4665f2f45536874f37810d730d01c21a46cf9b812f5e9aa818f6de1477d673462ecdf1e57828b6e1ec06d6170392a48cd
463865093aba7900ba8541e7682195fbdcf649a0d1ca03c42d0c34470829dbf7f74d928e561e806e61d0395a48e506b2f200
f19f3e1540738490b27dba474971d1c4a3c7edc2e07254909a8dd0925dd0d1c403d02ddcd3395a68e546701c107434f10074
0b83ced1422a57da461ef31542733ebd445b8230cf4f5b228fb8b627fc653db475a621b3b48d7c55d3341d4d3c00ddc2a073
b490ca15e8a9ec7d4c67987053b9c61f7e1ce0becff11d526d62e3641a320bf4ab9aa6e9e8e301e816069da385546e800db5
8a8e2e220fe81a069da38554ae9a97fc539f3e1ef8d8167ed4774623c63e40335645471f0f40b730e81c2da472239c4a041d
7d3c00ddc2a073b490ca8de0d822e8e8e601e816069da38554eefbc76d928e861e806e61d0395a48e53eb730116874f3e889
bb30d01c21a46c5fdfcd29345a7980b8857b3447082ddbf72f3b251dad3c00ddc2a073b490caa917f9fb794d711b7fbe3159
d633ae3d906ff26ba62153bdcaf76a9aa6a39507a05b18748e1652b90116a9283a587980f98e82cd5142ea36c01a28c10637
0f10df51a0691d9466032c74136470f378f09628c8b40a4ab111bedf7fc9e0e601de3bfa20d32afcabd8005baa0519dd3c00
dcc260d33a28cd0678a6acd8e8e601e416069ca384d46d040b134147370f40b730e81c2da472f5b6227b8aa453cbb684db81
333f375ae733deef5df7130a9d69c8acb7156f6a9aa6a39b470fdd8541e76821952b3790db19eeb2e6d4d1663de38ab4734d
dde796a0d3b487ff5fdf11e94c4366b9817c55d3341ddd3c00ddc23d9da38556ae3c9609773653741689035ce2b461498bd4
c3ec25f6fb9997fafe40261a12cb439937154da3d1ca03c42d0c344708295bf12f39f27f52b7b9a33be1c6768b67b8f99deb
46169d69c82cfe25af6a9aa6a39507a05b18748e1652b9d2b8709ae2719fde572e4b77833f87bbec7d9aef8df33ad1fac4d2
b6f04d45d368b4f200710b03cd1142ca96fb538693cdbec5c7dfe95bd8cad117fff0586ae6641a32737fca77354dd3d1ca03
d02d0c3a470ba95c716cf94dfb746ea993e25cf753a5016ee5aebe36e5d49986cce2d9f2aaa6693a5a793ca06b18748e1652
b90cbdcddb79ac7bf654597fcbbeedf16164fc43bdc72fd03ad39099a1dfd5344d472b0f40b7f083ced14228577ace4edb99
76eac536b8bf795ec2b0d21f42f1f470b69cc474a221b1749c7d53d1341a7d3c40dcc240738490b2e5bec247f75e36760edd
a7df91167dc67d8265117b6e2cac330d99b9b3f0bb9aa6e9e8e301e816069da385546e04f31241471f0f40b730e81c2da472
b96df66f9fc204714e8e49e1b6a72d678a1d06e37eb6bdf60ad79986ccdc36fb5d4dd37474f3e8a1bb30e81c2da472233448
177474f300740bf7748e165ab9115ae10b3a1a7a00ba8541e7682195fbdcdb42a0d1cd03c42d0c344708295b26ceef0aa7fc
952c676cb3744cd956673fb3ab5041d69986ccccfcaea6693a1a7a00ba8541e76821952bd0f139d992df474fcbfecbd67769
805bb8013dc37d4185969986cc02fdaaa6693a1a7a00ba8541e7682195cbd06b798b96ceacebbc9e213b7f293118bbd81766
996848ccc8af2a9a46a39b07885b18688e1052b6117c5b041ddd3c1ed0350c3a470ba9dc08e76d4147370f40b7f083ced142
2837c2155ad0d1cd03d02d0c3a470ba9dc0046728a8e861e806e61d0395a48e506f049547434f4e8a1bb30e81c2da472c506
743ed24def91ef04f6f07757daf13da79592cbbdbdd8c93464161bd057354dd3d1d003d02ddcd3395a68e532f49e5647ced9
8436da14fc8e231d8a73da36311df5258e9369c8ccd0ef6a9aa6a3a107a05b18748e1652b9fc30b02d789de29eb6394c2bae
6ca893a613eb75d4571a3ad390991f07beab699a8e861e806e61d0395a48e5ca63df7d0ed796392d770989e952b364ebe525
cf99eb23509d69c82c8f7d5fd5344d47430f40b730e81c2da4725f3fdd576874f300710b03cd1142ca36c0d24745474f0f40
b730e81c2da47203ac025474f4f47840d730e81c2da47275156035ff8aef13cfe35cc3497e4f87e512bea13336b7aaab0055
a621b3ae027c53d3341d6d3d00ddc20f3a470ba1dcb76635828b861ec06d6170392a48cd2a6e6b581b6e029729198dc4f115
4bb8d6744e661a322bf29b9aa6e968e801e816069da385546e8015078a0e76223d738b82cd5142eaf6f58e2e01d6db8980b6
067b2a2d8194ebeb75bcff42c1c904a8771454520021d5004e53820b4e26a0bda300d31a28bd1e4e26f8c43fbd3a505d8f44
8d7a80fd9f8a8d6626406e61c0394a48dd2af46fddc36d6c5a3336df0f87a2f9567c6f177d996e689569c8acd06f6a9aa6a3
9909a05b18748e1652b90116df293afa9900ba8541e7682195fbba3b8242a399c983b88681e60821651b6055a9a2a39909a0
5bf841e76821941b60e9b0a2a39909a05b18748e1652b901daec2a3a9a99f4d05d18748e165239011da662d7b9a71564f131
d9beffe2e351c15c130d8902d9ab681a8d4e26206ee11ecd1142cba63c6bd2e4255d4cf7f30c53b56d698d4754a2215139d6
38154da3d1c604c42d0c34470829db088635828e3626806e61d0395a48e546e8fe2fe8686302e816069da385544e411fd77a
1d6b32a63fe6259c5fcf531ddb2dd1fa4485ec54348d460f1310b730d01c21a46c698eafead345041fdbc2a8ef8c468e3d7f
aca84f1f0f7c6c0ba3be331a3976e5ee106e24c3119726c571e5415a44254fd777a62153b93b78354dd3d1c703d02d0c3a47
0ba99c727798c3bc38dc5bc7e726b1415d5e0aaaa0ef4c43a67277f06a9aa6a38fc703ba8641e76821951bc0dd41d1d1c703
d02dfca073b410ca0de0eea0e8e8e3d1437761d0395a48e594d1c11cdfcae65deffbb4c497a4dba42e542dd390a9ac0ebc9a
a6e9e8e301e816eee91c2db47203dc60283afa7800ba8541e76821951b608fb1a2a38f07a05b18748e1652b96c3a20ead34c
031fdbc2a8ef8c468efd69a6818ffdd32902f59dd1c8b1e78f15f56967818f6d61d4774623c72e1e525cfbeffced53ded717
26a5d3b2fe5635f16b99864cf190c2ad699a8e7616806e61d0395a48e53eb758106874d200710b03cd1142ca36c20944d0d1
49e3015dc3a073b490cae59f94a84f3b0b7c6c0b3fea3ba311631fe12725e8e868d18db70f83ced1422af77deb0f4947470b
40b77047e768e12857a1e7e44f9a9ed84cf1b9d199df456ffb32afebd9199788444362457e51d1341aed2c40dcc240738490
b27dfc145fa2d1ce02c42d0c34470829dbf72f6b241ded2c00ddc2a073b490ca55e865daae252d54dfd770e06df3949cb4b7
e9793b25330d9915fa4d4dd37474b400740b83ced1422a57df455eb1b5771a5fec963f9feb9ad6195ce14c33ef6bbbb15089
86c4fa26f24545d368b4b300710b03cd1142caf6fd2b6749473b0b40b730e81c2da472715620ebc356029f7a4751dd198b1c
79553ab9fca65775f79bbab452e917aea7675cbc752bad320d9955e937354db2c15602c47714685a07a5d90847d6bf64b095
78f09628c8b40a4ab101dabd294708da4af4b3b82ebc87abcf356f39dc8a38e1aec8d7aff4151a1d2d40dcc240738490b27d
efdb26e9e86b01e816069da38554aeba3cfc539fe6128f8fad61d4774623c7fef1e45aa2d1d702c42dfc40738410b2d5dfd3
158ef330cc74244ce1ac16bdd1d3d11fe60ae9b1e4fd83529986ccfa8b7a53d3341ddd2d00ddc2a073b490ca8d70e6147474
b700740b83ced1422a57a1d770d4efd943783dcaa8d20d405cc11c1b19dcd02ad39059a1dfd4344d47830b40b730e81c2da4
72dfcf09241d0d2e00ddc2a073b490ca7d6f9327e9e87101e816069da38554ae422f717978eab213ceadfb150e88b4832c1c
a173de3c7943ab4c4366857e53d3341d9d2e7ae82e0c3a470ba9dc003b77141d9d2e00ddc23d9da38556aeecd14a7dc39623
ddd5c677b3d1a220791514dbb7b9de34eb4c4366d9a3f5aaa6693a3a5d00ba8541e76821951bc1ff40d0d1e902d02d0c3a47
0ba9dcd7fb0f151a6d2e40dcc240738490b265e23d56f91df989cd521b93c401865fdf15cff8f507ad330d9999f95d4dd374
b4b97840d730e81c2da472053abdad9a92b7ce1a8ebc6b0b73a6b41378b95d1d0bb4cc346416e857354dd3d16403d02dfca0
73b410ca8df04d0b3a9a6c00ba8541e76821952b3fe86dfda5a1c401aee7116e3baffcfbdb6aa392fa8b969986ccf2937e55
d3341d4d3600ddc2a073b490ca95eb557ed6be97fd63c529229d66ebdba67abd929986cc72bd7a55d3341d4d3600ddc2a073
b490ca959949bad1ddf256d839dd249ce986b0f80caddbfdf051661a32cbcce4554dd37434d900740b83ced1422a37c01c54
d1d167a387eec2a073b490ca7dbdf952a1d16703c42ddca3394268d93e7e6923d168b201e216069a2384946d8067908a8e26
1b806e61d0395a48e5d49a8ae9b76ee10fe96718378a5ed7b52807869669c8546b2abc9aa6e960b201e63b0a364709a9db00
0b1d051b4c36407c4781a675509a0db09a5590c164e3c15ba220d32a28c5146fbc193af35c29cc94c39772ae72c5f29d687d
a2a2752a9ac482c30660efe8034b4bf0af5c03ac3f136474d800700b834deba0341b60f7a762a3c306905b18708e1252b7fa
5cfb13676ec5456f0de0b630b81c15a4669f4fb7041a8d357ae22e0c34470829db00b36a4547630d40b7704fe768a1951be0
c9aea2a3b706a05b18748e1652b901ba6f293a7a6b00ba8541e76821951be04190a2a3bd06a05b18748e1652b9d24eef571e
c4c631c5d36d18617a2f1ee6c067f222abd03ad39059dae9bdaa699a8ef61a806e61d0395a48e58a8bca964dc4d300975ff8
4ee2feae38c067f31e9d69c82c3e2aaf6a9aa6a3bdc603ba8641e7682195cbd0d3bafdc2493537738c3ea9d7142614f10fed
116d86d69986cc0cfdaea6693ada6b00ba851f748e1642b9118c54041d1d3600ddc2a073b490ca7dee2922d068af01e21606
9a2384942d37fa3dd7df190eb933fb7f94d65979b4d714b7caeee56b76320d99b9d1efbb9aa6e968af01e816069da385546e
044f1141477b8d1eba0b83ced1422a573c45ee5709b1d176b70064faa5d51fe1ccfb5ff114919986cce229f2aaa6693ada6b
00ba857b3a470bad5c868ea7d8b8b8237d11d92b7499f300cba9a836efd69986cc0cfdaea6693ada6b00ba8541e7682195cb
0df9cfa0fd7cfed2d4694a2b6c96396d3b0fa7dbfc17d58340671a327343fe77354dd3d16103d02d0c3a470ba95c865eba27
19d372edf126682f876559e29d9965a2213123bfaa681a8df61a206e61a0394248d912f1f64b0e097b6a0e32851972f8ddff
d27dfe74dc6d0113b29369c84ccc2f6b9aa6a3bd06a05b18748e1652b9b4aa56d5a7c7053eb68551df198d1c7b3ebaa6353f
e84edf7f7a3d7b9ef94477ac41afad9e469c44eb13f3d1f5aaa269341a5c3c886b18688e1052b6111c6b041d0d2e00ddc20f
3a470ba1dc085626828e0617806e61d0395a48e5f23572896338af74733dedb107f0ef3cb3334538e9c52546e53ce2641a32
f335f25d4dd37434b800740b83ced1422a97a1c3253cb5562a475f5d4f9326aceb2fbe43acfe2d3ad39099a1dfd5344d4783
8b1eba0b83ced1422a97e7bde1d2794e6130f9babd87ebca95efbfa6783312bf98f24d3b9986cc3cef7d57d3341d0d2e00dd
c23d9da385566e803b1c4547830b40b730e81c2da472f946b6ecf34e1b45e7650e179bf54c4fc4e7a53c2aaedeaf3ad39099
6f65dfd5344d47830b40b730e81c2da47219faec4eae73b8210bb722533a03cddbb94ddb6fdbaa7f8bce346466e877354dd3
d1e002d02d0c3a470ba95c7e38b5a4bbac29bdfa58f22eb0303f4f0f58c20f733bef97293ad1fac4fc68ea5545d36834b800
710b03cd1142ca36828f89a0a3c105a05b18748e1652b90116982a3a1a5c3ca06b18748e1652b9b2d6f25cd7d8362dbdf808
ffe2172611e921e9b2dd4eb565ada5cc346496b596af6a9aa6a3c105a05bf841e76821941b60ada5a2a3cd05a05b18748e16
52b9027dcc57b8ed99f31ea4b8de782e6f418e2b364fb9d6baaa56671a320bf4ab9aa6e9687301e816069da385546e805d20
8a0e36173d738b82cd5142ea36c09a5ac1d63b5d00b8067b30ad82546c8037e0ff72c1ec02b47714605203a15685cd2b2df2
bef3f0bfd35636a94ce7b64c4b5b6a2813ad4faca82f2a9a8482d30550ef28a8b4004aac0a9b5e59a64b67f842c2a925baf9
a5ebe86d03576955a621b3e2bea969928c3e1b006e61b0691d9466032c61516cf4d900720b03ce5142ea56a1db4dfc769433
68fa4ac24f2ff7e0bea155a621b342bfa9699a8e3e1b806e61d0395a48e506d87babe8e8b301e816069da38554eefb4e3392
8e561b0fe81a069da38554eefb7e55928e561b806ee1079da38550eeebf62b0a8d3e1b206e61a0394248d9be6f3228e9e8b3
d1437761d0395a48e506d807a0e868b501e816eee91c2db4720338aa283aba6d00ba8541e76821951ba173bba0a3db06a05b
18748e165239d5a33f5c68f66bf9c56b4bbd9eeef2f0be330d99aa47bf57d3341ddd3600ddc2a073b490ca651308519f9617
f8d816467d673472ec9f1bd80834ba6d80b88581e608216553c473f8cbf08fd634c0f2e05fed1b6b99864cc5ecd5344d47af
0f40b730e81c2da472ca9929fceed6503c5dd3b7308d5bb7f0cf14f49d69c854de4c5e4dd374f4fa00740b83ced1422aa71c
20ee9f799abb1dcbb15ca7dc3e76671a3295038457d3341dbd3e1ed0350c3a470ba9dcd77bc8141a8d3e40dcc20f34470821
9b9a0d951bd0342bbfd2b2f2ed923fe83bd390a966435e4dd37434fae8a1bb30e81c2da4729ff7ab176874f900710bf7688e
105ab60166fa8a8e2e1f806e61d0395a48e572bf7a519f3e1bf8d816467d673472ec4aebb4af2e9b8887e9e21ee6c697da9a
79271a1295d24e45d368b4f800710b03cd1142ca36c2d125e8683002e816069da38554ee6930828ffdd33d03f59dd1c8b10f
b0735fd1d16004d02d0c3a470ba9dc081610828e1e23806e61d0395a48e5067850a1e8e831f280ae61d0395a48e506e85aa0
e8e87002e8167ed0395a08e5069884293a3a9c74e3edc3a073b490ca7ddfc240d2d1e104d02ddcd1395a38ca7d7f22937434
3901740b83ced1422af7fdf440d2d1e704d02d0c3a470ba9dcc7b3408946931310b730d01c21a46c9fbbf708343a9c80b885
81e6082165fbbe7594a4a3c909a05b18748e1652b9efbbaa4b3a9a9c00ba8541e76821957b9a9ce063ffb0f14075672c72e4
9f9b43fd0b068713e0de51706911946059e17f6bc363e4f189258ada7a1c6accd987411874d0e5a39fc974e12d9cf4f7f81e
fb7f50c4097745be3f69483a1a7d00ba8541e7682195fbdc7c42a0d1e503c42d0c34470829db00fdaa141d3d461ed0350c3a
470ba95cb536f9a73e8d3ef0b12dfca8ef8c468cfdf3a34ba0d1e503c42d0c34470829db582e1f80fed3c202748e1652b901
5625293aba7c00ba8541e76821951b60fd99a2a3cb07a05b18748e1652b9bad230aa71e59fdc19bea264b91207b8cf613af8
eb566ec84c43665d69f8a6a6693aba7c00ba8541e7682195ab4b87af6b4a1dd5e3796639f7e5fc9dc9bf603e6343af6b6e0b
a555a621b32e1d7e53d3341d5d3e7ae82e0c3a470ba95c810ef7add77c2de919e4947628e40186c9cc91d686ded032d39059
a05fd5344d47970f40b7704fe768a1951b60278fa2a3cb07a05b18748e1652b9b2676bb9d26efe74f4adf36f0b3fc62b7d2b
61b8e9bea4decee84c4366d9b3f5aaa6693aba7c00ba8541e76821952bbbf3f256d0b41b76891b91cfb4e527fce15ef25136
e7c944eb13cbdebc37154da3d1e503c42d0c344708295bd98479e5dd5569bb60b8944ce13b489b8e9669ddc2bf99ee75233a
d390593661beaa699a8e2e1f0fe81a069da385546e80967f8a8e2e1f806ee1079da385506e846f5ad0d1e503d02d0c3a470b
a95cf94187a9f0765e795fc6543b4ea69fdf2f3f78aeb7753ad390597ed2af6a9aa6a3cb07a05b18748e1652b9cfbd8a041a
2d3e40dcc240738490b28d70851674b4f800740b83ced1422a37c25c4cd0d1e2a387eec2a073b490cad55b8d790e377d7993
e8761ee71abea8745b1097b3c78e4bf7c3039969c8acb71a6f6a9aa6a3cb07a05bb8a773b4d0cad59bca3536d6cc5db18fd2
a6248eefd9ac55261a12eb2de58b8aa6d1e8f201e216069a2384946d0087504547970f40b730e81c2da472033c83547470f9
00f31d059ba384d4edfb17ec8a0d2e1f20bea340d33a28cd06582f23c8e0f2f1e02d5190691594626a23d33cffc214714b4f
2c43567c15bfca3d3d77a621536d64f26a9a2483d10778efe8834cabf0af625faf48175874f9006d0b034c8ba004fb7e5190
64a3cb07905b18708e1252b70196742a3a7a7d00ba8541e7682195fb7ab7ad42a3d7474fdc8581e6082165abc4537aab94ce
2ccbb1cdcb152f96715e5c370c546295687d62257e51d1341a8d3e40dcc23d9a238496ed6b3f178546970f10b730d01c21a4
6c0334c3507474f900740b83ced1422a57a0a7f0adfcd6745bb7853bbcf80c23359ef94de556a032ab444362417e53d1341a
2d3e40dcc240738490b295c703536921940ebde90cf7f27beadbb1c4ae4af19758cfd83ad390591e0fbcaa699a8e161f806e
61d0395a48e54630731174b4f87840d730e81c2da472033cd95574b4f800740b3fe81c2d84725fb78e5468f4f700710b03cd
1142ca963b84fef6dc1e2d75768c8f2e62c3f3d4c2740ff73f7bdc0e9d919d4c4366ee10faaea6693a5a7c00ba8541e76821
952bbd60db229779d9d735763398d31f78add28986c4d209f64d45d368f4f700710b03cd1142ca3640cb5f45477f8f1eba0b
83ced1422a574c4d60ca70bf178d8d98f9cac2c9346416539357354dd3d1df03d02ddcd3395a68e54668e32de8e8ef01e816
069da385542e434ff9dd499a1c4ef18e27f5b34c5f516a8418572a66689d69c8ccd0ef6a9aa6a3bf07a05b18748e1652b9dc
9affc81678e97bc81d7943f134bef9d892d94c66d68986c4dc98ff5545d368f4f700710b03cd1142caf6b9e78440a3b90788
5b18688e1052b6ec687286e36f997ed928207a43fc42661aec7e859f5ebc41cac84ea621333b9abcab699a8ed622806e61d0
395a48e532f451fa2e274b88793bc3bf0a777f6980b917cc514e614ea6213343bfab699a8eee220fe81a069da385546e846f
5ad0d15d04d02dfca073b410ca65e86bdfd203abf4ebabeb78d2afaf3c7e2e3f689d6848ccc8af2a9a46a3b508885b18688e
1052b67c0adb72a3c7b47f6d8a3659e775a56775d3f13b7273e2720ed39986cc7c127b57d3341dad4500ddc2a073b490ca8d
60bf25e8682dd2437761d0395a48e506305a5374b41601740bf7748e165ab962a9b7875bdda047be8476cfa5e33be1d846a3
2eda76320d99c552ef554dd374b41601740b83ced1422a97ef33d6b4bae5979631a5174a71df5e5af2524ed9ffcf8cffffff
fc5dc334052d4400d7c2a07098a542192edeba8793c892ef71f2b9250f700e5fc43a45c3d60ca9330d99f926ea5d4dd374b4
1001740b83ced1422af7f5030285460b1110b730d01c21a46cf6bdb9ada2a38508a05b18748e165239fbdec658d1d142e401
5dc3a073b490ca65e8759af2f3b9ffe213c9ae77fe7c5cfb3ecd7723439d687d62b1ab7e53d1341afd4340dcc20f34470821
5b79947d374688ef149a7951b8b0a481d7e7d87fa4597988fd7735d350b40d016b0b03ca91400a36c0637b4547db1040b730
e81c2da472032cc85674b00de9995b146c8e1252b70196630bb6de3604c035d8836915a46203bc81fb970bb621a0bda30093
1a08b5caa2f36dfda521c48105d5d363a9b4467cbbcda5cbaa739969c82ccbce5fd534c905e710d0de5180690d945e85379c
41c399644f7b99f6780f139f36a6b3eb1c0efe73bf977ae94c4366e17d55d324199d4300dcc260d33a28cd06f076526c740e
01720b03ce5142eaf6f562098546db1010b730d01c21a46c23981d093ada8600ba8541e76821951b605b81a2a36dc803ba86
41e76821951b60dbb6a2a36d08a05bf841e7682194fb7a2f8542a36d08885b18688e1052b601f652283ada86f4d05d18748e
1652b9efdb5e493ada8600ba857b3a470baddcd7ad8c151a3d4340dcc240738490b20dd07455d1d13304d02d0c3a470ba9dc
08ee30828e9e21806e61d0395a48e54630381074742c01740b83ced1422aa72c71e6705b1d0dd8d200cf2d9c807ef2e8be13
ad4f5486384e45d368742c01710b03cd1142caa6bee6394cdab6fce2685f9e4d8364a621537dcd5e4dd374742c01740b83ce
d1422a37c20f5ad0d1b104d02d0c3a470ba9dc005d94151d1d4b1ed0350c3a470ba9dcd756160a8d8e25206ee1079a238490
6d801d9e8a8e8e253d7417069da38554eeeb899842a36309885bb8477384d0b20dd02653d1d1b104d02d0c3a470ba9dcd3b1
041ffba72307ea3ba391631fe0de46d1d1b404d02d0c3a470ba95cd65ad4a76d083eb68551df198d1cfbe7bf648146c71210
b730d01c21a46c036c6f5574742c01740b83ced1422af7f57559a1d1ae04c42d0c34470829db0077ce8a8e76250fe81a069d
a38554eeebedf90a8d5e25206ee1079a2384906d805b0b4547af926ebc7d18748e1652b9dc9c5cd4a75b083eb685bbface68
9cb1e78f15f5e9d7818f6d61d4774623c7fe7d8f0d4947bf0e40b730e81c2da4721fff92251afd3a40dcc240738490b2e5a3
4bd4a769063eb68551df198d1cfbf7b7e9928ea619806e61d0395a48e5be6fd323e9689a01e816069da385546e846f5ad0c1
b203cc77146c8e1252b711bee77fd9e0da01e23b0a34ad83d2ece1daf1f844edcf51a2a8adc7a1c63cc0bdaab2fba067483f
a3e8c26bb8dacecb2f875b1127dc1519e06188a2a36708a05b18748e1652b901bad7293ada8600ba8541e7682195fbfaf180
42a367c883b88681e60821651be16b1674742c01740b3fe81c2d847223fca0051d4d4b00ddc2a073b490ca8d655a02e83f1d
3940e76821951bcbb404d07f3a7280ced1422a57d7449d216d4a5bc8f7f9de351f976486dbf96b3e5a1b2c9969c8acaba2de
d4344d47d31240b730e81c2da47203ac7e5374342d01740b83ced1422a37c03a474547d3921eba0b83ced1422af7f51a5e85
46c71210b7708fe608a1651ba0e199a2a36309a05b18748e1652b9b2dde2dea916fd72b6d23a210e70b99bbf94ed1632d390
59b65bbcaa699a8e8e25806e61d0395a48e50a74d72c61b9baf53ce106e5b7cd57dc6a5ea065a621b340bfaa699a8ea62580
6e61d0395a48e5cacea97b23663cfad2f2f1dcff254c27d08ec0c93464969d53af6a9aa6a369c903ba8641e7682195fb6e3f
a082a25709585bf801e54820044b2f7b557dba85e0635b18f59dd1c8b117898ffc9f74ac1fdd4dc97c85135e98aacd754aa0
330d9945e857354dd3d12d04d02d0c3a470ba9dc0867104147c31040b730e81c2da47223b85b093a1a8600ba8541e7682195
2bb38275c9adb2e200cf250c715dd376fe682d973798d55981cc34649659c1ab9aa6e96818d2437761d0395a48e5cae42f79
816e79763aa74d0b677a72586633eb764ff365a621b34cff5ed5344d47c31040b7704fe768a195abd3fcedb7ecb9b7ce7694
cdede9b03cae2d7c65f78dbbcab33eaf4ef1ffae679a8b6e21c06d6170392a48cd06b87f5574740b01740b83ced1422a579f
54ec71577b1cde167ef2e13e734b5fcf321de9a9616bc8fe6f9e21af3ea5f8bb9e692ef88480f68e82cad1402a36c05347c1
069f1010df51a0691d946603acce1164f00979f09628c8b40a4ab1efd7302832f88480f78e3ec8b40aff2a36c0b21c4146ab
1000b730d8b40e4ab334a557d569d6810f6d619477c62247fef58a2f8546a70e10b730d01c21a46c2378b2083a9a75f4d05d
18748e1652b9afdfc329349a7580b8857b3447082ddb006f67141dfd3a00ddc2a073b490ca0db09559d1d1af03d02d0c3a47
0ba9dce7b62c028d7e1d206e61a0394248d946b0651174f4eb00740b83ced1422a3740df0d4547bf8e07740d83ced1422a37
c023094547bf0e40b7f083ced1422837c08b0a4547cb0e40b730e81c2da472033cb35774b4ec00740b83ced1422a97fb535e
5b3ee4f2a3f52bfce8c26927fd21b7e25fea6fdac93464e6fe94ef6a9aa6a36b07a05b18748e1652b90cbdd546e5714ceb6f
d9b77d4ea7ddfc9670b9bd1c9c4c4366867e57d3341d5d3b7ae82e0c3a470ba95c69385b5e76a776bf5798392ce15a1aafaf
f37ccfa34ac3599969c82c0d675fd5344d47d70e40b7704fe768a1951bc1aa44d0d1b503d02d0c3a470ba95c6e167db4763f
d311af2cd3943aa54ebfdbd12ff78a968986c4dc2afa5545d368b4ec00710b03cd1142ca36824f89a0a36b07a05b18748e16
52b9115ade0b3a1a7700ba8541e76821951bc1ce41d0d1b803d02d0c3a470ba95c86ae4fb8534ffeb25133cd98c3b4e99ae6
fdacd33127d39099a1dfd5344d47e38e07740d83ced1422a97a1d7749b171f4ec6019eeb2ffcf6d33c6a8a0bf8cf70162a33
1327d39099a1dfd5344d47e30e40b7f083ced1422897a197dcdb3b1d7cd588208f6fbde262fe6a62ab130d8919f95545d368
34ee00710b03cd1142ca36c2b12de868dc01e816069da385546e84b398a0a371470fdd8541e7682195cbd7ab7c973be52fa2
b4059e8af5d9192670db568db574a621335fafded5344d47e30e40b7704fe768a195cbd0fb11dbf386b34cbaa0e68ec0e95e
683aa770390d25caad869369c8ccd0ef6a9aa6a37107a05b18748e1652b9112c5a041d0d3d00ddc2a073b490ca158b96bbe1
733cfaba77f3e187185fcc2fbf3a07d59986cc62d1f2aaa6693a1a7a00ba8541e7682195cbf757e1fa794e6180f95e680f33
d52b5b514c57dc0d1e4e3ae5eced641a32f3fdd5bb9aa6e9e8e901e816069da385542e4387f3e9762cd9ceefd75d52e7e95e
95909965a2213123bfaa681a8d861e206e61a0394248d93271f871fdf6e32af7f9fb1ce62f71bb6abccfdfe6dff18b8f7732
b2ce346466e677354dd3d1d0e3015dc3a073b490ca7d6e5623d068e801e2167ea0394208d932f1def9f7cde1ec7ec6e7d1c9
60b8cc94ab67b14eb43e3113bfaa681a8db61e206e61a0394248d932f1998c19e6ecba11662e57748f4e7fc87677d354ed79
74a6213333bfab699a8eb61e806e61d0395a48e5ec7b6f6a45075b8f9eb945c1e62821751bc09148b0f5b61e00aec11e4cab
2015cbb4cd033dae678e3d7ac31792be8c70c609bfbcf87a2fd3ea4c4366a67d57d314176c3d407b47012635106a15dfa56d
fb1de79ace2a4bf41b4ded53e3c09efdd275a621b3382fbdaa69920bb61ea0bda300d31a28bd06784d25c868eb01e016069b
d6416956de46de6790b87e79df7fcb76a4669acb7496d34b7d1b29330d99e56de4ab9aa6d968eb01e416069ca384d46d8057
b08a8ece1e806e61d0395a48e5caa687fdb7c7472669a9476a44b2fed24478d9b6694fae5875d383cc3464964d0faf6a9aa6
a3b307a05b18748e1652b9015618283a3a7b3ca06b18748e1652b901363d283a3a7b00ba851f748e1642b9ba80e60ab3865f
6e04bc855f42a89d76e8ae65627c6f7ad59986ccba80e64d4dd37434f700740b83ced1422a57974ad5534bdc99112eaabf73
fa252bb47a9b772f8c979986ccba54ea4d4dd37434f7e8a1bb30e81c2da4727545dcfebbce332dd6bfd7eaa71de7612ab19e
71bbcabd244e651a32eb9ab837354dd3d1dc03d02ddcd3395a68e5eadac76d9fae5f7abbb26fb7cd6efc31eec7ef0897bb7b
7d98cc3464d6b58f6f6a9aa6a3bf07a05b18748e1652b9011aaf283afa7b00ba8541e76821951ba165baa0a3bf07a05b1874
8e1652b9efbbcf493afa7b00ba8541e76821954b9b22547d9a6ce0635b18f59dd1c8b10fb0bb48d1d16403d02d0c3a470ba9
dcd72d9e151a1d3640dcc240738490b28d701e117474d87840d730e81c2da4725fef04516874d800710b3fd01c21846c95f8
7eeebd9f539c3f9cb901515a4717a62e8d58241a122bf18b8aa6d168afd1137761a0394248d906682aa9e8e8b001e816eee9
1c2db472036cf8517474d800740b83ced1422af7e92667c545670fe0b630b81c15a46603ccf1151d9d3d00ddc2a073b490ca
55e8b8c4f7b8d29d57798593564586dfe3169786b62f59255a9f58915f54348d46531110b730d01c21a46c239cbf041d7d45
00ddc2a073b490caa90b73980e4ef1dd707ac470f7225257e69a69c8549766afa6693afa8a00ba8541e76821951b6033aea2
a3b508a05b18748e1652b9efbb6a4b3a5a8b3ca06b18748e1652b901fc29151ddd4500ddc20f3a470ba19c800e67f77d0d37
24e9acbbfdae73ddc2ad9882be330d9902daad699a8eee22dd78fb30e81c2da4720afad8ae73cf338a3097d8c3bdc12a6f24
6ba22151213b154da3d1d804c42ddca1394238b209e233fceee7f0933bf38566cfbbb114f29d69c814cc6e4dd374b4550174
0b83ced1422aa7a0cb7bae3d0df0fe252ae89a69c854d05e4dd374b45501740b83ced1422a970d02447d7a9be0635b18f59d
d1c8b18f7080093a3aab00ba8541e76821951be1001374745601740b83ced1422a27a07ff3117e825bea85bd877f3f8583e4
948f79ef4c43a680766b9aa6a3b30aa05b18748e165239057d1c679893a75beef3fcfdf670932daf9235d190a8909d8aa6d1
60ab02e03b0a304706295a3e8bfc5b1dc626f8cc3b8ae27a246ad40f6393c7276a0b9312456d3d0e35e601de4c294f121a9b
f493992ebcfc7fe1fe28dc10a6702be284bb22e97d85aa4f6b117c6c0ba3be331a39f6af9f702a34fa8a80b88581e6082165
fbfa29b642a3afc883b88681e6082165fbfc3b1668341501710b3fd01c21846c239c41041d4d4500ddc2a073b490ca0df014
5bd1d15404d02d0c3a470ba9dc58a62280fed33103748e1652b9019ab22a3a9a8a00ba8541e7682195abd0dd4b85df39c53d
d15b1ced1ac6b7afcbf6bb0f6f9969c8acd06f6a9aa6a3a908a05b18748e1652b90116de293a9a8af4d05d18748e1652b902
1d2e99d77c2de94dca1437671c7980e1e47becd7b9dd2b8875a621b340bfaa699a8ebe22806ee19eced1422b575610876bea
7c5d7bfa22f6b82b34fe06e300cbb2d7bb9da6ce34649615c4af6a9aa6a3af08a05b18748e1652b901d68a2b3afa8a00ba85
41e76821952bfb20ae63bed6333d780d33b6df146e088ef4ad4ceb163e61ba1f67eb4c4366d907f1aaa6693afa8a00ba8541
e76821951b60f387a2a3afc803ba8641e76821952b3b7ce2f2fddf3565a7826b5dd630374edb63e30ec2f0afeeed4c3ad390
5976f8bcaa699a8e0623806ee1079da38550ae409fbfe9b7aff927b786d34ef815e49da3275b81389986cc02fdaaa6693ada
9b00ba8541e76821952b8777f881cdcb96b6d485ff99a7fd8c5bdcb1a5b01cde32d390590eef57354dd3d1de04d02d0c3a47
0ba95c81dee7309198d34d509838847bbf6bcafb28af65fd9df11d6685969986cc02fdaaa6693ada9b00ba8541e76821952b
97ac6dda7ee192924eaeb179c69936abc6bd85f57d5ab962c944eb13cb05eb4d45d368f43601710b03cd1142ca566626dbfa
0b37b8e91e309c6bea49261e947797873233919986cc32337955d3341dbd4d7ae82e0c3a470ba95cd9db74c56eb2bfd4b523
5c63b6ed0ca7dcf44b0c17d9f3bcea394ce7599757f635bda8679a8bae26c06de19ecb51416b56ef337e7937ca7f29e9b7ce
c794f66bacb197ffbc4df761ad330d99f53ee34d4dd374f43601740b83ced1422a37c01da5a2a3b709a05b18748e1652b901
1e98283a389c80f98e82cd5142ea36c09a30c106871310df51a0691d9466032c4e1764703879f09628c8b40a4ab1cf8d28fe
c582bd0960efe8034b4bf0af5c0276effcdfc261b785e3f027df6cef9da55c4b14b05e459358f436016d0b034c8ba00453b8
b1ef4cf85779bded3a5ff3d45909cb4c43a602f66a9a66a3b30a905b18708e1252b76ce822ead3de041fdbc2a8ef8c468e7d
04a72041477b931eba0b83ced1422af7f5c943a1d1de04c42ddca3394268d9065838abe8686f02e816069da38554eeebb7be
0a8dde26206e61a0394248d9be5efbafd0e86d02e216069a238494adcee7aff8d23f0d2f2e66497d71d3fdd6314d73b89a35
3717916848acb3f917154da3d1d804c42d0c34470829db0077308a8ec6260fe81a069da38554aef66558c3b19797a86ffbb9
cfb1b17bfa56d6654ae7dbe6c1a9320d99b52fc39b9aa6e9686c02e8167ed0395a08e5067827a6e8686c02e816069da38554
6e8077628a8ec626806e61d0395a48e5ca53d55f38dac25fa631c51d75e7f1cb2df952cb9f3043ad972a9d69c82c4f555fd5
344d47631340b730e81c2da472a5f1dbfd44088dac6293b625b722bc1bbfc94c436669fcf6aaa6693a1a9bf4d05d18748e16
52b9e2e6b25fd3bea5478261325c5e4ec63fb42e6ec5cc45251a128b97cb9b8aa6d1e86a02e216eed11c21b46c2358b9083a
ba9a00ba8541e76821951bc1ca45d0d1d804d02d0c3a470ba95cb1720953fedffecb1dcd97257c3f6b1a5e5a417dfceab215
9967c82b462e2fea99e6a2a509705b185c8e0a52b3dc2dfd4c33c32ded07986a77b6d415391c8eeb7a4c4b39693b9986ccdc
2dfd5d4dd374b43401740b83ced1422a97cd00c24534dd06c4212df56555fe82f29d7d39ae75a221315b01bcaa681a8d7e26
206e61a0394248d946f07c1074f4337940d730e81c2da472197a8aed2de37ef6ffa634aab8e33d84cad4cbffeb8c77ffb5fd
f3af4d0f9e7625606ae1c7e01d54214c7ac2a6ead333041fdbc2a8ef8c468e3d7faca84fd70e7c6c0ba3be331a39f6fcb1a2
3e7d33fa8fedc2a8ef8c468e3d7dacaa4fe70a7c6c0bf7f59dd1e8b1e7e3750eb7e9a9df7efa41559b8974f02de1741abbd1
954357671a32f351fcaea6693a3a5700ba8541e7682195cbd0bf64acb1a7f52153b8650d53cedf964f9df36f3ac2b5a39e99
74a6213343bfab699a8ece15806e61d0395a48e5f255772d37edc942649dd7335caef20c2106b7302b2c175d996848ccd7dc
57154da3d1b602c42d0c344708295b26beb6630ed78af43c263eea5c8f30c74d0765d9cd5c8975a2f58999f85545d368f4ac
00710b03cd1142ca962792db796ce1fcb2e51fdbafac734c53bf701b1d4e02750d9d9369c8cc93c977354dd3d1b602d02d0c
3a470ba95c86beb203509ee29e535ec399467bdd2d7133b4ce346466e877354dd3d1b6e2015dc3a073b490ca0d709fa4e8e8
5c01e8167ed0395a08e58a254bb6cfc85685bf6a5c18ef5ea7625059a175a621b398b2bcaa699a8ee615806e61d0395a48e5
32f451b789a50186dbf5d897353da7080766da5a5656403b9986cc0cfdaea6693a9a5700ba8541e76821952bcf3e62cfddeb
58f2d197bf93f45a709ec331b94e676d06ef641a32cbb38f57354dd3c1bca2676e51b0394a48ddf203ae706393d6f9e5636f
fda53bf7fc68260cfd8c6b77cb132e9d69c8cc8fb8ded534c9d69b5700b8067b30ad8254ec6bdb9d7fa1e05c01d43b0a2a29
8090ea6bbb1d0105db0aa0de515069019458c569a7f3a90be33aa7256e048d7fd8cebc72b71ec43ad390599c765ed5344946
db0a00b730d8b40e4ab3e239539eb5a457416168db1edfe8c7d1c5059ccbb5dfaf8e75a621b378cebcaa699a8db615406e61
c0394a48ddcab68569ca4e407180cbb69e7b5c8a9c0678edfb34dfcd3374a2f58965d3c29b8aa6d1e85901e216069a238494
adbc68bafbb8c5975fed78fced79e0f52dd31f69565e31fd5dcd3414ad2ac0dac280722490827dbd6a5fa1d1a7e2415cc340
738490b20db06a5fd1d1a702d02dfca073b410ca8de04822e8e85301e816069da38554aeaedd6fa7962d4ceeb7d8b5e64887
65b8c94cdbe2eeb5fb2ad39059d7eebfa9699a8e3e153d7417069da385546e8045208a8e3e15806ee19eced1422b5797fb84
93f9f14b6f37b76b0ebfc7239d7fd62336068d0ee1f76a1f916848ac8b7d5e54348d46930a10b730d01c21a46c03ec565074
34a900740b83ced1422a3740f7124547930a40b730e81c2da47262816638d54d73dcc1927e89d53246add0bc330d996289a6
5bd3341d4d2a00ddc2a073b490ca3d4d2af0b17f3a30a0be331a39f601fa052b3a9a5400ba8541e76821951be1001374f4a9
00740b83ced1422a37c2a944d0d1a7e2015dc3a073b490ca7dbdf259a1d1a702c42dfc40738410b20dd0fa49d1d1aaa287ee
c2a073b490ca8dd0d15fd0d1aa02d02ddcd3395a68e5069810293a5a5500ba8541e76821954b9749559f9611f8d816467d67
3472ecf963457d9a36e0635b18f59dd1c8b18f705c0b3afa3600ba8541e7682195fbfa9cadd068da00e216069a2384946d80
f988a2a36903a05b18748e1652b9af9dc2141a1d1b40dcc240738490b20db0a952d1d1b1e1015dc3a073b490ca0db06b57d1
d1b101d02dfca073b410ca7ddf7f5bd2d1b1a11b6f1f069da38554ee5fe870baa9cbc2529380658dddc07e02ba651a32ff85
f66b9aa6a36903a05bb8a373b4709413d0f3718549f29ad61a0491c29def79aafe0a2dd1fa4481ec55348d46c70610b730d0
1c21a46c82381cffe156604f16f4c7fc5bd3761af5836e99864cc1ecd6344d47c70640b730e81c2da4724fc7067cec9f7604
a8ef8c468e5d695dde0de7cde4dbbc6d7b7c36acb4ae99864ca5b557d3341d1d1b00ddc2a073b490ca09e835f6520917b22b
0df0ba96e3da67b50fbf651a3205b45bd3341d1d1b00ddc2a073b490cae5034cd4a767023eb68551df198d1c7bd5fac88edc
5197b822ecdac3dc38eab2cd61fe14dfecdd5aab4c4366d5fa4d4dd374b04d00f31d059ba384d4ed619b80cffcc39800c5f5
48d4a81fb6098f4fd40609258ada7a1c6acc5fdf4229bb037a26f4d3892e3c8789c51a2734ff83224eb82b324063274547c7
0640b730e81c2da47203dc372a3a9a3600ba8541e7682195cbee18a23e9d131e1f5bc3a8ef8c468e7d20e70410ff690bf040
738410b20df0c240d1d13901d02d0c3a470ba95c3d8f7c6285abb8e89900dc160697a382d46c8067c88a8e9e09806e61d039
5a48e5eada89dfbaffd67483b187d3db39ede10af65fece5129f1dc493fbbd7842651a32ebea8937354dd3d13301d02d0c3a
470ba9dcd70b46141a0d1340dcc240738490b20db02e48d1d130a187eec2a073b490ca15e87415d9d2de8730f70e1791e94c
1398e517fe6a3ad6edbe3b979986cc02fdaaa6693a1a2600ba857b3a470bad5c59eb975f62ece5e82bdf511ce05ea73375ad
9fcc346496b57eaf6a9aa6a36102a05b18748e1652b9b10c1300fda71b00e81c2da472053a356ec9a7d9707adf7fcb76a4de
a35d5797022d330d9905fa554dd374344c00740b83ced1422af7ddea6c05459f84076b0d03ca91400a36964f02a0ff340178
d0395a08e5d22d8daa4fa7027c6c0ba3be331a39f6b19c0a00fd671b7ed0395a48e5c6722a00f49f9df841e76821951bc167
46d0d1ac00d02d0c3a470ba95cb956c453ccefc85d1ac3ecec8c6fcdd2c2f5b471214c94eb2dbbce3464966bc5ab9aa6e968
56d0437761d0395a48e506980a293a5a1600ba857b3a470bad5cddf6109f401e575e92b6869b92694f0b4e97eb8c5ebdf19e
bb6e7b509986ccbaede14d4dd374b42c00740b83ced1422a37c0ed8da2a36501a05b18748e1652b97a179bde2ca40792f7f3
c8f8d06c0db398359c85f6fbf646661a32eb7dec9b9aa6e9605900e63b0a364709a9db006bc0041b2c0b407c4781a675509a
7dfda05560c1afe0015ba2c0d21228b9be5e0425b0e05700d83bfac0d212fc2b975aff1466deb15b633a04e3aba570abb5a9
b7d92dd390a9d63f79354d92d1b200c02d0c36ad83d26c80c54f8a8d9605406e61c0394a48dd04f4710465f6fc003c5c3273
1f21c55c130d8902d9ab681a8d6e09206e61a0394248d9b24983a84fbf82fe63bb30ea3ba391634f1fabead334001fdbc27d
7d67347aec5fdb8e28343a0680b88581e60821651be0c4a5e8681a00e816069da385544e40cfe5e95f7a8d73a536d1db25d7
30de99864c01edd6344d47df0040b730e81c2da472ca2f690ae7f5f087345708f3fff9bac2fdbc7aa37e671a32955f9257d3
341dad0300ddc2a073b490ca7dbd5a44a1d137e0415cc340738490b255e2da072afede8ed20a2b8eef9e31556295687d6225
7e51d1341a4d0340dcc20f344708215bbd5b8c4dfce31d46ece6b8cccbb21ce97676dd8f23dcdc1ddbbdd04b661a32ebdde2
9b9aa6e9681a00e816069da385546e80e7028a8ea601806e61d0395a48e5ca13a0a9bc964b4f6ba633dcd2ef477a9db1874b
6b3c34eb294c671a32cb13a057354dd3d13400d02d0c3a470ba9dc084e09828ea6013d7417069da38554ae38251ce7396d67
ee3b33c70e02f39a463b8713fdbcfeb6fb5d89ce346416a78457354dd3d13700d02ddcd3395a68e5be6e38a6d0681a00e216
069a2384946d04a7044147d30040b730e81c2da472197a5ea7f88661cded6763b3d2d8b1338df6289b8e0bb4ce346466e877
354dd3d13a00d02d0c3a470ba95c69703b851fd816eef6e298f673997ee7919bc1ff6ec7a9d2e056661a324b83db57354dd3
d13a00d02d0c3a470ba95cee63bcfd7e7b9814e6feb37939c99a1e4087f963b8b8feae32177312ad4fcc7d8c5f55348d46eb
0010b730d01c21a46cc53a203a0fedf19d6c18d2dcfdf8c2b52dff45754bd09986cc621df0aaa6693a5a073ca06b18748e16
52b9ff55eb0030fdd995ff3178075508f3b40ec0c7fed9951ff59dd1c8b13fad03f0b17f76e5477d673472ec4feb80fe63ff
eeca8fface68e4d89fd601f8d83ffbe2f7f59dd1e8b1e78f15f5d9bc1f1fdbc2a8ef8c468e3d7faca8cff6f9f8d816467d67
3472ecf963457df6b0c7c7b630ea3ba39163cf2785352d139dd37ba829fc559e6fa55f78b9369569bb9369c8cca78a77354d
d3b18d3da05b18748e1652b9afbd72141a7bd883b88581e6082165fbfa12afd0d8c0fe415cc340738490b27d6ece20d0d8bd
1ec42dfc40738410b28d60ce20e8d8bd1ed02d0c3a470ba95cb136cbc75dfae1c5a5127338bba6efe4773f012cde662ad190
58cccdde54348dc6d6f5206e61a0394248d946b0a31074685ddf33b728d81c25a46ee556f43cc3a07ebfadfcd8ce29fcb334
cfbe920fd052570c3a9986cc722bfaaaa649b6be753d806bb007d32a48c532ed561efa2cf916b1dc49c42703e13b39c2d773
d61b6f9d69c8ccb4ef6a9ae242f77ad0de5180490d845a1936cc0ece79ff15a788729f949e034dd7145bcaede5cae4641a32
33ecbb9a26b9d0c01eb47714605a03a5d7082e0cff92b1813d805b186c5a07a5d9e74e0c028cddebc1dbc220736490a265e2
a336cc884f33aff84c31f6f1887e02f3eff8edf3524ed03ad19098895f55348dc6eef5206e61a0394248d932f1b5e5f6db79
1bc3ddcf218e36fc108e63a95daf9d4c4366667e57d3341ddbd803ba8541e7682195fbdca440a0b18dfd83b88681e6082165
2bef788efc9ff4b2ede85eaacf57b886fcb675ae8ba974a621b3bce37955d3341ddbd803ba851f748e1642b9017679293ab6
b107740b83ced1422a67df9bbd2b3ab6b1efa1bb30e81c2da47203ecdc54746c630fe816eee91c2db472032c625774ec640f
e816069da38554ae2e620fa79d734b76ec7bdfd46c3dc37d7c5c14742f3f919986ccba88fd4d4dd374ec640fe816069da385
54eeeb25370a8d6dec41dcc240738490b27dda4a4271b1813d705b185c8e0a52b3111a4f0b3ab6cf07740b83ced1422a97bb
f68bfa6c608f8f6d61d4774623c7fef58a3d85c6eef5206e61a0394248d9befd31092ef6ad7fe0d630b81c15a466036cfa51
746c5d0fe8167ed0395a08e52af49aa62869a3ce7a2fc988039cc217345dd3bdf856661a322bf49b9aa6e9d8babe87eec2a0
73b490ca7d7d5d5468ec5b0fe216eed11c21b46c03b4df5174ec5b0fe816069da38554eed9b71e1ffb674b78d4774623c75e
b53ee730015fd2e19fee78d6634fab3843f93d1c23d7bdec57661a32abd66f6a9aa663d77c40b730e81c2da4729fffa4041a
5be683b88581e60821651be072a1e8d8351fd02d0c3a470ba9dc00fde9141dbbe603ba8541e7682195fb7ac3a84263d77c10
b730d01c21a46c82383ec448ef84d30ff1d9a24e661a3205b35bd3341dbbe63fa06b18748e165239017d066de6682b9a0eca
29dc6e86d3c0aea0ef4c43a680766b9aa663d77c40b7f083ced14228a71ac8e7370b733e2cc37dc1765d9b6cfc7c671a3255
0379afa6693a76cdefc6db8741e76821951ba1dbb5a063d77c40b77047e768e12857a0e30bb4694e6dc7cf7dca96b16980fb
7a9db169cefdac40661a320bf4ab9aa6e9d8381fd02d0c3a470ba95c858e27f5333d810eff7dede1f29216a61d47ea9df3bb
5bb3e84c4366857e53d3341d1be703ba8541e7682195abd079d94e9c2c86b3ecb4a7bd33e90c143788ed5773a4959986cc0a
fda6a6693ab6ed07740b83ced1422a97a1af68cdb1ec6b1a5f691bbda6f1fda62b7ccedd8b5a261a1233f2ab8aa6d1d8b31f
c42d0c344708295b250e33a76d4df3c330695af2ed581c60f8a6cedf7added4575a621b332bfa9699a8e3dfb01ddc2a073b4
90ca15e8790e73d35fbacf0c3fbbdf326d7b5ac4744eb933ea5a8f6d9d69c82cd0af6a9aa6a36300a05b18748e1652b90abd
9e61aa107f6e67ffffa7fb9a5391ffcd33e455e0bfeb99e682570068ef28a81c0da46215b6bc57f92f2e7499b6f338d7d498
209c68788ba1330d9915f84d4d936c702a00f11d059ad6416956798f30ba29adee08d3ff6bd98e2ded220933c525b6d9bd37
ccea4c4366e57d53d324197c121ebc250a32ad82526c80fb66e57240ab847e0ed985e386dcd42ae17f50c4097745be7e79a0
d0e89300e216069a2384946d04730841479f0440b730e81c2da472033c1e517474697840d730e81c2da472d51ce29ffab44a
c0c7b6f0a3be331a31f6fcb1a23ecd0af0b12d8cface68e4d80778e5aae8685900e816069da385546e8473b6a0a36501a05b
18748e1652b901de99283a5a1600ba8541e7682195abd0f75e887b7b6bbcaa84d87a8509db799fb555a221b122bfa8681a8d
9605206e61a0394248d9ea72a8edb7c41e94718047598319c7b71ed776aeede7acf2accfab4ba1feae679a8b66053d6e1706
97a382d4acf667c93e1179aa9036edcde90f6152be85d9f0725fa164a2f589b53bcb8b8aa6d1e85400e216eed11c21b46ca5
394bdec234e7d765b11df85c1a8c1cf7438dd29c45661a324b739657354dd3d1a900d02d0c3a470ba95c59ce194ee4714f71
5a83b7addb74acbff48a21fceeebc39bb29c53661a32cb72ce57354dd3d1a900d02d0c3a470ba9dc000b77151d9d0a00ddc2
a073b490ca8d60cf20e86859f080ae61d0395a48e5065897aee8685900e8167ed0395a08e50a74b964a6a36f0f039df63935
9f08f3c5e33ac27cb55e9c75a621b340bfaa699a8e8609806e61d0395a48e5b24f83a84fcb027c6c0ba3be331a39f6b12c0b
00fd673f7ed0395a48e5c6b22c00f49ffdf841e76821951bcbb2a087febb1f3fe81c2da4720334ac5374b42c00740bf7748e
165ab932f54ddb24d7f414789be690bba45beee537951b933af5558986c432f57d53d1341afd0a40dcc240738490b2d5c97e
38afae7b9abc0459c22cf598d28ee03575a8d8e2d3b33adb579986cc3add7f53d3341dfd0a00ddc2a073b490cad51bba7843
7fe576f3e1c433a555a771807bb8d94c2fb8ee5b3a9569c8ac37756f6a9aa6835f0198ef28d81c25a46e032c39146cf02b00
f11d059ad64169f6e9ca74c104b3820769898249f32bad4473e45f1a4c7eb2b01fdbbe87fb20b526e94e34248ad6c85e4593
58302b00ec1d7d606909fe954bacc00a39e12a9ab794ede1123aa73b4eb502ebce34648a15586e4d9364342b00700b834deb
a0341b60a19d62a35901905b18708e1252376556505a27a5f758dbef3ad7ed9c4f057d671a32955d8157d3341dfd0a00ddc2
a073b490ca3dfd0afa8ffddb0a00f59dd1c8b10badc324748d8b5ae2a46c3fc38f619dd6537694bf330d99426bb7a6693aba
2500ba857b3a470bad5c366910f56959808f6d61d4774623c7aeb44e5b43d30533b6a80e1fb0fce4615d130d894a69a7a269
34fa1580b88581e6082165fbda7644a1d1ac00c42d0c34470829db00276c4547b30240b730e81c2da472ca9622bd67dfd27d
ce76c5ffb7efa7b41fb9330d99ca96c2ab699a8e7e050fe81a069da38554ee6b4f2c8546bf0210b7f003cd1142c836c05b5e
4547bf0240b730e81c2da472239834083afa1500ba8541e7682195abcf05c254780953e17493735ebf2bdc6da58755d3b25d
d7d4ad05979986ccfa5ce04d4dd374f42b00740b83ced1422a37c0332f4547bf821eba0b83ced1422a579e6e86d9ca6fdbb2
b740ecbb7ba62683d168a0ce91cbc34d99687d6279b6f9a6a269349a1580b8857b3447082d5b265ef630053fa6b4dc71c94b
e4a6dc52eab659c8c43ad1fac44cfcaaa269349a1580b88581e6082165cbbdb8d2eafbdc97354cc4cb5bc9f88775cbcfea0a
b14e3424e64e5caf2a9a46a35301885b18688e1052b6dc43f0b8e62b9c4497dcef2fbd9f3dd2d9362ea229abe5720f419d69
c8cc3d04dfd5344d47a70240b730e81c2da472a50d6898085e71e5611ce039974e70f10fd71aceb1fb59bf6727d39059da80
beaa699a8e4e05806e61d0395a48e532f41cbe8729eee38a63dac30f2eccce4bd3d2df1ea6c64b5defef641a3233f4bb9aa6
e9685600e816069da385542e77320ef7d0c9b12c7d11e1f6fe9ace333d879d8e351c985b5d84e5245a9f983b19bfaa681a8d
4e050fe21a069a238494ed7fd5a9004c7f9a003c06efa00a619e4e05f8d83f4d0050df198d1cfbd3a9001ffba70900ea3ba3
91637f3a15f41ffbb70900ea3ba391637f3a15e063ff3401e8eb3ba3d1637f3a15e063ff3401407d673472ec4fa7027cec9f
2600a8ef8c468efde954808ffdb30d3fea3ba39163cf1f2bead32b001fdbc2a8ef8c468e3d7faca8cf86fdf8d816467d6734
72ecf963457d76cd7f7c6c0da3be331a39f6018c71141d1be703ba851f748e1642b90c1d37cf84cb5e9ef6c42d60d7efb765
f786bb8d5f86d69986cc0cfdaea6693a36ce07740b83ced1422a9767156b69d198a63d6bdbf6371d3118667ed51f41261a12
f3ace25545d3686c9c0fe216069a2384942d139fd79a77adc631a515efe12292dbbf97bfa8c83ad3909999dfd5344d87c6f9
3d738b82cd5142ea96e7cbeb72a6674cd9be2159fd2e739ae825cfccd883b5cc9775a62133cf97dfd534c9d637ce07700df6
605a05a958f1820867d370b4e7632eadc4c9f3f8f46ee4f8d5073a32cf90579c205ed433458496f9e0bca34092f442a7010c
2004175ae683f68e024c6ba0f4cab7f7418a650ad38274fd3893c1eb919d66ae6b99ce70af58dd1775a62133dfdebfab6992
8c2df301dcc260d33a28cd32f19e1f9faef960cb3fb1f48469ba97a36460996848ccbcaf2a9a0663cb7cf0b630c81c19a468
c516613ee668027ae4af608f45f300d37ab2e5de2ae9641a328b2dc2ab9aa6e9d8351fd02d0c3a470ba95c79523717dfd3f4
2c2db5ff99934b647e561c3b9cd5477532d390599ed5bdaa699a8e5df301ddc2a073b490ca65e870cb183d397227fff03b9c
96f83c2d3d5a3cf3b3e4fa4deb4c4366867e57d3341d1be73fa06b18748e1652b9fc14baad85c4a689f8cc38fa3aac61865c
1e43eb4c4366f1097e55d3341d1be703ba851f748e1642b9b27e7c3a8f30ef4b630aff3347efd3341f9ee3943fce90ebb377
9d69c82cebc75fd5344dc7c6f9806e61d0395a48e5ca1b9670f19ce73dcd01d7e9572e3371b4cb7cc557eabffb158bcc3464
96772caf6a9aa663e3fc1eba0b83ced1422a37c0bb3445c7c6f9806ee19eced1422b37c05b5345c7c6f9806e61d0395a48e5
eafbf1bacb2d6e674d0bbf735f91f5e16de3641a32ebfbf137354dd3b1713ea05b18748e1652b911dc02041d7be703ba8541
e76821951b60bfbba263077d40b730e81c2da4725ff7ad55686c9f0fe216069a238494edd93e1f1ffb677b78d4774623c75e
855ea62ddcfda435b96b982c6e73dccf1a65792e9c939986cc2af59b9aa6e9d8411fd02d0c3a470ba95cd65ad46727fbc7c7
d630ea3ba39163fffca016686c630fe2167ea0394208d9eab92bbdc54e0742ecb3953c87d369e60833a97096b91728aa4443
623d73bda8681a8d3dec7be22e0c344708295b25de7fd779a69d57b1db44b4c84b96876bb82b0aa3fbedf72e279969c8accc
6f6a9aa6631b7b40b7704fe768a1951be0baace8d8c61ed02d0c3a470ba9dcb38d3d3ef6cf1eeda8ef8c468e7d84034cd0b1
8d3da05b18748e1652b901e6408a8e9dec01ddc2a073b490ca0de001a2e8d8c91ed02d0c3a470ba9dc00adda141d3bd903ba
8541e76821951ba8933d88ff6cd30e34470829db00fb35141d3bd93fa06b18748e165239d5c9beb4edd9d34199979fef72bf
c69d69c8549decbd9aa6e9d8c91ed02dfca073b410ca89a6eef3718589da9a1e1486eafb19ae3a6af7554bb43e51b474f72a
9a46631bfb6eb07d18688e1052b6119a9b0b3ab6b107740b77748e168e72b5eb75eeb6955adb9e6182b6cde1708c03fc6d73
9ea8ff57bb5eab4c4366ed7afda6a6693ab6b107740b83ced1422af76d0368c1c506f6c06d6170392a48cd0aee52160dfd17
5feaedf9b970fa152e8ff397ce346416e457354dd3b1813da05b18748e1652b90a7d84416d694cbf799bd3faebd4927cfd61
d5af9369c8acd06f6a9aa6630f7b40b730e81c2da472057a9d425a5cfa1d5dded7700fb0cca9af46749358b76dab8b929c4c
4366817e55d3341d7bd803ba8541e7682195abd0bf759db7b469e6b78653ea1a0ec734c0f8ec3bb648babf699969c8acd06f
6a9aa6630f7b40b730e81c2da472153a3de54f4d12c3cd507b93791e4b38c1aedbbd9f52671a322bf49b9aa6e9d0c91ecc77
146c8e1252b7821c46732d53f2738fafe9b77dbfcef4eb3bb7dc24f4be54c94c4366417e55d3241b3ad983f88e024deba034
1be1b8fe970c9dec1fbc250a32ad82526c80fb66d5849e9decfb6924c2b1b3f591c37d11277c17f9b4f58ee2621b7be0f6e1
8ecb51416a3642bb6f41c736f680eec31d9da385546e8067238a8e6dec1fd02ddcd1395a48e59e6decf1b17f76887fd47746
23c6fe6c638f8ffdb3473bea3ba391631fc19e41d0b18d3da0fb7047e76821951ba0c181a2631b7b40f7e18eced1422af7f5
cb4085c61ef620eec31d9a2384946d80b7448a8e6dec01dd873b3a470ba95c5da7138f80634fd791643e189712c5016e475c
c675ddbdaf75a621b3aed37953d3341d9bd9f7d00877748e1652b9ba22ab3cbb49d791e5dc97f3979a71ae7338db2efb752f
43d39986ccba22eb4d4dd374ec670fe83edce81c2db4725fb73155686c660fe23edca1394248d9ca6ac37c564dc7e17a1c7b
de761407b8d767b475b5a1cc346496d586af6a9aa663337b40f7e18eced1422a573af82f57483cb7ec1d31879bbd3ddcfdc5
0186e126f3a73ad5d69986ccd2c1ff554dd3746c660fe83edcd1395a48e546e8e02fe8d8ccfe01ddc21d9da38554eeeb1e2d
0a8d9dec41dc8781e60821641bc1b340d0b1933da0fb7047e76821951ba1bbbaa0631f7d40f7e18eced1422a57bee925cc21
623b957424ceebb1c5e658e9489cc2e9e78a1b29cb372d330d99e59b7e55d3341dfbe803ba0f77748e1652b9021d6ee3b7f3
ca4b9aa7daf23d0df0174e7fe7713f08d39986cc02fdaaa6693af6d107741feee81c2da47263f5d1efa1ff6e120f3a470ba9
5cb948ff7ef31c06992ea87bdc091b1b65a75944b8d2cc67fc57e5222d330d99e522fdaaa6693af6d107741f6e748e165ab9
321d2b5b5092c7531867382ca7f4d06e0927a2f3bceab95be7599757a6622fea99e662137de0f6e18ecb51416a566f32a29d
e671e52774eb74a45dcff9cabaa78da4f7334f9969c8ac37196f6a9aa663137d40f7e18eced1422a576f27cbdefe38c0fddc
e7635dd368d7f83b3ce66d6dbe682ad390596f27dfd4344d8726fa60eea21d9ba384d4ad3e36e81e3fffcef2ae30dd1ecc61
96b86c6d7b8fcc3464d6c7066f6a9a6443137d1077d10e4deba0341bc134e05f32b4d27ff0ded18e4caba014ab0fc2bef2e0
155868a50fd82e0a2c2dc1bf7229df803d9c51630798ff9225619a2d4c6a0958cb34642ae700afa64932b6d207701feed8b4
0e4ab3019a142b36b6d207721feee01c25a46e027a3dc2293d4c07d3a3e8786086db9eeda7a0ef4c43a680766b9aa6632b7d
40f7e18eced1422aa7d674deab77d21ea06339c22554da44dc99864cb5a6d3ab699a8e8dfc7b68843b3a470ba9dce7add605
1abbf883b80f373447082ddb00ab58151d3d0400dd873b3a470ba99c80bec21097f3975632edbf6bdba7659e95e14bcb3464
0a68b7a6693ada0800ba0f77748e165239f54dcf7bb8c999d394293e9fcb2d2de4375d330d99ea9bf66a9aa6a39300a0fb70
47e7682195fbbfcc9d49b2c4b69265e758855690c6be5941cd6b1db9ff71a12570f0ae8b9c9402324bfb7a9e2e27ce8d0812
20c1eb039cc5141d3b0900ba0d37748616523975bd9ae7eb9afc2f214e16c3dbefc1e1425eaf4aa643a6ba5e59359da66327
810eba861b3a430ba9dcafdfba5068ec2400e2360c34430821db001b48141d3b0900ba0d3774861652b901de705674ec2400
e836dcd0195a48e50668b6a7e8d84900d06db8a133b490ca953b42bfeaa1aed0d846a02546b841338490b20de0f7a2e8d849
00d06db8d2195a68e5f2039b6bf60bf9dcd73d6c520c7b4ea3b356ec75bececf96659de990991fd87caae9341d9b0900ba0d
3774861652b9043daddb150c1f535fe2c70f7d0abb05cb73d404ad331d3213f4b79a4ed3b19f00a0db7043676821954b1665
fedb169fa284db56737ab0b02543b5f0f6fdb44de5b55f23d3213359947dabe9341dfb0900ba0d3774861652b9e439b8edd7
325d6bdcb935f991c647c3e17b394ff93179fe4d1b990e99c975f05b4da7e9d84f00d06db8a133b490ca65c7d0630d2dcfa3
616bb8fd3c4f53fc264ed7b38f2d5b86aa4487c4ec19faa5a2d3686c2600e236dca0194248d99209ee7e6cd199389a22ae8d
6fde74fa09d5eebf8df9169191e990994c70bfd5749a8efd043ae81a6ee80c2da472c9e0d92f69b7f02663fc1e4ef3b5eddb
96fed84b978fe4efac135d9b98ec9d3f55741a8ddd0640dc868166082164fb0f3a2888c1b3950198da70337803550ad3b732
c0615fbb04a0be311a39f6be95417bd8e3b54b00ea1ba39163ef5b19e0b0af5d02dafac668f4d8fb560638ec6b9700d43746
23c7deb732c0615fbb04a0be311a39f6be95010efbda2500f58dd1c8b1f7ad0c70d8d72e01a86f8c468ebd6f6580c3be7609
407d633472ec7d2b83eeb02f3efda86f8c468ebd1cf64f7d3613c061db30ea1ba311634f8715f569e78fc3b6e1a6be311a39
f65f5fe2141aedfc41dc861b344308295b223e9b7d5f93fff24d979fe5c5a9d67efb39e7beacf98ea991e9909998bfd5749a
0e76fe2d731b6dd80c25a46e69f2e62f18c7115e730d235aee7d39e7298eee78cc93d2dc4d263a24a6a9dba78a4e82b55efe
a0adc14aa525907225543f5b9e8e391996afd7e957e57ef512a794cf3682842a131d1213eaa78a4e41c1ce1fa84db4a19202
08a9d21a640913e4f38e1bd9a63ddcd4bacef4053cfde7e1d78de5bd5123d32133ad42bed574920b76fea06da20d98d640e9
f5eb3597c0a2973f68db7003a6455082a575b5ff7a6fdb3927177a7fa54a6fa2873feed445a2bc326a643a64a675f5b79a4e
b3d1ce1fc86db881339490ba0d703341d1d1ce1fd06db8a133b490ca25e8bbbd78dc615ffd7c9fa9dfc012dee49e977cce32
321d3213f4b79a4ed3d1ce1fd06db8a133b490ca0d70af4cd1d1cebf83aee186ced0422a97a08fd2e9268ce9d8a6db2f6ab6
702d99c37968bdf7f26cc7c874c84cd0df6a3a4d473b7f40b761d0195a08e546e86120e868e70fe836dcd0195a48e506b8e9
afe868e7df4223dcd0195a48e5f236f63cbef8c4c9cf15f67b890e147ee280f3984e7448cc9bd8bf54741a8d5efe206ec315
cd1042cbf6d33dec8a8b2efec06dc30d97a182d46c801765151d5dfc01dd861b3a430ba9dc00cfa6151d5dfc01dd861b3a43
0ba95cd9d5fd4c8d9f368cf14b992e29f3e32226131d12cb9eee0f159d46a3853f88db7083660821651b601fbba2a38b3fa0
db7043676821951bc0a858d1b18700a0db7043676821954b76faa23e5dfc71d836dcd4374623c7fe73737981c606021d710d
3768861052b6d2b7e04f7dbaf8e3b06d18f58dd188b18ff04b167434f26fa1116ee80c2da47203b895283a1af903ba0d573a
430baddcafaf520a8d2efe206ec30d9a2184946d848f59d0b18700a0db7043676821951bab8700a05f0df241676821951be0
d2ace8d84300d06db8a133b490ca0d60ada5e8d84300d06db8a133b490ca956dcec1062b3e4af19f48586b9e6961109b42fb
ef627d6d41243a24964dce1f2a3a8dc60602206ec30d9a2184944dbd6497dd53d674ce0df73142ae7ac9ae643a64aa97ecac
9a4ed3b18700a0db70436768219513d07bb95317bf947e78e1f7275f5a78321d3205b459d3693af610e8a06bb8a133b490ca
a91791fce5f3da8f78c6f1a7a1e41524df5a78321d32d58b48564da7e9d84300d06d187486164239d143600dee85cb19df50
0f1ba3173f699f55b7889ae99029ba0898359da6631b8166bc0c3774861652b9e2a83f6f4b7a327a1ed9b431fce13f90e95e
a7ed718294890e89c54fff4345a7d1d84300c46df841338430642b76fac77c1da9e5a43fa5decb766cd113cbcf1f97233e6a
2bc832d321b358ea7fa9e9341d7b0800ba0d3774861652b911fce5051d3b0900ba0d3774861652b9115cc7051d3b0900ba0d
3774861652b9117cc7051d3b0900ba0d3774861652b9027dcda7ffcec54fa5b82aa70617fee7b71ec135bc40ab4c87cc02fd
a5a6d374ec2400e836dcd0195a48e532f49e7b1d84011ed9333b75b8d88e7bda9f67b146a6436686fe54d3693a761200741b
6ee80c2da472e1e694ac0f2f7f1cb58936d58db1c89117a5d33f5188e33af76b4eee3dfe22e3e7317770882f4aab4c87cca2
f4979a4eb2c1cb1fc44db441d33a28cd46f866fd2583977fc7fb441b32ad8252ece72ec9c283bf37f2af13b91ade6fbf6299
0fbf68f8dfb688116e8b8cd0bd40d0f576fe0df413269da185546e343bff06facdab9e74861652b911bcde055d6fe70fe81c
269da185546e8057d0155ddf4ca0817ec23d9da18550eed716570aadef63d0103f61a2194248d952fb0451bfef24d01cf609
b3be311a39f6df7641115c7d0f8106f70993cb50416a36c2b542d0f56d041ae8274c3a430ba9dc000f2f145ddf46a0817ec2
a433b490ca0d6082a8e8fa360215ba8649676821952b1b8596f002c619efb4873724fcaf206e4a5cc3db117e3d5f9f3bcb4c
87ccb251e84bcdbe8d4003fd66930f3a430bad5cd91276f859839f94c7f34c6aed10b794afd39277763d5bc254a64366d912
f6a566df49a0817eb3c9279da18554eea7dbfe1457df43a0c17dc2e43254909a655cffe38aefc084e19dfec27d859725c2f0
8e62265c8065a6436646fe54b3ef21d040bf19e493ced0422a378027b3a2eb7b08003a874967682195cbd0db7a454fec30c0
f53cfcccf44e1b6eb7f2827e8196990e9919fa53cdbe8d4003fde691dfd3195a08e546b01c17747d1b8106fa0993ced0422a
37428b0c41d7b71168a09f30e90c2da47219fa9cfc92728f4f48573f4e7fc25be3db224bd8631e4fc1055a663a6466e84f35
fb36020df49b473ee90c2da472237cbd055ddf46a0817ec2a433b490ca8d702213747d1b810a5dc3a433b490ca0d60afa5e8
fa36020df413069da185562e439fc739f9ab78fc2a861da1615c7180e7f3842d43cb4c87cc0cfda966df4ca0817e73ca279d
a18554aeccbad7730f5fbbb8d0bdfc69e63ad2efeff647388e60d55466dd2ad321b3ccbabfd4ec9b0934d06f5ef9a433b490
ca15687f350bfac7fb46e7b5cec71497c06b7d1dac40ab4c87cc02fda566df4ca081fe77bb7cb2194a48ddca92323e143db2
3d7afa84e2e9e7b8373f85aedd13449e6bf3ca72f2bd9e93545d1b8186b54409a515506a953b067e99795c5bda795b8c22c2
39e7c8a686cf0d0391e89058ee177ca8e82456d74300b0294a2c2d8192ebd78f890456d743a0812dd11e4b4bf057ae5fdf4d
16587d038186f609134c8ba004fbf56e4e05d6770f68789f30c90c19a468a249845f049c975f03c4abc8b95efe93b9ea4e65
99e990299a449835fbee010df49b353ee90c2da472626363ec3a73a6c9a01fe7761fe7aab6f8d544d7268a6d8d56c5be7540
257eb5c6279a2184946d04d36d41d7770f68a09f30e80c2db4720374c650747df78006fa0993ced0422a37c04e6545d7770f
68a09f30e90c2da472a3750f68a0dfacf1496768219553dd0366bf8edfd678177a5f7ab37199e990a9ba075835fbee010df4
9b353ee90c2da472bf3e752bb4be750088739868861052b6a15a0734c46fbef83d9a2184904dcc494e3f6d9ac3aa320cf079
5eaae6244fa643a698939835fbd6010df49b2f3ee90c2da472bf9e692bb4be6f4043fc8489660821651be0e1baa2ebfb0634
d04f9874861652b9725320ecb0386247962dbe07121e9086016e47f066b99f5bf73ad321b3dc18f852b36f1d50a15f7df149
67682195cbd071d5b3c54dd4e1a272efd319b78de5db93ebf67cd232d32133437faad9b70e68a0df7cf1416768a19573bf6f
deaee8fad6010df413269da185542e5b273d5d00c34de6c713dccfdba6292c4e32f15b9acba649efd5fa8e010deb9b1d3ea1
0c09a460c910ec6c4ccfe7730937dba778cfd54f1353f3f5fc5336321d329321d8b79a7dc78006facd0e9f74861652b9ec77
97ded88bef7d4cf73edfe145dc705199e76dbe8eeb793dcec874c8cc7e779f6af61d031ae837477cd2195a48e5b29d636a29
1ebbc4f995cf1a9640c9793299a76e65739091e99099ed1c3fd5ec9b0634d06f8ef8a433b490ca2533d6f35e8ff94c2fbb4c
c98d6f8e8b3e3fdaf4ffc85f6f23d32133d9b17eabd9370d00f4bf9be293ced0422a97fc85cb6ecb38c03c338e0f55263f81
9a667fd2c9d046a64366f217fe56b3ef1bd040bfb9e6f774861642b9ffb86f40c3f466c9cfc11ba85298bf7d039ac3be59f2
b3be311a39f6bf7d03ea618f374b7ed6374623c7feb76f4073d8374b7ed43746a3c7feb76f4073d8374b7ed6374623c7feb7
6f4073d8374b7ed6374623c7feb76f4073d8374b7ed6374623c7feb76f4073d8374b7ed6374623c7feb76f4073d8374b7ed6
374623c7feb76f000efbef96fcac6f8c468efd6fdf80e6b06f96fc7d7d633462ec7ffb0634877df3c5677d633472ece9b0a2
7e6fdedf1cf609b3be311a39f6745851bfb3cfaf477da2ac6e8c458e7c802bba60a3837e039c8300d32a48c512ed157b32ed
5734095fa6b020bfe24debe978bca112adce74c84cb4df6a3ac5d599e837b4254a30a981502bcd50b7cb2f04fd8711cdc1d3
9adf2fd6e2c0eee39aae3bcf608c44d726a6f9e9a78a4e42750efa0d6a89924a0ba0c44ab0fe1432c7f7ece3d4f998fd7c79
9ad3e22175ea2eb3719de8dac404fba9a29358bd837e43fb8409a6455082e58601fe0ae1cf5ef117762ccbb59e6bfcca5d7b
32bb2fed11549e435e6e17f0a15eef9ddfc0be19c313cbd0402a36c00a53d1f5def90df413269da185542e41efd92f377e0e
fe5717ef6d8565ffe4af1feb3617534d9de89098903f55ec8df31be2375778a2194248d9d22d23fff1c77da8d1e27d5aaff8
55dce29d8ed8ace359531b990e99e996d1b79abd713ea0ffdd159e74861652b904ed979dfe82b82733fff4e0748eef44a52d
a861b35386d6990e9909fa5bcdde38bf817e7385efe90c2d847209da9f5ce37b21e97ee5e3f214fa9d7a9d8e63299b3c8d4c
87cc04fdad666f9cdf40bfb9c293ced0422a97effe4ee7312f5bfcc9f9fff1bfba7389d386393871868944d908a8331d32f3
3de04f357be3fc0afd6a0c4f3a430ba9dc00ef2528bade3bbf817ec2a033b4d0cae58ddbdbe9e70e737c59773d83cbd6b9c6
1e524bfc1afa7f7f5ea393990e9979e3f6a79abd837e03fd660f4f3a430ba9dc00af4b2abade41bf817ec2a433b490cafd74
17b3e2eabdf31bdc274c2e4305a9d9002f3f2bbade3ebf817ec2a433b490ca0df0245ed1f5f6f90df413269da185546e04a7
7141d7dbe737d04f9874861652b9018c89155d6fdedf403f61d2195a48e546f8a4055defdf0fe81c269da185546e84dfb4a0
ebbb0734d04fb8a733b410caa5a605a27eefdf5f0f5bc3ac6f8c468efde756f602ad37ef6f889f30d00c21b46c516855bf77
d06f0efb8459df188d1c7b3aaca8df7bd837877dc2ac6f8c468e7d04e37e41d77bd837d04f9874861652b95fb73c5168bd81
7d43fc848966082165fba9a393e2eaadeb1bdc274c2e4305a999d8667bc511a559f97e6cfbbe4cb7da33ff243a248a4db656
c5debabe217ef365279a218494edd79bc7155aef5bdf103f61a2194248d906783142d1f5bef580ce61d2195a48e58485fb9c
5c54e37aebf073167f92bdf64340d74c874c61e16ed6ec7deb1be83753f69eced0422857a08f343b0b03dce22d85690aa7da
639bef65099e070fb4ca74c82cd05f6af6bef50ff4bb2f3be90c2da472c5b73e74db5ba3c1c1792d4b5a898401de693bd9f3
e6b1ce74c82ccef55f6af6d6f50df49b2f7b4b67686128575cdcd7fd38aff8f4f9f2e7d464a8ff4f6c2db06ef7149ad71717
7795e990595cdcbfd4ecadeb1be8375f76d2195a48e54670711774bd757d03fd8449676821951bc1d05cd0f5d6f50df41326
9da185546e049b6d41d75bd737d04f9874861652b902ed27fcc71caf28d7190c9b8393461ca05f24c4addf0fb4ca74c82cd0
5f6af6d6f50df49b2f3be90c2da47219fa58a77b9ed3efcfff00fd6af388f716aeee95279de8dac48cfca562ef5bdf10bf99
c213cd1042ca5688538390f8298425df326df1068a9fc9f82fe5bd3e0fc874a6436661fe52b377cd6fa0ffdd179e6c861252
b7821cde20cd230aad73c3ced974c68de7dbf9797f51673a6416e42f359d64eb5cf31be212259ad64169f6f32ff55facce32
1fb0294a2c2d81926b809b8ccaf01eaef9984636e1eb7f7c9d734ae15ac40837457e6d00a3d060994fe21a069a218494edb7
56d7820b66f9c4ad6170192a48cd8632cbefa1ffdd099e74861652b95fbf7facd0e0944fe21aeed00c21846cc9375ed4875d
3d0f5bc3a86f8c468ebdb3abe761df1cd959df188d1c7bf97c7fd89d53d1c1b49ed0350c3a430ba9dccf4fd5020d8ef524ae
61a0194248d90678d6a9e8e0584fe81a069da18554eed7cfdf141aecea41dc8481660821652bbb728a276fd845938d2da3f5
617963f64196990e996567ce979a4ed3c1ae9ed035dcd2195a68e506f0905474b0ab27740d83ced0422a37805ba8a283693d
a16b1874861652b9df7af40b2ed8d513b786c165a82035cbb8e9cab96787cb7df593a578d37dd9cb7387022c331d3233f2a7
9a4ed3c1aebe872e61d0195a48e5c6f0e8ff4307bb7a42d7704767682194cbd0e53d8c785af55387cbaf74e31ed7e4be7e3c
774374a6436686fe54d3693ad8d513ba8641676821951bc1cf5bd0c1ae9ed0350c3a430ba95c9ce6abfa308ce7616b18f58d
d1c8b167ad9bbb617e0ab34d87d729daf86f9b3f19f8b576b9c3a8331d32b3d69f6a3a4d07c37842d730e80c2da472239c4a
041d0ce301dd8441676821951bca309ed06f6ee8a033b4d0ca65e863be7de29c561fe124eb67a5f15339eef094f85ecba944
673a6466e84f359da683613ca16b1874861652b93c05f4b3a47bf653b3b8cc8ce6f67ee51d06786dc77e7ba91ecf3099e990
99a7809f6a3a4d07c37842d730e80c2da47265de1b6e11a5a7747e3d301dd7141a4c8739eab99fe7b53f0dda75a6436699f7
7ea9e9345d6b184fe6270a364309a95b414e2f7cc469f9b6f829fa993e93753e4367a5fbf945eb4c87cc82fca5a6936cad6d
3c899f28d0b40e4ab301bcf00459eb1cdff3e628c8b40a4ab101de2d1164ad793c799f6847a655f8abd800f7a20419fce309
5cc360d33a28cdcafdc61f36c9546cb09027720d03ce5042eaf6eb27070a0dfef124ae61a0194248d9ca5ebb1f36165674b0
90077413069da18554eed7db85151afce3495cc32d9a2184966d80de088a0efef184ae61d0195a48e594ddf47e1f47d82617
af2efeece367c987dc31fc643a642ac369aba6d374f08f27740d83ced0422af7eb8de10a0de6f124ae61a0194248d904f19c
6ffcc7efa19f23ad73d81ca2909f4c874cc16cd6749a0ee6f184ae61d0195a48e5d40fdaffee36ff477cf33abc6f73875db0
f2075d321d32d50fdaaae9341dfce37be812069da185546e04377541070b7942d7704767682194fb75231f8506ff7812d730
d00c21a46c651676dc7e54f18668d9d67ba5856e9a31d637f565a6436699877da9e9341d2ce4095dc3a033b490cafdfa1d35
8506ff7812d730d00c21a46cbf6eb5a6d0601e0fe2260c344308295bb94b50ac9bc28fcdcf4ed7fb4cfb0143eb1d5f627f7e
cb32d321b3dc25f852d3693a98c713ba865b3a430bad9cfb7dd76e4507f37842d730e80c2da472d94967f653fe3319092cd7
be5ffe8417e7c64bdcd21bfe9f195a673a6466279d4f359da6838b3ca16b1874861652b9043da54e5971a9b74c7efdb78709
b2ff236e445882555b86d6990e9909fa5b4da7e9e0224fe81a069da185542eb963655bff78636e5ee6699dfc17f28e7fec8b
5f16cde50d3d23d32133b9637dabe9341d5ce4095dc3a033b490ca25bbbb353e8fbfe2969ab81408bb3ef396b9e473f76f19
fff7ffbcd7709a026ef184ab615018cc52a1e4d758b6a547b3cc34f1f767d238ca708f3d3c3f4b9046a64366f26cfc56d369
3ab8c5f7d0250c3a430ba95c725b3d97fd5ca62b6eabf053e0e0ed185e940ba38dc68bcb9a3f5923d32133b9ad7eabe9341d
dce2095dc31d9da18550eebf748b27d39b113b076fa04a613ab7781ef6cd889df58dd1c8b1776ef138ecf166c4cefac668e4
d83bb7781ef6cd881df58dd1e8b1776ef13cec9b113beb1ba39163efdce279d8372376d6374623c7deb9c5f3b06f46ecac6f
8c468ebd738be761df8cd859df188d1c7be716cfc3be19b1b3be311a39f6ce2dbe3fecbf1bb1b3be311a39f6ce2d9e877d33
62efeb1ba31163efdce279d8372376d6374623c7deb9c5f3b06f56ecac6f8c468ebd738bc7615f1cd159dd188b1c793ca8a8
def8b5f39025d896d6e390634e1753bf34998f6d3ee2ac26bd4b30c54740935f249d53d9db6be4b9262f5d5ebfd4738aa875
6a27e7130592a4173a25f3f2b4d49c9247fe721e5798d8242feefdf46be06dcb132523d32133d9977fabe924576bd64eda27
0a30ad81d22bf15e79e75054bef850a73e47db312dc1d236f3ea4c87ccc4fbada69364f06b27700d834deba0344bf3ffedcc
efe4c6ef5ada3695e6b07ec57dfa39c1946f3f1a990e9969feffada6d36c706d27720d03ce5042ea96a0eff4354b4ef2fed7
b5860f228ef67e6eaf24689de99099a0bfd5749a0eaeed84ae61d0195a48e5fec365aca280413be16a181406b35428c15df9
06606a45e76794fe82bc2533f5e7f67782d4990e99c994fe5b4da7e960d0de439730e80c2da472e96ecceae7cffe321c0da6
fd3523bd5b167dc583dfe1b44de539b791e99099eec67cabe9341d0cda095dc31d9da185502e415f61a3c3141e2bc04a3c98
a93fb70c13b4ce74c84cd0df6a3a4d07837642d730e80c2da472e966a3ff12fa59de1a3f8865f56be1e9dee223ad39b40908
0e21e566a3ce74c84c371bbfd5749a0e06ed806ec2a033b490cae5dbcacd1d153f23f2e79ecb5f3bc21f736e91519e99e84c
87cc7c5bf9534da7e960d04ee81a6ee90c2db47203bc52a4e860d04ee81a069da185546e8077e5141d0cda095dc3a033b490
ca95e7634b58cdc4ee175bbed31dbb25aca9c9cb5e7759ca4c87ccf27cec4b4da7e960d34ee81a069da18554ae408737beef
f404fef43fc5b817320c307780991ee31199e99059a0bfd4749a0e36ed84ae61d0195a48e57efd9ab34283473b896b186886
1052b601fc65141d3cda095dc3a033b490ca0de036ace8e0d14ee81a069da185546e844f5ad0c1a3bd872e61d0195a48e546
f8a4051d3cda095dc31d9da18550aef368c7615f3dc859df188d1cfb005f3045079b7642d7704b6768a195eb6cda79d83707
74d6374623c7ded9b4f3b06f1ee4ac6f8c468e7d80f74d141d6cda095dc3a033b490cafddae246a1c1a69dc4350c34430829
db00f65c8a0e66ed84ae61d0195a48e506d8baace8e0d74ee81a069da185544e6c52f7df87743f387e2dfd4960bda743eecb
7f321d32c52675b3a6d374b06c27740d83ced0422aa72cdbf3bbf77bfc5e1eeb321da1edb0b26c2f990e99cab2ddaae9341d
2cdb7be812069da185544e58b62fa96d69dcb377ccd77a2c7edeb209e89ae990292cdbcd9a4ed3c1b29dd035dcd1195a08e5
32f4f37a79b09c7eac14c247943e937abb42663a6466e84f359da683657b3bde360c3a430ba95cf6479df30cf59fe0343ee5
cdb7f123eade9dd3990e99d921f5534da7e960d94ee81a6ee80c2d0ce58a7bf9bacdd3126f0c8766da975fe1c79d6c7ee6b6
1d4bf8d00ab4cc74c82ceee55f6a3a4d07cb7642d730e80c2da472c5bd7c3fe7758ed65ad71e9a829d573ce9845ea5d3eeb3
1f6899e99059dccbbfd4749a0e96ed84ae61d0195a48e546702f1774b06c27740d83ced0422a3782adb5a083653ba16b1874
861652b99f1b3d0b34f8b593b88681660821652bc47b6cb03dc7f1e5fedaf14b598df30ab2ca74c82ccc5f6a3a4d07cb7642
d730e80c2da47205fa886fa3c733ce1d3df497f802c1e5bf8e4bec13f140ab4c87cc02fda5a6d374ad653b999f28d80c25a4
6e61f9aeaab7a6e93ce61345713d1235ea11be597fc95adff49e374741a655508a0d70a742599ed337bd9dcb35e1d32f4f7d
3c856b1123dc14f9f59d0a8546df7410d730d00c21a46c23782e0b3abaa703ba864167682195fbb51d8842a3757a475cc240
338490b20d70434ad1d13d1dd035dcd1195a08e506f0405174f46e07740d83ced0422a37c04bd68a8ecef180ae61d0195a48
e57edbfd4170d1331eb8350c2e4305a959c1f52b0dff658b679b754eabecb8fb63ea6f3aca4c87cc82fca5a6d374b48d0774
0d83ced0422af7eb17e9151a3de3415cc340338490b2fd7a338542a3677c4bdc8481660821651bc06e4fd1d1331ed035dcd2
195a68e506700f5074f48c07740d83ced0422a378251bea0a3673ca06b1874861652b90c1d1d59b6f4a9ccd132ef8cd3a76c
69bb6ecf230499e99099a13fd5749a8ecef180ae61d0195a48e5f236cfb4e138dee15fc3868be0d31bed0dcea7b965dee629
331d32f336cf4f359da6a3737c075dc2a033b490ca65e8ed9ccf6d8e2688fe9b18969beb140dcf8f639976ffef4f7f0099e9
9099a13fd5749a8ecef180aee18eced0422837802388a2a3733ca06b1874861652b9013cae151d9de3015dc3a033b490caf5
cef138ecab2d3aea1ba391631f60c7b8a2a3733ca06b1874861652b911be60828ecef12d7413069da185546e805389a2a373
3ca06bb8a533b4d0ca0d70d15074748e07740d83ced0422af7d33e3a8a8b9ef1c0ad6170192a48cd0698022a3a78c683f989
82cd5042ea56a6fac9a32ddd0ef5ff3b6d73fc639dfcd5655aeab63099e8dac432d1ff50d1493018c603f789824b8ba0041b
c120ff2f190ce33bde1c05995641295616afe982193f0aff359c0f7f528dd7d2d57f10c154f6b9dd28331d32cbe2f54b4d27
c960180fde27da916915fe2a36c05b3d828c86f100ae61b0691d9466bfb6735460748b076f0d83cc90418a36c0a34f4547c3
7840d730e80c2da472bf7ef4a9d0e816df123761a0194248d90678f4a9e868180fe81a6ee90c2db4723f7d52a2b868150fdc
1a0697a182d42c2e54557d9ab5e3b0358cfac668e4d87ffd8059a1d1ac1dc4350c34430829db000f98151dcdda015dc3a033
b490caa56f97a84fbbf4eeb0258cfac668e4d80778c4abe868970ee81aeee80c2d8472bf3d71092e7aa503b786c165a82035
1ba04d8fa2a3573aa06b1874861652b901265e8a8e76e980ae61d0195a48e57e3dbf5668b44b6f899b30d00c21a46c033cf0
5474b44b07740db77486165ab9011e83293adaa503ba864167682195cbbe36e1f1d875a45d154bb1c00b038c8d71b7e59962
eb4c87ccec6bf3a9a6d374b44b07740d83ced0422a971d8cb6f41c3af6e7f467d82bbec1f74f701beaf66de84c87ccec60f4
a9a6d374b44b07740d83ced0422a97bcaafcd5d397d997f0939bfd6f31dd980a7fac97d7e978acaa74a26b139353d5a78a4e
a3d12b1dc4350c344308295bf2613b4a57957f82419c3fb5069ba3e8603f3d0bdfe4c3a6331d32930fdbb79a4ed3d1431dd0
350c3a430ba95cb2133c9b99e17434bd4aa72b9d71b77c1633321d32939de0b79a4ed3d143bd832e61d0195a48e5b23568b8
f7b9a4d6cad3b2e73733d368a3c7e399e7db46a64366b606fd54d3693a7aa803ba863b3a430ba1dc7feaa10ea6577b720cde
4095c2f41eea38ecab3d39ea1ba39163ef3dd4dbc3bedb93a3be311a39f6de431d877db5276feb1ba3d163ef3dd471d8577b
72d4374623c7de7ba8e3b0aff6e4a86f8c468ebdf750c7615fedc951df188d1c7befa18ec3beda93a3be311a39f6de431d87
7db527477d633472ecbd877a77d8177b72d4374623c7de7ba8e3b0aff6e45d7d633462ecbd873a0efb6a4f8efac668e4d87b
0f751cf6d5a01cf58dd1c8b1f71eeaed61df5cca51dd188b1c79e7a18e63feab61785b5a8f438e395d4cfd757ef3a790259a
09cfcdc436380cef5b98d0e7ebaace74c84c97d86f359de282933a689f28c0a40642ad345d5af3d3db38b0755ecf654fd39c
230437bf1acfb32599e89098264b9f2a3a09051b75a03e51506901945809d6ff88e773db633ff3f010623dfc6c367e0ae7be
fb556a79dc6f24ba3631c17eaae824163dd4415bc300d32228c1d2fc7f6fb64df97ae99df76491bd668bf73cffd7990e9969
feffada6d36cf45007720d03ce5042ea3696873aa05f8dc341676821954bcbdb6dbf16ff01c407b6933f7bcfdbbea65e65fe
17b885734ffe661b990e996979fbada6d374f45607740d83ced0422a97bdd5d30b21d1b0c47f10f379df6b327f9fbd4eabff
05e6c70c46a64366f656ff54d3693a7aab77d0250c3a430ba95cba75b3958b47bab7e2cfadfb1c6d75d34ed0f0d43edfbbd1
990e99e9e6cdb79a4ed3d15b1dd035dcd1195a08e512f4edbff5875f64c77b68d77e2cd33da73b4dc18e693e4b0b719de890
98903f55741a8dc6ea20ae61a0194248d9f26dc9b03af5a38c1d1757bfac58d7353e4d8d37d3d6fb2eed2f8c4c87cc7c5bf2
534da7e968acde423761d0195a48e5f20d68ff15dcce3b767158a77283388c365aff9fc754eec5ea4c87cc7c03fa534da7e9
68ac0ee81a6ee90c2db47203ec185774345607740d83ced0422a3796b13aa05f5dc341676821951bcb581dd0afaee1a033b4
90ca0db00553d1d1581dd0350c3a430ba9dcafdf055668345607710d03cd1042ca964c98457d5a9be3b0358cfac668e4d847
30bc1674b43607740d83ced0422a37c0c60b45476bf30eba8441676821952bd0fe8276a5bedbfb7c84079bdb1d461b7ab65d
53d863f640ab4c87cc02fda5a6d374b43607740d7774861642b9dedabc3decbb6f37ea1ba39163ffb5798342a3af39886bb8
453384d0b2f5bee638ecab6538ea1ba39163fff5366e85464b7510d730d00c21a46c03ec945374b45407740d83ced0422af7
eb0dc60a8d96ea20ae61a0194248d984bbb85fe34e7378c52ece514b5b83822c331d3285bbb859d3693a5aaa03ba86416768
21951346db73dcf1bcc52dbafe7ce7ffcdcfcba5d1f693e990298cb6cd9a4ed3d1521dd0350c3a430ba99c80de9bc665fea4
be4dd77149e6bde985561305b255d16934faa983b886816608216513c4579c2ac699d27e9ee951cea9884ba243a220b62a3a
8d4633f58eb88481660821651366ea6bd8d6efcfb7d1037cb9fddadaaf39956d7ecd74c81466ea664da7e968a60ee81aeee8
0c2d8472d94274aa4baed3ff331f4bbc8572dc93ffe1dddbf322ac4c7448cc06a25f2a3a8d4627f566b06d1868861052b64c
9c6f0f46e3d8d9ffd0fc6710dfbf3fb72d19bf3f0ec432d32133337faae9341d9dd4015dc30d9da185a15c31155f9778df24
0c704beef6f1aa725e53eabef97cce32d321b3988a7fa9e9341d9dd4015dc3a033b490ca1553f1fc550c03dc6f7f1dbdcf38
9738fd14263a0d94cbb3ce74c82ca6e25f6a3a4d47277540d730e80c2da472c513381af9474f7f7f623d96f58837482f7fe6
39f76b7ffa35c94487c4e208fca1a2d368b45107710d03cd1042ca568853bbebf83dbc52938ef828fe4a065bf3fe7ccc32d3
21b3307fa9e9341d6dd4015dc3a033b490ca8d60fd2ce8e8a40ee81a069da18554ae4087db526bbcf77ddd7113c2961cfdcb
1ee5fd8156990e9905fa4b4da7e9e8a40ee81a069da185546e2c277540bf389583cd5042ea36c297fb2f1b7cdc41fc4481a6
75509a25e7f8bfb5e1a4de1d3147515b8f438df9d7f7be94033a6dd4db895c133efe679beef002d9ffa288116e8a8ce04f2b
e8e8a40ee81a069da18554eeb71ec4828b1eeac0ad6170192a48cd7e7d074aa1d143bd232e61a0194248d9e21d5d559f2ee6
386c0d77f58dd188b1a7c38afaf411c7616b18f58dd1c8b10ff042bfa2a38f38a06b1874861652b9a4b5a84f3f6f1cb68651
df188d1cfb00b7b0151dfdbc015dc3a033b490cafdfa498542a39f37886b1868861052b6011e2d2b3a5a7ab7d04d18748616
52b901766b283a5a7a03ba865b3a430baddc00fb72141d2dbd015dc3a033b490cafddae74fa1d1cf1bc4350c34430829db00
7bce141dfdbc015dc3a033b490ca65683fd3f60bd8788f6a8bcbf9253e3e5aae296f232acc2ad12131237fa9e8341acdbc3b
e212069a218494eda716a58a8b36dec0ade18ecb50416896714373d6f02c210cef0a497e2d1d8d078ea7736b0696990e9919
f9534da7e968e30de81a069da185546e00ef5d45471b6f40d730e80c2da47223f8690b3a9a8803ba8641676821954bdee5a2
3e6dbc71d81a467d633472ec036cca5674b4f16ea19b30e80c2da47203fcaa141d6dbc015dc32d9da185562e7ec1547d1a69
e3b0358cfac668e4d807386d2b3ada6903ba864167682195fbf56444a1c14b1bc04f1460860c52b411ecc3ffb2c14e1bc44f
14685a07a5d900336c41063bed8e374741a655508a15dec3af36eef89c7fcbadaae2b350ff35cc2ee00fafca74c82cbc5f6a
3a49063b6df03ed18e4cabf057b101da400932da6903b886c1a675509a0d607ea7d8e8a80de41a069ca184d46d80fb218a8e
8eda80ae61d0195a48e50af431edfb1a2764db911df9e2f7702bdb961e6895e99059a0bfd4749a8ea6da2d7413069da18554
6e80db7d8a8ea6da80aee196ced0422b576eec36cfe7fc27b2f94b66f46f580fff9d5c97ad9eb565a64366b9b1fba5a6d374
b4d606740d83ced0422a573e693f150f9d73e200fd15c62f35e32687d55f43d7653ab6fa9b56990e99e593fe52d3693a1a7b
03ba864167682195fbb959ab40a3b137886b1868861052b65fbf78a8d0e8ea0de21a069a218494ad10efd77d9e71caf0bc07
104f37979f40f8f9d3b3ad4f673a6416e62f359da6a3a778075dc2a033b490ca152bf33ff5e9ea8dc3d67057df188d18fb00
8fc4141dbdbd015dc3a033b490cafdfc2422d068ec0de21a069a2184946d840ba4a0a3b137a06b1874861652b911a642828e
dede2d7413069da18554eed70f01151a8dbd415cc32d9a218496eda74f8714172dbd815bc3e03254909a25dc23ece5dccf78
aa59fd3c69f35fc73bf9bf3c1b5d13b0ce74c84cc8df6a3a4d474b6f40d730e80c2da472097adbe77b5ae73867590e7f45f3
038e864ef3ed97806778af2b43eb4c87cc04fdada6d374b4f406740d83ced0422a972c92aeb8237b8beef25e97c99f5db6f8
36ffbcddfe17b1f83f13b491e990994c92bed5749a8eaede80ae61d0195a48e59203981f853fab5e6772eb5a97ebf4d7f0e8
68342f614fd5bc14576f9de990991cc0bed5749a8eaede80ae61d0195a48e592c1ddb54fdbb6cc6bf29f6eb661865d62e11d
e432e335321d3293c1ddb79a4ed3d1d5bb832e61d0195a48e59273a39f1fa4218631adcdab467e9c7e9278dd79e66924ba36
3139377eaae8341a2dbd415cc31d9a218490ed3fb5f406d3ab5b36066fa04a617a4b6f1cf6d52d1bf58dd1c8b1f796deed61
dfddb251df188d1c7b6fe98dc3beba65b7f58dd1e8b1f796de38ecab5b36ea1ba39163ef2dbd71d857b76cd4374623c7de5b
7ae3b0af6ed9a86f8c468ebdb7f4c6615fddb251df188d1c7b6fe98dc3beba65a3be311a39f6ded2bb3bec8b5b36ea1ba391
63ef2dbd71d857b7ecaebe311a31f6ded21b877d75cb467d633472ecbda5370efbea978dfac668e4d87b4beff6b06fa6d9a8
6e8c458e3c5ddc9e5b89e1e214ce28fe03890b0abf382ff718d3754e673a64a64bdeb79a4eb2b586e2002ec1164cab2015cb
2d49a243f0149f6197164853b2c4f6eba8682d9c672f3ad32133b724f954d3292e188a83f689024c6a20d44ab057de6c1295
4f2f13fa01853fceec0f9d3f5a23d32133c17eabe924173cc541fb4401a635507a6593ede3bed7edd8d2d4b169223bf9ff36
fd93a7e33ad321339b6c7faae924196dc5015cc360d33a28cd0658802836da8a03b98601672821754b4b4dff8d5f263f1d88
bfb7f33caf7d3bd272e1befd17d0af50f37d6023d321332d35bfd5749a8eb6e280ae61d0195a48e512741e515cf37b59b2a9
435c0b3f26e8095a673a6426e86f359da6a3ad38a06b1874861652b9742765bec3abb24bbc513bcfc7b96ceb1a9b3d950d29
73719bd6990e99e94ecab79a4ed3d156bc832e61d0195a48e512f4915a5a453b6cbfb2b8b6235926cfddee079de89098903f
55741a8d9ee220aee10ecd1042c896ee12faef9a5fb89d7bba8977f96fe1b4acc902fcdace739f9e778075a64366ba4bf8ad
a6d374b41507740d83ced0422a37c0ad5145475bf116ba0983ced0422a37c04d7045475b7140d7704b6768a1951b6033b7a2
a3ad38a06b1874861652b95f3fcf5268f41407710d03cd1042ca561e5b6ed712f62884f1f91f5fdc2c1bbf93c7bd9d6bed2d
aef25c9b571e59bed7739a8b6ee2c0ad6170192a48cd067834ade8e8260ee81a069da18554eed79b7a141addc4415cc34033
8490b20db0d144d1d1cb1cd0350c3a430ba9dc08b6de828e5ee680ae61d0195a48e592c3b5a84f37f1eeb0258cfac668e4d8
7fedf2a3d068250ee21aeed00c21846cbf3670506874316f899b30d00c21a46c03f81a293a1a9903ba865b3a430baddc00e6
248a8e36ea80ae61d0195a48e50af4161f70461fffadb95db6c547a0cbf934aed7990e9905fa4b4da7e9e8a40ee81a069da1
8554eed7964e0a8d36ea20ae61a0194248d94630151774745207740d83ced0422a27dca69f136bd84f78a77b0a8af9497448
146ed35645a7d168a30ee21a069a2184944d10fbd34d7c2d2afe0cc3b773dee7ed52c84fa643a660366b3a4d471b7540d730
e80c2da47202faf445fd1c3cae48c22b44a9259f827e321d3205b459d3693a3aa903ba86416768219513bee2b39f4becdb12
575e87ffedf9c9f9b51f02ba663a640a5f71b3a6d3743453efa04b1874861652b90c7dcfbb5f6c1c7180bbff3af895757c70
971d939b974665a6436686fe54d3693a9aa903ba863b3a430ba15cf62dad6d80cff6dfa7e7d3caaea522cf212f7b967ea8e7
34179dd49b91b66170192a48cd326e9a18c6bba0feb2795e7e653f5ff1c3099bfdaffbe98aa0331d3233f2a79a4ed3d1491d
d035dcd0195a18ca1527f5fd9cd739bec97beda19552e85e1006781dfe84e34fb20fb4cc74c82c4eea5f6a3a4d47277540d7
30e80c2da472c5493dfd13fd728febdc2fbfa68e26c97e6e9cce3e055a663a641627f52f359da6a3933aa06b1874861652b9
e242bc4ef73c476f892b3ef008effb853fbaa9a74e746d62f120fe50d169343aa983b88681660821651bc17759d0d1491dd0
350c3a430ba95c325f16f5e9658ec3d630ea1ba391631fcbcb1cd0af56dda033b490ca65e873dace7823230cf03acf73f527
fc78d67b5e96cfd032d32133437faae9341dbcccc1fc44c1662821752b9ff3bd6c7e20f163b81bef9ad43164394a777a23d3
21b37cce5f6a3ac9062f73103f51a0691d9466237463f84b0627f58e374741a655508afdd6625b38a0d346bd9dc535e1fd7f
820d8597f27f51c40837457e7d3f46a1d1431dc4350c344308295bf25f16f5e9658ec3d630ea1ba391632fab467f999c97b4
7cf7d72e3fe5bfd3a3ac2df4f79e832d645935aa4c87ccb26afc52d3693ada9977d0250c3a430ba95c819ee3222fdea69ac2
53ff33bd56da5f1365a2436241fe50d169343aa983b8863b344308215befa48ec3beda84a3be311a39f6b19cd401fd6a520e
3a430ba9dc6f4fd5828b0eeec0ad6170192a48cd06785ea0e8e8e00ee81a069da18554eed78f3c151a1ddc415cc340338490
b20db08b40d1d1c1bd856ec2a033b490ca0de063a0e8e8e00ee81a6ee90c2db472033896293a3ab803ba8641676821951bc0
7d50d1d1c41dd0350c3a430ba9dc00969a8a8e26ee80ae61d0195a48e57eed9baad068e2de119730d00c21a46ca398b803f7
d5a1bce33254109a0db05957d1d1c41dd0350c3a430ba95c865e976d0e6f37c501fa956d7ca32d0e70d9e76d9bcf320bd399
0e9919fa534da7e968e20ee81a069da185546e2c137740bffaa383ced0422ad79bb8e3b0af0ee5a86f8c468e7d2c13f716fa
dda11c74861652b901be608a8e26ee80aee196ced0422b37c0a944d1d1421ed0350c3a430ba95c72ae17f569e28ec3d630ea
1ba391631fe05aa5e8e0e30ee6270a364309a95b468e3dbb97e883b486fdb5e1b71e7ff4d9d3f779b2ad331d3233f2a79a4e
b2c1c71dc44f14685a07a559fa66fdad0d27f5ee88398ada7a1c6acce5887d6d7899e3884fb4abadc7f177ccbf9e530b2c1a
9983b68601a6455082fd7442ada868610ed81a0696a181546c8065a2a2a38539a06b1874861652b9f42312f56922de1eb609
a3be311a39f664462bead3c61b87ade1b6be311a3df6016e7f283ada7803ba8641676821952bd0e1b95fda0db4f933cc714d
7bb4dc5eee738fafd63fb7ea65a6436681fe52d3693ada7803ba8641676821951be193167474f206740d83ced0422a37c227
2de868e60de81a069da185546e8006238a8e66de1d740983ced0422a37408f1145472b7140d77047676821941be08ebda2a3
9538a06b1874861652b9f2f54e6796f8c2983febeccb79c5e9e23a9f615579cff5fd7495e99059bede5f6a3a4d47377140d7
30e80c2da472697a20ead3cf1b87ad61d4374623c73ec2a944d0d1cfbb856ec2a033b490ca0dd0804bd1d1d21bd035dcd219
5a68e5dcef5be9293a1a7b03ba8641676821951bc0d344d1d1d81bd0350c3a430ba9dc00ee358a8ec6de80ae61d0195a48e5
9233d3bda5fb8dd13bc9af76fd37708fef1fcf7eba76c5fb6309dac874c84cde4cdf6a3a4d47636f40d730e80c2da472d983
cb2ffbe6fb88af1ccfb35f125deb1467aaf392b7d33c1e5c32d321337b707daae9341d8dbd015dc3a033b490ca25b735ffad
4f1eebff043b383f7599ee758bde707778ffc4cf6ef2d7dbc874c84c6e6bdf6a3a4d4763ef0eba8441676821954b7691feab
389fdb1e5fd10babdeb8377fcfa3bd8eb54c898c44d72626b3c84f159d46a3b137886bb843338410b225efd3fdf4f3866d8e
f66f53324d98d2e7e34ff6fec3cadf6c23cf3579c9f7f44b3da7b9e8f90ddc1a0697a182d4ec3ff0311783a7a138986a1883
3750a530bda1787bd877af6ed4374623c7de1b8ae3b0af5edd6d7d63347aecbda1380efbead58dfac668e4d87b43711cf6d5
ab1bf58dd1c8b1f786e238ecab5737ea1ba39163ef0dc571d857af6ed4374623c7de1b8ae3b0af5edda86f8c468ebd3714ef
0efbe2d58dfac668e4d87b43711cf6d5abbbab6f8c468cbd3714c7615fbdba51df188d1c7b6f288ec3beba75a3be311a39f6
74c2bffc2ccb4f29e39b2a935f45ecdbb625f3623ff75a96f378ae703ad321335d06bed5749a0e76e62d738d82cd5042eaf6
ebee2402acf532076d09b6545a022957327f9e4f7fde0cefbdc7cb71b6b28dbd72cec7ef34993feb4c87cc64fefcada6535c
f03207ed130598d440a89560c3bb197eed7f255ff5e6eef974a53baa5b7ef06f643a6426d86f359de4829739689f28c0b406
4aaf6c72bda4b77fd26ae159f487d582ff95adfe6756fc9e75a6436636b9fe54d349327a9903b886c1a675509a25e26df7c3
f19f46fc69f9f1ccdbbec6668cf3b4fa05f2bd6d65cda5331d3213f1b79a4eb3d1cb1cc8350c384309a95b82bef25380b862
b89b4e09f3fc7c3913b4ce74c84cd0df6a3a4d472f7340d730e80c2da472e99642fcaa4d77746bf32bfe69f5e7b7f4a92cf9
59417ee260643a64a65b0adf6a3a4d472f7340d730e80c2da47209dae7dc7e929d8cb7f77c73277c2fe7f832e974946d2e46
a6436682fe56d3693a7a9977d0250c3a430ba95c76f60eb615e17e7cb8a537f915dc3eed71274ebc5dbfdcfbf32aaece74c8
cccede9f6a3a4d473b7340d7704767682194cbf74643138c658bcfbbfcffcc7ef1bf5c7180c1db75bbcec77246673a64e67b
a39f6a3a4d473b7340d730e80c2da472097adff7653b93af7cd86eb5dce1f5b730da99d72b23d32133417fabe9341dedcc5b
e8260c3a430ba9dc003b7c151dedcc015dc32d9da185566e800d5d8a8e76e680ae61d0195a48e5ca43f2f439cce9369f9f19
fbf37cfc88567f0e3ae66dadf6de2ad321b33c24ff52d3693a3a9a03ba8641676821952bd0f7191f4fc475df322fcb72c4f9
e21a3d81fc94b9ee0c50990e9905fa4b4da7e9e86b0ee81a069da18554eee716df028da6e620ae61a0194248d90af17edde7
191f9f3f569b7187cae52fb46730a27f9055a6436661fe52d3693afa9a03ba8641676821951bc1e25bd0d1d71cd0350c3a43
0ba9dcaf5f3c5768343507710d03cd1042caf673af6d81463ff58eb88481660821651bc06b5bd1d1521dd035dcd1195a08e5
06f0da567474556fa19b30e80c2da472037867283abaaa03ba865b3a430bad5c813e6e3f358d0bbe7d6f6ee06df903aa7d56
64a6436681fe52d3693abaaa03ba8641676821951bc00947d1d1551dd0350c3a430ba9dcaf7d96141a5dd5415cc340338490
b20957f56b3fef293c3e0b033c967c7feb4156990e99c255ddace9341d5dd5015dc3a033b490ca09bbedb085d49f6be3db77
7eaeec27acc1fe4e40d74c874c61b76dd6749a8ec6ea80ae61d0195a48e504f411ef4a5ff1c7b76f7edabc86773015f493e9
9029a0cd9a4ed3d1581dd0350c3a430ba99c80f633c5f9da8f3bfe18afbcab511ae83f990e9902daace9341d8dd5015dc3a0
33b490ca0963f525bcb31eecedc268976df6279e708f4518ab3f990e99c258ddace9341d8dd53be812069da18554ae18ab97
9db97e80879f24def7195faff50b41bfc2dfefbaf959663a641663f52f359da6a3b13aa06bb8a333b410ca1563f5ec48e5c7
744d537ed5237e449d19a2ce74c82ce6ea5f6a3a4d477bf566bc6d1874861652b962af7eadebbc25e7e4f5b8ee759f96e80a
1e1a8bfa89e4f39bd6990e99c55efd4b4da7e968af0ee81a6ee80c2d0ce58ad378dd0470edb96d6e38d39e77dab1bb3ed032
d321b3388d7fa9e9341dedd5015dc3a033b490ca15cfdcf8f58bdfc4f3f44b81f5880b83b0e7e3f4338be726914c74482c8e
b91f2a3a8d466f7510d730d00c21a46c23b8040b3adaab03ba8641676821951bcb5e1dd0afcee5a033b490ca15e8ba17c2af
104277e825bec6155ebc5896adbe0da6331d320bf4979a4ed3d1dc1dd0350c3a430ba9dc58e6ee807ef52e079da185546e84
afb7a083b93b989f28d80c25a46e239cc6feb2c1dc1dc44f14685a07a5d9087d40fe92c1dcbde3cd51906915946203dc1c52
eeecb4786f67924d78f32bb4d0973d866b1123dc1419a0019da2a3cb3ba06b1874861652b9dee51d877db530477d633472ec
03dc9652747479efa04b187486165239051dadb25327eeddaf35f7656aee3e8a448744856c54741a8d2eef20aee10ecd1042
c836c24f4ad0d1631ed0350c3a430ba9dccf5b4508341acc83b88681660821651be0a991a2a3cd3ca06b1874861652b95f3f
065668f49807710d03cd1042ca36c0a60e45479b7940d730e80c2da472036c585274b4996fa19b30e80c2da47205da0f2b7c
1ce9ea993c22c2f8d6e3defc875537a6893cd7e615e0f77a4e73d1601eb835dc72192a68cd0670135174349807740d83ced0
422a37c09e4b4547837940d730e80c2da47203787d2a3a1acc03ba8641676821951bc0d355d1d163be832e61d0195a48e506
300d5674749a07740d7774861642b901368f2b3a3acd03ba8641676821951bc1085cd0d1691ed0350c3a430ba95c7608baae
79f65ffdf8fb0b8d73e3ded038c0f4457c6e3aea4c87ccec10f4a9a6d374749a07740d83ced0422ad73bcde3b0af36eaa86f
8c468e7d845f95a0a3d37c0bdd8441676821951be057a5e8e8340fe81a6ee90c2db47203fcaa141d9de6015dc3a033b490ca
0df0f55674f4b907740d83ced0422a37c2d75bd0c1e71ecc4f146c861252b7016644820d3ef7207ea240d33a28cd46f85eff
2583cb7ec79ba320d32a28c53a977d1cf1c54abeabadc7f177cc23fc88fe92d1681fc0350c36ad83d26c840b8460a3dd3e90
6b1870861252b70c7daeeb7ea48e9f5e97f9ba4ebfb20d03f4dffd7dba8ee769afce74c8ccd09f6a3a4d47bb7d40d730e80c
2da472197a5baff87a70bc4d751ebef41ddfdb0866a987ffd7c7965c673a6466e84f359da6a3d97f0bdd844167682195ebcd
fe71d85727fbb6be311a3df6ac75f3b47fddd66d3abc4e4bd4e5d99499b596990e9959eb4f359da6a3d93fa06b1874861652
b9014e258a8e66ff80ae61d0195a48e546f8a4051dcdfe015dc3a033b490caa55f95a84fbb7d1cb68651df188d1cfb00a76d
4547bbfd0eba8441676821951b60d2a7e868b70fe81aeee80c2d84720378642b3adaed03ba8641676821954bbf2a519f86f7
386c0da3be311a39f6019e20283adaed03ba864167682195fb697f29c545a3fd16b70983cb50416a36c0ef49d1d1681fd035
dcd2195a68e50658cc293a1aed03ba864167682195cbd7c8795bfd2f3ddeb2f0e3f1e7f8254dc8d725bcaced7f93e512a912
1d12f315f24b45a7d1e8b20fe21a069a2184942d3b491de99ff86b3b9abd54c13cddabb4cee502a9331d32b393d4a79a4ed3
d1651fd0350c3a430ba95c76d9bfa77d5ba6b8bd76beb7fc59843fba4d1c46a6436676d9ff54d3693abaec03ba8641676821
951bc0124fd1d1651fd0350c3a430ba95c323f5ce34ced8afe5ea917c0f4ec1d4ba687ff96f17fffcf7b0da729e8a6dfc195
30280c66a950322cdda6f5dab623b6400a6e6dd3751e7b32177d7c529261a9ce74c84c86a5df6a3a4d47437d40d770476768
21944beebbfe3b76addb127bb94df18e899f23441fca2374bada8a799291e8dac4e4befba9a2d368f4d407710d03cd1042ca
f69f7aea83e9d5ae1e833750a530bda77e7bd877bb7ad4374623c7de7beae3b0af76f56d7d63347aecbda73e0efb6a578ffa
c668e4d87b4f7d1cf6d5ae1ef58dd1c8b1f79efa38ecab5d3dea1ba39163ef3df571d857bb7ad4374623c7de7beae3b0af76
f5a86f8c468ebdf7d4ef0efb62578ffac668e4d87b4f7d1cf6d5aebeab6f8c468c3d9d02fdc4ea08af33c40b51defc1cdf00
9cf6fd9ee6fd3cf262c4c874c84c27c66f359da6a3a33fa06b1874861652b974a50bef7df9f541bc91392dc16d62c97d5ffc
dcc39f6bfc543a5fea74a64366bad67dabe9341dfb0900ba864167682195fb758720858666022d708d02cc90418a96e67047
756e9c8239e7eccfa97170d7b3e13f4de164a243629ac17daae82458db4c00b425d8526909a45cb9e3d53add5b5c9c04edfd
2433daa7843feec7fa2c77bc92990e99b9e3d5a79a4e71a19900689f28c0a40642ad041b5fd79de266e729bfaf1b9d95e7e9
69d69a5865a2436242fd54d14928741200ea130595164089959696d37ccc9b3f5fc61e1dc167629aeff8c8652e86d9f9576b
643a64a6a5e5b79a4e92b19300806b186c5a07a5595e4c579f2fbfd45dfdaf7b8ab778e6aea1ad4e7448cc4be92f159d0663
1b01f0d630c80c19a468d95c3e7c18fe17b5c75b1a73ee371bfe58aff93c8f25ff788d44d726666bf92f159d46630f0110d7
30d00c21a46cf92ed194ef5785215dfbb14cf79cc61a6e67cd7e5d5c6e12a94487c47c8fe84b45a7d1d84000c4350c344308
295bba19381fe7e94bc53b38cb1c2c9be735de949ec333bef50a0f16d3cd409de990996e067eabe9341d1b0874d0250c3a43
0ba99cfb7d9f5145c7060280aee18eced0422897bdf44fffa59ff7d8a7659dae73f653dfe87bb5847bd5f1614c86d6990e99
d94bff534da7e9d84000d0350c3a430ba9dc008f34141d1b08b4d04d1874861652b9fc5e57b8271d7a78c49b41f1b9626a19
be5c7eb1155e3c7ccc0f65a643667eafeb534da7e9d84000d035dcd2195a68e5ca1b7c3fecbdaee8d84000d0350c3a430ba9
5c794b33bc9a7c27bfd1f35ca668221c0698678ed3f3069fcc74c82c6f6a7ea9e9341d1b0800ba864167682195fbb997be40
63f70010d730d00c21a46c23d8ca0b3a361000740d83ced0422af7ebb7eb151abb0780b88681660821651bc1645dd0b17b00
a06b1874861652b9114cd6051d1b0800ba8641676821951bc02e43d1b18740075dc2a033b490ca15e8fd4cfb06f7ed986eff
41c40d3fe5a6ec554d51fee639e415e0f77a4e73b17b00706bb8e33254109a0de029afe8d83da0856ec2a033b490cafddaf9
46a1b17500886bb8453384d0b21583a36093bcc405fe7eece176d99d66125b67d0ab331d328bc1d1979a4ed3b17500a06b18
74861652b9015c9d141d5b0700ba86416768219513d077da40b0c559f251da172be827d32153409b359da663f70040d730e8
0c2da472ca483f7af46df1f4b3dde1dff6fd54366d35d3215319e95b359da663f70040d730e80c2da47202dacf1f62ebd3f8
630ca7a2799f3765af5e331d3205b459d3693a760f00740d83ced0422a27a0439b62bfec8917d4fd98fcbac44ba5cce96aa6
43a680366b3a4dc7ee0180ae61d0195a48e58491fe149cc09739beca7a4cc965f6b1f5d4990e99c248dface9341dbb0700ba
8641676821952bd047f2dc0a030caba0f0a8259c69fd92c8af83c27b690fb4ca74c82cd05f6a3a4dc7ee011d740983ced042
2a970d5ca7b4e33aba9f9fab5ff0cec7194e3ac7b5cda9c95886d6990e99d9c0f5534da7e9d83d00d035dcd1195a08e532b4
ff261ee7151fb15cd3bdfbff5be302c12f90d6ed9eeee3716396990e9919fa534da7e9d83da0196f1b069da185546eacee01
807eb5c66fe80c2d0ce5c6ea1e00e857777cd0195a48e506ea1e00e2576b7ca0194248d90af17c1dd315efe08455c2bd4cd9
2f7b39fcf8eea7b3bb91e9905998bfd4749a8edd03005dc3a033b490ca851d68b23efdfb71d81a467d633472ec23589c0b3a
faf703ba8641676821951be10b26e8e8df0fe81a069da18554ae9c3cd3de932bfeec833d78708d097ff8ebb99f2d4fcf7c5f
673a649693e7979a4ed3c1bf1fcc4f146c861252b711ae177fd9e0df0fe2270a34ad83d2ac4c0aca6b94e1b7f6f86f860b78
ec9ab1d5be1432d321b34c0abed474920cfefd1d6f8e824caba0141bc15e5d58efd3bfbf9dcb35e1f57f8ee39af614ae458c
7053442c619f673ae131a6ff3586d7d3652bb892e8902816b05645a7d168de0fe21a069a2184944ddd93f25fc76b9fd2e3ea
73bda6c5cf5a55cbbf9ae990a9ee4959359da663eb0040d730e80c2da472ea63f6279a23fc12c300af7d3dcef938e49d8a27
d321537dd0564da7e9d83aa0832e61d0195a48e506b8fba8e8d83d00d035dcd1195a08e50af4afbab52a34b60e00710d03cd
1042ca3642df1541c7ee0180ae61d0195a48e50678f0ade8d83d00d0350c3a430ba9dc009b3a141d1b0800ba864167682195
2bd0cd835a3f79dad66d8b53e4d58f6f5f97edf1d4d2990e9905fa4b4da7e9d84000d0350c3a430ba9dc4fbdf415175b07b4
b84d185c860a52b311bcf4051d1b0800ba865b3a430baddc085efa828e0d04005dc3a033b490ca8d602b2fe8d84000d0350c
3a430ba9dc00fe528a8e0d04005dc3a033b490ca0de024a6e8d840a0832e61d0195a48e592d9b8a84f0b7f1cb686bbfac668
c4d8d361457d9ae8e3b0358cfac668e4d80730e6537434d107740d83ced0422a3780c7a7a2a3893ea06b1874861652b9113e
6941470b7f40d730e80c2da472e95725ead344bf3d6c13467d633472ecc99259d4a78d3d0e5bc36d7d63347aece9b0a23e8d
e471d81a467d633472ece9b0a23eaddc71d81a467d633472ec03fc8a151dacdcc1fc44c1662821751be0372cd860e50ee227
0a34ad83d26c848ff82f19acdc3bde1c0599564129f66b5b3a81051779c03ed10e4b4bf057aefc6a6298805dc71abf78cbb4
9c875f70844f65891fcfb63c5bcf74a643667e35f1534d27c9e8220fe01a069bd6416936c0cb988a8d2ef240ae61c0194a48
dd327478e8196c0cc2988271d2191d41c24bc29b3fb35fcf7d019de8dac48cfca5a2d368b49007710d03cd1042ca965f34be
93e7673c899e619bd074468f57ffd16c7e9e323d3bed74a643667ed1f8534da7e96821df423761d0195a48e506f097547434
b007740db77486165ab9e4452cead3421e87ad61d4374623c79eb5bef2593d7e19fc5a3aec478bfe02f172bdcecf13129de9
9099b5fe54d3693a5ac803ba8641676821951be10b26e868210fe81a069da185546e844f5ad0d1c01ed0350c3a430ba9dc00
e74f454703fb0eba844167682195cb97c9259970c6afe23daffec2bddff18f79ba369f5a3c3e8d4c87cc7ca1fc54d3693a1a
d803ba863b3a430ba15c86bed24dbfd80379deb6db5fd5f7f8112de15d817d7d3c0b74a6436686fe54d3693a1ad803ba8641
67682195fbf5245fa1d13b1fc4350c34430829db00cb384547ef7c40d730e80c2da47203dc3c577474d06fa19b30e80c2da4
72c9ee55d4a7873d0e5bc36d7d63347aec59eb6492b6a5e7a1b33fb7f98560f4298936714bd83896b596990e9959eb4f359d
a6a3873da06b1874861652b90156938a8e36f680ae61d0195a48e5f2dce0d973181e93855f79dcebe1278dd314d6266562f0
92e6f2ace0bd9ad35074af076b0d03ca90400a36967b3da05fadd94167682195fbb5e19e42a3753d886b1868861052b65f3b
482a34fad977c4250c344308295bf2423d77bfccb8d35d3a3fb4322b0ba3bdd62b3c65c8f30123d32133b9a17eabe9341dfd
ec015dc31d9da185502e19de6e973fabf80f237cf5a6b5d9343f1db72f70ddf96c6d24ba363119de7eaae8341afdec415cc3
40338490b22533677fc538c35b5cd1b3b54ccae2f8c232fbf2b395fc311b990e99c9ccf95b4da7e968780fe81a069da18554
ee3f30f11783a79b7ecbd484317803550ad3bbe9e3b0af46f56d7d63347aecbd9b3e0efb6a548ffac668e4d87b377d1cf6d5
a81ef58dd1c8b1f76efa38ecab513dea1ba39163efddf471d857a37ad4374623c7debbe9e3b0af46f5a86f8c468ebd77d3ef
0efb62548ffac668e4d8d349e16c6ec44c7e70fedc7c1fa9f3caedaf4ebb5f52e4f383ce74c84ca78a6f359da6a3973fa06b
b8a333b410caa50b9e5fa8fa65d2195f8d7fced4b1b7ce391dfe3f59caaadfc874c84c97bc6f359da6a3973fa06b18748616
52b9015af5283a7af903ba8641676821954bd07e5171dce1bd9df841ccf3365fc14c27f879afebe4bfa77b81d6990e9909fa
5b4da7e960e7df32d728d80c25a46ebf9eb40bb0d6ce1fb425d8526909a45c2398beffe1829d3f689f28c0a40642ad04bbed
71811cd7875bb9611cfe58b7b4193d7fae3ad12131a17eaae82414ecfc81fa4441a505506225d82bbefbb4c5e7207e5a3df9
6fdf16df22f1a7153f4df127d3e2f5ae331d3213eeb79a4e92d1ce1fc0350c36ad83d26c04e373c146477f20d730e00c25a4
6e097a3dfd62d0afd1e24390d5afd5a67b4bb743cef58eefa897db663ad32133417fabe9341d4dfd015dc3a033b490ca0df0
b858d1d1d71fd0350c3a430ba95cbe25bcefcb76deb183e13a952d4ff17b79a55b9de59983ce74c8ccb7843fd5749a8ebefe
1d740983ced0422a37c0a64845475f7f40d770476768219473bf6f55ace8e8eb0fe81a069da18554aebc6316ac3f8f3b799f
faf39dff8fa6f470dbff27676834f2bc63a6321d32cb3b665f6a3a4d475fff16ba0983ced0422a57de232cd3bff0bef2e967
0cf7197d54a29760685cf9bc0e2d331d32cbbb845f6a3a4d475f7f40d7704b6768a1951bc1e25ed0d1d71fd0350c3a430ba9
5c794dd65f60ce2dce0ef7690e9e39777c36e34fb2ebec4bcccf2b9432d321b3bc26fba5a6d374f4f507740d83ced0422a57
a0e32a2fbeb0bd3ccbd9d01efef057f67dad4ea32ad121b1207fa8e8341a7dfd415cc340338490b20df0debba2a3af3fa06b
1874861652b911bccf051dadfd015dc3a033b490cafdd6fb5c70d1d41fb8350c2e4305a959bcb9aaead3561f87ad61d43746
23c73e82e5baa0a3ad7e075dc2a033b490ca0de018a2e868ae0fe81aeee80c2d8472bff699576874d66f899b30d00c21a46c
23f8cc0b3a9aeb03ba865b3a430baddc083ef3828ee6fa80ae61d0195a48e506b06e537434d707740d83ced0422a3780a797
a2a3b93ea06b18748616523901bd84bdbb67dab9ba4fd716b6782af3f19ae8da44816c55741a8dcefa20ae61a0194248d904
71d8bf7a4fe79a666deb7cfba5575dcdc94c874cc16cd6749a8ecefa80ae61d0195a48e504f4351feb346f714abe9fcb3cad
d37a2aebb69ae99029a0cd9a4ed3d1591fd0350c3a430ba99c70d6f79792f8203a7c2ac79c5b90ab4fba663a640a677db3a6
d37474d607740d83ced0422a97a1ef20c7b9473bf43db6b248cbcfe388af995c8f378cce74c8ccd09f6a3a4d4767fd0eba84
4167682195cbd6aab1bdfc7587d38c1f9a9f46f9f56eec03302d7e3de2179d8fb5aacc74c8ccd6aa9f6a3a4d47677d40d770
47676821942bcefa971fdeb42547df2b77790803dcf27de1a79d80cc74c82ccefa5f6a3a4d4767fd66bc6d1874861652b90c
edbf87f712362d85a799f3b6f979726cbd7efa3567f4247ba065a6436686fe54d3693a3aeb03ba861b3a430b43b9027dcde7
7ac63ba4575901863ffcb08f753d421f8002ad321d320bf4979a4ed3d1591fd0350c3a430ba9dc088ed8828ee6fa80ae61d0
195a48e57ede4e40a0d1591fc4350c344308295befac8fc3bedac6a3be311a39f611be5d828ecefa80ae61d0195a48e50a74
fa277efec7e53ffe393d6bf2679e73f68bfea7d590ce74c82cd05f6a3a4d47677d40d730e80c2da472e5e4b92efbb2c74dda
61ff4b7cad385aa45ff1b59ae9e9caac331d32cbc9f34b4da7e9e0ac0fe6270a364309a95b719a9fc23bc66b6a70b0fad5b5
9f98c656317bb6267d9065a64366719aff52d3493638eb83f889024deba0342bd3bfdcde215ebef67bd98e2dbe2a77fa1549
70447d1ec5ea4c87cc32fdfb52d3493238eb77bc390a32ad82526c80e670ca149fcefaed5cae092f61ede057e5315c8b18e1
a688ba5331878d7a677cc1649f96b0677f9b0e05fd643a64aa7b15564da7e968ae0fe81a069da185544ef53eccef11c54769
773447f0ff99827e321d3255ef43aba6d37434d707740d83ced0422aa7a0ffe3868f8a8ee6fa1d740983ced0422a3796b93e
a05fcde33b3a430ba1dc004f14141dfdf5015dc3a033b490ca0df08850d1d15f1fd0350c3a430ba9dcaf775628349aeb83b8
8681660821651bc1675ed0d15c1fd0350c3a430ba95cd92a957ca3e353cbfddc677f6589ab813518e0858701d5695e653a64
96ad525f6a3a4d47737d40d730e80c2da4726527a09f3b8476a3f1f776ded71dba50fe1377a86ea943e7b31350653a64969d
805f6a3a4d478bfd16ba0983ced0422a97a1c37bfbf3bdc46be8e4573c8b5ffcc59db87ee8fb7d6e8f5b8ece74c8ccd09f6a
3a4d478b7d40d7704b6768a1951bc1785dd0d1621fd0350c3a430ba9dc00de318a8e16fb80ae61d0195a48e57eed8ba4d0e8
af0fe21a069a2184946d00fb2b45477ffd0eba844167682195ebfdf571d857ebfaaebe311a31f6de5f1f877d358f477d6334
72ece9b0a23e1dee71d81a467d633472ec0378eb2a3a3adc03ba864167682195fbf90944a0d1de1ec4350c344308295b6f6f
df1ef6dd391ef58dd1c8b10ff0ed527434d707740db77486165ab9de5c1f877df5ad477d633472ecbdb93e0efbea1d8ffac6
68e4d8d361457dd8dbe3a84f14d58db1c891fffa072cc0e06d0fdc270a2e2d82126c808bbf2083b77dc79ba320d32a28c506
78375190c1de1ebc4fb423d32afc556c00175b41467b7b00d730d8b40e4ab34cbccf7e953cc73b7fcbe14fd8cb1d5e030eaf
c72eeb7586ed538558663a6466e24f359d66a3bd3d906b1870861252b70ced359f972d5a35f8ff9983414734ad98fd8470de
aef3699dae331d3233f4a79a4ed3d1e11ed0350c3a430ba95c7ea3fcb12e08a796cb5faafdf53a7c44f3e92f63fbe1ffa7bc
512e331d32f31be59f6a3a4d4787fb16ba0983ced0422a97a097794e7394787ed9d3fb73617cf565e1c42c131d1213f2a78a
4ea3d1de1ec435dca2194268d906704b507434d707740d83ced0422af7f38f59a0d1591fc4350c34430829db081fb3a0a3b3
3ea06b1874861652b9d4ba41d4a7b73d0e5bc3a86f8c468efde7df2e814663fb8eb88481660821651be05aa1e8686c0fe81a
eee80c2d847203cc0a141d8ded015dc3a033b490ca0d60b9a2e8e86d0fe81a069da185542e2f6be6e9dec30bc06180d7be5f
5ea26b8d53f2734effcfb2ac91990e997959f3a9a6d374f4b607740d83ced0422a37c05a4ed1d1dbbe856ec2a033b490ca8d
e0b72ee8e8ac0fe81a6ee90c2db472037cd28a8ecefa80ae61d0195a48e506f84d2b3a3aeb03ba8641676821951bc2591fac
afb6f180322490828de5ac0fe857e778d0195a48e5b2b561745cdce3fd403f03f7b9fb1c3d45e7f00ee3bc3caeeb46a64366
3637fc54d3693a9aeb03ba8641676821954b3e9de7ea4faafb7526bbd42bb8b6aed1e9c8ffe2a6e0adb0e793b691e990999c
3abfd5749a8efefa1d740983ced0422a37801faba2a3bf3ea06bb8a333b410ca25bbe12b9b948513ec54fc53e31f67be359a
af5446a64366b21bfe56d3693a5aec03ba8641676821951bc0585ad1d1621fd0350c3a430ba9dc7f6ab1df32bdbbd763f006
aa14a6b7d8c7615fddebdbfac668f4d87b8b7d1cf6d5bd1ef58dd1c8b1f716fb38ecab7b3dea1ba39163ef2df671d857f77a
d4374623c7de5bece3b0afeef5a86f8c468ebdb7d8c7615fddeb51df188d1c7bfa75966deeb129c77c6d7b788d395e35c2ab
3d8b9f14e6cd0146a64366facd7eabe9341d0dfe3be812069da185546e808b8fa2a3c13fa06bb8a333b410cae5363ecda677
3fbdad3e9afee2110d25cbcd4c23d32133b7f1f954d3693a1afc03ba8641676821954b13ca25afd1636ba1331a841ea9e98e
ff824ea7bf5416e7779de990992694df6a3a4d47837f40d730e80c2da472093adc5b3cee233e789dfce5f30cef26c69bcc73
78897f3acb82d8c874c84cd0df6a3a4d0783ff96b946c166282175fbb911fc5fb0d6e01fb425d8526909a45c7935bc1dfe73
f76798b840f75f393fd14fabf57d3bafe358ca19dbc874c8ccabe14f359de282c13f689f28c0a40642ad11ccd0ff72c1e31f
b44f14605a03a597fb7d5f6141468f7f00d730d8b40e4ab301365629367afc03b98601672821751be08d0145478f7f40d730
e80c2da4722398a00b3a7afc03ba8641676821951be02d2045478fff0eba8441676821951be07d2f45478f7f40d770476768
21942b6ff64527a4233b366eeb997ccfd6f5b8b773adefacaa3cd7e695b7fadeeb39cd45777fe0d630b80c15a46605377cd1
c2cee0b0e2f31fcdb21cd1ad72ed6eca1b990e9905f94b4da7e9e8eedf423761d0195a48e5ca2babc9d92cbe2ded7f827e40
4b7e2d6d3bb76d5dea4bba22d121b1bcb0faa1a2d368b4f607710db76886105ab642ec573bd7147b02f9d5c6116c18f674b6
3de6fd9e8ffbf99865a6436661fe52d3693a5afb03ba8641676821951bc1f35dd0d1da1fd0350c3a430ba9dc6fedcf05177d
fd815bc3e03254909a0d6032a0e8e8eb0fe81a069da185542e79be8bfa74d6c7616b18f58dd1c8b1ffdc095ca0d15c1fc435
0c34430829db0886ef828ed6fe80ae61d0195a48e57efe310b34fafa77c4250c34430829db081fb3a0a3af3fa06bb8a333b4
10ca0de049a3e868eddf423761d0195a48e50af47cdd73bc4db49f53982e0563f830be7d893d14abf5b94874482cc81f2a3a
8d465f7f10d7708b6608a1655316f7f1929dd622fbb1f9abd9742bb3f727d1215119dc1b159d46a3a93f886b186886105236
e530e56728d73ec5bbfc7eb9e5a7abebb59e0af9c974c8540e53564da7e968ea0fe81a069da185546e0003354547537f40d7
30e80c2da47203f8032a3afafa03ba86416768219513d07e79799f7b9a9efb89f9ee7f8aabead3f1243a240a64aba2d36834
f507710d03cd1042ca2688efe33c96f38ab744f7ebdef6699967d9b0e2c974c814cc664da7e968ea0fe81a069da18554ae40
6fc975df8fe95c1f5fddf0a5f4435cafba909389ae4d2cc81f2a3a8d46477f10d730d00c21a46cc5d1bfdceaf5e33b9e8e9f
e1f974b867141ee4156499e990591cfdbfd4749a8e8efe1d740983ced0422a572c5d73b3c07f4283ee29b7850ca35dba5998
ce74c82c96ae5f6a3a4d47477f40d77047676821942bbebdf1b6497c0aed27ffa7fff6cd73fc88f670e32c7c68055a663a64
16dfde2f359da6a3a37f33de360c3a430ba9dc08fecc828e8efe80aee186ced0c250ae38714fc7baad715fc5b5dfc7eebf88
77b21a4ecfa39fab95ce74c82c4edc5f6a3a4d47477f40d730e80c2da47223d88f0b3a3afa03ba8641676821951bc8d11fc4
af66f940338490b28df0310b3af61300740d83ced0422a57a0b33f7a1c60be0b1c473b6d47dc88ff5890cb4c87cc02fda5a6
d374ec2700e81a069da185546e8453b7a0633f0140d730e80c2da472bfbf484b3af41300f313059ba184d4ad4cc6d673d9b7
d41da6f9f7e9592c94a9d8df3c87bc32117bafe724153a0980f589024a2ba0d4fa7df72c45864e021d6f8e824caba014fbf5
a2597500601b817602d984677f5a98e6fb88e15ac4083745e23322559f46fe386c0da3be311a397621f4161d02d6b8fbd29f
ebba67ae32d32153486dd6749a8e46fe80ae61d0195a48e504f47fdd5153d1d1c8bf832e61d0195a48e5c632f207f4ab4b7d
476768219413d0ff75975c4547237f40d730e80c2da4723fdd52a0b868e10fdc1a0697a182d4acec9589374fe236717f4ebf
263f038fef14afc9e6659a9f5398cc74c82c7b65bed4749a8e2efe80ae61d0195a48e5ca3eb0b2a336fcd84e7faa39534fa0
752bed119f7d602ad321b3ec03fb52d3693abaf803ba8641676821952b1b1cebab2edbe267c37ece30a5fdd5fe2b1ade0879
beda32d321b36c70fc52d3693abaf803ba864167682195cbfb77e3e262897741d7f06cd82f70e3539d257f427399d7eb4c87
ccbc7ff7534da7e9e8e2df423761d0195a48e506d8a9ade8e8e20fe81a6ee90c2db47203ecc9577474f107740d83ced0422a
3780778aa2a38b3fa06b1874861652b901bc53141d8dfc015dc3a033b490ca0de07caae868e4df419730e80c2da47203d87d
293ab61100740d7774861642b911bede828e4d0c005dc3a033b490caf54d0c70d857877ed4374623c73ec0ab6b8a8e4d0c00
5dc3a033b490ca65e8f0243cbced9de66973b0e388edc0fdbac8afc5aef331199389ae4dccc85f2a3a8dc6260620ae61a019
4248d9fa2606ed61df1dfa51df188d1c7b72b817f5d9460087ade1b6be311a3df6be8d000efbead08ffac668e4d80730af53
746c6200e81a069da185546e800998a2430b05303f51b0194a48dd7e7dfa106068a100dc270a2e2d82126c800b8420430b85
8e374741a655508a0d30e3126468a100de27da916915fe2a36808bae20630b0500d730d8b40e4ab301cc92151b5b2800b986
0167282175fbb527b64263ff0410d730d00c21a46c03f84f2a3af64f68a19b30e80c2da47209fa8aef206cf1cedc7cfab9c3
15de8d087f6cfe62362dfecf0cad331d3213f4b79a4ed3b18502a06bb8a533b4d0ca25e8f3b8abf9e9b99c7e48536c6b35a7
3dafdb96ef0c18990e9909fa5b4da7e9d84201d0350c3a430ba95cf2d617f5d9c70087ad61d4374623c79eb58e13b339bef3
e27fefeb1d7a10c43fb6f2aa57d15a663a6466ad3fd5749a8e7d0c005dc3a033b490caf57d0c70d8579f7ed4374623c73ec2
f75ad0b19541075dc2a033b490ca0d6094a4e8d8ca00d035dcd1195a08e546e80e22e8d8ca00d0350c3a430ba95c9e874d79
e75b9c3325a78c295e4cfdc7b34ff359eee7ea44d726e659d8978a4ea3b18f01886b1868861052b601e6da8a8e7d0c005dc3
a033b490ca0d70df5ed1b18f410bdd844167682195cbd0576e8a180618763f9cc795ae6f69e7e7fcbc87a9331d3233f4a79a
4ed3b18f01a06bb8a533b4d0ca0dd08d4fd1b18f01a06b1874861652b9113e6941c73e0680ae61d0195a48e586e86300d657
937e40191248c1f2b57959b7794b2ff3370f8bc2456549b3d5e7da2c331d32f3b5f9534da7e9d8c700d0350c3a430ba95c9a
85edd1a96cbed22429cf9292ab663073f00bb17c8fc4c874c84cb3b06f359da6631f0340d730e80c2da4720338b02a3af631
e8a04b1874861652b9ecb5bb24eb82647a7c1cd1bb238d96f7ef8d4c87ccecb5fba9a6d374ec6300e81aeee80c2d8472c955
794b7b8ae33dd869bff23526fc71f909dc7e9f65be6d643a642657e56f359da6631f0340d730e80c2da472c93f7bbe567fe1
9ee284b8f4fe9e9207f47efa6bfa566e8819990e99c93ffb5b4da7e9d8c700d0350c3a430ba95c724a5fd286bcd85a213b1a
2c4bfc84e844ab131d12934ffaa78a4ea3b1cb414bdc848166082165fbffdfb9410d9e2d14c054c3ede00d542d4cdf420187
7ded4e80fac668e4d8fb160a38ec6b7702d4374623c7deb750c0615fbb13a0be311a39f6be85020efbda9d00f58dd1c8b1a7
efebbda7bb7cf10c529e77c633c872a68952fee6ca4487c4f41dfe54d16934766f00710d03cd1042ca96ae3e7e5071c51a4f
9ae9d6e7b6a79e3f4798846e77b9fae84c87cc74f5f956d3693a766fe8a04b1874861652b904ed27f3dbe6cf5857fc201a87
b0c90f3a58be9437868d4c87cc04fdada6d374ecde00e81aeee80c2d84726972e5c777eff7917a6af84b659aff85d9dfb4c5
c65dc569ccc874c84c93ab6f359da663f70640d730e80c2da47269ed3087bb8ed79d9a75f9d12cfee71fff98ef7359fd8ca1
3c8292890e8969e5f0a9a2d3686cdd00e21a069a2184942d11a719c176a715dce6bf7de7bac58f24f81684e7bfe5a188ce74
c84cccdf6a3a4d87d60d2d738d82cd5042eaf6eb9bb702ac6ddd00da126ca9b40452ae01b6a3fee542eb06d03e5180490d84
5a03bc3921b8d0ba01b44f14605a03a5d7084eef7fc9d8ba01c0350c36ad83d2acbc05951caa9321a7ff5f7fd9887facfe74
b3f879c4032c135d9b58de81fa50d16930f66d006f0d83cc90418af65b8f77c1c58e0dc0ad6170192a48cd46f0781774ecd8
00e81a069da185546e80b738151d3b3674d0250c3a430ba95c81bed6fd4a57917d7ee6f97e80fe37e8d7ed7bb5be94990e99
05fa4b4da7e9d8b101d035dcd1195a08e506b0795774ecdb00e81a069da18554ae406ff1564cfc54b6793e8fed8cee485bbc
59e34f7f8fe580cc74c82cd05f6a3a4dc7be0d2d7413069da18554eeb7def6828b4d1b805bc32d97a182d66c046f7b41c7a6
0d80ae61d0195a48e546f0e216746cda00e81a069da185542e19fa8bfa6c9e80c3d630ea1ba391634f8715f5d9be0087ad61
d4374623c7deb72fc0615fedf951df188d1cfb0886e7828e1d0c005dc3a033b490ca8df0731274ec6000e81a069da185542e
7dc1447db611e80e5bc2a86f8c468ebd1cf64f7d1af9e3b035dcd5374623c63ec2f75ad0d1c8bf856ec2a033b490ca0d64e4
0fe257a3fa16cd1042cb3690913f885f5dea81660821651bcbc81fd0af2ef5a033b490ca8d65e40fe857977ad0195a48e594
6b5cba7f105f9df233a6633996fbd928a8331d32956b9c55d3693a1af903ba86416768219513d0677e492a5ad01fcf6805f4
93e99029a0cd9a4ed3d1cb1fd0350c3a430ba9dc5fe8b3bd85e27f8ccb1aac3e54cb86b3bd2dd364fe85b66b3a4d472f7f40
d730e80c2da4720379f983f8d5a81e68861052b6eca43aa526aed1d0f6f43fb86df65fc7303e3ff0e0493a3f36c132d32133
bba97eaae9341dbdfc3be812069da185542e432f61dfc87224ebe2f822dc1a6f90fa187fd03ad32133437faae9341dbdfc01
5dc31d9da18550eef736c1928e5efecd78db30e80c2da4722318420b3a7af903ba861b3a430b43b9b1bcfc01fd6a540f3a43
0ba95c81bea6f998e3b9f53ae3763e3fc8f0c7fc9c750bb4ca74c82cd05f6a3a4d472f7f40d730e80c2da4722338db0b3ada
f903ba8641676821950bab77599f86fa386c0da3be311a39f6114e25828e86fa80ae61d0195a48e5ca95f2daae7bdad2973e
7b81c6abdab6a5263dcff440663a64962be5979a4ed3d1501fd0350c3a430ba9dcef5bfd483a18ea83f98982cd5042ea566c
e68fdd7fd9623387d37f1fa7f3f23ffb7fa263fab41ef37a3f4b1c99e990596ce6bfd474920db6fa207ea240d33a28cdc4ea
c69fdfd6fb487e187e1eb1fb1fc3a96e58d444d7268ab58d55d1492c78ea77b0390a2c2d81924bac5ffd47e0afa35b7c46ba
9fcb3cf973ed29ef543c990e9962fd6ad6a4adfe335cdaeab773b926ec0fe217ff470ad72246b82932402f3a4547537f40d7
30e80c2da472aa7bc2e340151e9b5e7ea5bd9db3fa5ed74c874cd53fc1aae9341d4dfd015dc3a033b490ca0d70234ed1d1d4
bf832e61d0195a48e5c632f507f4ab637d47676821941be0018aa2a3a93fa06b1874861652b9f260708a0f3ce2468025b4d5
083bc5c300a7e9f64befa56e9a5189ae4d2c8f053f54741a8dbefe20ae61a0194248d906d806a7e8e8eb0fe81a069da18554
ae6ceebc92f3fe3fd11ef35afd59272e05d6faca71d9dda9321d32cbf6ce2f359da6a3af3fa06b1874861652b96c719f8d30
e39653bfb2dc4ebf2e887e18a73ff1f9595346d679aec9cbf6f61fea39cd45477fe0d630b80c15a466036cd1567474f46fa1
9b30e80c2da472036cc2577474f407740db77486165ab96cb3f0bce9118c11fcba6fdae739be12e12731a9ed42715a90990e
99d96ce1534da7e9e8e80fe81a069da185546e00931c4547477f40d730e80c2da472f9759af39aae7d4dad15fc4f71f36971
6fc27cf23d6523d32133bf50f3a9a6d37474f407740d83ced0422af7eb7788141aedfc3be212069a2184946d848f59d0d1ce
1fd035dcd1195a08e546f8410b3adaf903ba864167682195fb9d438e82620f01b0d630a00c09a46003387c293af61000740d
83ced0422a37804fb4a2631b0140d730e80c2da47203985d293a363168a19b30e80c2da4727d13031cf6b53f405bdf188d1e
7bdfc400877d75e8477d633472ec034cb4151d9b1800ba8641676821954b5a8bfa682380a33e515437c622473ec08f49b0a1
9300889f28d0b40e4ab3012e13820c9d043ade1c059956412936c0bc4790a19300789f6847a655f8abd8afa7f3028b6d0440
5bc300d32228c106f0ec556c6c2300e41a069ca184d46d00776645c74e0280ae61d0195a48e506b0f75074ec24d0423761d0
195a48e5b203e4761eebbe2ff1e2b1cff3e9ffabb89fda7f0de7f3f4e7daccac135d9b98fd1fbf54741a8d6d04405cc32d9a
2184962d116f731e601ad2b2fbf36bb4799bd7f2782323eb4c87ccc4fcada6d3746c2300e81a069da185542e41afd1917389
164a73eab3b4a5d1863771a66d2a2f5d18990e9909fa5b4da7e9d8c400d0350c3a430ba9dc089fb4a063130340d730e80c2d
a4727d13031cf6d5a11ff58dd1c8b167adf7d83325f9bc96a6b15196a7654e965a253a2466a5bf54741a8d1d0c3ae212069a
2184946d00ef5c45c70e0680aee18eced042283742e31941c70e0680ae61d0195a48e5d214705a377fb23b9373f37dedab5f
7cace9f2f6f44d4e53409de99099a680df6a3a4dc7260680ae61d0195a48e57227293f35b98e698b4baee4b837a5b9db3379
cd9da464a26b137327a92f159d4663070310d730d00c21a46c032ce4141d3b18b4d04d1874861652b9019e3d2a3a76300074
0db77486165ab9016eea2a3a763000740d83ced0422a172706aa3e7b08e0b0358cfac668e4d807f855293a361300740d83ce
d0422a37c04d0345c7660280ae61d0195a48e506e863a5e8d84c00d0350c3a430ba9dcaf5d50151a3b0974c4250c34430829
5b32bbddf66b99aef54ac6b4d978397e0f27ffdbf3ff4df9988d4c87cc6476fbada6d374ec2400e81aeee80c2d8472c9d6f8
dcd326fdec799ff7f8c6d18661afcf52d2c874c84cb6c6df6a3a4dc74e0280ae61d0195a48e5522781f35e8ff9bcd2003734
7bc8ff8ffc491b990e99a993c0b79a4ed3b19300a06b1874861652b9e453ee2f1e8bbf92c5cbc9145eb7096f65275bf1e71d
f46454ae331d329353f9b79a4ed3b199400bdd844167682195fb75fb0485c6560320aee116cd1042cbf61fb44f1083671f03
30d530066fa04a61fa3e0638ec6b8b00d4374623c7fef32f8f40630b0510d730d00c21a46c039c23141d1b3800ba86416768
21951be06aa0e8d8c301d0350c3a430ba95cbeee375e8ad3e16722d792de6f9eaef04ae6bc9786da46a64366beee7faae934
1d7b3874d0250c3a430ba9dc00333c45c71e0e80aee18eced042289727f2777e653f4eb58f73d9d635353acf6f3f9457008d
4c87cc3c95ff54d3693af67000740d83ced0422a37c0ed5c45c7360e80ae61d0195a48e5f20686f24e795848fb33ccb2fa01
a705e679f8c0fd2cce75a64366dec2f0a9a6d37468e3d032d728d80c25a46e79b757d8a87ca5a5c6ba940e5e61accb9d9edc
953d483ad32133eff6fa54d349b6b69303804bb005d32a48c5f2db51d3b12cd71a5f9ff7ff9f65f557d338b06bba8f2bfe14
f2cb512ad12131bf1bf5a5a2535068e300d4270a2a298090aabcfcb6eef77646a3956d3ff7d9cf94e2a5640d9fc8e1a7bed5
f45e653a649697dfbed474920b6d1c40fb4401a635507a1527f4682b133f862c7c6c90936f38cfe7f3c9aa4487c4e283fea1
a29358ece100da1a0698164109f6ebb73915187b3880b7864166c820451bc01c5bd1b19303a06b1874861652b95fbb9e2a34
b67100710d03cd1042caf66bd75385c61e0e1d710903cd1042caa688fdc2670ad3ef7f626791f2d287422e990e998ad9aae9
341d7b3800ba863b3a430ba1dc00be038a8e3d1c005dc3a033b490ca0de06fabe8d8c3a1856ec2a033b490ca0df0492b3a76
7200740db77486165a3901edbf0ed31c5e3c88279dc92ff6aeb94e3465a643a680366b3a4dc74e0e80ae61d0195a48e57e7d
ea56686ce300e21a069a2184946d848f59d0b18904a06b1874861652b95fcf49141afb5780b88681660821651bc02f46d1b1
7b06a06b1874861652b9d4dc40d467ff0a1cb68651df188d1cfb58fd2b00fdda1a0274861652b9025d3bd7c6c6c5614b676a
a637c5bbc0b5d1914a746d6241fe50d16934b6cee8884b1868861052b6be75060efbda17a2ab6f8c468c3d1d56d467f38af6
b04d18f58dd1c8b10ff0a55674ec5f01e81a6ee90c2db472035c1c151d5b5800ba8641676821951b60b6abe8d8c202d0350c
3a430ba9dc00ce798a8e2d2c005dc3a033b490ca8dd5c202d0affd1940676821951bab8505a05ffb3380ced0422a37560b0b
40bfb668009da18554ae40dfb39f84c6677561ddb5cf7e647bfc5e9ec983f7b9c32c331d320bf4979a4ed3b18b05a06b1874
861652b9d2c5a2def33ffd3fe147173e94e39ee6edbeb7e7e42d131d124b0f8b0f159d46630b8b8eb88481660821651bab85
05a05ffb337474861642b9d2c2628f4f3ba21bfdfa5818fa011ecb76cfebf62c6175a643666961f1a5a6d3746c61d18cb70d
83ced0422a577ce7e3b6f068e278eda1a7afffbdc57632d7e12faff33e3f9fb4cc74c82cbef35f6a3a4dc7161680aee186ce
d0c250ae3436f053643f63881ef147eef39abe979bff09eecfcbad46a64366696cf0a5a6d3746c6101e81a069da185542e35
3610f5d9440287ad61d4374623c73e42e70c41c7261280ae61d0195a48e546e89122e8d8c202d0350c3a430ba95c397f5efe
a3d862b3856b3d2e7fee9be24bfdc16b66beaf697aa065a64366397f7ea9e9341d5b5800ba8641676821952b57ca753fce2b
4ec7af297c31ef356eea39c3d9fe9efce7f45c2955a64366b9527ea9e9341d5b5800ba864167682195fbf18448a2a17f0580
9f28c00c19a46862aabf861b2acb1977b884b724b32d8998ea3f990e9962aa6fd674920dfd2b40fc4481a675509aa9f59c9f
17cffe9339e3dacb9f66a287815ccf954c874cb59eb36a3a498616161d6f8e824caba01453fd3a5217a2d4466f3afc0f7fbd
76b568af990e99aa638755932d2c9ee1b285453b9743386c2c5b53b82d62849f223f6e5222d1d8bf02c46db841338490b28d
d0a944d0b17f05a0db70436768219513d0a17f5ee80f147736f98186395b79e1d0c874c814d0664da7e9d8bfa283aee186ce
d0422a37c0e61945c7fe15806ec3a033b410ca0db0154ed1b17f05a0db7043676821952b1b3cf7cb9f59e2ddfee7667f9820
ae97ffeffdd4787fbede32d321b36cf1fc52d3693ab6b000741b6ee80c2da4726517ef3caffe021a3f88dcec2b0d7059e66b
3dc34388b28b57653a64965dbc5f6a3a4dc71616806ec30d9da185542e432fd31ddd6ec200273f1b5f8eeb4ccea17e4912fa
6f3c676f99e99099a13fd5749a8e2d2c00dd861b3a430ba9dc083d1d041d1b5900ba0d3774861652b9fcf2813fb56ce71d4d
8dc2c6f9f472401860b40ff3a7da02ad331d32f3cb079f6a3a4dc746162d34c20d9da185546e80974c141d1b5900ba0d573a
430bad5cf60c9ae7f45248fc20f6d34f26d2eb4ef979cff67806a94487c4ec19f4a5a2d368ec6201e236dca0194248d9b251
66f4289be24a6f0eebdaffc7dcdbedec926bd7797dccabf80e2520dabb5845d6cfa11cc7028ce4c0f20e12c072804492250b
2dc9961407c8dd04c84de4ee32e72459e4e01a737d254bf05b0d6d75f7ecb9f8d633aa8ac5ff219936c31327f72127334066
3d28f3519981d3a18b05408fe181ced1822a57a1b5b7a567dae83549c747d26239ba4adb53f1d8efc341596280c48afca4c4
c0d1d0c20288c7f080e6084165b31525ac7cf491987eb68787f29daba1d75e0f1e2b8def72ecf1224f7b5eb2b5ddcc9a69bd
f2bd64886706c8ac078f3d2a33703af49100e8310c748e1644b98f5722040d4d2480780c0f688e1054b64f9fa8ccd0d04a02
88c7f080e60841657b83410aa1432b09801ec3039da30555ee052761313ab49200e8313cd0395a50e55e70be1ca3432b8911
1ac2039da305556eb692809ffdd6a5612cdfb91a7eed9f33a36150e85f01ac6378807224a082bde0b96274e85f01d06378a0
73b4a0cabdc15a81d0817b06300fd181cd5182eaf6c167fa472430cd00d0213a10717c26d50bbe0d840c4c3326de3b3a9071
1598629f6ef2102c70cc00d8210a585c821fe57a41339e90a16906008fe1818debc034fb740f9181a16306f08ee181cc9181
8af606f30842878e19003d86073a470baa5c3d027dd9ca61ec764dfbb9eb48bb35b9edc8aea833e0f508749a1920b31e81fe
a8ccc0e9d031638486f040e76841952bd06bdfc716d72c7f406e8c9df3bb1ef2fa5ffa1616669a1820b1203f2a31703474cc
00e231dcd11c21b86c2f381a89d1a16306408fe181ced1822a57a0977d499b3cf6f6be2d5b3901c66a9f789cc771dcc79d39
9901320bf4b33203a743c70c801ec3039da30555eed3a7d93234b4cb00e2313ca0394250d92a712a5f167bf4a4b52457b4d8
34702c073caec7fd6cd3cc009995f9519981d3a15907408fe181ced1822a57abb0d208b6850db1ae6b2887b1ad79cd578ef9
aec3686680cc5a893d2a33703af4eb98a07b78a073b4a0cabdc10d86d0a15f07408f61a073b420cabda059c2e8d0af03a0c7
f040e7684195fb74eb93a1a15907108fe101cd1182caf682f3e6191dfa7500f4181ee81c2da8722f18446674e8d731424378
a073b4a0cabd60dc8bd1a15f07408fe14ee768c1957b975f07407f6b8501748e1654b9f079df5846876e21003d86073a470b
aadc0b4638191dba8500f4181ee81c2da872b522bb0dcd74e84abeaf52c9eff6a9998f35e49901326b45f6a8ccc0e9d02d04
a0c7f040e76841957b81c528a343b710801ec3039da30555aeda301e4377480fc73bf58db47f598e4b9dbf5a1b946706c8ac
368c8fca0c9c0e0d4326e81e1ee81c2da872e538e9433a9bfb11ed0ce045ee4ad9ea647da365d9d22ef551817632036496e3
a49f9519381d1a8600f418063a470ba25c818e75fba09d617fa4b46ff2a9b1abbdf6bad6b642f3cc009905fa599981d3a161
08408fe181ced1822a574e4b57433b5da264877cd773fb63b945b5cb50efb4931920b39c96feacccc0e9d03004a0c7f040e7
6841952be7e2e73dd9d1fd7670ff36f8152dc7226d277946eb3bed6406c82ce7e23f2b33703a340c19a1213cd0395a50e5de
e08040e8d03304a0c770a773b4e0ca15e85d8fa997e68bdd8838d81c2ff994fa48bfb2159a6706c82cd0cfca0c9c0e4d4500
7a0c0f748e1654b9d954047ef65bd30c28dfb91a7aed45eb2b273bd1ca9e85e216709667a1ee48a935094f0c9058947e5462
e068682a02c46378407384a0b2557f0de962ead7db6e7e5a726991db3b2f7f6edb75a36cf5d7a0990132abbfc6a33203a743
5311801ec3039da30555ae7c31b28d016f36ddb1ec8327a81e802497b7b4f3759ccc0099e58bf1acccc0e9d05404a0c7f040
e76841952b6d83158c4ef41bbe96f1e1e5c205f14e6680ccd236785666e074682a3241f7f040e76841952bd097b41cd6e53c
ca35e9bba73300d662b3f5cd71ad77dac90c9059a09f9519381d9a8a00f418063a470ba2dc0b865f191d9a8a00f4181ee81c
2da8722fd8fac0e8d05404a0c7f040e76841957b81c306a303539191798c0e6c8e1254b717ece0226ca3a90800f76007e32a
50c5da26bd7e92553aa4da91f692cd5e6f391e511ae8cbbd33916606c86c9bf49e94191817f88a00ed101dc0a80644adb6f7
543e99d76a5d8ebcc54d1a478bcdec491be24cdbb5ecdd79826506c86c7b4f9f94192817f88a00ed101dc0b8064caf179c62
49c8d05a0480c7f0c0c675609a917de4ba746cd7b3adac1b267dec43ea197a26c49d192093ec2377cb0c9c0ddd4500790c0f
708e1254b7176c9e6774e82e02d06378a073b4a0ca3168f958ea06603b905acf0e8e392676b061cf0c90c9a0bd3203a74383
11801ec3039da305558e40e7289fb0335b8b296f475acefda42758de89614c24c85e8981a3a1c1c844dcc3039a230495cdfa
e9ac7cb4f8809f1dc350be7335e4da5f70402aa3438b0f801ec3039da30555ee0daf14a1438b8f111ac2039da305558e40af
51da69b998846569994adf232df44edf99013209b45b66e07468f101d063b8d3395a70e5d8a1c776de76b265d6e9d27fcaf9
a015c99d1920931d7aec9519381d5a7c00f4181ee81c2da8722f70b06174e8f201d06378a073b4a0cabde04c6f46872e1f00
3d86073a470baadc8b5c3e80f85b0b0b407384a0b2bdcbe503a0bf35d0003a470baadcc79d27081a1a8c00f1181ed01c21a8
6c8d381fe548949cf6451a0cd96692ea39ae52fddcc03fe605c86bbcdf971738175a8b00ee181eb81c15a866cd68e387f2d1
df63fad91e1eca77ae865efbecef013ffbad75c654be7335e4da677f8ff167bf37af80f29daba1d7fee9b788a1a1b907108f
e18ee608c1657b83a309a143730f801ec3039da30555ee0587cd313a34f700e8313cd0395a50e55e301cc4e8d0dc03a0c7f0
40e76841957b41378ed1a1b907408fe181ced1822af7f9c360291d9a7b00f4181ee81c2da872ccdc43cf9a5e632ccf65ac27
1213e89e192093997b7865064e87e61e003d86073a470baa5c83decbc9877a81ba19e5caf27db131957849b5b30d43332c33
4066837e5266e07468ee01d06378a073b4a0cab5b3ac75d6645bb43f2d5fd7753fa5ce29474f4b7ff7dcaef53640a0990132
db69d64fca0c9c0efd3d26e81e1ee81c2da872eff2f700e86fcd2b263a470ba25c3b9fde66906c25b72e1c3b2f1d7dffb2e6
8c7c56cfeb1ee3e5990132dbf9f44fca0c9c0efd3d86ebc5f040e76841957b97bf07407f6b5e31d0395a38cabdc17e81d0a1
bf07408fe181ced1822af706a30d4287ee22003d86073a470baa5c832e7f59a5b34ba75b2eccce419467e32887aedfd02c33
4066837e5266e074e82e02d06378a073b4a0cab577fa8cc776d8f2f5b39dca6cee59172efc73320364b677fa499981d3a1bb
08408fe181ced1822af77977264a87ee22003d86073a470baa5c858edb218de1721f867f5eeea66445267901f22af083f202
e7425f11c01dc30397a302d5ac36402f6d9a1ed95c40b2cd6c9cb6c9503e2bcb72ecf2df5b03946606c8ac0dd04765064e07
d622c03c4407364709aa1be95d5dbb3c78d23e349b43b9306944c5487b577766804cd2bb72cb0c940dac458078880e685c07
a6d9e757543032b0169978efe840c655608a7df20da67e20b3a9486f40f670ba74fdafcea2fe3a16e284c7425ee00fc9e866
5f9101fa0e239da305558ef98aecd232daacf1205fb2b22583cd7ede89011299ab8853e26c2a32107fe79881688e10543636
b5bfeb42e2cb4e2fc9cb99f48079363ad213c398c826f69d1267471120feb95d06a2394250d95e30dbcbe866479101fa0ecf
748e1644b9369c2f9f8bebb445d4d98ead48870d6aa47a2c4d1ffaa3990132db70fe933267479101fa3bbb0ca473b4a0ca35
4791db894c1ebc53d7acdb14cb36efada68901129b9fc88312673b9181f83baf0c447384a0b25562e9c9a56dbdca4492fc7d
49d1fe455a2e695dd6fe68d3c4302656e22725ce5e2203f177461988e60841652bc447199db2e74efa3647ded48ce24b4fb3
c8e5249e86cc33036416e66765ce5e2203f477461948e7684195ab3b0eb4bb9e52399f45b72c1eb635410f6b297bdb5aa5cd
13c39858f71b3c297136121988bf73c9403447082adb0b0e7c6174b3914887ee61a473b4a0cabde09c0446371b890cd07718
e81c2db8722fd837c5e8662f9101fa0e239da30555ae1e5096cdf8c3cea049ab4e9de5cd4e74ed6779d4f3c9586280c47a3c
d9931267239181f83ba30c447384a0b2155b0d52feece631fcec1dc6f29daba1d75e8f2db51d46f12c4f7c7de48b48ba0063
d9db6e0f273340663db6f45199b3970840ffdc2803e91c2da87205fa30bfa4584efbd9cfedd2e3bdec5f525b7a52a1796680
cc02fdacccd94b6480fece2b63a673b420cabda1f22474b39dc8007d8791ced1822af782e3a619ddec283240df61a473b4a0
ca7ddc3586a0cd762203f11d463447082adb0b4e8d6274b39dc8007d8791ced1822af7822620a39bed443a740f239da30555
ee051baa19dd6c663240df61a073b4e0ca150f1552feec2b32fcec1dc6f29daba1d75e7e96943f3b7b0c3f7b87b17ce76ae8
b57fd0e682404d861e036a8b2292230095eb0d4e173fb24dce1e03718b221ad78169f6826a9a904dce1ec05ba248c655608a
bde0285a4236997b0cbc2d3a9371157e54ec050e90846c36f71880ef30b2711d9866b527319c47168f5507ca165b2616cb12
9294dac9f83c334066ed493c2a73f6f71890bf33af40384709aa5bed330e4773c51ce51aa51760f7673bed93d94ec6e78961
4cac3dc62725cee61e03f177ce1588e6084165ab67555a775652f492ca7123a91c74aefb2396b4b45d804e6680cc7a56e5a3
3267738f0efdad7905d2395a50e53e6e8040d066738f81f80e039a230497ad9ecb74e881e6573d47324b75137569835e608a
a72d6bfeaae732d1cc0099f55ca64765cee61e03f477ce1548e76841957bc161548c6e36f718a0ef30d2395a50e5ac69cbca
9f1d36869fbdc358be7335f4daab918aae3894aace8e75b4f5e2ba52dc74a96307b7910acd0c90598d541e95393b6c0cd0df
d947209da30555ee0db622846e76d800e81a463a470baadc0bbe188c6e76d818a0eff04ce76841947bc16032a39b1d3606e8
3b8c748e1654b9027d16abbd728a7aef635943462a7cdd875ca1796680cc02fdacccd9646380fece4102e91c2da8722fe8d9
30bad9646380bec348e76841957bc1801ba39b4d363a740f239da30555ee0d2e04846e36d918a0ef30d0395a70e5ea7a82fb
2dd3d3f6723ea506b05d4eab9ab1d97f6c0b0a686680ccbaa2e05199b3c9c600fd9d8304d2395a50e52af4d0685bafe15427
e9ed9f295eba68be42d3cc009915fa5199b3c9c600fd9d8304d2395a50e5eae37ded515e426b9caec775dac269ab81163c74
d7c90c90591fef4765ce261b03f4770e1248e76841957bc33b4de866938d01fa0e239da30555eed3b3620c6d76d81888ef30
a2394250d9eaca914d5a4a7b31ea92b74edefaab34a35ac7b3b5c6686280c4ba6ee44989b3bd0610ffdc3b02d11c21a86c9f
1eb26068b3b7c6407c8767344708225be9bde7628a6a8b2e97b39ac2164fcde2b9d78ea57032036496defbb332676f8d01fa
3be308a473b4a0ca9593d2cfe603a117b89bf7c9d2ec4fb63def3ae1514e4ae7990132cb49e9cfca9cbd3506e8ef8c2390ce
d1822a57a097b554aa76647d596b926ce5bc9e937be856d5dae47632036416e86765cede1a1dfa5be308a473b4a0cabdc050
84d1cdde1a03f41d063a470bae5c812e27f62fe5aed493ae6dfe72919b224dd6d4a6519ccc009905fa5999b3b7c600fd9d6d
05d2395a50e50af43a1c8ab2d8229a68cba7f45db403829be3034d0c9058901f9538db7a0cc4dfd956209a230495ed0dcf36
a19b9d3d06e83b8c748e1654b9528bad7a0dc765f3594bd6b3bdd5ecd92e50eed026ef456d7a3a9901324b2df6acccd9d963
80feceb602e91c2da87205fa2a2fd856aee936e5d2f1dadb9fa340f3cc009905fa5999b3b3c700fd9d6d05d2395a50e54acb
64bfe2a5074558252337242fe76e8766483baa9d11585a263c33406669993c2b7376f600e89fdb56209da305552e7cde5899
d1cdce1e03f41d9ee91c2d88722f180c6474b3b3c7007d8791ced1822a57474be486688b49efcab634130abb5a34ae713203
64d6d1924765cece1e03f477c61548e7684195ab9b8bcab0b40d666c67b1c3dd6c0cebb8cf1ba89b8b686680ccbab9e85199
b3b34787fec6bd02d91c25a86ecdeb22c92bb7ecf584bab41dd96e89e45ce9d8d6fb58349617c6bce673f17d798152a1a7c7
805a8380c4f9a9566d43e4b6ea9e63db98badd1e8aba79513b03cbb5dc5be5686680ccb621f2499981714d9e1e036d8b2218
d580a8f5822daf846bf2f418685b14c1b8064cafcf1f4dc0c8664f8f01f80e231bd781694688f73d5dd256b03b93d391f3b9
6fccc4e44e0c904878bd1267438f81f73bb70a247364a0a27d7aab3e439bdd3c06e23b8c688e1054b6179c64c8e866378f01
fa0e239da30555ae795bfc50fe6ca9013f5bc358be7335f4da99d6b707a91d606f5dac959d0fda33036432adbd3267438f01
fa3bb78a99ced18228c74eb0896525977dccceadd80ad25ae4ce0c90c94eb0f1ca9c0d3d06e8efdc2a90ced1822af7824351
19dd6ce8d1a17b18e91c2da872cc23a7ed88fdb229e27869ef927e20efcc0099cc25c72b7336f418a0bf73ab003a470baedc
0b5a058c6e36f418a0ef30d2395a50e5988bc95e6db4bfecfc8a4bee4bba58d3af6706c8642e265e99b3a1c700fd9d5b05d2
395a50e55e70a435a39b0d3d06e83b8c748e1654b91718d931bad9d36380bec348e7684195fbf4c1fc0c6d36f41888ef30a2
394250d95e70920fa39b3d3d06e83b8c748e1654b917985232bad9d96380bec348e76841956bd08b39ead8e7a40d7dda93b8
2cd7baea886c636689614c6cc80f4a9c4d4580f8e77e1d88e6084165fbd15464f8d9effc3ae6f29dab21d7fea3a948ffd96f
4d33b07ce76ae8b5bf60088ad1cdbe2203f41d063a470bae5c83de2f69a1d90102ed705a5bea9beab5f68e15cd0c90d9a09f
9439fb8a0cd0df9966209da30555ee05ce538c6ef61519a0ef30d2395a50e55e30e8c8e8665f9101fa0e239da30555eed3c7
2932b4d9546420bec388e6084165fbf4c823439b1d4506e23b8c688e1054b6178cc331bad9516480bec348e76841957bc171
b08c6e761419a0ef30d2395a50e59aa3483feff190bfd482b6987fd47bd26a309a1820b1f9893c2871b61301e29f7b65209a
230495ed6d762203f4775e19339da30551eef307d253bad94ee4bede218c748e1654b937580f10bad94e6480bec3239da385
a3dc1b4c2608dd6c273240df61a473b4a0ca69f786963f1b7a0c3f7b87b17ce76ae8b537adb745f78698c9c85eed324b1d90
a4bacbf7965027334066d3fa4999b3a1c700fd9d5b05d2395a50e5def05611bad9d06380bec348e76841956b1f8d729e54b1
93a93b23ac824fd219c9977a86b58f06cb0c90d93e1a4fca9c0d3d06e8ef3c2b90ced1822af7e1b601459b3d3d06e23b8c68
8e1054b6d604b41d6cc51f6abb5d8374e2239fd7760e33ee2c318c89ad01f8a0c4d9d06320feb9650582393250d13e3f33c8
d826438f81b845118debc034fbbcfb29239b0c3d80b744918cabc0147b415b97397280ad0734e486b0dcbf3325a9fd7e8542
9cf050c8e78d79281dd87a20740f039da30555eed36b85181ad87a20710f039a2304958d10a7fbd1b3990ef9ac48af7767c8
7766804cc2ec9619381d387bccd02d0c748e1654b9178c38323a70f640e81e9ee81c2d88722f981a6374e0ec81d03d0c748e
1654b9b64ab9ae3ab01b51174edb32f269afaf931920b3ad547e5266e07460ee81d03d0c748e1654b917ac466774e0ef81d0
3d0c748e1654b9baefe0589618b32d2492c6cb11d765b31381d6355e7a06d33deec8330364d67d078fca0c9c0efc3d10ba87
81ced1822af7828352191d587c20740f039da30555ee0507b6333ab0f800e8210c748e1654b94f9f2dc9d0c0df03897b7844
7384e0b215e2daeab7cf8954a9a2cb56b63dc555b7b04a2dd00e4de59901320bf3b33203a7037f0f84ee61a073b4a0cabde0
4c384607161f08ddc340e76841959b2c3ef067bff3afc0f29daba1d7fe86078cd081c5c70cddc240e7684195fbb8810c4103
7f0f24eee109cd1182c8f682c3f7181df87b20740f039da30555ee0587ef313af0f740e81e063a470baadc1b4c4d081d587c
20740f039da30555ee059b8c191d587c20740f039da30555ae9e38583c0a6cb5c75af78ada111e5174ca4b3c5aff9d278631
b19e37f8a4c4c0d1c0df03888730a0394250d98ae704291f1c36f0677b782cdfb91a7eed93c306feec77e61558be7335f4da
df608840e8c0df03a17b18e81c2da8726fb0702174a3d10732df51607394a0babde0e450c2361a7d20f11d0534ae03d3ac9e
b567e7652dc96e47d93db0943ee75de3d6b3f668621813eb597b4f4a0c146b74f998616b14b0b8044caefa651a16b30c6779
eaa5adc509eefe32d1cc0099f5cbf4a8cc40c946970fe4bda3131957e147c5eab884f5b28e720ae029f765b962e95cebb070
3cf67b5882250648aca3124f4a0c140b2c3e90b687018c8bc0047b81ff2663038b0f44ee61807394a0bad5cec4b295ef885d
934ee8e4d376a946ddd9660702b6ce04cd0c90593b138fca0c9c0e5c3e10ba8781ced1822af7828e32a303970f801ec240e7
6841957bc1480ca303a30f84eee191ced1822bf72aa30f84fecec502e91c2da87205fa947e9cdc897a816556debea731de5b
440a34cf0c9059a09f9519381d187d20740f039da30555ae42eb96ef6b5f8b134939c9b43d978bbc8d473bc9dbc90c9059a1
1f9519381dd88c20740f039da30555ee55362308fd9d8706d2395a50e5de5091113ab01999a15b18e81c2da872d566e418ce
27d62343a41764633871d3c30f4f7d3eaacd08cd0c90596d461e9519381dd88c20740f4f748e1644b94f9f6bcdd0c0630489
7b18d01c21a86c2f980a6474e03182d03d0c748e1654b94f7722191a188c20710f039a230495ed73cea70c0a7c4580750803
94230115ec0dc60b840e7c4510ba87473a470baedc0b0e8d6474e02b82d03d0c748e1654b90a5d4c126c85feb6b6c3a7ed15
ac17db960df3cc009915fa519981d381af0842f730d0395a50e52ab47c26f5785a6d2c6c52cb5e52d196a1baf55e5655a169
6680cc0afda8ccc0e9c05704a17b18e81c2da87215bad89396f9143dddfcb07344f5514c595d00efa54934318c8915f94989
81a381a90812f730a0394250d93e57633328b01341d61e062847022a589d76dc9274f78e320c799d79bb96752bc374f78359
a71d696680cc3aedf8a8ccc0e9c05164866e61a073b4a0cabd60208cd181a90842f7f044e76841947b41a799d181a90842f7
30d0395a50e50af45a973d9a43c2711c674efb510e549766a25ceede8647786680cc02fdacccc0e9c05404a17b18e81c2da8
72e564fabdefb05af6ea10611e09e77d184e39989e2606482ce7d23f2a317034701401e2210c688e1054b6421ca55259740f
b05e539686ff9ab67a88fe5dfd14649e1920b3303f2b33703a701441e81e1ee91c2db872c56ae2b836b9c8b2497229b6a1d1
56fb4ae1e53fd4d688931920b3584d3c2b33703a701441e81e063a470baa9ccde9b3f2c1d6037fb687a17ce76ae8b5579f9e
62706c9fb445cf47dbaff32a9e3a475d42551f309e1920b3faf43c2a33703ab0f540e81e063a470baadca7eb1186069e1e48
dcc380e60841657bc1e782d181a70742f730d0395a50e55e306fc2e8c0d363866e61a073b4a0ca9526e0b116672c6b96dadc
fb719891907441ea7988b509c83303649626e0b33203a7034f0f84eee189ced18228f7b98e0d83022b0f64ed61807224a082
15d66ed7a87b9b36f94fdb69531b6bdd59df5e6527334066217e5666e07460e581d03d0c748e1654b9ba632c1d296d52917e
e9b238f9fb92a2fd8b34c5d3baac7d6b204d0c6362dd2ff6a4c4c0d1461f0f00ee51007364a0a2555eed641d871dd67a9fd5
aa6fdc26cfe276a82f4703a69901322bf1a33203651bdc3c10b8054730ae02558c6d5d5ff430a9e2382265e7bc2e17dba97f
270648641bd79d1203831aad3c10f58e0215158048f5e903e608d4e8e381a87714a8b8004c2c760243ad63f481cbfbb4439f
6706c86427307865064a063e1e08dcc3c0c675609a91b3e5d67238868d6eeff16c9376e46cb93b334026395bce2d337036b0
f240e41e06384709aadb870f16a16860e581c43d0c688e1054b6760a43ccb6c4e0abf8cf9a77a17e3b77e93f1fea207d9f82
4a330364b653189e9419381d587920740f039da305556eb2f2987ff6e73e1558be7335f4dadbcffe503e9869e0cff6f054be
7335e4dac99b2c158b34b60f3b164ca708ebe4337993efcc0099e44d76cb0c9c0ecc3410ba8781ced1822a47a06339fcacbc
77d25cb8e4cf6476ec44cf0c9049a0dd3203a703330d801ec240e7684195fbf4e9220c0d9c3490b887473447082edb1bec43
081d386920740f039da30555ee0df621840e9c3410ba8781ced1822ac78e54c9e7215d581baf49f2e22df24dd9d8c3dd3303
64b22355bc3203a703270d84ee61a073b4a0ca35e8681f537be3167d0b0e1bb92863bed23ceebd09921820b1213f28317034
b0d140e21e063447082adb1b4c25081d386920740f039da30555ae75912f3d3ec97ab37a245a3cb6b2c0e2daa5d996e5aedc
3d64921820b175901f9418381ad86820710f039a230495ad111f512af6d59ec318773d2ecb3e2cdbb9c77cc5fdba3fcf3433
4066637e5266e07460a381d03d0c748e1654b9069dec0b6ad5ea68bbb61d52e5ea294e77438c6606c86cd04fca0c9c0e9c34
66e816063a470baa5c83bea713a40f7fea48a099d86cf3464b9a1820b1213f2831703430f140e21e9ed01c21886c2f18f062
74602102d04318e81c2da8720dfa940eed66bdf91c775da7946c69c5662da62df7cf33cd0c90d9a09f9419381d588820740f
8f748e165cb9179c78c6e8c04204a17b18e81c2da8729f36ba6268e01f82c43d0c688e1054b64f8fe93234300f41e21e0634
47082adb0bbcfa181df88720740f039da305558ef987dccb1cb5cd2875ef96a429c3a0efcc0099cc41c42b33703ab01041e8
1e063a470baadc0bc649181d588820740f039da30555ae41efe57baa17a88b1a7457951d139b746f9574833a34cb0c90d9a0
9f9419381d588820740f039da30555ae1d13aee717459b0b959684b48fe37e68b3713f532c7d84af764e38cb0c90d90e0a7f
5266e074e0223243b730d0395a50e55e70b232a303171184eee189ced18228f72a179111fa7b8b0ca473b4a0cabdc1ef80d0
818b0842f7f040e768e128f706670b42072e2208ddc340e76841959b5c44f067bfb3c8c0f29daba1d75e7e96940f3e1ef8b3
3d0ce53b5743afbd3dd7763896ed3d39b3f68074e45d9f41f9e02c39e6fbac5d9e1920b33dd74fca0c9c0e7c3c10ba8781ce
d1822ad7aaed2defc76947299dd21a96ff6d367c7768d5732dd76d9bcc330364b66afb499981d3818f0742f730d0395a50e5
dee3e381c4df3955209a230495edf38d4f4a375a7920f31d05364709aa1b3b0edf4e08dbec60c7bcce4306343340263b0edf
2b3350b6d1ca0389ef28a0711d98669f77596564a395c7cc5ba340c655608ab5ea7a2be7cb7c95934b75bda18dda1d673dff
ef7e8b696680cc565d3f2913ad3ceecb452b8fb1353584e581d5c1b512ee8538e1a1900a1deb30b455ad4b926bdaec88a563
9d5c65796680cc0afda8ccc0e9d0ca03a07b18e81c2da872e44dbea4dbb31ea72d94975e60925e418cd494e7ce0c9049de64
b7ccc0e9d0cd03a07b18e81c2da8722f18186274e8e63141b730d0395a50e53e3de2c9d0d0ca03887b7842738420b27d7ad2
86a1a18f0710f730a0394250d9daaaec7e226b8a67d91d6afbdaa5ab97a51e5aeeb7996606c86cebb29f9419381dfa780074
0f039da30555ee05ebef191dfa7800740f039da30555ee731b2c1814da77006b0f0394230115ec053bc7191dda7700740f03
9da30555eed34e160c0dbd3b46e2210c688e1054b67a02d7706ed472c6b8aebb3943c5781df25d39db7a309e1820b19ebff5
a4c4c0d1d0bb03887b78447384e0b295fd70d28f3be479b391397de9ec7352aed686a8f6dcf6c3f1cc0099653fdcb33203a7
43ef0e80ee61a073b4a0ca15e8fd8a97b489d77281f6aaed36f0acb3e1fb1ed7b624dfc90c9059a09f9519381d7a7700740f
039da30555ee058727323a740e01e81e063a470baa5cddeeb966e9e69c4739b341e70cf5bdb30b5ccb8ee4bd6df7a4990132
eb76cf4765064e87ce2113740b039da30555ee05e708323a340f01e81e9ee81c2d8872f51c4173f958cdf42b960394921d2d
14b5fad5bd13f50bed6406c8ace7083e2a33703a340f01e81e063a470baadc0b4e766674681e02d03d0c748e1654b9171c25
c8e8d03c04a07b18e81c2da872ef320f01e86fcd3180ced1822af722f39091f87b5f0e407384a0b2bdc14e83d0a1750940f7
f048e768c195fbf46d6668e89a02c43d0c688e1054b662d642ca47df12f8d91e86f29daba1d7fe86a78bd0816f0930df5160
7394a0ba15a57f2c1d9c43e037ef2814ceaf845df5a70f7d2558e01c32c1d628607109985c2ff82c1132700e01de3b3a9171
157e54ec056d0f4286e62100dcc3c0c675609abda089c9d8d03c04907b18e01c25a86ed57b31d681aa3234b5e694e3514e13
5febe14d159a6706c8acee8b8fca0c9c0ecd4300ba8781ced1822af72ef39011fa7b670ca073b4a0ca7d7ac48fa1a1730810
f7f088e608c1657b83b502a143e71080ee61a073b4a0cabde03c504687ce2100ddc340e76841957bc159678c0e9d4300ba87
81ced1822af7f1179aa0a16d0810f730a0394250d95e506f333ab40d99a05b18e81c2da872e1c3eec80c0d3d4380b8872734
470822db0b0ef16374681b02d03d0c748e1654b9f079a3604687b62100ddc340e76841957b81d726a343e71080ee61a073b4
a0cad5d504a734fe4511ebf51dd2d13be463628f6259a617efbd163c3340665d53f0a8ccc0e9d04264841ec240e76841957b
83c702a1430b1180eee191ced1822b5717cbacd234cd870d67c893271da37c5dc5f676df6d3d577bbc796680ccba58e65199
81d3a1850840f730d0395a50e52af4212f9d8e5fe80526e9eb9dc7618b3dd674cf1557689a1920b3423f2a33703ab41001e8
1e063a470baa5c85966a5d5eb56c77453747d861127a813ade21bddffb68439e1920b3423f2a33703ab41001e81e063a470b
aadc1b1e6f42872e2200ddc340e7684195fbb86f0a41432f1120ee61407384a0b285cf7b80333af41299a05b18e81c2da872
e1c3c6c90c0d8d4480b8872734470822db0b9615303a341201e81e063a470baadc0bd647313a341201e81e063a470baa5c81
5e86034697f6654d6598633de292da494a4e6680cc02fdacccc0e9d04b64841ec240e7684195fbb8eb014143231120eee111
cd1182cb568875c5ad1e0b6ca6046587c41a6d34ab4d4a34649e1920b3303f2b33703a341201e81e063a470baa5c81964fa7
7c312f5b33bfc84b1f0f79038bbd8b3c88521f2db59fe16406c82cd0cfca0c9c0e6d4c00ba8781ced1822a57a0f37e5d5bda
6d3bccb20fa7d5c95b59ffaad03c334066817e5666e074686302d03d0c748e1654b9027d49af2ba5bd7af72cc535b21870d8
58a5ae4728d03c334066817e5666e074e86402d03d0c748e1654b952756779f8e43372d6fb501acbf669913762934e415bd9
4a130324968afb518981a3a18d0910f730a0394250d95e30e6cbe8d0c664826e61a073b4a0cabd60f29dd1a18d0940f7f044
e76841947b81cb2ca3433f1380ee61a073b4a0ca15e8a35c5f2c874e96a371ca289ed946c97f6c9d2b9e1920b3403f2b3370
3af43301e81e063a470baa5cdd399775abd56e2ba9929d037e969751ee49996abcb784d2cc009975e7dca33203a7034b9391
b94781cd5182ead63684e6a38c5f6579eee416e99fd1ede776564aeec76c93bc00796d3be8f7e5054a359a99006a0b8e489c
9f6ac576fbeab9a631d99443baf49f723ea8ebc59d192093edf6f5ca0c8c0bfc4c80f68e0218d580a8c52c3ea41e8d3a7aa3
f7649776a22d7c61b07766804c66f1e195192817589a00ed1d0530ae01d38b1cbeb1d81478b491aa7d89f57810c2db330364
92c337dc320325434b1300ee6160e33a30cddee07541d8d0d204907b18e01c25a86eed549998a4eade6c30e3dca37a35d819
3047390027b6017c9a1820b19d2af3a0c4c0d1d0d204887b18d01c21a86ced348a28cdee62ec2b6da1735d52b64ec0b1c89f
d9cfb4dd8707d1cc0099ed348a2765064e87962600ddc340e76841956bb7f93c97b459a74e5ef8b51c9d6bdda0e2097aafdf
e5990132db8d7e5266e07468a83241b730d0395a50e53efe6c1334f47201e21e9ed01c21886cad0a936fb5f4e4ec9648e766
39ced39a86fbbe4bd513b7ebfe22d3cc0099ad0a7b5266e074e8e502d03d0c748e1654b98fbb321134347219898730a03942
50d9d8b77938d53dd731aa7e6a12cd0c90c9becd5e9981d3a1970b40f7f048e768c1957b41038cd1a1970b40f730d0395a50
e53eeda2c7d0d0c805887b18d01c21a86cef327201e86f8d4a80ced1822ad77a8c977c482f1b999347af7e59f402f5147beb
05dd7d46961920b3f51a9f9419381d7ab900740f039da30555ae412fe63663e7b4adf73c8a1e65bf5cd2a95ffbe9f82c318c
890df9418981a3a1910b10f730a0394250d9de606b42e8d0cb05a07b18e81c2da8720d7a18b2903b94b694ec632a6d879c45
a9deada29901321bf4933203a7432f1780ee61a073b4a0cab5512f3d5ee82ade49c7b14aff5627fd75714f1982efc6633433
40661bf57a5266e074e8e53241b730d0395a50e55e30d4c7e8d0ce05a07b78a273b420ca35e8ebb0d1767b14d7b84a47c88e
90d89a4febfd4ed3cc00990dfa499981d3a19dcb083d8481ced1822ad7dee9d2ad2de7b5f52510dbbe57b3e9fb9526890112
db1bfda0c4c0d1d0cb05887b78447384e0b2bdc0b587d1a1970b40f730d0395a50e5de606043e8d0ce05a07b18e81c2da872
2f688b313a747401e81e063a470baa1ce967e8c6815d4ff4b7ea274b0738ee3bed5cdd990132494fc32d33703a747401e81e
063a470baa1c73078ddb9eb4e2b5e7523afce77ed2e98b3b318c89cc1bd429317034b47301e21e063447082adb0b0ea56574
68e702d03d0c748e165439325a227f64b59579d6fd4d51ba643adcc0acca5b66804c6655ee9519381ddab900740f039da305
55eed3639f0c0dbd5c26e216063447082adb0b4e996674e8e502d03d3cd1395a10e59a978b7d3f6db7eeb91ed2225673de2f
6bc8480beabcee3a8c6706c86c5e2e4fca0c9c0ebd5c86eb1dc340e76841957b83e705a1432f1780eee181ced1c251ae79b9
94bfcc6a653f8f7c4ad7decc2aa41b587c0c1a34cd0c90d9bc5c9e9419381d7ab900740f039da30555ee0d0636840e9d6400
ba8781ced1822a373bc9c0cf7e6b9302e53b5743affd0d6f15a143271980ee61a073b4a0ca7dde008cd2a1930c40f730d039
5a50e53edc36a068e82403c43d0c688e105436d6ee8de5b407eb8c9d75ab59a4edde96192093b57bbd3203a703271960bea3
c0e6284175fbbc15036303271920bea380c675609ab5019a0f99a4332cb09199606b14b0b8044cae77d9c8dc978b36326343
6e08ef367e194bb817e28487425eb07087d1a18d0c40f730d0395a50e558e7b5ac4a30c7eb3d4a35b3c6fd4cb4f3da320364
b2ceab5766e074682303d03d0c748e1654393207fcdfd45199a1a187cc44dcc280e60841656bb5d707fdd1191ddac800740f
4f748e1644b93675d2769dc88da87deb32fe5dce5fd741e93675c2320364b6a9932765064e874e3200ddc340e7684195ab9b
00d4d439453b4c743bb4bb776c7638cd6adb32e49fdb9de6990132eb26804765064e874e3200ddc340e7684195ab1be9a3bc
61477914a5c47cae69b70518d27c89e53f56689e1920b36ea57f5466e074e82403d03d0c748e1654b937b8ab103ab49401e8
1e063a470baadcc7dd55081afac900710f039a230495edd347b33134f49319898730a0394250d95eb0919ed1a1a50c40f7f0
48e768c1952bd09bee899506cb6a176817766f942db03fcbf8e33ffabe8cc029d03a06e07a18281c66aa50815b0f2967dbac
fa5c8ee390976e3fca5ed84baad3eb6a4e5f4e6680cc72479f9519381d5ac700740f039da30555ae40cbb76a5da4ed6bd774
aee50416ab58967ac254ed44399901320bf4b33203a743eb1880ee61a073b4a0cabdc11989d0a175cc04ddc240e76841957b
875fce0f74681d03d03d3cd1395a10e5ea77a91c80b7972f8734c1d3b6ede570163d26501b49edc344330364d62fd3a33203
a743eb1880ee61a073b4a0ca7dbaf9c1d0d03706887b18d01c21a86c855874d88e2bd94667b99e455ec0648b416392475407
9f1b32cf0c9059989f9519381dfac600740f039da30555ee054d6b4687be3100ddc340e7684195fbb405034343df98917808
039a230495ed054786323af48d01e81e1ee91c2db8726fb0332174681d03d03d0c748e1654b9d93a067ef65b671428dfb91a
7aed2f38ee8ed181750c30df51607394a0ba4dd631f09bdf38a440e1fc4ad855bfe12dfe910cdc6326de1a0532ae0253acd6
d37692e7769a9573ccd27c88a58dd4bd8f6b3dcd120324d67afa4989816281750cc0ded1098b4bf0a35cf5f41a3b5e7cb155
23f192beb3357eedab89cbdf9dcc0099f5f49a4765064a86d63100dcc3c0c675609abda0c1c5d8d03a06907b18e01c25a86e
9f6e573334f48d01e21e063447082adb0b8ec86474e81b33420f61a073b4a0cabda0cfc8e8d03a06a07b78a473b4e0ca7d7a
0097a1a16f0c10f730a0394250d9de609643e8d03706a07b18e81c2da872e1f3fec18c0e7d6300ba8781ced1822af78213f5
181d5ac700740f039da30555ee053e678c0ead6326e816063a470baa5c75b46b2b37f4737ae979fd97cd33dc0646b5d7c813
0324563fbb2725068e86d63140dcc3139a230491ed050727323ab48e01e81e063a470baadca75d65191afac600710f039a23
0495ed0d162a840e7d6300ba8781ced1822a57174c481f20c6bcd98a8ee5ac53697687e2a515df795b0fd0cc009975c1c4a3
3203a743df98117a08039da30555ee0dc61a840e7d6300ba87473a470baedc0bd603313af48d01e81e063a470baa5c85aecf
9b79bcc8cd49c7254d667b190ff9bfeb46a67961c8abc00fca0b9c0b1d6300b78781cb51816a66636eac7cf46c819fed6128
dfb91a7aed2f386f96d1a1670b40f730d0395a50e5de604944e8d0b605a07b18e81c2da872f573314c464b53b4ef3b96e679
39cbac8d39f2cc0099f573f1a8ccc0e9d0b665826e61a073b4a0cabd60a284d1a1730b40f7f044e76841947b815f22a343e7
1680ee61a073b4a0ca7d7a8486a1a16d0b10f730a0394250d93e7df63f4343cf96917808039a230495ed0546358c0e6d5b00
ba87473a470bae5cf1b5d0a33fe4137adacaa93d25e960970db852efdac2f53657e46406c82cbe16cfca0c9c0e6d5b00ba87
81ced1822af771771e82869e2d40dcc380e60841652bc4a71dfb7c96976f1f0e6c5b4e3bad4d3e3415996706c82ccccfca0c
9c0e3d5b00ba8781ced1822a575e68690b97892c7bf9b2b451e5eb6a8b41b58ad54345dbf24a9e1920b3bcd0cfca0c9c0e3d
5b00ba8781ced1822af782e5958c0e6d5b00ba8781ced1822a57a7ca8ee16baa27509f3abe62ffa2ae49a70e4dd7a9329a19
20b34e953d2a33703ab46d99a05b18e81c2da872a5f5d99748698b38968dee360aae1f99ed4cf7d2119e1920b3b43e9f9519
381ddab600740f4f748e1644b9170c4b313ab46d01e81e063a470baa5cdda6b665a96a52d95226fd817204b65ea06eec970f
4ebab760d2cc0099759bdaa33203a743db1680ee61a073b4a0ca7dfa843f86069e2d23708f0298230315ad6d396dbb4cf484
b6aca61897d537294da707f1cc00996dd3e9933203651b9d5b00b8054730ae02558c1c0ea0a7fbe8e146766aae743b8b913c
d961db33036492c301dc3203e302e716a0bda3004635206a11d8534f73daecc49b7ce86124693dd981d27762804482ea9518
2814d8b600ea1d052a2e00138bec17dff41c0ed1de0c47d67b368bec17bf33036492fde26e998192a16d0b00f730b0711d98
669f3ee98c81a1670bf0f63090393250d1da4108dbb1669bc538c67f5eeefbd48e41f8312f405e3b04e1fbf202e742c316c0
ed61e07254a09abde060374687862d00ddc340e7684195b32903563e5aa64c3fdbc250be7335f4da3ffb68112ef44b01dc1e
9eb81c158866adee2833186be96749335b9ae0e66d72a658ce7cb92b0f961920b3d51e4fca0c9c0efd5200ba8781ced1822a
f7069f2b4287962923f410063a470baa1cf92ec6fddaaebdaca796ea3d1f977afc92cfe29d18c644f255f44a0c1c0dfd5280
b887473447082edbbbfc5200fa5b3f10a073b4a0cabdc82f0588bf3503013447082adba7cfca6168689602c43d0c688e1054
b68fda6b322eb44901dc1e062e4705aad91b4c43081d3aa500740f039da30555ee0dae0a840e9d5200ba8781ced1822ad7a0
e52148871d92202f7b8e3a97614324727556f5dc1f669a1920b3413f2933703a744a01e81e063a470baa5c852e13ee7644a8
d4add272d2e5e27a816d7fc05d65d3cc009915fa519981d3a153ca04ddc240e7684195b3d63d2b1fbd4ae0677b782adfb91a
72ed4debda3ab53128e974493bd1ba405bdd7e731f1bc833036436ad9f9419381d7a958cd04318e81c2da872edad92975ded
35f402f7bac0d55e805d5e8cad5be5b1bc30e6b537eafbf202e742a312c0ede191cb51816bf6829a93d1a1510940f730d039
5a50e55ef08d6474685402d03d0c748e1654b9d60eda2f69fd9b3b76cec3ac7f5ab3ce4fa6fe68d3cc0099ad2df4a4ccc0e9
d0a804a07b18e81c2da8722fb0076474685402d03d0c748e165439d2ab49f721d266a0237d9255daacac5b736706c824fd1a
b7ccc0e9d0ab04a07b18e81c2da8726cd85eda2d3a2a6b86a447dd74440f30be330364b2817bafccc0e9d0ab04a07b18e81c
2da8722f38ee95d1a1570940f730d0395a50e5da80d4075d8b191dda954cd02d0c748e1654b9178c6a333ab42b01e81e9ee8
1c2d8872efb22b19a01f7871009da30555eef38775533ab42b01e81e1ee81c2d1ce59a73c75216e5e905e6ab0e03eb055e65
1bdd764f25d0cc0099cdb9e3499981d3a15d0940f730d0395a50e5de655702d0df3a81009da30555ee0d779ad0a1590a40f7
30d0395a50e5def04e133a344b01e81e063a470baadce73f59940ecd5200ba8781ced1822ad71a276d5b93ae06484bb92fda
90384a9379b907d8696680ccd638795266e074e89702d03d0c748e1654b977f9a500f4378e20c0e6284175fb7cc792b1815f
0a10df5140e33a30cd3ebcc08f61815fca045ba380c52560727ddedb8a5a9da05fcad8921cc2590754f7b3847b214e7828e4
f33517a543bf1480ee61a073b4a0cabdc0e194d1a15f0a40f730d0395a50e598d793d909271bc293d750fe29e76e7e443303
6432af27afccc0e9d03265826e61a073b4a0cab5816ef55db86cd7623ae4f18bcb61874c6d75a7db720f95d0cc00996da0fb
499981d3a1650a40f7f044e7684194fbe8c679c685662980dbc3c0e5a84035abfb0dee9377753ec9ce4f5fcb998f6baef663
6dbf01cd0c9059f71b3c2a33703a344b01e81e063a470baadc0bcecb6474689602d03d0c748e1654b917ec6b6674689602d0
3d0c748e1654b97a0edbbea4ed9456b85ed3224d42694195f396e2711e87de9a0acd330364d673d81e9519381dfaa500740f
039da30555aeba87a4bcda71975fb641734fba7e6db7ab3d96339e7a4c4a750fa1990132ab7bc8a33203a743cb94117a0803
9da30555ae7a4dd4dd5cc5f9a3fa83da5d91e6c32e177934cb149e1920b37a4d3c2a33703ab44c01e81e1ee91c2db8722fd8
d2cde8d04a05a07b18e81c2da87265afef752dc537d22e30d6c3d2ec02b76d59f3d11c659dcc009965afefb33203a7432b15
80ee61a073b4a0cad5bdbe758cb2dc959476ebd1db05ee75e2b8edf5a5990132eb5edf4765064e87562a00ddc340e7684195
7b837f0ca1432b9509ba8581ced1822a57a0cfeaa3576e44ad73cae7e55e2d50a0796680cc02fdacccc0e9d04a05a07b78a2
73b420cabde0585c4687562a00ddc340e7684195aba768663b86db3cb8526b1aebbff43372ea219a2c3140623d43f3498981
a3a1950a10f730a0394250d95e70b237a3432b1580ee61a073b4a0ca552b95b235d3bea1ab5c4fca4bb67e7e3bb6f7ee6df0
cc0099d54ae5519981d3a1950a40f730d0395a50e53e7d8a2643432b95917808039a230495ed059d4a4687562a00ddc3239d
a30557ae785d90f2d1cc047eb687a17ce76ae8b5cf6626f0b3df7a7540f9ced5d06baf07a52ec71ed764a73ac9dfa23c906b
e904e98121e93ceec5153c3340663d28f5519981d381990930df51607394a0ba55e41c2f6918958a66970b5caf652d064eeb
761ebae0a821d3cc009915f9519981b281950a10df5140e33a30cdde708b7f24032b9589b746818cabc014fb743d4db0c04a
0560efe884c525f851ae4ffb7b112cf45101da1e06302e0213ec0507e83036f45101e41e06384709aa5b1de62c67769523c5
8ff590c6d1b2dbcb968eb4a4f3ee333a990132eb30e7a33203a7432b1580ee61a073b4a0ca55f3984baa4e753ed16bd2536f
740d4fe9052cd2b55ef676e29b931920b3dac73c2a33703ab45219a18730d0395a50e5de604240e8d04a05a07b78a473b4e0
ca55bb091b6b2d6dc0354b87569a847653cad05cbe6eb709961820b19a4d3c29317034b45201e21e063447082adb0b6c8218
1d5aa900740f039da30555ce1ab8ac7c3433819fed6128dfb91a7aed9f1e916068e86402c43d0c688e1054b617cc8a313a74
3299a05b18e81c2da8727518a634d04a9bbbb741b59e8fcb9575cb701b86a1990132eb30cca33203a743331380eee189ced1
8228f769d34086864e2640dcc380e60841657b41df8dd1a1990940f730d0395a50e55e708022a343331380ee61a073b4a0ca
bde0487e4687662623f410063a470baadc672d2e0817da98006e0f8f5c8e0a5cb3b6b5bc1f349dacfdbde4a558071c7a2c9e
aec66c5bcb596680ccb6b5fc499981d3a18d0940f730d0395a50e55e707601a343331380ee61a073b4a0cabde01001468756
2a00ddc340e76841956bd06505b3ed80d7bf2f29dabf6c8b7cd11769acdfcc2c318c890df9418981a3a18f0a10f730a03942
50d92ab12d1cdaec2e48352bafdf6a5338ebb9a8a9523eef1d002c314062257e5262e06868a202c43d0c688e1054b6fa99ba
db45fa49d9d2b24b17dbe636d2bdd0b97ea6686680ccfa997a5466e07468a23241b730d0395a50e55ee0aec6e8d04405a07b
78a273b420cabd60ce93d1a1890a40f730d0395a50e5ea289cb40aad576b5d3d3db6369fb6ff266efbb6adf168bbac9dcc00
997514ee519981d3a18f0a40f730d0395a50e5ca72956b30315aae6d95d6f17514575a33158f6b5b82c633036496e52acfca
0c9c0ead5446e8210c748e1654b9172c36647468a502d03d3cd2395a70e55eb0048dd1a1950a40f730d0395a50e55eb0d890
d1a19b0a40f730d0395a50e5eab252694224b937e50257e985c5cd3eb072b38a65d7d99695d2cc00997559e9a33203a74337
1580ee61a073b4a0ca15e86d4fd2323dcff2fed97b17e5bb2ac15681fd2ce38fffe8fb3202a740d71480eb61a07098a94275
51a11edb2aaf54a951edb3b1354f6d790696a379603a990132eba2c24765064e87ae2900ddc340e76841957bc1f798d1a16b
ca04ddc240e7684195fbf41c374343cb1420eee109cd1182c8f6822d4c8c0e2d5300ba8781ced1822ad7bacb7d61857c3fca
a3687dbd4bde097923fad6539a1920b375989f9419381d5aa600740f039da30555aeeda8b6830cec1b929728dd5de9eee8d5
6e65ec7de9d60b34334066db53fda4ccc0e9c0356564ee51607394a0ba7dfaf06102365aa6006d0b8e545c022a17d963ab23
cfd2e3b321f79ca4431b9761bb29cd0c9049f6d8ba6506c6059629407b47018c6a40d422c7902dd2fc8b6b8c76958bb41992
5476d445e4ce0c90498e2173cb0c940b5c5380f68e0218d780e9f58623f17f2443d71400ee6160e33a30cdda592ed24d4da5
71704803573e9a79b3735796728ac3d65e5d9e1920b39de5f2a4ccc0d9d03805907b18e01c25a86e157a3de5f216730c3cad
5ba3cfa0756feb70dc7d3e13cd0c9059a11f9519381ddaa700740f039da30555ae388990f2d1c0047eb687a17ce76ae8b537
ad771b05b3239ca413697d477b32f4d89323a7db9d996706c86c5a3f2933703ab44f99a05b18e81c2da8720dba8e8ce8052e
b6bd63cbe5dcc5fb84cd06cd32036436e82765064e87262a00ddc3139da30551ae99a8e8b908f2d2d96bbfca1fd9b2cde448
f72b1ea7d4403734cd0c90d94c549e9419381d9aa800740f039da30555ae7d31f4084f79e4ac76b7f330b46fabe74b2ecb72
ecf2dfef2f06cb0c90d9be184fca0c9c0e4d5446e8210c748e1654b9179c4ecae8d04705a07b78a473b4e0cabdcb4705a0bf
f509013a470baa1cf35189cb75e5e22120f58f0ec0a6859e387c6706c8644e2a5e9981d3a1950a40f730d0395a50e53edd9b
616868a502c43d0c688e1054b67684f866dea3760cdb768f696b075b0fbe91fec83daf4d330364b623c49f9419381d1aaa00
740f039da30555ee0527e4333a345401e81e063a470baa5c835eb5136fabc5939d3219b36df9d9a6a5d94e6680cc06fda4cc
c0e9d05005a07b18e81c2da8722f1886627468a802d03d0c748e1654b90a2dd58a6d8ed00b5c74d1eb5ed6d2c84bb1db637a
bfd334334066857e5466e07468a83241b730d0395a50e53ebd5a85a1a1970b10f7f084e60841647bc36d2674e8e532420f61
a073b4a0cab5175a1a0eeb6547e4aaa9f079ad659a4eee4e5d4276bfd02c3340667ba19f9419381d3aba00740f8f748e165c
b91754dd8c0e1d5d00ba8781ced1822af7e1e349291adab900710f039a230495adb5c556117eb56d7cd2442da7dadb576691
3fb35d521fdd6d31961920b3b5c59e9419381ddab900740f039da305558e34b9a5afbd44dddd6e0fe5b21fbaf98b1d34dc33
03649246b75b66e07468e702d03d0c748e1654b94f9b543234f47201e21e063447082adb0bfad08c0ebd5c00ba8781ced182
2af769cb598686462e40dcc380e60841657b9791cb04fd8d4b09d0395a50e5da4cca472cb319175ab8006e0f4f5c8e0a44b3
178ceb333ab47019ae770c039da30555aef9d69cdb26176617b8ede7b565b942bd409def91bec27d1e3ecf0c90d97c6b9e94
19381d5ab800740f0f748e168e726f30f6207468e102d03d0c748e1654399d2da3e5a3890afc6c0f43f9ced5d06b2f3f4bca
471b13f8d91e86f29daba1d7fef973f1291dda9800740f039da30555aebdcc7d55afd4a9ab54b1d18ee1d359fa2da574de36
26343340667b999f9419381dda9800740f039da30555eef36600940e6d4c00ba8781ced1822a479a9dbb9da579da4afd9cce
ebd8e4b3c29a603d3340266976ba65064e073626c07c4781cd5182ead690eff332a4e7a3935a4739cdbe9d90dc1bda243140
62037e5062a060e06102b87714b8b8084cb0cf6f456464606332f1d6289071159862b5e63a5675bbb3aa552ef2b21d09a59e
6d3bab5acd45330364d69aeb51996863725f2eda988cada9219c7e93e5159107e85728c4090f85bc604d05a3431b1380ee61
a073b4a0cabd607901a3431b1380ee61a073b4a0ca91ea7a59ae753d6caf643aea8a676a3ad512032492cada2b317034f430
99885b18d01c21a86c6d68f723dee48c0bdd4b00b787272e4705a259dd612bdf123d94c42e491b8666ab2dffb2e698523cf2
3df2c5330364d61db68fca0c9c0e3d4c00ba8781ced1822a5777d8eee52f3bb8601f1ec878edcb7126e903b61db634334066
dd61fba8ccc0e9d0c304a07b18e81c2da8726573cbb5d42bd16b3ab3f4eca409a1ff6c171ad554a7ec6da1890112cbd69647
25068e86062640dcc380e60841657bc1796f8c0e0d4c00ba8781ced1822a57366e1ddb7948755e7d376c56742b7765316bd2
3dd717dac90c9059366e3d2b33703a343001e81e063a470baadc0bcefd6774686032420f61a073b4a0cabd607b2da3430313
80eee191ced1822b57b6d79ed2794ff245b36b328b4add5e6617b86c3aa7952bb4931920b36caf7d5666e074686002d03d0c
748e1654b9027ddc9e0b7a81c3113ccba9d6ac31b7a3c89ccc009905fa599981d3a1810940f730d0395a50e5ea46ea547a3e
b65bf458e44e48b1f6325eeb16cf4d0ffaae1ba9696680ccba91fa519981d3a1810940f730d0395a50e5ea9ee228cd60b9c0
cd2ef03caff5d8caf187ba9540fee8d63652f3cc0099e59d7e5666e07468603241b730d0395a50e55ef0c96274686002d03d
3cd1395a10e5de656002d0df1a74009da30555ee051bac191d7a9800740f039da30555ae36bc6d156a2a3ba04f9d59bb7239
19ba9e91bcdc2d6f9a1920b336bd1f9519381d7a9800740f039da30555eed387673234343001e21e063447082a5b3d18ff3e
9f56377ddb011547398e455a552253bc6761796680cc7a34fea33203a7430f93117a08039da30555ee0dee07840e3d4c00ba
87473a470bae5c85966fe6a1635736a4d10f7a97aac89c0c1af13769a1e27e5f5ae050689c02ac3d0c508e0454b0171c09cb
e8d0b605a07b18e81c2da872c52d86940fc629f0ab77144a77ae855ef90b8e7d236c609c02c47714d0b80e4cb322f28f6583
75c9f48b350a65f3eb60d7fc8667f94732702f01de3b3a9171157e54ec83f5d48f40e85b029c3d0c481c9f49f506272dc286
be2580dcc300e7284175ab662df7d9b37abae1514f4db706d15a0e91bfcd5a686680cc6ad6f2a8ccc0e9d0b704a07b18e81c
2da8722ff0036474e85b32420f61a073b4a0cabdc0ac85d1a16f0940f7f048e768c195abd0759cbd4c28c43ad36fdd3ae968
1dfbed38c013c39858919f9418381afa9600710f039a230495ed0db799d0a16f0940f730d0395a50e50af4de3694ebf55d7a
a8f7552e36b5833b2a334d0c9058901f9518381a5aa600710f039a230495ed055518a343d71480ee61a073b4a0cad5e37db7
746eeb51be25977c57a595b49517f13e38ad1eef4b330364d6e37d1f9519381dbaa64cd02d0c748e1654b94f9facc7d0d032
05887b7842738420b27daec5c9a0d02c05587b18a01c09a860f51441f994a86d8f19012ceb555ca1f452cd294a2ebc8df7f0
cc0099f514c14765064e87662900ddc340e76841957bc172114687662900ddc340e7684195fba86f08e3429b941177080397
a302d5ecf34b38291d9aa500740f8f748e165cb9b6ecab9dd2a9cbb4cced6329afe1d41071320364b6a55f4fca0c9c0ecd52
00ba8781ced1822ad776f42ed7b1677bf6925c5c5916af17b84b63511a307def36cd0c90d9f6f43e2933703a344b01e81e06
3a470baadca7b7313334744a01e21e063447082a5b7bb67589d46e7d1d6923d40df5f61ceee548e6fb85a6990132dbb3fda4
ccc0e9d02c05a07b18e81c2da8726fa8c5081dfaa500740f039da30555eeb30e57840b9d5226dc16062e4705aa596d8dacd7
b2e72395f5b4f14cd2e1bbacb964cb96565dbd5e5b23343340666d8d3c2a33703a744a01e81e9ee81c2d88729f6b63332834
4801d61e062847022ad80bfaca8c0e0d5200ba8781ced1822a173eecd8cbd0d01d65241ec280e6084165ab6603db2a1d9cad
2e34ac4626368013ef556ad56c806606c8ac66038fca0c9c0edd5100ba87473a470baedc0bd6ac323a744701e81e063a470b
aa5c815eab1566598b668681fb5116e05deb72c8c7bcdd699e1920b3403f2b33703a744701e81e063a470baadc7f4ba31042
812e2800d7c340e13053850a5cce512a91d35e33690297afa75d9f3c04d2d139da0da5890112cbfd7c5462e068688d02c43d
0c688e1054b66a6075e86687abd62c39ca973adb91e2f11e62aec83c33406635b07a5466e074688d02d03d0c748e1654b917
4c4e303ab44699a05b18e81c2da872755eb9b481ccd1765d875306e37ee5bcc4a3752178621813ebacf293120347436b1420
eee109cd1182c8560735cb12bbcdbaafab34bce51e94a314db60645b84c2330364d641cd4765064e87d62800ddc340e76841
957b81512ea3436b1480ee61a073b4a0cab561103bf4a38c59ec7581b4dea179b5224d0c90d886411e9418381af8a28cc03d
0a608e0c54b4171c6b46d8466b14006ec1118cab401523fb7693ad1addac6acdeb7c721dcd0c904976eeba6506c605d62840
7b47018c6a40d462d62883234fbef2bad9414604f6185d7e864c668de29519281758a300ed1d0530ae01d3eb05677a1132b4
4601e01e0636ae03d3ec5dd62880fcadef07c0394a50dd5e70ac17a3436b1480ee61a073b4a0cacdd628f0b3dffa7e40f9ce
d5d06b2f3f4bca477392e9675b18ca77ae865efb0b0ed26074684e02d03d3cd1395a10e5def032133a342701e81e063a470b
aadc1bea6c4287e62423f410063a470baadcbbcc4900fa5be78d91ced1822bf7828355191d9a9300740f039da305558e40eb
006b8cc9ce664f97fe53cec7c5a0efcc009904da2d33703a342701e81e063a470baa5c3b1efa4adb72d94895dc88738f6b59
3a2db5cd76d80537689a1920b31d0ffda4ccc0e9d09f04a07b18e81c2da872addf68be4ff6ca65e9ef1c7a9ca0cdef9b3ba2
5ee5dd71649901325bcff1499981d3a13f0940f730d0395a50e53e7a2810e3426712c0ed61e07254a09abdc1c880d0a13309
40f730d0395a50e5eadcb98dcbad362fb8e9172de7687be6d7ba72e73ec98d6706c8acf3e78fca0c9c0e9d4900ba8781ced1
822a57a1cb1c42aecb55b2bc01e9b061aadcaadf064d33036456e84765064e87ce2413740b039da30555ae429736b139d96d
fb72c815665ba4b5eaedd1637cdae3cd33036456e84765064e87e62400ddc3139da30551ae42cbbb2ecf402c2372f76917b6
6badaeb36cd034334066857e5466e074684e32420f61a073b4a0cabde09d6674684e02d03d3cd2395a70e55e309ccde8d09c
04a07b18e81c2da8721ffd42332e742601dc1e062e4705aa596b85e5539e441b6cbec79ab33d9171dd0ebdfcbb15c6320364
b656d8933203a743671280ee61a073b4a0cab5f6f62ef5ec667d0079e9faf945a91e4dd5ef32cd0c90d9dadb4fca0c9c0e9d
4900ba8781ced1822ac74ee0cc52df5c9b8dd8a5336fd217dc77da9dbc330364b23338bd3203a743731280ee61a073b4a0ca
1168a95ef7a46e7e7a8bb62349a7e8a4bdc93b318c89ccd7dd29317034742601e21e063447082adb0be630181d9a9300740f
039da30555ee0547ea323a342799a05b18e81c2da87275f873e9f3a487fc15f7752f36ca4b4cd7255fd736fac91203243667
e60725068e86fe2440dcc3139a230491ad8d726f7997368256abe77265f99f9e68a8930eeaebb65c7715c6330364b651ee27
65064e87fe24c3f58e61a073b4a0ca354b83d618542f91615fbd6e34bea25ce30d4d330364364b832765064e87fe2400ddc3
039da385a35c33af38e3b11d76eadad93ea7c536471e49e917ad3734cd0c90d9cc2b9e9419381dfa9300740f039da305556e
f627819ffdd6fa03ca77ae865e7bd57a8fe7ae0e44aacb59b682c66276b1cb1fbfee45b34e6680ccaaf5a33203a743771480
ee61a073b4a0cabde1012374e88e02d03d0c748e1654b977b9a300f4b7d61f40e76841957b973b0a407f6bfe01748e1654b9
cfcf1d513a704701e63b0a6c8e1254b7cf77ea181b18a400f11d0534ae03d3ac0d54b445ccba87ecd8cb61bbdacb2efb1574
78a30d55b0cc00996db0e2499981928141cac45ba340c655608a09efbff9facf5fb793880e9bf506d4b14ba5f3252d289db7
4d9bf44ee2578ceb6f744d8af4ccb62cf5cf9f7ffd2f5f7ff3f5db3f5cc35ffcfd57fcfa2badabbefeafafe5eb8fbe7efb2f
fffcbffcc73ffdf33ffea37ff1f5a77f5fd69c4903485a815f31efda8f8dd24029cbce742fbb546d1a971fda8fb6f06c5f75
87bdc4f36ff21157b99a1217663d174de3ba83425e9cb23849da98ab86937420d5a1a5864f6979d578cad2c3b8becaba1e01
5ca41712b3a408e15a8ac9bf598ea867556afcd0f359b71adfe5a13d4b7eda5791f4abac59915ae98a251ee57eca9595f515
524feb9653e1b473218eb64a63b90eb905163fe5b5dfb61a4ff13ca4f7a7f1432bbff45566efb76d9577c5e2f2626ca7c54f
3d7a71d733832dbee9857e953960db25dff237dd6b52a649f5b8f0abfcae7c43d6ab86e55b728884783912bf74dbd56a588b
bca1572966fb8d1de251e2f2f7356f756e6ecfb683096593f819f5aca822f36a7dfa32b975e8b977b1dd16ed2894f865339b
d36d3cf7dfe8b251bbceacd353bac7b3c4a5da92fed5f4949c8779cb5ff0b495f720e9b96ff23844b9136959d25edfb0b448
1578585ccf5d5eb65a4fe8deec5cf275def3ccb5c2dc773581b6b82ea53df71a8fd7aafb026394574e9affdb56bf29a2b9ee
e3d278d475a857fdc09e9bba5294f876e5b8d5d6861aaf6c352e5c57aa71dd0fb794b85caf4e95d52659924f569c7f576e9e
349f2a977c63e415a9ed56f9c72b95b87c68d2dd8c5eb51a28f1557a69f676693cca354b3f7dd2474f84d8f42097180f7b2d
d6b5c6e521d493e9352e6d565d385f3a64bac15f9f93b8eb2ea4abbc761297e68dce4d685c1d67d2567bad22f2ba6abedcdf
552bbc1adfe466c4dde23a371d531dccd8a592ba345fea31b911678bcb23709dd1e2d27aaaafd125cf95d465bbe66f66035c
af535feb537a141617c294531d133bb436bcac2a3cf594aab5c6f3b24b1d6ff15da45a531d383cf72b2f257f552bb036a028
2d77ddca1963d43550b9541bd766678c1c97c5733cd2b1d6b834f4f5d48d18f50825eb029421e84b4fbbdb2d7ea84950aef1
a85b802d2cdfd42b9562a2be2387de7679e24ff9872dd7f81a37fdd545ebfbb8ef25dd3615ebc32cefc1253fbae41a2e4753
68f8bca46a2875835455f21a6f165ef428f716dda3bce0259ce4d3532aaa53b7a5ab6167d4d6c7b59c8554e2429be5ba342e
15ea95f6fa4667a9e00fcdd7375adedcabbde9c2b75f165fe325efd65463c84b268f7451402798a5efa2c26be376910bdba7
0a49dfbdebd0250d5881495ca4aedf0f8bcbb5e51a97c7b02d5e6915a4bc7bf2195db654e3f2ce9c6b89a74377e44f15adc6
adfd37d5d712d73b5f717bfd2eeff075a8fcd3f740e34baa2fcaf0fd90b8ce20c43c7d6f347fbbbf37fdbba575877ac6c4e9
3b27f135eb198bd377d1ae7fd70d9d352e3f60b771e41dbebb1297df3f62fb7cb7eff4528e24dddaf75baa6e698a4c3a0fcd
00ed99693f61adf1d69cd0fb68e779fc6aad0fa912cf6bbeed12d78d907699d99cbf7391474facda92d6b3f254e90978a55e
d82f3d5b4957c6e2d32671915a5a76165737a152afe9060ff968dbed1d9f66896f7b962adbe2525f5c7699fba16e007a6214
be2c124f52d9c5f2ca496d144b75b7ef3a3caead397b439336266b5cde90642fcbf8e6ee7257a42799cbfb6f35560d9b29e1
3a550bbbdcace53ad7c3aa114959ea65ca9dbb742ecae279dd4aa5b36b0d1a9355d663652471916cb18fe258d9edab6ef6be
acf295ca519eb552694a5c3b7e7a955099ee51cf22d21a542be5559eeb1a3d742e304d55b5c4e5318b4bf914e847a43ce1fb
a267874a25307d0a763dc4491e85387d5276dbee168ff2094ae56bf4abed975ef540d9cbe297d40bbbe5dbcef194b7f95326
f12dae6a0d689f3e3dfcdfc2873cd5d6fbd1f0be9dc752d2f516265d0b5ce2426eb8d6cc917ea185a5e770960695b68ae2b1
da071d7e555b517bf94c9e6a995cb2b58da627841992b44ef61a95e6cd0f0258c3509702a95ed7a5eb782dac0d7af984151d
a5fed8eb8547f997ac5f079057c28b365da2dda6d346312c6ecb31d42a096f5fd6cda57a68bec5f593bbdba5276d0e694d6d
717955e4ddb7b87c37ac86b287e6923b96ac9c24faeabbb3d6b8b429edcd4ca2afde8fd5ca91e75e8ded7eb5d3d3b2bce1eb
355d8fc6ed874b5c9a22e55593ee921d2f57e2d779d50a396d6a027068450d3a485ca7385289cb27adb69493b6fa75a582a9
2fedd82a73d26d267a1ec2745752949a451a5225ff94e7bc7cb024be49bd91aee94148f2d8cbe5d40747ded3ad54ec490f32
3ba2953fb6d4246e58252e6d8fd261d8e4a94f56d9424350e3d2982b9779ea8b2435b8c5b51599d595c81aa067d4e3854a3c
ca135d1bc47783550fd495d6c8b59506f4b9e8fc7a8d471d95592d7ea937a0bd0e9bbe04d21890c76795aa48bed5b5fc5d3f
22abd680abb9829da5eadae479df75fa49e2f24d90466791d34ea59676f96a71b911e75af2f51ec9e569beb6a9f6bd606967
e658f5ad95f02e6dc8bcd6b81d6f59d2751b6569d56c727757350890b83ecdbaf5aec6a505bf1d25ae0f43e9ae6d51d9b75c
e2d2cc8fe5c5ddcc5a425e5d89ebce9474d5dfb51501faf55cf56b78ece5abba8a54f21db84afab1cb5db96a5cdef933499d
b66ac5a1834e3d9eb56228f1a374efd6b37c8535beaba143fd70485cda617a8b34beadb63de557339ed0c3b517cdd7dd1552
ef9f35be64690b9d169787c6b4b73f2a4d04cd3ef5f887b57ce3571d882b6d0ebb80ad3c227265b6644b0bd1eb55dfc412d6
da3e95f0aa0befcb6f4aa515ad79134d8e58de4fd3695737387913f47da80d1195551a105aef98d8b9e5eb6dc8baf73adacd
b96afbd36edbaaeb81a3de4c5d0cbcd5f8aaa769ac16979b7f9586a33c16f2795815b63c2cd7d2e2f2182dda33d2872b6df5
2329f14d1a0afa6d2ecfe259b1a2cee7e97ebb589ee9636f71793bb5928be5dd90c6598b8b6c5aaf495c5df2ca8bbbeaae3d
69d6caa3b3954ef3daf2933615e4d1b177585a2b478d6ffa95d0fc4b79d3d9e2f209c8da56b0aa431797d5b85c8ebd885605
49efe998ae47aa2c7993ca7b28b8ea39aaed58abe1e491d99b3cd93ee656514ab3749f54d3f8b56ff5eafb5dd10a57fee4d6
e2d2463eb5b7a07169be956193e16948c96ad9f2fa4b5cfaf6da888c56d1a7faa91c1eb6a4cdcc7ce4fac4467d7e35ddbe17
d795a7275ce272f5576a4fb89e74a39575f91e5d6b9ede1ff97e6925553e3612972efdaedfe2f25d3b73cb970e823d6ce53b
789c77fcb0ba2696efe651553eb4219314d7beb3c79ed2f4bb59776f08ef55e35167e7355fbfe3d2086f6faebe20c9c2d2ad
cf4b0b6b576d9f069506156cf0e82c8332a69a0ed0cc636eabf62de44b9fa6c1a0417c3b6beb3ceb239ee539958a6e1e0cb2
aa4f1ebc0d0783566ba41ed7515adbd1b6de557be645bb4c696a855be3559bfca5d19ef7545e096ba4ea87b936f2732c5589
c5d568736afc6bfb61d53ada3a0b795f6aab4cda03ba3f79993b17d67e908a22d6cec8729687d0da03f62996be8bb4808bca
51559327609fba3ad61a88b9758ca4f35d2e52977fe85f5307cb1a037a24b4858fb89766908e9a5d97f5aae556ad661d51c2
495715ef5377cfbeed526f5fa53b296f5ba94d259eb52688d6fd94eeec7ab57cf9b6afd69bd72e414ea519a1dff65d1a757a
95dabebcb6d21a2cdff6a8e37c4bad985aada6c3eddaec5874c9819433d7823a74a5edc8add623f286e8f0b12e19586dbc02
aa17092ffa801c35accb19e41ba1ab2aa43171b4ca4551e522af4327f0e499c3ba48c2729b96e18322afe6a5e1c50e18a90f
abbc75ebd765df59b5d229416d75894657d65357eb68a97eab7435cdfe7525ebc1ad77583b38e94b17ebc957f6ba0b11c9e5
c6e8c245730eafbfa7fb67e417e54b70d5e690d6933a9827d7acfe8b6b795606123d54502ac306280d3955febce44d3bd7a6
9dbc2ddaa33f4f3ddb7d2bedeb2ea986e5ae955e5cbf31a7ee4c9456c57dbfd6430f57d1b0a0ae116fbb8445f22db7a76111
c592652ff2712c83d6fda13aa5bedae591ddf1d93cd5b9a27e57ef075c8796a56a2a0da6fb6dd0e871d4afc0fd4a6901d2f0
2c8f810e545ee6098557d1de568dca8f15c0fe6e2b89bc1fa58aeb55c4a956b2f9289f231d7394875c1a3c2047af807455fd
5687427a75a5636d29e732f213ad09a9e3c9705f7a65a88bd4b561ff05e3e5972ee3d1d30cea94631d16b775ce71b94af63d
2aae613b56e20b06bf35ac03046dfebd8e7d5b58677ebf60e85bc3d27a3ddb88781df9d62b5997a30e34df03dc1a1659d7b6
2ea38e6f5f5a99ebe1e35f30bcad239efa892aa3e7ba146b97377a7c0786b16d894aed7a94af80ce88c9d3a42fddf01af581
ed4bdb21d23c6ce14557bcedf82a4ab5a18fb5be48e38baba3dda75a00e16b2e61f93eeb980c540a1296668a36d6a102d19f
ccd68f805a68b840fd736df854c30223af0c54655a0b1e7a8028d48745121bcc1a2ad57112c0e247ca6d124092a27699c74a
db26137425cb8195ff38f9a0eb6af4b0bfbdc6ad87358d25da4dd6eec4f4e91a2737242e95c776dd2b8af410d413bf8ce324
89c4e5939673aa71e9b1656d5c8edf638d4b9346cf5e84effa3839a343da9bb44cdaa48d0ef7d4d1bbbb3d314ef2e80bb5e8
1143f5c5d01d2a7580fd1ebd1b278be4bddcd2b6d4dfd5a6a3a89571f80e269df42845a9b8db6cbf9e82bbac384e07935a8b
5694fb52ae5faa0f01db70f80ee6c6a21dc37ed7193a166a7327c3f01dccbd69b3702b734b3abe240f43b5f2b3736bb77ace
f07d8abd744372194b6ce359129412e2de06bfca989835ba5b734ada7aabf6ebb6a1dc5f6d68571bbc71fa391df15db46f60
45cbe359da881a3eb743cf70913a3e6ae5dbc2657c58c2978ed4c71a56743d95ec681fae124ef9d2c342f54d3657c8129516
b63a9c48db5c1a605be93b1f52616da72efdcdbab3e1aae3347da45ac23a4c78b44357a427a7ab00a405bee99ac3927d0f83
8bba3ac976b5d171adabe53591b0bc75716fd9e725454bb63cf272f3af1aad43ec1a2d879ffd3a8ed4eb8cae5c4919d73aec
0d94efca971efaabe774d6681dee97e87ac569ce40e77ee531cd652cfe9e6290f0a67319a5089de011ce030bbee729b23698
a4466e3bfcebb4865c9cd430e77d3682f6d1441b00bce746f411947b716faf97bea7ce0583487d8645253dd73a5370dae754
bb587003fa3c8d7698d4bcaf4dcf675d16bae1cd95b00edeab51e7f828f449a0a4231ad2b46c93f3d26ad9759bcdf83c9d3a
b260eb18e0e1eb334c499bfad6bbfbd526a4a2366f123ed8b6b3385dc56ab0bf06a77696cfb20db9bf4af24d93ae5a2a0720
f757c93e75d2273bb08c4b8753e5f53ff117252c37374f572d511de9971b058c1ad643233714aa4f9b661dc5974e43bbbeeb
90122f545bbeb97247744c126e99846394ca7dc5fbae613d82f58267c72670a58bb9c353a9d19cb47e942a49e79b728b8a1a
a7452ff960dd53c0a70e407e656d3deee750b09ecafe950f6dfe1cdb755fdcaa37379f6a932d1fc409fcd271ab3db7f0a9b6
64bb86775d5cd1c2822bfa7de9b9a7ab1ea058c37572d9e646ae74b5bd5ffa1912e177a964b63dddbb9ecad4b21e239a8eab
cc6fde33c8bb3cb3f24e5749db4cb144f5bcc43a952bb76ed3f34f761ddbd2f3ba6bb44c1f6bf4b2c1865fc769623dc0f348
5769adf6d9e05dd709c5b5ce82de93bebbbe0cf13aa6c95d09cbf5d47aa8cfedee9bbef239b57de652716a5744c23ac216db
9efb3ab36ba76d5ed7b45066d7c75bbed4d39bab7365510714b15690b07c1096b39d5651ab965d1ecd432dd06bfd24af9cdc
51894a55b6a676e449adb5766d3bcad77b5a40b46b9b4747b6b09ed4b02879b5a3766a5d2be1fdb4713ea899257cea3a9c69
b194ce084a43b97ec0ee2f8186e5062ed3d4ac864f5da38b9f242dfb6c23757a6ad8b2ab13c09eb58b1deb3aa9fb732761ed
85d52afbfeee4af8cc6bae737d3a2d9b7572464f8b5cf658ae64d7553dbb7ab2e8918af2146d2d5c276b771d4f930f5b0bc7
b4e8889d8475a3ec3e4dedeea7fcd372947ea2969df45dd7b034b9ceab650b9b46e5c5d3a33377bc6c6d7de852ed362bad0b
6fe4628f723e5d9e26a5f500b9bb3dd30594047918ea98e0a1b3a5ba1c42c3492ea30cfdf5bb2061ada4a75b76681d97ea48
d8a12dd735c99ba0a79ae9d13419effba18337579babcf766092bce27a2298d4dd29e3c3a3c5ad5259b5f0a9c3789b86a5aa
4d65a8b43f815adc5ac76d242ae21c52fb49348abef7aa04793f16a9b125415ab4db362d56386cba279629d5fe2a1c3a177f
1d5b6b04481da6839e12d6add1fbb4c24e7f45608b52fd9d94b0742c635d5662cb6934c7c23671062fb6fcca9aaefeedadd5
8084b314b14c8b3e0eed082e4bf94e9c3a33b4eb04961e54a463271b563cda8c89bada056baf435782b591b55ed72958d41d
df58331edae3bbd669518d44a50b7c968a47c71cb28d32eae141d2f9ab8bfceeba586b269de36ed579a9b84f6d59e754bfbd
adea3fd57262ad0318f767e2d401bd5c27f1242aef92ce299d3a1573d6a9d5fb4ba36d2cddf7387dad54f4f358efd645fd12
9e36af57b02f9d32d1a3a7349a536b81f665597a668ebc4bdbb4baebd4e1286b0bfc3a2e12d3ece3aa515dea274f519e8abe
57a0e975c4ba7ca6af6353c2edfe410d5fda5345f04b9bb47bb92d5dffde85d7e743c7e9ef2ebc142ddf56786c247c183b3e
a9d6b3d7a340f12db8748c4cea831ddf300d9f3ad986efa38637f9b70bdf691b2bd013cdb162d01113f9c13cd52e973503e4
9b84159786e5a91535a152ecf05ac566f57069615d75bb63857be933b9eb2c3bd4ce4321baea71bfee42a42567ad97a1e61f
c635761bb2cb471bd6c83a74bdc1576518bdd05539eb716cd3608484758ea18e44d9a2ab533f7bc3176b188ad8d5c56d2bf3
32c3c882840f6970e7fc050308bbf680ebca94619c60d7d5de57bedad68f3a1ca0eb7de48189d3a09d7e69f54a0ae569afad
3c3ee37779e8f46b38d7f55543df5ec39bee50c15ac7261163d47521e5809c7db536354e2e0eabf0b643274163f9060fabf0
24ae13fe6bab03e535b165a0307939acc2db6cc076bb4f9652bfdc7d9a1b1d16e1a95b736ccd8ab23adad6acc31cebb0386f
b381ba65999a773a577b6895761fd876265bac0673b8c3a2bdcd3a1f31cf8bf636693e8ac6f5ebad8b2c6cc51fce1d0f8bf9
24be1d6d11e1b0986fd3b12c6dbae1b757e3cb76c575fa82eb9cb55413d7da16f3d9c4dd34953dace59378d606c23442a0f1
55b73d266c92d814babe0a7b8debb3ad034c30b53eace59378d6a34b32b68f34beaef5c51c17ed6d3aea9b96bba9d616e76d
76fc4d2eb771588467f1b6767f586db759073196a74dde5475ead0793bb8ce61159ef04a059ccbd3afef6a3e6d1d1be836ac
c25399b52a29cbf0745e6ed767156fefb00c4f9706d8f6e11a5fe4f3686fd7f8580dcbf3246e67a7b7b88eb5e479c9c0b00e
4fe25287c6526f0febede4759126dabef4157752476ed36b34acb8d3b74e070c2cdc17d06df6cdac5f8b7cd9e2e1e59c5ede
61619dc4e5fb58b7360c0be8d4765b57d99471bebe824ee397ed4cfc1556d059be34104b39f70aba4d1b7cba26a986e54ba2
a374d3e5b41574c2a4fb14ee75753aa6bb5c93063a1a106db40735cea5bf67a58ff74ac7d0a4856fcfe078cf4b7eb26528e3
b3a3e54bed7acecb50b20dbf9d799f96adf4ab5783a4b38e48649b11df4e5c12a36316ba5b1ad7d5a85c9bbcd92baec2e92a
ea2220e9e5d5657bbb4e40d96acc6185cf702b74118030d6a15ffd52eadb8bcb87b22d0857c3105c9e34dc699debdf535dd6
2a4f868e53e8f20758fe343c3136f9a9c363d313b68acef1d8aff6441e65e21457630d4fb0cef12ed270dca7075ee251ff68
8bb71767d545d44bdadb8257f9b66f7531595f3536bc80badc67d53bd25ee465b5171916a50d2fb82eaf91667ba9007540e0
d416e0b4e86da840563d086ab97a7d63eb8c704d9d3626aed3c6fbc7a57943b5a54bf9a466ab6b9c775dc3b6d832de7149e0
502dc643abef945afcda8e323f30ae381caaddb8ebb29bb3f6e27b356d2b17d37eb6b5d8ad5ab705b5a9bec8c36720ea1cae
4e8d4e9f8da81366729da9c6a51658f67905e4f0f989dae8d07edaf4198bba5e5b5e85b6545deac4688b6d471d86cfa1c6f5
13327d3da3f502ef2118fdc41fd30ea2f1db2c71b96bcbd97acc7af253ddf9d36fe2f08dd7f895db3c426f13ac7a1ac5729e
f780fa5926a3e0d919da16faac6df16c7dec5cc687f0d194268a569cdaa481477c68d2e8c2003d8f2ad7b81e5e64f9e32ba7
83f03adf3cbfd2439369d5aa6d8d77536a37e78aa9ca90f8a5a724cd2b1487a69a2e6e921aebbcf7cfc91b9436a8bf4e6d06
647585873a50c267b2b770a82f87c6e1aaade27cde07954a0b73a9f1bbd21ddb9eb696be0e590e2dd845bff9c7d5c62c45ad
4527d1a09e1f5bc83a88bb2f753bcdd0a28ee63255741f5be61a973fd9b6d3b46d36daf691d6756dc9f7ed3a124f974857e2
f7a61f09cb53d22e5f3b71cb654dc3f133d5b70e694b499e8875da7fb4e9a6c5bd8e695ebaa4e75ae7f653dfdca4ad5a7932
eeb3cae43db615cbf029d53d52a9ec40824fefb0a74a9b49d2f1c96dd640bed4b63c0b3ee197f5cd2e6b768e9ffcb11c7919
f5c3da7e57078fe7f6c470f5eae39cd67b6b96ce29a779bde5208dee8c3ad7da1dee42ea188220edd3ed38ed68edfcc3dd3b
7548b1ce2a8c775b07a8dadb4dfb6dbf86bf8465f68bfd097944aac7735b96af9bc2f4a404fba2f655fc8b8d59d4fd9fc36e
00891f527397c775d8542071790e8e1abff720e864b43c7d65a0aaef63d0eeeaa6d3bdd0e8b1e9f6ab4eb864eb2cda5a4e58
d6d75b32d2152e8306d86491f021fdb7656e83685c2bbed606696d0d8b1f752c67686b2c6a71b3b46f786f53e8d548d723df
db509276d3e72d40bdadb1e838f456a73687b6866ab09719f1a145a1d9ab3d9ed87228f723e5bbe5a0e791e7f9577bcb4157
0dd8c2f5da16d025ddd3a6a66177913c40d27ddfdaaea3435e841ce74d53d2d796bfb4a1012b1575f7d2928fba7ea32f491c
773b493358ed5c5a7b423a81da6d82b587d29e5857fbb3b8c670dc4ca59faeb38ca658fc9476c339ad31d44d5922b76e1980
b584fb6993e9d248813583da5cd1dd1517aea8e9ad185dfc93affa78e838d4658779c06a1ddd37b668c314d7f648f8d2a393
4e5c09a43317f245d6a584e3ba216d20ad490f8e8355461a963f2a39b026c99a539b2dde1ad63569d46678c7d581bb7eee16
ad94c6d5817a714bd495e1e332c0ceadcbcd74f7ca55c3fac9d121df71155abf3712def56b18f1ce6b765a97fb79886a9f34
95713f86a7564a4bac8fc9fd284b58def2f5686169f0abd270d5f77ba215eda9fb1d6bd8c66800fb7ed52e3de941ed956b58
6a4df50e03e5eeb755a2f2b55ce2d473d649d72d6faddb210d295d2f0d77e4ae38f4c3b36b3bbcd633ba4bf13cf1aef6ea47
e7d9cf33df9bd4740c7a9d1e8d5e89e9898752571fd31eb54b478197ba9250bb6969d5b34ae0b1eb35a78ea74bdd502b4edd
a5640b44e0d9bd6b5f89ca0f9f17d4e097aeccb449f2125477d4655a317b7f0424aa2dc0bdad2a2a5f8c4b5fdfbce616d511
9cf25ef5d7ad7f76e4e5940a2dd6ecfb2b75e957271f7560e1fea8e9523af9ac9f6d679a9e922dd538bcf8fdcba89bdcb339
afff3a7e4875c0fbcaa5d59e0e5d67b14febeec64d6cbab35e1a5f75d35bdfc4b6e88a5ca9b14a3cdb1660abe2c62a6bd8c4
b6e808d95957d4a7a44bacca4764acfa86cd6d123f7591e659e366b37d4e556bb20517b61f08aa62cdcfd27e8b538dae678d
48f3ff9cb7e58ebfab7b53a5f9df36e1c91fcdb6e26ffc928c5cda74d9ea36ad64cbe5d335afc01f75d3c570ea8f35ebace7
539f576e7179eacbaee2f1f3485b43da4e0affd68e56f9ed1fae5f7ff1f75febd7bf764f46b94f5f198f7fffeb311ccd6c5a
be647666cbbfa507bf8c87c4ccf9ce2ff0f363307f28ea67bfb028dfd75fd80930d128ff22fce710e5ef7fb04860d5a686bc
9947a9f0a32ec2b03f6d67bdfce95f877ff1bbf0db7f252fc2d7effe8394b57cfdeecfc2bffbfabd5ffee4f77ef93f7ef993
dfffe5eb97fffe977ff9cb297fff9dfcef4f7ff9db5ffefa97ff24ffffef7ff98fbffc83fcef6f7ff99bdffffaf75fbffbd7
e17ff85df837e1b77f18bfec389abf90a2fe4afe5764977f0f61d3ca389dba425a3b208a95750bb334fb84ab847e1d43d1d6
102ca7f59a863f0d71b9dfff21d47bdd74e8f75bd520bf3b84569d54f9b2037944be513aa97375bbcd290de1a86d6bfdfe47
fdd1aeda39a9b68c5ad85504a9e7a58a549f86fbd78790ffeb4977e0e978f695b5dbf1e4d723f9756de749dd24bf7bff7a0f
fde4d76ddff892b6d5f68f3cf8f1f5c71fcf7a7a82fcdcd9eff71df17f3aeb1cdf75e90923b61cf6c16f6fe36f8f77f0b8ca
b3940e5d6ca16310b514f9c37ac28a952b85dc85fede2fffe3ef7ffdeeaf42d66f712ebff07bbffc478badda5f93fedd71c7
ff53cd95c76319e37f6f711d2dcec9e6196bfc4f6b5ccf0718e37fd9cad1f5894339deeffe83c575459c7cec87f8ff5dcb9f
e35fad1c69013cb99ebfadf153f76eee5d87bf71aed3e3f5aef37faff93af63beaec95ef95c3b97efbafd67633ff274bd009
8878c421e1bedb9e327ff27bf527f1d6fef65f6ded4fc6c2d0fec05291b67c4443d2f6a1edc0b13f96bea4615cfee0ff7367
2e660b39147de76cfdc61cb6af227efd01167823fe7ff45e0d887f5d3550eb1c43fc275d59acba6c7a709c14365dd6fdab7f
f2fb9638d406ac1ad62d33638856c3ba6846f78147ab8675b86f2bb7e41f550f0f3faff33bd10e93814be8e1079791756030
e9b6015d47f88faa16872b396c79bdd4f178253dfce04a8e55d7f02ebab7eaa79aa49f5e891ed274d9fe0cb8921e7e7025ba
e0473a1cd2f8d66587fe95ec3fbd121d32971ea2aefc844b19e20fae45a71fa4036806233f53e5fcf9b5d8f6689d829baea5
c7f9b5d8908e9e6865d7a2e378f245d5553bdbcf1e96f8cd831b75adec7aeab40b5e4e8f3fb99c5dfb50ba65599b163fbd9c
153fac5ad81f68b19b2edeec2526dd6f9f537f27173b6b3aea0c870ed7d78aa6560d7f55ebdcbb361775b4477fe9890d2537
b7dcffd572f5382e35f0d36ae60f7eccbeebc2ffb97e3f92f46afb77e5673ffdf47bf08ffd62fd99d332f8c77e89ffae56b7
a776f6cffffa968197effdaec7fbab538ec7f55f6a393a653796f38f2dffff7474fef356cef6a4bdf3cffd551fbe8be9bfe2
bba8bdd4fbed2e6f52f90cdaeca36e72db165dcc6febae36ebeff4533f5bdfd6fa7d81f56e878a235dba9249c777a0cfd5c3
bfb65ef33fc7f504a91cb5ffa532e7fddc57155a171197bf8af368b3bbd2535aa7f34ac796828e35a7c356ae6d3a00b61eba
3b79d36d025b3df751c77497ace3dd3ac574c568c33949f7b0abe9a514227f6e53ab2d0beb92e8a8d3539b6d90291b87932e
a2d3b3c576a9db744cd156dba468c7ad59df4dcfccb111b96413d89b0e6f6daa9fee5cd3b0cedc5e8b2e4d4cbaa43296ddc4
490f40505b9e556bc92cf5ae757bb40b9857dd39685becceb59c3d99742dd771499b4fb7dea45dbdc22caca75065fdbe679d
aa5a96c292ccc64077a1dafc4a19484cc996bf5dd2c9ca3ae59d975492f51e5ebaf370d7d310cfb2a02865db387fa6dd964e
a632a52b15a86eb43de4c9d099fb73b3d9f4a4e3aa6a8c71e842553df2ca2677e4c6ea86d1e34cbaecf6bad26503554957ca
db3617dd2f1deb546c3a74e24107adacb910ebd047d2657abafcaa8c82c967bc8c8e259d43d5a977db5eb926fde468d8b627
eb817a657d8034515289eba8a79dd8a153d1d791cbeac0ac236e29d93cc5968bc5de69713de666b5412ab90fbaf1c82692c7
41886c672deca68b17ffe77c79d482e59ff0f64867d7a6daafdecc1e42ffdc6fbcd53efab64b834177d3cbbfc8affc7ba98d
feec9fe7fdd7657ffb591ac8c3fbafcb58aed35ef4f1fdd741e8d366c9c6d7df56b3e6757efd4f3d8ee6b041ccf1f5d7098a
b39c1da3efbf3c28ba3723e90ef7b36ce24dbae4e5d215215f497751ada7cd0b6b0d207f622f76cfbadf60db6b0db06ef2f4
d936d6755d63797975e9563af46592573dcb9596da45a7ec777dcef555976a683b4ac5a06b632f795fbeca2ab4a5d40b36ec
ad343a4d70d619c56443c48b2e3ad14d65d2273f4a5867c7e472ea42e1fd2a4ae959c5f18ca7ed9b9177a49c5e99749c31da
5afe4377e16d650e36e97eb4a47e94d6de4ed1e6ef6c4ce9b097e7ccb659ce261892ae22979ec96a133151d77a5db50e584f
5db4a1fd8b33d69918ad0316a9e4d63252ad3bc5d6b3d6017a5ae25a8e91b36ded7bad04163d16b12ca7d1d9cc73af95806e
dbb613b9746646ea561b59d435e5ebb9db7cbe4dad251b91ce7afae4b1dbba19ddeba99d8c1f5e769dca39d7f8b3f83fd7fb
a415807cd0ffdd97aea990df90c7e49ff04605ad13448673d326de30067987489df060ecfce7a3aa30442b85fe30d0fbd7ee
40af8eb23f1a29c6c4a18c9f96fdcd78b974d06a5d19f5a812ad3bb23c1cc90e632e9d9b4d77da58eb516e5069bd6223f27f
b326de1ff4bc3fb01d983f0cdd94d1a9cb4ef8d97f3664f4fff636e3d8add33948ed856ae374ff92d64103fde9b0fe9f9661
7d1dc4ffdb5f7efde5cf7ff933f9b77f90bfffbdfcffff4efef9f865fd25d5c1febfd3f8d42795468b3d84ea8312cd3c415e
ec6cff628fa62e9cfdbb3afeff9bf229eb7f26fceccfc8fdbc6cd7f0a18db3fa4ced35a2033f47add931f297c53d40aa753d
b9414f2d327b925ab4b9b5549787b5b807389961ca2ccd86edd44dcced7a92ba579550fbf9bca439645774b7ce4582f9ddd1
399cf2d8663d92bc3eb686f6435c914f1a1f1ee7f9318e3a317b5742513e0fcbf86c5807bdf48deac3b68f3d645dd81ab5fa
fdff7b3b93a65b72e33cdf756ef507bea51c211da230d4b0d46033423bc9ed15c30bbad512a9e8a282a64c85ffbd91035089
21eb426684b890badfcefbde7ccea983a980041d30cda3c5e7294f2592a7e627eebbaccf791bea4be8ff9069546eab0f3d53
7bff7bebdc5c42fe42422edc54bf27b52efd65fe97bf29f3cfcb9ddee999e6dfd675d36ed1fa2c53eeeda095d795bfe68702
e8da79eed79fb1593b5d0dcf6a46f32201f7aaf669feb6fc9db97f8ecd54f9df64928b051e4393e76f8adb1ef3ec45ff17eb
efffd9fc5becbfffd7c52bf7fbed62c3cf4f7345bf813f6936a87f2b3450c203d3dd6fe5d1dbdf8ad69ba6ff7939f8f29ba1
17d567bbd0bc77ede98fb985e417a13f7efb756e4dbfbefd91feffffc9ed26b6a5bfa336f4464db7a2fc91fc094366b03f24
9c94fa3436288fde7e485aa70f893fa05f9594beff41059c2ca6f34ffaa0fee6db6ff23ffdaf6fff377743f84f7ffccfeb74
9a0f09ab34e0c8113f3bacc0cfb2c71d53cf0c228e32f7412fe38aee33c357c0a15d7eed3fb29bfae6dfe68fe977f903fa43
feb0fe35f7dae5e39297f2edc7b4dc05aa8fe9f933f0f6679a8fc9533d83a3ff98362a6875f41f9392ffa31f136dabf7af4f
d6dbc7f453fe277ceaf089fa99627e5edead90a71a0e6f39a635979d266a793892a77e588b055f1715f567ade20ce2b84e5e
76501e8d4e7b1668ae5002f247b5edd5127b91dc36d03b389c3224da1a856b4479fad5cf18f0de24f902037e6bfc98e7a9d7
8edf74f99cf181b0c2729bfbeef2fffd69e4a9a5c38398befd302cf9f190bbdde4b22abcbaf24ab967c44bf7bea8867c9e14
d2ee5fb9da6d16084da0dcecb6e208f35c6f9b4ccb79521e69cede9a18723529c447c2b29227e587ffcd49b2b85130d2c51e
057916094d64615ef1eca125df1e5a6168f9cc3f525abf684d0cb99a14e83c84bcf0b43c26181c9e563ae9210c788ee7cc3f
830a3d8b8426b240af78f6d0926f0fad301e39e041017774bf734b5626051ac760f8abc7041d5d214ff9795cfdc173339579
12084d60415e708479b2b78da665bce592cfa23726865c4d0a3116de4cf48de092feee71eba9a306de6359a9c2fb1e0685f6
bb6e3daba4d9b3aaecb57c1d9eab5eb426865c4d0aabc77a089e7e6631e12ee633d017b35d797cbfc5adde3f3a8d8426b210
af78f6d0926f0fad301e195f74253ee8a34d0c599914e82d9c079e5bc704fdce174a52822944ac337354e859243491057ac5
13e6f9de369d968f3c50a4a5d0d6c490ab4981c68adc78498ea39b2f282d7a10b78d36f3a6bd42cf22a1892cd02b9e3db4e4
db432b8c47a65b2eb8089c3631646522d0e799a7f2819fbe3c430ab8bd8afec5e1d6fd8825a5047a1a094da4402f79c23cdf
dba6d372c23954df099b723529d0cfd001cb2a9ceea2fb512f17f902e058992781d00416e405c79e5892ed89158396715df7
ec3b2353ae268578c752389113c4a2a1f1e016378f630fbeb2b220cf22a1892ccc2b9e3db4e4db432b8c47a69b5722bf1b56
2686ac4c0a74a2aca8c5f1d7713aba7101dfeeee78d9e0166acf3c8d8426b240af78c23cdfdba6d3321e5a3cfb4ed894ab49
818e580827d134c11f5b6ed971a3152618b60b37e957e65920e8c082bce0d8134bb23db16278e4035f6773113f6d62c8caa4
10e7e118dde180f9e57955d8f1b012ce7a71f084afc72af22c129ac8c2bce209f37c6f9b4ecbd1fbd0f7c4965a2d0af29662
c21266344c727888f5a4a7901f41bc1ea820cf22a1892cc82b9e3d32a7db133f105a3df1c44ddf0d9b72f128c479608497e7
f02809df5beefbc94b1b74cd93ab2392692434918578c5b32796747b6445f1c8b86b032bad767086ac4c04fac8df7abc68d1
f0daea4a176e58e5e9fdf3639e46421329d04b9e30cff7b6e9b48c85f6af7eb268cad5a440635d032cdb8609467f5c9b0f91
12c4cb12a8e842819e45421359a0573c7b68c9b78756185ac6ba7957df119b723529d079dac32bef3434bc70ff2baddc1d1b
5e4f95e7c0157a16094d64815ef1eca125df1e5a613cf2459553afbe2336646552a06935e3e081d285d75bd1af6f3f021658
3eeb5ac1240e9ab802fc7d3f98677adb5c5ace5fda76f51d92295793821bb13cde49cd0d16e538e9c01c268807cef14a930a
3c8b8426b220af78f6d0926f0fad30aa8ce7645de2db019589216b93028d95a764c1c6056e64e921c493413b8e202af42c12
9ac802bde209f37c6f9b4ecbf1ec071e96580d0af0865b962f6a6a1c8f1ba8174dbb3f772c3a55792781d00416dc05c79e96
52ed596bfe5ac40244ce0f58862c168595f63d5d3cfe3db14260fe622835bcde60c3e21b057616094d64a15df1ec7125db1e
58413cf246b5675d374bb4646522d03b5e4d8e551769d0bfe16a11add49df1e4e95e5d269846421329d04b9e30cff7b6e9b4
4c97a2c601da90ab49813e5dee3269e9220ffbf1118c9c6dc4e24178ed6d859e45421359a0573c7b68c9b78756188f8ca5a7
fcf002c792954981a616d451d372e56700db57ca162b3ee35ed3da294f23a1892cd02b9e30cff7b6e9b41c5d0aee18a00db9
9a14e8dc49d3358eb44f32467fe13d4df812168b55ef58cdb740cf22a1892cd02b9e3db4e4db432b0c2de3052cae6f9d4db9
9a1468bc6a3a6e65ba932e8f77bf628271cb238a2dd5ae791a094d64815ef1eca125df1e5a613c3215cbe3f24cdac4909549
81a6f72ab4067b9ec11f69e785ca3394570e857916083ab0202f38c23cd9db46d3727e90fcb60dc4865c4d0af1869f80a7df
db892f0c9d7c3fbc4887b5390af22c129ac8c2bce2d9434bbe3db4c2d0f2857bd2fabed894ab4981ce1d2a1e78c09eecc8f3
bdb03bfaed7959e2a8d3a9592034810579c1b12796647b62c5f0c874792cdfedab4d0c59990871ca1dcbbed11add8977913b
aa17edf064adccf70af234129a48615ef28479beb74da7653c28bff5bd9229579302cdffa32f0277c1e1b915fa216eeea012
1cb5dd9e46421359a0573c7b68c9b78756188f8cfbbbcfb87553454b5626053a6031395ac1c13d72f89a3252b62ee2791af7
bca79a46421359a0573c619eef6dd36939e621fdd677c5a65c4d0ab4db71ebe5c6339e1d0fafd2d6072c484f2f4f6b67358d
8426b240af78f6d0926f0fad30b47ce2fa43df2b99723511e888bb380f9adbe73f8167f3b05217ed73da03d654ae03d06924
349102bde4d9434bbe3db4c278e41dab0587be899eabcaa220d359c59367b638cb3c784116df91634d8d67ea3c8d8426b220
af78c234dddb44d3eaee37ae1fd85a1872f128c4c1d32c8f27013c99a7a7f0ecdf524d23a1892cc42b9e3db1a4db232b8a47
c60dd9f1f2fdf769c8caa440bb83ee3fe049005e318a15e0b1aa6c4c11ff546dc3a691d04416e8154f98e77bdb745a0e27ed
60e8a10db99a0874fedec316794610f63c4c4af86bc3a383b8fdfe74cf5ad834129a48815ef2eca125df1e5a6168190b5bfb
be2f36e56a52a069199287c7c1f334801ec59d267ef199444f23a1892cd02b9e3db4e4db432b8c47c64abd912f04d12686ac
4c0a74cec979de869ac787f8d290de1ee6c941a4cb5a6a2b368d8426b240af78c23cdfdba6d3720a58526d8036e46a52a079
08c1f380fcb5e0bda5f42826bcd4832a5516e859243491057ac5b387967c7b6885a1e5cb5369c19ece90ab89407b5eade251
3196dfdb3cfd188f9c9cc396b73664d3486822057ac9b387967c7b6885f1c8d727cf4dfbed9c86aa2c0a321ea296f767b9e9
c1abd0133d8878dce0a032d9057916094d64415ef18469bab789a6d53dcf5387e18729178f421cd27e9cf496f874382bcacf
65a4ec72e38355556b1b368d8426b210af78f6c4926e8fac28aa8cd7c5e1f520ddf769c8daa4406f79c2e379c92acf007c1e
27d2f783e571f20401ab7416e859243491057ac513e6f9de369d96735fcf7735b426865c4d047a93fd3d9420cdf8b14a0e5f
b78b77e1a6da864d23a18914e825cf1e5af2eda1158696f38f84af28694d0cb99a146859cea15900de36bc47dae08417aeef
e1c0b3df057a16094d64815ef1eca125df1e5a613c329667d87ce8668e96ac4c0a34bef58f3cf351ffeccaa9918a3cc64113
5780bfef07f34c6f9b4bcbbbc35b13065c43ae260577c34b4fae93a73da7c73fc5e7b27cfee16185810a3c8b8426b220af78
f6d0926f0fad30b48c054efaad41b65c4d041a2f3ec3ba7194a0f77b6e6de8bdc38e57c8e301abda844d23a18914e825cf1e
5af2eda115c6237b3c51bdc56ede68c9caa440f3a206fdd88e83ae7ce2fce855135e9e52982781d00416e4054798277bdb68
5ace3d3c5fd3dc9a18723529c4585078f3d4d61cd9385299662cb41df3c829cf086afb358d8426b230af78f6d0926f0fad30
1e397c722b71c4be7b36646552a07343ea6935960a577b77d0727c1e53d04862ab0dd82c109ac082bce008f3646f1b4dcbc9
6f5ccea73531e46ac2c4276f9e0fd4dcec3c10a6fdc67899160e1d5c69c2e691d04432f39a670f2df9f6d00a43cb38928f7d
276ccad5a440efc9e1c1399af3a82332fbbe3bba77e1aad0b34868220bf48a670f2df9f6d00ae39123dd1818fb4ed8909549
818e582226518393fb51dcfaba7b4ad039971fceba577d1e094d64815ef18479beb74da7e5fdc0d9ea006dc8d5a4406ffcba
4566b78e6fa1c1722ab42e1deaf6d679243491057ac5b387967c7b6885f1c8789367ec57810c55590872ce2f0f7fe90777c4
94e7b9f969c4d97dfe5cf8b12caf6fe691d0440af292274cd3bd4d34ade2eec2d477c3a65c3c0af1bee3dd10344ac2636d39
25474f61dcf82692bd12cf22a1892cc42b9e3db1a4db232b0a2de7b1dd99fa3ec994ab4981ce89e0053e34bbad6f0ef13663
7a577c1e957916083ab0202f38f6c4926c4fac181e192b8144be0a559b18b23229c465d04ff3792cb871d03bc4dde35207dd
f15d906791d04416e6154f98e77bdb745a4ee71153ffb335e56a3281f678fba0dba9e5d9bdacc69f33e81a094de404daf4ec
a125df1e5a6168f9c202877d476ccad56406cd0f251d71c42a7ef8d2e38c53e812094de40cdaf2eca125df1e5a613cf2f109
290f76fa59a3212b930934568ac43a823480d83cd67f3c8e19730d041d3841b61c619eec6da369f9087861e5406cc8d56446
ccbb00b8c1717b4e0f6ff89b229748682267cc96670f2df9f6d00ae391b1ea82dbfb6778ae2a8b09b2934568fae9b94de67d
33e41a094de404d9f48469bab789a655acfdbef73346532e1e3362bc4332cf067006849758045aa19e129748682267c49667
4f2ce9f6c88a42cbf91ffcdef7c2a65c4d46e81dab42e26d60d8afe2f9cd84d73fa609f413094de4086d7bf6d0926f0fad30
1e192f1c937b24b589212b9319349ed44d3b2d63e191f4e654c23c129ac819b4e509f37c6f9b4ecb79d87e0ebf5c53ae2613
e81317a4036d724a076e8788fe9cb5da35109ac009b2e5d8134bb23db162a83256d4c4ea82dd84d190b5c98c78db03ae70d0
8f0f77b70787c54967c825129ac819b3e509f37c6f9b4ecbd16339f301da90abc904fad8790f2efdf8f6eebdcd3c129ac809
b4e9d9434bbe3db4c2d0f2e986221ca65a2d66c8787fef8157cb617a2e3f113ec5691b5623a1899c215b9e3d32a7db133f10
8f8a75a90f777413464b7e3c26c4fb1eb15434fe73c2cb91d3b98769fb5502a1099cf05a8e30cff5b6c9b48ca5c8fba249b6
5c4d66c45b4c21372f38cb4d31cf7df18aa469f35523a1899c315b9e3db4e4db432b0c2d5f29cf4fc34067c8d564029ddb72
77b9838e70a68867a3dce966b38b27129ac809b4e9d9434bbe3db4c27864bc87126fefeee80c5999cca037bc6831d1fbd214
f234ffdccf69fb55034107ce900d4798277bdb685ac677b9c73e101b72359910c7daa2627ef8aec55fc76c76f14442133961
363d7b68c9b78756188f1cf04ef034b4ce86ac4c66d09b3fe8e8082688a739e906a7297489842672066d79c23cdfdba6d372
c44de1d7006dc8d564029de7f2c1fb8d8eae26cfe526e26c8af14442133981363d7b68c9b78756185a3ee5c681d663ae568b
19321e4f768e5e3da4e74cfa0cb94442133943b63c7b644eb7277e201e357e7c0ab1af9e64c98fc784d8e77e2560f901cc8e
8b1084e9835d0341074e782d4798e77adb645ac6d3c47d11215bae2633e2cd5d1796d5a5fcfc41c5f7a70d588d842672c66c
79f6d0926f0fad301e397d5cfe859cdd6cd19295c9047acba3c2dc8dd1a1ddfc8d709198e9eca24642133981363d619eef6d
d36939ba10fae249b65c4d66d09bdbb62da74609e23fa5744c7be71a094de40cdaf2eca125df1e5a616819b7faf775846cb9
9a4ca05dbaf61d77206382270e5db67d9fce2f6a2434911368d3b387967c7b6885f1c878553696b3e9e80c5999cca01d1726
c4fc4a39b3e914a30442133843361c619eec6da36939e110beef854db99a8cc47801d171e6e6957e7c47381d9eda984d319e
4868224766dbb387967c7b6885a165aca837c2cdd56a3143767861fa15b976f7e936acd9386bc39e48682267c896678fcce9
f6c40fc4a31e7cd97b3f6334e4c763427ce6a625376f74303deede87039fcb19718d842672426c7ac23cdddb86d33255c3e8
3b6253ae26336887efc5a97648ccf3bf948789d7acfdaa81d004ce900dc79e5892ed8915c3239ff9cfedb1af9c64c9ca6442
7ce469f5b6273a011513dedc733e056ae791d0444e984d4f98e77bdb745a8e7962d81711b2e56a52a037daf141f939bcf8ed
a0ad0fb1af57310d8426b0202f38f6c4926c4fac18b47ce2d1d3be0b36e56a5288f17db0e7e9ce9e8efc0d5db44f31c66ebf
f63c129ac8c2bce2d9434bbe3db4c27864bcbd690f7df1244b5626027de065c817d5794abb93ad7a986048e1a0f153819e46
421329d04b9e30cff7b6e9b4bc6fc3176a88d5a0003fa5ceb19a5674b9e7dc28b9b668c33c107460c15d70ec6929d59eb5e6
5f45bcd7cb05e7ba69a2213f1685156baa061aff96e3b927aedb449fdb553c9b5d5708a691d04416da154f98677bdb6c5a0e
e71efad249b65c4d0a74c2dba831bdb8e7477d4fb4d42ee7d0d3f3666a12074d5c01febe5f8f2b99f6b80a40cbf913ddfb0a
42b65c4d0a6eee31aed3f1aa6bbda900efdf2b577b54e0592434910579c5b387967c7b6885f1c81bdd42e2ba29a2252b9302
1d3ceec0a31626ecf2e451822e7f41eeaa25d4e791d04416e8154f98e77bdb745a465fb70fd0865c4d0ab4c733c054c73f85
8d8f70d2f0dff58b9cd34868220bf48a670f2df9f6d00a43cb7859635f36c996ab498176346ca0e6e6a9de8f0996495f619e
05820e2cc80b8e3db124db132b8647f69f807bcbbaf992252b1321def3d48eb713273e1c10a8a6426e727333973bd03aa798
05421328c42b8e304ff6b6d1b48c7bfcfb3eca52ab45e1a5532fb4292de551f08145154e4a0f5f2085a754fe3c129ac842bc
e2d92373ba3df103f1a8783d215e0bd3b119f2e351888f3c247454c83f3d65eca969df527e32afbac2398d8426b210af78c2
3cdddb86d3721ea71f7df1205bae26059af615d31ca74e7112258887cbaeb396519a47421359a0573c7b68c9b78756185a3e
c3eeb6be1736e56a52a0e9b02615f24f7ac7569ec8e7bf06cbf755e859243491057ac5b387967c7b6885f1c8f1932726a1af
9864c9caa440d7135ef1a9e98ef96de1c23b77ea8c6216084d60415e708479b2b78da6e5ddf9bdaf1c64cbd5a410f319454e
70bbb0860235b07b1e38c518fcd3644f02a1092cc40b8e3db124db132b062de75fc7d5bf71b3e56a5288d548186b3e52fd04
4a70c365c1a77efa3c129ac8c2bce2d9434bbe3db4c278e494ad70d2ddd119b23211e8741df97ba043e6f1a9709e132c5b97
6abb3d8d842652a0973c619eef6dd369f9483ef5a5836cb99a14e8e7d84b3cf2709146c39460feddd101d60a3d8b8426b240
af78f6d0926f0fad301e79ff6c7900d0d70eb2646552a00fbcc99cf638443a1c715e342c0e713fce6b7faa89cf23a1892cd0
2b9e30cff7b6e9b49c367c5f3d401b723529d0f28e10137c4a7e6382a57048859e45421359a0573c7b68c9b78756185abe1c
dee43ed019723529d054cd699715763c6d44a346dc55138ff02c90cce240c715e0effbf5b892698fab001ef9c06b368fbe6a
90252b93821b52ce83de7dc774e4b1710874f027e4b68636033cc0b34868220bf28a27ccf3bd6d3a2def7b06e9fb6153ae26
055aeed6a1046b11ec9ca02f85fc2bf42c129ac802bde2d9434bbe3db4c278e4f3e370577fdf151bb23229d0db967f64549b
2f4a9d101e1be3961eac8353d785a691d04416e8154f98e77bdb745ac66ddf7d5365a9d54290a31cb5a7f47cfee51d27554a
095b7e1cf27754abdfcd23a18914e425cf1e99d3ed891f08ade26244e8bb61532e1e8558ee66c0ec9e52d0985dad64568867
91d04416e215cf9e58d2ed9115c5235f9fdc2ef8be5292252b9302cd23605e62cfffdf45aeed1fdc11c9a232cf02410716e4
054798277bdb685ac6aabe7dc9205bae268578f7741617f3938103558409b975e7e7b322cf22a1892ccc2b9e3db4e4db432b
0c2de35c2df4f34553ae26051adf9d70b9dce8439e0db94423637f1d89aea3a9eb42d34868220bf48a670f2df9f6d00aa3ca
1b5e018daf33bb9ed8d2b54dc1a62103559ecc63a42de261989d52f47b9e131caeb6dbd34868220bf68a271809df2f808d9e
5b8dd8570f7ad3ab8f90d30be1c8bdcbe6e5843d66891f10d68da9338d6924349142bee439904bc603b92251faf6d9b6f3e8
4b08d9baf21172efae3cc9f7f48ac51d6ef33b77323eff4a6982589bb46924349142bee40946c6f70b61a327e7fb77542f72
75116eb77b9fc70db4ca415759787aaf88ebc3bccc53b06781d0040af58ae300cdd90ecc0f4423679bd0974e7ad38b0d331f
57fef9e14c1f9bda2be561d4e568c9c3e789f0715ca5519bc7818a63e215bf0158521d881582d2fd273f3b7b5f46c8d6958f
305399604f7534429eeaee296db413d51fb5faaa504f23a18914ee254f3032be5f081b7d4f78afdd486ee9d547c8d55534e1
e43250546ab65cdf5be75bf3486822857cc97320978c077245a2f4f071b979ee0b29d9baf211723cf114377a839a4791b9d5
3b824b9425954377f56aa37924349142bee40946c6f70b61a3e72e32f58585def4ea23e4c1c7cded1baf50e3559f793ebc51
963e6d31cf9f4a0f3e8f842652c8973c0772c9782057248d7eba2d0fff47424baf3e42ce3b0592ac54cbba006699ca7eb742
3e8d842652c8973c0772c9782057244a8f9f0cb2f575956c5df908b993010565e98e9c66a2452f8fcd12be652e3df83c129a
48215ff20423e3fb85b0d1d34eb57207724baf3e4c9e7bd56dcbd308fa4d62397c7af94859f273594fa8cc23a18964f235cf
815c321ec81549a35f69c3fb9607424baf3e427e8480671ee93719af2d1b1f5499c0e3fe7377ee7557fb3c129a48215ff21c
c825e3815c91283d7df25fe2fa5a4bb6ae7c845c5dc614e8f44c3869f7bd8fb5ec9d904f23a18914f2254f3032be5f081bfd
c82d675f73e94daf3e429e87ed79d244634a2c86e5c3c587a73cd7433beb4c7c1e094da4902f790ee492f140ae4894be7f36
dc623c4c4b2d5df908395e3996f8b2f290fb9fb8e52f85e613a1dec826e4d3486822857cc9138c8cef17c2468ff9a94a43bf
6debd547c8b798a711b220e4f17c8de7af2778dc3496c7a0057c16084da070af380ed892ee80ad301a1dabe8f6f597def4ea
c3d8092ff23977de2944c76dcf8d170de8618dbe6e2c9c474213c9e06b9e03b9643c902b12a51fd84d8c4fb4212b17e13eb9
a9a5c5fcdcdde441ad4fd410f9fa324bb8a791d0440af79227cc13be6dbc46de433cf7a1c3b6f56223d88773db96e8f47ec8
996e1ecfad518a9b4c310bf634129a48c15ef21cb025e1815b8128fdfce4a1d8b60ffdb5a52b1f214f79f87e5cfc5ad29507
11b3dc4e3e7253faf17924349142bee40946c6f70b61a3e78142ec2b14bde9d547c8f9e66eda59e22fac49929f099a55d4fa
15023e0d041d28dc2b8e03b6a43b602b8c463ff2cf691fa6a0b65e7d04dba72337b83c8bb85409079f83a9186419becc23a1
8914f025cf815c321ec81589d2af0feeefeaab33d9baf211f21c9b0ede17e9cf944e1f777ab5e5e93a4dfc8f857c1a094da4
902f798291f1fd42d8e838e7e90b16bde9d587c9a35c0a4c8dee817719392ece9c9ba4b8bb9c6be9c5e791d04432f99ae740
2e190fe48aa4d1af6cb60f53505baf3e427eca6883b2cc3f05bccd09a7145bb9ebbbf4e3f3486822857cc97320978c077245
f2e81bbd63ed0ff458b27611ee9dff473fc25d1db7dbae3c993e63ee870bf734129a48e15ef28479c2b78dd7c8b8acd9176d
7ad38b8d60a7ed7261a3cd927ecf032a4fdb0230451fce23c550862ff3486822057bc973c09684076e05a2f42dffd198fa3a
46b6ae7c845cb683d2cf91df6a6d1b36447916bdf30271219f46421329e44b9e60647cbf10363aee003bfafefa45af3e425e
4f0663963861a4031ddb991c5dd157b0bf1306c2fc7db70158121d801540a3e34e92be7ad39b5e7d043837ad5b7e0ee97798
9f4e877beb764a72cbbfc1f3a8759de691d0440af692e7402e190fe48a44e93ecff162ec0b1ad9baf261f2502ef4a2e9e215
7c08813aa0eda82b074c3e8f842692c9d73cc1c8f87e216cf43ddb1d7d4ffda2571f213f4eac8bcdd3a680357573bb745296
3967bc4fb83ce7f3486822857cc97320978c077245a2f4f071b8c1b5efa94d5df90879e42d39f483cccfe449c5a828cb6e47
fd3c129a48215ff20423e3fb85b0d143ee13fb1a476f7af5117227578c52966a17c3b65f78d76badb9390f041d28dc2b8e03
b6a43b602b8c46cf7f431e1d8c78965e7d183bf734f92b71b42ee6f96cb18bf483dc8fc3e5395569dde681a003197bc971c0
9674076c85a1f4f8d9f0d4e8d0515bbaf2116c5af4e75add7e4b7c9c8572ac6bc3823d0b842650b0571cc148f77ec16bf414
a2efab1dbde9d587b19faf0093dcf882e18bb2cc23ebfc68d4267d1e094d2483af790ee492f140ae481afdca5dd3d94f3b5f
f4ea23e4eaba55355bc42cb1327db8f64a3e8d842652c8973c0772c9782057244a4f9fdc8e9c7db1275b573e4cee0e4994b2
c4db2ab16c1365e982f757aa9dd93c129a48265ff30423e3fb85b0d1f1d2aabee2d39b5e7d843cc433f883bf12dc3b9ea712
f4866b7b1649857c1a094da4902f790ee492f140ae4894be7fdc7e0caf822d59b910f799475129e29e2cca31ca523fe5d81e
1e3222a18924ee454f98277cdb788d9c1fa3fc608dd8965e6c043bf206059e41d4cb4530c58845c2bc4b057b1a094da4602f
790ed892f0c0ad401afd0c79383874d6b65e7d84dcc9a200cf9d7256d7c6cf24ae196cc75ec16781d0040af78ae3802de90e
d80a43e947ee2ef2d87798735abaf261ec938757bc3cf0cc9c30cbba50c8dcf3486822197ccd138c8cef17c246df715c34f4
d7b65e7d849c8a18455a01cc03786cf6229dc7de621e7fe1668e4a3e8d842652c8973c0772c9782057248d8ecbd37d49a837
bdfa30f9c16fa9f93758afd6a12c0f7e2120e37423129a48265ff31cc825e3815c9128fdfc843d0f92865ecbd2958f90d35b
9e8dd73e6974711cb4232d0fb9a41450219f46421329e44b9e60647cbf1036fa81c383a1dfb2f5eac3e4bc753e5e870c2aa9
f237f53dcf9a1993cf23a18964f235cf815c321ec81589d2afcf16863dd986aa3c84badc99846dcf959be29ddff16ddd11c9
79203481c2bce208d3646f13ac51d386d5b247604b1717064eb4877ce39492f4afb4e24d65aedc5ef6681b91d04432f29ae7
c02cf90ed48aa3d1af3cd577c3acd3d6ab8f90c73c5d4cc9730fbbc9712ffc973c55c785b1d29acf0341070af78ae3802de9
0ed80ae3d13dbe830be7b0f7ded2b58f60bbc06b43f47dec7c1f20f538583d387779e5925023129a48015ff20423e3fb85b0
d1f7fdd85cdf4fbfe8d587c9e986d6280f62d9974149d6b532069f064213c8dc4b8e03b6a43b602b0ca56f1f879771f6534e
53573e82bdc963c90f225e46b2d1b156de638ad5910af734129a48015ff20423e3fb85b0d1631e1cb9be937ed1ab0f93075a
22392f4e2c0fe54e7caf49ffe2b88c70219f474213c9e46b9e03b9643c902b9246cfed85eb4b49bde9d547c8a9f9f5b41767
e3bd2991bf2c4c1def763d0af934129a48215ff21cc825e3815c9128dd7fbcc337913da1a52b1f26f74f11d2cda710709441
bfcfee28f83c109a40e65e720423ddfb05afd1f1e2afadefac5ff4ea23d8fc468fea4d6c72b7142d0ee694f9f07ae59e4642
1329e04b9e03b9643c902b9246c76ae67d79a937bdfa30f9c6ef71a8bc489eaee6c17508bc062c0b85e54e3a23129a48265f
f31cc825e3815c91283d7c025e6f37f45b96ae7c843cf2820125b6f93c9e0c8e0ebbe62f8bcb3856f26924349142bee40946
c6f70b61a3e32ef8ad9f77bee8d547c8551becce9c67ee74f8bbbaf8ec4c19ba4d03a10914ee15c7015bd21db01586d2233e
4b675f6dcad6950f63bbddd15672fa41bac06721a9efd98ef338f095a670cf23a18964f0354f3032be5f081b3de146d1a1df
b2f5ea23e4db4e230e1c5c39bab31a6fabe65f27af7d57f26924349142bee439904bc603b92269f4136bda8f80865c5d88fb
b8f866c8e3a41cf1005efe1552e753a78fcc6d44421349dc8b9e0337273c603f1c4a4e7860efe86b4fd9fa6323d86ab3a5bb
705b29fe0cb9496ab6e41b91d0440af692271809df2f808dbee37deac3176bebd587c94fae14e5394b9a3985d218631de4a3
6ce03422a18964f235cf815c321ec81589d2f74f9e0386be1e93ad2b1f21971a33fc586e2e4f9ef7c80d512d422de4d34868
22857cc9138c8cef17c2468fdb9e27fa23b9a5571f26c7b21454dd13b33caf6b4fe1a282063949de682a4b30462434914cbe
e639904bc603b92269f4036fe818fa6b5baf3e421ef2af309d94182e0653dd47ee832f5cb4d85369dee691d0440af992e740
2e190fe48a44e9c767c3676b98825abaf261f25dd57cc446498efed1e8a3dc28c0e4f3486822997ccd138c8cef17c2464f7b
fe08872edbd6ab8f9097db2028cbfc7b0ce9e26d78ae166e16f26924349142bee439904bc603b92269f42b8ebb5c6cb9ba30
77e28a3c8152e4abba78b4e16a594cc69e064213c8d44b8e0334673b303f104a3e3fb9a5dafa8a4db6fed808b3947ba40cf1
d4089e06a40c6ba131619e05421328cc2b8e60647bbfd035fab1edb8e579a0b6f4ea23d8bc2c40256cdd290ba03cc0966ae4
a5419f47421329e04b9e03b9643c902b12a55f38103efa7a55b6ae7c983ca6338fb1c2c9597a3e1474f1d4828fc5c938dd88
842692c9d73cc1c8f87e216cf49867ba4355a717bdfa08b96cc2f294e526630dcad2e7f9c4e18ed2a0cf23a18914f225cf81
5c321ec81549a3e70633b79d23a1a5571f260f54bee13c394bcaaede28c1c46f11fff0cbef7b0c8492d940a8327ef480dbbd
f32fa7ef942d5dfb0861feade5e114559fc1f25297cf1373ee5e68ce184a911723129a48fe6ed73cc1c8f87e216cf43dff54
86d79e2f7af561722f473a39cb32aae409b377b9052a85008c486822997ccd7320978c077245a2f4ed93a7f77ea8d664eaca
47c83d5f5b1529cb7a5a1db36c0f8d1a91d0440af992271819df2f848d1eb0b246df37bfe8d587c9716f3aed43a12c533af2
6c995ed23b5c0abb922f078c8c486822997ccd7320978c077245d2e8f9033d8772552f7af511f2e0f0cc1fed2ccc6d6bfe75
e2774459d6eb62847c1a094da4902f790ee492f140ae4894ee3f1b5d49df135abaf2617239c27a729652a284be2cdc558aaf
394a5b3e8f842692c9d73cc1c8f87e216cf4842380be7f7ed1ab8f9087fca51c1baf02e22ff4dab84c66fedaf8d84469dbe7
91d0440af992e7402e190fe48aa4d1afe08fa15cd58b5e7d887cbf2ec7f7c6509665f18fb20cc1f97494f7e14624349144be
e839904bc603b922517af8e47f75c3965c53573e421e037e12819adea3ae87d0ea58a9bc25e4d3486822857cc9138c8cef17
c246cf9f705f59f545ae2ecc7d5e07bed9a22a156ee72147e2e5ef7a331473cf23a18964ee35cf819b131eb01f0e2547fc83
fb50aacad41f1bc1ce73a798bb5d6a757775991f5ef915f1929958b0a791d0440af692271809df2f808d8e953c877a4d2f7a
f511f25a5f08b3543708b993ae0fc2f38f423e8d842652c8973c0772c9782057248d7ee2fed77ec2f9a2571f263fd29e4773
5c88c6edea920287abbe5ba89b5e8c486822997ccd7320978c077245a2f4844582521a7ecc96ae7c845c5dd9e876550fd99d
5ccebd9cb23222a18914f2254f3032be5f081b7ddfd23914f079d1ab0f93efcf2d420eaf28c0956eea7bce7a091c834f03a1
0964ee25c7015bd21db01546a35f176d2f19f02cbdfa08761ed19c49be1b2a8ecb9d509e57d160a3b4e9b33868e2047ac16f
60965c0766c5a0f4fd134e1ffb1bc76d5df93073c2163eee549a01ab66d60b2c715190ff27d4f3486822997bcd138c8cef17
c2463ff6740ccfb2295717e1569541ddae8e79e2cde9baca8911094da4702f790edc9cf080fd7028f9c0e30c78d564c767e9
8f0d63c7fc5d04bff157934e29cb4229fa98c79447394169444213c9d86b9e60247cbf00367a0a783ff1086ee9d547c85dc8
4fdf4e7b2b1dbfc0e61b86dc592f7a14f26924349142bee439904bc603b92269f40b0fa90d3db5ad571f26c7f37e78c88dbf
1f7ebbb1f138430e1194e67c1e094d2493af790ee492f140ae48947e7e3c96691e7ecc96ae7c98dca73ca63a2e3a0ce3125e
99851b71685e812b257e2f97df1a91d04432f99a271819df2f848dbe9fdeefc3d4d3d6ab0f936f4772eedae94c904bb2f18e
b3ac873e997c1e094d2493af790ee492f140ae48947ee180300d65aa4c5df930b9e32d28f42ecbf1f8eabc78167dc8090a21
9f474213c9e46b9e60647cbf10367a8ce77020da96ab8b707bae3644c3683e161579fa78d45b5d857b1a094da4702f790edc
9cf080fd7034f2896fe086a9a7ad171bc24e57cccf654c74fba1cbff1cf63cc2e685b294ce3d94419b11083a90a0d71c0766
c9768056148f1ea9c6423cfaeedad2b50f639f52c09e06d4e5ed2daf09cb4843da7423129a48065ff30423e3fb85b0d1d385
959a47724baf3e4c7ee4b9716e7379ee189bf700f21fca373e8f842692c9d73c0772c9782057248d8e378d0c259b5ef4ea23
e434a088bc0a582a3df3bb1f9ffbdffc15a5423e8d842652c8973c0772c9782057244adf3e614f6118949bbaf261f29d2bfa
535e5c45cd274eb25edfcce0d3406802997bc9118c74ef17bc463ff0ec52df74bfe8d587b17396c7e98f835adca04ef9ba3c
9dda711c224dba11094d2483af790ee492f140ae4894ee3f584365a8d764eaca87c97313cfcb829425dd3e14f8cde67eeda7
3b2fd9cd6604820e64ee254730d2bd5ff01a3d6d69d8b266cbd545a0f94a5b3afbe7cabe144eb1adca6744421329d84b9e03
37273c603f1c8d7c5e745dd0c067e9c586b183dc89490f61ee5d0f9ff815cf8e229e9b10ea692034810cbde438304bb603b4
a2507af8e069a661b79aa92b1fc6f6090fa1043a0de53cbfc8a35678afb7b333f634109a40c65e720423ddfb05afd1b19138
fb49e78b5e7d189b17481c0f29a5aab99375e083dffc08f73c129a48065ff31cc825e3815c91283d7eb094c250a4cad4958f
90e36b3ccfb773e46ef5e4d3ec9425af929433944624349142bee40946c6f70b61a3c780b3dc91dcd2ab0f93bb800b609ece
0439aa92781cfca263c79b5063597f3002410732f792e3802de90ed80aa3d1f32fc60d759a5ef4ea43d8f14a5c8582722c55
bd29472ff5a7187b1e084d2061af390ed892ee80ad30949e3eb4d760e8a32d5df930f649e38bc46fae7d6e88b275e42cb733
f7b657294568444213c9e06b9e60647cbf10367a6e9f8732d1b65c5d84db6db8339d7bd7fcc34f31465ef5def31fcacd924b
857b1a094da4702f790edc9cf080fd7034f295db8fa14ad38b5e6c18fbd0a781b0e1a53312f82f89f6acf8200dba11094d24
63af790ed892f0c0ad4094be7f42a4e37a1da0a52b1f26dfe5720a1a666caa6205eebfe4573c423e8f842692c9d73cc1c8f8
7e216cf4236c69a850f5a2571f264f392ecfa0b8a7e1d7778e77a4a5dc00bafaa0cfe340c531f58adfc02cb90ecc8a41e907
1eca3b870a4da6ae7c9839caad9834c8d8d4092897e43f94e67c1e094d2473af798291f1fd42d8e8f13ab7a10cfc8b5e7d84
dce7d6e688fc62c35db9d9f19ec75ab92f72542bb4904f23a18914f225cf815c321ec81549a39ff81e66e8a96dbdfa307988
72c735657948ed19ca525eef94277d1e094d2493af790ee492f140ae48947e7ef2a070e09babca83a93d1797a256c7a9d330
2eb5d5a3e781d00432f392234c93bd4db046cde2e686a9a6ad8b0b036ff973a7cdb3945f5914a1fcfc15cf734fa5219f4742
13c9c86b9e03b3e43b502b0ea55f38bd8d43652a53573e428e1b4ea2e357b76580c159961370023e0d041d28dc2b8e60a47b
bfe0353a56cd190a53bde8d587b15db8b808c397a3ccdc75a2248b4af67f66c4fa9f61f8d30398243480a9441b9dee4619e6
92b65e7d32d8df7ffdfeeb177fe5bffef90f39ef7ffedabefe25a7777efd7bfeb75fa2a2ffaeb47d7051809a528f03cd5625
bfedebdfe1f7f9ffbaafbf74d92df8c07f7a3bb0d41a3e4c219c788f4bfafaf186bffe017ef1dfb6aff3eb877fc2cfe5eb87
7f845f7dfdf937f7edf3cdfd97affff9f5c3dfc17ffd01fe1e28259dca414b17bce5e549a5aa0ba9ecb9e1c6039a072de1bf
e7b2bde6725e785b2e9f647c72a9ea422e671e2ce1c390c72db9557bcfc5bfe6829712e1c8bf49a5880b99e0cd84b915bac2
71eddbfe9d54c27b2afec03d2cc7d667f3e82b09795c19c0bbcc1c3ebeef09c5f78470341ef6e8fb841e7d25215a95c34bb0
377cc3f39e50d209fd1ed0ee2fd138443c869ef6b0f3b398bf9fdcfa9d87063c3abf9fbefde1dbbf7dfbedb7fbdbaff3ffff
e9db3f7efbfaf64fdffe77feb71f49ffd76fbfd37f9bfffabbfc67cbaff9177ffbd31f7ffbe34ffff0cbbffefaf10f4d83a3
7fd454423ce1a928f5a39e35006f4d50f5f88f364dcafd3700ff3d33fc3ff78bd84d
>}

\EmbeddedPDFObject{application-1}{%
<< /Font \EmbeddedPDFRef{application-2} 0 R /XObject \EmbeddedPDFRef{application-3} 0 R /ExtGState
\EmbeddedPDFRef{application-4} 0 R /Pattern \EmbeddedPDFRef{application-5} 0 R /Shading
\EmbeddedPDFRef{application-6} 0 R /ProcSet [ /PDF /Text /ImageB /ImageC /ImageI ] >>
}

\EmbeddedPDFObject{application-2}{%
<< /F1 \EmbeddedPDFRef{application-7} 0 R /F5 \EmbeddedPDFRef{application-8} 0 R /F2
\EmbeddedPDFRef{application-9} 0 R /F3 \EmbeddedPDFRef{application-10} 0 R /F4
\EmbeddedPDFRef{application-11} 0 R >>
}

\EmbeddedPDFObject{application-3}{%
<< /M0 \EmbeddedPDFRef{application-12} 0 R /M1 \EmbeddedPDFRef{application-13} 0 R /M2
\EmbeddedPDFRef{application-14} 0 R /M3 \EmbeddedPDFRef{application-15} 0 R >>
}

\EmbeddedPDFObject{application-4}{%
<< /A1 << /Type /ExtGState /CA 0 /ca 1 >> /A2 << /Type /ExtGState /CA 1 /ca 1 >> >>
}

\EmbeddedPDFObject{application-5}{%
<<  >>
}

\EmbeddedPDFObject{application-6}{%
<<  >>
}

\EmbeddedPDFObject{application-7}{%
<< /Type /Font /Subtype /Type0 /BaseFont /BXGEVL+DejaVuSerif /Encoding /Identity-H /DescendantFonts
[ \EmbeddedPDFRef{application-16} 0 R ] /ToUnicode \EmbeddedPDFRef{application-17} 0 R >>
}

\EmbeddedPDFObject{application-8}{%
<< /Type /Font /Subtype /Type0 /BaseFont /CGMMQU+Cmex10 /Encoding /Identity-H /DescendantFonts [
\EmbeddedPDFRef{application-18} 0 R ] /ToUnicode \EmbeddedPDFRef{application-19} 0 R >>
}

\EmbeddedPDFObject{application-9}{%
<< /Type /Font /Subtype /Type0 /BaseFont /GYUURQ+Cmmi10 /Encoding /Identity-H /DescendantFonts [
\EmbeddedPDFRef{application-20} 0 R ] /ToUnicode \EmbeddedPDFRef{application-21} 0 R >>
}

\EmbeddedPDFObject{application-10}{%
<< /Type /Font /Subtype /Type0 /BaseFont /CGMMQU+Cmr10 /Encoding /Identity-H /DescendantFonts [
\EmbeddedPDFRef{application-22} 0 R ] /ToUnicode \EmbeddedPDFRef{application-23} 0 R >>
}

\EmbeddedPDFObject{application-11}{%
<< /Type /Font /Subtype /Type0 /BaseFont /CGMMQU+Cmsy10 /Encoding /Identity-H /DescendantFonts [
\EmbeddedPDFRef{application-24} 0 R ] /ToUnicode \EmbeddedPDFRef{application-25} 0 R >>
}

\EmbeddedPDFStream{application-12}{%
/Type /XObject /Subtype /Form /BBox [ -0.61237244 -0.61237244 0.61237244 0.61237244 ] /Filter [
/ASCIIHexDecode /FlateDecode ]
}{%
789c6d90510ac4300844ff3d452e3041633669afd06b14966561efffbb6d29d396f823e2e83cd5d257342db205686e56bc97
f413cdd64a55bf1535bb4dd6db5e7ad53ed5792b5577b5a3ebcc56a10a4ed003f4bd6c8f11322813418f93f0405164fbe8cf
7495808fd104110be34a18f7c6e32e048723f80fc24722f83842126e277e44def2076f985899
>}

\EmbeddedPDFStream{application-13}{%
/Type /XObject /Subtype /Form /BBox [ -0.61237244 -0.61237244 0.61237244 0.61237244 ] /Filter [
/ASCIIHexDecode /FlateDecode ]
}{%
789c6d90510ac4300844ff3d452e3041633669afd06b14966561efffbb6d29d396f823e2e83cd5d257342db205686e56bc97
f413cdd64a55bf1535bb4dd6db5e7ad53ed5792b5577b5a3ebcc56a10a4ed003f4bd6c8f11322813418f93f0405164fbe8cf
7495808fd104110be34a18f7c6e32e048723f80fc24722f83842126e277e44def2076f985899
>}

\EmbeddedPDFStream{application-14}{%
/Type /XObject /Subtype /Form /BBox [ -0.8660254 -0.8660254 0.8660254 0.8660254 ] /Filter [
/ASCIIHexDecode /FlateDecode ]
}{%
789c6d90d109c3300c44ff35851738a328ae54afd03502a504b2ff6f9312ae09d68f1127eb9e4e535945cb4bf6075a9fee6a
8fb28956b3ee315f44adadf5eefd90225a84ef924f36871dd2592dc22e38410fd0f76ffb1b21836d22e871126e2836f97df4
67b948c2c768828c8571258c7be3960b497024f7417a482417474ac225e247e42d5fc11e5a07
>}

\EmbeddedPDFStream{application-15}{%
/Type /XObject /Subtype /Form /BBox [ -0.8660254 -0.8660254 0.8660254 0.8660254 ] /Filter [
/ASCIIHexDecode /FlateDecode ]
}{%
789c6d90d109c3300c44ff35851738a328ae54afd03502a504b2ff6f9312ae09d68f1127eb9e4e535945cb4bf6075a9fee6a
8fb28956b3ee315f44adadf5eefd90225a84ef924f36871dd2592dc22e38410fd0f76ffb1b21836d22e871126e2836f97df4
67b948c2c768828c8571258c7be3960b497024f7417a482417474ac225e247e42d5fc11e5a07
>}

\EmbeddedPDFObject{application-16}{%
<< /Type /Font /Subtype /CIDFontType2 /BaseFont /BXGEVL+DejaVuSerif /CIDSystemInfo << /Registry
<41646f6265> /Ordering <4964656e74697479> /Supplement 0 >> /FontDescriptor
\EmbeddedPDFRef{application-26} 0 R /W \EmbeddedPDFRef{application-27} 0 R /CIDToGIDMap
\EmbeddedPDFRef{application-28} 0 R >>
}

\EmbeddedPDFStream{application-17}{%
/Filter [ /ASCIIHexDecode /FlateDecode ]
}{%
789c5d53cb6e833010bcf3153eb6878840c0241242aad20b873e54da539403d84b84540c32e4c0dfd7f61822152919cdecce
ee064dc273f95aaa6e66e1a71e4445336b3b25354dc35d0b620ddd3a154431939d983d73dfa2afc72034e66a9966ea4bd50e
419eb3f0cb14a7592fece9450e0d3d078cb1f0434bd29dbab1a79f7305a9ba8fe32ff5a466b60f8a82496acdb8b77a7caf7b
62a133ef4a69eaddbcec8cedd1f1bd8cc462c7239c240649d3580bd2b5ba5190efcd53b0bc354f119092ffea51025bd36efd
b1ed075c8057271f219fbcbc515405a8f055e165824c5e26c807ac381c21af3402c480032001a4000ec856bf1b97a0354930
ee4151c569893f2df1a7a5189c7a53eabb398ee0fe67ae1437714ce6b072dcc47113c74dfcb8fa310ecb33ff323de51280b7
c3dbb5c75932ecca382c2bc5ea0cabb374ed81e5045a7bcb465db541b3850bd0c8f13eb2130117e0d506644d82cd8a0df616
4471d7da64d0a5df85cfc6ae53b4fd41c661b42efbf903821bdac3
>}

\EmbeddedPDFObject{application-18}{%
<< /Type /Font /Subtype /CIDFontType2 /BaseFont /CGMMQU+Cmex10 /CIDSystemInfo << /Registry
<41646f6265> /Ordering <4964656e74697479> /Supplement 0 >> /FontDescriptor
\EmbeddedPDFRef{application-29} 0 R /W \EmbeddedPDFRef{application-30} 0 R /CIDToGIDMap
\EmbeddedPDFRef{application-31} 0 R >>
}

\EmbeddedPDFStream{application-19}{%
/Filter [ /ASCIIHexDecode /FlateDecode ]
}{%
789c5d90c16ac3300c86ef7e0a1ddb43715b28ec1002a3bde4b06e34ed69ece0d872302cb2519c43de7eb6d36530812d64fd
9ff82d796e2e0db908f283bd6e3182756418473fb146e8b077240e47304ec767556e3da8206482db798c383464bda82a90b7
d41c23cfb07935bec3ad0000f9ce06d9510f9bc7b95d9eda29846f1c9022ec455d83419bc6bda9705503822cf0ae31a9efe2
bc4bd89fe23e078463a90f8b25ed0d8e416964453d8a6a9fa286caa6a80592f9d77f529d5de5a797245fd2e792bf32f7abc8
23f27f577f7a624ed6ca528aa7ecc611ae7b0b3e642a9f1f6f9a736b
>}

\EmbeddedPDFObject{application-20}{%
<< /Type /Font /Subtype /CIDFontType2 /BaseFont /GYUURQ+Cmmi10 /CIDSystemInfo << /Registry
<41646f6265> /Ordering <4964656e74697479> /Supplement 0 >> /FontDescriptor
\EmbeddedPDFRef{application-32} 0 R /W \EmbeddedPDFRef{application-33} 0 R /CIDToGIDMap
\EmbeddedPDFRef{application-34} 0 R >>
}

\EmbeddedPDFStream{application-21}{%
/Filter [ /ASCIIHexDecode /FlateDecode ]
}{%
789c5d503d6bc33010ddf52b6e4c87a0244d8682318474f1d026d4ed6432c8d2c9086a49c8f2e07f1f9dd4bad003e971efee
dd17bf34af8d3511f82d38d962046dac0a38b93948841e0763d9fe00cac8f8e3e55f8ec2339ec4ed32451c1bab1dab2ae01f
2938c5b0c0e6ac5c8f4f0c00f835280cc60eb0f9bab4856a67efbf71441b61c7ea1a14ea54ee4df8773122f02cde362ac54d
5cb649f697f1b9788443f6f76524e9144e5e480cc20ec8aa5db21a2a9dac6668d5bff87351f57a4d3faa945ea02b7827fa74
ca34415730d3fd4ba609ba8277eaf25b8f1ad275d66de41c425a249f306f40b31b8beb95bdf3a4a2f700804a7e1d
>}

\EmbeddedPDFObject{application-22}{%
<< /Type /Font /Subtype /CIDFontType2 /BaseFont /CGMMQU+Cmr10 /CIDSystemInfo << /Registry
<41646f6265> /Ordering <4964656e74697479> /Supplement 0 >> /FontDescriptor
\EmbeddedPDFRef{application-35} 0 R /W \EmbeddedPDFRef{application-36} 0 R /CIDToGIDMap
\EmbeddedPDFRef{application-37} 0 R >>
}

\EmbeddedPDFStream{application-23}{%
/Filter [ /ASCIIHexDecode /FlateDecode ]
}{%
789c5d503d6fc32010ddf915372643441ca79b8554a58b872455dc4e51060c87855403c278f0bf2f1fa9abf624383ddebd3b
ded153fbd61a1d80be7b2b3a0ca0b4911e273b7b81d0e3a00da90e20b5084f946f317247681477cb14706c8db2a46980de22
3905bfc0e655da1eb70400e8d54bf4da0cb0f93c75e5a99b9dfbc2114d803d610c24aad8eeccdd858f08348b77ad8cbc0ecb
2eca7e2b3e168770c8b82a5f1256e2e4b840cfcd80a4d9c760d0a8188ca091fff8baa87ab596d7a93ca68ac1fd0f7c64b62e
f0f8645798d917cc30a57bc98f34f3a77b1a9f76b57a13b3f7d1565e68f6939c6883ebce9d754995ce3763758030
>}

\EmbeddedPDFObject{application-24}{%
<< /Type /Font /Subtype /CIDFontType2 /BaseFont /CGMMQU+Cmsy10 /CIDSystemInfo << /Registry
<41646f6265> /Ordering <4964656e74697479> /Supplement 0 >> /FontDescriptor
\EmbeddedPDFRef{application-38} 0 R /W \EmbeddedPDFRef{application-39} 0 R /CIDToGIDMap
\EmbeddedPDFRef{application-40} 0 R >>
}

\EmbeddedPDFStream{application-25}{%
/Filter [ /ASCIIHexDecode /FlateDecode ]
}{%
789c5d503d6bc33010ddf52b6e4c87a038359d8c21a48b877e10b753c8204b2723a82521cb83ff7d4e52ea420fa4c7bdbb77
5ffcdcbd76d644e09fc1c91e23686355c0d92d41220c381acbaa232823e3c3cbbf9c84679cc4fd3a479c3aab1d6b1ae0170a
ce31acb03b2937e0130300fe1114066347d87d9ffb42f58bf73f38a18d70606d0b0a35957b13fe5d4c083c8bf79da2b889eb
9e647f195fab473866bf2a2349a770f64262107644d61cc85a683459cbd0aa7ff1e7a21af496fe2228bdc0b5e02dd1a2ca74
826bc142d785ae1f744d3475f9ad971aa6eb6cdbc825045a249f306f90663716b72b7be7932abd3bab6a7e80
>}

\EmbeddedPDFObject{application-26}{%
<< /Type /FontDescriptor /FontName /BXGEVL+DejaVuSerif /Flags 32 /FontBBox [ -770 -347 2106 1110 ]
/Ascent 929 /Descent -236 /CapHeight 0 /XHeight 0 /ItalicAngle 0 /StemV 0 /FontFile2
\EmbeddedPDFRef{application-41} 0 R /MaxWidth 948 >>
}

\EmbeddedPDFObject{application-27}{%
[ 32 [ 318 ] 40 [ 390 390 ] 44 [ 318 ] 46 [ 318 ] 48 [ 636 636 636 636 636 636 636 636 636 ] 67 [
765 802 ] 76 [ 664 ] 84 [ 667 ] 97 [ 596 640 560 640 592 370 640 644 320 ] 108 [ 320 948 644 602 640
] 114 [ 478 513 402 644 565 ] 121 [ 565 527 ] 181 [ 650 ] 8211 [ 500 ] ]
}

\EmbeddedPDFStream{application-28}{%
/Filter [ /ASCIIHexDecode /FlateDecode ]
}{%
789cedcfe96a42410c06d00fb4d6adadad5bb575af4bf5fddfcfcbfd79a9205410e41c089924909949fea956a9eb792a73a3
88e7229a69a59d4eba79c96bded2fb73cb7b3e2a9d7ea51e5c7cc130a38cf39949a6f9ca776665779e459659655d9c37f9c9
36bbeccbc921bf57fdecd68e77b91500000000000000000000000000008047723a030b8203b0
>}

\EmbeddedPDFObject{application-29}{%
<< /Type /FontDescriptor /FontName /CGMMQU+Cmex10 /Flags 32 /FontBBox [ -24 -2960 1455 772 ] /Ascent
772 /Descent -2970 /CapHeight 0 /XHeight 0 /ItalicAngle 0 /StemV 0 /FontFile2
\EmbeddedPDFRef{application-42} 0 R /MaxWidth 1444 >>
}

\EmbeddedPDFObject{application-30}{%
[ 88 [ 1444 ] ]
}

\EmbeddedPDFStream{application-31}{%
/Filter [ /ASCIIHexDecode /FlateDecode ]
}{%
789c6360186280050000b60005
>}

\EmbeddedPDFObject{application-32}{%
<< /Type /FontDescriptor /FontName /GYUURQ+Cmmi10 /Flags 32 /FontBBox [ -35 -250 1048 750 ] /Ascent
706 /Descent -215 /CapHeight 706 /XHeight 443 /ItalicAngle 0 /StemV 0 /FontFile2
\EmbeddedPDFRef{application-43} 0 R /MaxWidth 970 >>
}

\EmbeddedPDFObject{application-33}{%
[ 77 [ 970 ] 85 [ 682 ] 185 [ 602 ] ]
}

\EmbeddedPDFStream{application-34}{%
/Filter [ /ASCIIHexDecode /FlateDecode ]
}{%
789c636018b480158dcf3620aea03e6001000a6b0010
>}

\EmbeddedPDFObject{application-35}{%
<< /Type /FontDescriptor /FontName /CGMMQU+Cmr10 /Flags 32 /FontBBox [ -44 -250 1009 750 ] /Ascent
706 /Descent -216 /CapHeight 706 /XHeight 453 /ItalicAngle 0 /StemV 0 /FontFile2
\EmbeddedPDFRef{application-44} 0 R /MaxWidth 500 >>
}

\EmbeddedPDFObject{application-36}{%
[ 48 [ 500 500 ] 51 [ 500 500 ] 94 [ 500 ] ]
}

\EmbeddedPDFStream{application-37}{%
/Filter [ /ASCIIHexDecode /FlateDecode ]
}{%
789c6360a0316061600392ec0cac34309b03000866001f
>}

\EmbeddedPDFObject{application-38}{%
<< /Type /FontDescriptor /FontName /CGMMQU+Cmsy10 /Flags 32 /FontBBox [ -29 -960 1124 779 ] /Ascent
707 /Descent -970 /CapHeight 707 /XHeight 0 /ItalicAngle 0 /StemV 0 /FontFile2
\EmbeddedPDFRef{application-45} 0 R /MaxWidth 777 >>
}

\EmbeddedPDFObject{application-39}{%
[ 106 [ 277 ] 161 [ 777 ] 164 [ 500 ] ]
}

\EmbeddedPDFStream{application-40}{%
/Filter [ /ASCIIHexDecode /FlateDecode ]
}{%
789c63601886808daeb6b182491600042f0010
>}

\EmbeddedPDFStream{application-41}{%
/Length1 13816 /Filter [ /ASCIIHexDecode /FlateDecode ]
}{%
789ccd7a7b7c14d5f9f773e6ccecfd327bcb269bdb6e36c992b0c92e1b4288041cc2fd2252400ad148800448022491fb2206
b5dc84106e864ba344a5d8a285401123a08046ad5a8bad2852b55a8917fcc5d4be524a3199bccf99dd4014dbdaf7fde37d77
3e67ce65ce9cf33ccff93ecf799eb3030400ac78e3c13d6af888913006e602103fb6268c9a78fb64fe6fff830fc96800eee0
a8c977147ec8fd633500bd8acfbfb86de894d1499ae917f0e58fb1cfa4db270742957f17330154fdf1f9d4d90b6656171606
96627d133ebf6bf6d2c5eefa9ca72e03a8bfc0fa8e39d573177cbefa6fad00da7bb1ffc6b9331755831a2fd069b06e983b7f
c59ca17f1c5486753b404a785ed9cc52c39f5fbc07c0bf059f0f98870d56b7fe13acff16eba9f3162c5efe6d86b308ebed58
ff607ed5ec99ef6efde82440167b5ebd60e6f26afe612107207b20d6dd0b672e288b7f3d5785f56948cf9ceaaa458bc7effe
603f40e0343e7faffa9eb2ea252b54570082bb0084bb81c9ca00911f8717857ed81682015866cf74108421c00f1f397e0ab8
66afb8673eb8e6de535609aef933172f0417ca187fdddd4a5f525976cf42607c2a357cc629b906387290f524af6083169cf8
c6b3ddcfc28ffefd77bda3ef4cff0fcf3f88e4f2dbf2db98a577cfec9e89ad33af3f574add95ff7794f4ccf2a3fa267727ff
17bd8dddc61fdf5bfe5afefa477726b8ea16b0820decb8c21ef0820ffa2026385c3b2b5e049fd8b066c78b620f86020f5e02
f6f4820a7bfb10f17df0d2e05b217c4ab0358077ee0766a3ca9dc7b751a9f03d0dce82fa8234e8119906308209cc205ea7c8
71d3085a88c10ba278ad864d202878752b78752b78752b784d5d3073f13c988534a182807a68d108370c1836690ade470d1d
85f771438bd87dd238bc4f19cadaa728e5eb188fe44a09b8f9e573678249b9b3f939e8413e2bab14ce23f827305ac9292611
fb0e50e49604fdb15f92225957afbe09119dc1114c90abf44c841cac272a33bba27c7e0c5fc044283080fa4385980750883e
d4e977224261e5debf9e3a99172ddff1238060c27197fee77e5014e9cbf2ebe55e637ca7bda8577df68d3287f6ebdb7a2c17
5c7fb50ff21fa348d505850a7a184a18f77cb407c555c47eda2578cf6288d3d6e1733f9640dbf09d7efdaef70b5eef17b8de
0f49e15d640bc3a09023ecc66a5224a7e7610e6745ea542a2aa8048ee357f72036fa9b3867442948e076db5476d94ef6a817
908b00518444e68ea40460ab06b00f6b44a9f3b01ef3244404454edd90819467c340180ee3610a9423929742189e71db1073
68db915af63c08c394e733613edc032b22cfbb2f767fd0fda7ee0bddef75ffbefbd5eed3dd47bb8f741fee3ed4bde4e3d85e
b4fcf02f62ff5ba2359332572451651500298bc8d21f4db8bf28926749adc81480ed3fccf60fc3341c9305d3f868b2619a12
4db8ef01b3ade598982ecfc754adc801901f0086372fa61598c2d1f99f8926dc87ddb64882be44078b10132df0065e6f6369
3c54400dac83262c7f0a2b95f633bc865dd87241607ddf86169281fdd85ec79100304d7d01c7c9c1679f62ff39586f420bc1
9e5fa36f2ad71efa26b70c383a114b1395379aa0850ee279fa6624296fbd0183e1381c6065e14d60a89a04e7f12ac4d1c7c2
49b8401e847de41d580575b048016622d1a1b656e0354b7847b9be86bb9499595b85f08eca8e3355209f2771f47d91769241
26d1123a874c430e3972808ec6d63550c197e0e553aee10a7f111e386e25ce1fe517dee4eee232781f3980f3b039dec4f10f
c060a4770e523a1a13c7e8a797a0893254c4096760ac7aacca4054ea55b87a1caca439648f2a114a60159d82234cc0b63a98
40cee32cb8b4c754024f395441b7d8cca58d296d967e32cdfddbe99e2cfff7aa6e51ed6e8689cdc615ee96eeee89d3f87861
7ab390d04cd334cd7c9af72fffeae15fb2fce3264e7337bf36627874d41125c3b16df2342cb21a3663fb88e159a0ac14ae3c
6a0fd3b8c7e5335c1d4ad4080992516d04558d49053a0be54ce2e5ced6503f103b7f979f1f24e99c45b4e639542c737275df
5cbdfacd9eb56bf7c86716900b643629251716ccaa906be527e47d726d05d3f4ad38f22a65e478c9a046db5d63d25bd4c09b
c473ada1f69e816db903ac1691f3e5c6b06ceb9e75ebf65cbe7a553e33ab82dc4fee22c5e4fe8a590b649fdc283f22fb70d4
b1f225f2325c440e1224036e1cbc56038215a816476db7e6e3b0e7d8b0ce216430c93111f5d8bba7bdacd9baedd51127d6a4
3df5f7221ce14cf745f20674e03e3af0a81155d6209eeb6c1dd76cfbc9b4a33a940c5f502016144cef0747e94c6d8941bcdc
1e792610aa57479e05f3fa0fc809c538ec2a6f4afa99f1f9791326e4e58f9f3770c2848179136e63f25dd77d91af47dcead1
abf34a362833a8caace5869258aa72d8b52ab33e16c7bdcc84d076b95dec08a68518f7de1445ba2960112127c4eedcc6ddbb
76ed66e91bf9afc4f2cd6522ca5f8f2413c96432854c949be5a7e55fcbcdb3c92eb29054915df23c79abbc4d9e875c5e4333
751929d0c17229812f452b4d4b55c26d54c5134155ae218b40abb6a02baa47b9b576b6f683407be872ab25df921f44d4a8f9
afd422ffd5f41432ae593765daf3c8888483ea41df7d7aa0256fe07449eb06370968033abe18fbe8953edda7237d4052fa78
723d1621372dc7e271102fe1e5b7c9fc87085fc1ffe9a507dbaf85158c6c44033a01a94c80db248fc10e66232ce64a5d6542
b97199c56e343841479c2a8d492712ab3e51bcdc7aaefd1ccacb929f8f2b1de8b8d221778848747e50d28fd48dd48f348cb4
8f7408881aaf6508c909f10e3baa28f1e4220db96408c9ed9fee4d51115575e58579978849fe7dac415ec7552de91c4c5e6e
58fdc843f58d7c0dd1e5177cf2e2fbf259273e6b91f72d20ad0ddf3eb4f3c066466f23ae6b2dd29b89f4f6d5269625db165b
cb9c65c9e68cc550965e46cbcd257ddd49da446f860ebc2a954b65b0ebc0caf545cadb6f909e1f59f7ce76f115c640076a01
a396014a9d441c76de9392eeb3e42461db0024b92fc98d149076b5f7d525e7ef7effca81c69637e4cfe5ab4b3e5a58ad3bf8
e8eec3c79a67840e1ddab26eddc3c66aa1c19ffdfc2fd73de3769d7fea0f1f857248ca961d87f7343f3367ebb60737aebe0f
f1792bcabd19f9e0703fda2ba571a55a9d9694ea74dadb88ce008b4c6a5158a42ba79c865781c548cd7a0d67152f9f2b40a8
14a0bc51f66d9d05e70ada4351b8880c2f02e2e548404b48f131b7d66d749b68041ae39a0d53ee1cd76c6437f3943b112956
0529cc698e20458ad513171948c6105ec5c5707d7403b881ba51dc689dba188a8945dbb3820c4a1e4b0ef786ccd1635dd3be
ee5acc6d7e8524ca17bf26d7a8abf333595541d216d0cb9dba4af97db65e0771bd56e06e920cbf96d2a82a71b9ab4c252e36
97d9cbd0782f8359861277a285d3258153a54db2e8cc9c1bf51241d6ca162aaa9dca1ac96d563482e39ab5531847d719712b
8ca02a4419c11e56ec618b3c5474c6cd3a45b84c3753336f16cc2ab3daac31c7995de678738239d19c64308399244332eebc
01a2623c239b0c12e0b0c377c1a060c0c2abbaa6f4df38f57f9158f9d36e98fb5e4d8df689ed8f1ef8d5dea20db9538577f6
c9535253e5cb5f7c25ff83adfee686b75e7af1ac348ceb60b66903caa441b14d69305b4a3180aacc0c658ee4f9492bbc65f1
e58e9274b72e2986aaac0e834a2bc4ead35124ed37ac15a2b6e35d944e504ad31ab4460b2f0aa24a545b343683cd289a44b3
285a2c19860ce384e409ee091e6b715a0fdd0e7b0ced61caab18394f28c6e9f06613347cdca375bb76d5d537ec7a6ef8d3f3
3f2254be74a94bfe9a1492b4db0fde59637860e18b2f8c3cfbe28b677ffbfcabef725f8f1b8bdc5d4493b882dc4dd6938349
4957ca2ae43fa1b89f46643f882bae817ba444ca76388e1202b7a1f95ea69e457017a6c0eb40c569994e5a9c0ccb97db7b60
2c4461acac342ea156595f2d5ebdd697b52b4bab65cf95f6a34e124bb9622dda398fc36bf1f00f76b53ccc65754d5842cf0b
55d7b60afd5ac85db8df32d9d745f785000c915252636999435b96642ecb2c4f2a09a6c61aac82cf2064812adea152f93cfa
2023f3728fe9e8d930deed607b728f541119b457396243bc29a9b9fd07e465a3d18b6e56f4f8de2d5bf6b294f1d0bd2b376e
5c79ef43f30fbc205fb9fa0ff9ef2f3cb5f3b1ab172e5c7d8cec6a3a7cb8e9b1c387e9d2f5bb76addfb06bd76bce37f7fef1
e2c53fee7dd3197f68fdd1b3678fae3fc4746b51f745611ff2920803a578d742a062adb9d4bed0504e4b922c66737aa20f9c
266d5a92c8a8ef859e573a8247c5e4aa640ecdb40ac94cf7a573b90c0bbcd312c1026e82c8469eb0af507ead63dbb599d50b
c83632264cec7b6bd23e7aedaf57aefcf5c9f78264e31fce97ceac268947d14d784a1a2afff2e411dc0c9f955be45fef7d14
715e8351cb6505090e182cc519d455a6593aa8d2aa89c38c5e88d5a84d058bde1383f4b585ceb5470d33db06db4341491b74
deee5ceddcebe449715a8e5b41ab170db1c59beb89d4c8b5e3b8dc79f2ebf207c7e5652fbcc04d220f9043f244747fc87832
00afd1d71a2a04a89497ca8fc94df2629419a329196932425072e9b56a5a25ccd2ae2655a04ad5e8cc061e523516ce638a92
24b6b14d0e77b990d81acc89cccdd085a51a9c6fd40bdc81956c82ca6f2dd7e48d6c4d308a100a717c33ac94e2b5829e56a9
671971789d59c39b55a6548dc102a92a4ec44dbfadb3b54dc4cd874d11ba2c17b486823da81715748b2046d17d24d9427a99
72d643d10b91f58c98367dc05262a9b734594e598462c55a334219c1961c0759774a1eb352c85e298f3955c15fbe444ae4a6
4bdf1a2a98359a831ab1117d731d4634f74981788f3e0e440309c76dd7d4db3c9bbd7549f5e99b6c2586b0898b13c0939562
f4bbcce0d30bae9498cc8cc8be1a0a2127166bc42768933bae7c26a2d070412df968b5a55431b32a73799fd3997cb8cf6972
9a3b4d5bf956a155d5ea688d69759e8e3d1d77da652c460e73afab900da9f6e49a108ce94c7d3c3d863757d9a769e0ab0d6b
d72f5e54d7b24fde37b8a9f2f05b1d1708b77dcdc207c41925cf4cfdf07332e12f4bc3550f6c2627bbfe58b168ecc8934dbf
383666c59ad259e73332de675c0f42ae5b54eca4c803a3a4348f60715048daa189dd6122615a03615389669335dea5cfb571
7e8785f741300541818c86982605dadb143b8c5c3bd1101ca9f5e2a69ba7489b116875d839a657c80251d889dadff6c3a47c
c0932b5baf5e6d5df9e400792729dfbea96eebd6ba4ddb2b04a9a26b41e90cf9f56bd7e4d76694eeaf204b4ebff6c9071f7e
fcfa29862bb64a89b84a2ec8955cb0db594f8d3b0d0f5beab59b6849bc29cbe002bb4b9519cf54fd77e71455ef6404ca6dc1
235509481b61a2845ce6d6c644b53c2ae93c3ed1f7e7e62fe4cf88eb8bb2df96553aea1e7c78d7ce86659b5de4d60347488e
0cdda45f20287fbd69f5e79f7df6c5bdaba298b91895deed527a0c2fa2f4e276189377a82d611b4a4f5d62dcc459fd0e91fa
881f72135cba4c26becbe8b414b4dd243e490b5ed15beb6df62aba1e5d782eb7bf35426dae87e6f632af7c4eb3bc8b89f19f
575f46311e24735ab66fdabc75eb661463c5fed21924efda35326046299751f1ed1be5a75efff8c30f3e792d4af55494a11d
e2618294c4691c61d0ec16ea0db1f5f073eb2643498246eb70413c71993313d862a3c3f8bb88289924c50e8665ab3b514aac
4da467e12c39cb9d759c8d1718723d5161e644e4dc9f850f6a45ca6a7e6ae7e78f3fbae2e7c93bfc5f9db9247f4a623ef91b
e1cea9b6aeafdd2662bc73fefdda55879a51d09da4bf7cfef0f1e75f788eadf8588c1f6a50c6095023a5ab63c0aa0b63c8c5
d7c7d58886b05dd368d21a491618f56a1f0cb09911a34174cf0bda4228e2a8f38466a5232477bcfb8ad811b23294d62711c5
8c38988d49542c48225e110bf26c099410eed624c2bc9f08401289e2ed393c8ae5cb89ec634e5a33934cbe7dee1492219f6f
ef7abab9b9f9b9df0c5b5b28c48d9b50b3754365e7c18a0a3aa972cd9163b1718a7d91f3f98b28f564f0c31c29c78f8e964d
4c70d30c082766ecb0691e551bea79ef8e38b431fc23898d59da2c7592df9d60f6416eba4b9d179399255e6e6bbf1cb131b8
1aeded6d0c3a7ffe2c0a1fe4ccaa40285bccaecd6ece4608118b1218fa72234edb60d21b4e68c17befdbb97c4ee7a5f2d3d3
4aac6b573dbef306aac83c79570faae8c9ce3b3ffd677afa47f34a9e6dadbf01b0fd95f2a628bc4e47568cefc0154b842ac9
efb2ab8dc4b94647c3424dbcd5444958b74904cb605b825f65f6db7cc65b78bf4be5033f174c620c86706f6e13db941da7a0
a3c319eaec88c62ac16381e4dae4e6647a63ed9294b54bc22bb276d1adc9120962a22b1661122b53f6379379394fcc6fdebf
4fe6f7ce08af5edba4989ba9f317befe6ed72b15151cbc547ffc88ec52f603e4822b112ea287e4803192936aad617b2342ce
045465f41b7c88354b10f7ecb65067db39c514863a514f105dce5ee88a512864e7df110a6dbd02765b0fa23caf0e0b640f2b
0c0486fdcfc1ae2f9b914aa1267bf8f0ec4061e1b77d1414e11883909e492855f4f824ab2aaca9e121cca151f1a9a95f08ea
228444c94022f4bd88d02944e8f08a1091767d62b2ef60d7a5c3688d2b2a9828a2faa6be1b711a845f4943dd2e9b9157f505
aa4e14d7505f38a331cd136bd70bea7036dde48dd1429c4d37d890e2776b5022aa5b9d83e37c0ebfedd63efe4cb72fd59f1e
ec175dd2b6ce9e65b5b0a391404101eaa433c462ce1062d7c9e2e65b43334255a12da1e6907083f47e0ae9fdf08aeeefb680
2ea00f3803b1677567f5679d67634f87b4c5e48717fca60a735b94ae7c5aee2f163eb96fbfcc359554adbcbf49d644734ec1
c8a3a5f7eea7d3962c3bfb070516694f3ca4c0a282f3ef5d1729d129e5cf5754afec413bca2b11aaa56cb5d1eee26914ef8d
ffe778b73aff0bb83b7e0cdc4b5630b823949eef85f6ad277e13453bdb132a900b160ff4956c506f50d5b3ad2096ba1cc3b5
b811dc382762562778d41d5715878e73ef98cad963f031ace2e6edd8b973c78e9dbbb6b31d35eed22512277f3ef2cbb6b64b
97dadabe9c8d565e2621f9f7b22cbfc9bc10792adf82b31bc00b77487e0296446ae477b88c61734a52d8addb61dfe42a713f
606e4c4df683da85b6cbe44f14c1670ca65e774698047b8c62e8fa967ab436ad398d39f88e5efe87af47683dfe494474645f
8f17d2f21dff84a90837ba3cea863c53b9ff867b22c7575646f7a8b791fe5878584ad1686d020173d80242cc26a3366c69d4
eb345c16887eabc667f09b8271917547b7ba5571ab43f22bb83589af28675ed240746689c889546401ad438c119d626cc025
8144244ea2122f09c31d528ce494621d2ca8649dddc4cda98a6f00254e014a1c5ed168b1c7d3f9be1ed0b72b8409e3d66f7d
a7eb4d86960107c259b3f90cf9caed134e36f7c0bcaad2148db2e811e431032e4843711b49f50c31e89392b944f710bd417f
9bdee0f6f031e4316812ea7d318f399a5c881f5f49a65b6ff0246a9c598ec421668dcf313295cb32c3884cc55f6e8bb83ed6
9e93a8ce7616525e612b1789824d5f6151addca7a71c71f7459b264df325fbdc3e8f2fc5e72dd215e98b0c45c6a2e4227791
a728659a776df25af75acfda94b5de86e4067783a721a5c1bb3f79bf7bbf677fca7e6fa2d4b7b6ef21fb21c7a19843ce43b1
87e24ed94f394ec59c729e8a3d1567c2fd3e2ff7faf14680f404acd1830186f064c2225a66bb1755f0750fddfbc08cdd6b36
ac7cf7f929c76757f0eb972c5e56b46edbce359fbf31b775d0d5654b8ba78f9cda2fcbffb339dbf6fb33bf98bb78d2a4613f
c9cc0ad455edfcb59f49147d6fea136a50e3e649894e1d88c67a3baf81b08a6bb497c4aa450bf16b051f8500ea5e0103ccef
2252c272ab282ba78d849d66e182c72a81522cc4f69cf03804b5a071a81d9a7475ba2610571b571fb7374e576c43f7c51239
bceae5a7a5e77ef8f8e3cf9122797fff91a30bb98d9acd5bf7fe8c9eac2463e563955dc5ebc6df3163ebdaba5f319af3d04e
bc8f51401c2c91529c1695c168e7d668635151a1515bc283df8914fbc92d26bfe833045c2c7e64ce6e5be4a0ae3dc4ecbdb2
99f79cd5ba14ca5de0ea39c0a88d6f8ee7ae23d9a52099fd871d45b2a75740a4ecf511bdb678b977c8b483f29e9cc72b0f3f
b99fcc50b677ded739ace26445d56be7b9fccaae875fdc72f237e42290ee6b68b36b910b13b448f93a2da7d96b3419d57b55
6a136643346af67184c0a986b0f31a3518340233e3b4114c46ad46ade20968cd7ace2c9efb5de4d8061973866e9cdc38436a
51f35524b11a4257a363d01dad52c7a87d6ab77ab27ab67aa95aadd168b4f17cbce0d03a746990c9670a99da4c5da6692004
c5d130ca38caf45398669a0bf34c4bf9a5c232d572cd72ed72dd72630242d54394185c1bc908b926af58412e1d207d49dfbd
a47e99bc02d7afb22b8d7bbfb2b27338309ed1ff643c27102a4d277b396236c527702ed86b369bf43a4ebbd7643669f6aa35
58d50cd16a08470415a71ec271e4364d2267760117e3409d366a55940bebf946eb23899cc3a44381a0b5d3980da8e2069a88
f13c3bc170e62ba796ed8a644474bd996d4345ff5722ea51750d3bb9559b98b08a9ce274524e96135e6d76881942860a45a3
f7997ce674711a99ccdd214c14ee504d544d564fd54dd5cf23a5e8ad95a84ad54b611959cedd47efe3170b8b552bb4cb0dcb
8d2b4c4bcc4bc4a596a5aea5f14b13faa0f884bcefca4f01548fa34c8e8f94177d479a71a170556076e86e557e41deb0970e
d3dff7c8f6db29f7de299adf183e6830d38e49b80fb4a38c8d5021a56887a00cb9211aade636bd96e7c2aa466d18b49c8e66
09ba2cb54100f6ffd3b9d69ef3b440872cb78aad371dffdd38eb3029ba60c22ba20b47cbb9395aae38273772d4e7500ea727
717bbafe5641e774cde10a2ba7d2392dc43bb573f719f91dc5e2e03e7b09e94b865a2927d1a972d90c3ce75aa3b5a0fe26d9
c20e68e4c2e607508f45b7dde437803f31a2cd767f8cc1670db233e950449f7b1cb9b6f60eb980edb721c5902bb63c7844f4
1072fddf98ef9f3e1fa9f5b040f1861efb6eec4a37b4f9a2a2cdfdb717e5dfed631bf07e72b7e2c2a04a0f9d872a9d9a2a37
5456f2dfde506b9ca4887c48f773b3d8fff54739403d0dfc5ef91f2e68438b5744c7771ee566adc17ecd5d9d74a27a1b7a3b
80eb2fdaacb61ca09c852c7aa9e9f8fe26b9b3a8b6abf34bba937cc20509edfa465edbd5d1f955e43db248bd81fdef6e437a
9b5f7a49bde1ca223673b36cc7113728233ad8805e0e07cc258b6ab7bfd1f4d69626f5862f3b7f21ff54b6cb656404e9e0d2
69e84bf81e25b922e5488ed566b538b8bada22b9b369fff12646888db392159c49eeea7a4b4eea2cff120857a7bcb741f96a
cbe2b504e9448514e5bb89deb4e4387050ea658372752f356d79abe98dedb58c9837bbde97adf209f273d24e7e45ef543cff
ee6e55264618b9b0547267f54bcb4aede70ff85207f4f72780da9e65a3c4e5b30dd00a7ebd3a38a0578c2d9e2b68bb116d4b
9ebe892e039f1a209a707caacd6bede3ac779b6b82b91c5587f5f1fec61c011ddd90e2fe63fe6ec12b0c3d21e6f192b49b82
6cc7f508fca6869e909c1eed9223f177cbd1c2b58525dca48973a6705caf0ae923bfc77ad013a47754feba12a8ffec465189
d9a33190e04349a4c02aa99f3a0162c5b00652f8fae49a1874e9756197a6d16e60a70e83cd36bfdaecd3fb411be7f05bd4be
7867d01bd9ad43e75ad1bf61ae7d87e2da47e29deb871092564c0da456a536a5f237fc36afa2e1ec2bade86ee7eb61362a98
9b4f21f8c0dd34bc72f766ee55a6255d7f5302c8e62f3ffde9b1e9828a4cbd63f3ee52dc0391e728e7252fffd615dfc3639c
12a52f91b20df11aa7396cd484a1c66161063eced868e3b5601206ab9905b0fa4da2c14762ed7e570c8b5b70edd91f70c89d
dc16b15ebd023ac6de31313990dcf4ef8297eb3cfde0010bf721a93bd475a9f9fc136b894fbe10e1eaaf2afbb5a92c507de0
91d29ef39592d75a41f97e0b343ae466106c92d202fe2c5f28ab3f2082f3fcb9bed45b7e00c005df03706bb4d4c682682935
82e1dcef62383b67c0204e88a0b85fbea044b13f84e36381c1c9fd9243f4ba99fbf788fe178b4c3fec7abbd7b952e4b089cb
5296f6e635a7276f86b6b2d617be0f0294d6488cf1afa2b406c00a49bc2e189d2298bc5e829174217f3fbf2f17e567634242
c32f69bc7e8f2f1d0dc307f911599953d2fbe572822a9c1daf6b4ceb7f432c921824dab0ae4fd85323d8acce7a738d827f16
e08642c1ff289a1fc607bdd0f5cecd52e9fbd7cb37e1e50725f25aeb771104d16fc1b8e9bbe70f2a9b3ac35cf07748d6281f
5afde13ed35f7af27fbed7758bf89ee6e3e83754911fbea75e20270246abff7cafb35a7cefa62fb9f278f6e50e1bfe00a68d
e81d24c3e398b662792ca6338213d661ba866923a6464cb7623a886903a6a759ae4a8645aa0aa81126629a0a76fe65982324
c3207ea592cfe14fc21cee0a8c6565c187e3d62b6990ba01dbb0ceff21d29fbe8de5785844f36010b6e7f16bbaaf29290f26
0939381e4011a6e6ef26ae8ee5c2211c6b6a2469f261b46608c3d0f55f164c8315f014b4935bc94ef22d57c5fd93cea3cff2
89fc02fe59fe7341124ea86caa22d5336a83ba5afda966b4668ba6557b97f69456d6ddad7b4bbf42ffb66180a1d938cb78c4
94621a609a649a615a69da69fa85e98c79a4b9454c168f59062bd2cde3b4d0177629df5f8ab09b7dddc6ebc93758e7956fdd
8ad8b74468b8005e50be8f63651618bd102db32f8ede8d9669af76be5759c0f29fa36515c4c015180655508d1cde03e580ee
382c56bef69b8db1b01b4210c42b074bb3b0871b0ab1cf625884e91e288399b000fcd83a061662ff6c2c0d85f978b9d157ec
196b91522bc3bc0cdf598af752eca983e158aac011a6c212ec311bfbcec451e62a3ddd5866e3bb71948578afc63eb370dc72
ece7c6f7ab70de99cab3ef8f335999a51c715988bdaaa012db59cb226cab52c60ae1ec39e805f47eefc65bd0f39570f77dca
b7a137fff214dd60dfaab32f84cd60897e171c83916e2c468e2e888774f041007b0e871188a45168bfc7e08e340ec6c3ed30
117e82f2980c77e0cc1800c174b80b8a61271c417f9a129e084445d44403a7e10cd1129d7ac9c2f2504e7e61341f16cd8747
f311d17c64241f1a64f9c8c2604f9e13cdfbb770aba58f5f95e92b36fa726baaf072297d59e25b53e94b46fae2199ff06229
3de3a3a78be9a955f4a49e9ed0d3e3cfd985e321fa9c9db684e8b3323d26d36764fa1b991e91e9e1e651c2e16bb479143d24
d383abe8af65fab4893e75c0203c65a7070cf45721facb52fa6412dd17a24f3c5e2a3c21d3c74be9630d26e1b134dab45c27
34a5d1bde3e8a3227d249b36ae4f121a65faf33da2f0f344ba47a4bb779984dd697417f6db65a2bb247e27beb8d34e77aee6
1b4cb441e21f4ea33b7e161476c874fb369bb03d8d6edb6a14b6d9e8b61622495a7eeb169db0d548b7b61090c6f05b7474cb
69bebe6a95507f826ebe5f2f6cb6d0cd125f87a5ba5be8a68d27844d32ddf850b1b0f104ddb89a7f68439af050317d48e237
205d1bd2e8fa7516617d125ddfd27d5aeae6d759e81a9c7a4d29fd59903e18431f68a0f7ebe9ead25261b54c6be78b426d1c
bd6f9549b82f445799e8bd2bcdc2bd36bad24cc30d7485852ed7d1654bddc2b26b74e9920461a99b2e49a08bf1a5c5497491
4cef91694db551a89169b591564b7cd52aba70c1106161255d3084ceaf3408f3453a7f355f69a095125f8153565ca3e5f34e
08e5329d37b758987782ce5bcdcf9d9326cc2da673257e4e1a2dc34e65d76869299deda0b3643a53a62533b2851299cec8a6
77cbb458a6778da377aea245329d3e9c4e93e94f653af504bd43a6934be9243bfd49884ebcdd2c4c5c456f37d30943a5423a
5e4fc796d231291a614c031d1da2a3a8288cb2d191563a82d30923e2e8f061366178251d56280ac36cb470a85e2814e95049
2b0cd553494b2526c7c9fcad0d74089f250cb98d0e2eb00b83c7d182413aa1c04e0b247e908ede926f156e29a6f9032d42be
950eb4d03c231d20d3dcfe762157a6fd736c427f3bcd09e9841c1b0df5d30a211d0d45d6a79f960603b14270380d643b8440
2c0d9ce6b3937442b68366afe6b3b4a5425603f5f7b50bfe71b42f32d1d74efb4a7c26929e594a33fa04858ca1b40f12d627
487d98f9649a7e0b4d33c60a69c534d56b155227532fbee6b552afc4a768a8c71d2b788aa93bd922b863a9fb349f8c93255b
68f26a3e494793243ed14b13cc343e95bae282826b328dc351e3823456a64e9cd429d318913aec76c15149ed369b60b753bb
c4db6cd48afdac27a805c56b91a988995848cd48bfb9819af09949a6461cc0184b8d126f90a91e2b7a696025d5611fdd2aaa
2da51ab545d0d8a9da4255424850ada202be2784288f83f1591407c5709d4ca62053d2424ad7d491beffdffee0ff3501fff6
875ed2ff0682798f06
>}

\EmbeddedPDFStream{application-42}{%
/Length1 1924 /Filter [ /ASCIIHexDecode /FlateDecode ]
}{%
789c75556d6c5355187ecffdecc7b6ae5bb77514e6ed2e0359efdd3ad6318773ebee6513443e2c8b699508776bb72e59b7b2
9665137ff8c38f6c0b89d1040d91c41043f841ccdd7e1034060163082a21b1318231f18f41889aa041b3c1587dcf5d5940e1
dc9ef37e3fef7b7ace7b2f10002886d78185ddbbf6346e5c7761691b801842edcefe94910617cc20ff134ea57f3c2bf5eedc
f53200b1a1fcee407a30c5b82f18287f88f3eae0f0e440d095cf01f0a7506e48268c785f6ee40e009b45ff4d4954b83ffab3
0bf14b515e9b4c6527aa5791eb0002c5f70c8ff61bd005efa33fc64351ca9848c31ff00dda6fa12c8d18a9c49ab742df0270
41c47f353d9ac99e7dc38158ec2f68df961e4ba4dd93939b513e43ebc549705f748745c0818494c387cad17c7e79cd9fb264
4c81160778a116d6e7ffb2341e92227b4933914929fc03b7e0129c8639380947e108bc090761145210873e88420476430bf8
9d1f38671c27ec115b89b8242e8a336246dcca7fc91fe10e7387b8035c1bfb337b86ddcaec25272d648d6223f2ef70c5c2fd
b880bb8ce80708575678cacbdca5ae92e222a7c36e13059e631902ca1c9e972e4df74dcbdac08baa32570fa0f74e98108856
b5c52cd26e11d3511f33895efa1f95aa04f06732f5aab2bc903e13646d96b0759ad423258db8c9d621954c4e970d293e1b0e
4ff746fd3866c1bf3ab63b4a19ae4eea499a625db761c20b51bfcfe4f5a7623e231edb13951be5055fa37c1373071e8f6ec4
1febc9d7699d37ff17dcb960a9dc65556daa721eb78d27a74bb4c8e96e93d58da1fd5bf66baad22d99acaccd1186d7c77559
9f96d16072b2e6359d74af88e6ed1e32c38664d56d76f990a7acd9e333b7c662262f6b0841646dc86464cd2caaf7aa8ad49d
dca22a9f811bf8fc390d8b9873f340a924cd32ecb4269bc4d07af00f0bf746cdf0fe18cda72a68d2dd3a96db04aad2a02a5f
a94aa7e9a957ad1b87d3dbfabdbecfd5fe37b036ebc2bd5677ecdc7d9abf7223e23c2ea8d6cd64e0e1813d12ae6524b68659
cd7a491553e92c176b344731f72ca7319d6b3a384dab2d0ec2e1336cfb9b6b0281fbf1f943e081470ddeaa87210c619de9b1
d1f8c1fe6c3631912dcd1c4ca58cecd0e8487c28931e362697bd45788976156747fe2cce659e608f9d2df00c94c0d7059ec5
8eba5ae0b9077c785805770abc0045c4033a76541a26610c8660109290c58e7d12b51b9036411b3e21ec6909343020833ec3
c86f87065c77a066182333189580714bd7851aea115941cb5852026902738ce31a47cf20626fb430b7637e8a24410f722385
085a15454a403f6a862c0bade2792b53022d2a4637a03585d204a205ad2c83d8c9c38836b682b6c3a28356f6cd0ff937617c
13345a95a8b013fdc691d2fdea0f78d133cacc83588d5c35b48683157646a8b1898ccdeb2e614b490d0187c0702ebb48b8b2
6a37e3a9b477147980efa8e440607c2e36802358deb2b1c51d925bfc15b5156e4f73859fca7eca657239e1442e776f26b778
2c370f9129e1b9a948646ae1345d85dc823a15a115dcbeeca415f82114562417cf7a7c76cfea62116c1c29f301c10a24bc48
4facaaaef2b878ceb6b6c4eee4898cd99b80a6af6d09b5dac9ba67486853b3bf5c5ea7900d84e08b4e247e4921adcdecdb1e
816bbb7b512816ed47e7dbf72d694bbea52f1677b01739eeee356e3d799afc4a3eed6d9fbf7db9d22e70bff19704ae4374de
3bffe3774d777ec077a4ed6ea02977ed5e8a360e59b9e2e261c838715a2a1e4f9eb7d40cde48012a91fb1c5eb1be0d6817d4
1b1df8c9a8ca5f59bcee3cfe00c84ab708ef9177282547c901aaba4f618029435027ffc8362b8c30bedfa398234233557d02
f02fb8bd8599
>}

\EmbeddedPDFStream{application-43}{%
/Length1 2724 /Filter [ /ASCIIHexDecode /FlateDecode ]
}{%
789c75567d4c1be7197fde7befcee70f6c631b0c38069bc3106c9f31e6bba1606c2074248c25909a9411cc371d1fe623e46b
5913755d0451956eb4cd14a55935a1ad7facd581b688b088b5cbd26991a6b68ab455daa6a9aab694a9fd67caa4b549ec3d77
b88c6eed6bdf3d9feff3fe9ee7def7b903020019700e28747efd7059c83a9ff94f005d176a3b0627e30998220aff67bcea07
17e65db379c7ff044072f0ca1e498c4e3299bf8e03304194df1f9d3835f289559c05e08b00e8bb63c3f1a163dff9ee190056
c0f9d563a8c8b8453b303efa43d1d8e4fc49c75bcc1480f63eca7b26a607e3e45d6613fdc3289b27e32713f009dc417f33ca
aea9f8e4b0f342e575b4bf86eb9d4e4ccfcda71ec1df503e89f66062763871645a9c43b913e562bc08e6a564680016943505
fc31480f4237dc48a5fecba5d653aba976d5777bf0e869840290a02fd99b9c4bddc58c16e945ee04b742cc84850fe01dd884
d7e01a5c852bf0322cc3f7e179780e9e8149188121e88543d0015f833688c0e3b00f6aa00a2ac00d05824e735f93cbdfe56f
f357b97f731f737fe4dee3dee1ae733fe7ce7367b8e35c2767643f64ffc22eb2fd3449efd37bf4f774935ea44fd35ce60453
c694904db24cf611276c218abf228edfc16fe136dc423c376103d6e13aacc11bf03344f713588157d3187fa8a2bc84282fc2
225c40accfc279380b67e0141c8739985191f7431f22ef840388bc159a117de30e7e3740383bcb66b5649a4dc60c835ea715
343cc75286807f0df750d4b534b02446468e48fe352f40b4eba40cbe98bdae4725f52a9175de1e9944cdffa392fc3efccb8c
57f26fdfc8800c626495504fc4d5ea1a8b0fc9d483d425b35131ee1a5a0d8797ba626e1cabe0ded3d3195318d6e36a1d9335
9e96b80cdf88b91d3217aded71c4877a0ec7c432f1534799b8856bfbbe3a7a7ce82b3d394fa471ebff26377eaaaa322df63a
c9ff16a68d3b28ea52402eb5c8341a1fef6fee8f48fe16974cc5c81a61b8e842548c2e89689059319223eb955c315a4ecbb8
1c8ebb54dc7293037985955b1d725b4f8fcc89110c41c4c8b8cc8811d9e0cd91fcae96b166c9bf0199c0a5de8c2088b54c0e
14ea72ad32742922ca241e69c58285bb6272b8bf47594ff2a3299a1945b8e520f90392ffb6e46f946d5e49ddf9789d5f9eaa
3e66aaff1750413d06dff6bcf2e6e734b591046180df3e1d0c7c71c4e10f649a7c14dea329a5459cdbbc872d8d19334c1732
a7f463153136169332aae0f9755aff9cd3e743fff4fcd419b0c1970d8d8a87c15d78194f6eadaa39aa9c66568b3cf688344f
f06c6fa679064fec9d344fa110de4ff3ec2e1f0ef2e0b334cf8381d8200ad390c033300be3300a63300f2ed88bda52a4e550
87bf4af0231fc11ce7d06702f97608e0fd206a2670e61cce1a860555d7841ac5e3d04eb439551a463a8c6b2ce07d083d8318
3ba4c66cc7f595482e3c6fd330959ea1a052220dc3206ac6558b82e280bad2305a249c1d40eb24fec6315a505d6514cff204
469bdd897650a5a3eaea8f7dc1bf1ce79743998a44c26e358d912535dfe82e2fe5591d4d6dd119ba881d00ede160411e63ca
d27a4b051d08a58526e0b3f6821634943593628bd649cee98bcd5c2119b5b0669e2933519faf1c7cbe20571820559596ea8a
10761123f1113ecb965d11aaaed9d668789e841ad02340c442e4af0e2c1f5c7ef842ecd99b1f5fe2dd5d474e7fb09fd7e9b2
d8f6643275e2ce8a375af4e4c5a3f691d7fb7f40c895a7ba17e61ca79f7c6a2931c8db75023b7e8a7167f7f415d7e5f75d56
76525db295bd8219344279786f5d25632135c4b3b75c9ba721ece33e07b5f23ad1956f3719b48c100900e14813e22e8746af
2f68adac210d44c5c86be836e81a046a4d2b559d9ac73e82b211e1179728f712aaf2750c91b465f97945849cb2b53df1d20c
212f6a42ce1c0fb36caa0f37e5b0193982bea8bbbf79dc5cb178e0dc13637c303fd7937cc6c0b2379c450579bc864dbabb1b
0e0c6c30380e394a5c7906f6d1da63ded8b584214373e9f58e48b8b33e3980be9829beaf853eee041c8670b82a9774d40769
7335d34af0a5ac7560b6fb9b9c866c4367fbfed6a670436d28e02df1e4e66466e82d52014378d26d56d36e0c047cc18ac2aa
06067325d5559525c5a22ab943762ba946926564c4420daf519e623eb1634d94ccdde929d5355ac22b152046a2b1a58d6336
824ff0968be8d89c924914f40219b7d357997f187349af2a9f2d771892f7584eab43078d607db4b5a326250f6f1b9c79f48d
e43d2cc337a3591aadd5c1e598cfb2924dc8a68283cb35d91f7c8b7dc9ae70466dd983ab36ada0a78ce0709ab75501f6c79f
7d64b773c283573893aecaed7058b062175277f95cac5804aac28140296364b5d5a43ee8a0c6229b01f75291889b9b616bed
66036bf1935029330825993c69c652e10812ccbf22d4c0d458b14e4a95089647dde258206aa4a2baef55eb7615c9aef2519e
48fb12edb19f7eef986d9dbd8639dbc84dad41d00bbf29b8619ffcd1af06677ef9b4e96db56068638ab3941a2af578db419c
ed0b4d6de7563a990fb3b5471c79568b96384d671f39fb7fb1dc35b8361fada58eedd2e4a99532ea020fff3ec1effacca1eb
70947d0fea34b7a08b3b8e5f0544edb483d821413df906ec92d9c8ddc4ef0216ad68e756926dcaa7546ae3615218d8156ca7
73d32479013b2d902b640620d9bb4dd9cb30c25830a89efbd2969f1e61ec5837708ddecf5f33ff01b5611734
>}

\EmbeddedPDFStream{application-44}{%
/Length1 2620 /Filter [ /ASCIIHexDecode /FlateDecode ]
}{%
789c75567d6c53d7153fe7be2f3b7112bf17db71be9ffd30a6b19d0f3b4e0809c4b193e08511520a691235210ef970b2a418
c8f8105dcbd6765419b095aa1362a38a1062a8dad823da34d43f58538654daa269a25dbb31a99aa2a96242ad3475532bc0d9
792fa1425b7bedfbcee73de777ef3df7be07080079701438e8d9f6444db870587e16c07a97b4ddbb67521938897ee2ff46bd
7df7815975ab7b5b1700aa24ff753c3331c3e4b75224dfa7fed1c4f4e1f1f5afae5f02103e07e04ea7c752a3bb4a5e3842fc
fbe4df902645de2e51a6f86f93bc263d337ba860896b27f938c965d37b76a7380bf71a006f23d931933a94812fe05db25f24
597d3a353356d65cff16d98f52bed9cc9efdb35cfe9493e424d91b33fbc6329ff59e1a21d9c0b78e3ad2bc8c19da8087f544
2df46344eba01e86969789ab85c82a07cbaf93af05720c9efc459040a0a705f2a11c0210cf0e64f72fdfa2d9ece7e68583c2
79bc097f87dbf001e870112ec079380dafc009780e66600ad23004bdb01dbaa10b9210878dd00c8d10a56c35e091b2d2a7c2
5de19aa00bf3c21161bb60e13fe62d5c96bbc315b3fbec73b6c45e67336c80f95925dec1457c119fc319f4c1a794f163caf9
17cafa47b80ed76011aec21b70057e030b70c944710ee6e1a7f032fc187e04c708cdf7e030ec873d846984f018589a098707
20e6723a0a15d95e909f67cbcdb15a2451e03986105ca07a48a87323735a7cbc37145ca80248ec38a443a0afa8a9df242d26
d173aafa754cd8ff47150a06e8afb3aa5070e581233a68f1cbc8f9e26aa79a4e8dea9c8fa8aaf3092da58e5e8ec5e676f479
a85d064f597f4f9fc1f03eb533ad4bbe8e940e8ff7794a7521b1bebf3435daff449f56a37d595aa3dda1dc816f8e9e1afd46
4fc1176fbdf37f835bbf3455b252d4140a2ed2b4a91a12aa0172ae43e712a9c9e1f6e17828d8a1ea9c165f4026240e24b4c4
9c46069dd7e26e3dd7982b4573774ceab1946ae2d6db4a893758bdb3544ff6f7eb8216a710a8c52775a6c5755b953b14543b
d2eda1e01b2083b0fc669c402cc8021854552f336e2eaee9988a77d282c576f4e9b1e17e235f2848a6849c20b875100a5687
82d743c156dd5115322b9ebaf393dbbfdf55d0f26fe02c66493fe33bfbe643ba7c210b16453444d13c118fb649b80cff420f
fe0adf8b55487ece9b5b6929e0fd83f979d2946d50da293cae7509838381bc289cb8c2b5bc581e083c8cb17c041cf075cd6a
6262c8a0108ac8c7051fb1be49c322c180714a79c3e32af5151ee91c5e5de5199dc0b757790e54f8f32acf3fe2234009fc67
95172107ed90a09acf50edefa3f94cd0799ca591eb48fb18d13a68a25f3d04898f438acec7244c13bf05aae9b99534d3e6a9
49c3181c30756da4313cb67f156dbf298d111da31c07e8394a9eb5143b6cc6dc42f98d482a7412f7f4ea08039511690c7693
66d2b41828be6d661a234b884657937586a2d6513c23c7047c972c29d23c8cb5d5a41366ee0d8f78d7d1d83aba630c1421ba
7df650d49039d7c4573e1cdd4cc06785835000a5b41e5db1b8dfcb7c8e5c2cb130ab5ac67cae0a1b6f7129cc6d672e478943
cc67cce202ab5de2f8ca7ce6c6a9228722946019b22945b6892c50c005ea2050551be122aea20a8c78c20d8d0d8d110ee588
d3138dc85ad4e3e40a3d85da5aff5acde3954469dcde2cb82d2f8d32fbd8318b5b68b6dfaf6966e79b9b1f3cf585ccef935b
0467c1d283fb4b054e0187ce965a44d1527a367b8e69984ea7b3a7d36966c364f6ca71b7cde63e4e15d2b37c8bbb24eca019
a9a0c5cadc7ec90616bb9fea4e62fc4e9b03849d651c88ccbb82b456f032b95e6988842b98d3c1447fb84189d6b3b59a5724
b9025d0ceefdb66df1d2b15f0e78daaefe7a5fa6a484de369992dcdaef84f02c962fe086db2ffce9b3e7f5eccd0fa3ffc4b2
f933c8df7a3ffbe0ccbe773e7c05084f1315f275c2e38262a889ad752373a1d7929323225f2cc8d622648a3d8749a860313e
6973950b584ac802add5d581dac228363463d8496ba7391d2e6305c30dd1a8cf2337b1c3527e45d18d1b87b63d786fdba1ac
1fff81db939c9e53747f3079a3aa28992caaba91641f18676d81100c110219d6c4ca7390475a7599d28355969053ac3890c7
a3808566da406d24aa4665518b861b2361975376789c5711a6cedd3b5a2ee52124dbf84824f97cae746f735bd288de4beb7d
91a26f819ed8661f162bedc0723966ad2c63c5be20554f14d728b92872da46da028e5764e6f25abb9a4410a2d85de7e2c5b2
02e0b678d7a1b839d90622bfd5c061d65023d50fcdb7be9a05d0bb11eb238de14d18ad37f646da84063a87a879d746eb0dc1
d8be7c945688e6ad26476353d1f0b9567b6ef0e0a833fc64eb4b8ab237b92bf8d450f7c92a5ee4ab376daa0ec5a67eb163d3
5c66f3d61ff6f4cfa79e792733bc78a43de269acecac7badb3fdddef6f981d6d2954ecd6e99f758e3cdbdc82180b0563b160
df0f12be8ee98ede535325f2e8cf07f6fe6ee25ba7c623ed95ebd5703badcb5dbe9b538413741fb963768107ae90d95a3904
0125b3f25aab6ad16f45c98f8e75e840e5b13b7c77001dd9bb819ba27971af345ad95e1ea087ea68814eeb5dd3648561baf3
8cc6e88b46a4da02f803bdd579b2925d389feda44f9cc6e50bcb272dca23c15643523d7c823f31289ec1bd00d98115ca9f83
71a650d05ce16bafeed516a39b6388720c3c7c65fc171571063e
>}

\EmbeddedPDFStream{application-45}{%
/Length1 2128 /Filter [ /ASCIIHexDecode /FlateDecode ]
}{%
789c75557b68535718ffce7d25695e4dd3e626b60d37bd8db1696e9fd65a2d6bbcb141d4765a8b344eb0b76dda6436b6b6b1
d83d70b075b04ad9647f0cc48d8d89880cb92dc3c9263ec68ab07fe6704c651326db60c260303aa8a524fbee697d6d7a9273
bfe7f97ddf77eff9ce01020076780358d8f5e29eda065fc0b617c08c4ae8eccf68a3f0157909f99f71b6f64f64a5ed81f6ad
00c48bf26783a34319c6f5b586f237386f0f0d4f0eee3ff16317007f03e51ba9a436d02b1f4330f614fa6f48a1c2915e9e07
30197895a94cf668c93b1c87f2c728970d8ff46bb04cde41ff1f502ecc684747491b2943fb2594a5435a2659fee97a5ccb75
23fe2ba323e3d9b7ba4bfe46d94878e7e858727444a96ac2f5bfa15c8193605d468536e0a00e8cc2ccc0207d193e81d3f9fc
632e7f2a3f90efa5be2b43c09f0dbc50090df91c40fed28a9a84880316e11eccc359380327e17d3801533009e390815ed80f
5dd009db4085cdd0024d10303b4ce3269f704c6813aa0407ff27ff137f9defe6bbf85ddcafdc3837c2e6d8bbec4dd6c72c90
bbe43a394282c486e8dfc1059883f318e323c47f135ea711fa601f74d308db3146fc618ca8a7a4d85de42a743aec366b81c5
6c12788e650844e6f0bbc6a4e9be69591ddcab44e6c200b1eea33a54f7882d094a5a29d10bc2099dc40affa35222d5f8d799
b0125979903e1d647596b041558a4b296d40678348259d8bc99a34301b8d4e77f70470cc42a02cb1abc760b8a0144fe9a660
bba6c3ee9e40a9cec736264ab581c49e1eb9567e505a2bdfc7d8d5cf47d7069eebc907d5b6fbff5bdcf680aa5c45628b12b9
8665e3578d494692d3ed3a1bd3d2bd5b7b5525d22ee9acacce11868f4dc4e4d8b48c069d9355af6e356a45346f7b5a8f6a12
cd5bdf528abcc1eaf1527d5b22a1f3b28a104456d33a23abba2dec5522527b6aab12f9125cc0e7afaa98c49c8b07834ad22c
c34eabb24e34358e2f2cdadda3477b13463c2582a6982b86e9d68312a95122f34aa44d2f0e2b7437e2fcc2d1ffcb0167eb3f
c0d2be84d7821f5e7d48f3f38b5de6656133ddb10c3c3d9a716fcfc31f512f5fc1973944b6acc366e7bbac1d1551b6a3a3d2
be1b662eb2ad53e5d5d5e8bbba36ff2a14c3b38689e6c21086b0f0171c70688786fac6b4fe83c9ec7072306bd3c6b3c9b1f4
f8c18c964d51ef7d46e77116e42fe35ce10976d4e5559e01077cbbcab3d8abb75779ee091f1ed6c0d22a8fdd488a21062330
8abd30066918821464418275a8ad425a8fddd002eb2182bc0a1af64b1a8691df0135f8ec40cd30eda214246182eab6a0c6f0
e87a84364ea524d224c698c0e7007ad6217603c5dc81f10d2409bb6f040eadae30b2329092d08f9a34b51859eca491926851
70750d5a33e83f89687534ca101c419b86911ea275503a44a36f7acabf1ed7d7432dcd44c113600491155a6fec092fe31bdd
cb0173560010c11b2df498ddc4e9603659590b4f7c4eb61a716aea2c646dd3fa66b2a1b1018f0ec164217e8254ae60f6e58e
731c3962f2178984e4de258709f1bafc391f7bc72c7036f696cfe3e60576b979b999e5852211a34d419c9d642fe239541b5d
d7581b0e957add4e1b14b82b094b1a1b22e1aa80e415f16c1278c746bfc5ca93669a45b8ce5d6124a118a9602222cf1ac453
8e2dbc9297c94e3c34ab50d04de9da260c7e4d0c969350ee4e795074da89e0aff079037e22d89d3eb93c778784ca659fd39e
5bf2cb1eafeccf2dd909171265c7cc715b500c155c219d3ec92f72b9b92b052131683b3ee3900d75ee7341f407bca4f34a81
f106cf915bfc0c5385f7a2182d2cc39d2f10aebe80059e3869eed52bb96f680cd0f7277feff2785cdc0785252585e4d61aa7
738d3111672107242e18f7ab3b6ab5721d66f208023f82db450b2e71d1da16bc2e639dcb9b0b9a2e18488bdbf189dbfef195
c496c23dbc2ca7846638c734c30235996006fb046807dbd0dd8305d8e06dec2362d8f9338b2f00584ee6e7977e372f3f01f6
a8abb93879cf402027c9617ac351cadc8441a608a9957fe671b03aa2787f9ec61878e70b9b4de701fe05a9249b84
>}


\ReserveEmbeddedPDF{window-1}
\ReserveEmbeddedPDF{window-2}
\ReserveEmbeddedPDF{window-3}
\ReserveEmbeddedPDF{window-4}
\ReserveEmbeddedPDF{window-5}
\ReserveEmbeddedPDF{window-6}
\ReserveEmbeddedPDF{window-7}
\ReserveEmbeddedPDF{window-8}
\ReserveEmbeddedPDF{window-9}
\ReserveEmbeddedPDF{window-10}
\ReserveEmbeddedPDF{window-11}
\ReserveEmbeddedPDF{window-12}
\ReserveEmbeddedPDF{window-13}
\ReserveEmbeddedPDF{window-14}
\ReserveEmbeddedPDF{window-15}
\ReserveEmbeddedPDF{window-16}
\ReserveEmbeddedPDF{window-17}
\ReserveEmbeddedPDF{window-18}
\ReserveEmbeddedPDF{window-19}
\ReserveEmbeddedPDF{window-20}
\ReserveEmbeddedPDF{window-21}
\ReserveEmbeddedPDF{window-22}
\ReserveEmbeddedPDF{window-23}
\ReserveEmbeddedPDF{window-24}
\ReserveEmbeddedPDF{window-25}
\ReserveEmbeddedPDF{window-26}
\ReserveEmbeddedPDF{window-27}
\ReserveEmbeddedPDF{window-28}
\ReserveEmbeddedPDF{window-29}
\ReserveEmbeddedPDF{window-30}
\ReserveEmbeddedPDF{window-31}
\ReserveEmbeddedPDF{window-32}
\ReserveEmbeddedPDF{window-33}
\ReserveEmbeddedPDF{window-34}
\ReserveEmbeddedPDF{window-35}
\ReserveEmbeddedPDF{window-36}
\ReserveEmbeddedPDF{window-37}
\ReserveEmbeddedPDF{window-38}
\ReserveEmbeddedPDF{window-39}
\ReserveEmbeddedPDF{window-40}
\ReserveEmbeddedPDF{window-41}
\ReserveEmbeddedPDF{window-42}
\ReserveEmbeddedPDF{window-43}
\ReserveEmbeddedPDF{window-44}
\ReserveEmbeddedPDF{window-45}
\ReserveEmbeddedPDF{window-46}
\ReserveEmbeddedPDF{window-47}
\ReserveEmbeddedPDF{window-48}
\ReserveEmbeddedPDF{window-49}
\ReserveEmbeddedPDF{window-50}
\ReserveEmbeddedPDF{window-51}
\ReserveEmbeddedPDF{window-52}
\ReserveEmbeddedPDF{window-53}
\ReserveEmbeddedPDF{window-54}
\ReserveEmbeddedPDF{window-55}
\ReserveEmbeddedPDF{window-56}
\ReserveEmbeddedPDF{window-57}

\EmbeddedPDFStream{window}{%
/Type /XObject /Subtype /Form /FormType 1 /BBox [ 0 0 398.438125 161.788125 ] /Resources
\EmbeddedPDFRef{window-1} 0 R /Filter [ /ASCIIHexDecode /FlateDecode ]
}{%
789ced594d531c3710bdcfafd0111f1052b73e0f39181353e5729272a0e2839d03068c21cb624c882bff3e4fda8f916647b3
ebe0532a076cb62db55eb7ba5befad0f8e2effba3ebffcf5f850bc38e90efa4fe70f9d1637f8b9124adce0e7abd0e2183f57
ddbec2c7db8e63908683268b8fb3f2a3765afa907f9da5d5d5e74f5df7b13b780e470fd877dc75c64a45f91fd9c868e159c3
1513f7a65969d286a4d5b0f51bd7a6a5735a38bf025c409701e07154b274ce4b1394215e9f5658b434e9b0ee10717fedeef1
a71209bf70243938e729fa18046919bc0ede8af3dbeef0b43b78a94510a71fbb9498d38bee9dd8d3f699f85d9cbeea7e3ced
de741942971264a2d6eb937b43f3e0c892ad651b8309769783496d1eac294aa3bd8fbe4f71616a1eaec920e94619e7b5a79d
4e1f095b3b95926782e94f2f4cedd3ad9514c933db60e32ea7f358ec51a70cea22f4ded23e3b38490e65a58c353be59dabc8
0b57c14aedad43954649cbfdd826236587d8bdf6b6f7e19938bde9b081b48aecedc2fbde59361be95db4bce8b7649eaf565b
62d3afbe1877f2359b9d0cc6bb0c6761bece66b44f4af256277f66334bd2b8bbdefc691c8958f936145d58990f5ed232dcd5
36e5551f569f8df1ed7beff7b23dc81ada6db67a59c7f1fe5936e352ee9173a5824857b6f8cd3a89da5a2c5e5dc7f9ad00be
fda3cb9bb3df1e4fcee60ffbbf7c985ddf3f5eee3fceaf151fbe104777b8df5575f52308ce2cbc995461464b3bb48e551929
54846569a23181a28ad4ae2f25fd486ff5e707853c2174aaceefadcdf33d52a6a2254c183f757c1869ae62022b23ad63c7b1
3abf3037016895961beb9ddf066132039a82b43eb8e06a08bdb90d812cee293a624f34d1e4281355b77972b19f9c6993ee12
7ff0961e9f8d37dcddb2c7837646f5ed79be34a717aae881c64468f86e34d2e5ca49d9b31f976ba32b9b6bdcda981d0dcc8d
d5a3281aa3ade1f9efa5d9ea2a1ba2ddfa0e4d87ba5bccc1edbdffc7d9e7cf67abcebf1743c6a2d1c394fe56e82527be5c8a
b7622e48bc1235a1eaac9611290c82a30c8c86d715ffb05eea686046a586946c8c6a273c6a530593f9cd9a31443ca0d67b9b
96166fb9d624c922b405675a3d749abd74a4169ebb93ee8dd83588778225f64b87aa57e2621850811d934e1be044443d7687
86401f2a2ab163e27020b47a0d1d6300c113d5c8b180950fd6f5c85362d59ad8d58836c9e4084b849f0d8a793b4a31b17227
7a5aad5bee6e7a5419ff829cea4c4dafca89543a0a6953731c3d2f67d166b9a1da78e98841238d4fe06dd830ce3a67478c05
d201429f5a66b91c97102b5ee46b8c3fbf2d414e9753019858aa90bbb282dc9b2bd0a539c3be42e12e4e6843c7e4773c2075
03f0af5f97e097a2a594446a29893abc1ca8deaa9e18454c1cca32294ceb8228767e838ca1c4a0a32d754c696a125b02a966
a59fa86428f5a7338594292cedb313a906d57b9a98618216632ac54c696a9eceda23f7f1896286312591bf52cc94a6f6e968
6552fe8962868349192cc44c61699fed232488fb1e62867c907846fe5733ff4935530eb1909ac55b9f7a1bc56686e6d106d7
a9c8d3c10ce9ae948b20046a52d48c14798102931fb43c30d7287af3340a9b98139a4e056b780ac5d888eb51a0e641d63257
2b40acadd318c0de02ae07654a610283de92898816c66961701fbd791a4574522be84b6314992914d399d01a74cde0624d0d
a3b04fe3d0da49a3004563204d8dde2de940d0d2c3d110c7dabc05063b89a189d28856b92918b621f70844db72ee6313d328
cb33b1e968fe389b89dbcbb3b93810f9f793a3d27166b54306b1b3aac0ad4af0ea246b9028701acd351d70e0d0daeb7c3b94
0a122f651216e0e118712ab390f5fb8d471a2f9acd8b8b8715920253d5e7d4f68fce98b0f886409434e033cd700c189a75e9
b5aec231e9cb58cf21d6e1601b90c4584563506f787bf3a828a2b104cf0623a48ac6437e47edecbf0dc64f05534abe321874
be4d5f3c0c82818e23763ad4d170ca36e5af328a688cc78c49eca28ac68166b351467f73345b445f5959e8280cfa98fbb0a8
2c875710d10e2acb48ed78c1d2abcaa2f45810f3a0b408c56cbd324fc69fcbac6b47813bf0680ff27514b803906432aebe03
239d713cec0f4c0187820ababe03dc4cc4b0fb1e31f86d31a41a306e10034b4be9ff0fea186cfaced19bc14d5895be69509e
ea18f076a0d92cefa6c04794d098c481a70d81743b2a90b0723775552f5cabf086cf2d2abcf2342dc30fa765781a01b47a9f
d63a9c207687565c11a8cda6b5adc4c9312830ef26c5cfefe6970fe2fdde4fe287cc196da29335f2b1815ce22f6579194129
c0cb183684f92a8ae3a9487650e6d70fb3b3f9c5c326fecd195ce2c7ab1d747aa96bf86b738dbe30ef0e5ea3255d4d0606d8
3fdc3dce2fcebefc2d6a79fba6fb073b1611d9
>}

\EmbeddedPDFObject{window-1}{%
<< /Font \EmbeddedPDFRef{window-2} 0 R /XObject \EmbeddedPDFRef{window-3} 0 R /ExtGState
\EmbeddedPDFRef{window-4} 0 R /Pattern \EmbeddedPDFRef{window-5} 0 R /Shading
\EmbeddedPDFRef{window-6} 0 R /ProcSet [ /PDF /Text /ImageB /ImageC /ImageI ] >>
}

\EmbeddedPDFObject{window-2}{%
<< /F2 \EmbeddedPDFRef{window-7} 0 R /F1 \EmbeddedPDFRef{window-8} 0 R >>
}

\EmbeddedPDFObject{window-3}{%
<< /F2-DejaVuSans-Oblique-kappa \EmbeddedPDFRef{window-9} 0 R /F2-DejaVuSans-Oblique-uni03BC
\EmbeddedPDFRef{window-10} 0 R >>
}

\EmbeddedPDFObject{window-4}{%
<< /A1 << /Type /ExtGState /CA 0 /ca 1 >> /A2 << /Type /ExtGState /CA 1 /ca 1 >> >>
}

\EmbeddedPDFObject{window-5}{%
<<  >>
}

\EmbeddedPDFObject{window-6}{%
<<  >>
}

\EmbeddedPDFObject{window-7}{%
<< /Type /Font /BaseFont /GCWXDV+DejaVuSans-Oblique /FirstChar 0 /LastChar 255 /FontDescriptor
\EmbeddedPDFRef{window-11} 0 R /Subtype /Type3 /Name /GCWXDV+DejaVuSans-Oblique /FontBBox [ -1016
-351 1660 1068 ] /FontMatrix [ 0.001 0 0 0.001 0 0 ] /CharProcs \EmbeddedPDFRef{window-12} 0 R
/Encoding << /Type /Encoding /Differences [ 104 /h ] >> /Widths \EmbeddedPDFRef{window-13} 0 R >>
}

\EmbeddedPDFObject{window-8}{%
<< /Type /Font /BaseFont /BXGEVL+DejaVuSerif /FirstChar 0 /LastChar 255 /FontDescriptor
\EmbeddedPDFRef{window-14} 0 R /Subtype /Type3 /Name /BXGEVL+DejaVuSerif /FontBBox [ -770 -347 2106
1110 ] /FontMatrix [ 0.001 0 0 0.001 0 0 ] /CharProcs \EmbeddedPDFRef{window-15} 0 R /Encoding <<
/Type /Encoding /Differences [ 32 /space 40 /parenleft /parenright 46 /period /slash /zero /one /two
/three 53 /five 55 /seven /eight /nine 61 /equal 65 /A /B 68 /D 76 /L /M /N 83 /S 87 /W 97 /a /b /c
/d /e /f 104 /h /i 108 /l /m /n /o 114 /r /s /t /u 119 /w 121 /y ] >> /Widths
\EmbeddedPDFRef{window-16} 0 R >>
}

\EmbeddedPDFStream{window-9}{%
/Type /XObject /Subtype /Form /BBox [ -1016 -351 1660 1068 ] /Filter [ /ASCIIHexDecode /FlateDecode
]
}{%
789c358dbb11c03008437ba660042320360be552d8fbb7310ea9f44ea70f895decd67911340e4d927056b44dd6bd3c1f56a4
085693f4047c52ed53b831067261f72091a45a9923ffdda487ee17ddb9181a
>}

\EmbeddedPDFStream{window-10}{%
/Type /XObject /Subtype /Form /BBox [ -1016 -351 1660 1068 ] /Filter [ /ASCIIHexDecode /FlateDecode
]
}{%
789c3d52498ec44008bbe715fe402476aaded3d2a80f3dffbf8e49327d323185c190e354c369d2f83dd403198dcf61f61f69
2e4c76a282ee0dcd242a31a0cbf162c6a14d76b0ea46d63d195b640d2af35eb1266122688359a302563ef03a5c75028f8d6e
78b39b24f910466e0823b62342583b4229f64c9afb1ba9b06b4d24ec6d82a403d57850ae3a465237331ac4bd9e04474c765c
2c5447d7d0da68451a27dd48ce4220ed797fef0b22e711e95470cccc3dd5d9325a4317d68c155c07b5164e3ed2c4d98f9593
0b88b58934c9150cd23e577e31ce8ae2160a53107444734193ec3f50f3dab77075f0a27742b0c5b5da9e36466f2365bcde2d
6eb43491ee1a71edb8c435791e5e799466f902def075cc4cf7fff0fd713ec7fbf8f903c1076c35
>}

\EmbeddedPDFObject{window-11}{%
<< /Type /FontDescriptor /FontName /GCWXDV+DejaVuSans-Oblique /Flags 96 /FontBBox [ -1016 -351 1660
1068 ] /Ascent 929 /Descent -236 /CapHeight 0 /XHeight 0 /ItalicAngle 0 /StemV 0 /MaxWidth 1350 >>
}

\EmbeddedPDFObject{window-12}{%
<< /h \EmbeddedPDFRef{window-17} 0 R >>
}

\EmbeddedPDFObject{window-13}{%
[ 600 600 600 600 600 600 600 600 600 600 600 600 600 600 600 600 600 600 600 600 600 600 600 600
600 600 600 600 600 600 600 600 318 401 460 838 636 950 780 275 390 390 500 838 318 361 318 337 636
636 636 636 636 636 636 636 636 636 337 337 838 838 838 531 1000 684 686 698 770 632 575 775 752 295
295 656 557 863 748 787 603 787 695 635 611 732 684 989 685 611 685 390 337 390 838 500 500 613 635
550 635 615 352 635 634 278 278 579 278 974 634 612 635 635 411 521 392 634 592 818 592 592 525 636
337 636 838 600 636 600 318 352 518 1000 500 500 500 1350 635 400 1070 600 685 600 600 318 318 518
518 590 500 1000 500 1000 521 400 1028 600 525 611 318 401 636 636 636 636 337 500 500 1000 471 617
838 361 1000 500 500 838 401 401 500 636 636 318 500 401 471 617 969 969 969 531 684 684 684 684 684
684 974 698 632 632 632 632 295 295 295 295 775 748 787 787 787 787 787 838 787 732 732 732 732 611
608 630 613 613 613 613 613 613 995 550 615 615 615 615 278 278 278 278 612 634 612 612 612 612 612
838 612 634 634 634 634 592 635 592 ]
}

\EmbeddedPDFObject{window-14}{%
<< /Type /FontDescriptor /FontName /BXGEVL+DejaVuSerif /Flags 32 /FontBBox [ -770 -347 2106 1110 ]
/Ascent 929 /Descent -236 /CapHeight 0 /XHeight 0 /ItalicAngle 0 /StemV 0 /MaxWidth 1342 >>
}

\EmbeddedPDFObject{window-15}{%
<< /A \EmbeddedPDFRef{window-18} 0 R /B \EmbeddedPDFRef{window-19} 0 R /D \EmbeddedPDFRef{window-20}
0 R /L \EmbeddedPDFRef{window-21} 0 R /M \EmbeddedPDFRef{window-22} 0 R /N
\EmbeddedPDFRef{window-23} 0 R /S \EmbeddedPDFRef{window-24} 0 R /W \EmbeddedPDFRef{window-25} 0 R
/a \EmbeddedPDFRef{window-26} 0 R /b \EmbeddedPDFRef{window-27} 0 R /c \EmbeddedPDFRef{window-28} 0
R /d \EmbeddedPDFRef{window-29} 0 R /e \EmbeddedPDFRef{window-30} 0 R /eight
\EmbeddedPDFRef{window-31} 0 R /equal \EmbeddedPDFRef{window-32} 0 R /f \EmbeddedPDFRef{window-33} 0
R /five \EmbeddedPDFRef{window-34} 0 R /h \EmbeddedPDFRef{window-35} 0 R /i
\EmbeddedPDFRef{window-36} 0 R /l \EmbeddedPDFRef{window-37} 0 R /m \EmbeddedPDFRef{window-38} 0 R
/n \EmbeddedPDFRef{window-39} 0 R /nine \EmbeddedPDFRef{window-40} 0 R /o \EmbeddedPDFRef{window-41}
0 R /one \EmbeddedPDFRef{window-42} 0 R /parenleft \EmbeddedPDFRef{window-43} 0 R /parenright
\EmbeddedPDFRef{window-44} 0 R /period \EmbeddedPDFRef{window-45} 0 R /r \EmbeddedPDFRef{window-46}
0 R /s \EmbeddedPDFRef{window-47} 0 R /seven \EmbeddedPDFRef{window-48} 0 R /slash
\EmbeddedPDFRef{window-49} 0 R /space \EmbeddedPDFRef{window-50} 0 R /t \EmbeddedPDFRef{window-51} 0
R /three \EmbeddedPDFRef{window-52} 0 R /two \EmbeddedPDFRef{window-53} 0 R /u
\EmbeddedPDFRef{window-54} 0 R /w \EmbeddedPDFRef{window-55} 0 R /y \EmbeddedPDFRef{window-56} 0 R
/zero \EmbeddedPDFRef{window-57} 0 R >>
}

\EmbeddedPDFObject{window-16}{%
[ 600 600 600 600 600 600 600 600 600 600 600 600 600 600 600 600 600 600 600 600 600 600 600 600
600 600 600 600 600 600 600 600 318 402 460 838 636 950 890 275 390 390 500 838 318 338 318 337 636
636 636 636 636 636 636 636 636 636 337 337 838 838 838 536 1000 722 735 765 802 730 694 799 872 395
401 747 664 1024 875 820 673 820 753 685 667 843 722 1028 712 660 695 390 337 390 838 500 500 596
640 560 640 592 370 640 644 320 310 606 320 948 644 602 640 640 478 513 402 644 565 856 564 565 527
636 337 636 838 600 636 600 318 370 518 1000 500 500 500 1342 685 400 1137 600 695 600 600 318 318
511 511 590 500 1000 500 1000 513 400 989 600 527 660 318 402 636 636 636 636 337 500 500 1000 475
612 838 338 1000 500 500 838 401 401 500 650 636 318 500 401 470 612 969 969 969 536 722 722 722 722
722 722 1001 765 730 730 730 730 395 395 395 395 807 875 820 820 820 820 820 838 820 843 843 843 843
660 676 668 596 596 596 596 596 596 940 560 592 592 592 592 320 320 320 320 602 644 602 602 602 602
602 838 602 644 644 644 644 565 640 565 ]
}

\EmbeddedPDFStream{window-17}{%
/Filter [ /ASCIIHexDecode /FlateDecode ]
}{%
789c3d503192033108ebfd0a9e0008b0fd9ebdb9b922f7ff36c2cea4d81583908c2884a820f9cb2a99a5f26323730aa0f23f
6283d46b847eb0c8f83cd516844b4ca33e8910907d46cc14cccd4e09d6bab8f33025e1d4702e0212a9fc763368a79050ce2f
a7fee2337a8faebc9d2ac5933a808c1b5d54c576708329b688da8c799e6d71c1164eb2d7f0f9ad8c2b977505da6c1367fe64
17ea92e70190c9643c2e9654855de4bae5b7b3b76470caa9752393c9200cdbd78c5c17ed30d567e03c43605153719c9b61e0
d2e388a0d3e7fccff81bbf6f39154eae
>}

\EmbeddedPDFStream{window-18}{%
/Filter [ /ASCIIHexDecode /FlateDecode ]
}{%
789c4d8e310ec43008047b5ec1072201c680ff738a5238ff6fcf984b740d3b6261c14590f0e8ab78137419f86110221453bc
412d364d684dd198173dee840bf6e69dd265357a94368e9d3541d77091199599875ea09c317fb562f497a3e1289cc4410ff1
2893c71fe4fe7e26bf3abf3175292e
>}

\EmbeddedPDFStream{window-19}{%
/Filter [ /ASCIIHexDecode /FlateDecode ]
}{%
789c3d514bb2c33008dbe7141cc1989f7d9ebe79d3457aff6d253ccd2668821192280b1912fc64b9d4dcf2a7d7f49298f2b9
6c1bc17d7928816f977420953de475852f510d899ca271ea54658728f1072f6c448f98b1430e8b14525a6ee10ed657af25ba
7f02eeebddd0310c31a58da0667a234f1307a9afd57458ab1059d48a95c34f759237a21dbcc8993d939e1404964cfc016b16
24609ef50822ba1f1554d4697d585aa2c2e103ce6bb41ec048a1581f84bd44e126a550b8b06d39b6a5623ea0c1b075ecae47
7b238498f44d771b7e3de8aae057c1863b3afcfa0a04bafa3470822305e3aed19ecdd8c9e049b20f3ed34feda311a96133b2
0451f26e9d6a6cd1eac80648a7f4ed1903f3f8ff027f296dcc
>}

\EmbeddedPDFStream{window-20}{%
/Filter [ /ASCIIHexDecode /FlateDecode ]
}{%
789c458f4b0ec3200c44f79c628e60c01f384faaaa0b72ff6d3da0b49bcc93ede0e7210d02b3fc842aa24dbc6a691ab086bb
f43e08ab686f041343cc9c77d4de711597a411f0ec37d79ddd8d9d247507274cfbfec76664878fb809f8a647804b98d7de4b
5a8fc12a9fb2e56ec62e541d7f38c3d9fa012f58a5eb431a7593f9844f8a666ef59053e5dd1a278f3aa9a560c84075c9f981
e934f7019e15924adc212c3228fafe0200ed3f2f
>}

\EmbeddedPDFStream{window-21}{%
/Filter [ /ASCIIHexDecode /FlateDecode ]
}{%
789c4d8d4b1280300843f79c8223d0caa7dec7715cb4f7dfda8863dd90c78404776561b3395c85a3ee7c147af601b1ca9d8a
b6051e31695a1f20d4697be34969568d3f09ee9b2c280dad78bc48b21472d1790397771edf
>}

\EmbeddedPDFStream{window-22}{%
/Filter [ /ASCIIHexDecode /FlateDecode ]
}{%
789c4dcf3b1203210806e0de5370046079799f4c26857bff36e01a4d239f323f8c842c80a09847f70b9c3bbca8a9e6fdaea2
0ca391c481b9a732b151a1d1d8784929800953eeb2de7ec31f3dd170fbd35cd0ad1fd4040bd975363c031b465afbd881b0e6
88c61263ac6e69062ea4839a397f39daa7bdbf130532e6
>}

\EmbeddedPDFStream{window-23}{%
/Filter [ /ASCIIHexDecode /FlateDecode ]
}{%
789c4dce410e03210805d0bda7f817680222a2f7699a2ea6f7df0e0cced8152f027e8629086de2c582210cab136f2efe42f8
45d18aa370b38d6ee6f2d683583a4a95bed44dc0da96724c3d692be7eec054764de44f7156fcd2962a793ed1d27592906e50
1e14e55b3e279edc2b00
>}

\EmbeddedPDFStream{window-24}{%
/Filter [ /ASCIIHexDecode /FlateDecode ]
}{%
789c35524b6e43210cdc738ab94025fc87f3bcaaea22bdffb6634816d138f0ecf9985c8189e5f812438aa25cf12d631b2cf0
d7a853f01ae2fb53c58478403221d321bb500bcf502d8440cb610b66d9f00cdbe7bff76fc373a194c7be1776b08113745f2c
e345573a1782b395227c2534bc5bd8ab6c767eaf5b6035617337757392e035945d161b12cd4a959a70297e433ef7d55e9dbd
84f03e66919c213391516877c9fb6708759706b9e3a46276918e729fcad9536dcb1bbba7b5961a820a8b894605a7b5c2aec2
9d0a43e7a7a2d36006ce04d3f4649364e31cdb48866abbfbe9d2ea209dae7ba2e4cc620ec22f631dbdca2974c0d39c1723db
e2a9385f1671d211fdfa8ab3b2de48dc54b9d3ced099181568b36faaf4de5f4f70261ef444f7a66753d4adcdca77a3dc7ca3
1cb7a7e28420e3dbb8cff7d25932ab7e6fc60c1b99e8bc95266faea9afb383a26f089317be96f3269ff13b7efe01f6698240
>}

\EmbeddedPDFStream{window-25}{%
/Filter [ /ASCIIHexDecode /FlateDecode ]
}{%
789c4d8fbb0dc440084473aaa08405965f3fd6e9025fffe961832c07689e9801012d0e5ca855b458d139f12070e3eafcc0e2
d213941435a54876dc1db65677b4aa8a3c7aeda880d38bdaa35843228c2451b43326a7e534f9b672b928326722485fd4b9e7
e4c1f65365a8df38e10b9f3f6a532d4b
>}

\EmbeddedPDFStream{window-26}{%
/Filter [ /ASCIIHexDecode /FlateDecode ]
}{%
789c3d52496ec4300cbbfb15fcc0005abdbc2745d1c3ccffafa594a6878084259b14953c138214bcd4917323ddf1a5c3cf86
4ec7a7992dc77bd8b687f96a66123076695522700d9d02b380f2493d7963ceaa14337e537194571c4b786e92084a78c26f25
c2355ca58887200e9c22cbea784ff0d9b6677fc8da357e466cb653fad32c8d3e7ba08708b1fa1fa42b925597e049ab4146a9
5211c7ab46b4cec568a49043ec3b29750e0e15a9cbd79813d32a456520ff1393981ccc03eb12a7db35b1c6448d67a2bc9e2d
5378b5ad62b7416752378b12e2e4e1442a457616b4158bae291a9befb0bf90995a36d3b3d831a1fc62b676b2979bd2606f6d
c3e96557e0479abc8bc4a92dab2987e1ed242a7113ad6230ba4f66657efa87b1b91acbd37de27c21952e9c3aebb012cc21c2
d16b3a86675db5baef5f801678c9
>}

\EmbeddedPDFStream{window-27}{%
/Filter [ /ASCIIHexDecode /FlateDecode ]
}{%
789c4d51496ec3300cbcfb15fc40007113a5f7b8087a70ff7fcd0c9db4056c702c52b3d033860cb12d0f75c93da4e6902f3d
5453d2e4a7418d25d781a13f80298091ff507837cd240a9411123bc4ca2475c8896b1b9429ee50c2eb75d7f308af46b13031
80a11059e8e404e2099c7968579b6423d25ac20955eb3bd34806127c939399a8c27a1ed425a28f47b6b14d5b306a78e87b09
832ce531c0b8c3be0bd6717df6721ddf3d611ed81191968a69411a0a70ba9b64627ab5621802240b9ca009105806b2446cd9
0c1bb0aabd06d0add5f50e4be4cec522d09a7d25acd3822320d19455b746d57bdd44f41073b4a9b0eec0a6f3d7c0b6477535
289cbf9118eff90230aa64da
>}

\EmbeddedPDFStream{window-28}{%
/Filter [ /ASCIIHexDecode /FlateDecode ]
}{%
789c3d51396e04310cebfd0a7e20800ecac77b2658a4d8fcbf0de5206946b4c4c3f2d43418caf0e18972a232f1e9a3a1d7c4
f72873b819b8166a8374a4e119b482a05a2d4db1ba3e23a22ef2a9b9a47e305dfda596ce8af27632c46c1b81644053dabe6c
caf2192d2f75daaeefd4015d9f91fadece9ca838c873543b82298ea239b587057816b8574f84b288f720eb0fe5be5a861429
762f59374117e4a22e46e9c56f2f552da779a3f0a994093f9ac4ef7d9570b45c899fbc35eaf444c8b73a62b8b8ad39bd7a9b
54a03d25e81015a5d3ee79b91e5b9ec4ec367d412ecc7ec7d228ee0f7afe7fd57b7c8dd70f9fd9560c
>}

\EmbeddedPDFStream{window-29}{%
/Filter [ /ASCIIHexDecode /FlateDecode ]
}{%
789c4d51396ec4300cecfd0a7e60019e3adee3204891fcbfdd196a37496170445173d02355544ae56121c34ce650f9b0abbc
a45c7e2ef600be1b286a46fdd66504362597c42ef194184bb6dc5780f851123a9bdac7a9f7e51a8d0ca2e662a6321cfd8923
ceeccec5e270725f0091b0a592a0e274d6449fcf0b1dd2551c015628eb4174c0187454462e5acc9dd28ec98808190b374453
17d20478ffa1c1a864f943bd8daf7ee2915851a355af9d4482dc210fa1d81b987663426a689bca09dbb43279e3b94fc760ac
96d8c68473233686c41ad81458c3bb9ea510d9c22c260c51f96637194830d19c7e44fcf5330068a256bbdac6d4b069f868db
a6c93bd28d789f4f05376445
>}

\EmbeddedPDFStream{window-30}{%
/Filter [ /ASCIIHexDecode /FlateDecode ]
}{%
789c3d514b6e053108dbe7145ca052cc2fc979a6aaba9877ffed33cc6b573816d886c451991253be6012ae1266f28d515049
bf06221add0ff2fc20ac14e416a8098ecb5e720d558a81fd47ccc5802ad730ced77babb88bcf2d19a49d5e3bc43901a478b2
46c9548842f7081cc19ce2cc49c2c3454f8d8209c4d23bb84ded5a01b2116880923d42d56b2c527c3307d6aaa239cba872a9
2ced50c0a6c3268fe435c8a83e17d1935db9caf2466ed11d4e397fb4e8e560bf3fa27f27bcc62f37cd0ef922e21d927e8af6
b4c3793bad4cd5acc3715d9ea53cab72296edd0cd354070e7dad73ae3ae7e2bab355eb63cae7fa77bce9fef30673f95a51
>}

\EmbeddedPDFStream{window-31}{%
/Filter [ /ASCIIHexDecode /FlateDecode ]
}{%
789c3552418e243108bbd72bf8c04a814048ded3abd15ee6ffd7b59dea43095702b681acb96cd86afbe3d36a1deb0cfbeb4f
ae657e8efd0a45b9654d8b83186d33867d9e1c6e338fcd95f7f3adf879a25d28662923c6796b7c5f16ef215646eae08628db
98e163a9a69b6420c96de27c45a821c5349ac02d5d299b3659ae0640f76de4f3fc7b329715ee7f2f3ab805c39a61e961ab58
3ff7b1b52155dbd6b97a8c6ceaa2402d3322f2ad0978208b9f215645e8a029a201f374d9533599744396c41f59a94e1d4638
a8f7044e98417c6be8962c720fd66f3fec6d9e898aadde30a13d2c0f9700fd0057d067698577cf1c3ee39d8dd0c409260643
3520442fc97a9b67eb7dd0232377725f8c77e1de3941a6439f436a7131881c806a3b24ee24921dcf9441df2dc391b7058e0d
e8607185213905d04e72094e671d86de0a0f84410300588ee7848f3b71e82d9c91caada340de7adbeca165965244246646ee
7c6beadd6c658995f1ca0841b7a6cb468d7a8d651f59cd7459ffee83bbf9f90ff09ba266
>}

\EmbeddedPDFStream{window-32}{%
/Filter [ /ASCIIHexDecode /FlateDecode ]
}{%
789cb330b6503050303430533034375530373652303135524831e402098198b95c30c11c30cbd8dc04c802c9225810d90c30
dbc8d418aa07c482e801190d518960416433b8d200514c1915
>}

\EmbeddedPDFStream{window-33}{%
/Filter [ /ASCIIHexDecode /FlateDecode ]
}{%
789c4d904b7203310844f73a0547e00f731ea7525e38f7dfa691ca1e6fa6df805a82b66262b2c4c78da992e947d6605ad1df
b2b64d2f9052a6905550f678f07f193d9679537190e945254ac6b6f5b1344f453d71026af0b44f471ab72529ba69b6353a77
0724d7bc18f5459e33c5746f0a9da6d60d3c9a6fd96551bde178d1fac079611fba09331e4a5712ec9f1d24b0940466940be4
81bd6447667ca24316f0ec8a238368a414d0d9d81a9e40cc8cb4dc4fdc361bbfd37eade7fafd0739ac4bcf
>}

\EmbeddedPDFStream{window-34}{%
/Filter [ /ASCIIHexDecode /FlateDecode ]
}{%
789c4d51496ec4300cbbfb15fa4000edb2df3345d143faff6b290533e8212063531229a72531eda04b8c220e951efa92156c
437f87a51fba97e4f9c7dcb9194a3d949481c9a4ba8149afa5c1e4859b4a6090b10fbe966d1de68e9330f203251fdc04dc58
c61831ce4135e91b3081b615e729893eef1e9bacce04d09383989efa9cc0cf8552d45cad97284a121592ddb1ad0d01a5aad3
20d39b1989e12f361d87d8a9b8fb0abc26a963a8cf3c00227935b19d6dcbc5bb77bb5312c1ae0233b3069f3cc3d0b81586cf
055e2a673bd0618a61846bcd8cc6cef430854b470245049749d51a462674b04ed3fbcbde83480dbb87f58bde9fb7bdd7cffa
fe0331f55ffc
>}

\EmbeddedPDFStream{window-35}{%
/Filter [ /ASCIIHexDecode /FlateDecode ]
}{%
789c4d90418ec4300804ef7e054fc0d060fb3d59adf690fdff751a349acc21e94a10a64c02a2e2c957ce94952a3f7360f2fb
bf224cee31cd1e58ba49ec78804df7b0695f045b45b60539c542057b8bad94d0946bb8d691107797e0e3793aaf01cb26e496
a0124eb017ac04a723e8c49950ed74d7aa905ab0eef0016dbdfdc92e00f94059bc2939c569cb4ba1e6b8950b5d1121be0e7d
9638b4f31a76769365b693718fb0591576fbe11f6ec1b13a6d9faaccf71aede80365d7fbbec7dff87d01cc004b50
>}

\EmbeddedPDFStream{window-36}{%
/Filter [ /ASCIIHexDecode /FlateDecode ]
}{%
789c45504112c42008bbfb0a9e405051dfd39d9d1edaff5f3750677bd064204462351595eabc6c0d19641f14329f2a7792d5
056a32740a506560c951608d6215d46708bd26b2e3bb4283542cec19db2ec613ae89544587cc9b84c23b38436c08b709f170
712aace73b81ecf0e5ac70935470b73db35d62799aee3047398bc1a41b734556922b8912f30712b20cb3973426cad69f74e6
b91eb79785fc2cdf1fb5233fa0
>}

\EmbeddedPDFStream{window-37}{%
/Filter [ /ASCIIHexDecode /FlateDecode ]
}{%
789c33363250305030b20413060ae666060a29865c4606a60aa6460ab95c2031202307cc3000d3300a2c6c68688a60981b58
40a4100c33b062a0690816587906571a007ce51753
>}

\EmbeddedPDFStream{window-38}{%
/Filter [ /ASCIIHexDecode /FlateDecode ]
}{%
789c4592cb75c5200c44f754a11284be508f7372b248fadf6624fcec059e01197425d8b68849039f2d935c95bee6f0b9c830
fe86ab9379907b922d23afc146d78889df45294c7a5be46abd46ea59c954f219b498b15710591264c8b80c39f7adca158173
a1dfd1141fc3d04879b403e9fc1a15fe38c4132c3699b2f21c16cccc9d62a18e48d0ee56f0f3710e6ef3792a44ce0b3507e8
0a036ad92a6b5704ae53075af09822505f8f76c0345ea3e8d3ed121c52dd9533b47aa93bbacb9ab39954fda6947d9c8435a5
98de9422d29482538ab2f45096ebd4b2f9354567f323bd3c455e63b18a3f5ee373d7de3aed712659ae2a88a2410fd7224155
ce5195709d6d7dabf50634f47e15c6c79dc6402bc21dd9f53e6a563781c6deafef1a3fe3fb1f968c7eb8
>}

\EmbeddedPDFStream{window-39}{%
/Filter [ /ASCIIHexDecode /FlateDecode ]
}{%
789c4d904112c3300803ef7e054fc00213fb3de9747a48ff7fada069d2438c32026b4db88b8a058fe821c34c1ebd79e7ff3b
cb801cad03b7f0d8a8387189d117053afe94234d60b2ad0b868acf29d8d8ad217b33cd2b5d8c81196ab1aaeecdf1c5f098bc
8f316b70d6e98c4c1f64ca70d5aa669a4e522460bde127b4f0e655cb708f5b24c5a98229465ae59739866421ab0fbadb2a1e
73adba37ac590a911d10708f394787a72dbe995b30dfaa62ae74fab9462cbd45d2d5be8ff66acf0fd5654b5c
>}

\EmbeddedPDFStream{window-40}{%
/Filter [ /ASCIIHexDecode /FlateDecode ]
}{%
789c3552496e03310cbbfb15fc40006bb7df93a2e8a1fdffb5940601125090b590d4a42536d2f01243e445b9e24b96e78169
e16fb9276c2b5c37f438ec2ab414ef65e9d0dc30632618df1a7c2f559f4892990a88064c2e5f8ec1b88d3f571db8876906c1
9927981208ffc9b6f7924a141bd56a9819593672f98d893cc85904b13935365fc20b71cea8892d8376aa5f1829eb837d92c2
dd1b57f8e076a08216d43e28eb1a29c41f675492b8219cfbca2646dd2f1da694c72d32743ba0a3bf4b223f112b85e484faea
40ae239ba6b2833c7a83d1adba0dd4d52ae9f221cd1ad365f768a73609fa9509e5613e077aaf9f65fb2239f76f697bcba82b
b2687ebbe2364e1eeebdd49d7444061fef27e2caaeb093d363a173c6a2e3b48653fb13e83d8d6419f164e8605738c5583c56
b6dbe4c9096efbc1aba34011d1bbc876e7f4a4cbdc929f202fdd539bfd474f6bfbfe07e9627d73
>}

\EmbeddedPDFStream{window-41}{%
/Filter [ /ASCIIHexDecode /FlateDecode ]
}{%
789c3550cb6dc5300cbb7b0a2df000eb6b7b9e3c14bd64ff6b492505028891458a544d932939e5a32e9926e92e5f1d3e553c
e41e1e87c077e24942a71c956b047a6a47229628de582d0f5f801c229cf0339b12b6f1428da8124a063077b05ec3280d645a
3da1c75e8ed66a15cd6c55d6670f113773824ec8696b14811e3569e109728ddfcec498f730cb465ae02bbe2345ea420bff38
87aec56235d106f0305930353726b6443ee602435b28c7bb71012ba292de47583d11db5f4ef268e8f0d614657db700f55a86
516d4e3db75e74494d9afe8fc1483f7fb155530b
>}

\EmbeddedPDFStream{window-42}{%
/Filter [ /ASCIIHexDecode /FlateDecode ]
}{%
789c4dcd410e80300804c03bafe00994c2d6fec7180ff5ff578118f50213925dd0c1c24d35a64de361ca7ba36679b86abbf2
22c5fc00df421972f447d011ea225511c2f8a9a2d9ff4232575f169d74dce05d1a1f
>}

\EmbeddedPDFStream{window-43}{%
/Filter [ /ASCIIHexDecode /FlateDecode ]
}{%
789c3d4fb90d03310ceb3d05173840cf49b2e74910a4b8ecdf86f205290cd1341fcb9740500b8746c075a152f0d4d170739f
61de90776d5d3a94843a1e632dccd56e9dd9c3c5c8129c3ac1c7d0738b6329f9b6674e745ed9f96f7becb252c3352ca9a9a2
c69146ad2a22b3dd6c0a61794dc6c79e775f239d7df81191edf1c9870e3958d4a14719ee9da47e8d7bbd6bbcc7eb0b2d6c31
38
>}

\EmbeddedPDFStream{window-44}{%
/Filter [ /ASCIIHexDecode /FlateDecode ]
}{%
789c3d4f3b0e433108db730a5f2012260192f3bcaaead0de7f2de4a99d7030fe646c8120884e330c12e18207db6ff5b99104
de8d9a3014b440d709c6c658b89a0a411128935cebcc215a4ca249435d988c2331f764cac475a14c3da23a044b1277878cb3
6c96314a817bbae6db36cb355b1be7e93bb9cebcf30a7139ea62ed23e1383d04dd333f179ddcff2f5fedd59e5fc6543073
>}

\EmbeddedPDFStream{window-45}{%
/Filter [ /ASCIIHexDecode /FlateDecode ]
}{%
789c2d4e5b0ac3300cfbcf29748182ed24757d9e8db29fddff77923b0828e8654dbf60a885c3272216dc4fbc7d90d98eaff0
2cb8192e69814abc86cb68e49775c0773552c97c988a7684ed27129eaad009160ad84f9a9fc9cb54f9e4de6a61fa1093bd4b
f5420a3cd80c63327050fbb98f718e9d81fa577fc6fd03b96828c9
>}

\EmbeddedPDFStream{window-46}{%
/Filter [ /ASCIIHexDecode /FlateDecode ]
}{%
789c4590411242210c43f79ca247a069a0701e1dc785de7f6ba8e377c1cf1b5ad2fc3297758ba90f8523c2eede0ad1ed5d14
bbdbab11f3221a7deaa43186b1eb70d9adc58671ca2f439a16dca5aa9c1e11d28dc30dd45bb82a40942f1cea9fa558fb5444
031a181a764125f09fd4b5037fe05ca773fe61f81694db45441ec2c9170625a26622b78dfecd9bb2a412792d25d657cf3f8e
22fab9f1721a3879c910656de0bcfd6df1d69eedf101f3c04403
>}

\EmbeddedPDFStream{window-47}{%
/Filter [ /ASCIIHexDecode /FlateDecode ]
}{%
789c3d924b6ec4300c43f739852e3080f58bedf34c5174d1de7fdb47a7e822a020c51249a93d6d58dff622a83bac33edc32f
32b1ed47e83decfbf2b1fea36dee6e1e6e2bccabec4e7b5f7e4fabb4186e5916750bde57ac13a453a592cb7a92cede36c372
0edb75c0435d4e546d0917bfa9d4349f4b9564e60e786d864c2680e78d2fcd6b714b727476d5e7b435cfefef0b6e1943a3d5
50b09c3441c5b259686f93b262021d11d5becf3c39125d0725c74f944e372cca12c61ff3f6b61abc1920aa6b6b8aa244c3f7
952bfea3811f187233331f5a85dbf409dece8638b8409104990d2345be794b179f7e78c3b789f8bc826e7f384e8528579281
ab34cad33e1ae990f28bcd66b06fb2e94709aec6d6ae931585dc919f0d1f1447ea1f4d0bf9a53d49cbfd9c021522565adc02
7ac422cefa58c160c0736ac1290925eac968892f1c87c5eb2c9575a52d54963dd7f8bebeaecf5f630f7fec
>}

\EmbeddedPDFStream{window-48}{%
/Filter [ /ASCIIHexDecode /FlateDecode ]
}{%
789c4dcc310ec0200c43d13da7f011200d86dca7aa3a84fbaf0d958abad86ffa3c888261398d86ae8eb3ca22bb638ae61684
68e1fbd607689eaa567f6a6d293b1b2b15f2456337436eb91e54a5173b
>}

\EmbeddedPDFStream{window-49}{%
/Filter [ /ASCIIHexDecode /FlateDecode ]
}{%
789c333636573000425d4b23056320dbdcc85221c590cbc8d412cccce58209e670995b8055e57019406998a21cae0cae3400
fd1e0e2d
>}

\EmbeddedPDFStream{window-50}{%
/Filter [ /ASCIIHexDecode /FlateDecode ]
}{%
789c3336b4503080c31443ae34001de60352
>}

\EmbeddedPDFStream{window-51}{%
/Filter [ /ASCIIHexDecode /FlateDecode ]
}{%
789c4d90bb0d03310c437b4fc1050eb0247fe7b9204871d9bf0de943e2547aa62c8252c98e0c9f382c10b3a08d8c8725cb03
a575bc137b824b506d12d4dba401d2fca7bb1b1cdb749ba8bbc9e24b33e08cd20adc0cd57126e76861b6a808ca75a8501eb6
de322265a6aa9483bfdb4478c5a4ec0de672d14e2286a0de0ba20fd906a738c20fe1389c7e7dddc06b5ef55cb144d68c15a6
2c26dd684099bb0f5f456b9cbf935de9959e1fd17040ac
>}

\EmbeddedPDFStream{window-52}{%
/Filter [ /ASCIIHexDecode /FlateDecode ]
}{%
789c3d523b6e43310cdb7d0a5e2080f5b7cf93a2e890de7f2da5041d1e444b3625522f2db1518987182215e58a2f59b790f7
e07789054a1452825283ee4499e0b9d499b1032d9f57b663e27319f91ab9914f135e4ab66425c894190856527562e4e94a23
75841cf829f8ddf0e88a732a5766d9ddf78195c1ee548ce85cf25f58169addac27080f2835b4220d9d283cb3d2482f3b5f54
b72169ab71e76db2e938a1f7ed08557286c9e8c6e3428ef3c8bcf84142da99d3fed1a6e7621492bee89afc239f7e121b97b0
787b372f7df484f2e30da51e065ae703ac1da064c91eb2a7e3d37e1164f24f1c471b695084fb4cedc2d7da15e3a816d4447d
c69edae7776f3b835e835c078dcb8d4cee20eb0c75d86ca2e579ef865eb7efc13e13533f9bc8cd8c0ad26af693b9e75f7024
5ded29f2fa6ca5633b6bef0c59f318f5f0e6b009674bde9748a46c466e7026ef95cee09ffff3b57ed6f71faee18468
>}

\EmbeddedPDFStream{window-53}{%
/Filter [ /ASCIIHexDecode /FlateDecode ]
}{%
789c4551498ec3300cbbfb157c82b5dbef4951f490f9ff752867d03924242c9922e5b4c4442efec216ca152f19a20b11819f
5176c8dd24571399859a06616b49418ad7b4700d15435941fd11d2a507af61f19cb874afc153a9d677622632389c13733e18
762a649e01a7bead0923ea9256ab09a51f635569dd5288ce8a5481adf770367e8954b38ef7cf1af30b61445d9bb24e663460
aa1cc92fc9372dcfc5012e0ef70d571adb7af0b17bd84a466434755aee60277cf68219801db9fd8468e4c2cc0f93cdf05c96
f4926377102f249f468c6a9311fe1ee41a9ff1fe0511b9524d
>}

\EmbeddedPDFStream{window-54}{%
/Filter [ /ASCIIHexDecode /FlateDecode ]
}{%
789c4d904b0e03210c43f79cc217a8447e04ce3355d5457bff6d4da6bfcdf8910c89cd704787262e62183d11b2709566e185
cf16aa458f939450ff7da053ddf4ab6b375c0291b06530818d04c71fcd5c7071967aedd31ca54753994592017e7dc2826551
c1bedb2784164ae7da0d92b3f368b4fe85d3a6acf9472af926096eecc1490cccc133f7624b0cd687ef44d64bb6d151e7ccbd
dfd602071ccd794d94ab9953d24a95018f4a7efad82ff7a3d3c8bddd5eac6143d2
>}

\EmbeddedPDFStream{window-55}{%
/Filter [ /ASCIIHexDecode /FlateDecode ]
}{%
789c4d8f390ec34008457b4ec111608001dfc78a5290fbb76136cb0d3c7d3e5b5847421e2154d0f8c29b418326fea0b320b3
62820ba1762fea662f1abe84d3bc68553d1e9f702d48b0b92d415ba0441409d1545a2923bbee963ae98135b819bfe8147d53
8b73a8b26fdf7923e10b9f3f99f32d4a
>}

\EmbeddedPDFStream{window-56}{%
/Filter [ /ASCIIHexDecode /FlateDecode ]
}{%
789c4d91510ec3300843ff730a2e5029264092f36c9af6d1ddff776649b74955f3140c36ad874b9543f928c443c531e58ea2
0839a6c9aba85330e42c1e62d109947f21e567d1167fb48af0b649e714a092cc7cdf35be7fb47a2ffb45abead5ae29c14460
ac734502f36863720c1345027d6e0533b7411384adb5a063018bd8abce9ea22903797696f82548744b31d3e668adc1cac7a4
66863e2e1a9940d9cf2e98c9ccd32d2d98f8486f20d5c1dddb06163df65daf5b9509766726cf61fa19a1b9d6fa0bb7f22c8f
37a12f504e
>}

\EmbeddedPDFStream{window-57}{%
/Filter [ /ASCIIHexDecode /FlateDecode ]
}{%
789c3550cb6dc5300cbb7b0a2e50c0fada9e2745d1cbdbff5a514e4f6622fea4b4c44426bec4106b62b9e25b86c986393ec3
b281cf400adc264412cff07048f13d1734ac5f4be7a490af92142362b5264539a14965d1338f83217c9fa1ef1f25b7183af3
d5c89676913cedcaf7e61031990c36a1e6766b1781deeaef2acff8edadb8e8a72817c96218c4149bd2bd21a13c88baa3bb52
5ac0b7a1a65102b2d3bb5d7dadb9413b5e8e01ab277656238fd38c98f66ac257bbf0da34e57b5388184b065b50b3efb10f5b
d293a5ffd7e04a3f7fc84753fe
>}

\end{document}